\documentclass{dune}
\pdfoutput=1

\usepackage{subfiles} 
\usepackage{authblk}

\usepackage[pdftex,bookmarks,hidelinks]{hyperref}
\usepackage{cite}
\graphicspath{ {graphics/} {common/}{common/graphics/}} %\sisetup{binary-units=true} %AH added for \bit and \byte 26may

\newif\ifdp
\newif\ifsp

\def\expshort{DUNE\xspace}

\def\explong{The Deep Underground Neutrino Experiment\xspace}

\def\thedocsubtitle{Single-Phase Vertical Drift Technology for\\ Far Detector Module 2 (FD2-VD)} 

\def\larmass{\SI{17.5}{\kt}\xspace} % full mass in cryostat
\def\cryostatht{\SI{17.8}{\meter}\xspace} % outer height of cryostat (Jim Stewart 5/2/19)
\def\cryostatlen{\SI{65.8}{\meter}\xspace} % outer length of cryostat (Jim Stewart 5/2/19)
\def\cryostatwdth{\SI{18.9}{\meter}\xspace} % outer width of cryostat (Jim Stewart 5/2/19)
\def\cryostathtinner{\SI{14.0}{\meter}\xspace} % inner height of cryostat (Jim Stewart 5/2/19)
\def\cryostatleninner{\SI{62.0}{\meter}\xspace} % inner length of cryostat (Jim Stewart 5/2/19)
\def\cryostatwdthinner{\SI{15.1}{\meter}\xspace} % inner width of cryostat (Jim Stewart 5/2/19)
\def\dunelifetime{\SI{20}{years}\xspace} % nominal operational life time of DUNE experiment
 
\newcommand{\nominalmodsize}{\SI{10}{kt}\xspace} % nominal FD module size 
\newcommand{\fdfiducialmass}{\SI{40}{kt}\xspace} % total Fid mass
\newcommand{\lntwo}{LN$_2$\xspace} 

\def\tpclength{\SI{58.2}{\meter}\xspace} % length of  TPC
\def\tpcwidth{\SI{13.5}{\meter}\xspace} % width of TPC
\def\maxdriftdist{\SI{6.5}{\m}\xspace} % max drift distance (checked)
\def\tgtdriftfield{\SI{450}{\volt/\centi\meter}\xspace} % target field strength
\def\pcbpaneldim{\qtyproduct[product-units=power]{1.494x3.366}{\m}\xspace} 
 
    \def\crudim{\qtyproduct[product-units=power]{1.496x3.370}{\m}\xspace} % latest 3V
\def\crpdim{\qtyproduct[product-units=power]{2.993x3.370}{\m}\xspace} % CRP LxW
\def\anodeplndim{\qtyproduct[product-units=power]{60.0x13.5}{\m}\xspace} % anode plane LxW

\def\channelspercrp{\num{3072}\xspace} % number of 3-view CRP readout channels 
\def\dge{$^\circ$\xspace} % easier to write!
\def\firstviewangle{{30$^\circ$}\xspace} % 3-view 1st induction view electrode angle wrt beam
\def\secondviewangle{$-$30\dge} % 3-view 1st induction view electrode angle wrt beam

\def\cooldown{cool-down\xspace}  
\def\coldbox{cold box\xspace} 
\def\Coldbox{Cold box\xspace} 
 
\newcommand{\rms}{RMS\xspace} % Might want this small caps?
\newcommand{\threed}{3D\xspace}
\newcommand{\twod}{2D\xspace}
\newcommand{\fdth}{feedthrough\xspace} %  easy to misspell!
\newcommand{\phel}{photoelectron\xspace} 
\newcommand{\frfour}{FR-4\xspace} 
\newcommand{\efield}{E field\xspace}

\newcommand{\numu}{\ensuremath{\nu_\mu}\xspace}
\newcommand{\nue}{\ensuremath{\nu_e}\xspace}

\newcommand{\anue}{\ensuremath{\bar\nu_e}\xspace}

\newcommand{\numubartonumubar}{
\ensuremath{\overline{\numu}\rightarrow\overline{\numu}}\xspace}

\newcommand{\ptoknubar}{\ensuremath{p\rightarrow K^+ \overline{\nu}}\xspace}

\def\argon40{${}^{40}$Ar}       
\def\Ar39{$^{39}$Ar}
\def\Cl40{$^{40}$Cl}
\def\K40{$^{40}$K}
\def\B8{$^{8}$B}
\newcommand{\lsim}{{\;\raise0.3ex\hbox{$<$\kern-0.75em\raise-1.1ex\hbox{$\sim$}}\;}}
\newcommand{\gsim}{{\;\raise0.3ex\hbox{$>$\kern-0.75em\raise-1.1ex\hbox{$\sim$}}\;}}
\newcommand{\beq}{\begin{equation}}
\newcommand{\eeq}{\end{equation}}
\newcommand{\bea}{\begin{eqnarray}}
\newcommand{\eea}{\end{eqnarray}}

\mathchardef\minus="002D

\DeclareSIUnit \c {$c$}
\DeclareSIUnit\magn{$\times$}
\DeclareSIUnit\min{min}
\DeclareSIUnit\hr{hr}
\DeclareSIUnit\hrs{hrs}
\DeclareSIUnit\week{week}
\DeclareSIUnit\month{mo}
\DeclareSIUnit\months{mos}
\DeclareSIUnit\year{yr}
\DeclareSIUnit\years{years}
\DeclareSIUnit\yr{yr}
\DeclareSIUnit\standard{std}
\DeclareSIUnit\str{sr}
\DeclareSIUnit\ppm{ppm}
\DeclareSIUnit\ppb{ppb}
\DeclareSIUnit\ppt{ppt}
\DeclareSIUnit\pe{PE}
\DeclareSIUnit\spe{SPE}
\DeclareSIUnit\pdm{PDM}
\DeclareSIUnit\ev{events}
\DeclareSIUnit\ct{counts}
\DeclareSIUnit\neutron{\mbox{$n$}}
\DeclareSIUnit\smp{samples}
\DeclareSIUnit\Sample{S}
\DeclareSIUnit\ch{ch}
\DeclareSIUnit\hit{hit}
\DeclareSIUnit\hits{hits}
\DeclareSIUnit\bin{(\mbox{5-PE}~bin)}
\DeclareSIUnit\sgm{\mbox{$\sigma$}}
\DeclareSIUnit\rms{RMS}
\DeclareSIUnit\keVee{\mbox{keV$_{e{\rm e}}$}}
\DeclareSIUnit\keVr{\mbox{keV$_{\rm nr}$}}
\DeclareSIUnit\eVee{\mbox{eV$_{\rm ee}$}}
\DeclareSIUnit\eVr{\mbox{eV$_{\rm nr}$}}
\DeclareSIUnit\ph{photon}
\DeclareSIUnit\el{\mbox{$e^-$}}
\DeclareSIUnit\pm{\mbox{PMT}}
\DeclareSIUnit\pixel{\mbox{pixel}}
\DeclareSIUnit\inch{''}
\DeclareSIUnit\foot{'}
\DeclareSIUnit\bit{bit}
\DeclareSIUnit\sample{samples}
\DeclareSIUnit\barn{barn}
\DeclareSIUnit\bara{bar}
\DeclareSIUnit\bar{bar}
\DeclareSIUnit\barg{barg}
\DeclareSIUnit\mlardepth{\mbox(meter~of~\LAr~depth)}
\DeclareSIUnit\Curie{Ci}
\DeclareSIUnit\PSI{psi}
\DeclareSIUnit\psia{psia}
\DeclareSIUnit\atm{atm}
\DeclareSIUnit\psf{psf}
\DeclareSIUnit\pcf{pcf}
\DeclareSIUnit\parsec{pc}
\DeclareSIUnit\cps{cps}
\DeclareSIUnit\slpm{\SI{}{\liter\per\minute}}
\DeclareSIUnit\rpm{rpm}
\DeclareSIUnit\mwe{\mbox{m.w.e.}}
\DeclareSIUnit\liveday{\mbox{live-days}}
\DeclareSIUnit\days{\mbox{days}}
\DeclareSIUnit\miles{\mbox{miles}}
\DeclareSIUnit\lumens{\mbox{lm}}
\DeclareSIUnit\degreeC{\mbox{$^{\circ}$C}}
\DeclareSIUnit\degreeF{\mbox{$^{\circ}$F}}
\DeclareSIUnit\electron{\mbox{$e^-$}}
\DeclareSIUnit\Euro{\mbox{\euro}}
\DeclareSIUnit\cph{cph}
\DeclareSIUnit\neq{neq}
\DeclareSIUnit\normal{\mbox{N}}
\DeclareSIUnit\USD{\mbox{\$}}
\DeclareSIUnit\Vpercm{\mbox{V/cm}}
\DeclareSIUnit\kV{\mbox{kV}}

\DeclareSIUnit \mm {\milli\meter}
\DeclareSIUnit \cm {\centi\meter}
\DeclareSIUnit \us {\micro\second}
\DeclareSIUnit \ms {\milli\second}
\DeclareSIUnit \pA {\pico\ampere}
\DeclareSIUnit \pC {\pico\coulomb}
\DeclareSIUnit \fC {\femto\coulomb}
\DeclareSIUnit \fF {\femto\farrad}
\DeclareSIUnit \pF {\pico\farrad}
\DeclareSIUnit \mV {\milli\volt}
\DeclareSIUnit \kV {\kilo\volt}
\DeclareSIUnit \V {\volt}
\DeclareSIUnit \GOhm {\giga\ohm}
\DeclareSIUnit \MOhm {\mega\ohm}
\DeclareSIUnit \ton {\tonne}
\DeclareSIUnit \kton {\kilo\tonne}
\DeclareSIUnit \kt {\kilo\tonne}
\DeclareSIUnit \Mt {\mega\tonne}
\DeclareSIUnit \eV {\electronvolt}
\DeclareSIUnit \keV {\kilo\electronvolt}
\DeclareSIUnit \MeV {\mega\electronvolt}
\DeclareSIUnit \GeV {\giga\electronvolt}
\DeclareSIUnit \km {\kilo\meter}
\DeclareSIUnit \kW {\kilo\watt}
\DeclareSIUnit \MW {\mega\watt}
\DeclareSIUnit \MHz {\mega\hertz}
\DeclareSIUnit \kHz {\kilo\hertz}
\DeclareSIUnit \mrad {\milli\radian}
\DeclareSIUnit \year {year}
\DeclareSIUnit \POT {POT}
\DeclareSIUnit \sig {$\sigma$}
\DeclareSIUnit\parsec{pc}
\DeclareSIUnit\lightyear{ly}
\DeclareSIUnit\foot{ft}
\DeclareSIUnit\ft{ft}
\newcommand{\driftd}{6.5\,m}

\newcommand{\PDelt}{many tens of ms}
\usepackage[toc]{glossaries}
\makeglossaries

\newcommand{\dshort}[1]{\glsentrytext{#1}}  % doesn't provide link
\newcommand{\dshorts}[1]{\glsentryshortpl{#1}}  % doesn't provide link
\newcommand{\dfirst}[1]{\glsfirst{#1}\glsunset{#1}}
\newcommand{\dfirsts}[1]{\glsfirstplural{#1}\glsunset{#1}}

\newcommand{\dword}[1]{\gls{#1}}
\newcommand{\dwords}[1]{\glspl{#1}}
\newcommand{\Dword}[1]{\Gls{#1}}
\newcommand{\Dwords}[1]{\Glspl{#1}}

\newcommand{\newduneword}[3]{
    \newglossaryentry{#1}{
        text={#2},
        long={#2},
        name={\glsentrylong{#1}},
        first={\glsentryname{#1}},
        firstplural={\glsentrylong{#1}\glspluralsuffix},
        description={#3},
        sort={#2}
    }
}

\newcommand{\newduneabbrev}[4]{
  \newglossaryentry{#1}{
    text={#2},
    long={#3},
    shortplural={{#2}\glspluralsuffix},
    longplural={{#3}\glspluralsuffix{}},
    name={\glsentrylong{#1}{} (\glsentrytext{#1}{})},
    first={#3 (#2)},
    firstplural={#3\glspluralsuffix{} (\glsentrytext{#1}\glspluralsuffix{})},
    description={#4},
    sort={#2}
  }
}

\newcommand{\newduneabbrevs}[5]{
  \newglossaryentry{#1}{
    text={#2},
    long={#3},
    plural={#4},
    shortplural={{#2}\glspluralsuffix},
    longplural={#4},
    name={\glsentrylong{#1}{} (\glsentrytext{#1}{})},
    first={#3 (#2)},
    firstplural={#4 (\glsentrytext{#1}\glspluralsuffix{})},
    description={#5},
    sort={#2}    
  }
}

\newduneword{dword}{DUNE Word}{A term in the DUNE lexicon}

\newduneword{nasa}{NASA}{U.S. National Aereonautics and Space Administration}

\newduneabbrev{nd}{ND}{near detector}{Refers to the collection of \gls{dune} detector components %or more  generally the experimental site
 installed close to the neutrino source at \gls{fnal}; also a subproject of \gls{usproj} that  includes  installation, infrastructure, and the cryogenics systems for this detector} %updated 19 nov 21

\newduneabbrev{fd}{FD}{far detector}{The \SI{70}{kt} total (\fdfiducialmass fiducial) mass \gls{lartpc} DUNE detector, composed of four \larmass total (\nominalmodsize fiducial) mass modules,  %or more generally the experimental 
  to be installed at the far site at \gls{surf} in
  Lead, SD, USA}

\newduneabbrev{sp}{SP}{single-phase}{Distinguishes a \gls{lartpc} technology by the fact that it operates using argon in its liquid phase only; a legacy \gls{dune} term now replaced by \gls{hd} and \gls{vd}} %11 may 21 {Distinguishes one of the DUNE far detector technologies by the fact that it operates using argon in its liquid phase only}

\newduneabbrev{dp}{DP}{dual-phase}{Distinguishes a \gls{lartpc} technology by the fact that it operates using argon 
 in both gas and liquid phases; sometimes called double-phase} % 1 sept2020; Anne added `double' after SSR mentioned that EU tends to use it

\newduneabbrev{pds}{PDS}{photon detection system}{The detector 
  subsystem sensitive to light produced in the \gls{lar} }

\newduneabbrev{hvs}{HVS}{high voltage system}{The detector 
  subsystem that provides the \gls{tpc} drift field}

\newduneabbrev{tpc}{TPC}{time projection chamber}{Depending on context: (1) A type of particle detector that uses an \efield together with a sensitive volume of gas or liquid, e.g., \gls{lar}, to perform a \threed reconstruction of a particle trajectory or interaction. The activity is recorded by digitizing the waveforms of current
  induced on the anode as the distribution of ionization charge passes by
  or is collected on the electrode. (2) TPC is also used in \gls{usproj} for ``total project cost''} 

\newduneabbrev{lartpc}{LArTPC}{liquid argon time-projection chamber}{A \gls{tpc} filled with liquid argon; %A class of detector technology that this technology forms 
the basis for the \gls{dune} \gls{fd} modules} %.   It typically entails observation of ionization activity by  electrical signals and of scintillation by optical signals}

\newduneabbrevs{apa}{APA}{anode plane assembly}{anode plane assemblies}{A unit of the \gls{sphd}
  detector module containing the elements sensitive to ionization in the \gls{lar}. 
  Each anode face has three planes of wires (two induction, one collection) to provide a 3D view, and interfaces to the cold electronics and photon detection system} 

\newduneabbrev{awg}{AWG}{American wire gauge} {U.S. standard set of non-ferrous wire conductor sizes}

\newduneabbrev{ufer}{Ufer}{concrete encased electrode} {U.S. National Electrical Code grounding method refered to as Concrete Encased Electrode}

\newduneabbrev{cro}{CRO}{charge readout}{The system for detecting
  ionization charge distributions in a %\gls{dp} 
  detector module} %11 may 2021 (no more DP module) still use?

\newduneabbrev{lro}{LRO}{light readout}{The system for detecting
  scintillation photons in a \gls{lartpc} detector module}

\newduneabbrev{shv}{SHV}{safe high voltage}{Type of bayonet mount
connector used on coaxial cables that has additional insulation 
compared to standard BNC and MHV connectors that makes it safer
for handling \gls{hv} by preventing accidental contact with the
live wire connector in an unmated connector or plug}

\newduneabbrev{fe}{FE}{front-end}{The front-end refers to a point that is
  ``upstream'' of the data flow for a particular subsystem. 
 For example the \gls{sphd} front-end electronics is where the cold electronics
  meet the sense wires of the TPC and the front-end \gls{daq} is where the \gls{daq} meets the output of the electronics}

\newduneabbrev{ }{DAQ RU}{DAQ readout unit}{The first element in the data flow of the \gls{daq}}

\newduneabbrev{cots}{COTS}{commercial off-the-shelf}{Items, typically hardware such as 
computers, that may be purchased whole, without any custom design or fabrication and 
thus at normal consumer prices and availability}

\newduneabbrev{i2c}{I2C}{Inter-Integrated Circuit}{I$^2$C or I2C is a synchronous, 
multi-master, multi-slave, packet switched, single-ended, serial computer bus widely used 
for attaching lower-speed peripheral ICs to processors and microcontrollers in short-distance, 
intra-board communication} %leave upper case

\newduneabbrev{spi}{SPI}{Serial Peripheral Interface}{The Serial Peripheral Interface is a 
synchronous serial communication interface specification used for short distance 
communication, primarily in embedded systems}%leave upper case

\newduneabbrev{miso}{MISO}{master in slave out}{The Master In Slave Out is a logic
signal on the \gls{spi} bus on which the data from the slave are transmitted once
a request from the master is received} %leave upper case

\newduneabbrev{mosi}{MOSI}{master out slave in}{The Master Out Slave In is a logic
signal on the \gls{spi} bus on which the data from the master is transmitted} %leave upper case

\newduneabbrev{uart}{UART}{Universal Asynchrous Receiver/Transmitter}{A universal 
asynchronous receiver-transmitter is a computer hardware device for asynchronous 
serial communication in which the data format and transmission speeds are configurable}%leave upper case

\newduneword{cr}{CR}{Capacitance-Resistance} %leave upper case

\newduneword{dc}{DC}{direct coupling} % I think these are ok lower case (anne)

\newduneword{ac}{AC}{Alternating Current; when used in the phrase ``AC coupling'' refers to a circuit element that filters out low-frequency components, such as constant offsets, leaving higher frequency signal components. The frequency filtering is determined both by a resistor and a capacitor}

\newduneabbrev{pll}{PLL}{Phase-Locked Loop}{A control system that generates an
output signal whose phase is related to the phase of an input signal}  %leave upper case

\newduneword{fifo}{FIFO}{First-In-First-Out} % leave in upper case

\newduneword{saci}{SACI}{\gls{slac} \gls{asic} Control Interface}

\newduneword{om3}{OM3}{Type of multi-mode fiber optic cable, typically capable of \SI{10}{Gbps} data transmission at lengths up to \SI{300}{m}}

\newduneword{om4}{OM4}{Type of multi-mode fiber optic cable, typically capable of \SI{10}{Gbps} data transmission at lengths up to \SI{550}{m}}

\newduneword{qfp}{QFP}{Quad Flat Package} % leave in upper case

\newduneabbrev{ams}{AMS}{analog and mixed signal}{Verilog-AMS is a derivative of the Verilog hardware description language that includes analog and mixed-signal extensions (AMS) in order to define the behavior of analog and mixed-signal systems}

\newduneabbrev{hepa}{HEPA}{High Efficiency Particulate Air}{The High Efficiency Particulate Air filters are a type of air filter that remove 99.97\% of particles that have a size greater than or equal to \SI{0.3}{\micro\meter}}  % leave in upper case

\newduneabbrev{uvm}{UVM}{universal verification methodology}{The Universal Verification Methodology is a standardized methodology for verifying integrated circuit designs}   % leave in upper case

\newduneword{lhc}{LHC}{Large Hadron Collider}

\newduneabbrev{lsb}{LSB}{least significant bit}{The bit with the lowest numerical value in a binary number}

\newduneabbrev{ldo}{LDO}{low-dropout regulator}{A low-dropout or LDO regulator is a \gls{dc} linear voltage regulator that can regulate the output voltage even when the supply voltage is very close to the output voltage}

\newduneabbrev{adc}{ADC}{analog-to-digital converter}{A sampling of a voltage
  resulting in a discrete integer count corresponding in some way to
  the input}

\newduneabbrev{inl}{INL}{integral non-linearity}{A commonly used measure of performance in \glspl{adc}. It is the deviation between the ideal input threshold value and the measured threshold level of a certain output code}

\newduneabbrev{dnl}{DNL}{differential non-linearity}{A commonly used measure of performance in \glspl{adc}. The DNL error is defined as the difference between an actual step width and the ideal value of one \gls{lsb}}

\newduneword{pnp}{PNP}{Type of bipolar junction transistor consistning of a
layer of N-doped semiconductor sandwiched between two layers of P-doped material}

\newduneabbrev{spice}{SPICE}{Simulation Program with Integrated Circuit Emphasis}{a general-purpose, 
open-source analog electronic circuit simulator. It is a program used in integrated 
circuit and board-level design to check the integrity of circuit designs and to 
predict circuit behavior} % Anne updated to abbrev oct 2022

\newduneabbrev{daq}{DAQ}{data acquisition}{The data acquisition system
  accepts data from the detector \gls{fe} electronics, buffers
  the data, performs a \gls{trigdecision}, builds events from the selected
  data and delivers the result to the offline \gls{diskbuffer}}

\newduneabbrev{iov}{IOV}{interval of validity}{Interval over which something is valid}

\newduneword{calci}{CALCI}{Calibration and Cryogenic Instrumentation}

\newduneword{detmodule}{far detector module}{The entire DUNE far detector design calls for segmentation into  four modules,  each with a total/fiducial mass of approximately \SI{17}{\kton}/\SI{10}{\kton}}
  
\newduneword{module}{module}{Many aspects of the DUNE far and near detectors are modular, so ``module'' must be understood in context.  It may refer to one of the four far detector modules,  distinct portions of a subdetector as in a ``field cage module,'' a software or electronics module,  e.g., a separate framework plug-in,  and so on}

\newduneword{detunit}{detector unit}{A portion of a \gls{detmodule} may be further partitioned into a number of similar parts.   For example, the \gls{sphd} \gls{tpc} is made up of \gls{apa}  units (and other elements)}

\newduneword{diskbuffer}{secondary DAQ buffer}{A secondary
  \gls{daq} buffer holds a small subset of the full rate as
  selected by a \gls{trigcommand}. 
  This buffer also marks the interface with the DUNE Offline}

\newduneabbrev{om}{OM}{online monitoring}{Processes that run inside
  the \gls{daq} on data ``in flight,'' specifically before landing on the
  offline disk buffer, and that provide feedback on the operation of
  the \gls{daq} itself and the general health of the data it is marshalling}

\newduneabbrev{dqm}{DQM}{data quality monitoring}{Analysis of the raw
  data to monitor the integrity of the data and the performance of the
  detectors and their electronics. This type of monitoring may be
  performed in real time, within the \gls{daq} system, or in later
  stages of processing, using disk files as input}

\newduneword{dumpbuffer}{DAQ dump buffer}{This \gls{daq} buffer
  accepts a high-rate data stream, in aggregate, from an associated
  portion of a \gls{detmodule} sufficient to collect all data likely relevant to
  a potential \gls{snb}}
\newduneabbrev{etf}{ETF}{Experiment Test Framework}{\gls{wlcg} testing middleware that runs grid jobs that actively test distributed sites' services and capabilities,  and reports back to monitoring services}

\newduneabbrev{etl}{ETL}{external trigger logic}{Trigger processing
  that consumes \gls{detmodule} level \gls{trignote} information
  and other global sources of trigger input and emits
  \gls{trigcommand} information back to the \glspl{mtl}}
\newduneabbrev{daqeti}{ETI}{external trigger interface}{Interface between \glspl{mtl} and external source and sinks of relevant trigger information}

\newduneword{trignote}{trigger notification}{Information provided by
  \gls{mtl} to \gls{etl} about \gls{trigdecision} %its 
  processing}

\newduneword{trigprimitive}{trigger primitive}{Information derived by
  the \gls{daq} \gls{fe} hardware that describes a region of space (e.g.,
  one or several neighboring channels) and time (e.g., a contiguous set
  of \gls{adc} sample ticks) associated with some activity}

\newduneword{externtrigger}{external trigger candidate}{Information
  provided to the \gls{mtl} about events external to a
  \gls{detmodule} so that it may be considered in forming
  \glspl{trigcommand}}

\newduneabbrev{daqoob}{OOB dispatcher}{out-of-band trigger command
  dispatcher}{This component is responsible for dispatching a \gls{snb} dump
  command to all \glspl{daqfer} in the \gls{detmodule}}

\newduneabbrev{mtl}{MTL}{module trigger logic}{Trigger processing
  that consumes \gls{detunit} level \gls{trigcommand} information
  and emits \glspl{trigcommand}. 
  It provides the \gls{etl} with \glspl{trignote} and receives back any
  \glspl{externtrigger}}

\newduneword{octant}{octant}{Any of the eight parts into which 4$\pi$
  is divided by three mutually perpendicular axes. 
  In particular in referencing the value for the mixing angle
  $\theta_{23}$}

\newduneword{trigcandidate}{trigger candidate}{Summary information derived
  from the full data stream and representing a contribution toward
  forming a \gls{trigdecision}}

\newduneword{trigcommand}{trigger command}{Information derived from
  one or more \glspl{trigcandidate}  that directs elements of a
  \gls{detmodule} to read out a portion of the data stream}

\newduneabbrev{tcm}{TCM}{trigger command message}{A message flowing
  down the trigger hierarchy from global to local context.  Also see \gls{tpm}}

\newduneabbrev{mlt}{MLT}{module level trigger}{The \gls{daq} component responsible for producing a \gls{trigdecision} that will be used to command the readout of a detector module}

\newduneword{trigdecision}{trigger decision}{The process by which
  \glspl{trigcandidate} are converted into \glspl{trigcommand}}

\newduneabbrev{tpm}{TPM}{trigger primitive message}{A message flowing
  up the trigger hierarchy from local to global context.  Also see \gls{tcm}}

\newduneabbrev{ipc}{IPC}{inter-process communication}{A system for software elements to exchange information between threads, local processes or across a data network.  An IPC system is typically specified in terms of protocols  composed of message types and their associated data schema}

\newduneword{daqdispre}{discovery and presence}{As used in the context of the \gls{ipc}, a system that provides mechanisms for a node on a communication network to learn of the existence of peers and their identity (discovery) as well as determine if they are currently operational or have become unresponsive (presence)}

\newduneabbrev{pubsub}{PUB/SUB}{publish-subscribe communication pattern}{An \gls{ipc} communication pattern where one element, the publisher, sends data to all connected elements, the subscribers.  Each subscriber may connect to multiple publishers.  A variant is PUB/SUB with topics where a subscriber may register an identifier, the topic, to limit the information received to just an associated subset}

\newduneabbrev{eb}{EB}{event builder}{A software agent that executes \glspl{trigcommand}  for one  \gls{detmodule} by reading out the requested data}
\newduneabbrev{daqdfo}{DFO}{data flow orchestrator}{The process by which trigger commands are executed in parallel and asynchronous manner by the back-end output subsystem of the \gls{daq}}

\newduneabbrev{daqubi}{UBI}{upstream DAQ buffer interface}{The process which provides read-only access to data residing in the upstream \gls{daq} buffers to processes on the network}

\newduneabbrev{cob}{COB}{cluster on board}{An ATCA motherboard housing four RCEs}

\newduneabbrev{rce}{RCE}{reconfigurable computing element}{Data processor located outside of the cryostat on a \gls{cob} that contains \gls{fpga}, RAM and \gls{ssd} resources, responsible for buffering data, producing trigger primitives, responding to triggered requests for data and synching \gls{snb} dumps}

\newduneabbrev{bow}{BOW}{Bump On Wire}{A working name for the front-end readout computing elements used in the nominal \gls{daq} design to interface the \gls{dp}  crates to the \gls{daq} front-end computers} %11 may 2021 (still needed? -check with Giovanna)

\newduneabbrev{atca}{ATCA}{Advanced Telecommunications Computing
  Architecture}{An advanced computer architecture specification developed for the telecommunications, military, and aerospace industries that incorporates the latest trends in  high-speed interconnect technologies, next-generation processors, and improved reliability, availability and serviceability} 

\newduneabbrev{utca}{$\mu$TCA}{Micro Telecommunications Computing Architecture}{The computer architecture specification followed by the crates that house charge and light readout electronics; used in the \gls{spvd} and \gls{dp} technologies} %11 may 2021 (still needed?)

\newduneabbrev{udp}{UDP}{user datagram protocol}{A simple,
  connectionless Internet protocol that supports data integrity
  checksums, requires no handshaking, and does not guarantee packet delivery}

\newduneabbrev{amc}{AMC}{advanced mezzanine card}{Holds digitizing
  electronics and lives in \gls{utca} crates}

\newduneabbrev{rf}{RF}{radio frequency}{Electromagnetic emissions
  that are within the (radio) frequency band of sensitivity of the detector
  electronics}

\newduneabbrev{fpga}{FPGA}{field programmable gate array}{An
integrated circuit technology that allows the hardware to be reconfigured to
execute different algorithms after its manufacture and deployment}

\newduneabbrev{fmc}{FMC}{FPGA mezzanine card}{Boards holding \glspl{fpga} and other integrated circuitry that attach to a motherboard}

\newduneabbrev{felix}{FELIX}{Front-End Link eXchange}{A
  high-throughput interface between \gls{fe} and trigger electronics
  and the standard PCIe computer bus}

\newduneword{daqpart}{DAQ partition}{A cohesive and
 coherent collection of \gls{daq} hardware and software working together to trigger and read out some portion of one detector module; it consists of an integral number of \glspl{daqfrag}. 
 Multiple \gls{daq} partitions may operate simultaneously, but each instance operates independently}

\newduneabbrev{fec}{DAQ FEC}{DAQ front-end computer}{The portion of one
  \gls{daqpart} that hosts the \gls{daqdr}, \gls{daqbuf} and
  \gls{daqds}.  It hosts the \gls{daqfer} and corresponding portion of the \gls{daqbuf}}

\newduneword{daqfrag}{DAQ front-end fragment}{The portion of one
  \gls{daqpart} relating to a single \gls{fec} and corresponding to an
  integral number of \glspl{detunit}.  See also \gls{datafrag}}

\newduneword{datafrag}{data fragment}{A block of data read out from a single \gls{daqfrag} that
span a contiguous period of time as requested by a \gls{trigcommand}}

\newduneabbrev{daqfer}{FER}{DAQ front-end readout}{The portion of a
  \gls{daqfrag} that accepts data from the detector electronics and
  provides it to the \gls{fec}}

\newduneabbrev{daqdr}{DDR}{DAQ data receiver}{The portion of the
  \gls{daqfrag} that accepts data from the \gls{daqfer}, emits
  trigger candidates produced from the input trigger primitives, and
  forwards the full data stream to the \gls{daqbuf}}

\newduneword{daqbuf}{DAQ primary buffer}{The portion
  of the \gls{daqfrag} that accepts full data stream from the
  corresponding \gls{detunit} and retains it sufficiently long for it
  to be available to produce a \gls{datafrag}}

\newduneword{daqds}{data selector}{The portion of the \gls{daqfrag}
  that accepts \glspl{trigcommand} and returns the corresponding
  \gls{datafrag}.  Not to be confused with \gls{daqdsn}}

\newduneword{daqdsn}{data selection}{The process of forming a trigger decision for selecting a subset of detector data for output by the \gls{daq} from the content of the detector data itself.  Not to be confused with \gls{daqds}}

\newduneabbrev{daqros}{DAQ RO}{DAQ readout subsystem}{The subsystem of the \gls{daq} for accepting and buffering data input from detector electronics}

\newduneabbrev{daqdss}{DAQ DS}{DAQ data selection subsystem}{The subsystem of the \gls{daq} responsible for forming a trigger decision based on a portion of the input data stream.  The majority subset of the \gls{daqtrs}}

\newduneabbrev{daqtrs}{DAQ TS}{DAQ trigger subsystem}{The subsystem of the \gls{daq} responsible for forming a trigger decision}

\newduneabbrev{daqbes}{DAQ BE}{DAQ back-end subsystem}{The portion of the \gls{daq} that is generally toward its output end.  It is responsible for accepting and executing trigger commands and marshaling the data they address to output storage buffers}

\newduneabbrev{daqtss}{DAQ TSS}{DAQ timing and synchronization subsystem}{The portion of the \gls{daq} that provides for timing and synchronization to various components}

\newduneabbrev{femb}{FEMB}{front-end mother board}{Refers to a unit of
  the \gls{sp} \gls{ce} that contains the \gls{fe} amplifier
  and \gls{adc} \glspl{asic} covering 128 channels}

\newduneword{asic}{ASIC}{application-specific integrated circuit}

\newduneword{lv}{LV}{low voltage}

\newduneabbrev{iceberg}{ICEBERG}{ICEBERG R\&D cryostat and electronics}{Integrated Cryostat and Electronics Built for Experimental Research Goals: a double-walled cryostat built and installed at \gls{fnal}  
for liquid argon detector R\&D and for testing of DUNE detector components}

\newduneword{coldadc}{ColdADC}{A newly developed 16-channels \gls{asic} providing analog to digital conversion}

\newduneword{coldata}{COLDATA}{A 64-channel control and communications \gls{asic}}
\newduneword{cryo}{CRYO}{(1) Integrated ASIC including \gls{fe} circuitry providing signal amplification and pulse shaping, analog to digital conversion, and control and communication functionalities for 64 channels; (2) acryonym for cryogenic systems and cryostat work scopes in \gls{lbnf}}

\newduneword{larasic}{LArASIC}{A 16-channel \gls{fe} \gls{asic} that provides signal amplification and pulse shaping}

\newduneword{cmos}{CMOS}{Complementary metal-oxide-semiconductor}

\newduneabbrev{enc}{ENC}{equivalent noise charge}{The equivalent noise charge is the input charge that corresponds to a \gls{s/n}$=1$}
\newduneword{sar}{SAR}{successive approximation register}

\newduneword{protodune}{ProtoDUNE}{Either of the two initial DUNE prototype detectors constructed at \gls{cern}. % and operated in  a CERN test beam (expected fall 2018). 
  One prototype implemented \gls{sp} technology and the other \gls{dp}}
  
\newduneword{protodune2}{ProtoDUNE-II}{The second run of a ProtoDUNE detector}  % #33

\newduneword{pdsp}{ProtoDUNE-SP}{The \gls{sphd} \gls{protodune} detector constructed at \gls{cern}  in \gls{np04}}

\newduneword{pddp}{ProtoDUNE-DP}{The \gls{dp} \gls{protodune} detector constructed at \gls{cern} in \gls{np02}}

\newduneword{wa105}{WA105 DP demonstrator}{The %\SI[product-units=power]{3x1x1}{m} 
3m$\times$1m$\times$1m WA105 \gls{dp} prototype detector at \gls{cern}}

\newduneword{rawevent}{DAQ event block}{The unit of data output by the
  \gls{daq}.  
  It contains trigger and detector data spanning a unique, contiguous
  time period and a subset of the detector channels}

\newduneabbrev{ssd}{SSD}{solid-state disk}{Any storage device that
  may provide sufficient write throughput to receive, both collectively and
  distributed, the sustained full rate of data from a \gls{detmodule}
  for many seconds}
\newduneabbrev{nvme}{NVMe}{Non-volatile memory express}{A specification for an interface to storage media attached via PCIe}

\newduneabbrev{hlt}{HLT}{high-level trigger}{This is actually a filter applied to data that has been triggered and aggregated in order to further reduce or characterize it}

\newduneabbrev{pid}{PID}{particle ID}{Particle identification}

\newduneword{readout window}{readout window}{A fixed, atomic and
  continuous period of time over which data from a \gls{detmodule}, in
  whole or in part, is recorded. 
  This period may differ based on the trigger that initiated the
  readout}

\newduneabbrev{zs}{ZS}{zero-suppression}{Used to delete some portion of a
  data stream that does not significantly deviate from zero or
  intrinsic noise levels. 
  It may be applied at different granularity from per-channel to per
  \gls{detunit}}

\newduneword{rc}{RC}{Depending on context, one of (1) resistive-capacitive (circuit), (2) run control, the system for configuring, starting and terminating the \gls{daq}, or (3) resource coordinator, a member of the \gls{dune} management team responsible for coordinating the financial resources of the project} % #33

\newduneabbrev{daqccm}{CCM}{DAQ control, configuration and monitoring subsystem}{A system for controlling, configuring and monitoring other systems in particular those that make up the \gls{daq} where the CCM encompasses \gls{rc}}

\newduneword{daqrun}{DAQ run}{A period of time over which relevant data taking conditions and \gls{daq} configuration are asserted to be unchanged. 
  Multiple \gls{daq} runs may occur simultaneously when multiple \glspl{daqpart} are active. 
  This term should not be confused with DUNE experiment or beam ``runs'' that typically span many \gls{daq} runs}
\newduneword{daqrunnum}{DAQ run number}{A monotonically increasing count that uniquely and globally identifies a \gls{daqrun}}

\newduneabbrev{snb}{SNB}{supernova neutrino burst}{A prompt 
  increase in the flux of low-energy neutrinos emitted in the first few seconds of a core-collapse supernova.  It can also refer to a trigger command type that may be due to this phenomenon,
  or detector conditions that mimic its interaction signature}

\newduneabbrev{snble}{SNB/LE}{supernova neutrino burst and low
  energy}{Supernova neutrino burst and low-energy physics program}

\newduneabbrev{snews}{SNEWS}{SuperNova Early Warning System}{A global
  supernova neutrino burst trigger formed by a coincidence of \gls{snb} 
  triggers collected from participating experiments}

\newduneabbrev{pps}{1PPS signal}{one-pulse-per-second signal}{An
  electrical signal with a fast rise time and that arrives in real
  time with a precise period of one second}

\newduneabbrev{sls}{SLS}{spill location system}{A system residing at
  the DUNE far detector site that provides information, possibly
  predictive, indicating periods of time when neutrinos are being
  produced by the \gls{fnal} Main Injector beam spills}

\newduneabbrev{wib}{WIB}{warm interface board}{Digital electronics
  situated just outside a FD cryostat that receives digital data
  from the \glspl{femb} (part of \gls{ce}) over cold copper connections and sends it to the \gls{rce}
  \gls{fe} readout hardware}

\newduneabbrev{gps}{GPS}{Global Positioning System}{A satellite-based system that provides a highly accurate \gls{pps} that may be used to synchronize clocks and determine location}

\newduneabbrev{ntp}{NTP}{Network Time Protocol}{A networking protocol that allows synchronizing of clocks to within a few \si{\milli\second} of a time standard on a local network and within a few tens of \si{\milli\second} over the Internet} 

\newduneword{ptp}{PTP}{Depending on context, either p-terphenyl, a \gls{wls} material, or Precision Time Protocol, a networking protocol that allows synchronizing of clocks to within a few \si{\micro\second} of a time standard on a local network}  % #33

\newduneabbrev{irig}{IRIG}{inter-range instrumentation group}{A standards body that defined a time-code standard for transferring timing information}

\newduneabbrev{nic}{NIC}{network interface controller}{Hardware for controlling the interface to a communication network.  Typically, one that obeys the Ethernet protocol}

\newduneabbrev{wiec}{WIEC}{warm interface electronics crate}{Crates mounted on the signal flanges that contain the \glspl{wib}}

\newduneabbrev{ptc}{PTC}{power and timing card}{Cards that provide further processing and distribution of the signals entering and exiting the \gls{sp} cryostat}

\newduneabbrev{ptb}{PTB}{power and timing backplane}{Backplane used to connect the \glspl{wib} and the \glspl{ptc} on the \gls{wiec}. Also connects the \gls{ce} flange on the cryostat penetration}

\newduneabbrev{sipm}{SiPM}{silicon photomultiplier}{A solid-state
  avalanche photodiode sensitive to single \phel signals}

\newduneabbrev{cisc}{CISC}{cryogenic instrumentation and slow controls}{Includes equipment to monitor all detector  components and  \gls{lar} quality and behavior, and provides a control system for many of the detector components}

\newduneword{fte}{FTE}{full-time equivalent. A unit of labor
  for the project. One year of work from one person}

\newduneword{art}{art}{A software framework implementing an
  event-based execution paradigm} %http://art.fnal.gov/

\newduneabbrev{sam}{SAM}{sequential
  access via metadata}{A data-handling system to store and retrieve
  files and associated metadata, including a complete record of the
  processing that has used the files}

\newduneword{artdaq}{artdaq}{A data acquisition toolkit for data transfer, aggregation and processing}

\newduneword{beamline}{beamline}{A sequence of control and monitoring devices used for the formation of a directed collection of particles; also subproject within \gls{usproj}} % #36

\newduneword{cdr}{CDR}{Depending on context, either ``conceptual design report,'' a formal project  document  that describes the experiment at a conceptual level, or ``conceptual design review,'' a formal review of the conceptual design of the experiment or of a component}  % #33

\newduneabbrev{cf}{CF}{conventional facilities}{Pertaining to
  construction and operation of buildings and conventional infrastructure, and includes  cavern excavation}

\newduneabbrev{cp}{CP}{charge conjugation and parity}{Product of charge conjugation and parity
  transformations} % updated to include `conjugation' 12/14/20 per Steve Manly

\newduneabbrev{cpt}{CPT}{charge, parity, and time reversal symmetry}{product of charge, parity
  and time-reversal transformations}

\newduneabbrev{cpv}{CPV}{Charge Conjugation-Parity Symmetry Violation}{Lack of
  symmetry in a system before and after charge conjugation and parity
  transformations are applied. 
  For \gls{cp} symmetry to hold,  a particle turns into its
 corresponding antiparticle under a charge transformation,  and a parity
transformation inverts its space coordinates,  i.e. produces the mirror image}

\newduneword{doe}{DOE}{U.S. Department of Energy}

\newduneabbrev{fra}{FRA}{Fermi Research Alliance}{A joint partnership of the University of Chicago and the Universities Research Association (URA) that manages and operates \gls{fnal} on behalf of the \gls{doe}}

\newduneabbrev{dune}{DUNE}{Deep Underground Neutrino Experiment}{A leading-edge, international experiment for neutrino science and proton decay studies; refers to the entire international experiment and collaboration} %updated 11 may 21

\newduneabbrev{esh}{ES\&H}{environment, safety and health}{A discipline and specialty that studies and implements practical aspects of environmental protection and safety at work} % The LBNF/DUNE ES\&H program complies with applicable standards and local, state, and federal legal requirements through the Fermilab ``work smart'' set of standards and the contract between Fermi Research Alliance and the DOE Office of Science (FRA-DOE)}

\newduneabbrev{ppe}{PPE}{personnel protective equipment}{Equipment worn to minimize exposure to hazards that cause serious workplace injuries and illnesses}

\newduneabbrev{odh}{ODH}{oxygen deficiency hazard}{a hazard that occurs when inert gases such as nitrogen, helium, or argon displace room air and thus reduce the percentage of oxygen below the level required for human life}

\newduneabbrev{feshm}{FESHM}{Fermilab Environment, Safety and Health Manual}{The document that contains \gls{fnal} 's policies and procedures designed to manage environment, safety, and health in all its programs}

\newduneabbrev{fscf}{FSCF}{Far Site Conventional Facilities}{The \gls{cf} at the DUNE far detector site, \gls{surf}, including all detector caverns and support infrastructure}%updated 11 may 21
  
\newduneabbrev{nscf}{NSCF}{Near Site Conventional Facilities}{The \gls{cf} at the DUNE near detector site, \gls{fnal}}

\newduneabbrev{fd1c}{FD1+C}{far detector module 1 + cryogenics}{The first far detector module to be built at \gls{surf}, including integration and installation, and all cryogenics infrastructure to support FD1 and \gls{fd2}; also a subproject of \gls{usproj}} %new 11 may 21

\newduneabbrev{fd2}{FD2}{far detector module 2}{The second DUNE far detector module to be built at \gls{surf}} %new 11 may 21

\newduneabbrevs{gut}{GUT}{grand unified theory}{grand unified theories}{A class of theories that unifies the electroweak and strong forces}

\newduneabbrev{lar}{LAr}{liquid argon}{Argon in its liquid phase; it is a cryogenic liquid with a boiling point of \SI{87}{K} and density of \SI{1.4}{g/ml}}

\newduneabbrev{lbl}{LBL}{long-baseline}{Refers to the distance between the 
  neutrino source  and the \gls{fd}.  It can also refer to the distance between the near and far detectors. 
  The ``long'' designation is an approximate and relative distinction. For DUNE, this distance  (between \gls{fnal} and \gls{surf}) is approximately \SI{1300}{km}}

\newduneabbrev{lbnf}{LBNF}{Long-Baseline Neutrino Facility}{Long-Baseline Neutrino Facility; refers to the facilities that support the experiment including in-kind contributions under the line-item project. The portion of \gls{usproj} responsible for developing the neutrino beam, the far site cryostats,  and far and near site cryogenics systems, and the conventional facilities, including the excavations } % #36, updated 11 may 2021
  
\newduneabbrev{lbnf-dune}{LBNF/DUNE}{LBNF and DUNE enterprise}{Long-Baseline Neutrino Facility/Deep Underground Neutrino Experiment; refers to the overall enterprise or program including \gls{usproj}, participating international projects, and the \gls{dune} experiment and collaboration} %updated 11 may 2021 (#35 09nov20)

\newduneword{usproj}{LBNF/DUNE-US}{Long-Baseline Neutrino Facility/Deep Underground Neutrino Experiment - United States; project to design and build the conventional and beamline facilities and the \gls{doe} contributions to the detectors. It is organized as a \gls{doe}/\gls{fnal} project and incorporates contributions to the facilities from international partners. It also acts as host for the installation and integration of the DUNE detectors} % new oct 2021

\newduneword{duneus}{DUNE-US}{Deep Underground Neutrino Experiment - United States; refers to the U.S. contribution to DUNE under the line-item \gls{usproj} project} % new 11 may 2021

\newduneabbrev{lbnc}{LBNC}{Long-Baseline Neutrino Committee}{The committee, composed of internationally prominent scientists with relevant expertise, charged by the \gls{fnal} director to review the scientific, technical, and managerial progress, plans and decisions associated with \gls{dune}}

\newduneabbrev{ncg}{NCG}{Neutrino Cost Group}{A group of internationally prominent scientists with relevant experience that is charged by the \gls{fnal} director to review the cost, schedule, and associated risks for the \gls{dune} experiment}

\newduneabbrev{mh}{MH}{mass hierarchy}{Describes the separation
  between the mass squared differences related to the solar and
  atmospheric neutrino problems (also written as \gls{mo})}

\newduneabbrev{mo}{MO}{mass ordering}{See \gls{mh}}

\newduneabbrev{mi}{MI}{Fermilab Main Injector}{An accelerator at
  \gls{fnal} that provides a beam of high-energy protons to the the \gls{usproj}  beamline} 

\newduneabbrev{pot}{POT}{protons on target}{Typically used as a unit
  of normalization for the number of protons striking the neutrino
  production target}

\newduneabbrev{qa}{QA}{quality assurance}{The process of ensuring that %The set of actions taken to provide confidence that 
the quality of each element meets requirements during design and development, and to detect and correct poor results prior to production} % #33

\newduneabbrev{qc}{QC}{quality control}{The process (e.g., inspection, testing, measurements) %An aggregate of activities (such as design analysis and inspection for defects) performed 
of ensuring that each manufactured element meets its quality requirements prior to assembly or installation} %adequate quality in manufactured products}  % #33

\newduneabbrev{sm}{SM}{Standard Model}{Refers to a theory describing
  the interaction of elementary particles}

\newduneword{tdr}{TDR}{Depending on context, either ``technical design report,'' a formal project  document  that describes the experiment at a technical level, or ``technical design review,'' a formal review of the technical design of the experiment or of a component}  % #33

\newduneabbrev{tp}{IDR}{interim design report}{An intermediate
milestone on the path to a full \gls{tdr}} % changed from ``technical proposal'' 6/6/2018

\newduneabbrev{ckm}{CKM matrix}{Cabibbo-Kobayashi-Maskawa
  matrix}{Refers to the matrix describing the mixing between mass and
  weak eigenstates of quarks}

\newduneabbrev{cl}{CL}{confidence level}{Refers to a probability
  used to determine the value of a random variable given its
  distribution}

\newduneabbrev{pmns}{PMNS}{Pontecorvo-Maki-Nakagawa-Sakata}{A type of matrix that describes the mixing between mass and weak eigenstates of
  the neutrino}

\newduneword{hnl}{HNL}{heavy neutral lepton} %  #37, 21 may 21
\newduneabbrevs{cpa}{CPA}{cathode plane assembly}{cathode plane assemblies}{The component of the \gls{sphd} detector module that provides the drift HV cathode}

\newduneword{fc}{field cage}{The component of a \gls{lartpc} that contains and shapes the applied \efield}

\newduneword{cpafc}{CPA/FC}{A pair of \gls{cpa} panels and the top and bottom \gls{fc} portions that attach to the pair; an intermediate assembly for installation into the \gls{spmod} }

\newduneabbrev{topfc}{top FC}{top field cage}{The horizontal portions of the \gls{sphd} \gls{fc}   on the top of the \gls{tpc}}

\newduneabbrev{botfc}{bottom FC}{bottom field cage}{The horizontal portions of the \gls{sphd} \gls{fc} on the bottom of the \gls{tpc}}

\newduneabbrev{ewfc}{endwall FC}{endwall field cage}{The vertical portions of the \gls{fc} near the end walls}

\newduneword{gp}{ground plane}{An electrode held electrically neutral relative to Earth ground voltage; it is mounted on the \gls{fc} to protect the cryostat wall}

\newduneword{gg}{ground grid}{An electrode held electrically neutral relative to Earth ground voltage; it is installed between the cathode and the \glspl{pd} in a \gls{dpmod} to protect the \glspl{pmt}, maintaining high transparency to light} %11 may 2021 (no more DP module) still used? Ask Jae Yu

\newduneabbrev{alara}{ALARA}{as low as reasonably
  achievable}{Typically used with regard management of radiation
  exposure but may be used more generally. It means making every
  reasonable effort to maintain e.g., exposures, to as far below the
  limits as practical, consistent with the purpose for that the
  activity is undertaken}

\newduneabbrev{ecal}{ECAL}{electromagnetic calorimeter}{A detector
  component that measures energy deposition of traversing particles (in the \gls{dune} near detector design)}

\newduneabbrev{hv}{HV}{high voltage}{Generally describes a voltage
  applied to drive the motion of free electrons through some media, e.g., LAr}

\newduneword{spmod}{SP module}{single-phase DUNE \gls{fd} module}  % to change? 11 may 2021 
\newduneword{vdmod}{vertical drift module}{vertical drift DUNE \gls{fd} module} %11 may 2021
\newduneword{dpmod}{DP module}{dual-phase DUNE \gls{fd} module} %11 may 2021 (no more DP module)
\newduneword{dsp}{DUNE-SP}{a single-phase DUNE far detector module} % 33 to change? 11 may 2021 (no more DP module)
\newduneabbrev{tcoord}{TC}{technical coordinator}{A member of the \gls{dune} management team responsible for organizing the technical aspects of the project effort; is head of \gls{tc}}

\newduneabbrev{tc}{TCN}{technical coordination}{The DUNE organization responsible for overall integration of the detector elements and successful execution of the detector construction project; areas of responsibility include general project oversight, systems engineering, \gls{qa} and safety}  % #33

\newduneabbrev{exb}{EB}{executive board}{The highest level DUNE
  decision-making body for the collaboration}

\newduneabbrev{tb}{TB}{technical board}{The DUNE organization responsible for
  evaluating technical decisions}

\newduneabbrev{rrb}{RRB}{Resources Review Board}{A part of \gls{dune}'s international project governance structure, composed of representatives of all funding agencies that sponsor the project, and of  \gls{fnal} management, established to provide coordination among funding partners and oversight of \gls{dune}}

\newduneabbrev{inc}{INC}{International Neutrino Council}{A highest-level international advisory body to the U.S. \gls{doe} and the  \gls{fnal} directorate on matters related to the  \gls{lbnf} and the  \gls{pip2} projects. This council is composed of representatives from the international funding agencies and  \gls{cern} that make major contributions the infrastructure}

\newduneabbrev{cc}{CC}{charged current}{Refers to an interaction
  between elementary particles where a charged weak force carrier
  ($W^+$ or $W^-$) is exchanged}

\newduneabbrev{dis}{DIS}{deep inelastic scattering}{Refers to interaction between
  elementary particles and a nucleus in an energy range where the
  interaction can be modeled as occurring between constituent quarks
  of one nucleon and resulting in no bulk recoil of the resulting
  nucleus} % 1sep2020 ArturA: swapped with {qe}

\newduneabbrev{fsi}{FSI}{final-state interactions}{Refers to
  interactions between elementary or composite particles subsequent to
  the initial, fundamental particle interaction, such as may occur as
  the products exit a nucleus}
  
\newduneabbrev{fsint}{FSI}{far site integration}{The scope of work at the \gls{fs} for the \gls{integoff}} % #36 - can't put it together with prev item

\newduneword{geant4}{Geant4}{A
  software toolkit for the simulation of the passage of particles
  through matter using \gls{mc} methods}

\newduneabbrev{genie}{GENIE}{Generates Events for Neutrino Interaction
  Experiments}{Software providing an object-oriented neutrino
  interaction simulation resulting in kinematics of the products of
  the interaction}

\newduneabbrev{mc}{MC}{Monte Carlo}{Refers to a method of numerical
  integration that entails the statistical sampling of the integrand
  function. 
  Forms the basis for some types of detector and physics simulations}

\newduneabbrev{qe}{QE}{quasi-elastic}{Refers to the 
  interaction of an elementary charged particle with a nucleus in an
  energy range where the interaction can be modeled as taking place with
  individual nucleons} % 1sep2020 ArturA: swapped with {dis}

\newduneabbrev{mou}{MoU}{memorandum of understanding}{A project management methodology that  \gls{usproj} uses to document agreement,s,  e.g., between \gls{fnal} and the Project, for how \gls{fnal}  will support the Project. More generally, a document
  summarizing an agreement between two or more parties} % #36 elaine

\newduneabbrev{pip2}{PIP-II}{Proton Improvement Plan II}{A \gls{fnal} project for
  improving the protons on target delivered delivered by the \gls{lbnf} neutrino production beam. 
  This is version two of this plan and it is planned to be followed by a PIP-III}
  
\newduneabbrev{sdsta}{SDSTA}{South Dakota Science and Technology
  Authority}{The legal entity that manages \gls{surf}, in Lead, S.D}
  
\newduneabbrev{sdsd}{SDSD}{Fermilab South Dakota Services Division}{A \gls{fnal}  division responsible providing host laboratory functions at SURF in South Dakota}

\newduneabbrev{firus}{FIRUS}{Facility Information Reporting Utility System}
 {Facility incident reporting systems, one at \gls{fnal} and at \gls{surf}, that monitors and reports the status of various fire, security and utility sensors} % #36

\newduneabbrev{bsi}{BSI}{building and site infrastructure}
 {The work package for outfitting of the \gls{lbnf} underground infrastructure}

\newduneabbrev{wbs}{WBS}{work breakdown structure}{An organizational
  project management tool by which the tasks to be performed are
  partitioned in a hierarchical manner}

\newduneabbrev{br}{BR}{branching ratio}{A fractional probability for a
  decay of a composite particle to occur into some specified set or
  sets of products}
\newduneword{bsm}{BSM}{beyond the Standard Model}

\newduneabbrev{dm}{DM}{dark matter}{The term given to the unknown
  matter or force that explains measurements of galaxy motion % motion of galaxies
  that are otherwise inconsistent with the amount of mass associated
  with the observed amount of photon production}
  
\newduneabbrev{bdm}{BDM}{boosted dark matter}{A new model that describes a relativistic dark matter particle boosted by the annihilation of heavier dark matter particles in the galactic center or the sun}

\newduneabbrev{cern}{CERN}{European Laboratory for Particle Physics}{The leading particle physics laboratory in Europe and home to the \glspl{protodune} and other prototypes and demonstrators, including the \glspl{mod0}} % per Steve Manly/Hiro Tanaka Dec 2020 (In French, the Organisation Europ\'{e}enne pour la Recherche Nucl\'{e}aire, derived from Conseil Europ\'{e}en pour la Recherche Nucl\'{e}aire)}

\newduneabbrev{dsnb}{DSNB}{diffuse supernova neutrino background}{The
  term describing the pervasive, constant flux of neutrinos due to all
  past supernova neutrino bursts}

\newduneabbrev{espp}{ESPP}{European Strategy for Particle Physics}{The
cornerstone of Europe's
decision-making process for the long-term future of the
field. Mandated by the \gls{cern} Council, it is formed through a broad
consultation of the grass-roots particle physics community, it
actively solicits the opinions of physicists from around the world,
and it is developed in close coordination with similar processes in
the USA and Japan in order to ensure coordination between regions and
optimal use of resources globally}

\newduneabbrev{gar}{GAr}{gaseous argon}{argon in its gas phase}
\newduneabbrev{gartpc}{GArTPC}{gaseous argon time-projection chamber}{A \gls{tpc} filled with gaseous argon} %; a possible technology choice for the \gls{nd}} AH rmvd 12/2020

\newduneabbrev{globes}{GLoBES}{General Long-Baseline Experiment
  Simulator}{A software package for simulating energy spectra of
  neutrino flux, interactions, and energy spectra measured after application of some
  model of a detector response)}

\newduneabbrev{snowglobes}{SNOwGLoBES}{SuperNova
Observatories with GLoBES} {From the official description: SNOwGLoBES is public software for computing interaction rates and distributions of observed quantities for \gls{snb} neutrinos in common detector materials} % (Anne thinks this was too long.)
\newduneword{l/e}{L/E}{length-to-energy ratio}
\newduneword{lri}{LRI}{long-range interactions}
\newduneabbrev{nc}{NC}{neutral current}{Refers to an interaction
  between elementary particles where a neutrally charged weak force carrier
  ($Z^0$) is exchanged}

\newduneabbrev{nh}{NH}{normal hierarchy}{Refers to the neutrino mass
  eigenstate ordering whereby the sign of the mass squared difference
  associated with the atmospheric neutrino problem is positive}

\newduneabbrev{ih}{IH}{inverted hierarchy}{Refers to the neutrino mass
  eigenstate ordering whereby the sign of the mass squared difference
  associated with the atmospheric neutrino problem is negative}

\newduneabbrev{no}{NO}{normal ordering}{Refers to the neutrino mass
  eigenstate ordering whereby the sign of the mass squared difference
  associated with the atmospheric neutrino problem is positive}

\newduneabbrev{io}{IO}{inverted ordering}{Refers to the neutrino mass
  eigenstate ordering whereby the sign of the mass squared difference
  associated with the atmospheric neutrino problem is negative}

\newduneabbrev{msw}{MSW}{Mikheyev-Smirnov-Wolfenstein effect}{Explains
  the oscillatory behavior of neutrinos produced inside the sun as
  they traverse the solar matter}

\newduneabbrev{nsi}{NSI}{nonstandard interaction}{A general class of
  theory of elementary particles other than the Standard Model}

\newduneabbrev{pfive}{P5}{Particle Physics Project Prioritization
Panel}{The Particle Physics Project Prioritization Panel (P5) was a
subpanel of the High Energy Physics Advisory Panel (HEPAP). It completed
its Report, a ten-year strategic plan for high energy physics in the
U.S., in 2014. This report included a recommendation that ``host a world-leading neutrino
program that will have an optimized set of short- and long-baseline neutrino oscillation experiments, and its long-term focus
is a reformulated venture referred to here as the Long Baseline
Neutrino Facility (LBNF)''}

\newduneabbrev{sme}{SME}{standard-model extension}{an effective field theory that contains the \gls{sm}, general relativity, and all possible operators that break Lorentz symmetry (Wikipedia)}

\newduneabbrev{susy}{SUSY}{supersymmetry}{Theoretical symmetry between a fermion and a boson}

\newduneabbrev{wimp}{WIMP}{weakly-interacting massive particle}{A
  hypothesized particle that may be a component of dark matter}

\newduneabbrev{ce}{CE}{cold electronics}{Analog and digital readout electronics that operate at cryogenic temperatures}

\newduneabbrev{crp}{CRP}{charge-readout plane}{An anode technology using a stack of perforated \glspl{pcb} with etched electrode strips to provide \gls{cro} in 3D; it has two induction layers and one collection layer; it is used in the \gls{spvd} \gls{fd} and \gls{dp} designs} % with projecting electrodes immersed in \dword{lar}}
\newduneabbrev{dram}{DRAM}{dynamic random access memory}{A computer memory technology}

\newduneabbrev{fnal}{Fermilab}
{Fermi National Accelerator Laboratory}{U.S. national laboratory in Batavia, IL.  It is the laboratory that hosts \gls{lbnf} and \gls{dune}, and serves as the experiment's near site}

\newduneabbrev{bnl}{BNL}{Brookhaven National Laboratory}{US national laboratory in Upton, NY}

\newduneabbrev{slac}{SLAC}{SLAC National Accelerator Laboratory}{US national laboratory in Menlo Park, CA}

\newduneabbrev{lbnl}{LBNL}{Lawrence Berkeley National Laboratory}{US national laboratory in Berkeley, CA}

\newduneabbrev{anl}{ANL}{Argonne National Laboratory}{US national laboratory in Lemont, IL}

\newduneabbrev{lanl}{LANL}{Los Alamos National Laboratory}{US national laboratory in Los Alamos, NM}

\newduneword{fs}{FS}{Depending on context, one of (1) the far site, \gls{surf}, where the DUNE far detector is located; (2) ``Full Stream'' relates to a data stream that has not undergone selection, compression or other form of reduction} % #36 (see change from item above)

\newduneabbrev{lem}{LEM}{large electron multiplier}{A micro-pattern detector suitable for use in ultra-pure argon vapor; LEMs consist of copper-clad PCB boards with sub-millimeter-size holes through which electrons undergo amplification}

\newduneabbrev{lng}{LNG}{liquefied natural gas}{Pertaining to natural gas in its liquid phase}

\newduneabbrev{mip}{MIP}{minimum ionizing particle}{Refers to a
  particle traversing some medium such that the particle's mean energy loss is  
  near the minimum}
\newduneabbrev{pd}{PD}{photon detector}{The detector
  elements involved in measurement of the number and arrival times of
  optical photons produced in a detector module} 

\newduneabbrev{pmt}{PMT}{photomultiplier tube}{A device that makes use
  of the photoelectric effect to produce an electrical signal from the
  arrival of optical photons}

\newduneabbrev{ppm}{ppm}{parts per million}{A concentration equal to one part in $10^{6}$}
\newduneabbrev{ppb}{ppb}{parts per billion}{A concentration equal to one part in $10^{9}$}
\newduneabbrev{ppt}{ppt}{parts per trillion}{A concentration equal to one part in $10^{12}$}
\newduneword{rio}{RIO}{reconfigurable input output}

\newduneabbrev{s/n}{S/N}{signal-to-noise}{signal-to-noise ratio}
\newduneword{ssp}{SSP}{\gls{sipm} signal processor}

\newduneabbrev{sbn}{SBN}{Short-Baseline Neutrino}{A \gls{fnal} program consisting of three collaborations, \gls{microboone}, \gls{sbnd}, and \gls{icarus}, to perform sensitive searches for $\nue$ appearance and $\numu$ disappearance in the Booster Neutrino Beam}

\newduneword{stt}{STT}{straw tube tracker}

\newduneword{wire board}{wire board}{At the head end of the APA in the \gls{sphd} \gls{tpc}, stacks of electronics boards referred to as ``wire boards'' are arrayed to anchor the wires.  They also provide the connection between the wires and the cold electronics} %?? long for a word. ??

\newduneabbrev{wls}{WLS}{wavelength-shifting}{A material or process by
  which incident photons are absorbed by a material and photons are
  emitted at a different, typically longer, wavelength}
  
\newduneabbrev{tpb}{TPB}{tetra-phenyl butadiene}{A \gls{wls} material}

\newduneabbrev{sft}{SFT}{signal feedthrough}{A cryostat penetration allowing for the passage of cables or other extended parts}

\newduneabbrev{sftchimney}{SFT chimney}{signal feedthrough chimney}{A volume above the cryostat penetration used for a signal feedthrough} %19 nov 2021  

\newduneabbrev{catiroc}{CATIROC}{charge and time integrated readout chip}{A complete read-out chip manufactured in AustriaMicroSystem designed to read arrays of 16 photomultipliers}

\newduneabbrev{wr}{WR}{White Rabbit}{A component of the timing system that forwards clock signal and time-of-day reference data to the master timing unit}

\newduneabbrev{mch}{MCH}{MicroTCA Carrier Hub}{A network switching device}

\newduneabbrev{wrmch}{WR-MCH}{White Rabbit \gls{utca} Carrier Hub}{A card mounted in \gls{utca} crate that recieves time syncronization information and trigger data packets over \gls{wr} network and disributes them to the \gls{amc} over \gls{utca} backplane} 

\newduneabbrev{wrtsn}{WR-TSN}{White Rabbit TimeStamping Node}{A unit on the \gls{wr} network that timestamps the trigger signals and sends out trigger data packets to \gls{wrmch}}

\newduneword{qap}{QAP}{quality assurance plan} %{A project management device for planning \gls{qa}}
\newduneword{ieshp}{IESHP}{integrated environmental, safety and health plan}%{Refers to the LBNF/DUNE project planning instrument}
\newduneword{dmp}{DMP}{data management plan} %{A project management device to state how the experimental data will be managed}
\newduneword{qam}{QAM}{quality assurance manager} %{The manager of \gls{qa} for the LBNF/DUNE project}

\newduneabbrev{dss}{DSS}{detector support system}{The system of rails suspended from the cryostat ceiling in a \gls{spmod} used to support the \gls{apa}s, \gls{cpa}s, and the \glspl{ewfc}} % update from MV 1sept2020 

\newduneabbrev{ddss}{DDSS}{DUNE detector safety system}{The hardware system responsible for the safety of the detector, implemented either via a \gls{plc} or via custom hardware protections} % update from MV 1sept2020 

\newduneabbrev{tco}{TCO}{temporary construction opening}{An opening in the side of a cryostat through which detector elements are brought into the cryostat; utilized during construction and installation}

\newduneabbrev{surf}{SURF}{Sanford Underground Research Facility}{The laboratory in Lead, SD, USA where the \gls{dune} \gls{fd} will be installed and operated; also where the \gls{usproj} \gls{fscf} and the \gls{fs} cryostat and cryogenics systems will be constructed} %  36

\newduneabbrev{uit}{UIT}{underground installation team}{An organizational unit responsible for installation in the underground area at the \gls{surf} site}

\newduneabbrev{cmgc}{CMGC}{construction manager/general contractor}{The contracted company hired to manage overall construction, used by \gls{lbnf} at the \gls{surf} site for the \gls{fscf} construction} % #36 old: The organizational unit responsible for management of the construction of conventional facilities at the underground area at the \gls{surf} site}

\newduneword{pdr}{PDR}{Depending on context, either ``preliminary design report,'' a formal project document  that describes the experiment at a preliminary level, or ``preliminary design review,'' a formal review of the preliminary design of the experiment or of a component} % #33

\newduneword{fdr}{FDR}{Depending on context, either ``final design report,'' a formal project document  that describes the experiment at a final level, or ``final design review,'' a formal review of the final design of the experiment or of a component} %  per #33, not sure if we need a `final design report' - yes per #36

\newduneabbrev{prr}{PRR}{production readiness review}{A project management mechanism by which the production readiness is reviewed}  % #33

\newduneabbrev{irr}{IRR}{installation readiness review}{A project management mechanism by which the plan for installation is reviewed}  % #33
\newduneabbrev{orr}{ORR}{operational readiness review}{A project management mechanism by which the operational readiness is reviewed}  % #33

\newduneabbrev{ppr}{PPR}{production progress review}{A project management mechanism by which the progress of production is reviewed}  % #33

\newduneabbrev{edms}{EDMS}{engineering document management system}{A computerized document management system developed and supported at the \gls{cern} in which some \gls{dune} project and collaboration documents, drawings and engineering models are managed} % #36  add `lbnf' before dune
\newduneabbrev{docdb}{DocDB}{Document DataBase}{A computerized document management system developed and supported at \gls{fnal} in which most \gls{lbnf-dune} documents are managed (docs.dunescience.org); some documents are maintained in \gls{edms}}

\newduneword{wrgm}{WR grandmaster}{White Rabbit grandmaster}

\newduneabbrev{larsoft}{LArSoft}{Liquid Argon Software}{A shared base of physics software across \gls{lartpc} experiments}
\newduneword{nova}{NOvA}{The \gls{nova} off-axis neutrino oscillation experiment at \gls{fnal}}
\newduneword{minerva}{MINERvA}{Neutrino cross sections experiment at \gls{fnal},  shut down in 2019}
\newduneword{microboone}{MicroBooNE}{A \gls{lartpc} neutrino oscillation experiment at \gls{fnal}}
\newduneword{sbnd}{SBND}{The Short-Baseline Near Detector experiment at  \gls{fnal}}
\newduneabbrev{nexo}{nEXO}{Enriched Xenon Observatory}{Experiment at Lawrence Livermore National Laboratory (U.S. national lab in Livermore, CA) searching for new physics with neutrinoless double-beta decay}
\newduneword{argoneut}{ArgoNeuT}{The ArgoNeuT test-beam experiment and \gls{lartpc} prototype at  \gls{fnal}}
\newduneword{icarus}{ICARUS}{A neutrino experiment that was located at the Laboratori Nazionali del Gran Sasso (LNGS) in Italy, then refurbished at \gls{cern} for re-use in the same neutrino beam from \gls{fnal} used by the \gls{miniboone} , \gls{microboone} and \gls{sbnd} experiments at \gls{fnal}}
\newduneword{atlas}{ATLAS}{One of two general-purpose detectors at the \gls{lhc}. It investigates a wide range of physics, from the measurements of the Higgs boson properties to searches for extra dimensions and particles that could make up \gls{dm}}

\newduneword{lbne}{LBNE}{Long Baseline Neutrino Experiment; (1) a terminated U.S. experiment that was reformulated in 2014 under the auspices of the new \gls{dune} collaboration, an internationally coordinated and internationally funded program, with \gls{fnal} as host; and (2) the former incarnation of \gls{usproj} } % #36

\newduneabbrev{lbno}{LBNO}{Long Baseline Neutrino Observatory} {A terminated European project that, during its six-year duration, assessed the feasibility of a next-generation deep underground neutrino observatory in Europe)}
\newduneword{wirecell}{Wire-Cell}{A tomographic automated \threed neutrino event reconstruction method for \glspl{lartpc}}
\newduneabbrev{wct}{WCT}{Wire-Cell Toolkit}{A software toolkit with data flow processing components for \gls{lartpc} noise and signal simulation, noise filtering, signal processing, and tomographic \threed ionization activity imaging}
\newduneword{ftslite}{F-FTS-lite}{Light-weight version of the \gls{fnal} File Transfer system used for rapid data transfers out of the online systems}
\newduneabbrev{fts}{FTS}{File Transfer System}{A file transfer system developed at \gls{fnal} to catalog and move data to permanent storage}

\newduneword{35t}{35 ton prototype}{A prototype cryostat and \gls{sp} detector built at \gls{fnal} before the \gls{protodune} detectors}

\newduneabbrev{mcr}{MCR}{main communications room}{Space at the \gls{fd} site for cyber infrastructure}

\newduneabbrev{cuc}{CUC}{central utility cavern}{The utility cavern at the 4850L of \gls{surf} located between the two detector caverns. It contains utilities such as central cryogenics and other systems, and the underground data center and control room}

\newduneabbrev{cfd}{CFD}{computational fluid dynamics}{High performance computer-assisted modeling of fluid dynamical systems}
\newduneword{vuv}{VUV}{vacuum ultra-violet}
\newduneword{tallbo}{TallBo}{A cylindrical cryostat at \gls{fnal} primarily used for developing scintillation light collection technologies for \gls{lartpc} detectors}

\newduneword{root}{ROOT}{A modular scientific software toolkit. It provides all the functionalities needed to deal with big data processing,  statistical analysis, visualization, and storage.  It is mainly written in C++ but integrated with other languages such as Python and R}

\newduneabbrev{eos}{EOS}{EOS}{The XRootD-based distributed file system developed by CERN}
\newduneabbrev{ehn1}{EHN1}{Experiment Hall North One}{Location at \gls{cern} of the \gls{np02} and \gls{np04} areas used for the \glspl{protodune} and for other test and prototyping activities for DUNE} 

\newduneword{led}{LED}{Light-emitting diode}
\newduneabbrev{rtd}{RTD}{resistance temperature detector}{A temperature sensor consisting of a material with an accurate and reproducible resistance/temperature relationship}
\newduneword{swc}{SWC}{Software \& Computing}
\newduneabbrev{las}{LAS}{LEM-anode Sandwich}{In the \gls{dp} technology, a \gls{lem} and its corresponding anode are mounted together in a module called a LEM-anode sandwich}

\newduneword{roi}{ROI}{region of interest}
\newduneabbrev{hpc}{HPC}{high-performance computing}{high-performance computing facilities; generally computing facilities emphasizing parallel computing with aggregate power of more than a teraflop}

\newduneword{comfund}{common fund}{The shared resources of the collaboration}
\newduneabbrev{ims}{IMS}{integrated master schedule}{A project management device consisting of linked tasks and milestones}

\newduneword{hvdb}{HVDB}{HV divider board}

\newduneword{hvft}{HVFT}{HV feedthrough}  % #33

\newduneword{sas}{SAS}{Another term for the materials airlock; a pass-through chamber used to ensure safe transfer of materials into a clean room, avoiding contamination in both directions}

\newduneabbrev{fea}{FEA}{finite element analysis}{Simulation of a physical phenomenon using the numerical technique called Finite Element Method (FEM), a numerical method for solving problems of engineering and mathematical physics}

\newduneword{fss}{FSS}{field shaping strips}
\newduneword{lvds}{LVDS}{low-voltage differential signaling}

\newduneword{esd}{ESD}{electrostatic discharge}%{ESD is the sudden flow of electricity between two electrically charged objects caused by contact, an electrical short, or dielectric breakdown. ESD can cause failure of electronic components such as integrated circuits}

\newduneabbrev{rp}{RP}{resistive panel}{Resistive panels form the constant potential surfaces for a \gls{spmod} \gls{cpa}; they are composed of a thin layer of carbon-impregnated Kapton and laminated to both sides of a \frfour sheet}

\newduneword{uhmwpe}{UHMWPE}{ultra-high molecular weight polyethylene}

\newduneword{cts}{CTS}{Cryogenic Test System}

\newduneabbrev{plc}{PLC}{programmable logic controller}{An industrial digital computer that has been ruggedized and adapted for the control of manufacturing or other processes that require high reliability, ease of programming, and process fault diagnosis} % update from Marco V 1sep2020

\newduneword{mppc}{MPPC}{\SI{6}{mm}$\times$\SI{6}{mm} Multi-Pixel Photon Counters produced by Hamamatsu\texttrademark{} Photonics K.K}

\newduneabbrev{sfp}{SFP}{small form-factor pluggable}{a particular standard for optical transceivers}

\newduneabbrev{minipod}{MiniPOD}{miniature parallel optical device}{a family of types of multi-channel optical transceivers}

\newduneword{ccc}{CCC}{configuration change command}
\newduneword{ccondc}{CCC}{code of conduct committee}

\newduneword{act}{ACT}{activation time stamp}
\newduneword{lcm}{LCM}{light calibration module}
\newduneword{lpm}{LPM}{light pulser module}
\newduneword{dac}{DAC}{digital-to-analog converter}
\newduneword{arapuca}{ARAPUCA}{A \gls{pds} design that consists of a light trap that captures wavelength-shifted photons inside boxes with highly reflective internal surfaces until they are eventually detected by \gls{sipm} detectors or are lost}
\newduneword{sarapu}{S-ARAPUCA}{Standard \gls{arapuca} design with different \gls{wls} coatings on both faces of the dichroic filter window(s) of the cell}
\newduneword{xarapu}{X-ARAPUCA}{Extended \gls{arapuca} design with \gls{wls} coating on only the external face of the dichroic filter window(s) but with a \gls{wls} doped plate inside the cell}
\newduneword{feb}{FEB}{front-end board}

\newduneabbrev{lsnd}{LSND}{Liquid Scintilator Neutrino Detector}{A scintillation detector and associated experiment located at Los Alamos National Laboratory}

\newduneabbrev{cvn}{CVN}{convolutional visual network}{An algorithm for identifying neutrino interactions based on their topology and without the need for detailed reconstruction algorithms}

\newduneword{pandora}{Pandora}{The Pandora multi-algorithm approach to pattern recognition} 

\newduneabbrev{pma}{PMA}{Projection Matching Algorithm}{A reconstruction algorithm that combines \twod reconstructed objects to form a \threed representation}
\newduneabbrev{bdt}{BDT}{boosted decision tree}{A method of multivariate analysis}
\newduneabbrev{cnn}{CNN}{convolutional neural network}{A deep learning technique most commonly applied to analyzing visual imagery}
\newduneword{pdg}{PDG}{Particle Data Group}

\newduneword{pci}{PCI}{Peripheral Component Interconnect}

\newduneword{labview}{LabVIEW}{Laboratory Virtual Instrument Engineering Workbench is a system-design platform and development environment for a visual programming language from National Instruments}

\newduneword{pcb}{PCB}{printed circuit board}

\newduneword{crio}{cRIO}{Compact Reconfigurable Input Output}

\newduneword{dcs}{DCS}{Distributed Communications System}

\newduneword{opc-ua}{OPC-UA}{OPC  Unified Architecture is a machine to machine communication protocol for industrial automation developed by the OPC Foundation. OPC stands for Object Linking and Embedding for Process Control}

\newduneword{cabangle}{Cabibbo angle}{A quark mixing parameter that governs the coupling of up quarks to strange quarks}
\newduneword{valor}{VALOR}{A neutrino oscillation fitting framework that is used by \gls{t2k}; the name stands for VALencia-Oxford-Rutherford, the original three institutions that developed it}
\newduneword{cafana}{CAFAna}{Common Analysis File Analysis}
\newduneabbrev{pca}{PCA}{principal component analysis}{A statistical procedure that uses an orthogonal transformation to convert a set of observations of possibly correlated variables into a set of values of linearly uncorrelated variables called principal components (Wikipedia)}
\newduneword{numi}{NuMI}{a set of facilities at \gls{fnal}, collectively called ``Neutrinos at the Main Injector.''  The NuMI neutrino beamline target system converts an intense proton beam into a focused neutrino beam}
\newduneword{gibuu}{GiBUU}{Giessen Boltzmann-Uehling-Uhlenback Project; a unified theory and transport framework in the MeV and GeV energy regimes for elementary reactions on nuclei }
\newduneabbrev{rpa}{RPA}{random phase approximation} {an approximation method commonly used for describing the dynamic linear electronic response of electron systems (Wikipedia)}%(from nu-osc 05)}
\newduneword{t2k}{T2K}{T2K (Tokai to Kamioka) is a long-baseline neutrino experiment in Japan studying neutrino oscillations}
\newduneword{mptdet}{MPT detector}{multipurpose tracking detector}

\newduneword{lariat}{LArIAT}{The repurposed ArgoNeuT \gls{lartpc}, modified for use in a charged particle beam, dedicated to the calibration and precise characterization of the output response of these detectors}

\newduneword{captain}{CAPTAIN}{Experimental program sited at \gls{lanl} that is designed to make measurements of scientific importance to \gls{lbl} neutrino physics and physics topics that will be explored by large underground detectors}

\newduneword{dayabay}{Daya Bay}{a neutrino-oscillation experiment in Daya Bay, China, designed to measure the mixing angle $\Theta_{13}$  using antineutrinos produced by the reactors of the Daya Bay and Ling Ao nuclear power plants}

\newduneword{nuwro}{NuWro}{neutrino interaction generator}

\newduneabbrev{neut}{NEUT}{neutrino interaction generator}{A neutrino interaction simulation program library for the studies of atmospheric and accelerator neutrinos}

\newduneword{minos}{MINOS}{A long-baseline neutrino experiment, with a near detector at \gls{fnal} and a far detector in the Soudan mine in Minnesota,  designed to observe the phenomena of neutrino oscillations (ended data runs in 2012)}

\newduneabbrev{efig}{EFIG}{Experimental Facilities Interface Group}{The body that provides a means to coordinate and discuss issues related to the interfaces within and between the facility and the DUNE detectors at both the \gls{fnal} and \gls{surf} sites} %The body responsible for the required high-level coordination between the project and collaboration in \gls{lbnf-dune}}

\newduneword{ashriver}{Ash River}{The Ash River, Minnesota, USA \gls{nova} experiment far site, used as an assembly test site for \gls{dune}} 

\newduneword{ipd}{project integration director}{Responsible for integration and installation of DUNE detector deliverables. Manages the integration project} %#35 09nov20 Responsible for integration and installation of \gls{lbnf} and \gls{dune} deliverables in South Dakota. Manages the \gls{integoff}}

\newduneabbrev{jpo}{JPO}{Joint Project Office}{The framework used to facilitate close and coherent coordination across all elements with many shared management resources; 
JPO functions include systems engineering, procurement, \gls{esh}, \gls{qa}, finance, project controls, risk management, compliance, internal reviews, partner agreement management, document management, and administrative support} % #35 09nov20 global project configuration and integration, installation planning and coordination, scheduling, safety assurance, technical review planning and oversight, development of partner agreements, and financial reporting}

\newduneword{ifbeam}{IFbeam}{Database that stores beamline information indexed by timestamp}

\newduneabbrev{marley}{MARLEY}{Model of Argon Reaction Low Energy Yields}{Developed at UC Davis, MARLEY is the first realistic model of neutrino electron interactions on argon for enegies less than \SI{50}{MeV}. This includes the energy range important for \gls{snb} neutrinos and also solar 8--boron neutrinos}

\newduneabbrev{es}{ES}{elastic scattering}{Events in which a neutrino
elastically scatters off of another particle}

\newduneabbrev{cno}{CNO}{carbon nitrogen oxygen}{The CNO cycle (for carbon-nitrogen-oxygen) is one of the two known sets of fusion reactions by which stars convert hydrogen to helium, the other being the proton-proton chain reaction (pp-chain reaction). In the CNO cycle, four protons fuse, using carbon, nitrogen, and oxygen isotopes as catalysts, to produce one alpha particle, two positrons and two electron neutrinos}

\newduneabbrev{sdwf}{SDWF}{South Dakota Warehouse Facility}{Warehousing operations in South Dakota responsible for receiving LBNF and DUNE goods and coordinating shipments to the access shaft (Ross Shaft) at \gls{surf}}

\newduneabbrev{wms}{WMS}{warehouse management system}{Commercial software package used to track shipments and interface to freight forwarders. This includes a database for shipping}

\newduneabbrev{dcdb}{DCDB}{DUNE construction database}{Database used by DUNE to track the history and testing of all parts of each \gls{detmodule}}

\newduneabbrev{aup}{AUP}{acceptance for use and possession}{Required for beneficial occupancy of the underground areas at \gls{surf} for \gls{lbnf} and \gls{dune}}

\newduneabbrev{bms}{BMS}{building management system}{A system provided by the \gls{cf} to manage the utility (cooling, ventilation, power, etc.) and fire/life safety systems. Separate systems are provided at \gls{surf} and at \gls{fnal}} % #36 Elaine  old: Part of the safety system at \gls{surf} that includes the fire and life safety system}

\newduneabbrev{fls}{FLS}{fire and life safety system}{Fire and life safety; systems designed with \gls{cf} to meet building/safety code compliance for safe facilities at \gls{surf} and at \gls{fnal}} % #36 

\newduneabbrev{sno}{SNO}{Sudbury Neutrino Observatory}{The Sudbury Neutrino Observatory was a detector built 6800 feet under ground, in INCO's Creighton mine near Sudbury, Ontario, Canada. SNO was a heavy-water Cherenkov detector designed to detect neutrinos produced by fusion reactions in the sun}

\newduneword{sk}{Super-Kamiokande}{Experiment sited in the Kamioka-mine, Hida-city, Gifu, Japan that uses a large water Cherenkov detector to study neutrino properties through the observation of solar neutrinos, atmospheric neutrinos and man-made neutrinos}

\newduneabbrev{id}{ID}{inner diameter}{Inner diameter of a tube}

\newduneabbrev{od}{OD}{outer diameter}{Outer diameter of a tube}

\newduneabbrev{rms}{RMS}{root mean square}{The square root of the arithmetic mean of the squares of a set of values, used as a measure of the typical magnitude of a set of numbers, regardless of their sign}

\newduneabbrev{orc}{ORC}{operational readiness clearance}{Final safety approval prior to the start of operation}

\newduneabbrev{gsc}{GSC group}{global safety coordination group}{DUNE group that evaluates applicable codes and standards, including international code equivalency, for the design, assembly, and installation of the \gls{fd}}

\newduneabbrev{ha}{HA}{hazard analysis}{A first step in a process to assess risk; the result of hazard analysis is the identification of the hazards present for a task or process}%different types of hazards}(anne)
\newduneword{har}{HAR}{hazard analysis report}

\newduneabbrev{tap}{TAP}{trip action plan}{A document required for any trip by a worker to the underground area at \gls{surf}, per that site's access control program; it describes the work to be accomplished during the trip} % ask Jim Stewart to check

\newduneword{em}{EM}{emergency management}
\newduneword{ert}{ERT}{emergency response team}

\newduneabbrev{ndk}{NDK}{nucleon decay}{The hypothetical, baryon number violating decay of a proton or a bound neutron into lighter particles}

\newduneabbrev{emi}{EMI}{electromagnetic interference}{Disturbance generated by an external source that affects an electrical circuit by electromagnetic induction, electrostatic coupling, or conduction}

\newduneabbrev{pe}{PE}{photoelectron}{An electron ejected from the surface of a material by the photoelectric effect}

\newduneabbrev{spe}{SPE}{single photoelectron}{A single photoelectron}

\newduneabbrev{fwhm}{FWHM}{full width at half maximum}{Width of a distribution measured between those points at which the distribution is equal to half of its maximum amplitude}

\newduneabbrev{gdml}{GDML}{geometry description markup language}{An application-independent, geometry-description format based on XML}

\newduneabbrev{xml}{XML}{extensible markup language}{A markup language that defines a set of rules for encoding documents in a format that is both human-readable and machine-readable}

\newduneabbrev{crt}{CRT}{cosmic-ray tagger}{Detector external to the TPC designed to tag TPC-traversing cosmic ray particles}

\newduneabbrev{sn}{SN}{supernova}{Event that occurs upon the death of certain types of stars}

\newduneabbrev{wg}{WG}{working group}{A group of persons working together to achieve specified goals}

\newduneabbrev{ctsf}{CTSF}{coating, testing and storage facility}{A facility where the the \gls{dp} photon detectors will be coated, tested, and stored} %11 may 2021 (no more DP module)- will other DP be tested? Is it at CERN?

\newduneword{rucio}{Rucio}{Data management system originally developed
by \gls{atlas} but now open-source and shared across \gls{hep}}
\newduneabbrev{doma}{DOMA}{data organization, management, and access}{data organization, management, and access efforts through the \gls{hsf}}

\newduneabbrev{hsf}{HSF}{HEP Software Foundation}{A foundation that facilitates cooperation and common efforts in high energy physics software and computing internationally} %updated 3/4/22 AH per P. Laycock

\newduneabbrev{wlcg}{WLCG}{Worldwide LHC Computing Grid}{Worldwide LHC
Computing Grid}
\newduneabbrev{osg}{OSG}{Open Science Grid}{Open Science Grid}
\newduneabbrev{sci}{SCI}{Scientific Computing Infrastructure}{Proposed
extension of the infrastructure component of \gls{wlcg} to other
experiments}
\newduneabbrev{csc}{CSC}{computing and software consortium}{DUNE
computing and software consortium}

\newduneword{dirac}{DIRAC}{Computing workflow management designed for LHCb and now used by many HEP experiments}

\newduneword{frp}{FRP}{fiber-reinforced plastic}
\newduneabbrev{hdpe}{HDPE}{high-density polyethylene}{A thermoplastic polymer made from petroleum commonly used to make plastic bottles}
\newduneword{hvps}{HVPS}{\gls{hv} power supply}
\newduneword{aisi}{AISI}{American Iron and Steel Institute}
\newduneword{ific}{IFIC}{Instituto de Fisica Corpuscular (in Valencia, Spain)}
\newduneabbrev{rsds}{RSDS}{radioactive source deployment system}{Proposed calibration system based on the deployment of
radioactive sources inside the \gls{dune} cryostat}
\newduneword{2p2h}{2p2h}{two particle, two hole}
\newduneabbrev{duneprism}{DUNE-PRISM}{\gls{dune} Precision Reaction-Independent Spectrum Measurement}{a mobile near detector that can perform measurements over a range of angles off-axis from the neutrino beam direction in order to sample many different neutrino energy distributions}
\newduneword{arcube}{ArgonCube}{The name of the core part of the \gls{dune} \gls{nd}, a \gls{lartpc}}

\newduneabbrev{citf}{CITF}{cryogenic instrumentation test facility}{A facility at \gls{fnal} with small ($<$\SI{1}{ton}) to intermediate ($\sim$\SI{1}{ton}) volumes of instrumented, purified TPC-grade \gls{lar}, used for testing devices intended for use in \gls{dune}}

\newduneabbrev{3dst}{3DST}{3D scintillator tracker}{The core part of the \threed projection scintillator tracker spectrometer in the near detector conceptual design}
\newduneabbrev{3dsts}{3DST-S}{3D scintillator tracker spectrometer}{The \threed projection scintillator tracker spectrometer  in the near detector conceptual design}
\newduneabbrev{mpd}{MPD}{multi-purpose detector}{A component of the near detector conceptual design; it is a magnetized system consisting of a \gls{hpgtpc} and a surrounding \gls{ecal}}
\newduneabbrev{hpg}{HPG}{high-pressure gas}{gas at high pressure to be used in a \gls{hpgtpc}} 
\newduneabbrev{hpgtpc}{HPgTPC}{high-pressure gaseous argon TPC}{A \gls{tpc} filled with gaseous argon; a possible component of the \gls{dune} \gls{nd}}

\newduneword{src}{SRC}{short-range correlated nucleon-nucleon interactions}
\newduneword{larpix}{LArPix}{ \gls{asic} pixelated charge readout for a \gls{tpc} }
\newduneword{arclt}{ArCLight}{a light detector for the \gls{arcube} effort}
\newduneword{fhc}{FHC}{forward horn current ($\numu$ mode)}
\newduneword{rhc}{RHC}{reverse horn current (\numubartonumubar mode)}
\newduneword{mwpc}{MWPC}{multi-wire proportional chamber}
\newduneword{na61}{NA61}{CERN hadron production experiment}
\newduneword{pdnd}{ProtoDUNE-ND}{a prototype \gls{dune} \gls{nd}}
\newduneabbrev{roc}{ROC}{readout chamber}{readout chamber for gaseous argon \gls{tpc}}
\newduneabbrev{iroc}{IROC}{inner readout chamber}{inner (radial) readout chamber for gaseous argon \gls{tpc}}
\newduneabbrev{oroc}{OROC}{outer readout chamber}{outer (radial) readout chamber for gaseous argon \gls{tpc}}

\newduneword{lux}{LUX}{Large Underground Xenon (LUX) dark matter detector at \gls{surf} }

\newduneword{mjdemo}{Majorana Demonstrator}{Experiment sited at \gls{surf} that  seeks to determine whether neutrinos are their own antiparticles}

\newduneword{lz}{LZ}{Experiment sited at \gls{surf} that  seeks to detect faint interactions between galactic dark matter and regular matter}

\newduneword{mu2e}{Mu2e}{An experiment sited at \gls{fnal} that searches for charged-lepton flavor violation and seeks to discover physics beyond the \gls{sm}}

\newduneword{pdsp2}{ProtoDUNE-SP-II}{A second test run in the single-phase ProtoDUNE test stand at CERN, acting as a validation of the final
single-phase detector design}  % #33 11 may 21-  may need an update Don't use!

\newduneword{osha}{OSHA}{Occupational Safety and Health Administration (USA Department of Labor) formed by the Occupational Safety and Health Act of 1970}
\newduneabbrev{pns}{PNS}{pulsed neutron source}{Calibration system based
on neutron capture gamma showers spread out in the whole detector}

\newduneabbrev{fv}{FV}{fiducial volume}{The detector volume within the \gls{tpc} that is selected for physics analysis through cuts on reconstructed event position}

\newduneword{p6}{P6}{software framework used to plan and status the resource-loaded schedule of activities associated with \gls{usproj}}

\newduneabbrev{evms}{EVMS}{earned value management system}{Earned Value Management is a systematic approach to the integration and measurement of cost, schedule, and technical (scope) accomplishments on a project or task. It provides both the government and contractors the ability to examine detailed schedule information, critical program and technical milestones, and cost data (text from the US DOE); the EVMS is a system that implements this approach}

\newduneword{core}{CORE}{CORE contributions are in either monetary units or labor hours. They can be technical components for the facility or experiment and the effort of the staff needed to produce, install, and test them;  major facilities for the experiment; or other products and services relevant for the completion of the facility or experiment} % 15 May - gina needs to `bless' this

\newduneabbrev{ahj}{AHJ}{Authority Having Jurisdiction}{An organization, office, or individual responsible for enforcing the requirements of a code or standard, or for approving equipment, materials, an installation, or a procedure (\gls{osha})}
\newduneword{cte}{CTE}{coefficient of thermal expansion}

\newduneabbrev{opc}{OPC}{open platform communications}{Open platform communications is a series of standards and specifications for industrial telecommunication} 
\newduneword{scada}{SCADA}{supervisory control and data acquisition}
\newduneword{ln}{LN$_2$}{liquid nitrogen}
\newduneabbrev{lapd}{LAPD}{Liquid Argon Purity Demonstrator}{Cryostat at \gls{fnal}  for long-term studies requiring a large volume of argon}

\newduneabbrev{pab}{PAB}{Proton Assembly Building}{Home of several \gls{lar} facilities at \gls{fnal} }
\newduneword{hep}{HEP}{high energy physics}
\newduneword{cms}{CMS}{Compact Muon Solenoid experiment; one of two general-purpose detectors at the \gls{lhc}. }
\newduneword{alice}{ALICE}{A Large Ion Collider Experiment, at CERN}
\newduneword{gpib}{GPIB}{general purpose interface bus}

\newduneabbrev{pfparticle}{PFParticle}{particle flow particle}{Each of the individual reconstructed particles in the hierarchy (or particle flow) describing the reconstructed event interaction}

\newduneabbrev{mcparticle}{MCParticle}{Monte Carlo Particle}{Individual true simulated particle}
\newduneword{au}{AU}{astronomical unit}
\newduneword{nufit}{NuFIT 4.0}{The NuFIT 4.0 global fit to neutrino oscillation data}

\newduneword{uhv}{UHV}{ultra high vacuum}
\newduneword{lps}{LPS}{laser positioning system}

\newduneword{unicamp}{UNICAMP}{University of Campinas, Sao Paulo, Brazil}
 
\newduneabbrev{fbk}{FBK}{Fondazione Bruno Kessler}{FBK is a research non-profit entity in Trento, Italy that partners in the development of technology with applications in various fields including High Energy Physics}

\newduneabbrev{enob}{ENOB}{effective number of bits}{The effective number of bits is a measure of the dynamic range of an \gls{adc} and its associated circuitry. The resolution of an \gls{adc} is specified by the number of bits used to represent the analog value, in principle giving 2N signal levels for an N-bit signal. However, all real \gls{adc} circuits introduce noise and distortion. ENOB specifies the resolution of an ideal \gls{adc} circuit that would have the same resolution as the circuit under consideration}
\newduneabbrev{sndr}{SNDR}{signal to noise and distortion ratio}{Also known as SINAD. Ratio of the \gls{rms} signal amplitude to the mean value of the root-sum-square of all other spectral components, including harmonics, but excluding \gls{dc} levels. It is a good indication of the overall dynamic performance of an \gls{adc} because it includes all components which make up noise and distortion}
\newduneabbrev{sfdr}{SFDR}{spurious free dynamic range}{Spurious free dynamic range is the ratio of the \gls{rms} value of the signal to the \gls{rms} value of the worst spurious signal regardless of where it falls in the frequency spectrum. The worst spur may or may not be a harmonic of the original signal}
\newduneabbrev{thd}{THD}{total harmonic distortion}{Total harmonic distortion is the ratio of the \gls{rms} value of the fundamental signal to the mean value of the root-sum-square of its harmonics} 
\newduneword{tvs}{TVS}{transient voltage suppression}

\newduneword{riskprob}{risk probabilities}{The risk probability, after taking into account the planned mitigation activities, is ranked as 
 L (low $<\,$\SI{10}{\%}), 
M (medium \SIrange{10}{25}{\%}), or 
H (high $>\,$\SI{25}{\%}). 
The cost and schedule impacts are ranked as 
L (cost increase $<\,$\SI{5}{\%}, schedule delay $<\,$2 months), 
M (\SIrange{5}{25}{\%} and 2--6 months, respectively) and 
H ($>\,$\SI{20}{\%} and $>\,$2 months, respectively)}

\newduneabbrev{lbls}{LBLS}{laser beam location system}
{Auxiliary calibration system providing an independent location measurement of the ionization laser beams direction}

\newduneabbrev{lsst}{LSST}{Large Synoptic Survey Telescope}{8.4 m telescope with 3.2G-pixel camera that will start taking data in 2023}
\newduneabbrev{ska}{SKA}{Square Kilometer Array}{International radio telescope array planned to start data-taking in 2027}
\newduneabbrev{hyperk}{HyperK}{Hyper Kamiokande}{260 kt water Cerenkov neutrino detector to begin construction at Kamiokande in 2020}
\newduneword{lhcb}{LHCb}{LHC experiment dedicated to forward physics}
\newduneword{belleii}{Belle II}{B-factory experiment now running at KEK}

\newduneabbrev{ldm}{LDM}{light-mass dark matter}{Refers to dark matter particles with mass values much lower than the electroweak scale, specifically below the 1~GeV level}
 
\newduneabbrev{bnv}{BNV}{baryon-number violating}{Describing an interaction where \gls{baryonnumber} is not conserved}

\newduneword{bugey}{Bugey}{Neutrino experiment that operated at the Bugey nuclear power plant in France}

\newduneword{minosplus}{MINOS$+$}{The successor to the \gls{minos} experiment, utilizing the same detectors and beam line, but operating at higher beam energy tune than \gls{minos}, parasitic with \gls{nova}}

\newduneword{baryonnumber}{baryon number}{A quantity expressing the total number of baryons in a system minus the number of antibaryons}

\newduneword{np04}{NP04}{%Experiment 
The CERN North Area in \gls{ehn1} intersected by the \gls{h4} hadron beamline,  the location of  \gls{pdsp} and \gls{pdsp2}; also used to refer to the 800\,t cryostat in this area}

\newduneword{np02}{NP02}{%Experiment 
The CERN North Area in \gls{ehn1} intersected by the \gls{h2} hadron beamline, the location of  the 800\,t cryostat used for \gls{pddp} and for \gls{spvd} tests and prototypes; also used to refer to the 800\,t cryostat in this area}

\newduneword{h4}{H4}{CERN North Area hadron beamline used for \gls{pdsp} and \gls{pdsp2}}  % #33

\newduneword{h2}{H2}{CERN North Area hadron beamline used for \gls{pddp} and \gls{spvd} prototypes and demonstrators}  % #33

\newduneword{ua1}{UA1}{UA1 (Underground Area 1) was a particle detector at \gls{cern}'s  Super Proton Synchrotron (SPS). It ran from 1981 until 1990, when the SPS was used as a proton-antiproton collider, searching for traces of W and Z particles in collisions. (CERN) The UA1 dipole magnet was reused in the NOMAD experiment and currently provides the magnetic field for the \gls{t2k} ND280 detector}

\newduneword{ssc}{SSC}{The Superconducting Super Collider was to be a huge underground ring complex beneath the area near Waxahachie, Texas, USA, that would have been the world’s most energetic particle accelerator. It was begun in 1990, but canceled by the U.S. Congress in 1993 (scientificamerican.com Oct 2013)}

\newduneword{daphne}{DAPHNE}{Detector electronics for Acquiring PHotons from NEutrinos is a custom-developed warm front-end waveform digitizing electronics module derived from the readout system developed at \gls{fnal}  for the Mu2e experiment}
 
\newduneword{nersc}{NERSC}{National Energy Research Computing Facility at \gls{lbnl}}

\newduneword{integoff}{integration project}{The \gls{doe} project element that organizes the onsite teams responsible for coordinating far detector installation and detector-facility integration activities at \gls{surf} as well as near detector installation activities at \gls{fnal}.  The integration project office includes overall LBNF/DUNE systems engineering, compliance and review offices} % #35

\newduneabbrev{sma}{SMA}{SubMiniature version A}{Connector interface for coaxial cables
with a screw-type coupling mechanism}

\newduneword{kloe}{KLOE}{KLOE is a $e^+ e^-$ collider detector spectrometer operated at DAFNE,  the $\phi$-meson factory at Frascati, Rome.  In DUNE it will consist of a \SI{26}{cm} Pb+scintillating fiber \gls{ecal} surrounding a cylindrical open detector region that is  \SI{4.00}{m} in diameter and \SI{4.30}{m} long.  The \gls{ecal} and detector region are embedded in a \SI{0.6}{T} magnetic field created by a \SI{4.86}{m} diameter superconducting coil and a \SI{475}{tonne} iron yoke}

\newduneabbrev{ro}{RO}{review office}{An office within the \gls{integoff} that organizes reviews} % #33
\newduneabbrev{doecd}{CD}{critical decision}{The U.S. DOE's Order 413.3B outlines a series of staged project approvals, each of which is referred to as a critical decision (CD)}

\newduneabbrev{lbnfspac}{LBNF/DUNE-US SPAC}{LBNF / DUNE-US Strategic Project Advisory Committee}{A committee charged by the host laboratory director to provide expert, independent advice on significant issues and strategies related to \gls{usproj} project organization, management, and risks} % #36

\newduneabbrev{sand}{SAND}{System for on-Axis Neutrino Detection}{The beam monitor component of the near detector that remains on-axis at all times and serves as a dedicated neutrino spectrum monitor}

\newduneword{4850l}{4850L}{The depth in feet (1480 m) of the access level for the DUNE underground area at SURF; called the ``4850 level''} % #33

\newduneword{apb}{APB}{authorship and publications board}
\newduneword{crb}{CRB}{collaboration resources board}
\newduneword{cube}{CuBe}{beryllium copper, used to make \gls{sphd} \gls{apa} readout planes}
\newduneword{drift}{drift}{(1) refers to electron drift under the influence of an electric field in a \gls{tpc}; (2) an excavated horizontal corridors (tunnels) in the underground areas at \gls{surf}}
\newduneword{exposure}{exposure}{The integrated detector fiducial mass times beam intensity; it is proportional to the number of interactions and is used to normalize cross sections in a data sample}
\newduneword{fr4}{FR-4}{Flame-retardant fiberglass-reinforced epoxy resin laminate used in making PCBs and other detector components}
\newduneword{g10}{G-10}{Non-flame-retardant fiberglass-reinforced epoxy resin laminate used in making PCBs}
\newduneword{kapton}{Kapton}{A polyimide plastic film that is stable over a broad range of temperatures and is resistant to radiation damage}
\newduneword{shaft}{shaft}{A vertical excavation at \gls{surf} connecting with the surface}
\newduneword{winze}{winze}{A vertical excavation at \gls{surf} connecting two drifts, not connecting to the surface}
\newduneword{ib}{IB}{institutional board; all institutions participating in DUNE are represented on this board}
\newduneword{irb}{IBR}{institutional board representative}
\newduneword{sc}{SC}{depending on context, either speakers committee or scientific computing} % #33
\newduneword{sac}{SAC}{spokespersons advisory committee}
\newduneword{eoc}{EOC}{education and outreach committee}
\newduneword{indico}{Indico}{Web-based meeting organization tool}
\newduneabbrev{htc}{HTC}{High Throughput Computing}{Computing facilities typically consisting of large numbers of commodity servers as opposed to a single large machine. Best suited for running large numbers of independent jobs in parallel, these facilities are what is usually meant by ``grid computing''}
\newduneword{dcache}{dCache}{A distributed, highly scalable (multi-PB) storage system, usable as both a standalone system and as a high-speed frontend to a tape storage system (such as \gls{pnfs} at \gls{fnal} )}
\newduneabbrev{ifdhc}{IFDHC}{Intensity Frontier Data Handling Client}{A multi-protocol tool for data transfer and file delivery in jobs. It is able to automatically select transfer protocols based on source and destination characteristics}
\newduneabbrev{ifdh}{ifdh}{Intensity Frontier Data Handling}{The actual command invoked when using \gls{ifdhc}, on the command line, e.g. ifdh cp source\_file dest\_file}
\newduneabbrev{pnfs}{PNFS}{Pseudo Network File System}{A file system often used in large storage systems. Typically interaction is very similar to a regular NFS volume, but there can be some subtle and important differences}

\newduneword{nde}{NDE}{non-destructive evaluation} % 33
\newduneword{psv}{PSV}{pressure safety valve}
\newduneword{pickling}{pickling}{steel pickling and oiling is a metal surface treatment finishing process used to remove surface impurities such as rust and carbon scale from hot rolled carbon steel}

\newduneabbrev{tof}{ToF}{time of flight}{The time a particle takes to fly between two visible interactions observed in the detector. If combined with the distance traveled by the particle, for example a neutron, it can be used for energy reconstruction}

\newduneword{pep4}{PEP-4}{TPC for the Positron Electron Project 4 Collider at Stanford}

\newduneword{ndlar}{ND-LAr}{\gls{lartpc} component of the near detector based on \gls{arcube} technology}

\newduneword{ndgar}{ND-GAr}{component of the near detector with a core gaseous argon \gls{tpc} surrounded by an \gls{ecal} and a magnet}

\newduneword{sfgd}{SuperFGD}{Super Fine-Grained Detector (SuperFGD) is a 3D granular plastic scintillator detector that adopts the same technology as \dword{3dst}. It will be installed in the \dword{t2k} \dword{nd280} system . The \dword{3dst} design will inherit in large part from the SuperFGD detector} %\cite{Abe:2019whr}

\newduneword{nd280}{ND280}{Near Detector 280, is the \dword{t2k} magnetized near detector} % \cite{Abe:2011ks}}

\newduneword{fee}{FEE}{front-end electronics}

\newduneword{mpgd}{MPGD}{MicroPattern Gas Detectors}

\newduneword{rmm}{RMM}{Resistive MicroMegas}

\newduneabbrev{stv}{STV}{Single Transverse Variables}{Kinematical variables obtained by projecting the neutrino interaction onto the transverse plane}

\newduneabbrev{ccqe}{CCQE}{charged current quasielastic interaction} {An interaction where a neutrino scatters from a nucleon, producing a charged lepton and converting a neutron to a proton or vice versa}

\newduneword{sis}{SIS}{shallow inelastic scattering}

\newduneabbrev{res}{RES}{resonant scattering}{The mode of scattering where the target nucleon is excited to a resonant state and decays, typically producing one or more pions}

\newduneabbrev{hadw}{W}{invariant mass of the hadronic system}{Refers to the invariant mass of the hadronic system formed during the neutrino scatter}

\newduneabbrev{agky}{AGKY}{Andreopoulos-Gallagher-Kehayias-Yang}{A model for hadronization of non-resonant inelastic neutrino reactions used in \gls{genie}. At low invariant hadronic masses, typically less than 2.3\,GeV/c$^2$, it is a KNO-inspired empirical model anchored on several bubble chamber measurements of neutrino-induced shower characteristics. For invariant hadronic masses between 2.3 and 3.0\,GeV/c$^2$, the model transitions linearly to a \gls{genie}-tuned version of PYTHIA, which is also used for the simulation of events at higher invariant masses} % update from Costas Andreopoulos sept2020
\newduneabbrev{clas}{CLAS}{CEBAF Large Acceptance Spectrometer}{A nuclear and particle physics detector located in the experimental Hall B at Jefferson Laboratory in Newport News, Virginia, United States. It is used to study the properties of the nuclear matter by the collaboration of over 200 physicists. Of particular relevance is its study of electron interactions with nuclei, including argon} 

\newduneabbrev{e4nu}{e4nu}{Electrons for Neutrinos}{A collaboration dedicated to using JLab's electron-scattering data to deliver improved neutrino-nucleus cross sections} 

\newduneabbrev{ldmx}{LDMX}{Light Dark Matter eXperiment}{The LDMX detector concept consists of a small precision tracker, and electromagnetic and hadronic calorimeters, all with near $2\pi$ azimuthal acceptance from the forward beam axis out to $\sim40^\circ$ angle. This detector would be capable of measuring correlations among electrons, pions, protons, and neutrons in electron-nucleus scattering at exactly the energies relevant for DUNE physics} 

\newduneabbrev{mec}{MEC}{meson-exchange currents} {An nuclear effect wherein pairs or larger groups of nucleons within a nucleus are bound together through the exchange of pions or other mesons. Neutrinos and other particles can scatter from these correlated pairs}

\newduneword{imt}{IMT}{Intranuclear momentum transfer}

\newduneword{miniboone}{MiniBooNE}{The Mini Booster Neutrino Experiment,  at \gls{fnal} , was designed to fully explore the LSND result}

\newduneabbrev{tms}{TMS}{The Muon Spectrometer}{A muon spectrometer for the Near Detector that will be installed for the initial running period of DUNE, before the \gls{mpd} detector component is ready} % #34 09nov20  08Apr22 Temporary -> The

\newduneabbrev{cvmfs}{CVMFS}{CERN VM File System}{A distributed file system designed for scalable, high-performance distribution of software to interactive and batch computers} %from tjunk sept2020 #21

\newduneword{kerberos}{Kerberos}{A strong authentication system used by the computing resources at \gls{fnal} and \gls{cern}} %from tjunk sept2020 #23

\newduneabbrev{mrb}{MRB}{Multi Repository Build System}{A \gls{fnal}-developed build system based on \dword{cmake} that allows development and builds of code from multiple repositories} %heidi sept2020

\newduneword{cmake}{Cmake}{CMake is an open-source, cross-platform family of tools designed to build, test and package software}%heidi sept2020

\newduneabbrev{ups}{UPS}{UNIX product support}{A software tool that sets up a consistent environment of versions of pre-installed products and their dependencies on UNIX-like platforms} % tjunk sept2020 #20

\newduneabbrev{upd}{UPD}{UNIX product distribution}{A tool for uploading and downloading pre-built software products between local systems and centralized software distribution servers.  UPD is not frequently used on DUNE because newer tools are more convenient} % tjunk sept2020 #22

\newduneabbrev{sso}{SSO}{single sign-on}{Used at \gls{fnal} to indicate that a group of services,  such as DocDB or the DUNE Wiki share common sign-in credentials and active sessions.  \gls{fnal}  services that say "Sign in with SSO username and password" mean to use your \gls{fnal} Services or federated username and password} % tjunk sept2020 #24

\newduneabbrev{vo}{VO}{virtual organization}{A database containing a list of member names, certificate distinguishing information, and a list of permissions members have to access computing grid and data resources} % tjunk sept2020 #27

\newduneabbrev{nas}{NAS}{network attached storage}{Disk storage that is available on computers but shared between them.  Relies on \gls{nfs} mounts rather than authenticated file transfer protocols.  Usually found on interactive servers to provide space for home directories, app and data storage} % tjunk sept2020 #25

\newduneabbrev{nfs}{NFS}{network file system}{Industry-standard mechanism for mounting disks over a network.  Provides regular UNIX file and directory access} % tjunk sept2020 #26

\newduneword{recombination}{recombination}{Electrons freed from Argon atoms will sometimes recombine with the positive argon ions, either the same ones from which they came or nearby ones.  Sometimes called ``quenching"} % tjunk sept2020 #29

\newduneword{elife}{electron lifetime}{The attachment of electrons drifting through liquid argon to impurity molecules such as oxygen or water is parameterized by an exponential as a function of time with a time constant called the electron lifetime} % tjunk sept2020 #30

\newduneword{gplane}{grid plane}{The uninstrumented plane of wires or electrodes on an anode plane %in an \gls{apa} 
facing the drift volume.   It shapes the signals and provides \gls{esd} protection} % tjunk sept2020 #31

\newduneword{gmesh}{grounding mesh}{A metal mesh attached to the SP \gls{apa} frame between the collection-plane wires and the space inside the frame where the \gls{pd} modules are installed.  It provides electric field uniformity so the collection-plane wires all have similar fields around them} % tjunk sept2020 #32

\newduneabbrev{bcr}{BCR}{baseline change request}{A DOE project change, part of the change management system process}

\newduneword{ncav}{North Cavern}{the location of two of the planned four DUNE far detector modules at \gls{surf}}

\newduneword{scav}{South Cavern}{the location of two of the planned four DUNE far detector modules at \gls{surf}}

\newduneword{semp}{SEMP}{systems engineering management plan}  % #36 replaced cmp 

\newduneabbrev{tpcost}{TPC}{total project cost}{The DOE terminology for the total budget and contingency for the entire \gls{usproj} project from CD-0 to CD-4}  % #36 replaced cmp 
\newduneabbrev{nsint}{NSI}{near site integration}{The scope of work at the near site for the \gls{integoff}} % #36 - can't put it together with other nsi

\newduneabbrev{opcost}{OPC}{other project costs}{The DOE project costs that support conceptual design, pre-operations commissioning, technical coordination, and power}

\newduneabbrev{moa}{MOA}{memorandum of agreement}{A project management methodology that documents an agreement between \gls{fnal} and the \gls{usproj}  Project for how \gls{fnal}  will support the project} %%%  MISSED THIS IN 36!!! ADDED AFTER
\newduneabbrev{croc}{CROC}{central readout chamber}{central (radial) readout chamber for the ND \gls{gartpc}} % from SManley 12/2/20 via email

\newduneabbrev{tki}{TKI}{transverse kinematic imbalance}{The imbalance among final-state particle momenta in the transverse plane to the neutrino direction; different aspects of the imbalance are sensitive to the detail of the nuclear effects in neutrino-nucleus interactions} % from SManley 25feb2021 via email

\newduneabbrev{iandi}{I\&I}{Integration and Installation}{One of the three project areas in the LBNF/DUNE-US \gls{doe} project, along with LBNF and DUNE-US} 

\newduneword{hmi}{HMI}{human-machine interface}

\newduneword{lhe}{LHe}{liquid helium}

\newduneword{echain}{energy chain}{mechanical machine elements used to carry and guide power to moving parts of machines or structures, as required for \gls{duneprism} to carry power, data, and utilities to and from each movable near detector component at any arbitrary position along its travel path}

\newduneabbrev{sphd}{FD1-HD}{horizontal drift detector module}{LArTPC design used in FD1 in which electrons drift horizontally to wire plane anodes (\glspl{apa}) that along with the front-end electronics are immersed in LAr} % updated 21 May 21

\newduneabbrev{spvd}{FD2-VD}{vertical drift detector module}{LArTPC design used in FD2 in which electrons drift vertically to PCB-based anodes at the top and bottom of the LAr volume,  with a cathode in the middle} % updated 21 May 21

\newduneabbrev{cru}{CRU}{charge-readout unit}{In the \gls{spvd} design an assembly of the \glspl{pcbp} plus adapter boards; two to a \gls{crp}} % 4 aug 2021 four --> two

\newduneword{mpv}{MPV}{most probable value}

\newduneword{pcbp}{PCB panel}{In the \gls{spvd} design, one of four \glspl{pcb} of size  %\SI{1.5x1.7}{m} 
1.5 $\times$ 1.7\.,m assembled into a \gls{cru}}

\newduneword{anodepln}{anode plane}{a planar array of charge readout devices covering an entire face of a detector module}

\newduneword{msps}{MSPS}{megasamples per second}

\newduneword{gpsdo}{GPSDO}{\gls{gps} disciplined oscillator}
\newduneword{tdaq}{TDAQ}{trigger and DAQ system} % 25feb2021 SPVD DAQ

\newduneword{nios}{NIOS}{network identity operating system} 

\newduneabbrev{corsika}{CORSIKA}{COsmic Ray SImulations for KAscade}{a program for detailed simulation of extensive air showers initiated by high-energy cosmic ray particles}

\newduneabbrev{scd}{SCD}{scientific computing division}{\gls{fnal}'s Scientific Computing Division}

\newduneabbrev{garsoft}{GArSoft}{Gaseous Argon Software}{A software toolkit similar to \gls{larsoft}, but targeted at the gaseous argon time projection chamber and calorimeter of \gls{ndgar}}

\newduneword{ndgarlite}{ND-GAr-Lite}{a temporary muon spectrometer consisting of the magnet and steel flux return of \gls{ndgar}, but with a simplified tracking chamber made with scintillating bars}

\newduneword{github}{GitHub}{a commercial web service providing code version management, storage,  and browsing via \gls{git}}

\newduneword{git}{git}{a distributed version-control system, commonly used to manage software}

\newduneword{xrootd}{xrootd}{a high-performance data system widely used in \gls{hep} to store and to distribute data to jobs.  It allows streaming of data}

\newduneabbrev{gpuaas}{GPUAAS}{GPU As A Service}{a technique that allows many non-GPU-enabled compute nodes to share a GPU resource by sending it work over the network and waiting for results to be returned}
\newduneword{sce}{SCE}{space charge effect}

\newduneabbrev{nest}{NEST}{noble element simulation technique}{Comprehensive simulation code modeling the excitation, ionization, and corresponding scintillation and electroluminescence processes in liquid noble elements}

\newduneword{bgr}{BGR}{A bandgap voltage reference is a circuit block in ASIC for providing stable reference voltages} %in electronics chapter

\newduneword{comsol}{COMSOL}{General-purpose simulation software based on advanced numerical methods (comsol.com)}

\newduneword{greyrm}{gray room}{ISO-8 clean room with a cleanliness level of 3.5M particles of 0.5 micron or less per cubic meter volume. ISO-8 clean rooms are referred to as grey rooms because at this level of cleanliness most standard clean room attire is not required.}

\newduneabbrev{pof}{PoF}{power-over-fiber}{a technology in which a fiber optic cable carries optical power, which is used as an energy source rather than, or as well as, carrying data; this allows a device to be remotely powered, while providing electrical isolation between the device and the power supply} %in prototype chap; got def from wikipedia (Anne)

\newduneword{ppc}{PPC}{photovoltaic power converter}% in PD
\newduneword{eelaser}{edge-emitting laser}{a laser in which  light is emitted from the edge of the substrate}% in PD
\newduneword{peek}{PEEK}{Polyether ether ketone, a colorless organic thermoplastic polymer}

\newduneword{fsm}{Finite State Machine}{a mathematical model of computation; it is an abstract machine that can be in exactly one of a finite number of states at any given time} % in DAQ
\newduneword{icecube}{IceCube}{South Pole Neutrino Observatory}
\newduneword{pde}{PDE}{photo detection efficiency} % in PD
\newduneword{pmos}{PMOS}{(see \gls{cmos}; PMOS is constructed with the p-type source and drain and an n-type substrate}% in PD
\newduneabbrev{poe}{PoE}{power-over-Ethernet}{systems that pass electric power along with data on twisted-pair Ethernet cabling} % in PD

\newduneabbrev{daqdts}{DTS}{DUNE timing and synchronization subsystem}{The portion of the \gls{daq} that provides for timing and synchronization to various detector systems}

\newduneword{tai}{TAI}{International Atomic Time}% in DAQ

\newduneword{larzic}{LARZIC}{The cryogenic amplifier \dword{asic} that is the principal component of the \gls{spvd} top drift \dword{fe} analog cards}

\newduneword{dpdfd}{DPDFD}{Deputy Project Director for far detectors}
\newduneword{bsws}{BSWS}{bearing sensor wire compression seal}

\newduneabbrev{ly}{LY}{light yield}{detected photons per unit deposited energy}

\newduneword{mtbf}{MTBF}{mean time between failures}

\newduneword{ingaas}{InGaAs}{Indium gallium arsenide is a room-temperature semiconductor commonly used as a high-speed, high-sensitivity photodetector for optical fiber telecommunications}

\newduneword{intlproj}{LBNF/DUNE Construction Project}{The international project to design and build the facilities and detectors for the \gls{lbnf-dune}; it includes the \gls{usproj} and projects at multiple international partners to manage the contributions from non-US institutions and funding agencies to design, build, and install the detector components}% new oct 2021

\newduneword{ddmp}{DUNE Management Plan}{DUNE Management Plan}

\newduneword{pd2hd}{ProtoDUNE-II-HD}{The second ProtoDUNE for the HD design, using NP04 to test installation, integration, and detector component performance}

\newduneword{mod0}{Module 0}{The final pre-production instance of a detector; for the DUNE \glspl{detmodule}, the \glspl{protodune2} in the 800\,t cryostats in \gls{np02} and \gls{np04} serve this purpose}

\newduneword{vdmod0}{FD2-VD Module 0}{The final pre-production \gls{protodune} instance for the DUNE \gls{spvd} \gls{detmodule},  it will use the 800\,t cryostat in \gls{np02} }

\newduneword{hdmod0}{FD1-HD Module 0}{The final pre-production \gls{protodune} instance for the DUNE \gls{sphd} \gls{detmodule},  it will use the 800\,t cryostat in \gls{np04} }

\newduneword{fsii}{FSII}{far site integration and installation}
\newduneword{fdc}{FDC}{Far Detector and Cryogenics Subproject}

\newduneabbrev{co}{CO}{compliance office}{a team of engineers from multiple partners that provides clear direction for designing and constructing components that will be used during integration}

\newduneword{fscfbsi}{FSCF-BSI}{\gls{usproj} subproject for far site conventional facilities, building and site infrastructure}

\newduneabbrev{poms}{POMS}{Production Operations Management System}{A workflow management system available for all DUNE users to submit and monitor grid jobs as well as view job log files}
\newduneabbrev{fifemon}{FIFEMON}{FabrIc for Frontier Experiments MONitoring}{Comprehensive suite of job and storage monitoring information available for most \gls{fnal} experiments, including DUNE}

\newduneword{larg4}{LArG4}{LArG4 is the replacement for LArsim/LArG4. LArG4 is based on \gls{artg4tk}}

\newduneword{artg4tk}{artg4tk}{artg4tk provides a general interface between \gls{geant4} and \gls{art}}

\newduneword{tde}{TDE}{top detector electronics}

\newduneword{bde}{BDE}{bottom detector electronics}

\newduneword{sigproc}{Signal Processing}{The goal of TPC signal processing is to reconstruct the distribution of ionization electrons arriving at wire planes from the digitized TPC waveform}
\newduneword{mcnd}{MCND}{More Capable Near Detector} % for ND TDR 9/26/22, Steve M

\newduneabbrev{compass}{COMPASS}{
Common Muon and Proton Apparatus for Structure and Spectroscopy }{a multipurpose experiment at \gls{cern}’s Super Proton Synchrotron (SPS)} %

\newduneword{vd}{vertical drift}{single-phase, vertical drift \gls{lartpc} technology}

\newduneword{hd}{horizontal drift}{single-phase,  horizontal drift \gls{lartpc} technology}

\newduneabbrev{esnet}{ESnet}{Energy Sciences Network}{The \gls{doe}'s dedicated science network} %

\newduneword{project.py}{project.py}{XML-based job configuration system developed by the \gls{microboone} collaboration} %

\newduneabbrev{ppfx}{PPFX}{Package to Predict the FluX}{\gls{fnal}-supported package that implements hadron production corrections to geant4 simulations and propagates uncertainties for the NuMI  and LBNF beam lines} %

\newduneabbrev{ml}{ML}{Machine Learning}{Machine Learning}

\newduneword{datalake}{data lake}{The not-yet-realized concept of a storage service with multiple levels of quality of service in which the end user can access data without knowing the data's source location} % from where the data are served}

\newduneabbrev{metacat}{MetaCat}{MetaCat}{Metadata Catalog, a modern replacement for the file description portion of the sam metadata catalog} %

\newduneword{g4lbnf}{g4lbnf}{LBNF neutrino beamline simulation program} %\cite{g4lbnf}} %

\newduneword{Geant4Reweight}{Geant4Reweight}{Framework for evaluating and propagating hadronic interaction uncertainties in Geant4} %\cite{Calcutt:2021zck}} %

\newduneword{geantnet}{G\'EANT}{G\'EANT interconnects Europe's national research and education networking (NREN) organizations with the high bandwidth, high speed and highly resilient pan-European backbone.
}

\newduneabbrev{nren}{NREN}{National Research and Education Network}{National level research computing network infrastructure} %

\newduneabbrev{vrf}{VRF}{Virtual Routing and Forwarding}{Networking overlays that provide  a logical routing infrastructure  that allows flexible traffic engineering} %

\newduneword{samvalue}{value}{A generic quantity describing a file in the \dword{sam} data catalog} %

\newduneword{samparameter}{parameter}{A user or experiment described quantity describing a file in the \dword{sam} data catalog} %

\newduneword{samproject}{SAM-project}{A server process running centrally that maintains a predefined list of files and delivers information about  their locations when asked by distributed processes. The project tracks success and failure of file processing} %

\newduneword{samconsumer}{SAM-consumer}{A client process that requests information about file locations from a \dword{samproject}, process the file and reports success or failure} %

\newduneword{samdataset}{SAM-dataset}{ A dynamic collection of files defined by queries to the \dword{sam} data catalog} %
\newduneword{metacatdataset}{MetaCat-dataset}{ A fixed but mutable collection of files defined by queries to the \dword{metacat} data catalog} %

\newduneword{samsnapshot}{SAM-snapshot}{A fixed collection of files corresponding to a \dword{sam} \dword{samdataset} at a particular point in time} %

\newduneabbrev{mql}{MQL}{\dword{metacat} Query Language} {A query language which supports queries of the \dword{metacat} data catalog, including parentage and logical functions such as union, join and subtraction} %

\newduneword{pd2}{ProtoDUNE-II}{\gls{protodune} test runs at CERN in 2022-2023; also called \gls{mod0}}

\newduneabbrev{ucondb}{uconDB}{Unstructured Conditions Database}{Unstructured conditions database developed for \gls{fnal} fixed target experiments} %\cite{bib:ucondb}} 

\newduneabbrev{json}{JSON}{JavaScript Object Notation}{Open standard data interchange format that uses pair-value pairs and maps well onto python data formats such as tuples and lists} 

\newduneabbrev{dqmdb}{DQMDB}{Data Quality and Monitoring Database}{Database storing the results of data-quality monitoring} % removed .

\newduneword{db}{DB}{database}

\newduneabbrev{vm}{VM}{virtual machine}{Emulator of a physical computer that allows multiple users to configure different operating systems  while sharing physical hardware}

\newduneword{enstore}{Enstore}{A mass storage system developed by \gls{fnal} that provides distributed access and management of data stored on tapes} % \cite{Litvintsev:2012dw}}

\newduneword{stashcache}{StashCache}{A distributed caching federation that enables opportunistic users to utilize nearby opportunistic storage} %\cite{bib:stashcache}}

\newduneabbrev{hdf5}{HDF5}{Hierarchical Data Format}{Data format widely used in \gls{ml}} %\cite{bib:hdf5}

\newduneword{glideinwms}{GlideinWMS}{A system of submitting pilot jobs to grid computing sites, inside of which user jobs run, presenting a uniform setup across many different sites} % KH 2022-10-15  \cite{sfiligoi2009pilot

\newduneword{mars}{MARS}{The MARS code system is a set of \gls{mc} programs for detailed simulation of coupled hadronic and electromagnetic cascades, with heavy ion, muon and neutrino production and interactions} % removed .

\newduneabbrev{fife}{FIFE}{Fabric for Intensity Frontier Experiments}{\gls{fnal} computing infrastructure for \gls{if} Experiments} % \cite{box2014fife}} \newduneword{postgres}{Postgres}{also known as PostgreSQL, Postgres is a free and open-source relational database management system used extensively for databases in \gls{hep}}

\newduneword{tr}{trigger record}{A data record produced by the \gls{dune} \gls{daq} system.  A Trigger Record can contain multiple interaction ``events'' or none} 

\newduneword{postgres}{Postgres}{also known as PostgreSQL, Postgres is a free and open-source relational database management system used extensively for databases in \gls{hep}}

\newduneword{fhicl}{FHICL}{Fermilab Hierarchical Configuration Language; a standard configuration language for the storage, communication, and manipulation of scientific parameter sets}

\newduneword{api}{API}{Application Programming Interface}

\newduneword{hwdb}{HWDB}{hardware database}   

\newduneword{s3}{S3}{The Amazon cloud-based commercial storage service} % removed .  %anne added

\newduneword{openstack}{OpenStack}{An open source cloud software used to deploy instances of containers}

\newduneabbrev{cric}{CRIC}{Computing Resource Information System}{a framework providing a centralized (and flexible) way to describe which resources are being used by the experiment and how}

\newduneabbrev{ccb}{CCB}{Computing Contributions Board}{a board made up of institutional representatives for larger countries and laboratories. It meets annually to negotiate collaboration contributions to computing infrastructure} 

\newduneword{garfield}{Garfield}{A simulation program developed at \gls{cern} for gaseous detectors } %\cite{Veenhof:1998tt}

\newduneword{datatier}{data tier}{Differing data types produced in a processing sequence, for example, raw data, reconstructed, derived analysis sample, histograms} 

\newduneabbrev{gpu}{GPU}{Graphical Processing Unit}{Specialized computing hardware optimized for image processing} 

\newduneabbrev{gpvm}{dunegpvm}{DUNE General Purpose Virtual Machine}{Centrally managed virtual Linux systems at \gls{fnal} with access to network attached and \gls{pnfs} storage. Used for small-scale data analysis and algorithm development} 

\newduneword{madx}{MAD-X}{framework that provides the de facto standard scripting language to describe particle accelerators, simulate beam dynamics, and optimize beam optics at \gls{cern}}

\newduneabbrev{cpu}{CPU}{Central Processing Unit}{A computing processor, when used as a unit of processing; generally refers to a single core} 

\newduneabbrev{fft}{FFT}{Fast Fourier Transform}{An algorithm that calculates the frequency components of a time-domain waveform in a computationally efficient manner} 

\newduneword{grain}{GRAIN}{In the \gls{sand} detector, a small cryostat containing \gls{lar} installed upstream of the straw-tube tracker inside the \gls{ecal}}

\newduneword{voms}{VOMS}{Virtual Organization Membership Service (in grid computing)}

\newduneword{cry}{CRY}{A cosmic ray shower library and software tool}

\newduneword{jwt}{JWT}{\gls{json} web token}

\newduneabbrev{ferry}{FERRY}{Frontier Experiments RegistRY}{a central database that keeps track of all scientific computing users at \gls{fnal}, the experiments and groups of which they are members, and the various capabilities they are allowed}

\newduneabbrev{perfsonar}{perfSONAR}{performance Service-Oriented Network monitoring ARchitecture}{a network measurement toolkit designed to provide federated coverage of paths and to help establish end-to-end usage expectations}

\newduneword{ipv6}{IPv6}{the most recent version of the Internet communications protocol that provides an identification and location system for computers on networks and routes traffic across the Internet}

\newduneword{monit}{MONIT}{a suite of tools using open source technology used in the monitoring of the \gls{cern} IT data center and \gls{wlcg} infrastructure. }

\newduneword{garana}{GArAna}{provides a facility for making an analysis ntuple from information stored in \gls{garsoft} data products}

\newduneword{ci}{CI}{continuous integration}

\newduneword{datadis}{Data Dispatcher}{communicates between running processing jobs and the data delivery systems.  It provides file location information and basic bookkeeping on file access and transfers} % removed .

\newduneabbrev{castor}{CASTOR}{CERN Advanced STORage manager}{a hierarchical storage system with tape and disk developed at \gls{cern}. It is being replaced by \gls{cta}} 

\newduneabbrev{cta}{CTA}{CERN Tape Archive}{a hierarchical storage system with tape and disk developed at \gls{cern}. It is replacing \gls{castor}} 

\newduneabbrev{sonic}{SONIC}{Services for Optimized
Network Inference on Coprocessors}{Framework for implementing machine learning algorithms on co-processors developed by \gls{fnal}}

\newduneword{jobsub}{JobSub}{The \gls{fnal} \gls{if} job submission client that supports user submission of complex workflows to HT-Condor} %\cite{box2014fife} 

\newduneword{hepcloud}{HEPCloud}{routes jobs to local or remote computing resources based on the policy for a particular experiment, workflow requirements, and cost and efficiency of accessing the various resources. It expands the resources available to include \gls{hpc} centers and commercial cloud resources}

\newduneword{wmsmon}{WMS monitoring}{Workflow Management System Monitoring}

\newduneword{redmine}{Redmine}{an open source code repository and issue tracking tool that was historically used by many of the \gls{fnal} general computing projects and \gls{if} experiments} 

\newduneabbrev{if}{IF}{Intensity Frontier}{refers to \gls{hep} experiments, particularly in the U.S., that rely on high luminosity instead of high energy for discovery.  Includes B-physics, neutrino, and muon experiments}

\newduneword{slack}{Slack}{a commercial business communication platform} 

\newduneword{servicenow}{ServiceNow}{a commercial enterprise workflow system used by \gls{fnal} for formal issue tracking and IT workflows such as user account preparation} 

\newduneabbrev{rse}{RSE}{Rucio Storage Element}{A storage element that is known to the DUNE \gls{rucio} instance} 

\newduneabbrev{coffea}{COFFEA}{Columnar Object Framework For Effective Analysis}{Columnar data analysis frame\-work developed at \gls{fnal}} %\cite{Smith:2020pxs}

\newduneword{nuisance}{Nuisance}{An open source C++ framework for studying neutrino interaction cross sections} %\cite{Stowell:2016jfr}} 

\newduneabbrev{highland}{HighLAND}{High Level Analysis at a Neutrino Detector}{Analysis framework developed by the \gls{t2k} collaboration} 
 
\newduneabbrev{raii}{RAII}{Resource Acquisition Is Initialization}{a C++ programming technique that binds the lifecycle of a resource that must be acquired before use (e.g., allocated heap memory, thread of execution, open socket, open file, locked mutex, disk space, database connection—anything that exists in limited supply) to the lifetime of an object}  
 
\newduneword{fts3}{FTS3}{File Transfer Service version 3, developed by \gls{cern}, and distinct from, the \gls{fnal} File Transfer Service} 
 
\newduneword{wfms}{WFMS}{Workflow Management System}
 
\newduneword{condmeta}{conditions metadata}{defined as the information necessary to understand the context of physics data, e.g., beam data or calibrations} 
 
\newduneword{cdb}{conditions DB}{The Conditions Database stores the \gls{condmeta} needed for data processing and analysis}

\newduneword{rntuple}{RNtuple}{The next generation of the \gls{root} I/O system} %\cite{Blomer:2020usr, ROOTTeam:2020jal}}
 
\newduneabbrev{lan}{LAN}{Local Area Network}{Computing network confined to a relatively small geographic area} 
 
\newduneabbrev{wan}{WAN}{Wide Area Network}{Computing network that interconnects \glspl{lan} across relatively broad geographic areas} 

\newduneword{layer3}{Layer-3}{Networking protocol  {https://en.wikipedia.org/wiki/Network\_layer}}
 
\newduneabbrev{rest}{REST}{Representational State Transfer}{Standard for web interfaces} %  \href{https://en.wikipedia.org/wiki/Representational\_state\_transfer}{https://en.wikipedia.org/wiki/Representational_state_transfer}}
 
\newduneabbrev{faq}{FAQ}{Frequently Asked Questions}{A software system for collecting and answering the most common questions about an activity} 
 
\newduneword{spack}{Spack}{Software package manager for UNIX and mac OS systems} 

\newduneabbrev{lxplus}{LXPLUS}{Linux Public Login User Service)}{The interactive logon service to Linux for all \gls{cern} users} 

\newduneword{dunesw}{dunesw}{The base DUNE software release} 

\newduneword{posix}{POSIX}{Portable Operating System Interface - Unix standard for operating system interfaces} 

\newduneword{nutools}{NuTools}{shared code for \gls{larsoft} + NOvA + other neutrino experiments that use \gls{art}.  Includes beam and event generators and re-weighting packages}

\newduneword{singlecube}{SingleCube}{A cubical time projection chamber 30 cm on a side with a single ND-LAr pixel readout tile, used for prototyping and tests}

\newduneword{hllhc}{HL-LHC}{High-luminosity \gls{lhc}} %anne added

\newduneabbrev{sof}{SoF}{signal-over-fiber}{a technology in which a fiber optic cable carries detector output that has been converted from an electrical to an optical pulse}

\newduneword{icd}{interface document}{Interface document}

\newduneword{ipmi}{IPMI}{Intelligent Platform Management Interface}

\newduneword{grafana}{Grafana}{add def}

\newduneword{trigact}{trigger activitation}{a correlation (or cluster) in time and space within a stream of \glspl{trigprimitive}}

\newduneword{cad}{CAD}{computer aided design}

\newduneword{gkvm}{GKVM}{Gava, Kneller, Volpe, McLaughlin Supernova model}

\newduneword{daqrou}{DAQ readout unit}{The basic component of the \gls{daqros}}

\newduneword{kpp}{key performance parameter}{(KPP) Defined by \gls{doecd}-2 (U.S.) as characteristic,  function, requirement or design basis that if changed would have a major impact on the system or facility performance, schedule,  cost and/or risk}

\newduneword{ftbf}{FTBF}{Fermilab Test Beam Facility}

\newduneabbrev{toad}{TOAD}{Teststand of an Overpressure Argon Detector}{A test of a high-pressure gaseous argon TPC in the Fermilab Test Beam Facility \gls{ftbf}}

\newduneabbrev{fcss}{FCSS}{field cage support system}{The support system suspended from the cryostat ceiling in the \gls{vdmod0} used to support the \gls{fc}} 

\newduneword{trigrec}{trigger record}{a collection of data and metadata from selected detector space-time volumes corresponding to a trigger decision}

\newduneword{solarization}{solarization}{Aging caused by exposure to UV-light, which can lead to increased attenuation}

\newduneword{cam}{CAM}{control account manager}

\newduneword{msds}{MSDS}{Manufacturer Safety Data Sheet}

\newduneword{darkside}{Darkside-50}{a dual-phase argon \gls{tpc}, developed for use in a search for direct evidence of dark matter}

\AtBeginDocument{\RenewCommandCopy\qty\SI}

\hypersetup{
    pdftitle={\expshort TDR \thedocsubtitle},
    pdfauthor={\expshort Collaboration},
    final=true,
    colorlinks=false,
    linktocpage=true,
    linkbordercolor=blue,
    citebordercolor=green,
    urlbordercolor=magenta,
    filecolor=black,
    pdfpagemode=UseOutlines,
    pdfborderstyle={/S/U},  
}

\begin{document}

% This should be \input first thing after \begin{document}
% I removed the stuff that the cover now covers

\pagestyle{titlepage}
\includepdf[pages={-}]{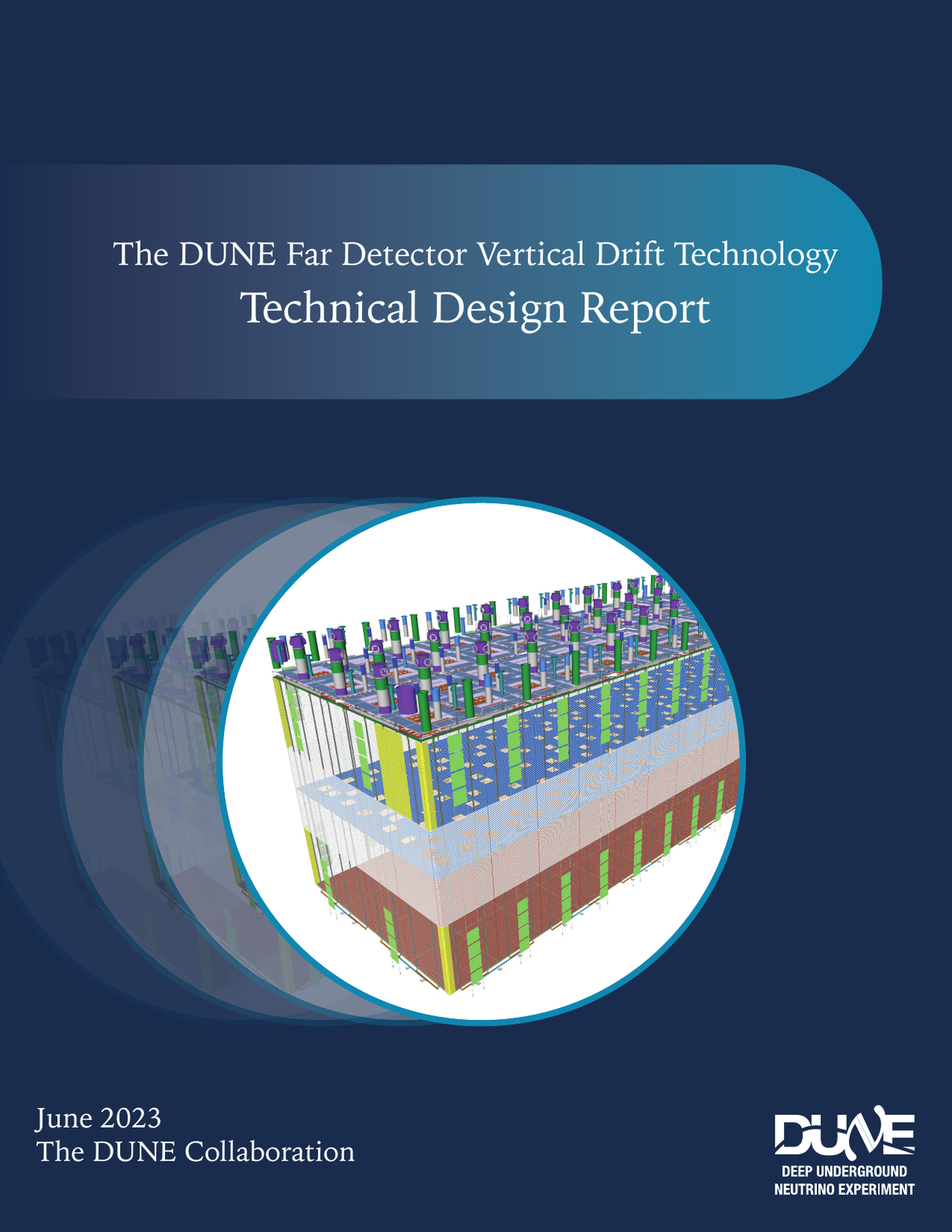}
\cleardoublepage

\vspace*{16cm} 
  {\small  This document was prepared by the DUNE collaboration using the resources of the Fermi National Accelerator Laboratory (Fermilab), a U.S. Department of Energy, Office of Science, HEP User Facility. Fermilab is managed by Fermi Research Alliance, LLC (FRA), acting under Contract No. DE-AC02-07CH11359.
  
The DUNE collaboration also acknowledges the international, national, and regional funding agencies supporting the institutions who have contributed to completing this design report.  
  }

\includepdf[pages={-}]{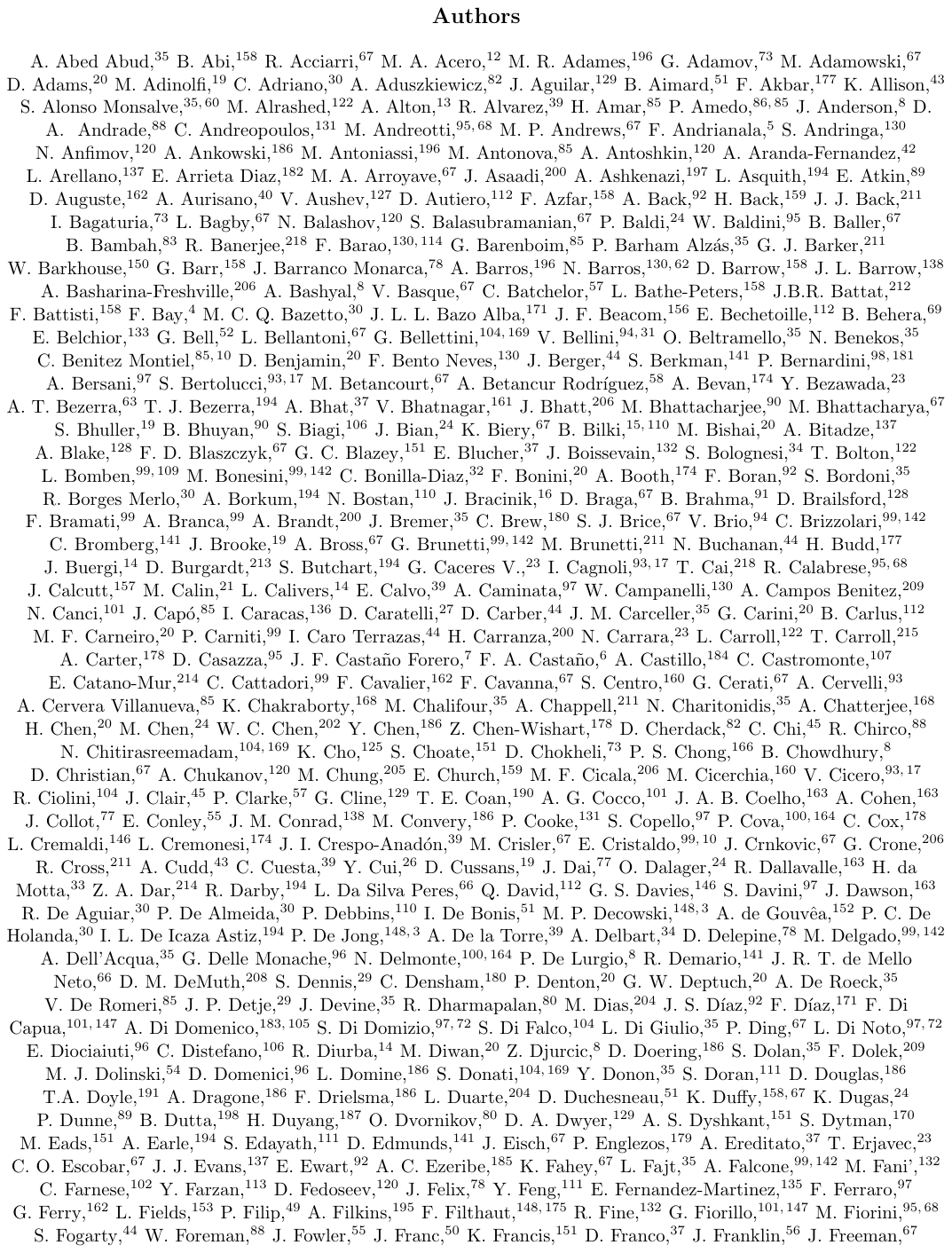}  

\renewcommand{\familydefault}{\sfdefault}
\renewcommand{\thepage}{\roman{page}}
\setcounter{page}{0}

\pagestyle{plain} 

%\clearpage

\textsf{\tableofcontents}
%\clearpage

\textsf{\listoffigures} 
\clearpage

\textsf{\listoftables} 
\clearpage

%For acronym list to appear just after TOC, TOF, TOT
%\printnomenclature
%\clearpage

\iffinal\else
\textsf{\listoftodos}
\clearpage
\fi

\renewcommand{\thepage}{\arabic{page}}
\setcounter{page}{1}

\pagestyle{fancy}

% Set how header/footers look
\renewcommand{\chaptermark}[1]{%
\markboth{Chapter \thechapter:\ #1}{}}
\fancyhead{}
%\fancyhead[RO,LE]{\textsf{\footnotesize \thechapter--\thepage}}
%\fancyhead[LO,RE]{\textsf{\footnotesize \leftmark}}
\fancyhead[R]{\textsf{\footnotesize \thechapter--\thepage}}
\fancyhead[L]{\textsf{\footnotesize \leftmark}}

\fancyfoot{}
\fancyfoot[R]{\textsf{\footnotesize  Technical Design Report}}
\fancyfoot[L]{\textsf{\footnotesize DUNE FD2 Vertical Drift LArTPC}}
\fancypagestyle{plain}{}

\renewcommand{\headrule}
{\vspace{-4mm}\color[gray]{0.5}{\rule{\headwidth}{0.5pt}}}

\newcommand{\jy}[1]{\textcolor{green}{#1}}
\newcommand{\jyc}[1]{\textbf{\textcolor{green}{(#1 --jy)}}}

% Not all main documents have any citations.
% When not built in "final" mode, add in one citation just to let the
% document build.
% If, after substantial editing a main document still lacks any
% citations then it should have its whole bibliography removed.
%\ifdefined\isfinal\nocite{}\else\nocite{CD0}\fi
\nocite{CD0}

% see also preamble.tex
%\input{common/acronyms}

\cleardoublepage

\chapter{Executive Summary} 
\label{ch:execsumm}

The \dword{dune} is designed to conduct a broad exploration of the three-flavor model of neutrino physics with unprecedented detail~\cite{DUNE:2020ypp}. Chief among DUNE’s potential discoveries is an unambiguous determination of the neutrino mass hierarchy and high-precision measurements of the parameters in the \dword{pmns} matrix, especially the \dword{cp} violating parameter $\delta$. Other goals include determining whether the 2-3 neutrino mixing is maximal and, if not, measuring the octant of $\Theta_{23}$; searching for sterile neutrinos; detecting a \dword{snb} in or near our galaxy should a supernova occur; and searching for proton decay.

To achieve these goals, \dshort{dune} will build on the strategies of
previous and current long-baseline neutrino experiments, such as 
%Super-Kamiokande~\cite{Super-Kamiokande:2002weg},
K2K~\cite{K2K:2001nlz},
MINOS~\cite{MINOS:2006foh},
T2K~\cite{T2K:2011qtm},
and NOvA~\cite{Adamson:2016xxw},
%current long-baseline neutrino experiments \dword{t2k} and \dword{nova}, 
for which horn-focused beamlines produce beams of almost entirely muon neutrinos. The beam's parameters are measured by a \dword{nd}, located just downstream of the beamline, and a \dword{fd}, some distance away. For DUNE, the \dword{usproj} project is building a beamline at \dword{fnal}, in Batavia, Illinois, that is designed to extract a proton beam from the \dword{mi} and transport it to a target area where the collisions generate a beam of charged particles. This secondary beam aimed towards the \dshort{fd} is followed by a decay region where the particles of the secondary beam decay to generate the neutrino beam. The beam intersects the \dshort{fd}, located  1300\,km away and 4850 feet (1,480\,m) underground at the \dword{surf} in Lead, South Dakota.

\Dword{lartpc} technology has been selected for the DUNE \dshort{fd}. 
This technology combines fine-grained tracking with total absorption calorimetry to provide a detailed view of particle interactions, making it a powerful tool for neutrino physics and underground physics such as proton decay and supernova-neutrino observation. 
It provides millimeter-scale resolution in \threed for all charged particles. Particle types can be identified both by their $dE/dx$ and by track patterns, e.g., the decays of stopping particles. 
The modest radiation length (14\,cm) is sufficiently short to identify and contain electromagnetic showers from electrons and photons, but long enough to provide good e/$\gamma$ separation by $dE/dx$ (one versus two \dwords{mip}) at the beginning of the shower. In addition, photons can
be distinguished from electrons emanating from an event vertex %by the flight path before their first interaction
by any gaps between the vertex and the start of the track. 
These characteristics allow the \dshort{lartpc} to identify and reconstruct signal events with high efficiency,  
while rejecting backgrounds, 
to provide a high-purity data sample~\cite{Adams:2013qkq}.

% RJW 24feb23: Added the following on the PDS.
In addition, taking advantage of the excellent scintillation properties of \dword{lar}, the \dshort{fd} will be instrumented with \dwords{pd} that will provide precise time information and additional calorimetric information.%, and a \dshort{pd}-based trigger for low-energy events.

The DUNE \dshort{fd} will be implemented as a set of four modules using independent cryostats, each containing about \larmass{} 
of \dshort{lar}. The cryostats, 
\cryostatlen long by \cryostatht wide by  \cryostatwdth high, have interior dimensions  \cryostatleninner by \cryostatwdthinner by \cryostathtinner. 
An 11.3\,m wide, 60.5\,m long mezzanine is installed 2.3\,m above each cryostat to house cryogenics and the \dword{daq} equipment. A second narrower mezzanine is provided for the 78 detector electronics racks. 

The first module, called the \dword{sphd}, will be located in the east end of the north cavern and the second, the \dword{spvd}, will be installed in the east end of the south cavern (see Figure~\ref{fig:cavern}). This arrangement will minimize interference between installation activities on the separate modules. A \dword{cuc} 
(190\,m long by  19.3\,m wide by  11\,m high) will house electrical, HVAC, internet, cryogenics, and other infrastructure for all the detector modules.

%$$$$$$$$$$$$$$$ 
\begin{dunefigure}
[Placement of detector modules in the north and south caverns]
{fig:cavern}
{Placement of detector modules at the east end of the north and south caverns at the \dshort{surf} 4850 foot level. The \dshort{cuc} is located between the detector caverns.}
\includegraphics[width=1.0\linewidth]{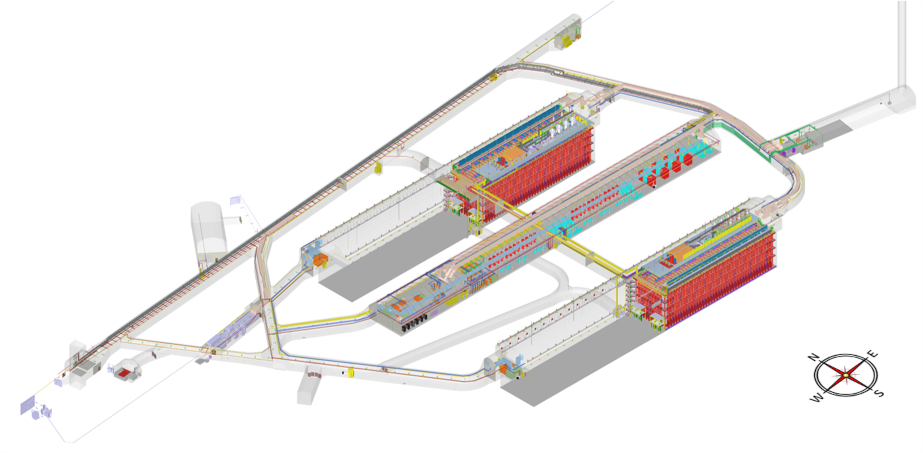}
\end{dunefigure}
%$$$$$$$$$$$$$$$ 

The first two far detector \dword{lartpc} modules differ in design. 
\dshort{sphd} uses \dword{hd} technology~\cite{DUNE:2020txw},  based on techniques used successfully by previous \dshort{lartpc} detectors, including \dword{icarus}~\cite{Amerio:2004ze}, 
\dword{microboone}~\cite{Acciarri:2016smi}, 
and \dword{pdsp}~\cite{DUNE:2021hwx}. 
%In \dshort{sphd}, i
Ionization electrons from charged particles will drift horizontally under the influence of an electric field (\efield) produced by vertically-oriented cathode and anode planes, with  the active volume surrounded by 
a \dword{fc}. 
The anode is formed by \dwords{apa} that feature planes of thin wires wrapped around and soldered onto an insulating frame. Two wire planes 
acquire induced signals as the ionization electrons drift past them, and the third wire plane 
collects the ionization electrons. 
A \dword{pds} implements \dword{xarapu}~\cite{Machado-XARA:2018} %, Brizzolari:2021akq} 
technology.

This \dword{tdr} describes the design for the second \dshort{detmodule}, \dword{spvd}, which implements \dshort{vd} technology. 
This design features a horizontal cathode plane placed at mid-height in the active volume of the cryostat, dividing the \dshort{lartpc} into two vertically stacked equal volumes, 
each 6.5\,m in height. 
The \dwords{anodepln} are constructed of perforated printed circuit boards (\dwords{pcb}) with etched electrodes forming a three-view charge readout.  
The top anode plane 
is placed close to the cryostat top, just below the surface of the \dword{lar}, and the other is located as close to the bottom of the cryostat as possible.
Ionization electrons will drift vertically, up or down, towards one of the \dshort{anodepln}s.  

The \dshort{spvd} design 
 offers a slightly larger instrumented volume ($60.0 \times 13.5 \times 13.0$\,m$^3$) compared to the \dshort{sphd} design and simpler, more cost-effective construction and installation due to its geometry and structure. 
The \dshort{lar} will be doped with a small quantity of xenon, which has no impact on the \dword{tpc} operation but significantly enhances the \dshort{pds} performance.
The \dshort{spvd} design will implement the same  \dshort{xarapu} \dshort{pds} technology as the \dshort{sphd} design.

\dshort{lbnf-dune} is committed to protecting the health and safety of staff, the community, and the environment, as described in the LBNF/DUNE Integrated Environmental, Safety, and Health Management Plan~\cite{edms-2808692}, as well as to ensuring a safe work environment for \dshort{dune} workers at all institutions and protecting the public from hazards associated with constructing and operating \dshort{dune}.  
Accidents and injuries are preventable, and the \dshort{esh} team will work with the global \dshort{lbnf-dune} project and DUNE collaboration to establish an injury-free workplace. All work will be performed so as to preserve  the quality of the environment and prevent property damage.

%%%%%%%%%%%%%%%%%%%%%%%%%%%%%%%%
\section{The DUNE Physics Program}
\label{sec:es:physics}

To reach the necessary precision on its measurements, DUNE will need to collect a few thousand neutrino interactions over a period of about ten years. The number of interactions is the product of (1) the intensity of the neutrino beam, (2) the probability that a neutrino will oscillate (approximately 0.02), (3) the interaction cross section, and (4) the detector mass. Currently, the
highest proton beam power that a %beam 
target can safely withstand is between 1 and 2\,MW, which limits the achievable neutrino beam intensity. This points to a required detector mass in the tens-of-kilotons range~\cite{DUNE:2020lwj}.

Central to achieving DUNE’s physics program is the construction of a detector that combines the many-kiloton fiducial mass necessary for rare event searches with sub-centimeter spatial resolution. The ability to accurately image the fine details of %events 
neutrino interactions will enable data analyses %that are able 
to select the  possibly rare, desired %events 
signals from a much larger %population of 
background. %events.  
As discussed in Chapter~\ref{ch:Phys}, the \dshort{spvd} \dshort{lartpc} will enable %event 
selection and analysis of events ranging from the MeV energy scale of solar and \dword{snb} neutrinos to the GeV energy scale of neutrinos produced from the 
neutrino beam  at \dshort{fnal}. 

A search for leptonic \dword{cpv} requires a study of \nue appearance in the 
mostly \numu beam. This analysis requires the capability to separate electromagnetic activity induced by \dword{cc} \nue interactions from similar activity arising from photons, such as photons from $\pi^0$ decay. Two signatures allow this differentiation: (1) photon showers are typically preceded by a gap prior to conversion, characterized by the 18\,cm conversion length in \dshort{lar}; and (2) the initial part of a photon shower, where an electron-positron pair is produced, has twice the $dE/dx$ of the initial part of an electron-induced shower. 

A search for nucleon decay, where the primary channel of interest is \ptoknubar{},
must identify kaon tracks as short as a few centimeters. It is also vital to accurately limit these possible nucleon-decay events to an origin within the fiducial volume, suppressing cosmic-muon-induced backgrounds. Here the detection of argon-scintillation photons is important for determining the time of the event. 

Detecting a \dshort{snb} poses different challenges, those of dealing with a high data rate while maintaining the high detector uptime required to ensure the capture of such a rare event.
The signature of a \dshort{snb} is a collection of MeV-energy electron tracks a few centimeters in length from \dword{cc} \nue interactions, spread over the entire detector volume. To fully reconstruct a \dshort{snb}, the entire detector must be read out with a data rate of up to 2\,TB/s, for 30\,s to 100\,s, including a 4\,s pre-trigger window.

%%%%%%%%%%%%%%%%%%%%%%%
\section{Far Detector Requirements and Specifications} \label{subsec:FD2ss}

The DUNE \dshort{detmodule}s are required to have fiducial mass in the multi-kiloton range. This large target mass, combined with mm-scale \dword{tpc} spatial resolution, will allow the separation of physics signatures of interest from the numerous backgrounds. The qualitative \dword{fd} physics requirements are listed here: 

\begin{itemize}
    \item  The \dshort{fd} shall enable identification of $\nu_\mu$ and $\nu_e$ charged current interactions, regardless of the final state; in particular, 
    \begin{itemize}
        \item 
        separation of muons
        from other charged particles at high efficiency using energy deposition and range;
        \item 
       separation of electromagnetic showers
        from  charged particle tracks at high efficiency using topological information;
        \item %rejection 
        identification of photon-induced showers based on observation of gaps between the start of the shower and the neutrino interaction vertex;
         \item   separation of 
        electron- and photon-induced showers using ionization density at the start of the shower;   
        \item 
        separation of protons from other particles using ionization density and range.  
    \end{itemize}     
    \item 
    The \dshort{fd} shall provide 
    sufficient energy resolution for electromagnetic showers to allow %precise 
    measurement of the electron neutrino (\nue) energy spectrum. 
    \item The \dshort{fd} shall be  modular to facilitate construction, installation and testing 1.5\,km underground (the \dword{4850l} of \dshort{surf}). 
    \item 
    The active volume shall be maximized.
\end{itemize}
%%%%

The specifications listed in Table~\ref{tab:specs:SP-FD2}
ensure that 
the \dshort{spvd} detector module
will satisfy the physics requirements. These specifications, most of which are common to both \dshort{spvd} and \dshort{sphd} designs, trace back to higher-level requirements on the DUNE experiment and ultimately to its science objectives~\cite{edms-198204}. 

% This file is generated, any edits may be lost.
\begin{footnotesize}
%\begin{longtable}{p{0.14\textwidth}p{0.13\textwidth}p{0.18\textwidth}p{0.22\textwidth}p{0.20\textwidth}}
\begin{longtable}{p{0.12\textwidth}p{0.18\textwidth}p{0.17\textwidth}p{0.25\textwidth}p{0.16\textwidth}}
\caption[Far Detector Specifications]{Specifications common to both far detector modules (labeled FD-1 through FD-30), and for FD2-VD (FD2-31 through FD2-35). %\fixmehl{ref \texttt{tab:spec:SP-FD2}}
} \\
  \rowcolor{dunesky}
       Label & Description  & Specification \newline (Goal) & Rationale & Validation \\  \colhline

   \newtag{FD-1}{ spec:min-drift-field }  & Minimum drift field  &  $>$\,\SI{250}{ V/cm} \newline ($>\,\SI{450}{ V/cm}$) %updated for FD2
   &  Lessens impacts of $e^-$-Ar recombination, $e^-$ lifetime, $e^-$ diffusion and space charge. &  ProtoDUNE \\ \colhline
    
   %  \newtag{FD-2}{ spec:system-noise }  & System noise  &  $<\,\SI{1000}\,e^-$ &  Provides $>$5:1 S/N on induction planes for  pattern recognition and two-track separation. &  ProtoDUNE and simulation \\ \colhline
  \newtag{FD-2}{ spec:system-noise }  & System noise  &  $<\,\SI{1000}\,e^-$ &  Provides $>$5:1 S/N on induction planes for  pattern recognition and two-track separation.\\ \colhline

  \newtag{FD-3}{ spec:light-yield }  & Light yield  &  $>\,\SI{20}{PE/MeV}$ (avg), $>\,\SI{0.5}{PE/MeV}$ (min) &  Gives PDS energy resolution comparable to that of the TPC for 5-7 MeV SN $\nu$s, and allows tagging of $>\,\SI{99}{\%}$ of nucleon decay backgrounds with light at all points in detector. &  Supernova and nucleon decay events in the FD with full simulation and reconstruction. \\ \colhline

   \newtag{FD-4}{ spec:time-resolution-pds }  & Time resolution  &  $<\,\SI{1}{\micro\second}$ \newline ($<\,\SI{100}{\nano\second}$) &  Enables \SI{1}{mm} position resolution for \SI{10}{MeV} SNB candidate events for instantaneous rate $<\,\SI{1}{m^{-3}ms^{-1}}$. &   \\ \colhline
    
   \newtag{FD-5}{ spec:lar-purity }  & Liquid argon purity  &  $<$\,\SI{50}{ppt}
   % updated for FD2
   \newline ($<\,\SI{30}{ppt}$) &  Provides $>$5:1 S/N on induction planes for  pattern recognition and two-track separation. &  Purity monitors and cosmic ray tracks \\ \colhline

  \newtag{FD-7}{ spec:misalignment-field-uniformity }  & Drift field nonuniformity due to component alignment  &  $<\,1\,$\% throughout volume &  Maintains APA, CPA,  FC orientation and shape. &  ProtoDUNE \\ \colhline

  \newtag{FD-11}{ spec:hvs-field-uniformity }  & Drift field nonuniformity due to HVS  &  $<\,\SI{1}{\%}$ in 99.8\,\% of active volume 
  %updated for FD2
  &  High reconstruction efficiency. &  ProtoDUNE and simulation \\ \colhline

  \newtag{FD-12}{ spec:hv-ps-ripple }  & Cathode HV power supply ripple contribution to system noise  &  $<\,\SI{100}e^-$ &  Maximize live time; maintain high S/N. &  Engineering calculation, in situ measurement,   ProtoDUNE \\ \colhline

%  \newtag{FD-13}{ spec:fe-peak-time }  & Front-end \newline peaking time  & \newline  \SI{1}{\micro\second} &  Vertex resolution; optimized for \SI{5}{mm} anode readout spacing. &  ProtoDUNE and simulation \\ \colhline
%    \fixme{check 5mm}
\newtag{FD-13}{ spec:fe-peak-time }  & Front-end \newline peaking time  & \newline  \SI{1}{\micro\second} &  Vertex resolution; optimized for approximately \SI{5}{mm} anode readout spacing.\\ \colhline
%   \newtag{FD-14}{ spec:sp-signal-saturation }  & Signal saturation \newline level  &  
%   \num{500000} e$^-$ \newline (Adjustable so as to see saturation in less than 10\% of beam-produced events) &  
%   Maintain calorimetric performance for multi-proton final state. &  
%   Simulation \\ \colhline
   \newtag{FD-14}{ spec:sp-signal-saturation }  & Signal saturation \newline level  &  
   \num{500000} e$^-$ \newline (Adjustable so as to see saturation in less than 10\% of beam-produced events) &  
   Maintain calorimetric performance for multi-proton final state. \\ \colhline    
   
  \newtag{FD-15}{ spec:lar-n-contamination }  & LAr nitrogen contamination  &  $<\,\SI{10}{ppm}$ &  Maintain \SI{0.5}{PE/MeV} PDS sensitivity required for triggering proton decay near cathode. &  In situ measurement \\ \colhline

   \newtag{FD-17}{ spec:cathode-resistivity }  & Cathode resistivity  &  $>\,\SI{1}{\mega\ohm/square}$ \newline ($>\,\SI{1}{\giga\ohm/square}$) &  Detector damage prevention.  &  ProtoDUNE \\
   \colhline

  \newtag{FD-18}{ spec:cryo-monitor-devices }  & Cryogenic monitoring devices  &   &  Constrain uncertainties on detection efficiency, fiducial volume. &  ProtoDUNE \\ \colhline

%  \newtag{FD-19}{ spec:adc-sampling-freq }  & ADC sampling frequency  &  $\sim\,\SI{2}{\mega\hertz}$ &  Match \SI{1}{\micro\second} shaping time. &  Nyquist requirement and design choice \\ \colhline
   \newtag{FD-19}{ spec:adc-sampling-freq }  & ADC sampling frequency  &  $\sim\,\SI{2}{\mega\hertz}$ &  Match \SI{1}{\micro\second} shaping time. \\ \colhline   
   
%  \newtag{FD-20}{ spec:adc-number-of-bits }  & \newline Number of ADC bits  & \newline  \num{12} bits &  ADC noise contribution negligible (low end); match signal saturation specification (high end). &  Engineering calculation and design choice \\ \colhline
  \newtag{FD-20}{ spec:adc-number-of-bits }  & Number of ADC bits  &$\ge{12}$ bits &  Makes ADC noise contribution negligible (low end); matches signal saturation specification (high end). \\ \colhline    
   
  \newtag{FD-21}{ spec:ce-power-consumption }  & Cold electronics \newline 
  power consumption (in-\dshort{lar})   &  $<\,\SI{50}{ mW/channel} $ &  No bubbles in \dshort{lar} to reduce HV discharge risk. %&  Bench test 
  \\ \colhline

  \newtag{FD-22}{spec:data-rate-to-tape}& Data rate to tape  &  $<\,\SI{30}{PB/year}$ &  Cost.  Bandwidth. &  ProtoDUNE \\ \colhline

  \newtag{FD-23}{ spec:sn-trigger }  & Supernova trigger  & Efficiency for a SNB producing at least 60 interactions with a $\nu$ energy $>$\,10 MeV in 12\,kt of active detector mass during the first 10\,s of the burst. &  $>\,$95\% efficiency for SNB within 20 kpc &  Simulation and bench tests  %\fixme{C. Cuesta, is 12 kton correct? the fiducial/total mass (10/17 kton) of the FD2-VD could be in this table (AH 20 Dec '22: comes from EB held reqs; let's not change it)} 
  \\ \colhline

  \newtag{FD-24}{ spec:local-e-fields }  & Local electric fields  &  $<\,\SI{30}{kV/cm}$ &  Maximize live time; maintain high \dshort{s/n}. &  ProtoDUNE \\ \colhline

%  \newtag{FD-25}{ spec:non-fe-noise }  & Non-FE noise contributions  &  $<<\,\SI{1000}\,e^- $ &  High S/N for high reconstruction efficiency. &  Engineering calculation and ProtoDUNE \\ \colhline
  \newtag{FD-25}{ spec:non-fe-noise }  & Non-FE noise contributions  &  $\ll\,\SI{1000}\,e^- $ &  High S/N for high reconstruction efficiency. \\ \colhline    
   
  \newtag{FD-26}{ spec:lar-impurity-contrib }  & LAr impurity contributions from components  &  $\ll\,\SI{30}{ppt} $ &  Maintain HV operating range for high live time fraction. &  ProtoDUNE \\ \colhline

  \newtag{FD-27}{ spec:radiopurity }  & Introduced radioactivity  &  less than that from $^{39}$Ar &  Maintain low radiological backgrounds for SNB searches. &  ProtoDUNE and assays during construction \\ \colhline

%  \newtag{FD-28}{ spec:dead-channels }  & Dead channels  &  $<\,\SI{1}{\%}$ &  Minimize the degradation in physics performance over the $>\,20$-year detector operation. &  ProtoDUNE and bench tests \\ \colhline
  \newtag{FD-28}{ spec:dead-channels }  & Dead channels  &  $<\,\SI{1}{\%}$ &  Minimize the degradation in physics performance over the $>\,20$-year detector operation. \\ \colhline
    
   \newtag{FD-29}{ spec:det-uptime }  & Detector uptime  &  $>\,$98\% \newline ($>\,$99\%) &  Meet physics goals in timely fashion. &  ProtoDUNE \\ \colhline
    
   \newtag{FD-30}{ spec:det-mod-uptime }  & Individual detector module uptime  &  $>\,$90\% \newline ($>\,$95\%) &  Meet physics goals in timely fashion. &  ProtoDUNE \\ \colhline
    
    \newtag{FD2-31}{ spec:fd2vd-anode-flatness }  & 
    CRP anode plane global flatness &  
    $<$\,20\,mm &  
    Maintains drift field uniformity of $<$\,1\% throughout each drift region &  \coldbox test
    \\ \colhline
    
        \newtag{FD2-32}{ spec:fd2vd-crp-transp }  & 
    \dshort{crp} minimum permeability & 
    $>$\,15\% &
    Allows efficient local heat dissipation and free \dshort{lar} circulation &  
   CFD simulation
    \\ \colhline
    
        \newtag{FD2-33}{ spec:fd2vd-crp-gap }  & 
   Gaps between \dshorts{crp} &  
   5\,mm for CRPs within superstructure; \newline 10\,mm between superstructures &  
   Minimizes loss of \dshort{fv} and allows space for cabling &  ProtoDUNE
    \\ \colhline

        \newtag{FD2-34}{ spec:fd2vd-shield-pln }  & 
    CRP shield plane on cathode-facing side &  
     &  
    Reduces impact on electronics from cathode discharge&   cold box test and ProtoDUNE
    \\ \colhline
    
        \newtag{FD2-35}{ spec:fd2vd-strip-width }  & 
  \dshort{crp} strip width and pitch  &  
  $<$\,8.5\,mm (ind.);\newline $<$\,5.5\,mm (coll.); \newline gaps 0.5\,mm &
  \dshort{s/n} consistent with 100\% hit reconstruction efficiency for \dshorts{mip}.  Spacing provides 1.5\,cm vertex resolution in $y$-$z$ plane. &  
   ProtoDUNE and bench tests
    \\  \colhline
     %incl to 35

\label{tab:specs:SP-FD2}
\end{longtable}
\end{footnotesize}

%%%%%%%%%%%%%%%%%%%%%%%%%%%%%%%%
\section{Vertical Drift Design}
\label{sec:es:design}

The \dword{spvd}  design (Figures~\ref{fig:vertical_drift} and \ref{fig:FD2-Detector}) has evolved from the \dshort{dune} \dword{dp} design~\cite{DUNE:2018mlo}, and in large measure from the experience gained with \dword{pddp}, which was installed and operated at the \dword{cern} in 2019-2022. The \dshort{dp} design, with one anode plane at the top of the drift volume, and one cathode plane at the bottom, offered many attractive features, such as simplicity of construction and installation, homogeneity of the active volume, and   
the prospect of increasing the signal by developing avalanches in the gas phase to compensate for the signal losses over the long (several millisecond) drift times. It also presented challenges, 
namely, it required cathode operation at higher voltages, and the presence of high voltages (up to a few kV) in the gas phase.  
The \dshort{spvd} design offers many of the same advantages while avoiding the challenges associated with the liquid-gas phase interface and taking advantage of the successes of the \dshort{pdsp} design features.

Electron lifetime values well above a few milliseconds (ms) have now been measured in 
large \dshorts{lartpc}, including both \dword{protodune} detectors, with lifetimes of \PDelt\ in \dshort{pdsp}. These results have exceeded expectations, lessening the motivation for the \dshort{dp} gas-phase signal gain. The experience accumulated by the \dshort{dune} collaboration in the construction, installation, and operation of \dshort{pdsp} and \dshort{pddp} 
represents a solid basis from which to evolve the design and construction of the second detector module, optimizing the collection of electrons drifting along the vertical detector axis in the liquid phase.  

The vertical drift design  simplifies the detector construction and installation, 
and reduces the overall detector costs 
relative to the well-established \dword{sp} horizontal drift design based on large wire plane assemblies. 
It %is a natural evolution of the \dshort{dp} design  as it 
uses most of  the same structural elements as \dshort{dp}, e.g., the \dwords{crp} that form the anode planes 
and the \dword{fc} that hangs from the cryostat roof and is constructed of modular elements that are easy to produce, transport, and install.  
The main difference between the \dshort{spvd} and \dshort{dp} designs is the removal of the extraction of ionization electrons to the gas phase and the subsequent charge amplification stage. %The \dshort{dp} anode is replaced by 
In the \dshort{spvd} \dshort{crp}s, a pair of perforated \dshort{pcb}s, etched with readout strips, now operate completely immersed in \dshort{lar}. 
The absence of \dword{hv} in
the %superficial liquid 
gas layer makes the \dshort{spvd} design much less sensitive
to environmental conditions, such as any instability of the \dshort{lar} surface or floating contaminants.

\begin{dunefigure}
[Conceptual vertical drift design with PCB-based charge readout]
{fig:vertical_drift}
{Schematic of vertical drift concept with PCB-based charge readout. Corrugations on cryostat wall shown in yellow; PCB-based CRPs (brown, at top and bottom with superstructure in gray for top CRPs); cathode (violet, at mid-height with openings for photon detectors); field cage modules (white) hung vertically around perimeter (70\% transparent portion in regions near anode planes); photon detectors, 
placed in the openings on the cathode and on the cryostat walls, around the perimeter in the vertical regions near the anode planes.}
\includegraphics[width=0.9\linewidth]{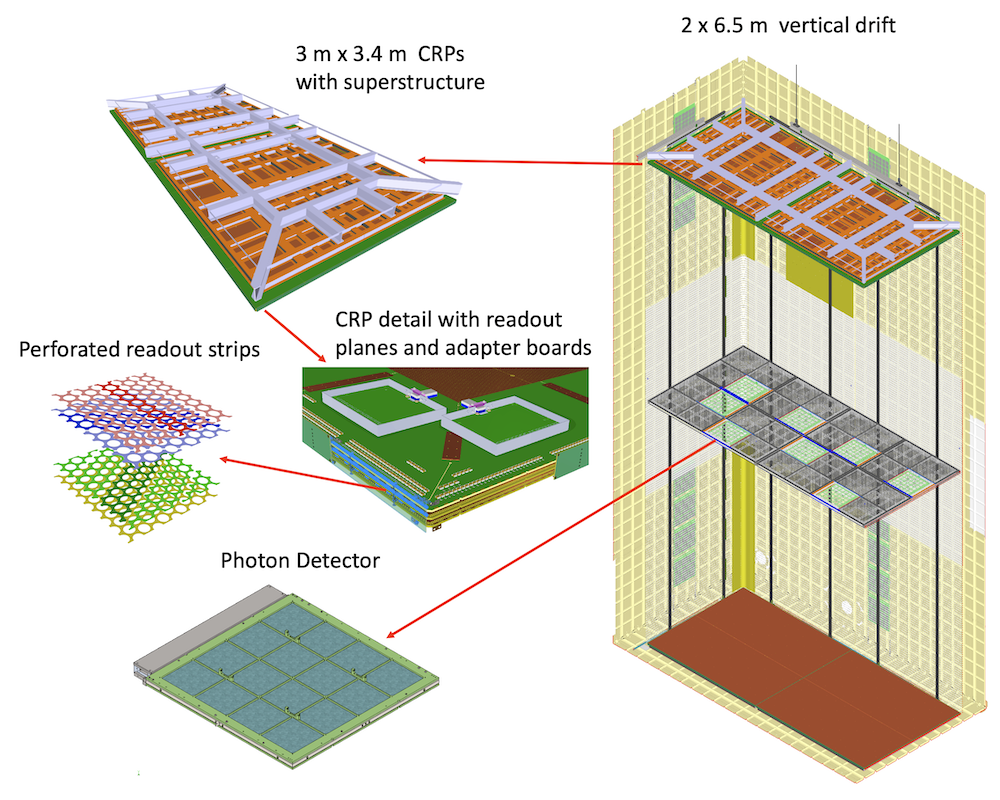}
\end{dunefigure}
\begin{dunefigure}
[Vertical drift detector]
{fig:FD2-Detector}
{Perspective view of the \dword{spvd} detector.} 
\includegraphics[width=1.\linewidth]{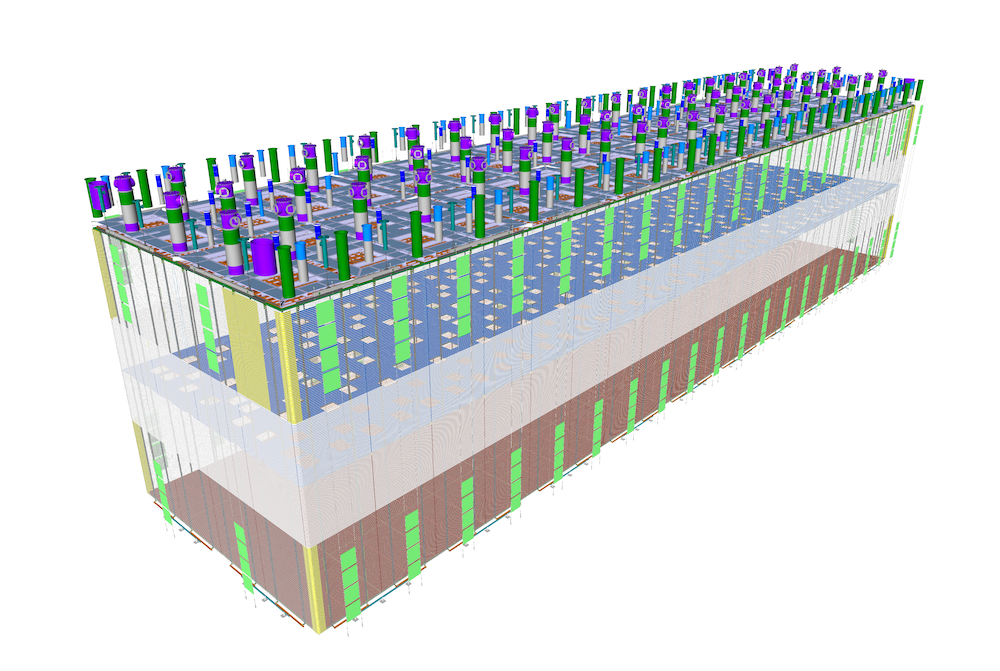}
\end{dunefigure}
The absence of gain in the \dshort{spvd} \dshort{crp}s results in a smaller, but more stable signal amplitude collected by the anodes. The signal reduction related to the unit gain in the perforated anodes is fully compensated by 
(1) increased strip pitch, 
(2) absence of extraction and collection inefficiencies present in the \dshort{dp}, and 
(3) charge collection on a single readout view rather than on %shared %among %strips etched 
two views oriented along orthogonal directions.

The \dshort{spvd} anode has a signal amplitude comparable to what a \dshort{dp} anode would have seen, but with much more reliability. Furthermore, considering that the 
drift length in 
\dshort{spvd} is %will be 
shorter than in the DP design, %will exploit a shorter drift length, 
the minimal signal level will at least a factor of four larger. 

Removing the cathode from the proximity of the cryostat floor maximizes the active volume by exploiting the total available vertical space of almost 13\,m. 
The \dshort{spvd} cathode is designed with a minimal thickness and hangs from the same top support structure of the detector that %also 
supports the top \dwords{crp}.  

Having subdivided the detector into two vertically stacked drift volumes of \driftd\ height each, the same \dshort{crp} geometry can be used for the top and bottom drift. 
The top \dshorts{crp}, %positioned 
suspended just under the liquid surface, %and suspended from superstructures, are 
are aligned parallel to this surface. 
This \dshort{anodepln} preserves full accessibility to the top drift electronics 
(\dword{tde}) via  \dwords{sftchimney}, as demonstrated in \dshort{pddp}. 
The bottom \dshorts{crp} are supported by feet on the cryostat floor, are mechanically  independent of the rest of the detector, and are not subject to movements.  
The bottom drift electronics (\dword{bde}) 
implements the same \dword{ce} as proven in \dshort{pdsp}, with some adaptations to  accommodate the \dshort{crp} geometry and modularity.

Before providing brief descriptions of each of the major subsystems in this chapter, the important features of the \dshort{spvd} design that make it a compelling candidate for the second \dshort{detmodule} are listed here:
\begin{itemize}
\item maximize the active volume;
\item high modularity of detector components;
\item simplified anode structure based on standard industrial techniques;
\item simplified cold testing of instrumented anode modules, for which  large cryogenic vessels are not required;
\item field cage structure independent of the other detector components;
\item extended drift distance;
\item reduction of dead material in the active volume, maximizing fiducial mass;
\item design allowing for improved light detection coverage; % and trigger efficiency;
\item simplified and faster installation and \dword{qa}/\dword{qc} procedures, not requiring large in situ infrastructures; and
\item cost-effectiveness.% for second module in light of restricted resources.
\end{itemize}

%%%%%%%%%%%%
\subsection{Charge Readout Planes (Anodes)}
\label{sec:vdd:crp}

The baseline design \dshort{spvd} anodes, illustrated in Figure~\ref{fig:supercrp2ch1}, are fabricated from two double-sided perforated \dwords{pcb}, 3.2\,mm thick, that are connected mechanically, with their perforations aligned, to form \dwords{cru}. 
A pair of \dshort{cru}s is attached to a composite frame to form a \dword{crp}; the frame provides mechanical support and planarity. 
The holes allow the electrons to pass through to reach the collection strips. 
Each anode plane consists of 80 \dshorts{crp} in the same  layout. 
The \dshorts{crp} in the top drift volume, operating completely immersed in the \dshort{lar}, are suspended  from the cryostat roof using a set of superstructures, and the bottom \dshorts{crp} are supported by posts positioned on the cryostat floor. 
%Similar to the \dword{dp} design, 
The superstructures hold either two or six \dshorts{crp}, and allow adjustment, via an externally accessible suspension system, to compensate for possible deformations in the cryostat roof geometry.

 The innermost face of a \dshort{crp}, i.e., the \dshort{pcb} face directly opposite the cathode, has a copper guard plane to absorb any unexpected discharges.
The reverse side of this \dshort{pcb} 
is etched with strips that form the first induction plane. The other \dshort{pcb} has strips on the side facing the inner \dshort{pcb} forming the second induction plane, %drift volume, 
and has the collection plane strips on its reverse side. 
The three planes of strips are segmented at about 7.5 and 5\,mm pitch, for induction and conduction planes, respectively, and are set at 60$^\circ$ angles relative to each other to maximize information in the charge readout from different projections. A potential difference of about 1\,kV %, completely confined in the liquid, (already said)
must be applied across each  \dshort{pcb} to guarantee full transmission of the electrons through the holes.

\begin{dunefigure}
[Exploded view of a top superstructure and CRPs]
{fig:supercrp2ch1}
{A top superstructure (green structure on top) that holds a set of six \dshorts{crp}, and below it an exploded view of a \dshort{crp} showing its components: the \dshort{pcb}s (brown), adapter boards (green) and edge connectors that together form a \dshort{cru}, and composite frame (black and orange).}
\includegraphics[width=0.99\textwidth]{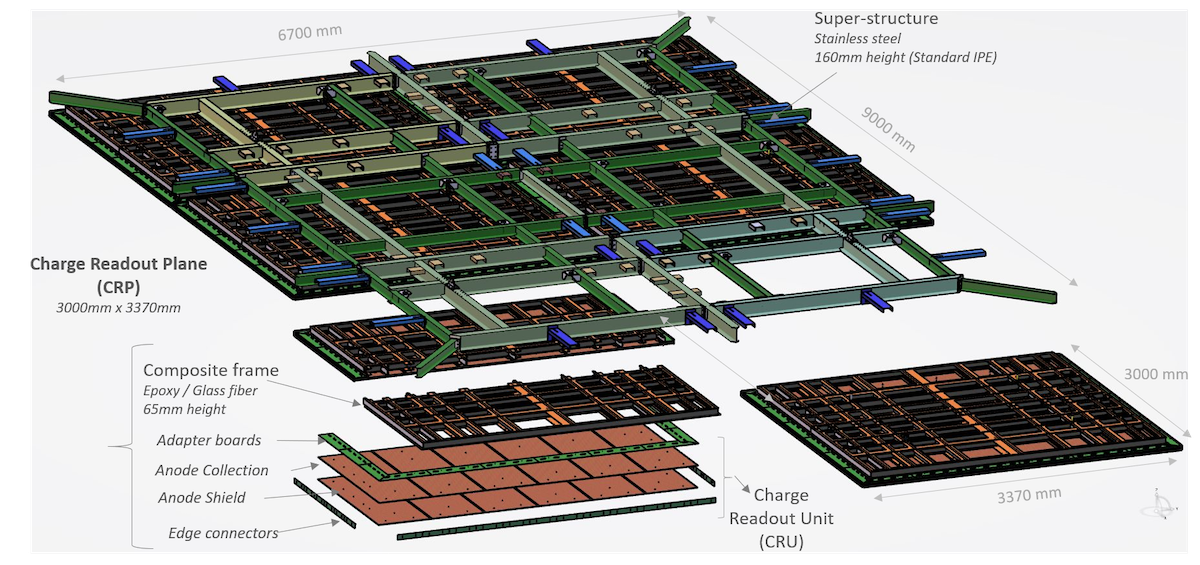}
\end{dunefigure}

%===================

%%%%%%%%%%%%
\subsection{Charge Readout Electronics}
\label{sec:vdd:tpcelec}

The \dshort{spvd} top and bottom drift volumes will implement different \dword{cro} electronics, top drift electronics (\dword{tde}) and bottom drift electronics (\dword{bde}), in order to take maximal advantage of 
placement of the top anode near the cryostat roof and to leverage the local amplification an digitization of the signal on the bottom anode to maximize the signal to noise.
The \dshort{tde}, based on the design used in \dshort{pddp},% and in the \dshort{dp} design, 
has both cold and warm components. The  \dshort{bde} optimizes signal to noise  % is implemented in the bottom drift volume, namely 
with the same \dword{ce} used in the \dshort{sphd} with signal processing and digitization on the \dshort{crp} in the \dshort{lar}.

The %\dword{spvd} 
\dshort{tde} design preserves access to the cryogenic amplifiers via \dwords{sftchimney}, without interfering with the detector operation, and keeps the digitization electronics completely accessible on the cryostat roof. 
Figure~\ref{fig:dp_croch1} illustrates the %top \dshort{cro} electronics (
\dshort{tde}. 

\begin{dunefigure}
[System architecture of the top drift readout electronics]
{fig:dp_croch1}
{System architecture of the top drift readout electronics.} 
\includegraphics[width=.9\textwidth]{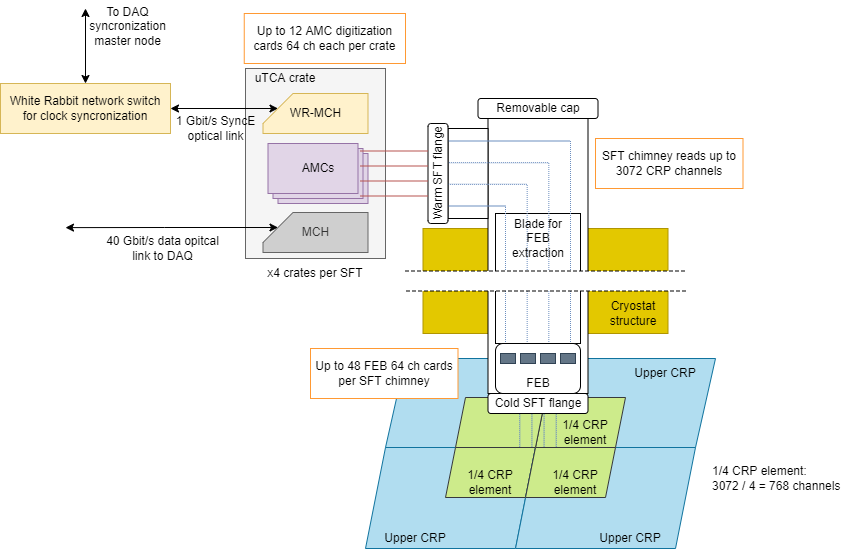}
\end{dunefigure}

The \dshort{ce} design shown in Figure~\ref{fig:system_overview_bottom_driftch1} amplifies and digitizes signals directly on the \dshort{crp} to maximize the signal to noise performance.
\begin{dunefigure}
[System architecture of the bottom drift readout electronics]
{fig:system_overview_bottom_driftch1}
{System architecture of the bottom drift readout electronics.} 
 \includegraphics[width=0.95\textwidth]{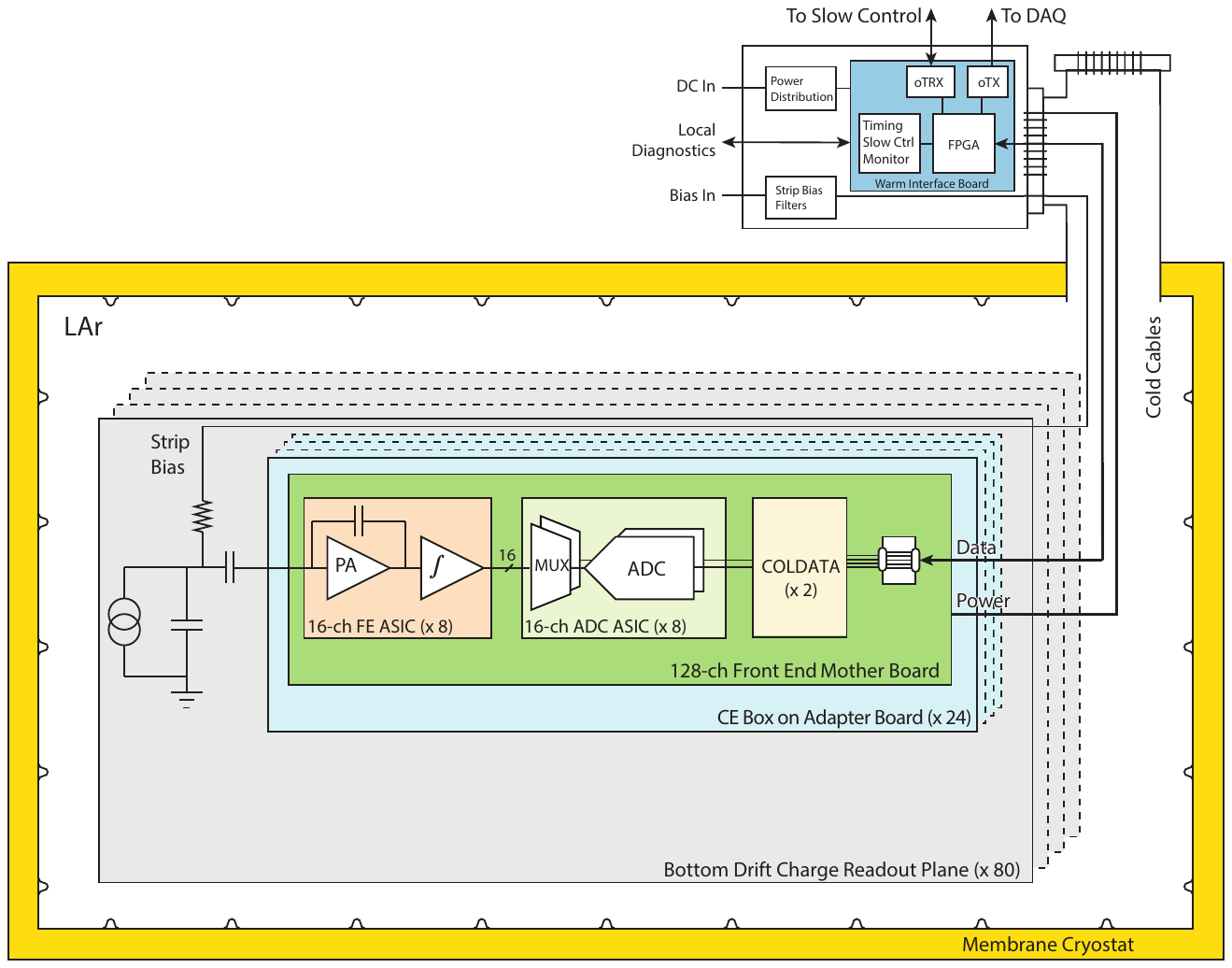}
\end{dunefigure}

%%%%%%%%%%%%
\subsection{High Voltage System and Drift Field}
\label{sec:vdd:hv}

The DUNE \dshort{spvd} detector module design, which includes two drift volumes of equal drift distance 6.5\,m and a nominal uniform \efield of 450\,V/cm, has a horizontal cathode plane placed at detector mid-height held at a negative voltage and horizontal \dwords{anodepln} (biased at near-ground potentials) at the top and bottom of the detector. The main \dword{hvs} components are illustrated in Figure~\ref{fig:vd_hvsch1}.

The \dshort{hvs} is divided into two systems: (1) supply and delivery, and (2) distribution. The supply and delivery system consists of a negative high voltage power supply (\dword{hvps}), \dshort{hv} cables with integrated resistors to form a low-pass filter network, a \dshort{hv} feedthrough (\dword{hvft}), and a 6\,m long extender inside the cryostat to deliver $-$294\,kV to the cathode. The distribution system consists of the cathode plane, the \dword{fc}, and the \dshort{fc} termination supplies. The cathode plane is an array of 80 cathode modules, with the same footprint as the \dwords{crp}, %modules, 
formed by highly resistive top and bottom panels mounted on fiber-reinforced plastic (\dword{frp}) frames. The modular \dshort{fc} consists of horizontal extruded aluminum electrode  profiles stacked vertically at a 6\,cm pitch. A resistive chain for voltage division between the profiles provides the voltage gradient between the cathode and the top-most and bottom-most field-shaping rings.

\begin{dunefigure}
[HV system components inside the cryostat]
{fig:vd_hvsch1}
{A birds-eye view of the \dshort{fc}, with one full-height \dshort{fc} column (highlighted in cyan) that extends the entire height, the \dshort{hv} feedthrough and extender (in the foreground), and the cathode (with one cathode module highlighted in cyan).}
\includegraphics[width=0.8\textwidth]{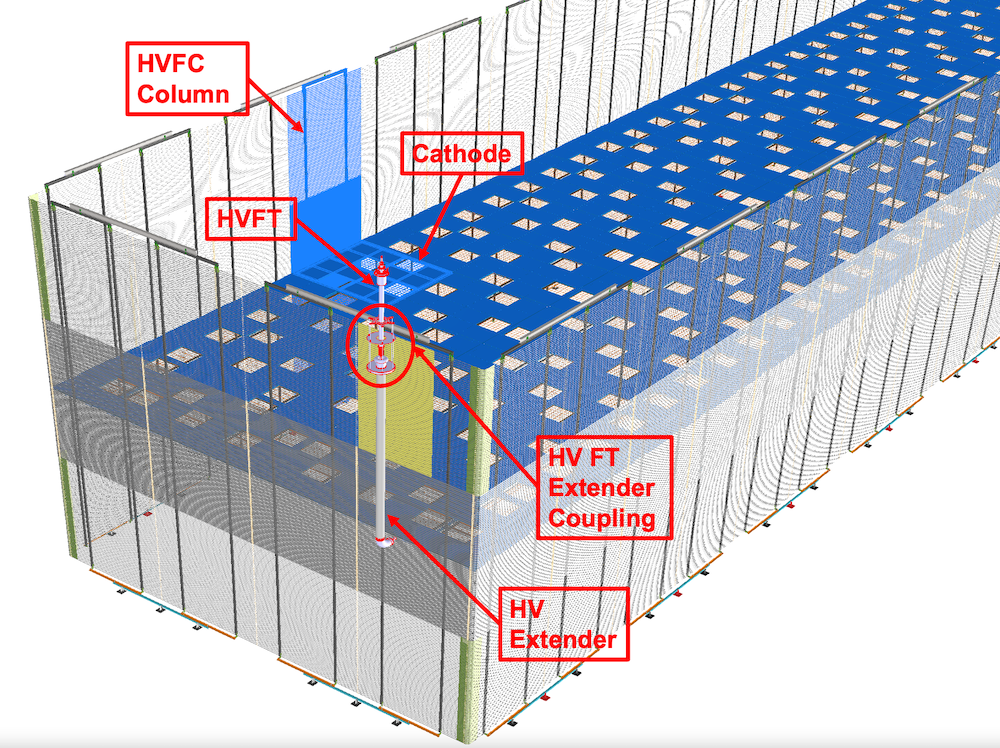}
\end{dunefigure}

In addition to the primary function of providing uniform \efield{}s in the two drift volumes, both the cathode and the \dshort{fc} designs are tailored to accommodate \dword{pds} modules (Section~\ref{sec:vdd:pds}) since it is not possible to place them behind the \dshort{anodepln}, as in the \dshort{sphd} design. Each cathode module is designed to hold four double-sided \dshort{xarapu} \dshort{pds} modules that are exposed to the top and bottom drift volumes through highly transparent wire mesh windows. Along the 
walls, the \dshort{fc} is designed with narrower profiles in the region within 4\,m of the \dshort{anodepln} to provide 70\% optical transparency to single-sided \dshort{pds} mounted on the cryostat membrane walls behind them, and conventional-width profiles within 2\,m of the cathode plane.  

%%%%%%%%%%%%
\subsection{Photon Detection System}
\label{sec:vdd:pds}

%\fixme{Missing ref to Arapuca - skip it?}
The \dshort{spvd} will implement  \dshort{xarapu}~\cite{Machado-XARA:2018, Brizzolari:2021akq} \dword{pds}. Functionally, an \dshort{xarapu} module is a light trap that captures wavelength-shifted photons inside boxes with highly reflective internal surfaces until they are eventually detected by \dwords{sipm}. An \dshort{xarapu} module has a light collecting area of approximately 600~$\times$~600\,{\rm mm$^2$} and a light collection window on either one face (for %single-sided 
wall-mount modules) or on two %(double-sided) 
faces (for cathode-mount modules). 
The wavelength-shifted photons are converted to electrical signals by 160 \dshorts{sipm} distributed evenly around the perimeter of the \dword{pd} module. Groups of \dshorts{sipm} are electrically connected to form just two output signals, each corresponding to the sum of the response of 80 \dshorts{sipm}.

Since their primary components are almost identical to those of \dshort{sphd}, only modest R\&D was required for the \dshort{spvd} \dshort{pds} modules. The primary differences were to optimize the module geometry and the proximity of the \dshorts{sipm} to the \dword{wls} plates.  Both of these are more favorable in \dshort{spvd}, leading to more efficient light collection onto the \dshorts{sipm}. 
As discussed in Section~\ref{sec:vdd:hv}, the design has the \dshorts{pd} mounted on the four cryostat membrane  walls
and on the cathode structure, facing both top and bottom drift volumes. This configuration produces approximately uniform light measurement across the entire \dshort{tpc} active volume. 

Cathode-mount \dshorts{pd} are electrically referenced to the cathode voltage, avoiding any direct path to ground. While membrane-mount \dshorts{pd} adopt the same copper-based sensor biasing and readout techniques as in \dshort{sphd}, cathode-mount \dshort{pds}  required new solutions to meet the challenging constraint imposed by \dword{hvs} operation. The cathode-mount \dshort{pds} are powered using non-conductive \dword{pof} technology~\cite{vasquez:ICTON-2019}, and the output signals are transmitted through non-conductive optical fibers (\dword{sof}), thus providing voltage isolation in both signal reception and transmission. \dshort{pof} is a well established technology, but extensive use in a cryogenic detector is a new application.

Converting electrical signals to optical signals in \dshort{lar} has been recognized as a critical aspect of the \dshort{spvd} \dshort{pds} design, and an aggressive R\&D plan to identify cold transceiver solutions operating at \dshort{lar} temperature was mounted and completed.

%%%%%%%%%%%%
\subsection{Trigger and DAQ}
\label{sec:vdd:tdaq}

The trigger and data acquisition (\dword{tdaq}) system is responsible for triggering on, receiving, processing, and recording data from the DUNE experiment. The main challenge for the DUNE \dword{tdaq} lies in the development of effective, resilient software and firmware that optimize the performance of the underlying hardware. The design is driven not only by data rate and throughput considerations, but also --- and predominantly --- by the stringent uptime requirements of the experiment.

The \dshort{tdaq} for DUNE has been designed and developed coherently by a joint 
consortium. The \dshort{tdaq} systems for the \dword{nd} and the different \dshort{fd} modules differ only in minor details to support the electronics and the data selection criteria for each. 

The \dshort{tdaq} has been subdivided into a set of subsystems. All the subsystems rely on the functionality provided by the \dword{daqccm} and \dword{dqm}, that provide the ``glue'' to the overall \dshort{tdaq}, transforming the set of components into a coherent system.

The \dshort{tdaq} is mainly composed of \dword{cots} components. A high-performance Ethernet network interconnects all the elements and allows them to operate as a single, distributed system. At the output of the \dword{tdaq} the high-bandwidth Wide Area Network (WAN) allows the transfer of data from \dshort{surf} to \dshort{fnal}.

%%%%%%%%%%%%
\subsection{Prototypes and Demonstrators}
\label{sec:vdd:proto}

The \dword{spvd} prototyping and demonstration program has followed a phased approach.
Commissioning of the first full-scale prototypes was achieved by the end of 2021. Further development continued in 2022--2023 with the procurement, validation, and installation of full-scale components into \dword{np02} for a \dword{vdmod0} validation test, which is expected to collect data in 2024. The timing of operation of \dword{vdmod0} is currently uncertain due to challenges in obtaining the needed$\sim$1~kton of \dshort{lar} needed to fill the cryostat.

%The \dshort{spvd} prototyping program is multi-phased, and 
This  multi-phased program has been taking place at both \dword{cern} and other facilities of appropriate scale for each test. The tests and demonstrations performed throughout 2021--2022 included tests of the perforated \dword{pcb} anode design, followed by optimization of the operating conditions and   readout. 
Following stand-alone tests of the \dshort{hv} system components, a full-scale, extended test of the \dshort{hv} distribution system was completed in the \dword{np02} cryostat at \dshort{cern} in 2022, with a new feedthrough and the 6\,m long \dshort{hv} extender, delivering 300\,kV to the cathode plane 6~m under the liquid surface. This test also validated the detection of cosmic muons across the full 6\,m anode-to-cathode drift length.

The first full-size \dshort{crp} module for \dshort{spvd}, equipped for both \dshort{tde} and \dshort{bde} readout, was successfully tested in a shallow cryostat (\coldbox) from fall 2021 through winter 2022.
This setup included %scintillation light 
\dshort{pds} modules installed on the cathode, and a membrane-mounted \dshort{pds} on the wall.  %gap. 
Figure~\ref{fig:bde_crp4_track} shows cosmic ray tracks detected with %this first 
a bottom \dshort{crp} module using the \dshort{bde} readout system in the \coldbox. The plots on the top show \twod images from the readout views, and the bottom plots show corresponding signal waveforms from channels in the induction views (left and center) and in the collection view (right).

\begin{dunefigure}
[Cosmic ray track from the prototype CRP-4 (BDE)]
{fig:bde_crp4_track}
{A sample cosmic ray track recorded using CRP-4 read out with \dshort{bde}. The individual channel waveforms shown are for strips that recorded ionization from multiple tracks.}
\includegraphics[width=0.95\linewidth]{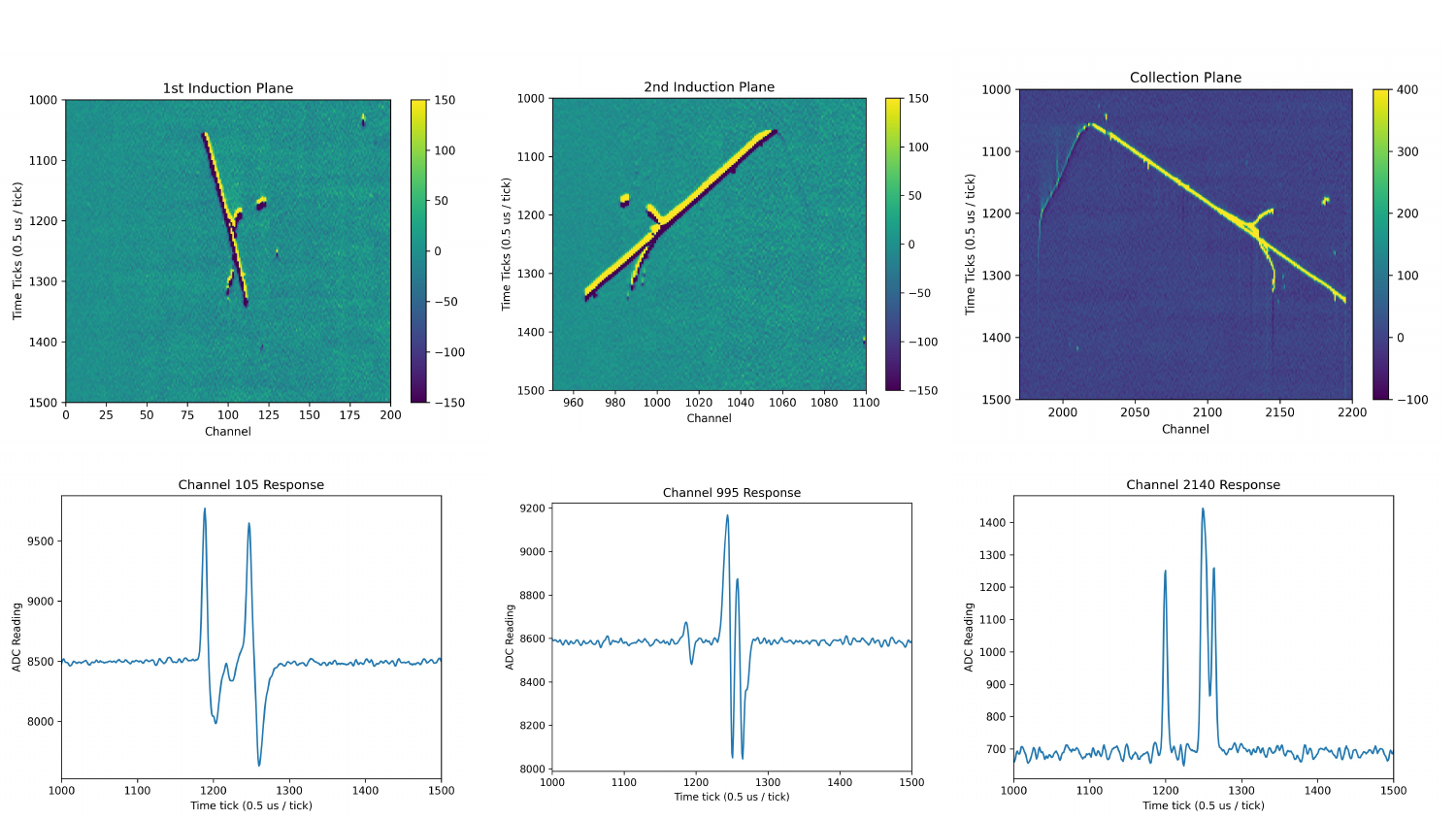}
\end{dunefigure}

%$$$$$$$$$$$$$$$ 
\begin{dunefigure}
[Installation of a \dshort{crp} in a cold box]
{fig:Module0-1-ch0}
{The \coldbox during installation of a \dshort{crp}. 
The cathode is visible on the bottom, with an \dshort{xarapu} \dshort{pd} installed.}
\includegraphics[width=0.9\linewidth]{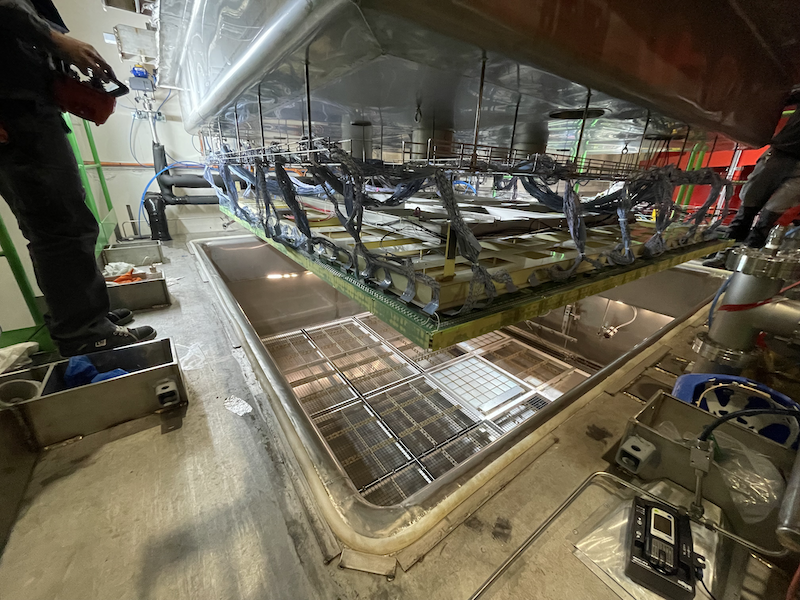}
\end{dunefigure}
%$$$$$$$$$$$$$$$ 

In 2022, the prototyping and validation activities included procurement of the additional four \dshort{crp} modules required for the \dword{vdmod0} program. The development of manufacturing, assembly, and \dword{qa}/\dword{qc} procedures for the \dshorts{crp} %modules 
has been a central part of this activity.

During this same period, validation activities have taken place for the \dshort{pds} system, both in dedicated setups and together with charge readout components in the cryostats at \dshort{cern}.  
Stability tests of the components in a cryogenic environment progressed in 2022 and %will extend into  
%is continuing in 
2023.
The risk associated with accidental discharges of the cathode \dshort{hv} 
has been studied with detailed simulations, and mitigated with design optimization.

%%%%%%%%%%%%
\subsection{Integrated Engineering and Installation}
\label{sec:vdd:iandi}

Detector integration is the responsibility of the \dword{usproj} \dword{fsii} team in cooperation with the \dshort{dune} consortium technical leads. It comprises activities ranging from \threed model integration, to integration with \dword{fscf}, to warehousing and logistics, to management and oversight of the underground installation. These activities require specialized engineering skills and involve oversight functions from groups within the project, e.g., the \dword{ro}, the compliance office, and safety professionals, and close coordination with the \dword{sdsd} and the \dword{sdsta} at \dshort{surf}.

Installation of the \dshort{spvd} components requires neither manipulation of extremely large or delicate objects (e.g., the \dshort{apa}s for \dshort{sphd}) nor significant assembly underground. This eliminates the need for a large cleanroom adjacent to the cryostat and large \coldbox{}es for component testing, which greatly simplifies the required infrastructure. Outside the cryostat a protected area (\dword{greyrm}) equivalent to an ISO-8 cleanroom, illustrated in Figure~\ref{fig:greyroom-ch0}, is needed for preparation and cleaning of components, and to rig them for transportation into the cryostat. The largest objects to be moved into the cryostat, the \dshort{crp} superstructures, are 9\,m by 6\,m.  The \dshort{greyrm} covers the entire \dword{tco} opening to protect the inside of the cryostat.

Much progress has been made in 2022 on the installation plans %procedures 
for %the components of 
\dshort{spvd}, %has made important progress in 2022, with 
including the design of tools and procedures to enable robust \dword{qc}, and %aimed at most 
safe, clean, and efficient installation. 
 Figure~\ref{fig:cleanspace-ch0} illustrates %an example of the 
 space allocation during a phase of detector installation.

\begin{dunefigure}
[\dshort{spvd} installation \dshort{greyrm}]
{fig:greyroom-ch0}
{The \dshort{greyrm} for the installation of the \dshort{spvd} detector module. The small structures to the left of the \dshort{greyrm} are the changing room and fenced areas for storage, and a small machine shop.}
\includegraphics[width=0.7\linewidth]{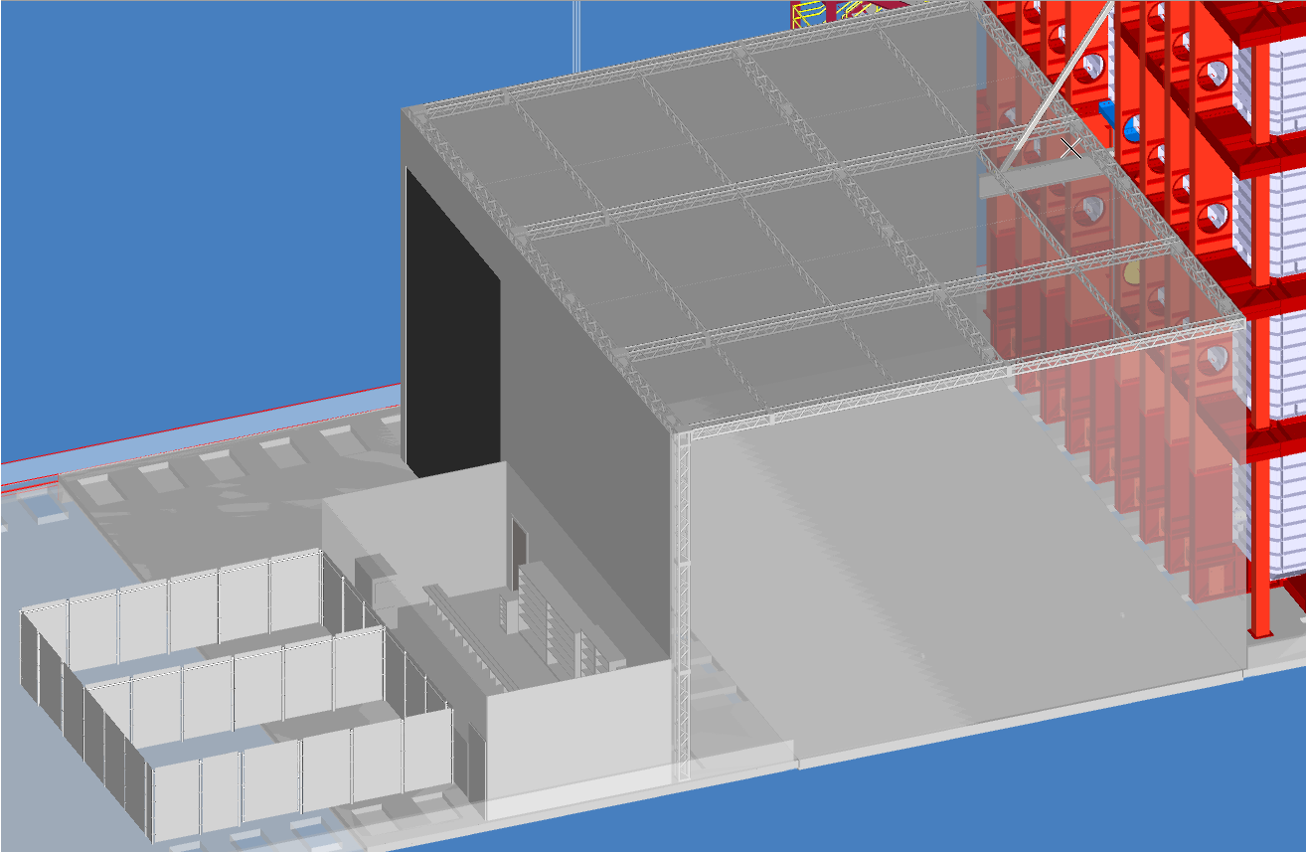}
\end{dunefigure}

\begin{dunefigure}
[Space allocation in the cryostat and \dshort{greyrm} during \dshort{crp} installation]
{fig:cleanspace-ch0}
{Top: Model of the main space allocations inside the cryostat %Image of the cryostat 
during week 8 of CRP installation. %, showing  
Bottom: The \dshort{greyrm}  
work area and space allocations.}
\includegraphics[width=0.7\linewidth]{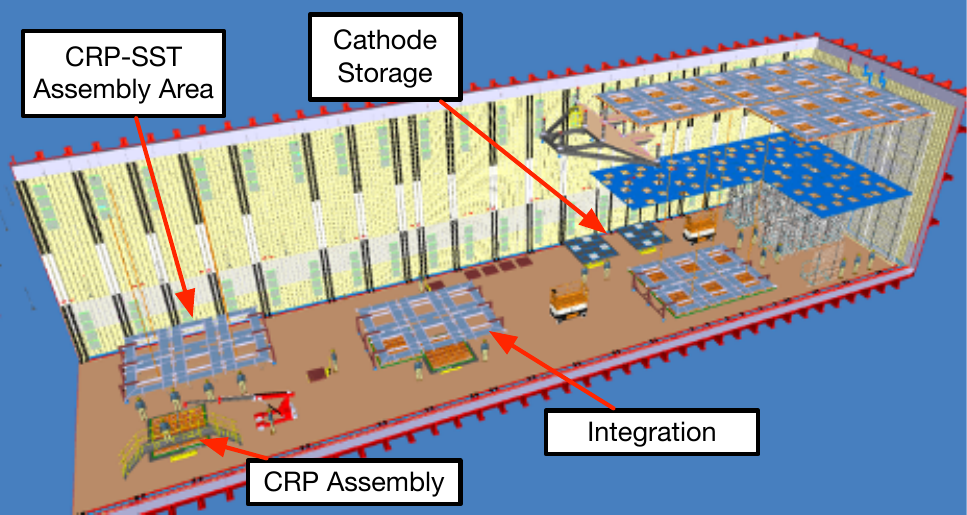}
\par\bigskip
\includegraphics[width=0.7\linewidth]{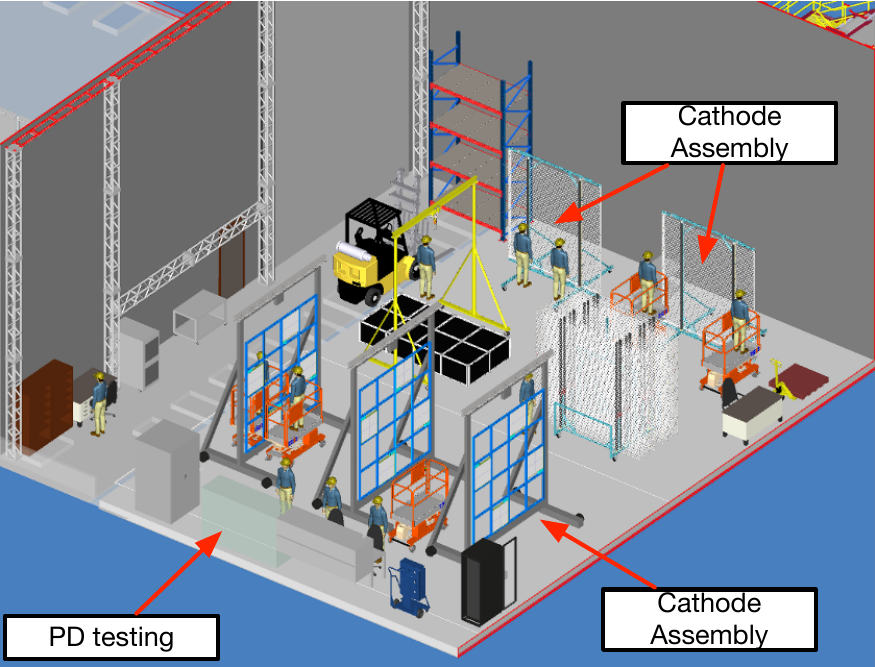}
\end{dunefigure}
\FloatBarrier
%%%%%%%%%%%%
\section{Project Organization}
\label{sec:vdd:proj}

The \dword{lbnf-dune} has been organized into five subprojects. Chapter~\ref{ch:project} outlines the roles and responsibilities of the various subprojects and their leadership, and of the DUNE collaboration.
In particular, the \dword{spvd} effort is part of the Far Detector and Cryogenics Subproject (\dword{fdc}), with a dedicated Deputy Project Director and a Technical Coordinator.
Construction of the \dshort{dune} detector components is carried out by consortia of collaborating institutions, who assume responsibility for detector subsystems. Each consortium plans, constructs, and participates in the installation, commissioning and operation of its subsystem. A total of nine \dword{fd} consortia have been formed to cover the subsystems required for \dshort{sphd} and \dshort{spvd}, two of which focus exclusively on \dshort{spvd} (\dshort{crp} and \dshort{tde}); %top drift electronics); 
most of the others have responsibility for subsystems common to both far detector technologies. 
Centrally organized support groups for cryogenics,  integration work and installation activities %will 
assist the consortia in their work on all phases of the project.

%%%%%%%%%%%%%%%%%%%%%%%
\section{Status of Project}
\label{sec:intro:projstatus}

%%%%%%%%%%%%%%%%%%%%%%%%%%
\subsection{Simulation and Physics Studies}
\label{sec:intro:simphys}

%RJW 23feb23: Tweaked to remove extraneous information.
Simulation of  \dshort{spvd}  was developed from the framework 
used for the \dword{dp} concept using the tools developed for \dword{sp}.
A large sample of simulated data with the 
baseline geometry implemented 
%in the prototypes developed 
for \dshort{vdmod0} %demonstrator, namely 
and CRP-2 through CRP-5 (Section~\ref{subsec:CRP_prototyping2022}), with strips oriented at -30$^{\circ}$, 30$^{\circ}$ and 90$^{\circ}$ 
was produced in summer 2022. 
As expected, %because of 
due to the similarity between the \dshort{spvd} and \dshort{sphd} readout designs, the samples of both have shown similar performance.

Simulation of the \dword{pds} has been used to optimize the location of the \dshort{pds} modules, resulting in the %reference
baseline design presented in this \dword{tdr}. Placing sensors on the cathode in addition to the walls significantly improves the level and uniformity of \dword{ly}, which 
will increase the performance of the detector for all interactions and in particular for low-energy neutrino physics, e.g., \dwords{snb} and the  high-energy tail of solar neutrinos. 
Current software activities include the integration of light simulation into the framework of \dword{larsoft}. 

\subsection{Prototyping} %The \dshort{vdmod0} Detector}
\label{sec:intro:mod0}

As discussed in Section~\ref{sec:vdd:proto}, the prototyping and validation activities that began %started in 
in 2021 and %continued in 
2022 will %be achieved with 
culminate in the \dshort{vdmod0} detector. The procurement and validation procedures for the detector components, which include full functionality and integration tests in the \coldbox setup, is well advanced.
Installation in the \dword{np02} cryostat is underway and will be completed in Q2 2023.

Figure~\ref{fig:Module0-ch1} shows a schematic of \dshort{vdmod0}, which includes 
two top and two bottom \dwords{crp},  and two cathode modules, separating the active volume into two drift regions, each 3.2\,m long. \dshort{pds} are mounted on the cathodes and on the cryostat membrane walls.

\begin{dunefigure}
[Model of the \dshort{mod0} detector]
{fig:Module0-ch1}
{A model of the \dshort{vdmod0} layout in the \dshort{np02} cryostat.}
\includegraphics[width=0.8\linewidth]{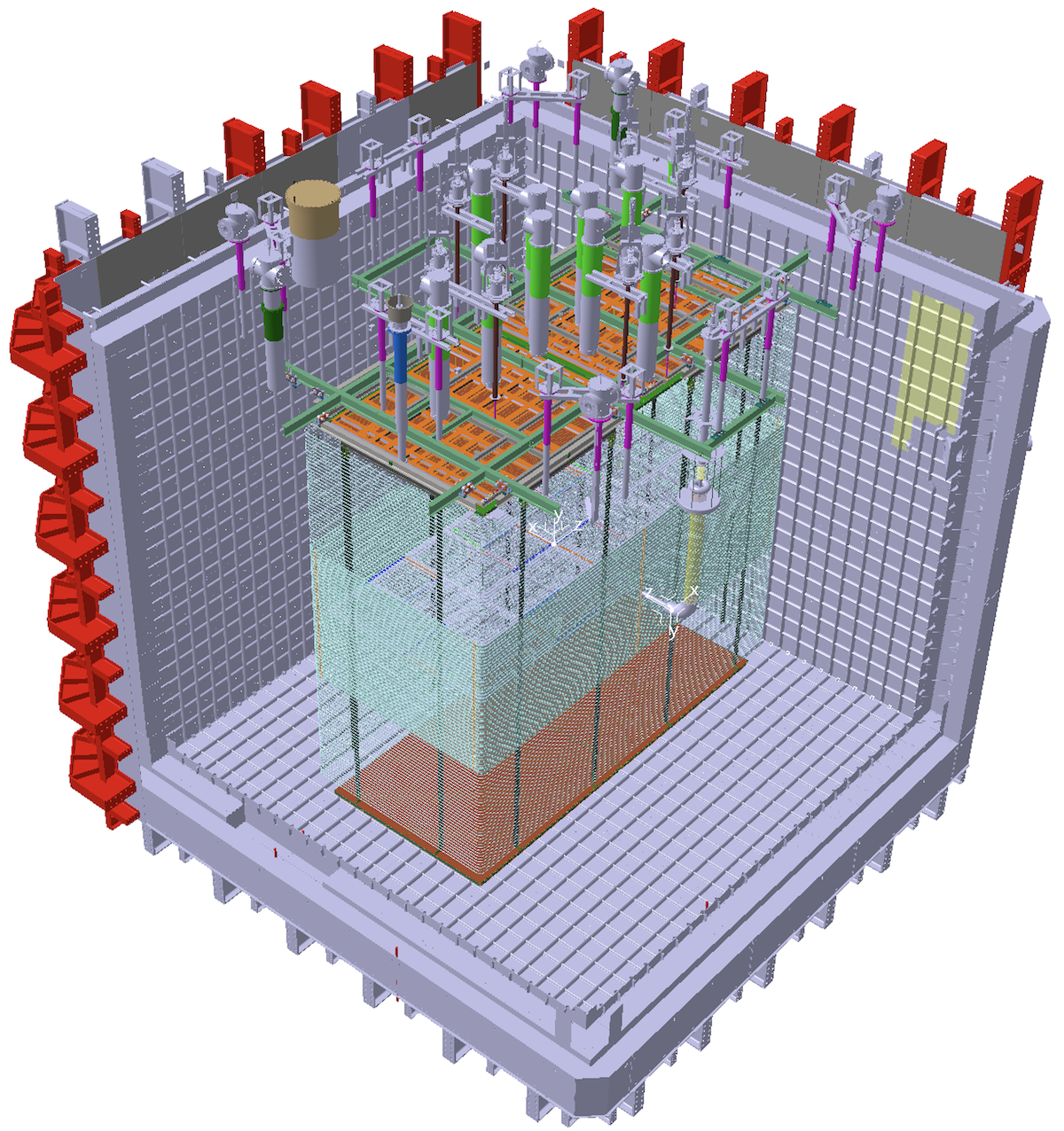}
\end{dunefigure}

\subsection{Path to Completion: Validation, Production and Installation}

%The schedule of the \dword{spvd} project 
The \dshort{spvd} schedule has been added to the LBNF/DUNE project schedule. R\&D and prototyping, including validation and optimization, have been completed; further validation will continue in 2023--2024, when the \dshort{vdmod0} detector %is tested as \dword{pd2vd} 
is installed and operated, respectively, at the \dshort{cern} Neutrino Platform, with operations dates still somewhat uncertain. The plan includes final design reviews in early 2023, based on this \dword{tdr}, 
and assessment of the detector performance for scintillation light and charge collection and readout.

The production of detector components is scheduled to start in 2024, after the \dwords{prr}, and complete in 2027. Transport to the far detector site,  \dshort{surf}, will take place in 2026--2027. 
Installation at the far site is scheduled to start with the warm cryostat structure.%s, in late 2026 for about a year.  
The detector assembly and installation will take place in 2027--2028.  Purge and fill of the cryostat is scheduled in 2029--2030.

%%%%

\chapter{Physics}
\label{ch:Phys}

%\tableofcontents

%%%%%%%%%%%%%%%%%%%%%%%%%%
\section{Introduction}

The \dword{dune} physics program is described in detail in the physics volume of the DUNE Technical Design Report (\dword{tdr})~\cite{DUNE:2020ypp} and in a series of journal articles~\cite{Abi:2020qib,Abi:2020kei,Abi:2020lpk}. The physics sensitivities presented in the TDR assume a modular \dword{lartpc} \dword{fd} with an initial fiducial mass of 20\,kt that is expanded to 40\,kt, a broadband neutrino beam beginning at 1.2\,MW beam power that gets upgraded to 2.4\,MW beam power, and a capable \dword{nd} to reduce systematic uncertainties. The TDR results are derived from a simulation and reconstruction of the 
\dword{sp} \dword{sphd} module design. It is implicitly assumed that all four \dshort{fd} modules will have at least the same performance as the \dshort{sphd}. In this chapter, the expected physics performance of DUNE including the \dword{sp} \dword{spvd} module design is presented, based on detailed simulations of the vertical drift detector.

Section~\ref{sec:ph:overview} reviews the DUNE physics program. For low-energy signals, such as solar and %supernova (SN) 
\dword{snb} neutrinos, the %Vertical Drift (VD) 
\dshort{spvd} design provides opportunities to improve the physics performance, primarily due to improved photon detection. The full simulation of the %Photon Detection System (PDS) for FD2-VD 
\dword{pds} and charge readout system for \dshort{spvd} 
%and the basic system requirements are 
are described in Section~\ref{sec:ph:pds}.
Studies of low-energy physics performance, including photon detection information, are presented in Section~\ref{sec:ph:le}. The reconstruction of high-energy signals in the vertical and horizontal drift designs is expected to be very similar. Section~\ref{sec:ph:lbl} describes the performance of the charge readout system for long-baseline neutrino oscillation physics.

\section{The DUNE Physics Program}
\label{sec:ph:overview}

DUNE has three primary physics goals:

\begin{itemize}
    \item measure the parameters governing %long-baseline 
    \dword{lbl} neutrino oscillations, including the neutrino mass ordering, the %CP 
    \dword{cp} violating phase $\delta_{CP}$, and the mixing parameters $\theta_{23}$ and $\theta_{13}$, and test the three-flavor oscillation paradigm;
    \item make astrophysics measurements with neutrinos from a %supernova burst
    \dshort{snb}, as well as other measurements with MeV-scale neutrinos;
    \item search for physics beyond the Standard Model (\dword{bsm}).
\end{itemize}

The motivation for this physics program is discussed extensively elsewhere~\cite{DUNE:2020ypp}. 

\subsection{Long-baseline Neutrino Oscillation Physics}
\label{sec:ph:overview:lbl}

DUNE measures neutrino oscillations as a function of neutrino energy over more than a full oscillatory period. The \dshort{lartpc} \dshort{fd} provides exquisite imaging capability, 
allowing efficient separation of $\nu_{\mu}$ and $\nu_{e}$ \dword{cc} reactions 
which are characterized by either a long muon track or an electromagnetic shower initiated by an electron. The \dshort{lartpc} is also sensitive to hadrons produced in neutrino interactions, enabling precise neutrino energy reconstruction over a broad range of energies.

Examples of the ultimate physics sensitivity of DUNE are given in Figures~\ref{fig:ph:lbl:2dresolutions} and \ref{fig:ph:lbl:mhcpv}. 
These sensitivity projections are based on a full simulation of the %Horizontal Drift far detector module. 
\dshort{sphd}. The exposures are quoted in kiloton-megawatt-years, the product of the \dshort{fd} fiducial mass in kilotons, the proton beam intensity in megawatts, and the number of years, including an accelerator up-time factor based on demonstrated \dword{fnal} performance. 
The principal characteristics of the \dshort{fd} that impact the physics reach of the experiment are:

\begin{itemize}
    \item fiducial mass; 
    \item reconstruction efficiency for $\nu_{\mu}$ and $\nu_{e}$ charged-currents, and rejection of backgrounds primarily %due to neutral currents;
    from neutrino \dword{nc} interactions with a $\pi^0$ in the final state;
    \item neutrino energy reconstruction in charged-current events; and
    \item %detector 
    systematic uncertainties related to, e.g., energy scales and particle responses.
\end{itemize}

\begin{dunefigure}
[Long-baseline (\dshort{lbl}) \twod resolutions]
{fig:ph:lbl:2dresolutions}
{DUNE's ultimate measurement capability for $\sin^{2}\theta_{23}$ and $\Delta m^{2}_{32}$ (left), and $\sin^{2}2\theta_{13}$ and $\delta_{CP}$ (right). DUNE can significantly improve on current measurements of the atmospheric mixing parameters, and measure the \dword{cp} violating phase, while achieving similar precision in $\theta_{13}$ to the current reactor measurements. The shaded area shows the current allowed region from global fits to data by NuFit~\cite{nufitweb}.}
  \includegraphics[width=0.45\textwidth,angle=0]{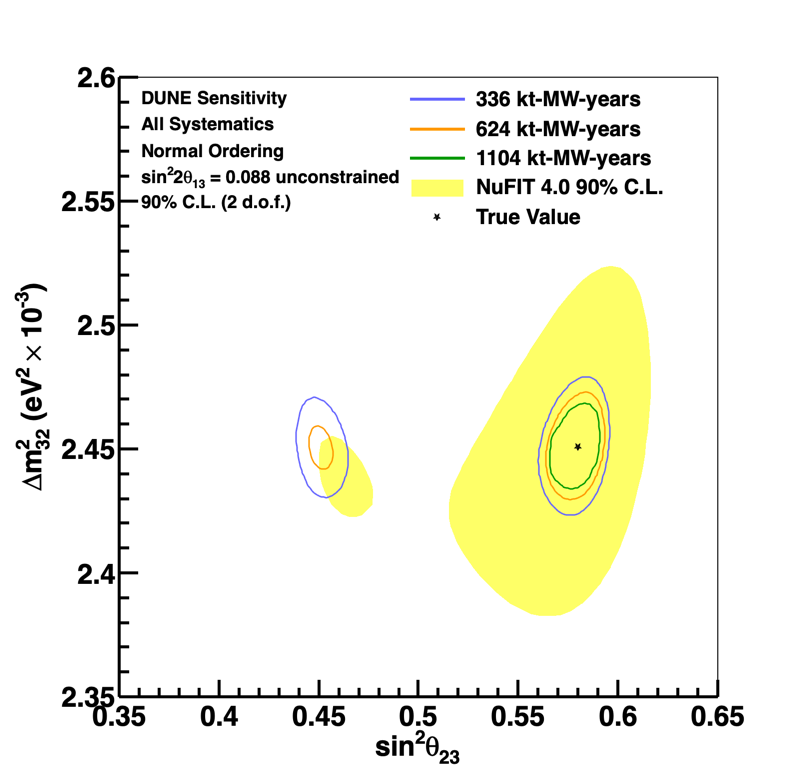}
  \includegraphics[width=0.45\textwidth,angle=0]{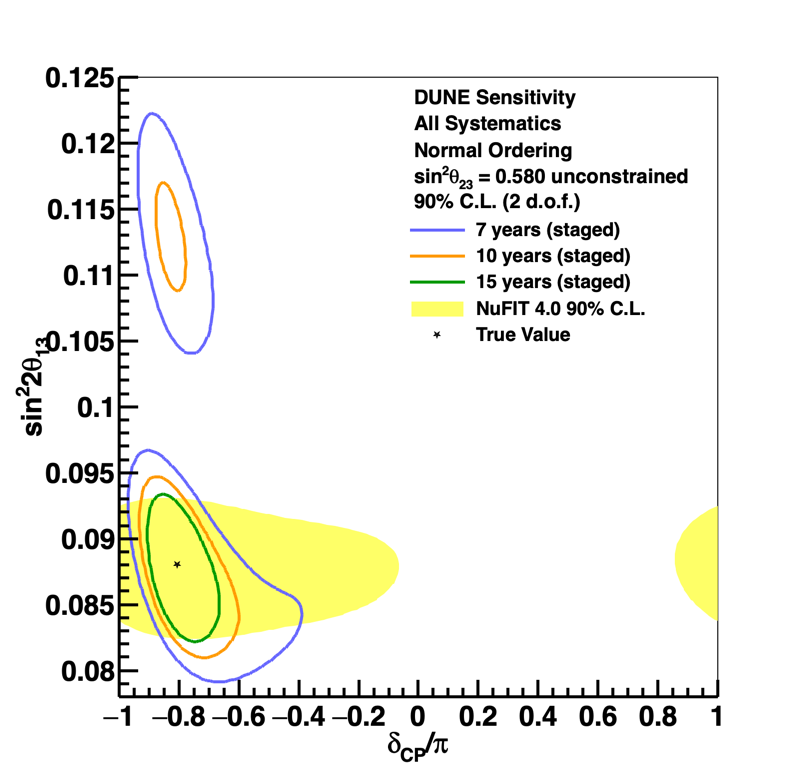}
\end{dunefigure}

\begin{dunefigure}
[LBL MH and CPV]
{fig:ph:lbl:mhcpv}
{The significance with which DUNE can resolve the neutrino mass ordering (left), and %CP violation
\dword{cpv} (right), assuming the true mass ordering is normal. The curves represent different assumptions about the value of $\delta_{CP}$, and the width of the bands correspond to the effect of using an external constraint on $\theta_{13}$ from reactor experiments.}
  \includegraphics[width=0.45\textwidth,angle=0]{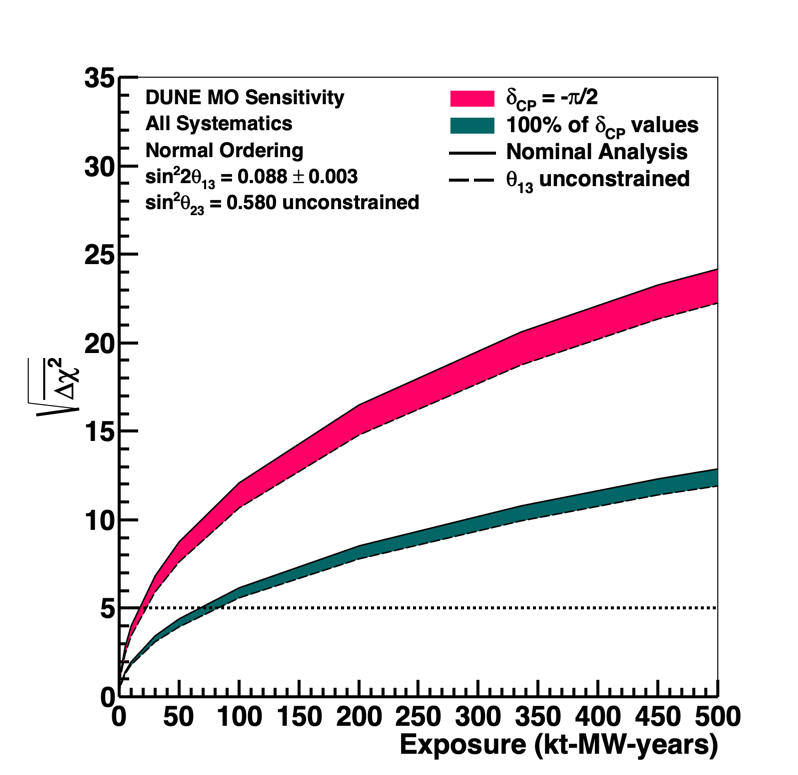}
  \includegraphics[width=0.45\textwidth,angle=0]{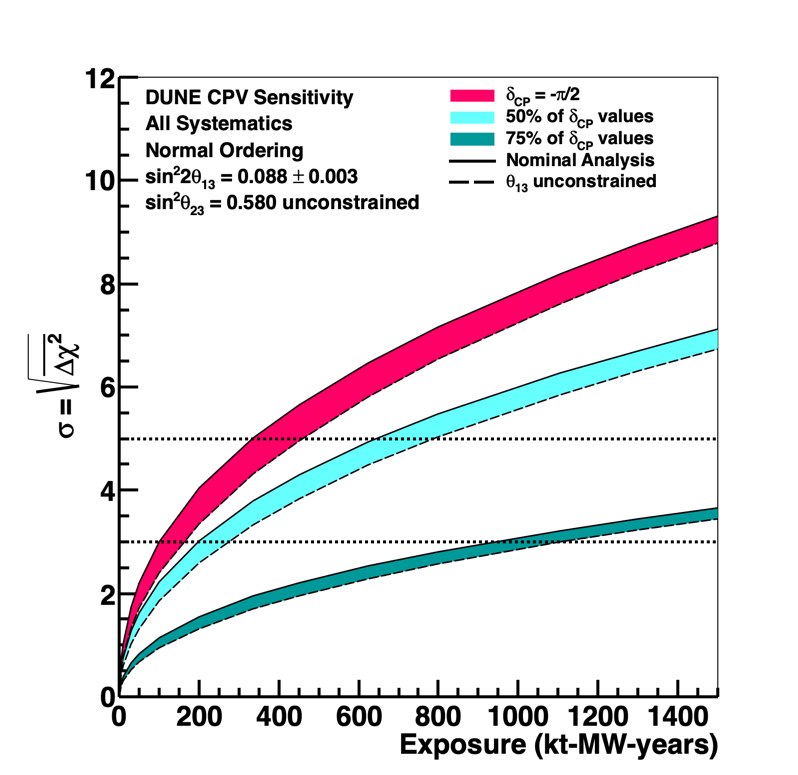}
\end{dunefigure}

Compared to the \dshort{sphd}, the \dshort{spvd} is expected to have the same or possibly greater fiducial mass for the same physical module size, 
therefore the expected \dshort{spvd} event sample is expected to be at least as large as that of the \dshort{sphd}.
The reconstruction efficiency based on the current \dshort{spvd} simulation is discussed in Section~\ref{sec:cvn}. The efficiencies are very similar to what was achieved with \dshort{sphd}, and they are expected to further converge as the \dshort{spvd} simulation and reconstruction continues to mature. The energy reconstruction for \dshort{spvd} is discussed in Section~\ref{sec:enureco}, and full simulations indicate that they are nearly identical for both technologies. 
Both detector modules are expected to have similar targeted
%bespoke \fixme{is this the word you want?} 
calibration programs, which will constrain energy scales and particle responses %about 
equally well. In summary, 
no significant differences between the performance of the \dshort{sphd} and \dshort{spvd} modules are  expected for long-baseline neutrino oscillations physics.

\subsection{Supernova Neutrino Bursts and Physics with Low-energy Neutrinos}
\label{sec:ph:overview:le}

The DUNE \dshort{spvd} is expected to have good sensitivity to neutrinos with energies above several MeV thanks to a \dword{pds} designed to achieve an efficient light collection and a large, unobstructed \dword{lar} volume that enables efficient rejection of background originating in the cryostat and electronics assemblies.  Charged-current interactions of \nue at MeV energies create short electron tracks in \dshort{lar}, potentially accompanied by gamma ray and other secondary particle signatures. Of proposed detector technologies at the multi-kt scale, the \dshort{lartpc} technology has optimal sensitivity to detecting \nue flavor low-energy neutrinos. This few-MeV regime is of particular interest for studying a galactic core-collapse \dword{sn} by measuring the associated burst of neutrinos, referred to as the \dword{snb}, released during the collapse which has been the primary focus of DUNE low-energy sensitivity studies. DUNE will also have sensitivity to neutrinos from other astrophysical sources, including solar neutrinos.

Each \dshort{sn} releases an intense source of neutrinos of all flavors. During a \dshort{sn} 99\% of the gravitational binding energy of the star ($\sim10^{53}$~ergs) is released as neutrinos and antineutrinos of all flavors, which play the role of astrophysical messengers, escaping from the \dshort{sn} core. In the event of a galactic \dshort{sn}, %explosion, 
DUNE data will probe the inner evolution of the core-collapse mechanism by studying the time and energy spectra of neutrinos arriving at DUNE.  \dshort{sn} neutrinos are emitted in a burst of a few tens of seconds duration~\cite{Huedepohl:2009wh}. There are three qualitative stages of in a supernova collapse, as shown in Figure~\ref{fig:ph:lep:SN-time}, which can be distinguished and studied with DUNE data.  These are:
\begin{enumerate}
    \item The neutronization burst -- a large pulse of \nue emission takes place in the first tens of milliseconds as electrons are captured on protons in the stellar core during the formation of a proto-neutron star.  A neutrino sphere forms around the proto-neutron star, inside of which the density is so large that neutrinos become trapped; at this point the neutronization burst of \nue emission quenches.
    \item The accretion phase -- during accretion, lasting from tens to hundreds of ms, neutrino emission is dominated by %infall 
    interactions of gas falling from the outer layers of the progenitor onto the outer extant of the proto-neutron star.
    \item The cooling phase -- after infall, the proto-neutron star cools over several seconds.  The neutrino opacity drops, allowing neutrinos to escape the core.  While trapped within the core, neutrino species thermalize %so that there is nearly luminosity equipartition between
    resulting in near equipartition of  neutrino species. % during the cooling phase.
\end{enumerate}

\begin{dunefigure}
[Supernova (SN) time-dependent flux parameters]
{fig:ph:lep:SN-time}
{Expected time-dependent flux parameters for a specific model for an electron-capture supernova~\cite{Huedepohl:2009wh}. No flavor transitions are assumed. The top plot shows the luminosity as a function of time, and the bottom plot shows average neutrino energy. $\nu_x$ stands for $\nu_{\tau}$ and $\nu_{\mu}$ including $\bar{\nu_x}$.}
  \includegraphics[width=0.7\textwidth,angle=0]{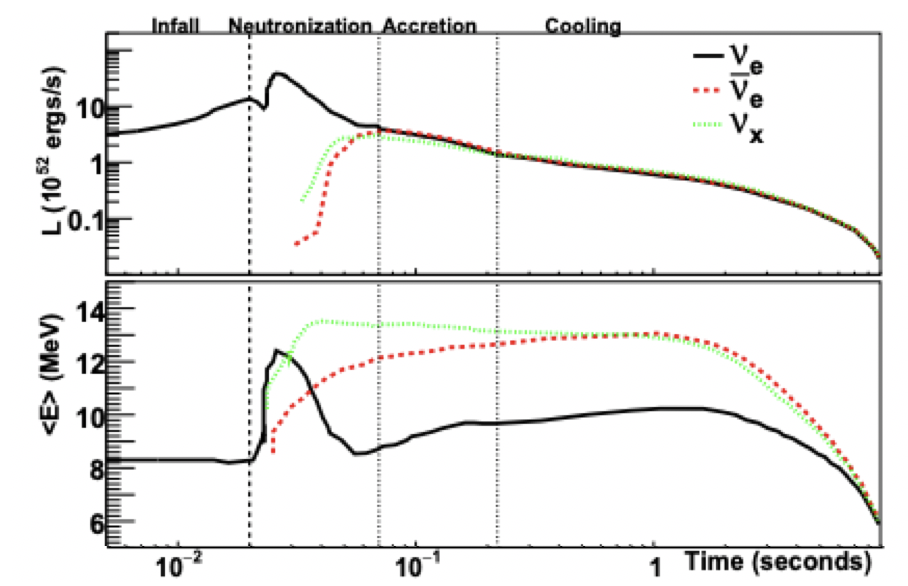}
  \label{fig:ph:le:tSNBTimeEvolutio}
\end{dunefigure}

%Among common detection media, 
\dshort{lar} uniquely offers sensitivity to the \nue component of a \dshort{snb} via the dominant \dword{cc} interaction: absorption of \nue on $^{40}$Ar, $\nu_e + ^{40}$Ar $\rightarrow e^- + ^{40}$K$^*$.  In \dshort{cc} interactions in argon, the final state $e^-$ is observed with any additional de-excitation products as the final-state $^{40}$K$^*$ decays to its ground state. Thus, DUNE 
will provide a distinct signature of the \dshort{sn} collapse in contrast to water and scintillator-based detectors, which are dominantly sensitive to \anue through inverse beta decay. Additional channels in argon include $\nu-e$ \dword{es}, $\nu-^{40}$Ar \dword{nc}  interactions, and $\bar{\nu}_e-^{40}$Ar \dshort{cc} interactions. Cross sections for the most relevant interactions are shown in Figure~\ref{fig:Xsection}.

\begin{dunefigure}
[Low-energy cross sections]
{fig:Xsection}
{Cross sections for \dshort{sn}-relevant interactions in argon as a function of neutrino  energy, from~\cite{Abi:2020lpk}.}
  \includegraphics[width=0.7\textwidth,angle=0]{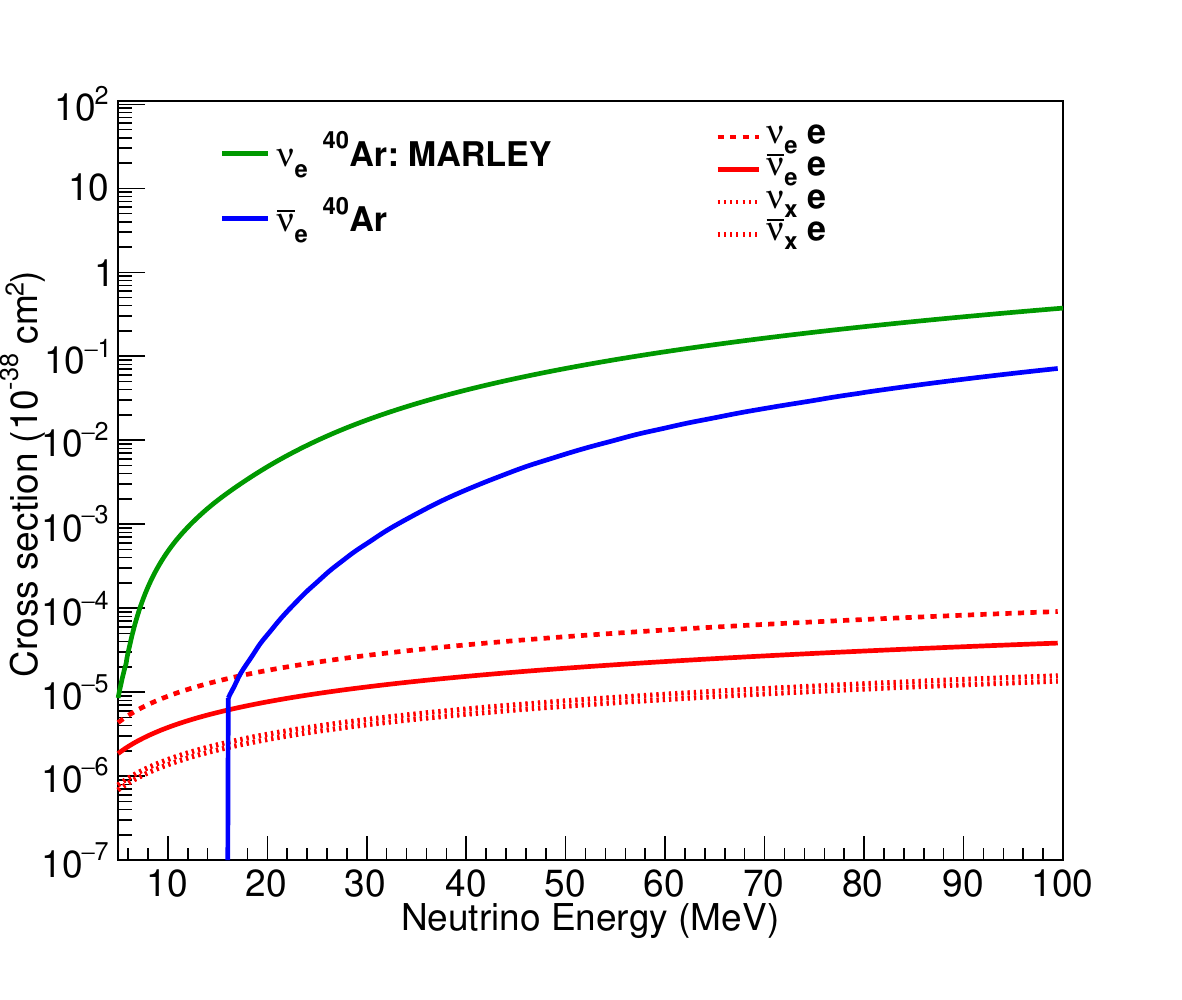}
\end{dunefigure}

The predicted event rate from a \dshort{snb} is calculated by folding together expected neutrino differential energy spectra and cross sections for the relevant channels using \dword{snowglobes}~\cite{snowglobes}. \dword{mc} simulated events are generated using the time and energy of incident neutrinos for a particular \dshort{snb} model using the \dword{marley}~\cite{marley,Gardiner:2021qfr} interaction model to simulate the dominant $\nu_e$ \dshort{cc} neutrino interaction and using \dword{geant4}-based detector models to simulate the DUNE \dshort{fd} detector simulation. Studies of simulated \dword{es} and \dword{nc} interactions are forthcoming. Standard \dshort{lartpc} algorithms are applied to reconstruct electron tracks. All visible energy from the event is used to calculate the incident neutrino energy. %calorimetrically. 
\dshorts{pd} are used to determine the time, and thus drift distance, of events. DUNE is also exploring methods for incorporating novel photon detector reconstruction and calorimetry to expand its low-energy physics reach.

Details of the simulation and reconstruction are provided in~\cite{DUNE:2020ypp,Abi:2020lpk}. Predicted event rates vary up to an order of magnitude among different \dshort{snb} models, and rates will scale as the inverse square of \dshort{sn} distance. As a benchmark, DUNE would observe $\sim$3300 \nue \dshort{cc}, 210 \dword{es} and 160 \anue \dshort{cc} events for a core collapse \dshort{sn} at 10\,kpc in the \dword{gkvm} model~\cite{Gava:2009pj}, assuming 40\,kt fiducial mass of argon for DUNE. The expected electron neutrino energy spectrum for three \dshort{sn} models is shown in Figure~\ref{fig:nue-CC-energy-spectrum} for events scattering in DUNE.

\begin{dunefigure}
[Interacted $\nu_e$-CC energy spectrum]
{fig:nue-CC-energy-spectrum}
{Probability density as a function of neutrino energy for $\nu_e$ \dshort{cc} events interacting in DUNE for three \dshort{snb} models with arbitrary normalization.}
  \includegraphics[width=0.5\textwidth,angle=0]{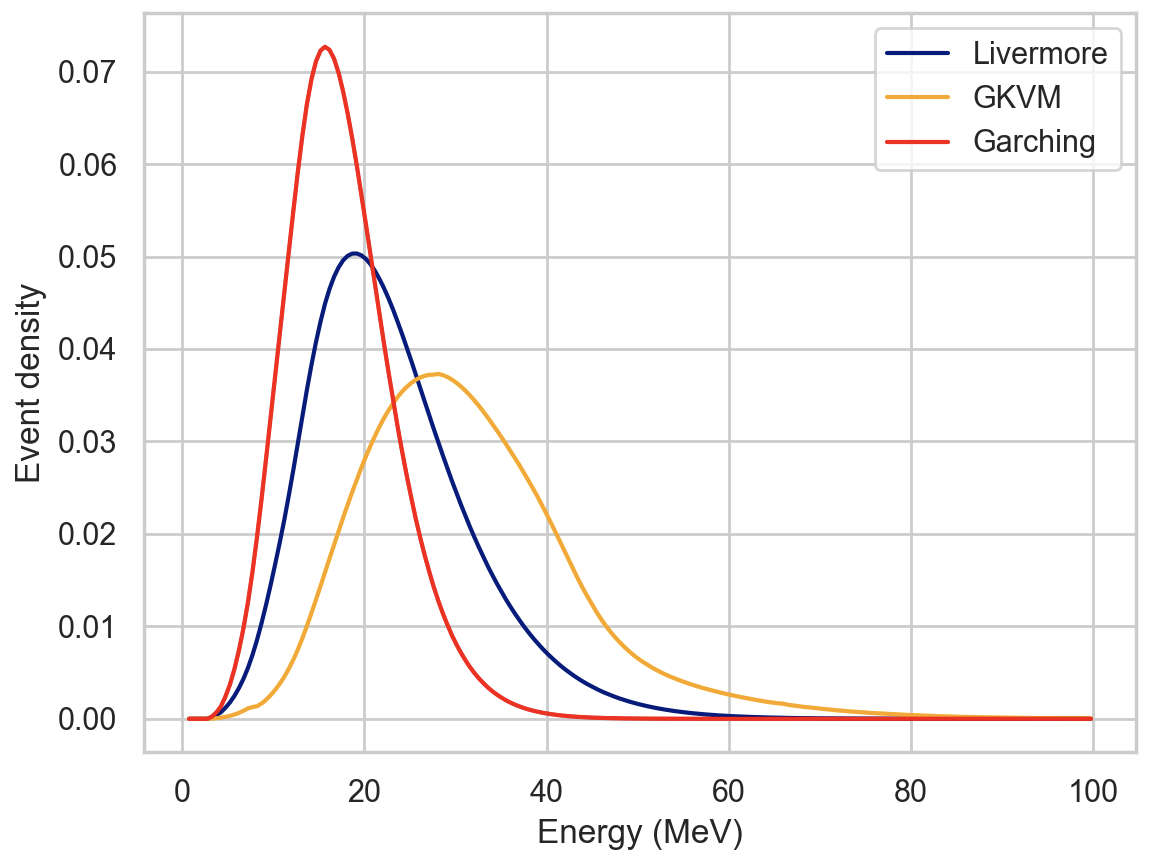}
\end{dunefigure}

Detection of \nue \dshort{cc} interactions from solar and other low-energy neutrinos is challenging in a \dshort{lartpc} because of relatively high intrinsic detection energy threshold, about 5\,MeV, and because radioactive backgrounds in the same energy regime can affect triggering capability. However, compared with other technologies, a \dshort{lartpc} offers a large \dshort{cc} event rate and unique potential to distinguish between \dshort{cc}, \dword{es}, and \dword{nc} interactions on an event basis due to sub-cm reconstruction performance. Furthermore, observed energy from the final state \nue \dshort{cc} interaction is much more tightly correlated with the incident neutrino energy on an event-by-event basis than the electron recoil spectrum from the \dword{es} channel that has been used for past solar neutrino observations such as in Super-Kamiokande~\cite{Super-Kamiokande:2016yck}. Due to this, DUNE can make more precise spectral measurements. Though background rates are large, the \dshort{lartpc} \dshort{spvd} detector allows for background reduction using fiducialization techniques. The solar neutrino event rate is also substantial in the DUNE \dshort{fd}, $\sim$100 per day, allowing samples of a few $10^5$ events after ten years of data collection. Detailed studies of solar neutrino detection capability are underway in DUNE along with subsequent physics sensitivity studies showing promise to improve on current measurements of oscillation parameters after applying fiducial and thershold restrictions. Similarly, DUNE can search for the  \dword{dsnb}~\cite{Beacom:2010kk} at energies just above the endpoint of the solar neutrino spectrum. As DUNE is primarily sensitive to the \nue component, it is the only experiment with sensitivity to the neutrino component of the \dword{dsnb}.

In summary, DUNE anticipates a broad low-energy physics program with sensitivity to the \dword{dsnb}, solar neutrinos, and \dshort{snb} neutrinos, including sensitivity to \dword{bsm} effects in these astrophysical neutrino samples, as well as participation in multi-messenger astronomy via early alert and pointing capabilities. DUNE has the capability to uncover a broad range of \dshort{sn} and neutrino physics phenomena, including sensitivity to neutrino mass ordering, collective effects, and potentially many other topics. 

In comparison with the \dshort{sphd}, the \dshort{spvd} presents two main advantages to perform low-energy physics searches. 
First, there is a larger free \dshort{lar} volume with no nearby components, so the 
geometric fiducial criteria, used to reject high-rate radiological backgrounds originating from detector components, encompasses a larger fraction of the total active \dshort{lar} mass than in the \dshort{sphd} design.   Secondly, an effort has been made to design the \dshort{pds} to achieve an efficient light collection, as presented in Section~\ref{sec:ph:pds}, which translates to %an 
improved reconstructed energy resolution and timing capabilities. Section~\ref{sec:ph:le} discusses the low-energy physics studies carried out to evaluate the physics reach of the \dshort{spvd} design. The \dshort{snb} triggering efficiency of the \dshort{spvd} is presented in Section~\ref{sec:ph:le:trigger} and the reconstructed energy resolution in Section~\ref{sec:ph:le:energy}. %It has to be noted that 
Importantly updated sensitivity for the \dshort{sphd} TDR studies are forthcoming to capture improvements to the background model allowing direct comparisons to studies shown in this document.

\subsection{Physics Beyond the Standard Model}
\label{sec:ph:overview:bsm}

Thanks to its deep underground location and large fiducial mass, as well as its excellent event imaging, particle identification, and calorimetric capabilities, the DUNE FD will be a powerful instrument for the search of phenomena beyond the Standard Model (BSM), including dedicated FD-only probes of the following topics \cite{Abi:2020kei}:
\begin{itemize}
\item \Dword{bdm}: The DUNE FD will be sensitive to hypothetical \dword{bdm} particles originating from various astrophysical sources in the universe, such as the galactic halo, the Sun, and dwarf spheroidal galaxies.
\item Baryon number violation (BNV): DUNE will search for nucleon decay and neutron-antineutron oscillations, baryon-number-violating processes whose existence is predicted by many physics theories beyond the SM. The LArTPC technology is particularly well-suited for the observation of proton and neutron decays into charged kaons.
\end{itemize}

Potential enhancements of the \dshort{spvd} design over the \dshort{sphd} may include a larger fiducial volume, improved energy resolution over the CP-optimized beam and atmospheric neutrino ranges, and improved timing resolution. These enhancements would further extend DUNE's sensitivity to the FD-only BSM probes mentioned above, as well as to BSM probes using a combination of the DUNE FD with the ND complex \cite{Abi:2020kei}, specifically:

\begin{itemize}
\item Search for active-sterile neutrino mixing and related phenomenology: DUNE is sensitive over a broad range of potential sterile neutrino mass splittings by looking for disappearance of \dshort{cc} and \dword{nc}  interactions over the long distance separating the \dword{nd} and \dshort{fd}, and can help distinguish between recently proposed more complex scenarios aimed at resolving the tension between long-baseline null results and short-baseline anomalies. These measurements will also probe other related phenomenology, such as neutrino propagation through large extra-dimensions. 
\item Searches for non-unitarity of the \dword{pmns} matrix: Through precision measurements of neutrino mixing parameters, DUNE will help assess the unitarity of the PMNS matrix, with any deviations requiring new physics explanations. Of particular relevance are FD measurements of atmospheric tau neutrino interactions, which would probe the least constrained sector of the PMNS matrix. 
\item Searches for \dwords{nsi}: 
DUNE is uniquely sensitive to \Dwords{nsi} affecting neutrino propagation through the Earth by leveraging its very long baseline and wide-band beam. If the DUNE data are consistent with standard oscillations for three massive neutrinos, DUNE will improve current constraints on $\epsilon_{\tau e}$ and $\epsilon_{\mu e}$, the magnitude of the \dword{nsi} relative to standard weak interactions, by a factor of 2 to 5, which may be further extended with improved FD energy resolution.
\item Searches for violation of Lorentz or \dword{cpt} Symmetry: \dword{cpt} symmetry, the combination of charge conjugation, parity and time reversal, is a cornerstone of any local, relativistic quantum field theory, and its potential violation would have wide-range repercussions in our understanding of particle physics. Through neutrino and antineutrino measurements with the DUNE ND and FD, DUNE can improve the present limits on Lorentz and \dword{cpt} violation by several orders of magnitude, contributing an essential test of these fundamental assumptions of quantum field theory.
\end{itemize}

%%%%%%%%%%%%%%%%%%%%%%%%%%
%\section{Photon Detector System in Vertical Drift}
\section{Simulations of FD2-VD}
\label{sec:ph:pds}

\subsection{Photon Detector Simulation}
\label{sec:ph:PDS-Simulation}

The \dshort{spvd} \dshort{pds}
uses large \dwords{xarapu} \cite{Segreto:2020jpd} distributed over five different planes (cathode and the four membrane walls) to collect scintillation light. An effort to optimize the \dshorts{pd}' locations taking into account the constraints present (e.g., mechanics, interface with other systems, installation) has been
undertaken to provide the most uniform detection \dword{ly}, i.e., how many photoelectrons will be detected by the \dshort{pds} with a given energy deposited in a given position within the volume \footnote{This is not the same definition as the scintillator light yield, which depends on the physical characteristics of the medium alone and is given in units of photons per energy deposited.}, possible across the detector volume.
Dependence on the drift direction is reduced due to the sensors placed over the walls along the $y$-$z$ plane, as described in Chapter~\ref{chap:PDS}. On the one hand, this solution aims to provide an improved energy reconstruction. On the other hand, it improves the position reconstruction and timing capabilities needed for triggering non-beam physics events, such as supernova burst or nucleon decay events, and fiducializing them along the drift direction.

A fully integrated \dshort{pds} simulation was developed within the \dword{larsoft} framework. In the simulation chain, after particles are generated and propagated using \dshort{geant4}, energy deposits along tracks are retrieved in order to estimate the number of electrons and photons generated at each step. The ionization %free 
charge is calculated based on the modified Birks or Box model with a correction at low \efield ~\cite{Marinho:2022xqk}. From the number of electrons and the total energy deposited, the number of photons is obtained, keeping the appropriate energy partitioning between charge and light. Since a very large number of scintillation photons (25000~ph/MeV at 500~V/cm) is produced in \dshort{lar}, it is very computationally demanding to individually propagate all photons using \dshort{geant4}. Instead, a fast simulation for photons is implemented, currently using a a generative neural network.

The fast photon simulation relies on a generative neural network to predict the fraction of photons (called visibility) reaching each \dword{pd} from a given point in the detector \cite{Muve2022}. This method is flexible enough to be used with any kind of boundary condition.
The generative model can be trained ahead of time using a full Geant4 optical simulation, with photons emitted from random vertices within the cryostat volume with isotropic
initial directions ($\mathord{\sim}10^6$ photons per point).  The output is saved to a computable graph and deployed in DUNE's LArSoft-based software stack. This procedure is followed for both 128\,nm photons (Ar scintillation) and 176\,nm photons (Xe scintillation) to account for the wavelength-dependent Rayleigh scattering and material reflectivity in the simulation of Xe-doped \dshort{lar} (see Section~\ref{sec:PD-Intro}).
When the computable graph is loaded into LArSoft, it quickly emulates photon transport by computing the visibility of each photon detector according to the photon emission vertex along the particle’s track.
This method is 20 to 50 times faster than the Geant4 simulation without introducing voxelization, which would limit the resolution of the simulation. While the computable graph is more accurate than previous `photon library'-based methods, it is not a perfect recreation of Geant4. The differences primarily occur in regions like corners with many reflections that typically do not contribute the majority of the observed light. 
The model inference also requires a relatively small amount of memory. The samples for ProtoDUNE-like and DUNE FD-like geometries show the required memory for the model inference is around 15\% of the Geant4 simulation. Furthermore, the memory usage does not directly scale with detector volume.

When simulating events, the computable graph %parameterization or photon library 
is used to assign the photons produced in the interaction to individual \dshorts{pd}. The time of arrival of the photons takes into account both the time profile of the scintillation process as well as the propagation time in the argon, including the effect of scattering. The emission times for Ar and Xe are treated separately. Argon emission follows a two exponential profile for fast and slow decays whereas the model used for Xe takes into account the time of the intermediate processes $\rm Ar_2^* + Xe \rightarrow ArXe^*$ and $\rm ArXe^* + Xe \rightarrow Xe_2^*$, and finally the $\rm Xe_2^*$ excimer decay time, with parameters obtained from preliminary ProtoDUNE-SP analysis.

The \dshort{pds} design presented in the \dshort{spvd} CDR~\cite{FD2-VD-CDR} was used for the studies discussed in this chapter. It does not include improvements implemented in this \dshort{tdr} such as the membrane coverage over the short walls, which increases uniformity in the \dshort{ly} closer to the end caps of the detector (see Section~\ref{subsec:PDS-Req}), nor the repositioning of the border cathode \dshorts{pd} farther inward, away from the \dshort{fc}, to mitigate the risk of electrical discharge. 
These changes, which will be implemented in the simulation software in early 2023, should only affect in a significant way events simulated in the $\sim$\,15\% of the volume closest to the cathode edges.
The studies presented in this chapter did not use the full \dword{tpc} volume; instead a subset of the top \dshort{spvd} drift volume was used, i.e., the full width (\tpcwidth) and drift length (\maxdriftdist), but only 21\,m of the full detector length (\tpclength).
Studies have shown that this volume is sufficient to characterize the light propagation anywhere in the full volume; 
disregarding the light that would be collected by more distant \dshorts{pd} has a negligible effect.

In the simulation, the aluminum \dshort{fc} profiles are included individually, as described in Section~\ref{subsubsec:FCsss}, and they are assigned a reflectivity value of 70\%. 
On the other hand, the anode is included as a single volume with an effective reflectivity that takes into account the metallic area of the perforated anode \dword{pcb} that could reflect light (about $\sim$40\% of the total area) and the wavelength. The relevant detector parameters and photon emission/propagation constants used in the \dword{mc} generation are listed in Table~\ref{tab:PD-Light-Parameters} (reflectivity assumed as specular for all materials listed). Values for liquid argon doped with Xe with a 10~ppm concentration take into consideration the latest ProtoDUNE measurements \cite{DUNE:2022ctp} showing that from the totality of photons emitted, 53\% will come from Xe dimmers deexcitation and of the 47\% that is emitted by Ar dimmers deexcitation, 35\% will be absorbed in the medium during propagation.

%$$$$$$$$$$$$$$$$
\begin{dunetable}
[Simulation input parameters and physical constants]
{rcl}
{tab:PD-Light-Parameters}
{Simulation input parameters and physical constants.}
{\bf Parameter} & {\bf Value} & {\bf Comment} \\ \toprowrule	
\dshort{lar} Photon Yield & 25,000 ph/MeV &  23\% Singlet (Fast) emission \\                (mip, \@ 500 V/cm) &       &  77\% Triplet (Slow) emission  \\ \colhline
Xe doping in Ar  &    10 ppm   & 53\% of total light emitted at 176~nm  \\
  &  & 35\% of light loss at 128~nm  \\\colhline
Rayleigh Scattering Length              & $\lambda_R(128\,\mathrm{nm}) = 1$\,m &  Ar light \cite{Abi:2020mwi, DUNE:2022ctp}   \\
            &  $\lambda_R(176\,\mathrm{nm}) = 8.5$\,m &  Xe light     \\\colhline
Absorption Length                   & $\lambda_{Abs}(128\,\mathrm{nm}) = 20$\,m &  3 ppm of ${\rm N}_2$ \\   &  $\lambda_{Abs}(176\,\mathrm{nm}) = 80$\,m &     \\
\colhline
\dshort{xarapu} Det. Efficiency      & $\epsilon_D=3\%$ & See Section~\ref{sec:PDS-LightColl} \\\colhline
Field Cage Reflectivity~\cite{Chepel:2013,Bricola:2007}& R=70\%   & Narrower profile in $\sim 60\%$ of longer walls\\\colhline
Cryostat Reflectivity & R(128\,$\mathrm{nm}$) = 30\%  & \\
 & R(176\,$\mathrm{nm}$) = 40\% & \\\colhline
Anode Reflectivity & R(128\,$\mathrm{nm}$) = 6\% & Assuming solid area of perforated PCB\\
 & R(176\,$\mathrm{nm}$) = 12\%  &  of 40\%\\

\end{dunetable}

After the generation and propagation of photons coming from energy deposits in the detector volume, the digitization of the light signal is performed assuming that each detected photon will generate a signal, called a single photoelectron (PE), with an idealized shape of a fast rising exponential and a slower exponential decay. The waveform is built by considering the relevant electronics characteristics. Details on the digitization parameters are given in Table~\ref{tab:PD-Digi-Parameters}.

%$$$$$$$$$$$$$$$$
\begin{dunetable}
[Input parameters for the optical digitization stage in the simulation]
{rc}
{tab:PD-Digi-Parameters}
{Input parameters for the optical digitization stage in the simulation.}
{\bf Parameter} & {\bf Value}  \\ \toprowrule	
Single PE rise time & 10 ns \\ \colhline
Single PE decay time  &    200 ns   \\\colhline
Single PE peak      & 10 ADC    \\ \colhline
Crosstalk probability   &   20\% \\ \colhline
Electronic channels   &   2 \\ \colhline
Baseline           &   500 ADC \\ \colhline
Electronics noise      & 2 ADC    \\ \colhline
Saturation      & 14 bits    \\ \colhline
Sampling frequency      & 62.5 MHz    \\
\end{dunetable}

The reconstruction of light signal includes the identification of optical hits in the recorded photon detector's waveforms, triggered by the criteria of signal above 1.5 \phel. Optical hits are later combined in optical flashes by a time coincidence criteria. Optical waveforms, hits, and flashes are the basic elements used for analyzing the light signal.

\subsection{Charge Readout Simulation}
\label{sec:ph:TPC-Simulation}

Despite the mechanical and electronic differences between the \dshort{sphd} and \dshort{spvd} detector modules, their underlying detection principles %of the two detectors 
are the same. This parity has been %utilized 
exploited in the software, where a model of %the 
\dshort{spvd} %detector 
has been implemented in the \dshort{larsoft} framework with a nearly identical simulation and reconstruction workflow to %the 
\dshort{sphd}. % detector. 
Because of this similarity, many of the advanced reconstruction techniques that were developed for the \dshort{sphd}~\cite{DUNE:2020ypp} have been reused by the \dshort{spvd}, with some retuning and reoptimization required.  The simulation and reconstruction for \dshort{sphd} has been described extensively elsewhere~\cite{DUNE:2020ypp} and will only be briefly summarised here.  The overall simulation and reconstruction for both detectors can be broken down into the following steps:
\begin{enumerate}
    \item \textbf{Event generation:} Simulation of neutrino interactions within the detector volume, relying on a neutrino interaction generator such as \dword{genie}.
    \item \textbf{Particle propagation:} Simulation of the trajectories and interactions of the neutrino interaction's final state particles through the detector volume using \dshort{geant4}.  The number of electrons ionized from the liquid argon and the number of scintillation photons produced from recombination are calculated at the end of this step.
    \item \textbf{Detector response:} The signal simulation of the raw output of the detector, as would be expected from the \dword{daq}.  This step involves transportation of the ionization electrons to the anode plane, and simulation of the charge readout's response to those electrons.
    \item \textbf{Signal processing:} The first step of the reconstruction is to remove detector effects (such as the shaping time of the electronics) and noise from the raw waveforms, recovering the true charge waveforms observed on the readout channels.
    \item \textbf{Hit reconstruction:} A search for, and fits to, peaks in the charge waveforms.  These fitted peaks are called hits.
    \item \textbf{Pattern recognition:} The process of clustering waveform hits, and matching those clusters across readout planes to form 3D representations of particles.  This is typically handled by a dedicated package such as \Dword{pandora}~\cite{Acciarri:2286065} (see section~\ref{sec:pandora}).
    \item \textbf{High level reconstruction:} Advanced reconstruction techniques that are applied to upstream reconstruction to deduce particle and neutrino-level properties.  These include neutrino flavor tagging with a \dword{cvn} (see section~\ref{sec:cvn}) and neutrino energy reconstruction (see section~\ref{sec:enureco}).
\end{enumerate}

Despite the general software similarly between \dshort{sphd} and \dshort{spvd}, direct comparisons are, in some cases, complicated by advances made in the software %development 
since the most recent full end-to-end \dshort{sphd} production. For the benchmark performance plots in previous design reports~\cite{DUNE:2020ypp}, the neutrino interaction generator was \dshort{genie} version 2.12, and a simplified model was used to simulate the detector response. Since then, the \dshort{larsoft} framework has moved forward to \dshort{genie} version 3.0, and a more realistic detector response model is simulated through the \dword{wct}, in addition to several other incremental improvements to algorithms.  These recent additions are included in the \dshort{spvd} samples that are shown in detector performance comparisons throughout this chapter, but those same additions are not included in \dshort{sphd} benchmark being compared to.

\subsection{Signal Processing with Wire-Cell Toolkit}
\label{sec:wirecell}

\dword{wct} is a software suite based on the \dword{wirecell} tomographic reconstruction principle. It provides algorithms for the \twod convolution-based \dshort{lartpc} signal simulation and signal processing.
This simulation agrees with real detector data better than the previous 1D version. 
It is used and validated for wire readouts in the \dword{pdsp} and the \dword{microboone}  experiments~\cite{MicroBooNE:2018swd,MicroBooNE:2018vro}. Initial validations for the \dword{pcb} readout was done using a 50L prototype.
As a reverse process, the first step in the \dshort{lartpc} charge reconstruction is the \dword{sigproc}, which extracts original ionization electron distributions from digitized \dshort{tpc} waveforms. The \dshort{wirecell} \twod deconvolution based \dshort{lartpc} algorithm is a core port of \dshort{spvd}'s simulation and reconstruction, and has been validated using real detector data.
A similar detector response model is currently being deployed for \dshort{sphd}.

%%%%%%%%%%%%%%%%%%%%%%%%%%
%\section{Low-energy Physics with Vertical Drift}
\section{Low-energy Physics Performance}
\label{sec:ph:le}

The design and optimization of the \dshort{pds} in terms of detection coverage, detection efficiency, and timing capabilities may allow for the enhancement of the DUNE physics reach. A high \dshort{ly} is highly appealing for low-energy physics as it entails improvements in the triggering efficiency and energy resolution.

As a window into this possibility, the \dshort{tpc} and \dshort{pds} trigger efficiencies and energy resolutions for low-energy neutrino events are studied in Sections~\ref{sec:ph:le:trigger} and~\ref{sec:ph:le:energy}, respectively. 

The \dword{mc} simulation for \dshort{snb} events in \dshort{larsoft} employs the \dword{marley} %(Model of Argon Reaction Low Energy Yields) 
generator~\cite{marley,Gardiner:2021qfr}. \dshort{marley} simulates neutrino-nucleus interactions in \dshort{lar} in the tens-of-MeV range by selecting an initial excited state of the %residual 
final state $^{40}$K$^*$ nucleus and sampling an outgoing electron direction %using 
according to the $\nu_e$\dshort{cc} differential cross section. After simulating the initial two-body $^{40}$Ar($\nu_e$, e$^-$)$^{40}$K$^*$ reaction for an event, \dshort{marley} also handles the subsequent nuclear de-excitation.

Background generation follows the \textit{BxDecay0} package\footnote{https://github.com/BxCppDev/bxdecay0}, a C++ library providing simulated nuclear decays that is integrated into \dshort{larsoft}.  A radiological model was developed for the \dshort{sphd} and adapted for the \dshort{spvd} design. It includes radioactive decays in the \dshort{lar} bulk ($^{39}$Ar, $^{42}$Ar, $^{85}$Kr, $^{222}$Rn and their decay chains), the cathode (drifted $^{42}$K from \dshort{lar}'s $^{42}$Ar decays, $^{40}$K and the $^{238}$U decay chain), the \dwords{crp} ($^{60}$Co and the $^{238}$U decay chain), the \dshort{pds} ($^{222}$Rn decay chain), and external sources (gammas and neutrons from surrounding rocks). Table~\ref{tab:bkg} summarizes the different activities considered. This background model is much more detailed than the one considered in the \dshort{sphd} \dshort{tdr}, although even further improvements are planned, as for instance, a more accurate model of the cavern gammas. 

\begin{dunetable}
[Backgrounds]
{lc}
{tab:bkg}
{Activities taken into account in the radiological model. The numbers are coming from estimations or dedicated measurements.}
 \bf Component         & \bf Activity (mBq/cm$^3$)\\ \toprowrule
$^{39}$Ar in LAr &			1.41 	\\ \colhline
$^{42}$Ar and $^{42}$K in LAr & 	0.128 $\times 10^{-3}$ \\ \colhline
$^{85}$Kr in LAr		&	0.16  \\ \colhline
$^{222}$Rn chain in LAr 	&1.395 	$\times 10^{-3}$ \\ \colhline
$^{40}$K in cathode		&	9.1 	\\ \colhline
$^{238}$U chain in cathode		&	0.113 \\ \colhline
$^{60}$Co  in anode		&	0.361  \\ \colhline
%$^{238}$U  chain in anode		&	0.095 $\times 10^{3}$ \\ \colhline
$^{238}$U  chain in anode		&	95  \\ \colhline
$^{222}$Rn  chain in PDS		&	0.021 	\\ \colhline
External neutrons &		7.6	$\times 10^{-3}$   \\ 
(rocks, concrete walls, etc)&		                \\ \colhline
%Cavern gammas &	0.064 		$\times 10^{3}$ 
Cavern gammas &	64 		\\ 
\end{dunetable}

After the generation stage, events pass through the \dshort{geant4} stage %for 
to estimate the number of electrons and photons reaching each detector channel, % to be estimated, 
with special %care of 
attention to neutron transport and electromagnetic physics ($QGSP\_BERT\_HP\_EMZ$ physics list), and the digitizer stage, where the corresponding recorded signals will be generated. Finally, the reconstruction stage provides input for the physics analysis.

\subsection{SNB Trigger Efficiency}
\label{sec:ph:le:trigger}

As %each 
a \dfirst{sn} is such a rare and unpredictable event, it is vital that DUNE  trigger on each galactic \dshort{snb}. %using the increased physics activity within the active volume. 
The \dshort{snb} trigger will be limited by radiological backgrounds within the \dshort{lar} volume and, relying on an increase in physics activity within the active volume. 
%In order to trigger on a \dshort{snb}, t
The detector must be able to detect and reconstruct events in the range 5–100\,MeV, and special triggering and \dword{daq} requirements %that 
must take into account the short, intense nature of the burst, as well as the need for prompt propagation of information %in a 
worldwide. % context. 
The trigger is required to achieve 95\% efficiency for a neutrino burst from a \dshort{sn} at 20\,kpc.  A \dshort{snb} is expected to last approximately 30 seconds, but can last as long as a few hundred seconds; a large fraction of the events are expected within approximately the 1-2 seconds of the burst. 

The DUNE detector systems must be configured to provide information to other observatories on possible astrophysical events (such as a galactic \dshort{sn}) in a short enough time to allow global coordination. This interval should be less than 30 minutes, and preferably on a few-minute timescale. Any \dshort{snb} trigger observed in DUNE will be forwarded immediately to \dword{snews}~\cite{snews} %, so that they can begin 
to allow observation of the evolution of the event.

In DUNE, the trigger on a \dshort{snb} can be done using either \dshort{tpc} or \dshort{pds} information. In both cases, the trigger scheme exploits the time coincidence of multiple signals over a timescale matching the typical \dshort{sn} luminosity evolution. A redundant and highly efficient triggering scheme is envisaged. %We describe here a 
This section describes a trigger design study based on the \dshort{tpc} and another on the \dshort{pds}. 

\subsubsection{SNB Trigger Efficiency with the TPC}
\label{sec:ph:le:triggerTPC}

With the planned \dshort{daq} computing resources, the reconstruction available for real-time triggering of a \dshort{snb} in DUNE will be limited. The \dshort{snb} trigger is based on \dwords{trigprimitive}, reconstructed \dshort{tpc} hits that include basic information such as start-time, time-over-threshold, channel number, and pulse-integral (a value proportional to the charge collected in that channel).  Clusters are formed from nearby \dshort{trigprimitive}s and the detector is triggered if the number of clusters observed in a sliding 10 second time window passes above a certain threshold.  

The algorithm to trigger a \dshort{snb} should be robust and reliable. To determine the number of clusters in the \dshort{tpc}, the \dshort{trigprimitive}s are sorted by time and channel number. \dshort{snb} \dword{mc} samples with low-energy neutrino interactions were generated and combined to create a stream of \dshort{trigprimitive}s to understand DUNE's potential to trigger on a burst.  Clustering is performed to group nearby hits within 20 ticks (10~$\mu$s) and two channels (in order to keep the algorithm robust against dead channels). If at least six \dshort{trigprimitive}s are clustered together, the cluster is taken as a neutrino candidate. The mean number of \dshort{trigprimitive}s reconstructed together in a cluster is proportional to the neutrino energy.  This algorithm has $>50\%$ efficiency for identifying neutrinos with $\geq16$~MeV of energy with modest efficiency at lower energies (see Section~\ref{sec:tdaq:trigger}).  The number of clusters accumulated in a sliding 10 second window is then calculated to determine whether the detector should trigger.

An important constraint of DUNE regarding the \dshort{snb} triggering is the number of fake triggers the \dshort{daq} can handle. 
%We assume a false trigger rate of once per month 
As in previous DUNE studies~\cite{DUNE:2020lwj} and required by \dword{snews}, the maximal tolerable rate of false triggers is assumed to be one per month. Using the radiological model previously described and summarized in Table~\ref{tab:bkg}, a rate of background events was found to be around 1\,Hz, using a selection of six \dshort{trigprimitive}s per cluster. Neutrons from the rocks are the main source of remaining backgrounds, since their capture in Ar atoms produces a gamma cascade that is similar to a \dshort{snb} neutrino interaction. A requirement of 25 clusters in a ten second interval produces about one fake trigger per month.

With a requirement of 25 clusters, each with at least six trigger primitives, reconstructed within 10 seconds, the \dshort{snb} triggering efficiency is depicted in the left plot of Figure~\ref{fig:SNB_TPC_trig_eff}, calculated with Poisson statistics. In this plot, the efficiency is given in terms of true interactions in 10\,kt. Using the Garching model~\cite{Huedepohl:2009wh}, depicted in Figure~\ref{fig:ph:lep:SN-time}, the efficiency in terms of the \dshort{snb} distance is shown on the right plot of Figure~\ref{fig:SNB_TPC_trig_eff}. This model gives the time dependency of the interactions, which is taken into account when setting a ten second window for triggering. Figure~\ref{fig:SNB_TPC_trig_eff} shows that the trigger requirement is met with the \dshort{tpc} signal of the \dshort{spvd} and the Garching model, which is the most conservative of the three considered in this document.

\begin{dunefigure}
[TPC SN burst (SNB) triggering]
{fig:SNB_TPC_trig_eff}
{\dshort{snb} triggering efficiency using the \dshort{tpc} signal. Left: In terms of the real number of interactions. Right: Converted to the distance from an \dshort{snb}.}
  \includegraphics[width=0.48\textwidth,angle=0]{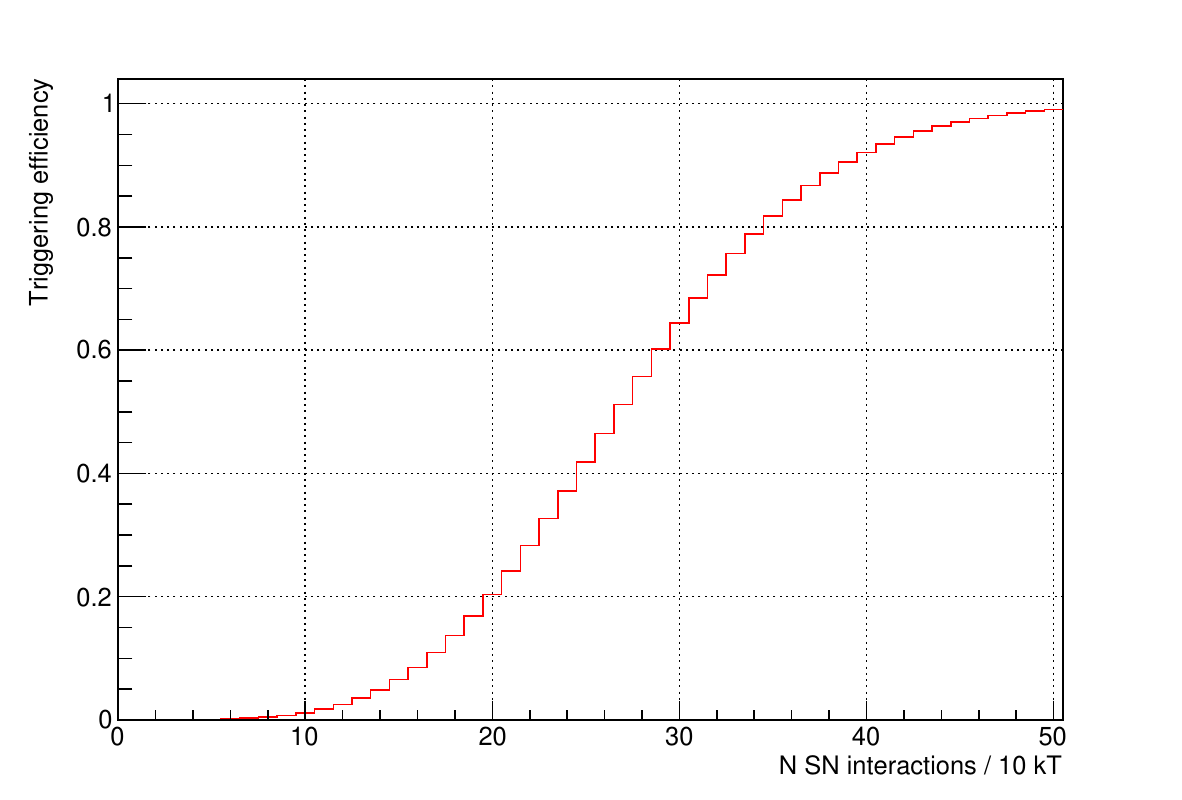}
  \includegraphics[width=0.48\textwidth,angle=0]{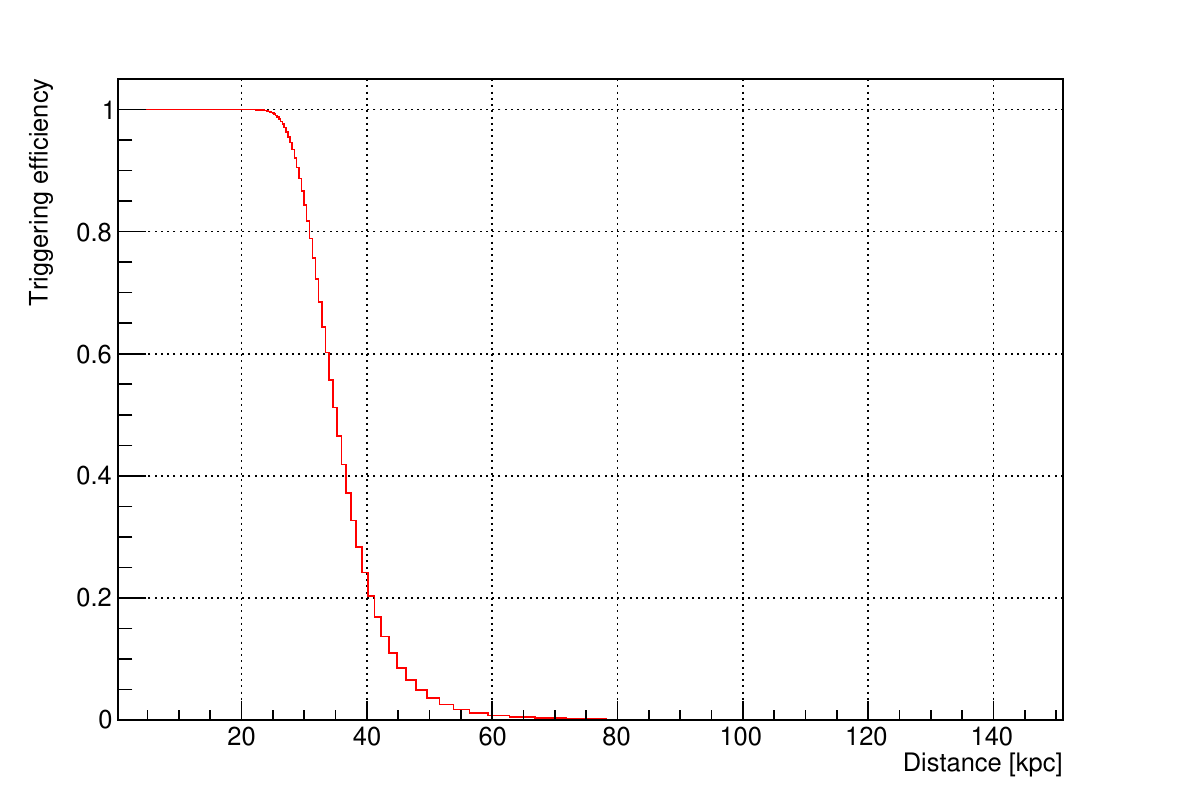}
\end{dunefigure}

\subsubsection{\dshort{snb} Trigger Efficiency with the \dshort{pds}}
\label{sec:ph:le:triggerPDS}

The \dshort{pds}-based \dshort{snb} trigger uses optical flashes that are clustered according to trigger primitive hits in the optical detectors reconstructed nearby in time and space.
For the \dshort{pds} trigger study, a real-time algorithm provides the trigger primitives by searching for optical hits and optical clusters based on time and spatial information.

\dword{mc} samples were generated using %the 1$\cross$8$\cross$14 \dshort{spvd} geometry. 
a subset of the \dshort{spvd} geometry consisting of one drift gap height, full width, and 21 m length along the detector long side.
A \dshort{snb} simulation was run using the \dshort{spvd} geometry producing $10^6$ \nue-\dshort{cc} events with the \dshort{marley} event generator. %according to a flat distribution in the 4-100\,MeV range
A sample of {$10^6$} simulated exposures, each using a 8.5\,ms exposure time, was

produced with the radiological model, including the radioactive decays listed in Table~\ref{tab:bkg}. The \dshort{pds} trigger study was carried out for three different core-collapse \dshort{sn} neutrino flux models: Livermore, Garching, and \dword{gkvm}, as detailed in~\cite{Abi:2020lpk}.

First a clustering algorithm was developed. For this, %we define 
an optical hit is defined as a peak in the digitized signal arriving in one of the optical channels that is above 1.5\,PE. A cluster is defined as a collection of hits that present certain correlations in time and space, such that %we believe 
they %were 
are believed to have all been caused by the same underlying neutrino event inside the detector. The parameters %we 
explored to optimize %our 
the clustering algorithm are:

\begin{itemize}
    \item Maximum cluster duration -- the maximum time difference between the earliest and the latest hit in the cluster.
    \item Maximum time difference between consecutive hits.
    \item Maximum spatial distance between neighboring optical channels detecting the hits.
    \item Minimum hit multiplicity -- the minimum number of hits required to classify a collection of hits as a cluster.
\end{itemize}

As an orientation, the optimal parameters were around 300~ns for the maximum cluster duration, 200~ns for the maximum hit time differences, 250~cm for the maximum hit spatial distances, and 10 for the minimum hit multiplicity. 

Second, %we developed 
a trigger algorithm was simulated. For a given set of parameters, %we run
the clustering algorithm is run on the whole set of %supernova 
\dshort{snb} and background events. %We train a 
A classifier (boosted decision tree), trained on the clusters produced by this set of parameters, may be used to discriminate between signal neutrino and radiological background events that is trained based on the cluster parameters.

Next, for a fixed neutrino energy spectrum, %we run our 
the clustering algorithm is run on 1000 subsamples of %supernova 
\dshort{sn} events and corresponding backgrounds, using %. We use 
a burst time window of one second, which was found close to optimal for identifying a \dshort{snb}. %Our 
This classifier is then applied to remove $>50\%$ of the background while keeping most ($>90\%$) of the \dshort{snb} signal.

%For each of these subsamples we perform a
A $\chi^2$ test is performed on each of these subsamples, comparing the signal with the expected background. %The variable in which we perform the comparison is h
Hit multiplicity (number of hits in a cluster) is the variable used to perform the comparison, as this %we found it 
was found to have the highest discriminating power. The trigger is activated successfully if the significance of the $\chi^2$ test is above a critical value indicating a trigger frequency of %1/
one per month, a requirement for limiting data rates with DUNE and \dword{snews}. The trigger efficiency for this set of parameters is estimated as the number of successful \dwords{trigact} over the total number of subsamples.

%We 
The next step is to iterate through a wide set of parameters and select those %which give us 
that yield the higher trigger efficiency for a particular neutrino spectrum. For the three %mentioned %supernova neutrino 
\dshort{snb} flux models considered, the efficiency values are presented in Table~\ref{tab:PD-trigger-eff} and Figure~\ref{fig:PD-trigger-eff}.

\begin{dunetable}
[PDS trigger efficiencies]
{lcccc}
{tab:PD-trigger-eff}
{\dshort{pds} trigger efficiencies for three different \dshort{sn} models and a range of distances for a 12\,kt \dshort{spvd} module. The rightmost column indicates the distance at which efficiency falls below 95\% .}
          & Eff. at 10 kpc & Eff. at 15 kpc & Eff. at 20 kpc & 95\% thresh. (kpc) \\ \colhline
Livermore & 1                    & 0.954              & 0.492          & 16.12             \\ \colhline
Garching & 0.758            & 0.066             & 0.006          & 9.24              \\ \colhline
GKVM     & 1                     & 1                   & 1               & 28.56           \\ 
\end{dunetable}

\begin{dunefigure}
[PDS trigger efficiencies]
{fig:PD-trigger-eff}
{\dshort{pds} trigger efficiencies for three different \dshort{sn} models and a range of distances for a 12\,kton FD-2 VD module.}
  \includegraphics[width=0.7\textwidth,angle=0]{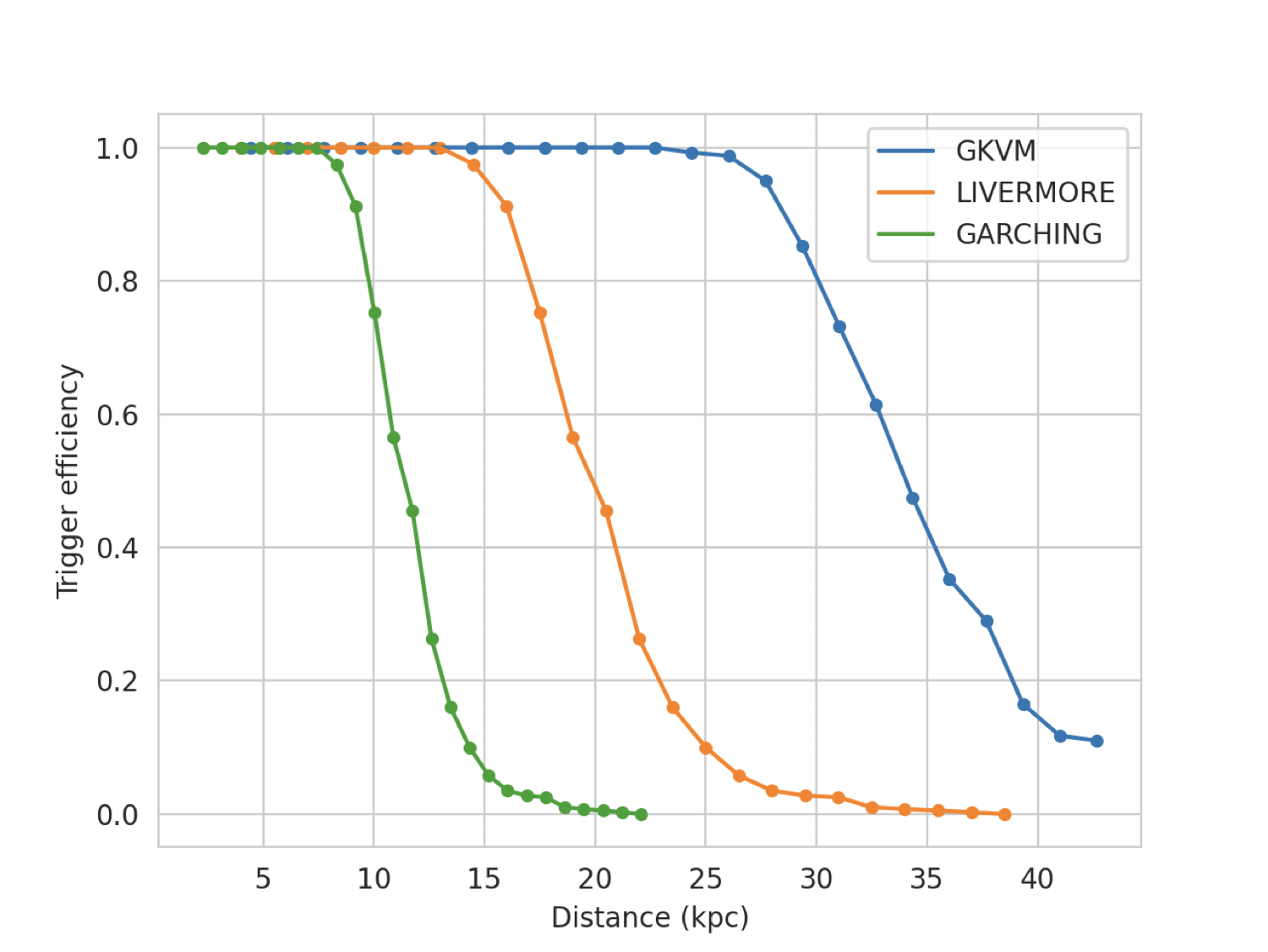}
\end{dunefigure}

DUNE's current \dshort{pds} trigger sensitivity to a core-collapse \dshort{sn} is found to be in the 5-20\,kpc range, the precise value being highly dependent on the \dshort{snb} flux model. The \dshort{spvd} \dshort{pds} should yield a highly efficient trigger for a \dshort{sn} occurring anywhere in the Milky Way when combined with other subsystems, and the proposed trigger strategy meets the requirements, although it has a dependency on the model. 
For a direct comparison with the \dshort{sphd}, a new study is in progress updating the simulation parameters and the radiological model.

\subsection{Energy Resolution with TPC and PDS}
\label{sec:ph:le:energy}

A simulation has been performed to assess the energy resolution for neutrino interactions in the 5 to 30\,MeV energy range. For this, \dshort{marley} was used to generate monoenergetic signal samples with ($\mathcal{O}(10^4)$ events in 1\,MeV steps).  The full DUNE simulation and reconstruction chain was also run as previously described. 

From the sample without backgrounds, a calibration was performed to transform the charge read by the \dshort{tpc} into the energy deposited by the interacting neutrino. This was done by summing the charge of all the hits, after electron drift lifetime correction, as long as the true neutrino vertex was at least 15\,cm from any surrounding wall-like structure so that the full event topology is contained.

To compute the visible energy in a given event, an algorithm that takes the local information from \twod trajectories, stitching together nearby \twod hits was used to form reconstructed clusters. Next \threed track information was produced by taking the \twod clusters %formed 
and matching %clusters 
them in the three \twod projection strip planes to build the tracks. A calorimetric sum of the hits associated with the highest charge track and surrounding clusters
was performed (after electron lifetime correction) both for signal-only and signal-and-background generation in order to evaluate the impact of radiological background on the low-energy detector performance.

Figure~\ref{fig:tpc-le-eres} shows both the tracking performance and energy resolution attained for the signal-only sample. The tracking efficiency grows with energy and presents a plateau at $\sim$90\% and the energy resolution is comparable to the \dshort{sphd} module. A closer look at track reconstruction at low energy with the \dword{pcb} technology will be taken in order to evaluate possible improvements in this metric. The presence of backgrounds does not affect the signal tracking efficiency and brings only a small degradation on the energy resolution, especially on the lower energy end.

\begin{dunefigure}
[MARLEY tracking efficiency and energy resolution for low-energy events]
{fig:tpc-le-eres}
{Left: Tracking efficiency for \dshort{marley} \dshort{cc} $\nu_e$ scattering on $^{40}$Ar events. Right: Energy resolution for low-energy events, defined as the \dword{rms} of the distribution of the fractional difference between the reconstructed and the visible energy with respect to visible energy, as a function of neutrino energy for reconstructed \dshort{tpc} tracks corrected for drift attenuation.}
\includegraphics[width=.48\textwidth]{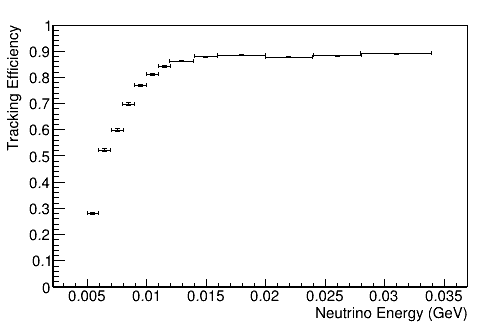}
\includegraphics[width=.51\textwidth]{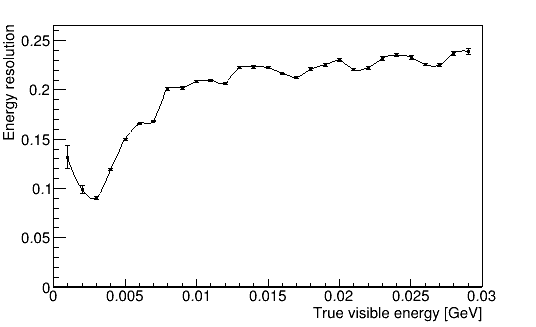}
\end{dunefigure}

The \dshort{pds} should be able to provide a calorimetric energy measurement for low-energy events, complementary to the \dshort{tpc} energy measurement%(see Table~\ref{tab:specs:SP-FD2})
. Achieving \dshort{pds} energy resolution comparable to the \dshort{tpc} will allow taking %full 
advantage of the anti-correlation between the emission of light and charge signals.
The \dshort{pds} energy performance at low energy was studied for MARLEY-generated neutrino events simulated in the \dshort{spvd} detector geometry. %The events were generated uniformly spread throughout the \dshort{tpc}
Neutrino interactions with energies between 4\,MeV and 100\,MeV were uniformly generated within the \dshort{tpc}, and analysis was performed by selecting events in the central region of the detector along the beam direction to avoid border effects from the cryostat's end caps. 

A precise knowledge of the \dshort{ly} as a function of position  e.g., from a dedicated calibration, will enable calorimetric energy reconstruction from photon counting. Since the \dshort{ly} of a detector is position-dependent, the energy reconstruction was carried out by correcting the total observed number of photoelectrons using the simulated neutrino's true position in the detector. In future analyses, a reconstructed position based on PDS and TPC signals can be used. The energy resolution was obtained as the standard deviation of the Gaussian fit to the distribution of the relative difference of visible energy to the reconstructed deposited energy.

Figure~\ref{fig:PDS-E-resolution} shows the energy resolution with statistical-only uncertainties when using collected light from simulated low-energy neutrinos. 
The fit presents the expected behavior, consisting of the ``noise,'' ``stochastic,'' and ``constant'' terms added in quadrature ($\sqrt{p_0{}^2+(p_1/\sqrt{E})^2 + (p_2/E)^2}$).
The stochastic term ($\propto p_1$) corresponds to the intrinsic statistical spread in the number of photons detected given by Poisson statistics, whereas the noise contribution ($\propto p_2$) is due to the cumulative electronic noise of the ARAPUCA cells’ readout chain (simulated baseline and dark noises). The constant term alone ($p_0$) 
provides an energy resolution of the PDS at higher energy of about 10\%. The largest contribution to this number comes from the calibration procedure currently being used. It relies on averaging the \dword{ly} over a large region ($0.25 \times 1 \times 2~$m$^3$) in order to obtain the total expected number of photons landing on a given \dword{pd}. Further exploitation of the computable graph photon simulation, combined with rigorous calibrations should significantly reduce this term and improve the energy resolution.

\begin{dunefigure}
[\dshort{pds} energy resolution for the \dshort{spvd}]
{fig:PDS-E-resolution}
{\dshort{pds} energy resolution with statistical-only uncertainties for the \dshort{spvd}. The fit shows a function of the form $\sqrt{p_0{}^2+(p_1/\sqrt{E})^2 + (p_2/E)^2}$, with $p_0$, $p_1$, and $p_2$ representing the constant, stochastic and noise coefficients respectively.}
\includegraphics[width=.49\textwidth]{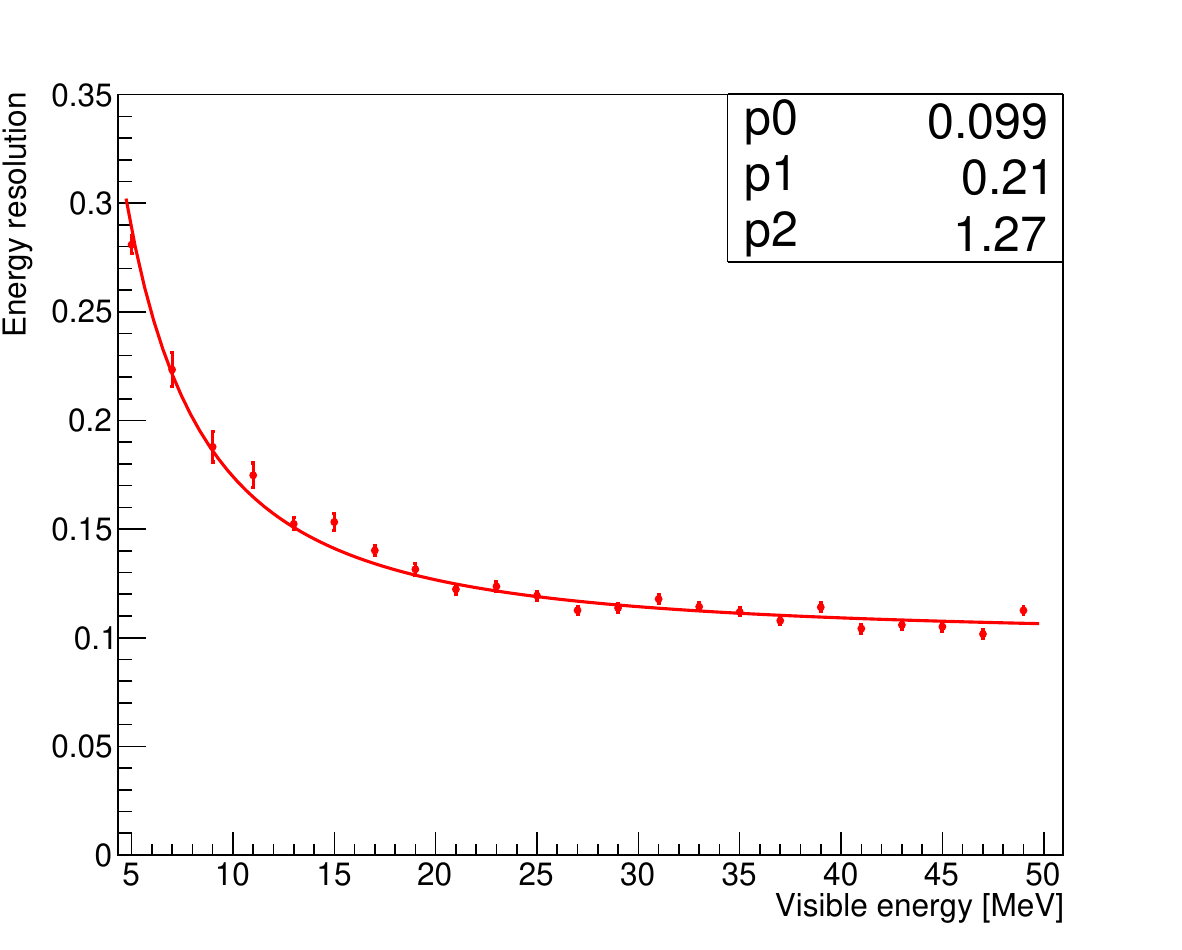}
\end{dunefigure}

%%%%%%%%%%%%%%%%%%%%%%%%%%
\section{Long-baseline Oscillation Physics Performance}
\label{sec:ph:lbl}

The \dword{lbl} %long-baseline 
neutrino oscillation physics sensitivity of DUNE is described in the \dshort{sphd} TDR~\cite{DUNE:2020ypp} and in two related journal articles~\cite{DUNE:2020jqi, DUNE:2021mtg}. 
The analysis described in those references is based on simulations of the \dshort{sphd}. 
The performance of the \dshort{fd} for the oscillation analysis depends on three main factors, all of which are expected to be very similar in \dshort{sphd} and \dshort{spvd}: the reconstruction efficiency and background rejection for \dshort{cc} \numu and \nue signals, the neutrino energy estimation for these signal events, and the residual calibration uncertainties. %We can compare t
The two detector designs can be compared in these three areas, without completely repeating the %long-baseline
\dshort{lbl} analysis, which requires tens of millions of CPU core hours, largely dominated by the processing time of the $\sim \!\! 1,000,000$ throws for each of the $\sim \!\! 150$ systematic uncertainties.

The inputs to the oscillation analysis are selected samples of \numu and \nue \dshort{cc} candidates, in \dword{fhc} and \dword{rhc} beam modes, with a neutrino energy estimate for each event. The reconstruction stages required to %do this 
categorize the event type and estimate its energy
are signal processing (using \dshort{wirecell}, described in Section~\ref{sec:wirecell}), hit reconstruction, \threed event reconstruction (using \dshort{pandora}, described in Section~\ref{sec:pandora}, event selection (using a \dshort{cvn}, %Convolutional Visual Network, 
described in Section~\ref{sec:cvn}), and neutrino energy reconstruction (based on \dshort{pandora}, described in Section~\ref{sec:enureco}). The hit reconstruction, which finds signal peaks on the strip waveforms, is identical to the \dshort{sphd} and is not described here.

The performance benchmarks presented here are based on a simulation of six million neutrino interactions in \dshort{fhc} beam mode. \dshort{fhc} results are shown. 
Comparisons of \dshort{rhc} events were not possible due to a technical difficulty with the tape system at Fermilab that prohibited accessing the relevant files for \dshort{sphd}.
These \dshort{fhc} comparisons %are to 
used a \dshort{sphd} simulation that was based on a simplified detector response and a reconstruction that has undergone years of optimization and is substantially more mature than its \dshort{spvd} counterpart.   For \dshort{spvd}, some areas of the simulation and reconstruction have undergone a first round of optimisation (sections~\ref{sec:wirecell}, \ref{sec:cvn} and \ref{sec:enureco}) while other areas (section~\ref{sec:pandora}) rely on tunings taken from \dshort{sphd}, with \dshort{spvd} tuning planned for the future. In all cases, the preliminary \dshort{spvd} performance is expected to converge with that of \dshort{sphd} once the additional realism is added to the \dshort{sphd} simulation, and the \dshort{spvd} reconstruction undergoes further development and optimizations.

\subsection{3D Event Reconstruction with Pandora}
\label{sec:pandora}

\dshort{pandora} is a multi-algorithm reconstruction suite~\cite{Acciarri:2286065}, with specific design features to perform pattern recognition on plane-based \dwords{lartpc}.  The overall aim of \dshort{pandora}'s reconstruction is to read in the reconstructed \twod hits from a \dshort{lartpc}, and output fully \threed representations of the particles emanating from the neutrino vertex. Due to the similarities between the two detector modules coupled with \dshort{pandora}'s detector-agnostic nature, the \dshort{spvd} instance of \dshort{pandora} was able to inherit all configurations and tunings from the \dshort{sphd} and achieve competitive performance out of the box.  

Figure~\ref{fig:lbl_pandora_momeff} shows \dshort{pandora}'s reconstruction efficiency of the simulated leading lepton in \dshort{cc} interactions, shown as a function of the lepton's true momentum. The efficiencies are shown for both the \dshort{spvd} and \dshort{sphd}. % detectors. 
Performance is similar for both detector modules, with only a small relative deficit in efficiency for the \dshort{spvd}, especially for electrons. Figure~\ref{fig:lbl_pandora_angulareff} similarly shows \dshort{pandora}'s reconstruction efficiency of the leading muon, but as a function of the muon's angle in the anode plane.  While the performance is again competitive between the two detectors, Figure~\ref{fig:lbl_pandora_angulareff} also shows that both detectors have consistent angular acceptance, even when the muon is aligned with the direction of the wires/strips in a given anode plane (about $\pm30^\circ$ and $90^\circ$ for both detectors).  This is due to the three-plane design of both detector designs: when the readout quality of one plane is compromised because of the co-linearity of the lepton with that plane's readout, the other two complementary planes provide enough information to achieve \threed reconstruction~\cite{Brailsford:2021htz}.

\begin{dunefigure}
[Pandora reco efficiencies of leading lepton (CC interactions) in \dshort{sphd}/\dshort{spvd}]
{fig:lbl_pandora_momeff}
{\dshort{pandora}'s reconstruction efficiencies of the leading lepton from \dshort{cc} neutrino interactions in \dshort{sphd} and \dshort{spvd}. %the vertical drift detector.  
The efficiencies are shown as a function of the leading lepton's true momentum.  Left: $\mu^-$.  Right: $e^-$.}
\includegraphics[width=.49\textwidth]{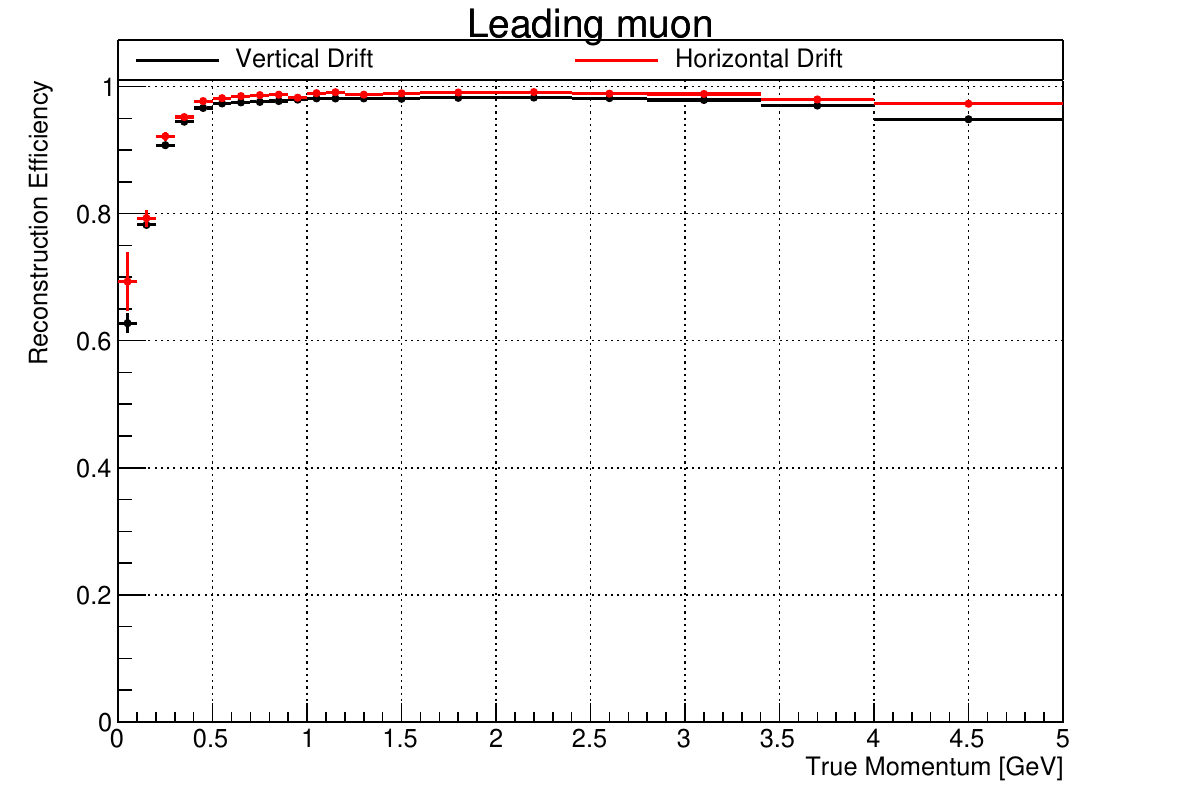}
\includegraphics[width=.49\textwidth]{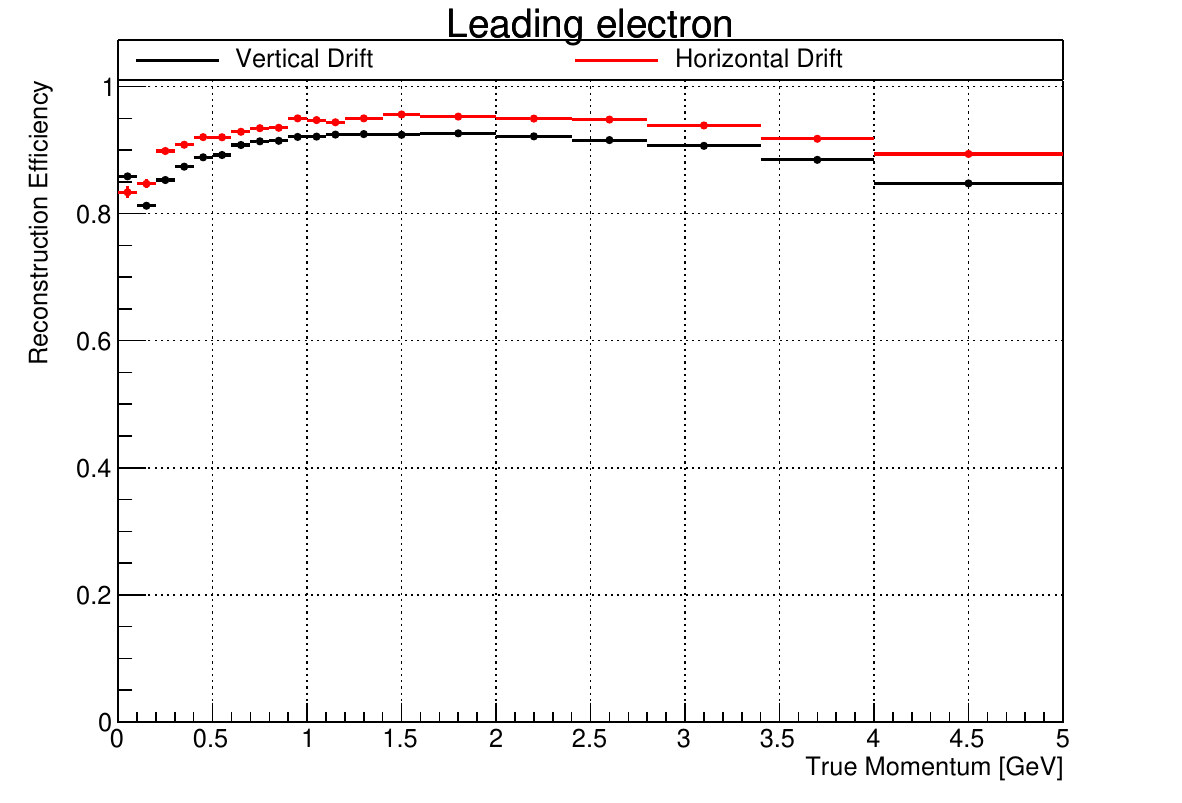}
\end{dunefigure}

\begin{dunefigure}
[Pandora reco efficiencies of the leading $\mu^-$ (CC interactions) in \dshort{sphd}/\dshort{spvd}]
{fig:lbl_pandora_angulareff}
{\dshort{pandora}'s reconstruction efficiencies of the leading $\mu^-$ from \dshort{cc} neutrino interactions in \dshort{sphd} and \dshort{spvd}. %the vertical drift detector.  
 The efficiencies are shown as a function of the muon's angle in the plane of the anode.}
\includegraphics[width=.75\textwidth]{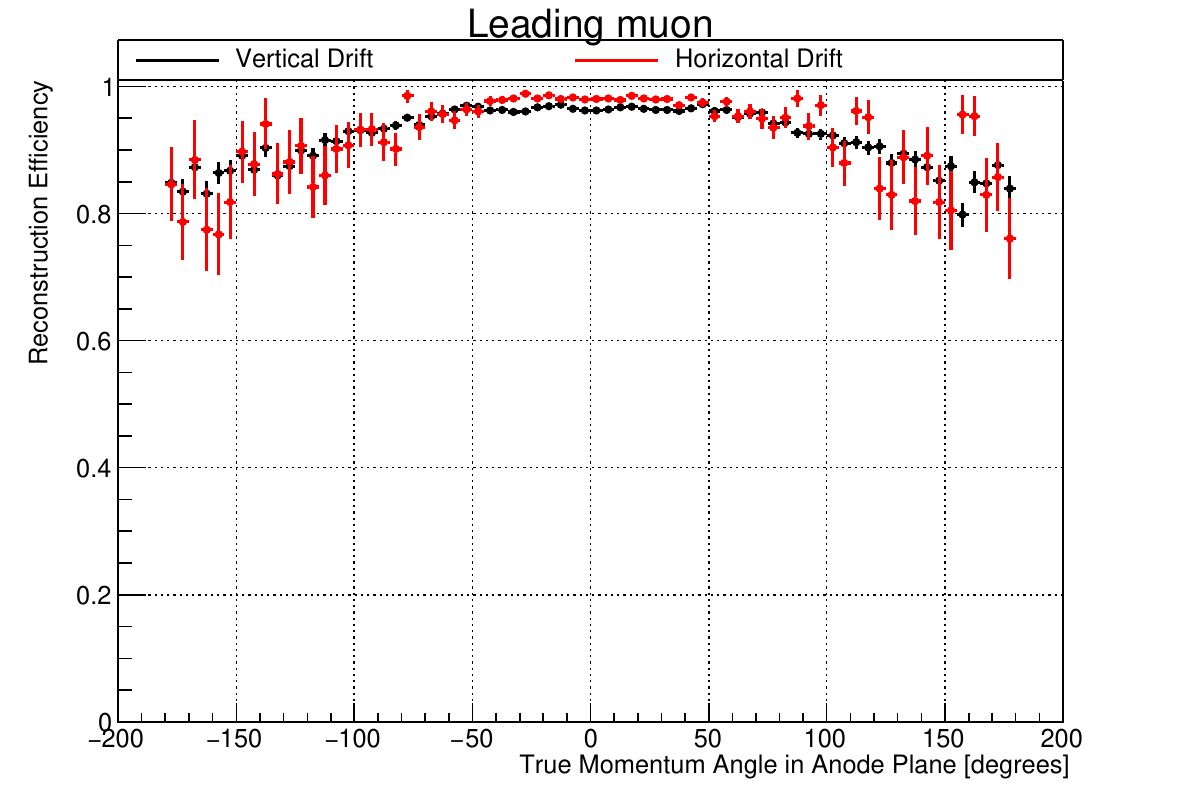}
\end{dunefigure}

\subsection{Neutrino Event Selection with the Convolutional Visual Network}
\label{sec:cvn}

The \dshort{cvn}~\cite{PhysRevD.102.092003} is the primary means of flavor-tagging a neutrino interaction in the DUNE far detector, and was used extensively in the DUNE oscillation sensitivity analysis with \dshort{sphd}~\cite{DUNE:2020ypp}. The \dshort{cvn} is a a deep learning-based image recognition technique that can exploit the fine-grained detail observed in a \dshort{lartpc} to maximize selection power. The technical configuration of the \dshort{cvn} has been described extensively elsewhere~\cite{DUNE:2020ypp, PhysRevD.102.092003}. The \dshort{cvn} was retrained for  \dshort{spvd} using approximately three million simulated neutrino interactions. The simulated neutrinos were processed through the \dshort{spvd}'s reconstruction chain up to the hit reconstruction, where the reconstructed hits provide the primary input to the \dshort{cvn}.

The primary output of the \dshort{cvn} is a set of scores that describe how likely the observed neutrino interaction is to be $\numu$ \dshort{cc}, $\nue$ \dshort{cc}, $\nu_\tau$ \dshort{cc}, or \dword{nc}. The \dshort{cvn} score distributions for \dshort{spvd} are shown in Figure~\ref{fig:lbl_cvn_score}, displaying a good degree of separation between signal and background.  For comparative purposes, the equivalent scores for \dshort{sphd} are shown in Figure~\ref{fig:lbl_cvn_score_hd}.

\begin{dunefigure}
[\dshort{cvn} score distributions for respective signal and backgrounds, FHC mode, \dshort{spvd}]
{fig:lbl_cvn_score}
{\dshort{cvn} score distributions for their respective signal and backgrounds in \dshort{fhc} beam mode for \dshort{spvd}.  Left: $\nu_\mu$ \dshort{cc} score.  Right: $\nu_e$ \dshort{cc} score.}
\includegraphics[width=.49\textwidth]{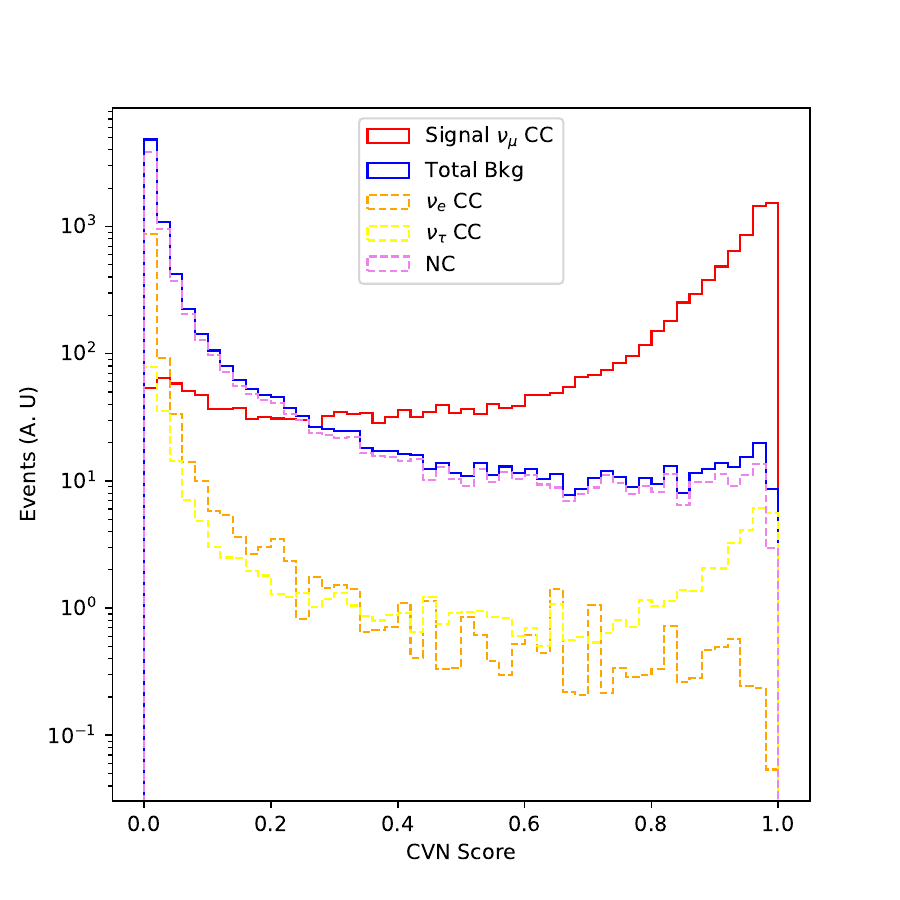}
\includegraphics[width=.49\textwidth]{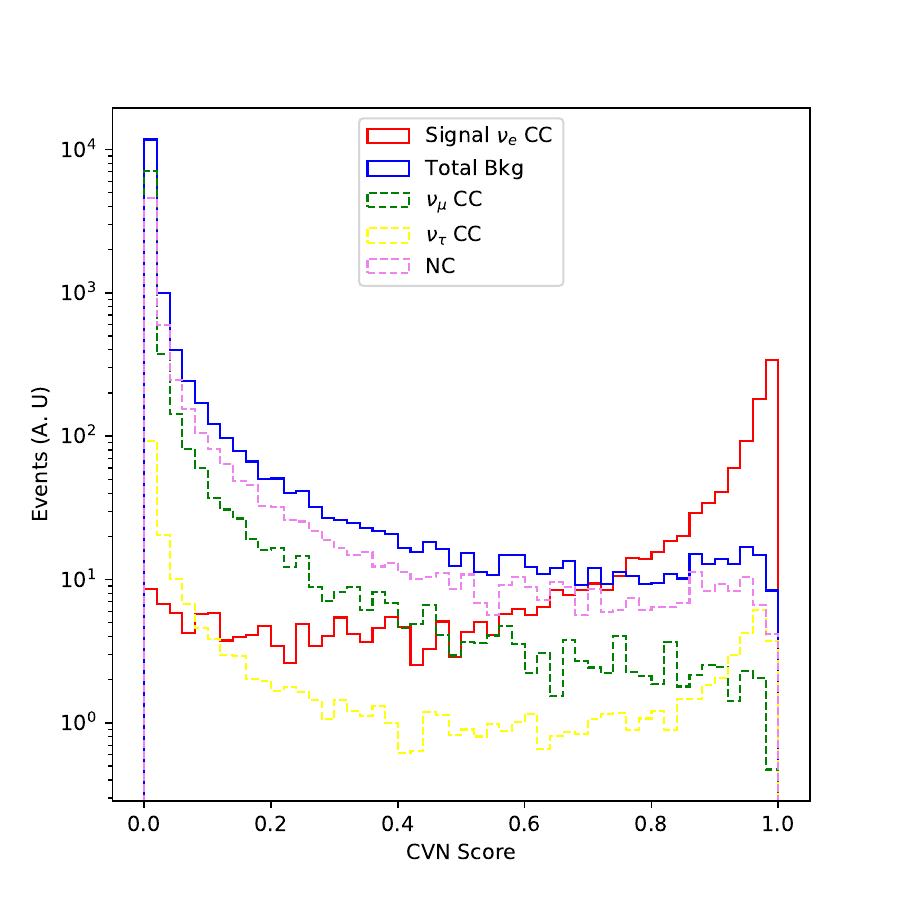}
\end{dunefigure}

\begin{dunefigure}
[CVN score distributions for respective signal and backgrounds, FHC mode, \dshort{sphd}]
{fig:lbl_cvn_score_hd}
{\dshort{cvn} score distributions for their respective signal and backgrounds in \dshort{fhc} beam mode for \dshort{sphd}.  Left: $\nu_\mu$ \dshort{cc} score.  Right: $\nu_e$ \dshort{cc} score.}
\includegraphics[width=.49\textwidth]{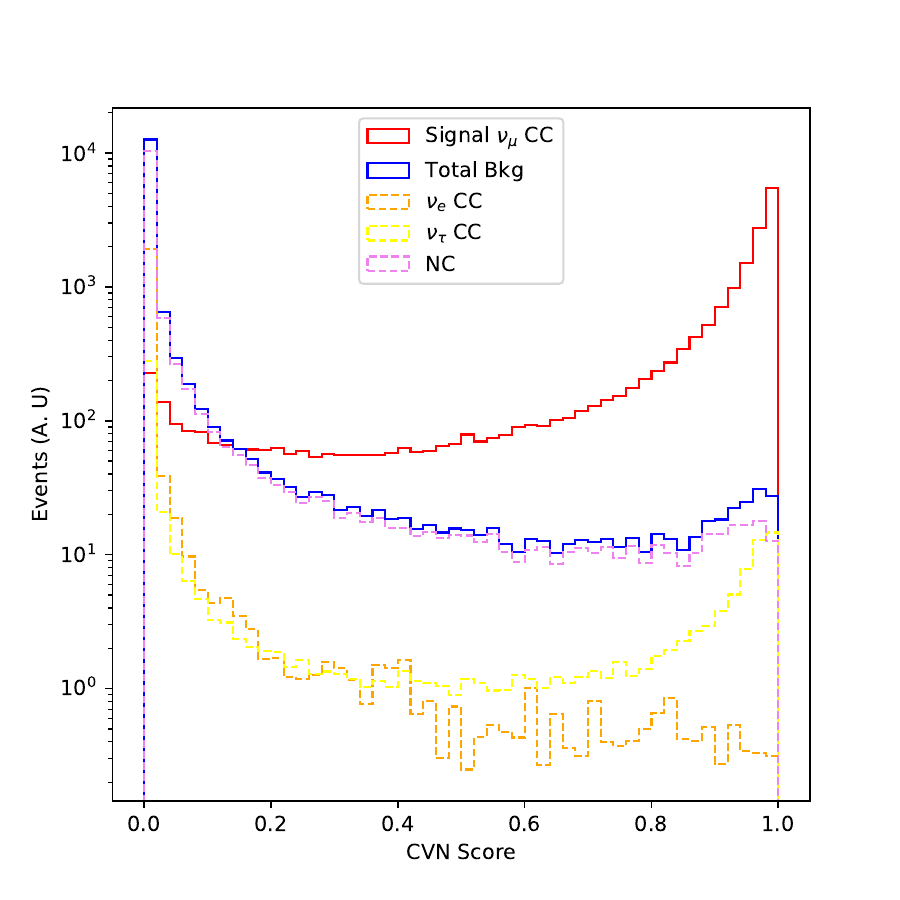}
\includegraphics[width=.49\textwidth]{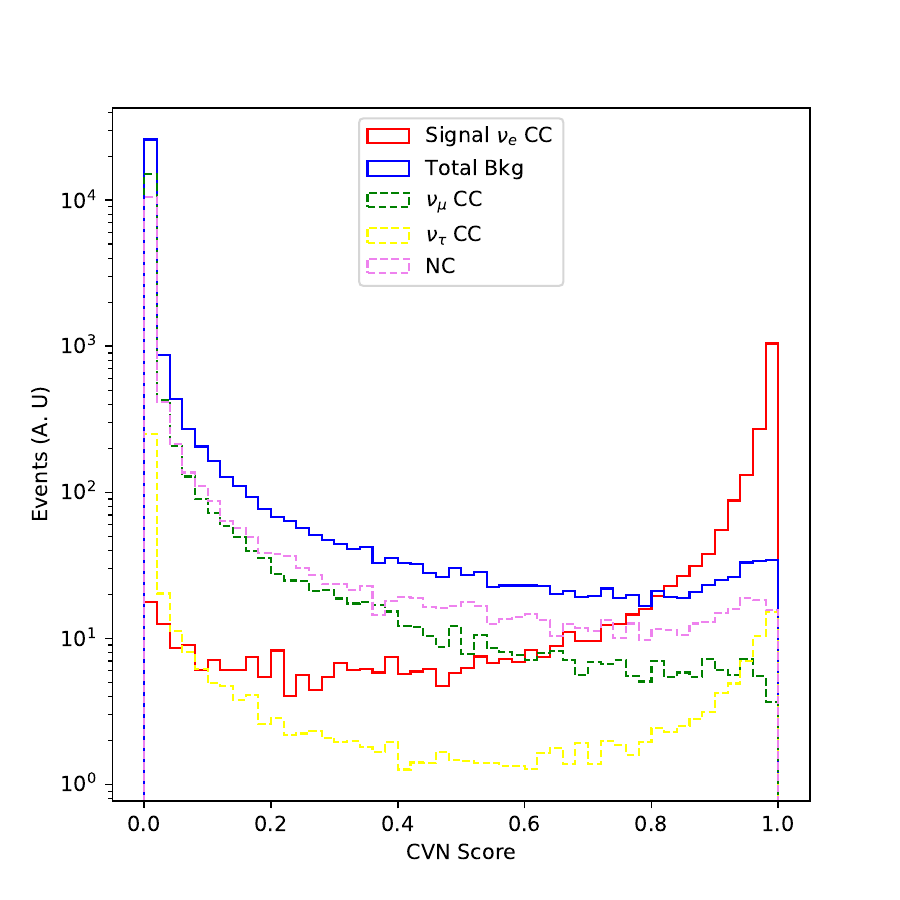}
\end{dunefigure}

The trained \dshort{cvn} was used to develop a neutrino selection, where a tuned cut was applied to the \dshort{cvn} score. The cut was separately tuned to maximize the product of selection efficiency %times 
and purity for $\numu$ \dshort{cc} and $\nue$ \dshort{cc} interactions. This procedure was followed in both the \dshort{sphd} and \dshort{spvd} detector simulations.  Figure~\ref{fig:lbl_cvn_eff} shows the resulting selection efficiency as a function of true neutrino energy for the $\numu$ \dshort{cc} and $\nue$ \dshort{cc} selections while  Table~\ref{table:lbl_cvn_eff_pur} shows the overall selection efficiencies and purities. The overall selection performance is similar between the two detector designs, with the most notable difference being a small divergence in the $\nue$ \dshort{cc} purity. The \dshort{spvd} instance of the \dshort{cvn} is much more recent and is still under active development and tuning, and it is expected that these small differences will converge over time.  

The cut on the \dshort{cvn} score can also be tuned to achieve equal purities between the \dshort{sphd} and \dshort{spvd} versions, where the \dshort{sphd} cut is optimized and the \dshort{spvd} cut is forced to meet the same selection purity. With the current version of the \dshort{spvd} \dshort{cvn} tuning, this gives an $8\%$ decrease in average $\nue$ \dshort{cc} efficiency. As the purities are identical, all of the background uncertainties would be the same in a %long-baseline 
\dshort{lbl} sensitivity analysis. This selection would therefore result in the same sensitivities, but with 8\% additional required exposure.  This does not account for the increased fiducial volume size of \dshort{spvd}, which would reduce the required exposure. Given the relative immaturity of the \dshort{spvd} reconstruction, this estimated increase in exposure time should be taken as a worst-case scenario.

\begin{dunefigure}
[Neutrino selection efficiencies in FHC beam mode using the CVN]
{fig:lbl_cvn_eff}
{The neutrino selection efficiencies in \dshort{fhc} beam mode using the \dshort{cvn}.  Left: $\nu_\mu$.  Right: $\nu_e$.}
\includegraphics[width=.49\textwidth]{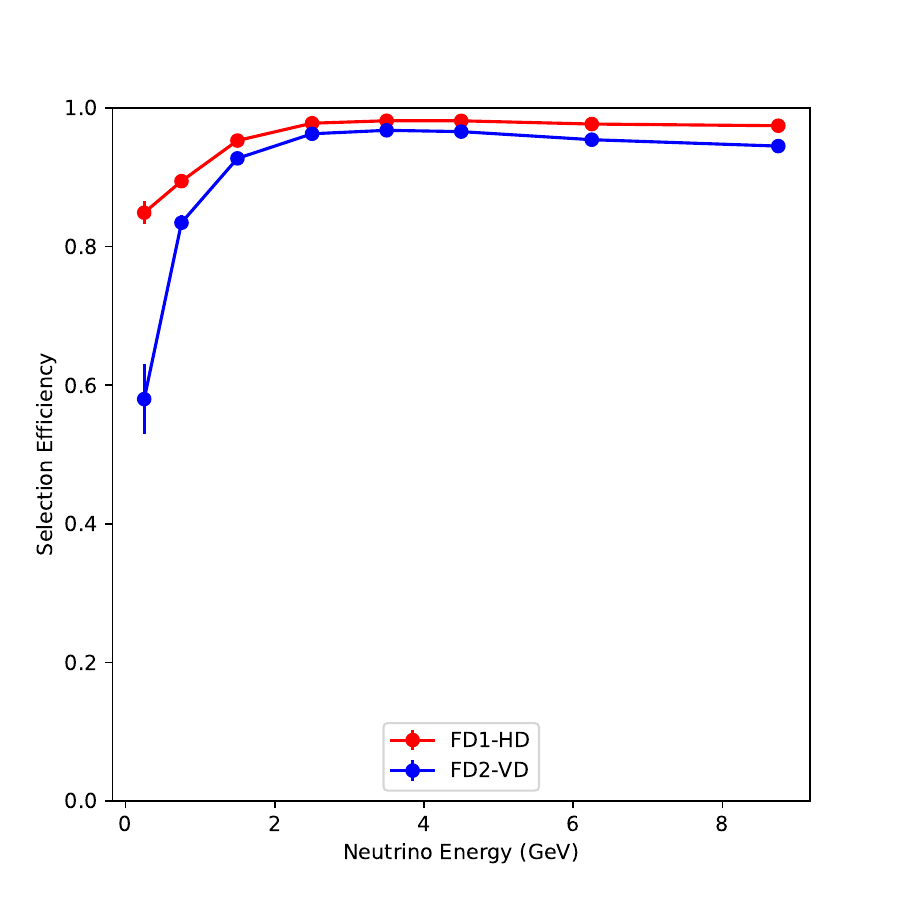}
\includegraphics[width=.49\textwidth]{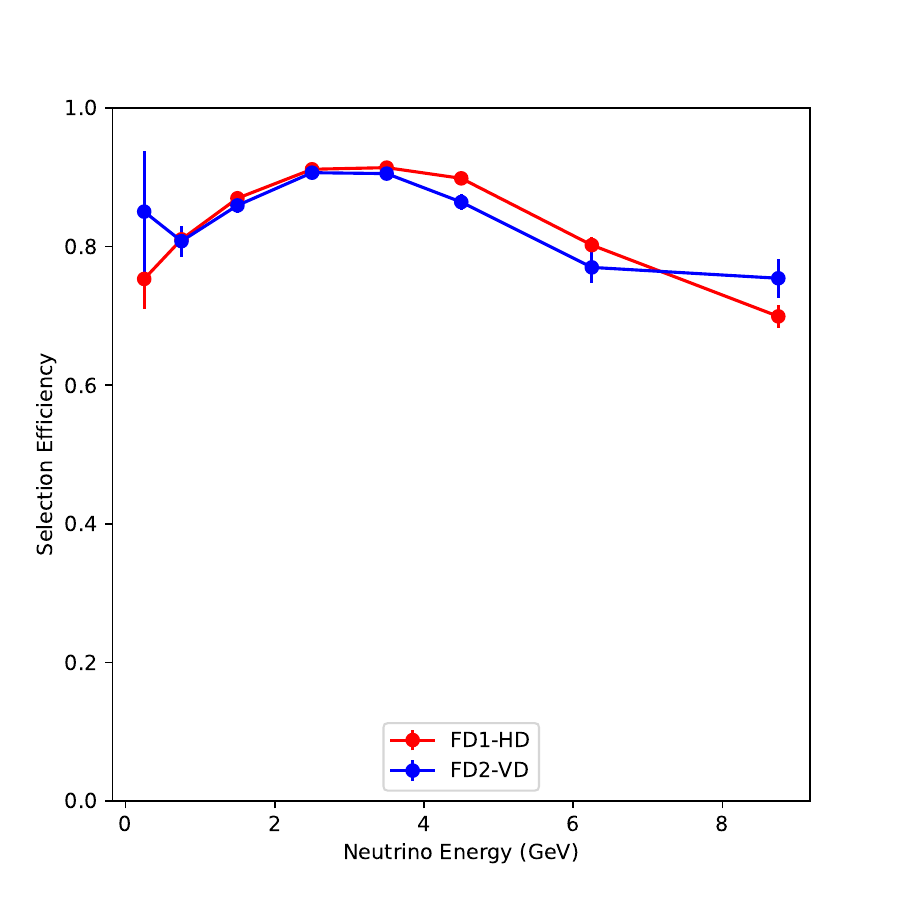}
\end{dunefigure}

\begin{table}[!b]
\begin{center}
\begin{tabular}{ l c c c c } 
 \hline
 & $\nue$ CC efficiency & $\nue$ CC purity & $\numu$ CC efficiency & $\numu$ CC purity \\
 \hline
\dshort{spvd} & 84\% & 83\% & 94\% & 93\%\\
\dshort{sphd} & 86\% & 88\% & 97\% & 94\% \\
 \hline
\end{tabular}
 \caption[Overall neutrino selection efficiencies and purities using the \dshort{cvn}]{The overall neutrino selection efficiencies and purities using the \dshort{cvn}.  The results for the \dshort{spvd} and \dshort{sphd} detectors are shown.}
 \label{table:lbl_cvn_eff_pur}
\end{center}
\end{table}

\subsection{Neutrino Energy Reconstruction}
\label{sec:enureco}

The neutrino energy reconstruction is based on the algorithm developed for \dshort{sphd}~\cite{DUNE:2020ypp}. The method uses the \threed reconstruction provided by \dshort{pandora} to identify, and separate out, the leading lepton and hadronic system. The lepton is always taken to be the longest primary reconstructed track or the highest-charge primary reconstructed shower, depending on whether the neutrino interaction is being reconstructed under a $\nu_\mu$ or $\nu_e$ hypothesis. All of the reconstructed hits not associated with the reconstructed lepton are assigned to the hadronic system. The neutrino energy is then calculated as

\begin{equation}
E_{\nu} = E_{\textrm{lepton}}^{\textrm{cor}} + E_{\textrm{hadron}}^{\textrm{cor}},
\end{equation}

where $E_{\textrm{lepton}}^{\textrm{cor}}$ and $E_{\textrm{hadron}}^{\textrm{cor}}$ are the reconstructed energies of the leptonic and hadronic systems respectively.  The muon energy is %reconstructed by ranging 
estimated by range or multiple Coulomb scattering, depending on whether the muon is fully contained in the detector.  Both the electron and hadron energies are estimated using reconstructed hit-based calorimetry.  $E_{\textrm{lepton}}^{\textrm{cor}}$ and $E_{\textrm{hadron}}^{\textrm{cor}}$ are independently corrected using a simulation-based calibration curve to minimize the energy reconstruction bias; this calibration curve was retuned for the \dshort{spvd} detector.

The fractional energy residuals for the three %hypotheses
event categories are shown in Figure~\ref{fig:lbl_enu_residual}, and the overall energy resolutions (the width of the Gaussian fits to the fraction energy residuals) are summarized in Table~\ref{table:lbl_enu_resolutions}. The full reconstructed neutrino energy spectrum is used in all oscillation studies; the Gaussian fit results given are merely a convenient way to quantify the observed resolution. Reconstructed energy performance is nearly identical between \dshort{sphd} and \dshort{spvd}.

\begin{dunefigure}
[Reconstructed neutrino energy residuals for different reconstruction hypotheses]
{fig:lbl_enu_residual}
{The reconstructed neutrino energy residuals for the different reconstruction hypotheses with overlaid Gaussian fits to the center region of the distributions. Left: $\nu_\mu$ hypothesis with contained muon. Middle: $\nu_\mu$ hypothesis with exiting muon. Right: $\nu_e$ hypothesis.}
\includegraphics[width=.32\textwidth]{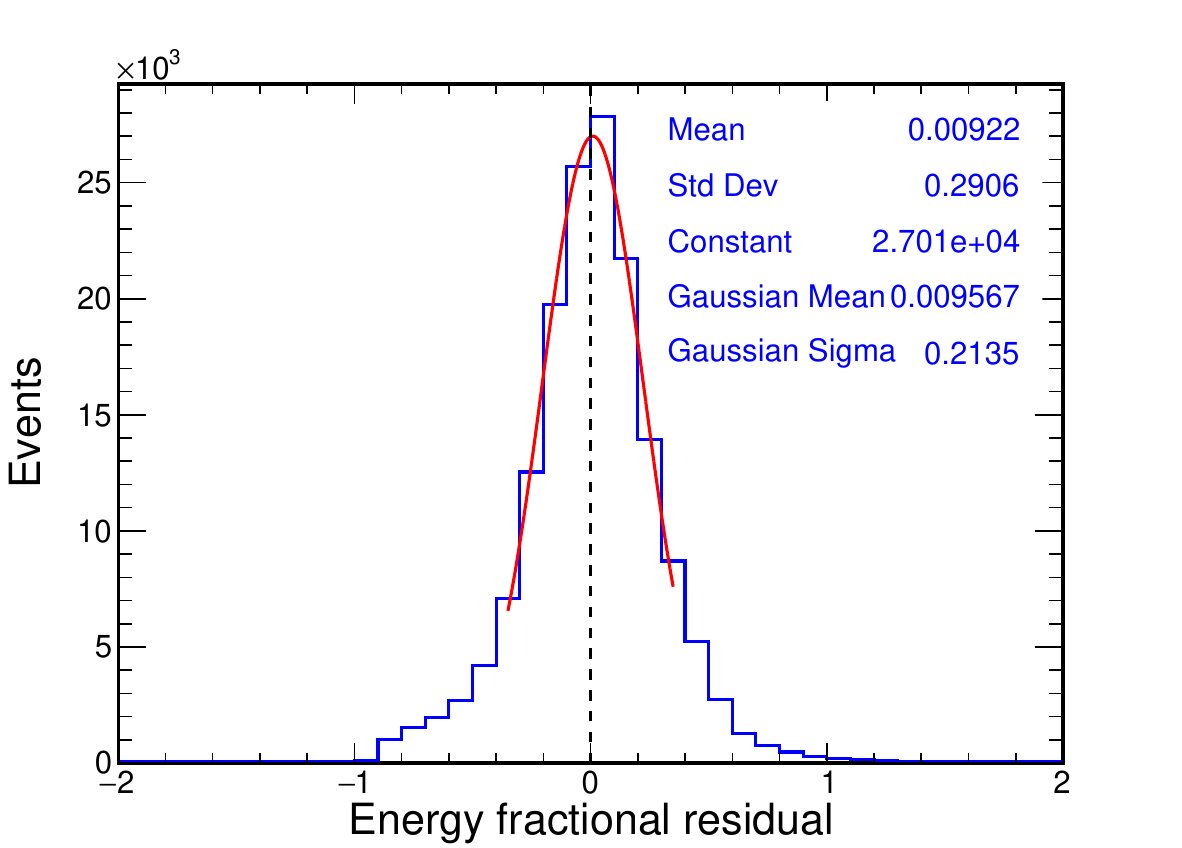}
\includegraphics[width=.32\textwidth]{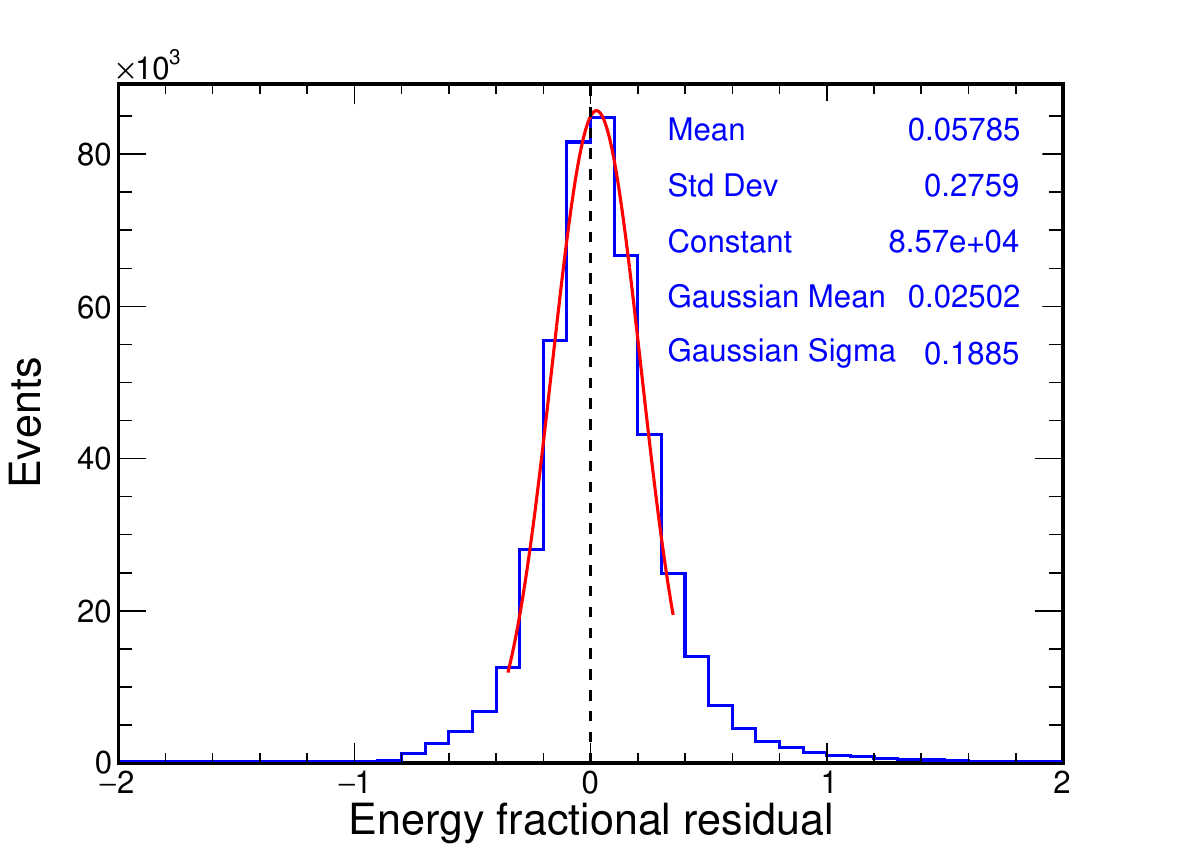}
\includegraphics[width=.32\textwidth]{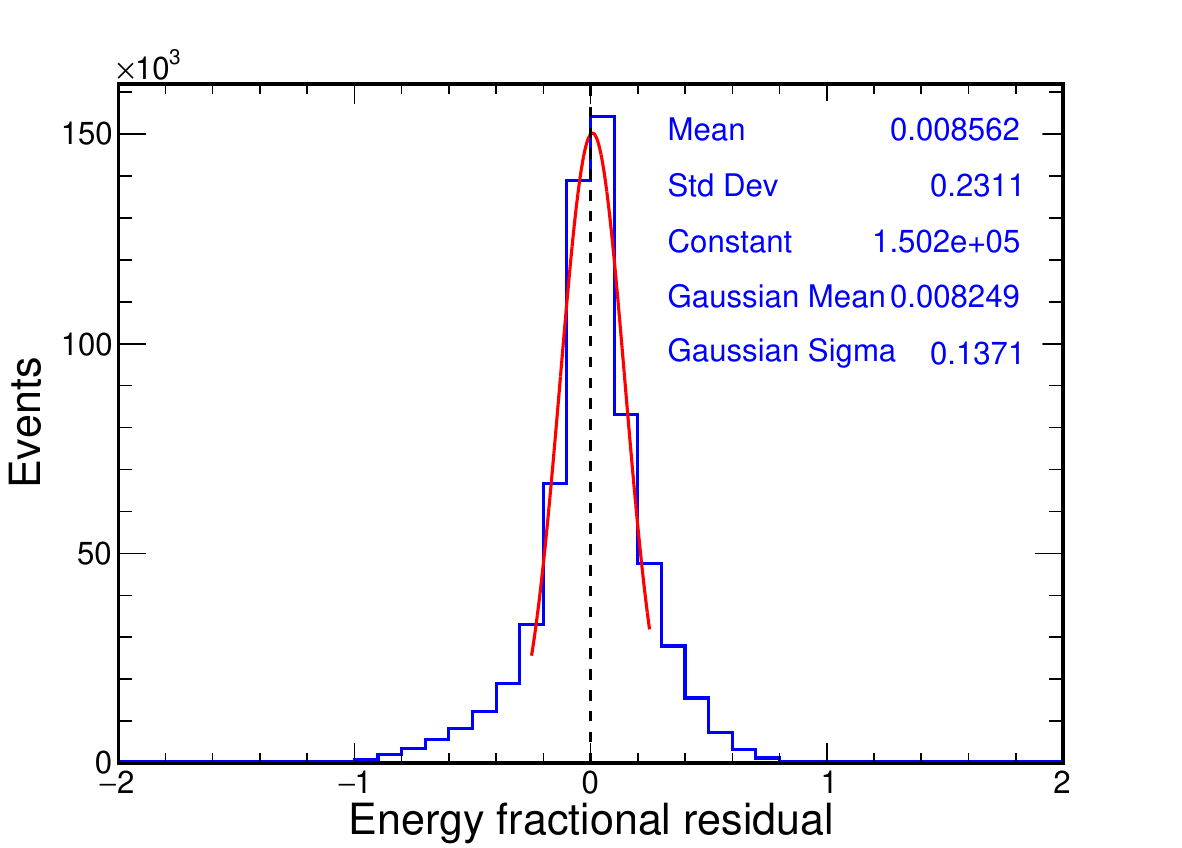}
\end{dunefigure}

\begin{table}
\begin{center}
\begin{tabular}{ c c c } 
 \hline
 Event hypothesis & Vertical Drift & Horizontal Drift \\ 
 \hline
 $\numu$ CC with contained $\mu$ track & 21\% & 18\% \\ 
 $\numu$ CC with exiting $\mu$ track & 19\% & 20\% \\
  $\nue$ CC & 14\% & 13\% \\
 \hline
\end{tabular}
 \caption[Reconstructed neutrino energy resolutions for the three event hypotheses]{The reconstructed neutrino energy resolutions for the three event hypotheses. The energy resolutions for both  \dshort{spvd} and \dshort{sphd} are shown.}
 \label{table:lbl_enu_resolutions}
\end{center}
\end{table}

\dshort{spvd} will have a targeted calibration program 
%\fixme{again - right word, bespoke? -- changed bespoke to targeted}
that is currently under %investigation
development, and will be very similar to that for the \dshort{sphd}. It is expected that detector uncertainties that are relevant for %long-baseline 
\dshort{lbl} oscillation sensitivities, such as energy scales and particle responses, will be at the same level in both detector modules. A dedicated hadron and electron test beam run will occur for both \dshort{sphd} and \dshort{spvd}, so any hadronic response uncertainties that are constrained by test beam data will be very similar. To the extent that calibration uncertainties are uncorrelated between \dshort{sphd} and \dshort{spvd}, it may be possible to decrease the overall systematic uncertainty because specific effects will impact only the portion of the overall \dshort{fd} sample from a given module. Uncertainties on particle propagation in LAr, for example, will be fully correlated between \dshort{sphd} and \dshort{spvd}. Any increase in sensitivity due to uncorrelated detector uncertainties is expected to be very small.

\subsection{PDS Channel Saturation for High Energy Beam Events}
\label{sec:chsat}

One of the \dshort{pds} requirements is the ability to retrieve information contained in the photon signal even for extreme events, e.g., when higher energy interactions ($\mathcal{O}$(10\,GeV)) occur near the planes instrumented with \dword{xarapu} devices. Estimates were performed for the fraction of \dshort{pd} electronic channels that can saturate using the \dshort{larsoft} \dword{mc} simulation with neutrino beam events. These are meant to ensure the 14-bit \dshort{pds} readout design satisfies the needs of the experiment and to %verify if 
determine whether signal saturation effects are a concern for the \dshort{spvd} module in some particular cases.

Two data sets of $\rm 10^5$ events each of $\nu_{\mu}$ and $\nu_{e}$ beam neutrinos were used. The optical waveforms were obtained for all \dshort{pds} readout channels in the simulation taking into account all physical effects and parameters described in Section~\ref{sec:ph:PDS-Simulation}. The number of channels in which the signal saturates ($\rm amplitude > 2^{14}$ \dshort{adc}) and the total number of valid waveforms in each event were calculated. A waveform is considered valid if its amplitude %goes above 
exceeds a threshold of 1.5\,PE. Figure~\ref{fig:pds-saturation} (left) shows the distribution of the registered valid waveforms' peak amplitudes where the maximum value peak indicates the number of times a channel signal was saturated over all analyzed data events.

\begin{dunefigure}
[Peak amplitude distribution and fraction of events that saturate]
{fig:pds-saturation}
{Left: Distribution of the channels peak amplitude. Maximum values corresponds to the maximum amplitude of $\rm 2^{14}~ADC$. Right: Fraction of events as function of the fraction of its valid channels per event that saturate above a certain minimum level required.}
\includegraphics[width=.49\textwidth]{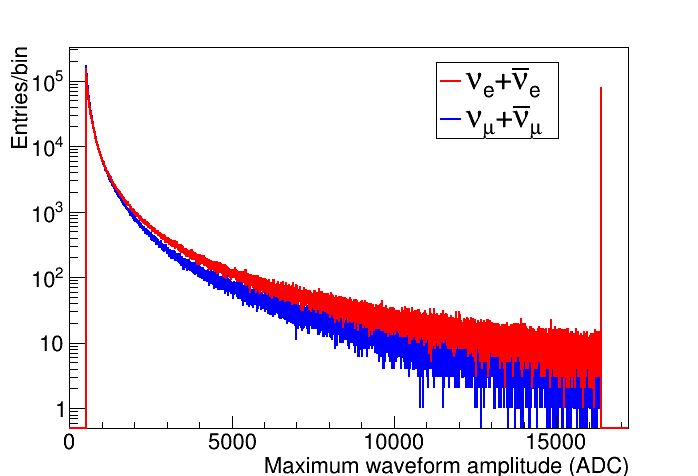}
\includegraphics[width=.49\textwidth]{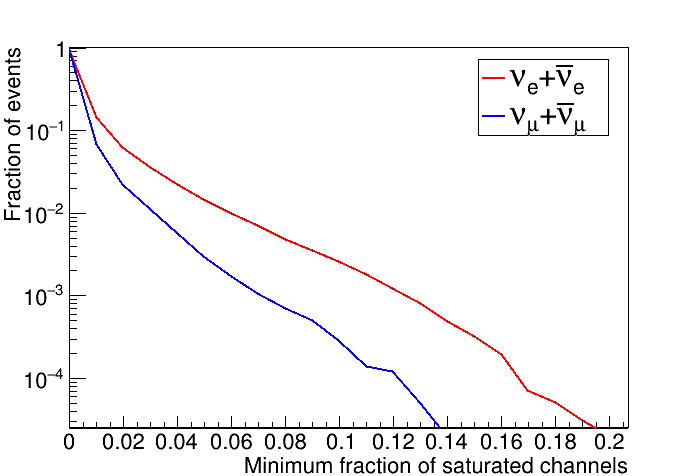}
\end{dunefigure}

Figure \ref{fig:pds-saturation} (right) shows the fraction of beam %muon and electron neutrino 
\numu and \nue events %which have 
in which a fraction of %its 
the valid channels saturate %d 
above a certain minimum level. The obtained fractions of events are kept at levels that satisfy the \dshort{pds} requirements. 

The small fraction of saturated channels per event increases with the incoming neutrino energy, with a slightly higher probability for primary interaction positions closer to the cathode. The effect on the \dshort{pds} energy reconstruction capability  was evaluated by calculating the decrease in the hit area caused by saturation. For %muon (electron) neutrino 
\numu (\nue) events the average fraction reduction of the summed signal integrals was estimated to be $\rm \sim 2(3)\%$ per event and an average of $\rm 15(16)\%$ for each saturated channel. These numbers indicate \dshort{pds} energy calibration and reconstruction should not be significantly affected by saturation effects and that these rare situations representing extreme events do not compromise the requirement of up to 20\% of saturated \dshort{pd} channels in a given beam event (Table~\ref{tab:PD-VD-Requirements}). We note that the dynamic range simulated in our studies, up to approximately $2^{14}/10\simeq 1600$~\dwords{pe}, is lower than our 1--2000~\dwords{pe} specification (see Chapter~\ref{chap:PDS}), so an even lower fraction of saturated events is expected.

\subsection{Conclusions}

Despite the relative infancy of the \dshort{spvd} simulation and reconstruction, in addition to some known differences between the \dshort{sphd} and \dshort{spvd} software, the performance is similar between the two detector designs for GeV neutrino interactions relevant for \dshort{lbl} oscillation physics. The small differences in performance will be investigated and are expected to converge over time. The \dshort{sphd} simulation and reconstruction will be updated and a new simulation campaign will be run as part of this effort.  

The \dshort{lbl} oscillation sensitivities, including the sensitivity to observe %CP violation
\dword{cpv}, depend on the selection efficiency, energy reconstruction, and detector uncertainties, all of which are very similar in both detector configurations. Hence, the oscillation sensitivities obtained in the \dshort{sphd} TDR analysis~\cite{DUNE:2020ypp} are also valid for a configuration where the \dshort{fd} contains both \dshort{sphd} and \dshort{spvd} modules.

An update to the \dshort{lbl} oscillation sensitivity estimates in DUNE is currently underway and expected to be available in 2024. The primary improvement, relative to the previous estimates, is a more realistic treatment of the \dword{nd}, as its design is now at a stage where realistic simulations are possible. However, as part of this update, it is planned to also include full simulation and reconstruction samples of the \dshort{spvd} module in addition to those of the \dshort{sphd}. Given the analysis and results presented in this chapter, it is not expected that this will have any significant impact on the results.

%%%%

\chapter{Charge Readout Planes}
\label{sec:AR}
%\tableofcontents
%%%%%%%%%%%%%%%%%%%%%%%%%%%%%%%%%%%%%%
\section{Introduction}
 \label{sec:AR:intro}
 
Since the \dword{lartpc} technology was first proposed~\cite{rubbia77}, 
intense R\&D and novel ideas have fueled %its 
continuous evolution. 
Major developments in most of the \dshort{lartpc} core components have led to substantial improvements in 
 %their 
 performance and stability. 
These include new \dword{pd} technologies, improvements in the \dword{hv}  and \dword{ce} systems, resistive cathode designs, and modular \dwords{fc}. 
Charge collection has changed very little, however, and multi-layer wire planes remain the standard anode technology.

Only in the last few years have new anode ideas been proposed and partially developed to replace the conventional wire planes. A modified version of the traditional \dshort{lartpc} chamber has been under development since 2010 in a \dword{dp} configuration~\cite{Cantini:2014xza}, in which the usual wire arrays are replaced by multiple \dword{pcb} \dword{lem} planes, which are derived from Gas Electron Multiplier (GEM) detectors~\cite{Sauli:2016eeu}. Leveraging the advancements in \dshort{lem} technology, 
this dedicated R\&D program has been developing an anode technology using perforated \dshort{pcb}-based \dword{cro} with projecting electrodes immersed in \dword{lar}~\cite{Baibussinov:2017ogy}.

The \dword{spvd} anode design will implement \dshort{pcb}-based charge readout using stacked, perforated \dshort{pcb}s with etched electrode strips. The \dshort{pcb}s are attached to a composite frame to form \dwords{crp}; the frame provides mechanical support and maintains the required planarity. Figure~\ref{fig:supercrp2} provides an overview of the \dword{anodepln} structure and Table~\ref{tab:complist} lists its components and their quantities and sizes. As discussed in Chapter~\ref{ch:execsumm}, the \dshort{spvd} detector module is split vertically into two drift volumes, with \dshort{crp}s at the top and bottom of the cryostat forming the \dshort{anodepln}s, both of which are immersed in \dshort{lar}, and a horizontal cathode plane placed midway between them. The \dshort{crp}s in the top drift volume are collected into sets of ``superstructures'' that are
suspended from the cryostat roof and the bottom \dshort{crp}s are supported by posts positioned on the cryostat floor.  The top drift volume electronics components are mounted  on signal feedthrough flanges on the cryostat ceiling. The \dshort{crp}s at the bottom have integrated \dshort{ce} attached to them.

\begin{dunefigure}
[Exploded view of a top superstructure and CRPs]
{fig:supercrp2}
{A top superstructure (green structure on top) that holds a set of six \dshort{crp}s, and below it an exploded view of a \dshort{crp} showing its components: the \dshort{pcb}s (brown), adapter boards (green) and edge connectors that together form a \dshort{cru}, and composite frame (black and orange). Dimensions are given.}
\includegraphics[width=0.99\textwidth]{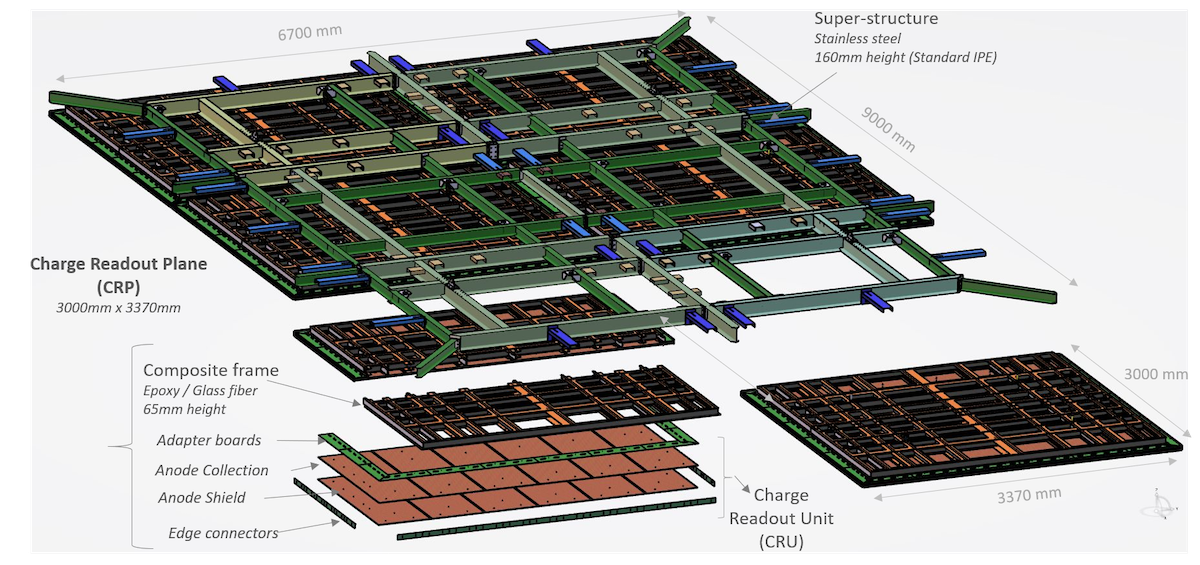}
\end{dunefigure}

The \dshort{anodepln} design for the \dshort{spvd} provides three-view charge readout and is constructed of two stacked,  perforated  \dshort{pcb}s with biased strip electrodes on one or both faces. The three sets of electrode strips are set at different angles relative to each other, as shown in Figure~\ref{fig:3V_anode_layout}, to provide charge readout from different projections. A \dshort{pcb}-based anode is less expensive and less delicate than a wire-based anode plane and can be produced more rapidly using commercially available tools. It therefore offers attractive potential reductions in cost, schedule, and risk. The design of the \dshort{crp}s incorporates the lessons learned from \dword{pddp}~\cite{DUNE:2018mlo}, intense R\&D activities from 2019 to 2021, and ongoing prototyping activities. \dshort{crp}s are modular structures for purposes of fabrication, assembly, and transportation; they are constructed from two half-size \dshort{anodepln} pieces, called \dshort{cru}s, and a composite frame.

To protect against possible risk from cathode charge injection, the \dshort{pcb} view facing the drift volume is a shield plane with no readout.  The back face of this \dshort{pcb} becomes the first induction plane, the front face of the second anode \dshort{pcb} is the second induction plane, and its back face (facing either the cryostat roof  facing 
or the bottom membrane) 
is the collection plane.

\begin{dunefigure}
[Basic electrode strip configuration for the three-view top and bottom CRUs]
{fig:3V_anode_layout}
{Illustration of the basic electrode strip configuration for a top (right) and bottom (left) \dshort{cru}s. Each 
\dshort{cru} is constructed from a one-view (induction-1, magenta) \dshort{pcbp}, overlaid with a two-view (induction-2, blue and collection, green) \dshort{pcbp},  
and readout electronics adapter boards (along the edges) that host cable connectors for the top \dshort{anodepln},
or \dshort{ce} readout modules for the bottom. 
The lines at half-height indicate electrical discontinuity for the collection plane strips. 
Strip pitches and adapter boards are not drawn to scale.}
\includegraphics[width=.85\linewidth]{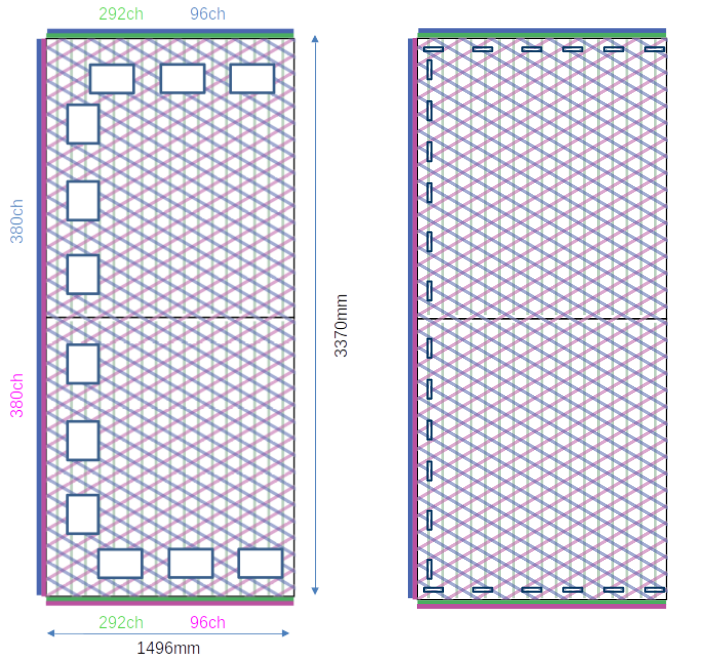}
\end{dunefigure}

\begin{dunetable}
[Anode plane component list with quantities and sizes]
{p{0.2\textwidth}|p{0.23\textwidth}|p{0.23\textwidth}|p{0.2\textwidth}}
{tab:complist}
{Reference three-view anode plane component list with quantities and sizes.}
Component & Sub-components & Quantity & Size \\
\toprowrule
\dshort{pcb} segment  & (none) &  12 per \dshort{cru} (glued and stacked), 3840 total &  in 6 different flavors and 3 sizes. See Section \ref{subsec:3V}\\ 
\colhline
\dshort{pcb} panel  & \dshort{pcb} segment &  2 per \dshort{cru} (stacked), 640 total &   \pcbpaneldim \\ 
\colhline
View (also called ``layer'') & electrode strips on \dshort{pcb} &  1 set of parallel strips per \dshort{pcb} panel side &  See Table~\ref{tab:anode_parameters} \\ 
\colhline
Adapter board & 4-layer \dshort{pcb} + bias capacitors and resistors & 12 per \dshort{cru}, 3840 total  & in 7 different flavors and sizes for top and bottom  \\
\colhline
Edge cards & 4-layer \dshort{pcb} + small connectors & 24 per \dshort{cru}, 7680 total  & in 3 different flavors and sizes for top and bottom  \\
\colhline
\dshort{cru}  & 2 \dshort{pcb} panels + 12 adapter boards + 24 edge cards& 2 per \dshort{crp}, 320 total &  \crudim\ \\ 
\colhline
\dshort{crp} & 2 \dshort{cru}s + composite frame  & 80 per anode plane, 160 total &  \crpdim \\ 
\colhline
Top superstructure &  6 or 2 \dshort{crp}s  & 16 (top anode plane only, twelve 6-\dshort{crp}, four 2-\dshort{crp}) & 9.0$\times$6.7\,m and 3.0$\times$6.7\,m  \\ 
\colhline
Anode plane & 80 \dshort{crp}s  & 2  &  \anodeplndim \\ 
\end{dunetable}

Electronics adapter boards, an interface between the \dshort{anodepln}s and readout electronics for strip biasing and charge readout, are attached to the two-\dshort{pcb} stack. Vertical interconnection between the \dshort{anodepln}s and adapter boards are done via small PCBs called ``edge cards.'' Small connectors on the edge cards are in contact with the strips on the \dshort{pcb} end and with the electrical pads on the adapter boards at the other end. Anode \dshort{pcb}s, adapter boards, and edge cards are connected to a composite frame that serves as the mechanical support structure.  Together, the perforated \dshort{pcb}s, adapter boards, and edge cards form a \dshort{cru}, and two \dshort{cru}s plus a composite frame form a \dshort{crp}. Each \dshort{spvd} \dshort{anodepln} is a grid of 80 (20 by 4) \dshort{crp}s that spans the horizontal area of the detector.

\dshort{pcb}-based readout in \dshort{lar} has well defined induction and collection signals, and is well suited for a vertical drift, which enables a larger, unobstructed active volume with a longer drift distance than DUNE's \dword{sphd} offers. Bias voltages that direct electron trajectories to the holes enable uniform, fully localized and sharp signals both for collection and induction signals.
%. This effect is more evident for the induction signals, but also applies to the collection signals. 

The \dshort{pcb}-based design is well suited to capturing the induction signal for tracks that have a large dip angle (i.e., are nearly parallel to the \efield direction). Induction signals in \dwords{lartpc} are bipolar with positive and negative lobes, and the effect of signal cancellation between the lobes grows as the dip angle increases. 
The \dshort{pcb} anode design aims to enhance the asymmetry of the induction signal by pairing a short high lobe formed inside the holes with a long low lobe formed in an extended open low-field region. This asymmetry will reduce the cancellation effect, enhancing signals from tracks with large dip angles. 

%%%%%%%%%%%%%%%%%%%%%%%%%%%%%%%%%%%%%%%%%%%%%%%%%%%%%%%%%%%%%%%%%%%%%%%%%%%%%%%%%%%%%%
\section{Specifications}
\label{sec:crp-specs}

The principal specifications for the \dwords{crp} concern the anode 
planarity, the transparency to liquid flow, and the geometrical structure configurations. Table~\ref{tab:reqs:FD2-CRP} lists these and the other specifications developed for the \dshort{crp} design to meet overall detector performance requirements.  

\begin{footnotesize}
\begin{longtable}{p{0.12\textwidth}p{0.18\textwidth}p{0.17\textwidth}p{0.25\textwidth}p{0.16\textwidth}}
\caption{CRP anode plane specifications} \\
  \rowcolor{dunesky}
       Label & Description  & Specification \newline (Goal) & Rationale & Validation \\  \colhline
  \newtag{FD-2}%was: SP-APA-2}
  { spec:anode-active-area }  & Active area  &  Maximize total active area. &  Maximize area for data collection  &  ProtoDUNE  \\ \colhline

  \newtag{FD-6}%SP-APA-6}
  { spec:anode-bad-channels }  & Missing/unreadable channels  &  $<$1\%, with a goal of $<$0.5\% &  Reconstruction efficiency &  ProtoDUNE \\ \colhline
  
 \newtag{FD-7}{ spec:misalignment-field-uniformity }  & Drift field nonuniformity due to component alignment  &  $<\,1\,$\% throughout volume &  Maintains anode, cathode,  FC orientation and shape. &  \dshort{vdmod0} \\ \colhline
    
     \newtag{FD2-31}{ spec:fd2vd-anode-flatness }  & 
    CRP anode plane global flatness &  
    $<$\,20\,mm &  
    Maintains drift field uniformity of $<$\,1\% throughout each drift region &  \coldbox test
    \\ \colhline
    
        \newtag{FD2-32}{ spec:fd2vd-crp-transp }  & 
    \dshort{crp} minimum permeability & 
    $>$\,15\% &
    Allows efficient local heat dissipation and free \dshort{lar} circulation &  
   CFD simulation
    \\ \colhline
    
        \newtag{FD2-33}{ spec:fd2vd-crp-gap }  & 
   Gaps between \dshorts{crp} &  
   5\,mm for CRPs within superstructure; \newline 10\,mm between superstructures &  
   Minimizes loss of \dshort{fv} and allows space for cabling &  ProtoDUNE
    \\ \colhline
    
        \newtag{FD2-34}{ spec:fd2vd-shield-pln }  & 
    CRP shield plane on cathode-facing side &  
     &  
    Reduces impact on electronics from cathode discharge&   cold box test and ProtoDUNE
    \\ \colhline
    
        \newtag{FD2-35}{ spec:fd2vd-strip-width }  & 
  \dshort{crp} strip width and pitch  &  
  $<$\,8.5\,mm (ind.);\newline $<$\,5.5\,mm (coll.); \newline gaps 0.5\,mm &
  \dshort{s/n} consistent with 100\% hit reconstruction efficiency for \dshorts{mip}.  Spacing provides 1.5\,cm vertex resolution in $y$-$z$ plane. &  
   ProtoDUNE and bench tests
    \\  \colhline
    
           \newtag{FD2-36}{ spec:fd2vd-gap-pcb }  & 
    Vertical gap between the \dshort{pcb}s &  >\SI{8}{mm}
     &  
    Allows >5$\mu$s drift time between the two \dshort{pcb}s for better separation of the signal and ensures a safe running environment at voltages lower than the component ratings. &   Prototyping and bench tests
    \\ \colhline
 
           \newtag{FD2-37}{ spec:fd2vd-bias-volt }  & 
   Anode plane bias voltages &  <2\,kV
     &  
    Beyond this value the size and cost of capacitors shoot up and choices plummet. &   Prototyping and bench tests
    \\ \colhline
   
          \newtag{FD2-38}{ spec:fd2vd-hole-sep }  & 
   Minimum wall thickness between holes &  0.5\,mm
     &  
     This is for mechanical properties of the perforated \dshort{pcb}; given the full electron transparency, the hole size does not have any effect on charge collection. &   Prototyping and bench tests
    \\ \colhline
   
\label{tab:reqs:FD2-CRP}
\end{longtable}
\end{footnotesize}
Regarding FD-7, a maximum overall deformation on an individual \dshort{crp} of 10\,mm  allows the \efield amplitude to remain within 1\% of the nominal value, and limits the drift line deflection to 3\,mm.

In FD2-33 the horizontal gaps of 5\,mm  and 10\,mm as listed refer to measurements at warm. These gaps allow for positioning the \dshort{crp}s  and accommodating the cables that support the cathode structure. Due to the different shrinkage properties of the frames and \dshort{pcb}s when cooled to \dword{lar} temperature, these gaps become about 4\,mm and 20\,mm at cold, respectively.

Regarding FD-35, vertical position tolerance of < 4\,mm between two adjacent \dshort{crp}s keeps the drift line distortion around the \dshort{crp} borders below 6\,mm.
%%%%%%%%%%%%%%%%%%%%%%
\section{Anode Plane Design} 
\label{subsec:3V}

As discussed in Section~\ref{sec:AR:intro}, the \dword{anodepln} design for \dword{spvd} features three layers of readout channels (two induction views and a collection view) plus a shield layer, and is constructed by stacking two double-sided anode \dwords{pcb}. The configuration is illustrated in Figure~\ref{fig:3V_anode_layout}. %In addition to the 
The readout strips on the three views are all at %holding readout channels with 
different angles relative to each other. The shield layer, % plane which does not have readout strips, 
which faces the cathode, provides additional safety for the readout electronics; 
when connected to a sufficiently large capacitor bank ($\mathcal{O}($$\mu$$F)$), it could block up to 95\% of the capacitive coupling between the cathode and the first induction plane. %The minor disadvantage of t
This additional layer of electrode %is the 
requires an increase in overall bias voltage differential across the entire anode stack.
The back side of the cathode-facing \dshort{pcb} %panel becomes 
is the first induction plane (induction-1), with strips %having strip angles 
running diagonally %on the CRP 
at \firstviewangle.
The front side of the second \dshort{pcb} %panel 
is the second induction plane (induction-2), with induction strips %running with an angle -30\dge. 
at \secondviewangle. Its back side is the collection plane, %on which the electrode 
with strips running perpendicular to the beam (90\dge) to offer the best charge measurement for the beam events. 

Dimensional details of the \dshort{pcb} layout and schematics of a cross-section of the anode \dshort{pcb} stack are listed in Table~\ref{tab:anode_parameters} and illustrated in Figure~\ref{fig:3V_hole_pattern}. The perforation pattern on %the second %anode 
 both \dshort{pcb}s is arranged such that the narrow gaps between two strips always bisect a row of holes. 
  The hole pattern on the cathode-facing \dshort{pcb} %is set to 
 aligns with the pattern on the second \dshort{pcb} to improve electron transparency and liquid flow. The strip pitch is 5.1\,mm for the collection view and 7.65\,mm for the induction views.
The gap between strips on the three views is set to 0.5\,mm. Strip lengths are 1.68\ for collection, and up to 1.72\,m for the induction views. The numbers of readout channels are 476, 476 and 584 per \dword{cru} for induction-1, induction-2 and collection, respectively, %reaching a 
totaling 3072 channels on %the entire 
a \dshort{crp}. 
The detailed design drawings for the perforated PCBs can be found at~\cite{edms2670965}. 

%$$$$$$$$$$$$$$$$$$$$$$$$$$$$$$$
\begin{dunetable}[Key parameters for the baseline anode design]
{|p{.34\textwidth}|l|l|l|}
{tab:anode_parameters}
 {Key parameters for anode design}
             & \multicolumn{3}{c|}{Three-view configuration}  \\
\rowtitlestyle    Parameter    & Induction 1      & Induction 2  & Collection      \\    \toprowrule
    Strip length [m]      &   up to 1.74  &   up to 1.74    &   1.68    \\ \colhline
   Strip pitch [mm] & 7.65 & 7.65 & 5.1
 \\ \colhline
    Strip gap [mm]       &   0.5  &   0.5    &   0.5    \\ \colhline
    Unit capacitance [pF/m]      &   103  &   103    &   81 \\ \colhline
    Total capacitance [pF]      &   up to 177  &   up to 177    &   135    \\ \colhline
    Number of strips per \dshort{cru}      &   476  &   476    &   584    \\ \colhline
    Number of readout channels per \dshort{crp}   & \multicolumn{3}{c|}{\channelspercrp} \\ \colhline
    Strip angle w.r.t. beam       &   \firstviewangle  &   -30\dge    &   90\dge    \\ \colhline
    Bias voltage [V] for a shield plane bias at -1500 V      &   -500  &   0    &   1000    \\  \colhline
    Hole diameter [mm]    &  2.4  &  2.4     &     2.4  \\  \colhline
    Inter-\dshort{pcb} gap within \dshort{crp} (at room temp.)  [cm] % was mm, Anne changed 1/27
    &    \multicolumn{3}{c|}{1}    \\  
 
    \end{dunetable}

Given the smaller separation between the strips on the \dshort{pcb} and higher dielectric constant of the \frfour, the anode \dshort{pcb} has higher capacitance per unit length relative to the wire-based charge readout.  
Electrostatic %finite element 
\dword{fea} calculations predict a capacitance per unit length of about $\sim$100\,pF/m for the strips immersed in \dword{lar}, versus 20\,pF/m for the \dword{sphd} \dword{apa} wires. The reference \crudim \dword{cru} segmentation keeps the detector capacitance and, as a consequence, its contribution to the electronic noise, similar to that of the APAs.

The %charge readout plane 
\dword{crp} and the bias voltages, illustrated in Figure~\ref{fig:3V_hole_pattern}, are designed to first focus the drifting electrons into the cathode-facing \dshort{pcb}'s perforations, then defocus them as they travel toward the second \dshort{pcb}, and refocus them through its perforations, as shown in Figure~\ref{fig:3VElectronPath}. The shapes of the signals from the induction and collection planes are well adapted to the shaping and sampling frequency of the readout electronics. 
As mentioned %above 
in Section \ref{sec:AR:intro}, the asymmetric bipolar signals on the induction planes are expected to enhance measurement of tracks having a large dip angle relative to the readout plane. 

\begin{dunefigure}
[Anode PCB hole pattern and board stack]
{fig:3V_hole_pattern}
{Left: details of the hole and strip pattern on the anode \dshort{pcb}s. The pitches and gaps between electrodes are given; the electrode widths are 7.65\,mm (induction-1, blue), 7.65\,mm (induction-2, green), and  5.1\,mm (collection, red). 
Right:  Illustration of the anode layers, adapter board and edge card stack. Bias voltages for each plane are shown.}
\includegraphics[width=.45\linewidth]{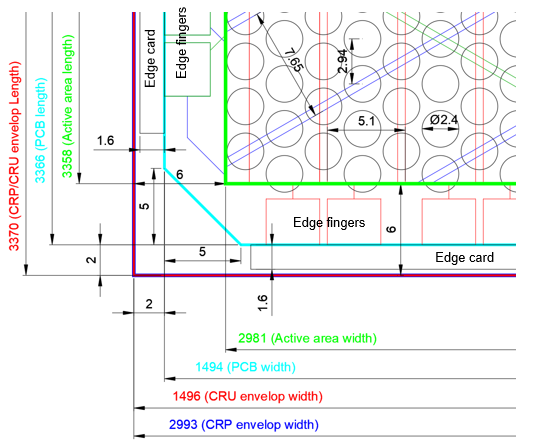}
\includegraphics[width=.45\linewidth]{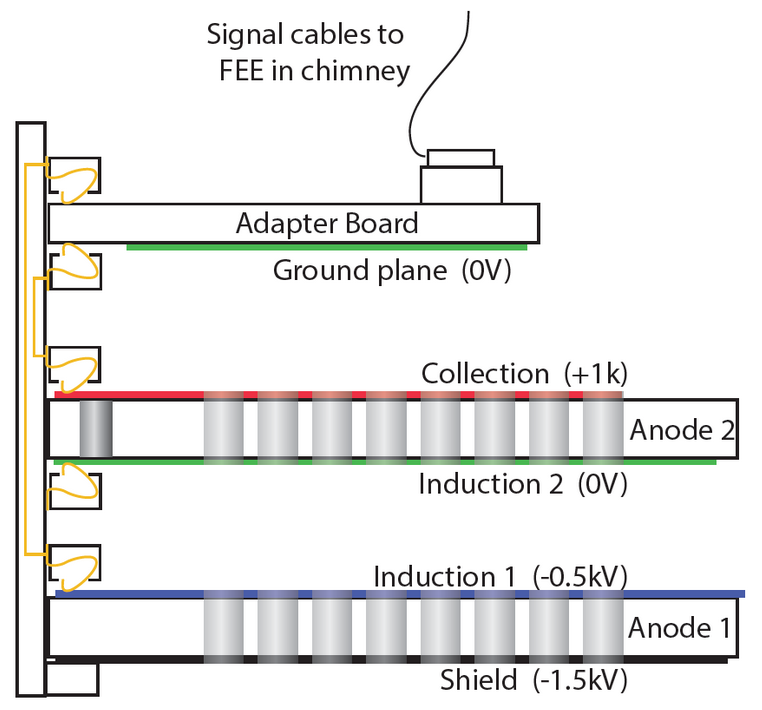}
\end{dunefigure}

\begin{dunefigure}
[Drift path in the anode plane]
{fig:3VElectronPath}
{Field lines in the two-\dshort{pcb} design of the perforated anode plane, illustrating the path of ionization electrons from a track segment. The study used 3.2\,mm PCB thickness, 2.4\,mm hole size and 10\,mm gap between the PCB panels.}
\includegraphics[width=.45\linewidth]{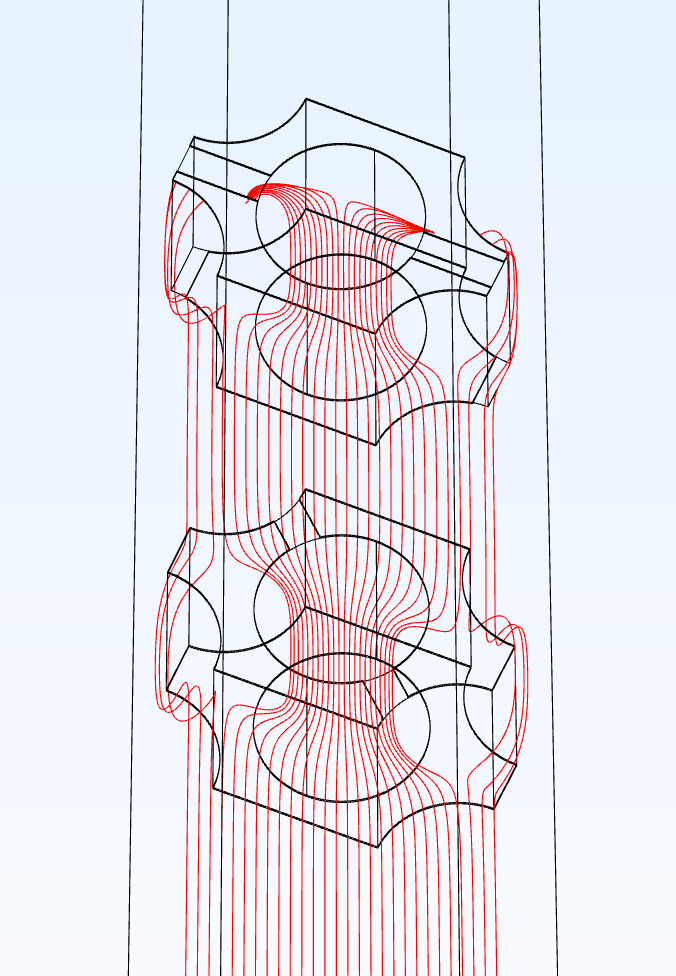}
\end{dunefigure}

With the chosen perforation pattern and thickness of the \dwords{pcbp}, and external fields of $\sim$500\,V/cm, the voltage differential required for complete charge extraction through the holes is estimated to be about 1\,kV. The minimum transfer field between the two anode \dshort{pcb}s should be equal to the drift field (\tgtdriftfield).  The separation between the stacked \dshort{pcbp}s in a \dshort{cru} is determined by the stack height of the \dword{peek} connectors used between them.  Both the separation between the \dshort{pcbp}s and the gap between the inner \dshort{pcbp} and the adapter board is 10\,mm.  Since the distance between the two \dshort{pcbp}s impacts the transfer field, which in turn affects the electron transfer, the voltage differential between the two induction planes is set accordingly.

Figure~\ref{fig:3V_anode_layout} shows the strip layout for both the top and bottom \dshort{cru}s. Each \dshort{pcbp} is constructed by joining six \dshort{pcb} segments (12 segments total for a \dshort{cru}, 24 for a \dshort{crp}). The segments are commercially produced, standard 2-layer, 3.2\,mm thick perforated \dshort{pcb}s with dimensions given in the figure. Each \dshort{pcbp} requires three different ``flavors'' of \dshort{pcb} segment, with the copper traces running at the correct angles. As shown in Figure~\ref{fig:3V_sixSegments}, the middle four segments (segments 2-5) are identical, but segment 1 and segment 6 are different. 
Each segment has %so called 
``half-lap'' joint surfaces (one for segments 1 and 6, two for segments 2-5), where a half-thickness of the \dshort{pcb} is removed for 2-cm width along its long edge. 
The half-lap portions of the \dshort{pcb} segments are %glued 
epoxied together to form a \dshort{pcbp}. 

\begin{dunefigure}
[PCB panel and PCB segment details]
{fig:3V_sixSegments}
{Each \dshort{pcbp} is constructed by gluing six PCB segments together. There are three different flavors with different widths, half-lap joints and copper patterns: segment-1, four identical middle segments and segment-6.}
\includegraphics[width=\linewidth]{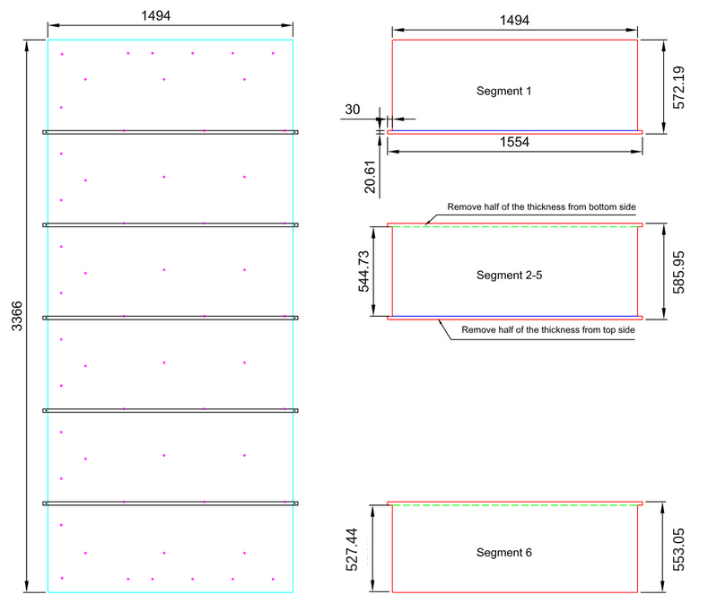}
\end{dunefigure}

Once the \dshort{pcb} segments are mechanically bonded to form a \dshort{pcbp}, electrodes on the induction-1, induction-2, and collection views are bridged by screen-printing conductive ink patches onto them. 
 Figure~\ref{fig:pcb_jumpers} is a concept drawing of the \dshort{pcb} ``half-lap'' edge-to-edge bonding method and the electrical bridging with conductive  silver ink patches. %Conductive ink is a mixture of silver and epoxy. 
 The ink is applied %to the joint 
 at the surface over the joint using a special mask  is 
 then treated under $\sim$120$^{o}$C for $\sim$3 hours for polymerization. Once polymerized, it allows electrical continuity between the patched surfaces.

\begin{dunefigure}
[PCBs bonded edge-to-edge to create PCB panel]
{fig:pcb_jumpers}
{The \dshort{pcbp} are constructed from smaller \dshort{pcb} segments using ``half-lap'' joint bonding technique (yellow and green overlap). Electrodes on one (induction-1) or both (induction-2 and collection) sides of a bonded \dshort{pcbp} are bridged by screen printing conductive ink patches on the \dshort{pcb} (red).}
\includegraphics[width=.6\linewidth]{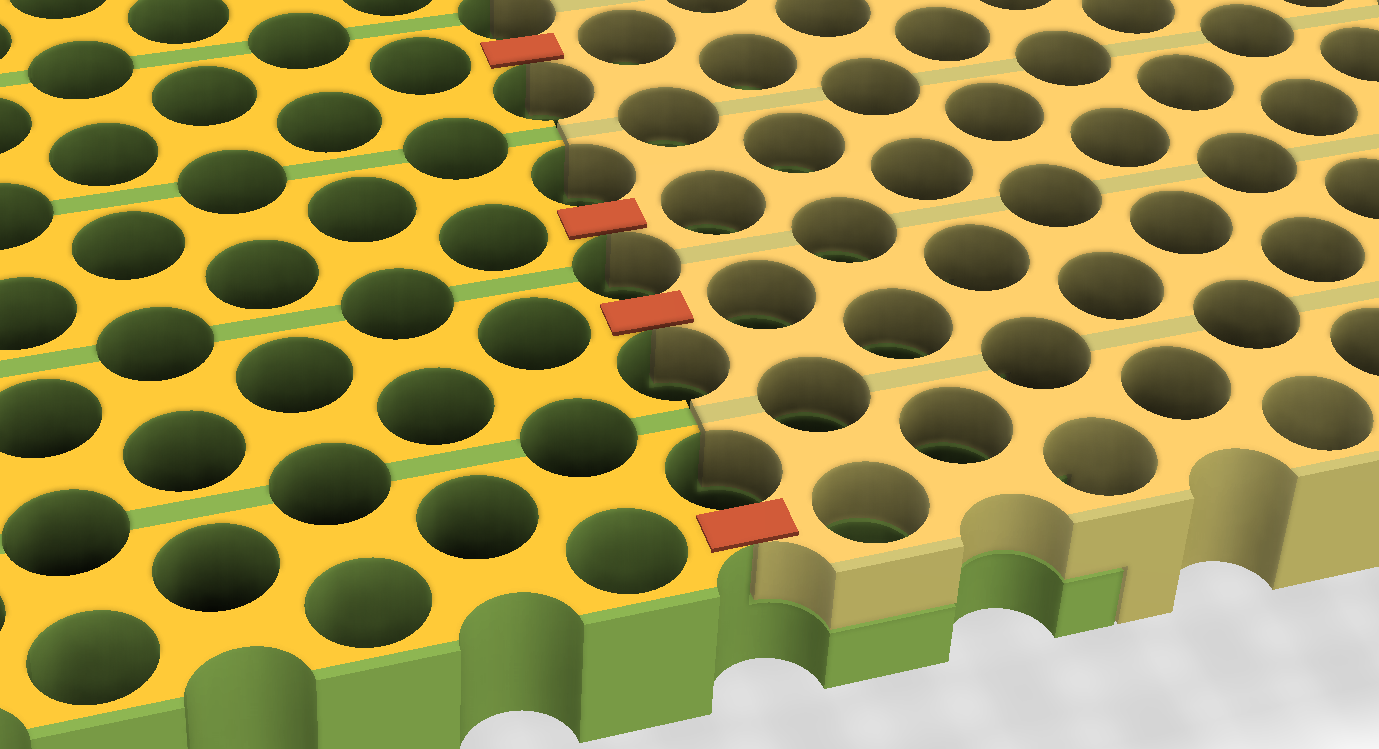}
\end{dunefigure}

The interface between the \dshort{anodepln}s and readout electronics is implemented 
via adapter boards that link the readout pads along the edges of the \dshort{cru}s to the readout cables (top \dshort{crp}s) or to the \dword{ce} modules (bottom \dshort{crp}s). 
In addition to being an interface to the readout electronics, adapter boards also provide an interface for biasing the %PCB planes
\dshort{pcb} views. Adapter boards are four-layer \dshort{pcb}s that host current-limiting resistors, \dword{hv} AC coupling capacitors, connectors for readout electronics, noise filters on the bias line, and various bias and charge readout lines on different layers of the board. Each \dshort{cru} has 12 boards installed around its periphery.  Adapter boards for the  top and bottom \dshort{crp}s have different designs to accommodate the different %needs of the two 
readout systems. 
As shown in Figure~\ref{fig:CRP_adapterBoards}, both top and bottom \dshort{crp}s have 12 adapter boards, of seven distinct designs. %flavors.  
Design models, assembly and manufacturing details, electrical specifications for all flavors of the top and bottom CRP adapter boards can be found respectively at~\cite{edms2717223, edms2765576}.

\begin{dunefigure}
[Top and bottom CRP adapter boards]
{fig:CRP_adapterBoards}
{Top and bottom \dshort{crp} adapter boards. The top image shows an inset with details.} % as an example.  }
\includegraphics[width=\linewidth]{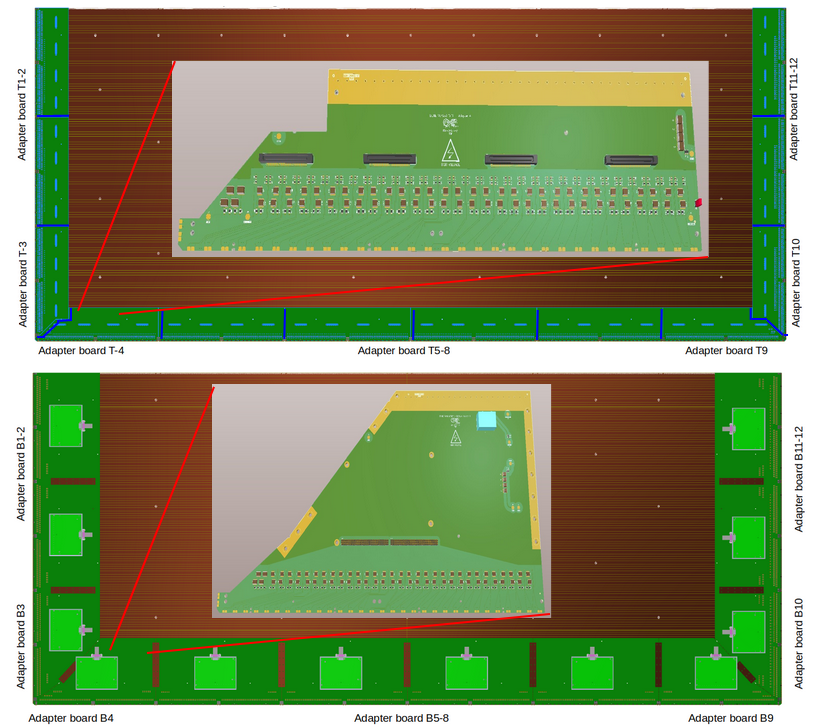}
\end{dunefigure}

Vertical interconnection between the \dshort{pcb} %layers 
views and adapter boards is done via edge cards. Figure~\ref{fig:CRP_edgeCards} shows %an example of 
how edge cards are attached to the %PCB anode.
\dshort{crp}. Edge cards are small four-layer \dshort{pcb} boards with connectors soldered on them. There are 24 edge cards plugged and secured on %the anode-adapter board assembly. 
a \dshort{cru}. Connectors on the edge cards are in physical contact with the pads on the anode strips and adapter boards. 
As well as biasing the readout strips, they bring the signal from strips to the readout electronics. Edge cards of identical designs are %identical 
used for the top and bottom \dshort{crp}s. There are three %flavors of boards where each kind 
designs, each responsible for different sets of channels to be connected to the adapter boards. Details for each design and electrical scematics can be found at~\cite{edms2785495}. As shown in Figure~\ref{fig:CRP_edgeCards}, to secure the edge cards in place and ensure their electrical contact with pads, % during the operation, 
they are mechanically supported by \threed printed guides and brackets. The guides are glued onto specific locations on the boards to maintain %ensure 
a 1\,cm %proper 
gap between the \dshort{pcbp}s and adapter boards. Brackets %secure 
bolt the cards %by bolting them 
to the adapter boards and to the shield plane.

\begin{dunefigure}
[CRP edge cards]
{fig:CRP_edgeCards}
{Top: Details of anode assembly and edge card installation. Two layers of \dshort{pcbp} are visible with an adapter board on top of them. Edge cards are plugged into the side % of anode assembly is visible. T
(blue/transparent edge card %view 
on the left shows the details of the connectors,
also a green one on the right). FEMBs (bright green rectangles) and capacitors %(small brown/yellow rectangles) 
on the adapter boards are also visible. %on the edge card (blue). 
Bottom left: An edge card showing connectors with yellow pins. % are visible. 3D 
\threed printed guides (black) and brackets are also shown. Bottom right: \threed printed guide, its location for gluing onto the edge card, and  dedicated %teeth 
notches on %the PCB planes are shown.
the adapter board and \dshort{pcb}s are shown.}
\includegraphics[width=\linewidth]{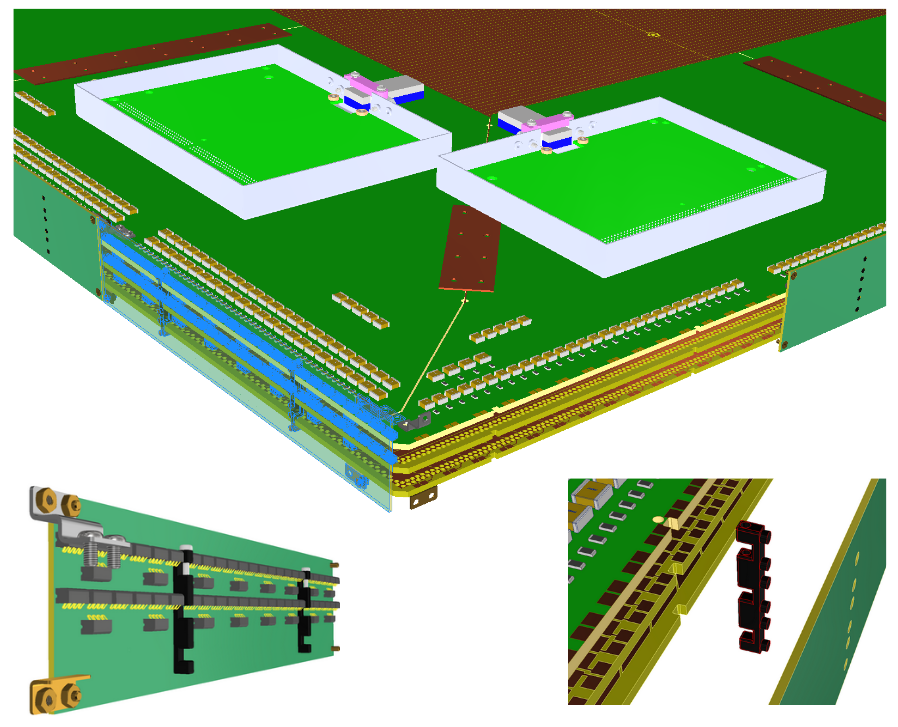}
\end{dunefigure}

\section{CRP Mechanical Support Structure} %Composite Structure and 
%Anode Plane Support}
\label{sec:invar-frame}

A \dword{crp} is  designed to have the thermomechanical stability and planarity
to 
meet the \efield uniformity requirement so as to ensure optimal ionization charge detection and collection.

Figure~\ref{fig:crp}  illustrates a full \dshort{crp} mechanical support, which is a composite frame made of two  parts that fit together. Each part is designed to attach and support a complete CRU; the top and bottom \dshort{crp}s use a similar composite structure and concept. A full CRP is built from %the junction of 
two half composite structures holding a CRU each. 
The right-hand image shows a detail of the composite structure at the edge close to the two half-\dshort{crp}s junction parts.

\begin{dunefigure}
[CRP illustration showing its two component CRUs]
{fig:crp}
{Left: Illustration of a complete top \dshort{crp} with its composite frame (orange skins and gray profiles) coupled to the \dshort{cru}s underneath. Right: Detail of the composite structure close to the \dshort{crp} junction parts that are used to link the two \dshort{cru}s.} 
\includegraphics[width=0.5\textwidth]{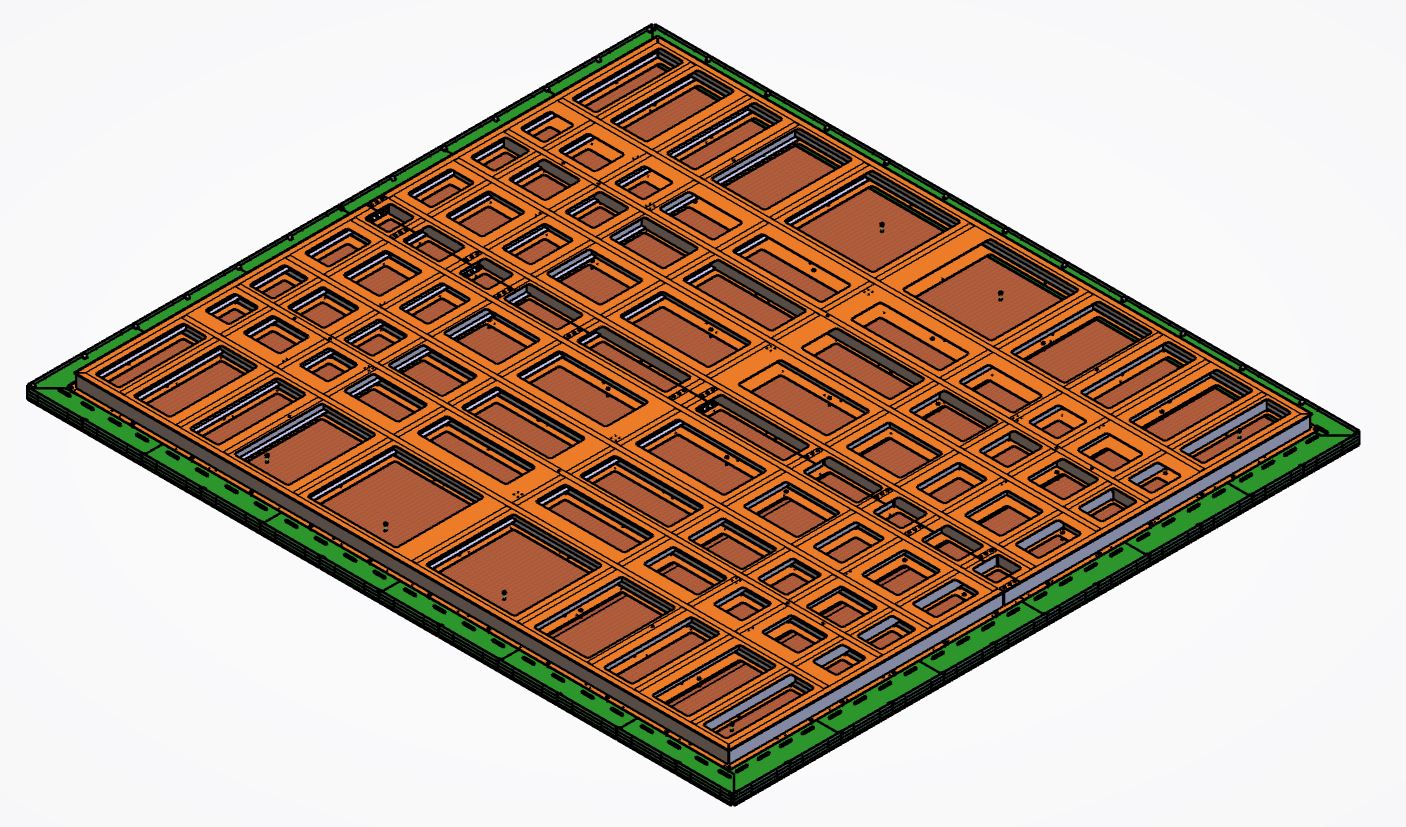}
\includegraphics[width=0.450\textwidth]{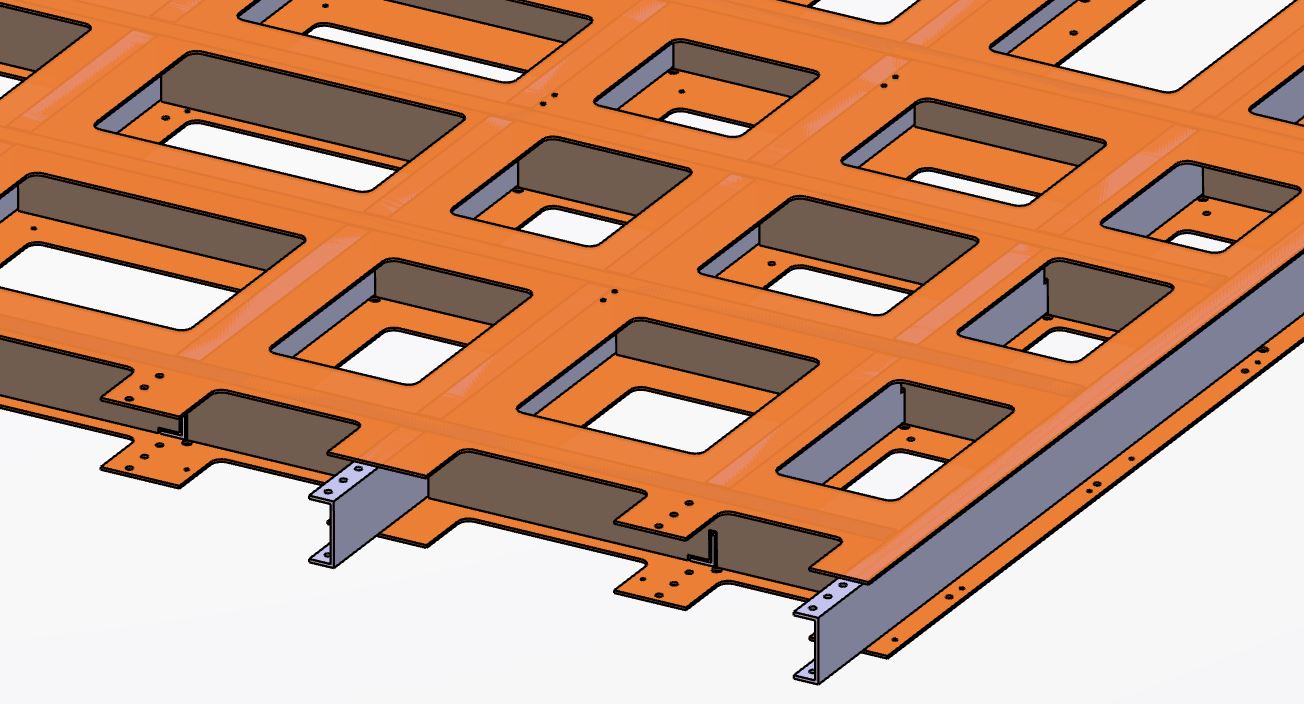}
\end{dunefigure}

The \dshort{crp} dimensions (Table~\ref{tab:complist}) take into account the largest size \dword{pcb} that the industrial manufacturers can provide and  maximize the active detection area in the cryostat. The full \dword{spvd} module will implement 80 \dshort{crp}s each on the top and bottom (160 total). The top \dshort{crp}s are suspended from the cryostat roof and the bottom \dshort{crp}s are supported by posts placed on the cryostat floor. 

The rigid composite frame for a \dshort{crp} is made of two layers (``skins'') made of perforated (water-jet-cut) glass-reinforced epoxy laminate material, each \SI{2.4}{mm} thick, separated by imbricated (overlapping) U-shaped profiles made of Durostone\textregistered{} EPGM Epoxy (\dword{frp}), of transverse dimensions \SI{60}{mm} $\times$ \SI{23}{mm} $\times$ \SI{3}{mm}. 
The frame is composed of two identical sections (half-frames)of dimensions \SI{3.3}{m}$\times$\SI{1.56}{m} for a bottom half-\dshort{crp} and \SI{3.2}{m}$\times$\SI{1.5}{m} for a top one. 
A \dshort{cru} is connected via attachments to the composite half-frame for mechanical support (Figure~\ref{fig:crpspacer}). 
The advantage of having the structure split into two parts that are easily connectable, allows building, testing and transport of the smaller half-\dshort{crp}s from the production sites to \dword{surf}. 
Each full-size \dshort{crp} is assembled from two half-\dshort{crp}s only after transport into the cryostat. 

 At several positions in the top \dshort{anodepln} frames, G10 spacers are used in place of the \dshort{frp} profiles, reducing the weight.  
The anode \dshort{pcb}s and adapter boards are connected to the composite frame via a number of suspension points using machined pins and spacers made of \dshort{peek} material.   

\begin{dunefigure}
[\dshort{crp} spacer and anode attachment to the composite frame]
{fig:crpspacer}
{Details of the different spacers (light blue and light green) and attachment screws used to link the two anode layers and the adapter boards to the composite frame.} 
\includegraphics[width=0.73\textwidth]{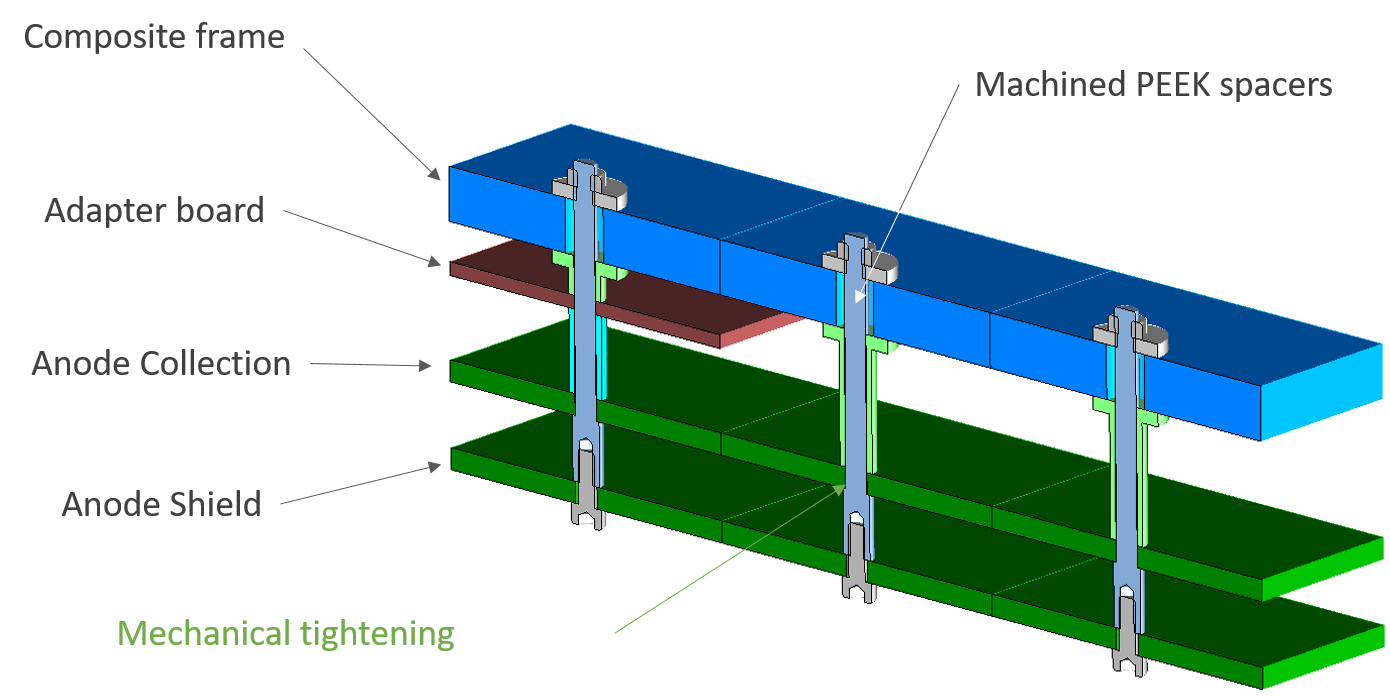}
\includegraphics[width=0.24\textwidth]{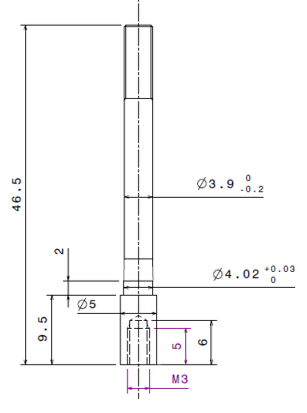}
\end{dunefigure}

\begin{dunefigure}
[\dshort{crp} glass epoxy skin]
{fig:compositeskin}
{Left: One glass-epoxy skin after being  machined and before being glued to %U-shaped fiber glass 
\dshort{frp} profiles. Right: the final composite frame after its construction, showing G10 rods used in place of profiles in several locations.} 

\includegraphics[width=0.48\textwidth]{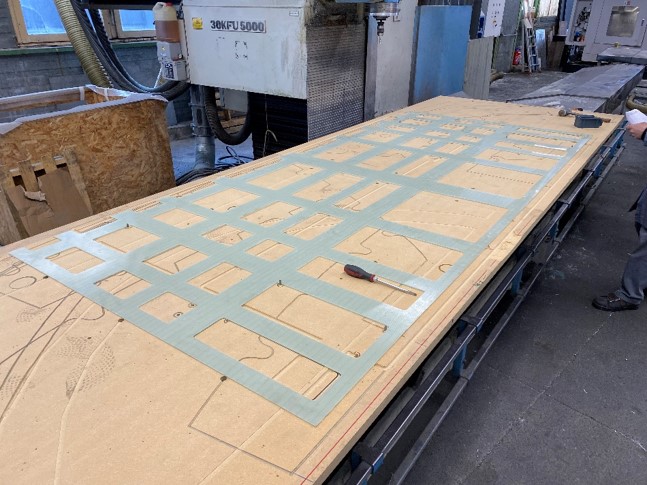}
\includegraphics[width=0.48\textwidth]{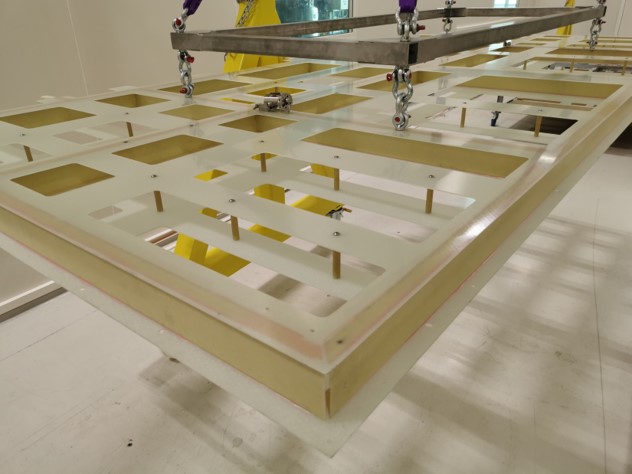}
\end{dunefigure}

The two fiber glass-epoxy water-jet-cut skins are fabricated from prepreg  with two 0\dge/90\dge layers and two +45\dge/-45\dge layers for the glass-fiber orientation of the composite material.  The skins are produced between two metallic plates before being cured in an oven under vacuum. Then they are water-jet cut to create the openings. Figure~\ref{fig:compositeskin} (left) shows one skin of a top \dshort{crp} after the complete process.  The right-hand image shows the final composite frame.

 The glass-reinforced epoxy laminate material for the skin of the frame was chosen to match the thermomechanical behavior of the  \dshort{cru}s to (1) avoid over-stress from differential thermal contraction, and (2) control the (horizontal) spacing between \dshort{cru}s. The coefficients of thermal contraction of both the perforated anodes and the composite frame material have been measured to be the same within \num{1e-6}K$^{-1}$. 
 The holes in the structure serve to reduce the weight, 
 allow for \dshort{lar} flow across the \dshort{anodepln}s, and allow access to the adapter boards and connectors in order to connect the %different 
readout electronics.

The height of the composite frame is optimized to retain the necessary stiffness while keeping its weight under \SI{100}{kg}. The weight for a top \dshort{crp} frame is well under that at \SI{35}{kg}. A bottom frame must be stiffer to support the electronics boxes installed along the periphery; this is accomplished by using twice the number of profiles, which increases the weight to \SI{90}{kg}. 

Figure~\ref{fig:pcbframe} shows a portion of a \dshort{crp} composite frame  with a detail of the U-beam profile interconnections at the crossing points. Figure~\ref{fig:pcbframe2} shows how the two half-frames are coupled. 
Twelve junctions are distributed along the long \dshort{crp} side, made by imbrication of protruding U-shape profiles of one half under the fiber glass skins of the other half, and securing with six  screws.  

\begin{dunefigure}
[\dshort{crp} composite frame profiles]%mechanical support structure]
{fig:pcbframe}
{CAD rendering of  \dshort{crp} composite frame profiles %inner part and  details of the structure components made of two 
with a fiber glass epoxy layer glued to one side.} %U-shaped fiber glass profiles.} 
\includegraphics[width=0.65\textwidth]{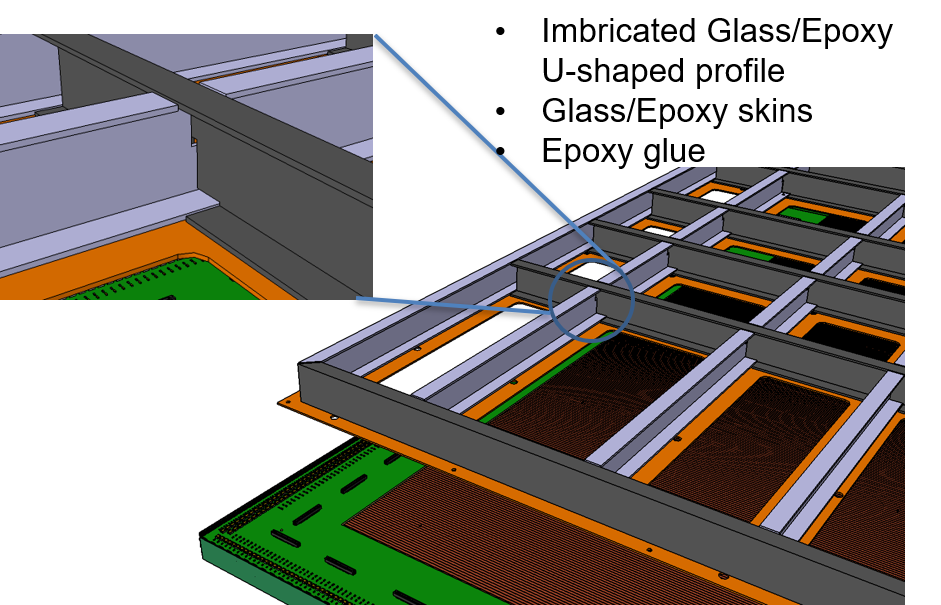}
\end{dunefigure}

\dword{fea} calculations of both kinds of composite frame %structures 
have been performed. They included the full weight of the %anodes
\dshort{crp}, the adapter boards, the electronics, and the positions of the suspension for the top or the feet for the bottom. The results showed that the maximum deformation expected under gravity (without buoyant forces) of the structure over the full \dshort{crp} size at warm is less than 1.3\,mm for the two composite geometries described above.

\begin{dunefigure}
[\dshort{crp} composite frame junction system] %mechanical support structure]
{fig:pcbframe2}
{Left: CAD rendering of the \dshort{crp} composite frame junction system to couple two %half-CRPs
\dshort{cru}s; Right: photo of the first assembly of 2 half \dshort{crp}s during the CRP-2 prototype production.} 

\includegraphics[width=0.65\textwidth]{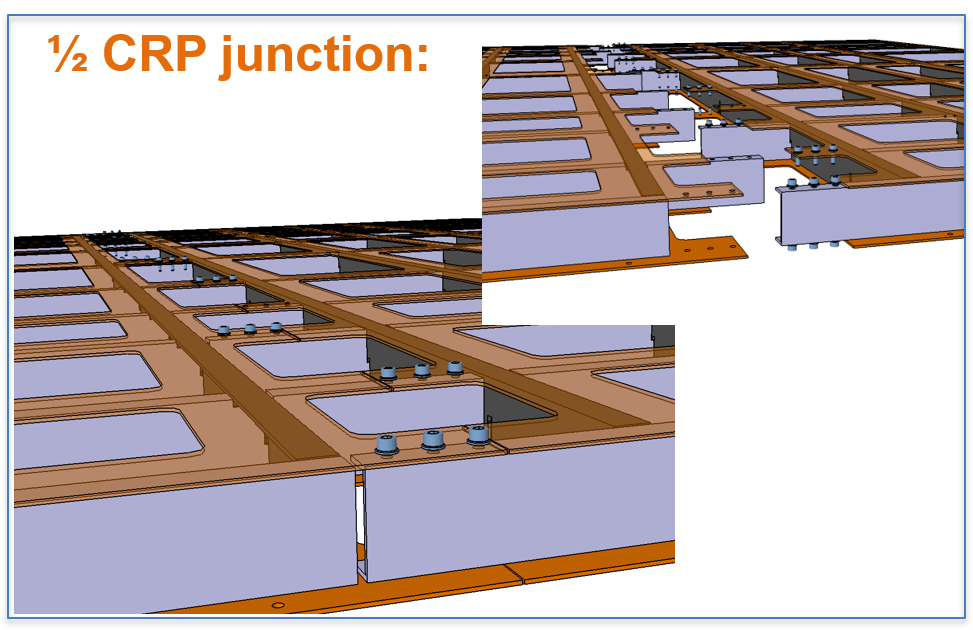}
\includegraphics[width=0.33\textwidth]{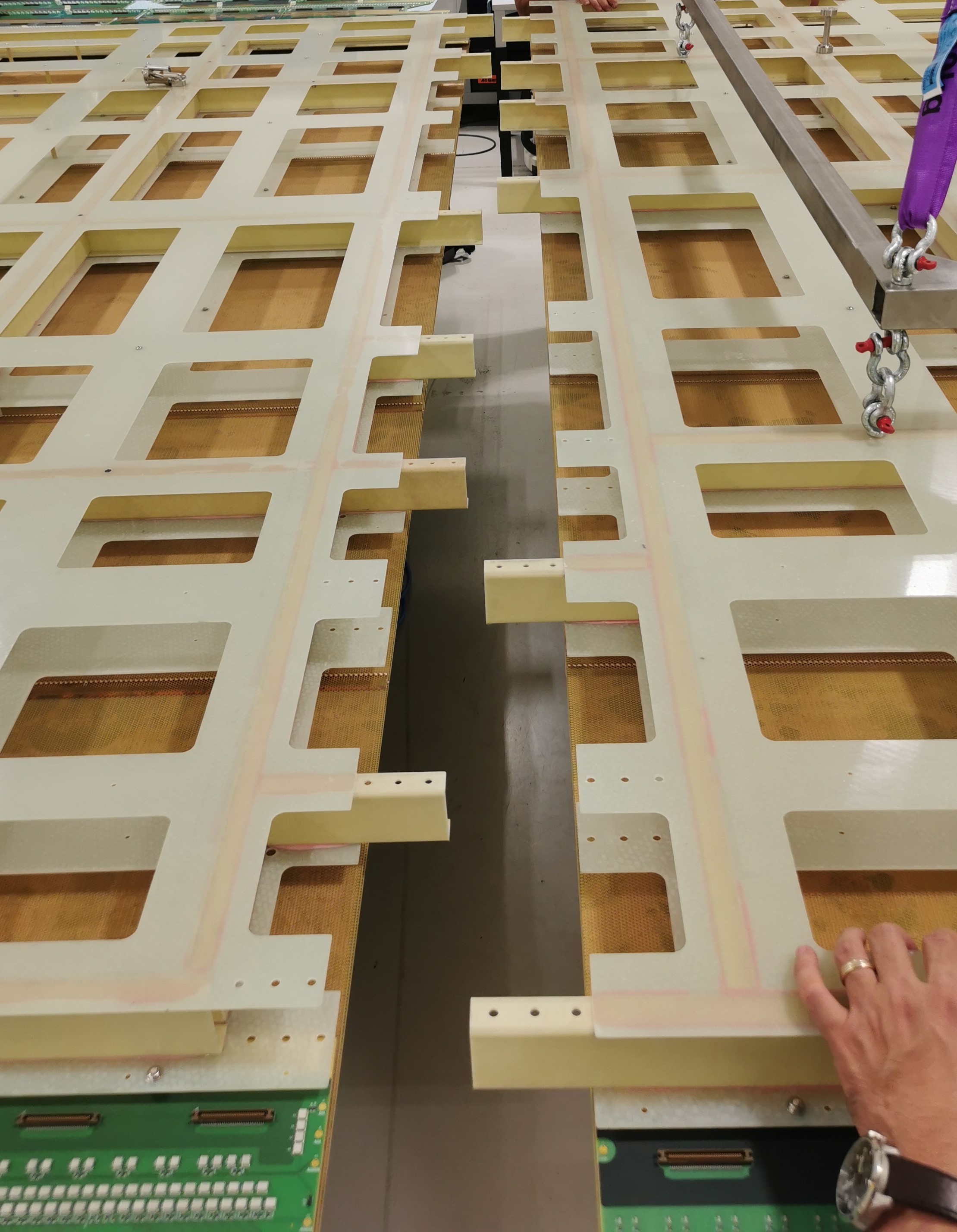}
\end{dunefigure}

The baseline design specifies 61 suspension points per \dshort{cru} between the \dshort{cru} and the composite frame, based on \dshort{fea} calculations, 
with a higher density along the borders to keep the relative deformation of the anode 
\dshort{pcb} layers under 1\,mm over their length, while keeping the overall \dshort{cru} planarity to about 2\,mm.  This requirement ensures that the connections between the anode strips and the adapter boards can be made easily and with sufficient precision. 

The two stacked anode \dshort{pcb}s are vertically separated by 10\,mm and the adapter boards are 11\,mm 
from the nearest anode layer. The \dshort{cru}s are 
attached to the composite frame at the 61 suspension %attachment 
points using machined pins and spacers made of \dshort{peek}  material. 
Figure~\ref{fig:crpspacer}  shows the design  for  attaching the different \dshort{pcb} layers to the composite frame using the spacers and screws. The spacers will be inserted (cryo-fitted) into the supporting holes made in the \dshort{pcb} (4\,mm diameter) with a force appropriate 
to prevent them from turning when the screws are tightened. This mechanical tightening allows a quick, accurate, and clean assembly with no gluing.

\section{Top CRP Superstructures and Suspension}
\label{sec:CRPMss}

The \dwords{crp} that compose the top \dshort{anodepln} are suspended from the cryostat roof via a set of 16 ``superstructures'' (SST), 12 large (holding six \dshort{crp}s) and four small (two \dshort{crp}s), that serve to minimize the number of suspension penetrations required 
(Figure~\ref{fig:supercrp-pattern}). The superstructures also support the cathode modules \SI{6.5}{m} below, using the same suspension 
pattern. %%%% anne to here
Each superstructure is suspended from four points through dedicated feedthroughs in the cryostat roof, represented by white dots in the figure. This will allow finely spaced vertical adjustments to compensate for possible deformations of the cryostat roof geometry after cooling and filling with \dshort{lar}, and will maintain 
the planarity and horizontality of the entire \dshort{anodepln} within the required \SI{20}{mm}. %The design foresees 
The susperstructure will be fully immersed in the \dshort{lar} with a nominal position corresponding to a few cm below the liquid surface.  The position of the top %part 
of the superstructure with respect to the liquid level will be %given 
monitored by capacitive level meters located at %the SST 
its corners, to % able to give 
an absolute position precision of the order of a mm. 

\begin{dunefigure}
[CRP superstructure pattern in the cryostat]
{fig:supercrp-pattern}
{Layout of the 16 superstructures %in the FD2 cryostat 
for the top \dshort{crp}s. The four superstructures at the ends each hold two \dshort{crp}s, the other twelve each hold six, for a total of 80. The white dots indicate the positions of the suspension feedthroughs.}
\includegraphics[width=0.98\textwidth]{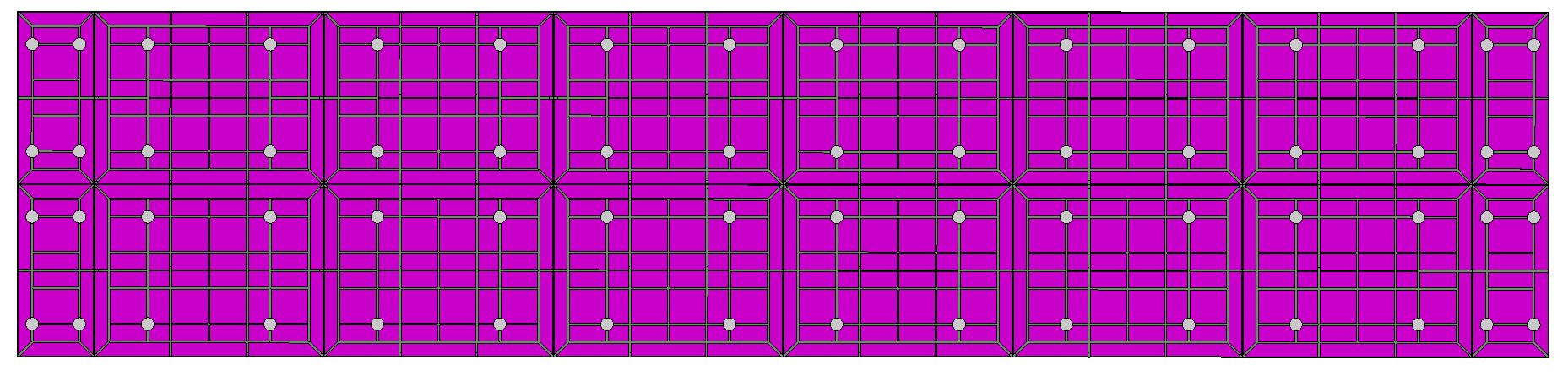}
\end{dunefigure}

\begin{dunefigure}
[Dimensions and positions of the superstructures]
{fig:supercrp-patternv2}
{Dimensions and positions of the 16 superstructures and the \dshort{crp}s for the top \dshort{anodepln}. The gap between \dshort{crp}s within the same superstructure is \SI{5}{mm} in both horizontal dimensions, and \SI{10}{mm} between \dshort{crp}s in adjacent superstructures.
}
\includegraphics[width=0.98\textwidth]{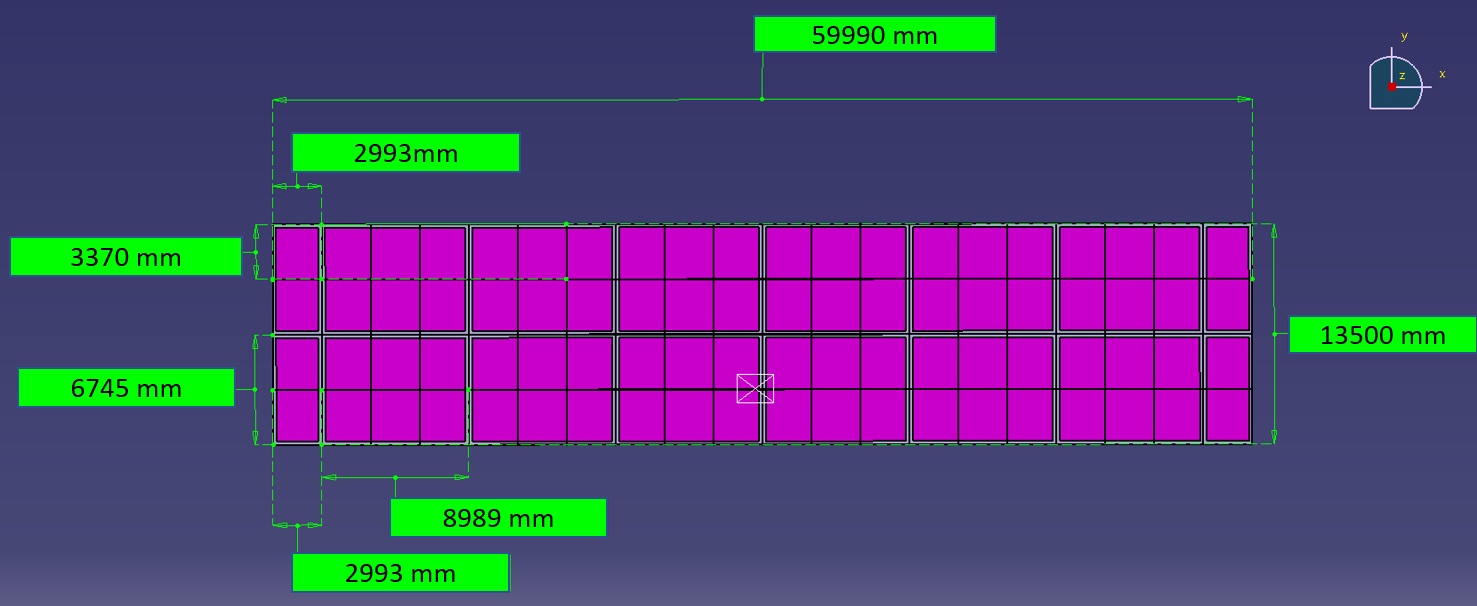}
\end{dunefigure}

The gap between \dshort{crp}s within the same superstructure is \SI{5}{mm} in both horizontal dimensions, and \SI{10}{mm} between \dshort{crp}s in adjacent superstructures.
Given the \dshort{crp} dimensions of {\qtyproduct[product-units=power]{3.370x2.993}{m}}, 
the large %SST frame 
superstructure dimensions are \qtyproduct[product-units=power]{8.989x6.745}{m} and the small %SST 
are \qtyproduct[product-units=power]{2.993x6.745}{m} 
as illustrated in Figure~\ref{fig:supercrp-patternv2}.
The slope of the vertical deformation of the roof is expected to be smaller 
near the %extremities of the 
cryostat ends than closer to the center, therefore the %groupings of 
smaller, two-\dshort{crp} superstructures in this location %for the  at the cryostat ends 
are designed to take into account the potential need to perform differential height adjustments  along the first and last three meters along the beam direction. 
During the filling of the cryostat, parts of the roof will deform (downwards) up to 10\,mm in the vertical direction, mostly near the center. %at the central part.

During operation
the vertical position of the roof is expected to change up to $\pm$1\,mm due to variations between the internal and external pressures (again, mostly near the center). %at the central part, while at the periphery will be affected much less).
The \dshort{crp} superstructures also support the cathode supermodules, which have similar dimensions (Section~\ref{subsubsec:CAsss}). For the large superstructure ({\qtyproduct[product-units=power]{9x6.7}{m}}), 
the cathode modules are attached by means of twelve  suspension cables, ten attached at the \dshort{crp} superstructure extremities and two near the center. For the small superstructures ({\qtyproduct[product-units=power]{3x6.7}{m}}), 
the cathode modules are attached by six cables, four at the extremities and two near the center, as shown in Figure~\ref{fig:supercrp-cathode}. The cables are made of Dyneema fibers\footnote{In \dshort{lar}, the ropes are expected to stretch by $\sim\,14\,mm$ due to thermal expansion at low temperature (Dyneema has a negative \dword{cte} of about $-10^{-5}/K$). 
Dyneema\textregistered, \url{https://www.dsm.com/dyneema/en_GB/home.html}}, a high-strength composite material.
\begin{dunefigure}
[CRP superstructure suspension cables for cathode]
{fig:supercrp-cathode}
{Model of the cable locations on the superstructures (SST) used to suspend the cathode supermodules, % to the Large superstructure (Left) and to the small superstructure (Right) 
after attachment of the \dshort{crp}s to the superstructure. 
}
\includegraphics[width=0.90\textwidth]{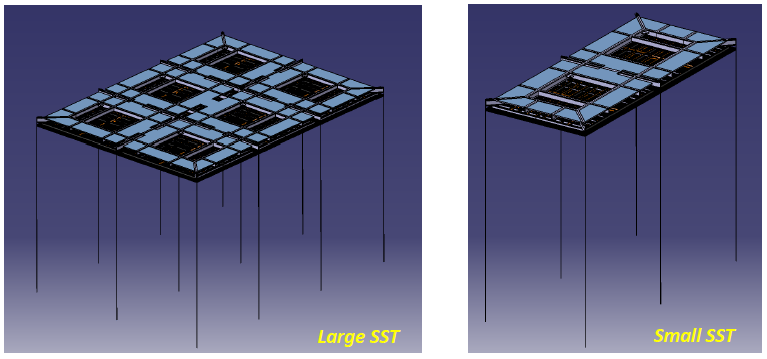}
\end{dunefigure}

The baseline  design for the \dshort{crp}-supporting superstructure %is mainly 
consists of a stainless steel frame composed of S6$\times$12.5, S4$\times$7.7 and C4$\times$4.5 standard US profiles. 
The superstructure will also be used to support people to allow cabling of the top CRPs at the cold flanges below the chimneys (after the superstructures are raised close to the cryostat roof). Catwalks and safety systems are added to the stainless steel frame for this purpose. 
Figure~\ref{fig:supercrp} shows a superstructure frame designed for a group of six \dshort{crp}s ({\qtyproduct[product-units=power]{9x6.7}{m}}) 
where the catwalks (dark areas) and the beam profiles are visible. 
\begin{dunefigure}
[CRP superstructure before and after assembly]
{fig:supercrp}
{(Left) \dshort{crp} superstructure modules (metallic frames) before assembly and (right) after assembly  with the four cables used to suspend it. The six \dshort{crp}s to which it will attach are not shown. The yellow arrows indicate the 12 points of attachment of the cathode suspension ropes.  
}
\includegraphics[width=0.45\textwidth]{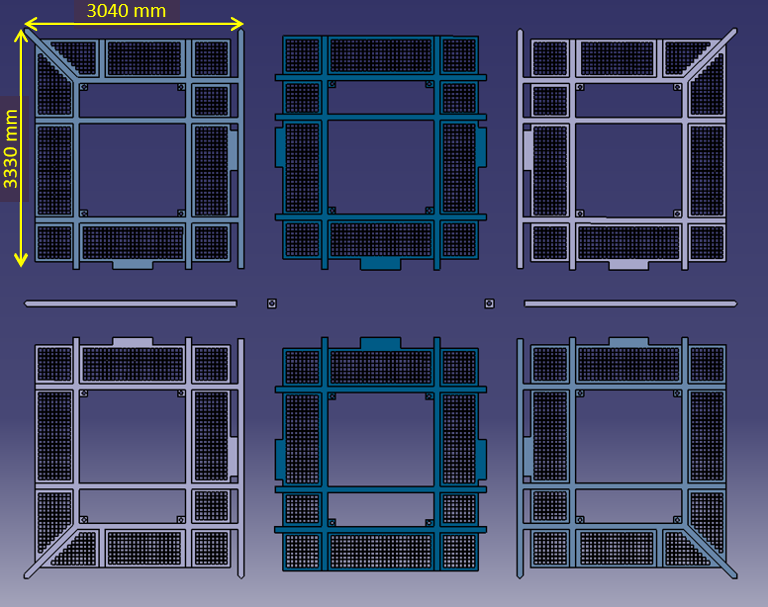}
\includegraphics[width=0.54\textwidth]{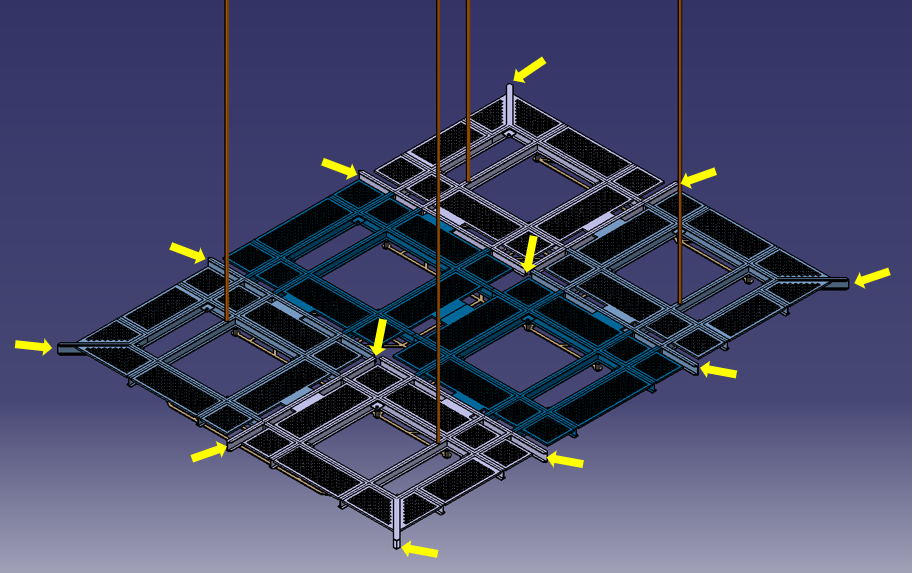}
\end{dunefigure}

Each superstructure is assembled underground at \dword{surf} from %multiple 
either six or two stainless steel modules. %of dimensions (\SI{3x3.4x0.25}{m}), 
The modular design allows use of %. The size allows the modules to be transported in a 
standard transport to the site and down the Ross shaft in the cage. %to be brought down in the cavern using the Ross cage.
Assembly, during which the modules are bolted together,  will take place inside the cryostat. The four %oblique 
extensions at the superstructure corners are used to connect to the cathode suspension cables.  
Several configurations have been studied and \dshort{fea} calculations have been performed with the 
appropriate constraints applied on the superstructure as it is suspended by four points.  The calculations have yielded maximal deformations of the order of 5\,mm given the full charge loading from the six \dshort{crp}s (assuming 200\,kg each), %and the 
an attached cathode %structure below 
supermodule (600\,kg), and the catwalks at warm. 
Figure~\ref{fig:supercrpdetails} shows a closeup view of a large superstructure corner with \dshort{crp}s attached, catwalk grids and the beam extensions to attach the cathode modules.
\begin{dunefigure}
[CRP superstructure corner extensions]
{fig:supercrpdetails}
{Details of the corner extensions of a top superstructure (foreground and at right, green) with  \dshort{crp}s attached below. The catwalk and the stainless steel beams are visible. 
}
\includegraphics[width=0.95\textwidth]{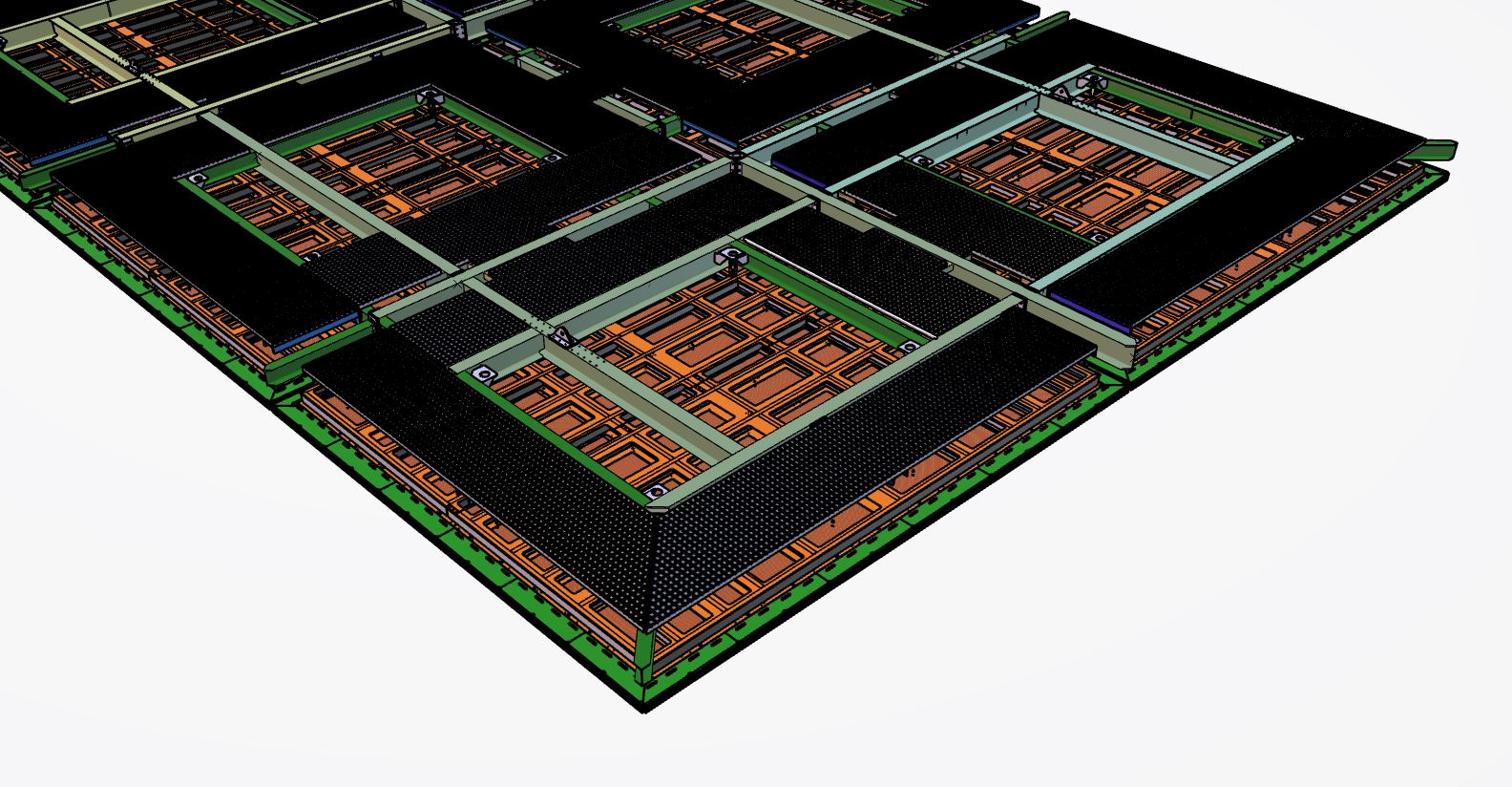}
\end{dunefigure}

%The design of the structure is a result of optimization. 
In the optimized superstructure design, %keeps avery low deformation change 
the change in deformation when the suspension anchoring points are shifted  by up to  250\,mm is very low. This amount of shifting is expected on some of the suspension cables %to take 
taking into account the cryostat roof's main structure beam positions. % at the top. %The resulting effect on the SST 
This deformation is negligible compared to the centered suspension positions. 
Additional calculations have been done assuming the %also by shifting the 
cathode suspension points and %also the CRP 
the \dshort{crp} attachment points are shifted by a few cm. %The effects 
These shifts add no more than a mm to the global deformation.

Each %individual 
\dshort{crp} is attached at four points to a superstructure via decoupling devices that allow %the 
transverse displacement of the %different 
two structures relative to each other during \cooldown without impacting the planarity or the mechanical behavior of the \dshort{crp} assembly. 
The decoupling system 
defines a central fixed point on the \dshort{crp} towards which all its components shrink.

The decoupling system implements %There are 
three different types of mechanical links between the metallic and composite frames. % for this decoupling system.
One is a solid vertical bar attachment that defines the fixed position of the \dshort{crp} with respect to the superstructure.
The other links are made with double-ball joints, which allow free displacement in any direction during differential thermal shrinkage 
while keeping the vertical positions of both structures fixed. 
 To prevent the \dshort{crp} from  rotating around the fixed point, a guiding rail constrains one of the double-ball joints  to move in a chosen direction.  Complete sets of junction components between the \dshort{crp} composite and metallic frame have been built for the three recent prototype \dshort{crp}s %built since 2021 
and successfully tested in the \coldbox.
Figure~\ref{fig:crp-decoupling} illustrates the three types of mechanical links and the principle for a single \dshort{crp} attached to a metallic superstructure, and shows the final components built and used to couple the  \dshort{crp} composite prototypes to the metallic frame for the \coldbox tests in 2022. The height of each connecting decoupling system is \SI{92}{mm} with a diameter of the support of \SI{37}{mm}. All pieces are made of stainless steel.

\begin{dunefigure}
[CRP decoupling system]
{fig:crp-decoupling}
{Illustration of the decoupling system and the three types of links designed to compensate for differential thermal contraction of the top CRP superstructures made of fiber glass composite and stainless steel. The photos show the final components built and validated with the \dshort{crp} prototyping.}
\includegraphics[width=0.99\textwidth]{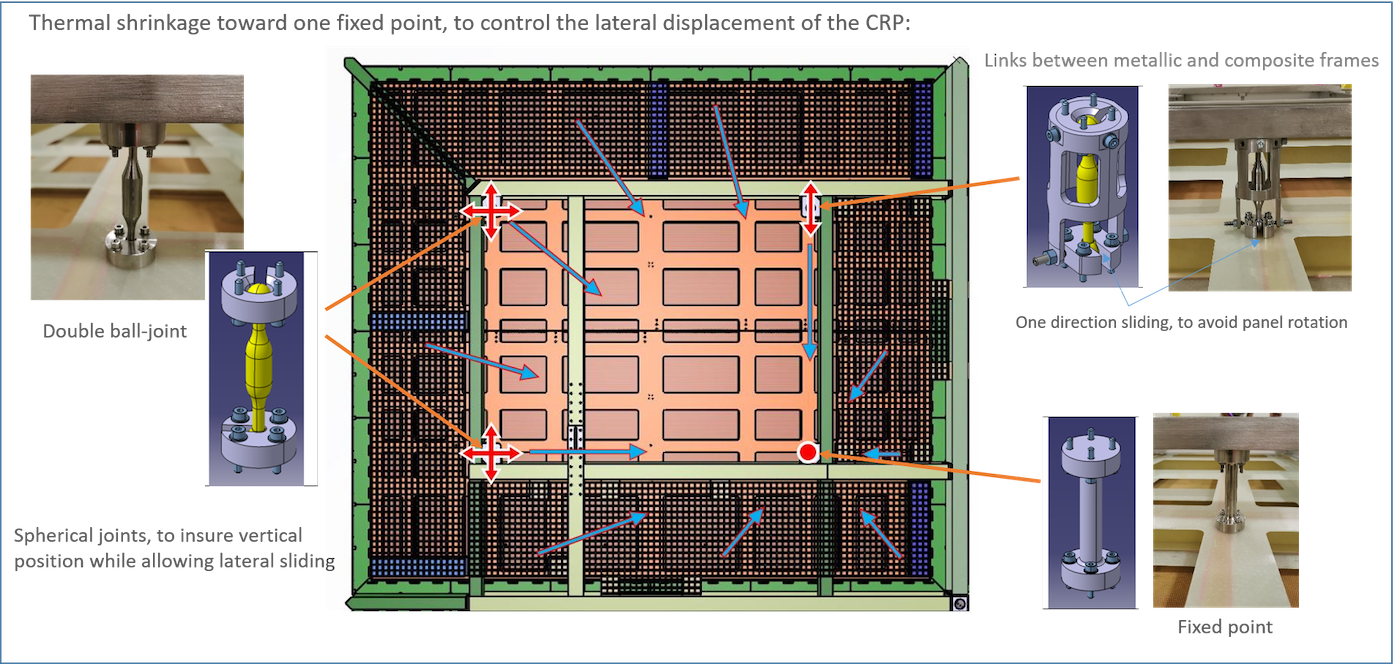}
\end{dunefigure}

Each  \dshort{crp} superstructure is hung by four cables from the top of the cryostat and attached to a 
motorized system that allows vertical adjustment after the cryostat is filled. The principle of the suspension system is similar to that used to control the \dword{pddp} \dshort{crp}s
in which the suspension feedthroughs are arranged geometrically such that their barycenter coincides with that of the \dshort{crp} superstructure, and an automated system is used to suspend the structures at the required position and adjust them precisely in the cryostat. 

Figure~\ref{fig:spft} shows the design planned for the suspension feedthroughs, including the bellows and the motors to be installed on top of the cryostat. The blue part on left image linking the two pieces of cables acts as a  mechanical end-of-stroke to limit the movement downwards and to provide a mechanical safety stop.
The photograph on the right shows one of the twelve similar suspension systems built and installed on the \dshort{pddp} cryostat roof.

\begin{dunefigure}
[Views of suspension feedthrough system for a CRP superstructure]
{fig:spft}
{Diagrams and photo of suspension feedthrough system for a \dshort{crp} superstructure.}
\includegraphics[width=0.75\textwidth]{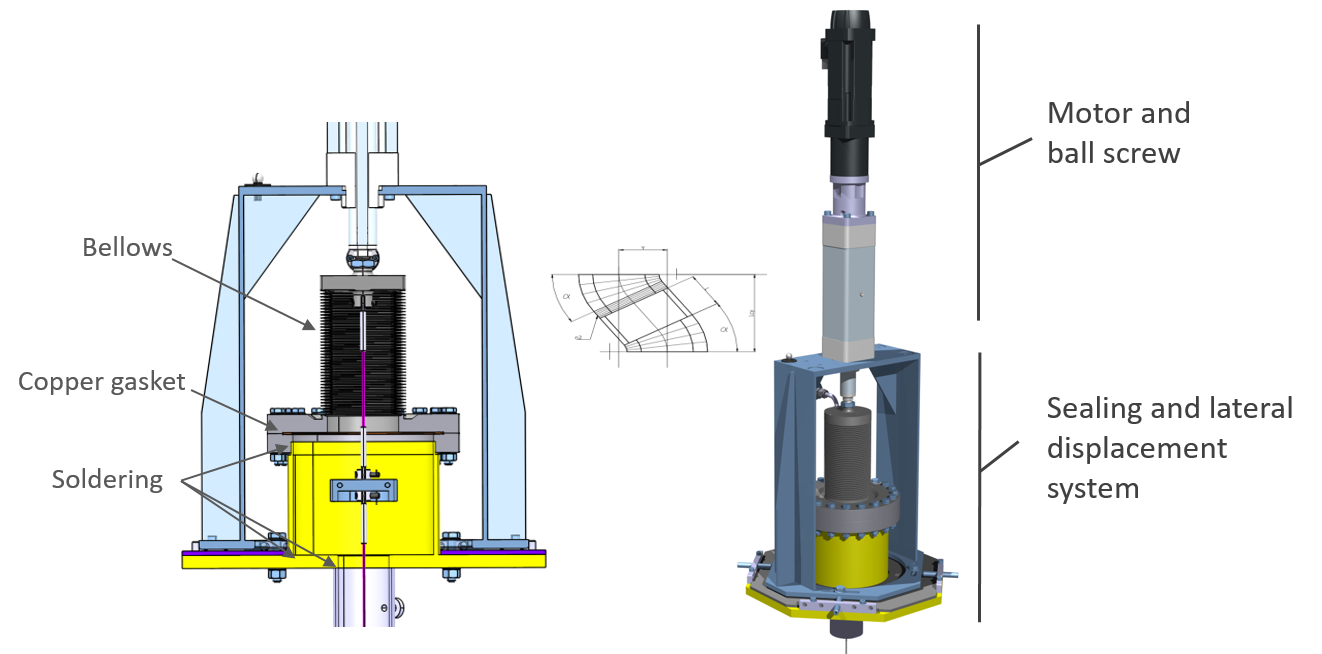}
\includegraphics[width=0.24\textwidth]{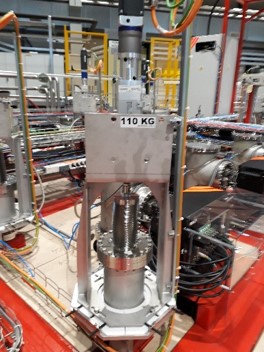}
\end{dunefigure}

Using long pushing screws at the level of the adjusting table on the cryostat roof, the transverse position of the suspension cable can be translated by a few cm, allowing correction (using metrology) of the superstructure %. This position to its nominal position inside the cryostat % measurement 
before raising it.

The size of the bellows in the suspension feedthrough determines the maximum vertical 
displacement, and the baseline 
design bellows enables a range of about  $\pm$\SI{40}{mm}.
To allow for maintenance or replacement of the bellows, the system incorporates a mechanical stop and a simple device to obstruct the chimney, closing it off.
To allow for precise surveying of the feedthrough position during installation, a special slot is available at the top of the suspension feedthrough for mounting a laser tracker target.

 \section{Bottom CRP Support System}
\label{subsec:CRPBss}
 
The bottom  \dwords{crp} are identical to the top ones, each composed of two \dwords{cru} plus a composite frame. On the bottom \dshort{anodepln}, each \dshort{crp} is supported about 160\,mm above the membrane surface by four %posts, or feet 
dead-weight load bearing supports, as shown on Figure~\ref{fig:crpbottom}, positioned inside flat portions of the cryostat inner membrane. There is no intermediate metallic structure, as for the top. 
The number of posts and density have been 
optimized to take into account  the weight distribution of the 
\dword{ce} cards and boxes, as well as the need to use the composite structure as a stress relief for the cabling of the bottom electronics.
The height of the %posts 
supports can be finely adjusted during installation to set the horizontality and planarity of each \dshort{crp}.

\begin{dunefigure}
[Bottom CRP supported on the membrane by posts]
{fig:crpbottom}
{Illustration of a bottom \dshort{crp} 
supported about 160\,mm above the membrane flat surface by four posts.}
\includegraphics[width=0.95\textwidth]{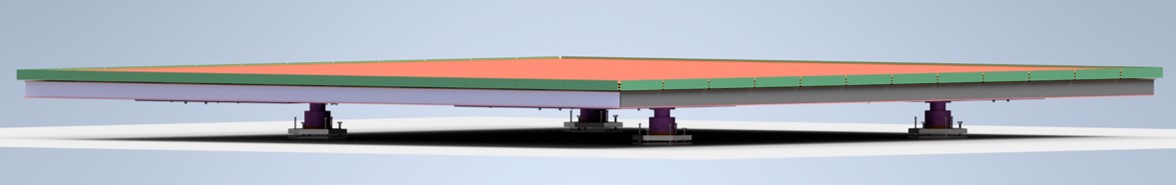}
\end{dunefigure}

Each support is attached to the \dshort{crp} composite  frame  using an adapter plate made out of \frfour (Figure~\ref{fig:crpbottom2}).  The support component consists of an aluminum post, a sliding plastic connection, and a stainless steel foot, which together span the 15.9\,cm gap between the \dshort{crp} and cryostat floor.  A low-friction sliding plastic connection allows the support to displace in-plane with the \dshort{crp} as thermal contraction occurs.  %Contact is made with the membrane floor of the cryostat with a 
The stainless steel foot %.  This same material contact 
matches the thermal contraction properties of the cryostat membrane, % of the foot and floor, 
ensuring a slip-free interface.  
The support feet and sliding connections are mounted to the aluminum post using a custom designed
 centering mechanism.  The purpose of the mechanism is to keep the foot and 
sliding connection locked and centered during installation.
One  connection among the four is fixed in order for the CRP to contract towards this point, and one 
connection is guided along one direction to prevent any possible rotation of the CRP during thermal contraction.
The two other connections are free to slide in any direction. The centering system allows for travel within a 7\,mm radius of the center.  
This is sufficient for the maximum anticipated thermal contraction displacement of 6.3 mm. 
A completely fixed version of the centering mechanism  is used  for the fixed support configurations.

\begin{dunefigure}
[Bottom CRP support components]
{fig:crpbottom2}
{The bottom support design consists of four components: the adapter plate out of G10/\frfour, and aluminum post, a low-friction plastic sliding connection, and a type 316 stainless steel foot.}
\includegraphics[width=0.65\textwidth]{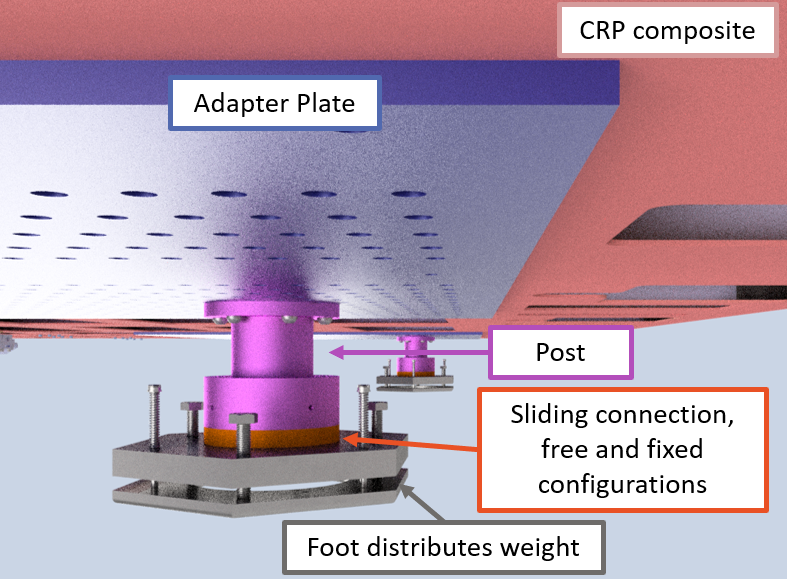}
\end{dunefigure}

Adapter plates are used for mounting the bottom support to the underside of the %CRP composite panel. 
\dshort{crp}s, two per \dword{cru}. Made out of \frfour glass composite, these plates match the thermal contraction of the \dshort{crp} structure 
and provide the needed rigidity.  The adapter plate is 3/8 inch thick, although that value may change based on material availability and analysis of the plate deflection under load.

Twelve M4 bolts are used to attach each adapter plate %The adapter plates are bolted 
to the composite frame %structure 
just inside of the \dshort{crp}  corners, as seen in Figure~\ref{fig:crpbottom3}. The plates span the U-shaped profiles of the composite structure at three horizontal and vertical positions.  
Two adapter plates will be bolted to each \dshort{cru} at the %CRP factory.
production site

\begin{dunefigure}
[Bottom CRP adapter plate attachment]
{fig:crpbottom3}
{Adapter plates are attached to the \dshort{cru}s at the \dshort{crp} factory.  Bottom supports are attached to the adapter plate using a six-holed bolt pattern}
\includegraphics[width=0.7\textwidth]{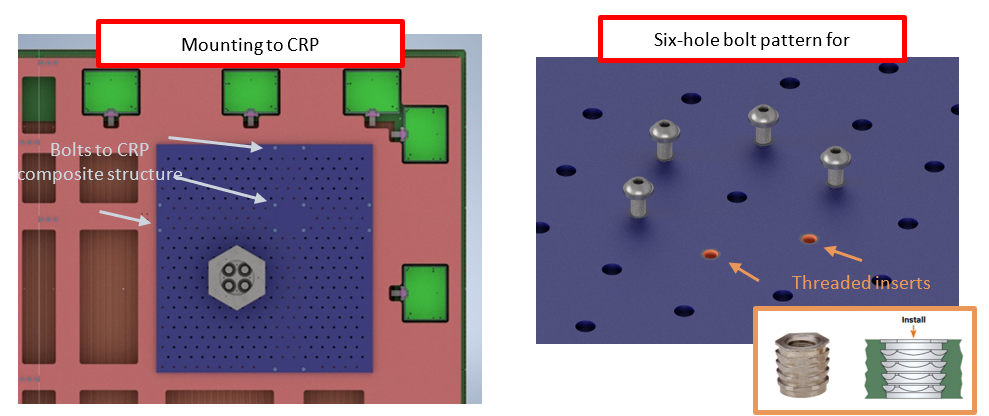}
\end{dunefigure}

%Additionally, t
The support design features height adjustment for leveling of the \dshort{crp} and a centering mechanism that fixes the free-sliding feet during %the CRP 
installation. % process.

The adapter plates are perforated with a field of 5.5\,mm holes, which %.  These holes 
provide a universal plane for mounting all required bottom support positions and %.  The holes also provide open area for 
allow  flow of both \dshort{lar} %flow and the escape of 
and vapor/bubbles during the filling process.
Six stainless steel M6 bolts are arranged in a ring pattern at the desired support mount location %are used 
to attach the aluminum support post to the adapter plate. The six chosen holes have threaded inserts for the bolts. 
Since the mount positions are different for each \dshort{cru}, %so 
care must be taken %at the CRP factory 
to ensure the correct positioning of inserts and labeling of \dshort{cru}.  The inserts are press fit into the adapter plate holes from the reverse side so that when supports are bolted on, the inserts can be further tightened. % into the adapter plate.  
Belleville washers are used %in conjunction 
with the six bolts, maintaining bolt tension during the thermal contraction of the aluminum post's bolting flange.  Contraction of the support post's bolt ring relative to the adapter plate is accounted for and aids in the tightening of the bolted connection.

The foot of the %bottom 
support is made out of two type 316 stainless steel plates.  The thicker 3/4 inch plate provides contact for the sliding mechanism and material for mounting the centering mechanism hardware as well as hardware for height adjustment of the foot.  A thinner 3/8 inch plate is attached to the thicker plate using retaining bolts and pre-tensioned springs.  This thinner plate has filleted bottom edges so as not to mar the membrane floor with which it will be in contact.  
The distance between the two plates can be adjusted $\pm$10\,mm using three adjustment screws.  The adjustment screws use hex heads so that they can be easy accessed and adjusted from the side, since when installed on a \dshort{crp} they can’t be accessed from above.  Independent height adjustment of each bottom support allows for the in-place leveling of \dshort{crp} during installation within the cryostat.

The horizontality of each bottom \dshort{crp} and the overall planarity will be adjusted using the adjustable supporting feet during the installation and verified with metrology measurements.

\section{Installation Procedures and Tooling} 

 The \dword{crp} installation procedures have been developed and presented at the %preliminary design review 
 \dword{pdr} in May 2022.  This section summarizes these procedures.
 
\subsection{Top Anode Plane} %CRP plane}

All top  \dshort{crp} installation work %is foreseen 
will take place inside the cryostat %with the exception of 
except the suspension and lifting systems, which are installed on top of the cryostat. The steps are detailed in Section~\ref{sec:detcompinst}.

The installation process includes the top \dshort{crp} signal cabling to the \dword{tde} chimneys after the cathode module attachment and the raising of the complete system 
%\fixme{a superstructure + CRPs? => DD answers: yes it is the full assembly}
to a height called ``cabling position,'' which allows  access for people doing the cabling on  top of the superstructure. This  intermediate position is
defined by a distance of \SI{800}{mm} between the top of the superstructure %frame 
and the cryostat membrane. When the cabling is completed and tested on a SST, it is raised to the final operating position defined by a distance of \SI{490}{mm} between the top of the superstructure  and the cryostat membrane.  At that moment the long cables   used with the electrical winches to raise the whole SST are replaced by the final suspension cables on the suspension feedthroughs.

There are 96 signal cables per \dshort{crp} for a total of 7680 cables connected to 105 \dshort{tde} chimneys.
The cabling process also includes the connection of the bias cables (three per \dshort{crp})  for a total of 240 cables to the cold \dword{hv} filter boxes 
sitting on the composite frame.  The cables are routed on top of the superstructures to \dword{fc} support feedthroughs.

The main steps of the top \dshort{crp} installation also shown in Figures~\ref{fig:CRPcabling} to~\ref{fig:topcrp} are:
\begin{itemize}
\item Install circular cable trays with pre-routed cables on a temporary jig (Figure~\ref{fig:CRPcabling} left, center)
\item Assemble superstructures;
\item Assemble \dshort{crp}s from %1/2CRPs 
half-\dshort{crp}s extracted from a transport cradle (Figure~\ref{fig:SSTassembly} left);
\item Install \dshort{crp}s under superstructures and perform metrology measurements  (Figure~\ref{fig:SSTassembly} center, right);
\item Install the suspension system and  lifting winches;
\item Install the elevated workstation;  %platform
\item Raise the superstructure together with the cathode supermodule (Figure~\ref{fig:topcrp} left);
\item Cable the \dshort{crp}s when superstructures are at cabling positions under the roof (Figure~\ref{fig:CRPcabling} right, Figure~\ref{fig:topcrp} right); and 
\item Raise the superstructures to their nominal positions. 
\end{itemize}

\begin{dunefigure}
[CRP signal cabling]
{fig:CRPcabling}
{%Illustration of the 
\dshort{crp} signal cabling steps: Left: pre-route the cables on a jig with the circular cable tray to be attached (middle) to the bottom part of the chimney. The cables on their cable trays are installed before the \dshort{crp}s are raised to the cabling position. Right: when the superstructures are at \SI{800}{mm} from the cryostat membrane, a team of trained people %are going on the superstructure catwalk to 
transfer the pre-routed cable connectors from the jig to the \dshort{crp} adapter board connectors. }
\includegraphics[width=0.95\textwidth]{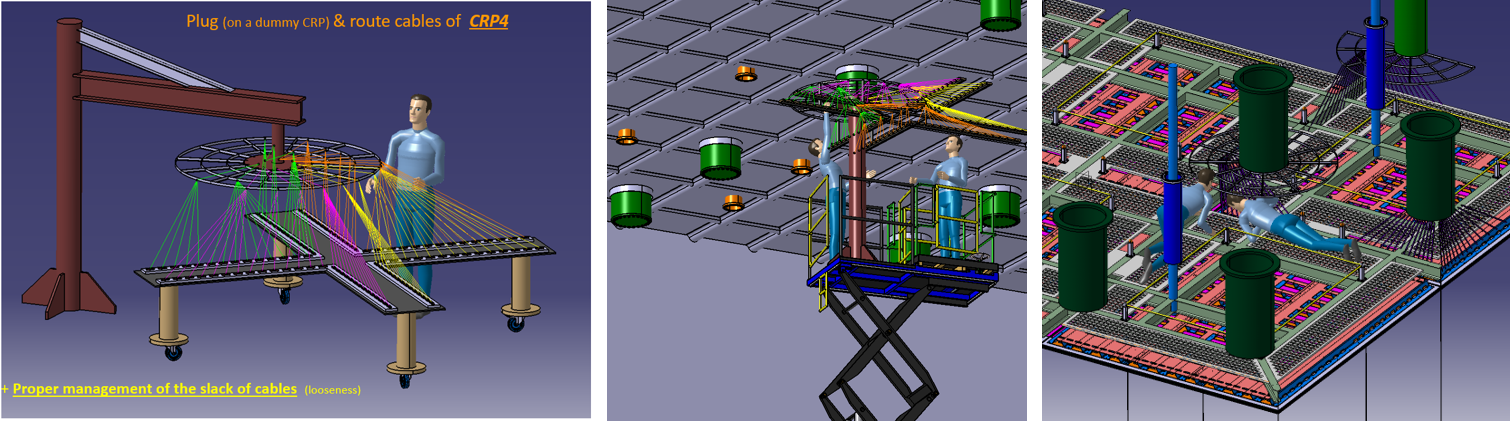}
\end{dunefigure}

\begin{dunefigure}
[Top CRP and superstructure assembly]
{fig:SSTassembly}
{%Illustration of the a
Assembly of a top \dshort{crp} from %2 half CRP 
two half-\dshort{crp}s on a dedicated table %used to roll 
that rolls under the superstructure %frame 
for attachment to the decoupling systems. The last picture on the right shows a complete superstructure with six \dshort{crp}s installed %waiting 
ready to be raised.}
\includegraphics[width=0.95\textwidth]{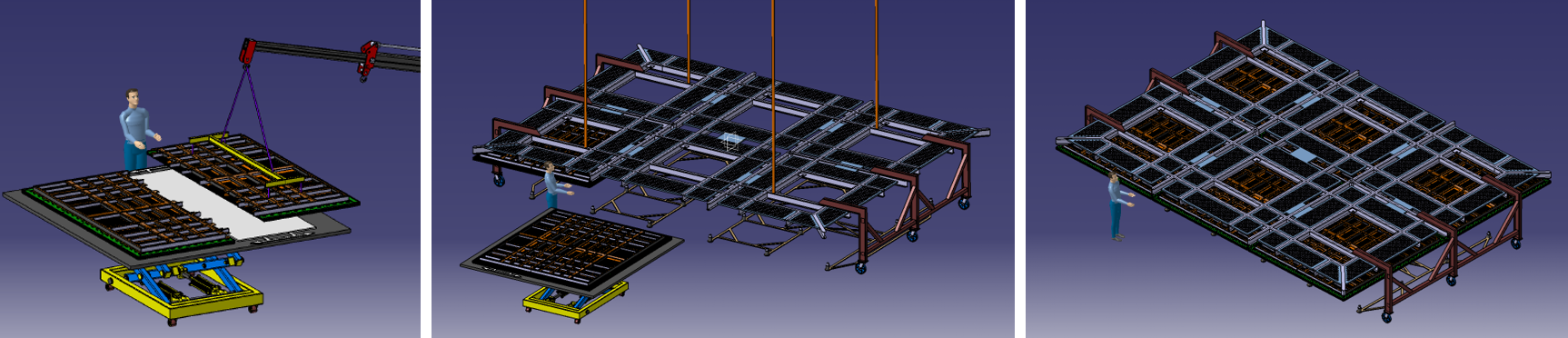}
\end{dunefigure}

\begin{dunefigure}
[Top CRP superstructure installation]
{fig:topcrp}
{%Illustration of CRP 
Left: Illustration of \dshort{crp} superstructure installation with the %access platform 
elevated workstation for cabling and three installed superstructures (two small and one large) with the cathodes suspended. Right: View from above  of two large SSTs connected to  the elevated workstation  to perform the \dshort{crp} cabling .}
\includegraphics[width=0.95\textwidth]{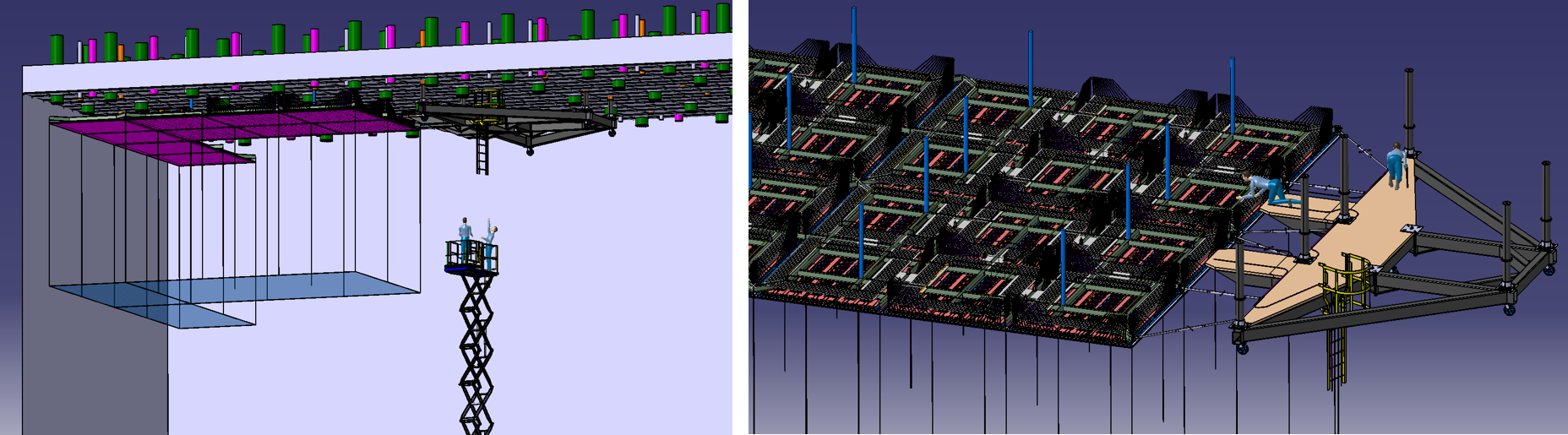}
\end{dunefigure}

The \dword{qc} %measures 
processes to perform during \dshort{crp} installation are:
\begin{itemize}

\item  Visually inspect \dshort{crp};
\item Power the \dshort{crp} anode bias circuit to control the complete electrical chain with low voltage; 
\item Verify metrology of superstructure assembly;
\item Check the electrical connectivity of the \dshort{crp} connections to the signal feedthroughs  after cabling together with \dshort{tde} consortium; and
\item Check the electrical connectivity of the anode bias cables to the biasing system on the cryostat roof together with \dshort{hv} consortium. 
\end{itemize}

\subsection{Bottom Anode Plane} %CRP plane}

Bottom \dshort{crp}s will be installed anode-side up using a system that lifts them from below using tines attached to a lifting carriage that is hoisted by a crane (Figure~\ref{fig:BotCRPLifting}.  The combined weight of the \dshort{crp} and lifting system is restricted to %is \fixme{can be? => DD answer: it is the case} no more than 
510 kg, and the center of mass of the system is %considerable far in front of the crane being used, 2.3 meters minimum. 
at least 2.3\,m in front of the crane.  This distance is needed when lifting the \dshort{crp} from the short edge and when rotating a \dshort{crp} below the crane.  Lightweight construction of the tines and carriage system is therefore necessary 
to keep the total lift weight down. 
\begin{dunefigure}
[Crane lifting bottom CRP]
{fig:BotCRPLifting}
{Counterbalanced crane lifting a bottom \dshort{crp} using the tine and carriage system to pick up the \dshort{crp} from below the adapter plates.}
\includegraphics[width=0.95\textwidth]{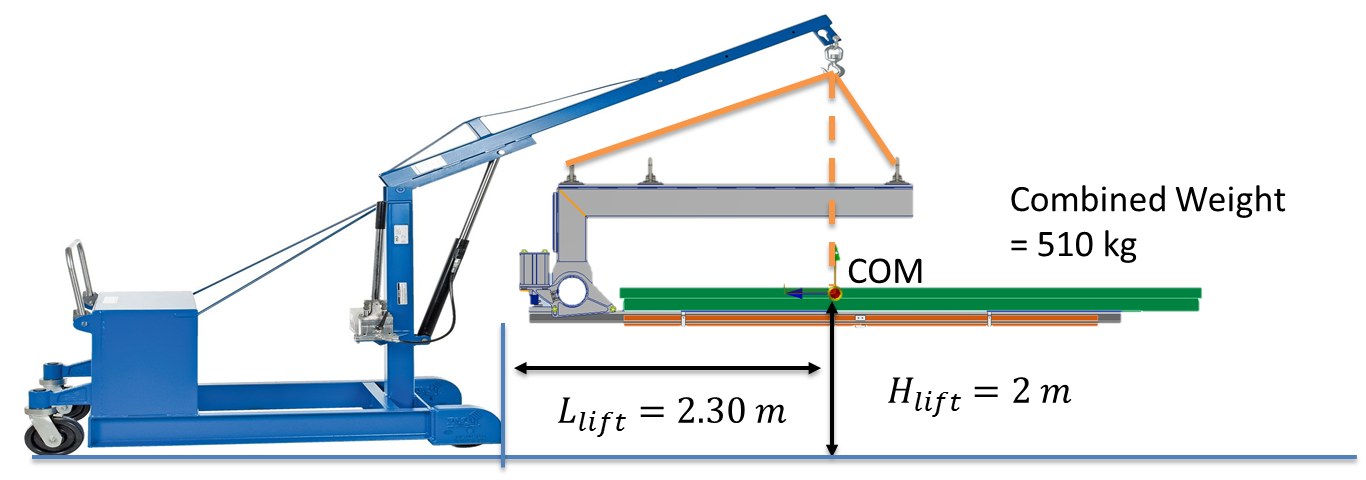}
\end{dunefigure}

The %whole 
system consists of three components: the \dshort{crp} and its attachments, the tines, and the carriage as shown in Figure~\ref{fig:BotCRPInstall}
\begin{dunefigure}
[Bottom\dshort{crp} lifting system components]
{fig:BotCRPInstall}
{Components of the bottom \dshort{crp} lifting system. The tine guiding system, called ``tine cage,'' located below the \dshort{crp} is shown as the orange bars on the right figure. They are attached to the adapter plates in blue.}
\includegraphics[width=0.95\textwidth]{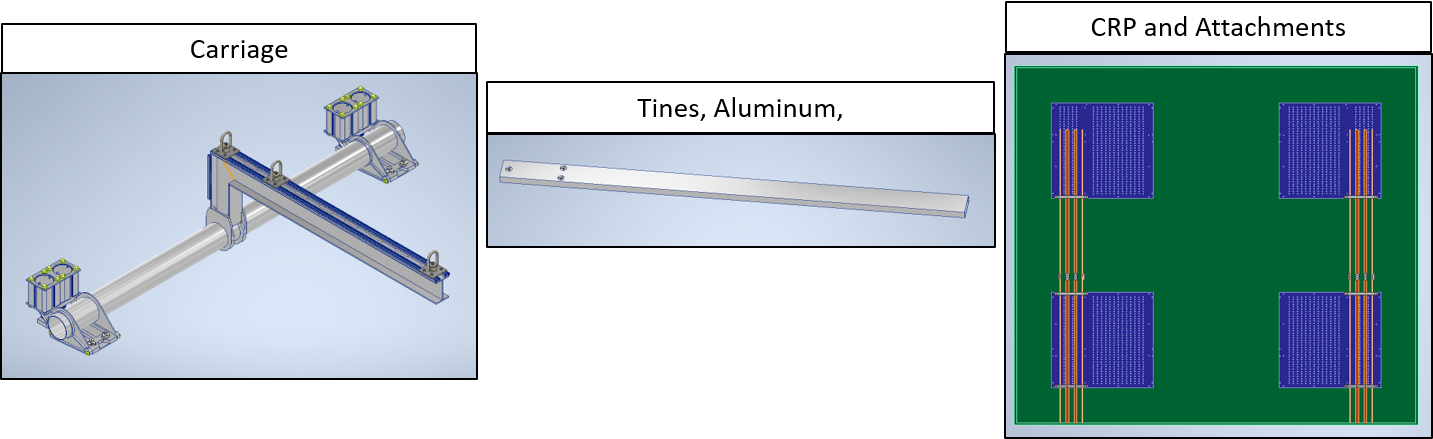}
\end{dunefigure}
Tines are inserted and extracted from under the \dshort{crp} as individual units.  %Being able to 
Detachable from the lifting carriage, the tines can be maneuvered manually underneath the \dshort{crp}, %being 
guided and slung into place by a tine cage shown in Figure~\ref{fig:tinecage}.  The cage ensures that the tines pick up from the adapter plates and don’t collide with under-\dshort{crp} wiring harnesses.  The tines feature a stop bar that contacts the side of the \dshort{crp} structure, avoiding contact with the grounding plane and edge boards.  This prevents the tines from being over-inserted.
\begin{dunefigure}
[\dshort{crp} tine cage]
{fig:tinecage}
{Details of the tine cage in orange. The tines are placed and guided underneath the CRP using the tine cage.}
\includegraphics[width=0.95\textwidth]{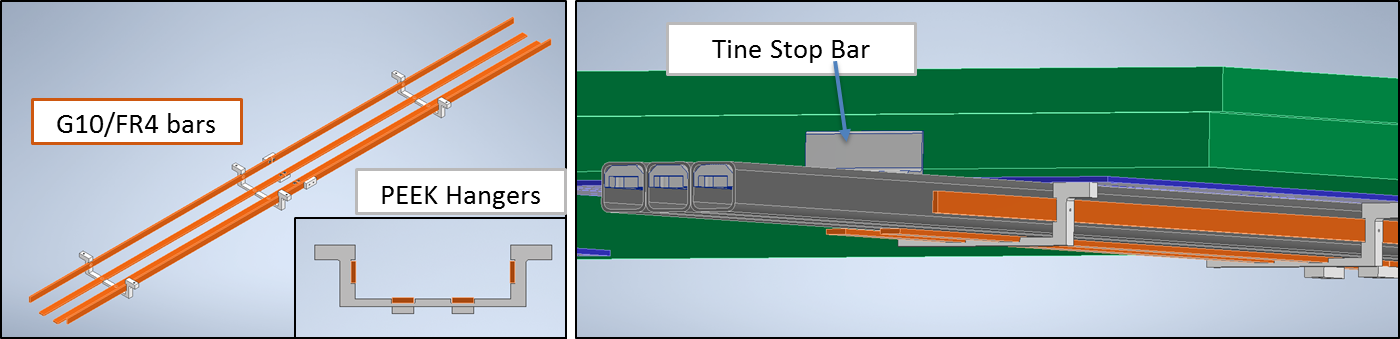}
\end{dunefigure}
The tine cage consists of \dshort{peek} hangers and G10/\frfour bars that run the length of the tine insertion.  The cages will be installed within the cryostat as the \dshort{crp} is being worked on on the assembly table, the hangers will be %being 
attached to the adapter plates in areas where the bottom supports are not attached.  The tine cages will remain underneath the \dshort{crp} after installation.  %Being made of G10/FR4 t
They will shrink at the same rate as %along with 
the \dshort{crp} structure during filling. % as LAr is filled.

The main steps of the bottom \dshort{crp} installation inside the cryostat are:
\begin{itemize}
\item Assemble \dshort{crp}s from half-\dshort{crp}s;
\item %Form one 
Outfit each \dshort{crp} with four support feet and two patch panels;
\item Assemble installation truss on membrane floor at location of \dshort{crp} installation;
\item Lift \dshort{crp} onto installation truss;
\item Attach cables to patch panels and test electronics;
\item Survey and adjust position of \dshort{crp};
\item Level %ing 
screws on bottom support feet;
\item Raise \dshort{crp} from the installation truss;
\item Break down and remove the installation truss from below;
\item Lower \dshort{crp} to floor (anode side up); and
\item Align to neighboring \dshort{crp} and/or reference bar using edge bump blocks. 
\end{itemize}

The \dword{ce} cables are connected to the patch panels located under the \dshort{crp} before being lowered down to the floor. To allow this cabling, the \dshort{crp} is positioned with the lifting system on the installation truss as shown in Figure~\ref{fig:CRP-Truss}.   The installation truss will be used also to survey and level the \dshort{crp}.

The truss is assembled on the membrane floor out of bolt-together components and uses fixed support feet for \dshort{crp}-like load transfer to floor.
The \dshort{crp} is placed on top of installation  \SI{1.22}{m} above the membrane floor.

\begin{dunefigure}
[CRP truss system with bottom CRP]
{fig:CRP-Truss}
{Left: Truss system with a bottom \dshort{crp} sitting on it for cabling of the \dshort{ce} to the \dshort{crp} patch panels. Right: Bottom view of the CRP showing the position of the four patch panels in blue.}
\includegraphics[width=0.55\textwidth]{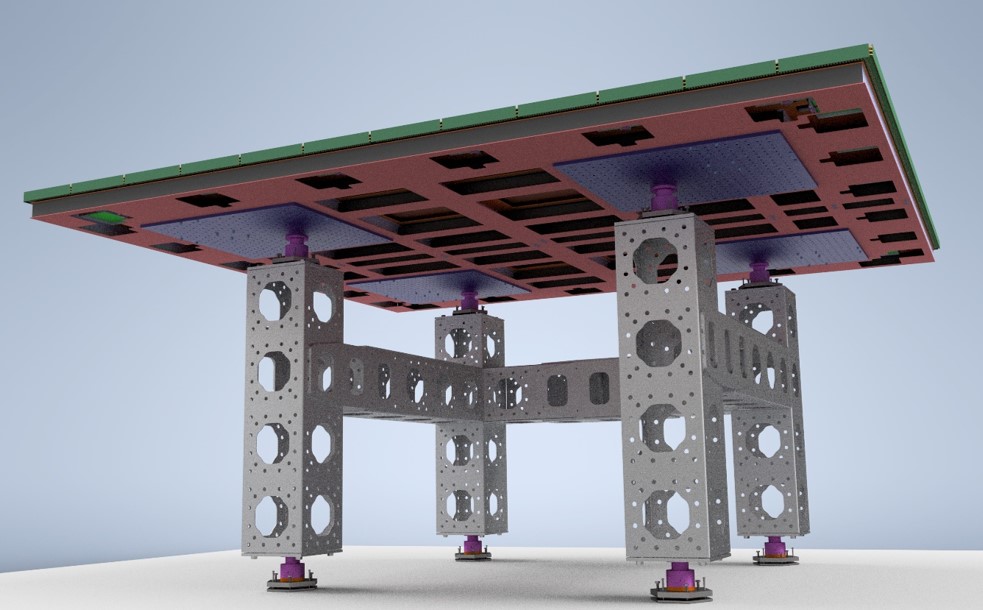}
\includegraphics[width=0.4\textwidth]{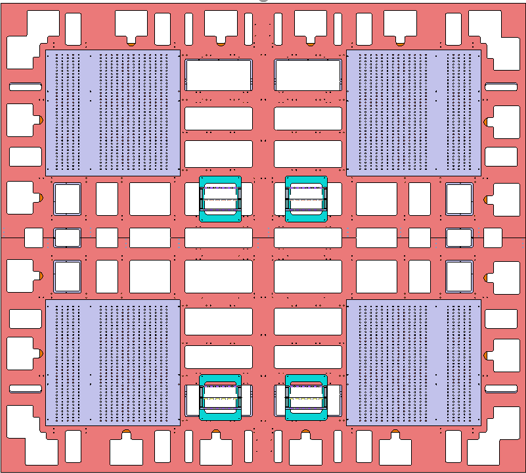}
\end{dunefigure}

%%%%%%%%%%%%%%%%%%%%%%%%%%%%
\section{Prototyping and Validation} % Plans} %

The first step in prototyping was a proof-of-principle of the perforated \dword{pcb}-based charge readout. It was successfully demonstrated at small scale at the 50L \dword{tpc} setup at \dword{cern} for both %. The first tests with a 
two-view anode and later three-view anode readout. As described in detail in~\cite{FD2-VD-CDR}, %the FD2-CDR, 
further tests at the 50L \dshort{tpc} led to improvements in the three-view geometry and detector components, %pieces 
and optimization of the anode operation. %the working conditions of the perforated anodes.

%After the successful validation campaign at small scale, 
Next, \dword{crp} prototyping moved to %the next step in 2021 when 
the first full-scale \dshort{crp} in 2021, which was built and tested at \dshort{cern} to validate the design and construction procedures. %At the time of constructing the first full scale \dword{crp}, it had the 
It was of the design  described in the \dword{spvd} \dword{cdr}~\cite{FD2-VD-CDR} and details are given in Section~\ref{subsec:FirstFullCRP}.

While the first full scale \dshort{crp} was undergoing testing, the \dshort{crp} design %had been evolved 
was still evolving to better address the physics needs, to facilitate production and assembly, and to reduce the cost of various sub-components. 
This document describes the evolved design. With the new \dshort{crp} design in place, four new full-scale \dshort{crp}s (CRP-2 through CRP-5) were constructed in 2022. CRP-2 and CRP-3 were built as top \dshort{crp}s while CRP-4 and CRP-5 as bottom \dshort{crp}s. These four CRPs will be installed into the \dword{vdmod0} in 2023. Production details for these \dshort{crp}s, as well as warm and \coldbox tests, are described in Section~\ref{subsec:CRP_prototyping2022}.

%%%%%%%%%%%%%%%%%%%%%%%%%%%%
\subsection{First Full-scale CRP Prototype (2021)} % with the FD2-CDR design}
\label{subsec:FirstFullCRP}

As shown in Figure~\ref{fig:CRP_fullscale_first}, the first full-scale \dshort{crp} had a hybrid setup where one half of the \dshort{crp} was equipped for \dword{bde} and the other half %was equipped 
for \dword{tde}. %the top drift electronics (TDE).

\begin{dunefigure}
[\threed model of the first full-scale CRP]
{fig:CRP_fullscale_first}
{\threed model of the first full-scale \dshort{crp} that was tested in the \coldbox. Perforated \dshort{pcb} anodes, adapter boards, composite frame and support structures are shown in exploded view. This \dshort{crp} had a hybrid setup in which both readout electronics designs (\dshort{bde} and \dshort{tde}) were integrated.}
\includegraphics[width=\linewidth]{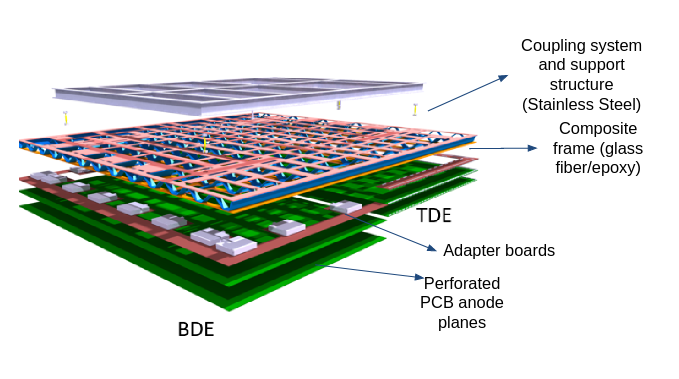}
\end{dunefigure}

The perforated \dshort{pcb} had a three-view layout with a shielding layer and the induction-1 view on the first \dword{pcbp}, and the induction-2 and collection views on the second \dshort{pcbp}. Strip orientations were (48\dge, 0\dge, 90\dge) for the induction-1, induction-2 and collection views, respectively. Two sets of adapter boards were produced to accommodate the needs of \dshort{bde} and \dshort{tde}. Interconnection between the anode planes and adapter boards was done using surface-mounted connectors and pins.

The \dshort{crp} support structure, composed of a composite frame, coupling system and metallic structure, was designed specifically for this \dshort{crp} to be compatible with both types of readout electronics. The \dshort{crp} design was finalized in April 2021 and it was fully assembled over the following six months. 

%The assembly of the first \dword{crp} was performed 
This \dshort{crp} was assembled in the clean room at building 185 at \dshort{cern}. Following the reception of the \dshort{pcb} segment from the manufacturer, \dword{qc} and cleaning %on the pieces were 
was performed and %they 
the components were prepared for gluing. Six segments were glued together to form a \dshort{pcbp} with dimensions of 1.7\,m $\times$ 3\,m. Silver printing and \dshort{qc} tests both at room and cryogenic temperatures followed the gluing. % PCB segments. 
Once the %PCB panels 
\dshort{pcbp}s were ready, surface-mount connectors and pins were soldered onto them. Then the panels were stacked together and the adapter boards %were 
installed. The anode and adapter board assembly were then connected to the composite frame and metallic structure.  

Once assembled, \dshort{bde} was installed on one of the \dshort{cru}s %at the clean room in Bld-185 at \dword{cern} 
and all channels were tested. % to be alive. 
The \dshort{crp} was then moved to \dword{ehn1} and attached to the \coldbox lid. Before warm and cold testing, %installation of 
the \dshort{tde}, \dshort{crp} bias filter and cables, as well as the cabling of the \dshort{bde} and \dshort{tde} were %performed at EHN-1. 
installed. The \dshort{crp} %then moved into 
was then put into the dedicated testing area prepared inside the \dword{np04} clean room for warm tests, during which % During the warm tests, 
noise levels of each readout electronics were measured. Different grounding configurations and bias line filtering options were also tested. Once the characterization of the \dshort{crp} was finalized at the room temperature, it went into the \coldbox for cryogenic testing. Figure~\ref{fig:CRP_coldboxInsertion} shows the \dshort{crp} being moved to the \coldbox  before it was closed. %on the left image and final moments before he \coldbox was closed on the right image.

\begin{dunefigure}
[First full-scale CRP \coldbox insertion]
{fig:CRP_coldboxInsertion}
{Left: \dshort{crp} moved from warm testing area to be inserted into the \coldbox. Right: %Final moments before \dword{crp} is fully inserted into the \coldbox.
The \coldbox just before it is closed.}
\includegraphics[width=\linewidth]{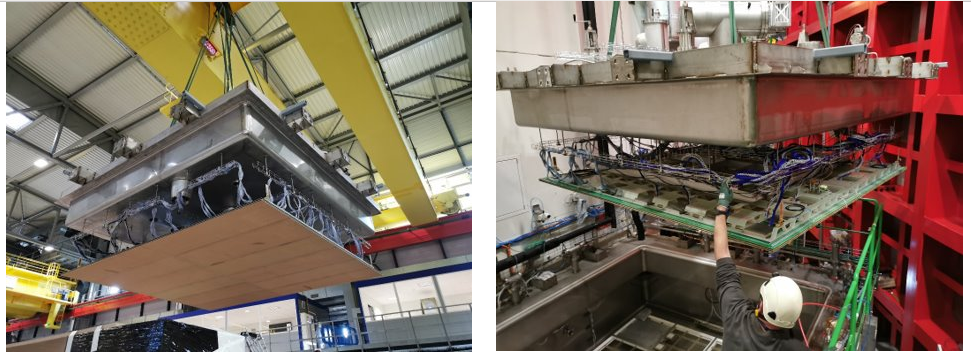}
\end{dunefigure}

There were two test runs %running periods 
with this \dshort{crp} in the \coldbox; 
the second %phase of running 
run included improved grounding for the \dshort{bde}.
During the two runs, millions of triggers were collected by the two readout chains. Figure~\ref{fig:CRP_firstTracks} shows
an example of a cosmic track recorded by the \dshort{bde}. 
%few example of tracks and related waveforms detected by the \dword{crp}. 
Details from the \coldbox runs and \dshort{tde} analysis of the collected data is presented in Section~\ref{subsubsec:topelec:cold-box}.
\begin{dunefigure}
[First cosmic track in full-scale \dshort{crp} in the \coldbox.]
{fig:CRP_firstTracks}
{Cosmic ray track recorded by the first \dshort{crp}. Induction and collection views are shown.}
\includegraphics[width=\linewidth]{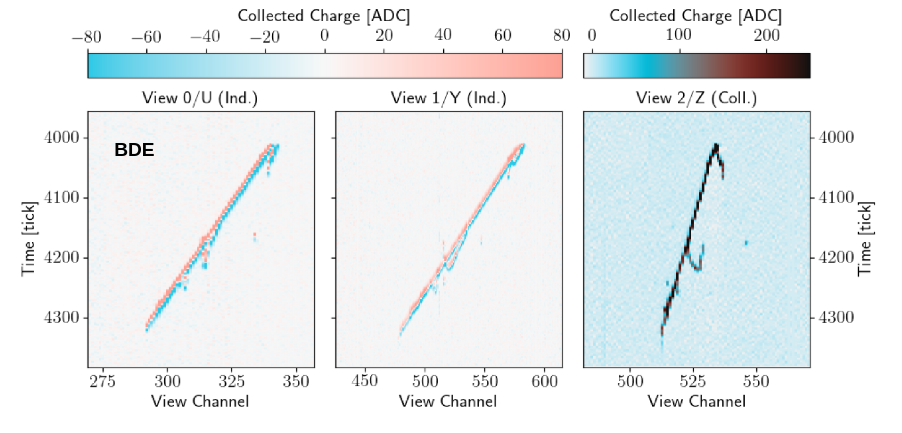}
\end{dunefigure}

The \dshort{crp} was very stable during the \coldbox runs.  Problematic channels, defined as those with with no signal, or partial signal due to lost connectivity in one of the silver print junctions, or higher noise, were quantified as 1.3\% of the channels.  % were defined as the problematic channels. 
They were monitored during the \coldbox runs and no additional degradation was observed. Results from the \coldbox runs demonstrated that the full-scale \dshort{crp} behaved well and in line with expectations from the 50L \dword{tpc} tests.

%%%%%%%%%%%%%%%%%%%%%%%%%%%%
\subsection{CRP Production and Testing (2022)} 
\label{subsec:CRP_prototyping2022}

The anode \dwords{pcb} were produced using 
3.2\,mm thick \frfour as the base material and are perforated with 2.4\,mm holes.

A bare %finish
\dshort{pcb} board has a 35\,$\mu$m thick copper image applied on both sides. In order to evaluate the finishing options, CRP-4 used a silver coating over the copper. % plating. 
Given that six \dshort{pcb} segments are glued together to form a \dshort{crp} panel, milling on the ``half-lap'' joint is one of the important steps in the \dshort{pcb} production. A gap or an uneven surface between two glued \dshort{pcb}s could create difficulties for the silver printing.  

The thickness of the board was measured before milling, and half of the measured thickness plus 50\,$\mu$m was removed on the half-lap joint. The tolerances on this process are $+0/-50$\,$\mu$m.

At the prototyping stage, production of perforated PCBs for a CRP %takes 
has been taking 4-6 weeks. Upon receiving the PCB segments they have been visually inspected for any possible production problems, % is performed. Then, 
then the segments have been cleaned, washed and dried %for the construction.
prior to gluing.
Preparation for gluing starts with covering the half-lap regions with a special tape %. The purpose of the tape is 
to prevent the epoxy from spreading to unwanted regions. 
Then epoxy was applied to the half-lap surface using a roller.

Adjacent \dshort{pcb} segments were aligned by using two sets of alignment holes with metallic pins inserted into them. Once all six panels were aligned and glued, they were covered with a vacuum cloth and sealed under vacuum for drying. Drying takes approximately one day. Once dry, the assembly was inspected. % for any possible problems or if small corrections are needed.
 Figure~\ref{fig:CRP_PCB_gluing} shows the %details of PCB 
gluing process. 
\begin{dunefigure}
[PCB gluing details]
{fig:CRP_PCB_gluing}
{Left: Preparation of PCB segment for epoxy application. Middle: Six PCB segments are glued, aligned and ready to be sealed. Right: Glued assembly is under vacuum for drying. }
\includegraphics[width=\linewidth]{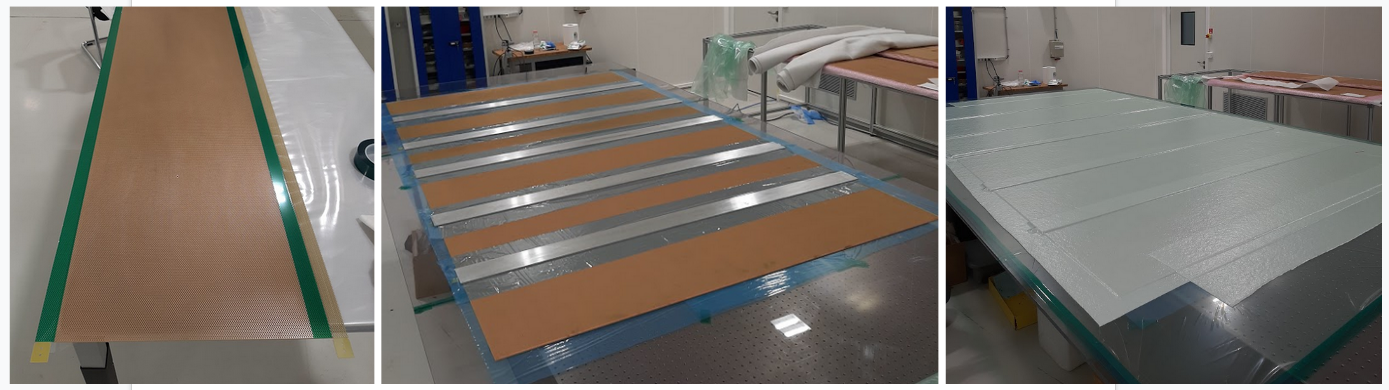}
\end{dunefigure}

Next, the electrical connections between them were made by applying 
silver ink, a mixture of 
%made from a combination of 
silver and epoxy, to the joint points using a special screen. 
Then it was treated under $\sim$120$^{\circ}$C for $\sim$3 hours for its polymerization. 
Figure~\ref{fig:CRP_silverPrinting} shows the silver ink application using the screen, epoxy treatment with heat, and the result. % of silver printing.
\begin{dunefigure}
[PCB silver-printing details]
{fig:CRP_silverPrinting}
{Left: Screen for the silver ink. Silver ink is visible (gray) next to the dot pattern, ready to be applied. Middle: Treatment of the silver print at high temperature for polymerization. Right: Result of silver printing.}
\includegraphics[width=0.8\linewidth]{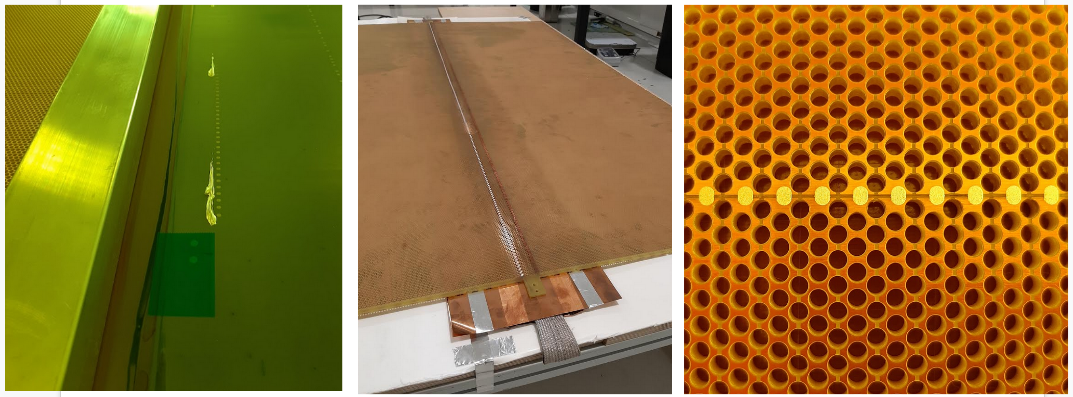}
\end{dunefigure}

%Once the silver ink is polymerized, 
The initial electrical continuity and isolation tests were performed at room temperature, followed by cold tests in a large cryogenic bath, then drying and cleaning before assembly. 
it is ready for  \dshort{cru} assembly. 
Figure~\ref{fig:CRP_coldTest_dry} shows a picture from testing PCB panels in LAr and drying cleaned panels for the assembly. 
\begin{dunefigure}
[PCB panel cold testing and drying before assembly]
{fig:CRP_coldTest_dry}
{Left: Cryogenic test of the glued and silver printed PCB panel. Right: Drying in the clean room after final cleaning prior to assembly.} % and getting ready for the \dword{crp} assembly.}
\includegraphics[width=0.8\linewidth]{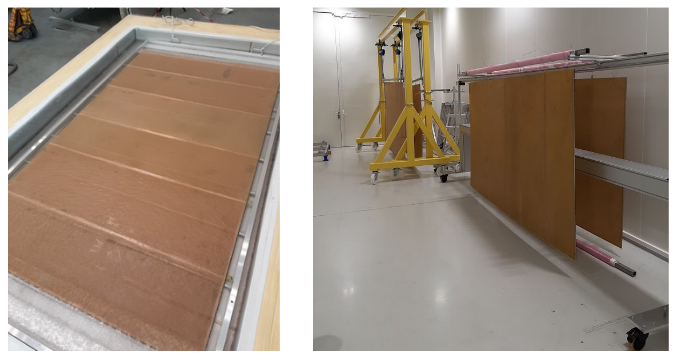}
\end{dunefigure}

The \dshort{crp} assembly started by inserting \dword{peek} support screws to the induction-2/collection plane. 
Each pin was kept at cold before being cryo-fit to the special holes on the PCB.
Once the support pins were installed, the induction-2/collection plane was placed on top of the shield/induction-1. The two planes were aligned using the corners and support screws, then %. Once the alignment is achieved, 
the shield/induction-1 board was connected to the pins using nylon screws. 

At this point the PCB assembly was ready for the adapter board installation. The adapter boards sit on the spacers that are installed on the \dshort{peek} pins, located at the periphery of the PCB assembly. 
Once the \dshort{pcbp}s were stacked and the adapter boards  installed, the assembly was attached to the composite frame using the \dshort{peek} support screws. 
Upon receipt of the composite frame from the manufacturer, 
the two halves were first visually inspected and cleaned. 
For this prototyping stage, the composite frames were tested in a large cryogenic bath to check for any mechanical issues or flatness deviations. 
Figure~\ref{fig:CRP_compositeFrame} shows the composite frame received from the manufacturer,  its attachment to the %half \dword{crp} 
\dshort{cru} and connecting the two halves into the \dshort{crp}. 
\begin{dunefigure}
[CRP attachment to the two halves of composite frame] % and its attachment to the CRP]
{fig:CRP_compositeFrame}
{Top Left: Two halves of the composite frame as received from the manufacturer. Top Middle: Half composites are attached to the anode assembly and they are ready to be connected. Top Right: Connecting two half units. Bottom: Composite frames are connected and assembly is ready for the edge card installation. }
\includegraphics[width=0.8\linewidth]{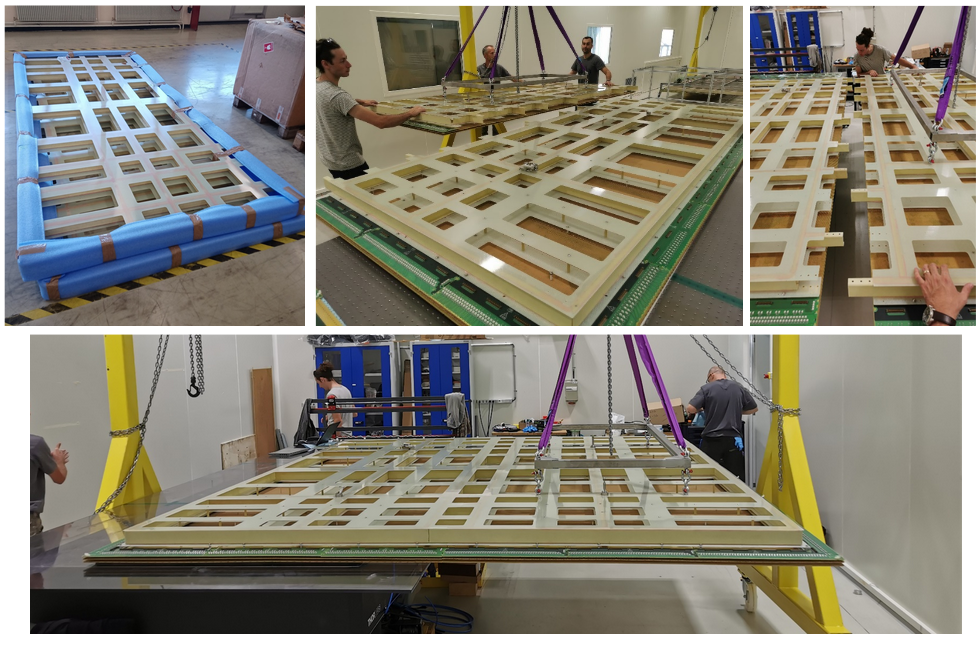}
\end{dunefigure}

As the final step of the \dshort{crp} assembly,  the vertical interconnection between the anode layers and adapter boards was done by installing the edge cards, which are plugged into the three sides of the anode assembly. 
Figure~\ref{fig:CRP_assembly_details} shows the final \dshort{crp} assembly. The process of constructing a CRP takes roughly a month for a group of three people: three weeks for PCB panel preparations and a week for the final assembly. 
\begin{dunefigure}
[Details of the finalized CRP assembly]
{fig:CRP_assembly_details}
{Finalized top \dshort{crp} assembly where the different components are shown. One edge card at the (left) corner is removed to reveal %show the details of 
the \dshort{pcbp} adapter board stack. U-beam and skin structure of the composite frame are visible.  Adapter boards and edge card are also shown. } % in smaller pictures.}
\includegraphics[width=0.8\linewidth]{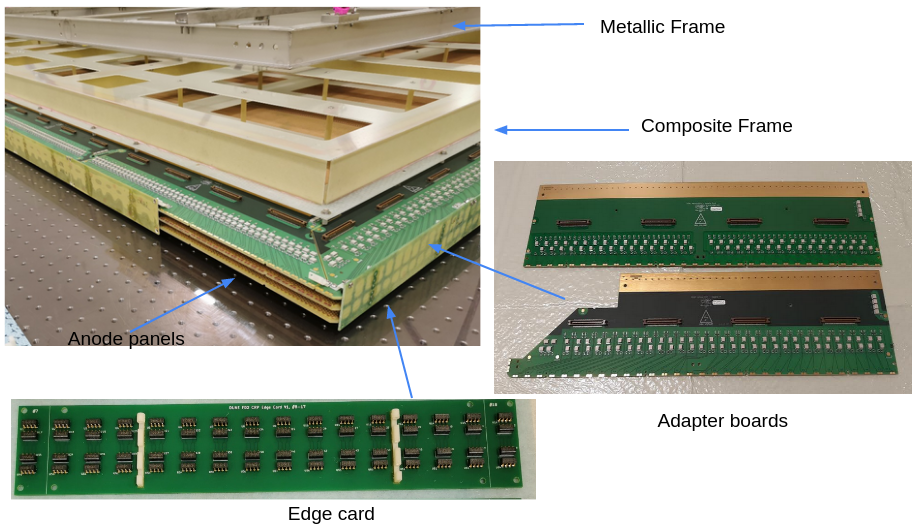}
\end{dunefigure}

%%%%%%%%%
\subsubsection{CRP-2} % Production and Testing}
\label{subsubsec:CRP_2_production}

CRP-2 was constructed %during the first three weeks of 
in July 2022. 
Despite careful preparations, the results using the silver printing mask were not satisfactory and several channels required manual corrections. Taking into account preparation time, difficulty, and unsatisfactory results, using a screen for silver printing was discarded. It was done manually for the remaining two PCB panels of the CRP-2. 

Adapter boards for CRP-2 and CRP-3 were produced %partially by 
both by the CERN PCB workshop and a commercial company\footnote{Tecnomec Srl}.
%Similar to the adapter boards, e
Edge cards for CRP-2 and CRP-3 were also produced %together, %. Production of the edge cards were done 
by a commercial company\footnote{PCBWay}. The adapter boards and edge cards 
were assembled at the CERN electronics assembly lab. 
The composite structure of CRP-2 was delivered in early May %on May 3rd 
in two parts. They were bolted together and the composite assembly was finalized by the following week. 

Figure~\ref{fig:CRP_assembly_crp2} shows the finalized CRP-2 along with some of the people involved in the construction. Upon finalizing the assembly in the clean room at CERN building 185, CRP-2 was put into the transportation box and shipped to EHN1 for warm and cold testing (Figure~\ref{fig:CRP_EHN1_operations_crp2}, top left). At EHN1, CRP-2 was attached to the \coldbox roof and secured to the required height (Figure~\ref{fig:CRP_EHN1_operations_crp2}, top right). Then, the TDE signal cables were connected to the adapter boards on one end and to the readout chimney on the other end. After finalizing the level meter installation and necessary cabling, CRP-2, %together with 
secured to the \coldbox roof, was moved to the Faraday cage for warm testing. Figure~\ref{fig:CRP_EHN1_operations_crp2} shows the CRP plus \coldbox roof at EHN1 and their transport to the dedicated Faraday cage for warm testing. 

\begin{dunefigure}
[CRP-2 with the team involved in construction]
{fig:CRP_assembly_crp2}
{Finalized CRP-2 assembly and some of the people involved in the construction.}
\includegraphics[width=.8\linewidth]{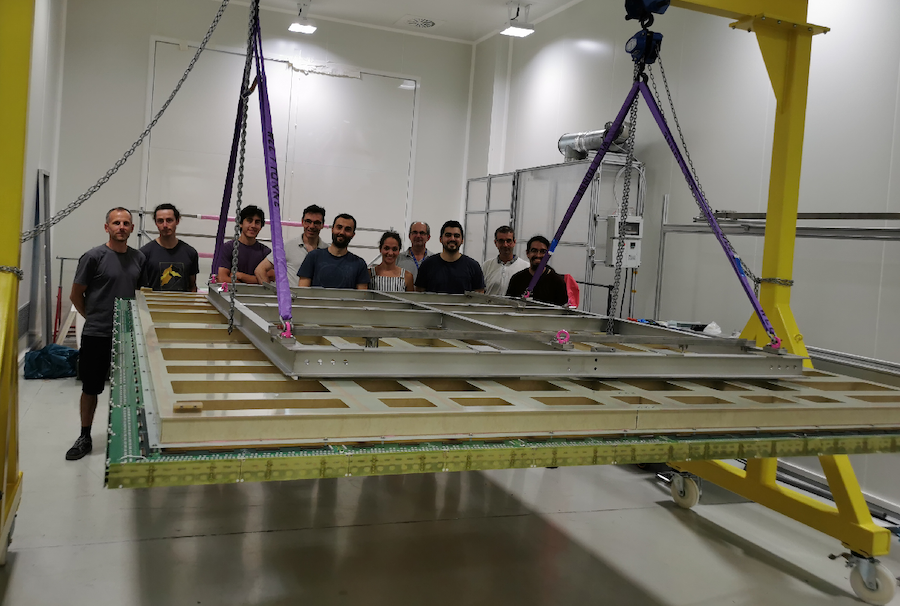}
\end{dunefigure}

\begin{dunefigure}
[CRP-2 transport and preparations at EHN1]
{fig:CRP_EHN1_operations_crp2}
{Top Left: CRP-2 inside the transport box is being loaded to a truck for the transport to EHN1. Top Right: At EHN1, CPR-2 is attached to the \coldbox roof and raised to the final height. Bottom Left: The \coldbox roof with the attached CRP-2 is on its way to the Faraday cage. Bottom Right: The \coldbox roof positioned at its final position on the Faraday cage room.}
\includegraphics[width=\linewidth]{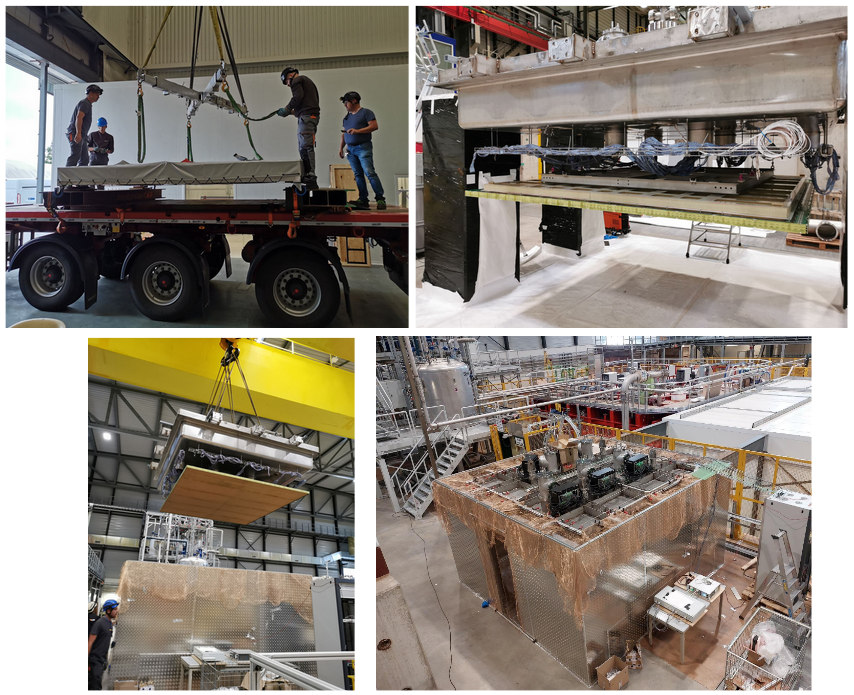}
\end{dunefigure}

Once inside the Faraday cage, the  biasing circuit for the CRP-2 was finalized. 
Cold filters were installed, and HV cables were routed and attached to the HV flange on top of the \coldbox roof. 
A voltage $\sim$50\,V was applied to each bias line to check for any leakage current, and none was found. %Once the CRP-2 was in its final configuration, t
The \dword{tde} group then made extensive warm noise measurements and identified four channels with very low noise and 10 channels with a slightly higher noise levels. 

%After the validation in Faraday cage, t
The assembly was moved to the \coldbox for testing.
%Data taking was started right 
Very soon after the filling, %, and soon after
very clean samples of tracks at low noise conditions were collected. %More than 1M triggers were collected during the \coldbox runs with CPR-2. 
Figure~\ref{fig:CRP_crp2_tracks} shows few examples from the detected cosmic ray tracks. Analysis of the collected data is presented in Section~\ref{subsubsec:topelec:cold-box}.

\begin{dunefigure}
[Examples from the tracks recorder during the CRP-2 \coldbox test]
{fig:CRP_crp2_tracks}
{Two examples from the cosmic ray tracks recorded with CRP-2 during the \coldbox run. }
\includegraphics[width=\linewidth]{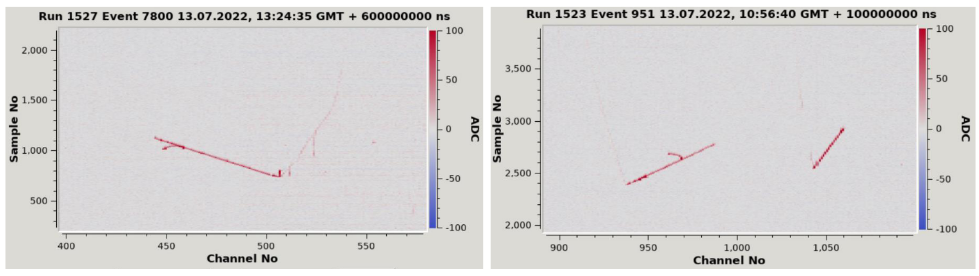}
\end{dunefigure}

Detailed analysis %of the \dword{crp} 
confirmed, however, that the abnormal noise observed at warm was also present in the data collected in the \coldbox.
In fact, additional channels with lower noise or partial signal appeared. Most of these newly found problematic channels were close to strips located on the induction-2 plane on one \dshort{cru} only. 
In early September 2022, this \dshort{cru} was disassembled, and visual inspection and electrical measurements %were performed on the induction-2 layer. In the non-responsive regions, 
revealed electrical discontinuity on the silver joints in the non-responsive regions. % were measured.
The silver joints %on the problematic induction-2 channels 
were then manually fixed, remeasured, and validated. 

%After finalizing the inspections and fixes 
CRP-2 was re-assembled at the end of October and capacitance measurements confirmed that the problems on induction-2 were no longer present. %CRP-2 then shipped to EHN-1 and cold-tested from November 2nd to November 6th. 
During the second \coldbox runs in early November, CRP-2 worked as designed and tests were concluded successfully. More than 1M triggers were collected during the \coldbox runs with CPR-2.

\subsubsection{CRP-3} % Production and Testing}
\label{subsubsec:CRP_3_production}
PCB gluing for the CRP-3 production was finalized at the end of July 2022 and %. After the summer break, 
silver printing was concluded at the end of August. 
After QA/QC tests at room and cryogenic temperatures, % and silver printing,  
assembly took place at the end of September 2022. See Figure~\ref{fig:CRP_crp3_assembled}.

\begin{dunefigure}
[Finalized CRP-3 assembly inside the clean room at CERN building 185]
{fig:CRP_crp3_assembled}
{Finalized CRP-3 assembly inside the clean room at CERN building 185 and some of the \dshort{crp} assembly team.} %working on the assembly.}
\includegraphics[width=0.8\linewidth]{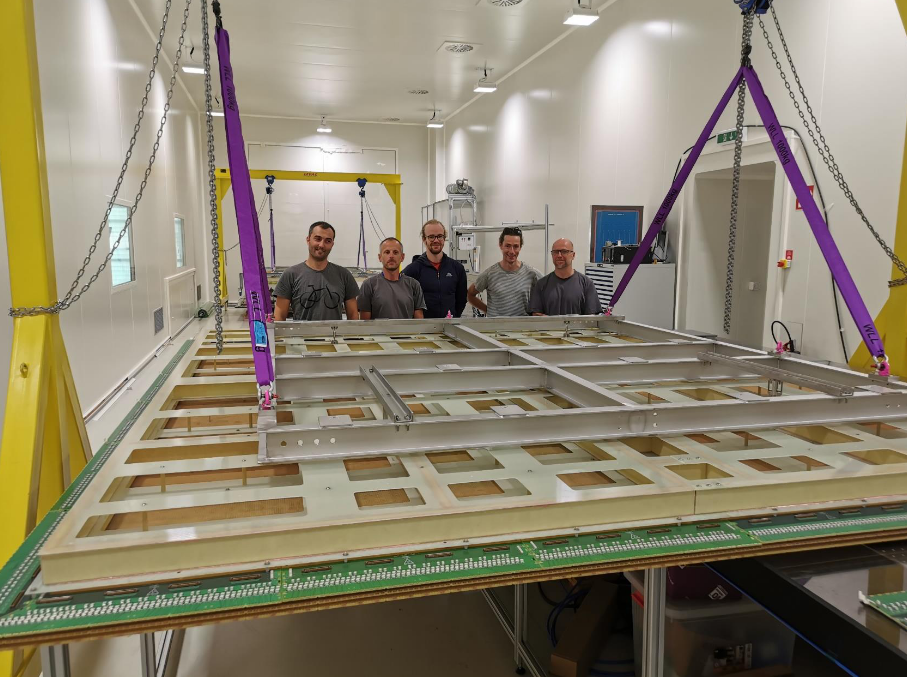}
\end{dunefigure}

Before the \coldbox test, capacitance tests revealed contact problems that were then found to be due to inadequate mechanical support. 
The problem was quickly resolved by adding mechanical supports to the edge cards. Two other channels were found to have breaks in the silver joint, and two with shorts.

Following Faraday cage tests, the \coldbox test was performed in mid-October, and both CPR-3 and TDE readout %were active soon after the filling, responding 
responded as expected; %where 
clean cosmic tracks were visible immediately. 
No major issues were observed during the tests. See Figure~\ref{fig:CRP_crp3_hitmap}. Analysis of the collected data with CRP-3 is presented in the Section~\ref{subsubsec:topelec:cold-box}.

\begin{dunefigure}
[Hit map of CRP-3 for one of the runs during the \coldbox tests]
{fig:CRP_crp3_hitmap}
{Hit map of CRP-3 during one of the runs in the \coldbox.}
\includegraphics[width=\linewidth]{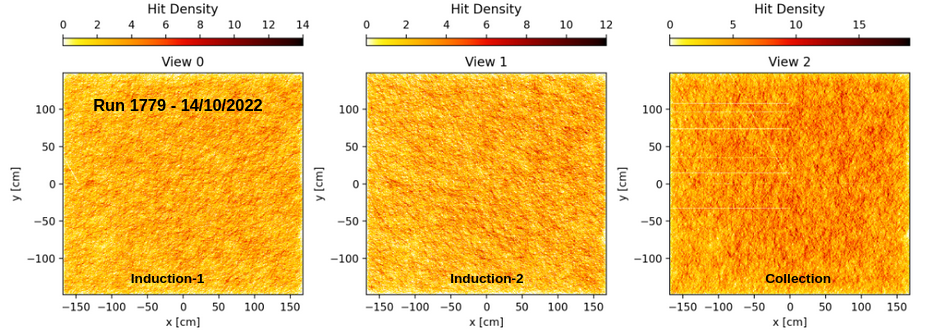}
\end{dunefigure}
After the \coldbox tests, CRP-3 was moved back to 185 to be stored in the clean room until its installation in Module-0.

\subsubsection{CRP-5} % Production and Testing Status}
\label{subsubsec:CRP_5_productionTesting}
CRP-5 was split into its two halves, CRP-5a, to be assembled in the U.S., and CRP-5b, to be assembled at CERN. The \dshort{pcb} preparation and \dshort{qc} for both 
were finalized at \dshort{cern} by the middle of July 2022. The two PCB panels for CRP-5a, half composite frame, and various small components were packed and shipped in early August to Yale University for assembly. 
In addition to the  assembly steps followed for CRP-2 and CRP-3, a thin copper grounding layer was also attached to the CRP-5a composite frame.

Figure~\ref{fig:CRP_crp5a_yale} shows various photos from %the Yale setup for 
the CRP-5a assembly at Yale. After assembly, CRP-5a %then was put into a shipping crate and moved 
was shipped to BNL for extensive testing.
\begin{dunefigure}
[CRP-5a assembly at the Yale Wright Laboratory]
{fig:CRP_crp5a_yale}
{Top Left: Two PCB panels are stacked together using the PEEK spacers. Top Middle: Installed adapter boards on the PCB panel assembly. Top Right: The half composite frame with the copper grounding plane attached to it. Middle Left: Attaching composite frame to the anode assembly. Middle Center: Installation of edge cards. Middle Right: CRP-5a is ready to be moved to BNL. Bottom: Team that assembled the CRP-5a. }
\includegraphics[width=0.9\linewidth]{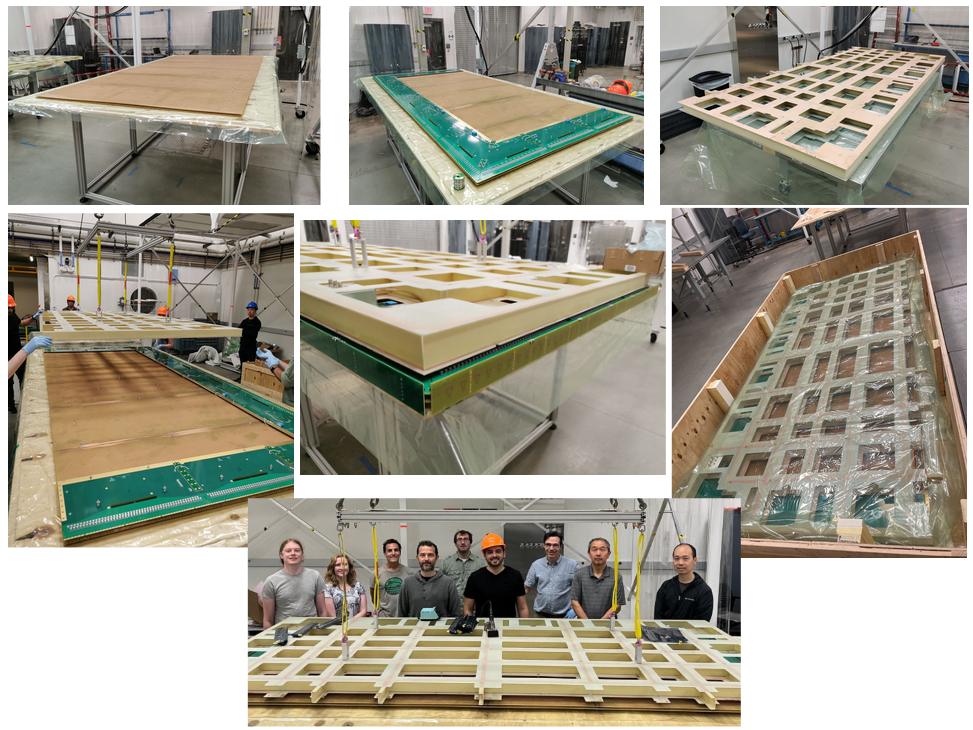}
\end{dunefigure}
At BNL, once the safety procedures were approved and reviews were passed, CRP-5a was moved out of the shipping crate, FEMBs were installed, cabling was finalized, and warm tests were conducted by late November. Cold tests were started immediately afterwards. %Then, following the warm tests, at the end of November cold tests were started. 
For the cold tests at BNL, a dedicated \coldbox was built and certified. At the time of writing,  the cold tests are ongoing. 

Following the extensive cold-tests at BNL, CRP-5a will be shipped to CERN where it will be connected to %the second half CRP (
CRP-5b. CRP-5b was assembled at CERN at the end of November, and FEMBs were installed and tested in December.
The full CRP-5 will be tested in the \coldbox at CERN in early 2023.

\subsubsection{CRP-4} % Production Status}
\label{subsubsec:CRP_4_production}
CRP-4 is the second bottom \dshort{crp} prototype.  
The PCB gluing and silver printing were done at CERN in 2022 and the assembly will be completed at Yale University by January 2023. 

As shown in Figure~\ref{fig:CRP_crp4_perforatedPCBs}, perforated PCBs for CRP-4 has a layer of silver coating over the bare copper.  
The silver coating is not costly and may address possible oxidation issues given the long-term storage that may be required.
Figure~\ref{fig:CRP_crp4_perforatedPCBs} shows %pictures of the 
perforated PCBs with silver coating and PCB panels built from them.
\begin{dunefigure}
[Silver-coated perforated PCBs for CRP-4]
{fig:CRP_crp4_perforatedPCBs}
{Top and Bottom Left: Close up images from the silver coated perforated PCBs. Top Right: Panels for CRP-4; One on a low table (foreground), %the table on the floor, 
one on the optical table (rear), and a stack of two on the workbench (right). Bottom Right: %Closeup image of one of the 
A PCB panel for CRP-4.}
\includegraphics[width=0.8\linewidth]{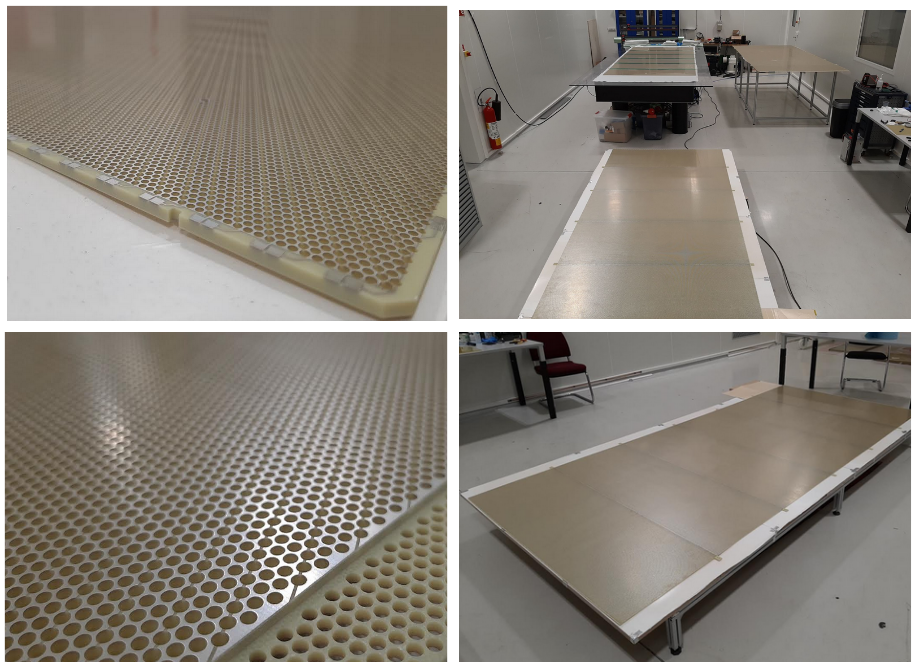}
\end{dunefigure}

In early December, upon finalizing all the necessary preparations at CERN, four PCB panels, %layers, 
the composite structures, and various small components were shipped to Yale for assembly.
After the assembly and integration of cold electronics, CRP-4 will be cold-tested at a dedicated test stand at Yale, then %. After the cold-tests it will be 
shipped to CERN, possibly for another \coldbox test, and later its integration into \dword{vdmod0}.

%%%%%%%%%%%%%%%%%%%%%%%%%%%%%%%%%%%%%%%%%
\section{Interfaces} %added by anne 30 sep
\dword{crp} consortium has interfaces with the 
%BDE, TDE, HV 
\dword{bde}, \dword{tde}, and \dword{hv} consortia and with the %I$\&$I 
\dword{fsii} group. Table~\ref{tbl:crp-interfaces} %contains a brief summary of 
summarizes %all of 
the interfaces, % between the \dword{crp} consortium and other consortia, 
with references to the current versions of the interface documents. % in EDMS. 
More details follow. 

\begin{dunetable}
[\dshort{crp} interfaces]
{p{0.15\textwidth}p{0.8\textwidth}}
{tbl:crp-interfaces}
{\dshort{crp} interfacing systems (linked to interface documents) and interface descriptions.}
Interfacing System & Description  \\ \toprowrule
\href{https://edms.cern.ch/document/2618995/1}{\dshort{bde}} & Mechanical (connections of CE boxes, patch panel and cable routing) and electrical (bias voltages, \dshort{femb}--copper plane connection, grounding 
scheme) \\ \colhline

\href{https://edms.cern.ch/document/2618998/1}{\dshort{tde}} &  Signal cables, cable routing, cable installation and circular cable trays \\ \colhline

\href{https://edms.cern.ch/document/2619003/1}{\dshort{hv}} &  Mechanical (support of cathode modules) and electrical (bias connections on the \dshort{fc} flange, bias power supplies, warm cables for all \dshort{crp}s, as well as the cold cables for the top \dshort{crp}s)\\ \colhline

\href{https://edms.cern.ch/document/2648559/1}{Installation} &  Infrastructure: Design files, \twod drawings, safety analysis and approval, fabrication of 16 supermodule structures, rigging equipment, tooling and hardware. Installation: Engineering notes, procedures, tooling, storage space and rigging equipment. Bottom and top \dshort{crp} installation.  \\ 
\end{dunetable}

The \dshort{crp} and \dshort{bde} consortia share mechanical and electrical interfaces. Mechanical interfaces include installation of \dwords{femb} to the \dshort{crp} adapter boards, installation of patch panels to the composite frame and cable routing for the \dshort{bde}. Electrical interfaces %composed of 
include biasing the \dshort{crp} planes and \dshort{femb}-copper plane grounding connections. Design, production and installation of the adapter boards and composite frame, installation and testing of \dshort{femb}s, cable routing to the patch panel are \dshort{crp} consortium responsibilities. \dshort{bde} consortium is responsible for the design and production of bottom \dshort{crp} bias connections and warm filter on the \dshort{bde} flange, procurement and installation of the cold \dshort{hv} cables, final \dword{ce} cabling inside the cryostat. In addition, \dshort{bde} is responsible for %to provide 
the signal and power cables for the \dshort{femb}s, and the test stand for the bottom \dshort{crp} factories.

The interfaces between the \dshort{crp} and \dshort{tde} consortia include signal cables, cable routing, cable installation and their support via circular cable trays. The \dshort{crp} consortium is responsible for the design and production of the top adapter boards, procurement of the flat signal readout cables, installation of the flat cables to the cold flanges inside the cryostat, design and production of the circular cable trays to dispatch the cables from the cold flange to the \dshort{crp} adapter boards. %Meanwhile, 
The \dshort{tde} consortium is responsible for the installation of the chimneys and  providing mechanical attachment points for these cable trays to the chimneys.

The interfaces between the \dshort{crp} and \dshort{hv} consortia include supporting the weight of %all cathode modules 
the cathode under the \dshort{crp} superstructure, providing bias connections for the top \dshort{crp}s on the \dshort{fc} flange, procuring \dshort{crp} bias power supplies, providing warm bias \dshort{hv} cables for all the \dshort{crp}s and cold \dshort{hv} cables for the top \dshort{crp}s. The \dshort{crp} consortium is responsible for the design of the support structure and %its 
the \dword{qa}\dshort{qc} testing of all its components, including the anchor points of the cathode suspension ropes. The \dshort{hv} consortium is responsible for the complete model of the cathode module, 
including detailed weight distribution, rope selection and the design of the connection between the ropes and the superstructure anchor points. In addition, \dshort{hv} %consortium 
is responsible for implementing \dword{shv} feedthroughs to allow the top \dshort{crp} bias voltage channels to be brought out of the cryostat through the \dshort{fc} support penetrations, and procurement of the bias power supplies and warm cables for all \dshort{crp}s, and the cold cables for the top \dshort{crp}s.

The \dshort{crp} and %I$\&$I 
\dword{fsii} group share infrastructure and installation interfaces. Infrastructure interfaces include design files, \twod drawings, safety analyses and approvals, fabrication of 16 supermodule structures, rigging equipment, tooling and hardware. The \dshort{crp} consortium is responsible for providing approved engineering notes and \twod drawings of the two types of \dshort{crp} superstructures, %. \dword{crp} %consortium is also responsible to provide 
safety analyses and approvals of the superstructure design, assembly procedures, testing and \dshort{qc} information for the superstructures. In addition, it will provide the design of the motorized suspension system and related hardware for the \dshort{crp} support structure as well as the automated level adjusting system and related hardware. % are the \dword{crp} responsibilities. I$\&$I group 
\dshort{fsii} is responsible for fabricating the 16 superstructures (four small and 12 large), and providing any crates or boxes, tooling, hardware (nuts and bolts) needed to assemble  the superstructures. % In addition, I$\&$I is responsible to 
\dshort{fsii} will also provide the rigging equipment needed to move the superstructures underground and into the cryostat, and the related drawings.

The installation interfaces between the \dshort{crp} consortium and the \dshort{fsii} group include bottom and top \dshort{crp} installation, provision of engineering notes, procedures, tooling, storage space and rigging
equipment. \dshort{crp} consortium is responsible for providing temporary winches, roof flanges, and cables to lift the \dshort{crp} to position in the cryostat. Providing engineering notes, safety related documentation, installation procedures, cleaning procedures, parts and tooling are also \dshort{crp} consortium responsibilities. %I$\&$I 
\dshort{fsii} %is responsible to 
will provide any crates or boxes needed to move the superstructures, 
storage space in the cavern area for 20 half-\dshort{crp}s storage/transport crates, a gantry crane that meets \dshort{crp} specifications to be used to lift the half-\dshort{crp} 
onto the assembly table. The \dshort{crp} consortium provides the \dshort{crp} assembly table.

For the top \dshort{crp} installation, the \dshort{crp} consortium is responsible for providing any required tooling % needed to install the \dword{crp} 
including the %access platforms
elevated workstation. \dshort{crp} personnel are responsible for %accessing the \dword{crp} superstructure for 
cabling the \dshort{crp} from the superstructure. The survey plan for \dshort{crp} installation will be a joint responsibility. For the bottom \dshort{crp} installation, \dshort{crp} consortium %is responsible to 
will provide procedures, and design and construction of the bottom \dshort{crp} support structure. %I$\&$I group 
\dshort{fsii} will provide commercial equipment to place the \dshort{crp}s in position.

%%%%%%%%%%%%%%%%%%%%%%%%%%%%%
\section{Production and Schedule} 
\label{sec:crp:prod}

The production model for the \dwords{crp} takes into account the experience gained from the procurement of components for the \coldbox tests (Sections~\ref{subsec:CRP_prototyping2022} and~\ref{subsec:FirstFullCRP}), which will go into \dword{vdmod0}.  A timetable for procurement and assembly has been developed that feeds into the schedule outlined in Chapter~\ref{ch:project}.
%\fixme{check section refs - correct?}

The production of the 3840 \dword{pcb} segments is planned between October 2024 and October 2026. Vendor/s will be selected 
%in 2023 in 
accordance with \dword{cern}'s procurement process. The %expected 
delivery rate is required to be 50 PCB segments per week to allow a six month  ramping-up period.
The \dshort{pcb} segments will be delivered to CERN where the gluing, silver printing and \dword{qc} will be performed. Multiple batches of PCB panel shipments from CERN to assembly sites are expected. This phase is scheduled from October 2024 to January 2027.

The production and assembly of the adapter boards and edge cards will be performed by the industrial contractors between %, in the period from 
January 2024 and June 2026.

Components for \dshort{crp} assembly, namely composite frames, decoupling system, cables, and connectors for readout for the \dword{tde} will be manufactured by industry between November 2024 and July 2026.

Assembly %factories 
sites will be set up at collaborating institutions, two in the EU and two in the U.S.
Preparatory work (including design and manufacturing of tooling, cold test bench installation, test and site validation) will begin in 2024, with production starting at a lower rate in early 2025. Full-speed production of one \dshort{crp} per week at each site is expected starting in June 2025 in the EU for top \dshort{crp}s, and from April 2026  in the U.S. for bottom \dshort{crp}s. This assembly rate is based on one shift per day, five days per week, as validated in the assembly of %\dword{crp} \#1. 
CRP-1. A second shift per day could be added %implemented %in case of need. 
if needed. The assembly operation is expected to be completed in August 2026 for the top \dshort{crp}s
and July 2027  for the  bottom \dshort{crp}s. %, respectively.

Manufacturing the top superstructures and suspension feedthroughs will be done by industry contractors between November 2024 and December 2026. 
The %automatization 
automation system for the suspension of the top superstructures will be developed the collaborating institutions. 
The supports for the bottom \dshort{crp}s will be produced industrially between September 2025 and October 2026. Installation tooling will be manufactured at the same time at  collaborating institutions. 

The production schedule is summarized in Table~\ref{tab:crp-prod}.
 
 \begin{dunetable}
[CRP production timetable]
{p{0.19\textwidth}|p{0.50\textwidth}|p{0.21\textwidth}}
{tab:crp-prod}
{\dshort{crp} Production Timetable. Items where industrial procurement/manufacture is not indicated will take place at collaborating institutions. }
Item  & Tasks & Dates \\
\toprowrule

Perforated anode PCBs &
\textbullet Construction of panels and drilling of holes &  Oct 2024 -- Oct 2026 \\ 

\colhline

Interface boards (adapter boards and edge cards) & 
\textbullet Production of four-layer \dshort{pcb}s  (industry) & 
Oct 2024 -- Aug 2025  \\ 
&   \textbullet Procurement of capacitors, resistors and connectors&   \\ 
\colhline

\dshort{pcb} anode plane  assembly& \textbullet PCB gluing, silver printing and QC & Jan 2025 -- Dec 2026\\ 
&\\ 
\colhline

Top \dshort{crp} Interface board assembly
&\textbullet Adapter board assembly (industry) & Oct 2024 -- Jan 2026\\ 
&   \textbullet Edge card assembly (industry) & 
Oct 2024 -- Jan 2026\\ 
\colhline

Bottom \dshort{crp} Interface board assembly
&\textbullet Adapter board assembly (industry) & Mar 2025 -- Jun 2026\\ 
&   \textbullet Edge card assembly (industry) & 
Mar 2025 -- Jun 2026\\ 
\colhline

Top Composite Frame production & 
\textbullet Fabrication & 
Nov 2024 -- Jul 2026 \\ 
\colhline

Bottom Composite Frame production & 
\textbullet Fabrication & 
Feb 2025 -- Oct 2026 \\ 
\colhline

Top \dshort{crp} assembly & 
\textbullet Assemble the anode layers
 & 
Jun 2025 -- Jul 2026 \\ 
&   \textbullet Connect adapter and interconnection boards
& \\
&   \textbullet Connect the mechanical frame 
 & 
\\ 
\colhline

Bottom \dshort{crp} assembly  & 
\phantom{ai}Same sequence as for Top & 
 Apr 2026 -- Jun 2027 \\ 
  & & \\
\colhline

Top \dshort{crp} super- & 
\textbullet Production of large superstructure components  & 
Nov 2024 -- Jul 2026 \\ 
structure  & \phantom{ai}(industry) & \\
\colhline

Top suspension feedthroughs & 
\textbullet Production of the 64 FTs components (industry) & 
Jan 2025 -- Sep 2026 \\ 
 &
\textbullet Superstructure suspension automation system  & May 2025 -- Dec 2026 \\
\colhline

Bottom mechanical support & 
\textbullet Production (industry) & 
Sept 2025 -- Oct 2026 \\ 
 &
\textbullet Tooling for installation  & Sept 2025 -- Oct 2026 \\

\end{dunetable}

A more detailed schedule for production and installation of the \dshort{spvd} is found in Figure~\ref{fig:crp_schedule}.
\begin{dunefigure}[Key \dshort{crp} milestones and activities toward 
\dshort{spvd}]{fig:crp_schedule}{
Key \dshort{crp} milestones and activities toward  the \dshort{spvd} in graphical format (Data from~\cite{docdb22261v28}).}
\includegraphics[width=0.95\textwidth]{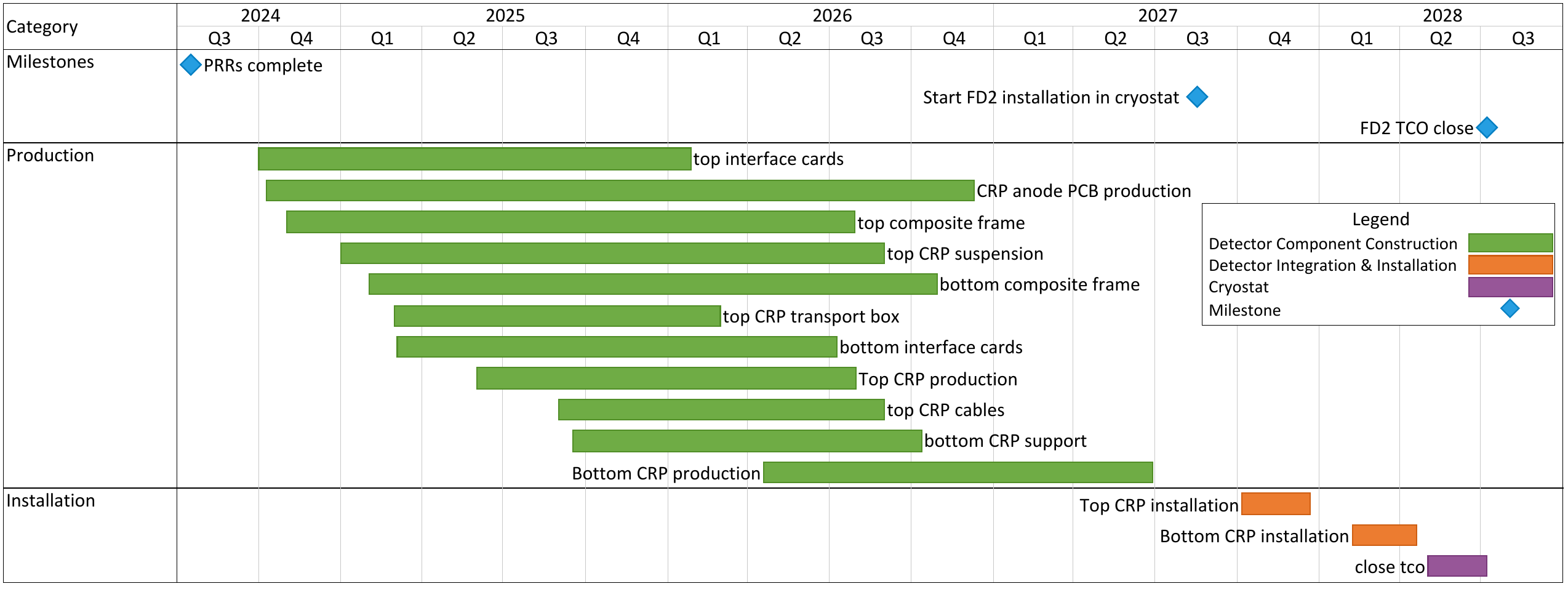}
\end{dunefigure}
%%%%% 

\section{Consortium Scope and Organization} 

\subsection{Scope }
The scope of the \dword{crp} system, provided by the \dshort{crp} consortium, covers the procurement of materials, fabrication, warm and cold testing, \dshort{qa}/\dshort{qc}, delivery, and installation of all
 components needed to complete the top and the bottom \dshort{crp} anode planes.

The %scope includes the 
following subsystems and items are included: 
\begin{itemize}
\item  Production of the  anodes;
\item  Production of the  adapter boards and edge cards;
\item  Production of the composite frames and of the mechanical links between the anode layers, adapter boards and structures;
\item Production of the superstructures for the top \dshort{crp} plane;
\item Production of the mechanical support systems for the bottom \dshort{crp} plane;
\item Production of the suspension system and position control for the top \dshort{crp} superstructures; % of the top plane;
\item Production of the \dshort{hv} distribution system associated with the top \dshort{crp}s for biasing; % the anodes;
\item Production of the level meter system associated with the top anode; %\dword{crp};
\item Production of the cabling of the \dshort{crp}s to the cold flange of the top electronics;
\item Production of the transportation and storage boxes for the \dshort{crp}s;
\item Setting up of the anode plane assembly site;
\item Construction and setting up of \dshort{crp} %factories;
production sites at collaborating institutions 
\item Assembly of the \dshort{crp}s %in the factories;
at the factories;
\item Testing of the assembled \dshort{crp}s at the %factories;
%production sites;
factories;
\item Placing the \dshort{crp}s into transport boxes and delivering them to \dword{surf}; 
\item Design and fabrication of specific tooling for installation in the cryostat;
\item Installing and cabling the \dshort{crp}s in the cryostat.
\end{itemize}

The scope also includes R\&D, design validation for the detector components, fabrication, integration, and testing of prototype \dshort{crp}s in various cold test setups that % already 
started in  2021 and %to 
that will be pursued with \dshort{vdmod0} in 2023. (Chapter~\ref{ch:mod0}).

\subsection{Institutional Responsibilities}
\begin{dunetable}
[\dshort{crp} institutions]
{p{0.28\textwidth}p{0.12\textwidth}p{0.60\textwidth}}
{tbl:crp-institutes}
{Institutions participating in the \dshort{crp} consortium}

Institution & Country & Detector \& Deliverables \\ \toprowrule

European Organization for Nuclear Research (CERN) & Switzerland & 
Anodes, anode panel assembly and test \\ \colhline
Brookhaven National Laboratory & USA & Anodes, Interface boards, CRP assembly, installation, bottom CRP transport frames \\ \colhline
Chicago University &  USA & CRP assembly \\ \colhline
LAPP Annecy IN2P3 & France  &  CRP mechanical design, assembly, composite support frame design and procurement, superstructures design, automation system, installation\\ \colhline
LPSC Grenoble IN2P3 & France  &  CRP assembly, top CRP transport frames, installation, level meters\\ \colhline
University of Wisconsin &  USA & Bottom CRP support feet system design and procurement, installation  \\ \colhline
Yale University &  USA & CRP assembly \\ 
\end{dunetable}

%%%%

\chapter{Charge Readout Electronics}
\label{ch:tpc-elec}

%\tableofcontents
%%%%%%%%%%%%%%%%%%%%%%%%%%%%
\section{Introduction}
\label{sec:tpc-elec:intro}

The \dword{tpc} \dword{cro} encompasses the hardware systems necessary to amplify, digitize, and transmit the \dshort{tpc} ionization charge signals out of the \dshort{dune} \dword{spvd} detector module. 
This includes the \dword{fe} electronics (amplifiers, digitizers, digital controllers), power and data cabling, related cryostat feedthroughs, and power supplies.

The \dshort{cro} as presented here does not include the electronics associated with the detection and recording of \dword{lar} scintillation photons, called the \dword{lro}. The \dshort{lro} is discussed in Chapter~\ref{chap:PDS}.

The \dshort{spvd} top and bottom drift volumes will implement different \dshort{cro} electronics in order to take maximal advantage of the different configurations of the two drift volumes. The top volume front-end (\dshort{fe}) electronics (``top drift electronics'' or \dword{tde}), based on the design developed for the proposed \dword{dp} \dshort{detmodule}~\cite{DUNE:2018mlo} design and used in \dword{pddp}, has both cold and warm components.
Because of the long cable paths from the bottom \dwords{crp} to the cryostat roof, the \dword{ce} \dshort{cro} solution chosen for the bottom drift volume (\dword{bde}) is the same as that used in the \dword{sphd} design~\cite{DUNE:2020txw,Abi:2020mwi}
and validated in \dword{pdsp}. %similar to 
 This \dshort{ce} design was developed to 
to collect signals %at %the bottoms of 
throughout the full depth range of the \dshort{sphd} \dwords{apa}, and therefore to operate in the conditions and depth of the \dshort{spvd} bottom \dword{anodepln}.

The \dshort{spvd} concept has developed %quite naturally 
out of the \dshort{dp} design implemented in \dshort{pddp}. The top drift volume is very similar to a  \dshort{dp} detector with the \dshort{crp} suspended from the cryostat roof. 
This scheme naturally allows using the existing \dshort{dp} electronics, which was designed to read anode strips and performed well in \dshort{pddp}. In particular, this  \dshort{spvd} design preserves 
 straightforward access to the cryogenic amplifiers via \dwords{sftchimney} at any time, without interfering with the detector operation, and keeps the digitization electronics completely accessible on the cryostat roof. Accessing the cryogenic electronics in the \dshort{sftchimney}s  is a simple operation, which was demonstrated 
 %and used 
 in \dshort{pddp}. 
 This accessibility will also allow DUNE to take advantage of technological evolution and potential cost reductions, as has already been witnessed, for example regarding the \dword{utca} digital components with the development of a 40\,Gbit/s bandwidth capability since 2020.

Detector integration aspects are optimized according to the characteristics of the %top and bottom \dword{cro} 
\dshort{tde} and \dshort{bde} configurations. The bottom drift \dshort{crp}s lie close to the cryostat floor,
supported by posts as described in Section~\ref{subsec:CRPBss}. The top drift \dshort{crp}s hang from the cryostat roof, supported by superstructures, from which the cathode modules are also suspended. The layout of the top \dshort{crp}s is  designed to facilitate access to the electronics via the cryostat roof. The weight of the %top \dword{fe} electronics 
\dshort{tde} is supported by the \dwords{sftchimney} and does not affect the mechanical structure of the top drift \dshort{crp}s. Since the top electronics is inside the \dshort{sftchimney}s, no heat dissipates directly onto the \dshort{crp} structures from the electronics components.
In addition, the separation between the \dshort{crp}s and the %electronics in the top readout 
\dshort{tde} will introduce fewer constraints on the installation schedule and simplify \dword{qc} procedures.

Because of the different layouts, risks are different between the top and bottom \dshort{cro} electronics. In particular, 
the top \dshort{crp}s would be exposed to any bubbles and/or dust contamination that may float towards
the liquid surface during the detector filling. Either bubbling or dust could cause sparking to occur in the anode plane.  Although this risk is low, the accessibility of the electronics  provides definitive mitigation for
for this risk %these issues 
and guarantees effective functioning of the %detector 
\dshort{tde} over a very long life span.

The bottom drift (cold) electronics (\dshort{bde}) components are located below the bottom \dshort{crp}s; the heat dissipation in this area was carefully studied~\cite{voirin-sim-rpt}, and no issues of local heat accumulation arose. The \dshort{bde} is designed with a requirement of greater than 30 years of operational lifetime. The prototype \dshort{bde} operated in \dshort{lar} at \dshort{pdsp} for over a year without any degradation in performance. A \dword{mod0} %second ProtoDUNE 
run with the final electronics is being planned. In addition, dedicated accelerated aging studies on the \dwords{asic} and other key components are ongoing to ensure long-term reliability.

Adopting different solutions for the top and bottom drift volumes also takes advantage of  
 a dedicated international R\&D program started in 2006 that has provided resources and non-U.S.  funding for the \dshort{tde}. 
\dshort{pddp} demonstrated this technology at the level of 10k channels, a scale roughly 1/20 that of \dshort{spvd}.

The scope of the \dshort{crp} electronics \dshort{cro} system includes the selection and procurement of materials for, and the fabrication, testing, delivery, and installation of the system. 

%%%%%%%%%%%%%%%%%%%%%%%%%%%%
\section{Requirements}
\label{sec:tpc-elec:spec}
The \dword{tpc} electronics system is designed to
produce a digital record representing the waveform of the current produced by charge induction
and collection on the \dword{pcbp}  strips.

The requirements/specifications for the top and bottom drift electronics are listed in Table~\ref{tab:specs:just:SP-ELEC}, approved by the DUNE Executive Board, and are the same as commonly defined in the DUNE design reports for the \dword{sphd} and \dword{pddp} electronics. 

% This file is generated, any edits may be lost.
%\begin{footnotesize}
%\begin{longtable}{p{0.14\textwidth}p{0.13\textwidth}p{0.18\textwidth}p{0.22\textwidth}p{0.20\textwidth}}
%\begin{longtable}{P{0.12\textwidth}P{0.18\textwidth}P{0.17\textwidth}P{0.25\textwidth}}
%\caption{Specifications for SP-ELEC \fixmehl{ref \texttt{tab:spec:SP-ELEC}}} \\
%\caption{\dword{cro} Electronics Requirements}
%  \rowcolor{dunesky}
%       Label & Description  & Specification \newline (Goal) & Rationale \\  \colhline
%  \rowcolor{dunesky}
%       Label & Description  & Specification \newline (Goal) \\  \colhline

%\input{generated/req-SP-ELEC-01.tex} % 1
%\input{generated/req-SP-ELEC-02.tex} % 2
%\input{generated/req-SP-ELEC-03.tex} % 3
%\input{generated/req-SP-ELEC-04.tex} % 4
%\input{generated/req-SP-ELEC-05.tex} % 5
%\input{generated/req-SP-ELEC-06.tex} % 6
%\input{generated/req-SP-ELEC-07.tex} % 7
%\input{generated/req-SP-ELEC-08.tex} % 8

\begin{footnotesize}
%\begin{longtable}{p{0.14\textwidth}p{0.13\textwidth}p{0.18\textwidth}p{0.22\textwidth}p{0.20\textwidth}}
%\begin{longtable}{p{0.12\textwidth}p{0.18\textwidth}p{0.17\textwidth}p{0.25\textwidth}p{0.16\textwidth}}
\begin{longtable}{p{0.08\textwidth}p{0.22\textwidth}p{0.22\textwidth}p{0.36\textwidth}}
\caption{\dshort{cro} electronics requirements} \\
  \rowcolor{dunesky}
       Label & Description  & Specification \newline (Goal) & Rationale\\  \colhline

   %  \newtag{FD-2}{ spec:system-noise }  & System noise  &  $<\,\SI{1000}\,e^-$ &  Provides $>$5:1 S/N on induction planes for  pattern recognition and two-track separation. &  ProtoDUNE and simulation \\ \colhline
  \newtag{FD-2}{ spec:system-noise }  & System noise  &  $<\,\SI{1000}\,e^-$ &  Provides $>$5:1 S/N on induction planes for  pattern recognition and two-track separation.\\ \colhline
     % 2
   
%  \newtag{FD-13}{ spec:fe-peak-time }  & Front-end \newline peaking time  & \newline  \SI{1}{\micro\second} &  Vertex resolution; optimized for \SI{5}{mm} anode readout spacing. &  ProtoDUNE and simulation \\ \colhline
%    \fixme{check 5mm}
\newtag{FD-13}{ spec:fe-peak-time }  & Front-end \newline peaking time  & \newline  \SI{1}{\micro\second} &  Vertex resolution; optimized for approximately \SI{5}{mm} anode readout spacing.\\ \colhline % 13
%   \newtag{FD-14}{ spec:sp-signal-saturation }  & Signal saturation \newline level  &  
%   \num{500000} e$^-$ \newline (Adjustable so as to see saturation in less than 10\% of beam-produced events) &  
%   Maintain calorimetric performance for multi-proton final state. &  
%   Simulation \\ \colhline
   \newtag{FD-14}{ spec:sp-signal-saturation }  & Signal saturation \newline level  &  
   \num{500000} e$^-$ \newline (Adjustable so as to see saturation in less than 10\% of beam-produced events) &  
   Maintain calorimetric performance for multi-proton final state. \\ \colhline     % 14
   
%  \newtag{FD-19}{ spec:adc-sampling-freq }  & ADC sampling frequency  &  $\sim\,\SI{2}{\mega\hertz}$ &  Match \SI{1}{\micro\second} shaping time. &  Nyquist requirement and design choice \\ \colhline
   \newtag{FD-19}{ spec:adc-sampling-freq }  & ADC sampling frequency  &  $\sim\,\SI{2}{\mega\hertz}$ &  Match \SI{1}{\micro\second} shaping time. \\ \colhline    % 19
   
%  \newtag{FD-20}{ spec:adc-number-of-bits }  & \newline Number of ADC bits  & \newline  \num{12} bits &  ADC noise contribution negligible (low end); match signal saturation specification (high end). &  Engineering calculation and design choice \\ \colhline
  \newtag{FD-20}{ spec:adc-number-of-bits }  & Number of ADC bits  &$\ge{12}$ bits &  Makes ADC noise contribution negligible (low end); matches signal saturation specification (high end). \\ \colhline     % 20
   
  \newtag{FD-21}{ spec:ce-power-consumption }  & Cold electronics \newline 
  power consumption (in-\dshort{lar})   &  $<\,\SI{50}{ mW/channel} $ &  No bubbles in \dshort{lar} to reduce HV discharge risk. %&  Bench test 
  \\ \colhline
     % 20
   
%  \newtag{FD-25}{ spec:non-fe-noise }  & Non-FE noise contributions  &  $<<\,\SI{1000}\,e^- $ &  High S/N for high reconstruction efficiency. &  Engineering calculation and ProtoDUNE \\ \colhline
  \newtag{FD-25}{ spec:non-fe-noise }  & Non-FE noise contributions  &  $\ll\,\SI{1000}\,e^- $ &  High S/N for high reconstruction efficiency. \\ \colhline     % 25
   
%  \newtag{FD-28}{ spec:dead-channels }  & Dead channels  &  $<\,\SI{1}{\%}$ &  Minimize the degradation in physics performance over the $>\,20$-year detector operation. &  ProtoDUNE and bench tests \\ \colhline
  \newtag{FD-28}{ spec:dead-channels }  & Dead channels  &  $<\,\SI{1}{\%}$ &  Minimize the degradation in physics performance over the $>\,20$-year detector operation. \\ \colhline
     % 28

\label{tab:specs:just:SP-ELEC}
\end{longtable}
\end{footnotesize}

Below is %the list of requirements that have been  
more information on the items (numbers correspond to Table~\ref{tab:specs:just:SP-ELEC}):

\begin{itemize}
    \item {FD-2 System noise $<$ 1000 $e^-$: total system noise seen by each \dshort{crp} strip should be less than 1000 enc of noise. It is expected that random noise on the \dshort{fe} amplifier will be the dominant contribution to the total system noise.}
    
    \item{FD-13 Front-end peaking time of $\sim$ \SI{1}{\us}}
    \item{FD-14 Signal saturation level $\sim$ 500,000 electrons: the largest signals correspond to events with multiple protons produced in the primary event, in particular, when the trajectories of one or more of those particles are parallel to the strips, causing the charge over a long path length to be collected within a short time period.}
    
    \item{FD-19 ADC sampling frequency $\sim$ \SI{2}{\MHz}: this value is chosen to match \SI{1}{\us} shaping time (the approximate Nyquist requirement).}
    
    \item{FD-20 Number of ADC bits $\ge 12$: the lower end of the ADC dynamic range is driven by the requirement that the ADC digitization not contribute to the total electronics noise. The upper end of the ADC dynamic range is defined by the signal saturation level. The two requirements with the specification for the total electronics noise results in the 12 bit ADC digitization requirement.}
    
    \item{FD-21 Cold electronics power consumption in \dshort{lar} $<$ \SI{50}{\mW}/channel: \dshort{ce} power consumption must be low enough to prevent the occurrence of local heat accumulation. 
    } 
    
    \item{FD-25 Non-FE noise contributions $\ll$ 1000 $e^-$: noise contribution from all non-FE noise sources should be much lower than the system noise requirement.}
    
    \item{FD-28 Dead channels $<$ \SI{1}{\%}: detector components shall be sufficiently reliable so as to ensure that the number of dead channels does not exceed \SI{1}{\%} over the lifetime of the experiment.}
\end{itemize}

%%%%%%%%%%%%%%%%%%%%%%%%%%%%

\section{Top Drift Readout} 
\label{sec:TAROss}

\subsection{System Overview}

The top drift charge-readout system (\dword{tde}) is the outcome of a long R\&D process, launched before DUNE and aimed at optimizing cost and performance. It is based on the design used in \dword{pddp}~\cite{DUNE:2018mlo}. 
Minor modifications and some optimizations were made to the design in 2021 to adapt it for the top anode plane of the \dword{spvd} module, which has a total of 245,760 readout channels.

The  \dshort{spvd} \dshort{tde} was extensively tested in 2021 and 2022 in %various 
a series of \coldbox \dshort{crp}  tests, %along with 
including tests of the two %final 
top drift  \dshort{crp}s that will be used in \dword{vdmod0}, and which 
match the final \dshort{crp} channel layout and the final layout of the associated \dshort{tde} readout system adopted for \dshort{spvd}.
Figure~\ref{fig:dp_cro} %schematically 
illustrates the architecture of the \dshort{tde} system. Each strip of the top drift \dshort{crp}s is read by a system of front end (\dshort{fe})  cryogenic analog amplifiers  that connect to external \dshort{fe} digital warm electronics through \dshort{sftchimney}s in the cryostat roof.

%$$$$$$$$$$$$$$$  
\begin{dunefigure}
[System architecture of the top drift readout electronics]
{fig:dp_cro}
{System architecture of the top drift readout electronics.} 
\includegraphics[width=1.0\textwidth]{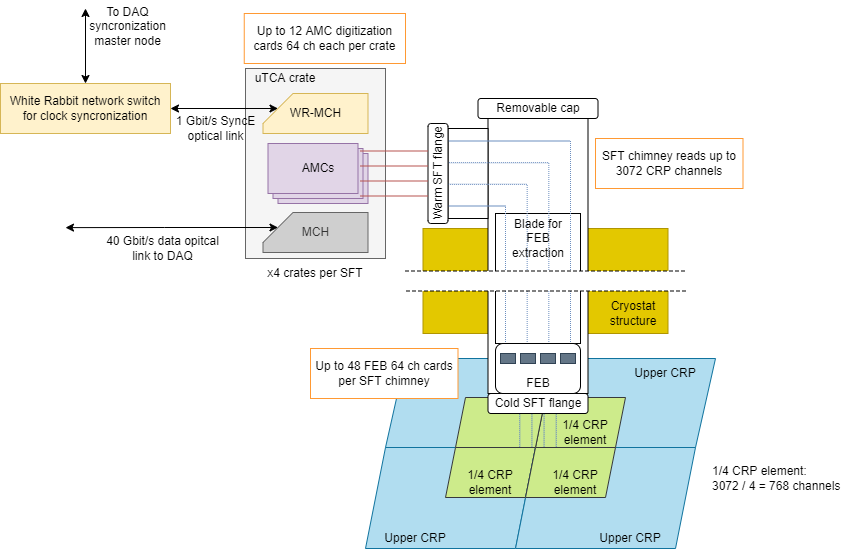}
\end{dunefigure}
%$$$$$$$$$$$$$$$  

The analog amplifiers are implemented in \dshort{asic}s located on \dshort{fe} cards that are bonded to extractable blades, allowing servicing or replacement via a hot swapping procedure. Each card plugs into a cold flange at the bottom of its \dshort{sftchimney}. %  with 48- or 24-card capacity.
Each \dshort{fe} card hosts four \dshort{asic}s, each of which reads 16 channels. %Thus there are 64 channels per \dword{fe} card in total.   

The \dshort{fe} cards operate at the bottom of the \dshort{sftchimney}s in close proximity to the \dshort{crp}s. The %N2 
nitrogen-filled inner volume of the %SFT chimneys 
\dshort{sftchimney}s is separated from the cryostat and the external environment by a pair of vacuum-tight feedthrough flanges that dispatch signals and slow controls. 
The analog electronics is thereby completely shielded from the environment and operates at cryogenic temperatures. 

The \dshort{sftchimney} cold flanges define the interface to the \dshort{crp}s.   The adapter boards mounted on the \dshort{crp}s  are connected with flat cables to the connectors on the cold flange side facing the cryostat.

The \dshort{fe} cards, mounted on 2\,m long blades, glide along the guiding rails inside \dshort{sftchimney}s to plug into  dedicated connectors mounted on the inner %(chimney volume) 
side of the cold flange (within the chimney volume). The \dshort{fe} card \dshort{asic}s amplify the input signal and produce analog differential output signals carried by flat cables running along the blades up to warm flanges
located near the top of the  \dshort{sftchimney}s. The warm flanges seal the chimney from the outside environment. They pass slow control and low-voltage lines to the \dshort{fe} cards inside and bring out the differential analog signals.

The \dshort{sftchimney}s are designed explicitly to allow access to the \dshort{fe} cards during detector operation. A top cap is removed to open the chimney inner volume, allowing extraction and re-insertion of the blade-\dshort{fe} cards assembly from the cryostat roof. During access,  nitrogen flow is activated to prevent humidity from entering the \dword{sft} inner volume. 
Moreover, replacement of a \dshort{fe} card can be executed via hot swapping without switching off the rest of the electronics, and 
without interfering with the detector operation.

A system of this design was successfully deployed and operated in 2019-2022 in \dshort{pddp}, where the \dshort{fe} cards were installed in groups of 10 in smaller-radius  \dshort{sftchimney}s. 

The \dshort{spvd} \dshort{tde} system contains 105 \dshort{sftchimney}s of two types. One type (positioned along the long edges of the cryostat), of which there are 21 on each side (42 total), each hosts 24 \dshort{fe} cards (SFT24).  The  other 63 (three rows of 21) each host 48 cards (SFT48).   
Except for the SFT48s at the ends, these larger ones are positioned above each intersection of four \dshort{crp}s, and read 3072 channels from the eight \dshort{cru}s. The end SFT48s read out half as many channels (1536). The SFT24s each read out four \dshort{cru}s (1536 channels), except for those at the ends, which read out two (768 channels).   Table~\ref{tab:table_summary_td} lists the counts. 

The external warm flanges on a %chimney 
\dshort{sftchimney} connect the amplified signals to the \dshort{fe} digitization system, located on the cryostat roof.  
Installing the digital electronics on the cryostat roof allows use of inexpensive \dword{utca}  electronics, a standard that is commonly used in the telecommunications industry. 

The amplified signals from \dshort{fe} cards pass through the warm flange to the digitization cards (\dword{amc}).  The \dshort{amc}s  are hosted in \dshort{utca}  crates and are connected to the warm flanges with shielded VHDCI cables. Each \dshort{utca} crate can host up to 12 \dshort{amc} \dshort{fe} digitization units of 64 channels each. Therefore an SFT48 interfaces to four \dshort{utca} crates, except for those at the end, which interface to two.

Each \dshort{utca}  crate also hosts a \dword{mch} switch that supports 40 Gbit/s bandwidth via an Ethernet optical link connection to the back-end \dword{daq}. \dshort{amc}s send their data to the \dshort{daq} via the \dshort{mch} by using 10\,Gbit/s XAUI Ethernet lanes in the crate backplane. 
%Considering 
Given the 12 bit dynamics and 2\,MHz continuous sampling of the signals, the occupancy of the link, assuming 12 \dshort{amc}s operating simultaneously in each \dshort{utca} crate, corresponds to about 60\% of the maximal \dshort{mch} bandwidth. No data compression is thus required.

The design of the \dshort{daq} system on the receiving side of the  Ethernet links  is common to both \dword{fd} modules. The \dshort{fe} digitization system is coupled to a local timing distribution system based on the \dword{wr} standard, which is connected to the DUNE timing system.

Each  \dshort{utca}  crate contains a \dword{wr} end-node 
(\dword{wrmch}) that serves as the local synchronization source to the global reference clock for all \dshort{amc}s hosted in the same crate. The \dshort{utca} crate 
contains a \dshort{wr} slave end-node card, a \dshort{wr} Lite Embedded Node (LEN) (Seven Solutions OEM WR-LEN), as a mezzanine card. The WR-LEN is a %an OTS 
\dword{cots} component that runs on customized firmware to enable it to decode the trigger timestamp data packet received over the \dshort{wr} network, in addition to the \dshort{wr} synchronization packets. This firmware, developed for the integration of the WR-LEN in the \dshort{wrmch}, also allows for the distribution of both the timing and synchronization signals on dedicated lines of the crate backplane to the \dshort{amc} %as well as of 
and the time-stamping data packets. The \dshort{wrmch}s are connected to the global timing %Grand Master
\dword{wrgm} via a network of dedicated switches.

\begin{dunetable}
[Top drift charge readout electronics: unit counts]
{lr} 
{tab:table_summary_td}
{Top drift charge readout electronics: unit counts for final \dshort{crp} and readout system configuration.}
\textbf{Item} 
& \textbf{Quantity} 
\\ \toprowrule
			3.0\,m x 1.7\,m CRUs in the top drift & 160\\
			\colhline
			Anode channels per CRP 
			& 3072 \\
			\colhline
			Channels per FE card or AMC card   
			& 64\\
			\colhline
			FE cards or AMC cards per CRP 
			& 48\\
			\colhline
			Number of SFT  
			&  105\\
			\colhline
			FE card slots per SFT 
			& 48 or 24\\
			\colhline
			\dshort{utca} crates  
			&  320\\
			\colhline
			\dshort{amc} cards per crate  
			&  12\\
			\colhline
			WR-MCH  
			&  320\\
			\colhline
			40 Gb/s data links  
			& 320\\
			\colhline
			Anode channels in the top drift 
			& 245,760\\
\end{dunetable}

%%%%%%%%%%%%%%%%%%%%%%%
\subsection{Analog Cryogenic Electronics}
\label{subsubsec: UACEsss}
The \dshort{spvd} top drift cryogenic analog \dshort{fe} electronics and the \dshort{sftchimney}s are similar to those implemented in \dshort{pddp}~\cite{DUNE:2018mlo}. 
Figure~\ref{fig:tde_analog_chain}  shows a synopsis of the analog chain including the \dword{larzic} \dword{asic} and the analog stage (\dword{adc} buffer ADA4940) integrated in the \dword{amc} card before the \dshort{adc}. The  \dshort{larzic} \dshort{asic} integrates a cryogenic  Charge Sensitive Amplifier (CSA) and a differential output buffer stage acting as low-pass filter. 

%$$$$$$$$$$$$$$$  
\begin{dunefigure}
[System architecture of the \dshort{tde} electronics]
{fig:tde_analog_chain}
{Synopsis of the \dshort{tde} analog chain including the \dshort{larzic} \dshort{asic} and the analog stage (\dshort{adc} buffer) present in the \dshort{amc}.} 
\includegraphics[width=.9\textwidth]{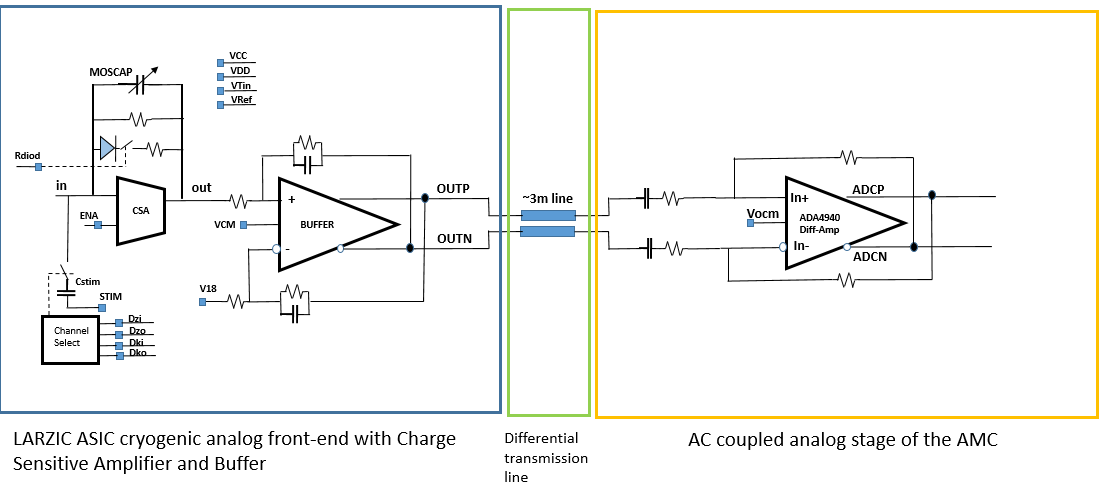}
\end{dunefigure}
%$$$$$$$$$$$$$$$  

\subsubsection{Cryogenic \dshort{asic} amplifier }
\label{subsubsec: ASICsss}

The cryogenic Charge Sensitive Amplifier (CSA)  \dshort{asic} (\dshort{larzic}) is the principal component of the \dshort{fe} analog cards. Its design is based on \dword{cmos} 0.35\,$\mu$m technology, for which R\&D began in 2006. The chips are produced 
at CMP\footnote{Circuits Multi-Projets\textregistered, \url{https://mycmp.fr/}} and AMS Full Service Foundry \footnote{ams\textregistered, \url{https://ams.com/full-service-foundry}}. 

This technology remains fully exploitable for production on the timescale of the DUNE \dshort{detmodule}s. Each \dshort{asic} contains 16 amplifier channels with differential line buffers and has a power consumption of 11\,mW/channel. The main characteristics of the  \dshort{larzic} \dshort{asic} are listed in
Table~\ref{tab:table_summary_larzic}).

\begin{dunetable}
[\dshort{tde} \dshort{larzic} \dshort{asic} specifications]
{p{0.4\textwidth}p{0.4\textwidth}}
{tab:table_summary_larzic}
{Top drift charge cryogenic \dshort{larzic}  \dshort{asic} main characteristics and specifications}
Label & Description
\\ \toprowrule
			Technology
			& CMOS 0.35\,$\micro$m \\
			\colhline
			Channels per ASIC   
			& 16\\
			\colhline
			Integrated components per channel
			& Charge sensitive amplifier + differential buffer \\
			\colhline
			Peaking time  
			&  1 us\\
			\colhline
			Operation temperature
			& Typically around 110\,K at the bottom of the SFT chimneys, the LARZIC/\dshort{fe} card can operate as well at LN2 temperature\\
			\colhline
			Power consumption 
			&  11 mW/channel \\
			\colhline
			ENC
			&  <400 electrons at cold, < 600 electrons at warm\\
			\colhline
			Conversion factor 
			&  14 mV/fC (including all the entire TDE analog chain ASIC + ADC buffer)\\
			\colhline
			Calibration  
			& Integrated charge injection system with embedded capacitors and the possibility of activating single channels or groups of channels\\
			\colhline
			Crosstalk 
			& <1\%\\
			\colhline
			Unipolar dynamics (ASIC)
			& Linear up to 400 fC, max signals up to 1200fC with dual-slope regime (used in dual-phase)\\
                        \colhline
			Bipolar dynamics (ASIC)
			& Linear up to +-200 fC\\
                        \colhline
                        Bipolar dynamics (ASIC+ADC buffer)
			& Linear up to +-80 fC\\
\end{dunetable}

The CSA has linear gain for input charges of up to 400\,fC and a logarithmic response in the 400–1200\,fC range. This double-slope behavior is obtained by using a MOSCAP capacitor in the feedback loop of the amplifier that changes its capacitance above a certain signal threshold. The MOSCAP yields a lower gain for input charges larger 400\,fC. The feedback circuit includes also a selectable branch (Rdiod) 
with a resistor in series to a diode, acting %on 
over the discharge time.  The activation of this branch in the feedback circuit guarantees similar discharge times in the dual-slope regime for signals smaller or exceeding the 400\,fC threshold by keeping the RC of the feedback circuit at about 500\,ns in both cases. 

The possibility for 
a dynamic range larger than 400\,fC was mainly developed for the \dword{dp} application (with \dword{lem} gain of at least 20) and is of %low interest 
much less importance for the \dshort{spvd}, where the dynamics of the entire analog chain has been re-optimized to work in bipolar mode. In this configuration, the \dshort{asic} features a linear regime up to $\pm$200\,fC, 
where the linearity is at the 1\% level. The %signals dynamics 
signal dynamic range is then limited to $\pm$80\,fC by the \dshort{adc} dynamics after accounting for the  \dshort{amc} analog stage.

The  \dshort{larzic} output signals are transmitted on differential lines at \SI{120}{\ohm} to the  \dshort{amc} analog input stage. The buffer integrated in the \dshort{larzic} produces positive and negative differential signals (Figure~\ref{fig:tde_diff_out}) separated by an adjustable positive offset (V18). These signals are then processed
by the differential amplifier embedded in the  \dshort{amc}, which is AC coupled at the end of the transmission line. 

Five independent DC supply voltages are supplied for the operation of the \dshort{larzic} \dshort{asic}: VCC (supply voltage of the differential buffer), VDD (supply voltage for the output stage of the CSA), VRef (reference voltage used by the cascode 
stage of the CSA), V18 (bias voltage applied to the buffer circuit to separate the signal branches), and %V$\_T_{in}$ 
V$_{T_{in}}$
(supply voltage for the input transistor of the CSA). These voltages are typically 3.3, 3.3, 1.4, 1.8, and 2.2\,V respectively. The \dshort{asic} can be completely disabled (switched off) with an external enable control level (ENA). 

%$$$$$$$$$$$$$$$  
\begin{dunefigure}
[Examples of differential output signals from the  \dshort{larzic} \dshort{asic} ]
{fig:tde_diff_out}
{Examples of differential output signals from the  \dshort{larzic} \dshort{asic} } 
\includegraphics[width=.3\textwidth]{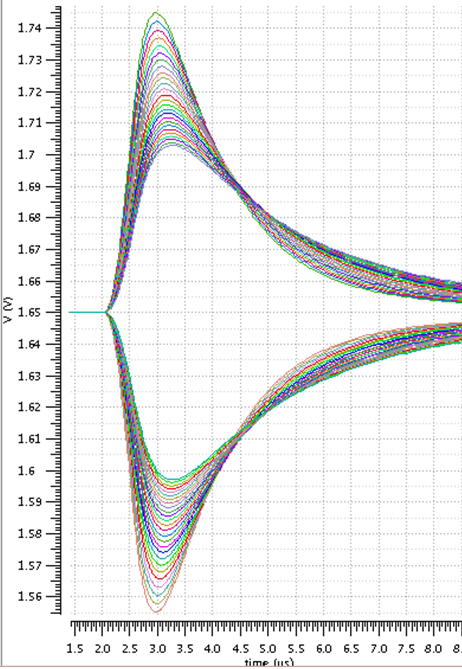}
\end{dunefigure}
%$$$$$$$$$$$$$$$  

Figure~\ref{fig:tde_larzic_connections} shows the \dshort{larzic} \dshort{asic} chip die with its electrical connections  together with an enlarged view of a single channel layout.

%$$$$$$$$$$$$$$$  
\begin{dunefigure}
[\dshort{larzic} chip die with connections and single-channel layout]
{fig:tde_larzic_connections}
{Top: \dshort{larzic} \dshort{asic} chip die with its electrical connections. Bottom: 
enlarged view of a single-channel layout ($1100\,\mu m \times 170\,\mu m$)} 
\includegraphics[width=.45\textwidth]{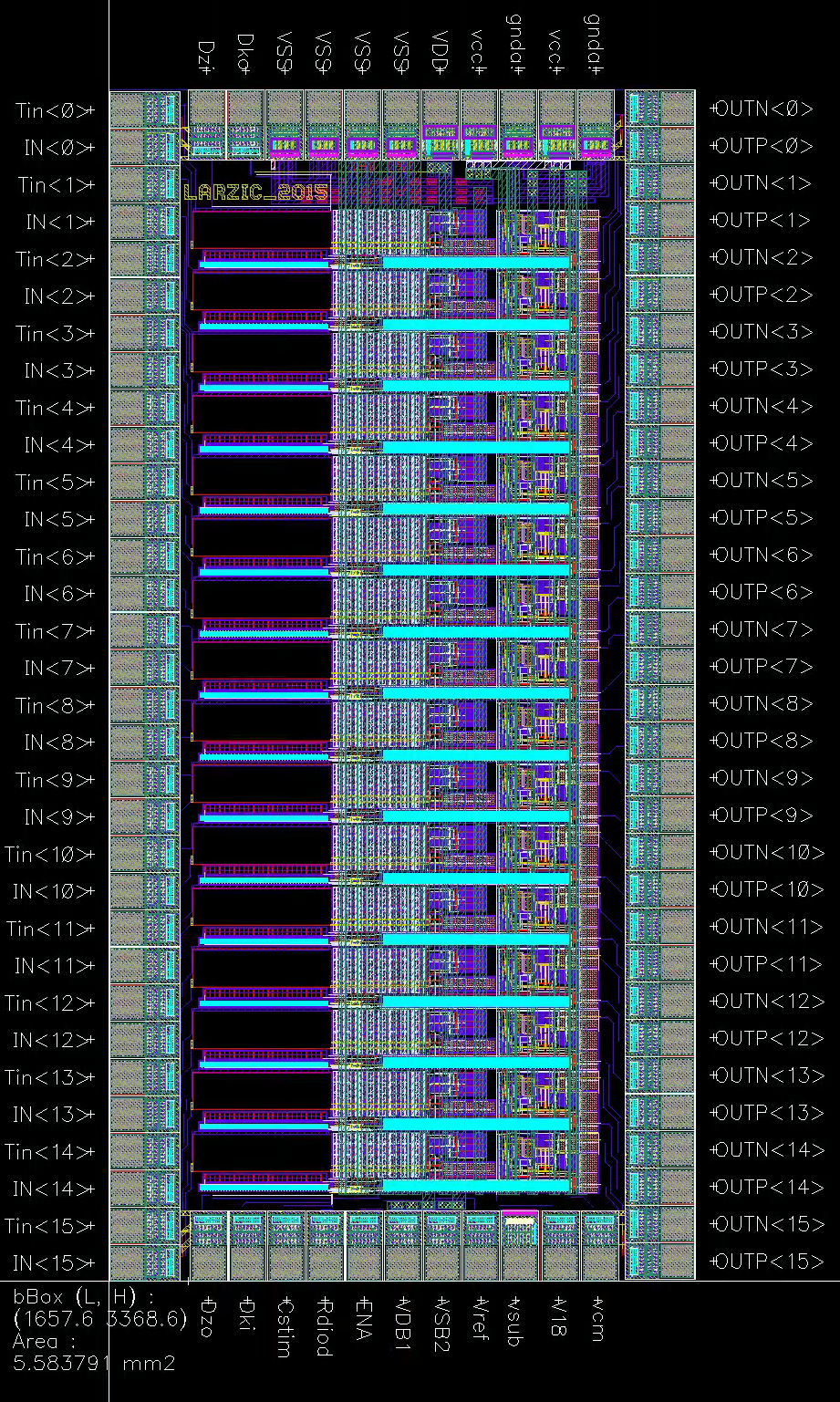}
\includegraphics[width=1\textwidth]{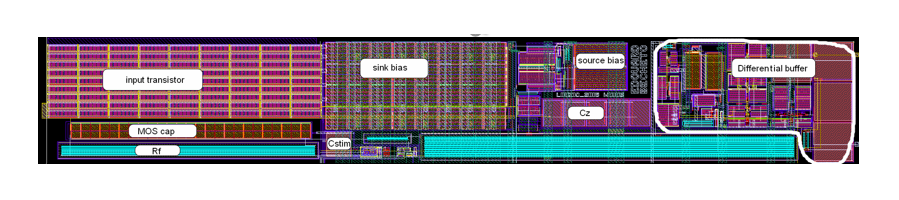}
\end{dunefigure}
%$$$$$$$$$$$$$$$  

The  \dshort{larzic} is equipped with a charge injection system with embedded capacitors (C$_{stim}$ = 1\,pF). A slow control system allows selection of injection from an external signal injection source connected to the stimulus (STIM) line for single channels or for a desired configuration of channels (Figure~\ref{fig:tde_cis}). This configuration is defined with a \dword{spi} bus including input and output data and clock lines (Dzi, Dzo, Dki, Dko). The %possibility of activating the
option to activate injection single channels 
has also allowed measurement of the crosstalk, which was found to be negligible (<1\%).

%$$$$$$$$$$$$$$$  
\begin{dunefigure}
[Details of the \dshort{larzic} charge injection and channel selection systems]
{fig:tde_cis}
{Details of the \dshort{larzic} charge injection system and channel selection system} 
\includegraphics[width=.8\textwidth]{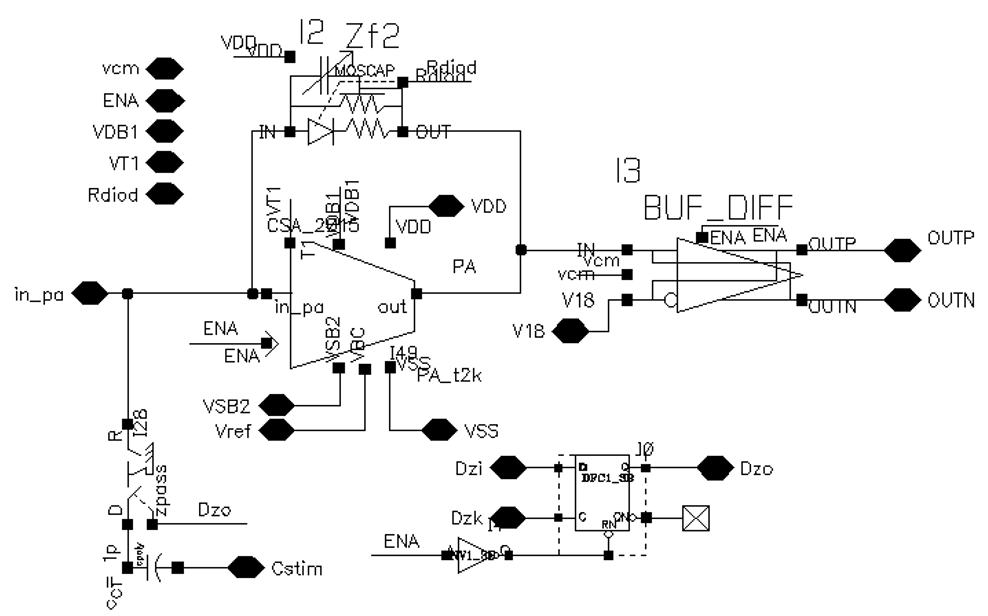}
\end{dunefigure}
%$$$$$$$$$$$$$$$  

Multiple \dshort{larzic} \dshort{asic} chips on the same \dshort{fe} card can be daisy-chained on the \dshort{spi} bus using the input and output clock and data connections as shown in Figure~\ref{fig:tde_daisy_chain}.

%$$$$$$$$$$$$$$$  
\begin{dunefigure}
[SPI bus daisy-chaining of multiple \dshort{larzic} \dshort{asic}s]
{fig:tde_daisy_chain}
{SPI bus daisy-chaining of multiple \dshort{larzic} \dshort{asic}s} 
\includegraphics[width=.5\textwidth]{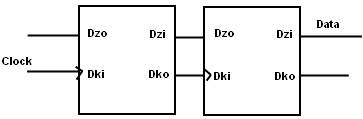}
\end{dunefigure}
%$$$$$$$$$$$$$$$  

%%%%%%%%%%%%%%%%%%%%%%%%%%%%%%%%%%%%%%%

\subsubsection{Cryogenic \dshort{fe} cards }
\label{subsubsec: AFEBsss}

The \dword{fe} cards amplify bipolar signals that are propagated with differential analog lines to the digitization system located in the \dword{utca} crates on the cryostat roof. 

Each cryogenic \dshort{fe} card holds four amplifier \dshort{asic}s  and a few passive discrete components  for total of 64 readout channels. Common input low-voltage supply lines (VCC, VDD, VRef, V18 and V$_{T_{in}}$) are distributed on the \dshort{fe} card  to the four \dshort{asic}s. Banks of blocking capacitors suited to work at cryogenic temperatures further filter the power lines, which are previously filtered by the low-voltage filtering and distribution system connected to the power supply so as to distribute the power to the chimneys.  

Similarly, the configuration control signals (ENA, Rdiod) and the charge injection signal STIM are connected to the \dshort{asic}s. The  \dshort{asic}s are daisy-chained for the clock and data \dword{spi} bus signals, which define the global charge injection configuration of the 64 channels. Pull-up resistors mounted on the \dshort{fe} card define the default configuration of the \dshort{fe} card, in the absence of external ENA levels which otherwise have to be applied to the \dshort{fe} cards via the warm flange.

Two independent input connectors, each handling 32 signal channels, bring the CRP signals to the two groups of two  \dshort{asic}s. The output connectors include a 68-pin connector for the output of the second group of 32 channels. The output of the first group of 32 channels is on a larger output connector of 80 pins, which is also used for the connection of the low-voltage and control lines, and the external charge injection signal. 

The  \dshort{fe} card (Figure~\ref{fig:feb_dp}) originally included some front-end components, which in the case of the  \dword{dp} were \dword{hv} rated and used to polarize the  \dword{crp} anode strips (2.2\,nF decoupling capacitors and \SI{1}{G\ohm} resistors, rated up to 3\,kV).

%$$$$$$$$$$$$$$$ 
\begin{dunefigure}
[FE cryogenic  amplifier card] 
%for the DP configuration]
{fig:feb_dp}
{FE cryogenic  amplifier card for the \dshort{dp} (left)
and for the \dshort{spvd} (right)  configurations.}
	\includegraphics[width=.5775\textwidth]{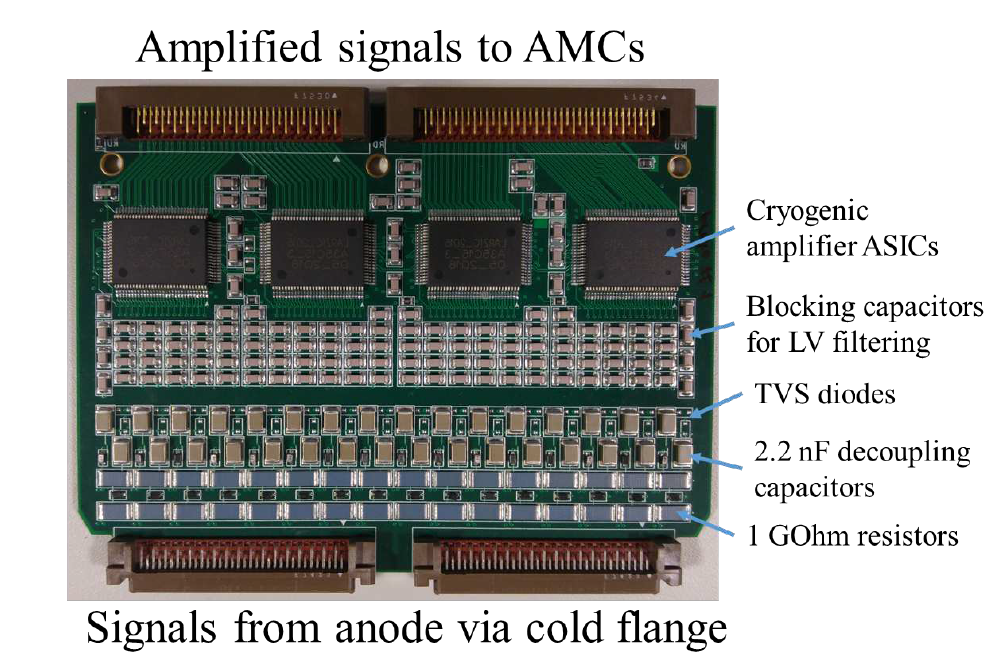}
 \includegraphics[width=.395\textwidth]{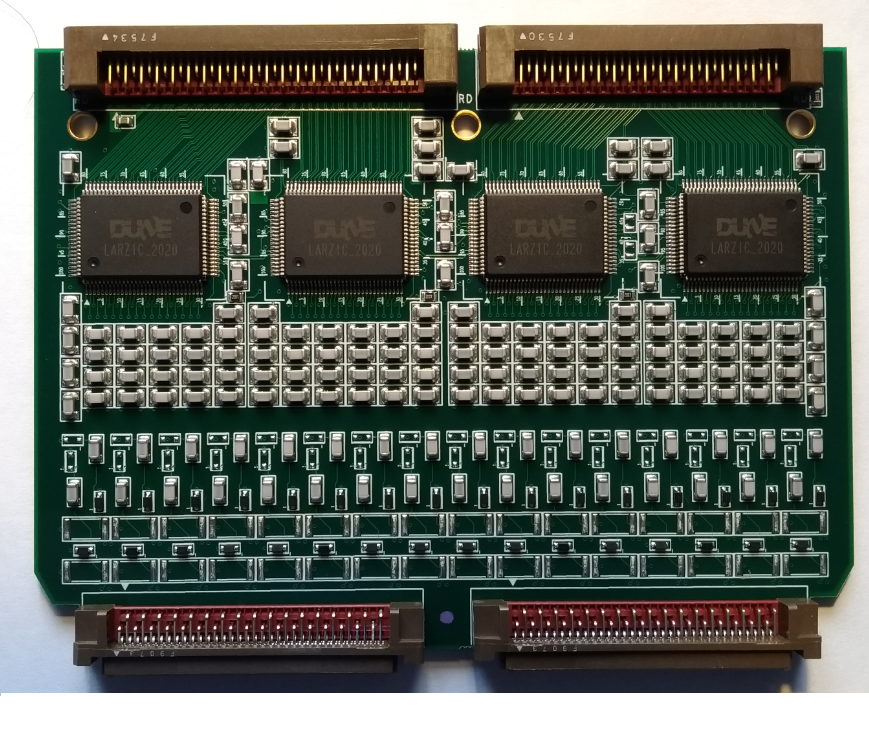}
\end{dunefigure}
%$$$$$$$$$$$$$$$ 

This configuration has been simplified for the \dshort{spvd} since these polarization components are mounted on the adapter boards themselves and no longer needed on the \dshort{fe} cards.

An \dword{esd} protection device (TVS diodes Bourns CDSOD323-T08LC) is also included in each input stage and is used to protect the amplifiers against discharges coming from the detector. The particular device was selected after %a long campaign aimed at 
extensively checking the performance of different \dshort{esd} components, 
subjecting them to many discharges of a few kV and a stored energy similar to the one of the \dwords{lem}. 

These components proved to be very efficient in the \dshort{dp} to protect against %very 
frequent \dshort{lem} sparking. In general the  \dshort{fe} card and \dshort{asic} design are very robust --- for several years the  \dshort{fe} cards have been  manipulated with bare hands with no particular precautions, without %ever observing 
causing any observable damage. 
Very minor modifications have been made to the configuration of the \dshort{fe} cards for their use in \dword{spvd}:

\begin{itemize}
\item The  \dshort{spvd} design has decoupling capacitors and biasing resistors as part of the anode system on the \dshort{crp}'s adapter board rather than on the input stage of each amplifier channel. %, as was done in \dword{pddp}. 
\item The \dshort{spvd} uses a simplified, customized version of the \dshort{fe} cards (Figure~\ref{fig:feb_dp}) %Figure~\ref{fig:dp_FE_card_VD}) 
where the  decoupling and biasing components, rated for operation %in \dword{pddp} 
at several kV, have been removed.
\item The \dshort{fe} cards still host \dshort{esd} protection components with diode pairs, which are very effective since they were originally designed for the \dshort{dp} application to withstand \dshort{lem} discharges at 3\,kV. 
\end{itemize}

%%%%%%%%%%%%%%%%%%%%%%%%%%%%%%%%%%
\subsubsection{\dshort{amc} Analog Stage }
\label{subsubsec: AASsss}

As previously described, the \dword{amc} cards include an analog stage with a differential amplifier (ADC buffer ADA4940-2) before the \dword{adc}, see Figure~\ref{fig:amc_analog_stage}.

%$$$$$$$$$$$$$$$ 
\begin{dunefigure}
[Schematic of the analog stage on the \dshort{amc}]
{fig:amc_analog_stage}
{Two-channel schematic of the analog stage on the \dshort{amc} with differential amplifiers connected to the signal transmission lines.}
	\includegraphics[width=.9\textwidth]{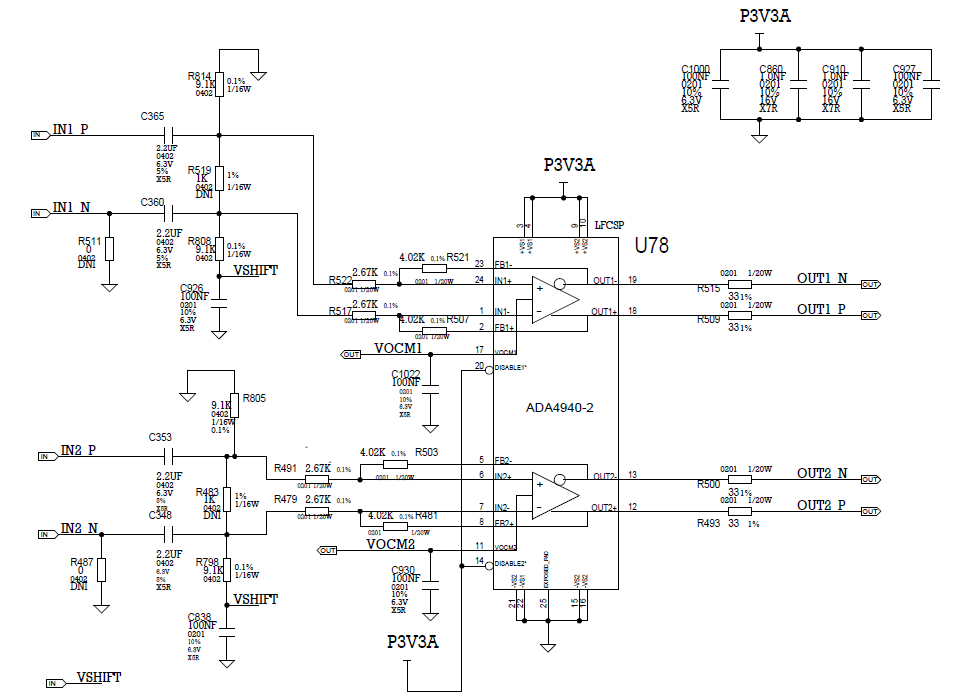}
\end{dunefigure}
%$$$$$$$$$$$$$$$ 

This stage is AC coupled to the signal transmission lines from the \dword{fe} card. For the \dword{dp} application, where the anodes had two identical collection views, the detector signals were %always 
unipolar and the analog stage was set with an external offset (VSHIFT) in order to fully exploit the ADC dynamics. 
For \dshort{spvd}, VSHIFT is set at ground and the baseline is naturally placed at half of the \dshort{adc} range (2048 counts) allowing for equivalent dynamics for the digitization of signals of both polarities.

%%%%%%%%%%%%%%%%%%%%%%%%%%%%%%%%%%
\subsubsection{Response Uniformity }
\label{subsubsec: AREUsss}

The \dword{tde} analog chain features excellent channel response uniformity. Extensive calibration campaigns were 
systematically performed for the different production runs, demonstrating high uniformity. 

A dedicated calibration card was developed for the first production run for \dword{pddp} to test and calibrate the analog \dshort{fe} cards (see Figure~\ref{fig:feb_calib_card}). The calibration card hosts a bank of 32 (QuadTech 7600) air varicaps, 
mechanically trimmed and calibrated individually, 
that are tuned to <2\% within the nominal target value of 1.1\,pF. 

%$$$$$$$$$$$$$$$ 
\begin{dunefigure}
[Charge injection system with varicaps connected to \dshort{fe}]
{fig:feb_calib_card}
{32-channel charge injection system with calibrated varicaps, with a \dshort{fe} card connected to it via one of its input connectors.}
	\includegraphics[width=.6\textwidth]{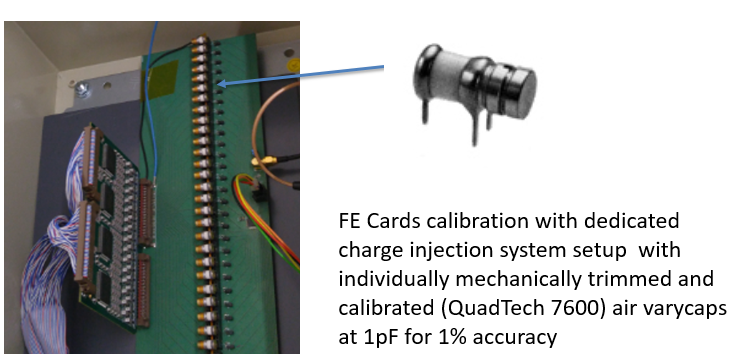}
\end{dunefigure}
%$$$$$$$$$$$$$$$ 

A known charge can be injected into a given \dshort{fe} card channel  by pulsing %the 
its corresponding capacitor with an external pulse generator. The channel selection is performed by a 32:1 analog MUX controllable via \dword{spi}. For %this version of the injection card 
this initial version of the injection card,
the design featuring a single 32-channel output connector, only half of the channels %of a \dword{fe} card 
could be tested at the same time, so a %. The 
full test of each \dshort{fe} card required %was performed in two shots by 
swapping the input connector of the card and retesting. 

These tests resulted in (see Figure~\ref{fig:feb_calib_results}) a remarkable 2\% %channels 
uniformity across channels. 
A tiny systematic effect that accounts for this dispersion was found among the two blocks of 32 channels corresponding to the same injection card. The effect is related to the accuracy of the injection system itself. This result indicates that the intrinsic response uniformity at the level of the \dshort{larzic} \dshort{asic} is even better than 2\%. 

%$$$$$$$$$$$$$$$ 
\begin{dunefigure}
[\dshort{pddp} \dshort{fe} card calibration results]
{fig:feb_calib_results}
{Superimposed calibration results for all \dshort{fe} cards produced to equip \dshort{pddp} as a function of the channel number on the card (1-64)}
\includegraphics[width=.8\textwidth]{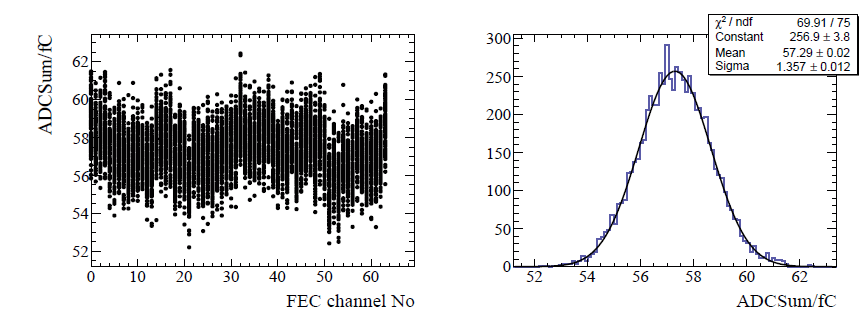}
\end{dunefigure}
%$$$$$$$$$$$$$$$ 

The calibration card %provides 
allows for %a complete test of 
testing continuity of the entire analog chain, starting from the input connectors of the \dshort{fe} card. The injected signals follow the same path as the signals coming from the \dshort{crp}. A complementary technique %allows performing the 
enables \dshort{asic} calibration via charge injection through its internal calibration circuitry, as mentioned in Section~\ref{subsubsec: ASICsss}.

This testing technique has been used throughout the prototyping program and will be used in \dshort{spvd} to perform periodic calibration checks once the \dshort{tde} system is installed. The \dshort{asic} calibration uses an \dshort{spi}-like bus (clock and data lines) that can activate one or more channels of the \dshort{fe} card for injection. A calibration pulse, delivered to the warm flange and passed by a dedicated conductor on the flat cables, generates the injection charge. The protocol consists of passing 64 bit codes at 1 bit per clock cycle to enable (1) or disable (0) any combination of the 64 channels.

The direct \dshort{asic} injection tests will: %should:
\begin{itemize}
\item check correct channel mapping. This can by done by addressing the channels via the \dshort{spi} bus for two possible patterns (1 channel over 8 enabled and 1 channel over 64 enabled); and

\item measure the linearity of the response and compare it to the reference template.
\end{itemize}

%Comparisons of t
The two calibration methods, charge injection via the \dshort{larzic} \dshort{asic} or via the injection card connected to the \dshort{fe} card input, have %shown 
yielded equivalent and excellent results, channel-by-channel, at the 1\% level, enabling 
flexibility in %the application of 
the calibration policy. The %possibility of addressing 
ability to address single channels in different configurations %, in both calibration methods 
has also allowed measurements of the crosstalk, which is < 1\%.

The calibration %test with 
method that uses the external pulsing card %is 
has proved preferable at the level of the \dword{qc} tests during production since it allows for %complete 
integrity tests of the %integrity of 
all connections in the \dshort{fe} card along the same signal path as for signals coming from the \dshort{crp}. 
The calibration card was explicitly developed for this purpose.

 In 2022 a new version of the calibration card that introduced %many 
 several simplifications was designed and  successfully tested. This design is also  more compact and easier to produce. In particular, \dword{pcb}-embedded capacitors replace the trimmed varicaps, and by including two 32-channel output connectors, %the calibration card can now be used to test 
 a full \dshort{fe} card can be tested in one run. %in a single shot.

%%%%%%%%%%%%%%%%%%%%%%%%%%%%%%%%
\subsubsection{TDE Readout Chain Optimizations for Vertical Drift}
\label{subsubsec: ADVOsss}

The \dshort{tde} readout electronics, originally developed for the \dshort{dp} detector design, %has been successfully operating on the 3x1x1 demonstrator 
operated successfully in the \dword{wa105} (2016-2018) and in 
\dshort{pddp} (2019-2022). 
The configuration was optimized in 2021 for 
\dshort{spvd} by adapting the signal dynamics to the bipolar signals produced by the two induction views. 
As discussed above, the modifications included the removal of the \dword{hv}-rated decoupling and biasing components on the \dshort{dp} anode \dshort{fe} cards, with equivalent components now present on the anode adapter boards. 
The analog electronics was already bipolar except for the baseline shift applied at the \dword{adc} buffer, which was removed setting VSHIFT at ground.
The \dword{amc} analog stage gain was increased to 4, yielding a total conversion factor of the analog chain of 14\,mV/fC, which matches better expected signals ranges.  
New productions for all elements of the chain were  carried out after these modifications.  

%the Vertical Drift application. 
The new configuration %has been 
was extensively tested %since 
starting in summer 2021 at \dword{cern}, %in the Vertical Drift CRPs tests campaign 
first with the tests in a dedicated integration facility,
%\dword{crp}  there were no CRPs in the integration facility !!
%and 
then with the first \dshort{crp} in the \coldbox, and finally in late 2022 %. As previously mentioned, this program was concluded in 2022 
with the tests of the two %final 
top-drift \dshort{crp}s built %in order to equip Module-0. 
for \dword{vdmod0}.

Figure~\ref{fig:mip_tde_pulser} shows examples of the measured response to charge injection signals in bipolar mode for the equivalent charge released by a minimum ionizing particles.

%$$$$$$$$$$$$$$$ 
\begin{dunefigure}
[\dshort{tde} response to bipolar charge injection signals]
{fig:mip_tde_pulser}
{\dshort{tde} response to bipolar charge injection signals. The amplitude of the square wave pattern produces injected charges equivalent to \dwords{mip}  for both polarities.}
	\includegraphics[width=.8\textwidth]{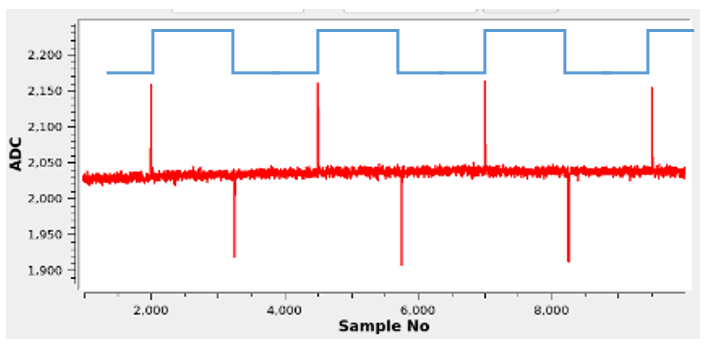}
\end{dunefigure}
%$$$$$$$$$$$$$$$ 

Additional developments %concerned as well 
centered on the digital electronics chain, with the successful conversion of the readout  system to the 40\,Gbit/s standard at %the level of 
the \dword{utca} crates \dword{mch} connectivity, as described in Section~\ref{subsubsec:UTCsss}. 

%%%%%%%%%%%%%%%%%%%%%%%%%%%%%%%%%
\subsection{Signal Feedthrough Chimneys}
\label{subsubsec: USFTsss}

The \dword{fe} cards are mounted on sliding blades that can be inserted into or extracted from the \dwords{sft} (see Figure~\ref{fig:dp_FE_card}). 

\begin{dunefigure}
[FE cryogenic amplifier card mounted on a \dshort{sft} sliding blade]
{fig:dp_FE_card}
{A \dshort{fe} cryogenic amplifier card mounted on a \dshort{sft}  sliding blade.}
	\includegraphics[width=.7\textwidth]{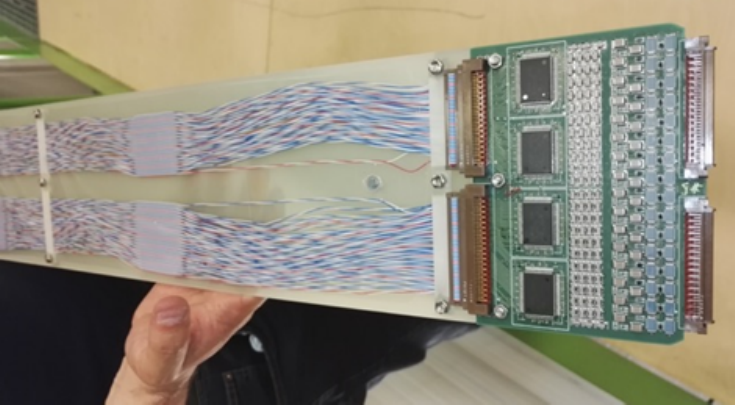}
\end{dunefigure}

The \dshort{sftchimney}s, ultra-high-vacuum-sealed on the top and bottom, enable access to the  \dshort{fe}  analog electronics for repair or replacement while the detector is operational, without affecting the inner cryostat volume. This capability was well demonstrated during the operation of \dshort{pddp} (see Figure~\ref{fig:dp_chimney_access}).  

\begin{dunefigure}
[Access to a FE cryogenic amplifier card in a SFT chimney]
{fig:dp_chimney_access}
{Access to a \dshort{fe} cryogenic amplifier card in a \dshort{sftchimney}.}
	\includegraphics[width=.4\textwidth]{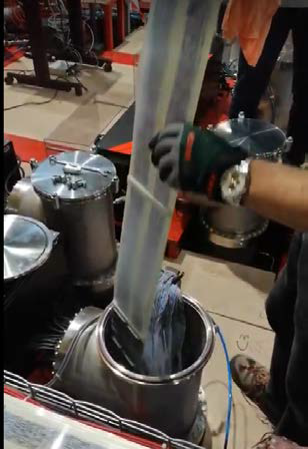}
\end{dunefigure}

%In addition, t
The metallic structure of the \dshort{sftchimney}s acts as a Faraday cage, isolating the  \dshort{fe}  \dshort{asic}s from environmental interference. 
The vacuum-tight feedthrough flanges at both ends of the \dshort{sftchimney}s dispatch the signal and slow control lines to and from the \dshort{fe} electronics inside it. The bottom (cold) feedthrough flange isolates the inner volume of the detector from the chimney volume and interconnects the signals from the \dshort{crp} to the analog \dshort{fe}  cards.  %A picture representing  is shown in  
Figure~\ref{fig:cold_flange} illustrates how the \dshort{fe} cards (shown dismounted from their blades) are plugged into the top side of a cold flange. 

\begin{dunefigure}
[\dshort{fe} cards plugged into the top side of a cold flange]
{fig:cold_flange}
{\dshort{fe} cards (seen from the chimney side and dismounted from the blades) plugged into the top side of a \dshort{pddp} cold flange. }
	\includegraphics[width=.5\textwidth]{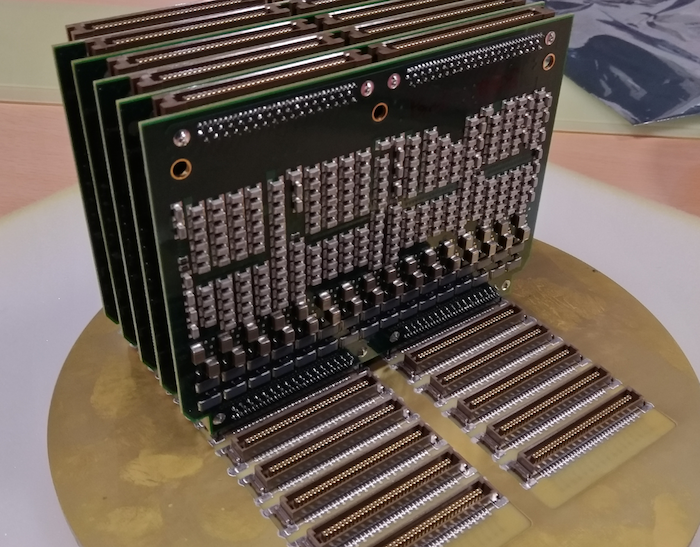}
\end{dunefigure}

The (warm) feedthrough flange at the top seals the chimney from the outside environment, passes the low-voltage and control lines to the enclosed \dshort{fe} electronics, and feeds out the differential analog signal lines from the \dshort{fe} amplifiers (see Figure~\ref{fig:warm_flange}). 

\begin{dunefigure}
[Warm flange of a SFT chimney in \dshort{pddp}]
{fig:warm_flange}
{Warm flange of a \dshort{sftchimney}  in \dshort{pddp}}
	\includegraphics[width=.7\textwidth]{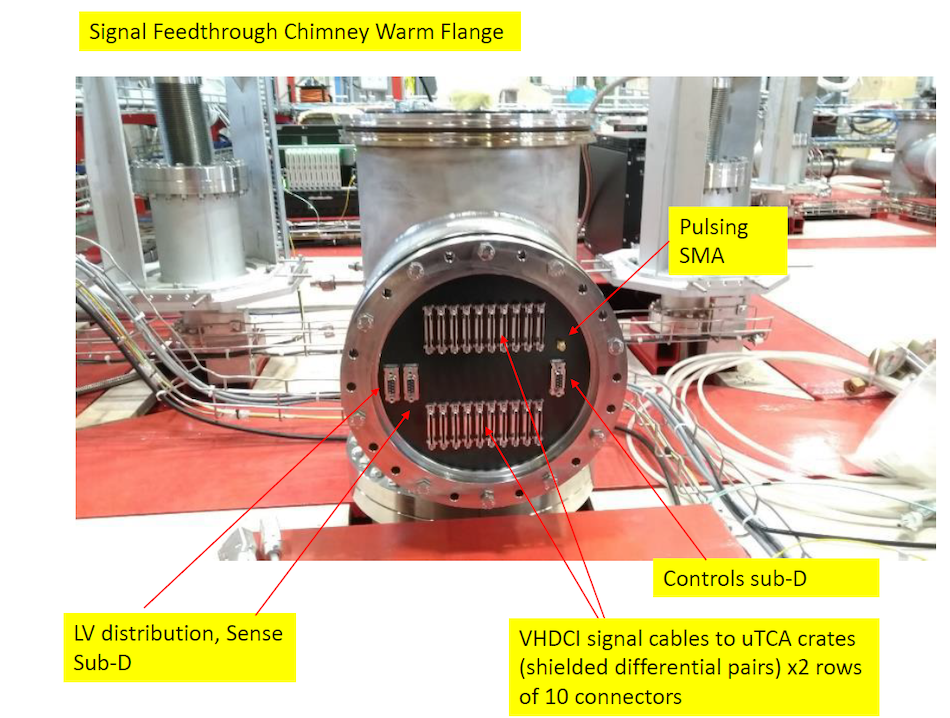}
\end{dunefigure}

Twenty power supply and filtering/distribution units are distributed along the cryostat on the detector electronics mezzanine  to feed the \dshort{sftchimney}s. A similar number of calibration/control boxes are also arranged on the mezzanine.

Based on \dshort{pddp} experience and optimization of the cryostat design, the 
63 roof penetrations (three rows of 21) for the inner \dshort{spvd} chimneys (those not along the sides) 
%\dword{spvd} roof penetrations have been 
were increased to 526\,mm diameter and the chimneys have been configured to host up to 48 \dword{fe} cryogenic cards, for a maximum of 3072 readout channels per chimney. 
The chimneys located along the cryostat long sides %along the long side 
(21 penetrations per side), 
host only 24 \dshort{fe} cards each % instead of 48 cards and to 
and fit into smaller penetrations of 381\,mm diameter. This arrangement accommodates a total of 105 chimneys %  on the cryostat roof 
(see Figure~\ref{fig:cryostat_chimneys}), reading 245,760 channels.

\begin{dunefigure}
[Cryostat roof layout with locations of the SFT chimneys]
{fig:cryostat_chimneys}
{Cryostat roof layout including the location of the top drift \dshort{sftchimney}s.}
\includegraphics[width=.9\textwidth]{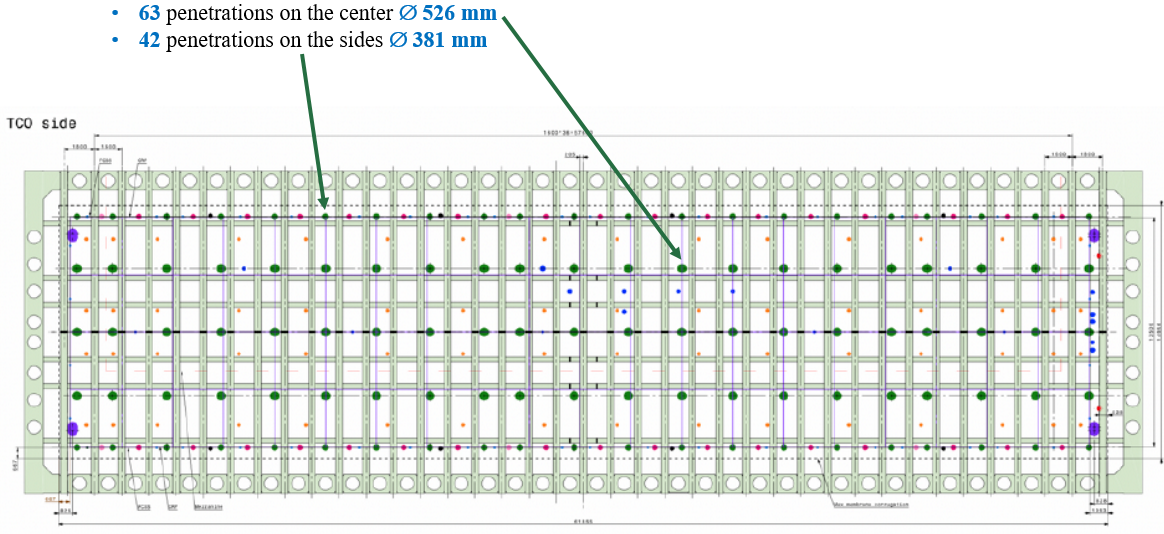}
\end{dunefigure}

Both %the 24 cards and the 48 cards 
chimney designs %exploit 
use the same warm flanges, each serving %each one 
a set of 24 cards. The larger chimneys have two warm flanges in a V-shape configuration. Figure~\ref{fig:new_chimneys} shows \threed models of the chimneys, including the warm flanges, as well as a bottom view with the cold flange removed % which allows seeing 
to show the location of the \dshort{fe} cards.

\begin{dunefigure}
[CAD models of the %larger 
\dshort{spvd} chimneys] %containing  24/48 cards designed for \dword{spvd}]
{fig:new_chimneys}
{CAD models of the %larger 
\dshort{spvd} chimney designs for  24 and 48 cards. %designed for \dword{spvd}, 
In the views at left, %the body of 
the %two 
chimneys %has been made 
are transparent %in order to be able to see 
to show the blades inside and the \dshort{fe} cards plugged into the cold flange at the bottom. % of the chimney. 
The two projections at %the 
right are %correspond to a 
bottom views where the cold flange has been removed %in order to appreciate 
to show the position of the \dshort{fe} cards (in green).}
	\includegraphics[width=.8\textwidth]{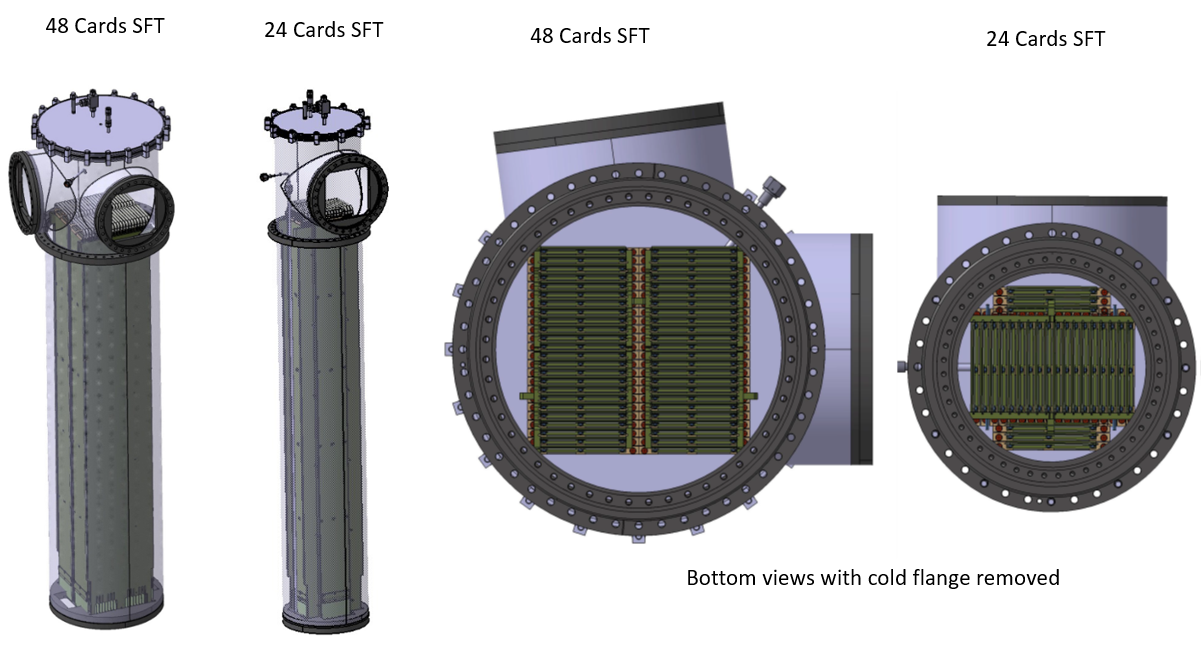}
\end{dunefigure}

The design of the 24/48 card chimneys was developed in 2022. %, including 
Thermal simulations were done and prototypes for the two sizes are being built. These prototypes will be tested and characterized in a dedicated test-bench environment, in parallel with \dshort{vdmod0} operation. %This is naturally due to the fact that 
The \dword{np02} cryostat is already equipped with 10-card chimneys from \dshort{pddp} and its roof structure cannot be modified to host the new larger chimneys.

%%%%%%%%%%%%%%%%%%%%%%%%%%%%%%%%%%%%%
\subsection{Digital Front-end}
\label{subsubsec: UADEsss}

 The warm digital electronics, located on the cryostat roof, digitizes the analog signals coming from the  \dshort{sftchimney}s and transmits them to the \dword{daq} system.  

%%%%%%%%%%%%%%%%%%%%
\subsubsection{\dshort{amc} Digitization Boards}
\label{subsubsec: USFTsss}

Each \dword{utca} crate can host up to  12 \dword{cro} \dwords{amc}. The configuration planned for the \dshort{tde} readout includes 12  \dshort{amc}s in each crate.  Each card (see Figure~\ref{fig:dp_amc}) has eight  Analog Devices AD92574\footnote{Analog Devices, \url{www.analog.com}} \dword{adc} chips, two dual-port memories,   and a \dword{fpga} (Altera Cyclone V\footnote{Altera Cyclone\textregistered, \url{https://www.intel.com/content/www/us/en/programmable/b/cyclone-v.html}}) on board. 

\begin{dunefigure}
[A front-end \dshort{utca} digitization card (CRO-AMC)]
{fig:dp_amc}
{A front-end \dshort{utca} digitization card (\dshort{cro} \dshort{amc}s): the two VHDCI connectors on the right attach to the cables that bring analog signals from the warm flange to the \dshort{amc}.}
		\includegraphics[width=.7\textwidth]{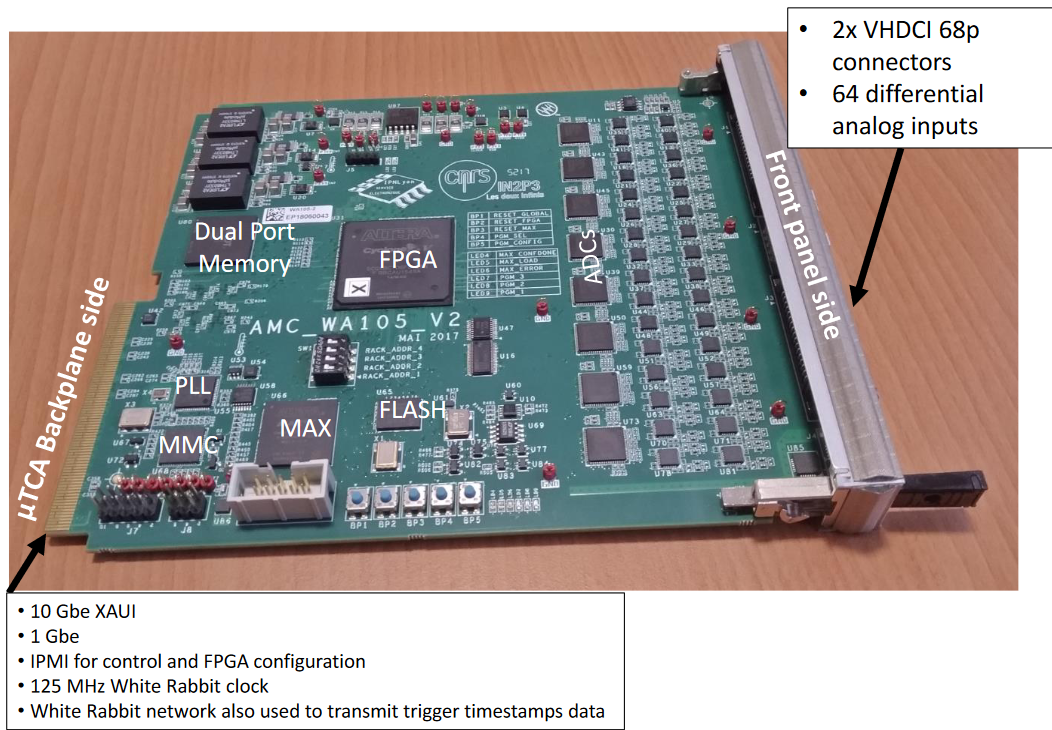}
\end{dunefigure}

The components were chosen to meet functional design requirements and technical criteria such as cost, chip footprint, power consumption, and ease of use~\cite{DUNE:2018mlo}.

The \dshort{fpga} provides a \dword{nios} virtual processor that handles the readout and data transmission. Given the programming flexibility provided by the firmware, which can be flashed to all \dshort{amc}s via the crate network connection, this digital  \dshort{fe} stage can also analyze and compress the data before transmitting them over the network. Figure~\ref{fig:amc_schematics} shows a schematic diagram of the  \dshort{amc} digitization board.

\begin{dunefigure}
[Schematic diagram of the \dshort{amc} digitization board]
{fig:amc_schematics}
{Schematic diagram of the \dshort{amc} digitization board.}
	\includegraphics[width=.9\textwidth]{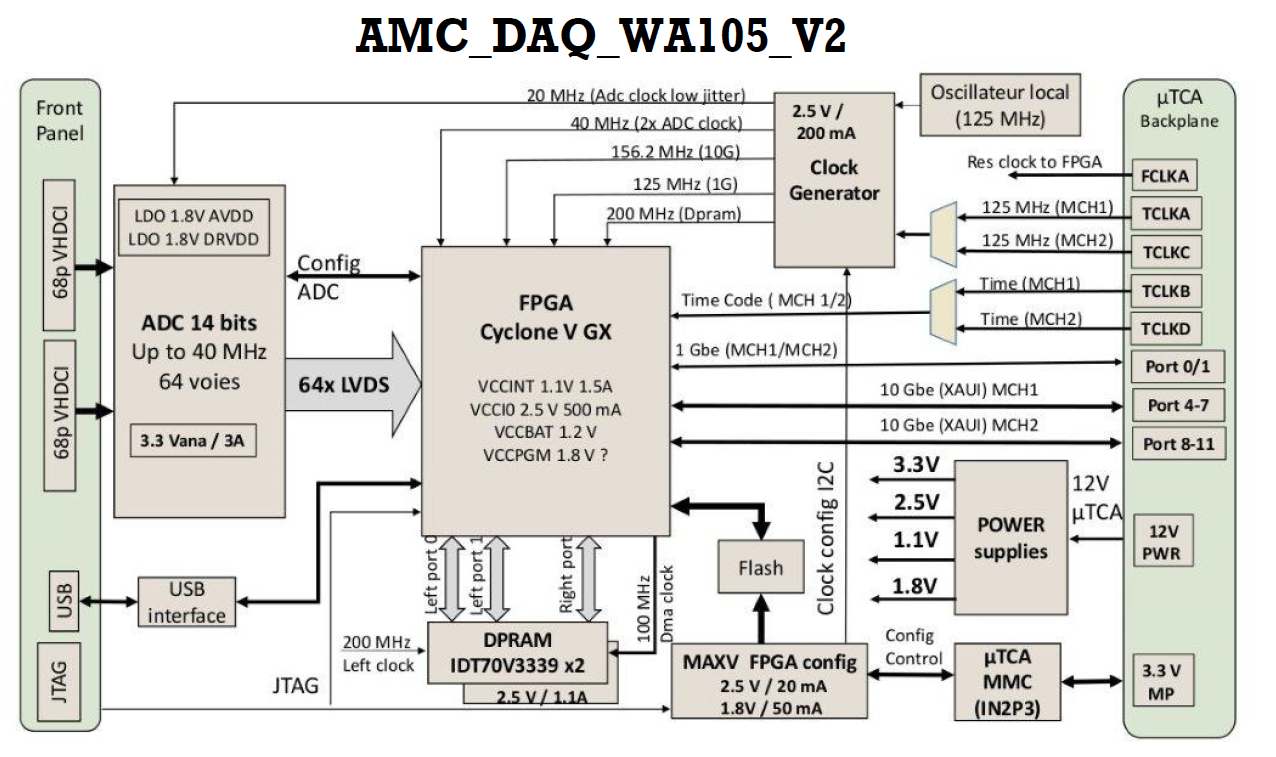}
\end{dunefigure}

The \dshort{amc} generates a continuous stream of data for each readout channel. Each \dshort{amc} has 64 channels and digitizes the signals coming from one analog \dshort{fe} card.  The \dshort{amc}s can be programmed in different operating configurations, offering quite a bit of flexibility: sampling can be performed either at  2.5\,\dword{msps} or at  2.0\,\dshort{msps}; data can have 12\,bit  or 14\,bit dynamics; data compression may be implemented or not; the \dshort{amc}s can acquire drift windows based on external triggers  (external trigger mode) or can simply continuously stream  all the sampled data to the \dshort{daq} system.  

For the final \dshort{amc}s output, data acquired at a given resolution/sampling rate are down-sampled in the \dshort{fpga}. This capability provides flexibility in the operating conditions. For instance, when operating in the 2.5\,\dshort{msps} sampling mode, data are acquired at a sampling rate a factor of eight higher,  then down-sampled in the \dshort{fpga} to 2.5\,MHz. Similarly, data with 12\,bit  dynamics are produced by selecting only the twelve most significant bits from each digitized 14\,bit sample. In the running configuration adopted for \dshort{pddp}, final sampling was  done at 2.5\,\dshort{msps}, 12\,bit. These 12\,bit data could then be  losslessly compressed using an optimized version of the Huffman algorithm, or kept  uncompressed, and in both cases buffered and organized into frames for network transmission in external trigger mode.  The configuration planned for \dshort{spvd} operation is simpler; it is based on a 2.0\,\dshort{msps} final sampling rate,  12\,bit  dynamics, and continuous streaming with no data compression applied. 
 
For the \dshort{spvd}, the \dshort{adc} dynamics has been optimized to deal with bipolar signals by removing the level offset used in \dshort{pddp}. 
Each  \dshort{amc} card has 10\,Gbit/s individual connectivity to the backplane of the \dshort{utca} crate.  Each \dshort{utca} crate contains a network switch, \dword{mch}, through which data collected from the hosted \dshort{amc} cards are sent to the \dshort{daq}. The timing synchronization of the \dshort{amc}s is achieved via a \dword{wrmch} module (also housed in the crate) connected to a \dword{wr} network, which ensures the last stage of the timing distribution at the level of the \dshort{fe} units. The \dshort{mch} and \dshort{wrmch} require one optical fiber link each. 

\dshort{pddp} functioned with \dshort{mch}es operating at 10\,Gbit/s. In this initial design, where the connectivity of each crate was limited to 10\,Gbit/s, 
lossless Huffman data compression in the \dshort{amc}s was used to ensure the necessary bandwidth for continuous streaming from ten \dshort{amc} cards to the  \dshort{daq} system. 

Starting in 2021, the readout system for the  \coldbox tests used the upgraded  40\,Gbit/s connectivity on each \dshort{utca} crate. The \coldbox tests of the top-drift \dshort{crp}s in 2021-2022 were conducted by reading the \dshort{amc} in external trigger mode, without compression, similar to  the operation of \dshort{pddp}. The new uncompressed operation mode for the \coldbox  however profited of the final 40\,Gbit/s connectivity, which does not have bandwidth limitations.

The \dshort{tde} baseline design has relied since the beginning on Ethernet links and data transmission in UDP packets. Beyond its original use in the \dshort{tde}, in 2022 the \dshort{daq} design evolved to adopt this readout scheme %everywhere in DUNE .
for the DUNE detectors.

Continuous streaming firmware for the \dshort{amc}s was developed by the \dshort{tde} in 2022, %accordingly to 
using the UDP data format agreed %with 
to by the \dshort{daq} consortium at the Preliminary Design Reviews (\dwords{pdr}).

%%%%%%%%%%%%%%%%%%%%%%%%
\subsubsection{\dshort{utca} crates}
\label{subsubsec:UTCsss}

In 2020 the commercial \dword{mch} technology had evolved, without increased costs, to 40\,Gbit/s connectivity. The 40\,Gbit/s links provide enough bandwidth to comfortably support continuous data streaming from each crate (768 readout channels) without the need for any data compression or zero skipping. This is the baseline operation mode for the top drift electronics.

The use of 40\,Gbit/s data links for each crate was extensively tested in April-July 2021 and it has been systematically implemented since the first \dshort{crp} \coldbox test in fall 2021. It %was then 
has now been validated as the reference design for \dshort{spvd}.   Figure~\ref{fig:crate_40} shows a \dshort{tde} \dshort{utca} crate equipped with a  40\,Gbit/s \dshort{mch}.

\begin{dunefigure}
[\dshort{utca} crate with a  40\,Gbit/s \dshort{mch} ]
{fig:crate_40}
{\dshort{utca} crate with a  40\,Gbit/s \dshort{mch}.}
\includegraphics[width=.6\textwidth]{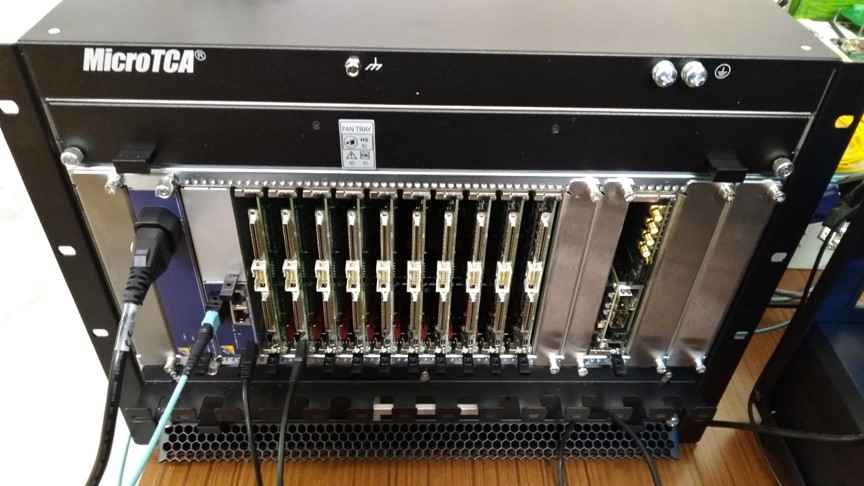}
\end{dunefigure}

In order to fully exploit the new \dshort{mch}s, the \coldbox %benefited since 2021 of 
also implemented a dedicated  fiber network infrastructure in 2021 (see Figure~\ref{fig:fiber_infra}) supporting a total bandwidth up to 240 Gbit/s and up to five \dshort{utca} crates for the full \dshort{tde} tests on CRP-2 and CRP-3.
Customized optical patch cables were developed %for the connections of 
to connect the crates to this infrastructure.  
 This was the first large-scale \dshort{utca} 40\,Gbit/s system to operate anywhere in the world. %wide.

\begin{dunefigure}
[\Coldbox fiber network infrastructure for 40\,Gbit/s bandwidth ]
{fig:fiber_infra}
{\Coldbox fiber network infrastructure supporting a system of five \dshort{utca} crates with 40\,Gbit/s \dshort{mch} for the tests of the full  \dshort{crp} \dshort{tde}.}
	\includegraphics[width=.8\textwidth]{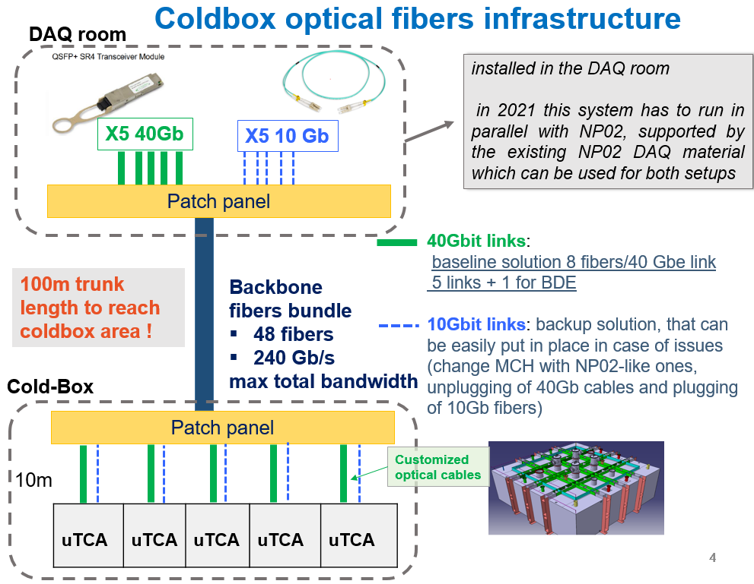}
\end{dunefigure}
 
%%%%%%%%%%%%%%%%%%%%%%%%%%%%%
\subsubsection{Front-end Timing Distribution System}
\label{subsubsec:UATSsss}

The  last stage of the %top drift 
\dshort{tde} timing distribution at the level of the \dshort{fe} units uses a \dword{wr} network that combines the synchronous 1\,Gbit/s Ethernet (SyncE) technology with the exchange of \dword{ptp} (V2) packets to synchronize clocks of distant nodes to a common time while automatically compensating for the propagation delays.  The system is fed by a  high-stability \dword{gps} disciplined oscillator (GPSDO) providing a clock reference signal to be distributed over the physical layer interface of the \dshort{wr} Ethernet network. 

The network topology uses specially designed switches that have standard IEEE802.1x Ethernet bridge functionality with  additional \dshort{wr}-specific extensions to preserve clock accuracy. Time and frequency information is distributed to the nodes on the \dshort{wr} network via optical fibers. 

The \dshort{wr} protocol automatically performs dynamic self-calibrations to account for any propagation delays and keeps all connected nodes continuously synchronized to sub-nanosecond precision, independently of the network topology and changes in environmental conditions.  The \dshort{wr} Ethernet network %, by being an Ethernet network, 
can be also used to transport data other than the \dshort{ptp} packets used for synchronization, %and 
which occupy a very small bandwidth,  
e.g., to transmit the data containing the \dshort{wr}  timestamps of external trigger signals.  The \dshort{amc}s use these timestamps to select the drift-window data samples starting at the the time of the external trigger. 

The first \dshort{wr} switch connected to the DUNE timing system acts as the \dshort{wr} network grandmaster (\dword{wrgm}), and it is connected via 1\,Gbit/s optical links through a cascade of secondary \dshort{wr} switches to the \dshort{wr} end-node slave cards (see Figure~\ref{fig:wr_mch}). A  \dshort{wr} end-node slave card is present in each \dshort{utca} crate (\dword{wrmch}) and it is used to keep the \dshort{amc}s synchronized to the reference time signals and to distribute timestamp data to them.

\begin{dunefigure}
[\dshort{wr} end-node slave card]
{fig:wr_mch}
{\dshort{wr} end-node slave card.}
	\includegraphics[width=.8\textwidth]{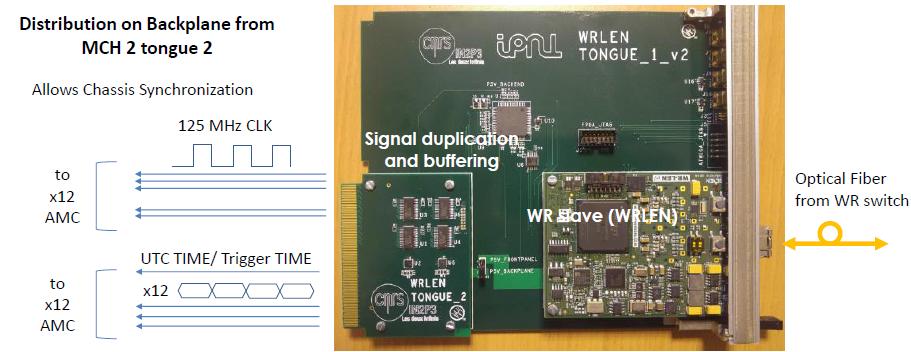}
\end{dunefigure}

The \dshort{wr} end-node slave card is a \dshort{utca}  board into which the WLREN (commercial \dshort{wr} end-node) mezzanine is plugged.
%The \dword{wr} end-node
This card occupies the second \dshort{mch} slot and  provides power to the WRLEN via standard \dshort{utca} facilities. It %then 
delivers the signals WR$_{clock}$ (125 MHz) and WR$_{DATA}$ for synchronization to each  \dshort{amc} via dedicated lanes on the \dshort{utca} crate backplane.

 %%%%%%%%%%%%%%%%%%                
\subsection{Performance in \dshort{pddp}}
\label{subsubsec:topelec:perf-np02}

 The elements of the \dshort{tde} readout chain %have already 
 operated reliably in the \dword{wa105} (2016-2017) and \dword{pddp} (2019-2022). These elements are very similar to those planned for the \dshort{spvd} \dshort{tde} apart from the minor modifications previously described. 
 
 All the components had undergone very rigorous \dword{qc} procedures before installation  and demonstrated reliability over %these
 extended time periods with no failures. The \dshort{wr} time distribution chain in \dword{np02}, for instance, was in continuous operation with no interruptions from 2019 to 2022. 
 Figure~\ref{fig:tde_pddp} shows the charge readout system on the \dshort{pddp} cryostat roof and Figure~\ref{fig:events_pddp} shows some cosmic ray events that it %has 
 read out. %were read out by it. 
 During \dshort{pddp} operations, 100\% of the channels were active.

 \begin{dunefigure}
[Charge readout system on the \dshort{pddp} cryostat roof]
{fig:tde_pddp}
{Charge readout system on the \dshort{pddp} 
 cryostat roof.}
\includegraphics[width=.7\textwidth]{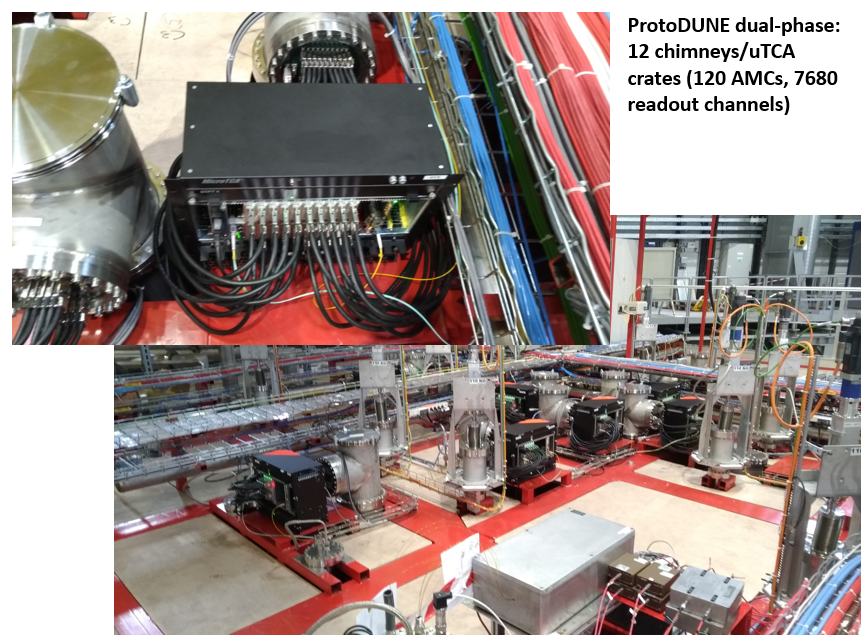}
\end{dunefigure}

\begin{dunefigure}
[Cosmic ray events acquired with \dshort{pddp}]
{fig:events_pddp}
{Cosmic ray events acquired with \dshort{pddp}}
\includegraphics[width=.8\textwidth]{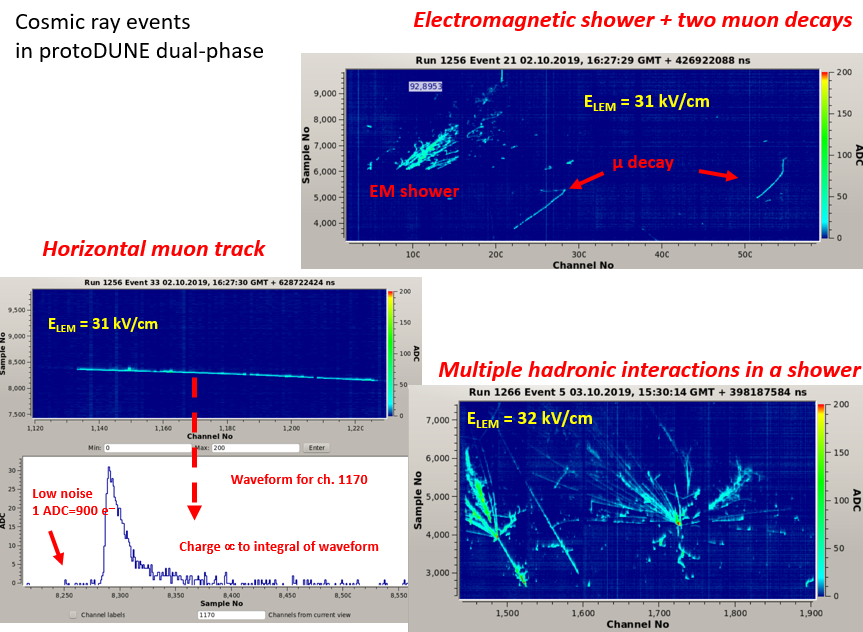}
\end{dunefigure}

 The \dshort{pddp} cards operated with a typical intrinsic noise around 600 electrons.   Overall noise levels in \dshort{pddp} were dominated by environmental conditions inside the cryostat related to grounding imperfections of the slow-control connections at their dedicated cryostat flanges (\dshort{hv}, temperature probes, cameras, and \dshort{crp} instrumentation). Strip  capacitance for the \dshort{spvd} application is lower than %the capacitance expected for 
 in the \dshort{dp} design. 
 The electronics is well suited to reading strips in this capacitance range.

 The \dshort{pddp} electronics, despite a higher gain (from the \dword{lem} micro-pattern detectors operating in the gas phase), received lower-amplitude input signals than those that have been produced by the \dshort{crp}s developed for \dshort{spvd}. During \dshort{pddp} data collection, the gain was limited to 6. Several factors conspired to erode its  signal amplitude: that fact that two collection views shared the signal, a finer strip pitch, and inefficiencies related to the extraction of the electrons into the gas phase and their collection on the anodes.  The performance of the \dshort{tde} in the 2021 and 2022 \coldbox tests (Section~\ref{subsec:CRP_prototyping2022}) has in fact surpassed that of \dword{pdsp} in terms of noise and signal amplitude.

\dshort{pddp} operated in two runs: Run1 (July 2019 - September 2020) and Run2 (September 2021 - March 2022). Before Run2 the cryostat was emptied to replace the HV extender delivering 300\,kV to the cathode, because it had developed a short circuit to the \dshort{fc} at the beginning of Run1. 
Run2 included the stability tests of the HV distribution system designed for \dshort{spvd}. Operating the \dshort{tpc} at the design voltage of 300\,kV  provided the opportunity to 
observe 6\,m long tracks of cosmic ray muons crossing the entire %gap 
anode-to-cathode drift length.  For this demonstration, the \dshort{pddp} readout electronics was used with an anode of 1\,m$^2$ active surface, equipped with a grid for the extraction of ionization electrons into the gas phase, but without \dwords{lem} for gas amplification.  In this configuration of charge-sharing among readout views and using narrower strips, the signal amplitude was approximately four times smaller than in the \dshort{spvd} configuration.
Data for this observation were collected with an electron lifetime in \dshort{lar} of about 2 to 3\,ms, to be compared to a maximum drift time of 3.7\,ms in \dshort{pddp}.

Despite these limitations, thanks to the good performance and the low noise achieved by the \dshort{tde} electronics, a large sample of clean cosmic ray tracks was collected (over 1M events and 120\,TB of data) and successfully reconstructed.  
Figure~\ref{fig:6m_tracks} shows examples of cosmic ray events collected in these conditions. 
They represent the operation of a \dshort{lartpc} over the largest drift distance %gap 
used so far.

\begin{dunefigure}
[Cosmic ray tracks acquired with ProtoDUNE-DP demonstrating 6\,m drift]
{fig:6m_tracks}
{Cosmic ray tracks acquired with \dshort{pddp} demonstrating 6\,m drift}
	\includegraphics[width=.8\textwidth]{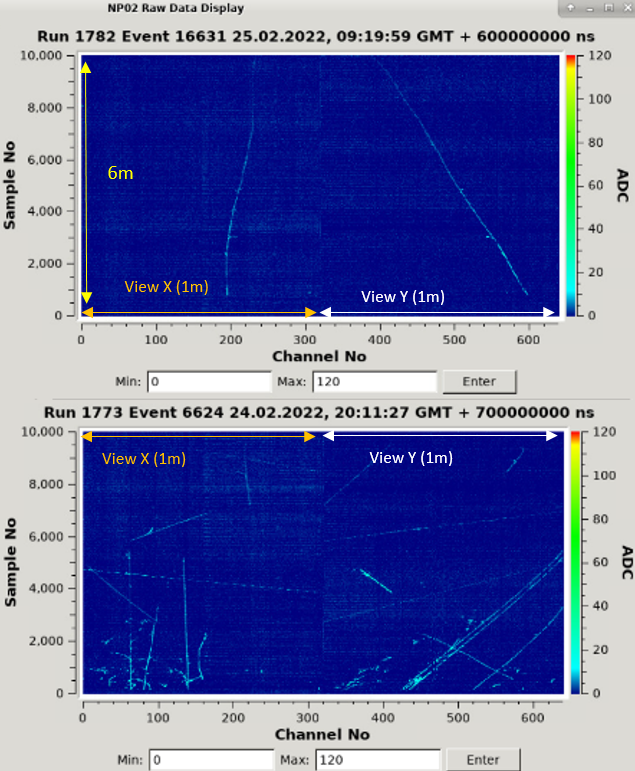}
\end{dunefigure}
        
 %%%%%%%%%%%%%%%%%%                
\subsection{\Coldbox Tests of the \dshort{tde}} 

\label{subsubsec:topelec:cold-box}

Starting in fall 2021, the \dshort{tde} readout chain has undergone  a very intensive campaign of qualification tests along with the \dshort{crp} prototypes both at warm in a Faraday cage and in the \coldbox.
In the first tests, CRP-1 was equipped with both \dshort{tde} and \dword{bde} electronics. 
A modified setup, the CRP-1b, included grounding improvements and the installation of an isolation transformer. 

 The first \dshort{crp} with the final channel layout and strip orientations (CRP-2) was built for the top drift layout and tested in the Faraday cage and in the \coldbox in June-July 2022 (Section~\ref{subsubsec:CRP_2_production}). The second ``final'' module (CRP-3, Section~\ref{subsubsec:CRP_3_production}) was also tested in both the Faraday cage and the \coldbox in September-October 2022. CRP-2 was tested for a second time in the \coldbox in October-November 2022 after some corrections to the anode boards. %silver-printed joints. 
 These tests concluded the qualification campaign of the \dshort{crp}s and \dshort{tde} %that have been built to equip 
 for the top drift volume of \dword{vdmod0}.

 During this campaign, the \dshort{crp}s were mounted to the \coldbox roof and the assembly was used in both the warm and cold tests. 
 The \dshort{tde} readout chain was installed and dismounted about 20 times throughout this process with no deterioration. This provided an excellent opportunity to develop and validate installation procedures, and also demonstrated the robustness of the components. A graphical timeline of all these milestones and achievements is presented in Figure~\ref{fig:timeline_tde}.

 \begin{dunefigure}
[Timeline of the 2021-2022 \dshort{tde} tests] %campaign in 2021-2022]
{fig:timeline_tde}
{
Timeline of the 2021-2022 \dshort{tde} tests.}
	\includegraphics[width=.95\textwidth]{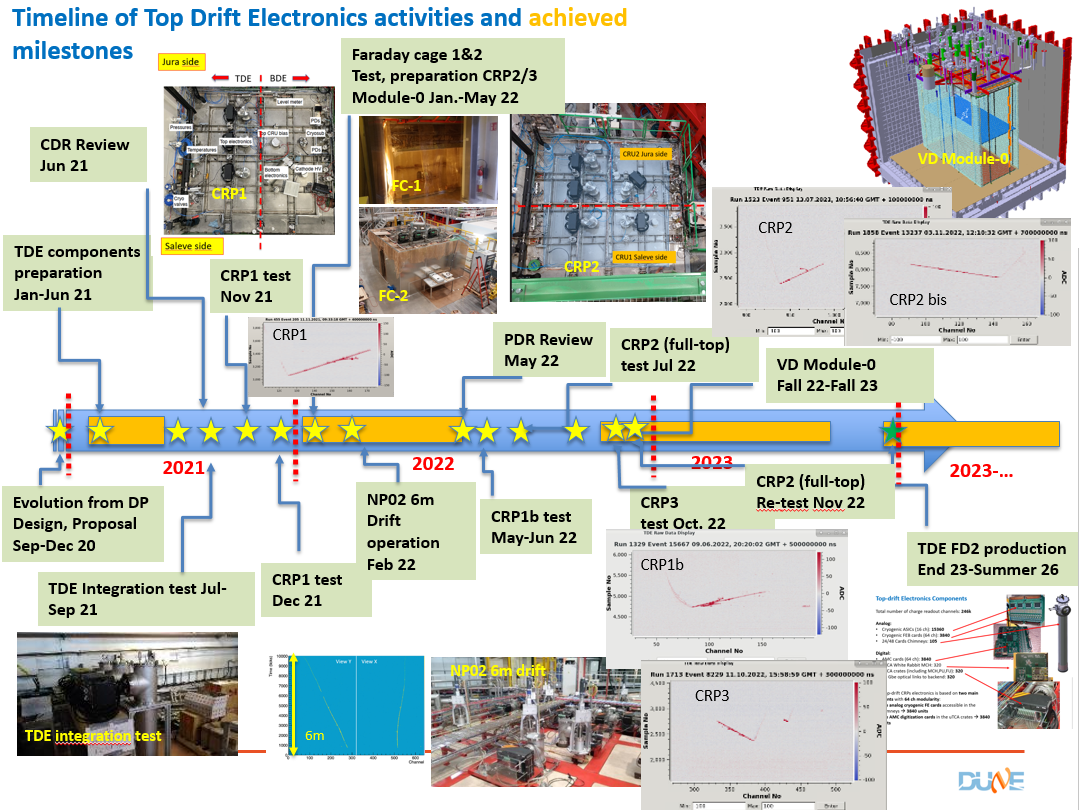}
\end{dunefigure}

 After defining the required adaptations relative to the \dshort{dp} design, the new components for the \dshort{tde} \coldbox test, listed below, were procured in spring 2021 and extensively tested before installing them in the TDE integration test at \dshort{cern} a few months later. %by the beginning of summer 2021.
 The following lists the modifications:

 \begin{itemize}
\item New production of \dshort{asic}s, 
\item  New front-end cards, 
\item  %Modifications to  
Modified digitization cards for bipolar signals
\item  %Production of n
New timing cards and a new \coldbox-dedicated timing distribution network, independent of infrastructures in the \dshort{np02} experimental area,
\item  New  \dshort{mch}s in \dshort{utca} crates at 40 Gbit/s and associated fiber network infrastructure,
\item  New low-voltage generation and distribution system, independent of \dshort{np02},
\item New calibration system, also usable for \dshort{spvd},
\item  %Setting up n
New \dshort{daq}/network system setup, %for \coldbox
\item  %Production of ne
New cold/warm flanges,
\item  %Production of 
VHDCI cabling and inner chimney cabling, and
\item  %Dedicated production of 5 
Five ten-card mini-chimneys, optimized for the \coldbox roof thickness.
\end{itemize}

 A full integration test of these elements was %then 
 performed at warm, without the \coldbox roof and without \dshort{crp}s, %(which was still under construction), %since 
 starting in July 2021 in a dedicated area at \dword{ehn1}  (see Figure~\ref{fig:integration_test}).

 \begin{dunefigure}
[TDE integration test at EHN1 (summer 2021)]
{fig:integration_test}
{\dshort{tde} integration test at \dshort{ehn1} in summer 2021}
	\includegraphics[width=.7\textwidth]{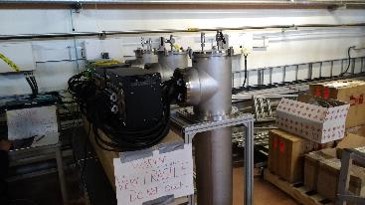}
\end{dunefigure}
 
 This integration test validated the entire readout chain and, despite the rudimentary grounding scheme of this test facility  (connected to the building ground),  
 demonstrated a low noise level of 2.5 \dword{adc} \dword{rms}, %that was 
 in agreement with expectations at warm with no strips connected, 
 and the absence of coherent noise %present 
 (see Figure~\ref{fig:integration_fft}).
 \begin{dunefigure}
[\dshort{fft} noise spectrum during the integration test]
{fig:integration_fft}
{\dword{fft} noise spectrum during the integration test showing the absence of coherent noise peaks}
	\includegraphics[width=.5\textwidth]{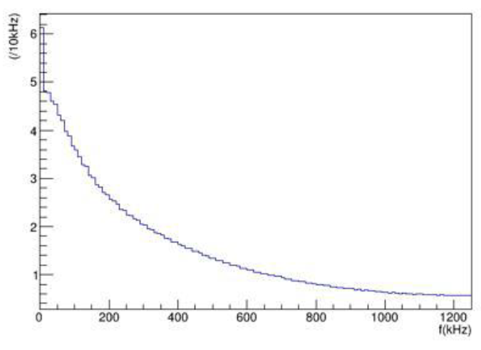}
\end{dunefigure}

The integration test started with a single chimney. In September-October 2021 it was extended to three chimneys, and all channels were tested with the \dword{daq} and shown to be active.  The integration setup was dismounted at the end of October 2021 and the components were moved to the \coldbox area  where they were installed on the \coldbox roof  (see Figure~\ref{fig:crp1_tde}).

 \begin{dunefigure}
[\dshort{tde} installation on the \coldbox for the test of CRP-1]
{fig:crp1_tde}
{\dshort{tde} installation on the \coldbox cryostat roof for the test of CRP-1}
	\includegraphics[width=.7\textwidth]{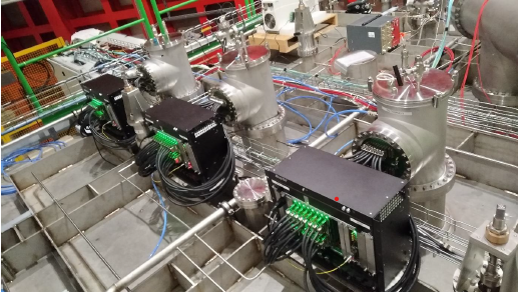}
\end{dunefigure}

 After validating the installation and the cryogenics system,  the \coldbox was filled with \dshort{lar} on 11 November 2021 (Run-1, covering the period 11-19 November).  After filling, the anode biases and the cathode \dshort{hv} were activated. Clean tracks were immediately visible, indicating good performance of the \dshort{tde}. 

Figure~\ref{fig:TDE_events} shows examples of raw cosmic ray data events acquired with the \dshort{tde}, %These are  display images 
without treatment for coherent noise removal (CNR). Waveforms for the three views highlight
%are also shown in order to appreciate 
the different signal shapes for the two induction views (bipolar) and the collection view (unipolar). 

\begin{dunefigure}
[First raw data event displays acquired during \coldbox Run-1]
{fig:TDE_events}
{Raw cosmic ray data event displays of the first events  acquired with the \dshort{tde} during \coldbox Run-1}
	\includegraphics[width=.8\textwidth]{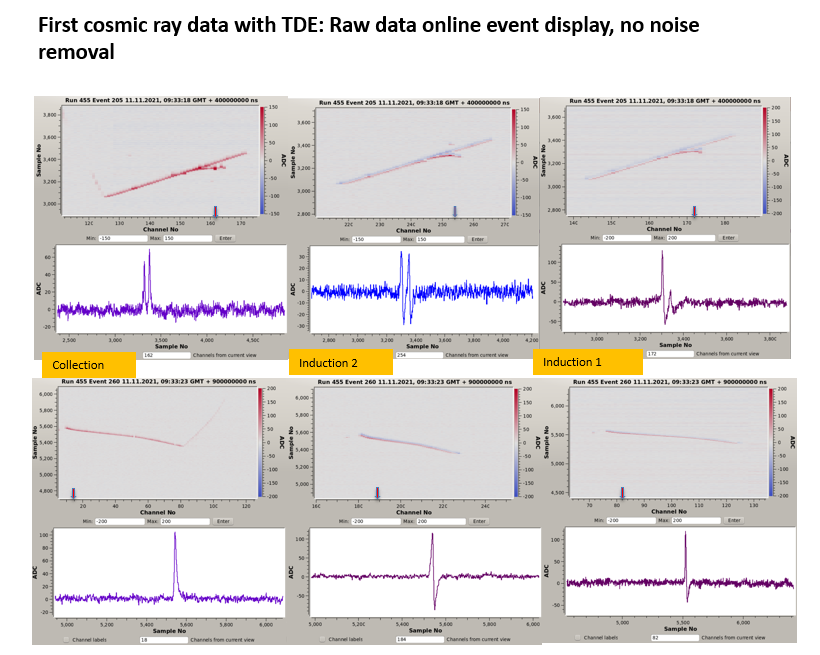}
\end{dunefigure}

 A large cosmic ray data sample that can be used %, exploitable 
 for \dshort{crp} performance studies was then acquired with the \dshort{tde}.  Run-1 collected 0.74M triggers, corresponding to 20\,TB, and with Run-2, this statistic climbed to 1.6M triggers. Already from Run-1, results demonstrated detection performance well in line with expectations from previous %preliminary 
 tests performed in a 50 liter \dshort{tpc} setup.   
 
 Although the \coldbox grounding system during these runs left some %small 
 low-level coherent noise that affected the electronics, 
 the standard DUNE Coherent Noise Removal (CNR) procedure, shown in  Figure~\ref{fig:CNR_example}, %could easily 
 was able to remove it easily. 
 
  A slight worsening of the coherent noise conditions for the half-\dshort{crp} readout with the \dshort{tde} was observed in Run-2. %the second cold-box run. 
  This was traced back %after the end of the run 
  to a bad contact %which 
  that had developed on the bias connection on two anode adapter boards serving half of the induction-2 view. 
 Once CNR was applied offline, noise levels %are 
became comparable to %the results 
those obtained in \dshort{pdsp}.

 \begin{dunefigure}
[Example of application of the Coherent Noise Removal (CNR) procedure]
{fig:CNR_example}
{Example of application of the Coherent Noise Removal (CNR) procedure to event 3257 of Run 483. Plots  with tracks in the three CRP views, produced with and without CNR, show on the horizontal axis the channel number, on the vertical axis the time ticks. The color code corresponds to the signals amplitude in ADC counts. The plot at the top right shows the wave-forms of for all channels of induction 1 for this event before and after CNR}
	\includegraphics[width=.9\textwidth]{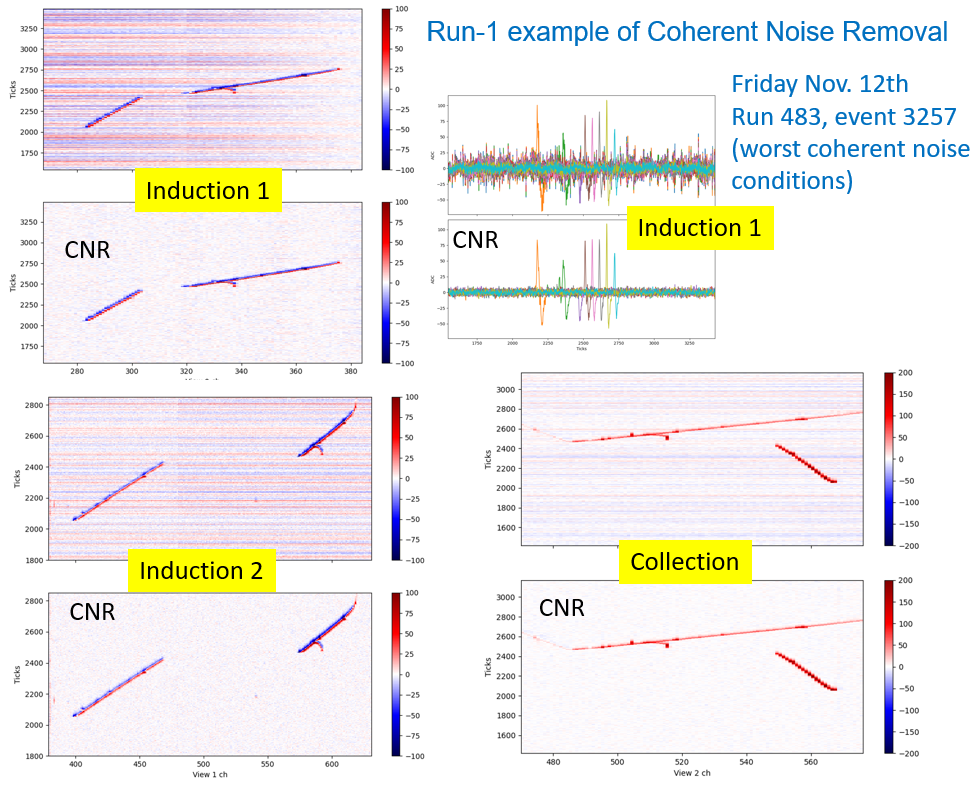}
\end{dunefigure}

Figures~\ref{fig:noise_noCNR} and~\ref{fig:noise_CNR} show  summaries of the noise performance for the three views of CRP-1 for the \dshort{tde} (Run-1 and Run-2) and the \dshort{bde} (Run-2) before and after the CNR procedure, respectively.

\begin{dunefigure}
[Noise performance for the three views of CRP-1 before CNR]{fig:noise_noCNR}
{Noise performance for the three views of CRP-1 before CNR}
\includegraphics[width=.7\textwidth]{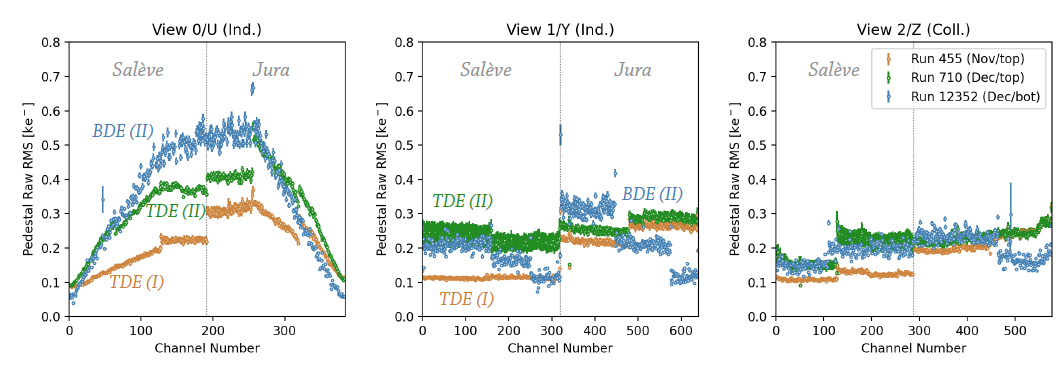}
\end{dunefigure}

\begin{dunefigure}
[Noise performance for the three views of CRP-1 after CNR]
{fig:noise_CNR}
{Noise performance for the three views of CRP-1 after CNR. The red %reference 
line represents the noise level in \dshort{pdsp}.}
\includegraphics[width=.7\textwidth]{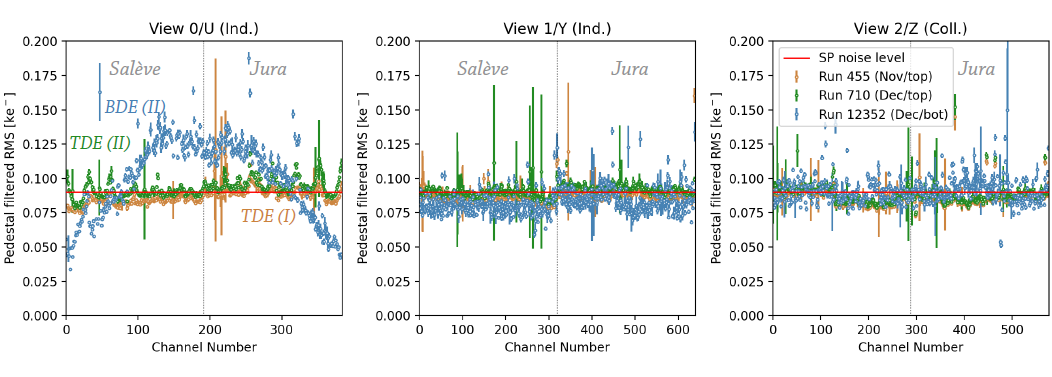}
\end{dunefigure}

These first \coldbox tests showed that the noise levels were %indeed 
in line with expectations, with an \dshort{rms} corresponding to 2.7\,ADC counts (666 electrons) for a capacitance of about of 200\,pF.
The coherent noise contamination conditions %in the \coldbox setup 
greatly improved in spring 2022 with the installation of an isolation transformer and other %little 
minor setup improvements, % of the setup 
and better shielding of the flanges used for the slow control services. 
This was demonstrated in subsequent \coldbox tests of CRP-1b, CRP-2 and CRP-3.

Besides reduced coherent noise, the test of CRP-1b %also 
allowed comparison of the signal to CRP-1 over several months. Figure~\ref{fig:crp1_stability} shows superimposed $dQ/ds$ and $dE/dx$ curves for the November (black) and June (red) CRP-1 runs. 
The Landau distributions correspond to the nominal values/shapes and show a very good detector resolution. The curves corresponding to the two different periods are practically indistinguishable, demonstrating the stability of the detector and of its readout system. 

\begin{dunefigure}
[Calorimetric response stability for CRP-1/CRP-1b ]
{fig:crp1_stability}
{Calorimetric response stability for runs taken several months apart with CRP-1/CRP-1b.}
	\includegraphics[width=.6\textwidth]{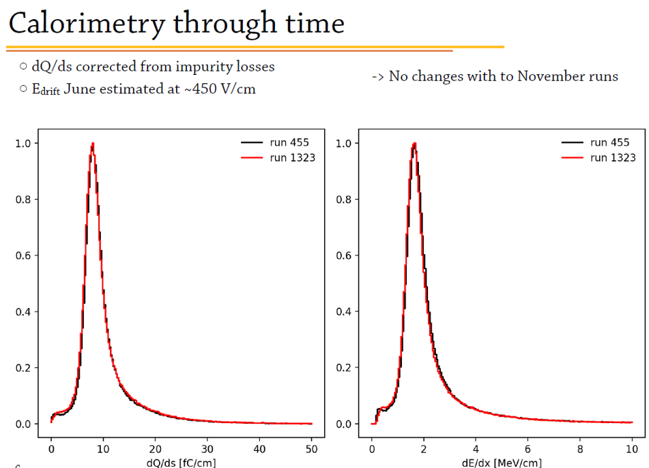}
\end{dunefigure}

Figure~\ref{fig:crp2_tests} shows the Faraday cage and \coldbox installations %in occasion of 
for the tests performed on CRP-2  in June-July 2022. This first full vertical drift \dshort{crp} %was involving for its readout 
used 48 \dshort{fe} readout cards %hosted by 
installed on five 10-card chimneys. The chimneys were connected to five %crates with 
48-\dshort{amc} crates.

\begin{dunefigure}
[\dshort{tde} readout chain in the Faraday cage and the \coldbox for CRP-2 tests]
{fig:crp2_tests}
{Left: \dshort{tde} readout chain installed in the Faraday cage. Right: \coldbox setup for the CRP-2 tests in July 2022}
	\includegraphics[width=.8\textwidth]{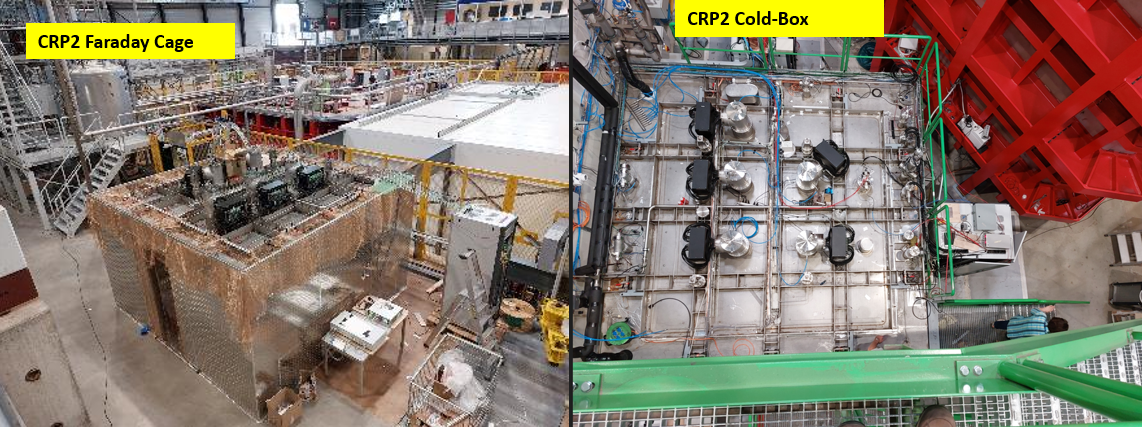}
\end{dunefigure}

Given the grounding improvements already performed on the \coldbox in spring 2022, cosmic ray tracks were collected for the full CRP tests in very clean coherent noise conditions.
Figure~\ref{fig:crp23_events} shows two 
%very clean 
cosmic ray tracks acquired with CRP-2 in July 2022 and with CRP-3 in October 2022,  respectively.  The raw data from the collection view is shown with no noise treatment. 
 
 \begin{dunefigure}
[Cosmic ray tracks collected during CRP-2 and CRP-3 \coldbox runs]
{fig:crp23_events}
{Examples of cosmic ray tracks collected during the \coldbox runs for CRP-2 (left)  and CRP-3 (right).}
	\includegraphics[width=.9\textwidth]{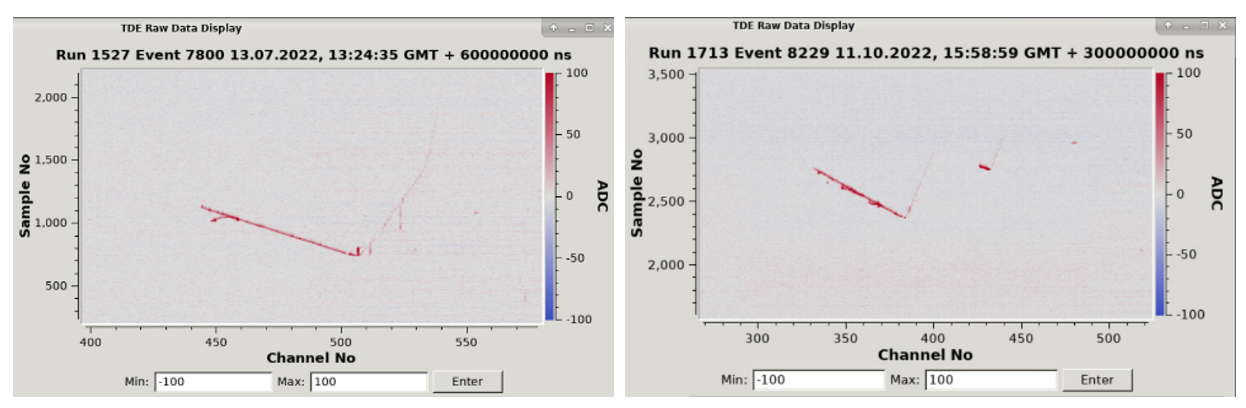}
\end{dunefigure}

The noise levels for CRP-2 an CRP-3 before and after noise reduction are summarized in Figure~\ref{fig:crp23_noise}. Compared to the  \coldbox runs of CRP-1 in 2021, % one can notice the large 
the significant reduction of the coherent noise contamination is evident. 

\begin{dunefigure}
[Noise levels before and after CNR for CRP-2 and CRP-3]
{fig:crp23_noise}
{Noise levels before (top) and after (bottom) CNR for the operation of CRP-2 and CRP-3.}
	\includegraphics[width=.8\textwidth]{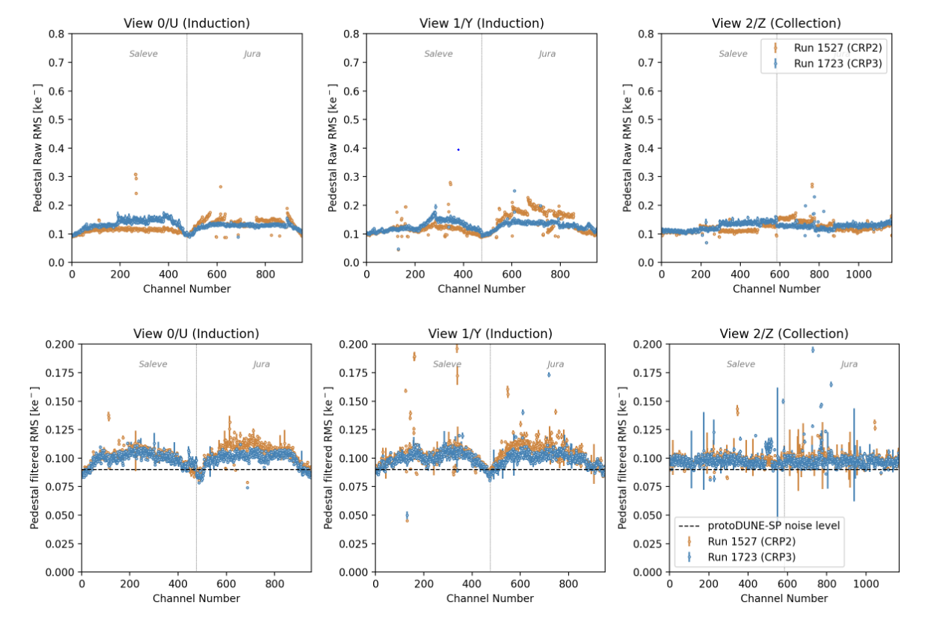}
\end{dunefigure}

Due to the good performance of the \dshort{tde} in the \coldbox runs, it was possible to collect a high statistics sample of cosmic-ray events. These were used to finely map and characterize the response of CRP-2 and CRP-3, which will be installed in \dshort{vdmod0} in early 2023.

 %%%%%%%%%%%%%%%%%%
\subsection{\dshort{tde} Services on the Cryostat Roof}
\label{subsubsec:topelec:services}

A \threed model of the cryostat roof (Figure~\ref{fig:chimneys_cryostat}) shows the position of the chimneys with respect to the steel I-beam structure. %cryostat roof. 
The %48 cards 
larger chimneys have two warm flanges, each  connected to two \dword{utca} crates via the VHDCI cables. The %24 cards 
smaller chimneys have a single warm flange. 

\begin{dunefigure}
[Model of cryostat roof I-beams structure and chimneys]
{fig:chimneys_cryostat}
{\threed view of the cryostat roof I-beam structure and the position of both the larger and smaller %48 cards  and 24 cards 
chimneys.}
	\includegraphics[width=.6\textwidth]{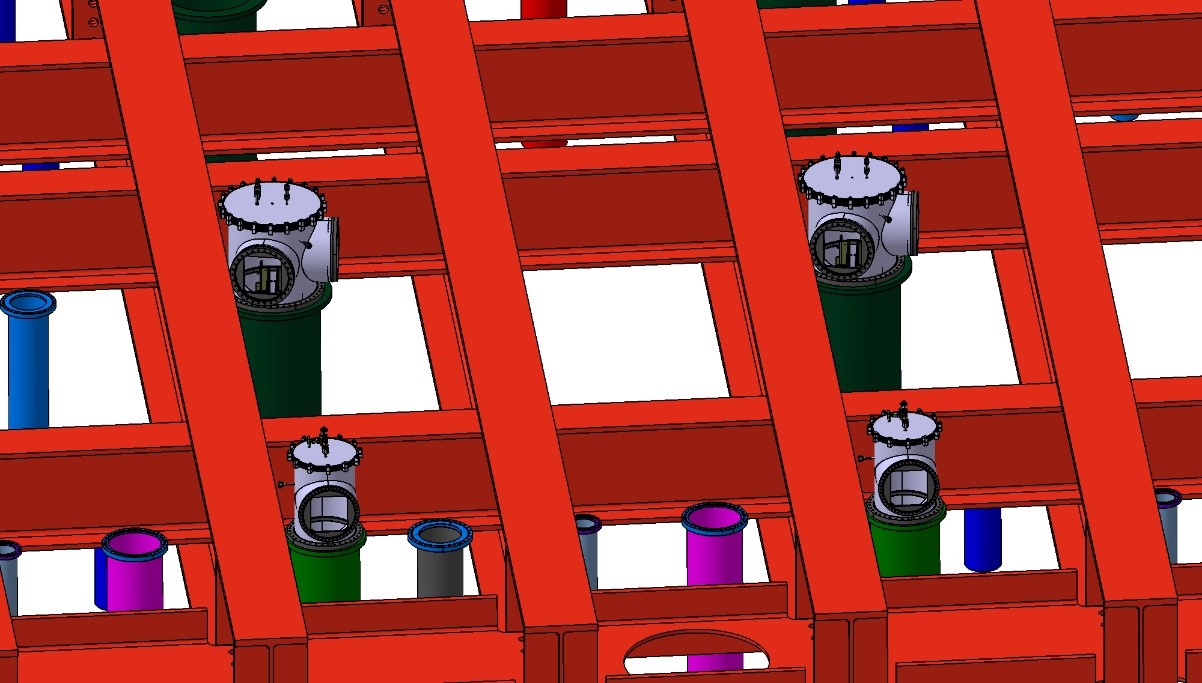}
\end{dunefigure}

The positioning of the chimneys with respect to the I-beams %of the steel structure %of the cryostat 
allows positioning the crates directly on the top surface of the  I-beams and to cable them to the warm flanges via the front connections of the \dword{amc} cards.  Each \dshort{utca} crate %needs to be connected 
requires connection to a 110V AC power cord, % and to 
a \dword{wr} fiber, and a data fiber. Each warm flange %needs to be connected 
requires connection to two low-voltage cables, a cable for controls, and a pulsing cable. A single sense cable is used for two warm flanges.

Figure~\ref{fig:crates_cabling} illustrates the VHDCI cable connections of a large (48-card) chimney to the four \dshort{utca} crates, and the position of the cable trays for the service cables and the flanges.

The 24-card chimney connections are similar, but connect to only two \dshort{utca} crates to a single flange.

\begin{dunefigure}
[Large (48-card) chimney cabling to \dshort{utca} crates]
{fig:crates_cabling}
{Top and side views of a larger (48-card) chimney with  the VHDCI cabling to four \dshort{utca} crates, including the integration of the cable trays on the I-beam structure.} 
	\includegraphics[width=1\textwidth]{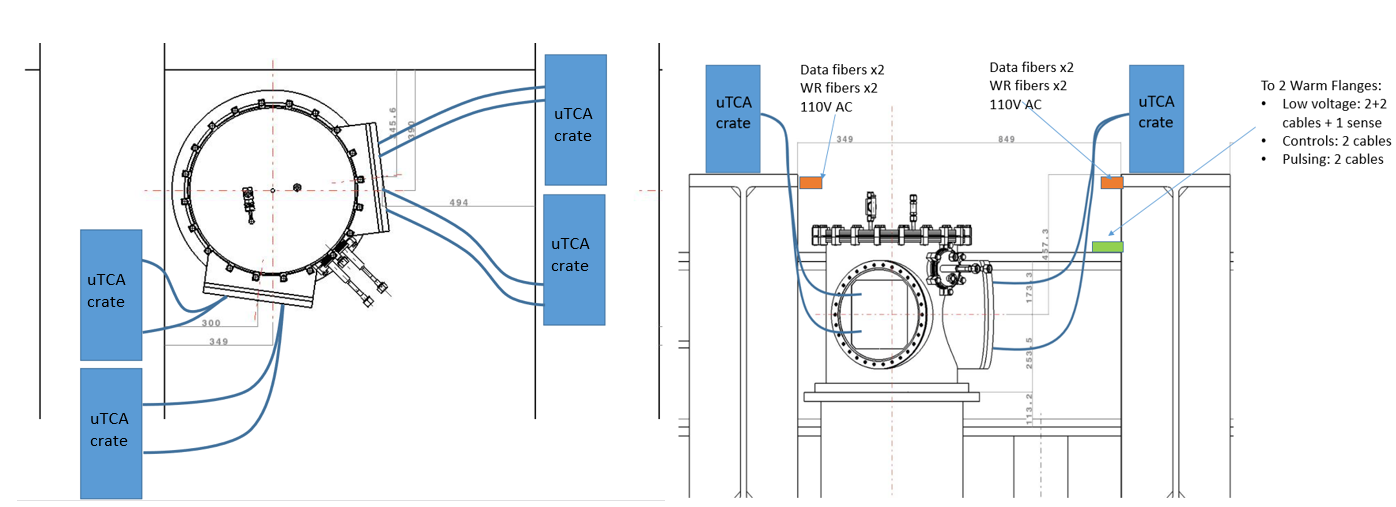}
\end{dunefigure}

The infrastructure needed to support the \dshort{tde} installation and operation on the cryostat roof includes the following items:

\begin{itemize}
\setlength\itemsep{1em}
\item	 110\,V AC power network to supply the power to the \dshort{utca} crates. Each TDE crate (hosting 12 AMC card, the MCH, the WR-MCH and the power and the cooling units) has a typical power consumption %typically lower than 
< 500\,W.

\item $N_2$ network to purge and fill the chimneys before the cryostat \cooldown and for accessing a blade. 
The network is made of plastic pipes with the $N_2$ at about 10\,mbar overpressure. %Large operation 
Experience comes from NP02 and the \coldbox tests. After the initial chimney purging (typically equivalent to 10 volumes) the $N_2$ circulation is closed, until an % new 
access is needed in cryogenic conditions. In that case the $N_2$ flow must be started before opening the chimney cap %in order 
to %avoid 
keep humidity %to enter in 
from entering the chimney. %After the access and the closure of 
Once the chimney is closed again, purging should continue for ten equivalent volumes %. The internal volume of a chimney is 
(10 $\times$ 0.46 m$^3$). The $N_2$ flow should %allow 
complete purging in a few hours.  Each chimney has a plastic input pipe and an output valve to directly vent  the $N_2$ into the environment %air 
during purging.

\item The \dshort{wr} network for the timing distribution. The WR-MCH in each \dshort{utca} crate is connected with a single fiber to a Level 1 (L1)  WR switch. Given the ports' topology on the switches, 19 L1 \dshort{wr} switches are needed to connect the 320 crates. These switches will be %hosted 
installed in the detector mezzanine racks and should be evenly distributed along the 60\,m length of the detector. The 19 uplinks of the L1 switches are then aggregated in a patch panel located %at half of the 
in the middle of the mezzanine (by length) 
%in a 
which is connected to a 24-fiber trunk cable length
going to the DAQ room (see Figure~\ref{fig:wr_ancillary}). In the DAQ room the 19 links are connected to two cascaded Level 2 (L2) switches, the first one  of which acts as \dword{wrgm} 
%grandmaster %switch and it 
and is connected with its 1\,PPS and 10\,MHz inputs to the DUNE timing system. A \dshort{wr} time-tagging unit in the DAQ room is %also 
connected to the second switch and it ensures signal time-tagging from the DAQ. Each \dshort{wr} switch requires about 200\,W max AC power and occupies 1U.

\item The low-voltage generation and distribution system for the TDE \dshort{fe} includes twenty units, each capable of serving four larger chimneys %of 48 cards 
(192 \dshort{fe} cards in total). Given  that the TDE readout system includes 3840 \dshort{fe} cards distributed in 24- and 48-card chimneys, a total of 20 low-voltage units is needed. These should also be evenly distributed %in 
along the mezzanine racks along %the 60m of 
the detector length (see Figure~\ref{fig:lv_ancillary}). A %48 cards 
large chimney, for instance, will have its two warm flanges connected with four nine-wire shielded cables to the low-voltage distribution unit, where a single cable  serves a group of 12 \dshort{fe} cards. A %24 cards 
small chimney will have its warm flange connected via only two low-voltage cables. A low voltage unit is composed of a power supply (Wiener crate MPOD Micro 2 LX 800 W with two modules MPV 4008I) and a low-voltage filtering and distribution box (see Figure~\ref{fig:lv_boxes}). An additional nine-wire low-voltage shielded cable can be connected to a warm flange to the low-voltage unit for voltage sensing. Each %48 cards 
small chimney warm flange has two sense connectors corresponding to the two independent low-voltage cables, but only one sense connector is connected via a low-voltage cable to the distribution box, with the other one kept in case verifications are needed. Each low-voltage unit requires max 1\,kW AC power, and occupies 3U$+$4U.

\item Each chimney warm flange has one sub-D 9 connector for the controls and one SMA connector for charge injection. The controls connector will be connected (with the same kind of cables used for the low voltages), together with a coaxial cable for pulsing, to a calibration unit. The calibration units will also be %hosted 
installed in the mezzanine racks. A calibration unit could serve 16 flanges. %11 
Eleven calibration units will be then needed %in order 
to connect the 105 chimneys. These calibration units should also be distributed evenly along %in 
the mezzanine racks % evenly along the 60m length of the detector 
(see Figure~\ref{fig:calibration_ancillary}). They will require AC power %smaller than 
< 100\,W per unit and occupy 2U.

\item Each \dshort{utca}  crate is connected %with 
via an optical fiber data patch cable to a patch panel on the cryostat roof. The patch cable aggregates eight optical fibers to establish a 40\,Gbit/s connection between the MCH and the DAQ system. 
The \dshort{daq} consortium is responsible for defining the positions of the patch panels on the cryostat roof and it is responsible of providing these items and the corresponding trunk cables going to the DAQ room.
\end{itemize}

\begin{dunefigure}
[Layout of the \dshort{wr} switch positions on the mezzanine]
{fig:wr_ancillary}
{Layout of the \dshort{wr} switch positions (blue boxes) on the cryostat roof with the uplinks connected to the patch panel at the mezzanine center (orange box).}
	\includegraphics[width=.9\textwidth]{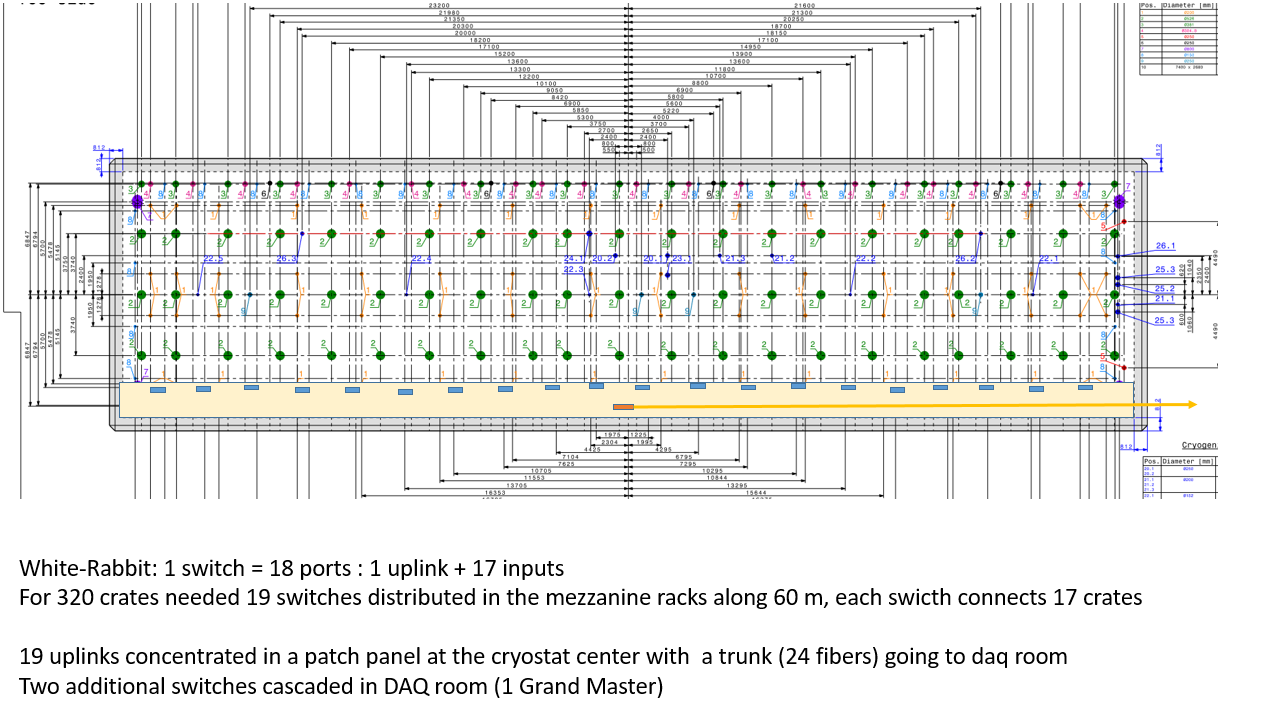}
\end{dunefigure}

\begin{dunefigure}
[Layout of the LV power supplies and distribution boxes on mezzanine]
{fig:lv_ancillary}
{Layout of the low-voltage power supplies and distribution boxes positions on the cryostat roof mezzanine  (blue boxes).}
	\includegraphics[width=\textwidth]{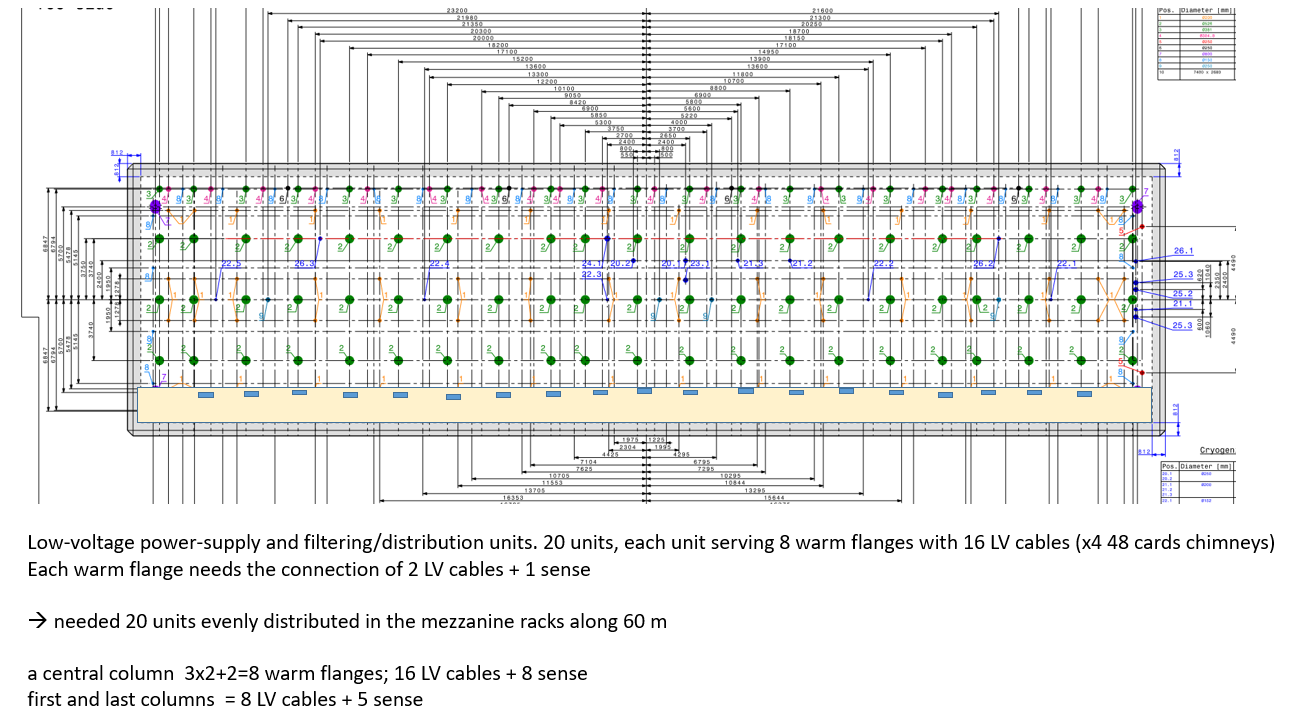}
\end{dunefigure}

\begin{dunefigure}
[Rack mountable LV power supply and distribution box] % to be installed on the cryostat mezzanine]
{fig:lv_boxes}
{Rack mountable low-voltage power supply and distribution box to be installed on the cryostat mezzanine.}
	\includegraphics[width=.9\textwidth]{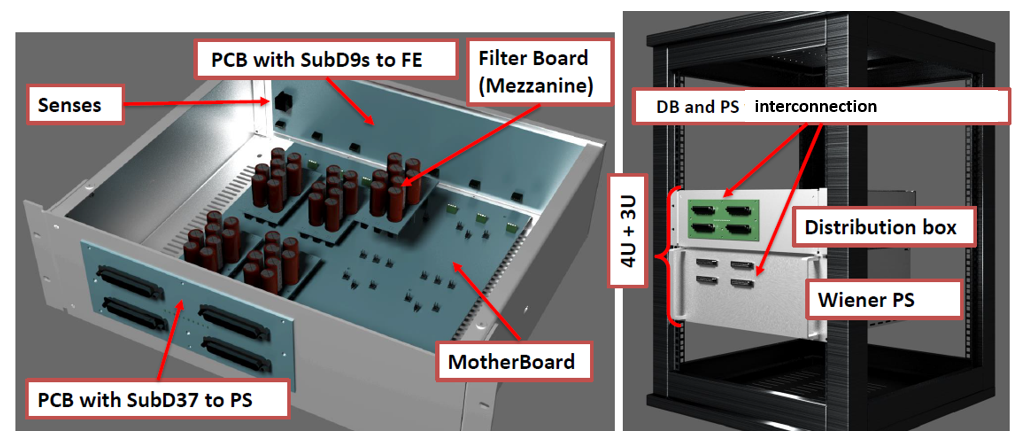}
\end{dunefigure}

\begin{dunefigure}
[Layout of the calibration box positions on mezzanine]
{fig:calibration_ancillary}
{Layout of the calibration box positions on the cryostat roof mezzanine  (blue boxes).}
	\includegraphics[width=.9\textwidth]{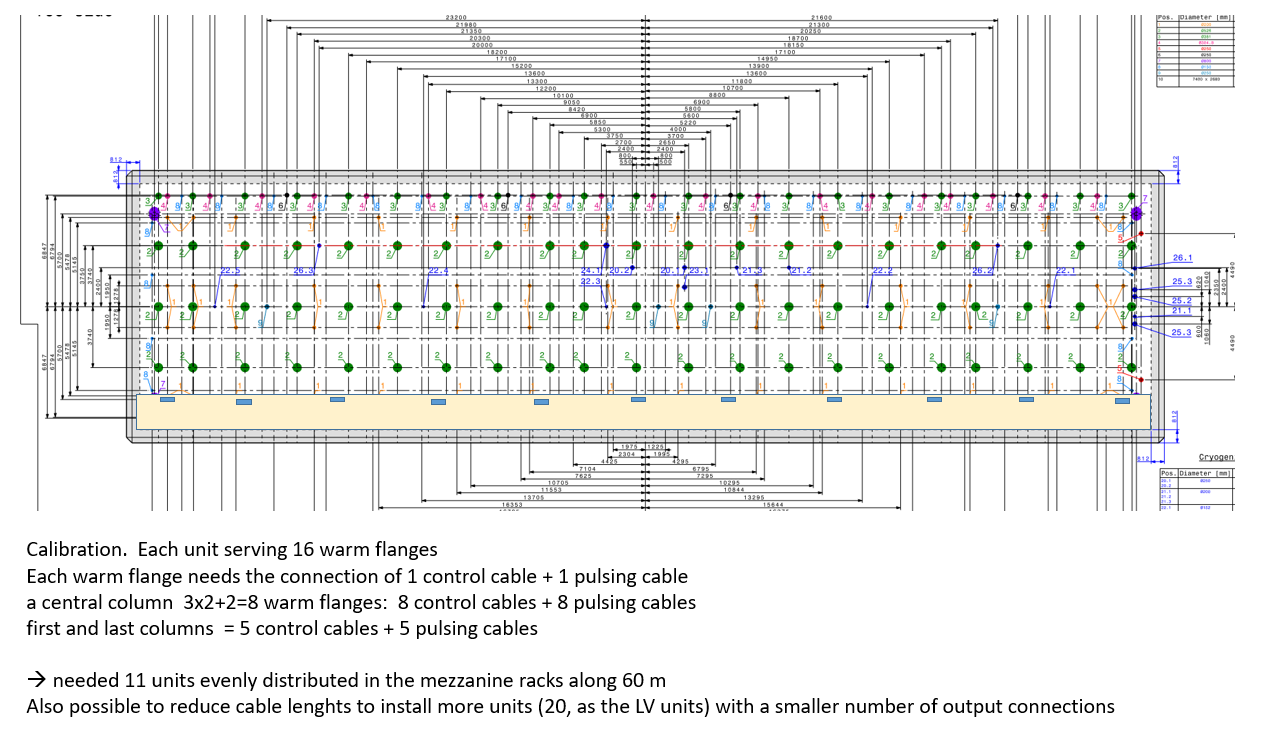}
\end{dunefigure}

 %%%%%%%%%%%%%%%%%%
\subsection{Grounding}
\label{subsubsec:topelec:grounding}

An overview of the TDE grounding plan, which relies exclusively on the cryostat ground, %aspects 
is available in~\cite{edms2737071}. 

The \dshort{tde} readout chain has the front-end analog electronics (\dshort{fe} cards) completely enclosed in a Faraday cage, the %signal chimney
\dword{sftchimney}. Each %signal 
chimney, which is at the cryostat ground potential,  is made of stainless steel and %it 
is in tight mechanical contact with the flanges of the cryostat penetrations.

%The c
Contact between the cold flange %printed circuit board 
\dword{pcb} and the mechanical ring of the flange is required %in order ensure the 
for cryostat tightness and for electrical contact to ground, which  
is effected by the pressure produced by multiple bolts used to close the flange ring. 

The chimneys/cryostat are %then 
the grounding reference for the entire TDE readout chain. 

The %good level of 
secure grounding at the flanges (for which the design of the warm flange \dword{pcb} is an important aspect) and the exclusive use of shielded cables for all connections external to the chimneys together 
prevent any connections from injecting noise. Another advantage of the \dshort{tde} design is that there are no components of the readout chain inside the cryostat, so it is not possible to induce noise on any other systems.

Both the cold and warm flange \dwords{pcb} in the chimneys have a ground plane %which goes 
that is in tight contact with the flange mechanics (and hence the cryostat ground). The \dshort{fe} card grounds are connected to these ground planes, and therefore %By design of the chimney and of its PCB flanges, 
all possible connections are well grounded at the %level of the 
flanges. 

On the warm flange the connectors for the VHDCI connectors, the SMA connector for pulsing and the sub-D connectors for the low-voltage supplies and the control signals are %well referred to the
connected to the flange/cryostat ground. Only shielded cables are used: 
VHDCI cables to send the differential analog signals to the \dwords{amc}, shielded multi-wire cables for the connections of the low voltages and the controls, and coaxial cable for the charge injection pulsing. The shields of all cables are grounded at the cryostat warm flange.

The chimneys receive the signals %collected by the perforated anodes via 
from the adapter boards mounted on the \dshort{crp}s %by 
via flat cables connected at %the bottom of the 
cold flanges at the bottom of the chimneys.  

These flat cables connect the adapter board to the ground at the cold flanges. 
Each flat cable has 68 wires, 32 of which are used to read the 32 channels from the adapter board. The remaining 36 wires (34 ground wires $+$ two unused channels) are used for grounding.

To avoid grounding loops over the \dword{crp} area, the adapter board ground planes are not interconnected among different boards. Each board is only connected via its flat cables to a chimney.
The anode adapter boards  just ensure the routing of the strips signals to the flat cable connectors for the connection to the cold flange and the HV polarization of the strips.

 The polarization voltages are applied to the anode strips via a network of biasing and decoupling passive components on the adapter boards. These are connected to a HV cable going to a filter and distribution box on the \dshort{crp} frame. There are two boxes on each \dshort{crp} daisy-chained and, connected via a single shielded coaxial cable to a dedicated flange on the cryostat. The cable shield is grounded only on its box end; it is not connected to the adapter board ground plane. Several adapter boards are daisy-chained for the distribution of a HV bias voltage so that only the first one of the chain is connected to the filtering board and a single HV cable is used. However the adapter boards are only daisy chained for the HV potentials but not for the ground, %for the reasons explained before. 
 as explained. The \dshort{crp} biasing is in the scope of the CRP consortium. 

The \dshort{amc}s that digitize the differential signals brought by the VHDCI cables from the warm flange are referred via the shield of the VHDCI cables to the cryostat ground and not to the \dshort{utca} crate chassis ground. The analog stage in the \dshort{amc}s is AC coupled to the twisted pairs present in the VHDCI cables. Each \dshort{utca} crate is connected to the power network via an isolation transformer.

The low-voltage distribution system  includes a power supply and a filtering and distribution box that suppresses the power supply ripple. The power supply (Wiener crate MPOD Micro 2 LX 800 W with two modules MPV 4008I) is connected to an isolation transformer.  The grounding of the low-voltage distribution system to the cryostat ground is ensured by the shielded cables going to the warm flanges and the ground connectors (three ground wires per dub-D 9 connector, the other wires bring the five supply voltages, one voltage VCC, which is associated to higher currents, uses two wires). The low-voltage cables are routed on cable trays on the cryostat surface on top of copper foils connected to the cryostat ground.

The grounding scheme in  Figure~\ref{fig:tde_grounding} summarizes the description of the grounding aspects in the TDE readout chain presented above.

\begin{dunefigure}
[TDE grounding scheme]
{fig:tde_grounding}
{TDE grounding scheme.}
	\includegraphics[width=.9\textwidth]{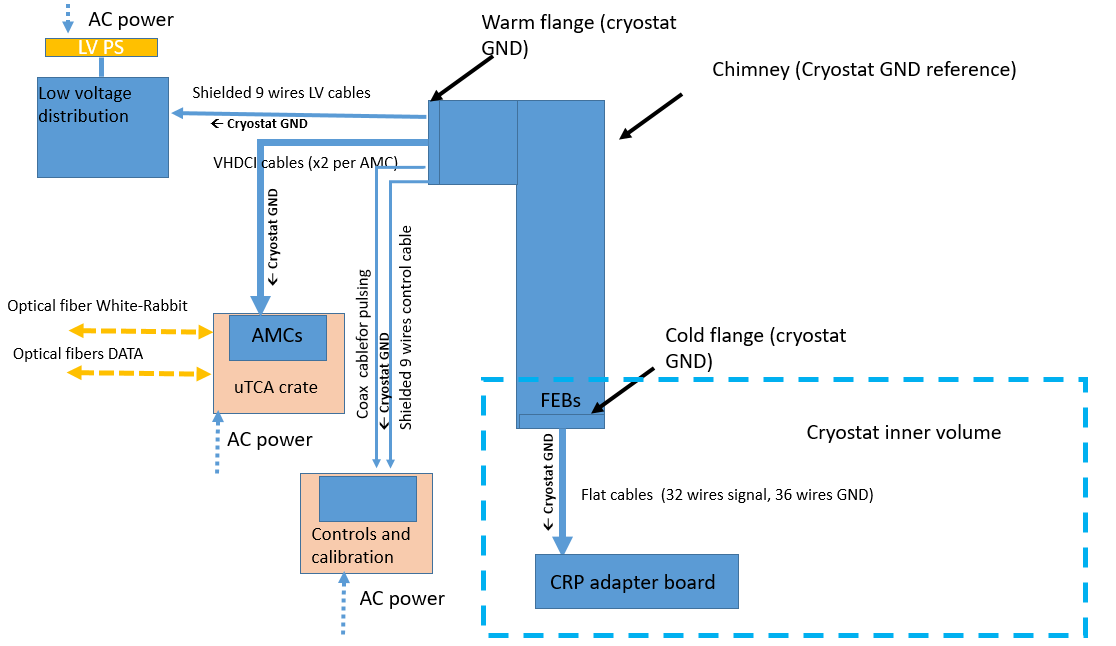}
\end{dunefigure}

The accessibility of the cryogenic electronics allows disconnecting the \dshort{fe} cards from the flat cables to check if any coherent noise has infiltrated the analog chain from the external connections due to grounding issues. This  also provides a very good handle for checking the grounding of the entire chain.

This operation was performed several times in \dword{pddp} and in the \coldbox. The results have always been consistent with the measurement of the intrinsic noise of the amplifiers; i.e., absence of any form of coherent noise coming from the readout system itself. These checks have %been useful to 
shown that coherent noise, observed before  disconnection of the \dshort{fe} cards,  was either originating from grounding issues on the \dshort{crp} or %it was  
entering the cryostat from another path and being picked-up by the \dshort{crp} anode strips.

%%%%%%%%%%%%%%%%%%                
\subsection{Production}
\label{subsubsec:topelec:prod:assembly}

The \dword{tde} readout chain  includes %different items concerning 
components for the analog and digital front-end stages as well as for the timing distribution system and the \dwords{sftchimney}.  
The main components are listed in Table~\ref{tab:prod_tde}, and shown in Figure~\ref{fig:dp_cro}.

\begin{dunetable}
[%Items 
Components to be produced for the \dshort{tde}]
{lr} 
{tab:prod_tde}
{%Items 
Components to be produced for the top drift charge readout electronics} %: unit counts.}
\textbf{Item} 
& \textbf{Unit Counts} %Quantity} 
\\ \toprowrule
			\dshort{larzic} \dshort{asic} (16 ch.) & 15360\\
			\colhline
			Cryogenic \dshort{fe} cards (64 ch.) & 3840 \\
			\colhline
			24-card (smaller) chimneys   & 42\\
			\colhline
			48-card (larger) chimneys & 63\\
			\colhline
			\dshort{amc} cards  &  3840\\
			\colhline
			\dshort{utca} \dshort{wr} MCH & 320\\
			\colhline
			\dshort{utca} crates (including MCH, PU and CU)  &  320\\
			\colhline
			40\,Gbe Ethernet optical links to back end  &  320\\
\end{dunetable}

The main elements of the readout system are the \dword{fe} cards and the digitization \dword{amc} cards, both of which require %, which need both to be installed in 
3840 units. The production yield, given past experience, is close to 100\%;  %given 
some rare welding defects, which once identified, 
have always been fixed. %after their identification.  

The analog \dshort{fe} cards are located in the chimneys and remain accessible throughout the lifetime of the experiment. The rest of the system components (\dshort{utca} crates with \dshort{amc}) are located in the ambient temperature of the detector cavern at the cryostat roof or in the vicinity and thus remain accessible. Working with these components at CERN over the course of several years, no replacements have been needed under normal conditions.
However, in order to ensure compliance with the requirement of maintaining fully functioning channels over the long term, a sufficient stock of spare components will be be maintained on site.

It is therefore planned to order spares in appropriate quantities, according to the characteristics of the system and possible exposure to risks. The number of spares will range from 5\% (e.g., for the  \dshort{fe} cards) and 3\% (for the \dshort{amc}s),  to at least 20\% for the \dword{larzic} \dword{asic}. %are foreseen to be produced with 
Other %infrastructural elements of the system as 
components, e.g., the chimneys/PS or timing system elements, will have 2\%  spares.

Within the time requirements dictated by the \dshort{spvd} installation schedule, the \dshort{tde} production organization aims  to minimize production risks by conducting QC tests as each production batch is delivered.

The production is organized as follows:
\begin{itemize}
\item The production of the main components: \dshort{fe} cryogenic cards, \dshort{amc} cards, \dshort{utca} crates, cables, and chimneys can take place in parallel. %completely parallelized

\item 
The minimal production and testing for all the components is of the order of one year, and could be done in parallel, starting a year before the items are needed for installation in South Dakota. However, given that \dshort{spvd} installation is set to start in 2027, a three-year production time will be implemented, allowing staggered production throughout the years 2023 through 2026.

This arrangement will smooth the spending profile over three years, and provide better flexibility and risk mitigation with respect to production delays, production non-conformities and manpower involved in the QC.

\item  
Deliveries will be planned in monthly batches, starting one to two months after the beginning of the production. The \dshort{qc} activities will be performed upon delivery of each batch allowing for earlier detection and correction of any problems and for regular follow-up with the production team. It will also limit any delay to a single batch or a fraction thereof, and therefore to a single month, which can be easily re-absorbed on the production schedule.  

\end{itemize}

Relationships with vendors, component design, other aspects of production, and the \dword{qa}/\dshort{qc} procedures have been honed since 2016 via the development of the \dword{wa105},  \dword{pddp},  the vertical drift  \coldbox tests, and finally  the \dword{vdmod0}, procuring %for 
a total of about 10k channels over this time.  Some components, e.g., the \dshort{fe} cards and the \dshort{amc}s have been %produced by
procured from
different companies at different stages.

The milestones in Table~\ref{tab:tope-prod} include periods for production and \dshort{qc} from 2023 to 2026. The ``ready for installation'' milestone is set three months before actual installation.

 \begin{dunetable}
[\dshort{tde} production timetable]
{p{0.35\textwidth}p{0.30\textwidth}}
{tab:tope-prod}
{Top drift electronics production timetable}
Item  &  Dates \\
\toprowrule
\dshort{amc} digitization cards &  June 2024 -- July 2025   \\ \colhline
Cryogenic \dshort{asic}s and \dshort{fe} cards & March 2024 -- Dec 2025   \\ \colhline
Timing end nodes &  Jul 2024 -- July 2025  \\ \colhline
\dshort{utca} crates &  Jul 2024 -- July 2025 \\ \colhline
\dshort{sftchimney}s &  October 2024 -- April 2026 \\ \colhline
Cabling &  February 2025 -- July 2026 \\ \colhline
Ready for installation &  September 2026 \\
\end{dunetable}

%%%%%%%%%%%%%%%%%%  

\subsection{Quality Assurance and Control}
\label{subsubsec:topelec:qaqc}

It is crucial that the \dshort{spvd} \dword{cro} electronics chain be fully understood and completely debugged. 
The \dshort{qa} plan must ensure that the number of functioning channels %in the system delivered and installed 
satisfies the DUNE requirement (<1\% of non-functioning channels) for the lifetime of the experiment. 
Testing policies were %accurately 
defined and optimized during the previous \dshort{tde} production phases. 

\dshort{qc} procedures applied at the production companies (e.g., optical and electrical tests) are carefully defined in the contracts. However, experience has shown that any change in the production lines, personnel, or execution of the \dshort{qc} procedures can lead to problems during  production, and that additional \dshort{qc} procedures are required upon receipt of components.

The \dshort{tde} \dshort{qc} %policies,
procedures, in place since 2014, %described below, 
are very %strict and imply 
thorough and involve full operation tests of the %different 
components. %In all cases, the  procedures 
They have reliably detected problems;  for example, systematic issues in production of electronics cards have been prevented via early validation of prototype cards prior to mass production. The immediate identification and correction of production anomalies has resulted in the attainment of % reaching the level of 
zero defects at installation, and the maintenance of %. They then  resulted in achieving and maintaining 
nearly zero malfunctioning channels %reliably 
over several years of operation at \dshort{cern}.

All components in the readout chain undergo full-functionality and operation tests at the single channel level, completely emulating the use of the components in the final detector layout.  For %other 
connection components, such as the flanges, an individual channel-by-channel continuity test is also %separately 
required.

The \dshort{tde} \dshort{qc} tests are %based 
done on a test bench setup that emulates the full detector readout chain. The setup, while quite sophisticated, is minimally configured, requiring only a single readout \dshort{utca} crate. It can therefore be replicated in several %centers 
locations at moderate cost. 
A new test bench, shown in Figure~\ref{fig:dup_tb}, was constructed in 2022 in a \dshort{qc} center at LP2I Bordeaux. It is a replica of the test bench used for several years at IP2I Lyon. 
%Another important point is that t
These analog readout tests, which are conducted using a calibration system with an external pulsing card (Section~\ref{subsubsec: AREUsss}),  
enable very %good 
high accuracy (at the $\sim$1\% level).

In the absence of anomalies, the very high uniformity response of the electronics is also at the 1\% level. This testing procedure enables immediate detection of small problems related to bad connections or to malfunctions at the %level of the 
\dshort{fe} cards %and 
or successive elements %of 
in the analog chain.  

\begin{dunefigure}
[Duplicated test bench system at a second QC center ]
{fig:dup_tb}
{Duplicated test bench system at a second QC center}
	\includegraphics[width=.7\textwidth]{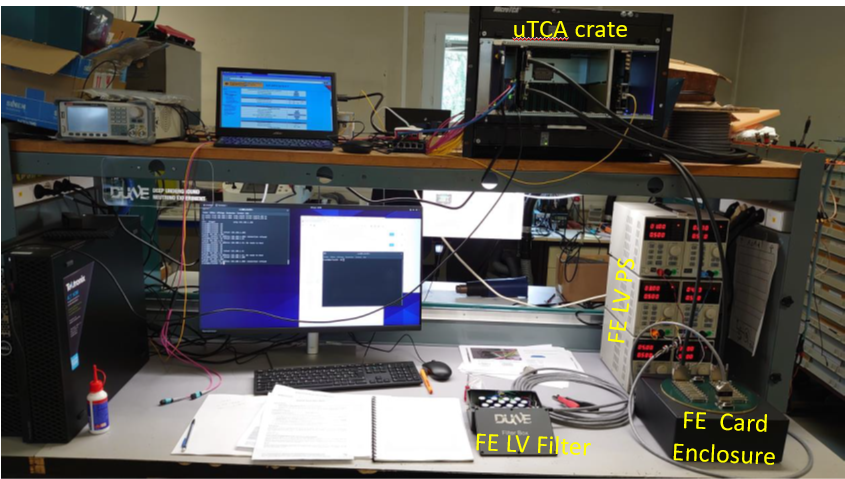}
\end{dunefigure}

A new version of the calibration charge-injection board with PCB-embedded capacitors, shown in Figure~\ref{fig:new_injection}, was successfully designed and tested in 2022. It is more compact and easier to produce, and will enable multiple \dshort{qc} tests to be done in parallel. % to parallelize the \dword{qc}, a

\begin{dunefigure}
[New version of the calibration card]
{fig:new_injection}
{New version of the calibration card, with embedded PCB capacitors, used in the QC tests.}
	\includegraphics[width=.9\textwidth]{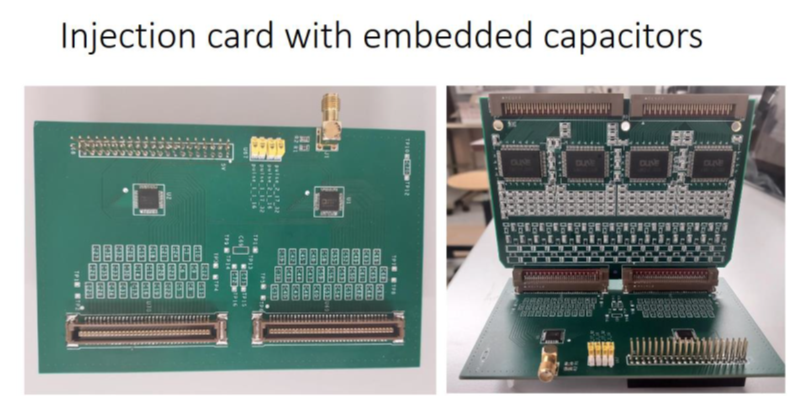}
\end{dunefigure}

The \dshort{qc} tests will be pipelined with production in order to quickly identify and correct any potential systematic defects in the manufacturing. To the extent possible, the production will be organized %as close as possible 
in monthly batches so as to both mitigate production risks and ensure continuous checks on %real-time follow-up of 
the production process.

The post-delivery \dshort{qc} procedures will be performed at test sites. The test reports for each element will be logged in a common database, following the DUNE hardware database requirements. %In order to %easy 
To facilitate the execution of the tests and any follow-up, %of the different components 
each component will be 
labeled with its unique identifier in the %hardware 
database. The test benches %chains 
will be equipped with bar-code readers interfaced to the \dshort{qc} software for 
%\fixme{automatic?} 
the immediate identification of components and transmission of  test %outcomes 
results to the database.

The dedicated test-bench setup is schematically illustrated in Figure~\ref{fig:tde_test_bench}.   
The setup %represents a miniaturized 
is a miniature version of the \dshort{spvd} \dshort{tde} system, which will enable full functionality tests of the various elements of the both the analog and digital chains.

\begin{dunefigure}
[Production QC test bench scheme]
{fig:tde_test_bench}
{Production \dshort{qc} test bench scheme.}
	\includegraphics[width=.8\textwidth]{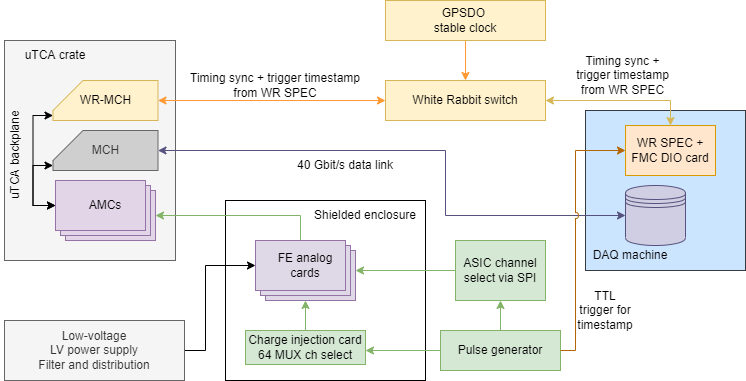}
\end{dunefigure}

It includes:
\begin{itemize}
\item The hardware for \dword{wr} trigger timing and synchronization;
\item The low-voltage system to power the \dword{fe} analog cards and the voltage/current monitoring;
\item One computer for \dword{daq} and online analysis;
\item \dword{fe} analog cards (to be tested or used as support to test other elements of the chain) located in a shielded enclosure with a warm flange feedthrough;
\item A \dword{utca} crate with \dwords{amc} (to be tested or used used as support to test other elements of the chain); and
\item Systems to inject calibration charge either via external injection capacitors or internal calibration circuitry on the pre-amplifier \dword{asic}.
\end{itemize}

In triggered readout mode, the trigger time stamp information is used by the \dshort{amc} to output the data sequence that matches the trigger timing, which, for %example in the case of 
the QC tests, is related to the timing of the charge injection pulse distributed to the analog front-end cards.

The construction schedule, which is driven by the availability of the detector for installation rather than by the production and testing time of the electronics components, 
very conservatively assumes the following:

\begin{itemize}
\item The testing time %assumed 
is typically shorter than the production time.
\item The span of the production and testing activities over several years %with 
allows the possibility of reabsorbing delays.
\item 
The time needed to execute the tests using the two test benches is %of the order of 
a few days per month. % including two testing chains
\item  
It would be easy to expand the test-bench including multiple \dshort{utca} crates operating in parallel; for instance going from one crate to two or four crates, depending on needs. 

\item Buffer time is available at the end of the production runs before installation.
\end{itemize}

The \dshort{fe} cards have been and are currently comfortably tested at a rate of at least five per hour per test bench, %. This rate has been very comfortably achieved by 
testing a single card at a time, % in the test-bench setup. This would mean testing 
or about 40 \dshort{fe} cards per day. 
It would be easy to test up to 10 cards %tested 
in an automated way at the same time on a test bench, i.e., testing %.   This implies the possibility of testing 50 cards per hour or 
up to 400 cards %in a single day.
a day.

By taking a conservative margin this shows that the 400 cards of a monthly batch could be tested over two days relying on a single test site equipped with a single test-bench. This time shrinks to a single day by including two sites (baseline assumption) or by operating two test benches in parallel.

\dshort{amc} boards are tested in batches of 12 on a single \dshort{utca} crate bench. 
Testing a batch, including various insertion manipulations, takes at most one hour, enabling the testing of up to 96 cards per day on a single test bench, and the entire monthly production batch of 320 in %3.3 (seems too precise!)
less than 3.5 days (or half that if two test setups are used).   This time can be shortened further by running multiple crate benches in parallel connected to the same \dshort{daq} system.

The tests of the \dshort{utca} crates and of the \dshort{wr} end nodes can be performed at the same time (and with complete \dshort{daq} operation) since these two elements are coupled. Production of \dshort{wr} end-nodes is faster than that of crates, but testing activities can be combined in a single test. 

The \dshort{wr}  end-node components are pretested and there have never been problems.  On the basis of this coupled test scheme, one hour per site should be sufficient to test a new \dshort{utca}-\dshort{wr} system.  A monthly batch of 46 units can therefore be tested in three days assuming two sites.  Again, this time can be shortened by running more than one crate in parallel. % in a single site.
Figure~\ref{fig:testing_time} summarizes these \dshort{qc} activities.

\begin{dunefigure}
[Summary of the QC tests duration]
{fig:testing_time}
{Summary of the QC tests duration for monthly batches of components.}
	\includegraphics[width=.98\textwidth]{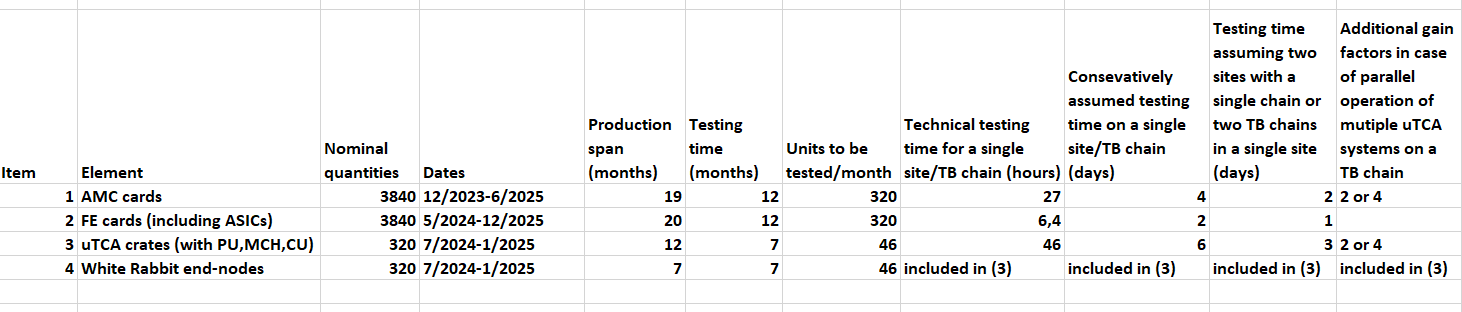}
\end{dunefigure}

The chimneys will be tested at warm for vacuum tightness at the level of a few $10^-3$ mbar. The chimneys should be leak-tight down to a few $10^-9$\,mbar l/s. 

A randomly selected sub-sample of chimneys will  
be tested at cryogenic temperatures. After the target temperature is reached at the cold flange ($\sim$100K) a new round of helium leak tests is performed to ascertain that it is still leak-tight to $10^-9$\,mbar l/s. 

A full set of blades will be installed to check the behavior of the guiding system at cryogenic temperature. The continuity of the signal lines for warm and cold flanges will be  verified for each channel. This procedure can be automated with 64:1 MUX and 1:64 DEMUX stages connected at the input and the output of a given signal chain (e.g., warm flange inner side / outer side). By enabling a single input channel at a time, a sweep over the output channels will verify that only the corresponding output channel is active. Presence of signals on any other channel would imply a resistive contact that would need to be fixed.

%%%%%%%%%%%%%%%%%%                
\subsection{Installation, Integration, and Commissioning}
\label{subsubsec:topelec:install:integration}

%Start here modified text:
The installation, integration and commissioning procedures for the \dshort{tde} were repeated many times during the development and operation of \dshort{pddp} and the \coldbox tests in 2021-2022, demonstrating the absence of significant risks, for both material and personnel. 

The installation in \dshort{spvd} has been studied together with the \dword{fsii} team, and besides the installation of detector components, the related documentation includes cabling layout, power requirement, and schedule. 

The predecessors to the \dshort{tde} installation work are:
\begin{itemize}
\item Cryostat roof completed (walking surface available);
\item Cryostat detector mezzanines in place;
\item Cryostat penetrations welded and leak tested;
\item Mezzanine racks, power to racks, and safety system in place;
\item Power and fibers for the \dshort{utca} crates in cable trays; and
\item \dshort{daq} ready to readout the \dshort{utca} crates (i.e.,\dshort{tde} commissioning, which may %also exploit 
be accomplished using a portable system).
\end{itemize}

The TDE installation is then subdivided into three phases:
\begin{itemize}
    \item Mezzanine Racks: Installation of all TDE electronics ancillary modules in the racks on the mezzanine (low-voltage generation and power distribution systems, level 1 switches. Calibration and control units). Cabling. Powering and testing of all these units.

    \item Chimneys: Insertion of all 105 chimneys in the cryostat penetrations. Connection to the penetration flanges interfacing to the chimneys followed by leak testing. Positioning of the mechanical stands/mounts of the \dshort{utca} crates. Connections to the N$_2$ distribution network for chimney purging.

    \item TDE electronic components: installation of the blades with the FE cards in chimneys. Positioning of the \dshort{utca} crates on their stands. Cabling of the \dshort{utca} crates to the warm flanges. Commissioning.
\end{itemize}

The installation and integration of TDE is to a large extent independent of CRP installation activities, although proper coordination is needed. To avoid possible damage to the electronics related to the CRPs and interfering with cabling activities inside the cryostat, the blades in already-equipped chimneys can be lifted from the cold flanges and connected back once the CRP cabling is completed. However, complete and activated chimneys are required in order to test CRP cabling. Purging and filling of the chimneys with \lntwo is not required during commissioning phase, but must be done before the cryostat \cooldown.

%%%%%%%%%%%%%%%%%%                
\subsection{Interfaces}
\label{subsubsec:topelec:interfaces}

The \dshort{tde} system has important interfaces with the \dword{crp} and \dword{daq}  consortia and with \dword{fsii}, which are defined and documented as cited: 
\begin{itemize}
\item \dshort{crp}-\dshort{tde}~\cite{tdeinterfacecrp}
\item \dshort{daq}-\dshort{tde}~\cite{daqinterfaceTDE} 
\item  Installation-\dshort{tde}~\cite{tdeinterfaceinstall}
\end{itemize}
              
Each document describes the interfaces between two consortia to complete the design, fabrication and installation of the far-detector.

The interface with the \dshort{crp}s  %similarly to the \dword{dp} design, 
is defined at %the level of 
the chimneys' cold flanges. The cold flanges are part of the \dshort{tde} system and everything below the flanges, including the flat cables used for the \dshort{crp} connection 
is the responsibility of the \dshort{crp} consortium.

The flat cable connectors on the cold flange and the matching cables are well defined. The flat cables have a length of 2.2\,m apart some exceptions in particular chimney locations. A circular cable tray designed and produced by the \dshort{crp} consortium for the routing of the flat cables is attached to the bottom of each chimney. The \dshort{tde} consortium provides the fixation points on the cold flange for it. 

At the production level: the \dshort{crp} consortium fabricates the anode adapter boards and provides the flat cables; the \dshort{tde} consortium provides the chimneys and cold flanges. 

At the installation level: the chimneys and cold front-end electronics are installed by the \dshort{tde} consortium; the flat cable connection to the cold flange and the circular cable trays installation is performed by the \dshort{crp} consortium.

The \dshort{tde} data readout happens via 320 fiber connections (40\,Gbit/s) between the \dshort{mch} on \dshort{tde} side and the back-end on the \dshort{daq} side, involving the following: % aspects:

\begin{itemize}

\item  The \dshort{tde} system includes 320 \dword{utca} crates, with 12 \dword{amc} cards/crate and 64 channels per \dshort{amc}.

\item  In each crate data are collected by an \dword{mch}, having a 120\,Gb/s  maximal connectivity, when exploiting three ports operating at 40\,Gb/s each  (this total 120\,Gb/s  bandwidth corresponds to 10\,Gbit/s connectivity at the level of a single \dshort{amc}, supported for all the 12 cards in the crate).

\item The \dshort{mch} has a MPT-CXP 24 connector, supporting up to three 40\,Gb/s ports.

\item Assuming a sampling rate 2\,MHz, 12 (16) bits, the  raw data flow is 17 (23)\,Gb/s  (with 12 \dshort{amc} cards operating in continuous streaming mode). 

\item For the \dshort{tde} application only one port of the \dshort{mch} is connected, ensuring  a 40\,Gbit/s connectivity, with a factor of two safety margin on the needed bandwidth.

\item Since just a single \dshort{mch} port is used, customized MTP patch cables have been designed and produced. These accommodate a group of eight fibers corresponding to the selected port~\cite{edms-2737105}. 

\item The other optical cable patch end is terminated with a standard MTP-12F connector with only two groups of four fibers connected (standard 40\,Gbit/s connectivity). It can be plugged into a QSFP+ transceiver in any commercial network card.

\item The 40\,Gbit/s system and these patch cables have already been tested and validated  during \coldbox data taking in 2021. Cables needed for \dword{vdmod0} have been procured.  

\end{itemize}

The \dshort{tde} consortium has responsibility for the data link aggregation into the optical cables for \dshort{mch} connection and for the procurement of customized optical patch cables. 

The \dshort{daq} consortium is responsible for the connection to the \dshort{daq} room via patch panels distributed on the cryostat roof close to the crates and with connected trunk cables going to the \dshort{daq} room. 
Optimization criteria will drive the definition of the patch panels' position.

The \dshort{tde} readout is based on UDP  packet in JUMBO frames. This readout was already operated with \dword{wa105}, \dword{pddp}, and the \dword{vd} \dshort{crp} \coldbox tests in external trigger mode. A firmware to work in continuous data streaming mode has been developed. 

The data format from \dshort{tde} to  \dshort{daq} is based on an evolution of \dshort{pddp} data format, agreed with the \dshort{daq} consortium and detailed in~\cite{EDMS2737054}. %\fixme{reference not accessible}
The data packets structure includes: a header followed by the associated \dword{adc} samples. The header includes information on the channel identification, \dword{tai} timestamp, global \dshort{daq} timestamp, error flags, and so on.

The timing distribution of the \dshort{tde} is using \dword{wr}  components. There is a dedicated  \dshort{wr} end node in each of the 320 \dshort{utca}, connected to network including 19 \dshort{wr} switches. The \dword{wrgm} is integrated in the DUNE central timing system which provides the 1\,PPS and 10\,MHz signals. 

The \dshort{tde} and \dshort{daq} consortia are jointly responsible of the integration of the  \dshort{wrgm} switch with the DUNE timing system. A DUNE timestamp counter will be  maintained by the \dshort{amc}s as well as the TAI one. This counter is reset/aligned to the timestamp of the \dword{daqdts} signal.  

The \dshort{tde}-installation interface document includes several aspects which are also reported in the \dshort{tde} installation section of this chapter. 

Additional technical and engineering details (installation of dedicated cable trays close to the chimneys, racks allocation on mezzanine and cables routing, the \dshort{daq} fiber patch panels on the cryostat roof, the path of the network for the $N_2$ distribution to the chimneys ) are in progress. % to be finalized. 

%%%%%%%%%%%%%%%%%%                
\subsection{Safety}
\label{subsubsec:topelec:safety}

This section describes the basic requirement for materials and personnel safety during the \dword{tde} installation activities. All the work will be performed entirely on the cryostat roof.

For what concerns material safety, the \dshort{tde} \dword{fe} cards are very robust and can be manipulated bare hands without \dword{esd} issues. They are resistant to repeated (tens of thousands) discharges at a few kV. 

In order to avoid any risks, the \dshort{tde} concept allows lifting and unplugging the blades from the cold flanges should any welding activities occur inside the cryostat during installation. 

The chimneys are the heaviest  objects ($\sim$200\,kg). The tooling for the rigging on the chimneys  was designed by the I\&I team and the \dshort{tde} experts. Points where the load can be applied during rigging without damaging the chimneys have been defined. For risk mitigation, the chimney installation happens before the installation of fragile components as the \dword{utca} crates. 

There are no particular safety rules for the protection of people apart the standard ones for the manipulation of loads and work underground.

%%%%%%%%%%%%%%%%%%                
\subsection{Management and Organization}
\label{subsubsec:topelec:mgmt}

%%%%%%%%%%%%%%%%%%                
\subsubsection{Institutions}
\label{subsubsec:topelec:consortium}

The \dfirst{tde} consortium is responsible for delivering the electronics used for the readout of the %Charge Readout Planes (
\dwords{crp} for the top-drift volume of \dword{spvd}. 

The production of the system elements in the \dshort{tde} consortium is foreseen to be funded by the France-IN2P3 DUNE project, with possible contributions from Japanese and U.S. groups for % what concerns 
testing and installation at SURF.

The list of institutions contributing to the \dshort{tde} consortium is shown in Table~\ref{tab:TDE:institutions}.

\begin{dunetable}
[\dshort{spvd} \dshort{tde} electronics consortium intitutions ]
{ll}
{tab:TDE:institutions}
{Institutions participating in the \dshort{spvd} \dshort{tde} consortium}
Institution  \\ \toprowrule
IN2P3 IJCLAB Orsay \\ \colhline
IN2P3 IP2I Lyon \\ \colhline
IN2P3 LP2I Bordeaux  \\ \colhline
Iwate University \\ \colhline
KEK \\ \colhline
NITKC \\ \colhline
Southern Methodist University (SMU)  \\
\end{dunetable}

Institutional responsibilities are summarized below:

\begin{itemize}
\item{\dword{larzic} \dword{asic} design and procurement: IN2P3 IP2I, IN2P3 LP2I}
\item{\dword{fe} cards design and procurement: IN2P3 IP2I, IN2P3 LP2I}
\item{\dword{sftchimney} design and procurement: IN2P3 IJCLAB}
\item{\dword{amc} cards design and procurement: IN2P3 IP2I, IN2P3 LP2I }
\item{\dword{utca} crates design and procurement: IN2P3 IP2I}
\item{timing system end-nodes design and procurement: IN2P3 IP2I, IN2P3 LP2I}
\item{LV PS and filtering and distribution boxes design and procurement: IN2P3 IP2I, IN2P3 LP2I}
\item{VHDCI cables design and procurement: IN2P3 IP2I, IN2P3 LP2I}
\item{Calibration units design and procurement: IN2P3 IP2I, IN2P3 LP2I}
\item{\dword{qc} testing:  all institutions }
\item{Installation and commissioning:  all institutions }
\end{itemize}

%%%%%%%%%%%%%%%%%%                
\subsubsection{Milestones}
\label{subsubsec:topelec:milestones}

The milestones of the \dword{spvd} \dword{tde}  consortium are listed in Table~\ref{tab:TDE:timeline}.

\begin{dunetable}
[Milestones of the \dshort{tde} consortium]
{p{0.75\textwidth}p{0.20\textwidth}}
{tab:TDE:timeline}
{Milestones of the \dshort{tde} consortium.}
Milestone & Date \\ \toprowrule
Test of first full top drift \dshort{crp} (CRP-2) & July 2022     \\ \colhline
Test of second full top drift \dshort{crp} (CRP-2) & October 2022     \\ \colhline
Start of Module-0 installation & October 2022     \\ \colhline
Completion of top drift \dshort{crp} tests & November 2022     \\ \colhline
Completion of  Final Design Review & March 2023     \\ \colhline
Completion of Production Readiness Review & End of 2023 \\ \colhline
Start of  \dshort{amc} production & June 2024 \\ \colhline
Start of \dshort{fe} cards production & October 2024 \\ \colhline
Start of \dshort{utca} crates production & July 2024 \\ \colhline
Start of timing system production & July 2024 \\ \colhline
Start of cabling production & February 2025    \\ \colhline
Start of chimneys production & October 2024 \\ \colhline
End of timing system production and QC & July 2025 \\ \colhline
End of  \dshort{amc} production and QC & July 2025 \\ \colhline
End of \dshort{utca} crates production and QC & July 2025 \\ \colhline
End of \dshort{fe} cards production and QC & December 2025 \\ \colhline
End of cabling production & July 2026    \\ \colhline
End of chimneys production and \dshort{qc} & April 2026 \\ \colhline
Delivery to SURF completed & September 2026 \\ \colhline
Completed \dshort{tde} installation and commissioning & December 2027 \\ 
\end{dunetable}

\FloatBarrier

%%%%%%%%%%%%%%%%%%%%%%%%%%%%%%%%%%%%%%%
\section{Bottom Drift Readout} 
\label{subsec:BAROss}
%%%%
\subsection{System Overview}
\label{subsubsec:BASOsss}
%%%%

The \dword{cro} electronics for the bottom drift volume is  based on, and therefore similar to, the successfully deployed \dword{pdsp} \dword{ce}~\cite{DUNE:2020txw}, and on the \dshort{asic}s developed for the \dword{sphd} detector module. 
The same \dshort{asic}s will be used for both detector technologies.

The system architecture is shown in Figure~\ref{fig:system_overview_bottom_drift}.
Anode signals are input to \dwords{femb} mounted on the edges of the \dshort{crp}s.  The anode signals are amplified, shaped, and input to \dwords{adc}.  The digitized signals are multiplexed in groups of 32 channels, and output on serial links operating at \SI{1.25}{Gbps}. Short (\SI{2.5}{m}) power and signal cables connect the \dshort{femb}s to patch panels near one edge of each half-\dword{crp}.  Long (\SI{25}{m}) power and signal cables are connected to the patch panels and routed up the sides of the cryostat to penetrations distributed along the top of the cryostat near each of the long sides, where they are connected to \dwords{wib} through feedthrough flanges.  The \dshort{wib}s further concentrate the data and output information from groups of 256 channels over optical links at roughly \SI{10}{Gbps} to the \dword{daq} system.

\begin{dunefigure}
[TPC readout electronics architecture in the bottom drift]
{fig:system_overview_bottom_drift}
{System architecture of the TPC readout electronics in the bottom drift volume.}
	\includegraphics[width=1\textwidth]{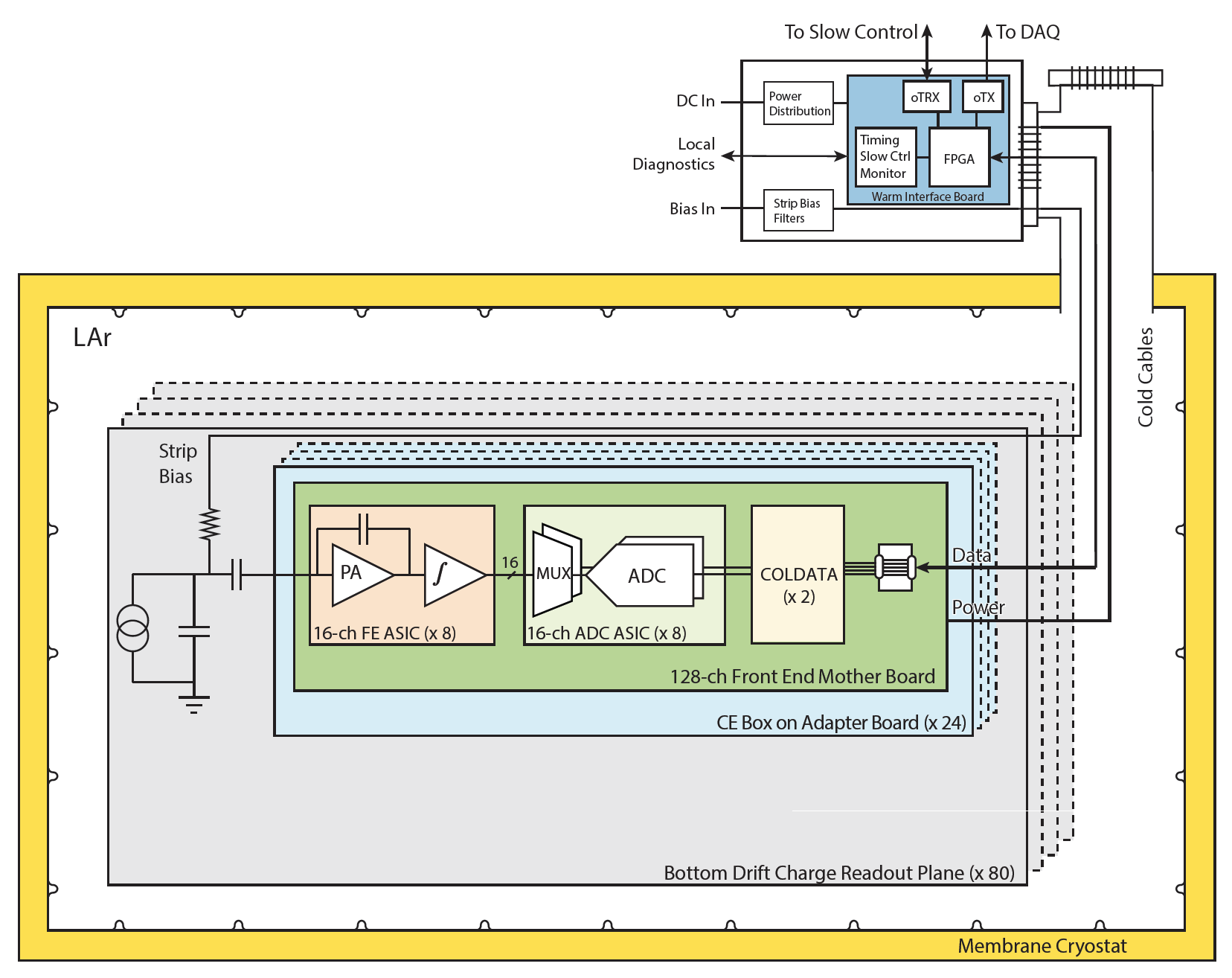}
\end{dunefigure}

\subsection{Specifications}
\label{sec:bde-specs}
In addition to the high level requirements listed in Table~\ref{tab:specs:just:SP-ELEC}, the BDE consortium has defined a number of engineering specifications, listed in Table~\ref{tab:specs:BDE-ELEC}.  Some of these are derived from the requirements and others represent design choices.
% This file is generated, any edits may be lost.
\begin{footnotesize}
%\begin{longtable}{p{0.14\textwidth}p{0.13\textwidth}p{0.18\textwidth}p{0.22\textwidth}p{0.20\textwidth}}
%\begin{longtable}{p{0.12\textwidth}p{0.18\textwidth}p{0.17\textwidth}p{0.25\textwidth}p{0.16\textwidth}}
\begin{longtable}{p{0.1\textwidth}p{0.22\textwidth}p{0.2\textwidth}p{0.36\textwidth}}
\caption{BDE specifications} \\
  \rowcolor{dunesky}
       Label & Description  & Specification \newline (Goal) & Rationale\\  \colhline

  \newtag{FD-CE-1}{ spec:num-FE-baselines }  & Number of baselines in the front-end amplifier  &  \num{2} &  Use a single type of amplifier for both induction and collection strips  \\ \colhline
    
   \newtag{FD-CE-2}{ spec:gain-FE-amplifier }  & Gain of the front-end amplifier  &  $\sim\SI{10}{mV/fC}$ \newline (Adjustable in the range \SIrange{5}{25}{mV/fC}) &  The gain of the FE amplifier is obtained from the maximum charge to be observed without saturation and from the operating voltage of the amplifier, that depends on the technology choice. \\ \colhline
    
   \newtag{FD-CE-3}{ spec:syncronization-CE }  & System synchronization  &  \SI{50}{ns} \newline (\SI{10}{ns}) &  The dispersion of the sampling times on different strips of a CRP should be much smaller than the sampling time (500 ns) and give a negligible contribution to the hit resolution.   \\ \colhline

  \newtag{FD-CE-4}{ spec:FEMB-femb-ch }  & Number of channels per FEMB & 128 & Design \\ \colhline
    
   \newtag{FD-CE-5}{ spec:FEMB-data-link }  & Number of links between the FEMB and the WIB  &  \num{4} at \SI{1.25}{Gbps}  &  Balance between reducing the number of links and reliability and power issues when increasing the data transmission speed. \\ \colhline

  \newtag{FD-CE-6}{ spec:number-FEMB-per-WIB }  & Number of FEMBs per WIB  &  \num{4} &  The total number of FEMB per WIB is a balance between the complexity of the boards, the mechanics inside the crate holding the WIBs, and the required processing power of the FPGA on the WIB.\\ \colhline

  \newtag{FD-CE-7}{ spec:WIB-data-link }  & Data transmission speed between the WIB and the DAQ backend  &  \SI{10}{Gbps} &  Balance between cost of the FPGA on the WIB and reduction of the number of optical fiber links for each WIB. \\ \colhline

\newtag{FD-CE-9}{ spec:double-pulse-resolution }  & Ability to resolve two tracks  & Time corresponding to $\sim\SI{5}{mm}$  & Should be about the same as the strip pitch and the longitudinal diffusion of electrons drifting to the CRP \\ \colhline
\newtag{FD-CE-10}{ spec:saturation recovery }  & Recovery from a very large input pulse & Monotonic recovery without nonlinear behavior & Recovery from saturation must be predictable so that dead time is known.\\ \colhline
\newtag{FD-CE-11}{ spec:crosstalk }  & Maximum allowable crosstalk  & $< 1\%$ ($<0.1\%$)  & Some crosstalk can be mitigated in signal deconvolution. At $0.1\%$ the effect of crosstalk can be ignored \\ \colhline
\newtag{FD-CE-12}{ spec:input_cap }  & Input capacitance to front end ASIC  & \SI{120}{pF} to \SI{210}{pF} &  Cold electronics should be optimized for the specified range of capacitance. \\ \colhline
\newtag{FD-CE-17}{ spec:PDS_Cables }  & Provision for PDS cables  &  &  Enough space  must be provided in the cryostat penetration for the PDS and CALCI cables. \\ \colhline
\newtag{FD-CE-18}{ spec:DNL }  & Differential NonLinearity (DNL) & Absolute value $<1$ (12-bit) LSB&  The minimum requirement ensures no missing (12-bit) codes; the goal is as low as is practical.  \\ \colhline
\newtag{FD-CE-19}{ spec:INL }  & Integral NonLinearity (INL) & Absolute value $<1$ (12-bit) LSB & Should not contribute to energy resolution   \\ \colhline
\newtag{FD-CE-20}{ spec:ENOB }  & Equivalent Number of Bits (ENOB) & $>10.3$ & Pulse amplification and digitization should not contribute to energy measurement uncertainty.  \\ \colhline
\newtag{FD-CE-21}{ spec:WIBs_per_WIEC }  & Number of WIBs in each Warm Interface Crate (WIEC)  &  \num{6} &  Design choice; one WIEC per CRP  \\ \colhline
\newtag{FD-CE-22}{ spec:PTCs_per_WIEC }  & Number of Power and Timing Cards (PTCs) in each Warm Interface Crate (WIEC)  &  \num{1} &  Design choice; one WIEC per CRP  \\ \colhline
\newtag{FD-CE-23}{ spec:Timing }  & WIB timing and clocks  &   &  WIBs must receive clock signals from the DUNE Timing System through the Power and Timing Card and provide the \SI{62.5}{MHz} clock to the \dwords{femb}.  \\ \colhline
\newtag{FD-CE-24}{ spec:WIBthroughput }  & WIB data processing  &   &  WIB must be able to receive high-speed TPC data from \dwords{femb}, format the data, and transmit it over optical fibers to the \dword{daq} \\ \colhline
\newtag{FD-CE-25}{ spec:WIBSC}  & WIB interface to Slow Control  &   &  WIB must be able to interface to the Slow Control system  \\ \colhline
\newtag{FD-CE-26}{ spec:WIBCal}  & Calibration of readout electronics  &   &  WIB must be capable of performing calibration of the FEMBs via charge injection to the front end ASICs. \\ \colhline
\newtag{FD-CE-27}{ spec:PTCclk}  & PTC timing distribution  &   &  The PTC must transmit clock and control signals from the timing system to the \dwords{wib} and multiplex the return signals \\ \colhline
\newtag{FD-CE-28}{ spec:PTCpower}  & PTC power distribution  &   &  The PTC must step down the 48V from the LVPS to 12V and distribute power to the \dwords{wib}. \\ \colhline
\newtag{FD-CE-29}{ spec:PTCmon}  & PTC monitoring  &   &  The PTC must be able to gather WIB status and transmit relevant information to the DUNE Detector Safety System (DDSS). It must also be able to receive inhibit/enable signals from the DDSS and transmit them to the WIBs  \\ \colhline
\newtag{FD-CE-30}{ spec:PTCripple}  & PTC output voltage ripple  & <\SI{15}{mV}  &  The ripple on the 12V provided to the \dwords{wib} must not be larger than \SI{15}{mV}.  \\ \colhline
\newtag{FD-CE-31}{ spec:Clockjitter}  & WIB clock jitter  & <\SI{180}{ps}  &  The \SI{62.5}{MHz} clock provided to the \dwords{femb} by the \dwords{wib} is used to generate the ADC sampling clock. ADC sampling clock jitter adds effective noise to the digitization. Assuming ENOB of $\sim 11$ and a signal bandwidth of 350 kHz, the required signal to noise ratio is $\sim$\SI{68}{dB}. This translates to a sampling clock jitter of \SI{180}{ps}. \\ \colhline
\newtag{FD-CE-35}{ spec:ADC_overflow}  & ADC overflow protection  & & When the input signal exceeds the upper or lower limit of the ADC range, the output should be fixed at the maximum or minimum value. \\ \colhline

\label{tab:specs:BDE-ELEC}
\end{longtable}
\end{footnotesize}
\subsection{System Design}
\label{sec:fdsp-tpcelec-design}
%\fixme{AH starting review 12/1}

This section describes the overall
system design of the \dword{bde}, starting in
Section~\ref{sec:fdsp-tpcelec-design-grounding} with a description of the
grounding and shielding scheme adopted
to minimize the overall noise in the %detector
\dword{spvd} module, followed in 
Section~\ref{sec:fdsp-tpcelec-design-bias} by a discussion of the bias
voltage distribution system.
Section~\ref{sec:fdsp-tpcelec-design-femb} describes the \dwords{femb}, including
the design of the \dshort{asic}s that will be used.
%In 
Section~\ref{sec:fdsp-tpcelec-design-infrastructure} 
discusses the infrastructure for the \dshort{bde} inside the cryostat,
including the \coldbox{}es that shield the \dshort{femb}s, the
cold cables, and the cryostat feedthroughs. 
Section ~\ref{sec:fdsp-tpcelec-design-warm} describes the electronics mounted on the warm side of the feedthroughs, and section ~\ref{sec:fdsp-tpcelec-design-services} describes the services that provide the low-voltage power and the bias voltage to the \dshort{bde}.

%%%%%%%%%%%%%%%%%%%%%%%%%%%%%%%%%%%
\subsubsection{Grounding and Shielding}
\label{sec:fdsp-tpcelec-design-grounding}

The overall approach to minimizing the system noise 
%in the  \dword{sphd} 
relies on enclosing the sensitive \dwords{anodepln} in a nearly hermetic 
Faraday cage, then carefully controlling currents flowing into or 
out of that protected volume through the unavoidable penetrations 
needed to build a working detector. Done carefully, this can %result in avoiding 
avert all unwanted disturbances that result in detector noise. 
Such disturbances could be induced on the anode strips by 
changing currents flowing inside the cryostat or even on the cryostat 
walls as, for instance, a temperature-sensing circuit that acts as a 
receiving antenna on the outside of the cryostat and a transmitting 
antenna in the interior. % of the cryostat. 
In addition, unwanted signals 
might be injected into the electronics either in the cold or just 
outside the cryostat by direct conduction along unavoidable power 
or signal connections to other devices. This approach to minimizing
the detector noise by using appropriate grounding and shielding procedures
is discussed in detail in~\cite{Radeka:1998as}. 
It results in the 
following set of requirements that need to be respected during the
design and the construction of the  \dword{spvd}:
\begin{itemize}
\item{The \dword{crp} copper ground plane shall be connected to the common of
all the \dword{fe} \dshort{asic}s;}
\item{All electrical connections (low-voltage power, bias voltage,
clock, control, and data readout) from one \dshort{crp} shall lead to a 
single signal feedthrough (\dword{sft});}
\item{All \dshort{crp}s shall be insulated from each other;}
\item{The common of the \dshort{fe} \dword{asic} and the rest of the 
\dword{tpc} readout electronics shall be connected to
the common plane of the \dword{femb};}
\item{The return leads of the \dshort{crp} strip bias lines and any shield
for the clock, control, and data readout shall be connected
to the common plane of the \dshort{femb} at one end and
to the flange of the \dshort{sft} at the other end; these shall be 
the only connections of the \dshort{crp} ground plane to the cryostat;}
\item{The mechanical supports of the \dshort{crp} shall ensure that the \dshort{crp} is electrically isolated from the cryostat;}
\item{The last stage of the readout strip and shielding plane bias filters shall be
connected to the common of all the \dshort{fe} \dshort{asic}s and therefore
to the \dshort{crp} ground plane.}
\end{itemize}

To minimize system noise, the \dshort{tpc} electronics cables for each \dshort{crp} 
enter the cryostat through a single \dword{ce} flange.
%, as shown in Figure~\ref{fig:connections}.
This creates, for grounding purposes, 
an integrated unit consisting of a \dshort{crp} ground plane, \dshort{femb}
ground for all \num{24} \dshort{ce} modules, a \dshort{tpc} flange, and 
warm interface electronics. The input amplifiers on the 
\dshort{fe} \dshort{asic}s have their ground terminals connected to 
the \dshort{crp} ground plane. All power-return leads and cable shields are connected to both the ground plane of the \dshort{femb} and to the 
\dshort{tpc} signal flange.

The only location where this integrated unit makes electrical contact 
with the cryostat, which defines the detector ground and acts as a 
Faraday cage, is at a single point on the \dshort{ce} \fdth board in 
the \dshort{tpc} signal flange where the cables exit the cryostat. 
Mechanical support of the \dshort{crp}s is accomplished using 
insulated supports. To avoid structural ground loops, the \dshort{crp}s 
are electrically insulated from each other.

Filtering circuits for the \dshort{crp} strip bias voltages are 
locally referenced to the ground plane of the \dshort{femb}s 
through low-impedance electrical connections. This approach 
ensures a ground-return path in close proximity to the 
bias-voltage and signal paths. The close proximity of the 
current paths minimizes the size of potential loops to further 
suppress noise pickup.

%%%%%%%%%%%%%%%%%%%%%%%%%%%%%%%%%%%
\subsubsection{Distribution of Bias Voltages}
\label{sec:fdsp-tpcelec-design-bias}

The \dword{crp} includes two perforated \dwords{pcb} with strips on both sides. The strips are biased such that electrons pass through the holes in the \dshort{pcb}s unimpeded until they are collected on the final strip layer on the back of the second %printed circuit 
board. This condition is referred to as ``transparency.''

The filtering of strip bias voltages and the \dword{ac} coupling 
of wire signals passing onto the charge amplifier circuits is 
done on adapter boards that connect to edge boards that mount on the edges of the \dshort{crp}s and connect to the \dshort{crp} strips (see Section~\ref{subsec:3V}). Each adapter board includes \dword{rc} filters 
for the collection and first induction plane strip wire bias voltages, while the second induction plane 
strips have a floating voltage. %connect directly to ground.
In addition, each board 
has pairs of bias resistors and \dshort{ac} coupling 
capacitors. The coupling capacitors block \dword{dc} levels while passing \dshort{ac} 
signals to the \dshort{femb}s. On the \dshort{femb}s,
clamping diodes limit the input voltage received at the amplifier
circuits to between $\SI{1.8}{V}+U_D$ and $\SI{0}{V}-U_D$, where $U_D$
is the threshold voltage of the diode, approximately \SI{0.7}{V} at \dword{lar} temperature.
The amplifier circuit has a \SI{22}{nF} coupling capacitor at the
input to avoid leakage current from the protection clamping diodes.
Tests of the protection mechanism have been performed by discharging
\SI{4.7}{nF} capacitors holding a voltage of \SI{1}{kV} (\SI{2.35}{mJ} of
stored energy). The diodes have survived more than 250 discharges at \dword{ln}
temperature.

%%%%%%%%%%%%%%%%%%%%%%%%%%%%%%%%%%%
\subsubsection{Front-End Motherboard}
\label{sec:fdsp-tpcelec-design-femb}

Each \dword{crp} is instrumented with \num{24} \dwords{femb}.
Each \dword{femb} receives signals from \num{128} strips.
The \dshort{femb} contains eight \num{16}-channel 
\dword{larasic} chips, eight \num{16}-channel 
\dword{coldadc} \dwords{asic}, and two \dword{coldata} control and 
communication \dshort{asic}s.
%(see Figure~\ref{fig:system_overview_bottom_drift}).
The \dshort{femb} also contains regulators that produce the voltages 
required by the \dshort{asic}s and filter those voltages. The \dshort{larasic} 
inputs are protected by two external
diodes as well as internal diodes in the chip.

\begin{dunefigure}
[FEMB for DUNE FD2-VD]
{fig:DUNE_vd_femb}
{Bottom drift electronics \dshort{femb}: Four \dshort{larasic} front-end \dshort{asic}s (at the bottom of the picture) and four \dshort{coldadc} \dshort{asic}s are mounted on the visible side of the printed circuit board.  Another four \dshort{larasic}s and four \dshort{coldadc}s are mounted on the other side of the printed circuit board.  Two \dshort{coldata} \dshort{asic}s (the larger parts near the top of the picture) concentrate data from the \dshort{coldadc}s and transmit data to the \dword{wib} using \SI{1.25}{Gbps} serial links.}
\includegraphics[width=0.8\linewidth]{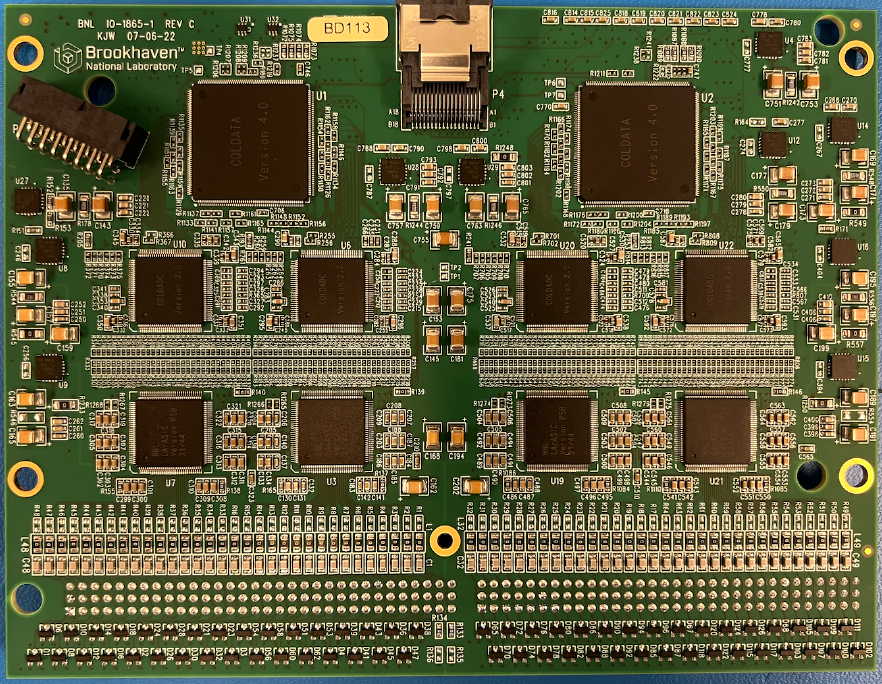}
\end{dunefigure}

The functionality of the \dshort{femb} for \dshort{dune} is very similar to that of the \dshort{femb} used in \dword{pdsp}. The design has changed slightly to accommodate the new \dshort{asic}s (fewer voltage regulators are
required) and features have been added to allow the analog values of the ColdADC reference
voltages to be measured. The \dshort{femb} used in \dshort{pdsp} consisted of an ``analog motherboard’’ and a ``digital daughterboard’’ which included an \dword{fpga}.  The \dshort{dune} \dshort{femb} is implemented on a single printed circuit board and is often referred to as a ``monolithic FEMB.’’ The \dshort{dune} \dshort{femb} for FD2-VD (shown in Figure 4.50) is identical to the \dshort{femb} used in FD1-HD except that it uses a 36-pin miniSAS connector instead of a Samtec connector for data I/O.  This allows the use of miniSAS cables on the \dshort{crp}s.  The miniSAS cables are less expensive and more flexible than the Samtec data cables, but the insertion loss is significantly greater on the miniSAS cables, so they cannot be used for the full run between the \dshort{femb}s and the \dshort{wib}s.

The power consumption of the \dshort{femb} \dshort{asic}s depends on whether the \dshort{larasic}s are operated in single ended or differential mode.  In single ended mode, the \dshort{asic}s consume $\sim$\SI{29}{mW/channel}.  In differential mode, this increases to $\sim$\SI{34}{mW/channel}.  Including the power consumption of the voltage regulators, the power consumption in differential mode is slightly less than \SI{45}{mW/channel}.

The noise measured with \SI{150}{pF} at the input is shown as a function of peaking time in Figure~\ref{fig:bde-femp-noise}.

\begin{dunefigure}
[FEMB noise]
{fig:bde-femp-noise}
{Noise as a function of peaking time for \SI{150}{pF} input capacitance, measured at room temperature (red) and in liquid nitrogen (blue). For these measurements the \dshort{larasic} gain was set to \SI{14}{mV/fC} and single ended output was used.}
\includegraphics[width=0.8\linewidth]{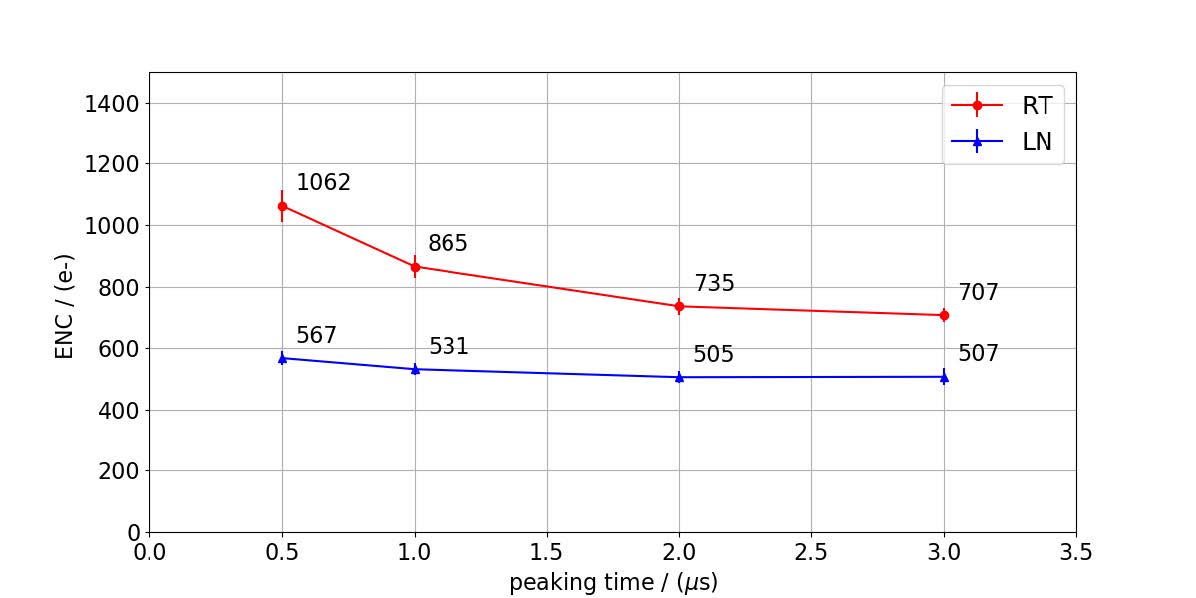}
\end{dunefigure}

All the discrete components mounted on the \dshort{femb} have been
characterized for operation in \dshort{lar}. %In some cases 
Some of the components (resistors,
capacitors, diodes) %the components 
used on the \dshort{pdsp} \dshort{femb}
belong to the same family of components already used for other boards
operating in a cryogenic environment, namely the boards used for the 
\dword{atlas} accordion \dshort{lar} calorimeter, providing relevant information
on the lifetime of these components, which is discussed later in 
Section~\ref{sec:fdsp-tpcelec-qa-reliability}. %There we 
That section also discusses
procedures for the measurement of the lifetime of discrete
components that have been adopted in recent years to demonstrate
that the \dshort{tpc} electronics can survive in \dshort{lar}. 
These types of measurements have been performed already for 
other neutrino experiments that use \dword{lartpc}
technology.

In the case of custom \dshort{asic}s, appropriate steps must be taken prior 
to starting the layout of the chips. Both \dshort{coldata} and 
\dshort{coldadc} are implemented in a 
\SI{65}{nm} \dword{cmos} 
process. The designs were done using cold transistor models 
produced by Logix Consulting\footnote{Logix\texttrademark{} Consulting, http://www.lgx.com/.}.  Logix made measurements of 
\SI{65}{nm} transistors (supplied by \dword{fnal}) at \dshort{ln} 
temperature and extracted and provided to the design teams \dword{spice}
models valid at \dshort{ln} temperature.  These models were used in 
analog simulations of \dshort{coldata} and \dshort{coldadc} subcircuits.  
In order to eliminate the risk of accelerated aging due to the hot-carrier
effect~\cite{Hot-electron}, no transistor with a channel length
less than \SI{90}{nm} was used in either \dshort{asic} design.
A special library of standard cells using \SI{90}{nm} channel-length 
transistors was developed by members of the University
of Pennsylvania and \dshort{fnal} groups. Timing parameters were
developed for this standard cell library using the Cadence Liberate
tool\footnote{
\href{https://www.cadence.com/content/cadence-www/global/en_US/home/tools/custom-ic-analog-rf-design/library-characterization/liberate-characterization.html}{Cadence Liberate\texttrademark{}.}
} 
and the Logix \dshort{spice} models. Most of the digital logic
used in \dshort{coldadc} and \dshort{coldata} was synthesized 
from Verilog code using this standard cell library and the Cadence Innovus tool\footnote{
\href{https://www.cadence.com/content/cadence-www/global/en_US/home/tools/digital-design-and-signoff/hierarchical-design-and-floorplanning/innovus-implementation-system.html}{Cadence Innovus\texttrademark{}}.}.
Innovus was also used for the layout of the synthesized logic.
The design of \dshort{larasic}
is implemented in a %\dword{tsmc} see above
\SI{180}{nm} 
\dshort{cmos} process. The models used during the design of \dshort{larasic} were obtained by extrapolating the
parameters of the models provided by the foundry, which are 
generally valid in the \SIrange{230}{400}{K} range.  After the design was complete, simulations using cold models provided by Logix Consulting were used to verify the design.

%%%%%%%%%%%%%%%%%%
\subsubsection{\dshort{larasic} Front-end \dshort{asic}}
\label{sec:fdsp-tpcelec-design-femb-fe}

\begin{dunefigure}
[FE ASIC block diagram]
{fig:larasicblockdiagram}
{\dshort{larasic} block diagram, showing 16 channels and shared circuitry including a Band Gap Reference (BGR) circuit.}
\includegraphics[width=0.99\linewidth]{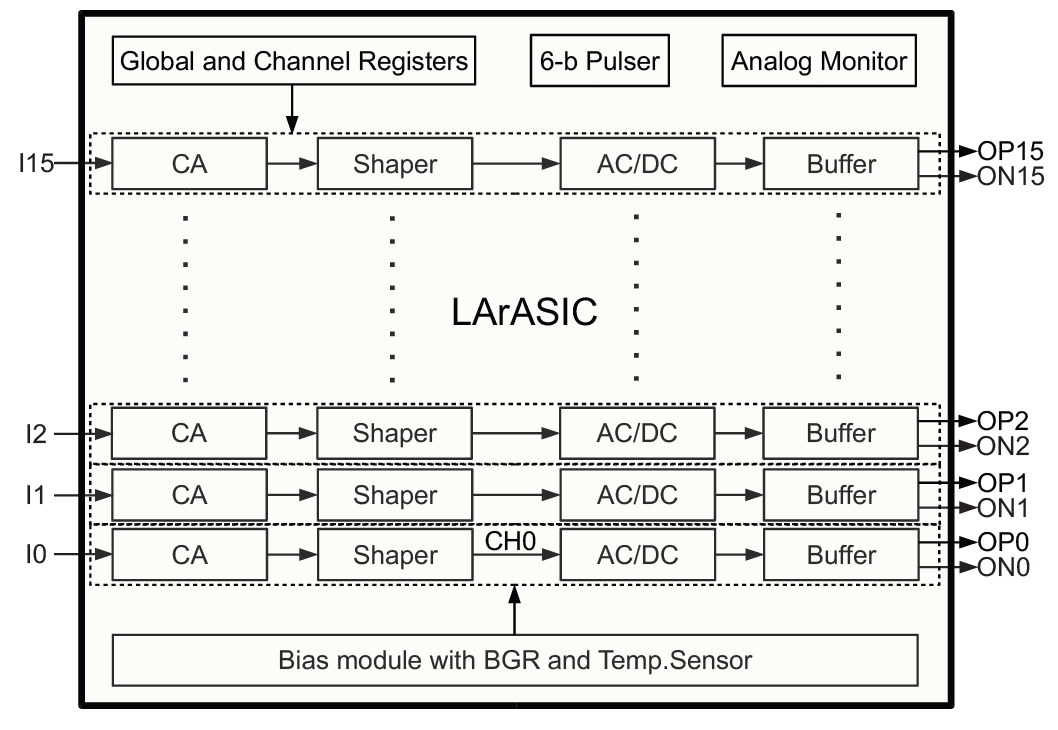}
\end{dunefigure}

Each \dshort{larasic}~\cite{DeGeronimo:2011zz} receives 
signals from 16 \dshort{crp} strips, amplifies and shapes the signals, and outputs the signals to a \dshort{coldadc} for digitization.  \dshort{larasic} is designed to minimize noise for an input capacitance of approximately 150 pF.
Figure~\ref{fig:larasicblockdiagram} shows a block diagram of \dshort{larasic}, and 
a simplified block diagram of one channel of \dshort{larasic} is shown in Figure~\ref{fig:larasicchannel}. \dshort{larasic} is implemented using a %\dword{tsmc} see above instances
\SI{180}{nm} \dshort{cmos} process and was designed by engineers from \dword{bnl}.

\begin{dunefigure}
[FE ASIC channel schematic]
{fig:larasicchannel}
{Channel schematic of \dshort{larasic}, which includes a 
two-stage charge amplifier and a \num{5}$^{th}$ order semi-Gaussian shaper.}
\includegraphics[width=0.99\linewidth]{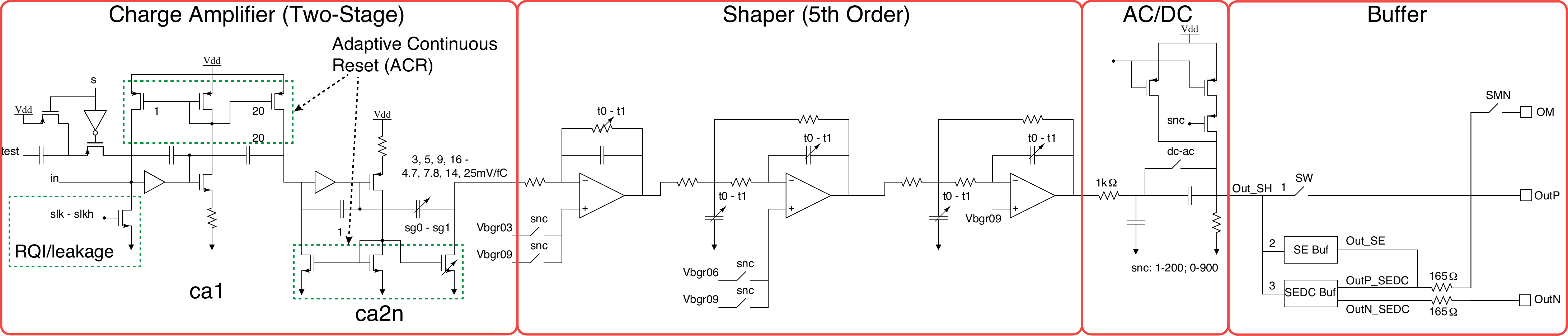}
\end{dunefigure}

A two-stage charge amplifier with adaptive continuous reset is followed by a fifth-order shaping amplifier with complex conjugate poles that converts charge to voltage and performs shaping to maximize the \dword{s/n} ratio and provide an anti-aliasing filter for the \dword{adc}.  The output of the shaper can be AC or DC coupled and an optional buffer can drive a long cable (for debugging purposes) or can provide a differential output.

Each channel can be independently configured by setting the gain, peaking time, and output baseline (assuming DC coupling).  A bandgap reference circuit produces all of the required internal bias voltages and a six-bit digital to analog converter (\dword{dac}) can provide a voltage level that can be used to test the channels using a charge injection capacitor. Figure~\ref{fig:larasicpulse} shows output pulses from test inputs for a variety of gain and shaping time settings.

\begin{dunefigure}
[LArASIC pulse]
{fig:larasicpulse}
{\dshort{larasic} output pulses recorded in response to test pulses input using the internal \dword{dac}.  The black waveform is shown for a channel set to 14 mV/fC gain, 1 microsecond shaping time, and the 900 mV baseline intended for induction channels.  The colored waveforms illustrate the various gain and shaping time settings with a channel set to the 200 mV baseline intended for collection channels.}
\includegraphics[width=0.99\linewidth]{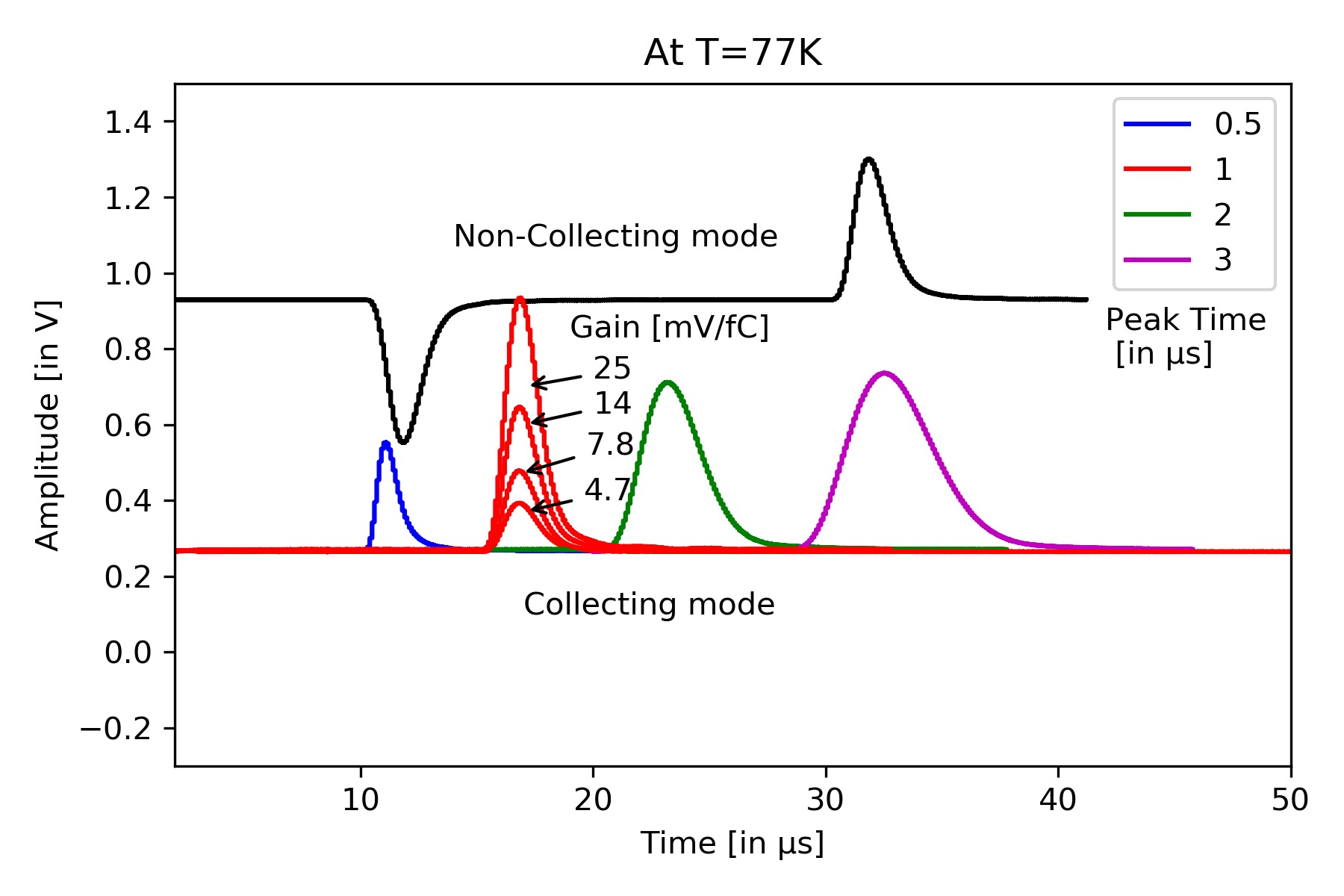}
\end{dunefigure}

A test output can be used to monitor the output of one channel, the bandgap reference voltage, or a voltage proportional to temperature.

%%%%%%%%%%%%%%%%%%
\subsubsection{\dshort{coldadc} \dshort{asic}}
\label{sec:fdsp-tpcelec-design-femb-adc}

\dshort{coldadc} is a low-noise \dshort{adc} \dshort{asic} designed to digitize
\num{16} input channels at a rate of $\sim\SI{2}{MHz}$. 
\dshort{coldadc} requires two external clocks, the digitization clock and a master clock which is used by the \dshort{coldadc} digital logic and must be 32 times the frequency of the digitization clock.  DUNE will use a \SI{62.5}{MHz} master clock, so the digitization frequency will be \SI{1.953}{MHz} (1/512 ns). For convenience, the digitization clock is referred to as \SI{2}{MHz} and the master clock as \SI{64}{MHz}.

\begin{dunefigure}
[ColdADC block diagram]
{fig:COLDADC_Block_Diagram}
{\dshort{coldadc} block diagram.}
\includegraphics[width=0.7\linewidth]{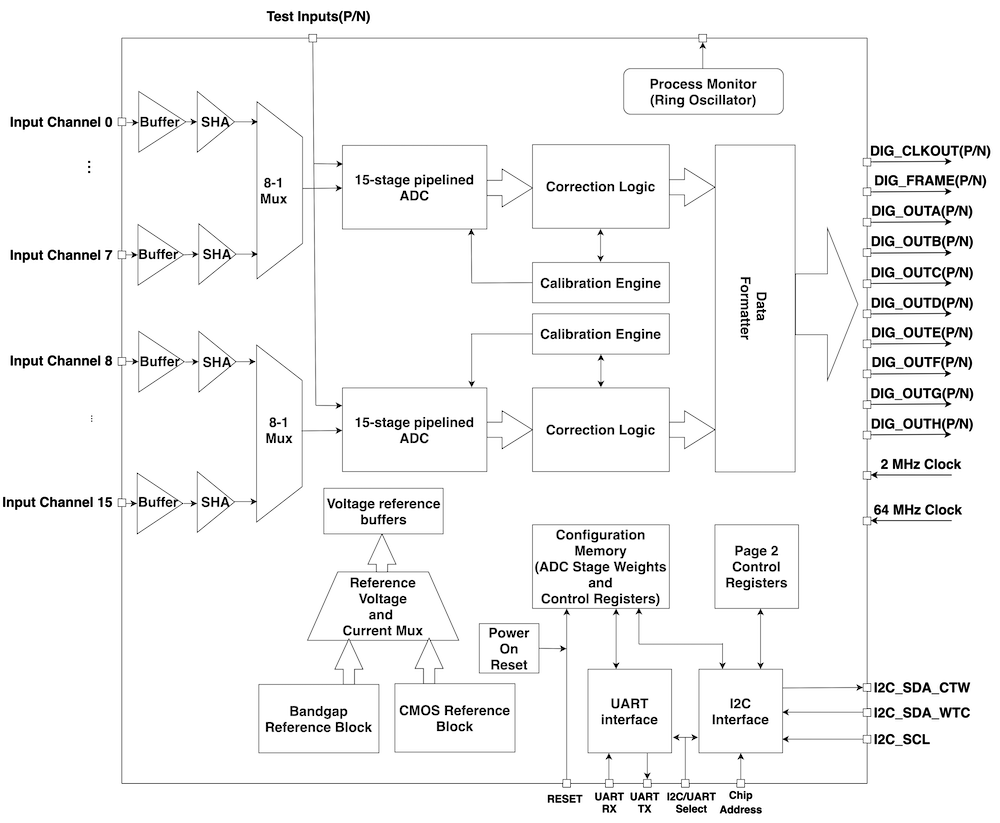}
\end{dunefigure}

\dshort{coldadc} is implemented in a %\dshort{tsmc} see above instances
\SI{65}{nm} \dshort{cmos} technology and has been designed by a team of engineers
from \dword{lbnl}, \dword{bnl}, and \dword{fnal}.  The \dshort{asic} uses a conservative,
industry-standard design including digital calibration.  Each \dshort{coldadc}
receives \num{16} voltage outputs from a single \dshort{larasic} chip.  The voltages
are sampled, multiplexed by eight, and input to two \num{15}-stage pipelined \dwords{adc}
operating at \SI{16}{MHz}. The \SI{16}{MHz} clock is generated internally in
\dshort{coldadc} by dividing the \SI{64}{MHz} clock by four,
and shares its rising edge with the \SI{2}{MHz} clock. 
The \dshort{adc} uses the well known pipelined architecture
with redundancy~\cite{PipelinedADC}.  Digital logic is used to correct non-linearity
introduced by non-ideal amplifier gain and offsets in each pipeline
stage~\cite{CalibrationCorrection}, and an automatic calibration procedure is
implemented to determine the constants used in this logic.  The \dshort{adc} produces
\num{16}-bit output which is expected to be truncated to \num{12} or \num{14} bits.

The \dshort{adc} is highly programmable to optimize performance at different
temperatures.  Many circuit blocks can be bypassed, allowing the performance 
of the core digitization engine to be evaluated separately from the ancillary 
circuits. A block diagram of the chip is shown in Figure~\ref{fig:COLDADC_Block_Diagram}. 
%Each of the major blocks is described below.
A detailed circuit description is given in reference~\cite{Grace:2021ybz}.

\dshort{coldadc} is designed to be very low noise so that it will contribute a negligible amount to the noise of a digitized waveform even when LArASIC is operated at its lowest gain setting.  At liquid nitrogen temperature, the noise (measured either with open channels or with a constant voltage at the inputs) is $\sim$\num{0.75} 14-bit \dshort{adc} counts with differential inputs and $\sim$\num{1.5} 14-bit \dshort{adc} counts with single ended inputs. The noise is quite uniform channel to channel.

The static linearity of \dshort{coldadc} has been measured both using a very slow linear ramp and using a sine wave input, with consistent results.  Typical results are shown in Figure~\ref{fig:rawINLDNL}. The differential nonlinearity (DNL) is small and there are no missing codes, but the integral nonlinearity (INL) is larger than was expected and has a characteristic shape. It is  believed this is due to dielectric absorption in the capacitors used in the switched capacitor circuits of the sample and hold amplifiers and in the pipeline \dshort{adc}s~\cite{Grace:2021ybz}.

The INL varies systematically channel to channel and to a lesser extent chip to chip and can be corrected with a polynomial function.  The result of fitting a third-order polynomial to the INL of the \dshort{adc} channel depicted in Figure~\ref{fig:rawINLDNL} and using the fit function to correct data from the same channel taken on a different day (after multiple temperature cycles) is shown in Figure~\ref{fig:correctedINLDNL}.

\begin{dunefigure}
[ColdADC raw INL and DNL]
{fig:rawINLDNL}
{Typical static nonlinearity of a \dshort{coldadc} channel differential inputs at \SI{77}{K}. The INL shows a marked polynomial shape that suggests it could be corrected with a cubic (or another odd-order polynomial)}
\includegraphics[width=0.70\linewidth]{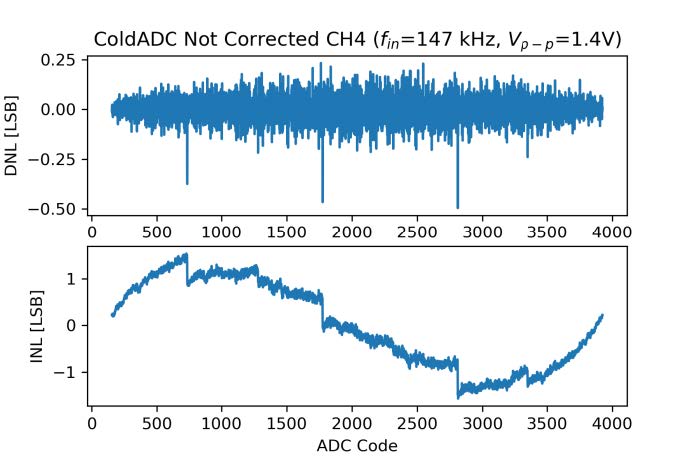}
\end{dunefigure}

\begin{dunefigure}
[ColdADC corrected INL and DNL]
{fig:correctedINLDNL}
{Post-correction static nonlinearity of a separate measurement (on a different day) of the same \dshort{coldadc} channel shown in Figure~\ref{fig:rawINLDNL}. The INL is improved by use of a cubic fit to the INL shown in Figure~\ref{fig:rawINLDNL}.}
\includegraphics[width=0.70\linewidth]{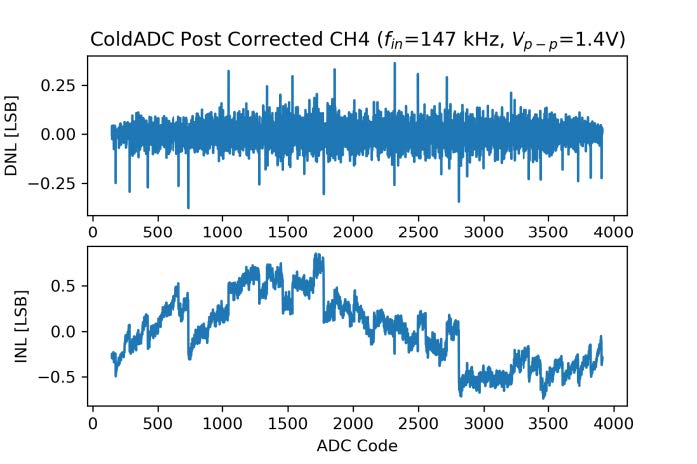}
\end{dunefigure}

The structure in the INL that is attributed to dielectric absorption is very stable with time.  Linearity correction has been found to be stable for at least several months, but in practice the transfer function of each channel will be measured in situ from time to time using the \dshort{larasic} test input.

\begin{dunefigure}
[ColdADC FFT ch4 dif 39kHz]
{fig:FFTch8dif_39kHz}
{Typical frequency spectrum of ADC output codes (with differential inputs at \SI{77}{K}) after polynomial correction. The input frequency was $\sim$\SI{38.6}{kHz}}
\includegraphics[width=0.60\linewidth]{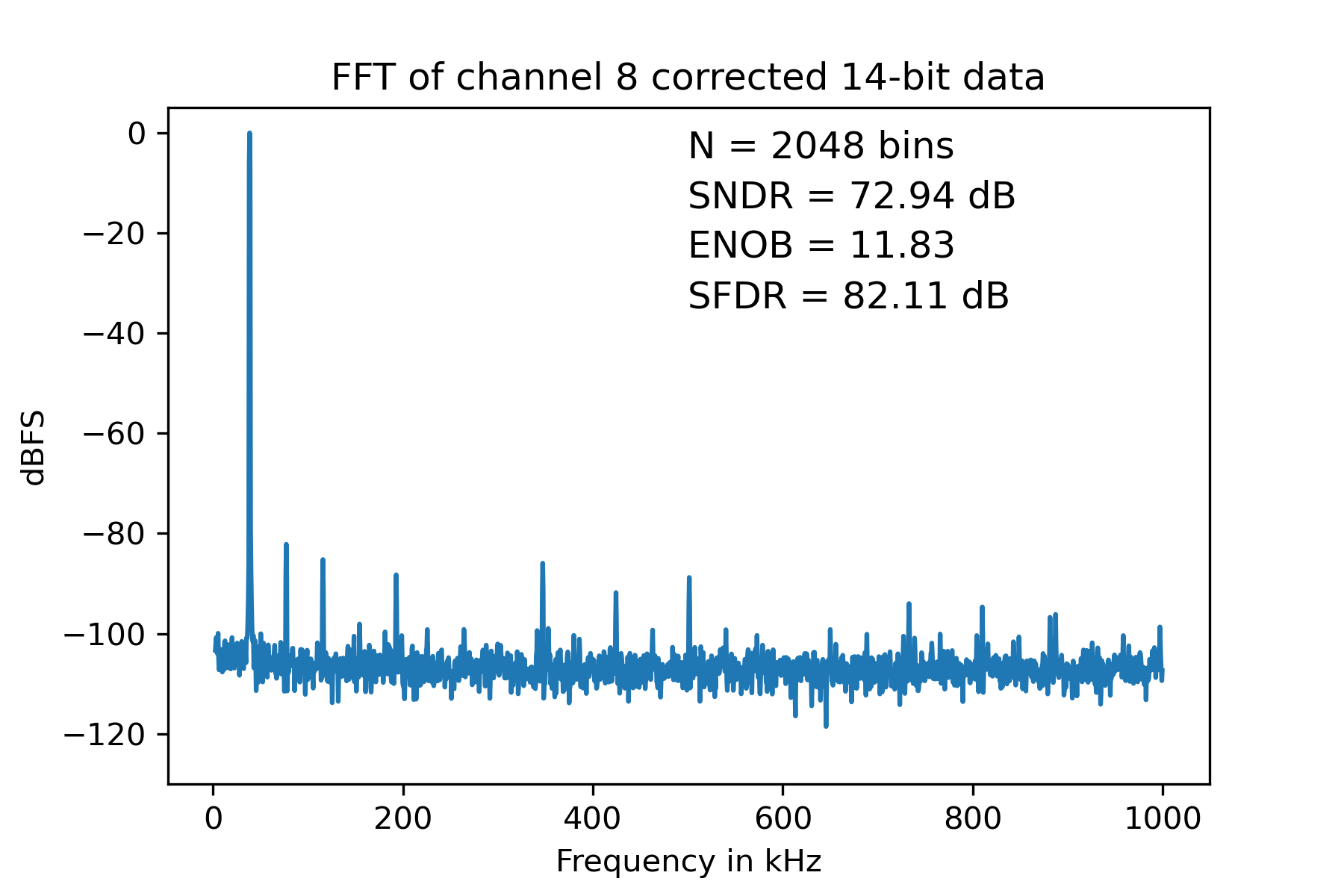}
\end{dunefigure}

The dynamic linearity of \dshort{coldadc} has been measured using sine wave inputs. Figure~\ref{fig:FFTch8dif_39kHz} shows a typical Fourier-transformed output spectrum collected at \SI{77}{K} using differential inputs. The measured dynamic performance as a function of input frequency is shown in Figure~\ref{fig:ENOBvsFreq_ch8}. The \SI{-3}{dB} bandwidth of \dshort{larasic} is $\sim$\SI{410}{kHz} when the shortest shaping time of \SI{0.5}{{\micro}s} is selected and $\sim$\SI{205}{kHz} when the shaping time is \SI{1}{{\micro}s}.

\begin{dunefigure}
[ENOB vs freq ch8]
{fig:ENOBvsFreq_ch8}
{Typical \dshort{enob} as a function of input sinewave frequency (differential input, \SI{77}{K})}
\includegraphics[width=0.60\linewidth]{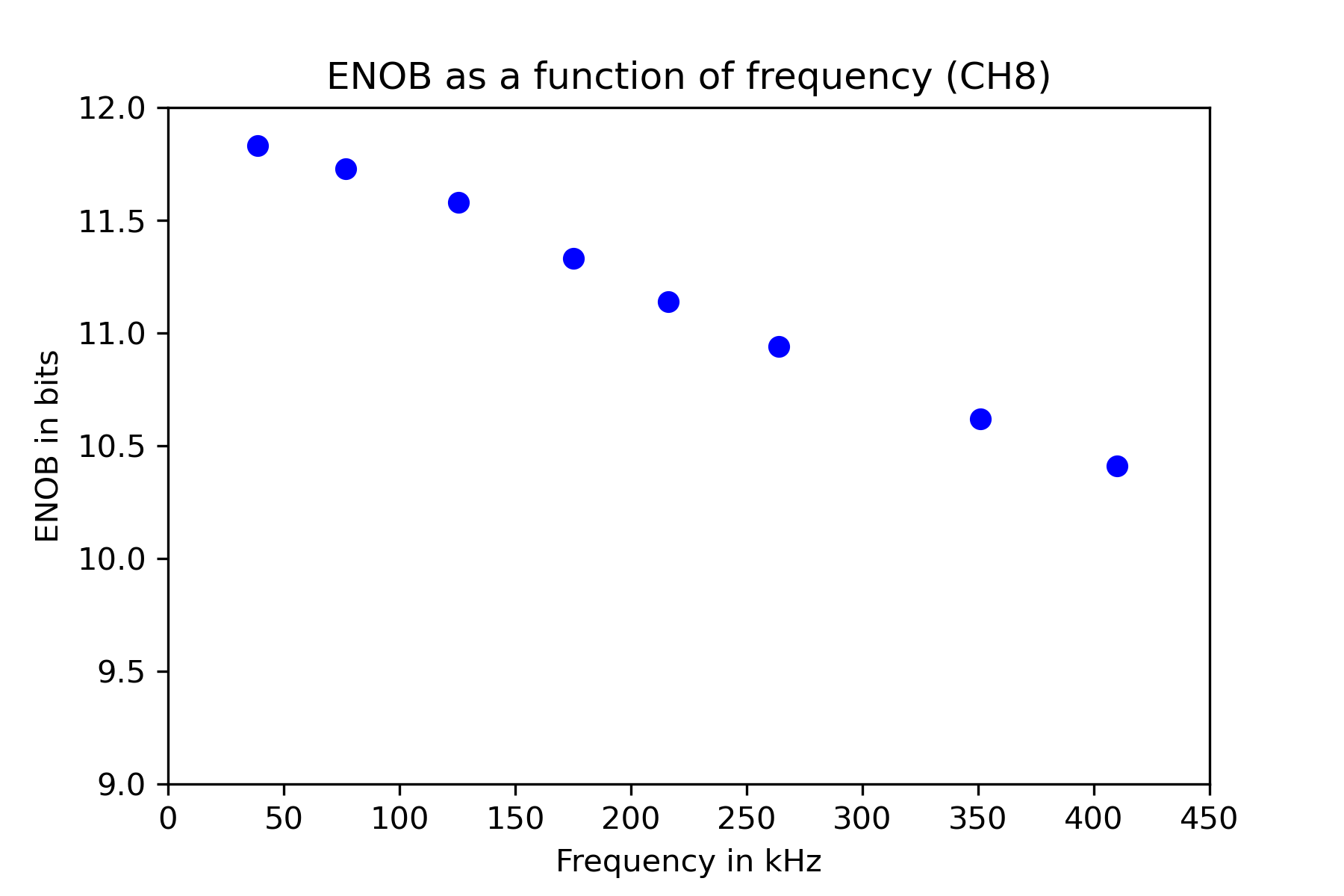}
\end{dunefigure}

\dshort{coldadc} exceeds its \dword{enob} requirement of \num{10.3} bits across the required input frequency range. 

\begin{dunefigure}
[ColdADC crosstalk]
{fig:ADCCrosstalk}
{Channel-to-channel crosstalk in ColdADC measured with differential inputs at \SI{77}{K}}
\includegraphics[width=0.70\linewidth]{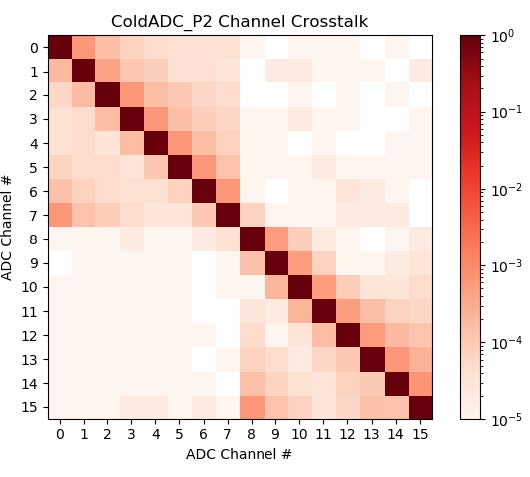}
\end{dunefigure}

The crosstalk of \dshort{coldadc} has also been measured by injecting a signal into one channel at a time and recording the \dshort{adc} codes from all 16 channels (see Figure~\ref{fig:ADCCrosstalk}). The peak crosstalk is $\sim$\num{0.06}$\%$ and is observed in the next channel in the multiplexing cycle, indicating that most of the crosstalk is associated with the multiplexer.  This level of crosstalk is small enough that no correction is expected to be needed in the analysis of DUNE data.

As detailed in Table~\ref{tab:ColdADCpower}, \dshort{coldadc} dissipates approximately \SI{332}{mW} in a typical configuration (differential inputs with input buffers bypassed) at \SI{77}{K}.

\begin{dunetable}
[ColdADC power dissipation]
{cccc}
{tab:ColdADCpower}
{\dshort{coldadc} power dissipation at \SI{77}{K}.}
\textbf{Power Domain} &\textbf{Description} &\textbf{Voltage (V)} &\textbf{Power Dissipation (mW)}\\ \toprowrule
VDDA2P5 & Analog Circuitry & \num{2.25} &\num{286} \\ \colhline
VDDD2P5 & ADC Digital Logic and Switches & \num{2.25} &\num{11.9} \\ \colhline
VDDD1P2 & Digital Logic & \num{1.10} &\num{1.3} \\ \colhline
VDDIO & ESD Ring, LVDS, CMOS I/O & \num{2.25} &\num{32.6} \\ \colhline
Total &  &  &\num{331.8} \\ \colhline
Per Channel & & &\num{20.7} \\
\end{dunetable}

%%%%%%%%%%%%%%%%%%
\subsubsection{COLDATA \dshort{asic}}
\label{sec:fdsp-tpcelec-design-femb-coldata}

\dshort{coldata} is a control and communications \dshort{asic} designed to control four \dshort{larasic} front end \dword{asic}s and four \dshort{coldadc} \dshort{asic}s and to merge data from four \dshort{coldadc}s. It transmits data to a \dword{wib} over two \SI{1.25}{Gbps} links. \dshort{coldata} receives commmands either from a \dshort{wib} or from another \dshort{coldata} \dshort{asic}.  It either responds to commands directly (if they are intended for it) or relays the commands to their destination and relays responses from the destination back to the \dshort{wib}.

\begin{dunefigure}
[ColdDATA block diagram]
{fig:coldata_block_diagram}
{\dshort{coldata} block diagram.}
\includegraphics[width=0.90\linewidth]{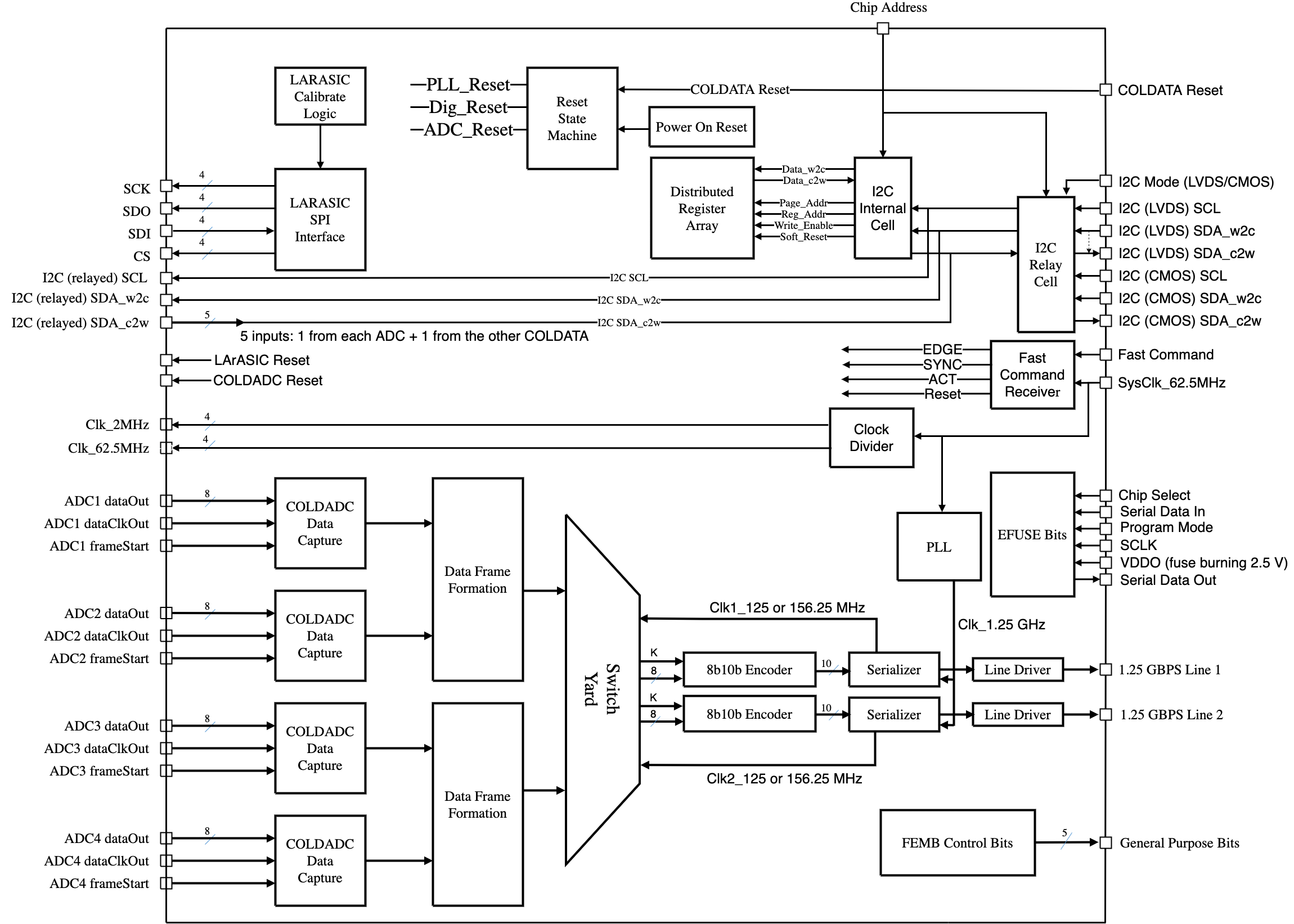}
\end{dunefigure}

A block diagram of \dshort{coldata} is shown in Figure~\ref{fig:coldata_block_diagram}. 
%Each of the major functional blocks is described below. 
\dshort{coldata} is implemented in a  
\SI{65}{nm} \dword{cmos} technology and was designed by a team of engineers from \dshort{fnal}, Southern Methodist University, and \dshort{bnl}.

\dshort{coldata}'s control communication protocol is based on \dword{i2c}~\cite{bib:I2C} protocol.  The major difference between \dshort{coldata} ``\dword{i2c}'' and standard \dshort{i2c} is that \dshort{coldata} uses two separate SDA (Serial Data) lines, rather than one bidirectional line. 
This is
motivated by the fact that \dshort{coldata}'s control communications must travel over long cables between the \dshort{wib}s and the \dshort{femb}s.  Consequently, for these signals, canonical \dshort{i2c} signaling is replaced by differential \dword{lvds}.  Since \dshort{lvds} is not amenable to bidirectional communication, the bidirectional SDA (serial data) line is replaced by two data lines, one from Warm to Cold (I2C-SDA-w2c) and one from Cold to Warm (I2C-SDA-c2w).  To reduce power dissipation on the FEMBs, single ended \dshort{cmos} signaling rather than differential \dshort{lvds} is used for ``\dshort{i2c}'' communication on the FEMBs. The \dshort{coldata} data sheet contains a full description of the protocol that is implemented.

Time-sensitive commands are sent from the \dshort{wib} to \dshort{coldata} in the form of DC balanced 8-bit ``Fast Command'' words. Fast Command words are bit-aligned with the rising edge of the \SI{62.5}{MHz} system clock and captured by \dshort{coldata} using the falling edge of the system clock. An example of a Fast Command is the ``Edge'' command, which is used to move the rising edge of the ADC sample clock to the next rising edge of the \SI{62.5}{MHz} system clock.  This command, along with another feature of \dshort{coldata} that allows the time delay on the cable between the \dshort{wib} and the \dword{femb} to be measured, is used to synchronize the sample time of all \dshort{coldadc}s.

The \SI{1.25}{Gbps} hybrid-mode Line Driver is designed with current-mode transmitter equalization and voltage-mode pre-emphasis to drive 25-35 meter long twin-axial cables.  The current-mode transmitter equalization circuit uses a finite element response (FER) 
%\fixme{anne changed from (FIR)} 
filter to distort the data pulse to compensate for the large frequency-dependent signal loss over a long twinax cable with low dynamic power consumption.  The voltage-mode main driver and pre-emphasis circuit uses source-series-terminated (SST) output stages to provide a large output voltage swing and low static power consumption.  The Line Driver is highly programmable and can also be operated in a pure current-mode or a pure voltage-mode.  The pure current-mode with no equalization or pre-emphasis is appropriate for use with short cables.  Figure~\ref{fig:coldata_eyemeasurements} shows eye diagrams for a PRBS7 pseudorandom bit pattern driven (at room temperature) over \SI{25}{m} and \SI{35}{m} lengths of Samtec twinax cable.  PRBS-7 has approximately the same degree of imbalance as is provided by the 8b10b encoding that is used by \dshort{coldata}.  The cable insertion loss is less significant at cryogenic temperature, so the eyes become even more open when the cables are cold.

\begin{dunefigure}
[ColdDATA links with long cable]
{fig:coldata_eyemeasurements}
{PRBS7 eye measurements made at room temperature using \dshort{coldata} with a \SI{25}{m} length of Samtec twinax cable (top) and a \SI{35}{m} length (bottom). The eye height is approximately 150 mV with a \SI{25}{m} cable and slightly more than 100 mV with a \SI{35}{m} cable.}
\includegraphics[width=0.70\linewidth]{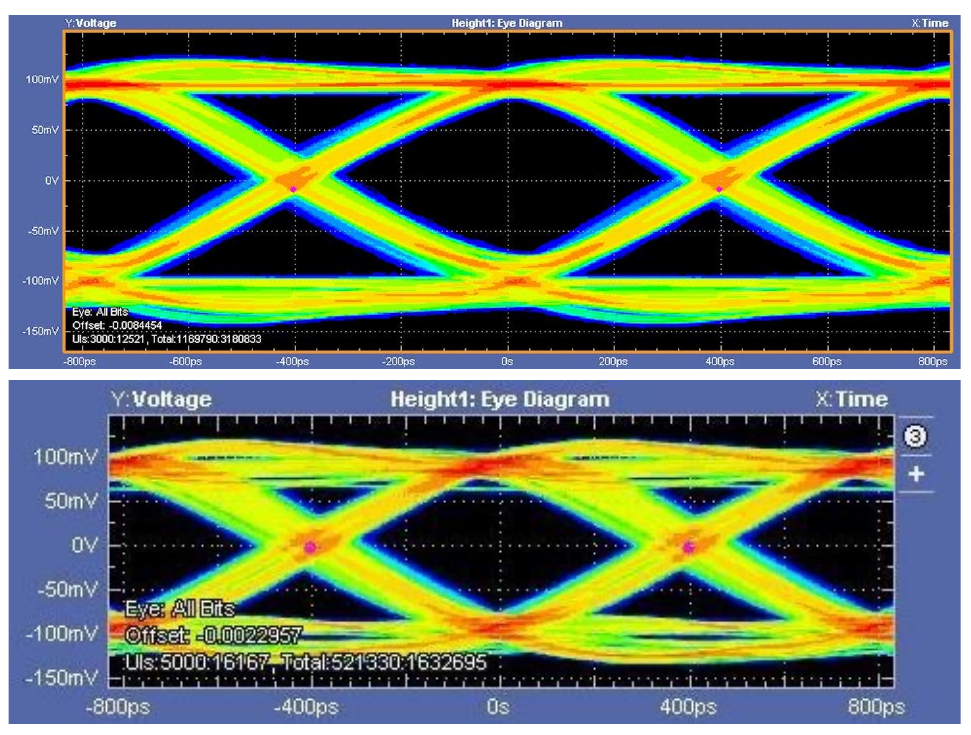}
\end{dunefigure}

%%%%%%%%%%%%%%%%%%%%%%%%%%%%%%%%%%%
\subsubsection{Infrastructure Inside the Cryostat}
\label{sec:fdsp-tpcelec-design-infrastructure}

Each \dword{femb} is enclosed in a \dword{ce} box 
which is connected to the ground of the \dshort{femb} and
provides mechanical support for the \dshort{femb} and cable strain relief
(see Figure~\ref{fig:ce-box}). 
A \SI{0.25}{inch} copper braid provides
the electrical connection between the \dword{ce} box and the \dword{crp} 
copper reference plane, as discussed %defined 
in Section~\ref{sec:fdsp-tpcelec-design-grounding}.

\begin{dunefigure}
[Aluminum CE Box for DUNE]
{fig:ce-box}
{Picture of an open \dshort{ce} box (left) and a \dshort{ce} box with the lid on (right) mounted on a \dshort{crp}.}
\includegraphics[width=0.9\linewidth]{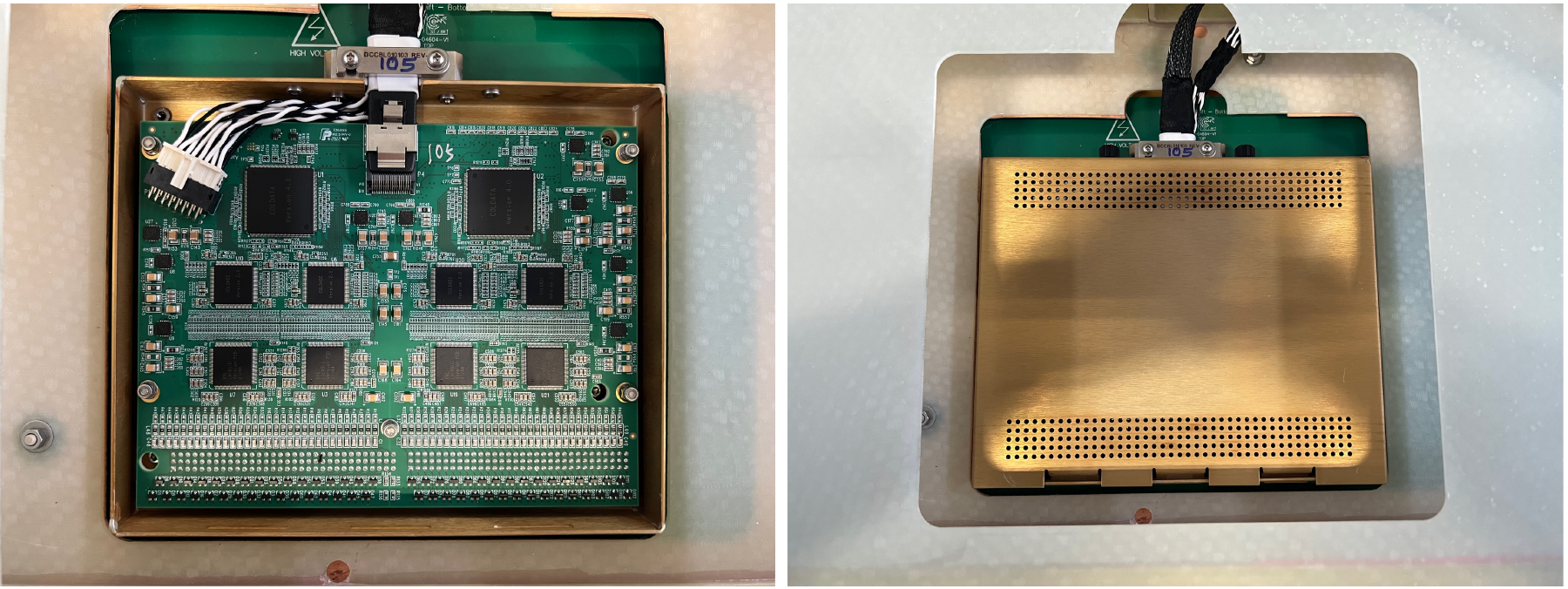}
\end{dunefigure}

Twelve \dshort{femb}s are mounted on each half CRP and connected to patch panels (also mounted on the half CRPs) by \SI{2.5}{m} long power cables and \SI{2.5}{m} long miniSAS data cables as shown in Figure~\ref{fig:ce-boxs_on_CRP}.  \SI{25}{m} long power and data cables are connected to the patch panels.  These cables are \SI{3}{m} longer than the power and data cables used in \dword{sphd} but otherwise identical.  The long cables will be routed from the patch panels on the \dshort{crp}s along the floor of the cryostat to the long walls.  They will be supported by ``rope ladders'' on the side of the cryostat and routed to penetrations located close to the long side of the cryostat. 

\begin{dunefigure}
[FEMB CE boxes mounted on CRPs and cabled to patch panels]
{fig:ce-boxs_on_CRP}
{Cable routing between \dshort{femb}s and patch panels on a CRP.}
\includegraphics[width=0.9\linewidth]{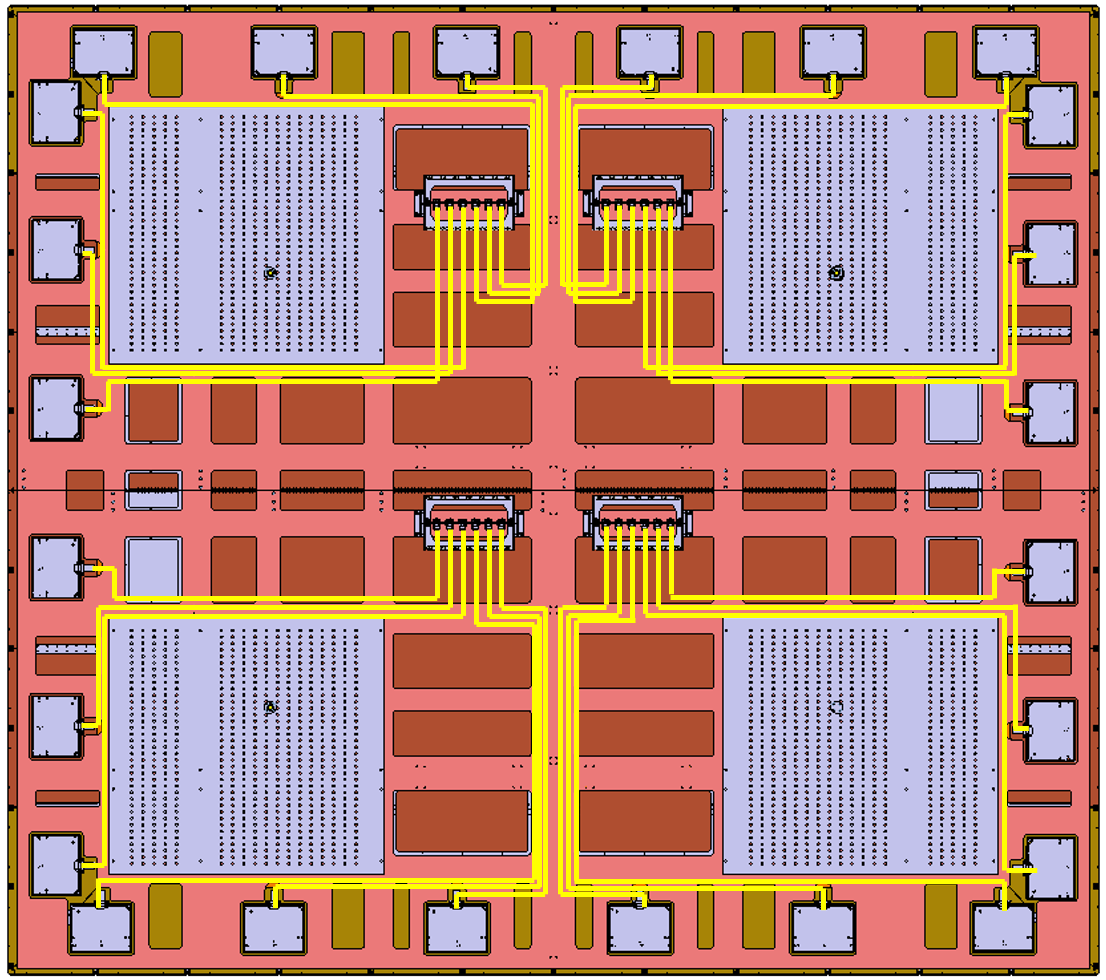}
\end{dunefigure}

%%%%%%%%%%%%%%%%%%%%%%%%%%%%%%%%%%%
\subsubsubsection{Cold Cables and Cold Electronics Feedthroughs}
\label{sec:fdsp-tpcelec-design-ft}

All cold cables originating inside the cryostat connect to the outside 
warm electronics through penetrations in the top of the cryostat.
A fixture holding four threaded rods and a T-shaped spool piece are mounted on top of the cryostat penetration ``crossing tube.'' The threaded rods support a cable strain relief fixture positioned at the bottom of the crossing tube inside the cryostat (see Figure~\ref{fig:tpcelec-signal-ft}). \dword{bde} cold signal and power cables connect to feedthrough boards mounted on flanges on two of the three ports of the spool piece.  The third port is for photodetector cables and cables associated with cryogenic instrumentation and calibration systems.

\begin{dunefigure}
[TPC cold eletronics \fdth]
{fig:tpcelec-signal-ft}
{\dshort{bde} cryostat \fdth. Warm Electronics Interface Crates mount on the two horizontal flanges of the spool piece. The top flange is for photodetector cables and fibers. A fixture at the bottom of the crossing tube provides strain relief for the cable bundles.}
\includegraphics[width=0.9\linewidth]{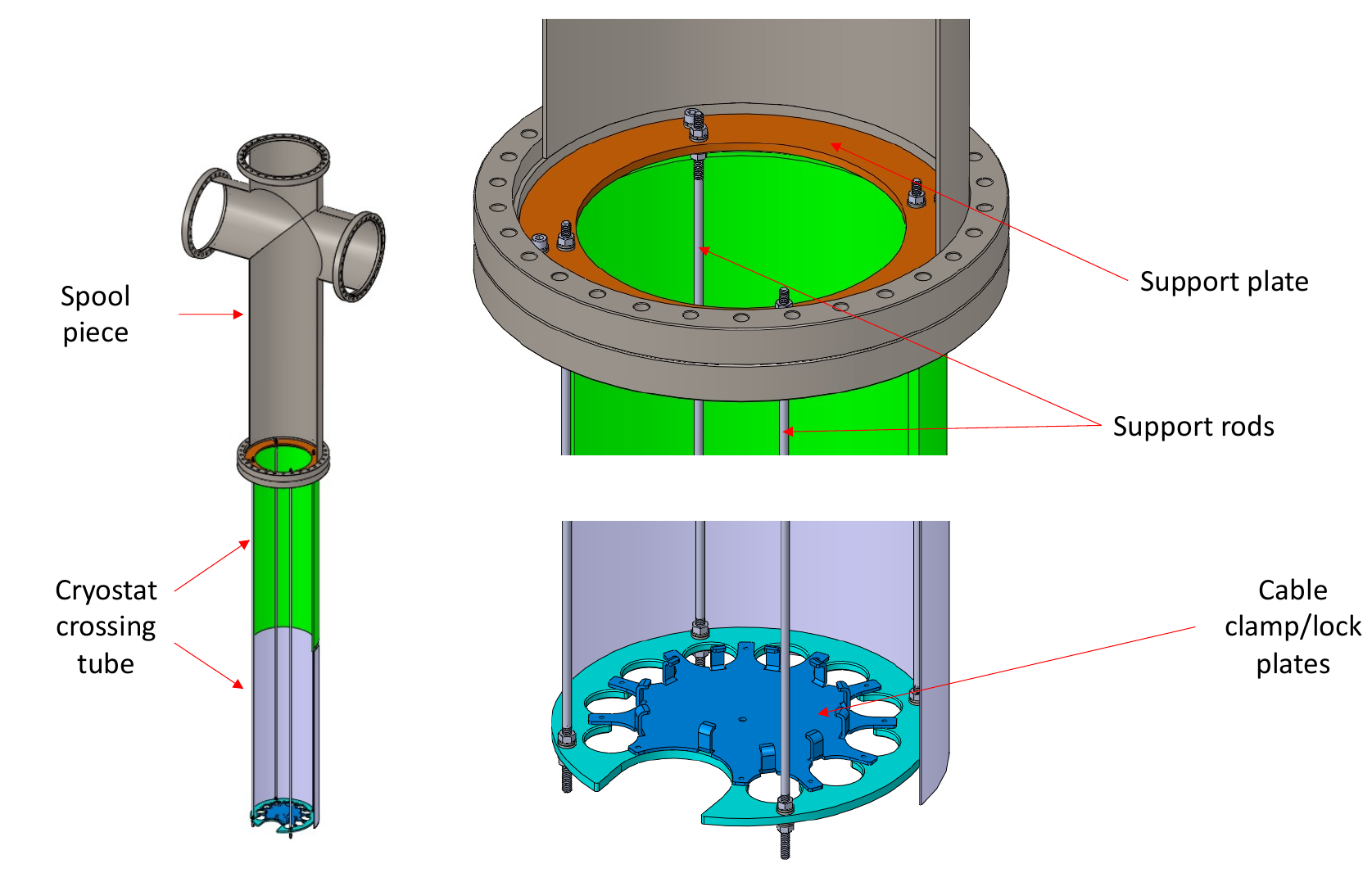}
\end{dunefigure}

Data and control cable bundles send system clock and control signals 
from the signal flange to the \dshort{femb}s and
high-speed data from the \dshort{femb}s to the signal flange. Each of the \num{24}
\dshort{femb}s on one \dshort{crp} connects to a flange via one data cable bundle.
Ten low-skew shielded 
twin-axial cables transmit the following signals
between the \dshort{wib} and the \dshort{femb}:
\begin{itemize}
\item four differential \SI{1.25}{Gbps} data links (two from each \dshort{coldata});
\item one \SI{62.5}{MHz} system clock (sent from the \dshort{wib});
\item one fast command line;
\item three \dshort{i2c}-like control lines (clock, data-in, and data-out);
\item one multipurpose pair of lines (described below in the section on the \dshort{wib}).
\end{itemize}

The \dword{lv} power is passed from the flange to the
\dshort{femb} by bundles of \num{16} \SI{20}{AWG} twisted-pair wires, with half of the wires serving as power feeds and the other half as returns. 
The bulk of the power consumed on a \dshort{femb} is used by the analog section of \dshort{coldadc}, which operates at \SI{2.25}{V}.  Most of the balance of the power is consumed by \dshort{larasic}, which operates at \SI{1.8}{V}.  For this reason, four wire pairs are devoted to supply voltage to the \dshort{coldadc} low dropout (LDO) linear regulators and two pairs are used to supply voltage to the \dshort{larasic} LDOs.  One pair is used to supply voltage to the \dshort{coldata} LDOs and one pair provides the \SI{5}{V} required by the LDOs themselves.

The cable plant for one \dshort{crp} in the \dshort{lar} also includes
the cables that provide the bias voltages applied to the
shield, first induction, and collection plane strips.
%$X$-, $U$-, and $G$-plane strips. 
The voltages are supplied
through three of eight \dword{shv} connectors mounted on the signal flange.
RG-316 coaxial cables carry the voltages from the signal flange to
a filter box mounted on the \dshort{crp}.

%%%%%%%%%%%%%%%%%%%%%%%%%%%%%%%%%%%
\subsubsection{Warm Interface Electronics}
\label{sec:fdsp-tpcelec-design-warm}

The warm interface electronics provide an interface between the 
\dword{ce} and the \dword{daq}, timing, and slow control systems. 
\dwords{wiec}) are attached directly to the \dshort{ce} 
flanges.  %The 
A \dword{wiec}, shown in Figure~\ref{fig:tpcelec-flange}, 
contains one \dword{ptc},
 six \dwords{wib}, 
 and a passive \dword{ptb}
that fans out clock signals and \dword{lv} power from the \dshort{ptc} to the 
\dshort{wib}s. The \dshort{wiec} provides a Faraday shield and 
robust ground connections from the \dshort{wib}s to the detector ground 
(Section~\ref{sec:fdsp-tpcelec-design-grounding}). Only optical
connections are used for the communication to the \dshort{daq} and the
slow controls.

\begin{dunefigure}
[Exploded view of the cold electronics signal flange for ProtoDUNE-SP]
{fig:tpcelec-flange}
{Exploded view of the \dshort{ce} signal flange.}
\includegraphics[width=0.9\linewidth]{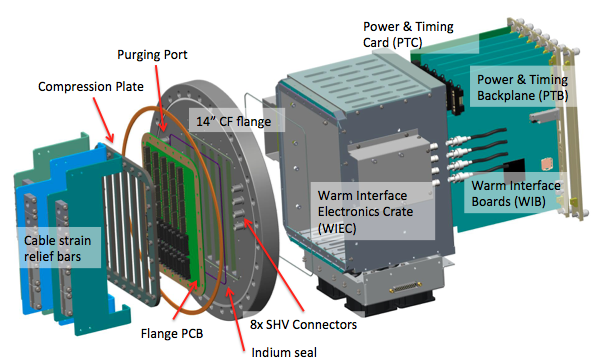}
\end{dunefigure}

\subsubsubsection{Power and Timing Card}
The \dshort{ptc} receives \SI{48}{V} \dshort{lv} power and steps the \SI{48}{V} down to \SI{12}{V}.  Filtered \SI{12}{V} is output from the \dshort{ptc} to the \dshort{wib}s on point-to-point connections on the \dshort{ptb}. The \dshort{wib}s in turn power the \dwords{femb} 
(see Figure~\ref{fig:tpcelec-wib-power}).
The \dshort{ptc} also contains a bidirectional fiber interface to the timing system.  The PTC fans out the encoded clock signal to the \dshort{wib}s via the \dshort{ptb}.

\begin{dunefigure}
[Low voltage power distribution to the \dshort{wib} and FEMBs]
{fig:tpcelec-wib-power}
{\dshort{lv} power distribution to the \dshort{wib} and \dshort{femb}s.}
\includegraphics[width=0.75\linewidth]{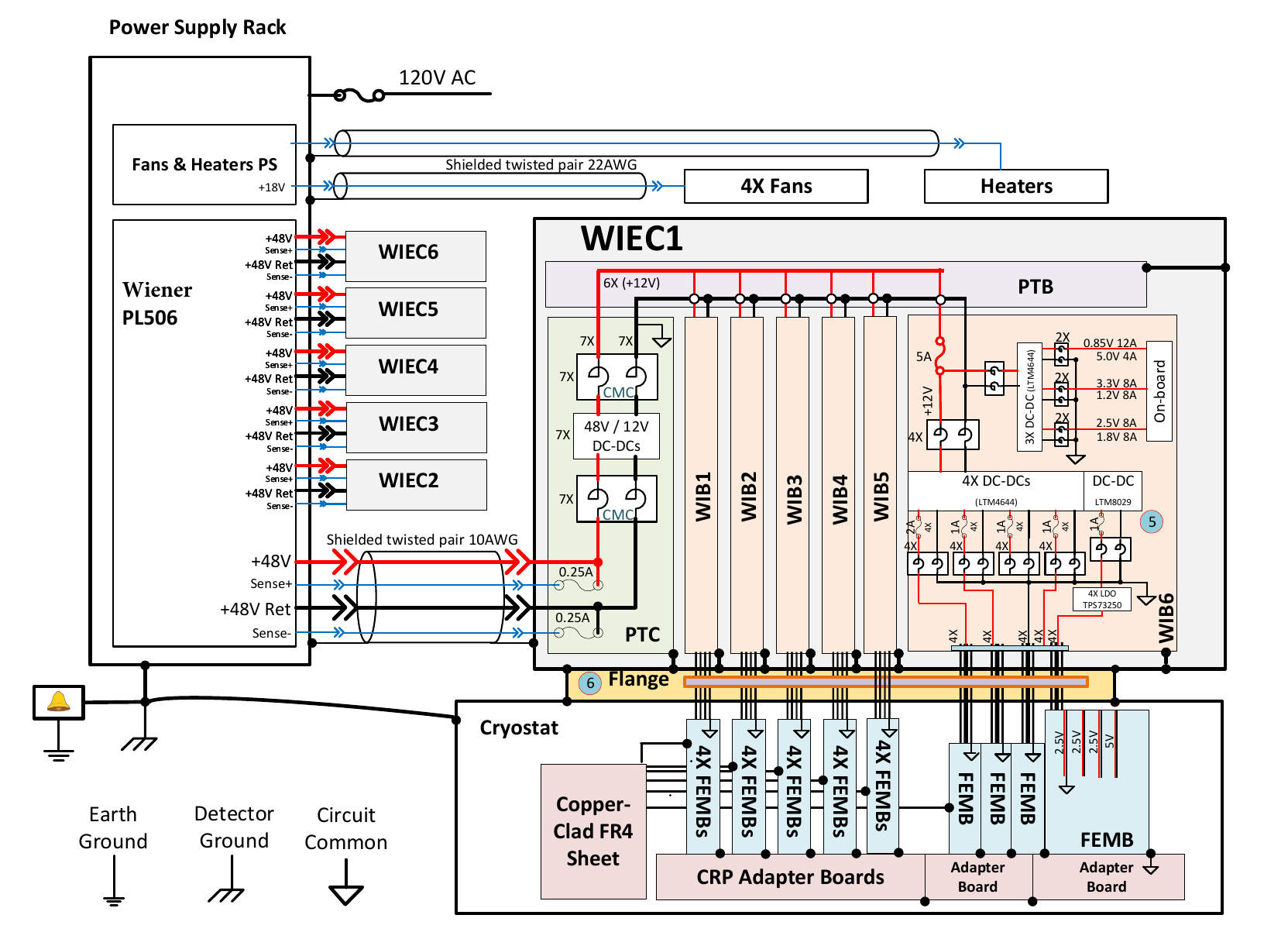}
\end{dunefigure}

\subsubsubsection{Warm Interface Board}

\begin{dunefigure}
[Warm interface board]
{fig:tpcelec-dune-wib}
{\dshort{wib} block diagram including the fiber optic connections to
the \dshort{daq} backend, slow controls, and the timing system, as well as the
data readout, clock, and control signals to the \dshort{femb}s.}
\includegraphics[width=0.75\linewidth]{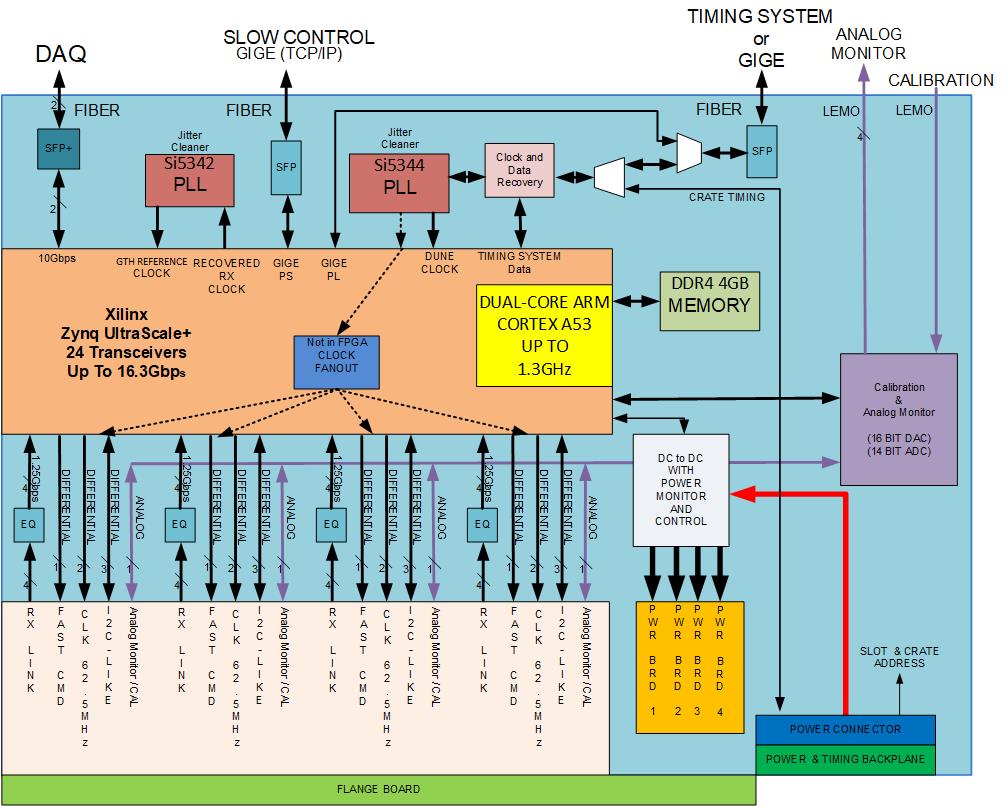}
\end{dunefigure}

A block diagram of the \dshort{wib} is shown in 
Figure~\ref{fig:tpcelec-dune-wib}.
The \dshort{wib} receives the system clock and control signals from the \dshort{ptc} that allow it to synchronize the \dshort{femb}s and appropriately time stamp data received from the \dshort{femb}s.
Each \dshort{wib} provides power to and controls four \dshort{femb}s and receives high-speed serial data on \num{16} \SI{1.25}{Gbps} links from the \dshort{femb}s.  It reformats these data and transmits data to the \dshort{daq} using two \SI{10}{Gbps} optical links.  A \SI{1}{Gbps} Ethernet interface provides connectivity between the \dshort{wib} and the slow controls system and also the configuration, control, and monitoring component of the data acquisition system.

The \dshort{wib}s are attached directly to the \dshort{tpc}
\dshort{ce} \fdth on the signal flange. 
The \fdth board is a multilayer \dshort{pcb} with
surface mount connectors on both sides electrically connected to one another using offset blind vias so there are no holes through the \dshort{pcb}.
Cable strain relief for the cold cables is 
provided from the back end of the \fdth.

All \dshort{wib} functions are controlled by a Xilinx Zynq Ultrascale+ system-on-a-chip.  Time-critical functions are implemented in the programmable logic and most other functions are implemented in software.  Communication between software and firmware is done via a memory map (``REG'' in Figure~\ref{fig:wiblogic}).  The software framework for the \dshort{wib} is based on a client server model.  A server process (written in C) running on the \dshort{wib}'s ARM cores provides a network API for client software running on other machines, either as part of DUNE \dshort{daq} or in a simpler bench-top setting.  An OPC/UA server is also implemented to allow DUNE Slow Controls to access the \dshort{wib} server.  The \dshort{wib} logical blocks and their interconnections are illustrated in Figure~\ref{fig:wiblogic}.

\begin{dunefigure}
[Warm interface board software and firmware]
{fig:wiblogic}
{\dshort{wib} logical blocks and interconnects.}
\includegraphics[width=0.7\linewidth]{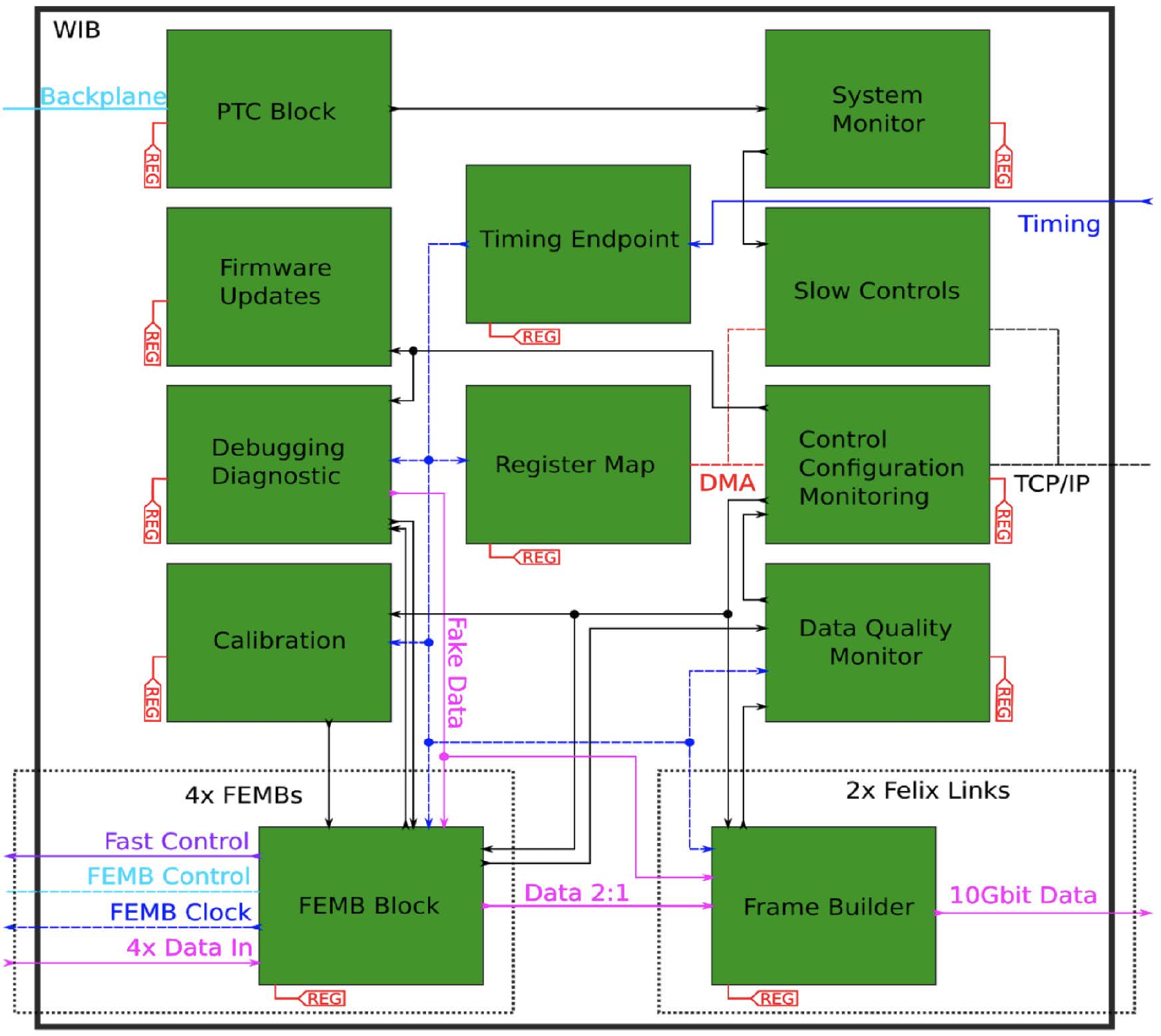}
\end{dunefigure}

%%%%%%%%%%%%%%%%%%%%%%%%%%%%%%%%%%%
\subsubsection{Services on Top of the Cryostat}
\label{sec:fdsp-tpcelec-design-services}

Each PTC receives \SI{48}{V} from a Wiener PL506 power supply installed on the top of the cryostat.
Four wires are used for each \dword{ptc} module; two \SI{10}{AWG}, shielded, twisted-pair 
cables for the power and return; and two \SI{20}{AWG}, shielded, twisted-pair 
cables for the sense. The primary protection is the over-current 
protection circuit in the \dword{lv} supply modules, which is set 
higher than the $\sim\SI{10}{A}$ current draw of the \dword{wiec}. 
Secondary sense-line fusing is provided on the \dshort{ptc}.  
The power cable shields can be connected at one or both ends of the cable.  In tests of the first full-scale \dword{crp} in June 2022, good noise results were obtained with the shields connected at both ends.  This will be verified in %ProtoDUNE-II 
\dword{vdmod0} running.

Switching power supplies controlled by the slow controls system provide
power to the heaters (\SI{12}{V}) and the fans (\SI{24}{V}) that are 
installed on the \dword{ce} flanges. Temperature sensors mounted on the
flanges, and power consumption and speed controls from the fans are 
connected to the interlock system that is part of the \dword{ddss}, in 
%\dword{ddss}, in
addition to being monitored by the slow controls system.

RG-58 coaxial cables connect the strip bias voltages from bias voltage
supplis provided by the high voltage consortium to the standard \dword{shv} connectors that are machined directly 
into the \dshort{ce} \fdth and insulated from the low voltage and 
data connectors.

Optical fibers are used for all connections between the \dshort{wiec}s 
and the \dshort{daq} and slow 
control systems.

To support the electronics, fan, and heater power cables, as well 
as optical fibers on top of the cryostat, cable trays are installed
below the false flooring on top of the cryostat. 
All the necessary \dshort{lv} supplies and %, in addition to 
the bias
voltage supplies are installed in these racks. Patch panels for
the optical fiber plant used for the control and readout of the
detector are also installed on the detector mezzanine.

%%%%%%%%%%%%%%%%%%%%%%%%%%%%%%%%%%%
\subsubsection{Summary of Differences between \dshort{bde} and \dshort{sphd} TPC Electronics}
The \dword{bde} are almost identical to the \dword{sphd} TPC electronics.  The only differences are:
\begin{itemize}
    \item {\dshort{bde} \dshort{femb}s uses miniSAS data connectors rather than Samtec data connectors.}
    \item{The \dshort{bde} uses longer cold cables than are used in \dshort{sphd} and patch panels on the \dshort{crp}s. The patch panels allow the \dshort{femb}s to be installed on the \dshort{crp}s and cabled to the patch panels at the \dshort{crp} factories, and also allow the long cold cables to be installed into \dshort{spvd} before the \dshort{crp}s are installed.}
    \item{Short mini-SAS data cables carry signals between the \dwords{femb} and the patch panels.}
    \item{The routing of cold cables in \dshort{sphd} and \dshort{spvd} is different.}
    \item{The method of strain relieving the cold cables at the bottom of the cryostat crossing tubes is slightly different in \dshort{sphd} and \dshort{spvd}.}
    \item{Six \dshort{wib}s are used in each \dshort{spvd} \dword{wiec} and five \dshort{wib}s are used in each \dshort{sphd} \dshort{wiec}.}
\end{itemize}
\subsection{Performance with \dshort{crp}}
As detailed in Section ~\ref{subsec:FirstFullCRP}, the first full-scale \dword{crp} prototype was built so that one half of the \dshort{crp} could be read out using \dword{bde} and the other half \dword{tde}.  Since monolithic \dwords{femb} were not yet available, \dword{protodune}-style \dshort{femb}s were used for tests of this \dshort{crp}.  Tests were carried out in two phases.  During the first phase, the \dshort{bde} suffered from a relatively high level of coherent noise.  The second phase of running included improved grounding and the \dshort{bde} performed well with very little coherent noise.

\begin{dunefigure}
[Cosmic ray track from CRP-4 (BDE)]
{fig:bde_crp4_track}
{A sample cosmic ray track recorded using CRP-4 read out with \dshort{bde}. The individual channel waveforms shown are for strips that recorded ionization from multiple tracks.}
\includegraphics[width=0.95\linewidth]{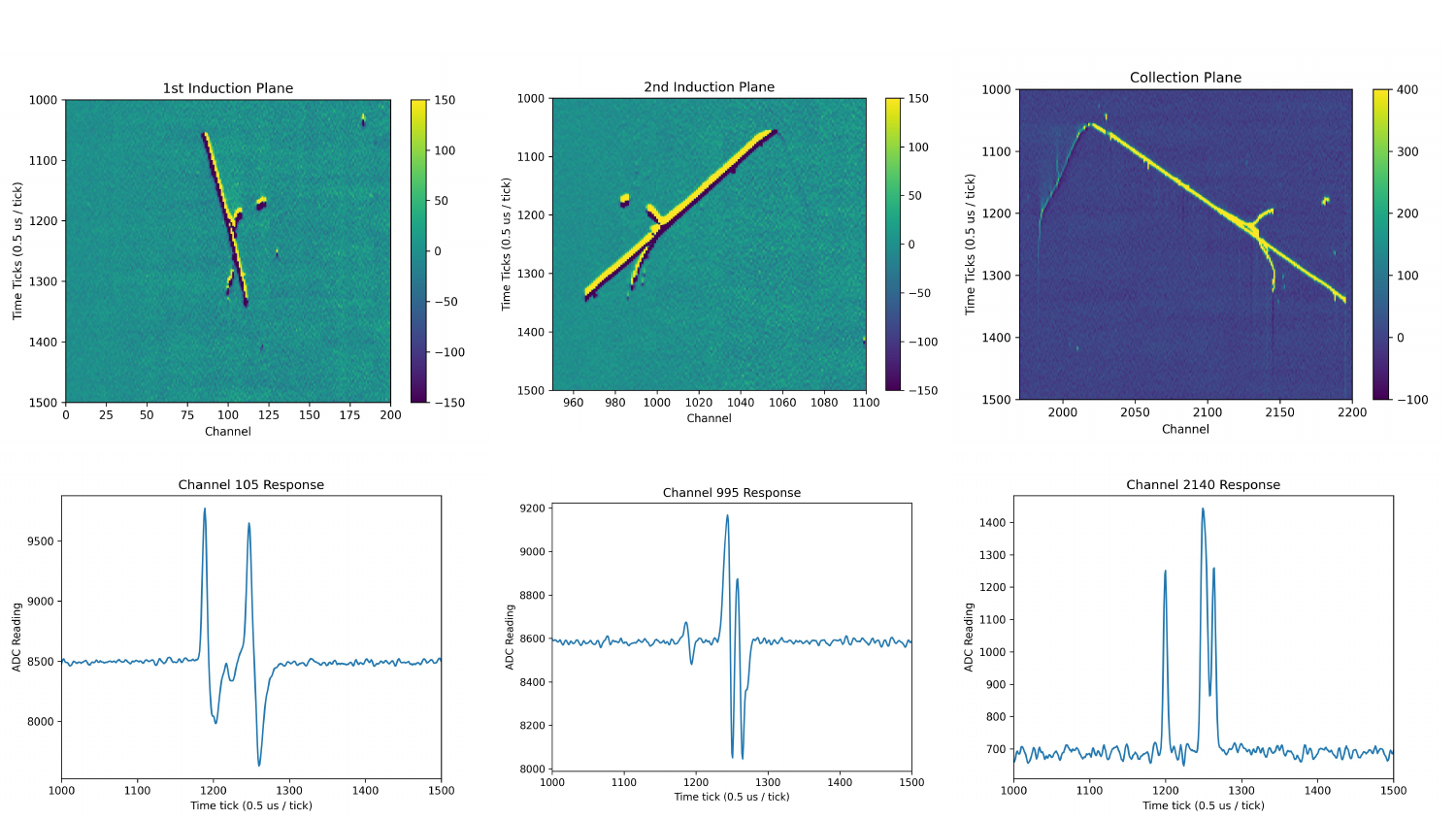}
\end{dunefigure}

One half of \dshort{crp}-5, instrumented with 12 monolithic \dshort{femb}s, has been tested in LN2 in a cold box at BNL. The noise performance was in line with expectations.  The average noise (before coherent noise removal) for full length strips from the first and second induction layers is $\sim$\SI{585}{e^-} and $\sim$\SI{560}{e^-} and the average noise for collection strips is $\sim$\SI{465}{e^-}.  Recently, \dshort{crp}-5 and \dshort{crp}-4 have been tested in the cold box at NP02 using the final \dword{bde}.  The measured noise in these cold box runs was also in line with expectations and very little coherent noise was observed.  A sample event taken with \dshort{crp}-4 (without coherent noise removal) is shown in Figure~\ref{fig:bde_crp4_track}.
\subsection{Quality Assurance} % and Control} 
\label{sec:fdsp-tpcelec-qa}

%The goal of the \dword{qa} plan is 
DUNE aims to maximize the number of 
readout channels in \dword{spvd} %the detector 
that achieve the performance specifications % for the detector 
discussed in Section~\ref{sec:tpc-elec:spec}.
Particular care must be used for
the \dword{ce} components that will be installed inside the cryostat, since they cannot %will not be able to 
be repaired or replaced once the cryostat is closed.

A complete \dword{qa} plan starts with ensuring that the designs of all
detector components fulfill the specification criteria, considering
also system aspects, i.e. how the various detector components interact
among themselves and with the detector components provided by other 
consortia. %We discuss validating the design in 
Section~\ref{sec:fdsp-tpcelec-qa-initial} discusses design validation. 
The other aspects of the \dshort{qa} plan involve documenting the 
assembly and testing processes, storing and analyzing the information
collected during the \dword{qc} process, training and qualifying 
personnel from the consortium, monitoring procurement of 
components from external vendors, and assessing whether the
\dshort{qc} procedures are applied uniformly across
the various sites involved in detector construction, integration,
and installation. The \dword{tpc} electronics consortium plan involves
having multiple sites using the same \dshort{qc} procedures,
many of which will be developed as part of system design tests during the \dshort{qa} phase,
with the possibility of a significant turnover in the personnel
performing these tasks. To avoid problems during most of the
production phase, %we plan to emphasize 
training as well as documentation
of the \dshort{qa} plan will be emphasized. Reference parts will be tested at
several sites to ensure consistent results. At
a single site, some parts will be tested repeatedly to ensure
that the response of the apparatus does not change and
that new personnel involved in testing detector components are 
as proficient as more experienced personnel.

%%%%%%%%%%%%%%%%%%%%%%%%%%%%%%%%%%%
\subsubsection{Initial Design Validation}
\label{sec:fdsp-tpcelec-qa-initial}

As described in Section~\ref{sec:fdsp-tpcelec-design}, three \dwords{asic} 
have been developed for the \dshort{dune} \dword{fd} \dshort{tpc} readout 
(\dword{larasic}, \dword{coldadc}, and \dword{coldata}). 
As each new prototype \dshort{asic} was produced, the groups responsible for the \dshort{asic} design performed the first tests of 
\dshort{asic} functionality and performance. Some of these tests used bare chips mounted directly on a printed circuit board 
and wire bonded to the board while others used package parts.  The goal of these tests was to determine 
the extent to which the \dshort{asic} functions as intended, both at room 
temperature and at \lntwo temperature.  For all chips, these tests 
included exercising digital control logic and all modes of operation. Tests 
of \dshort{fe} \dshort{asic}s included measurements of noise levels as a function 
of input capacitance, baseline recovery from large pulses, cross-talk, linearity, 
and dynamic range. Tests of \dwords{adc} included measurements of the effective noise levels and 
of differential as well as integral non-linearity. Tests of the \dshort{coldata} \dshort{asic}
included verification of both the control and high-speed data output links using 
cables with lengths up to \SI{35}{m}.
After the initial functionality
tests, the \dshort{asic}s
were mounted on \dshort{femb}s so measurements could be repeated with varying input loads and
with real \dwords{apa}. 

%%%%%%%%%%%%%%%%%%%%%%%%%%%%%%%%%%%
\subsubsection{Cryogenic Tests}
\label{sec:fdsp-tpcelec-qa-cryo}

Tests of \dshort{asic}s and \dshort{femb}s in a cryogenic environment
are performed in \lntwo instead of \dshort{lar} for cost reasons, ignoring
the small temperature difference. These tests can be performed immersing
the detector components in a dewar containing \lntwo for the duration
of the tests. Condensation of water from air can interfere with
the tests or damage the detector components or the test equipment,
particularly during their extraction from the \lntwo. A test dewar
design developed by Michigan State University, referred to as the
\dword{cts} (Cryogenic Test System), has been developed to avoid
this problem. Several \dshort{cts} units
were deployed at \dword{bnl} during the \dword{pdsp} construction
and used for the \dshort{qc} on the \dshort{asic}s and \dshort{femb}s
for \dshort{pdsp}. Later they were also used to perform similar tasks
during the construction of the electronics for \dword{sbnd}.
Several other \dshort{cts} units have been deployed to institutions involved in
developing \dshort{asic}s to test the first prototypes of \dshort{asic}s
and \dshort{femb}s for the \dshort{dune} \dshort{fd} as well as for \dshort{qc} tests. 

Two \dshort{cts} units 
in operation at \dshort{bnl} are shown in Figure~\ref{fig:CTS}.

\begin{dunefigure}
[The Cryogenic Test System]
{fig:CTS}
{Cryogenic Test System (CTS): an insulated box is mounted on top of a commercial \lntwo dewar.  Simple controls allow the box to be purged with nitrogen gas and \lntwo to be moved from the dewar to the box and back to the dewar.}
\includegraphics[width=0.4\linewidth]{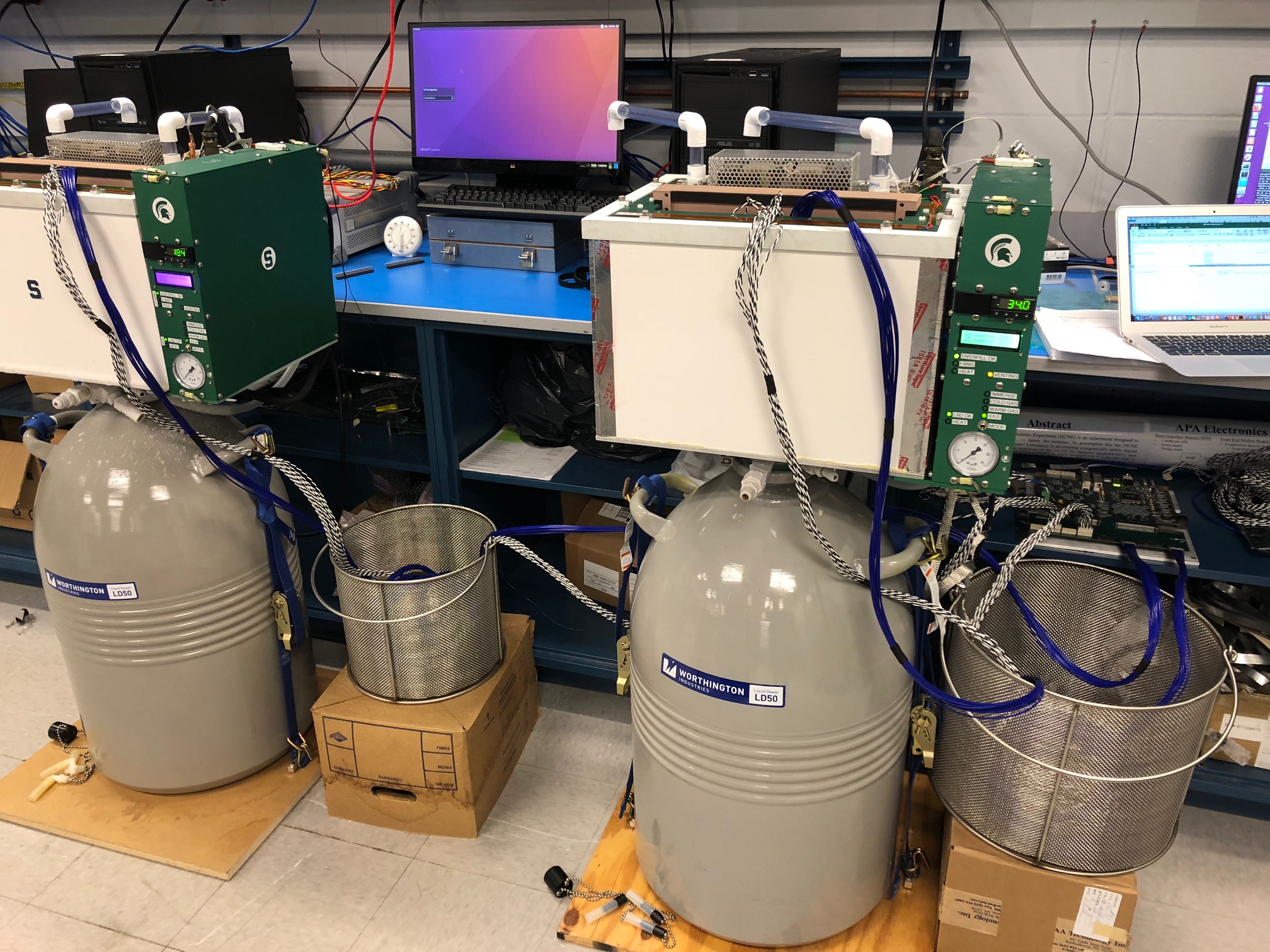}
\end{dunefigure}

%%%%%%%%%%%%%%%%%%%%%%%%%%%%%%%%%%%
\subsection{Reliability Studies}
\label{sec:fdsp-tpcelec-qa-reliability}

The \dword{tpc} \dword{ce} system of the \dshort{dune} \dword{spvd} \dshort{fd} must meet 
stringent requirements, including a very small number of failures ($<\num{1}$\% of the total number of
channels) for components installed 
on the detector inside the cryostat during the \dunelifetime of 
detector operation.

A few \dword{hep} detectors have operated without intervention for a 
prolonged period, with few readout channel losses, in extreme 
conditions that are similar to those in the \dshort{dune} \dshort{fd} cryostats:
\begin{itemize}
\item The NA48/NA62 liquid krypton (LKr) calorimeter has 13,212 channels 
of JFET pre-amplifiers installed on the detector. It has been kept at LKr temperature 
since 1998. The total fraction of failed channels is $<$ 0.2\% in more than \num{24} years of operation.
\item The ATLAS \dshort{lar} accordion electromagnetic barrel calorimeter has 
approximately 110,000 readout signal channels, with up to seven connections and different 
circuit boards populated with resistors and diodes inside the cryostat. This
calorimeter has been cold since 2004, for a total of \num{18} years of operation. So far, the
number of readout channels that have failed is approximately 0.02\% of the total channel count.
\item The ATLAS \dshort{lar} hadronic endcap calorimeter has approximately 35,000 GaAs
pre-amplifers summed into 5,600 readout channels that are mounted on cold pre-amplifier
and summing boards. The ATLAS \dshort{lar} hadronic endcap 
calorimeter \dshort{ce} have been in cold since 2005, with
$<$ 0.1\% of the channels failing during 17 years of operation. 
\end{itemize}
Neither NA48/NA62 nor the ATLAS \dshort{lar} hadronic endcap calorimeter
uses \dword{cmos} electronics; however, the procedures used in the construction and \dshort{qc} of \dwords{pcb} and 
for the selection and \dshort{qc} of connectors and discrete components mounted
on the \dshort{pcb}s are directly relevant for the \dshort{dune} \dshort{fd}.

A list of reliability topics to be studied (including some that have already been completed and others that have started) for the \dshort{tpc} electronics operated 
in \dshort{lar} environment are:
\begin{itemize}
\item For \dword{cots} components, accelerated lifetime testing, a methodology 
developed by \dword{nasa}~\cite{nasa_nepp} will be used to verify the expected
lifetime of operation at cryogenic temperatures. A \dshort{cots} \dword{adc} has undergone
this procedure and has been qualified as a solution for the SBND experiment~\cite{Chen:2018zic}.
\item The custom \dshort{asic}s incorporate design rules 
intended to minimize the hot-carrier effect~\cite{Li:CELAr,Hoff:2015hax},
which is recognized as the main failure mechanism for integrated circuits
operating at \dshort{lar} temperature.  Extensive lifetime studies of individual transistors used in \dword{larasic} were performed early in its development~\cite{Li:CELAr,Ma:2015}.  The lifetime of \dshort{coldadc} has been verified using the accelerated lifetime testing methodology, and a similar verification of the lifetime of \dshort{coldata} will be completed before the production order for \dshort{coldadc} and \dshort{coldata} is placed.
\item Printed circuit board assemblies are designed and fabricated to survive 
repeated immersions in \lntwo.
\item In addition to the \dshort{qa} studies noted above, a very detailed
and formal set of \dshort{qc} checks of the production pieces will be required in order to ensure
a reliable detector. The \dshort{qc} plans for the 
BDE
detector components are discussed in Section~\ref{sec:fdsp-tpcelec-production-qc}.
\end{itemize}

\subsection{Production and Quality Control} %Assembly}
\label{sec:fdsp-tpcelec-production}

This section,  discusses the production and assembly plans,
including the plans for the spares required during the detector
construction and for operations, for
procurement, assembly, and quality control.

%%%%%%%%%%%%%%%%%%%%%%%%%%%%%%%%%%%

\subsubsection{Spares Plan}
\label{sec:fdsp-tpcelec-production-spares}

The bottom anode of the \dshort{spvd} will consist of 80 \dwords{crp}.  This means that at least 1920 \dwords{femb} with the corresponding bundles of cold 
cables will be required for the integration.
\dword{apa}s). 
To have spare \dshort{femb}s, the \dword{tpc} electronics consortium plans to
build 2100 \dshort{femb}s, about 10\% more than required to instrument 80 \dshort{crp}s. If more spares are needed during the
\dword{qc} process or during integration, additional 
\dshort{femb}s can be produced quickly as long as any components that have 
long lead times are on hand. For these components, it is planned to keep on hand a
larger number of spares. The \dshort{asic}s require a long lead time; a plan for those spares is
discussed below.

In the case of \dword{larasic}, the number of spare chips is driven by concern that it may soon become impossible to have devices fabricated using the 
\SI{180}{nm} \dword{cmos} process. Consequently, it was decided to hold a \dword{prr} for \dshort{larasic} and purchase chips.  The review committee recommended the immediate purchase of 250 wafers using the mask set for the engineering run of \dshort{larasic}.  Each wafer contains 310 final-design \dwords{larasic}, so this is a sufficient number for both of the first two DUNE far detector modules, even making pessimistic assumptions of chip yield.

In the case of \dword{coldadc} and \dword{coldata}, the number of spares expected is driven by the fact that these two designs are implemented on the same wafer and that production wafers must be purchased in batches of 25.
The expected number of chips per wafer is about
850 for \dshort{coldadc} and 275 for \dshort{coldata}. 

Assuming an overall yield of $85$\% ($5$\% loss in dicing and packaging and approximately $10$\% failure rate in the \dshort{qc} testing), 34 wafers are required for the \dword{sphd}.
Since wafers must be fabricated in batches of 25, 50 wafers will be purchased for \dshort{sphd}.  %We expect 
It is planned to purchase another batch of 25 wafers for \dword{spvd}. 

In general, for other components, %we 
it is planned to procure between 5 and
10\% additional components for spares for the construction of the 
\dshort{spvd} module.% We will need m
More spares will be needed for components that have
a larger risk of damage during integration and 
installation.

The components on top of the cryostat (power supplies, bias
voltage supplies, cables, \dwords{wiec} with their \dword{wib}
and \dword{ptc} boards) can be replaced while the
detector is in operation. For these components, additional spares may be required
during the \dunelifetime operation period of \dshort{spvd}.
The plan is to purchase 10\% additional components for spares.

%%%%%%%%%%%%%%%%%%%%%%%%%%%%%%%%%%%
\subsubsection{Procurement of Parts}
\label{sec:fdsp-tpcelec-production-procurement}

The construction of the detector components for \dshort{dune} requires many large procurements that 
must be carefully planned to avoid delays. For the \dwords{asic}, the 
choice of vendor(s) is made at the time the technology used in designing 
the chips is chosen. For almost all other components, several vendors 
will bid on the same package. Depending on the requirements of the funding
agency and of the responsible institution, this may require a lengthy
selection process. The cold cables used to transmit data from the
\dshort{femb}s to the \dwords{wib} represent a critical case. In this case
a technical qualification, including tests of the entire chain (from the \dshort{femb}
to the receiver on the \dshort{wib}) is required.

%%%%%%%%%%%%%%%%%%%%%%%%%%%%%%%%%%%
\subsubsection{Assembly}
\label{sec:fdsp-tpcelec-production-assembly}

The \dshort{tpc} electronics consortium plans to minimize
the amount of assembly work at any one of the participating
institutions. 
For instance, all printed circuit boards will be assembled by commercial fabricators.
One of the few exceptions is the assembly of \dshort{wiec}s, since this involves the integration of the power and timing backplane and installation on a feedthrough flange.

%%%%%%%%%%%%%%%%%%%%%%%%%%%%%%%%%%%
\subsubsection{Quality Control}
\label{sec:fdsp-tpcelec-production-qc}

Once the \dwords{crp} are installed inside the cryostat, 
only limited access to the detector components will be available to the \dshort{tpc} electronics
consortium. After the \dword{tco} is closed, no access to detector
components will be available; therefore, they should be constructed to
last the entire lifetime of the experiment (\dunelifetime). This
puts very stringent requirements on the reliability of these
components, which has been already addressed in part through 
the \dword{qa} program discussed in Section~\ref{sec:fdsp-tpcelec-qa}. The
next step is to carefully apply stringent \dword{qc} procedures for  
detector parts to be installed in the detector.
All detector components installed inside the cryostat will
be tested and sorted before they are prepared for integration
with other detector components prior to installation.
A preliminary version of the \dshort{qc} process is being used as %we fabricate 
parts are fabricated for use in \dword{vdmod0}.  These processes will be further developed and fully documented before the \dshort{prr}s for each specific part.

Some of the requirements for the \dshort{qc} plans can
be laid out now based on the lessons learned
from constructing and commissioning the \dword{pdsp}
detector. A small but significant fraction ($\sim 4\%$) of the \dshort{larasic} chips produced for \dshort{pdsp} passed \dshort{qc} tests at room temperature but failed when immersed in \lntwo.  Based on that experience,% we are planning 
it is planned to test all \dshort{asic}s in \lntwo before they are mounted on the \dshort{femb}s.  If the fraction of \dshort{asic}s that fail cold after passing \dshort{qc} tests at room temperature proves to be very low, %we will reconsider 
this plan will be reconsidered.

Based on experiences at \dshort{pdsp},
discrete components like resistors and capacitors
need not undergo cryogenic testing before they are installed
on the \dshort{femb}s.

\dword{asic} testing is performed with dedicated test boards that
allow tests of the functionality of the chips.
The dedicated test boards reproduce the
entire readout chain where the input to the \dword{fe} amplifier
or to the \dword{adc} is replaced by an appropriate signal generator,
and some parts of the backend may be replaced by a simple \dword{fpga}
that is directly connected to a computer. Tests of the \dshort{femb}s
can be performed by connecting them directly to a standalone \dword{wib},
as discussed in Section~\ref{sec:fdsp-tpcelec-design-warm}. Given
the large number of \dshort{asic}s and \dshort{femb}s required for
one \dshort{dune} %\dshort{fd} \dword{sp} detector, 
\dword{detmodule},
corresponding \dshort{qc} activities wil be distributed among multiple institutions
belonging to the \dshort{tpc} electronics consortium. Up to six
test sites are needed for the \dshort{asic}s plus an additional
five sites for the \dshort{femb}s, with each test 
equipped with a cryogenic system such as the \dword{cts}. All tests
will be performed following a common set of instructions.

The choice of distributing the testing activities among multiple
institutions has been made based in part on the experience gained
with \dshort{pdsp}, where all associated testing activities were concentrated
at \dword{bnl}. While this approach had some advantages, like the direct availability
of the engineers that had designed the components, a strict conformance
to the testing rules, and a fast turn-around
time for repairs, it also required a very large commitment of
personnel from a single institution. Personnel from other institutions
interested in the \dshort{tpc} electronics participated in the test
activities but could not commit for long periods of time. For
this reason, %we are planning 
it is planned to distribute the \dshort{qc} testing
activities for \dshort{asic}s and \dshort{femb}s among multiple
institutions belonging to the \dshort{tpc} electronics consortium.
It should be noted that this approach is used in the \dword{lhc}
experiments for detector components like the silicon tracker 
modules where both the assembly and \dshort{qc} activities take
place in parallel at multiple (of the order of ten) institutions.
To ensure that all sites produce similar
results, %we will emphasize 
emphasis will be placed on training experienced personnel
that will %overview 
oversee the testing activities at each site,
and% we will have 
a reference set of \dshort{asic}s and \dshort{femb}s
%that 
will be initially used to cross-calibrate the
test procedures among sites and then to check the
stability of the test equipment at each site.

Criteria will be developed for the acceptance of \dshort{asic}s and \dshort{femb}s.  
The procedures adopted for detector construction will evolve from the experience gained with \dword{pdsp} and the \dwords{mod0}. The \dshort{qc} procedures will be reviewed before the \dword{prr} that triggers the beginning of production. During production, the results of the \dshort{qc} process will be reviewed at regular intervals in production progress reviews. 
 The yields and acceptance rate of the production will be centrally monitored and compared among different sites. 
In case of problems, the failures will be analyzed and root cause analyses will be performed. If necessary, production and/or the test program will be stopped at all sites while issues are being investigated, 
followed by changes in the procedures if necessary. All data from the \dshort{qc} process will be stored in a common database.

Since the number of \dshort{asic}s to be tested is very large,
%we are developing 
a robotic system 
is being developed at Michigan State University that includes a pick-and-place robot and a modified version of the \dshort{cts}.  

After assembly, 
each \dshort{femb} will be tested in \lntwo using
the current \dshort{cts} design. 

In the test, each channel of the \dshort{femb} is connected to a \SI{150}{pF} capacitive load. This allows
connectivity checks for each channel as well as measurements of
the waveform baseline and of the channel noise level. Calibration 
pulses will be injected in the front-end amplifier, digitized,
and read out. 
The test setup requires one \dshort{wib} and
a printed-circuit board similar to those used on the cryostat
penetration, allowing simultaneous testing of four \dshort{femb}s.
A standalone \SI{12}{V} power supply is required, and the readout
of the \dshort{wib} uses a direct Gb Ethernet connection to
a PC. The setup used for \dshort{asic} testing is similar.
In both cases, the data can be processed locally on the PC,
and the results from the tests and calibrations are then stored 
in a database. The plan is to have the capability to retrieve  
these test and calibration results throughout the entire life
of the experiment. As in the case of \dshort{asic} testing,
%we will monitor 
the test results will be monitored to ensure that all
sites have similar test capabilities and yields and to
identify possible problems during production.

The tested \dshort{femb}s will be installed on \dshort{crp}s and cabled to patch panels mounted on the \dshort{crp}s at the \dshort{crp} factories.
The \dshort{bde} consortium will provide a \dshort{wiec} with six \dshort{wib}s, and a \dword{ptc} to each \dshort{crp} factory.  The \dshort{bde} consortium will also provide the necessary firmware, software, and training so that \dshort{crp} consortium members can verify the successful installation and cabling in warm and cold tests.

Stringent requirements must be applied to the cryostat penetrations in order to ensure that the liquid argon is not contaminated by nitrogen or oxygen.
%in order to avoid argon leaks. 
The cryostat penetrations 
have two parts: the first is the crossing tube with its spool pieces,
and the second one is the three flanges used for
connecting the power, control, and readout electronics to the
\dword{ce} and \dword{pds} components inside the
cryostat. On each cryostat penetration there are two flanges for
the \dshort{ce} and one for the \dword{pds}. The crossing
tubes with their spool pieces are fabricated by industrial vendors and pressure-tested
and tested for leaks by other vendors. The flanges are assembled
by institutions that are members of the \dshort{tpc} electronics and \dshort{pds} consortia; the
flanges must undergo both electrical and mechanical tests to ensure their
functionality. Electrical tests comprise checking all of the
signals and voltages to ensure they are passed properly between the two sides of the
flange and that there are no shorts. Mechanical tests involve 
pressure-testing the flange itself, including checking for leaks. Further
leak tests are performed after the cryostat penetrations are installed
on the cryostat and later after the \dshort{tpc} electronics and \dshort{pds}
cables are attached to the flanges. These leak tests are
performed by releasing helium gas in the cryostat penetration and
checking for the presence of helium on top of the cryostat. Similar
tests were performed during the \dshort{pdsp} installation.

All other detector components that are a responsibility of
the \dshort{tpc} electronics consortium can be replaced, if necessary,
even while the detector is in operation. Regardless, every component
will be tested before it is installed in \dword{surf} to ensure
smooth commissioning of the detector. The \dshort{wiec}s will be
assembled and tested with all of the \dshort{wib}s and \dshort{ptc}
installed. Testing requires a slice of the \dword{daq} back-end,
power supplies, and at least one \dshort{femb} to check all 
connections. All cables between the bias voltage supplies and
the end flange, as well as all of the cables between the low-voltage power
supplies and the \dwords{ptc} will be tested for electrical
continuity and for shorts. All power supplies will undergo a
period of burn-in with appropriate loads before being installed
in the cavern. Optical fibers will be tested by measuring the
eye diagram for data transmission at the required speed. All
test equipment used for qualifying the components to be installed
in the detector will be either transported to \dshort{surf} or duplicated
at \dshort{surf} in order to be used as diagnostic tools during operations.

\subsection{Installation, Integration, and Commissioning}
\label{sec:fdsp-tpcelec-IandI}

The installation of \dword{bde} components consists of four steps:
\begin{itemize}
\item{Install power supplies and \dword{ddss} components in the detector mezzanine.}
\item{Install 40 cryostat penetrations (spool pieces and crossing tubes)and 80 \dwords{wiec}.}
\item{Install \SI{25}{m} long cold cables from the \dwords{wiec} to the bottom of the cryostat.}
\item{Connect the cold cables to the patch panels on the bottom \dwords{crp} as they are installed.}
\end{itemize}

The installation of power supplies and \dword{ddss} components in the detector mezzanine can take place as soon as the mezzanine is available and will be the first step in \dword{bde} installation.

The %I\&I 
\dword{fsii} team will provide the transport crate and mobile gantry used in the installation of the cryostat penetrations. They will also do the installation of the penetrations and spool pieces.  A \dword{bde} crew will install the flanges, \dwords{wiec}, \dwords{wib}, and \dwords{ptc}.  After each \dword{wiec} is installed, the \dword{bde} crew will verify that the \dwords{wib} and \dword{ptc} are functional.

Cold cables will arrive at \dword{surf} pre-assembled into cable bundles fastened to rope ladders wound onto large reels. Each reel will hold the cables associated with one \dword{wiec}. %I\&I 
The \dword{fsii} team will install the cables and two \dword{bde} teams will test the cables after installation.  The rope ladder will be attached to the cryostat wall and the cable bundles will be hoisted through a penetration using a lift.  A \dword{bde} team on the cryostat roof will strain relief the cables and connect them to a warm electronics flange.  A second \dword{bde} team will connect \dwords{femb} to the cables on the floor of the cryostat where a \dword{crp} will be installed.  The \dword{bde} team on top of the cryostat will verify that every \dword{femb} is powered properly and can be read out, and then the \dwords{femb} will be disconnected.

The half \dword{crp}s will be moved into the cryostat by %I\&I 
\dword{fsii} and \dword{crp} consortium members.  Two half \dwords{crp} will be joined to form a full \dword{crp}, moved into position, and placed on a support truss that will hold the \dword{crp} approximately \SI{1.2}{m} off the floor.  \dword{bde} crew members will connect cold cables to the \dword{crp} patch panels and perform readout tests to verify that all of the cables are connected properly and the full readout chain works. The %I\&I 
\dword{fsii} crew will then remove the truss and lower the full \dword{crp} into its final position.

Quality control tests will be performed at each step of the installation.  After the \dwords{wiec} are installed, test patterns will be read out from each \dword{wib}.  After the cold cables are installed, four (data cable,power cable) pairs at a time will be connected to \dwords{femb} inside the cryostat and read out through a \dword{wib} on top of the cryostat.  After each half \dword{crp} is transported to the clean room, a \dword{bde} crew will connect cold cables to the \dword{crp} patch panels and use a test system to verify that all \dword{femb} channels are working.  Any \dword{femb} with a dead channel will be replaced before the half \dword{crp} is moved to the cryostat.  Finally, after a full \dword{crp} is positioned, it will be read out through the normal readout chain.  If any dead channel is observed and the \dword{crp} installation schedule allows, the \dword{crp} will be raised back onto support trusses and the faulty \dword{femb} will be replaced.
\subsection{Interfaces}
\label{sec:fddp-tpcelec-interfaces}

Table~\ref{tab:CEinterfaces} contains a list of all of the interfaces
between the \dword{bde} consortium and other consortia or groups,
with references to the current version of the interface documents in EDMS.

\begin{dunetable}
[TPC electronics system interfaces]
{p{0.25\textwidth}p{0.65\textwidth}}
{tab:CEinterfaces}
{\dword{tpc} electronics system interfaces. }
Interfacing System & Description 
\\ \toprowrule

\href{https://edms.cern.ch/document/2618995}{CRP} & Mechanical (connections of CE boxes and 
cable routing) and electrical (bias voltages, \dword{femb}--copper plane connection, grounding 
scheme)  \\ \colhline

\href{https://edms.cern.ch/document/2088713}{DAQ} & Data output from the \dword{wib} to the \dword{daq} back-end, clock signal distribution,
controls and data monitoring responsibilities 
\\ \colhline

\href{https://edms.cern.ch/document/2726647}{HV} & Grounding, bias voltage distribution, installation and testing 
\\ \colhline

\href{https://edms.cern.ch/document/2618994}{PDS} & Electrical (cable routing, installation, and grounding scheme), cold flange 
\\ \colhline

\href{https://edms.cern.ch/document/2694691}{Installation} & Integration and installation activities at \dshort{surf},
equipment required for  \dword{tpc} electronics consortium activities, material handling 
\\ 
\end{dunetable}

\subsection{Safety}
\label{sec:fdsp-tpcelec-safety}

The leadership of the \dword{tpc} electronics consortium will work with the \dword{lbnf-dune}
\dword{esh} manager and other relevant responsible personnel at the
participating institutions to ensure that all consortium members receive the appropriate training for the work they
are expected to perform and that all preventive measures to minimize
the risk of accidents are in place. Where appropriate,
\dword{bde} will adopt the strictest standards and requirements from among
the different institutions. Hazard analyses will be performed,
and the level of \dword{ppe} will be determined
appropriately for each task. The scientists in charge of the \dword{bde} activities at each site will be responsible for monitoring the training of personnel at their site.

\dword{esh} plans for the activities to be performed in various
locations, including all universities, national laboratories,
and \dword{surf}, are discussed in the 
review process (Preliminary Design, Final Design, Production 
Readiness, Production Progress) that takes place during the construction of the detector.

It is also important to  protect the detector
components during testing, shipping, 
integration, installation, and commissioning.  The most important risk during construction is damage 
induced by \dword{esd} in the 
electronics components, followed by  mechanical damage to 
parts during transport and handling. Appropriate 
preventive measures will be documented and enforced during all phases,
%\fixme{I made up "documented and enforced" (Anne)}
and \dword{qc} procedures and integration tests will ensure full functionality of the components and system.

To avoid unsafe conditions for the \dword{bde} 
during operations, the \dword{ddss} will include hardware interlocks 
to prevent operating or even powering up
detector components unless conditions are safe
both for the detector and for personnel. Interlocks will be
used on all low-voltage power and on bias voltage 
supplies, including inputs from environmental monitors both
inside and outside the cryostat.

\subsection{Management and Organization}
\label{sec:fdsp-bde-management}

\subsubsection{Institutions}
\label{sec:fdsp-bde-inst}

Table~\ref{tab:SPCE:institutions} lists the institutions participating in the joint \dword{sphd}/\dword{spvd}  \dword{bde} consortium.

\begin{dunetable}
[\dshort{spvd} \dshort{bde} \dshort{tpc} electronics consortium intitutions ]
{ll}
{tab:SPCE:institutions}
{Institutions participating in the joint \dword{sphd}/\dword{spvd}  \dword{tpc} electronics (\dword{ce}) consortium}
Institution  \\ \toprowrule
%Boston University \\ \colhline
Brookhaven National Laboratory \\ \colhline
%% University of Chicago \\ \colhline
University of Cincinnati \\ \colhline
Colorado State University  \\ \colhline
%% Columbia University \\ \colhline
University of California, Davis \\ \colhline
\dword{fnal} \\ \colhline
University of Florida \\ \colhline
University of Hawaii \\ \colhline
Iowa State University \\ \colhline
University of California, Irvine \\ \colhline
Lawrence Berkeley National Laboratory \\ \colhline
Louisiana State University \\ \colhline
Michigan State University \\ \colhline
University of Pennsylvania \\ \colhline
%University of Pittsburgh \\ \colhline
SLAC National Accelerator Laboratory \\ \colhline
Stony Brook University \\
\end{dunetable}

Institutional responsibilities for \dword{spvd} \dword{bde} are summarized below:
\begin{itemize}
\item{\dword{larasic} design and procurement: BNL}
\item{\dword{larasic} QC: BNL, Michigan State U., Stony Brook}
\item{\dword{coldadc} design and procurement: LBNL, Fermilab, BNL}
\item{\dword{coldadc} QC: LBNL, LSU, UC Irvine}
\item{\dword{coldata} design and procurement: Fermilab, BNL}
\item{\dword{coldata} QC: Fermilab, UC Irvine}
\item{\dword{asic} test stand development: BNL, LBNL, Fermilab}
\item{\dword{femb} test stand development: BNL}
\item{Development of cryogenic and robotic test systems: Michigan State U.}
\item{\dword{femb} design and procurement (including \coldbox{}es): BNL}
\item{\dword{femb} QC: BNL, Fermilab, Cincinnati, Iowa State, U. Florida}
%\item{FEMB mechanical supports: Colorado State U.} \fixme{same for FD2?}
\item{Cold cable specification and procurement: BNL}
\item{Cable tray and cryostat penetration design and procurement: BNL}
\item{Flange design and procurement: BNL}
\item{\dwords{wib} design, procurement, and QC: BNL, U. Penn}
\item{\dword{wib} firmware and software: U. Florida, U. Penn, BNL}
\item{\dwords{ptc} design, procurement, and QC: U. Penn}
\item{\dword{ptc} firmware and software: U. Penn, U. Florida}
\item{Interface to \dword{ddss}: Fermilab}
\item{Low voltage and bias power supplie specification and procurement: Fermilab}
\item{Warm cable and fiber procurement: Fermilab}
\item{Support for \dword{crp} integration tests using the \coldbox{}es at NP02: BNL, LBNL, LSU, UCI, Fermilab}
\item{Installation at SURF: All institutions in the consortium}
\item{Detector monitoring during installation at SURF: All institutions}
\item{Detector commissioning: All institutions}
\end{itemize}

\subsubsection{Milestones}
\label{sec:fdsp-bde-milestones}

The milestones of the \dword{spvd} \dword{bde} \dword{tpc} electronics consortium for 2022 and beyond are listed in Table~\ref{tab:SPCE:timeline}.

\begin{dunetable}
[TPC electronics consortium schedule]
{p{0.75\textwidth}p{0.20\textwidth}}
{tab:SPCE:timeline}
{Milestones of the \dword{bde} TPC electronics consortium.}
Milestone & Date \\ \toprowrule
Start of \dword{vdmod0} CE installation & November 2022     \\ \colhline
Completion of \dword{vdmod0} CE installation & May 2023     \\ \colhline
Completion of \dword{bde} Final Design Review & June 2023     \\ \colhline
Completion of Production Readiness Review & February 2024 \\ \colhline
Start of \dword{coldadc}/\dword{coldata} \dword{asic} production & November 2024 \\ \colhline
Start of cold cable procurement & October 2024 \\ \colhline
Start of \dword{femb} production & October 2024 \\ \colhline
Start of production of \dwords{wiec}, \dwords{wib}, and \dwords{ptc} & June 2025 \\ \colhline
Start of production of cryostat penetrations & April 2025 \\ \colhline
Completion of \dword{asic} QC & February 2026 \\ \colhline
Completion of cold cable QC & September 2026 \\ \colhline
Completion of \dword{femb} QC & July 2026 \\ \colhline
Completion of \dword{wiec}, \dword{wib}, and \dword{ptc} QC & November 2026 \\ \colhline
Completion of cryostat penetration QC & March 2026 \\ \colhline
Start of power supply, penetration, and WIEC installation at SURF & September 2026 \\ \colhline
Start of cold cable installation inside FD2 cryostat & July 2027 \\ \colhline
End of cold cable installation inside FD2 cryostat & September 2027 \\ \colhline
Start of \dword{spvd} bottom \dword{crp} installation inside FD2 cryostat & February 2028 \\ \colhline
End of \dword{spvd} bottom \dword{crp} installation inside FD2 cryostat & April 2028 \\ 
\end{dunetable}

A more detailed schedule for production and installation of the complete \dshort{spvd} drift readout electronics is found in Figure~\ref{fig:cro_schedule}.
\begin{dunefigure}[Key \dshort{cro} milestones and activities toward 
\dshort{spvd}]{fig:cro_schedule}{
Key \dshort{cro} milestones and activities toward  the \dshort{spvd} in graphical format (Data from~\cite{docdb22261v28}).}
\includegraphics[width=0.95\textwidth]{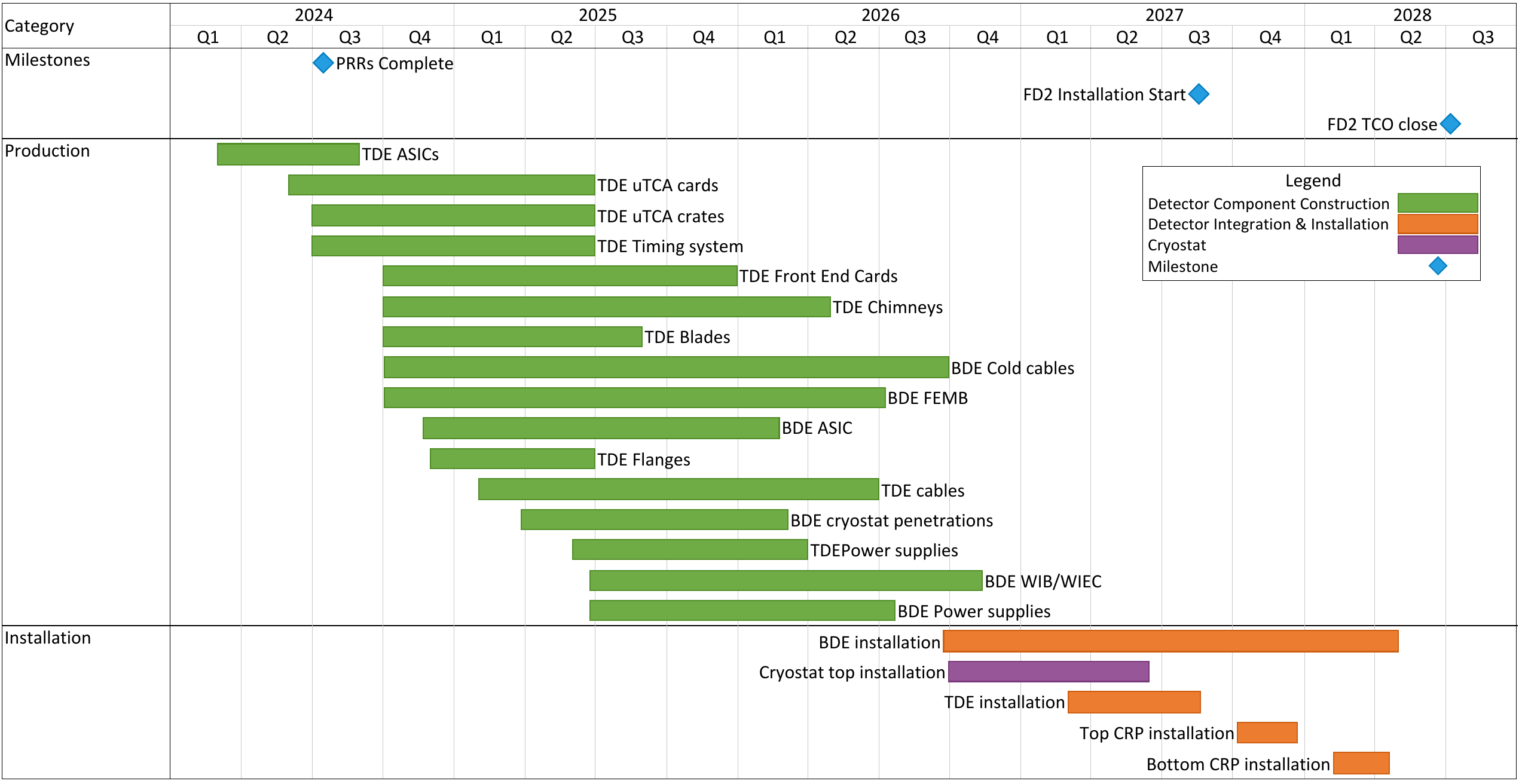}
\end{dunefigure}

%%%%

\chapter{High Voltage System and Drift Field}
\label{ch:DFS}
%%%%%%%%%%%%%%%%%%%%%%%%%%%%%
\section{Introduction}
\label{ch:DFS:intro}

A \dword{lartpc} uses an electric field (\efield) to drift ionization electrons 
through \dword{lar} to an anode sensor plane.  
The system that drives these electrons consists of an equipotential cathode plane biased at a negative \dword{hv}, parallel to and a distance away from the anode plane, and an interior \efield that is maintained uniform by a \dword{fc} system that surrounds the detector active volume.  

The DUNE \dshort{lartpc} detector module design for  \dword{spvd} encompasses two drift volumes of equal (maximum) drift distance 6.5\,m and has a nominal uniform \efield of 450\,V/cm. A flat horizontal cathode plane %is placed 
spans the detector at mid-height and is held at a negative voltage.  The two horizontal \dwords{anodepln} (biased at near-ground potentials) span %are located at 
the top and bottom of the detector, as described in Chapter~\ref{sec:AR}. 

The \dword{hvs} is divided into the supply and delivery system, and the distribution system. The supply and delivery system consists of a negative high-voltage power supply (\dword{hvps}), \dshort{hv} cables with integrated resistors to form a low-pass filter network, a \dshort{hv} feedthrough (\dword{hvft}), and a 6\,m long extender inside the cryostat to deliver $-$294\,kV to the cathode inside the \dword{tpc}. The distribution system consists of the cathode plane, the \dshort{fc}, and the \dshort{fc} termination supplies.  Figure~\ref{fig:vd_hvs_schematic} is a schematic circuit diagram of the high-level components of the \dshort{hv} system.

The \dword{spvd} cathode plane is 60\,m long, 13.5\,m wide, and 6\,cm thick.   It is tiled from an array of 20$\times$4 cathode modules with highly resistive top and bottom panels mounted on fiber-reinforced plastic (\dword{frp}) frames. The purpose of the %cathode with 
high-resistivity surfaces is to slow the voltage swing on the cathode in case of a discharge, thereby reducing the peak current injection into the front-end electronics connected to the anode readout strips (Section~\ref{subsec:3V}). 

The \dshort{fc}, in conjunction with the cathode and \dshort{anodepln}s, defines the detector drift volumes and is designed to ensure uniformity of the \efield in the vertical direction.
The \dshort{fc} features 48 columns of four-unit modules, and extends
%extending the \dword{fc} 
the full 13\,m height and the full 148\,m cryostat perimeter. 
These modules consist of horizontal field-shaping electrodes (extruded aluminum profiles) that are stacked at regular intervals and interconnected by resistive voltage-divider chains to maintain a uniform vertical voltage gradient. 
To ensure a uniform response of the wall-mounted portion of the \dword{pds}  (see Chapter~\ref{chap:PDS}), the \dword{fc} in the region where the \dwords{pd} are installed is designed for 70\% optical transparency. 

\begin{dunefigure}
[HVS schematic]
{fig:vd_hvs_schematic}
{A high-level schematic circuit diagram of the \dshort{hv} system in \dshort{spvd}.   Each box marked FCM represents a \dshort{fc} module.  The resistance on the FCMs is provided by high voltage resistor divider boards (\dshort{hvdb}s). $R_O$ is the internal 17\,M$\ohm$ output resistor of the power supply. 
%$R_{f1}$ and $R_{f2}$ 
$R_{1}$ and $R_{2}$ are the two filter resistors included in the  \dshort{hv} cable terminations, near the power supply and near the \dshort{hvft}, respectively. }
  \includegraphics[width=0.9\textwidth]{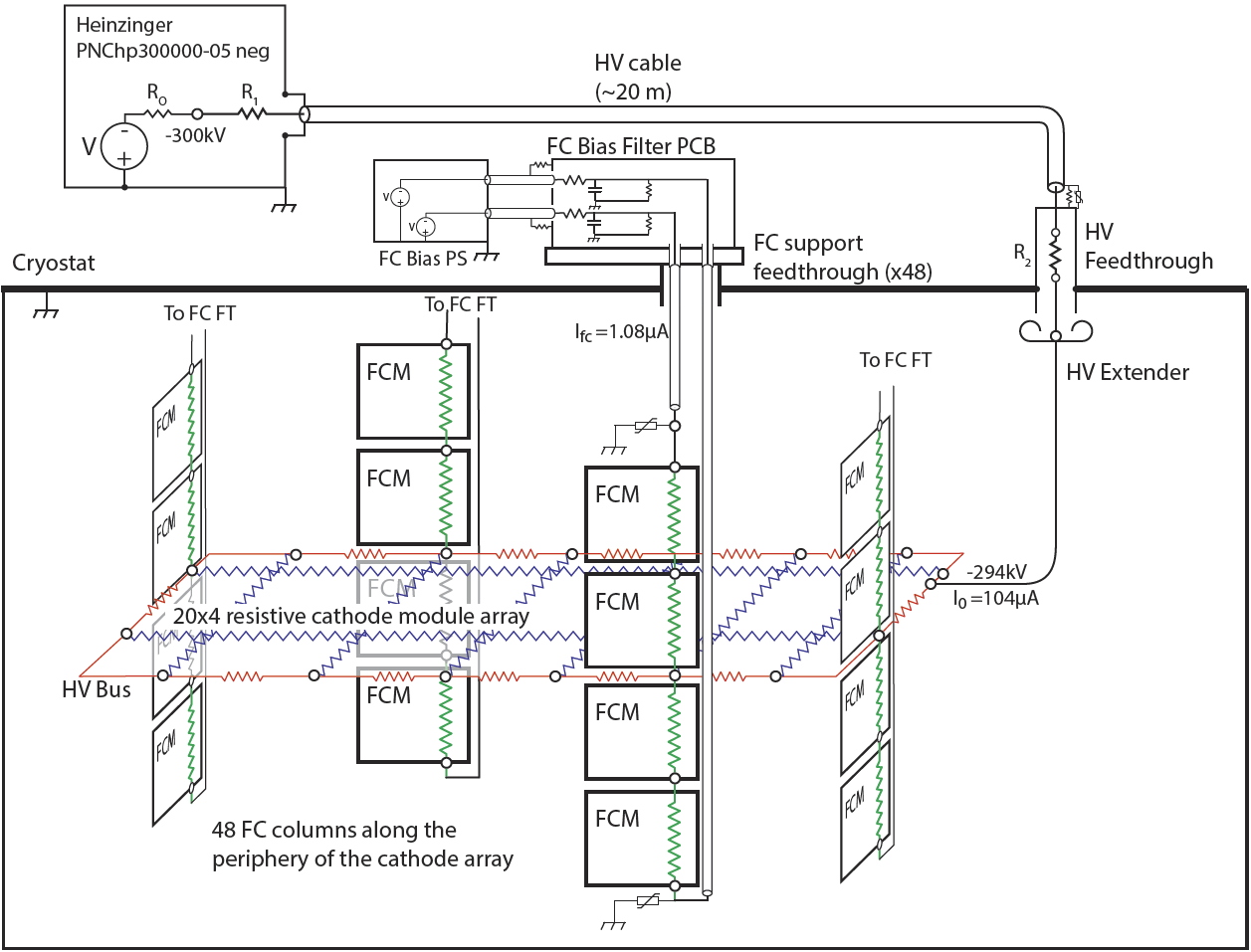}
\end{dunefigure}

Figure~\ref{fig:vd_hvs} gives a birds-eye view of the %key 
\dword{hvs} with each of the key \dshort{hv} components clearly indicated.

The baseline design presented in this chapter reflects significant improvements and optimization gained through extensive R\&D and testing activities. Some %parts 
aspects of the \dword{spvd} \dshort{hvs} design are still %in evolution 
evolving to provide the flexibility to adopt potential further improvements.
The majority of the design elements presented in this chapter will be tested in the \dword{vdmod0}, as described in Chapter~\ref{ch:mod0}.
\begin{dunefigure}
[HV system components inside the cryostat]
{fig:vd_hvs}
{A birds-eye view of the \dshort{fc}, with one full-height \dshort{fc} column (highlighted in cyan) that extends the entire height, the \dshort{hv} feedthrough and extender (in the foreground), and the cathode (with one cathode module highlighted in cyan).  The \dshort{fc}  profiles are mechanically and electrically independent.  
The figure distinguishes between the 70\% optical transparency portion 
that spans the vertical ranges $2.5<y<6.3$\,m and $-6.3<y<-2.5$\,m and the standard portion.} 
\includegraphics[width=0.95\textwidth]{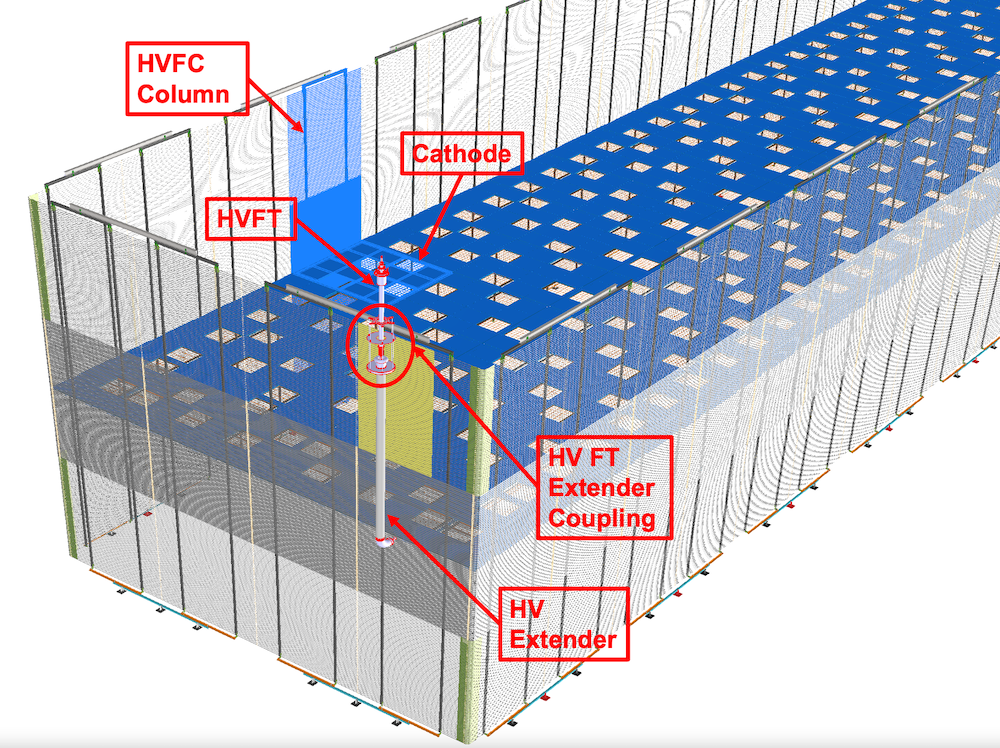}
\end{dunefigure}

%%%%%%%%%%%%%%%%%%%%%%%%%%%%%
\section{%High Voltage System Design 
Requirements and Specifications} 
\label{sec:HV:spec}

The high-level design requirements and specifications 
of the \dword{hvs} are as listed in Table~\ref{tab:specs:SP-HV}. 
All are common to both \dshort{fd} modules except FD2-HV-3, which is specific to \dword{spvd}.

\begin{footnotesize}
\begin{longtable}{p{0.12\textwidth}p{0.18\textwidth}p{0.17\textwidth}p{0.25\textwidth}p{0.16\textwidth}}
\caption{HV specifications %\fixmehl{ref \texttt{tab:spec:SP-HV}}
} \\
  \rowcolor{dunesky}
       Label & Description  & Specification \newline (Goal) & Rationale & Validation \\  \colhline

   \newtag{FD-1}{ spec:min-drift-field }  & Minimum drift field  &  $>$\,\SI{250}{ V/cm} \newline ($>\,\SI{450}{ V/cm}$) %updated for FD2
   &  Lessens impacts of $e^-$-Ar recombination, $e^-$ lifetime, $e^-$ diffusion and space charge. &  ProtoDUNE \\ \colhline

  \newtag{FD-11}{ spec:hvs-field-uniformity }  & Drift field nonuniformity due to HVS  &  $<\,\SI{1}{\%}$ in 99.8\,\% of active volume 
  %updated for FD2
  &  High reconstruction efficiency. &  ProtoDUNE and simulation \\ \colhline

  \newtag{FD-12}{ spec:hv-ps-ripple }  & Cathode HV power supply ripple contribution to system noise  &  $<\,\SI{100}e^-$ &  Maximize live time; maintain high S/N. &  Engineering calculation, in situ measurement,   ProtoDUNE \\ \colhline
    
   \newtag{FD-17}{ spec:cathode-resistivity }  & Cathode resistivity  &  $>\,\SI{1}{\mega\ohm/square}$ \newline ($>\,\SI{1}{\giga\ohm/square}$) &  Detector damage prevention.  &  ProtoDUNE \\
   \colhline

  \newtag{FD-24}{ spec:local-e-fields }  & Local electric fields  &  $<\,\SI{30}{kV/cm}$ &  Maximize live time; maintain high \dshort{s/n}. &  ProtoDUNE \\ \colhline
    
   \newtag{FD-29}{ spec:det-uptime }  & Detector uptime  &  $>\,$98\% \newline ($>\,$99\%) &  Meet physics goals in timely fashion. &  ProtoDUNE \\ \colhline

  \newtag{FD-HV-1}{ spec:power-supply-stability }  & Maximize power supply stability  &  $>\,\SI{95}{\%}$ uptime &  Collect data over long period with high uptime. &  ProtoDUNE \\ \colhline
    
   \newtag{FD-HV-2}{ spec:hv-connection-redundancy }  & Provide redundancy in all \dshort{hv} connections.  &  Two-fold \newline (Four-fold) &  Avoid interrupting data collection or causing accesses to the interior of the detector. &  Assembly QC \\ \colhline
    
   \newtag{FD2-HV-3}{ spec:hv-optical-transparency }  & Provide optical transparency to the photon detectors on the membrane  &  $\geq$~70\%  &  Enable photons detectors outside the field cage to collect light with efficiency similar to those mounted in the cathode. &  VD Module-0 \\ \colhline

\label{tab:specs:SP-HV}
\end{longtable}
\end{footnotesize} 

All these specifications were met during \dword{pddp} operation. 
In particular, the \dword{hv} stability was tested at \SI{-300}{kV} on the cathode (about 500\,V/cm in the 6\,m drift) %\dword{fc}) 
during the long-term stability test performed from fall 2021 through early 2022 in the \dword{np02} cryostat at the \dword{cern} Neutrino Platform.

%%%%%%%%%%%%%%%%%%%%%%%%%%%%%
\section{HV Delivery System}
\label{subsec:HVss}
This section describes the baseline design of the \dword{spvd} \dword{hv} delivery system, which has been derived from the system developed for the \dword{dp} \dshort{lartpc} proposal~\cite{edms2084051}.  The \dshort{spvd} design %also 
has undergone a design optimization process, and reflects the lessons learned from the long-term \dshort{hvs} stability test, %performed in the fall of 2021 through March 2022 
where the new concept of the \dshort{hv} extender and its coupling to the \dword{hvft} were extensively tested at \SI{-300}{kV}. This provides confidence that the design  
improves long-term reliability and %long-term 
stability at the \dshort{spvd} operation voltage.

%%%%%%
\subsection{\dshort{hv} Power Supply and Cable}
\label{subsubsec:SCsss}

Low-noise $-$300\,kV  HV power supplies (\dword{hvps}) and dedicated $>$300\,kV cables are commercially available. The baseline design uses the Heinzinger PNChp300000-05~\footnote{Heinzinger 
 \url{https://www.heinzinger.com/en/applications}} as the \dshort{hvps} of choice. %The $-$300\,kV \dword{hvps} from Heinzinger, PNChp300000-05 
It has a residual ripple of $10^{-5}$ (corresponding to 3\,V at the maximum voltage of 300\,kV) at an absolute precision in nominal voltage of $\pm$\SI{50}{mV}, with a typical frequency of 30\,kHz.  The cable is Dielectric Sciences type 2236~\footnote{Dielectric Sciences \url{https://catalog.dielectricsciences.com/item/all-categories/wire-cable-2/item-1260?plpver=1001}}, which is designed and certified to operate at 320\,kV DC. 

Given the 100\,$e$ noise ceiling requirement %from 
on the \dshort{hvs}, %we can derive 
the upper limit of the ripple voltage on the cathode is found by dividing the noise charge by the capacitance between the cathode and the longest  readout strip on the first induction plane. Using a %pessimistic 
conservative assumption that the coupling is a simple parallel plate capacitor, and factoring in the shielding effect of the shield plane in front of the first induction plane (calculated from a \dword{fea} using the perforated anode \dword{pcb} geometry), the cathode voltage ripple limit is found to be 15\,mV.  
As a consequence, at $-$300\,kV output voltage and the typical HVPS frequency of 30 kHz, the required noise reduction from 3\,V to 15\,mV is a factor 200 ($\sim$46\,dB).
Note that this number is much greater than that for the \dword{sphd} (0.9\,mV) for several reasons: a larger drift distance, a smaller area %sustained 
covered by a \dword{crp} readout strip compared to that of an \dword{apa} wire (%mostly 
due to the wires' length), and the presence of a much more effective shielding electrode. 

The additional \dword{rc} filtering is achieved with a resistor at the output of the \dword{hvps} and the capacitance of the \dshort{hv} cable. Assuming  cable lengths of at least \SI{20}{m}, calculations and experience confirm that a resistance as low as a few \si{\mega\ohm} yields the required noise reduction.

As shown in Figure~\ref{fig:ps_filter_ft_schematic}, the Heinzinger PNChp300000-05 features a 17\,M$\ohm$ output resistance $R_O$, with %the voltage drop across this resistance 
an internally compensated voltage drop.
The primary %ripple 
low-pass filtering, aimed at reducing  the \SI{30}{kHz} voltage ripple on the output of the power supply, can be achieved by using this output resistance %and a 
plus the 2\,nF capacitance %due to 
from the 20\,m long \dshort{hv} cable ($\sim$100\,pF/m). 
The corner frequency of this low-pass filter is 4.7\,Hz, hence the noise attenuation at 30\,kHz is expected to be $\sim$76\,dB (well above the minimum required noise reduction of $\sim$46\,dB).
This minimal layout of the ripple noise filter was implemented in the HV delivery system of the \dword{np02} \dshort{hv} long-term stability run: as expected, no %ripple induced by the 
\dshort{hvps}-induced ripple  was ever observed on the \dword{crp} read-out electronics.

No additional characteristic frequencies are expected from the HVPS. Very low frequency oscillations (e.g., 50\,Hz), if present, would be difficult to identify as they would appear as tiny baseline variations during the full drift time of 4.2\,ms. The preamplifier bandwidth %would also contribute 
is expected to further attenuate the baseline fluctuations.

To further increase the ripple filtering capability a \dshort{hv}-rated resistor $R_{1}$ in a custom housing is inserted in the \dshort{hv} cable termination (\dshort{hvps} side) (Figure~\ref{fig:ps_filter_ft_schematic}). 
A second $R_{2}$ is installed in the opposite cable termination (\dshort{hvft} side); it serves as a current-limiting resistor to restrict sudden energy dumps from the \dshort{hvps} and the long cable into the \dshort{tpc} in the event of a discharge. % inside the cryostat. 

To ensure a drift field of 450\,V/cm over the 6.5\,m drift, the voltage to be provided to the cathode is 294 kV. Hence, the maximum voltage drop across the ripple filter resistors is %at most 
6\,kV.
Based on the total current drawn by the entire \dword{fc} ($\sim$104\,$\micro$A), the sum of $R_{1}$ and $R_{2}$ can be as high as 57\,M$\ohm$. 
The actual value of each resistor can be tuned, but the baseline %we 
assumes equal values of about 25\,M$\ohm$.  
This filtering scheme, with similar power supplies, has been used successfully in other \dword{lartpc} experiments, such as \dword{microboone} and \dword{icarus} and, more recently, it has been demonstrated in \dword{pdsp} and \dword{pddp}. Figure~\ref{fig:ps_filter_ft_schematic} shows the \dshort{hv} supply circuit.

\begin{dunefigure}[Power supply photos and schematic of \dshort{hv} delivery system to the cryostat]
{fig:ps_filter_ft_schematic}
{Left: The \SI{-300}{kV} Heinzinger power supply. %An example (Credit: CERN). 
Right:  A schematic showing the \dshort{hv} delivery system to the cryostat. 
Two filter resistors %$R_f$ 
are included in the HV cable terminations: $R_{1}$ sits near the power supply, and $R_{2}$ near the HVFT. The 10\,k$\ohm$ ground-loop breaking resistor $R_G$, placed at the HVFT termination, is also shown.}

\begin{minipage}{\textwidth}%{6in}
  \centering
 $\vcenter{\hbox{\includegraphics[width=0.22\textwidth]{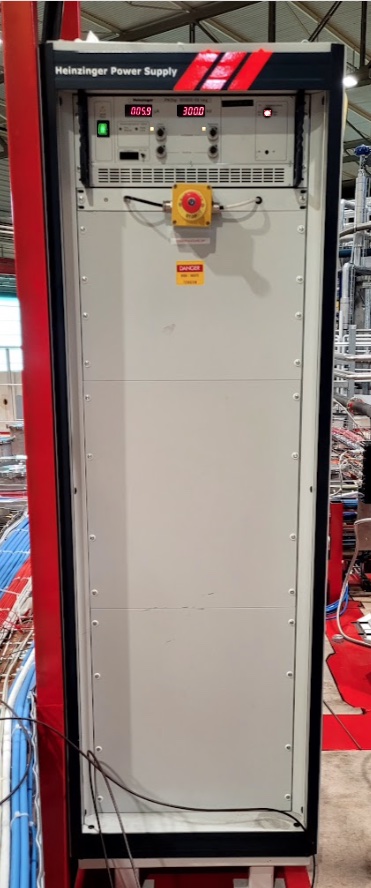}}}$
 \hspace*{0.03\textwidth}  $\vcenter{\hbox{\includegraphics[width=0.72\textwidth]{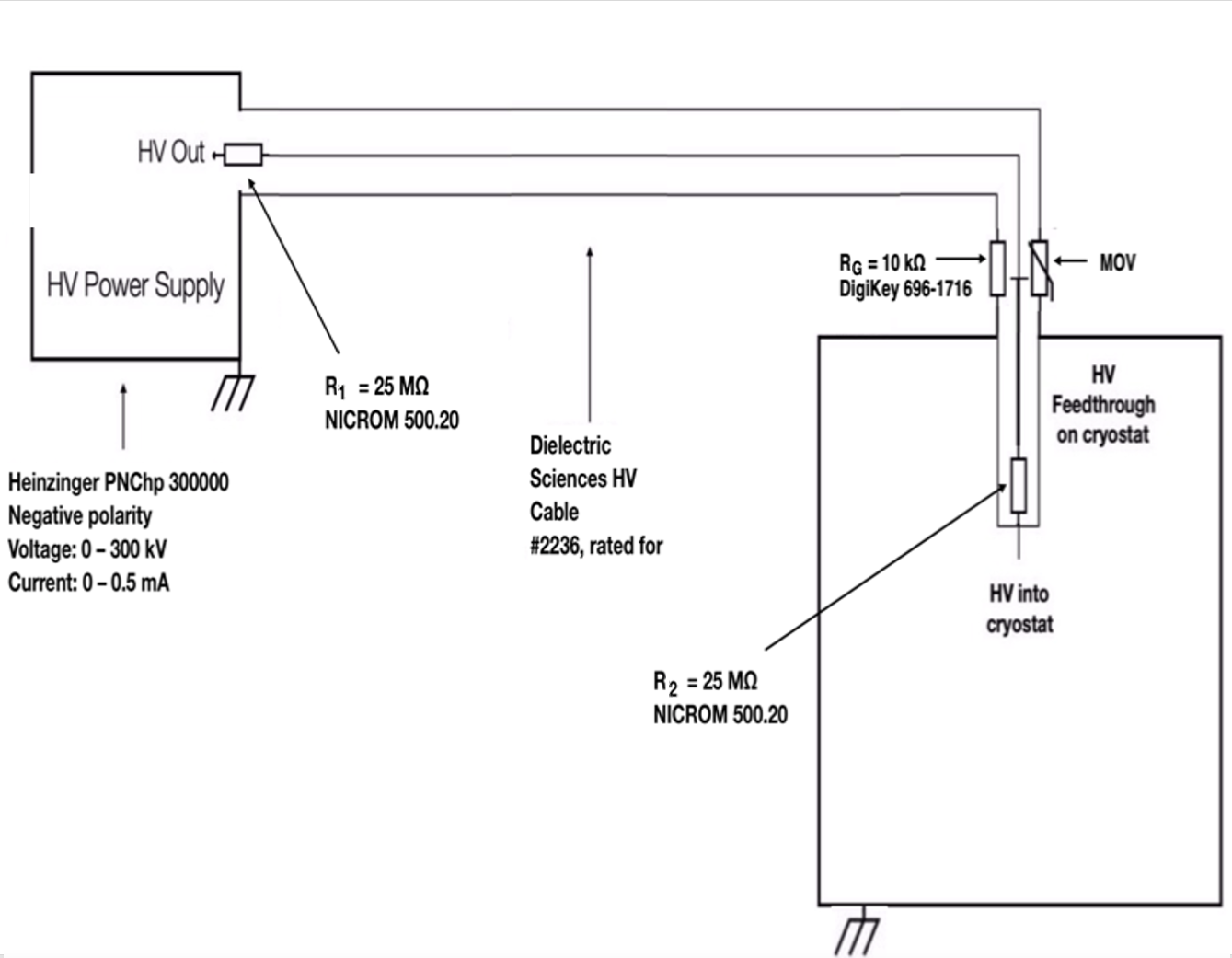}}}$
\end{minipage}
\end{dunefigure}

The %current plan for the 
cylindrical filter resistors %have a cylindrical design. They 
are located in the termination tips of the \dshort{hv} cable, which in turn are inserted into the \dshort{hvps} and \dshort{hvft} cable receptacles. The resistor %should shall be selected to 
must withstand a large over-power condition. Resistors withstanding up to 150\,kV are commercially available and were used in \dshort{pddp}. The filter layout allows for easy and rapid replacement of the resistors in case of damage. %,  for which other designs have used transformer oil or \dword{uhmwpe}. 
A %ground 
breaking circuit to avoid ground loops is implemented on the cable termination shield at the level of \dshort{hvft} plug. 

%%%%%%%
\subsection{\dshort{hv} Feedthrough}

%%%%%%%%%%%%%%%%%%%%

The HV feedthrough (\dword{hvft}) %will be 
is based on the successful \dword{icarus} design~\cite{Amerio:2004ze}, %Icarus-T600}, 
and scaled to hold to \SI{-300}{kV}.  
%and successfully tested in \dword{pdsp} and \dword{pddp}. 
The voltage is transmitted by a stainless steel center conductor.  On the warm side of the cryostat, this conductor mates with a cable end.  Inside the cryostat, the end of the center conductor has a spring-loaded tip that ensures contact to the receptacle cup mounted on the HV extender, delivering \dword{hv} to the cathode and the \dword{fc}. The \SI{40}{mm} diameter center conductor of the \fdth is surrounded by an ultra-high molecular weight polyethylene (\dword{uhmwpe}) insulator cylinder. The insulator is surrounded by a tight-fitting stainless steel ground tube.  A %\SI{25}{\centi\meter}  
Conflat industry-standard flange is welded onto the ground tube for attachment to the cryostat. 
%At the bottom 
On the under (cold) side of the \dshort{hvft}, the \dshort{uhmwpe} cylinder extends beyond the edge of the outer ground shield by about 20\,cm, along the length.  This exposed cylindrical surface is corrugated with grooves to increase surface path length between the protruded center conductor below and the grounded shield above, as depicted in the top left and top right photos in Figure~\ref{fig:NP02-extender-installation}. The central conductor is equipped with a spring-loaded tip (10\,cm excursion) to ensure electrical contact with the connected electrode. 

\begin{dunefigure}
[Components of the full-scale \dshort{hv} long-term stability test in the \dshort{np02} cryostat]
{fig:NP02-extender-installation}
{Components of the full scale \dshort{hv} long-term stability test in the \dshort{np02} cryostat. Top: two views of the doughnut and the sphere of the extender head with the \dshort{hvft} tip connected. The gold colored ring in the top %right 
left photo is the sleeve to confine the gas bubbles produced by the heat input from the HVFT and evacuate them to the gas phase close to the HVFT ground skin. Bottom left: the tip of the extender in front of the \dshort{fc}. Bottom right: Detail of the extender connection to the \dshort{fc}.}
\includegraphics[width=.95\textwidth]{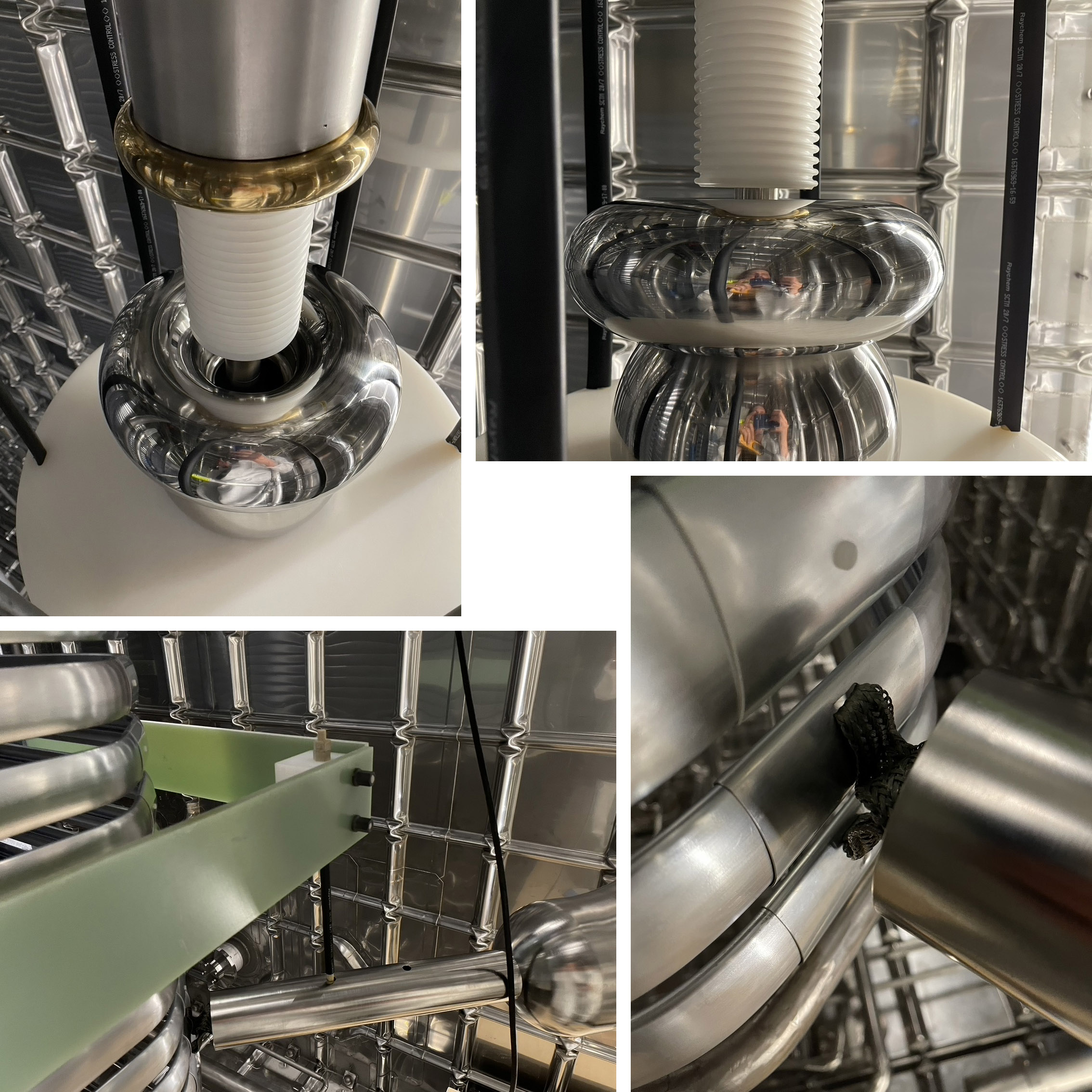}
\end{dunefigure}

A doughnut ($\sim$20\,mm annular thickness) is added to the bottom edge of the outer stainless steel cylinder ground shield of the \dshort{hvft}, as shown in Figures~\ref{fig:filterAndFeedthrough} and~\ref{fig:NP02-extender-installation} to minimize the local \efield strength. 
On the top (warm) side of the \dshort{hvft}, a receptacle with a depth of $\sim$80\,cm and a diameter of 38\,mm receives the cable (Dielectric Science type 2236) from the \dword{hvps}.

\begin{dunefigure}[Photo and drawing of a \dshort{hvft}]{fig:filterAndFeedthrough}
{Photograph and drawings of a \dshort{hvft}. The photograph shows the \dshort{pdsp} installation with the ``cold'' \dshort{hv} tip %touching 
in the doughnut cup that is connected to the top of the (vertical) cathode; no extender is necessary for the \dshort{hd} detectors.
A \threed view and vertical cross sections of the \dshort{hvft} are on the right, illustrating  %. Detailed zoomed views of 
the cable insertion and the \dshort{hv} tip touching the doughnut cup. The \dshort{hv} cable (brown) is inserted about 1\,m into the \dshort{hvft}. The exposed \dshort{uhmwpe} insulator (aqua) %at 
near the ``cold'' \dshort{hvft} tip is about 40\,cm long and is corrugated. The insulator diameter is 150\,mm. The diameter of the inner conductor is 40\,mm. As in the \dshort{sphd} \dshort{hvft}, the spring-loaded tip of the \dshort{hvft} makes contact with the center part of the doughnut cup, which is part of the \dshort{hv} extender.} 
\begin{minipage}{\textwidth}%{6in}
  \centering
 $\vcenter{\hbox{\includegraphics[width=0.20\textwidth]{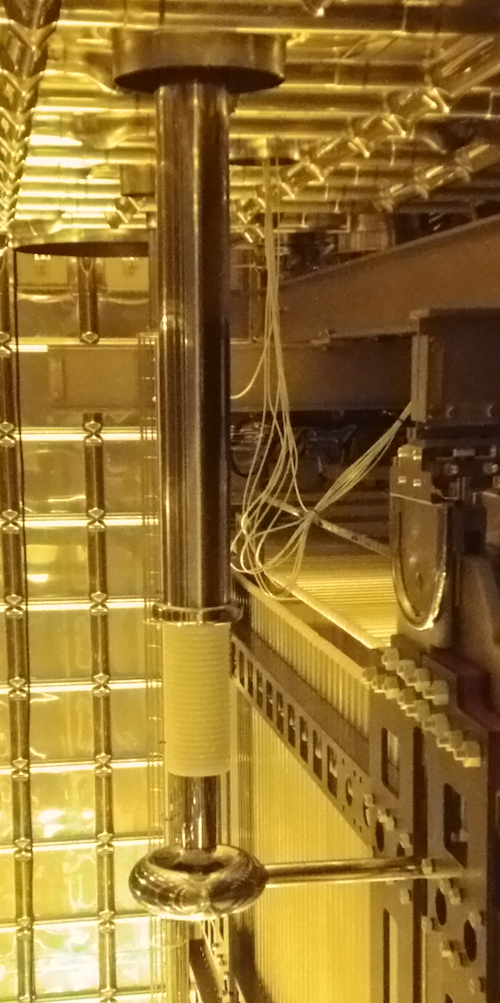}}}$
 \hspace*{0.001\textwidth}  $\vcenter{\hbox{\includegraphics[width=0.75\textwidth]{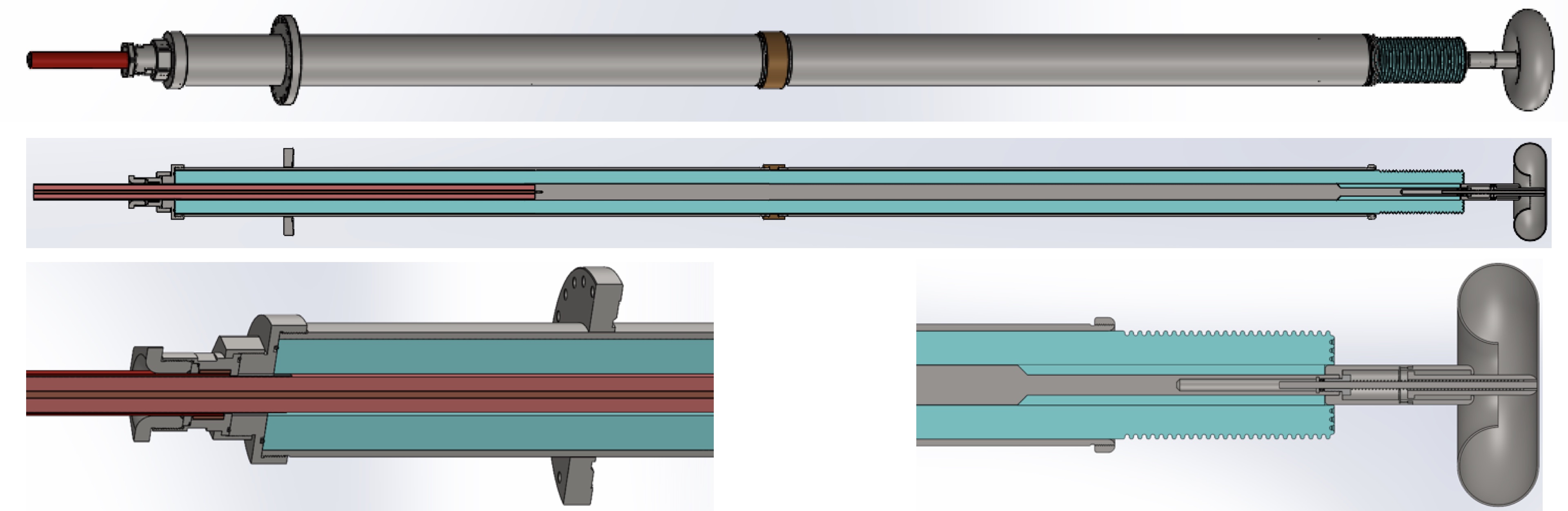}}}$
\end{minipage}
\end{dunefigure}

Three \dshort{hvft}s were built for  %\dword{np02} and \dword{np04} cryostats
\dshort{pdsp} and \dshort{pddp}, and all functioned properly at \SI{-300}{kV} on a short-term basis in dedicated test stands.
One of these \dshort{hvft}s was also successfully used in the second part of the %\dword{np02} (
\dshort{pddp} \dshort{hv} stability run  (for more than three months), where the detector was operated at the nominal voltage of \SI{-300}{kV} delivered from the \dword{hvps} under very stable conditions. 

Following the long-term operation in both %\dword{np04} (
\dshort{pdsp} 
and the two %NP02 (
\dshort{pddp} runs, it was found that the \dshort{hvft} was affected by ice formation at the cable receptacle. The cable tip was %sitting \SI{\sim 80}{cm} above the cryostat insulation, 
reaching the depth of the inner cryostat membrane, 
thus %experiencing quite a 
subject to a quite low temperature. Investigation showed that although
the cable receptacle was flushed continuously with dry N2, it was insufficient to prevent ice formation over the long term.
The formation of ice did not affect the performance of the \dshort{hv} distribution, given its good insulating properties. However, it would prevent %s the possibility of extracting 
potential extraction of the cable, which might be required at some point %the case in DUNE FD over 
during the decades-long operation of \dword{spvd}.
%
%For the reason above,  a
An improved version of the \dshort{hvft} %, {\it a.k.a.}UCLA \dword{hvft}, 
was initially used in the %NP02 (ProtoDUNE-DP) 
\dshort{pddp} \dshort{hv} stability run. This \dshort{hvft}, %with a similar concept as 
conceptually similar to the original ones, but one meter longer on the warm end to keep the cable receptacle in the warm section of the \dshort{hvft}s,  was designed and built at UCLA. 
%The UCLA \dword{hvft} 
It was extensively tested in the U.S. up to \SI{-200}{kV} %after manufacturing 
 and was then %successfully 
 tested at \dword{cern} at \SI{-300}{kV} prior to its use %during the VD 
 in the \dword{hvs} long-term stability test. % in the fall 2021 through March 2022.
The UCLA \dshort{hvft} performed stably at \SI{-300}{kV} for more than two months in the early part of the stability test, at which point %until 
a failure due to impurities found in the %PE
\dword{uhmwpe} insulation cylinder required its replacement by one of the original \dshort{hvft}s for the remainder of the test at \SI{-300}{kV}.

To prevent a similar failure, % similar to the UCLA \dword{hvft}, 
a new \dshort{hvft} was designed as shown in Figure~\ref{fig:filterAndFeedthrough} for %the future operation of 
\dword{vdmod0}. % VD \dword{mod0}.
This \dshort{hvft} will be the prototype %on the \dword{hvft}'s of 
for both \dword{sphd} and \dshort{spvd}; its components were built in 2022 and, as of this writing, the \dshort{hvft} has been assembled via cryogenic insertion. % in March 2023. 
Validation in \dshort{lar} at 300\,kV will be carried out in Q2 2023. 
It %will be 
is similar to the %latest 
UCLA design but %it will be 
longer (\SI{4}{m}), with the ``warm'' side of it that contains the cable receptacle extending above the cryostat roof by about \SI{1}{m}.  It is also thicker (\SI{150}{mm} instead of \SI{100}{mm}) to improve the insulation reliability. X-ray inspection of the \dshort{uhmwpe} purity will be performed before assembling the parts. % of the \dword{hvft}. 
The inner conductor will also be thicker (40\,mm diameter) to limit the \efield strength at the conductor surface. 

Given the thickness of the \dword{fd} cryostat outer structure and insulation layers, and the depth of the gas ullage ($\sim$150\,cm in total), 
the bottom rim of the \dshort{hvft} ground shield will be immersed in \dword{lar} to a depth of at least 40\,cm. This depth minimizes the risk of gas pocket formation below the ground shield doughnut from heat dissipation by the metallic components of the \dshort{hvft} itself; %pockets 
gas bubbles generally form closer to the surface (in the top $\sim$20\,cm of depth). To further mitigate this %possible 
potential issue, holes are drilled in the \dshort{hvft} outer shield cylinder a few \,cm above the doughnut at the tip,  allowing any gas bubbles formed inside to easily escape.
%\fixme{Concerning the lines below, was the SBND feedthrough really tested?  SP}

The new design is illustrated in Figure~\ref{fig:filterAndFeedthrough}. Detailed technical drawings of the \dshort{hvft} can be found at~\cite{edms-2593486}. The \dshort{hvft}s are constructed by the same company that %successfully 
produced those for \dword{icarus}, \dword{protodune} and \dword{sbnd}.

\subsection{\dshort{hv} Extender and its Coupling to \dshort{hvft}}
\label{subsec:hv-extender-coupling}

An ``HV extender'' is introduced in the \dword{spvd} design to feed the  cathode 
with the \SI{-300}{kV} \dword{hv}. It consists of a \SI{20}{cm} diameter, \SI{6}{m} long polished and passivated stainless steel pipe that 
%the pipe 
is terminated at the top with a coupler to the \dword{hvft} and at the bottom with a $90^{\circ}$ %bend 
elbow for the connection to the cathode.
The \dshort{hv} extender is supported by an insulating \dword{uhmwpe} disk suspended from the cryostat roof using six insulating \frfour rods. The extender exploits the surrounding \dshort{lar} as the dielectric insulator.

The \dshort{hvft} and the extender will be located on the side of the cryostat opposite the \dword{tco} (Section~\ref{ch:IEI:cryostat}), where the distance between the \dword{fc} profiles and the flat face of the cryostat membrane is 1.1\,m. The knuckles and corrugations of the membrane with small radii of curvature could be covered by a vertical band of flat conductive sheet to further smooth the \efield in case future tests indicate that this is necessary. The internal cryogenic pipes on this end of the cryostat will be placed away from the \dshort{hv} extender. 
A \dword{fea} has shown that a conductive cylinder at $-$300\,kV with a diameter between 200\,mm and 400\,mm placed in the center of a 1\,m gap between two grounded parallel planes will have a maximum surface \efield of 17\,kV/cm.  This is well below the 30\,kV/cm \efield limit required for safe operation in \dshort{lar}.

\begin{dunefigure}
[HVFT+extender assembly connection and interface]{fig:hvft-extender}
{Left: Schematic diagrams of the optimized metal tube extender assembly showing the $90^\circ$ elbow connection to the cathode, with details of the coupling to the \dshort{hvft} and to the support disk. Right: (top) \threed model of the extender  with HV tip inserted into doughnut and (bottom) detail of the head and support disk.}
\includegraphics[width=0.9\textwidth]{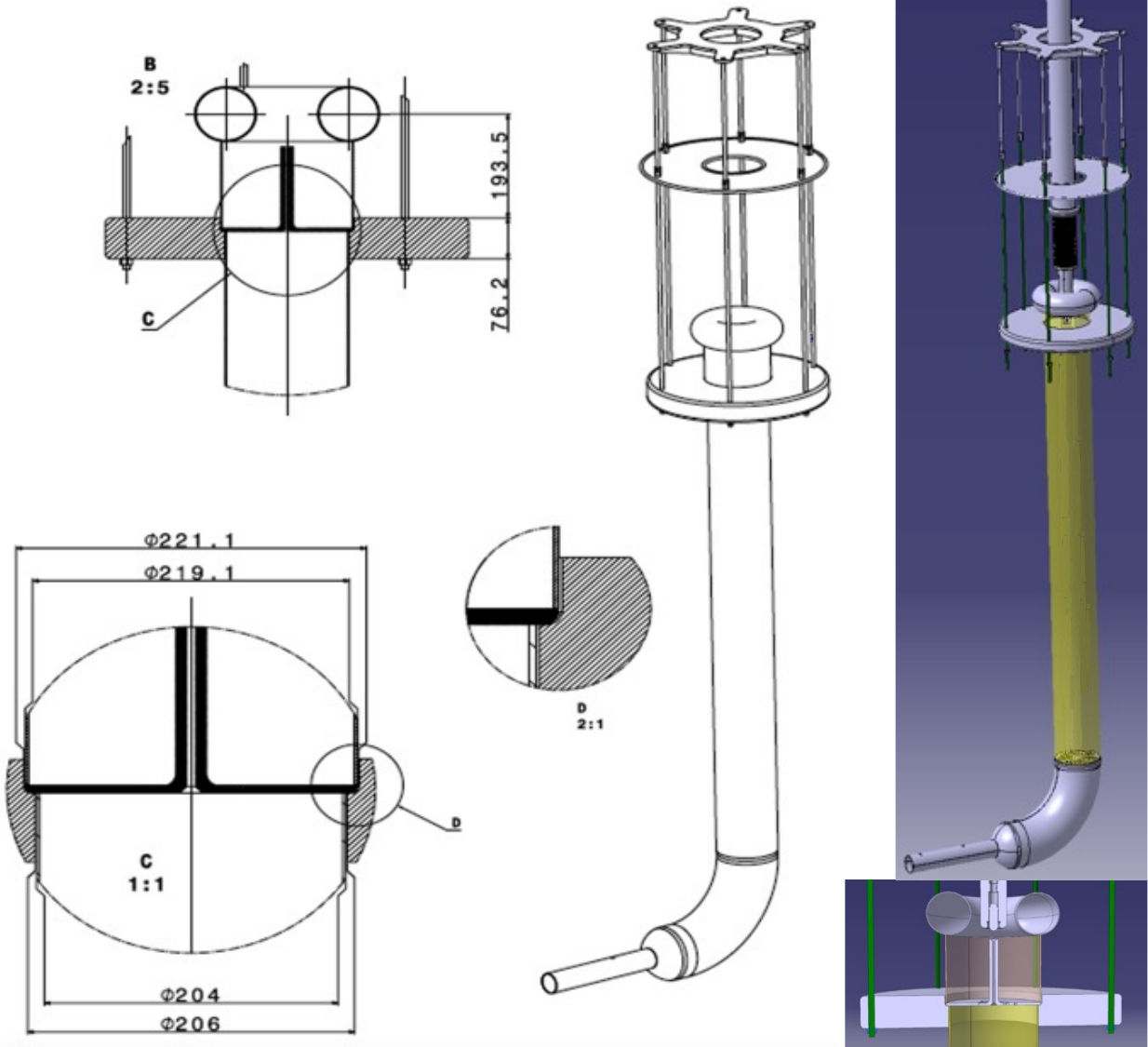}

\end{dunefigure}

Figure~\ref{fig:hvft-extender} shows schematic diagrams of the %various 
components of the optimized \dshort{hv} extender. 

The new version of the \dshort{uhmwpe} support disk 
%\fixme{this was NOT tested in np02? -  I understand that it was not [SP].}
has a notch at the extender tube weld joint, as illustrated in the
figure (shaded areas), avoiding the critical triple-point of the previous design~\cite{FD2-VD-CDR}  (where three different materials, LAr, \dshort{uhmwpe} and stainless steel meet with $<90^{\circ}$ angles).
Triple-point locations 
are subject to potential instabilities because the insulating surfaces could get charged up by free charges moving in LAr under the \efield in the contact region; if the angle between the insulating material and the metal surfaces is larger than $90^{\circ}$, the charging up is highly suppressed.
The notch in the new \dshort{uhmwpe} disk is required to support the extender 
and fully contains the weld joint inside the insulating plate, reducing this possibility.
%This \dshort{hv} extender design is being realized. 
The extender head and the support disk will be tested together with the new \dshort{hvft} before installation in the upcoming \dshort{vdmod0}. 

%%%%%%%%%%%%%%%%%%%%%%%%%%%%%%%%%%%
\section{HV Distribution System}
\label{subsec:VDss}

The two main components of the \dword{hv} distribution system are the cathode plane and the \dfirst{fc}. Together with the top and bottom anode planes, they define a uniform drift \efield in each of the two active volumes. The nominal bias voltage on the anode shield plane is $-$1.5\,kV.  Therefore, in order to achieve the goal of 450\,V/cm drift field, the cathode voltage needs to be set to $-$294\,kV for the 6.5\,m drift distance.

The cathode plane is composed of 80 cathode %unit 
modules identical in horizontal area to the \dwords{crp}. Since each cathode %unit 
module will be adjacent to others % cathode modules 
at the same voltage, their outer edges are not directly exposed to a high \efield region.  
The \dshort{fc}, on the other hand, designed to cope with high \efield outside the active volume, fully surrounds and shields the cathode plane.

The bias voltage to the cathode is delivered from the \dword{hvft} through the 6\,m extender to a \dshort{hv} bus that is mounted on the inside of the \dshort{fc} at mid-height.  This bus connects to the resistor divider board (\dword{hvdb}) chains on all the \dshort{fc} columns and to the outer edges of the cathode plane to provide a constant voltage to the cathode plane, independent of the current flowing through each branch along a \dshort{fc} column. Each chain of  resistive \dshort{hvdb}s on the anode 
ends is brought out of the cryostat to a separate bias power supply that provides adjustable voltage for fine-tuning the drift field near the anode planes and %measures 
monitors current. % for monitoring purposes. 

In addition to the primary function of providing uniform \efield{}s in the two drift volumes, both the cathode and the \dshort{fc} designs are tailored to accommodate \dfirsts{pd}, according to the configuration described in Chapter~\ref{chap:PDS}. 
In the %\dshort{spvd} \dshort{hvs} 
baseline design, each cathode module is designed to hold four double-sided \dword{xarapu} modules 
that are exposed through highly transparent wire mesh windows to the top and bottom drift volumes. 
\dshort{pd}s will also be mounted along the full perimeter of the cryostat walls,  limited to the vertical ranges $2.5<y<6.5$\,m and $-6.5<y<-2.5$\,m  (vertically separated from the cathode plane by 2.5\,m above and below). The \dshort{fc} is designed to provide approximately 70\% optical transparency, at normal incidence, over this vertical range.

%%%%%%%%%%%%%%%
\subsection{Cathode}
\label{subsubsec:CAsss}
%%%%

\begin{dunefigure}
[%Conceptual design 
Diagram of a cathode module with integrated \dshort{pd} units]
{fig:cathode}
{%Conceptual design of a %unit 
Diagram of a cathode module. The cathode module frame is constructed of \dshort{frp} beams. One double-sided \dshort{pd} module (blue) is integrated into each of four dedicated openings. Each \dshort{pd} is covered on both sides by a metallic mesh with high optical transparency while the rest of the top and bottom faces of the cathode frame is reinforced with cross ribs that support % covered with 
perforated resistive panels (not shown) on both sides. For the modules placed along the cryostat perimeter, the \dwords{pd} that would be located close to the cryostat wall are moved one slot away from it (see Section~\ref{sec:PDS-LightColl-discharge}).}
  \includegraphics[width=0.8\textwidth]{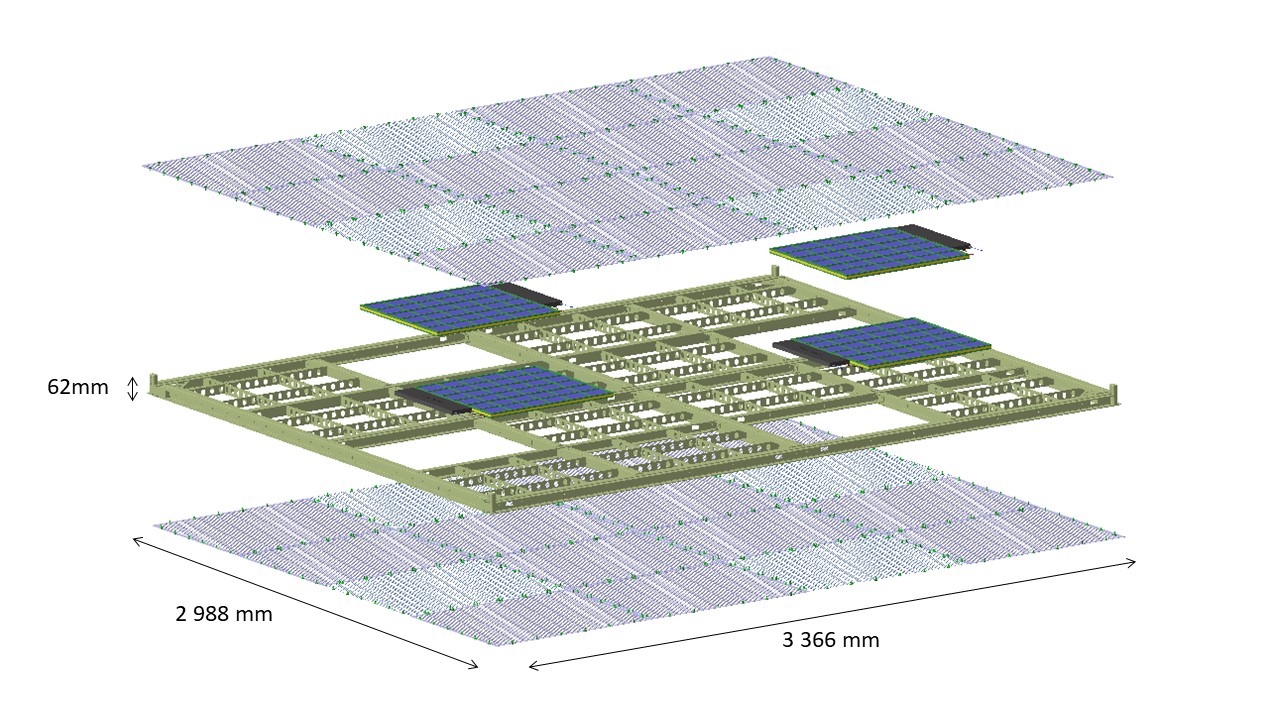}
 
\end{dunefigure}

The cathode plane, while modular like both the \dword{protodune} cathodes, requires an all new design due to the unique requirements that the \dword{spvd} design imposes on it, such as integrating the \dwords{pd} into it. A %conceptual design of a %unit 
cathode module is %shown 
illustrated in Figure~\ref{fig:cathode}.

\begin{dunefigure}
[View of the two half frames that form a cathode module frame] %structure] %the cathode]
{fig:cathode_half_frame}
{Two identical half frames are assembled to form a complete cathode module frame. } %structure.}
  \includegraphics[width=0.8\textwidth]{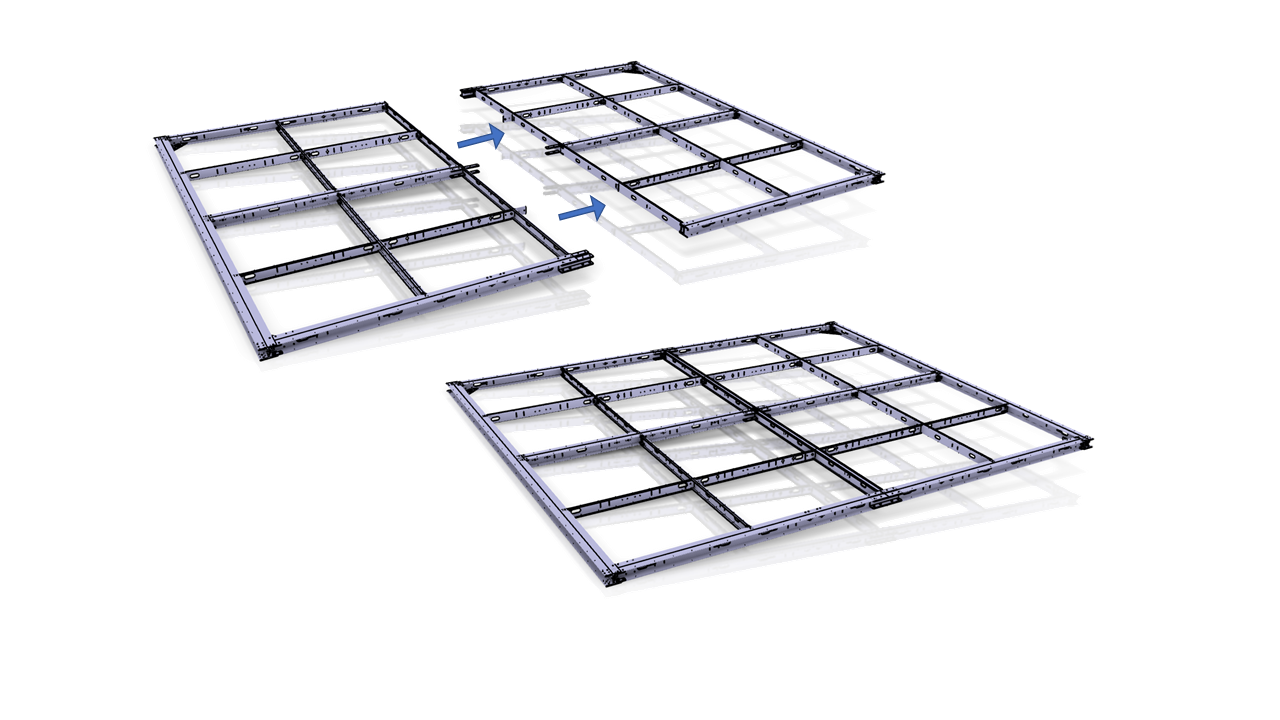}
\end{dunefigure}

A %unit 
single cathode module has the same footprint as a \dword{crp} module, with dimensions of 2.988\,m $\times$ 3.366\,m $\times$ 62\,mm. The weight of a %unit 
cathode module, including the integrated \dwords{pd}, is required to be less than 150\,kg (15\,kg/m$^2$) in air to minimize deformations of the \dword{crp} superstructure from which it hangs. The current estimate is 128\,kg. %In order to ease the 
To facilitate transportation, the frame consists of two identical half-frames (Figure~\ref{fig:cathode_half_frame}) that, after transport underground, are joined in the %clean room 
\dword{greyrm} in front of the cryostat, and then assembled into a %unit 
cathode module.

To remain below the maximum %bending 
deformation constraint of 20\,mm across the entire surface of each cathode module in \dshort{lar}, the cathode is constructed from \dword{frp} I-shaped and C-shaped beams (Figure~\ref{fig:cathode_beam_sizes}).

The dimensions of the overall frame are: 2948.3\,mm $\times$ 3326\,mm; this % in order to 
accommodates the \dshort{fc} supports, allowing use of % with 
the same frame design for both perimeter and inner %all 80 
cathode modules. While the cathode frame is slightly smaller than the \dshort{crp} footprint, the perforated resistive panels that deliver the \efield  will be machined to exactly %fill the CRP footprint
match it since %. As 
the shapes of the perforations have no impact on the mechanical behavior and they can be more easily machined than the frame. This %, it 
simplifies the production and reduces the overall cost.

\begin{dunefigure}
[Dimensions of the I-shape and C-shape beams used for cathode frame construction]
{fig:cathode_beam_sizes}
{Dimensions of the I-shape and C-shape beams used for cathode frame construction.}
  \includegraphics[width=0.9\textwidth]{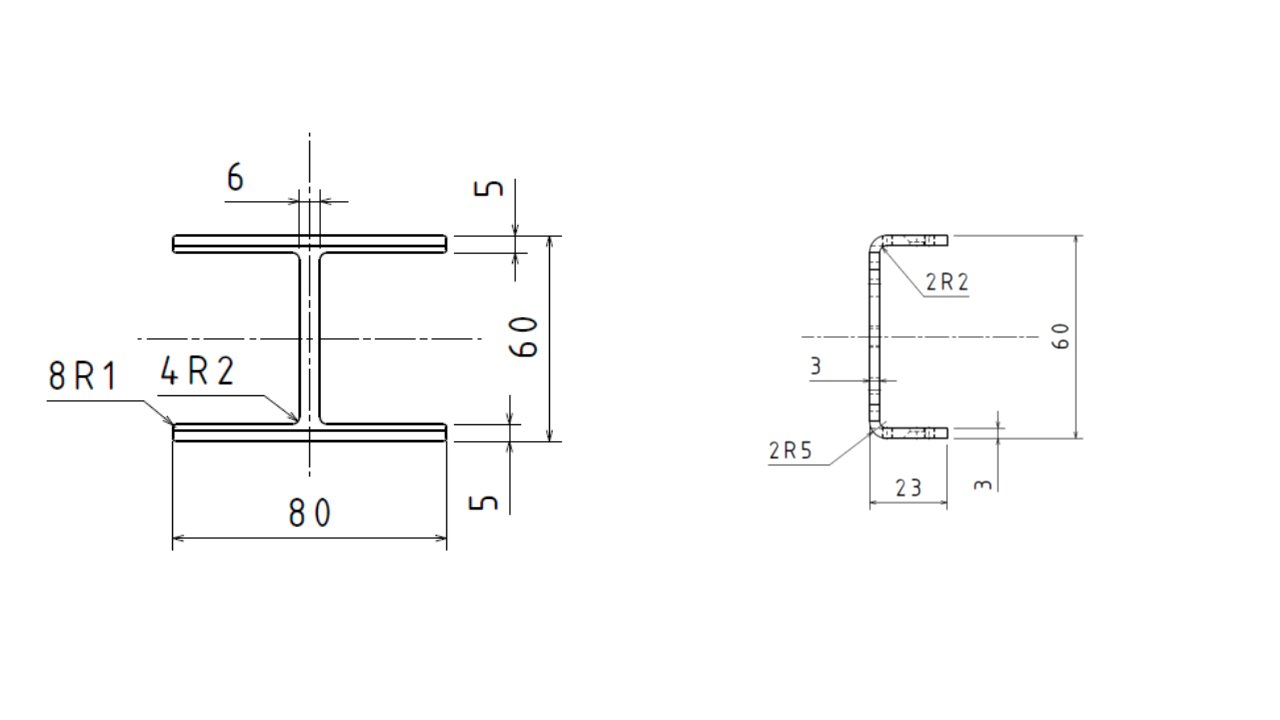}
\end{dunefigure}

According to the simulations that correctly predicted the measurements made on the prototype currently used in the \dword{np02} \coldbox  (with beams of 50\,mm depth), the expected deformations of the frame are given in Table~\ref{tbl:deform-numbers}:
\begin{dunetable}
[Expected cathode deformation]
{lcc}{tbl:deform-numbers}
{Expected cathode deformation}
Position on Cathode & In Air, z deformation (mm) & In \dword{lar}, z deformation (mm)\\ \toprowrule
Center & -26.9 & -8.5 \\ \colhline
Middle of Long Side & -14.3 & -4.5 \\ \colhline
Middle of Short Side & -10.4 & -3.3 \\ 
\end{dunetable}
These numbers are applicable for a single cathode module as well as for a six-module cathode supermodule configuration, described below (see Figure~\ref{fig:cathode_suspension_to_CRP}), and assume placement of all the \dshort{pd}s at the center of each cathode module, which would cause the most deformation; i.e., worst case. 

\begin{dunefigure}
[Cathode supermodule suspended from CRP superstructure]
{fig:cathode_suspension_to_CRP}
{Conceptual view of a six-module %unit 
cathode supermodule suspended from a \dshort{crp} superstructure. Left: side view. Right: top view overlaid with a \dshort{crp} superstructure and the embedded \dshort{pd} \dshort{xarapu} modules in blue squares.} 
  \includegraphics[width=0.8\textwidth]{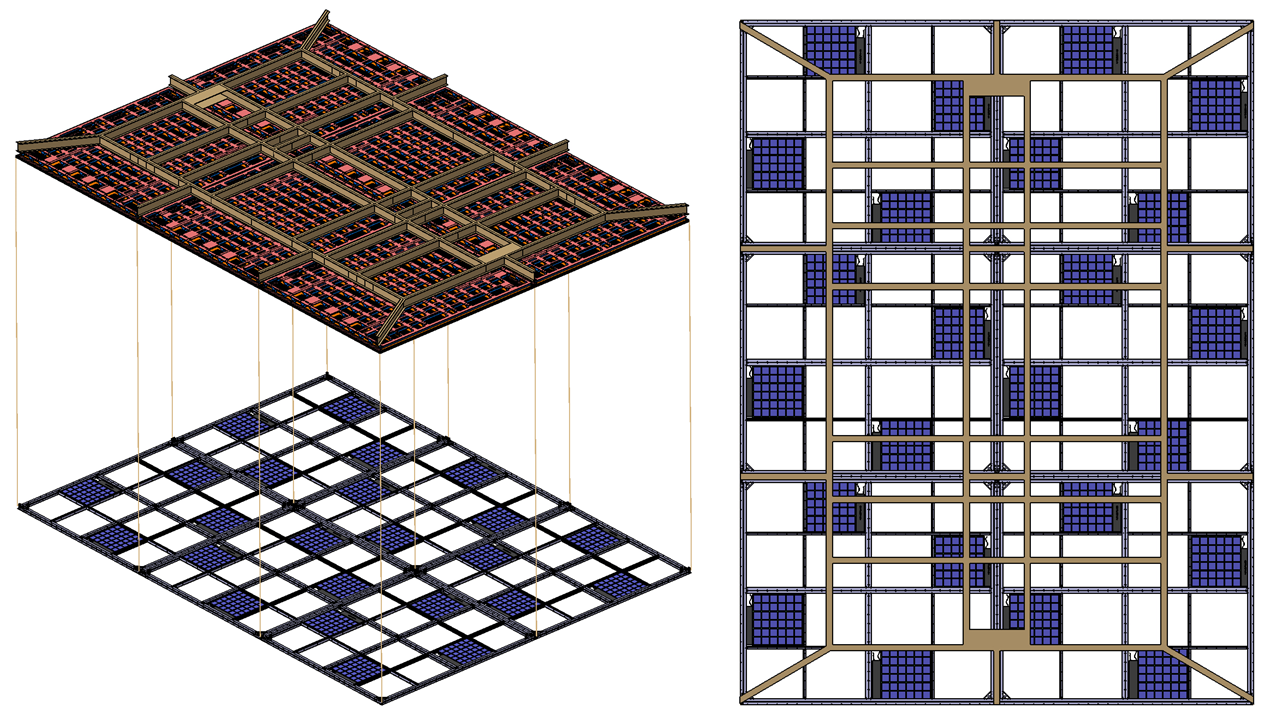}
\end{dunefigure}

After production, the deformation of the frames will be measured with the expected loads (using meshes and dummy \dword{pd} modules)  %in order 
to check compliance with the requirement (FD-11 in Table~\ref{tab:specs:SP-HV}). 
The measurements will be done before and after a plunge in liquid nitrogen in order to check the impact on them of sudden cooling %at 
to cryogenic temperatures. A visual inspection of the beams and connections will also be done to check for any non-conformity.

The frame for each cathode module has 16 openings, four of which are dedicated to \dshort{pd}s. The four openings are all of slightly different sizes ({\qtyproduct[product-units=power]{635x730}{mm}}, 
{\qtyproduct[product-units=power]{635x812}{mm}}, 
{\qtyproduct[product-units=power]{696x730}{mm}} and 
{\qtyproduct[product-units=power]{696x812}{mm}}), 
all of which can accommodate the active part of the \dshort{pd} module, which is {\qtyproduct[product-units=power]{621x621}{mm}}. %, fitting in all openings.

The four openings %each of which 
containing the double-sided \dshort{pd} modules are covered on both sides by stainless steel wire-mesh panels of optical transparency above 86.5\% at normal incidence. A mesh of this type that satisfies all the DUNE requirements is commercially available. A set of meshes has been produced and installed on the cathode prototype in the \coldbox and has exhibited no issues throughout the duration of the tests.

To ensure the planarity of the perforated resistive panels and the wire meshes, which are quite flexible due to their high transparency, additional supporting cross ribs are provided in all cathode frame openings; these do not change the mechanical properties of the frame. The cross ribs in the four openings for the \dshort{pd} are integrated into the \dshort{pd} mechanical design (see Figure~\ref{fig:cathode_crossribs}), and have been successfully tested in the \coldbox.  
The cross-ribs in the remaining 12 openings %of each cathode  
support Vetronite EGS 619 AS %perforated 
resistive panels from Von Roll\footnote{Von Roll \url{https://www.vonroll.com/en/}} (which is like G10) on both sides. The panels are perforated to reduce their weight and allow \dshort{lar} flow.
The square shaped perforations ({\qtyproduct[product-units=power]{25x25}{mm}} 
spaced by \SI{2.5}{mm}) are machined by water-jet cutting. 
Small hooks hold the panels.

The resistivity measurement of the Vetronite is \SI{16} {\mega\ohm}/sq, in agreement with the requirements. %An 
An optical transparency (due to the perforations) %for the resistive mesh 
of $75\%$ has been achieved while maintaining good mechanical properties and %. It will also facilitate 
achieving %good 
sufficient \dshort{lar} flow across the cathode plane.  
The use of resistive material both reduces the peak current flow along the cathode in the event of a \dshort{hv} discharge and slows down the associated voltage swing, which reduces %the 
any potentially dangerous charge injection into the readout electronics. The \dshort{pd}s would be protected from a dangerous large spatial voltage gradient across their surface %during a \dword{hv} discharge 
by the stainless steel meshes that cover them.

By analyzing the spatial distribution of the hits recorded in the \coldbox, %we have been able 
it has been possible to measure the small variations in \efield intensity (at the level of few percent) close to the cathode surface. 
This non-uniformity is due to the perforated structure and to the plates supporting the \dshort{pd} electronics. The %effect is 
measurements are in good agreement with the \dword{comsol} simulations of the cathode performed in-house.

The pitch of the mesh wires covering the \dshort{pd}s is 25\,mm, chosen to avoid significant distortions of the \efield. As shown in Figure~\ref{fig:cathode_electrical_field_close_surface}, obtained from a \dshort{comsol} simulation of the cathode, the \efield component along the drift direction is within 1\% of its nominal value beyond 5\,cm from the cathode plane (top plot). The amplitude of the transverse component of the \efield is smaller than 1\% of the longitudinal one beyond 3\,cm (bottom plot).

\begin{dunefigure}
[Electrical fields close to the cathode surface]
{fig:cathode_electrical_field_close_surface}
{\efield (normalized to the nominal %one
value) close to the cathode surface. The top figure shows the uniformity of the \efield component along the vertical direction, while the bottom %one presents 
shows the relative amplitude of the component in the horizontal plane. The error bars give the range of  variation %for 
at a given distance. The red lines indicate the 1\% specification.} % is shown with the red lines.}
  \includegraphics[width=0.8\textwidth]{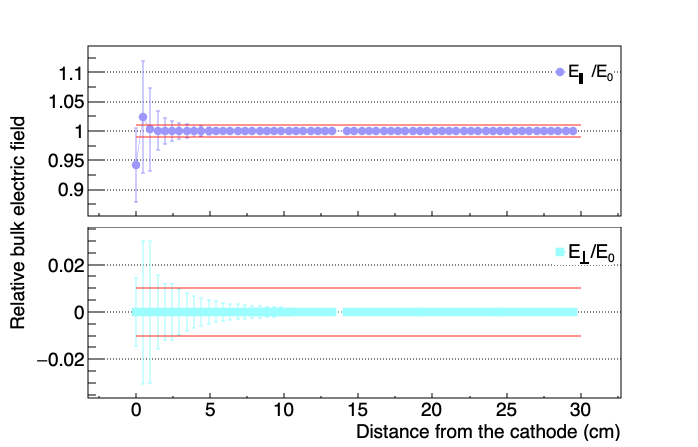}
  
\end{dunefigure}

For a given specification on the drift field uniformity, %homogeneity 
the simulation also allows computation of the fraction of the drift volume for which the specification is satisfied. Figure~\ref{fig:cathode_specif_homogeneity} shows that the difference between the nominal drift field and that generated by the cathode is less than 1\% over $\sim$ 99.4\% of the drift volume.

\begin{dunefigure}
[Volume within specification vs field homogeneity specification]
{fig:cathode_specif_homogeneity}
{Volume within specification vs field homogeneity specification}
  \includegraphics[width=0.8\textwidth]{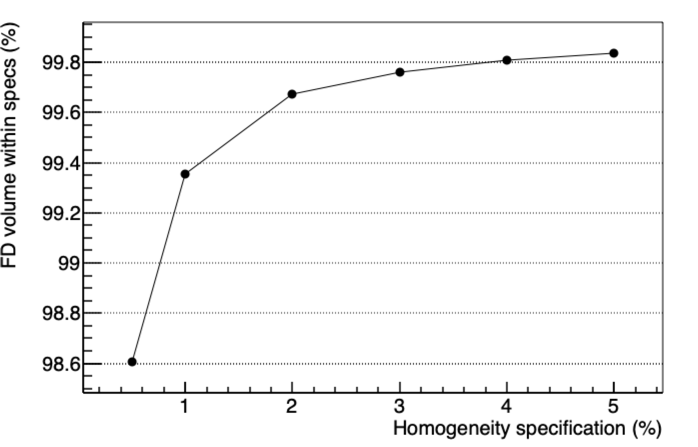}
\end{dunefigure}

\begin{dunefigure}
[View of a portion of a cathode module] 
{fig:cathode_crossribs}
{View of a portion of a cathode module. The opening in the foreground (and the one in the right rear) hosts a \dshort{pd} module (not visible in the image) covered on both sides by a stainless steel wire mesh (shown as light gray), with its electronics box at left. The openings in the left rear and on the right show the cross ribs and the (75\%) perforated resistive panels, one on each side; the green bar is the fiber bundle.} 
  \includegraphics[width=0.8\textwidth]{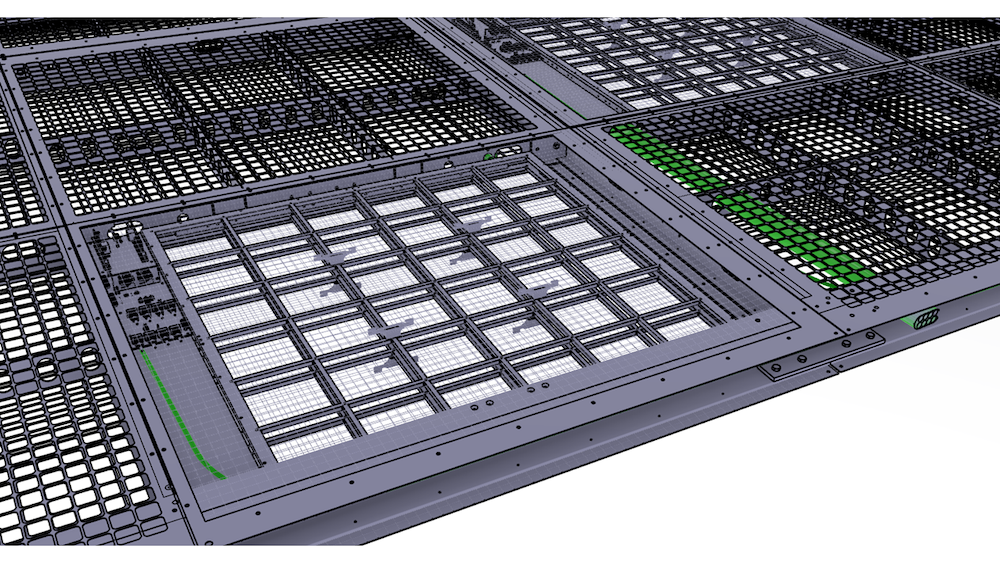}
\end{dunefigure}

To maintain the entire cathode at the same voltage, adjacent resistive panels %or wire-mesh  panels 
are interconnected at regular intervals both laterally and %through the thickness of the frame 
%on 
across opposite sides of the cathode 
to ensure good electrical connections with sufficient redundancy.  The top and bottom interconnects of the wire meshes also offer some electrostatic protection to the \dshort{pd} modules mounted in between. Similar electrical interconnects between adjacent cathode modules and to the neighboring \dword{fc} modules are made through flexible hookup wires. 

\begin{dunefigure}
[Frame opening of \dshort{pd} fibers and cables]
{fig:cathode_pd_fiber_routing}
{View of the frame openings %foreseen 
designed for \dshort{pd} fiber and cable routing} 
  \includegraphics[width=0.8\textwidth]{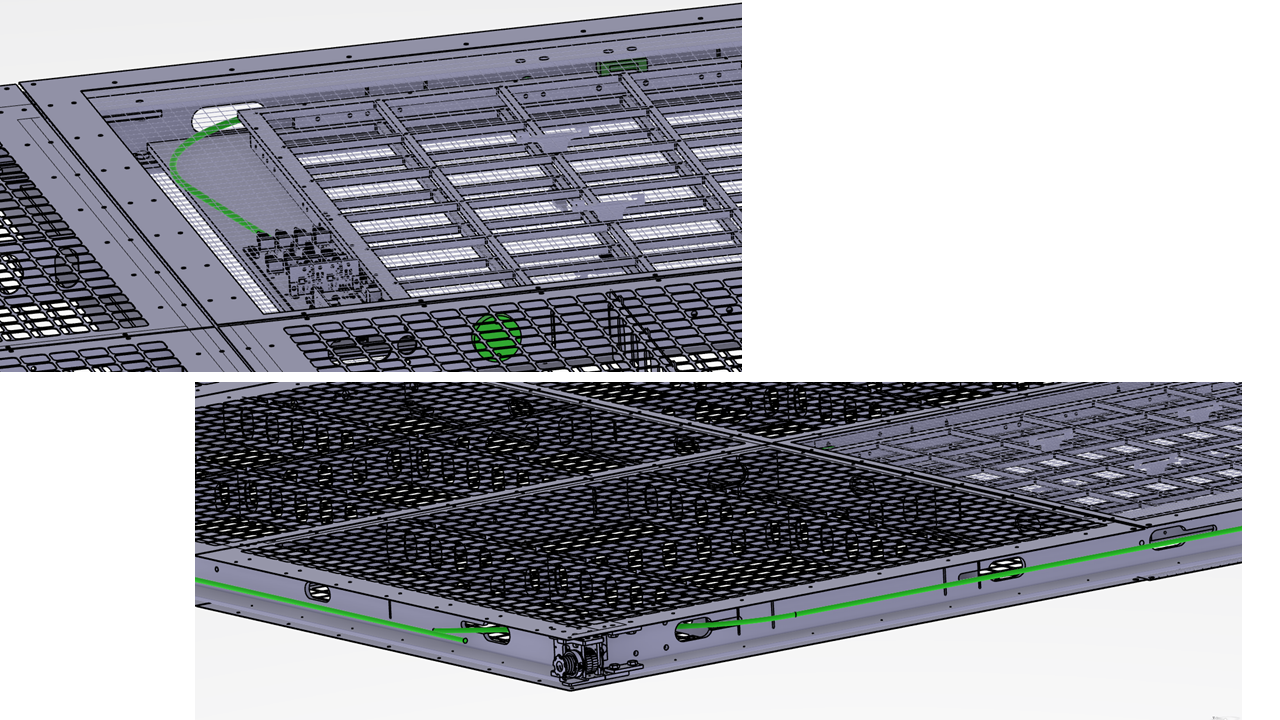}
\end{dunefigure}

To reduce the amount of work in the underground environment, mounting hardware and local cables and fibers are pre-installed into the cathode module at the cathode assembly site. The beams in the frame %has been designed with 
have several openings %in the beams 
(Figure \ref{fig:cathode_pd_fiber_routing}) %to insure the possibility to 
for routing the \dshort{pd} cables and fibers from any location %of the  \dword{xarapu} module 
in the cathode frame. 
The \dshort{pd} modules are installed into the cathode module shortly before the cathode is attached to the \dword{crp} superstructure once in the cryostat. 
The optical fibers that carry power and signal for the \dshort{pd}s, described in Section~\ref{subsec:PoFsss}, are routed from the cable trays on the bottom of the cryostat, up the \dword{frp} box beams on the bottom half of the \dshort{fc} below the cathode, and along the sides of the cathode frame to the designated cathode module. Section~\ref{sec:detcompinst} provides details on cable and fiber routing.

The cathode plane consists of 80 cathode %unit 
modules total, assembled into a set of 16 supermodules (12 six-module %unit 
and four two-module), 
matching the modularity of the \dword{crp} superstructures (see Section~\ref{sec:CRPMss}) to simplify the installation. 
The six-module 
supermodules are in $2 \times 3$ formation, and the %second type consists of 
two-module %cathode modules 
supermodules are $2\times 1$. % formation.
To form the entire cathode, the 12 six-module  supermodules are arranged  $2 \times 6$, each with the three-cathode-module (long) side oriented lengthwise along the cryostat  
and two (2) two-cathode-module supermodules %arranged in 
along each end wall arranged with their long sides against the end wall.  

Each six-module cathode supermodule is suspended from the ten \dshort{crp} superstructure extensions along its perimeter and %at 
two junctions near its center using 12 non-conductive cables (``ropes''), as illustrated in Figure~\ref{fig:cathode_suspension_to_CRP}. 
The %cables 
ropes are made of 3.35\,mm diameter Dyneema DM20, produced by Corderie Lancelin\footnote{Corderie Lancelin \url{https://lancelin.com/}}, which was chosen because it has one of the highest strength-to-weight ratios 
%and an absence of creeping 
and, critically, according simulations done with the simulation tool of the Dyneema provider, maintains a very stable length over ten years of use at cryogenic temperatures. 
 The suspension configuration limits the %cathode 
distortion to about 8.5\,mm in \dshort{lar} and 26.9\,mm in air across the entire surface of each cathode supermodule.

Figure~\ref{fig:cathode_suspension_system} presents a conceptual view of the suspension system. % composed of five components which are described hereafter. 
Each of the ten long ropes (about 6\,m) that support the cathode supermodule from the top of the \dshort{crp} superstructure is attached %thanks 
to a top adjusting device (TAD). % on the top of the superstructure that supports the \dword{crp}. 
At the bottom of the long rope, short ropes (one, two or four according to the position of the rope and the number of connected cathode  modules) will be attached to the long rope via a shackle, and %. Then, the short rope are 
connected to the cathode supermodule frame using the length adjusting device (LAD).

\begin{dunefigure}
[Cathode suspension system]
{fig:cathode_suspension_system}
{Conceptual view of the cathode suspension system composed of five elements: the top adjusting device (TAD), a long rope (about 6\,m), a shackle, short ropes (1, 2 or 4) and the length adjusting device (LAD).} 
  \includegraphics[height=12cm]{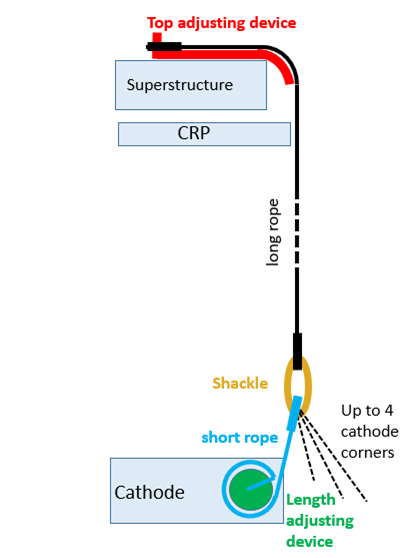}
\end{dunefigure}

The TAD shown in Figure~\ref{fig:cathode_TAD} %is mounted on the top side of the \dword{crp} superstructure. This 
is the anchor point of the rope. The TAD provides an approximate large-scale tuning of the rope length as well as  
the precise positioning of the %\SI{6}{m} long cable 
rope %that 
as it goes through a $20\,\times 20 $\,mm square space at the corner between the \dword{crp} %supermodules. 
superstructures. 
\begin{dunefigure}
[Cathode top adjusting device]
{fig:cathode_TAD}
{Model of the cathode top adjusting device (TAD)  on the \dshort{crp} superstructure} 
  \includegraphics[width=0.8\textwidth]{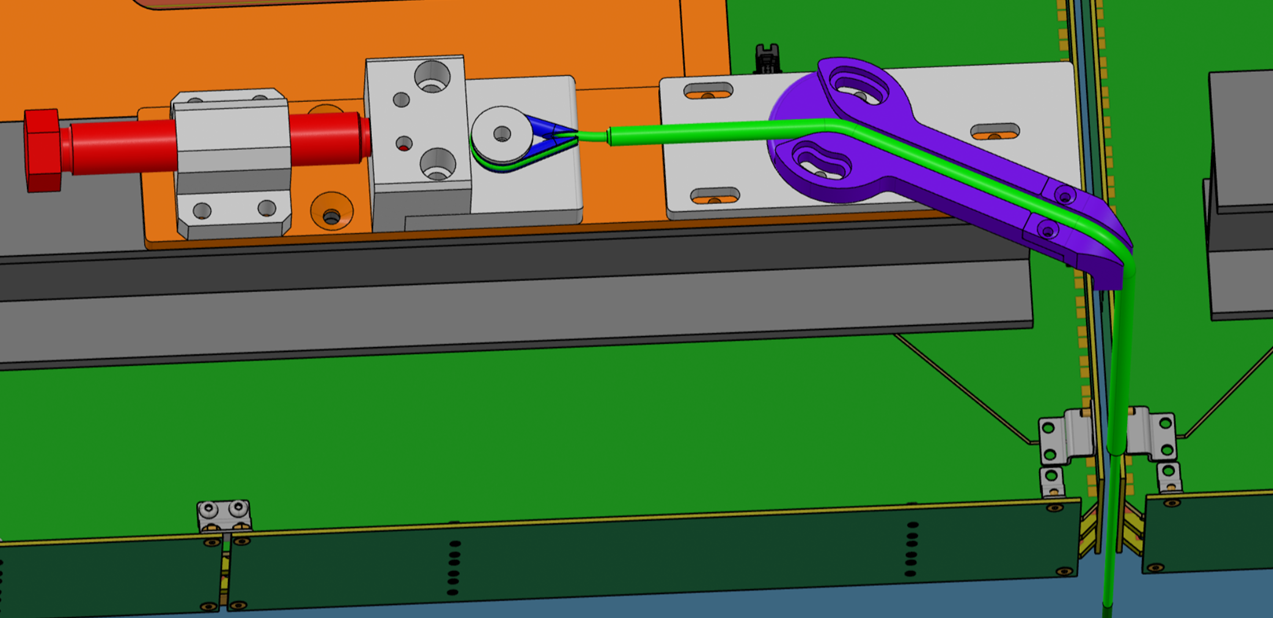}
\end{dunefigure}

 Uniformity of the rope production, in particular the reproducibility of the rope lengths after load, has been validated with a joint R\&D program between Physics Laboratory Irène Joliot-Curie, Paris (IJClab) and the rope producer. The dispersion on these lengths has an impact on the tuning range of the TAD and the LAD.

Several ropes were extensively tested at room temperature in late 2022. Some preliminary behavior tests at cryogenic temperatures have also been performed, and more recently quantitative measurements were carried out on a dedicated test bench. 
The rope stretches quite linearly with respect to the load, with a measured elongation coefficient of about 0.33\,mm per kilogram. Measurements made over the course of a few months show reproducibility to about 1\,mm, which is the precision of the current test setup. In \dshort{lar}, the ropes are expected to stretch by $\sim14\,$mm due to thermal expansion at low temperature (Dyneema has a negative \dword{cte} of about $-10^{-5}/K$). The buoyant forces, however, will reduce the load, so the rope length  will depend on the number of supported cathodes, at about $\sim$12.5\,mm per cathode. 
Table~\ref{tbl:rope-lengths} presents the estimated length difference for the three possible cases (rope supporting one, two or four cathode supermodules):

\begin{dunetable}
[Length differences of ropes supporting the cathode]
{cc}
{tbl:rope-lengths}
{Estimated length differences of ropes supporting the cathode}
 & Length Difference (in mm) \\
\rowtitlestyle Number of supported & between installation and running conditions \\
\rowtitlestyle cathode supermodules & (+ in up direction) \\ \toprowrule
1 & -1.5 \\ \colhline
2 & 11 \\ \colhline
4 & 36 \\ 
\end{dunetable}

\begin{dunefigure}
[Model of test bench for rope characterization]
{fig:cathode_rope_testbench}
{A conceptual view of the test bench under construction to be used for the quantitative characterization of the ropes (elongation under load and thermal elongation)} 
  \includegraphics[width=0.8\textwidth]{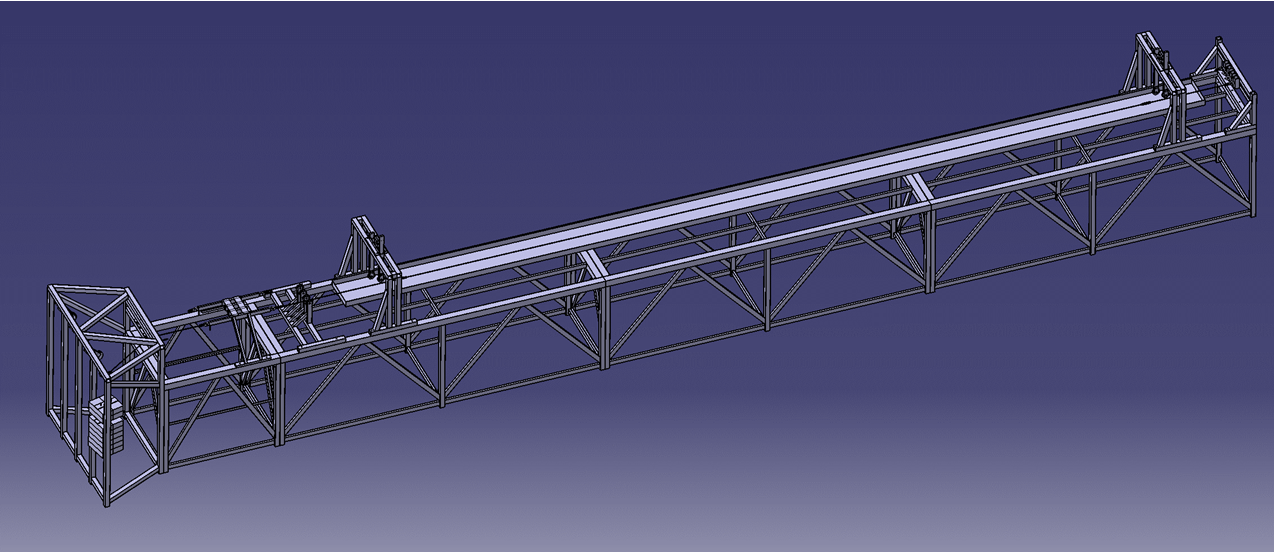}
\end{dunefigure}

Since there is no way to adjust the rope lengths once the cryostat is closed, it is essential to know precisely the rope's elongation behavior due to load and thermal factors. 
A dedicated bench (Figure~\ref{fig:cathode_rope_testbench}) is under construction to measure each of the 168 \SI{6}{m} long ropes and the 192 \SI{50}{cm} long ropes needed to support the full set of cathode supermodules.
The coupling between the long and the short ropes is performed through the standard Dyneema shackle. %produced by Corderie Lancelin. 

\begin{dunefigure}
[Cathode length-adjusting device (LAD) ]
{fig:cathode_LAD}
{A model of the cathode length adjusting device (LAD) placed in the cathode supermodule frame.} 
  \includegraphics[width=0.8\textwidth]{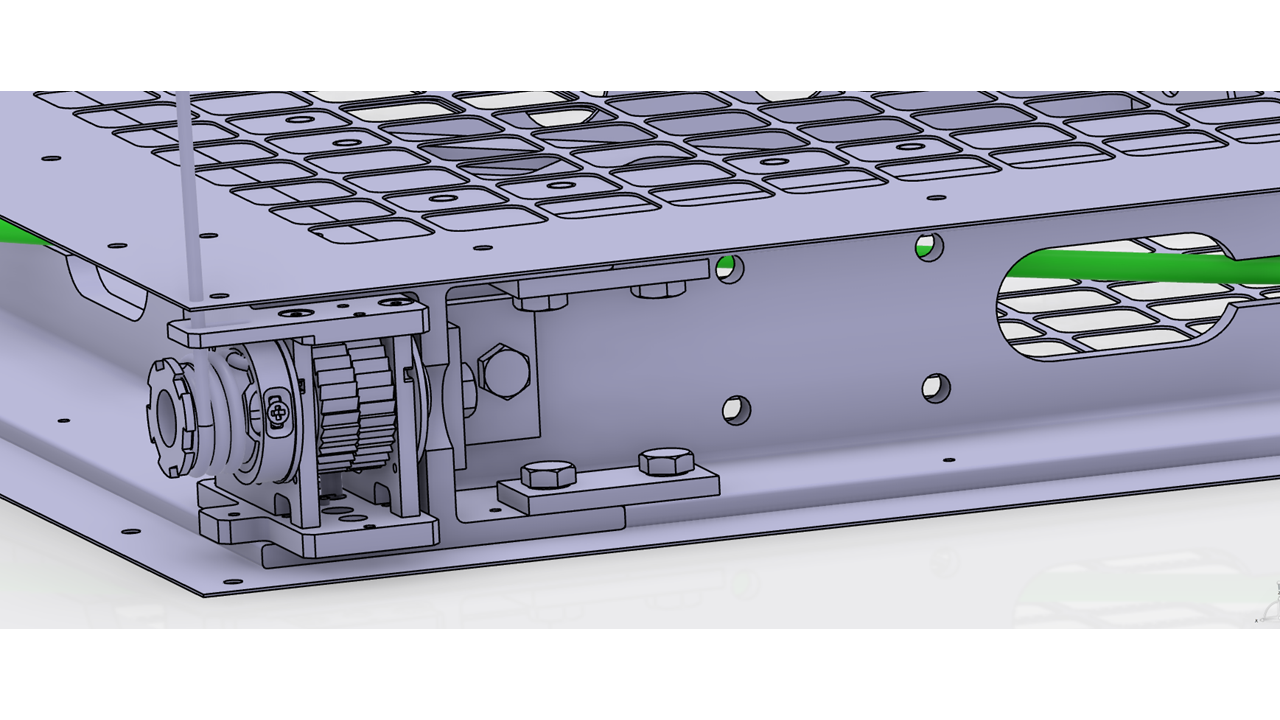}
\end{dunefigure}

For the connection to the cathode and the fine-tuning of the rope length, a dedicated device has been designed called the length-adjusting device (LAD), shown in Figure~\ref{fig:cathode_LAD}. The LAD is installed inside the I-shaped beam of the cathode frame to ensure the correct transmission of the stress. A system of two gears allows length tuning with a 1.5\,mm precision and a range of $\pm$5\,cm.

\begin{dunefigure}
[Cathode length-adjusting device tests ]
{fig:cathode_test_LAD_Rope}
{A photograph of the cathode LAD and a Dyneema rope under test in air. The system is loaded with a 200\,kg weight. (The measuring stick shows inches.)} 
  \includegraphics[width=0.65\textwidth]{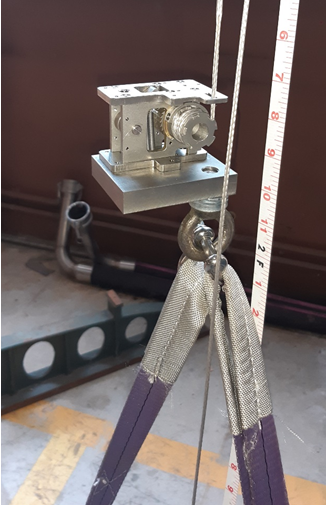}
\end{dunefigure}

A LAD prototype was produced that successfully passed load tests in air. Figure~\ref{fig:cathode_test_LAD_Rope} shows the prototype connected to a Dyneema rope with a load of 200\,kg. No measurable deformation has been observed after one week of stress under the load. The LAD was also tested in a cryogenic bath and showed no deformation. The next step is to perform a load test in liquid nitrogen once the test bench illustrated in Figure~\ref{fig:cathode_rope_testbench} becomes available. 

All elements described in this section (frame, resistive panel and metallic mesh, ropes, the TAD, the LAD, etc.) will be produced for \dword{vdmod0} 
with their current designs, which are expected to be final. %the final one. 
Some of the tools needed for installation will also be produced for and tested during %Module-0 
\dshort{vdmod0} installation. Due to the limited size of \dword{np02} and to the different \dword{crp} and cathode support structure under the cryostat ceiling, 
the actual full assembly test of the \dshort{crp} superstructure hanging six cathode modules will require a dedicated mock-up to tune the final installation procedure and the associated tools.

%%%%%%%%%%%%%%%. 
\subsection{Field Cage}
\label{subsubsec:FCsss}
%%%%
%anne to here
\dword{pddp} demonstrated the simplicity of installing a \dword{fc} in a vertical drift \dword{tpc}, and \dword{pdsp}, whose \dshort{fc} had identical design fundamentals, operated successfully.  The lessons learned in  \dshort{pdsp} have led to an updated \dshort{fc} design for the \dword{sphd}, in which all insulating materials on the side of the \dshort{fc} facing the grounded cryostat membrane wall are removed to improve \dword{hv} stability. The updated \dword{spvd} design, illustrated in Figure~\ref{fig:xparent-fc-overview-new}, implements these features, % of the \dword{sphd}, 
and aims to achieve both improved performance and an even simpler \dshort{fc} assembly.

\begin{dunefigure}
[Field cage with 70\% optical transparency]
{fig:xparent-fc-overview-new}
{Left: A corner view of the \dshort{fc} baseline design with 70\% optical transparency along the four cryostat walls, starting at 2.5\,m away from the cathode plane.
%in both directions for 4\,m. 
Right: \dword{fea} calculation of the \efield (V/m) of the corner without the additional insulation shield. The maximum value on the color scale is 1.4$\times 10^6$\,V/m.} 
  \includegraphics[width=1.0\textwidth]{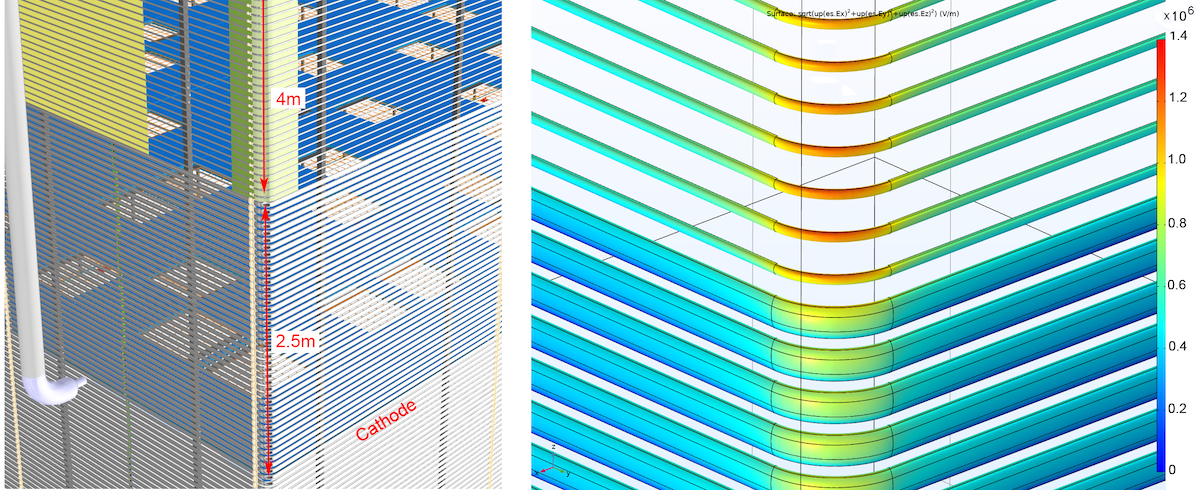}
\end{dunefigure}
\subsubsection{Geometry, Mounting and Support} 

The \dshort{fc} modules for the long walls are 3.0\,m wide $\times$ 3.24\,m high, and for the endwalls, 3.38\,m wide $\times$ 3.24\,m high, the widths of their aluminum electrode profiles matching the \dwords{crp}' edges.  
Each \dshort{fc} module consists of 54 extruded aluminum profiles stacked vertically at a 6\,cm center-to-center spacing. The profiles are mounted on two %5-cm-tall 5-cm-wide 
 $5\times5$\,cm$^2$ square cross section \dword{frp} box beams of length 3.24\,m, spaced 1.8\,m apart from each other. A subset of the profiles  are fixed on both box beams at regular intervals to define the spacing between the box beams. The rest of the profiles are fixed only on one beam and are allowed to slide on the other (see discussion below on shrinkage). 
On the side of each endwall profile that meets one of the four corners of the cryostat, the profiles are bent 90$^\circ$ with a 10\,cm bending radius to connect to the adjoining long-wall profiles and to avoid charge buildup on the insulating caps at the corners.
Figure~\ref{fig:field_cage} shows the proposed \dshort{spvd} \dshort{fc} long wall module design with the \dwords{hvdb} shown in green. 
The weight of a \dshort{fc} module various between 35kg for the thin profile long wall type and 60 kg for the mixed profile end wall type.

\begin{dunefigure}
[Field cage module with bent profiles]
{fig:field_cage}
{A long-wall \dshort{fc} module to be placed at a corner. Its profiles are bent to $90^\circ$ on the corner side to minimize the local field and to align with the \dshort{fc} module along the adjoining end wall. Note the 12 thin profiles at the bottom of the figure; 
a module containing only thin profiles will be installed below the one in the picture and the two together will be part of the \dshort{fc}  of the bottom drift region.
On each module, six \dshort{hvdb}s are mounted %symmetrically 
along a vertical line passing through its center of gravity. %of the \dshort{fc} module. 
}
  \includegraphics[width=0.8\textwidth]{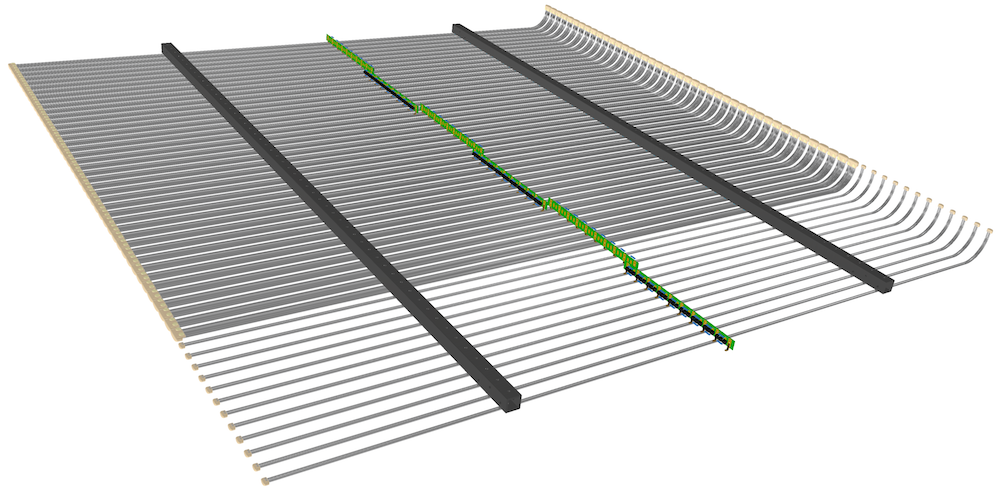}
\end{dunefigure}

\begin{dunefigure}
[Field cage support structure]
{fig:fc_support}
{Top: Top \dshort{fc} support beam and yoke structure for a \dshort{fc} supermodule. A stainless steel I-beam (cyan) is suspended by two lift rods that are each lowered through a \dshort{fc} 
support penetration in the cryostat roof.  Each end of the beam is connected to an aluminum yoke (orange) at a single pivot point.  The two ends of the yoke are attached to the two box beams of a \dshort{fc} column.  Bottom: Bottom \dshort{fc} stabilizer.  One stainless steel angle beam (cyan), and two aluminum angles (orange) mirror the beam and yoke structure at the top of the \dshort{fc} column to ensure parallel lateral shrinkage of the column. The aluminum angles have clearance holes in the middle, and slots at the ends that slide along the vertical pins mounted on the stainless steel angle. The three feet are glued onto the membrane floor to prevent lateral sway %of the field cage column 
while allowing contraction in the vertical direction. The middle foot (red) only allows vertical motion of the stainless steel angle while the other two feet allow its horizontal contraction. 
}
  \includegraphics[width=0.9\textwidth]{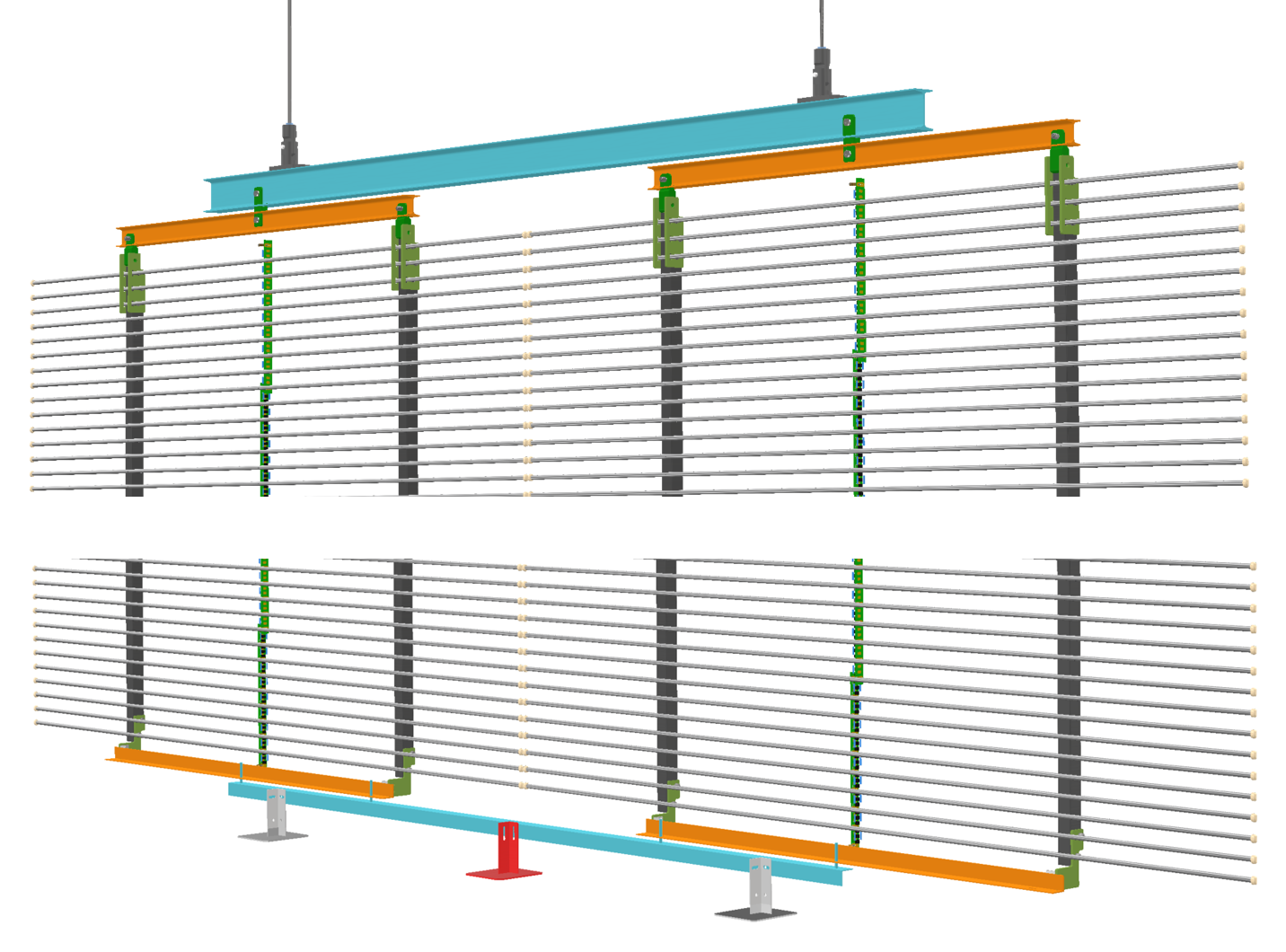}
\end{dunefigure}

Two columns of four \dshort{fc} modules each form a \dshort{fc} supermodule,  
6.0\,m(W) $\times$ 13\,m(H) for the long walls and 6.76\,m(W) $\times$ 13\,m(H) for the endwalls.  
Each long wall has 10 adjacent supermodules and each endwall has two; together these 24 supermodules form the \dshort{fc} that surrounds the two active drift volumes.
Each \dshort{fc} supermodule hangs from a stainless steel I-beam shown in Figure~\ref{fig:fc_support} with two lifting rods, each going through one of the 48 roof penetrations provided for support of the \dshort{fc}, as described in Chapter~\ref{ch:IEI}.    
Each mechanical feedthrough flange on top of these penetrations, described in Section~\ref{subsubsec:FCFT}, are designed to allow the \dshort{fc} lift rods 
to be pulled up and anchored on the flange. The \dshort{fc} support flange has a mechanism to allow the 
supermodule to be moved laterally by a few centimeters to avoid interference with the \dword{crp} during the \dshort{fc} installation.

The cryostat roof is expected to deform downward by up to  13\,mm after the cryostat is filled with \dword{lar}, and upward by about 6\,mm  
before safety relief valves open~\cite{EDMS2217676}.  
Because of this, a pair of 
\dshort{fc} supermodule support points on the roof could have a height differential up to 3.5\,mm, or an equivalent rotation of 1\,milliradians, which translates into a 13\,mm lateral swing at the bottom of a 13\,m tall \dshort{fc} column. To maintain perpendicularity of each \dshort{fc} column and to ensure even load balance on each \dshort{fc} module,  each top module of the two \dshort{fc} columns is connected to the \dshort{fc} support  beam through an intermediate aluminum yoke at a single point located directly above the center of gravity of the assembled \dshort{fc} column. At the bottom of each column, a stabilizer prevents the \dshort{fc} column from lateral movement while permitting overall shrinkage %of the field cage 
in the vertical direction. The bottom stabilizer %has a 
construction mirrors that of the top 
support structure.  With its middle foot glued down to the membrane, % flat, 
which has little lateral movement during \cooldown, it will keep the columns vertical even if the \dshort{fc} construction is slightly asymmetrical in weight distribution.    

\subsubsection{Shrinkage during Cool-down}

At \dword{lar} temperature, the bottom of the \dword{fc} will shrink up by about 23\,mm from a combination of \dword{frp} and stainless steel structure thermal contractions.  This movement needs to be accounted for when setting  the bottom \dword{crp} plane height.  The top of the \dshort{fc} moves more or less in sync with the top \dshort{crp} since they both mostly shrink at the rate of the stainless steel support structures.   
Along the long walls of the cryostat, the (horizontal) gaps between \dshort{fc} columns on adjacent supermodules will increase by about 20\,mm due to shrinkage of both the stainless steel support beams and the aluminum profiles.
The gaps between \dshort{fc} modules on the same supermodule will see a smaller increase of about 3\,mm, however, due only to the difference in the shrinkage between the two metals. 
The corresponding two values along the end walls will be 10\% greater due to the larger span of both the beams and profiles.  Perpendicular to the \dshort{fc}, the (horizontal) gaps between the \dshort{fc} and the anode/cathode will increase by about 7\,mm. The 20\,mm increase in \dshort{fc} module gaps is expected to cause E field non-uniformity exceeding the 1\% limit in small volumes of LAr near the \dshort{fc} gaps at the outer edges of the active volume. The total fraction of the impacted LAr is negligible compared with the active mass of the TPC.

Since most of the \dword{spvd} components are oriented horizontally and occupy relatively narrow slices of vertical space, they are 
quite insensitive to vertical temperature gradients.
The \dshort{fc}, however, is an exception. 
The amount of contraction in the lengths of the aluminum \dshort{fc} profiles at different heights will vary due to the vertical temperature gradient during the initial \cooldown and \dshort{lar} filling period. To avoid excess stress buildup in the \dshort{fc} modules, most profiles are fixed on only one of the two \dshort{frp} box beams in a \dshort{fc} module and can slide on loosely tightened but secured slip nuts. 
To maintain the box beam spacing, a subset of profiles at large vertical intervals are fixed on both box beams.  The interval is determined by the expected temperature gradient (from CFD studies and \dword{protodune} results) above the \dshort{lar} surface, and the maximum stress the box beam and aluminum profiles can tolerate.  This will be finalized as part of the ongoing \dshort{fc} stress analysis.
This configuration, in which the profiles are fixed  only on one box beam, increases the \dshort{fc} modules' tolerance to a higher vertical temperature gradient, and therefore reduces some constraints on the cryogenics system for cryostat \cooldown and filling. 

\subsubsection{\efield Uniformity and Field Cage Transparency} \label{subsubsec:TranspFC}
The \dword{fc} profiles are positioned 5\,cm away from the boundary of the anode planes.  This buffer zone ensures a sufficient \efield uniformity at the boundaries of the sensitive volume, avoiding the highly modulated pattern very close to the discrete \dshort{fc} electrodes.  

The 5\,cm vertical gap between the cathode and the \dshort{fc} is in the transition region between the two drift fields in opposite directions. Due to the lack of well-defined electrodes in this gap, a non-uniform \efield extends into the active volume.
To improve the field uniformity, a set of aluminum strips is mounted on the \dshort{fc} profiles at the cathode height, effectively extending the cathode surface to the \dshort{fc}. Figure~\ref{fig:cathode_edge} shows the position of the field-shaping strip and the improved field uniformity at the edge of the cathode plane. The additional field-shaping strips essentially form a 5\,cm$\times$5\,cm channel between the \dshort{fc} and the outer perimeter of the cathode plane. 

\begin{dunefigure}
[\efield uniformity near the cathode-FC intersection]
{fig:cathode_edge}
{\efield uniformity near the cathode and \dshort{fc} intersection. A field-shaping strip is installed in the gap between the outer edge of the cathode and the \dshort{fc} electrodes to extend the cathode surface to the \dshort{fc}.  The  \efield lines inside the \dshort{tpc} are shown in gray, and the color gradient represents the \efield strength (V/m) in a logarithmic scale.  
The dark blue half-ellipses closest to the cryostat wall (left) represent the \dshort{fc} profile cross sections. The calculations were made with a drift field of 500\,V/cm, which can be scaled linearly down to the goal drift field value of 450\,V/cm.}
\includegraphics[width=0.9\textwidth]{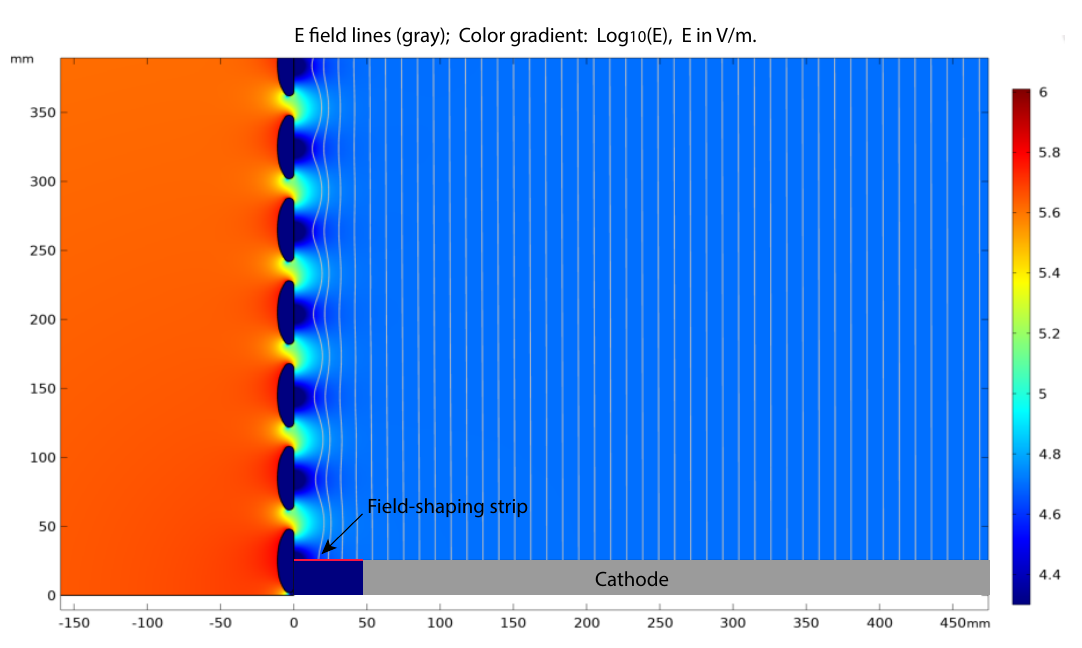}
\end{dunefigure}

The middle two rows of \dshort{fc} profiles at the cathode level form a \dword{hv} bus (see Figure~\ref{fig:hv-bus}) that distributes the cathode bias voltage to all cathode modules as well as all the \dshort{fc} columns. Unlike what has been done for \dword{sphd}, this \dshort{hv} bus has interleaved resistive elements to retard large voltage swings of the cathode modules in case of a discharge to reduce the risk of damages to the cathode mounted PD modules.
\begin{dunefigure}
[Cathode \dshort{hv} bus]
{fig:hv-bus}
{A schematic diagram illustrating the interconnects between \dshort{fc} profiles (blue), HV bus resistors, and the resistive (black) and conductive (light gray) mesh panels on the cathode modules in a corner of the cathode plane. The two rows of \dshort{fc} profiles at the cathode level serve as conductive segments of the HV bus, interconnecting the outer edges of the cathode modules. Custom resistor modules are mounted between two adjacent profile segments to form a complete loop around the cathode. }
\includegraphics[width=0.9\textwidth]{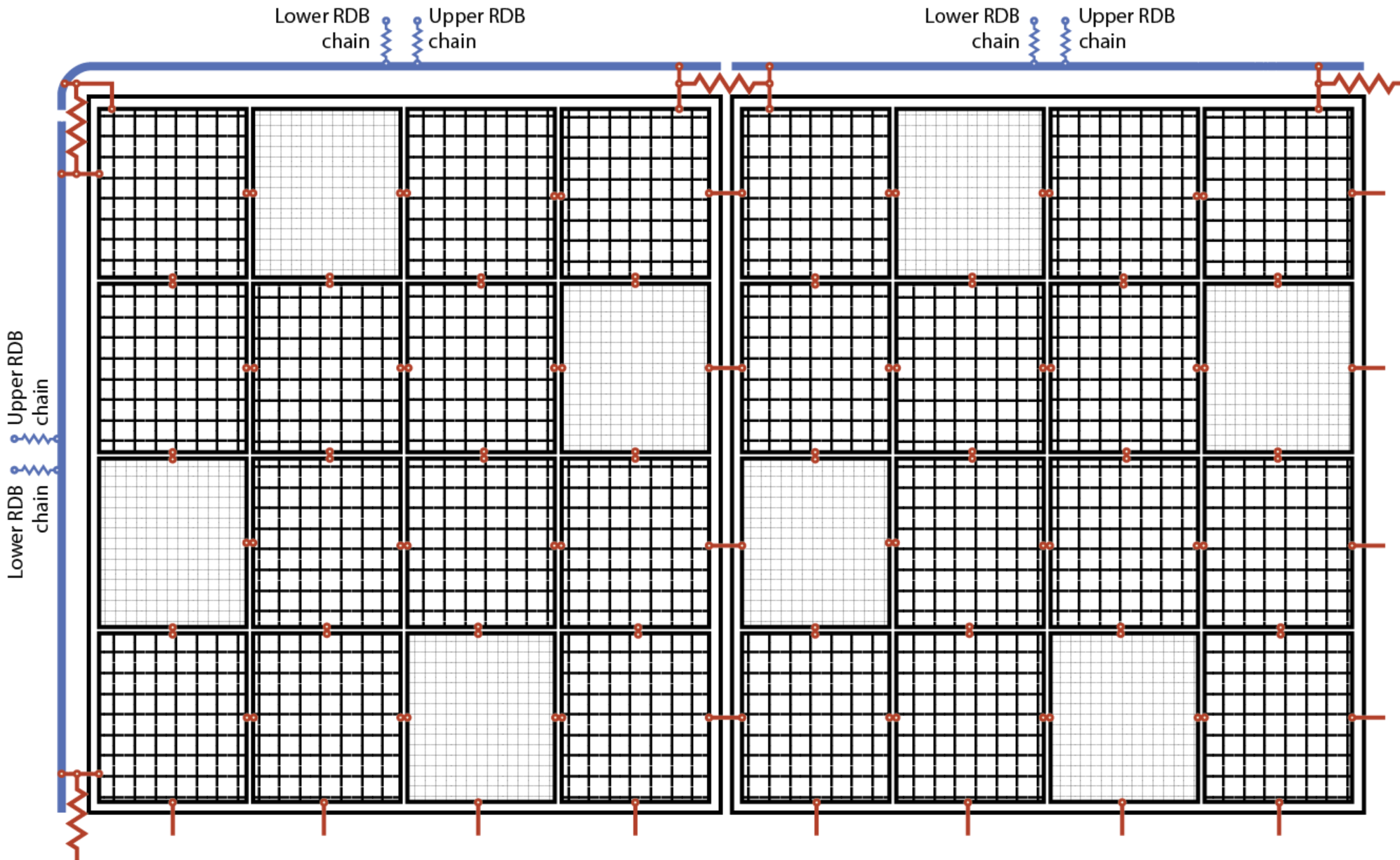}
\end{dunefigure}

In the \dword{pds} baseline design, a large fraction of the \dshort{pd}s are mounted on the cryostat wall to collect photons transmitted through the \dshort{fc}. A highly transparent \dshort{fc} is therefore desired to allow as much light as possible to reach these \dshort{pd}s. This could be accomplished either using narrower \dshort{fc} profiles or a larger profile spacing. Both of these options will increase the surface \efield on the profiles toward the grounded membrane wall, which increases the risk of \dshort{hv} instabilities on the \dshort{fc}. Additionally, increasing the profile spacing would require a deeper buffer zone for the \efield to reach good uniformity. Therefore, using narrow profiles at the same spacing has been selected as the design choice.

The profiles have elliptical surfaces facing the outside of the \dshort{tpc}, yielding a relatively low \efield on the surface. 
The narrow profile of 15\,mm width at 6\,cm spacing provides a maximum 70\% optical transparency, taking into account all the other \dshort{fc} structures.  The cross section of this profile is shown in the top left of Figure~\ref{fig:xparent-narrow-profile}.  A comparison of this and the conventional profile (labeled ``FD1 profile'') is shown on the right. 

The lower left of Figure~\ref{fig:xparent-narrow-profile} shows the \efield distribution surrounding the narrow profile when it is positioned as the 
profile immediately next to the cathode plane. The peak \efield at this position is 13\,kV/cm, well below the 30\,kV/cm limit imposed on all electrodes.  
Nevertheless, it is about 50\% higher than that of the conventional 46\,mm profiles at the same position.  To mitigate this potential risk of higher \efield on the \dshort{fc} surface, the narrow profiles will only be used to cover \dshort{fc} surfaces further away from the cathode plane where the bias voltages of the profiles are much lower.  For this reason, only the profiles in the range $\pm$2.5\,m to $\pm$6.5\,m away from the cathode plane (taken as $y=0$), namely within the 4\,m of the top and bottom \dwords{anodepln}, have 15\,mm width, while the remaining profiles are the standard 46\,mm version, in the \dshort{fc} baseline design.  This configuration accommodates the \dshort{pds} baseline design, described in Chapter~\ref{chap:PDS}.

\begin{dunefigure}
[Profiles for 70\% transparent \dshort{fc}]
{fig:xparent-narrow-profile}
{Top left: a cross section of a narrow aluminum profile. Bottom left: local \efield of a narrow profile when positioned next to the cathode plane. Right: 
cross sections of the standard and narrow profiles, with dimensions.}
\includegraphics[width=0.8\textwidth]{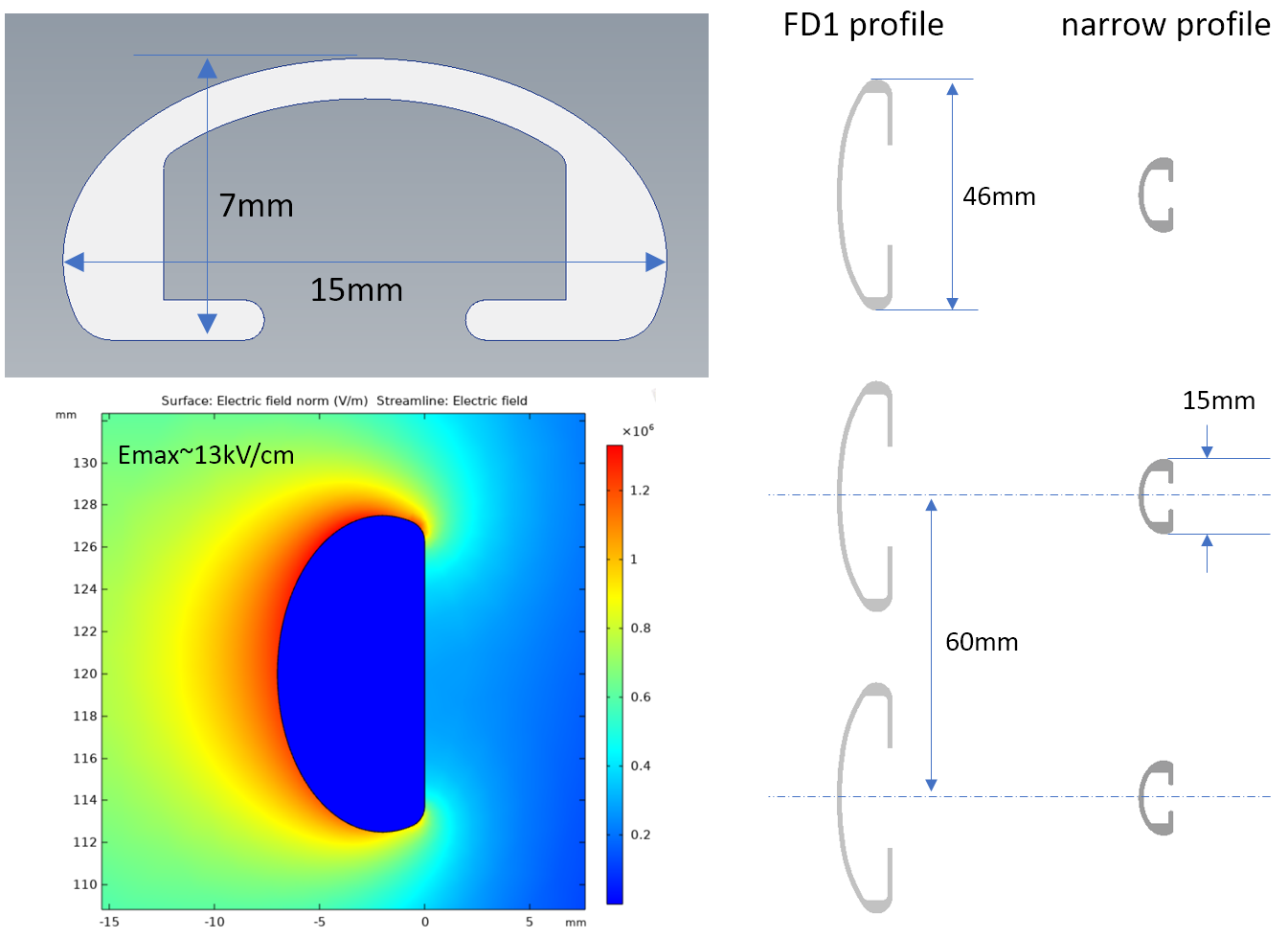}
\end{dunefigure}

At the $90^\degree$ bend at the corners of the \dword{fc}, the standard 46\,mm width profiles have a maximum local \efield of 17\,kV/cm, which is higher than in other areas on the \dshort{fc}, but much lower than the 29\,kV/cm value that would result from $90^\degree$-bent narrow profiles. 
A \threed \dword{fea} of the \efield at the corner section of the \dshort{fc} at and above the cathode plane is shown in Figure~\ref{fig:xparent-fc-overview-new}, right.

The mixing of the narrow and standard profiles introduces a discontinuity in the electrode coverage distribution, and results in an increase in \efield non-uniformity at the transition region. 
See Figure~\ref{fig:xparent-fc-efield} (left). % (right). 
Modifications to the standard \dword{hvdb} resistance are needed to largely eliminate this local degradation; %for this correction scheme whose 
the implementation details are described %in detail 
in Section~\ref{subsubsec:HVDB}. The \dword{fea} shows that the least drift line distortion is achieved when the voltage steps beyond the third narrow profiles are about 0.7\% higher than those of the standard profiles. Without this slight increase, there is a subtle divergence of \efield lines near the edges of the \dshort{fc} up to 2\,cm over the 6.5\,m full drift, although the \efield amplitude remains within the requirement. To implement this small correction, we need to use about 40\% of resistors with 0.7\% lower value on the divider boards for the wide profiles. Since the tolerance on these resistors is +/- 1\%,  we plan on sorting all resistors after their cold test at the QC stage and dividing the resistors into two groups with their mean values approximately 0.7\% apart.

\begin{dunefigure}
[\dshort{fea} of E field at the transition region of the 70\% transparent \dshort{fc}]
{fig:xparent-fc-efield}
{Comparison of the drift field uniformity near the transition region of the field cage profiles. Left: All \dshort{fc} profiles are biased with a linear voltage gradient of 3000\,V (for a 500\,V/cm drift field); Right: A simple correction scheme that eliminates most of the field non-uniformity near the transition region. \efield values are in units of V/m. The active volume starts at $x>$50\,mm.  A number of  \efield lines (vertical), and equipotential lines (horizontal) are also shown. The calculations were made with a drift field of 500\,V/cm, which can be scaled linearly down to the goal drift field value of 450\,V/cm.
The region colored in yellow (orange) corresponds to a deviation in the range $+1$ to $+2$\%  (beyond $+2$\%) in the \efield intensity.
The light blue (dark blue) regions correspond to deviation from the nominal \efield in the range $-1$ to $-2\%$ (beyond $-2$\%).
}  
\includegraphics[width=0.95\textwidth,height=0.4\textwidth]{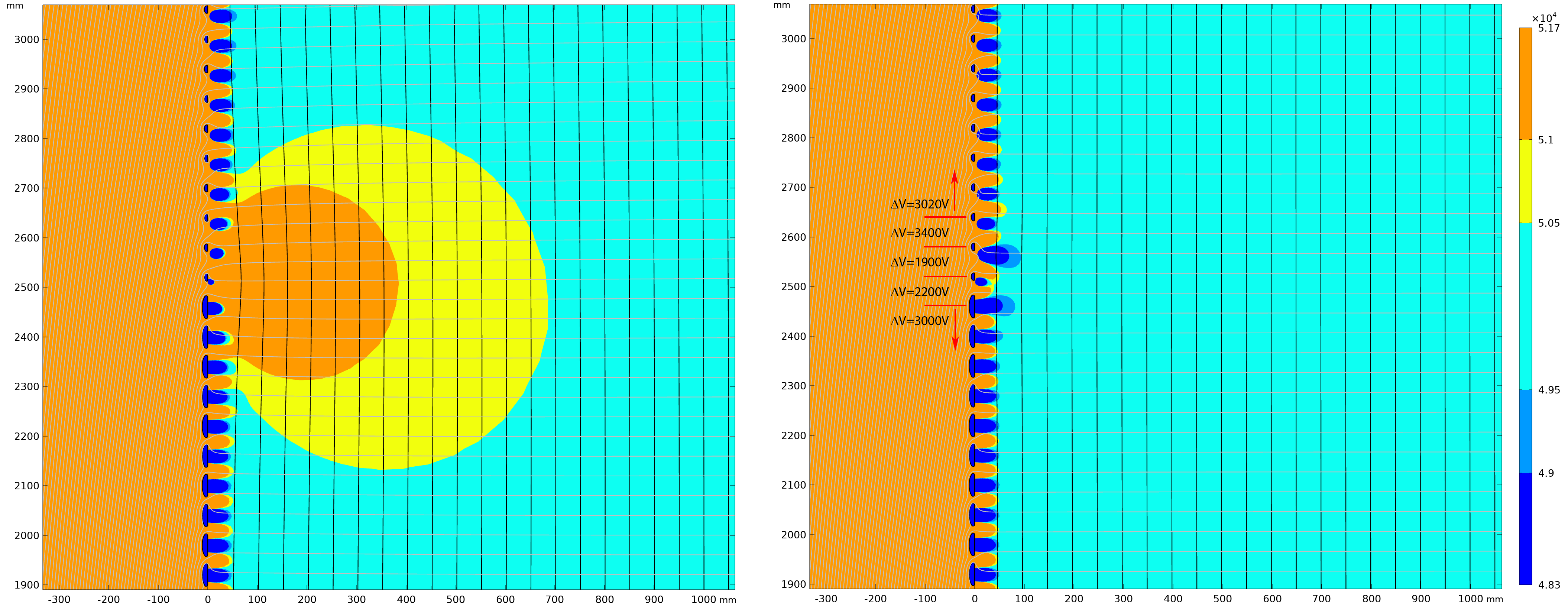}
\end{dunefigure}

The distance of the \dshort{fc} profiles from the surface of the flat part of the membrane is %estimated to be 
$\sim$700\,mm.  Once the height of the ``knuckles'' in the membrane (the intersections of the vertical and horizontal corrugations) and the depth of the cable trays for the bottom \dshort{crp}s are taken into account, this distance is reduced to $\sim$620\,mm.  Extrapolating from the \dshort{pdsp} \dshort{hv} experience, this is a safe separation for operation at $-$294\,kV on the cathode. The hydrostatic pressure from the $\sim$7\,m of \dshort{lar} also helps the stability of the field, suppressing the formation of gas bubbles at the cathode level. 

\subsection{High Voltage Divider Boards (\dshort{hvdb}) }
\label{subsubsec:HVDB}
The resistive chain for voltage division between the \dword{fc} aluminum profile electrodes provides the voltage gradient between the cathode and the top-most and bottom-most field-shaping rings. This chain is critical because it determines the uniformity of the \efield inside the active volume of the \dword{tpc}.
The chain is constructed with a column of high voltage divider boards (\dwords{hvdb}) to provide voltage divisions to each column of the \dshort{fc}, independently.
The \dshort{hvdb}s are \dwords{pcb} with nine stages, each of them consisting of  
two 5\,G$\ohm$ resistors in parallel, for intra-board redundancy and to keep the overall current of the system low, and three varistors of threshold voltage 1.7\,kV, 
which will act as a 5.1\,kV voltage clamp in case of a sudden discharge, and protect the resistors from failure due to excessive voltages across them~\cite{Asaadi_2014}.
Given a $\sim$2.7\,kV voltage differential between each stage, the total expected current is $\sim 1.08\,\mu$A along each \dshort{hvdb} chain on a 6.45\,m high \dshort{fc} column.
With a total of 48 columns surrounding the active volume, each with its own resistor chain, the total current per active volume is $\sim52.0\,\mu$A. 
The two drift volumes are in parallel, so the \dshort{fc} for the entire \dshort{spvd} will draw $\sim104.0\,\mu$A. 

Figure~\ref{fig:vd-rdb} shows a section of a \dshort{hvdb} board mounted on a section of the \dshort{fc} in the 70\% optical transparency region. 
The boards can be mounted either on the profiles with each of the electrical pads making  direct contact as in the \dshort{sphd} field cage, or through a set of right-angle metal brackets such that the boards are edge-on to the \dshort{fc} surface and the electrical pads making contact with the profiles indirectly through the bracket.

\begin{dunefigure}
[Resistor divider board]
{fig:vd-rdb}
{Illustration of the resistor divider board mounted edge-on to the \dshort{fc} to improve the light transmission to the wall-mounted \dshort{pd}s.}
\includegraphics[width=0.5\textwidth]{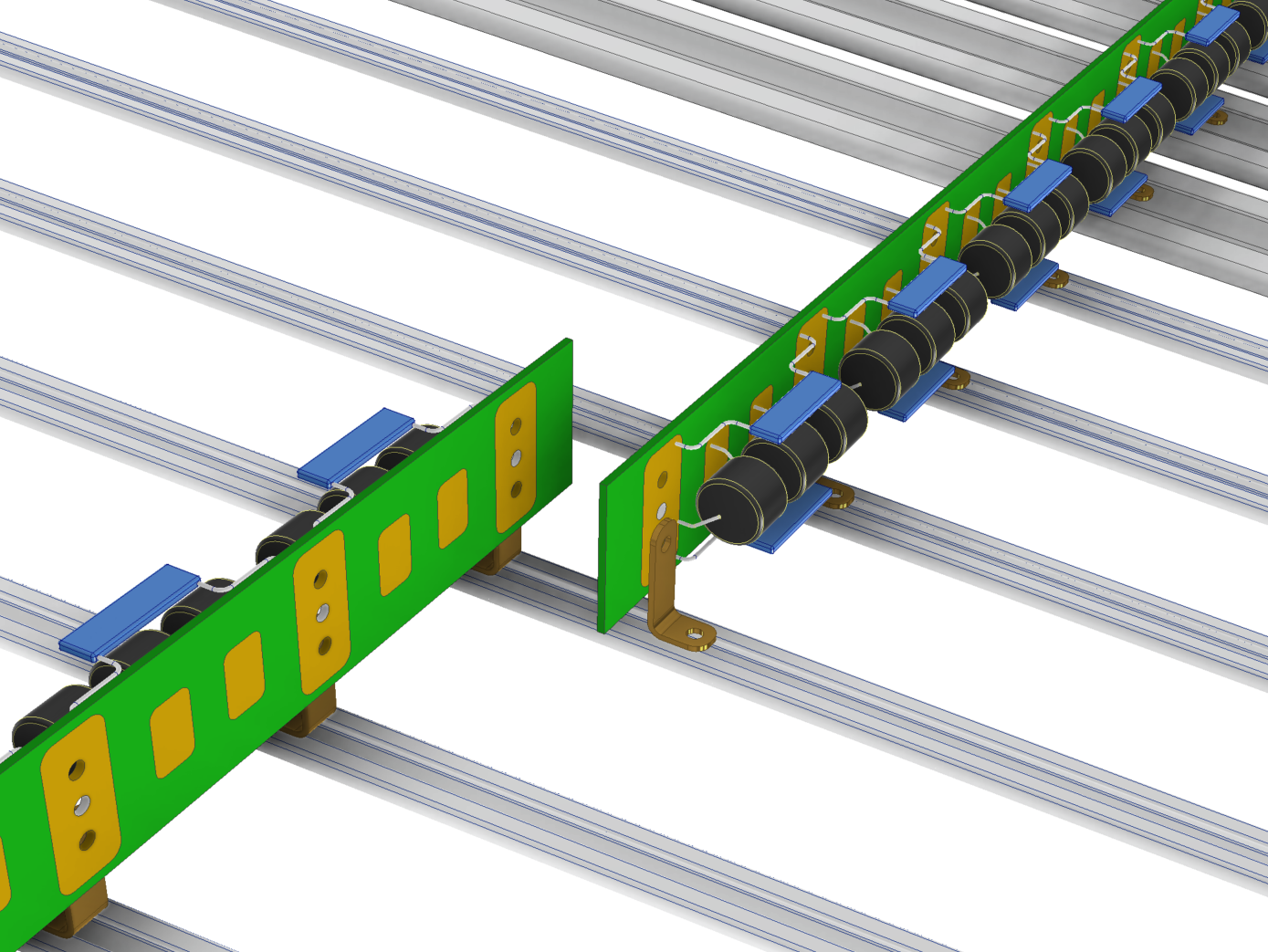}
\end{dunefigure}

To implement the local correct scheme over the wide-to-thin profile transition region, preliminary analysis has shown (see Figure~\ref{fig:xparent-fc-efield}) that the voltage drops in this region need to be:  $V_n=73\%\,V_s$, $V_{n+1}=63\%\,V_s$, and $V_{n+2}=113\%V_s$, where $n$ is the gap between the last wide and the first narrow profiles, and $V_s$ is the nominal voltage between adjacent profiles. The resistor values across these gaps must be adjusted accordingly by adding a special, short correcting \dword{pcb} over this transition region. 
This strategy keeps all \dwords{hvdb} uniform but enables finer adjustments. % 

The \dshort{fc} voltage divider chains for both the top and bottom active volumes are terminated through wired connections outside of the cryostat to a set of power supplies that can control the termination bias voltages and can measure and record the current flowing through each \dshort{hvdb} chain.  The voltage adjustment provides the capability for fine-tuning the drift field near the edges of the \dwords{crp}, and recording the current has proven to be a valuable diagnostic tool. 
Provision for transmitting these signals has been made on the \fdth for the bottom anode plane readout, as described in Section~\ref{subsubsec:FCFT}. %the next section.

\subsection{Field Cage Support Feedthroughs}
\label{subsubsec:FCFT}

The \dword{fc} is suspended from the roof of the cryostat independently of the top anode and the cathode structures.  A total of 48 dedicated penetrations are available on the roof of the cryostat: 20 along each long wall, and four along each end wall to support the 24 
\dshort{fc} supermodules.

As shown in Figure~\ref{fig:hvs-fcft}A, the vertical crossing tube has a 150\,mm \dword{od} and 10\,mm wall thickness to support half the weight of a 
supermodule ($\sim$600\,kg). Directly on top of the crossing tube is an oversized base flange (300\,mm diameter) to serve as a platform for the lateral adjustment of the \dshort{fc} anchor point. A slider plate with an O-ring groove underneath it is placed on top of the base flange.  This slider plate has a keyhole-shaped opening and a hemispherical cradle to allow the ball end of a \dshort{fc} support lift rod to pass through from below and sit on top.  This slider plate can be pushed around by the four jacking screws on the base plate while maintaining a tight seal by the bottom O-ring. A very slim lifting adapter is attached to the threaded portion at the top of the \dshort{fc} lifting bar.  
\begin{dunefigure}
[Field cage support feedthrough]
{fig:hvs-fcft}
{Conceptual design of a \dshort{fc} support penetration and the associated feedthrough flanges for both mechanical support and electrical connections. Left (A): exploded view of the components of the \dshort{fc} support penetration. Right (B -- E) views of the top feedthrough flange at different stages of the installation. }
\includegraphics[width=0.75\textwidth]{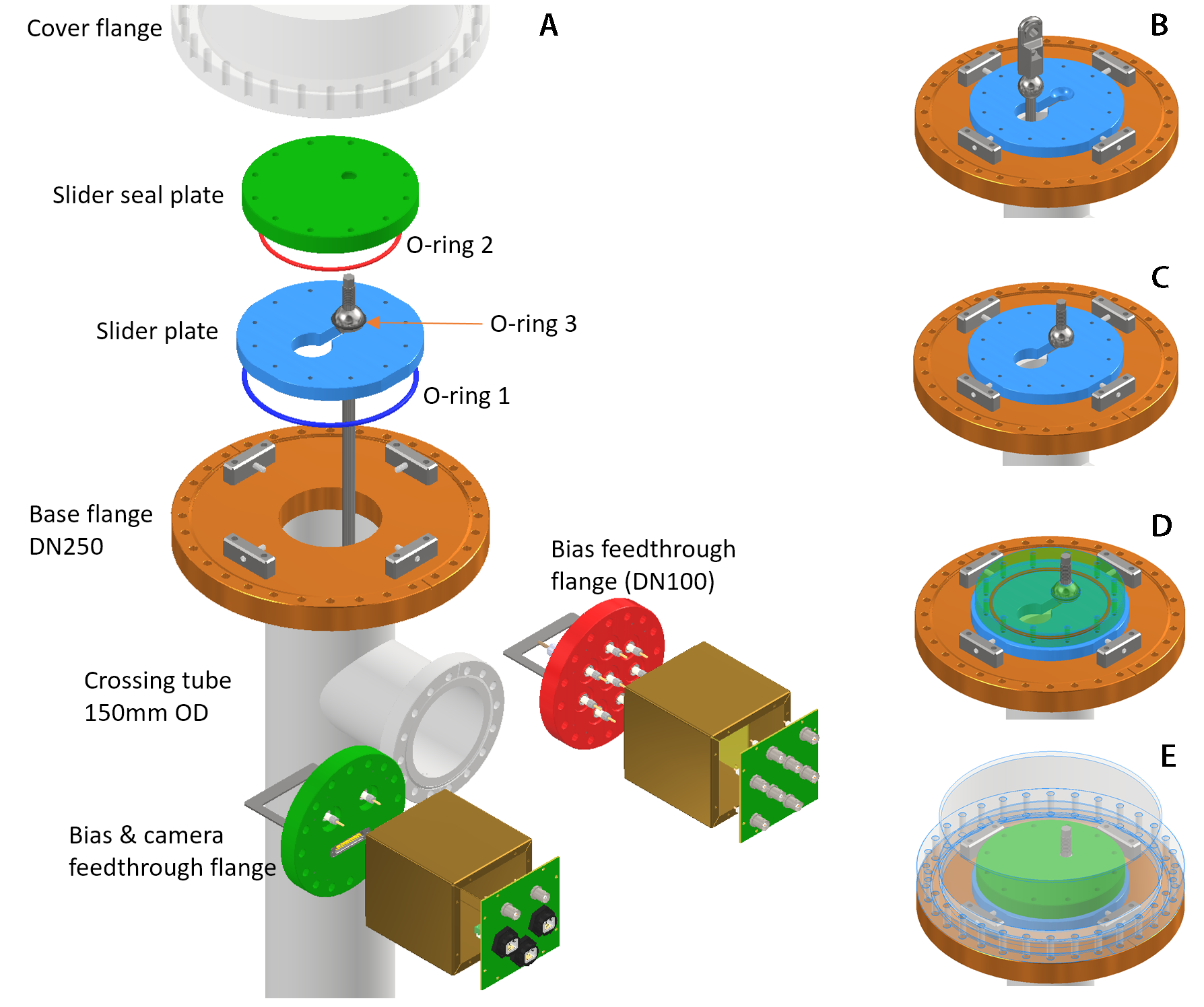}
\end{dunefigure}

A winch mounted on a tripod above the penetration is used to pull the \dshort{fc} lift rod through the larger opening of the keyhole on the slider plate. 
This large opening is oriented away from the interior of the \dword{tpc} to provide an offset of about 5\,cm during the lifting of the field cage supermodule to provided added clearance between the field cage and the cathode or top CRPs.  Once the ball end of the \dshort{fc} lift rod emerges through the keyhole (Figure~\ref{fig:hvs-fcft}B), it is settled into the cradle, and the weight of the \dshort{fc} is transferred to the base flange (Figure~\ref{fig:hvs-fcft}C).  Before the final adjustment of all \dshort{tpc} component positions, the \dshort{fc} support feedthroughs can be left in this state (with the cover flange loosely placed for protection).

To provide a leak-tight seal of the opening in the slider plate, a slider sealing plate is added on top.  This plate has an O-ring groove at the bottom face (for O-ring 2 in Figure~\ref{fig:hvs-fcft}A), and another inside a hemispherical cavity (for O-ring 3).  Combined with O-ring 1, these three O-rings provide a hermetic seal at the top of the penetration (Figure~\ref{fig:hvs-fcft}D).  In this state, the \dshort{fc} support rod can still be adjusted in all directions. The vertical adjustment is done by rotating the lift rod against a threaded connection at its bottom end. 
During normal operation, the cover flange is sealed against the base plate. The volume inside the cover flange can either be evacuated or purged together with the crossing tube below.

As shown in Figure~\ref{fig:hvs-fcft}, the crossing tubes of \dshort{fc} support penetrations have electrical flanges on their side, designed to bring out the \dshort{fc} termination bias voltage cables, the cold camera data cables and the top anode bias voltage cables.

Two types of signal feedthrough flanges are designed for these ports.  Along the long walls of the cryostat, two \dshort{fc} termination lines and six top anode bias lines are needed.  A DN100 Conflat flange with eight SHV feedthroughs are used here (see the red flange in Figure~\ref{fig:hvs-fcft}A).  Along the end walls, there are two \dshort{fc} termination lines per penetration, but no anode bias lines. Cameras will be placed inside the cryostat for monitoring of the \dshort{hv} components. A different flange with two SHV and one DB25 connector is used for the camera signal (see the green flange at the bottom of Figure~\ref{fig:hvs-fcft}A). The filter boards inside a shielded enclosure are mounted on the signal feedthrough flange to remove noise pickup inside the cryostat. 

The top anode bias voltage lines are routed through the \dshort{fc} support penetrations and secured to the side ports before the \dshort{fc} installation.  The bottom \dshort{fc} termination cables are embedded inside the cable trays mounted on the cryostat wall, running from floor to  ceiling.  The upper ends of the cables are pulled through the \dshort{fc} support penetrations while the rest of the cable bundles are installed into the cable trays.  The top \dshort{fc} termination cables are tied to the \dshort{fc} lift rods at  installation, and carefully detached as the lift rods pass through the top flange.  Continuity checks are made at every step of the cable connections.

\subsection{Field Cage Assembly and Installation Procedure and Tools}
\label{subsubsec:fc-tools}

Detailed \dword{fc} unit module assembly and \dshort{fc} super module installation procedures have been developed and were presented at the 
\dword{pdr} in May 2022. The assembly procedure  includes mounting of the \dword{hvdb}. This section briefly describes these procedures and the tools required for them.

\begin{dunefigure}
[Field cage assembly and installation tools]
{fig:hvs-fc-tools}
{Tools for the \dshort{fc} (FC) assembly and installation process - (A) FC module assembly station;  (B) FC module installation cart; (C) FC module storage cart; (D) tripod for raising \dshort{fc} supermodule assembly} 
\includegraphics[width=0.95\textwidth]{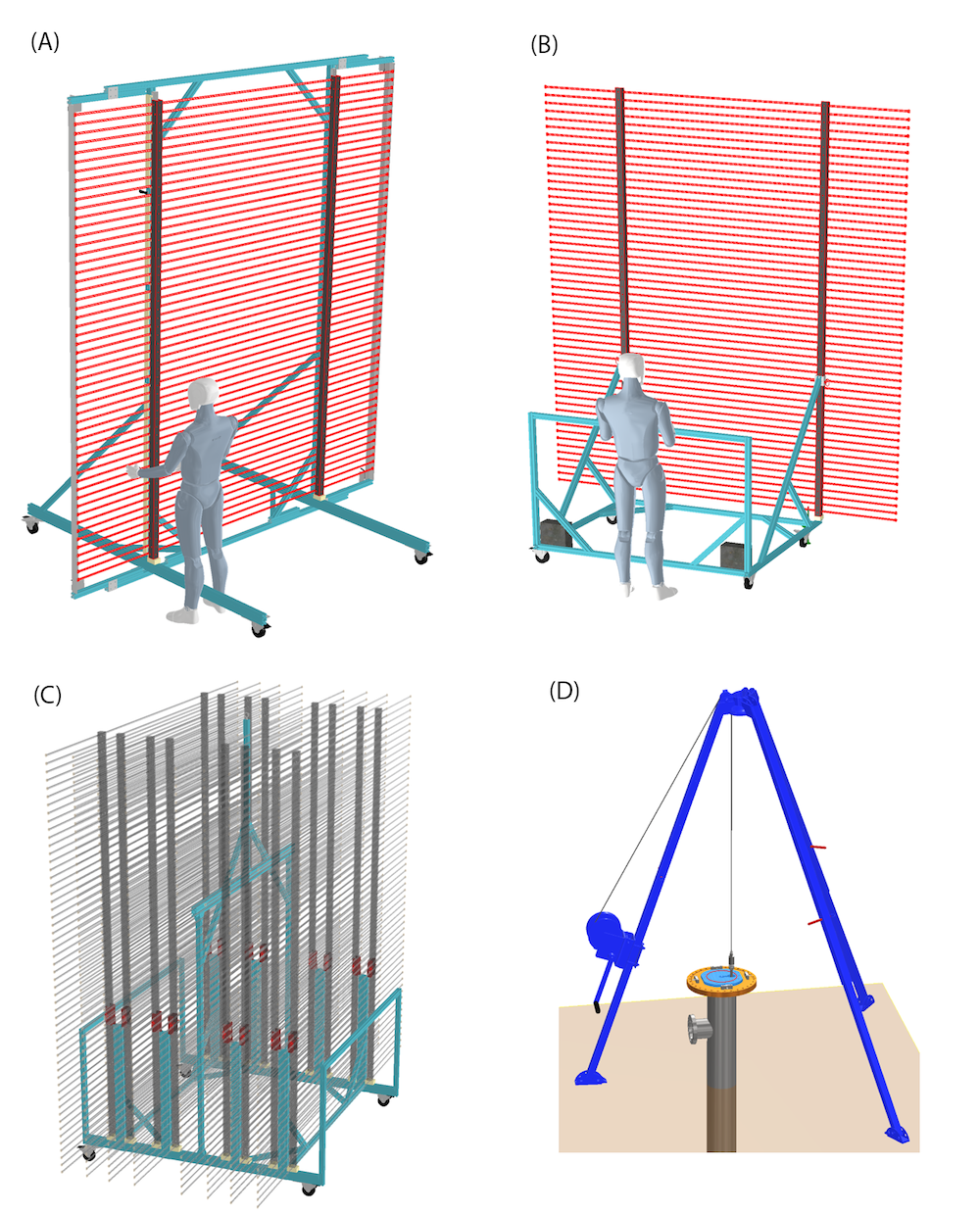}
\end{dunefigure}

Figure~\ref{fig:hvs-fc-tools} shows %various 
several tools designed for the \dshort{fc} installation process.  Each of the 3.24\,m tall \dshort{fc} %unit 
modules will be assembled vertically on the assembly station (A) %depicted in Figure~\ref{fig:hvs-fc-tools}.A  
to minimize stress and deformation of the profiles %at the time of pick up 
as the module is lifted after %the completion of the 
assembly. 
The storage cart (C) accommodates eight completed \dshort{fc} modules, i.e., enough for one supermodule. Once it is full, 
the storage cart will be transported into the cryostat for installation of the supermodule.
Each module is moved to % picked up out of the storage cart and put on 
the installation cart (B) %shown in Figure~\ref{fig:hvs-fc-tools}.B and be 
and positioned under the stainless steel I-beam hanging on the cables attached to the 
installation tripod (D) %shown in Figure~\ref{fig:hvs-fc-tools}.D 
on the cryostat roof.
The two modules in the same row of a 
supermodule are installed, then %at the given height.
the completed row is raised by about 3.5\,m for the installation of the next row.
This process is repeated until all four rows are installed.

%

%%%%%%%%%%%%%%%
\section{Design Validation}
\label{sec:fdsp-hv-protodune-lessons-design}

After the \dword{pddp} run, the most critical item of the vertical drift design was determined to be the \dshort{hv} delivery system, in particular the \dshort{hv} extender and its coupling to the \dword{hvft}. 

%\textbf{Extender} 
\textbf{Delivery System}. The overall design of the extender was highly simplified with respect to the \dshort{pddp} version, and it was the subject of the dedicated R\&D campaign in spring/summer 2021 in small scale cryostats (\dword{iceberg} at \dword{fnal} and the two-ton vessel at \dword{cern}). The main outcome of the R\&D was a further optimization of the extender-to-\dshort{hvft} coupling to minimize the \efield at the extender surface and the charging up on insulating elements such as the exposed \dshort{hvft} \dword{uhmwpe} and the extender support disk. The material of the extender support disk was also selected to be \dshort{uhmwpe} against the \frfour because of its homogeneity and  better dielectric rigidity.  Operation at \SI{-300}{kV} in pure argon was  achieved and maintained for the full test duration (several days) in these R\&D campaigns.

The validation of the \dshort{hv} delivery system was performed during the \dshort{np02} \dshort{hv} long-term stability run in 2021-2022 in which operation at \SI{-300}{kV} in ultra pure \dword{lar} was demonstrated (up-time > 99.9\% over two periods of about two months each of continuous operation). At the end of this %the \dword{np02} long term stability 
run, cosmic rays tracks were also recorded with the \dword{dp} \dword{crp} readout system, demonstrating that \SI{6}{m} drift distances are feasible and that the newly designed \dshort{hv} delivery system  introduces no detectable noise from residual ripple or leakage currents.

The down-time intervals of the \dword{hvs} in this %NP02 long term stability 
run occurred at a rate of about one every two hours for few seconds each (< 0.1\% of the time on average); they have been identified as %being 
\dshort{hv} current glitches located at the junction between the extender support \dshort{uhmwpe} disk and the metallic head of the extender-to-\dshort{hvft} coupler.  These glitches are most likely due to instability of the charged-up layer on the insulating surface of the support disk. While they were of limited duration with no visible degradation over the %NP02 long term stability 
test run, a design optimization to further reduce the down-time was introduced and discussed in Section~\ref{subsec:hv-extender-coupling}.  Tests in the two-ton cryostat for this new design are in progress. 

\textbf{HV \fdth and HV cable}. The \dshort{hv} cable and the \dshort{hvft} experienced a failure during the \dshort{np02} long-term \dshort{hv} stability run. 
In-depth investigations have been carried out in collaboration with the \dshort{cern} experts of the ``Materials, Metrology and Non-Destructive Testing'' group with tools such as the electron microscope, computerized tomography, and X-rays. 
A crack in the cable insulation was most likely caused by pre-existing mechanical stress due to handling issues (e.g., bending the cable more acutely than its  prescribed limit or making a small cut during the preparation of the cable termination). This incident is feeding into udpates to the cable installation procedure.

Regarding the \dshort{hvft}, the investigation points to pollution of the \dshort{uhmwpe} insulating cylinder by extraneous filaments during extrusion as the most probable explanation. Cracks progressively developed along these filaments and eventually opened a path for current flow.  (Note that this problem occurred only in one \SI{300}{kV} \dshort{hvft} out of four available.) The mitigation consists of selecting very pure extruded \dshort{uhmwpe}. A certificate will be required; CT examination of a sample of the \dshort{uhmwpe} cylinder will also be an option. Since X-rays can be used to detect the filaments in \dshort{uhmwpe}, the full  cylinder can be examined this way.

\textbf{\dword{hvft} length}. During the long-term runs of the original 300\,kV \dshort{hvft}s built for \dword{np02} and \dword{np04}, %were affected by 
ice formed at the cable receptacle, which reached the depth of 
the cryostat roof inner membrane and was subject to a quite low temperature.
Ice formed as a consequence, 
despite continuous flushing of the cable receptacle with dry nitrogen. 
This was confirmed in the \dshort{np02}
stability run when the failed \dshort{hvft} was replaced by the one previously used on the \dword{pddp} run. Ice formation does not affect the \dword{hvs} performance, given its good insulating properties, however it prevents extraction of the cable, which might be required in \dshort{spvd} over the decades-long operation.

The new \dshort{hvft} initially installed in \dshort{np02} for the long-term \dshort{hv} stability run was about 80\,cm longer on the warm side %such that 
to keep the cable receptacle temperature %was always 
above 0 \degree{}C, avoiding freezing and frost buildup, which caused instability during \dshort{np04} operation. %For this reason, 
This allowed the cable to be easily extracted when the failure occurred after about two months of operation. The design of the \dshort{hvft}s %newly %being built 
under construction maintains the %feature of having a 
``warm'' cable receptacle, keeping it fully above the cryostat insulation skin. This results in a \dshort{hvft} length longer than 3\,m; at present the design is 4\,m, which % as it 
is %to be 
feasible with the extruded UHMW-PE cylinders commercially avaialable. 

\textbf{Cathode}. The challenging features of the \dshort{spvd} cathode design that require dedicated testing are the \dword{frp} frame and the %hanging 
suspension system. A prototype of the cathode  module in 1:1 scale, has been built and installed in the \dshort{np02} \coldbox for the \dword{crp} test program, with integrated  \dword{xarapu} cells and electronics. The cathode module was %sitting 
placed on feet resting on the \coldbox floor. % of the %NP02 
Although some details of the design were not final, % in some details 
(the \dword{frp} frame was made from C-shaped beams glued together instead of H-shaped beams, %the thickness of 
that were \SI{5}{cm} thick instead of \SI{6}{cm}, and a metallic %mesh 
rather than resistive perforated panel for the first semester of tests; resistive meshes are currently in operation) %instead of a resistive one), 
the cathode operated properly and stably up to \SI{15}{kV} as planned in all the % many CRP's 
test runs (from November 2021 %up to now, 
through November 2022). % with no issues at all and very stably. 
No impact %of the possible sag 
on the field uniformity from any possible sag has been measured. 

Based on these first results, the final design of the cathode is proceeding as defined, with the first final prototypes to be built for %NP02 Module-0. 
\dword{vdmod0}. 

Investigations are ongoing at %Physics Laboratory Irène Joliot-Curie, Paris (
IJClab concerning the behavior of the Dyneema rope used in the suspension system. 
The preliminary creep data %are being 
acquired on several ropes % several have been tested, 
show a reproducible and predictable elongation within %some 
a few cm precision, limited by the setup. A cryogenic and more precise (mm precision) setup at IJClab is undergoing validation and will be used for %Module-0.
\dshort{vdmod0}.

\textbf{Field Cage} The design of the \dshort{fc} based on C-shaped aluminum profiles and supported by \dshort{frp} beams is well established, and its validity was confirmed by the operation of \dword{pdsp} and more recently in the \dshort{np02} long-term \dshort{hv} stability run. 
The configuration in which narrower cross-section %of the 
aluminum profiles are used along portions of its height (closer to the anode planes) has %been 
undergone preliminary tests in \dshort{np02}. It has a $\SI{2}{m} \times \SI{2}{m}$ window integrated into the existing \dword{dp} \dshort{fc}. No issues were recorded, but due to the \SI{1}{m} distance of the \dshort{fc} to the membrane (instead of 70\,cm) and the location of the window close to the anodes, the nominal \efield expected around the profiles close to the cathode was not reached. 

A dedicated test is planned with a mini-\dshort{fc} in the 
50 liter cryostat at \dshort{cern}. %It is worth noting, however, that t
Its field cage %of the 50 liter TPC 
is made of aluminum profiles %with shape 
similar to %the ones 
those of the 70\% transparent design. With a distance from the cryostat vessel as close as \SI{3.5}{cm} and a \dshort{hv} as high as \SI{25}{kV} on the cathode, the \efield at the surface of the aluminum profiles can be as high as in the %far detector 
\dshort{spvd} design close to the cathode.

%%%%%%%%%%%%%%%
\section{Production, Handling and Quality Control}
\label{sec:fdsp-hv-protodune-lessons-prod}

The production and handling of \dword{hv} components must be approached with 
great care to avoid scratching and potentially compromising the electrical components. 
Part production %should 
is to be carried out so as to avoid introducing sharp edges wherever possible.
The aluminum field-shaping profiles are particularly prone to scratches and must be packaged and handled so as to avoid direct contact with other profiles and materials.
Kapton strips are used to separate the profiles from the \dword{frp} of the \dword{fc} frames as they are inserted in order to protect against scratching or removal of the profile coating.
Any scratches found in the \dshort{frp} beams are covered with epoxy to prevent fibers from escaping into the \dword{lar}.

\Dword{qc} tests have been conducted on \dshort{hv} modules and individual components at every step:  
parts procurement, production, integration, and installation, and will continue to be conducted as production for \dshort{spvd} proceeds. 
Documented procedures, including \dword{qc} procedures, come with checklists that must be completed for component parts at each step. %  as production proceeded
%forms.  
Printed copies of the checklists completed in the procurement and production stages were and will continue to be included as travelers in shipping crates.  
To ensure that nothing %was 
is compromised during transport, \dshort{qc} tests %were 
are repeated on individual components and assembled pieces after shipping. 

Resistance between steps on the resistor divider boards %were 
is measured and verified to be within specification both after %their 
production and after %they were shipped 
delivery to \dshort{cern}.
Once the resistor divider boards %were 
are mounted onto an assembled \dshort{fc} module, the resistance between adjacent profiles %was 
is measured to verify sound electrical connection.
%In a similar way, 
These \dshort{qc} checks of connections between cathode modules and between the cathode and \dshort{fc} modules, that have been completed for the received parts, will also % need to 
be performed after installation. %\fixme{in both \dshort{vdmod0} and \dshort{spvd}?}

\dshort{qc} tests on the \dshort{hv} components of \dshort{protodune} required many measurements %to be made 
with several different testing devices.  Extrapolating these measurements to the scale of %DUNE 
\dshort{spvd} will require development of dedicated tools %so that 
to optimize the \dshort{qc} process %can be made more efficient and optimal 
at each step.
%

%%%%%%%%%%%%%%%%%%%%%%%%%%%%
\section{Interfaces} 
The major interfaces of the \dword{hv} system are shown in Table~\ref{tbl:hv-interfaces}.  They are all %in a 
well advanced, % stage, 
although not fully finalized. The interface with the \dword{pds} is most critical since %in the part this interface 
it concerns the routing of the fibers along the \dshort{fc} modules and within the cathode modules, which may require some design changes to accommodate it.

\begin{dunetable}
[\dshort{hv} interface links]
{p{0.1\textwidth}p{0.15\textwidth}p{0.7\textwidth}}
{tbl:hv-interfaces}
{\dshort{hv} interface descriptions and links to full interface documents.}
System & EDMS \# & Description  \\ \toprowrule

CRP & \href{https://edms.cern.ch/document/2619003/1}{2619003}  & Field cage to CRP clearance, CRP bias power supplies \& cables \\ \colhline

BDE & \href{https://edms.cern.ch/document/2726647/1}{2726647}  & Bottom CRP bias voltage feedthroughs\\ \colhline

PDS & \href{https://edms.cern.ch/document/2619007/1}{2619007} & Mounting of the PD modules on the cathode, routing of the PD fibers off the cathode, providing transparency in the field cage for wall mounted PDs. \\  \colhline

I\&I & \href{https://edms.cern.ch/document/2648558/3}{2648558} & Installation requirements and procedures. \\

\end{dunetable}
%%%%%%%%%%%%%%%%%%%%%%%%%%%%%
\section{Production Timetable}
\label{sec:hv:prod}

A timetable for \dword{hvs} component procurement and assembly has been developed, as shown in Table~\ref{tab:hvs-prod} and feeds into the schedule outlined in Chapter~\ref{ch:project}.

\begin{dunetable}
[HV system production timetable]
{p{0.3\textwidth}|p{0.35\textwidth}|p{0.2\textwidth}}
{tab:hvs-prod}
{HV system production timetable}
Deliverable  & Production type & Dates \\
\toprowrule

\dshort{fc} profiles &
commercial &  Q2 2024 -- Q3 2025 \\ 
\colhline

\dshort{fc} divider boards  &
\textbullet component selection: in-house
 &  Q1 2024 -- Q4 2025 \\ 
production &   \textbullet board production: commercial &   \\ 
\colhline

\dshort{fc} box beams and  &
\textbullet fabrication: commercial
 &  Q2 2024 -- Q1 2025 \\ 
hardware & \textbullet \dshort{qc}: in-house & \\
\colhline

cathode frame fabrication &
commercial
 &  Q2 2024 -- Q3 2026 \\ 
\colhline

cathode frame integration and \dshort{qc} &
in-house
 &  Q1 2025 -- Q3 2026 \\ 
\colhline

\dshort{hvps} and \dshort{hvft} &
commercial
 &  Q2 2024 -- Q2 2025 \\ 
\colhline

remainder of \dshort{hv} delivery system &
commercial and in-house
 &  Q2 2024 -- Q4 2024 \\ 
\end{dunetable}

%%%%%%%%%%%%%%%%%%%%%%%%%%%%%
\section{Organization and Management}
\label{sec:hv:orgs}

\subsection{Institutional Responsibilities}
\label{sec:insts}

The \dword{hvs} joint consortium %aims at 
is responsible for the design, construction and assembly of the \dshort{hv} systems for both %FD1-HD and FD2-VD. 
\dword{sphd} and \dword{spvd}. It currently comprises several US institutions and \dword{cern}. One French institution is participating %but limited to 
in \dword{spvd} only. All the participating institutions are involved in construction and operation of %ProtoDUNE-HD and ProtoDUNE-VD.
\dword{hdmod0} and \dword{vdmod0}.

%As for ProtoDUNE, 
\dshort{cern} is %heavily 
committed to a significant role in terms of funding, personnel, and the provision of infrastructure for R\&D and detector optimization for the \dwords{mod0}. %the \dwords{protodune}. 
Moreover, \dshort{cern} will be responsible for a significant fraction of %\dword{sphd}
\dshort{spvd} subsystem deliverables.

%At present, i
In the current \dshort{hvs} consortium organization as listed in Table~\ref{tbl:hv-institutes}, each institution is naturally assuming the same responsibilities that it assumed for %ProtoDUNE-HD and ProtoDUNE-VD.
\dword{pdsp} and \dword{pddp}. The consortium organizational structure includes a scientific lead (from \dshort{cern}), a technical lead (from \dword{bnl}), technical design report (TDR) editor (from UTA), and a \dshort{hvs} design and integration lead (from \dword{anl}).

The successful experience gained with %the ProtoDUNE-HD detector 
\dshort{pdsp}, \dshort{pddp} and the recent R\&D with the NP02 Coldbox has demonstrated that the present \dshort{hvs} consortium organization and the number of institutions are appropriate for the construction of the \dshort{hv} system 
for both the \dshort{sphd} and the \dshort{spvd}.

The consortium is organized into working groups (WG) that address the design and R\&D phases of development, and the hardware production and installation.

\begin{itemize}
\item  WG1. Design optimization for \dshort{sphd} and \dshort{spvd} modules: assembly, system integration, detector simulation, physics requirements for monitoring and calibrations.
\item  WG2. R\&D activities, R\&D facilities.
\item  WG3. \dshort{sphd} \dword{cpa}: Procurement of resistive panels, frame strips, electrical connections of planes; assembly, \dword{qc} at all stages, and shipment of these parts.
\item  WG4. \dshort{spvd} cathode and suspension system: material procurement; construction, assembly, shipment to %ITF,
\dword{sdwf},
%quality assurance (QA), QC.
\dword{qa} and \dshort{qc}.
\item  WG5. \dshort{sphd} top/bottom and endwall \dword{fc} modules, %HD-EndWall modules , VD-FC 
\dshort{spvd} \dshort{fc} modules: procurement of mechanical and electrical components, assembly and shipping to %ITF.
\dword{sdwf}.
\item  WG6. \dword{hv} supply and filtering, %HV 
power supply, and cable procurement, R\&D tests, filtering and receptacle design and tests.
\end{itemize}

Taking advantage of identified synergies, some activities of the \dshort{sphd} and \dshort{spvd} working groups are merged: \dshort{hv} feedthroughs, voltage dividers, aluminum profiles, \dword{frp} beams, and assembly infrastructure.

\begin{dunetable}
[\dshort{hv} institutions]
{p{0.28\textwidth}p{0.12\textwidth}p{0.60\textwidth}}
{tbl:hv-institutes}
{Institutions participating in the HVS consortium}

Institution & Country & Detector \& Deliverables \\ \toprowrule

European Organization for Nuclear Research (CERN) & Switzerland & 
FD1 \& FD2: System design, HV R\&D, HV distribution system \& monitoring, components procurements (FC, CPA, HV), ProtoDUNE installation \& operation\\ \colhline
Argonne National Lab & USA  & FD1: System design \& analysis, CPA production and installation;  FD1 \& FD2: QA/QC\\ \colhline
Brookhaven National Laboratory & USA & FD1 \& FD2: System design \& analysis, project management, interfaces, cold cameras\\ \colhline
Kansas State University &  USA & FD1: HV bus \& interconnects, GP monitoring system\\ \colhline
Louisiana State University &  USA & FD1: Resistive divider boards, FC termination boards, end-wall FC production \& installation\\ \colhline
SUNY Stony Brook University &  USA & FD1: Top field cage production \& installation\\ \colhline
University of Texas Arlington &  USA & FD1: Bottom field cage production \& installation; FD2: field cage production \& installation   \\ \colhline
College of William and Mary - Virginia &  USA & FD1: CPA production \& installation; FD2: Resistive divider boards, FC assembly \& installation.
\\ \colhline
Laboratoire de Physique Irène Joliot-Curie &  France & FD2: Cathode design,  construction, assembly \& tests  \\ 
\end{dunetable}

%%%%%%%%%%%%%%%%%%%%%%%%%%%%
\subsection{High-level Schedule}
\label{sec:fdsp-hv-org-cs}

Table~\ref{tab:HVsched} lists the most high-level milestones for the design, testing, production, and installation of the \dword{spvd} \dword{hvs}. Dates in this tentative schedule are based on the assumed start of installation of the %VD \dword{spmod}
\dshort{spvd} at \dword{surf}. The dates for the \dshort{hvs} production of a 
\dshort{spvd} are %also 
included as a reference.

 The production scenario 
 for the schedule presented in Table~\ref{tab:HVsched} assumes 
 one factory site for the cathode construction and two for the \dword{fc} components, namely UTA and W\&M with the field cage profiles supplied by CERN. Given the present starting date % for installation underground of
 for \dword{spvd} installation, this assumption is fully compatible with the time available after the operation of %the Module -- 0 prototype.
 \dword{vdmod0} and with a reasonably large float in the production schedule.
 \begin{dunetable}
[HVS consortium schedule]
{p{0.65\textwidth}p{0.25\textwidth}}
{tab:HVsched}
{High level milestones and schedule for the production of the \dshort{hvs} of  \dshort{spvd}} 
Milestone & Date   \\ \toprowrule

%\rowcolor{dunepeach} 
Complete \dshort{spvd} \dshort{hvs} Final Design Review & April 2023 \\ \colhline
Complete \dshort{hvs} installation of \dshort{mod0} & April 2023     \\ \colhline
\dshort{mod0} commissioning & Fall 2023      \\ \colhline
 %\rowcolor{dunepeach} 
 Post \dshort{mod0} Review & Jan -- March 2024     \\ \colhline
 %\rowcolor{dunepeach} 
 \dshort{spvd} \dshort{hvs} production readiness review & March -- April 2024      \\ \colhline
\dshort{spvd} \dshort{hvs} component production & April 2024 -- Aug 2026 \\ \colhline
%\rowcolor{dunepeach}
Start of  %\dshort{detmodule} \#2 TPC 
\dshort{spvd} \dshort{hvs} \dshort{tpc} installation& July 2027      \\ \colhline
%\rowcolor{dunepeach}
End of  %\dshort{detmodule} \#2 TPC 
\dshort{spvd} \dshort{hvs} \dshort{tpc} installation& April 2028      \\ 
\end{dunetable}

A more detailed schedule for production and installation of the \dshort{spvd} is found in Figure~\ref{fig:tasks}.
\begin{dunefigure}[Key HVS milestones and activities toward 
\dshort{spvd}]{fig:tasks}{
Key HVS milestones and activities toward  the 
\dshort{spvd} in graphical format (Data from~\cite{docdb22261v28}).}
\includegraphics[width=0.95\textwidth]{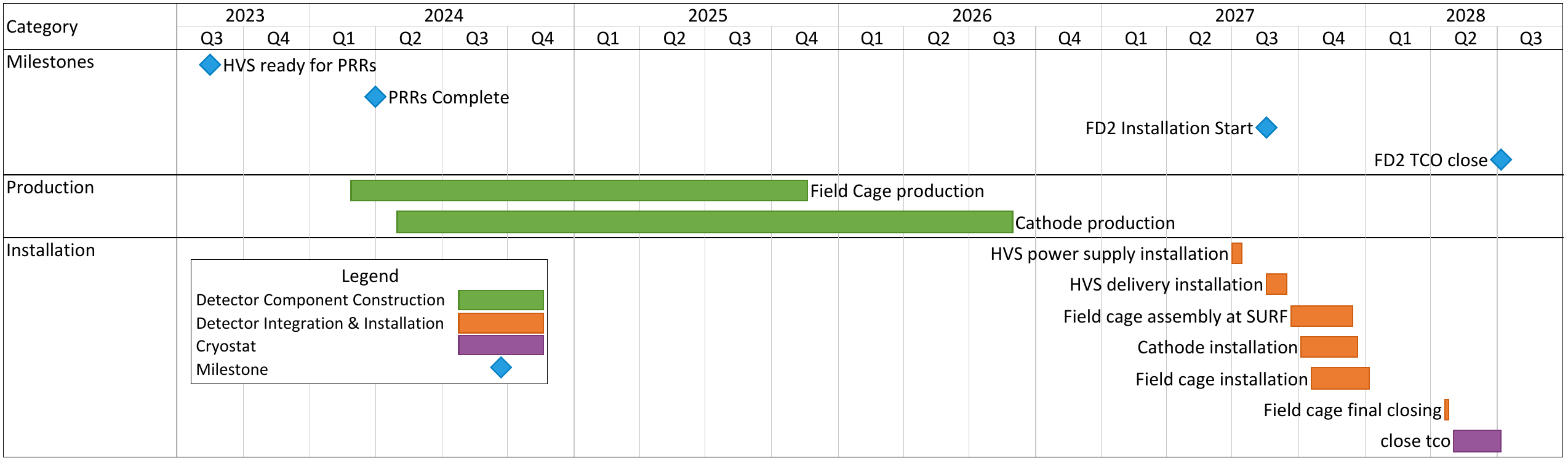}
\end{dunefigure}

%%%%

\chapter{Photon Detection System}
\label{chap:PDS}
% Editors: R.Wilson, M.Sorel

\section{Introduction}
\label{sec:PD-Intro}
%{\color{blue} Content: Cavanna}\newline
%{\color{blue} Edit: Wilson}

Energy deposition from the passage of high energy charged particles in \dword{lar} yields both free charge from ionization and fast scintillation light. 
In fact, \dshort{lar} is an excellent scintillator producing $\sim$25,000 photons per MeV of energy deposited by a \dword{mip} in the presence of an electric field (\efield)  of 500\,V/cm. Scintillation light provides information on three key detection aspects of the experiment: event triggering, event (and sub-event) 
precision time reconstruction, and event energy reconstruction as discussed in Chapter~\ref{ch:Phys}.

The mechanism for light production is quite well understood. Ionized and excited argon atoms combine on a picosecond timescale to produce $\rm Ar_2^*$ singlet or triplet excimer states that rapidly decay (about 25\% with a time constant of $\rm \tau_s = 6$~ns, and 75\% with $\tau_t = 1.5~\mu$s) yielding a characteristic scintillation of $\rm 128~nm$ wavelength \cite{Doke1981, Hitachi1992}. In the environment of \dshort{lar} doped with a small amount of xenon, the slow argon scintillation component is almost entirely shifted to $\rm 175~nm$ \cite{Wahl2014, KUBOTA1993}. This has the effect of increasing the scattering length for that component to approximately eight times higher than for $\rm 128~nm$ light with the impact that a higher amount of light is collected by the detector for sources beyond $\sim$4\,meters~\cite{Babicz2020}. 
%RJW 23feb23 
Xenon doping is assumed for the \dword{spvd} and this effect is included in all relevant simulations.

The presence of nitrogen impurities in \dshort{lar} affects both scintillation light production and propagation. 
Nitrogen affects the light output of argon through non-radiative collisional reactions that destroy the argon excimers before deexcitation, thus quenching scintillation light production. This process particularly affects the long-lived triplet excimer states, effectively reducing the amplitude of the slow component of the argon scintillation light~\cite{WArP:2008rgv}. In argon doped with xenon, the vast majority of the argon light that would otherwise be quenched by nitrogen impurities (via $Ar_2^{\ast}+N_2 \to 2Ar+N_2$) can be recovered (via the competing process $Ar_2^{\ast}+Xe \to ArXe^{\ast}+Ar$). The impact of nitrogen on scintillation light propagation is that it can absorb VUV photons, reducing argon transparency. Here again, xenon doping helps to mitigate this detrimental effect of nitrogen, since the photoabsorption cross-section by nitrogen is higher at 128~nm than at 175~nm~\cite{nitrogen}.

\Dwords{pd} are implemented in \dword{lartpc} experiments to exploit the information provided by scintillation light and thereby both improve and expand the capabilities of the apparatus.
The physics motivation for the system capabilities is presented in Section~\ref{sec:ph:pds}.
In the \dshort{spvd} proposal for the second DUNE far detector module, the implementation of a robust \dshort{pds} is an important feature of the design.

The \dshort{spvd} design will implement the same  \dword{xarapu} \dword{pds} technology as the \dword{sphd} design~\cite{Abi:2017aow} but with a modified configuration. Functionally, an \dshort{xarapu} module is a light trap that captures wavelength-shifted photons inside boxes with highly reflective internal surfaces until they are eventually detected by \dwords{sipm}. The wavelength-shifted photons are converted to electrical signals by \dshorts{sipm} distributed evenly around the perimeter of the \dword{wls} plate. 
Figure~\ref{fig:ARAPUCA-module-VD-photo} shows a full-scale  \dshort{xarapu} \dword{pd} module prototype.

\begin{dunefigure}
[\dshort{spvd} \dshort{pds}  detector module]
{fig:ARAPUCA-module-VD-photo}
{
 Full-scale prototype of the \dwords{xarapu} with (left) and without (right) the electronics cover. 
 }
  \includegraphics[width=0.4\textwidth]{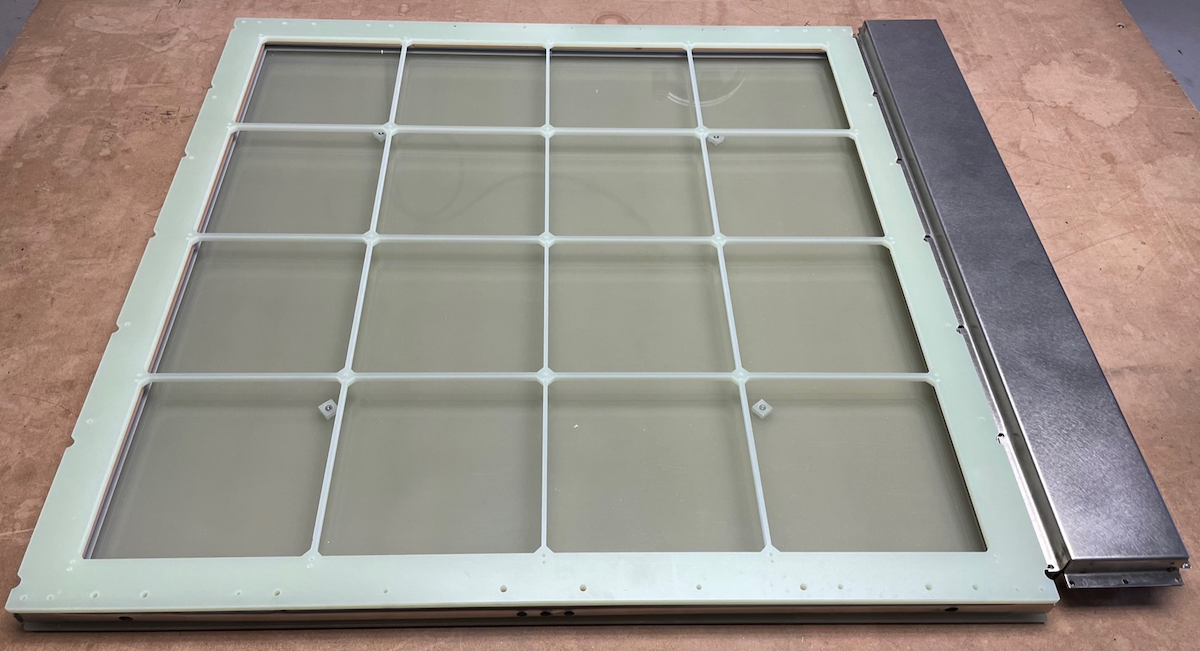}
  \includegraphics[width=0.4\textwidth]{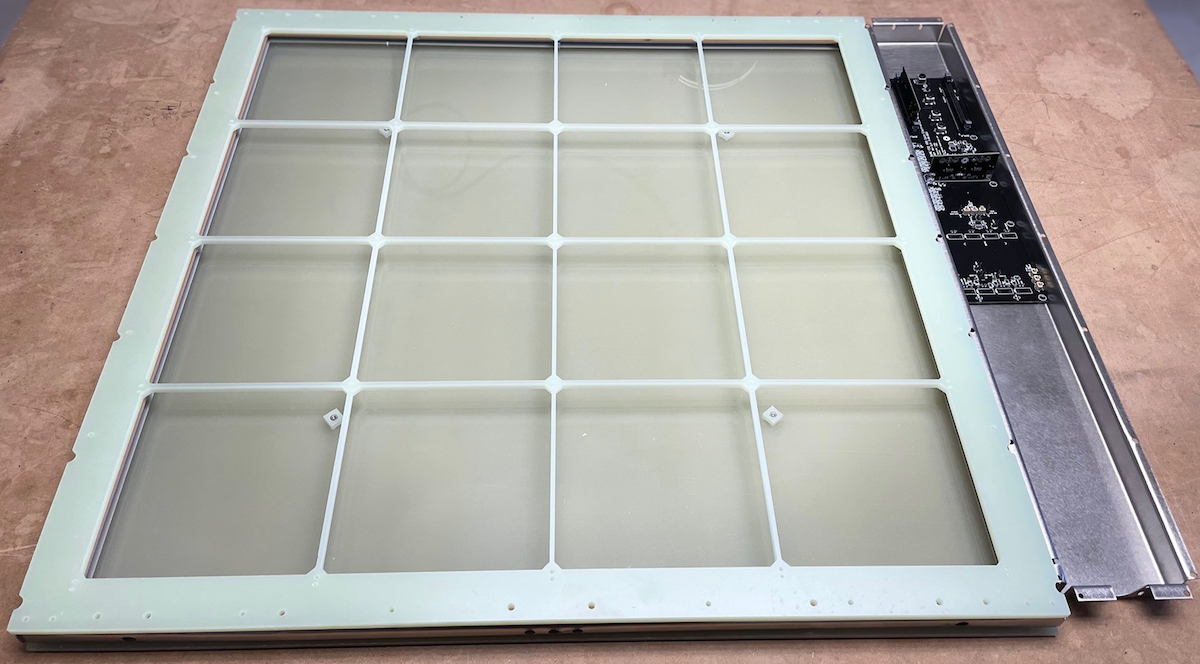}
\end{dunefigure}

 In the  vertical drift \dword{tpc} configuration, even though the \dword{crp} structure is perforated, it is effectively opaque to light and therefore does not allow for \dword{pd} installation at the anode (ground) side of the \dshort{tpc} volume. 
 This has the consequence that the \dshorts{pd} can only be installed on the cathode plane, on the %four 
 \dword{fc} walls or on the cryostat membrane walls behind the \dshort{fc}, provided that the latter is sufficiently transparent to light. 
 
 In the \dshort{spvd}  \dshort{pds} design, \dshort{xarapu}s are mounted on all four membrane walls (at ground potential) and within the cathode plane structure, as shown in Figures~\ref{fig:PD-Inside-VD-Mount-Concept}
 %, \ref{fig:Membrane_PDS_Layout} 
 and~\ref{fig:cathode-photon-modules}. 
 Cathode-mount \dshorts{pd} are at the cathode voltage, so there can be no conductive path to ground. 
 
 While membrane-mount \dshorts{pd} will adopt the same copper-based sensor biasing and readout techniques as \dshort{sphd}, cathode-mount \dshorts{pd} require new solutions to meet the challenging constraint imposed by operation in a \dword{hv} environment. The cathode-mount \dshorts{pd} are powered using non-conductive \dword{pof} 
 technology\cite{vasquez:ICTON-2019}, and the output signals are transmitted through non-conductive optical fibers, i.e.,  \dword{sof} technology. This solution provides % thus providing 
 voltage isolation in both signal reception and transmission. \dshort{pof} is a well established technology, but its extensive use in a cryogenic detector will be a new application. 
 
 \begin{dunefigure}
[\dshort{xarapu} modules mounting positions concept]
{fig:PD-Inside-VD-Mount-Concept}
{Perspective view of \dword{xarapu} modules locations on the horizontal cathode plane  and on the vertical cryostat membrane walls behind the \dshort{fc}.}
  \includegraphics[width=1.0\textwidth]{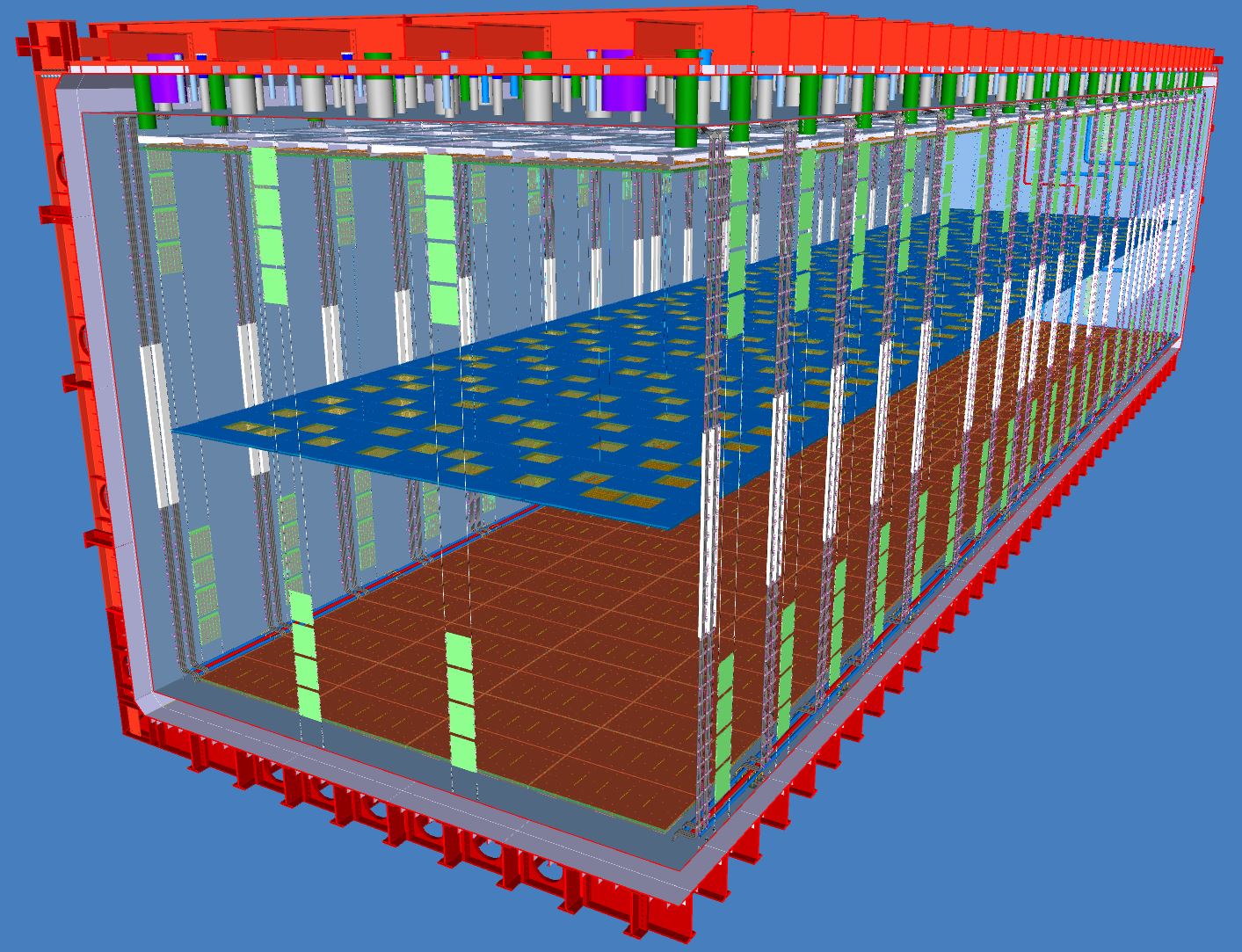}
\end{dunefigure}

\begin{dunefigure}
[Central cathode module showing the PD modules placement]
{fig:cathode-photon-modules}
{A cathode plane module showing the placement of the four \dshort{spvd}  \dshort{pds}  \dshort{xarapu}s in the central modules (blue squares). In the cathode plane modules adjacent to the \dshort{fc}, there are no PD modules abutting the \dshort{fc} to minimize potential discharge effects. 
The cathode module has a resistive skin over most of the surface, except for a conductive mesh of 90\% optical transparency over the \dshort{xarapu}s.} 
\includegraphics[width=0.7\textwidth]{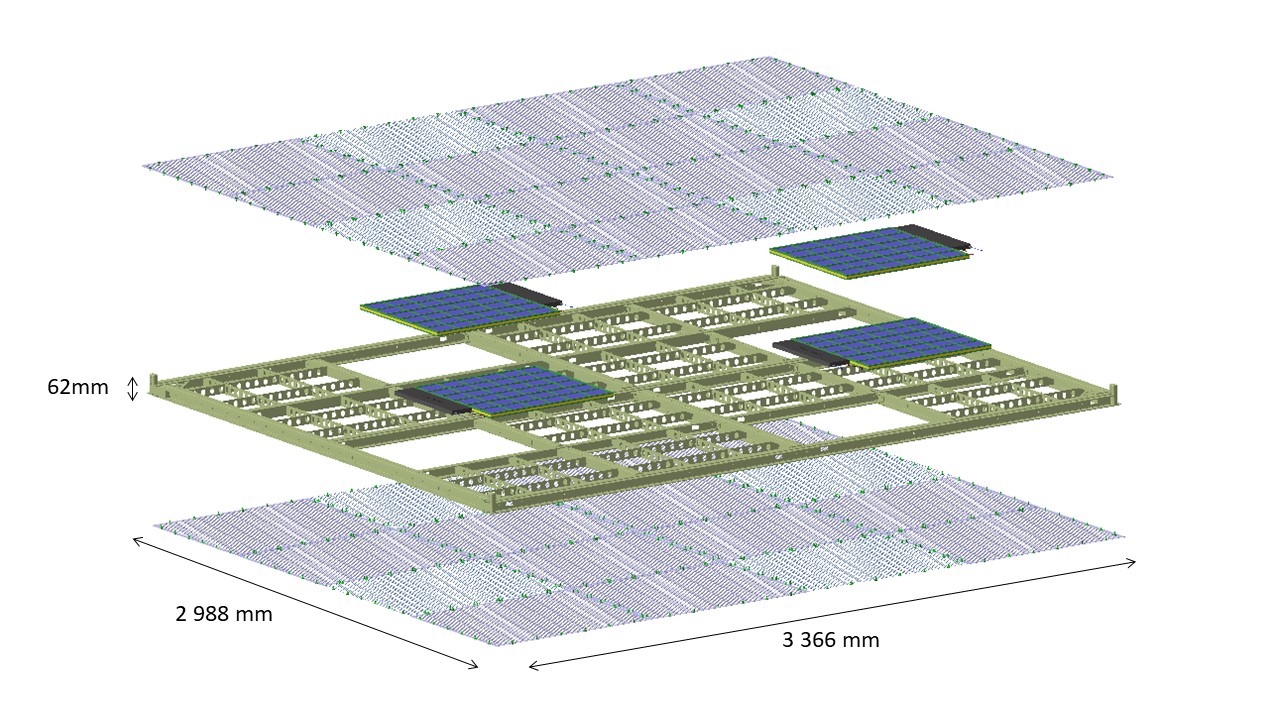}
\end{dunefigure}

Another important difference of the \dshort{spvd} design with respect to \dshort{sphd}, which has smaller optical volumes,  is the doping of \dshort{lar} with xenon at the level of $\mathcal{O}$(10\,ppm)\footnote{Although the baseline \dshort{sphd} design does not call for xenon doping, the cryogenics system allows the option for addition of a xenon injection port as a potential mitigation for nitrogen contamination.}. Results from large-volume measurements with \dword{pdsp} and \dword{pddp} are summarized in Section~\ref{subsec:PDS-xenon-protodune}. This feature will ensure improved light detection uniformity (see  Figure~\ref{fig:LY_maps_baseline}) 
and will make the system more resilient to nitrogen contamination. In particular, the ratio of minimum-to-average \dword{ly} in the $z$=0 plane of the active volume improves from 0.16 for pure argon to 0.51 for Xe-doped argon.

As shown in Section~\ref{subsec:PDS-Req}, the \dshort{spvd} \dshort{pds} configuration, with uniformly distributed \dshort{pd} coverage across the cathode and partial  coverage on all four membrane walls near the anode planes, produces significantly better overall light yield and light yield uniformity across the entire \dshort{tpc} active volume when compared to the \dshort{sphd} design for a comparable cost.

%%%%%%%%%%%%%%%%%%%%%%%%%%%%%%%%%%%%%%%%%%%%%%%%%%%%%%%%%%
\section{Design Specifications, Performance and Scope}
\label{subsec:PDS-Req-scope}

\subsection{Specifications and Performance}
\label{subsec:PDS-Req}

The detector specifications for the \dshort{spvd} \dshort{pds} are the same as for the \dshort{sphd} system. 
%(see Table~\ref{tab:specs:FD2-VD-EBheld}). 
Selected specifications, those with a direct connection with physics performance, are reproduced here in Table~\ref{tab:PD-VD-Requirements}. The proposed system will exceed these requirements, as summarized in Table~\ref{tab:PD-VD-Performance} and described below, providing a robust  margin against unexpected degradation and may enhance the system performance for some physics studies.

\begin{dunetable}
[PDS specifications]
{p{.03\textwidth}p{.20\textwidth} p{.20\textwidth} p{.45\textwidth}}{tab:PD-VD-Requirements}
{Key \dshort{spvd} \dshort{pds} specifications. 
}
{\bf No.} & {\bf Description} & {\bf Specification} & {\bf Rationale}\\ \toprowrule
1&Light~yield %(SP-FD-3)
& >20~PE/MeV~(avg),    & Gives PDS energy resolution comparable to that of the TPC for 5-7 MeV supernova (SN)\\   
      &  & >0.5~PE/MeV~(min) & $\nu$'s, and allows tagging of > 99\% of nucleon decay backgrounds with light at all points in detector.\\ \colhline  
2& Time~resolution %(SP-FD-4) 
&< 1\,$\mu$s \newline Goal: < 100\,ns & Enables 1\,mm position resolution along drift  direction.\\  \colhline
3& Spatial localization in plane perpendicular to drift %(SP-PDS-2)
&< 2.5\,m    & Enables accurate matching of \dshort{pd} and \dshort{tpc} signals.\\ \colhline 
4& Single PE pulse height divided by baseline noise RMS %(SP-PDS-14)
& $>4$ & Signal-to-noise sufficiently high to keep data rate within electronics bandwidth limits and to ensure efficient trigger.\\ \colhline
5& Dark rate per electronics channel %(SP-PDS-15)
& $<1$\,kHz & Dark noise sufficiently low to keep data rate within electronics bandwidth limits and to ensure efficient trigger. \\ \colhline
6& Fraction of beam events with saturating channels %(SP-PDS-16)
& $<20\%$ & Sufficient dynamic range is needed to reconstruct the energy calorimetrically, but a small amount of saturation can be mitigated. \\ 
\end{dunetable}

\begin{dunetable}
[PDS performance]
{p{.03\textwidth}p{.20\textwidth} p{.20\textwidth} p{.45\textwidth}}{tab:PD-VD-Performance}
{\dshort{spvd} \dshort{pds} 
estimated performance and basis for estimates (see text for details).}
{\bf No.} & {\bf Description} & {\bf Estimated Performance} & {\bf Basis for Estimate}\\ \toprowrule
1  &Light yield %(SP-FD-3) 
& $\simeq$39~PE/MeV~(avg), & Geant4-only simulations. \\   
   &                      & $\simeq$16~PE/MeV~(min)  & \\ \colhline
2& Time resolution %(SP-FD-4) 
&$\simeq$4\,ns & \dshort{pdsp} cosmic-ray data. \\  \colhline
3& Spatial localization in plane perpendicular to drift & $\simeq$0.75\,m~at~400~MeV & \dword{larsoft} simulations of nucleon decay events. \\ \colhline
4& Single PE pulse height divided by baseline noise RMS& 6 & Cold box tests at CERN and ganging tests in standalone facilities.\\ \colhline
5& Dark rate (DCR) per electronics channel & $\simeq$0.2\,kHz & Standalone tests at 77~K for 80 ganged \dshort{sipm}s (verified for both vendors). \\ \colhline
6& Fraction of beam events with saturating channels & 2\% of beam events with $>5$\% saturated channels & \dshort{larsoft} simulations of beam neutrino interactions. \\ 

\end{dunetable}

The light yield (\dword{ly}, detected photons per unit deposited energy)
is the most important \dshort{pds} specification in Table~\ref{tab:PD-VD-Requirements} (first row). Both the average \dshort{ly} from light emission anywhere in the detector volume, and the lowest detectable light level from the dimmest part of the detector, are relevant. For \dshorts{pd} located on a single detection surface, the \dshort{ly} variation across the volume would present a large gradient along a perpendicular axis to this plane. Placing photosensors along the four membrane walls in addition to the cathode, as in this system design, significantly reduces the \dshort{ly} non-uniformity along the drift ($y$) direction.

A dedicated study using a Geant4-only %\dword{mc} 
simulation has been performed to determine the \dshort{ly} map of the reference \dshort{pds} design. 
Compared to the \dword{cdr} \dshort{ly} studies that also relied on a \dword{geant4}-only simulation, several simulation assumptions affecting the detector geometry, the light production in \dshort{lar} and the reflectivity of detector materials have been updated. The result is a realistic simulation, based on essentially the same optical assumptions included in the end-to-end \dshort{larsoft} event processing chain discussed in Section~\ref{ch:Phys}. The advantage of the standalone \dshort{geant4} simulation over the \dshort{larsoft} simulation framework is a much faster turnaround for \dshort{ly} studies for different \dshort{pds} layout geometries or optical simulation assumptions.

In the \dshort{geant4}-only simulation used to estimate the \dshort{ly}, the full FD2 geometry is simulated, including the most relevant detector components from the \dshort{pds} point of view. In particular, the geometry of the thin ($\sim$70\% transparent) and thick (optically opaque) \dshort{fc} profiles, as well as the location of all cathode-mount and membrane-mount \dshort{xarapu}s, is accounted for. The average light transmission of the conductive mesh that covers the cathode-mounted module windows is taken to be 90\%. 

For scintillation light production in \dshort{lar}, we use 12,700 photons/MeV at 175~nm, plus 7,300 photons/MeV at 128~nm, as inferred from \dword{pddp} Xe-doping data (at 10~ppm Xe). For light propagation in \dshort{lar}, an absorption length of 80~m (20~m) and a Rayleigh scattering length of 8.5~m (1~m) are assumed at 175~nm (128~nm). This absorption length corresponds to a level of N$_2$ impurities in the \dshort{lar} of about 3\,ppm, see \cite{Jones:2013bca}. The reflectivity at 175~nm and 128~nm of the anode PCB, the \dshort{fc} profiles, and the membrane cryostat wall materials are accounted for in the simulation. Finally, motivated by prototype measurements and simulations~\cite{Souza:2021pfq, Brizzolari:2021akq, Palomares:2022xjc, Segreto:2020jpd}, the \dshort{xarapu} detection efficiency is set to the target value of %$\epsilon_D= 
$3\%$ at both 175~nm and 128~nm. In the light yield plots (Figure~\ref{fig:LY_maps_baseline}) the detection efficiency is a scale factor, as discussed below. 

\begin{dunefigure}
[Reference configuration light yield map]
{fig:LY_maps_baseline}
{(Left) Map of the \dfirst{ly} in the central $(x,y)$ transverse plane at $z=0$ for the reference configuration. (Right) Map of \dshort{ly} in the central $(z,y)$ longitudinal plane at $x=0$, close to the detector boundary at $z=30$~m in front of one of the membrane short walls, for the same \dshort{pds} configuration.

}
  \includegraphics[width=0.49\textwidth]{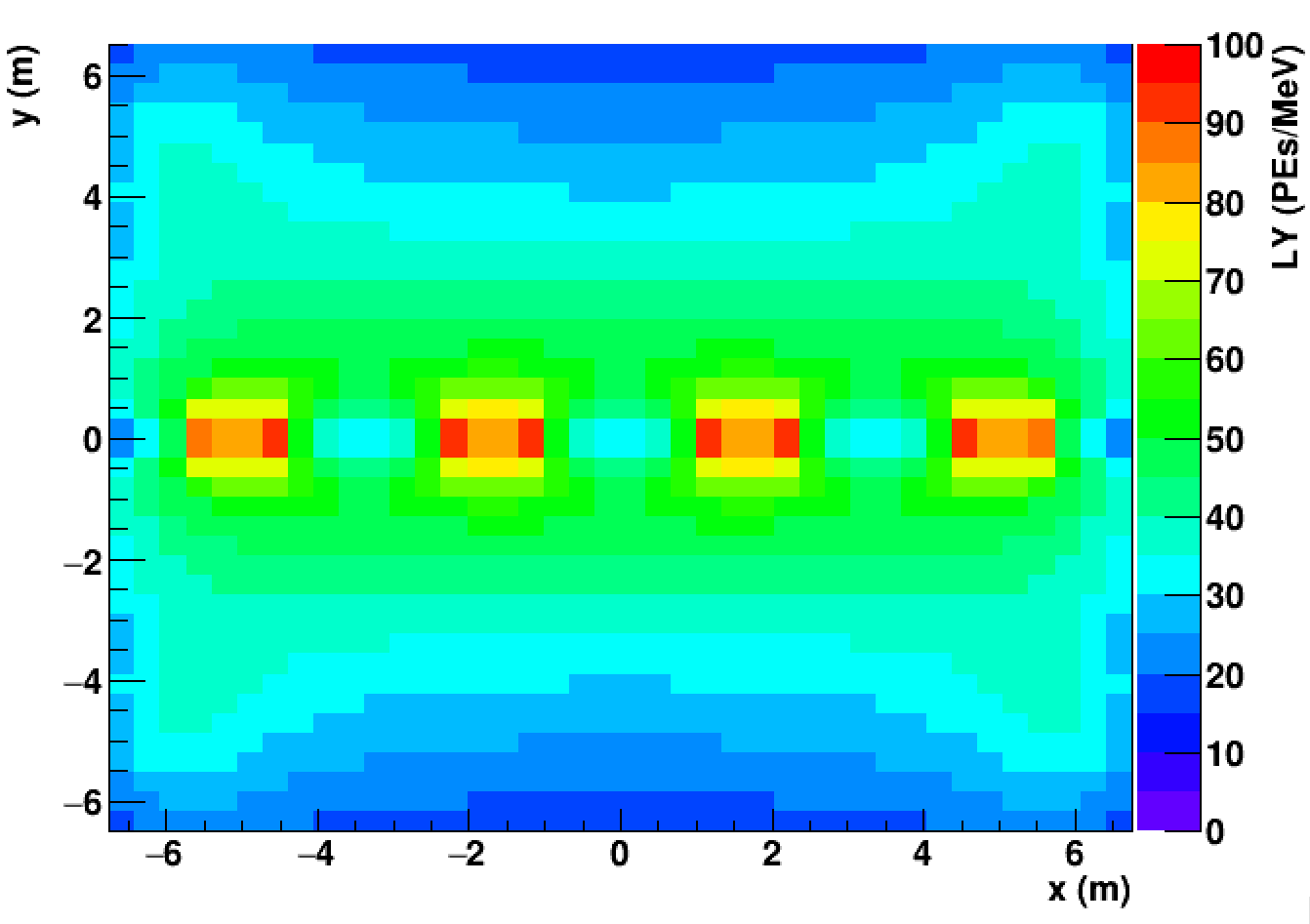} \hfill
  \includegraphics[width=0.49\textwidth]{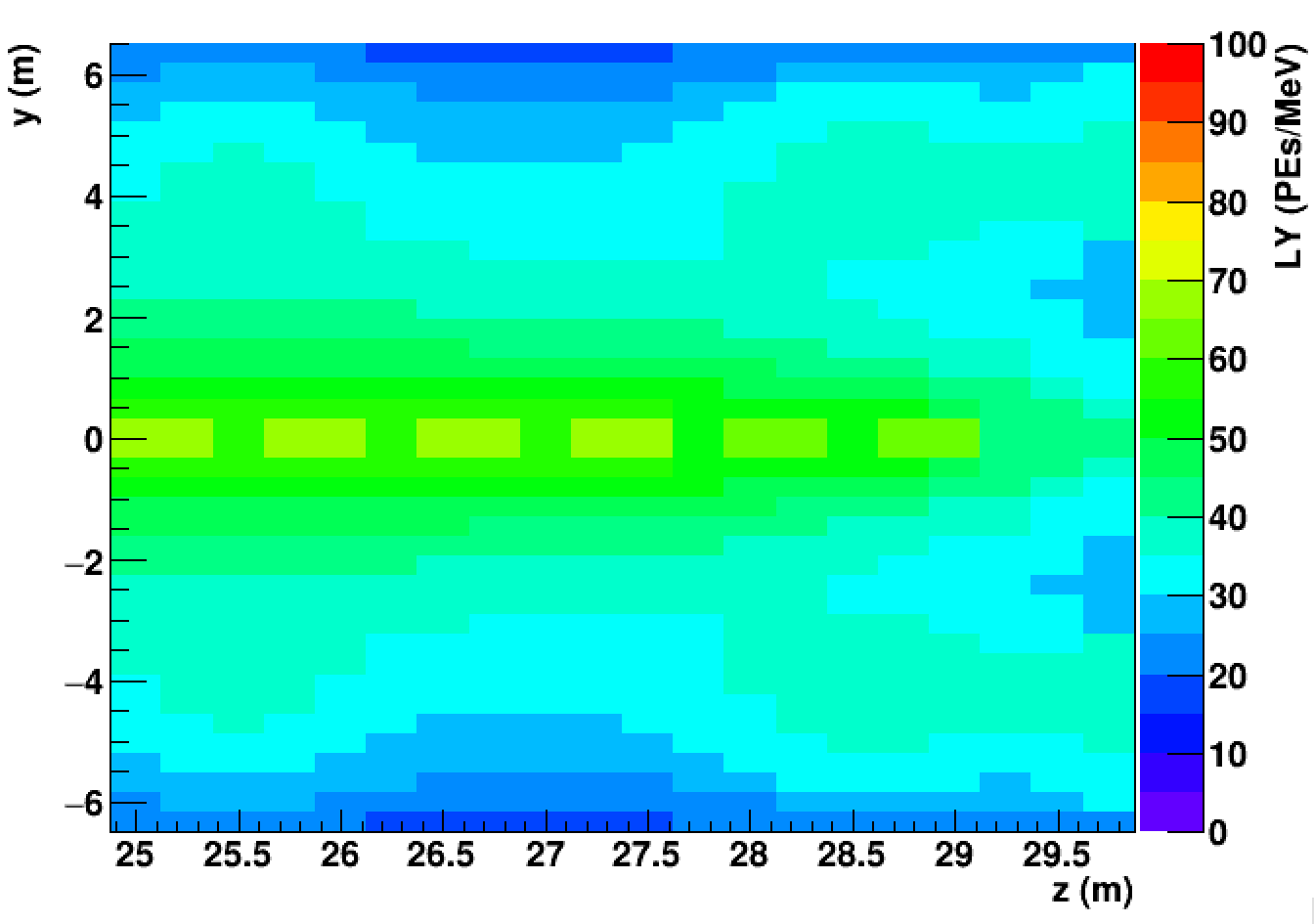}
\end{dunefigure}

\begin{dunefigure}
[Improved spatial uniformity of response by instrumenting membrane short walls]
{fig:LY_at_border}
{Values for average (left) and minimum (right) \dshort{ly} in $(x,y)$ planes as a function of the $z$ direction, obtained close to the detector boundary at $z=30~$m. Light yield values for two detector configurations are shown: reference configuration including \dshort{xarapu}s on the membrane short walls in blue, without short wall \dshort{xarapu}s in red.}
  \includegraphics[width=0.49\textwidth]{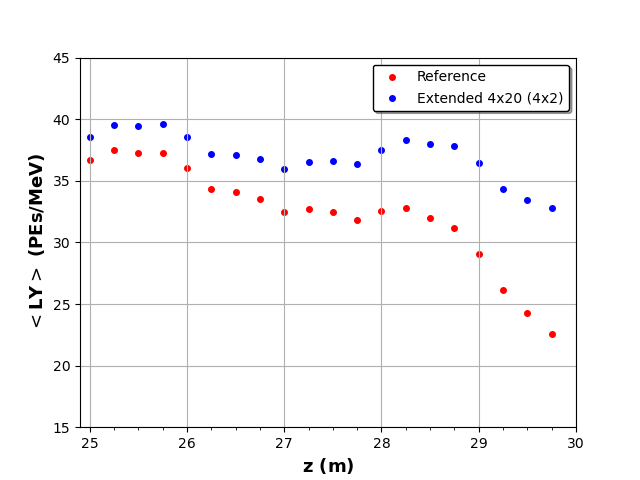} \hfill
  \includegraphics[width=0.49\textwidth]{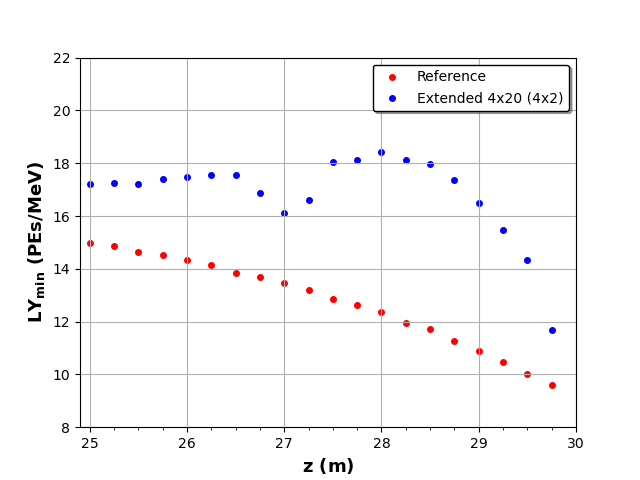}
\end{dunefigure}

The light yield \dshort{ly}$(x,y)$ for the central transverse plane at $z=0$ inside the target volume was evaluated and is shown in Figure~\ref{fig:LY_maps_baseline}~(left).\footnote{We use the \dshort{spvd} coordinate system: $z$ is parallel to the Far Detector Module's long dimension, aligned with the incident neutrino beam; $y$ is the vertical drift direction; and $x$ is along the module's narrow horizontal dimension. The origin is at the geometric center of the module.}  
The $(z,y)$ projection of \dshort{ly} for $x=0$ and near the detector end-wall at $z=30$~m is shown in Figure~\ref{fig:LY_maps_baseline}~(right). As is evident in both projections, the \dshort{ly} is highest near the cathode plane at $y=0$, and lowest near the non-instrumented anode planes at $y=\pm 6.5$~m. The average \dshort{ly} value for events in the $0<z<3$~m detector portion near the central transverse plane at $z=0$ is $\langle \mathrm{LY}\rangle \simeq 39~{\rm PE/MeV}$, as reported in Table~\ref{tab:PD-VD-Performance}. Boundary conditions close to either ends of the detector volume at $z=\pm 30$~m also affect the \dshort{ly}. 

As shown by the red data points in Figure~\ref{fig:LY_at_border}, a clear decrease for both average (left) and minimum (right panel) \dshort{ly} per $z$ slice would be observed along the $z$ direction at both detector ends if no \dshort{xarapu}s were installed on the membrane short walls. The presence of 32 \dshort{xarapu}s in the reference design on the end walls ensures a more uniform \dshort{ly} response along $z$ near the detector boundaries, as shown by the blue points in Figure~\ref{fig:LY_at_border}.
The goal of this modest (5\%) coverage increase, compared to the option with non-instrumented end walls, is to enlarge the FD2 fiducial volume for analyses relying on \dshort{pds} information.

These values significantly exceed the \dshort{ly} specifications of Table~\ref{tab:PD-VD-Requirements}. They are also much better than the \dshort{sphd} \dshort{pds} values, particularly with regard to the spatial uniformity of the detector response. While these simulation results assume our target \dshort{xarapu} detection efficiency of 3\%, light yield estimates for the \dshort{spvd} \dshort{pds} scale linearly with efficiency. Light yield specifications would thus be met even under the worst-case \dshort{xarapu} detection efficiency that would satisfy requirements, that is 2\% (see Sec.~\ref{sec:PDS-LightColl}). As discussed in Table~\ref{tab:PD-VD-Requirements} and in Section~\ref{ch:Phys}, this high and uniform light yield will ensure a \dshort{pds} energy resolution comparable to that of the TPC at \dword{snb} energies, \dshort{snb} event triggering using \dshort{pds} information alone, and efficient tagging of nucleon decay backgrounds anywhere in the \dshort{lar} active volume.

Preliminary time resolution results for an \dshort{xarapu}-based \dshort{pds} have been obtained using the 12 \dshort{xarapu} channels in APA6 of \dword{pdsp} at CERN. Photons coming from the same cosmic-ray muon track are detected by two separate (nearby) channels. \dshort{tpc} crossing cosmic-ray muon tracks nearly parallel to APA6 are used, such that differences in photon arrival time to the two nearby channels become negligible. To reduce the effect of the 6.67\,ns sampling time, the waveforms' rising edges are fitted to extract 
the time at which the amplitude crosses one \dword{pe} for each channel. The difference in time measured by the two channels provides a direct estimate of the single-channel time resolution, measured to be about 4\,ns. The time resolution of the optical flashes, being a collection of time-coincident and space-correlated optical hits, is  better than 4\,ns.\footnote{The current \dshort{pds} reconstruction assumes a 1.5 \dshort{pe} threshold to reconstruct an optical hit, and a 3.5 \dshort{pe} threshold to construct an optical flash.}
Even though this result was obtained with \dshort{pdsp} data, the similar \dshort{pds} technology and sampling time ensure that a similar time resolution is expected in \dshort{spvd}, exceeding the time resolution requirement by a large margin.

Initial results on spatial resolution in the \dshort{spvd} and in the plane perpendicular to the drift direction were obtained with a Geant4-only simulation of point-like energy deposits at \dshort{snb} energies (tens of MeV) and a pseudo-reconstruction relying on the barycenter of the light pattern, see \cite{Paulucci:2021sqn}. A spatial resolution in the transverse plane of about 0.5\,m was obtained. These results have now been confirmed with full \dshort{larsoft} simulations of nucleon decay events. 

For this study, the simulation and reconstruction of $10^5$ events has been carried out. The events include GENIE-generated $p\to K^+\bar{\nu}$ decays, one decay per simulated event, uniformly distributed in the \dword{lar} volume and depositing about 400~MeV of energy on average. They also model the detector activity due to radiological backgrounds, according to the model in Table~\ref{tab:bkg}. The radiological model introduces a rate of reconstructed optical flashes of about 200~kHz, or about 800 flashes throughout the 4~ms long event time window. For the position reconstruction study, we define the signal-like flash in each event as the flash of highest charge among those with a timing consistent (within $\pm$1~$\mu$s) with the nucleon decay time. 

As in our earlier study, the reconstructed position of the nucleon decay candidate event in the plane perpendicular to drift is reconstructed using a simple barycenter algorithm, using position and detected charge information from \dshorts{pd} located on the cathode. The reconstructed position is then compared to the true position of the MC simulated nucleon decay vertex. A spatial resolution of about 75~cm is obtained. This expected performance is significantly better than the 2.5\,m spatial resolution required.

The single-\phel pulse height divided by the baseline noise \dword{rms} (or \dword{s/n} ratio, in the following\footnote{The \dshort{s/n} ratios quoted throughout the chapter refer to the smallest possible signal of interest for one \dshort{pd} channel, which is a single-\phel pulse. Most physics event categories in DUNE will typically produce much larger signals.}) and the dark rate per electronics channel (rows 4 and 5 in Table~\ref{tab:PD-VD-Requirements}), have been studied with \dshort{pdsp} \dshort{pds} data, with \dshort{cern} \coldbox data, and with dedicated setups. 
Since the initial \coldbox runs in August 2022, light leakage from \dword{pof} receivers and fibers 
have been reduced to negligible levels.  
The \dshort{s/n} ratio measured in a \coldbox run in February 2023 for a PD module with \dshort{pof} and \dshort{sof} readout was 5.9, as shown in Figure~\ref{fig:pds_charge_spec}, significantly higher than the specification of 4. 
 
The dark count rate at 77\,K of the \dshorts{sipm} under consideration for \dshort{spvd} (see Section~\ref{sec:PDS-Photosens+Bias}) has been measured to be about 60\,mHz/mm$^2$, corresponding to about 0.2\,kHz per electronics channel, which is well within requirements.

Table~\ref{tab:PD-VD-Requirements} specification No. 6, the maximum fraction of beam events with saturating channels that impacts energy resolution, has been demonstrated in full \dshort{larsoft} simulations of beam electron neutrino interactions and assuming a 14-bit dynamic range, see Section~\ref{ch:Phys}. It was found that only a 2\% (0.2\%) of events have more than 5\% (10\%) \dshort{pds} channels overflowing the ADC range. Also, this dynamic range induces a mean reduction in the determination of the total deposited energy per event, estimated from the reconstructed charge sum over all optical hits, of only 2\%. In other words, \dshort{pds} channel saturation level in \dshort{spvd} is expected to be within requirements. It is also lower compared to the saturation studies presented in the \dshort{sphd} \dword{tdr} \cite{DUNE:2020txw}, which assumed a 12-bit digitizer.

\subsection{\dshort{pd} Consortium Scope}
\label{subsec:PDS-Scope}

The \dshort{dune} \dword{pd} Consortium (see Section~\ref{sec_OrgMan}) will provide a photon detector system for \dshort{spvd} that meets the performance requirements established by the DUNE collaboration as presented in Table~\ref{tab:PD-VD-Requirements}. The scope of the consortium activity includes selecting and procuring material for, and the fabrication, testing, delivery and installation of light collectors (\dshort{xarapu}), photosensors (\dshorts{sipm}), electronics, and a calibration
and monitoring system. 
The reference design components 
%for the \dshort{spvd} \dshort{pds} 
are listed in Table~\ref{tab:pds-config-scope}.

\begin{dunetable}
[\dshort{pds} reference configuration]
{p{0.13\textwidth}p{0.25\textwidth}p{0.51\textwidth}}
%{lll}
{tab:pds-config-scope}
{\dshort{pds} reference configuration}
Component  				& Description 						& Quantity		\\ \toprowrule
Light collector 		& \dshort{xarapu}							& 352 single-sided modules (160 per long-dimension wall, 16 per short-dimension wall), 320 double-sided modules (cathode plane); \num{672} total. Light collection area 3600~cm$^2$/module/side \\ \colhline
Photosensor 			& Hamamatsu and FBK \SI{6}{mm} \dshort{sipm}$\times$\SI{6}{mm} & 160 \dshorts{sipm} per module; \num{107520} total	\\ \colhline
\dshort{sipm} signal summing		& 5 groups (in series) of 4 \dshort{sipm}s (in parallel) per flex \dshort{pcb}				& 8 flex \dshort{pcb}s per module; \num{5376}  total	\\ \colhline
Readout electronics		& Analog signal conditioning circuit & 2 channels/module; \num{1344}  total	\\ \colhline
Calibration and monitoring	& Pulsed UV via \dshort{fc}-mounted fibers & 15 fibers per roof penetration; 20 penetrations per side; \num{600}  total		\\
\end{dunetable}

%(Table~\ref{tab:penetrationtable})
Although the configuration of the \dshort{spvd} and \dshort{sphd} led to structurally different solutions for the \dshort{pds}, many of scientific and technical issues are similar. 
A common consortium facilitates the sharing of information and helps to exploit the similarity of these two detectors, as appropriate.

%%%%%%%%%%%%%%%%%%%%%%%%%%%%%%%%%%%%%%%%%%%%%%%%%%%%%%%%%%%%%%%%%%%%%%%%%%%%%%%%

\section{Photon Detector System Overview}
\label{sec:PD-Overview}

Scintillation light detection in the \dshort{spvd} module is based on the \dword{xarapu} technology developed for the \dshort{sphd} 
module, with the photocollector %geometrical configuration 
design modified to match the different \dshort{spvd} mechanical and electrical constraints. %, see Figure~\ref{fig:ARAPUCA-module-VD}. 
The \dshort{pds} consists of a large number of %modest-sized 
photon detection units distributed on or behind the surfaces delimiting the \dshort{lartpc} active volume. 
The basic unit is the \dshort{xarapu} detector module with a light collecting area of approximately 600~$\times$~600\,{\rm mm$^2$}. 
The modules have light collection windows on either one (single-sided) or two (double-sided) faces, depending on the location in the active volume.
The wavelength-shifted photons are converted to an electrical signal by 160 \dwords{sipm} distributed evenly around the perimeter of the module. Groups of \dshorts{sipm} are electrically connected to form just two output signals, each corresponding to the sum of the response of 80 \dshorts{sipm}.      

The \dshort{spvd} \dshort{pds} topology consists of 320 double-sided \dshort{xarapu} modules evenly distributed across the cathode plane (cathode-mount modules) and 352 single-sided modules mounted behind the \dshort{fc} onto the cryostat walls (membrane-mount modules), as illustrated in Figure~\ref{fig:PD-Inside-VD-Mount-Concept}.  The 352 membrane-mount modules includes 32 modules on the cryostat end-walls\footnote{These were not part of the Conceptual Design Report design, but were added to provide better light-yield uniformity throughout the active volume.}.

The cathode-mount \dshort{xarapu}s  are embedded in the mechanical frames of the central cathode plane, as illustrated in Figure~\ref{fig:cathode-photon-modules}. They will have two optical surfaces, one facing upward into the top \dshort{spvd} volume, the other facing downward into the bottom volume.
Operating photodetectors on a \dword{hv} surface requires electrically floating photosensors and readout electronics, i.e., power (in) and signal (out) transmitted via non-conductive cables. This is achieved with the use of \dshort{pof} technology and optical transceivers for communication via \dword{sof} technology. 

Membrane-mounted \dshort{xarapu}s will be arrayed in columns along all four walls of the cryostat. They will face inward from behind the \dshort{fc}, spanning the vertical ranges $3.3<y<6.3$\,m and $-6.3<y<-3.3$\,m %in the top and bottom detector halves,
as measured from the cathode plane, respectively, where the \dshort{fc} is approximately 70\% transparent (Section~\ref{subsubsec:FCsss}). They will be supported on hanging vertical mechanical structures, %arranged in columns, 
each holding 
four modules mounted near the top anode plane and four near the bottom. 
There will be 20 columns on each of the long walls and two columns on each short wall. % of the cryostat. 
The column structures are electrically referenced to detector ground and thus do not require \dshort{pof} and optical transceivers; standard transmission through copper cables will be used for both power and signal, leveraging the \dshort{sphd} experience.

A conceptual diagram of the %signal and power optical fiber
the \dshort{sof} and \dshort{pof} routing for the cathode-mount modules and the copper-based routing for the membrane-mount modules is shown in Figure~\ref{fig:PD-power-signal-routing}.

\begin{dunefigure}
[Conceptual diagram of power and signal routing]
{fig:PD-power-signal-routing}
{Conceptual diagram of power and signal routing. One of the 80 cathode four-module groups (see Figure~\ref{fig:cathode-photon-modules}) is shown on a greatly expanded scale; fibers pass through holes in the cathode modules, down along the \dshort{fc} and over to the membrane wall, en route to penetrations on the top surface of the cryostat. More detailed drawings are provided in Section~\ref{sec:PDS-CathodeModule-FiberRouting}.
}

\includegraphics[width=0.8\textwidth]{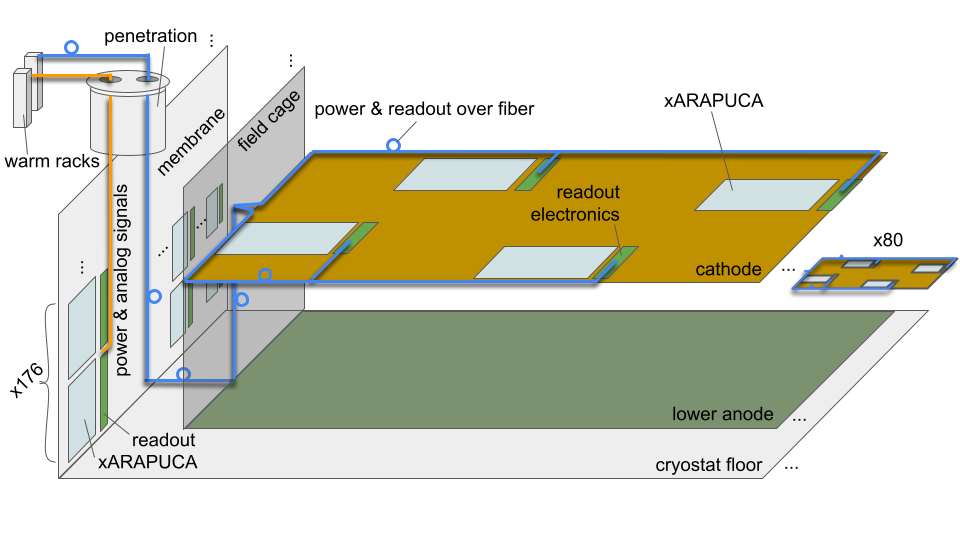}
\end{dunefigure}

To optimize the light collection uniformity, the \dshort{lar} is doped with $\mathcal{O}$(10\,ppm) xenon, which like argon, is a high-yield scintillator. 
Several studies have demonstrated its beneficial light production properties when used as a low concentration dopant~\cite{KUBOTA1993, Hofmann_2013, Wahl2014, Akimov_2019}.
Motivated by these studies done on smaller LAr volumes, DUNE tested xenon doping in the \dword{protodune} detectors, and concluded that it will
provide greater light yield uniformity and allow photon detection recovery from accidental nitrogen contamination for multi-meter drift-path TPCs 
(see Section~\ref{subsec:PDS-xenon-protodune}).

A concentration of $\mathcal{O}$(10\,ppm)  is sufficient for the Ar triplet scintillation component to be fully transferred to the xenon emission wavelength (175\,nm), which reduces the Rayleigh scattering rate of the wavelength-shifted light and so increases collection efficiency for light emitted at a far distance from the \dshorts{pd}.
It also mitigates the risk due to inadvertent nitrogen contamination of the \dshort{lar}. 
The time response of the \dshort{pd} modules to scintillation light will be used to monitor the xenon content in \dshort{lar}. \dshort{pdsp} data have demonstrated that the shape of the average \dshort{pd} modules’ waveforms is very sensitive to the xenon concentration.

\begin{dunetable}
[Summary of \dshort{spvd} \dshort{pds} scintillation medium and reference design technology]
%{rl}
{p{.13\textwidth}p{.80\textwidth}}
{tab:PDS-Summary}
{Summary of \dshort{spvd} \dshort{pds} scintillation medium and reference design technology.}
	Item	& \textbf{Description} \\
\toprowrule
Liquid argon  & Ar+Xe (10 ppm). Residual impurity: [N$_2$]$<1$ ppm, [O$_2$]$<50$ ppt.\\ 
Scintillator &Absorption length: $\lambda_{abs}\ge 30~$m.\\ 
    &Photon yield (\dshort{mip}, nominal \efield=500 V/cm): 25,000 ph/MeV.\\ 
    &(Ar-Xe) Energy transfer reaction:  Ar triplet scintillation component fully \\
    & transferred to Xe $\Rightarrow Y_{ph}({\rm Ar})\simeq {\rm 6,000~ph/MeV}$, $Y_{ph}({\rm Xe})\simeq {\rm 19,000~ph/MeV}$. \\
    & Rayleigh scattering length: $\lambda_R({\rm Ar})\simeq 1$m, $\lambda_R({\rm Xe})\simeq 8.5$m.\\
    \colhline
Photon  & \dshort{xarapu} (light trapping by dichroic filters and two-stage \dshort{wls}). \\
collector &Sensitive to both Ar light (128\,nm) and Xe light (147 and 175\,nm).\\
    &Detection efficiency $\epsilon_D\simeq 3\%$ (detected PEs per photon impinging upon the\\
    &optical surface of the detector).\\
    \colhline
Photosensor &\dword{sipm} %(Si avalanche photodiode micro-cell array), 
$6\times 6~{\rm mm}^2$ area, single-photon sensitive.\\
    &Requirements at nominal voltage: PDE(\@450 nm)$>$35\%, dark count rate\\
    &$<$200 mHz/mm$^2$, cross-talk $<$35\%, afterpulse $<5$\%, gain $>2\times 10^6$. \\
    &\Dword{ce} read-out (active ganging, shape and noise filtering). \dword{s/n}$\ge 4$ \\
\end{dunetable}
%$$$$$$$$$$$$$$$$$

The basic properties of the scintillation medium, photon collectors, photon sensors and associated electronics for the \dshort{spvd} \dshort{pds} are summarized in Table~\ref{tab:PDS-Summary}.

The \dshort{spvd} \dshort{pds} design of 320 
%\fixme{not 352?} that is the membrane 
cathode mounted \dshort{xarapu}s provides 105.6\,m$^2$ %\fixme{115 here and below - check} 
photodetector coverage out of the 810\,m$^2$ cathode area, for 13.0\% optical coverage.
The assemblage of 320  \dshort{xarapu}s mounted on the long walls of the cryostat membrane gives 105.6\,m$^2$ coverage 
%out of the 1560\,m$^2$ \dshort{fc} line-of-sight area, 
which is about 6.8\% of the \dshort{fc} area. 
Finally, the optical coverage on the cryostat membrane short walls is 10.6\,m$^2$ out of 351\,m$^2$ for 3.0\%. Figure~\ref{fig:PD-Inside-VD} (left) is a \threed rendering of the \dshort{spvd} showing the top two rows of \dshorts{pd} \dshort{xarapu}s  (blue) on one long wall and those in the cathode frame. The view is from the opposite cryostat wall, behind the \dshort{fc}, in the top drift volume.

\begin{dunefigure}
[Mounted \dshort{xarapu} modules; view through the \dshort{fc}]
{fig:PD-Inside-VD}
{\dshort{xarapu} detector modules (indigo) on the cathode plane and on the \dshort{fc} walls -- view through $\sim$70\% transparent \dshort{fc} into detector volume (left); view of the membrane-mount  \dshort{xarapu} detector modules from inside the active volume (in this close-up view, the \dshort{fc} bars are visible in front of the modules) (right).}
  \includegraphics[width=0.5\textwidth]{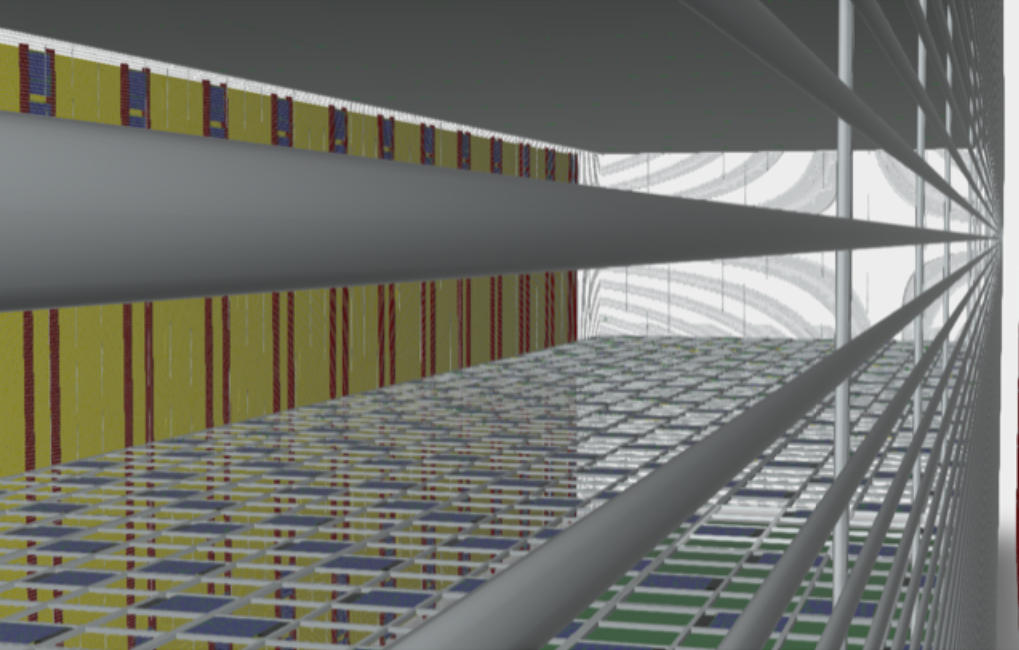}  
  \includegraphics[width=0.48\textwidth]{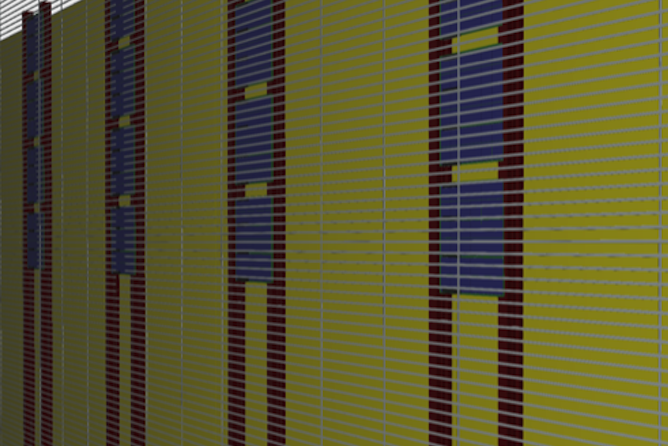}
\end{dunefigure}

The \dshort{fc} electrodes in the foreground are $\sim$70\% transparent to light, allowing transmission to the \dshort{pd} detector modules behind it. 
This configuration provides the \dshort{spvd} \dshort{pds} with a photon detection performance that meets the experiment's requirements, as presented in Chapter~\ref{ch:Phys} and Section~\ref{subsec:PDS-Req-scope}.

\begin{dunetable}
[\dshort{spvd} photon detector system components] 
{rrl}
{tab:PD-VD-components}
{\dshort{spvd} \dshort{pds} components.} 
			\textbf{Item}          & \textbf{Quantity} & \textbf{Detector surface}\\
			\toprowrule
			\dshort{xarapu} modules  & 320 double-sided  & Cathode plane\\
			                        &  320 single-sided  & Membrane long walls \\
			                     &  32 single-sided    & Membrane short walls \\
			\colhline
			Dichroic filters & 17,856 &  \\
			\dshort{wls}  plates       & 672  &  \\
			Photosensors (\dshorts{sipm}) & 51,200  & Cathode plane\\
			                                   &  51,200  & Membrane long walls \\
                                               &  5,120  & Membrane short walls \\
			\colhline
			Signal channels  &  640     & Cathode plane   \\
			                                 & 640  & Membrane long walls \\
			                                 & 64  & Membrane short walls \\
			\dshorts{sipm} per channel   &   80      &     \\
			Optical Area &   105.6\,m$^2$  $\times$ 2  & Cathode plane  \\
			             &   105.6\,m$^2$  &  Membrane long walls \\
%			             &   11.52\,m$^2$  &  Membrane short walls \\
			             &   11.5\,m$^2$  &  Membrane short walls \\
			Active coverage    & 13.0\% & Cathode plane\\
			& 6.8\% & Membrane long walls \\
			& 3.0\% & Membrane short walls \\
\end{dunetable}

\section{Light Collectors}
\label{sec:PDS-LightColl}

The \dshort{xarapu} developed for the \dshort{sphd} has been adopted for use in \dshort{spvd}.
It offers compact and flexible design geometry and has a very small impact on the \dshort{lartpc} active volume. The \dshort{xarapu} module has an excellent ratio of optical to inactive surface areas (68\,\%), a good \dword{pde}, a moderate fabrication cost, and is relatively easy to integrate into the \dshort{tpc} layout.
Table \ref{tab:PD-VD-components} itemizes the components of the \dshort{pds}.

\begin{dunefigure}
[\dshort{spvd} \dshort{pds} detector module]
{fig:ARAPUCA-module-VD}
{(Left) Dimensioned drawing of a \dshort{spvd} \dshort{pds} detector module. (Right) Exploded view showing major design elements.  The double-sided module concept to be used in the cathode is shown (with dichroic filter windows on both sides of the module). For the membrane mount modules, the windows on the side facing the cryostat wall are replaced with a reflector-coated G-10 sheet. [A photograph of a module that will be tested in the \coldbox at \dshort{cern} is shown in Figure~\ref{fig:ARAPUCA-module-VD-photo}.]}
  \includegraphics[width=0.45\textwidth,angle=0]{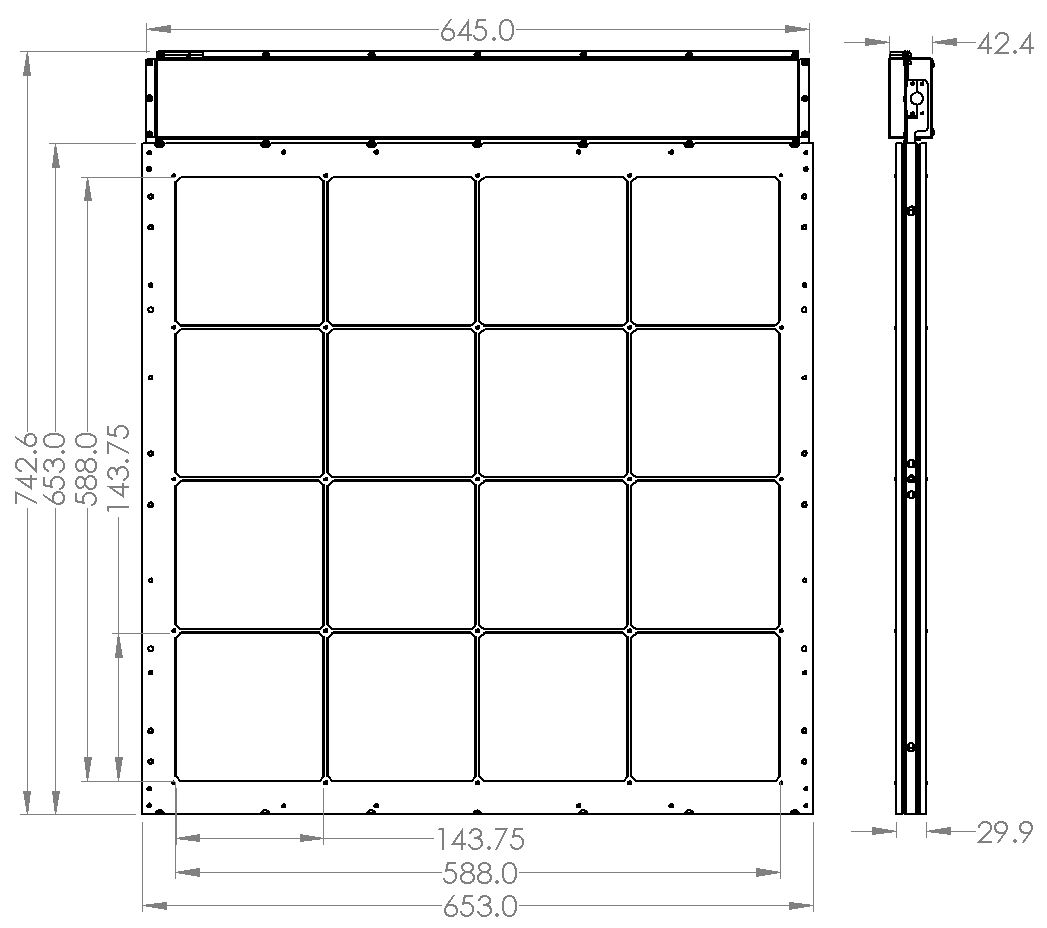} % \hfill
  \includegraphics[width=0.45\textwidth]{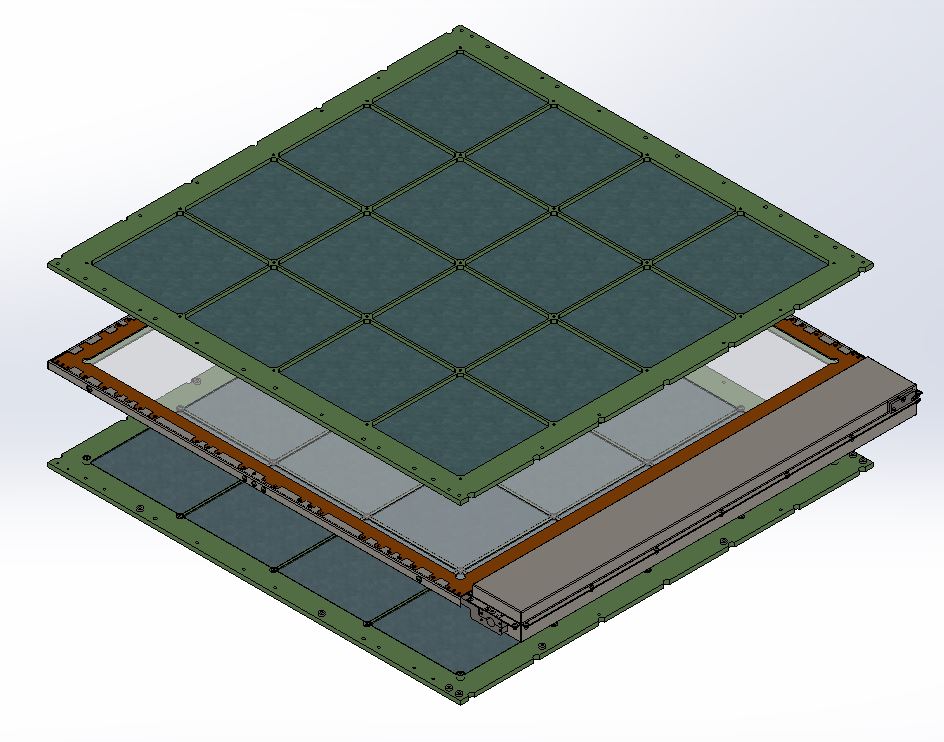}
\end{dunefigure}

The basic unit of the \dshort{pds} is a module as depicted in Figure~\ref{fig:ARAPUCA-module-VD} with dimensions provided in Table~\ref{tab:PD-module_summary}. 

It is a tile-shaped thin module with overall dimensions
\qtyproduct[product-units = power]{653 x 653 x 29.9}{\mm} plus an attached electronics box \qtyproduct[product-units=power]{645x89.6x42.4}{\mm}, bringing the total outside dimensions to \qtyproduct[product-units=power]{742.6x653x42.4}{\mm}. 
\dshort{pd} modules are composed of three basic design elements:  
\begin{itemize}
\item one (single-sided) or two (double-sided) dichroic window frame assemblies composed of 16 \dword{ptp}-coated dichroic filters mounted in \dword{g10} frames -- the dichroic window dimension establishes the active area of the module as \qtyproduct[product-units=power]{575x575}{\mm}, $0.33{\rm \,m^2}$;
\item one frame and Faraday cage shielding assembly consisting of a \dword{wls}-doped acrylic plate onto which the \dshorts{sipm} are coupled (a three-sided conducting box protects the \dshorts{sipm} from induced currents due to a potential cathode HV discharge, see Section \ref{sec:PDS-LightColl-discharge}); and
\item one cold electronics/\dword{pof}/\dword{sof} 
electronics enclosure, mounted to one side of the basic module following assembly.
\end{itemize}

\begin{dunetable}
[PD basic unit: \dshort{xarapu} module]
{rcc}
{tab:PD-module_summary}
{PD basic unit: \dword{xarapu} module.}

	Item/Parameter	& \textbf{Quantity} & \textbf{Dimensions}\\
\toprowrule
			Light collection module area   &    1   & \qtyproduct[product-units=power]{653x653}{\mm} = $0.43\,{\rm{m^2}}$ \\
		     (excluding electronics box)   &       &  \\
			Module thickness &  1  & $29.9$~{mm} \\
			Weight &    1   & $\sim8.6{\rm ~kg}$\\ \colhline
			Active Area &  2 (two-sided)   & 
			$0.33{\rm \,m^2}$ per side\\ 
			\colhline
			Dichroic filters & 
			16 $\times$ 2 sides & \qtyproduct[product-units=power]{143.8x143.8}{\mm}\\
			\dshort{wls} plate & 1 &	\qtyproduct[product-units=power]{607x607x3.8}{\mm} \\ 
		
			\dshorts{sipm} 	& 160 & \qtyproduct[product-units=power]{6x6}{\mm} \\	\colhline
		
			Read-out channels  & 2 &    \\
			\dshorts{sipm} per channel   & 80 & \\

\end{dunetable}

The square modular geometry optimizes the ratio of the light collecting area to the \dword{sipm} area and the module mechanical design allows use of the same structure for both the cathode-mount and membrane-mount configurations.  In the case of the single-sided membrane-mount modules, the rear side (facing the cryostat wall) window frame assembly is removed and replaced by a single sheet of \dshort{g10} coated with 3M Vikuiti\textsuperscript{TM} reflective material.

The \dshort{wls} plate acts as a secondary shifter, serving three main functions: providing the wavelength shift required to trap the photons within the \dshort{xarapu} volume,  matching the \dshorts{sipm} wavelength sensitivity, and as a light guide -- the optical-grade surfaces enable efficient uniform light collection over a large area.
Prototype \dshort{wls} plate have been manufactured at a casting reactor at UniMIB, while for the large scale production, the Glass to Power s.p.a. company (G2P)\footnote{Glass-to-Power - Italy \url{https://www.glasstopower.com}.} G2P has been identified as an industrial partner.

The mechanical properties of acrylic plates have been tested extensively in both the \dshort{sphd} and \dshort{spvd} \dshort{pds} geometries.  ProtoDUNE-SP (the Horizontal Drift ProtoDUNE) included 38 acrylic bars (externally coated with TPB) and showed no indication of surface cracking after more than a year of continuous immersion in LAr. Each of the 20 Glass to Power bars installed in \dword{hdmod0} were been submerged in LAr and inspected following testing, with no signs of surface degradation. Finally, full-size \dshort{spvd} WLS plates have been tested through multiple cooling cycles (up to 4 cycles for the V1 module) with no observed degradation. \dword{vdmod0} will provide the final long-scale validation of these full-size bars.

A spring-loaded \dshort{sipm} mounting system (see Figure~\ref{fig:ARAPUCA-module-corner}) makes a dynamic connection between the \dshort{wls} plate and the \dshort{g10} frame to allow for  relative thermal contraction of the \dshort{wls} plate and frame.

\begin{dunefigure}
[\dshort{wls} plate centered in the frame by spring-loaded dynamic mounting system]
{fig:ARAPUCA-module-corner}
{\dshort{wls} plate centered in G-10 frame, held by spring-loaded dynamic mounting system. Coil springs (16 total, three visible in this image) maintain physical contact between the \dshort{wls} plate and the \dshort{sipm} face throughout cool-down. Screws (seen in this figure) passing through the springs are used to allow controlled application of force while the \dshorts{sipm} are installed, then removed following installation.}
  \includegraphics[width=0.7\textwidth, angle=0]{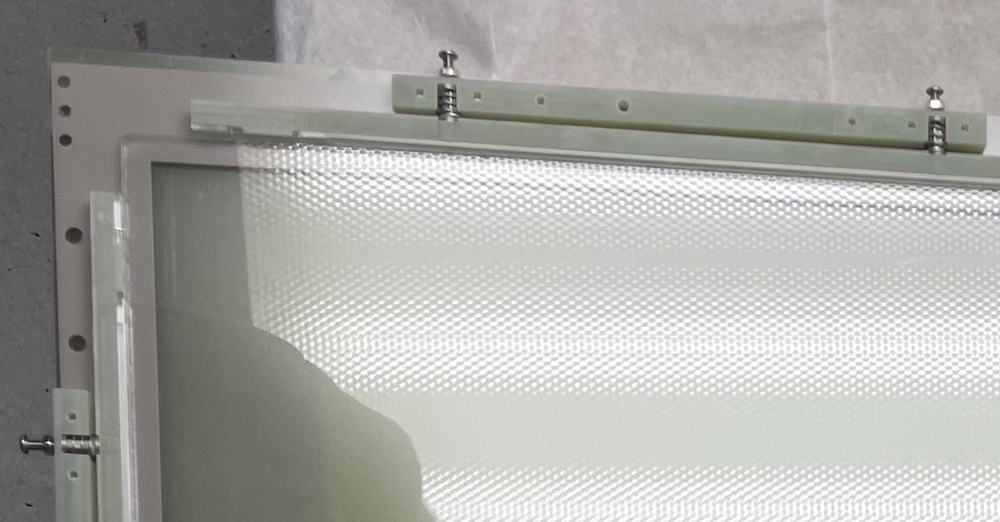}
\end{dunefigure}

The \dshorts{sipm} have dimensions 6$\times$6\,mm$^2$ and are mounted in a series of five adjacent units of four \dshorts{sipm}, each passively ganged on one flexible Kapton \dshort{pcb} strip (i.e., one 30\,cm flex circuit has 20 \dshorts{sipm}; see Figure~\ref{fig:ARAPUCA-module-flexipcbs}, left). 
The \dshort{sipm}s are positioned symmetrically with respect to the mid-plane of the plate\footnote{Simulation of two additional \dshort{sipm} geometries with the same active area 
(4$\times$9\,mm$^2$ and 3$\times$12\,mm$^2$) 
in the \dshort{sphd} PD module bar configuration showed no substantial difference in the detection efficiency that would justify a custom geometry for the \dshort{sipm}.}.

Each flex circuit mates through a connector to a shielded twisted-pair cable 130~cm long that passively routes the flex circuit signal to the module interface board where a stage of signal shaping and buffering prepares the signal for readout. There are eight such flex circuits along the perimeter of the \dshort{wls} plate, for a total of 160 \dshorts{sipm}. The \dshorts{sipm} are grouped into two readout channels of 80 \dshorts{sipm} each; two channels offer a level of redundancy for the module.

The strips of \dshorts{sipm} are held against the \dshort{wls} plate edges by an array of low-strength coil springs.  
The use of a spring-loaded backing plate and flexible \dshort{pcb}s for mounting (Figure~\ref{fig:ARAPUCA-module-flexipcbs}) the \dshorts{sipm} combined with dynamic supports for the \dshort{wls} plate inside the \dshort{g10} provides accommodation for the roughly 1\% relative thermal contraction between the \dshort{wls} plate and the \dshort{g10} frame during \cooldown. A Vikuiti reflector adhered to the Kapton flexi PCB covers the plate edge between adjacent \dshorts{sipm}.

A possible enhancement of the reference PD module design to enhance the efficiency of photon extraction from the \dshort{wls} plates onto the \dshorts{sipm} would be to include cutouts in the \dshort{wls} plates at the location of the \dshorts{sipm}. The performance and cost-effectiveness of three different cutout shapes and associated production costs are being evaluated as part of the final module evaluation, with a final decision as to their inclusion decided prior to the \dword{prr}.

\begin{dunefigure}
[\dshort{sipm}s mounted to flexible \dshort{pcb}]
{fig:ARAPUCA-module-flexipcbs}
{20 \dshorts{sipm} mounted to flexible \dshort{pcb} (left).  Flexible \dshort{pcb} positioned against a dummy \dshort{wls} plate (white block) with spring-loaded dynamic mount (right).}
  \includegraphics[width=0.45\textwidth, angle=0]{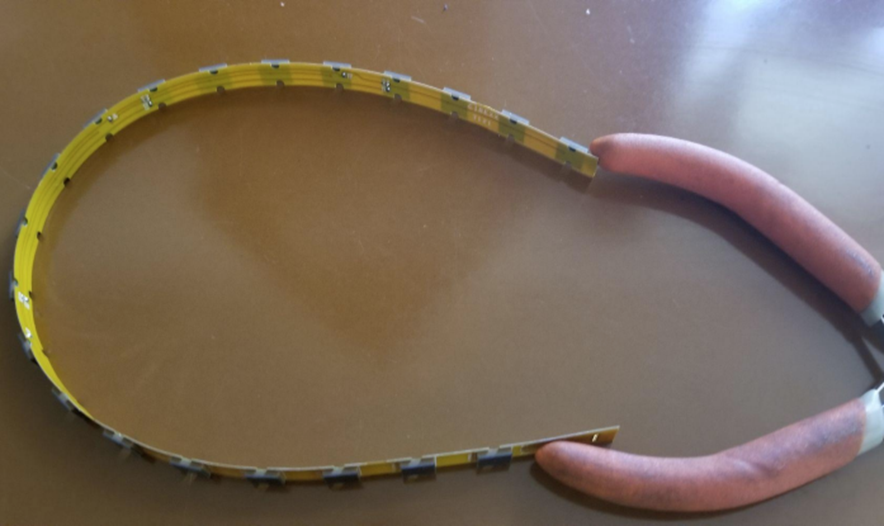}  
  \includegraphics[width=0.4\textwidth, angle=0]{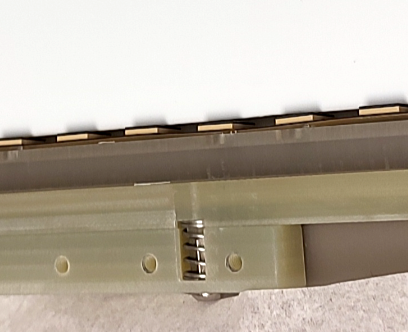}
\end{dunefigure}
% renamed PDS-module-flexipcbs-02.png 9/21/23 Anne - perl script didn't pick it up

The required \dword{pde} for the PD modules is 2\% with a target value of 3\%, similar to the best achieved by the \dshort{sphd} \dshort{xarapu}~\cite{Souza:2021pfq, Brizzolari:2021akq, Palomares:2022xjc, Segreto:2020jpd}. 
Measurements of \dword{pde} have not yet been made on full-sized \dshort{spvd} \dshort{pd} modules but the fundamental design of the \dshort{xarapu} is the same as for \dshort{sphd}, so we expect similar performance.  Nonetheless, there are differences that directly affect the \dword{pde}, most significantly the ratio of photosensors per light collection area is reduced by a factor of 2.4 compared to \dshort{sphd}.
Simulations indicate that this factor is compensated for by the more efficient \dshort{wls} plates, spring-loading of the \dshort{sipm} to \dshort{wls} plate mounting to ensure mechanical contact, and reflective \dshort{xarapu} frames. 
An additional enhancement may come from inclusion of light-concentrating cutouts in the edge of the \dshort{wls} plates that simulations indicate increases the photon collection efficiency - validation studies will be completed prior to design baselining in summer 2023.

\subsection{Cathode Modules HV Discharge Protection}
\label{sec:PDS-LightColl-discharge}

Having the cathode modules mounted in the cathode plane exposes them to a risk of damage from a discharge event of the -300~kV \dword{hv} system.

Two independent studies, one at \dword{bnl} and one at \dword{fnal},  agreed that the original design presents an unacceptable risk of damage to the \dshort{xarapu} modules' electronics from a discharge, especially for the configuration in which multiple modules share power and signal, 
but that an open faced conductive enclosure around the \dshort{sipm} flexible PCBs and readout cables, together with a complete enclosure of the cold readout electronics, provides sufficient risk mitigation.

As a result of these studies, the \dshort{xarapu} module design has been modified to include a three-sided Faraday cage composed primarily of copper-clad \dword{fr4} surrounding the \dshort{sipm}s, flexible readout boards, and readout cables for all PD modules. Additionally, the readout electronics, and \dshort{pof} %power over fiber 
in the case of the cathode mount modules, are enclosed in the Faraday cage.  Details of this cage can be seen in Figure~\ref{fig:ARAPUCA-module-faraday}. The opening in the Faraday cage is ${\rm 593.4\times 593.4\,mm^2}$, matching the coverage of the dichroic filter windows.

This design will be tested and improved in test benches that will emulate the cathode discharge, with progressively increasing charge injection amounts to identify weaknesses. In \dword{vdmod0} we will test the design on both shared/distributed power and signal system and independently powered \dshort{xarapu}s. At the end of the \dshort{vdmod0} run, discharges will be induced from maximum cathode HV.

The final design will be validated in a \coldbox run in 
in advance of the Production Readiness Review."

\begin{dunefigure}
[Module frame with discharge shield for \dshorts{sipm}]
{fig:ARAPUCA-module-faraday}
{Copper-clad G-10 frames and formed stainless-steel external covers provide a 3-sided discharge shield for \dshorts{sipm}.}
  \includegraphics[width=0.65\textwidth, angle=0]{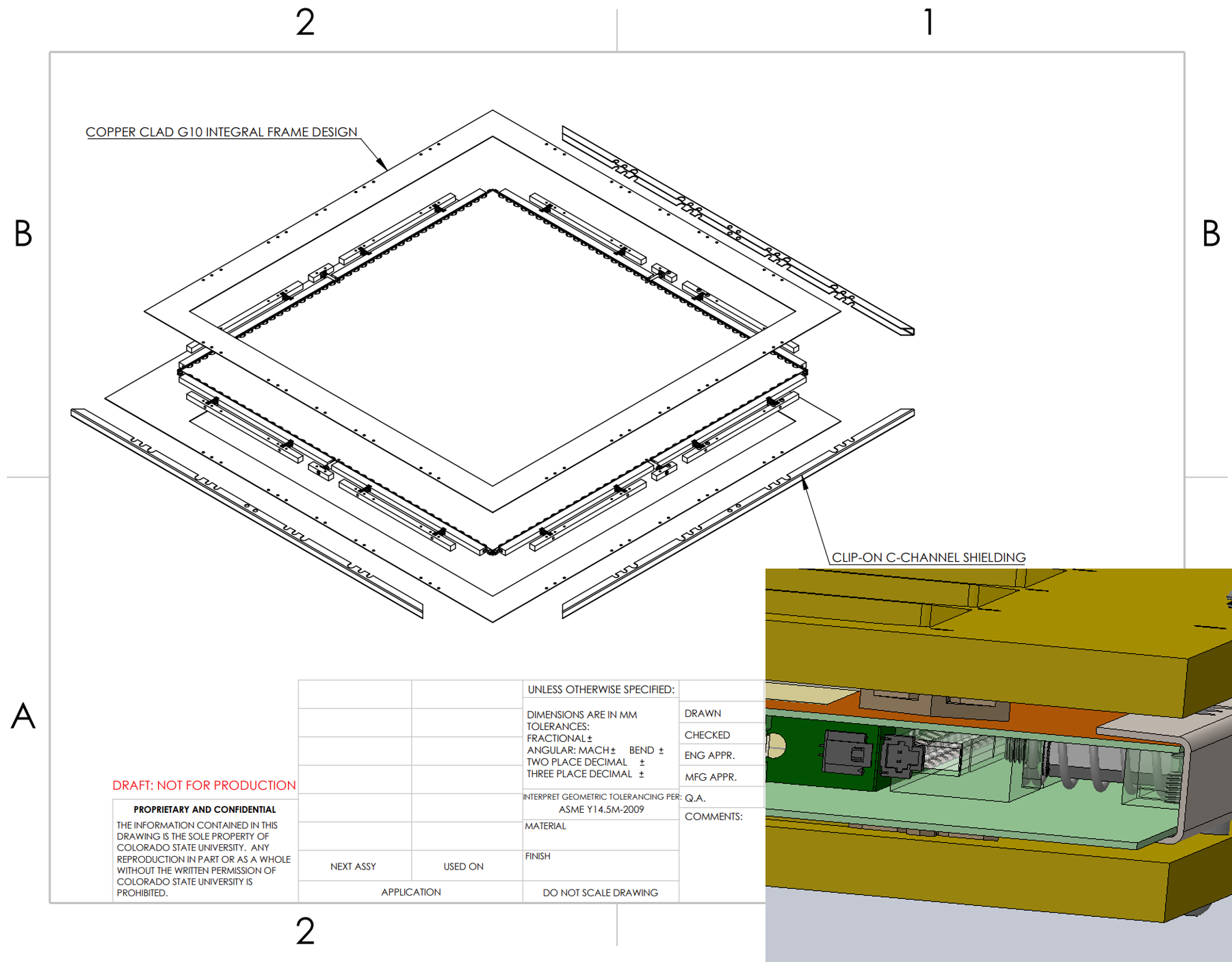}
\end{dunefigure}

\subsection{Membrane Modules Suspension System}
\label{sec:PDS-LightColl-fixation}

The membrane-mount portion of the \dshort{pds} is composed of 20 columns of 8 \dshort{pd} modules on each of the two long walls of the cryostat, plus 2 columns of 8 \dshort{pd} modules on each of the two short walls, for a total of 352 modules. Each \dshort{pd} column is supported by two suspension lines. The suspension lines are made of three connected stainless steel pieces: a top rod bar (5\,mm diameter, 3.7\,m length) fixed on the cryostat roof, a central tube (12\,mm diameter, 6\,m length) in the central part of the detector, and a bottom rod bar (5\,mm diameter, 3.7\,m length) fixed on the cryostat floor. The top and bottom rod bars have eye bolts to be fixed on the M10 membrane bolts welded on the cryostat. The central tube has a larger (12\,mm) diameter in order to avoid inducing large field gradients in the high voltage region near the cathode plane. The central tube is fixed between the two rod bars by shackles. There is a jaw to jaw straining screw and spring on the bottom rod bar, in order to compensate for differences between nominal and real line dimensions (75\,mm adjustment range), to pre-load the spring, and to absorb the thermal expansion. 

Each \dshort{pd} module has four attachment points, two on each suspension line, as shown in Figure~\ref{fig:ARAPUCA-module-fixation} (left). The fixation points are made of wire rope grips pre-positioned along the 5\,mm rod bars. Figure~\ref{fig:ARAPUCA-module-fixation} (right) shows a zoomed image of one fixation point. Signal cables are routed along the rod bars toward the cryostat roof and floor, and then exit on the \dword{bde} cables tray. The estimated weight of each \dshort{pd} module column, including 8 \dshort{xarapu} modules, electronics, cables, fixation elements, and rod bars/tube lines, is about 110\,kg. Several prototype suspension lines have been produced.

\begin{dunefigure}
[Fixation system for membrane-mount \dshort{pd} modules.]
{fig:ARAPUCA-module-fixation}
{Fixation system for membrane-mount \dshort{pd} modules. (Left) Each \dshort{xarapu} module is supported using four fixation points on two vertical suspension lines. (Right) Detail of one fixation point.}
  \includegraphics[width=0.30\textwidth]{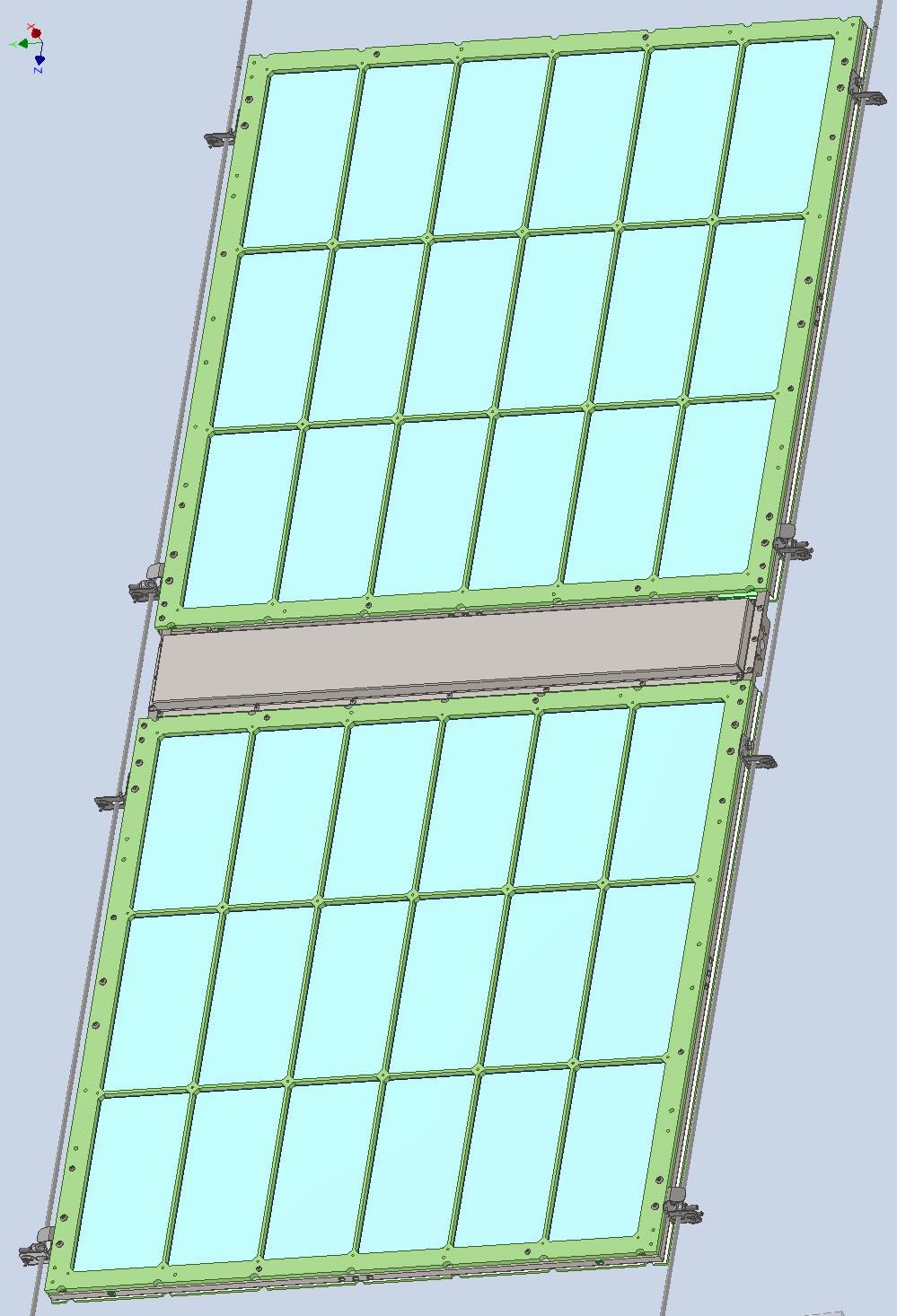} %\hfill
  \hskip 0.3in
  \includegraphics[width=0.25\textwidth]{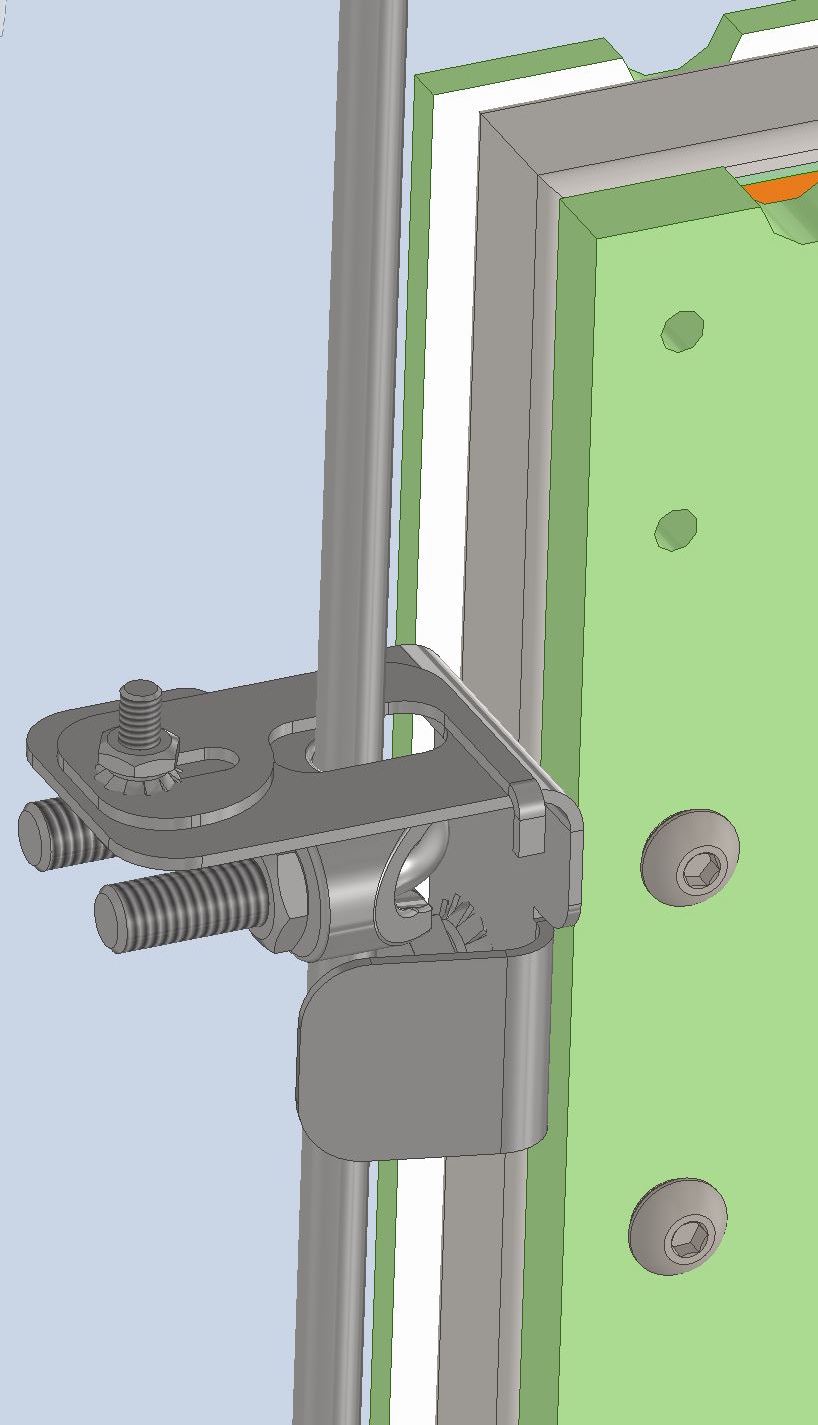}
\end{dunefigure}

\section{\dshort{pds} Module Signal} % and Sensor Biasing}
\label{sec:PDS-Photosens+Bias}

\subsection{Silicon Photosensors}
\label{subsec:PDS-sipm}

The photosensors of choice for the \dshort{spvd} are \dwords{sipm}. This choice is driven by the use of the \dshort{xarapu} as the underlying technology for photon collection and the experience leveraged from the \dshort{sphd} module development.
The sensors developed in collaboration with the two main vendors of cryogenic \dshorts{sipm} (Fondazione Bruno Kessler (FBK) and Hamamatsu Photonics (HPK)) fulfill the sensor specifications listed in Tables~\ref{tab:PD-VD-Requirements} and \ref{tab:PDS-Summary}. 
Several candidate devices from these vendors have undergone extensive evaluation by the DUNE team over several years. 
The typical breakdown voltage for Si-photosensors at cold temperature is in the few tens of volts ($\sim42$\,V for Hamamatsu \dword{mppc} and $\sim27$\,V for FBK \dshort{sipm}). They will be operated at +3 to +5\,V overvoltage, with a typical \dword{pde} at room temperature of 45\% at 430\,nm. 

The \dshort{sipm} development needed for the \dshort{spvd} \dshort{pds} was modest.
Thanks to the experience gained in the construction of %ProtoDUNE-HD (ProtoDUNE-SP Run II)
\dword{hdmod0}, we know that the failure rate is well below 0.5\% and the dark count rate is marginal compared with the expected contribution from $^{39
}$Ar and other radiological backgrounds\footnote{Our background simulation (where $^{39}$Ar is the dominant contribution) yields a flash rate of about 200~kHz for a flash reconstruction threshold of 3.5 photoelectrons. This is  much larger than the flash rate induced by \dshort{sipm} dark counts alone, and for the same 3.5 PE threshold.}.
 
For optical coupling of the \dshorts{sipm} to the \dshort{wls} plate, both gluing and non-adhesive positioning were investigated. 
The results indicate that 
a spring loading system (see Section \ref{sec:PDS-LightColl}) provides an optimal solution. 
This solution was successfully validated in  laboratory test-stands and during the \coldbox tests.
The associated risks related to the thermal stress 
are mitigated by the flexible boards that host the \dshorts{sipm}.

The baseline \dshorts{sipm} choice is to use the S13360-5075HD-HQR (HPK) for the membrane modules and the NUV-HD-CRYO triple-trench (FBK) for the cathode modules. This solution benefits from the lower operating voltage of the FBK sensors to be biased by the \dshort{pof}.  
Both types of sensors have been tested with the \dshort{pof} system and achieved the expected performance.

As a consequence, the choice of \dshort{sipm} models is mostly driven by cost-effectiveness and the performance of the cold electronics described in Section~\ref{sec:PDS-Electronics}. 
The \dshorts{sipm} do not represent a significant risk source for the design of the \dshort{spvd} module.

\subsection{Signal Ganging}
\label{subsec:PDS-ganging}

The \dshort{pd} module \dshorts{sipm} are mechanically pressed against a %\SI{607x607}{mm} 
\dshort{wls} plate and electrically connected to a flexible Kapton(R) \dshort{pcb} where 20 \dshorts{sipm} are passively ganged to sum their electrical response. 
Summed signals from four flex \dshort{pcb}s are fed into the amplifier stage (active ganging, biased at +5~V), so a single electronic channel 
provides a readout of the combined response of 80 ganged \dshorts{sipm}; there are two such channels per \dshort{xarapu}. 
Both passive and active ganging stages are cold. This double-stage ganging solution, adopted for both cathode-mount and membrane-mount modules, represents an extension of the scheme adopted for the \dshort{sphd} \dshort{pd} electronics where eight 
boards of six \dshorts{sipm} are summed in a single-stage cold amplifier. Commercial components are used for the ganging signal conditioning circuit.

A flexible \dshort{pcb} is host to the passive ganging of 20 \dshorts{sipm} of the \dshort{spvd} \dshort{pds}. Five groups of four \dshorts{sipm} connected in parallel are connected in series, as shown in the equivalent circuit of Figure~\ref{fig:PDS_PassiveGanging}. 
%This circuit has the 
advantage of allowing all the \dshorts{sipm} to be biased at the same voltage while keeping the overall capacitance of the system low, thus ensuring a short
This circuit has the advantages of ensuring the same potential on the surface of the \dshort{sipm}s, and a small capacitance, hence a lower noise and generally faster response time.
The ganged signal is read in differential mode in order to increase the \dword{s/n} ratio.

Figure~\ref{fig:PDS_Flex_Boards} shows four flexible \dshort{pcb}s ganged to make a single \dshort{pds} channel (there are two such channels per module) that have been used successfully in the \dshort{cern} \coldbox.

\begin{dunefigure}
[Equivalent circuit for the passive ganging of 20 \dshorts{sipm}]
{fig:PDS_PassiveGanging}
{Equivalent circuit for the passive ganging of 20 \dshorts{sipm} of the \dshort{spvd} \dshort{pds}.} 
  \includegraphics[width=0.60\textwidth]{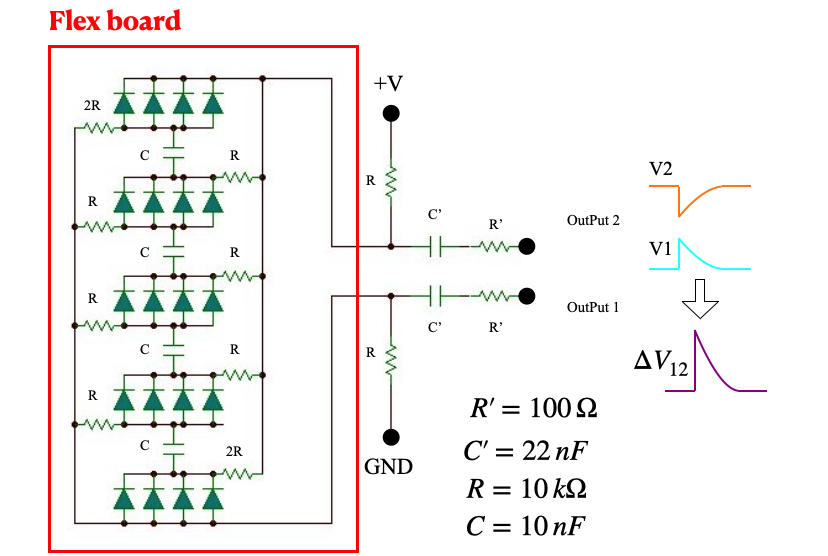} 
\end{dunefigure}

\begin{dunefigure}
[PDS SiPM flexible boards]
{fig:PDS_Flex_Boards}
{Four flexible boards with the 80 ganged \dshorts{sipm} of a single \dshort{spvd} \dshort{pds} channel. The \dshort{xarapu} light collector to which the four flexible boards are coupled is not shown in this figure.} 
  \includegraphics[width=0.75\textwidth]{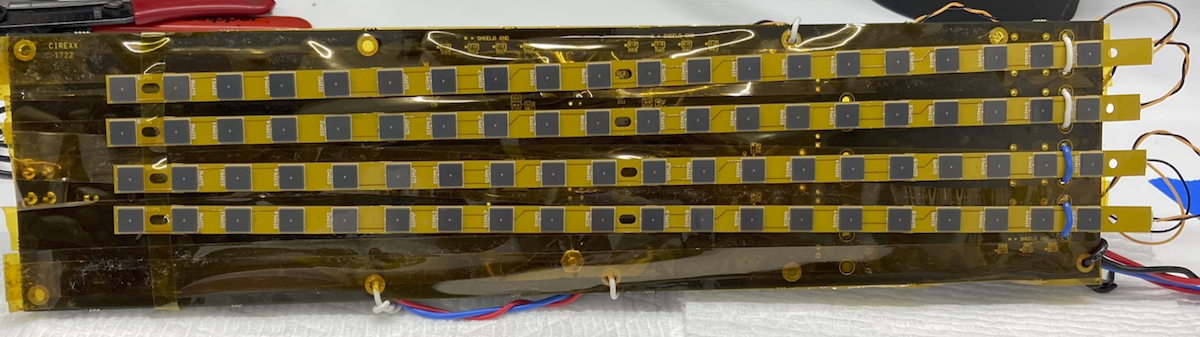}
\end{dunefigure}

\section{Sensor Biasing and Signal Readout}
\label{sec:PDS-Electronics}

\subsection{Membrane Mount Modules}
\label{sec:PDS-membrane-elec}

The  \dshort{spvd} design calls for 56,320 \dshorts{sipm} and 704 signal transmission electronics channels (two signals per light collector module) to read out membrane mount modules.

The membrane-mount \dshort{xarapu}s and associated readout electronics are at ground, so the readout solution developed for \dshort{sphd} \dshort{pds}, with power and signal transmitted over conductive cable, is a viable solution with minor changes and is adopted as the reference design. In parallel, a study is in progress to determine whether a more effective solution is to adapt the board containing the signal summing and amplifier stage developed for the cathode modules for use with the membrane system; power and signal would be over copper for the membrane modules. This would have the benefit of maximizing the similarity for the first stage signal processing of the subsystems.

The warm electronics (\dword{daphne} digitizer), and associated interface to \dword{daq}, are adopted from the existing \dshort{sphd} \dshort{pds} solution \cite{DUNE:2020txw}.

\subsection{Cathode Mount Modules}
\label{sec:PDS-cathode-elec}

The cathode-mount modules will float at the cathode \dword{hv}, requiring power and signal transmission via non-conductive cables.  Liquid immersion, low temperatures, long maintenance-free 
lifetime, and \dword{hv} isolation requirements combine for a particular challenge for the electronics.

\subsubsection{Power-over-Fiber for Active Devices on the Cathode}
\label{subsec:PoFsss}

Power-over-fiber (\dword{pof}) is a  power delivery technology that delivers electrical power by sending laser light through robust, lightweight, non-conductive optical fibers to a remote photovoltaic receiver or photovoltaic power converter (\dword{ppc}) to power remote sensors or electrical devices.  This innovative technology provides three major benefits: (1) noise immunity, (2) voltage isolation, and (3) spark-free operation.

\begin{figure}[!tbp]
  \centering
   \includegraphics[width=0.75\textwidth]{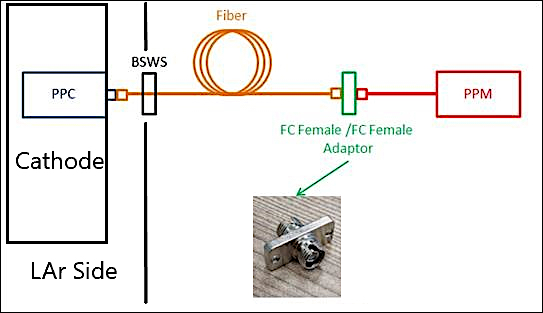}
    \caption[Block diagram of a single \dshort{pof} system]{A block diagram of a single \dfirst{pof} system with the vacuum feedthrough (\dword{bsws}) separating the \dshort{lar} and warm sides.} 
    \label{fig:PoF-BlockDiagram} 
%  \end{minipage}
\end{figure}

 The current \dshort{spvd} design consists of discrete \dshort{pof} units that include a laser transmitter module on the warm side, a fiber link, and a \dshort{pof} receiver on the \dshort{lar} side. Figure~\ref{fig:PoF-BlockDiagram} shows the basic block layout of a single \dshort{pof} system. The laser transmitter is made of photonic power modules (PPMs), which are laser modules with fiber pigtails. The fibers are passed through a bearing sensor wire compression seal (\dword{bsws}) rubber cork feedthrough. The \dshort{pof} receivers, called \dwords{ppc} or optical photovoltaic converters (OPCs), deliver power to readout electronics and bias generation circuits.

 The first silicon-based transmitter control unit (PPM module) consisted of a switching power supply, laser interlocks, controls and five Class 4, 971\,nm wavelength, laser transmitters that each had fused fibers. This compact power housing unit is shown in Figure~\ref{fig:cathode_pof} and has been fully tested. A final, improved, design of the transmitter module that includes both local system communication capability and remote control is underway. 
 An improved system will be deployed that includes eight, 2~W power capable and 808~nm wavelength, gallium arsenide (GaAs) lasers per control unit; the operational laser power target is 0.4~W. The control unit will have the same basic design, but the laser's compact footprint allows up to eight lasers per box.

\begin{dunefigure}
[Photograph of a \dshort{pof} transmitter module]
{fig:cathode_pof}
{Photograph of the fully tested \dword{pof} transmitter module showing five 971\,nm lasers.}
\includegraphics[width=0.7\textwidth]{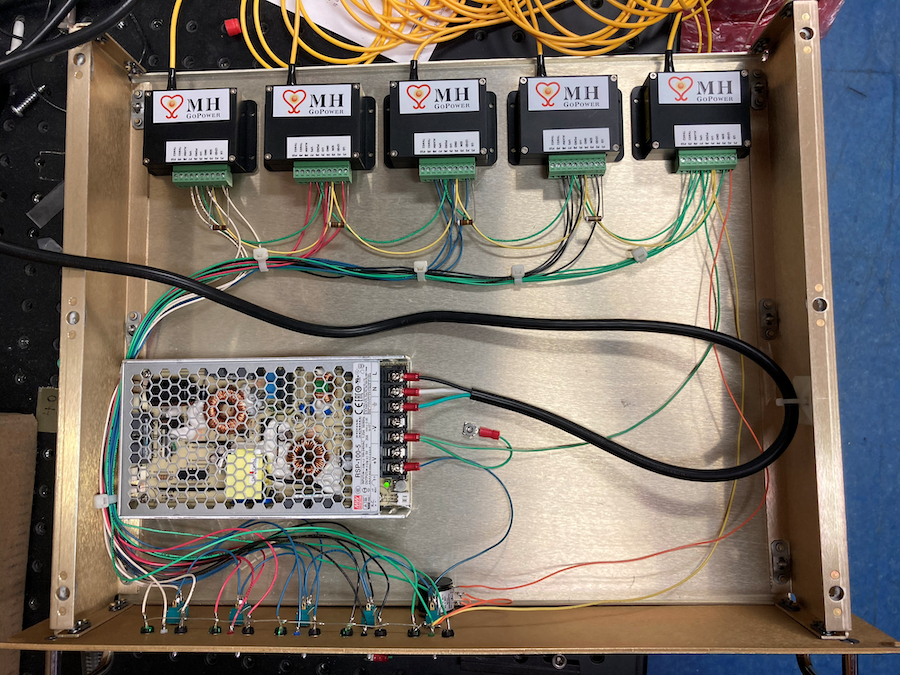}
\end{dunefigure}

The \dword{bsws} feedthrough leak rate with optical readout and \dshort{pof} fibers has been tested at \dshort{fnal} in a stand-alone test bed with silicone potting by pumping against a small sealed volume to $10^{-6}$ Torr and conducting leak tests with a leak checker. The results were satisfactory (e-10 atm-cc/sec). The \dword{bsws} flange prototype with silicone potting has also been successfully tested in the NP02 \coldbox test bed, demonstrating the ability to reach the required \dshort{lar} purity. The full-scale \dshort{spvd} \dshort{pds} flange with epoxy potting will be tested in \dword{vdmod0} allowing for final design validation well ahead of DUNE \dshort{spvd} installation.
 
 The readout electronics system voltage is 6~V and 450~mW power. To deliver the low voltage for the readout electronics, GaAs \dword{ppc} receiver units~\cite{fahrenbruch:IEEE-photov-2019} are deployed in parallel to provide sufficient current supply and redundancy. The voltage and power needs of the readout electronics must be closely monitored during \coldbox and \dshort{vdmod0} operation, as changes in power delivery requirements
 %for example to optimize redundancy or signal-to-noise, 
 could imply changes to the \dshort{pof} full system topology and a possible re-optimization for performance and cost. Compared to the silicon option, the GaAs \dword{ppc} systems have higher efficiency at \dshort{lar} temperatures at a similar, if not lower, cost. In \dshort{lar}, the commercial silicon options deliver 200~mW at 12~V, while the commercial GaAs options deliver 200~mW at 6.5~V or 400~mW at 6.5~V. Specifications of the single \dword{ppc} modules that have been considered are summarized in Table~\ref{tab:Cathode\dshort{sipm}PoFs_power}. Because of its higher efficiency, the GaAs technology is our chosen option for \dshort{spvd} \dshort{pof}. 

\begin{table}[htbp]
\caption[Power estimates for \dshort{pof}  cathode \dshort{sipm} system]
{Power estimates for \dshort{pof} cathode \dshort{sipm} systems. Numbers refer to individual PPC modules.}
\centering
\begin{center}

\begin{tabular}
{rrrcccc}
\hline
\rowtitlestyle
{\bf Del. Power} & \bf Type & \bf Wavelength & \bf Current & \bf Voltage & \bf Usable Power & \bf  Eff. \\
\rowtitlestyle
(W)$^{*}$ &  & (nm) & (mA) & (V)$^{**}$ & (W) & (\%) \\ \toprowrule
0.6 &GaAs warm & 808 & 70 & 5.5 & 0.40 & 65\\
0.4 &GaAs cold & 808 & 30 & 6.5 & 0.20 & 50 \\
\colhline
0.8 &Si warm & 971 & 70 & 5.5 & 0.40 & 50  \\
1 &Si cold & 971 & 16 & 12.0 & 0.20  & 20 \\
 \hline
\end{tabular} \\
\footnotesize
{$^*$ The power delivered is not all converted to usable power; e.g. for Si cold, the efficiency is about 20\% in LAr. \\
$^{**}$ Each PPC module voltage can vary about 3\%. }
\label{tab:Cathode\dshort{sipm}PoFs_power}
\end{center}
\end{table}

\begin{figure}[!tbp]
  \centering
%  \begin{minipage}[b]{0.42\textwidth}
\includegraphics[width=0.75\textwidth]{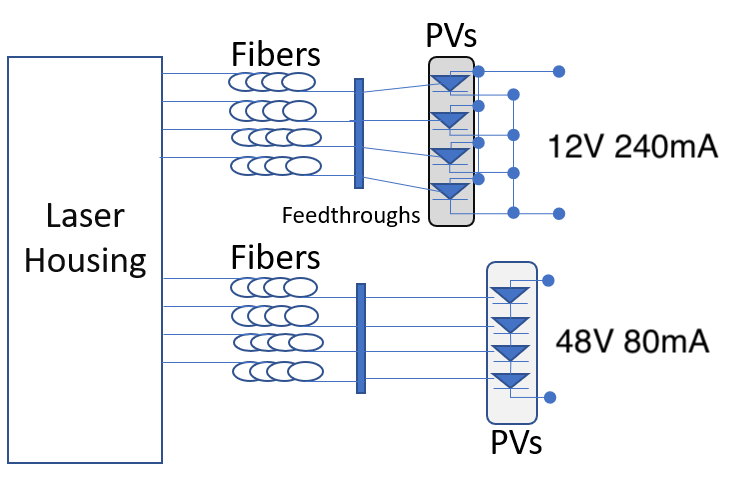}
    \caption[Basic block layouts for \dshort{pof} power systems]
    {\dshort{pof} receiver options. The photovoltaic (PV) components are housed in a small metal \dshort{fc}-style %connectorized 
    unit called a \dword{ppc} or OPC (optical photovoltaic converter). The series and parallel connections shown are for illustrative purposes only, assuming each unit is 12\,V/80\,mA. The actual number of photovoltaic components used will depend upon the semiconductor choice. For example, GaAs \dwords{ppc} have lower voltage but higher current compared to silicon-based ones. }
    \label{fig:PoF-BatteryBlockDiagram}
%  \end{minipage}
\end{figure}

 All \dword{ppc} systems can be assembled in parallel or series sets, like batteries, in combination with Zener diodes, to achieve the required voltage and current. Figure~\ref{fig:PoF-BatteryBlockDiagram} shows the basic block layouts of arranging \dshort{pof} power systems. The \dshort{pds} cathode-mounted module readout electronics motherboard, DUNE Cold Electronics Motherboard (DCEM), as shown in Figure~\ref{fig:pds_dcem}, has been designed to accommodate the \dword{ppc} units on G10 standoffs, with a heat sink with through-hole vias to provide power connection to the signal conditioning and bias generation circuitry. Up to four \dwords{ppc} can be established in parallel on the motherboard \dshort{pcb}. The parallel mounting of three GaAs \dwords{ppc} on the readout electronics motherboard is shown in Figure~\ref{fig:pds_dcem}.
 
\begin{dunefigure}
[DUNE cold electronics motherboard for the \dshort{pds}]
{fig:pds_dcem}
{DUNE Cold Electronics Motherboard (DCEM) for the \dshort{spvd} \dshort{pds}. The labels indicate the main components. The board has been used in \coldbox tests at CERN.}
  \includegraphics[width=\textwidth]{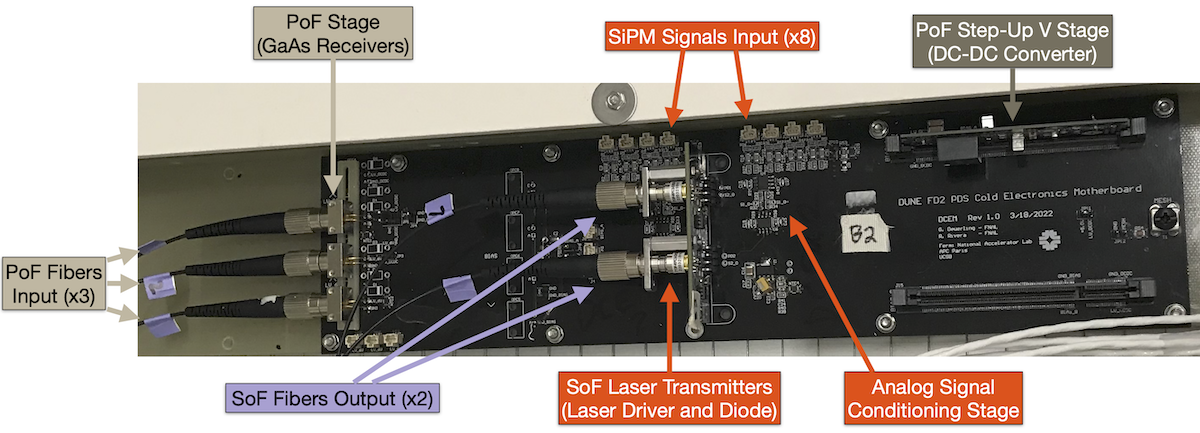}
\end{dunefigure}

While sensor biasing and signal readout electronics will operate in cold, validation in warm conditions after installation is a requirement of the system, see Section~\ref{sec:fdsp-pd-installqc}. 
 The \dshort{sipm} bias is established with a DC-to-DC step-up converter to generate a bias voltage derived from the low voltage delivered by the redundant, parallel GaAs \dword{ppc} units. The step-up converter is designed as a daughtercard to the readout electronics motherboard. The step-up converter can be tuned separately to the desired bias voltage for the \dshorts{sipm} which is 20~V to 50~V in \dshort{lar} depending on the \dshort{sipm} type, see Section~\ref{subsec:PDS-sipm}. A critical constraint for \dshort{sipm} bias comes from the risk of high voltage discharge by the cathode; that is, the collapsing of the -300~kV on the cathode, which could occur due to arcing at the HV feedthrough. If not for this risk, \dshorts{sipm} bias voltage could be shared among multiple \dshort{xarapu}s, thus reducing the cost and complexity of the \dshort{spvd} power distribution. 

 However, as discussed in Section~\ref{sec:PDS-LightColl-discharge}, a conductive distributed power system elevates the risk of damage to the readout electronics to unacceptable levels. Although a well-designed electrical shielding should provide sufficient protection, the safest option is to power each \dshort{xarapu} independently in addition to shielding. 
 To optimize costs, generating the bias through the use of \dword{ppc} units in series was abandoned in favor of local bias generation using DC-to-DC step-up conversion.

A concern for the use of \dshort{pof} in the vicinity of very sensitive photodetectors is related to photons leaking from the high intensity system.
%system, which could stimulate signals from the \dshorts{sipm}, 
The selected black-jacketed fibers\footnote{MH GoPower 62.5~$\mu$m core diameter, 1.5mm OD, 40~m length, black PTFE coating.} were found to reduce the leakage of IR photons to undetectable levels using high-sensitivity thermal cameras. The \dshort{pof} fibers are also routed through dedicated black polytetrafluoroethylene (PTFE) tubing (sloped and periodically slit to allow trapped gas to escape) and silicon potting may be employed at the terminations, providing further optical isolation.

%\fixme{DWW> Added text to describe tube venting}

 Tests at the \dshort{cern} \coldbox have shown that black-jacketed fibers and potted terminations are successful at preventing light leakage. 
 %This aspect of the system will be verified during 
 \dshort{vdmod0} operation will confirm that the power delivery does not introduce an unacceptable background. More than one type of conduit, tubing, or braided sleeving will be deployed at \dshort{vdmod0} to compare background measurements for different solutions.

\subsubsection{%Cathode Mount Modules 
Signal Transmission}

Analog optical transmission of the signal has been selected as the reference design based on numerous factors\footnote{A digital optical transmission was investigated but could not be validated on the appropriate timescale.}. A similar parallel development by the DarkSide experiment showing promising results served as proof that analog transmission in \dshort{lar} is possible, and therefore achievable within the constrained timescale of the development. The simplicity of design reduces risk associated with long-term cold qualification and affords a straightforward approach to \dshort{pd} module redundancy by increasing the number of analog transmitter channels per \dshort{pd} module.

As shown in Figure~\ref{fig:analog_transmission_schematics}, the analog signals on the cathode are transferred, via a cold optical transceiver, over optical fiber to the external \dshort{daphne} warm electronics for digitization and recording (the same system as the membrane modules). The \dshort{spvd} topology calls for 51,200 \dshorts{sipm} and 640 signal transmission electronics (two signals per light collector module) on the cathode.

\begin{dunefigure}
[Readout electronics for cathode-mount PD modules]
{fig:analog_transmission_schematics}
{Readout electronics for cathode-mount \dshort{pd} modules -
optical transmission of the analog signals to digitizers outside the \dshort{lar} cryostat. 
The green box represents the cold readout components for one analog readout channel. The differential input on the left of the figure corresponds to 80 ganged \dshorts{sipm}, hence two analog optical transmitters and two receivers per \dshort{pd} module are required.}
  \includegraphics[width=1.0\textwidth]{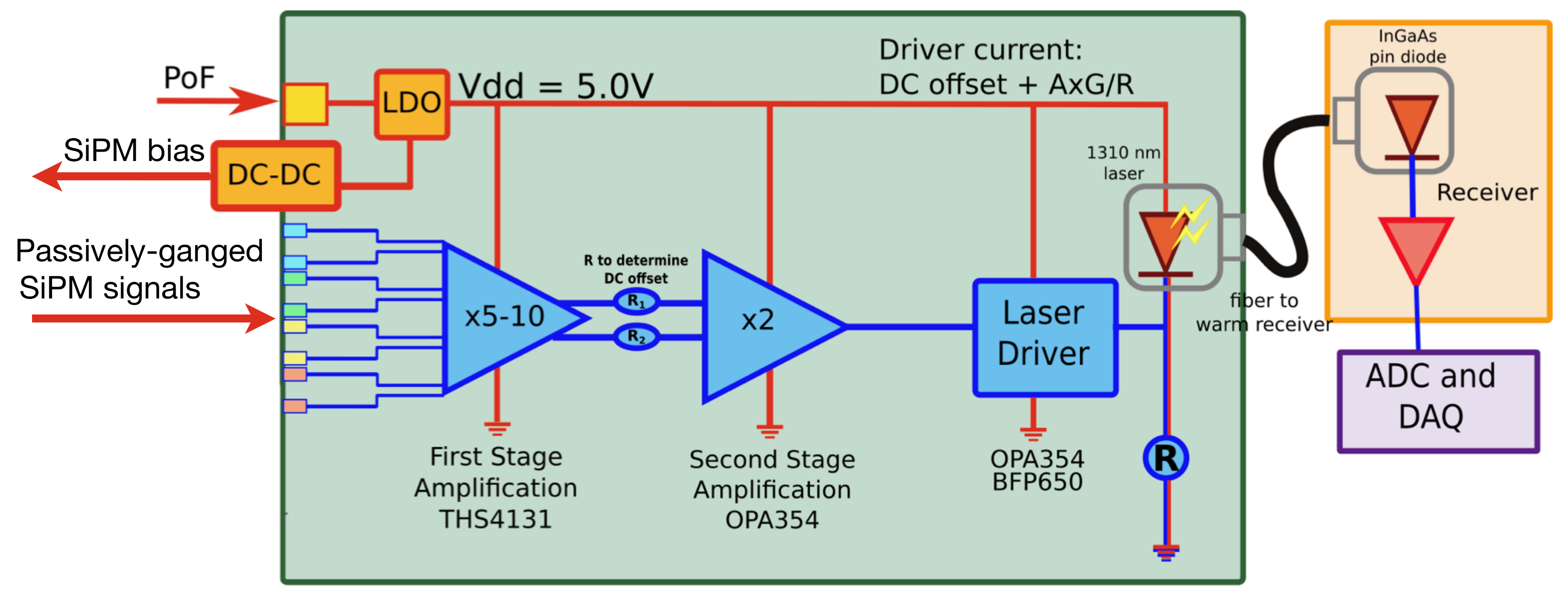}
\end{dunefigure}

Commercial transceivers are operated and certified only for temperatures greater than 233\,K. Converting electrical signals to optical signals at \dshort{lar} temperatures is thus recognized as a critical aspect of the readout of the \dshort{pds}. Analog optical transceivers have been developed to operate at such temperatures previously~\cite{atlaslar:1999}. In particular, the DarkSide Collaboration has developed a prototype transmitter for their experiment that shows satisfactory results \cite{Consiglio:2020fgk}. 

An analog optical transceiver has been developed for \dshort{spvd}. These transceivers are based upon commercially available discrete analog components, and the light source can be either a \dshort{led} or a laser. A laser is used in the \dshort{spvd} \dshort{pds}, owing to its higher optical output efficiency in cold. Compared to DarkSide, the \dshort{spvd} optical driver has no major radiopurity constraints but must serve a detector with a larger dynamic range. 
The design aims to reduce risks by implementing a minimum topology of discrete components that are known to have adequate behavior in cold, while attempting to match as well as possible the dynamic range of the signals coming from the \dshort{xarapu} and minimizing power consumption.

The signals of each analog channel (80 \dshorts{sipm}) are first amplified (signal shaping stage), then converted to a current signal (laser driver). Both stages are based on high-bandwidth operational amplifiers. The bandwidth of the transmitter circuit is one of the key aspects for properly transmitting the ganged signal of the \dshorts{sipm}. With a rise time of around 50\,ns, a bandwidth of at least 20\,MHz is desirable in order to transmit the signal without significant deformations. In addition, the expected signal level for the single \phel{} is about 30-50\,$\mu$V and hence requires amplification on the transmitter board. An appropriate balance between gain and bandwidth must be reached, given that a configuration with larger gain reduces the bandwidth capabilities of the circuit components.
%\fixme{SS Edits 06mar23: end} 

Although it has been found that operation in cold reduces the performance of the driver with respect to 
operation at room temperature, an adequate bandwidth of 50\,MHz in cold has already been achieved. The circuit drives current through an \dword{eelaser} coupled to an optical fiber. Edge-emitting lasers show satisfactory behavior, with a threshold lasing current reduced to only a few milliamps in cold and with good linearity.

The linearity of the electrical-to-optical conversion and transmission for the analog signals in the dynamic range of expected signals is a crucial feature of the \dshort{sof} system, and has been tested in detail with satisfactory results. Commercially available Fabry-P\'erot laser diodes have been shown to work well as an optical signal source at cryogenic temperatures, with a lower threshold current in cold and the same efficiency as in warm across its range of operation. The custom high bandwidth laser driver with fixed DC offset, to set the working point of the laser diode just above its threshold current in cold, enables the laser to efficiently operate in linear regime. The response of a sample of Fabry-P\'erot laser diodes during acceptance tests in cold is shown in Figure~\ref{fig:pds_sof_linearity}.

\begin{dunefigure}
[Linearity of response for the \dshort{pds} \dshort{sof} laser diodes.]
{fig:pds_sof_linearity}
{Linearity of response for a sample of six Lasermate Fabry–P\'erot laser diodes from acceptance tests in deep \dshort{lar} at \dword{fnal} \dword{pab}. The optical power transmitted through the fiber is shown as a function of excitation input current (2.1~mA threshold). This test is for 500\,$\micro$W devices over 40\,m of 62.5\,$\micro$m diameter multimode fibers, and read out via an optical power meter.}
\includegraphics[width=0.65\textwidth]{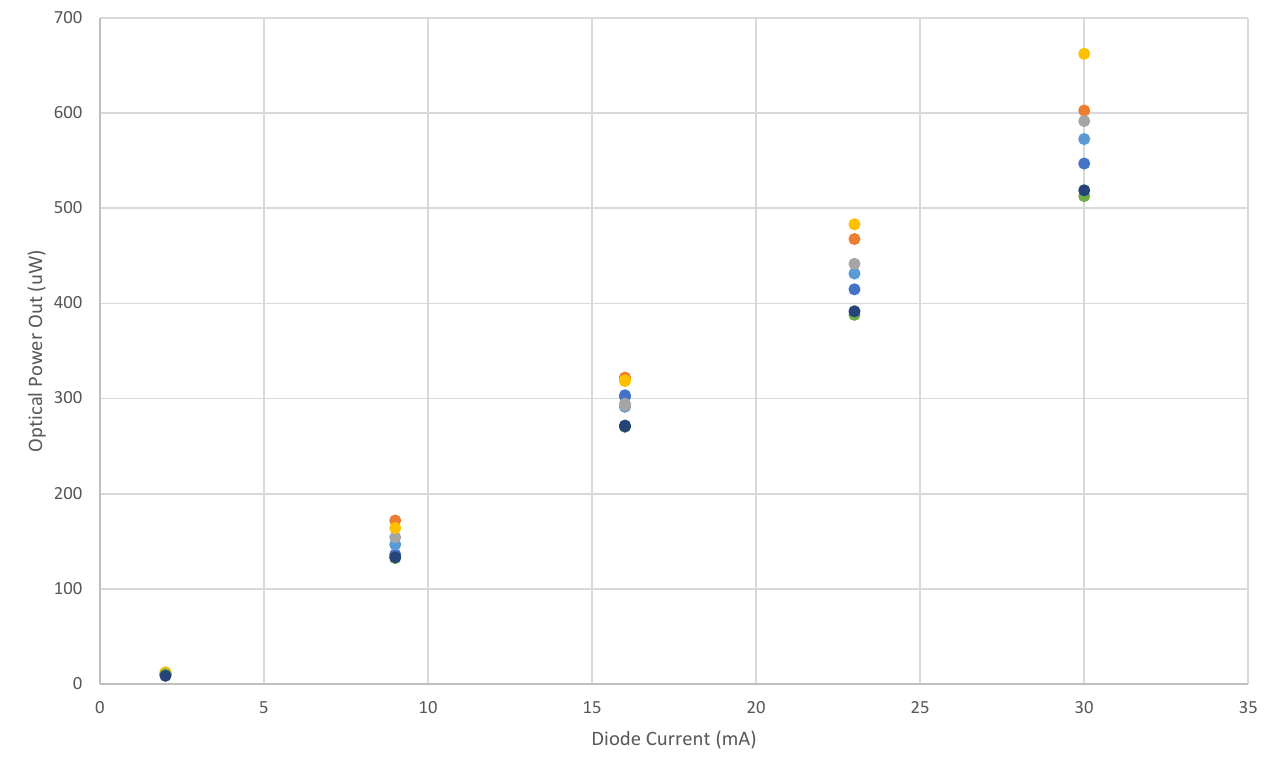} 
\end{dunefigure}
%\fixme{SS 10/03/2023 end}

An annotated photograph of the DCEM %for the \dshort{spvd} \dshort{pds} 
is shown in Figure~\ref{fig:pds_dcem}. One such board per \dshort{xarapu} will be used. As shown by the labels in the figure, this single integrated board operating at \dshort{lar} temperatures includes circuitry for \dshort{pof}, %power-over-fiber, 
analog signal amplification and conditioning, delivery of \dshort{sof} %signal-over-fiber 
and \dshort{sipm} bias.

Power dissipation from the DCEM is within the specification from the \dword{hv} group that limits the heat load to less than 1\,W/cm$^2$ in order to avoid bubbles. We estimate a power dissipation of 400\,mW across the two, 50\% efficient, \dword{ppc} modules needed per \dshort{xarapu} \dshort{pof}. The local heat load from the \dword{ppc} modules does not exceed 200\,mW/cm$^2$ at any location, and is hence within specifications. The DCEM board itself uses an additional 400\,mW of power, yielding a maximum local  heat load of 130\,mW/cm$2$, also within specifications. Total power consumption across the entire cathode mount \dshort{xarapu} system is about 300\,W.

At the other end of the signal fibers and outside the cryostat, the analog optical receivers will operate at room temperature. The photodiodes to be used are \dword{ingaas}, since they have the optimal sensitivity to the 1310\,nm light being used in the transmitter on the DCEM. The radius of the photosensor should be well adapted to the fiber core diameter; currently 300\,$\mu$m is planned, since it would be appropriate for any of the fiber core diameters used, and it is a commercially available standard size. These devices have a responsivity\footnote{Output signal (typically voltage or current) of the detector produced in response to a given incident radiant power falling on the detector.} power-to-current conversion that depends on the incident light wavelength; for 1310\,nm light this is close to 0.8\,A/W.

The photodiode is followed by a fast, low-noise amplification circuit. A carefully selected high-gain transimpedance operational amplifier is the basis of this stage. The gain should be such that the smallest signals expected from the transmitter, the single \phel{}s (SPE), are correctly digitized by the ADC that follows. %following after. 
This is a 14-bit ADC currently used in the DAPHNE digital electronics board, with a 1~V maximum input. The goal is to adapt the gain of the entire transmitted/receiver path to achieve both a good S/N in SPE signals and the readout of large signals. The analog signal targets a 1-to-2000 \phel dynamic range, the upper bound being dictated by %\dshort{pds}-based energy reconstruction of beam events. 
the range of signals expected in beam-induced neutrino events.

Simulations have been performed to determine the occurrence of ADC saturation events for such a dynamic range. ADC saturation within specifications was achieved, see Section~\ref{subsec:PDS-Req-scope}. Knowing the DC level of the transmitter signal is important, since it allows for checking that the transmitter circuit is working and functioning correctly. For this reason, DC coupling in the receiver and digitizer stage is desirable.

A commercial solution for the analog signal receiver, the \href{https://www.koheron.com/photonics/pd100-photodetection?ref=PD100-DC}{Koheron PD100}, has been in use since the first prototype.
Despite its excellent noise performance and robustness, its gain  
is too large to be used in combination with an \dword{adc} like that in \dshort{daphne}, and %. In addition, due to this high gain configuration, it has a fairly low 
its power input saturation point of 600~$\mu$W is too low. An in-house receiver is being developed that will allow the receiver stage to be fully configurable and adapted to the specific needs of this application. It will also enable a cost reduction in the final construction, and will better integrate with \dshort{daphne}. %We will install our 
The final design will be installed in \dshort{vdmod0}, final optimizations will be made based on 
%30-year lifetime studies and 
the installation experience during summer 2023.
Since the receiver is external to the cryostat, it can be optimized even at a later time.

\subsubsection{%Cathode Module 
Power and Signal Fiber Routing}
\label{sec:PDS-CathodeModule-FiberRouting}

Figure~\ref{fig:pds-fiber-routing-detail-01} provides a \threed rendering of the routing of the \dshort{pds} signal and power optical fibers for the cathode mount modules. From the module, they run internally to the cathode grid over to the edge next to the \dshort{fc} vertical supports. From there, they run down to the cryostat floor into the cables trays provided for the bottom \dword{crp} detector electronics. These run up to
the \dshort{pds} flanges (Figure~\ref{fig:pds_flange_8c6}) on 32 of the penetrations that also carry feedthroughs for 4 CAT-6 cables, each of which serves 2 membrane-mounted PD modules, while the remaining 8 flanges will each carry 8 CAT-6 feedthroughs.

\begin{dunefigure}
[Cathode module optical fiber routing]
{fig:pds-fiber-routing-detail-01}
{Cathode module optical fiber routing: Internal to cathode plane to the edge (top left); from the cathode plane edge to the \dshort{fc} vertical supports then down to the floor (top right); (bottom left) transfer into bottom detector electronics cables conduit on the membrane walls; up the membrane walls and over to flanges on the cryostat top.}
  \includegraphics[width=0.45\textwidth, angle=0]{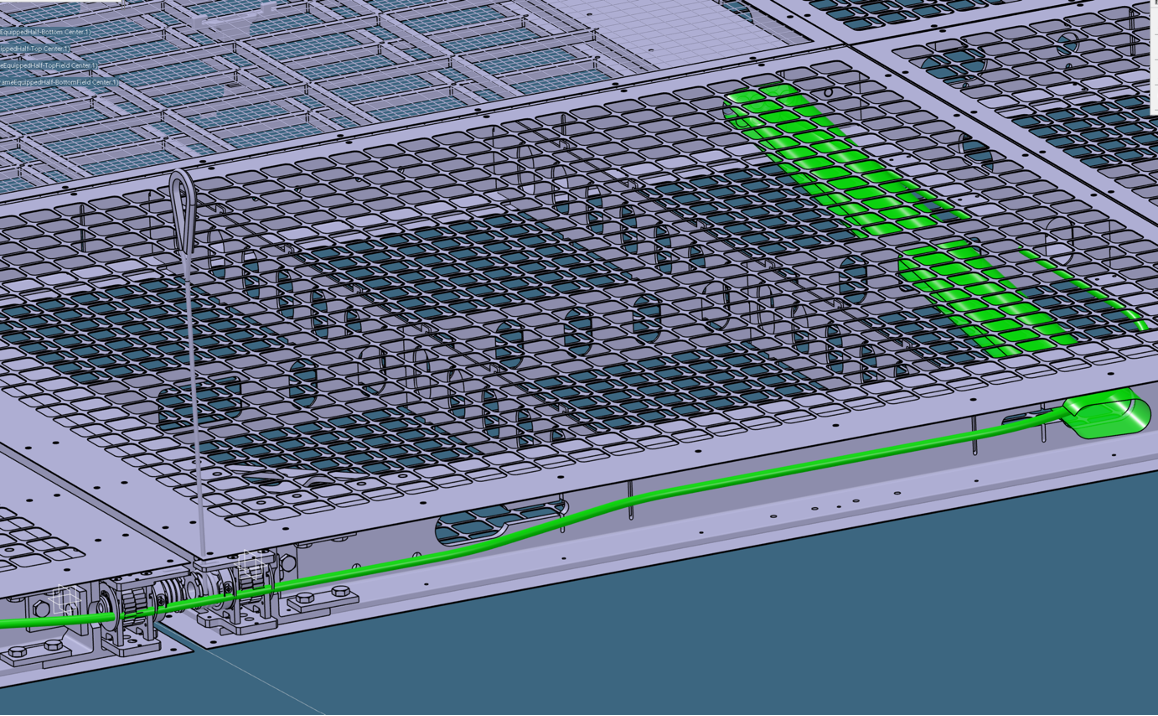}
  \includegraphics[width=0.51\textwidth, angle=0]{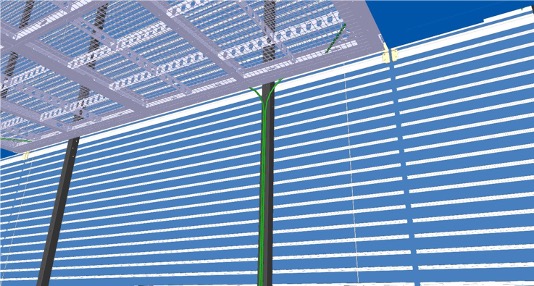}
  \includegraphics[width=0.49\textwidth, angle=0]{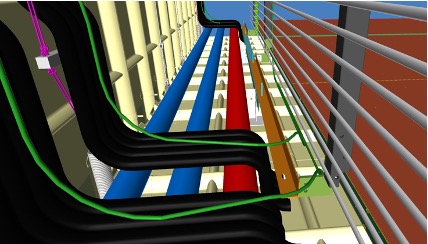}
  \includegraphics[width=0.24\textwidth, angle=0]{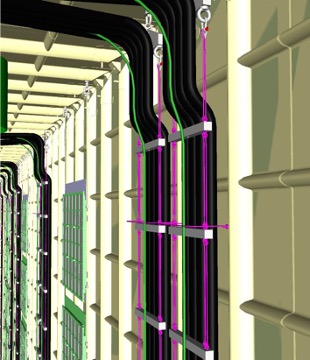}
\end{dunefigure}

\begin{dunefigure}
[\dshort{pds} flange with 8 cat6 feedthroughs]
{fig:pds_flange_8c6}
{\dshort{pds} flange with 3 feedthroughs for light response monitoring fibers, (top-left), 8 feedthroughs for fibers for cathode-mounted \dshort{pds} modules (left), and  feedthroughs for CAT-6 cables (right).  Of the 40 flanges, 32 will have 4 CAT-6 feedthroughs, with the remaining 8 flanges having 8 CAT-6 feedthroughs as shown.}
\includegraphics[width=0.4\textwidth]{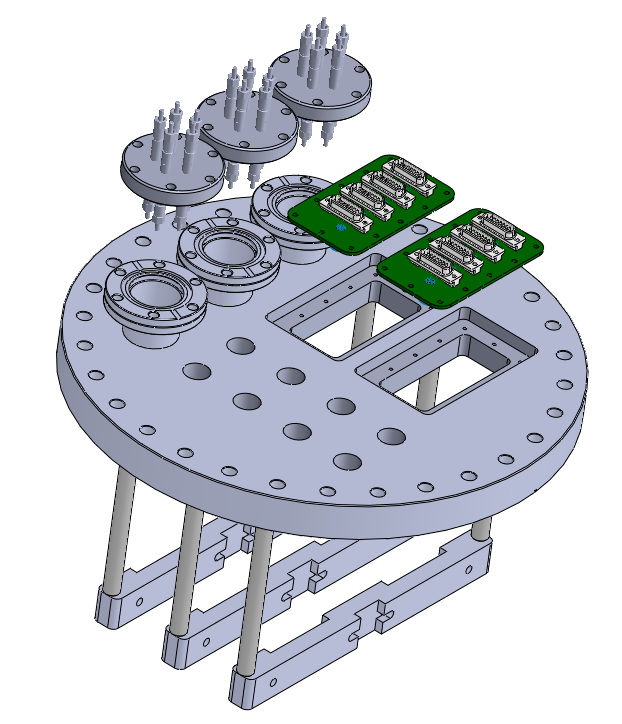} 
\hspace{2pt}
\end{dunefigure}
%was FD2_PDS_Flange8Cat6.png (my perl script didn't pick it up (AH 9/21/23) 

%%%%%%%%%%%%%%%%%%%%%%%%%%%%%%%%%%%%%%%%%%%%%%%%%%%%%%%%%%%%

\section{Light Response Monitoring and \dshort{sipm} Calibration}
\label{sec:PDS_Calib_Mon}

The \dshort{pds} will incorporate a pulsed UV-light system to calibrate and monitor \dshort{pd} response over time. The calibration system produces UV light flashes with a variable pulse amplitude, pulse
width, repetition rate, and pulse duration. The calibration data recorded by \dshort{pds} will be used to characterize and calibrate the \dshort{pd} gain, crosstalk, time resolution, channel-to-channel timing, and PDS stability over time. Examples of potential time instabilities to be monitored include dissolution of \dword{ptp} coatings over time, or drifts in the power of the readout electronics lasers.

The system design, which has both warm and cold components, is very similar to that designed for the \dshort{sphd} \dshort{pds} and operated in \dword{pdsp}~\cite{Abi:2020mwi} where all the primary components of the system have been validated. 
A significant difference between \dshort{pdsp} and \dshort{spvd} is the typical distance between the point-like light source locations in the cold volume and photon detectors. 
However, illumination of a large area ($6\times6$~m$^2$) of PMTs from a fiber placed at the top of the \dshort{fc} in \dword{pddp} demonstrated that there will be sufficient intensity for these longer distances.

Based on \dshort{pdsp} prototyping and operational experience, the system hardware for \dshort{spvd} calibration and monitoring system consists of both warm and cold components.

Fibers with diffuse fiber-end points mounted on the 
support structure around \dwords{crp}' perimeter above the \dshort{fc} strips %(will add figure here!) 
and at locations behind the \dshort{fc} strips will serve as point-like light sources, used to illuminate the \dshorts{pd} mounted at the cathode and on the cryostat membrane walls. % behind the \dshort{fc}. 
Quartz fibers transport light from the optical feedthrough (at the cryostat top) to the diffuse fiber-end points. In this configuration the light emission points are about 6.5\,m away from the cathode surface.
%located conveniently against \dshorts{pd} that observe the calibration light. 
Warm components of the system include electronics boards with controlled pulsed-UV source (currently 275\,nm and/or 365\,nm) and warm optics. Cold and warm components are interfaced through an optical feed-through, optimized for the number of calibration fibers and for the size of the cryostat flange. 
Figure~\ref{fig:pds_flange_8c6} %\ref{fig:pds_feedthrough_flange} 
illustrates the top, horizontal flange,  which provides eight fiber feedthroughs each serving up to eight \dshort{pof} and \dshort{sof} fibers for one cathode-mount PD module, three optical feedthroughs serving up to 15 Response Monitoring System diffusers.

Most of the design of the \dshort{sphd} calibration system will be reused for \dshort{spvd}. The primary differences with respect to the \dshort{pdsp} system are the number and location of the fibers and its diffuse end-points, the lengths of the optical fibers, and the addition of a monitoring photodiode within the \dshort{led} light source. 
The diffuse fiber end-points will be Figure~\ref{fig:concept_vd_UV_calib}. 
attached at support structure around the \dshort{crp} perimeter at locations indicated in Figure~\ref{fig:concept_vd_UV_calib}. Each of these point-like diffuse sources will illuminate the cathode and membrane-mount \dshort{xarapu}s using pairs of fibers. In each pair, one fiber will be pointing inward toward the cathode-mount \dshort{pd} modules and one fiber will be pointing outward toward the membrane-mount modules. 
The calibration system will have 24 pairs  close to the top \dshorts{crp}, and another 24 pairs closer to the bottom \dshorts{crp}. These fibers run to the flanges at the top of the cryostat, where the electronics modules with light sources will be located. % on the cryostat top.

\begin{dunefigure}
[Location of UV calibration light sources on the anode plane.]
{fig:concept_vd_UV_calib}
{Location of the calibration system %point-like UV light sources produced by 
optical fibers with diffusers at the ends. They are  installed along the perimeter of the \dshort{crp} (viewed from above) and illuminate the \dshorts{pd} mounted on the cathode and cryostat walls. }
\includegraphics[width=1.00\textwidth]{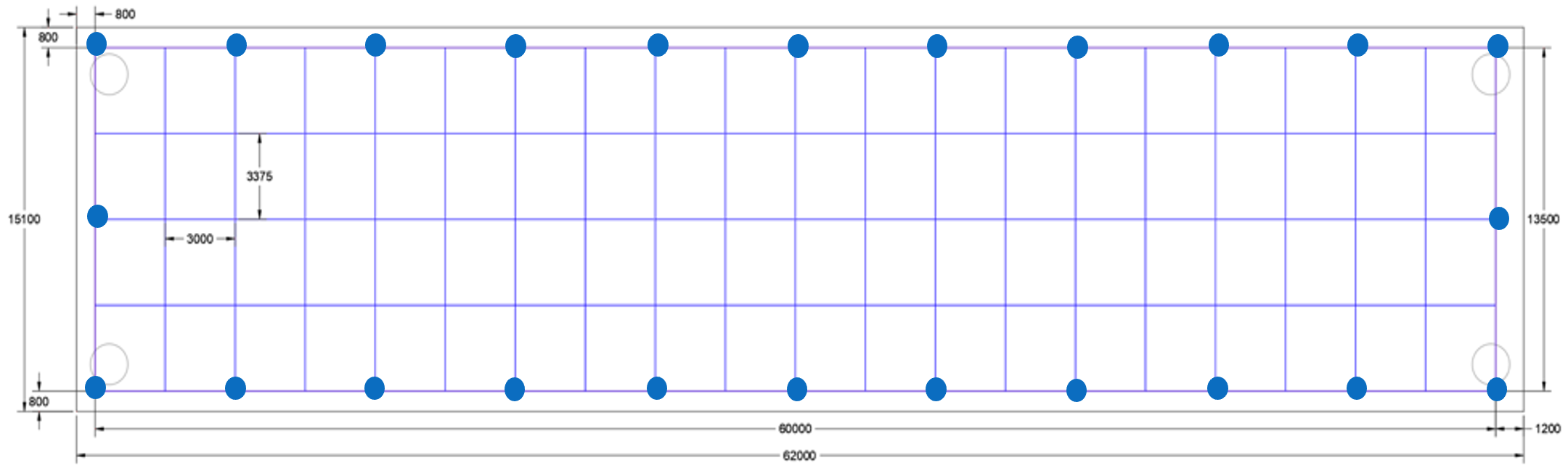}
\end{dunefigure}

The optical fiber, fiber routing scheme and optical feedthrough components, as well as light source electronics, have been designed for \dshort{sphd} and will be tested in \dword{hdmod0}. %the phase II of \dshort{pdsp}.
For \dshort{spvd}, the diffuse light point design and distribution will be optimized for light coverage of \dshort{pd} units located across the cathode and on the cryostat membrane walls. % behind the \dshort{fc}. 
Design options include %a possibility to 
use of a bare fiber end as a diffuse light point, and/or  installation of a small quartz light diffuser at the fiber end to provide a more uniform illumination (see Figure~\ref{fig:pds_diffuser}). 

\begin{dunefigure}
[\dshort{pds} diffuser and bare fiber]
{fig:pds_diffuser}
{Diffuser (6.0\,mm diameter) attached to bare fiber (0.7\,mm diameter).}
\includegraphics[width=0.40\textwidth]{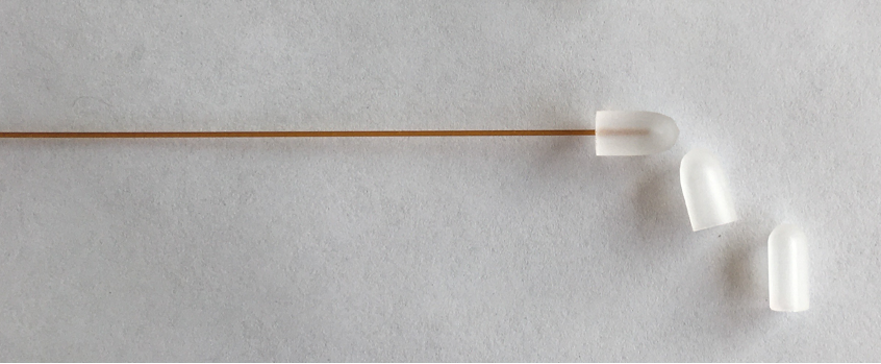}
\end{dunefigure}

The %ProtoDUNE-VD 
\dshort{vdmod0} %(also called \dword{pd2vd}) 
will be used to validate the design and evaluate the performance of the calibration %is used to prototype and qualify the calibration 
system for \dshort{spvd}. The \dshort{vdmod0} system provides ten calibration fiber-end diffuse ends. 
Calibration light is diffused from six light emission points attached to upper \dwords{crp} support structure with four diffuse-end points to illuminate cathode \dshorts{pd}, and two diffuse-end points to illuminate membrane wall \dshorts{pd}.
There are four light emission points attached to lower \dwords{crp} support structure with two diffuse-end points to illuminate cathode \dshorts{pd}, and two diffuse-end points to illuminate membrane wall \dshorts{pd}. A test will be performed to determine if the light emitting points at fibers-end may diffuse light via either bare fiber-end or with the designed diffuser unit attached to it as shown in Figure~\ref{fig:pds_diffuser}. One of the top fiber-end emission points will be instrumented with the diffuser unit.

Multiple quartz fiber types will be tested for the light transport and diffusion characteristics in \dshort{vdmod0}: two types of fibers with different \dword{solarization} resistance  of the quartz core (200 and 400\,$\mu$m core diameter), and the third type already selected for
\dshort{hdmod0} (600\,$\mu$m core diameter, which is expected to have a negligible solarization effect). 
Two fiber jacketing options are considered depending on fiber installation route constraints: PTFE jacket, or stainless-steel jacket.

The system has no active components within the cryostat. The active system component consists of a 1U rack-mount light calibration module (LCM) at warm temperature. The LCM electronics system generates light pulses that propagate through a quartz fiber-optic cable to the diffusive fiber-end to distribute the light across the PDs. The current calibration module design consists of a field programmable gate array (FPGA) based control logic unit coupled to an internal LED pulser module (LPM) and an additional bulk power supply. 

The LPM uses multiple digital outputs from the control board to control the pulse amplitude, pulse multiplicity, repetition rate, and pulse duration. Analog-to-digital converter (ADC) channels internal to the LCM are used to read out a reference photodiode used for pulse-by-pulse monitoring of the LED light output. The output of the monitoring diode is available for monitoring and for normalizing the response of the \dshorts{sipm} in the detector to the calibration pulse. 
Hardware components for the far detector (FD) modules will be costed based on the prototyping cost and expertise gained with \dshort{vdmod0}.

Complementary to the monitoring system, a calibration of the absolute energy scale as a function of position within the detector will be pursued. The goal is to derive the position-dependent \dword{ly} in the detector, in units of detected photoelectrons per unit energy deposit, from the data themselves. Several options are under consideration, including cosmic ray tracks, radioactive sources (at fixed positions or diffused into the \dshort{lar} volume), pulsed neutron generators, and ionization lasers.

\section{Design Validation}
\label{sec:PDS_validation}

This section summarizes the results of the R\&D and prototyping plan carried out for the \dshort{spvd} \dshort{pds} following the Conceptual Design Report and the on-going design validation program.

%%%%%%%%%%%%%%%%%%%%%%%%%%%%%%%%%%%%%%%%%%%%%%%%%%%%%

\subsection{Xenon doping in \dshort{pdsp}}
\label{subsec:PDS-xenon-protodune}

A xenon-doping campaign was carried out in \dshort{pdsp} a few months after an incident with a gas recirculation pump had let an unknown amount of air inside the detector volume.  Oxygen and water were efficiently  removed by the purification loop, but nitrogen cannot be removed by the purification filters.
A residue of nitrogen ($\sim5.4$~ppm in mass) remained in the \dshort{lar} with the result that the scintillation light was effectively quenched. 
Although it was not originally foreseen to do xenon testing in the presence of nitrogen contamination, its presence presented the opportunity to demonstrate light recovery due to xenon doping and a large scale. 

The doping run consisted in doping argon with up to 18.8~ppm of xenon, in steps. The amount of detected light increased as a function of concentration up to around 16~ppm, after which the gain flattened. This trend was consistent across the three types of light detectors in \dshort{pdsp} \dshort{pds}. The results were corroborated by measurements taken with two prototype \dshort{xarapu} detectors inserted for this run in the non-active TPC region behind one of the APAs. %APA6. 
A paper on this work is in internal review. 

Initial small-scale tests and the procedure used for \dshort{pdsp} demonstrated that the doping of xenon and the mixing with argon must happen in gas phase, before condensation of the fluid. Indeed, since xenon liquefies at 165~K and solidifies at 161~K, creating a solution with liquid argon must be performed with extreme care, in order to avoid its freezing.
Several mixing ratios were tested, showing that the Ar/Xe ratio must be above $10^3$, to avoid xenon solidification on the walls of the argon condenser. This \textit{freeze-out} effect can be easily observed since, at the highest xenon concentrations, the pipes of the condenser get clogged and the argon recirculation stops.

For \dshort{pdsp}, the xenon injection point was placed on the gas recirculation system; on the line collecting the chimney boil-off, after the argon gas purification filter but some distance before the condenser. This allows for full mixing within the gas flow. The maximum xenon mass flow rate was set to 36~g/h, to be well within the Ar/Xe ratio limit mentioned above; this corresponds to 50~ppb/hour in the \dshort{pdsp} detector. Based on a numerical \dword{cfd} simulation of the LAr flow within the ProtoDUNE cryostat, the xenon injected at this rate is expected to be uniformly distributed in LAr within few hours.

Here we summarize the main results of the successful run:
\begin{itemize}
    \item Since one of the two \dshort{xarapu} detectors was fitted with a quartz window, which makes it insensitive to 128~nm photons, it is possible to disentangle the contribution of the xenon photons alone from the total emitted light.  Comparison of the signal from the two detectors clearly shows that the amount of light above the quartz window cut-off, ``xenon light'', increases with xenon concentration. The ratio of the xenon-to-total argon light collected by the two detectors, as a function of xenon concentration, is shown in Figure~\ref{fig:PD-xarapuca-pdsp-ratio}.
    \item The data shows that the effect of xenon is simlar across the several detector technologies which are at different locations in \dshort{pdsp}.  This indicates good uniformity of the doping throughout the TPC volume. Time stability of the mixture is observed at the level of few weeks after the last doping event (afterwards, the run was stopped). 
    \item Light attenuation curves show that the amount of detected light increases more further from the detectors, demonstrating the usefulness of the doping for very large volume TPCs (Figure~\ref{fig:06_LightRecoveryAra}). This is  even more evident in a subsequent xenon-doping campaign in the \dword{pddp}, which features a larger drift distance~\cite{DUNE:2022ctp}.
    \item In the \dshort{pdsp} configuration, the increase of light with xenon concentration starts flattening out at around 16.6~ppm.
    \item No detectable deterioration of the charge collection performance, i.e., of the imaging capability of the TPC, was observed during the doping campaign.
\end{itemize}

 \begin{dunefigure}
[Xenon light sensitive \dshort{xarapu} signal in \dshort{pdsp} following xenon injections]
{fig:PD-xarapuca-pdsp-ratio}
{Fraction of light collected by the xenon light-only sensitive X-ARAPUCA in \dshort{pdsp}, with respect to total light collected by the other device: $\frac{\text{Xe}}{\text{Ar+Xe}}$. The ratio increases with the doping concentration and reaches a plateau around 0.65 for xenon concentration greater than \SI{16.1}{ppm}. The red points correspond to data collected with the nominal TPC electric field (500~V/cm), while black points refer to data with no electric field. Shaded areas indicate xenon injection periods.}
  \includegraphics[width=0.95\textwidth]{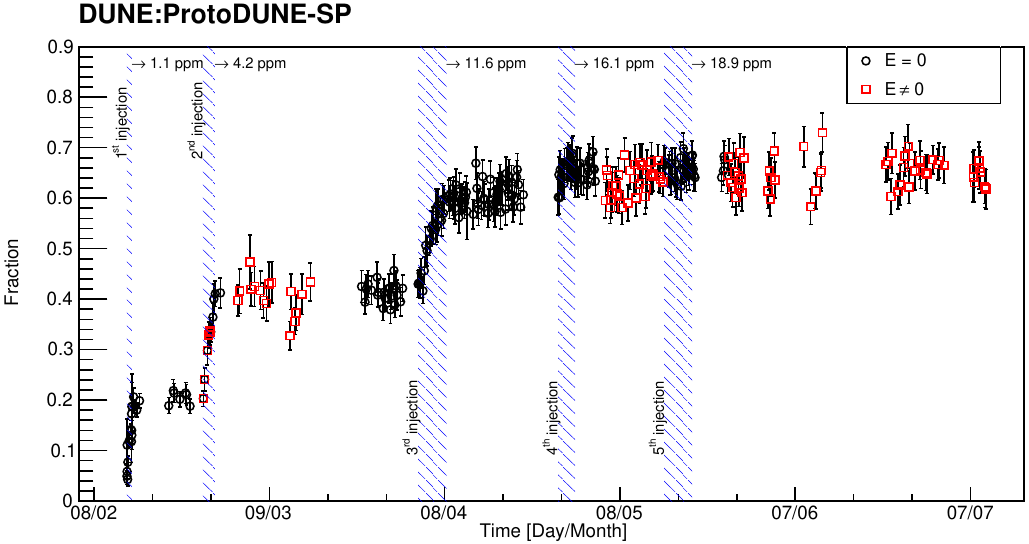}
\end{dunefigure}

 \begin{dunefigure}
[Light recovery following xenon injection in \dshort{pdsp}]
{fig:06_LightRecoveryAra}
{Light recovery is demonstrated through attenuation curves after xenon injection in the nitrogen-polluted \dshort{pdsp}. Data collected with the non-beam side \dword{arapuca}. The left plot shows the collected light versus distance from the detector; the right plot shows the ratio of collected light with nitrogen (black points) and with nitrogen+xenon, with respect to non-polluted LAr (blue points). Data refer to runs with no active electric field. The right plot shows how the increasing concentration of xenon enhances light recovery further from the photon detectors.}
  \includegraphics[width=0.6\textwidth]{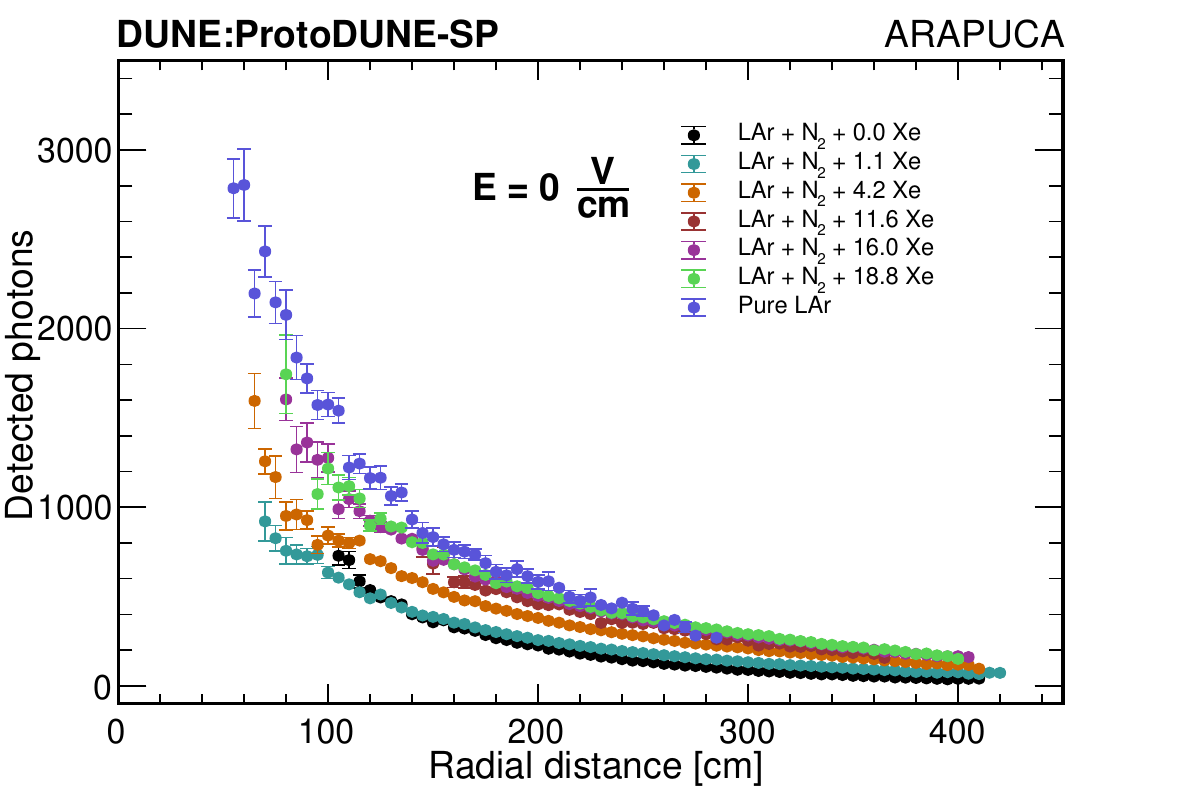}
  \includegraphics[width=0.6\textwidth]{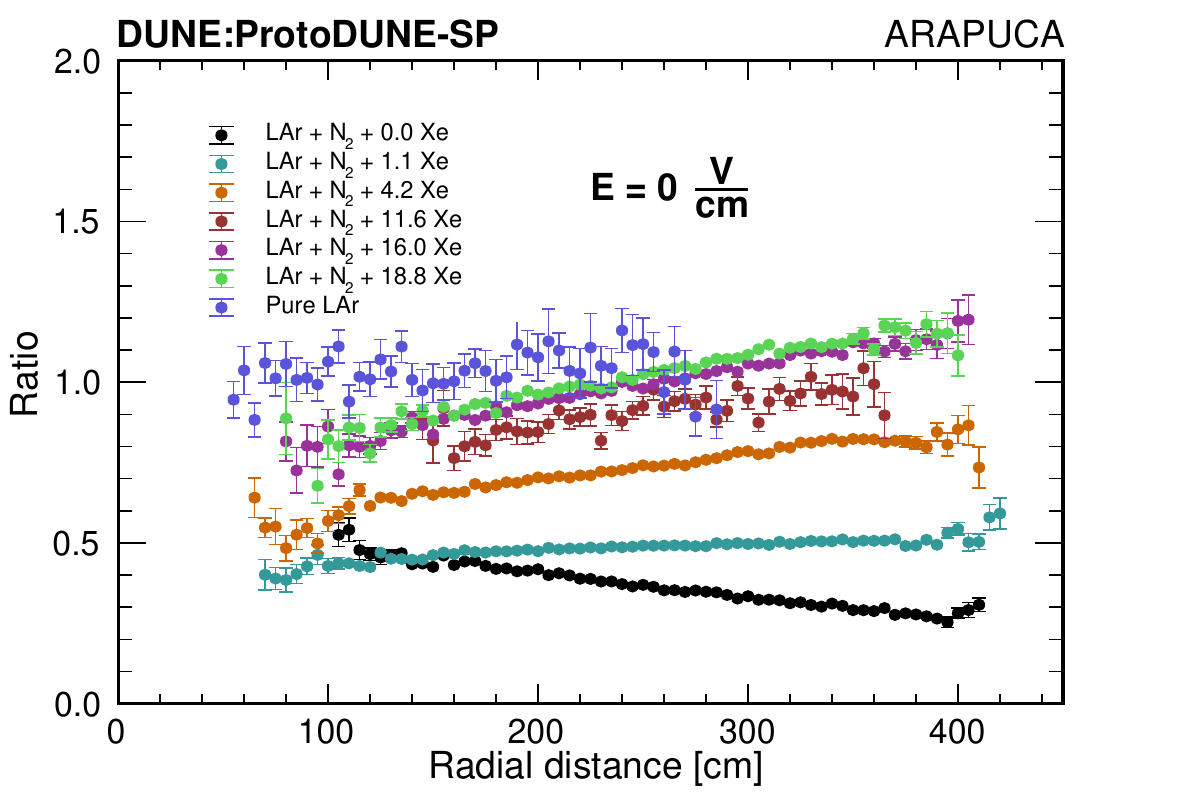}
\end{dunefigure}

%\fixme{RJW - add DP xenon citation} done
In addition to the \dshort{pdsp} run, \dshort{pddp}  was partially emptied and refilled with the doped argon transferred from \dshort{pdsp}. The amount of nitrogen was increased in order to reproduce the conditions of ProtoDUNE SP. The measurements taken in  \dshort{pddp} also show an increase in the amount of collected light, which is becoming more and more evident with increasing distance from the detectors (due to the larger Rayleigh Scattering length). This result is published in the overall paper on the performance of the \dshort{pddp} light detection system~\cite{DUNE:2022ctp}.

The results of the two campaigns cannot be directly compared due to the significantly different TPC configurations and light collection technologies. However, both demonstrate that xenon doping will provide greater light yield uniformity and allow photon detection recovery from accidental nitrogen contamination for multi-meter drift-path TPCs.

%%%%%%%%%%%%%%%%%%%%

\subsection{\dshort{xarapu} Module Optical and Mechanical Development}
\label{subsubsec:X-ARAPUCA Module Developmnet and Validation}

The \dshort{xarapu} concept of the \dshort{spvd} \dshort{pds} module is the same as that  of \dshort{sphd}. Though the design benefited greatly from the \dshort{sphd} development through to the final design, an R\&D program was pursued in those aspects most impacted by the significantly different geometrical layout of the \dshort{spvd} modules.

The development of the \dshort{spvd} module focused on:  

\begin {itemize}
\item Production of large area \dshort{wls} plates of $607\times607\times3.8$~mm$^3$ dimensions compliant

with cryo-resilience, optical grade, and thickness requirements;% (INFN-MiB);

\item \dshort{sipm} mounting onto flexible kapton \dshort{pcb} (instead of rigid FR4 \dshort{pcb} pin coupled to FR4 routing boards) utilized for both the \dshort{sipm} mechanical support and the signal routing;% (UCSB,INFN-MiB,NIU,FNAL);
\item Spring-loaded mounting of the \dshort{wls}/\dshort{sipm} photon collector structure inside a \dshort{g10} frame, to compensate for their differential thermal shrinking (about 1\%);% (Iowa,CSU,INFN-MiB); 
\item Optimization of the dichroic filters optical density and reflectivity at specific angle of incidence (maximimized for 45$\deg$, the average angle of the Lambertian emission angle distribution) and wavelengths;% (Spain, Italy);
\item Development of larger size (202$\times$97.5 mm$^2$ and 143.75$\times$143.75 mm$^2$) dichroic filters to minimize the surface area of the non-active holder frame profiles of the dichroic filters.
\item Optimization of \dword{ptp} coating on large size filters;% (Brazil).
\item \dshort{sipm} coupling via optical grade epoxy to a \dshort{wls} plate forming an integrated readout structure; 
\item Laser cutting of the \dshort{wls} tile edges with cutouts of rectangular/elliptical shape, to improve the photon collection as they exit the lightguide and to mitigate \dshort{sipm} gluing.
\end{itemize}

%%%%%%%%%%%%%%%%%%%%%%%%%%%%%%%%%%%%%%%

\subsection{Cryogenic PD Testing Facility for \dshort{vdmod0}}
\label{sec:PDS_cryo_testfac}

In late 2022 a cryogenic facility was constructed at the Neutrino Platform facility at \dshort{cern} EHN1 building to test full size %$60\times60$~cm$^2$ 
\dshort{xarapu} modules and read-out electronics, see Figure~\ref{fig:pds_mini-coldbox}.
This facility provides final functionality check and validation in cold prior to deployment in the \dshort{vdmod0} cryostat and detector integration. 
The $4\times4$~m$^2$ %delimited 
test area is located at CERN building 887/R-291 Sal\`eve side, in front of the ground floor \dshort{pds} Lab barrack, 
where the assembly of the \dshort{xarapu} PD modules is be performed.

A 70-cm diameter $\times$ 120-cm tall open air Dewar contains a (non-purified) \dshort{lar} bath, which can be sealed with a custom-made lid that ensures light tightness and free exit for gaseous argon boil-off. A trolley with a winch allows insertion and extraction of a PD module. %in and out of the liquid. 
To minimize the usage of liquid argon and ensure safe warm-up and fast operation turnaround, a stainless steel box that can accommodate the \dshort{xarapu} module is inserted inside the Dewar and filled with \dshort{lar}.  

At the end of a test, when extracting the module while it is still cold, nitrogen gas flowing under the box contained by a polyethylene skirt, prevents frost condensing on the modules. Cables and optical fibers routed via dedicated feedthroughs on the lid allow for cold electronics readout during test and calibration with a LED pulsed flasher installed inside the Dewar. Temperature sensors and a slow control system are used to monitor the \dshort{lar} level and temperatures at different heights inside the inner vessel. 

All 16 \dshort{xarapu} modules for \dshort{vdmod0} were tested, and qualification data acquired, analyzed, and stored.

\begin{dunefigure}
[\dshort{pds} mini-cold box test facility at CERN]
{fig:pds_mini-coldbox}
{Schematic (left) and Dewar photos (right) of the mini-\coldbox at CERN, for testing  \dshort{xarapu} modules prior to installation in \dshort{vdmod0}.
}
\includegraphics[width=0.33\textwidth]{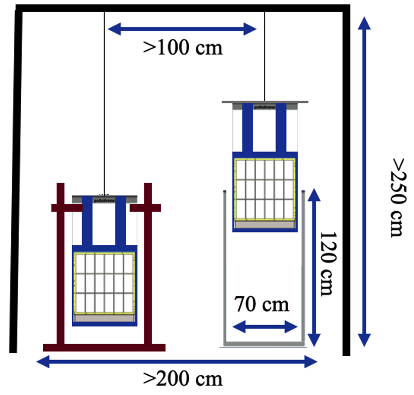}
\includegraphics[width=0.16\textwidth]{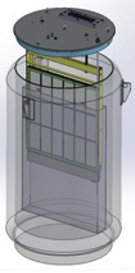}
\includegraphics[width=0.48\textwidth]{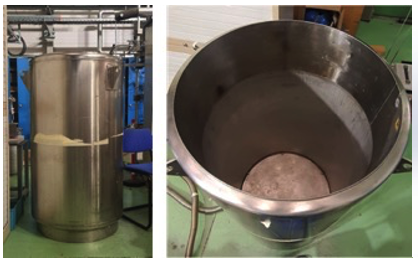}
\end{dunefigure}

%%%%%%%%%%%%%%%%%%%%%%%%%%%%%%%%%%%%%%%%%%%%%%%%%

\subsection{PD Module Validation}
\label{subsubsec:X-ARAPUCA Module Validation}

During 2021 and 2022, all the components were developed and prototypes fabricated. Integration techniques and procedures were developed and tested.

Fully-instrumented \dshort{xarapu} modules have been tested both at the Proton Assembly Building at \dword{fnal}, which serves as a test-bed for cryogenic electronics and \dword{pof} development, in a test facility at CERN commissioned in later 2022 (described in Section~\ref{sec:PDS_cryo_testfac}), and in the \coldbox demonstrator, in \dword{np02} at \dshort{cern}, starting in summer 2021 and continuing through to 2023, see Table~\ref{tab:PD-VD-modproto}.

Three prototype versions (\dshort{xarapu}.V1, \dshort{xarapu}.V2, and \dshort{xarapu}.V3) have been deployed and operated in the \coldbox demonstrator, as can be seen in Figure~\ref{fig:pds_coldbox}. The V4 and V5 prototypes were validated in the EHN1 FD2 cold box ahead of \dshort{vdmod0} installation: these versions integrate different technologies of electronics shielding, \dshort{sipm}-to-\dshort{wls} coupling, and \dshort{wls} edge shaping. An independent module mechanics design committee set up by the \dshort{spvd} \dshort{pds}  consortium management evaluated the designs and down-selected from the two options (V4 and V5) to a single option (essentially V5) that is  deployed in \dshort{vdmod0}. These value engineering efforts are aimed at   increased manufacturability, reduction of fabrication costs, and optimization for shipping.  

\begin{dunetable}
[PD module prototypes]
{p{.10\textwidth}p{.65\textwidth}}{tab:PD-VD-modproto}
{\dshort{pd} module prototypes. }
{\bf Version} & {\bf Description} \\ \toprowrule
V1 & First full-size FD2 PD module. 3-sheet  filter frame design, coil springs for pressing \dshort{sipm}s against \dshort{wls} plate. \\  \colhline
V2 & Changes primarily to dichroic filter frames, incorporating leaf springs to press \dshort{sipm}s against \dshort{wls}, Single-sheet filter frame with spring-loaded filter plates.\\  \colhline
V3 & Evolution of V2 design, making improvements to filter frame assembly. \\  \colhline
V4 & Final version of leaf-spring \dshort{sipm} mounting design with spring-loaded filter windows in single-piece frame. Evaluated as part of Module-0 mechanical design process. \\  \colhline
V5 & Final version of coil-spring \dshort{sipm} mounting design with Teflon washers mounting filter windows in single-piece frame. Evaluated as part of Module-0 mechanical design process. \\ 
\end{dunetable}

\begin{dunefigure}
[CERN \coldbox facility with three \dshort{xarapu} prototype modules]
{fig:pds_coldbox}
{Top view of the \coldbox demonstrator at CERN, showing three \dshort{xarapu} prototypes installed.
%at the time of writing. 
The first prototype is mounted on the cryostat wall, while the second and third prototypes are mounted on the cathode.}
\includegraphics[width=0.65\textwidth]{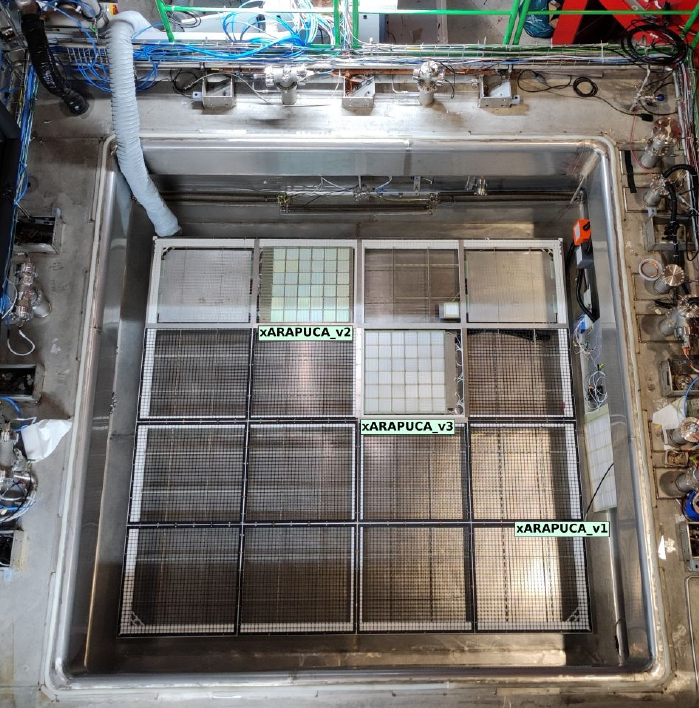}
\end{dunefigure}

\subsection{Cathode-Mount Module Cold Electronics Validation}
\label{subsubsec:DCEM Board Validation}
Numerous tests of the electronics developed for the \dshort{sipm} signal read-out of the cathode-mount PD modules V1 to V5 have been performed in the \coldbox, starting from the earliest prototypes (Dec. 2021) to the latest fully integrated final design board (Dec.22-Feb.23 - DCEM board shown in Figure~\ref{fig:pds_dcem}). In this section first results are reported on the read-out electronics response characterization validating the final design of the DCEM board. 

For the analysis of the electronics response, a pulsed LED calibration system is implemented in the \coldbox. It consists of a board that drives a 310~nm LED flasher, into a quartz fiber ending into a light diffuser placed in the LAr bath inside the \coldbox. The intensity and the width of the LED pulse can be selected, down to about 20~ns minimum duration.

The DCEM board in its final configuration is made of three main stages, the \dshort{pof} stage (with two OPCs, optical-to-electrical power converter, and a bias circuit transforming the OPC low output voltage to the high \dshort{sipm} bias voltage), the signal amplification and conditioning stage, and the 2-channel \dshort{sof} transmitter (laser diode and its driver for signal transmission over optical fiber). The board, with input cable connections from the \dshort{sipm} flex boards, is placed within a Faraday shielding metallic box along one side of the \dshort{xarapu} module. 

The first key milestone achieved is the demonstration of a signal-to-noise ratio (SNR) above the requirement of 4. Figure~\ref{fig:pds_charge_spec} shows the charge spectrum obtained with minimum LED pulse amplitude illumination, with peaks representing the charge distribution for 1-PE to 6-PE, above the first noise (0-PE) peak. From the multi-Gaussian fit, the ratio of the 1-PE peak position to the width of the 0-PE peak SNR=5.9 is obtained. 
This SNR is obtained operating the \dshort{xarapu} module (V4) on the cathode at high voltage, during TPC operation. The DCEM read-out board is powered by \dshort{pof} laser power via optical fiber and \dshort{sipm} electrical signals are converted into optical signals and transmitted by \dshort{sof} laser diode via optical fiber. Signal and noise levels were stable during the cold box run.     

\begin{dunefigure}
[Charge distribution from \dshort{pd} cathode module \coldbox test with \dshort{pof} and \dshort{sof}]
{fig:pds_charge_spec}
{Photoelectron charge distribution for a full cathode module, obtained with the data from February 2023 \coldbox test with \dshort{pof} and \dshort{sof}. An SNR of $\sim$~5.9 was obtained.}
\includegraphics[width=0.40\textwidth]{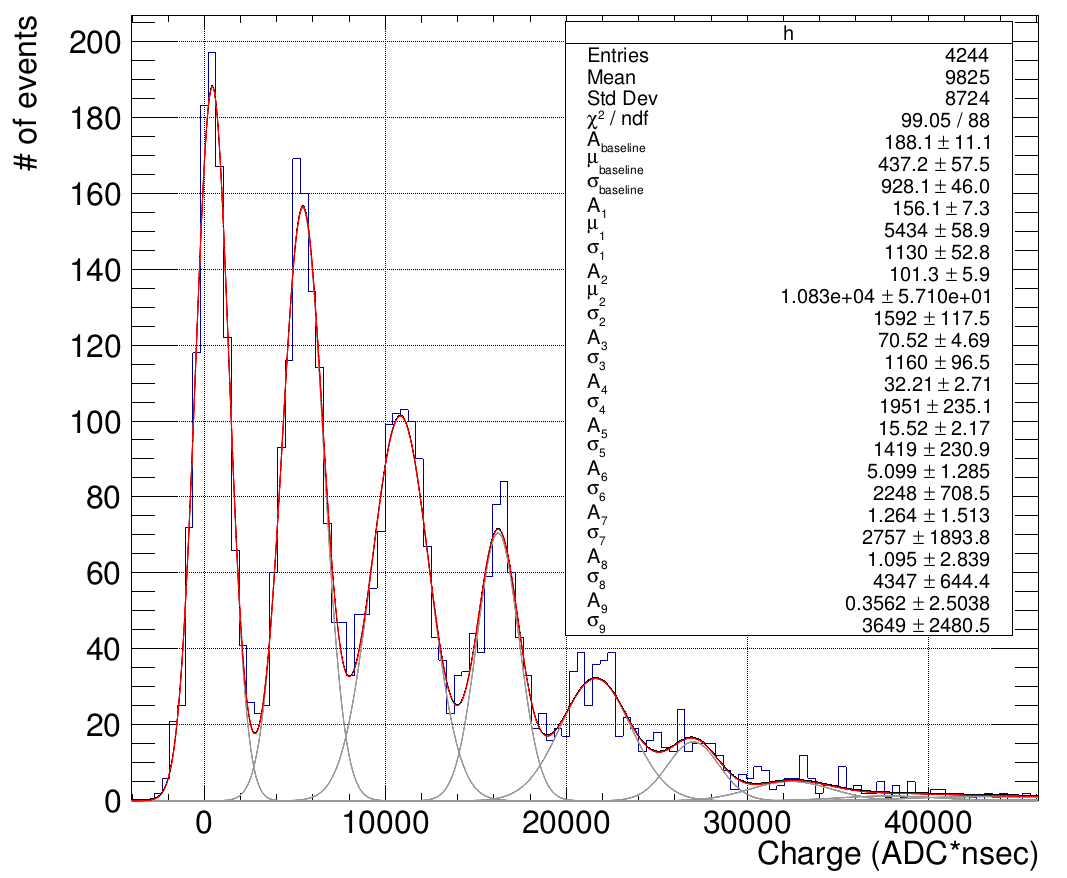}
\end{dunefigure}

The impulse response function of the readout electronic chain, \dshort{sipm} signal summing, cold DCEM analog signal conditioning and warm signal optical-to-electrical conversion (commercial Koheron PD100 board) and digitization, is obtained by $\sim$20~ns narrow LED flasher pulsing. Figure~\ref{fig:pds_avg_signal} shows the average of 500 signals obtained with fixed amplitude LED pulse. The relevant characteristics are highlighted on the plot, namely the rise time and discharge time measured between the 10\% and 90\% of the pulse amplitude, of 40~ns and about 400~ns respectively. The rise-time obtained is that expected from 80-ganged \dshort{sipm}s, indicating that the bandwidth of the transmitter  is adequate. 

\begin{dunefigure}
[Average signal from \dshort{pd} cathode module \coldbox test with \dshort{pof} and \dshort{sof}]
{fig:pds_avg_signal}
{Average of 500 pulses obtained from the cathode modules. The rise time measured between 10\% and 90\% amplitude is $\sim$ 40~ns, while the discharge time is around 380~ns. }
\includegraphics[width=0.5\textwidth]{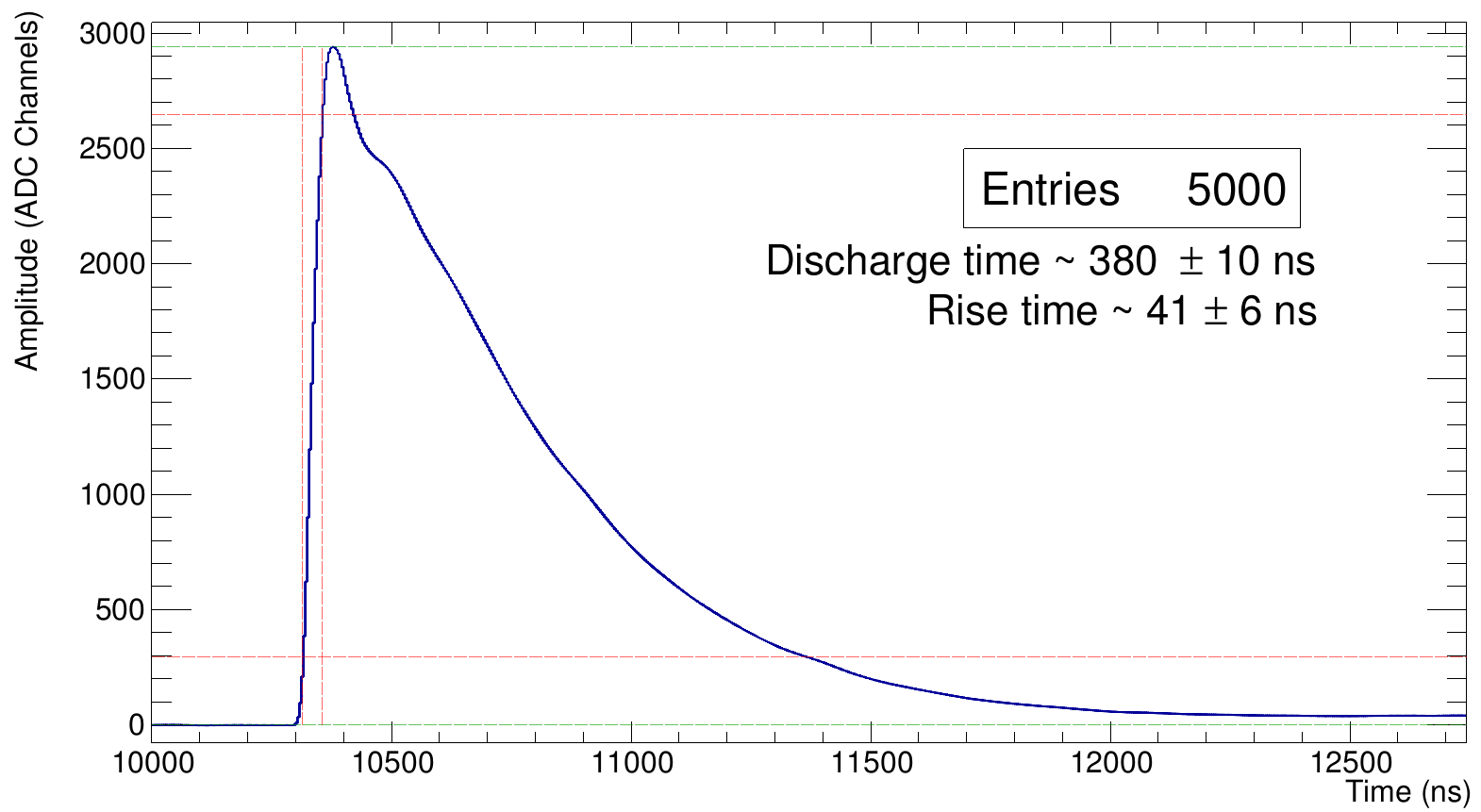}
\end{dunefigure}

Linearity of the \dshort{sof} electronics was first demonstrated with lab test-bench measurements over a limited range of signal amplitude, corresponding to few tens of PEs. Using data acquired at the \coldbox, it has been possible to evaluate the linearity of the entire readout chain, from the photosensor to the warm receiver, for a much larger dynamic range. 

Signal waveforms were recorded with LED calibration flasher pulsed from minimum to maximum amplitude settings. The linear correlation between amplitude of the signal waveform (pulse height expressed in ADC counts) and signal waveform integral 
%(signal charge expressed in number of PEs, after 1-PE to ADC*ns charge calibration) 
(after charge-to-PE calibration is applied)
for various LED flasher amplitude setting is shown in Figure~\ref{fig:pds_linearity}. The pulse amplitude of the recorded signals is limited to a maximum of about 1.5~V ($\sim~13000$ ADC), since the warm receiver is saturated at this point.
The linearity shown in Figure~\ref{fig:pds_linearity} demonstrates that the signal maintains the expected shape without distortions over the dynamic range of interest. 
A slight non-linearity of the LED calibration system signals appears at higher LED pulse amplitude settings
%show slight shape distortion (smaller widths), 
that does not occur for scintillation signals from cosmics 
%show no shape distortion 
for all amplitudes.
This residual non-linearity of the LED calibration system has to be taken into account.

\begin{dunefigure}
[Linearity of \dshort{pd} waveform amplitude and integral]
{fig:pds_linearity}
{Linear growth of signal amplitude as a function of the number of PE, estimated from the integral of the SPE charge. The groups of data correspond to LED voltage levels. The saturation is due to the warm receiver's maximum input limit.
%\fixme{SS edit 13/03 updated file for this figure}
}
\includegraphics[width=0.6\textwidth]{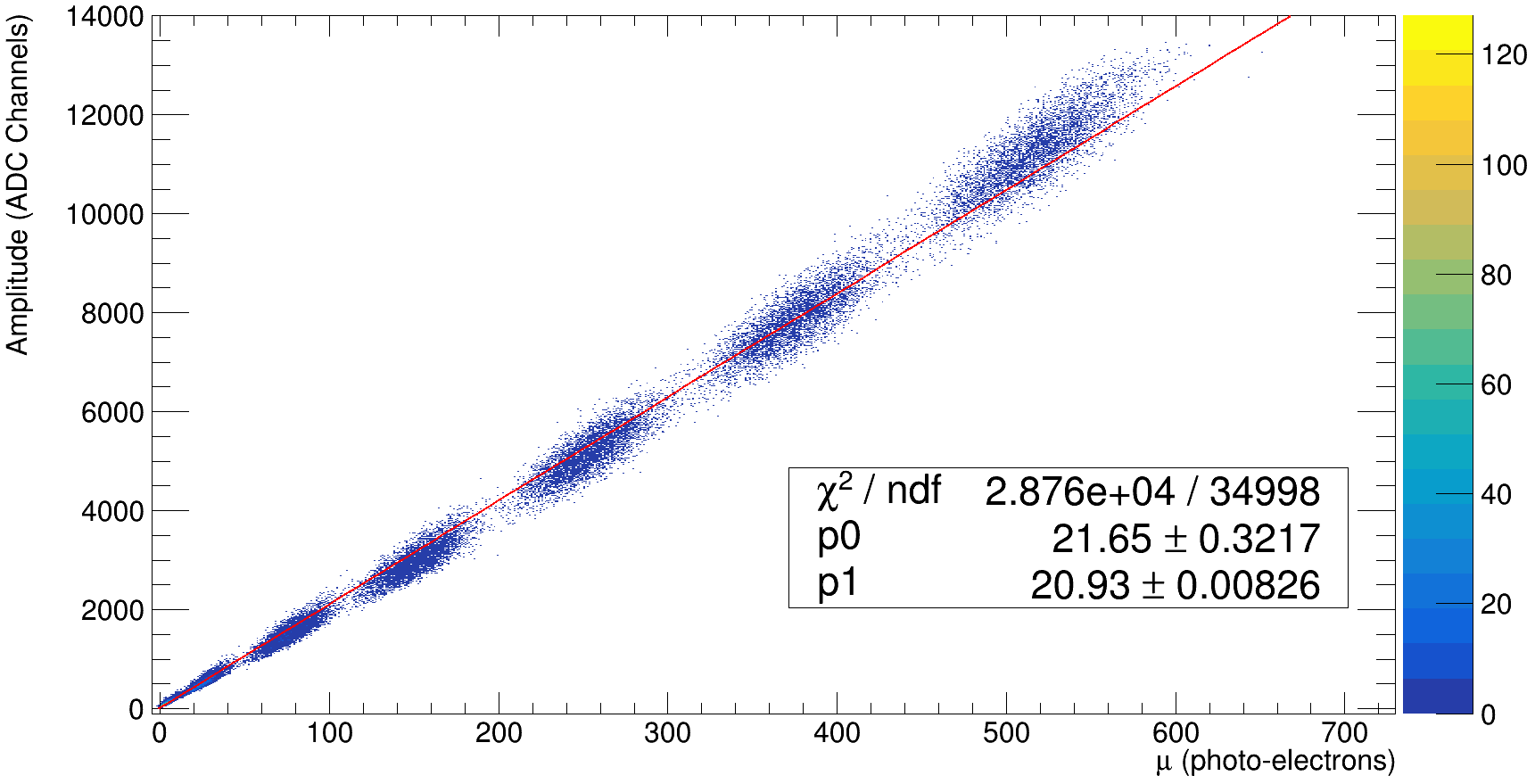}
\end{dunefigure}

The full dynamic range was evaluated in a special channel of a DCEM board with lower light output also tested in \coldbox for V4 \dshort{xarapu} module read-out, without the saturation limitation of the warm receiver (whose gain will be adjusted for the in-house receiver under development to replace the commercial Koheron board). The resulting measurement, going up to almost 2000 PE equivalent signal amplitude, is presented in Figure \ref{fig:pds_dynamic_range}.

\begin{dunefigure}
[\dshort{sof} dynamic range]
{fig:pds_dynamic_range}
{The dynamic range of the system has been demonstrated from the SPE level to over 2000 PE.
%\fixme{SS edit 13/03 updated file for this figure}
}
\includegraphics[width=0.6\textwidth]{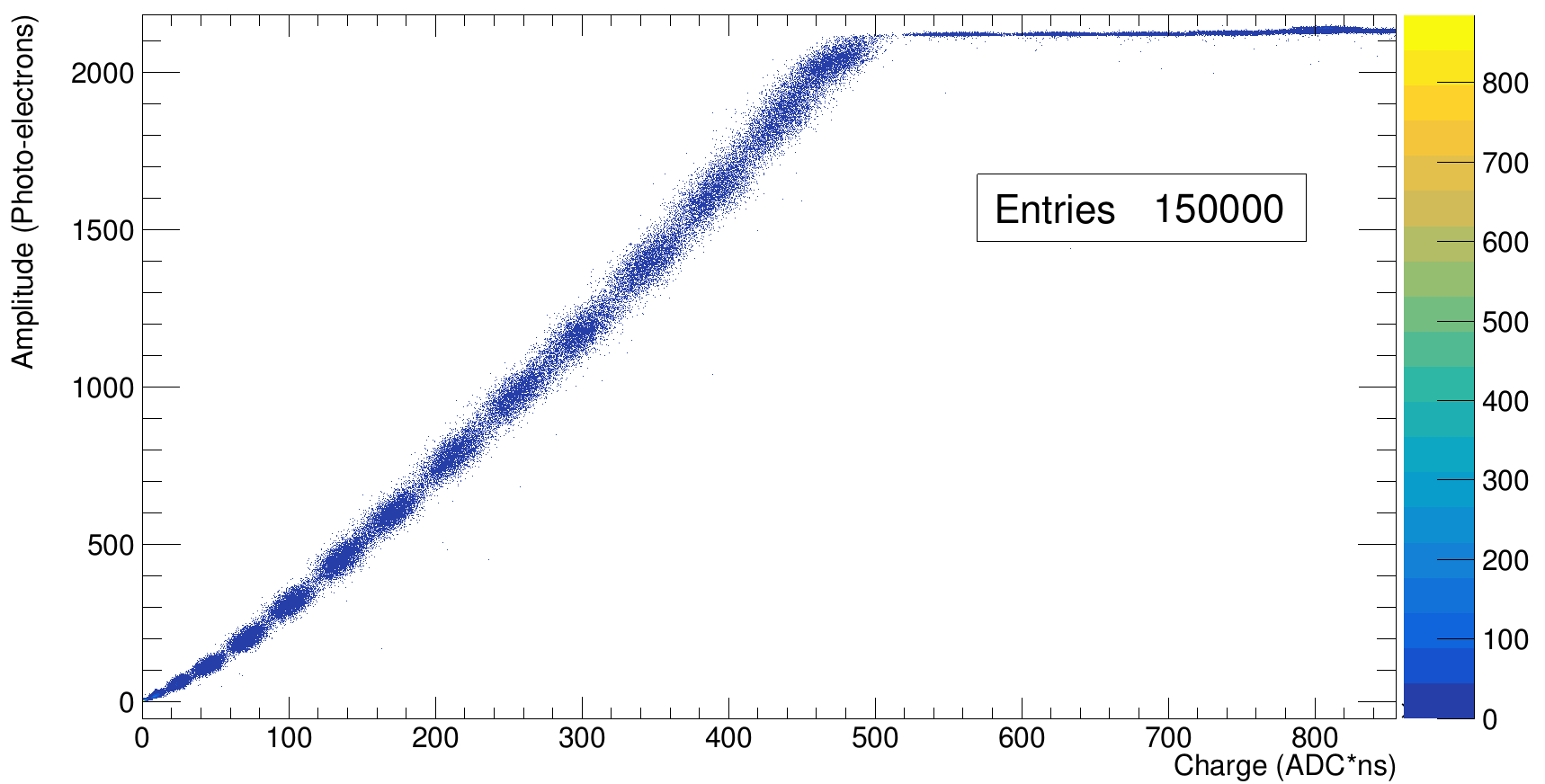}
\end{dunefigure}

The \dshort{xarapu} detectors have a very small, but non-zero, sensitivity to the 808~nm infrared light of the \dshort{pof} laser. This means that the \dshort{pof} fibers and connectors at OPC receivers must be adequately shielded. Potential light leakage is effectively mitigated by potting the OPC receivers and connectors on the DCEM boards with electronic grade silicone paste and also by the metallic Faraday boxes surrounding the electronic boards. Light leakage from the \dshort{pof} fibers becomes negligible when black jacketed fibers are bundled in the protective black tubes used for fiber routing from PD modules on the cathode to the feedthrough flanges at the top of the cryostat. 
Lab tests have demonstrated optical noise due to \dshort{pof} light leakage decreases to sub-PE levels with these measures. 

The level of detectable light leakage at the \coldbox was evaluated by measuring the rate of single PE signals with an independent \dshort{xarapu} module mounted on the wall (V1). Data were taken while both the cathode \dshort{xarapu} modules (V4 and V5) with \dshort{pof} lasers were turned ON and OFF. No difference was found in the number of detected single-photon signals with \dshort{pof} ON compared to \dshort{pof} OFF (no laser light injected in \coldbox). 
The noise distribution evaluated from baseline fluctuations from the V1 membrane-mount \dshort{xarapu} module on the wall before and after turning on \dshort{pof} lasers is shown in Figure~\ref{fig:pds_Pof_noise} and demonstrates no significant noise increase. It is concluded the light leakage has been reduced to inconsequential levels. 

\begin{dunefigure}
[\dshort{xarapu} baseline noise with \dshort{pof} lasers OFF and ON]
{fig:pds_Pof_noise}
{Distribution of baseline noise for membrane V1 \dshort{xarapu} (DCEM v1.0 coaxial cable readout) with \dshort{pof} OFF and ON for February 2023 \coldbox data.}
\includegraphics[width=0.4\textwidth]{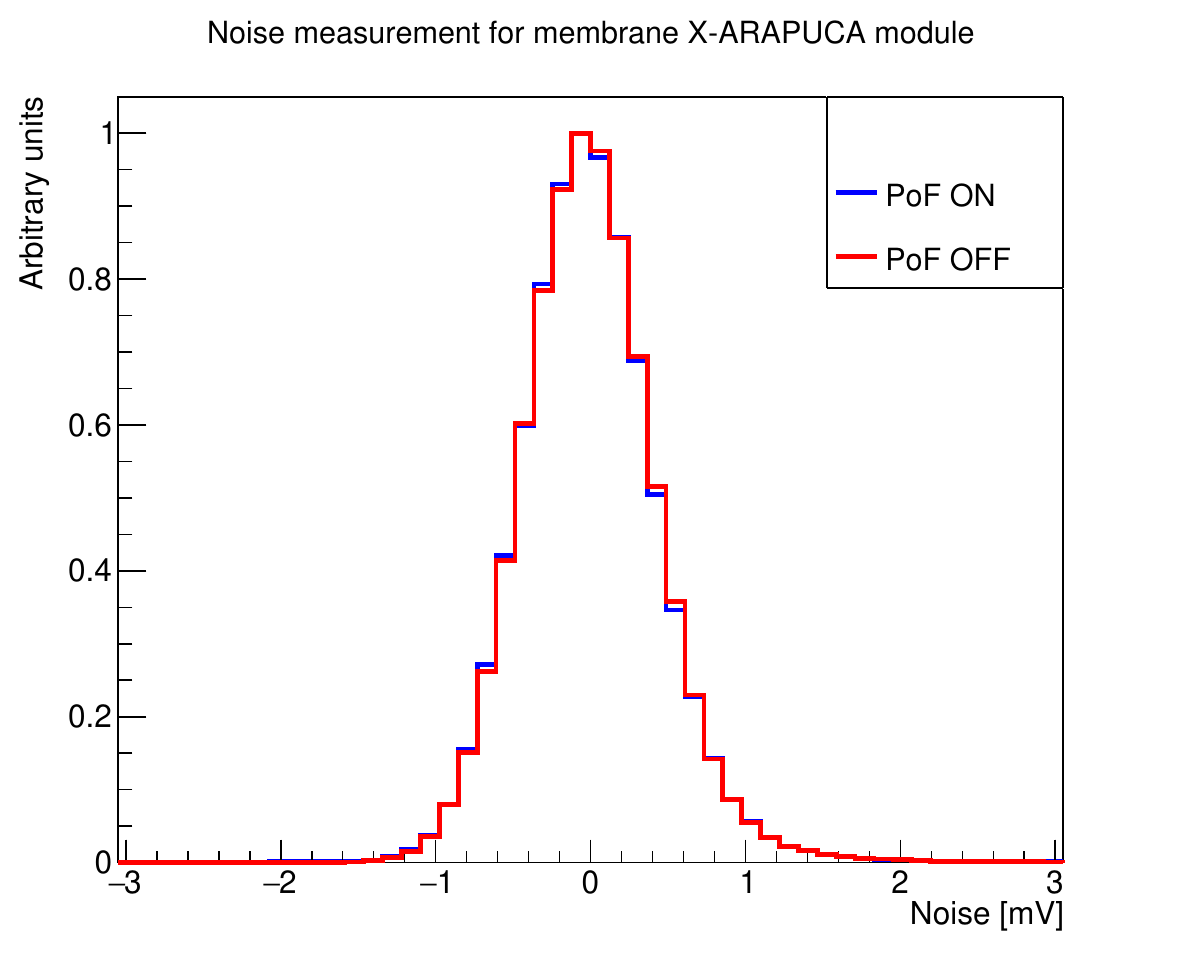}
\end{dunefigure}

Interactions between the PDS system and the rest of the detector were evaluated. The noise levels on the PDS system were compared with the cathode HV voltage ON (10~kV) and OFF. Figure~\ref{fig:pds_HV_noise} shows the Fourier transform of the data collected from a cathode module; the peak around 30~MHz corresponds to the frequency domain of the signals detected. No difference was observed due to the state of the HV. 
The TPC CE also evaluated the noise level with PDS on and off, finding no differences. 

\begin{dunefigure}
[Fourier transform of \dshort{xarapu} baseline signal with cathode HV OFF and ON]
    {fig:pds_HV_noise}
    {Fourier transform plot of the baseline signal of the PDS readout with the high voltage of the cathode ON and OFF. No difference is found.}
    \includegraphics[width=0.6\textwidth]{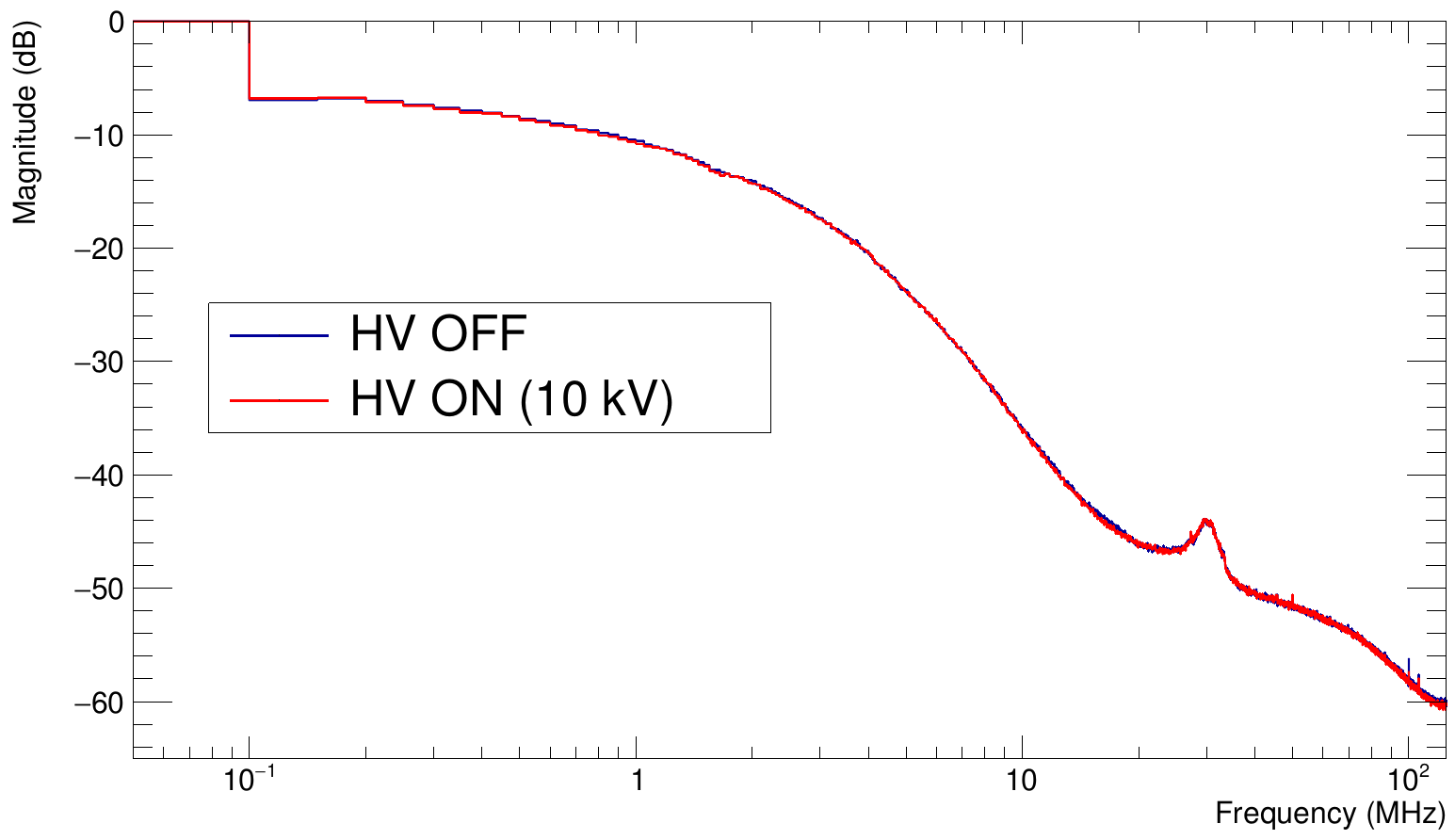}
\end{dunefigure}

Lastly, the time resolution of the PDS system is evaluated by comparing the measured time of arrival of signals from two channels for the same \dshort{xarapu}; the result is shown in Figure~\ref{fig:pds_time_res}. The convoluted time dispersion of both channels is found to be of the order of a few nanoseconds. It should be noted that the result is preliminary in that a fixed threshold is applied and with no correction for time-walk. 

\begin{dunefigure}
[Time difference between two \dshort{xarapu} module channels]
    {fig:pds_time_res}
    {Histogram of the time difference between signals from the two channels of one \dshort{xarapu} detector. }
    \includegraphics[width=0.6\textwidth]{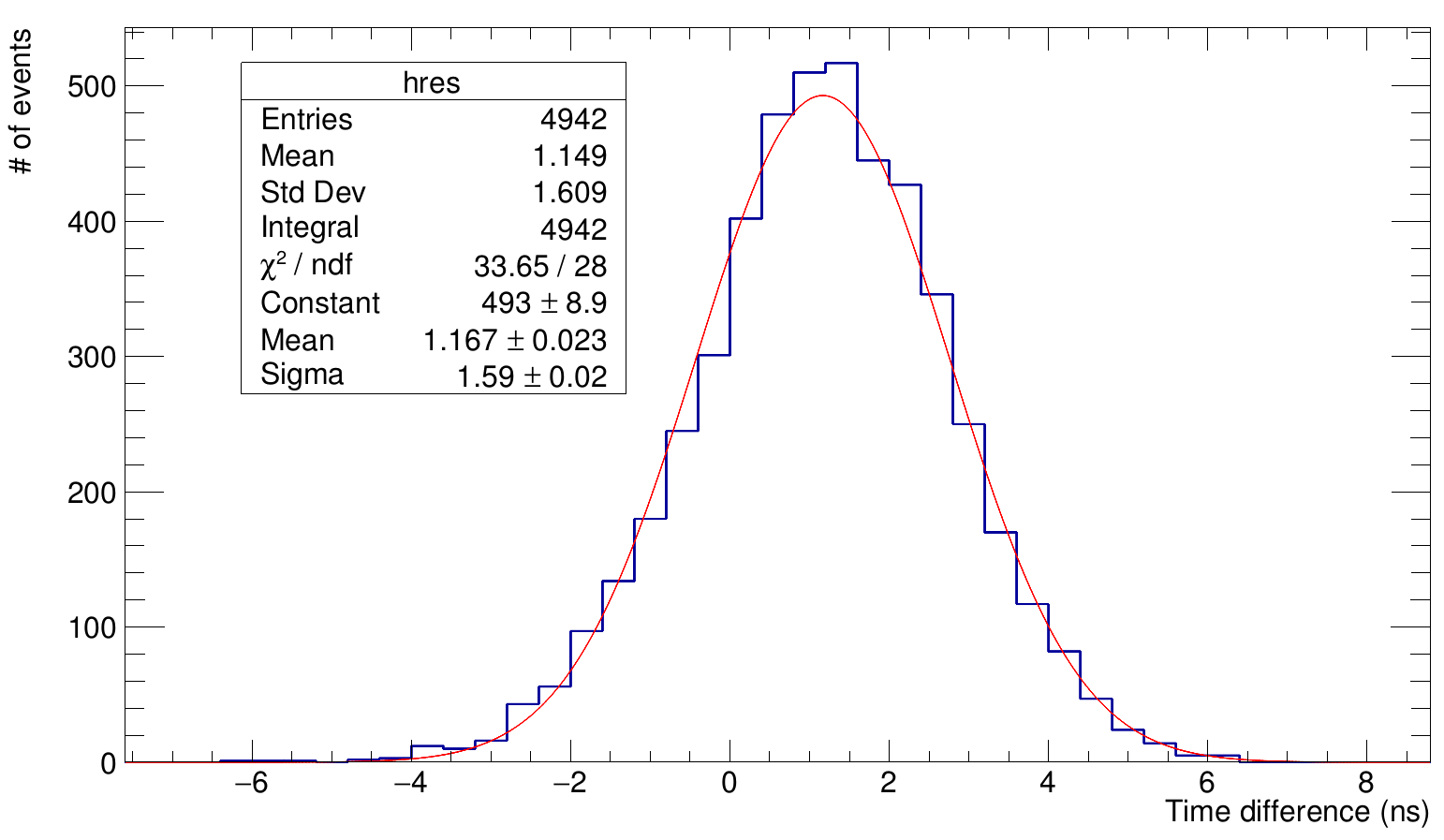}
\end{dunefigure}

%%%%%%%%%%%%%%%%%%%%

\section{Production and Assembly}
\label{sec:PDS_Prod_Assbly}

The \dshort{spvd} \dshort{pds} production phase will be launched after a successful \dword{prr}, which 
should be completed in early 2024.

Production procurement will be split into three phases: an initial 10\% phase to establish the \dword{qa}/\dword{qc} procedures with vendors and \dshort{spvd} \dshort{pds} assembly centers, followed by two 45\% time-sequential phases. This sequence of fabrication stages is initially envisioned to allow for an initial 10\% production startup phase followed by two main production phases to control cost envelopes over fiscal years. It also provides a handle for multiple sites and/or multiple vendors, as a secondary benefit. Once production begins, the schedule drivers for \dshort{spvd} \dshort{pds}  are procurement handling, vendor lead-time, and production site throughput rate.  
No schedule logic that interfaces to other subsystems will be present until installation begins.

The reference design calls for using several similar or identical components for \dshort{spvd} \dshort{pds} as used in \dshort{sphd}: photosensors, dichroic filter plates, \dshort{wls} plates, and membrane-mount conductive cables. Dealing with the same vendors as the \dshort{sphd} should mitigate production delay concerns on these components. Schedule risk is significantly mitigated by a two-vendor strategy for both photosensor and \dshort{wls} plate fabrication. Multiple vendors of high-quality dichroic filters have also been identified.

Planning for module assembly is based on the experience with the \dshort{pdsp} \dshort{pds}. 
The modules were designed to be factorizable into multiple free-standing sub-assemblies, thereby increasing the number of assembly sites that can operate independently. \dshorts{sipm} already mounted to flexible \dshort{pcb}s will be procured from vendors, simplifying the logistics chain while reducing the number of required \dshort{qc} tests.  
Dichroic filter plates will be procured, tested and assembled into sub-assemblies as shown in Figure~\ref{fig:ARAPUCA-module-VD}.  These sub-assemblies will be assembled into \dshort{pd} modules at multiple production sites.  
Studies of assembly rate for \dshort{spvd} \dshort{pd} modules will be conducted as part of the \coldbox and \dshort{vdmod0} validation efforts. Preliminary tests at CSU, NIU and at \dshort{cern} all suggest that module assembly times (from a complete set of components) of one module per day using a two-person assembly team are reasonable. Particular care will be taken to understand the impact of coupling the \dshorts{sipm} to the \dshort{wls} plates, which is likely to have the most significant impact on production rate.

The reference plan is for 672 detector modules for the \dshort{spvd} \dshort{pds}. Potential production sites at collaborating institutions have been identified. 
At least two production sites will be selected, possibly more, depending on interest from collaborators and the ability to meet production requirements. These production sites will be supplied from multiple sub-assembly sites, the number of which will be determined by collaborator interest and the required production rate.  
It is planned to launch production orders over three years beginning in 2024 and for fabrication of core components (SiPM, WLS plates, dichroic filters etc.) with a ramp up during summer of 2024.  
After the ramp-up period, the target production rate for modules is eight per week. 
In the reference schedule, detector module assembly begins in early 2025 and should be completed by Q2 2027.

For cold electronics readout, commercial vendors will fabricate custom \dshort{pcb}s. 
On-board components will be qualified for long-term cold operation following a peer-reviewed process. 
\dshort{pof} assembly is currently planned to involve custom assembly by collaborating institutions on the transmitter side and for individual receiver mounting to the cold readout electronics motherboard. There is an opportunity to explore turnkey vendor assembly of the \dshort{pof} rack mount transmitter units and cold receiver units. The \dshort{pof} fiber optical power converter receiver unit is fabricated by commercial vendors, as is the \dshort{sof} transmitter unit.

The readout electronics in warm is a custom module assembly of commercial digitizer and aggregator electronics.

%%%%%%%%%%%%%%%%%%%%%%%%%%%%
\section{Interfaces} %added by anne 30 sep

Tables~\ref{tbl:pds-interfaces} summarizes the \dshort{pds} interfaces to the other \dshort{spvd} systems. Each interface is discussed in more detail in the following sections.

\begin{dunetable}
[\dshort{pds} interface links]
{p{0.15\textwidth}p{0.8\textwidth}}
{tbl:pds-interfaces}
{\dshort{pds} interface descriptions and links to full interface documents.}
Interfacing System & {\bf Description}  \\ \toprowrule

\href{https://edms.cern.ch/document/2619004/1}{\dshort{crp}} & Response Monitoring System diffuser placement (potential, TBD), routing of \dshort{pds} fibers at bottom of cryostat. \\ \colhline

\href{https://edms.cern.ch/document/2618994/1}{\dshort{bde}} &  Shared use of cable trays and cryostat penetrations.\\ \colhline

\dshort{tde} &  None.\\  \colhline

\href{https://edms.cern.ch/document/2619007/1}{\dshort{hvs}} &  Mechanical and electrical contact between cathode and \dshort{pd} modules, fiber routing in cathode, fiber routing along \dshort{fc}, \dshort{fc} transparency, \dshort{pds} fiber diffuser placement, \dword{hvs} camera lighting.\\  \colhline
\href{https://edms.cern.ch/document/2088726/4}{\dshort{daq}} &  High-speed data links, timing signals.\\  \colhline

\href{https://edms.cern.ch/document/2648555/1} {\dshort{iandi} and \dshort{sc}} &  Membrane-mounted \dshort{pd} module support, fiber installation process, grounding, rack infrastructure, \dword{lv} power, \dshort{pds} readout configuration, power supply control and monitoring, \dshort{pds} Response Monitoring system control and monitoring. \\\colhline

\dshort{calci} &  None.\\  \colhline

\href{https://edms.cern.ch/document/2145146/3} {\dshort{swc}} &  Software and databases to support data-taking and offline analysis.\\

\end{dunetable}
%\section{Systems Interfaces}
\label{sec:PDS_Interfaces}

\subsection{Charge Readout Plane (\dshort{crp})}
Previously, a potential interface between the \dshort{pds} and \dword{crp} was identified as arising from the possible installation of \dshort{pds} Response Monitoring System optical fibers on the central long axis of each \dshort{crp}.  As of the drafting of this report, it is believed that the optical fibers placed between the \dshort{fc} and cryostat wall will provide sufficient coverage of all \dshort{pds} modules in the system, thus eliminating this interface.  This will be verified with the FD2 \dshort{vdmod0} run scheduled to take place in 2023, and the potential interface revisited accordingly.

The precise routing of bundles of fibers for the \dshort{pds} %power-over-fiber and signal-over-fiber 
\dshort{pof} and \dshort{sof} systems for cathode-mounted modules is a potential mechanical interface with the \dshort{crp}. The interface will be further defined to ensure that the fiber bundles at the bottom of the cryostat are neither damaged by nor impede the placement of the \dshort{crp} during its installation.

\subsection{Bottom Drift Electronics (\dshort{bde})}
The interfaces between \dword{bde} and \dshort{pds} stem from overlapping cable and fiber paths for the two systems. 
As described in Section~\ref{sec:fdsp-tpcelec-design-ft}, 
the two systems share 40 penetrations through the cryostat roof, each consisting of a four-way cross-shaped spool piece.  Two vertical flanges indicated in Figure~\ref{fig:tpcelec-signal-ft} %\fixme{which figure?} 
serve \dshort{bde}, with the top horizontal flange serving \dshort{pds}.  
 
The design elements required for the routing of BDE and PDS cables and fibers through the penetrations, and the responsibilities for finalizing them, are laid out in the Interface Control Document.

The routing of cables and fibers in shared cable trays along the cryostat membrane presents another interface between PDS and BDE.

\subsection{Cathode Plane Assembly and High Voltage System (\dshort{hvs})}
\label{subsec:VD-PDS-HV-Interface}
The mechanical and electrical contact between the cathode and \dshort{spvd} \dshort{pds} as described in Sections~\ref{subsubsec:CAsss} and \ref{sec:PD-Intro} presents several interfaces between the \dshort{pds} and \dword{hvs}. 
The electrical contact with the cathode will be a critical consideration due to field uniformity and the risk of cathode high voltage discharge as described in Section~\ref{sec:PDS-LightColl}.
The reference design envisions that each cathode module contains four \dshort{pd} modules with an independent set of optical fibers to provide power to each module, eliminating the risk of discharge damage resulting from transient potentials between modules.  An alternative mitigation using custom-designed balun circuits as part of a conductive distribution of \dshort{sipm} bias voltage between \dshort{pd} modules on a single cathode module is under investigation, targeting a demonstration at Module-0.  Individual \dshort{pd} modules are further protected from the effects of discharge by a conductive mesh covering each side of the cathode openings which house \dshort{pd} modules and by a Faraday cage enclosing the electrical components of each cathode-mounted \dshort{pd} module as shown in Figure~\ref{fig:ARAPUCA-module-faraday}.  

Mechanical supports connecting cathode-mounted \dshort{pd} modules to the cathode structure have been designed and deployed successfully in \dshort{cern} \coldbox prototyping runs, and will be updated as the \dshort{pds} mechanical design in finalized.  Both the dry and buoyant weight of cathode-mounted \dshort{pd} modules must remain below their respective maxima determined by cathode flatness specifications.  The cathode suspension system must be calibrated in accordance with the PD module weight; 12kg is the dry weight limit per \dshort{xarapu}.

As shown in Figure~\ref{fig:PD-power-signal-routing}, the \dshort{spvd} \dshort{pds}
fiber routing from the cathode modules is expected to go along the cathode to the \dshort{fc}, down along the \dshort{fc}, from the anode over to the membrane wall where cathode cable routes will join \dshort{pds} membrane module cable routes up to penetrations at the cryostat roof. 

%\fixme{RJW: Removed reference to splicing.}
The process of fiber installation represents an interface with \dshort{hvs} in the placement of fibers along the \dshort{fc} cable tray, the proximity of fiber handling,
and the installation of fibers in the cathode.  
%Remove reference to splicing as that is not currently an option.
The fiber installation in the cathode will take place  \textit{in situ}.

The light yield specification (Table~\ref{tab:PD-VD-Requirements}) implies a design interface with \dshort{hvs} through the transparency of the \dshort{fc} and its impact on light collection by the membrane-mounted \dshort{pd} modules. All membrane-mounted PD modules are positioned within the region of 70\% FC transparency. 

The \dshort{pds} response and monitoring system has a potential interface with the \dshort{hvs} through the possible mounting of light sources (diffusers and fibers) on \dshort{fc} elements, respectively. This interface will be updated as the monitoring system design is finalized after the operation of \dshort{vdmod0}.

A final interface with \dshort{hvs} is presented by infrared LEDs providing illumination for inspection cameras in \dshort{hvs} scope.
%(Section~\ref{}).  
The resulting risk of damage to powered \dshort{sipm}s requires a system, to be designed, to prevent operation of \dshort{sipm} bias while the camera LEDs are in use.

\subsection{Data Acquisition (\dshort{daq})}
The \dshort{spvd} \dshort{pds} interfaces with the \dword{daq} system through high-speed links and timing signals connected to a layer of warm readout electronics that is in \dshort{spvd} \dshort{pds} scope. The DAQ high-speed links and timing signal cables and fibers are not in \dshort{spvd} \dshort{pds} scope, nor is the supporting infrastructure such as racks, rack power, and rack monitoring. % are not in \dshort{spvd} \dshort{pds} scope. 
The \dshort{spvd} \dshort{pds} warm readout electronics will handle extracting timing signals and distributing them to the detector readout electronics, as well as aggregating and formatting detector data in the \dshort{daq} event structure.
%(i.e., interface with \dword{felix} \dword{pci} cards). 
%\fixme{RJW: removed FELIX reference.}

\subsection{Cryogenics Instrumentation and Slow Control (\dshort{sc})}

\dword{sc} has several interfaces with \dshort{pds}:
\begin{itemize}
    \item \dword{sc} control (configuration) and monitoring of the \dshort{daphne} modules used for \dshort{pds} readout.
    \item SC control (voltage level, current limits, and on/off state) and monitoring (current draw, sense voltage, on/off state) of low-voltage direct-current power supplies powering the DAPHNE crates.  
    \item SC control and monitoring of power distribution units (PDUs) controlling distributing of AC power to \dshort{pds} direct-current power supplies and \dshort{pof} transmitter units.  
    \item SC control and monitoring of the flashers for the PDS response monitoring system.
\end{itemize}
These interfaces will require implementation of software modules to reflect configuration options yet to be determined.

\subsection{Facility, Integration and Installation Interfaces (\dshort{iandi})}
The \dshort{pds} has an interface with \dword{iandi} in the design and execution of the installation and process.  The completion of routing of fibers for \dshort{pof}, \dshort{sof},
and the Response Monitoring System will require coordination with other systems via \dword{iandi}, as well as careful consideration for Class-4 laser safety protocols during testing of \dshort{pof} functionality in place.  

The reference voltage of the cold electronics for the 352 membrane-mounted \dshort{pd} modules will be tied to detector ground at the cryostat penetration.  The reference voltage of the 320 cathode-mounted  \dshort{pd} modules will be connected via a 1 M$\Omega$ resistor to the conductive mesh that covers each cathode cell housing a \dshort{pd} module.

The \dshort{pds}-\dword{iandi} interface also includes 40 27u racks on the cryostat roof, close to the penetrations, housing \dshort{pof} transmitter boxes and \dshort{daphne} readout cards.  An additional 4 \dword{lv} power supply racks serving PDS will be located among the detector racks, approximately 25~m from the penetrations.

\subsection{Calibration and Monitoring}
While early versions of the \dword{calci}-\dshort{pds} \dword{icd} %interface document 
anticipated light sources potentially capable of damaging powered \dshort{sipm}s and lasers capable of damaging \dshort{wls} components in \dshort{pds}, the scope of \dshort{calci} does not currently include such components.

\subsection{Physics, Software and Computing}
The interfaces of \dshort{spvd} \dshort{pds} with physics, software, and computing are identical in form to those for \dword{sphd}, impacting
\begin{itemize}
    \item {Operations}
    \item {Data products}
    \item {Database schema and management}
    \item {Software design}
    \item {Computing power.}
\end{itemize}

The \dshort{pds} system is expected to contribute a very small fraction of the data that is selected to be transferred to and stored at Fermilab.
Data products, algorithms, and database schemas provided by \dword{swc} for \dshort{spvd} \dshort{pds} data will require different treatments of geometrical information for \dshort{sphd} and \dshort{spvd}. \dshort{pds} will be responsible for ensuring that the technical geometry description of the \dshort{spvd} \dshort{pds} remains up-to-date.

%%%%%%%%%%%%%%%%%%%%

\section{Transport and Handling}
\label{sec_TandH}

\label{sec:fdsp-pd-install}
%\metainfo{Content: Onel, Kemp, Warner}

A storage facility near the \dword{fd} site (the \dword{sdwf}) will be established to allow storage of materials for detector assembly until needed.  Transport of assembled and tested PD modules, electronics, cabling, and monitoring hardware to the \dshort{sdwf} is the responsibility of the \dshort{pd} consortium.

Following assembly and quality management testing at \dshort{spvd} \dshort{pds} construction centers, the \dshort{pd} modules with associated cold readout electronics will be packaged and shipped to an intermediate cryogenic testing facility for final full-chain testing including operation in LN2 to validate detector performance. Following this, the modules will be stored in their shipping containers in the \dshort{sdwf}.  Cables, warm readout electronics, and monitoring hardware will be shipped directly to the \dshort{sdwf} and stored until needed underground for integration.

Packaging plans are informed by the \dshort{sphd} \dshort{pds} experience.   \dshort{xarapu}s will be packaged in groups of 4 modules (matching the installation pattern), 
%\fixme{DWW> I have edited the crate size}
approximately \SI{1}{m} $\times$ \SI{1}{m} $\times$ \SI{50}{cm}.  These shipping boxes may be gathered into larger crates to facilitate shipping.  The optimal number per shipment is being considered.

Documentation and tracking of all components and \dshort{pd} modules will be required. % during the full logistics process. 
Well-defined procedures are in place to ensure that all components/modules are tested and examined prior to, and after, shipping. These procedures will be presented during the FD2 PDS Final Design Review and posted in \dshort{edms}.

Information coming from testing and examinations will be stored in the \dshort{dune} hardware database.  Each \dshort{xarapu} will be labeled with a text and barcode label, referencing the unique ID number for the \dshort{xarapu} contained, and allowing linkage to the hardware database upon unpacking prior to integration in cathode modules and membrane-mount support systems underground.

Tests have been conducted and continue to validate environmental requirements for photon detector handling and shipping. The environmental condition specifications for lighting (no exposure to sunlight), humidity ($<$50\% RH at 70 $\deg$F), and work area cleanliness (Class 100,000 clean assembly area) apply for surface and underground transport, storage and handling, and any exposure during installation and integration underground.

Details of \dshort{pd} integration into the cathode and installation into the cryostat, including quality management testing equipment, tests, and documentation are included in Chapter~\ref{ch:IEI}.

\section{Quality Assurance and Quality Control}
\label{sec_QAQC}

The \dshort{qa} and \dshort{qc} programs as well as the design-phase quality assurance and validation plans 
%for the \dword{fd} 
are based on our experience with the \dshort{pdsp}. The \dshort{spvd} \dshort{pds} quality assurance program is focused on final specifications and drawings, and developing a formal set of fabrication procedures including a detailed set of \dshort{qc} procedures. 

During fabrication, integration into the detector, and detector installation into the cryostat, the \dshort{spvd} \dshort{pds} \dshort{qc} plan will be carefully followed, including incoming materials and other inspection reports, fabrication travelers, and formal test result reports entered into the \dshort{dune} \dshort{qa}/\dshort{qc} database.

Steps in this process are detailed below.

\subsection{Design Quality Assurance}
\label{sec:fdsp-pd-designqa}

\dshort{pd} design \dshort{qa} focused on ensuring that the detector modules meet the following goals:
\begin{itemize}
\item Physics goals as specified in the \dshort{dune} requirements document;
\item Interfaces with other detector subsystems as specified by the subsystem interface documents; and
\item Materials selection and testing to ensure non-contamination of the \dshort{lar} volume.
\end{itemize}

\dshort{qa} for full system prototypes is underway at multiple cryogenic test sites during the design to ensure the module design achieves requirements.  In particular, a series of tests is ongoing at the \dshort{cern} \coldbox at NP02, as detailed in Section~\ref{sec:PDS_validation}.

In addition, the lifetime of all electrical components that will be located inside the cryostat must be established.  Most electronics failure mechanisms can be characterized by a (positive) activation energy and are greatly suppressed at cryogenic temperatures.  One notable exception is the hot electron effect~\cite{Hot-electron} that can limit the lifetime of NMOS transistors.  For this reason, all CMOS circuits must either be designed to mitigate this damage mechanism, or be operated at reduced bias voltage.
Another exception is damage to components caused by material CTE mismatch; this mechanism is especially important for capacitors~\cite{nasa_nepp}.

The circuit used to read out cathode-mounted \dshort{pds} units includes a CMOS operational amplifier that may be difficult to qualify for resistance to the hot electron effect.  High priority will be given to qualifying this part~\cite{Chen:2018zic} or finding a replacement part.
A number of capacitor failures have occurred during \coldbox tests.  High priority will also be given to establishing a procedure for selecting and qualifying capacitors.

The \dshort{pds} consortium will perform design and fabrication of components in accordance with applicable requirements as specified in the relevant \dshort{lbnf}-\dshort{dune} \dshort{qa} plan. If an institution (working under the supervision of the consortium) performing the work has a previously-existing documented \dshort{qa} program meeting PD consortium requirements, work may be performed in accordance with their own existing program.
As part of the final design review and \dword{prr} process, the reviewers will be charged to ensure that the design demonstrates compliance with the goals above.

\subsection{Production and Assembly Quality Assurance}
\label{sec:fdsp-pd-prodqa}

The \dshort{pds} will undergo a \dshort{qa} review for all components prior to completion of the design and development phase of the project.  The \dshort{vdmod0} test will represent the most significant test of near-final \dshort{pd} components in a near-\dshort{dune} configuration, but additional tests will also be performed.  The \dshort{qa} plan will include, but not be limited to, the following areas:

\begin{itemize}
\item Materials certification (in the \dword{fnal} materials test stand and other facilities) to ensure materials compliance with cleanliness requirements;
\item Cryogenic testing of all materials to be immersed in \dshort{lar}, to ensure satisfactory performance through repeated and long-term exposure to \dshort{lar}.  Special attention will be paid to cryogenic behavior of fused silica and plastic materials (such as filter plates and wavelength-shifters), \dshorts{sipm}, cables and connectors, optical fibers and connections, and all electronics and optics operated cryogenically.  Testing will be conducted both on small-scale test assemblies (including small cryostats and dewars at institutions throughout the consortium) and full-scale prototypes (including mechanical testing at the large CDDF dewar at CSU, and operational testing at \dshort{cern} \coldbox{}es, \dword{iceberg} at Fermilab, and other large cryostats available to the consortium).
\item Mechanical interface testing, beginning with simple mechanical go/no-go gauge tests, followed by installation into the \dshort{vdmod0} system, and finally full-scale interface testing of the \dshort{pds} into the final pre-production \dshort{tpc} system models; and
\item Full-system readout tests of the \dshort{pd} readout electronics, including trigger generation and timing, including tests for electrical interference between the \dshort{tpc} and \dshort{pd} signals.
\end{itemize}

Prior to beginning construction, the \dshort{pds} will undergo a final design review and \dword{prr}, where the planned \dshort{qa} tests will be reviewed, and the system declared ready to move to the production phase.

\subsection{Production and Assembly Quality Control}
\label{sec:fdsp-pd-prodqc}

Prior to the start of fabrication, a manufacturing and \dshort{qc} plan will be developed detailing the key manufacturing, inspection, and test steps.  The fabrication, inspection, and testing of the components will be performed in accordance with documented procedures. 
For example, vendors will provide the I-V curve of each SiPM at room temperature and test 10\% of each lot (1 lot is about 100 flexi boards) at 77~K. The PDS consortium will test the boards before and after three thermal cycles and check the I-V curve in reverse bias and the single SiPM response to LED light of the flexi-boards.

The work will be documented on travelers and applicable test or inspection reports. Records of the fabrication, inspection and testing will be maintained. When a component has been identified as being in noncompliance to the design, the nonconforming condition shall be documented, evaluated, and dispositioned as: \textit{use-as-is} (does not meet design but can meet functionality as it is), \textit{rework} (bring into compliance with design), \textit{repair} (will be brought to meet functionality but will not meet design), and \textit{scrap}. For products with a disposition of accept, as is, or repair, the nonconformance documentation shall be submitted to the design authority for approval.

All \dshort{qc} data  (from assembly and pre- and post-installation into the cryostat) will be directly stored to the \dshort{dune} database for ready access of all \dshort{qc} data.  Monthly summaries of key performance metrics (to be defined) will be generated and inspected to check for quality trends.

Based on the \dshort{sphd} \dshort{pds} model, we expect to conduct the following \dshort{spvd} \dshort{pds} production testing:

Prior to shipping from assembly site:
\begin{itemize}
\item Dimensional checks of critical components and completed assemblies to ensure satisfactory system interfaces;
\item Post-assembly cryogenic checkouts of \dshort{sipm} mounting flex \dshort{pcb}s (prior to assembly into \dshort{pd} modules);
\item Module dimensional tolerances using go/no-go gauge set;
\item Room temperature scan of complete module using motor-driven \dshort{led} scanner (or UV \dshort{led}  array), with the final electronics in place. The readout electronics for cathode-mount \dshort{pd} modules will be modified to allow them to be tested at room temperature; and
\item DAQ tests using DAPHNE: communication check of a test pattern generated inside DAPHNE, confirmation of \dshort{sipm} bias voltage settings, and a check of the digitization of known injected charges. 
\end{itemize}

Following shipping to the US reception and checkout facility but prior to storage at \dshort{sdwf}:
\begin{itemize}
\item Mechanical inspection;
\item Room temperature scan (using identical scanner to initial scan); and
\item Cryogenic testing of completed modules (in CSU CDDF or similar facility).
\end{itemize}

\subsection{Installation Quality Control}
\label{sec:fdsp-pd-installqc}

\dshort{pds} pre-installation testing will follow the model established for \dshort{sphd} \dshort{pds}.  Prior to installation in the cathode and on the membrane walls, \dshort{pd} modules will undergo a room temperature scan in a scanner identical to the one at the \dshort{pd} module assembly facility and the results compared to results collected during fabrication.  In addition, the module will undergo a complete visual inspection for defects and a set of photographs of selected critical optical surfaces taken and entered into the \dshort{qc} record database.  

Following installation into the cathode, an immediate check for optical fiber continuity will be conducted.
Spare fibers will be installed in every feedthrough, so if a bad fiber is found at this point, a spare will be used (before the cathode is raised).
Following the mounting of the \dshort{pd} module on the membrane wall, an immediate check of cable continuity will be conducted.  During this test, the \dshort{pds} system will undergo a final integrated system check for expected warm response for all channels, electrical interference with other electronics,  compliance with the detector grounding scheme, and power consumption.

%%%%%%%%%%%%%%%%%%%%

\section{Safety}
\label{sec_safety}

Safety management practices will be critical for all phases of the \dshort{spvd} \dshort{pds} assembly, and testing.  Planning for safety in all phases of the project, including fabrication, testing, and installation will be part of the design process.  The initial safety planning for all phases will be reviewed and approved by safety experts and the DUNE safety management team in the Project Office as part of the Final Design Review and the \dwords{prr}.  All component cleaning, assembly, testing,  and installation procedure documentation will include a section on safety concerns relevant to that procedure and will also be reviewed during the design reviews.

Areas of particular importance to the \dshort{spvd} \dshort{pds} include:
\begin{description}
\item[Hazardous chemicals and cleaning compounds:]  All potentially hazardous chemicals used (particularly \dshort{wls} chemicals such as \dword{ptp} used in filter plate coating) will be documented at the consortium management level, with materials data safety sheets (MSDS) and approved handling and disposal plans in place.

\item[Class-4 laser hazards associated with the \dshort{pof} %power-over-fiber 
installation:] Class 4 is the highest and most dangerous class of laser; by definition, a class 4 laser can burn the skin, or cause devastating and permanent eye damage as a result of direct, diffuse or indirect beam viewing. These lasers may ignite combustible materials, and thus may represent a fire risk. Class 4 lasers must be equipped with a key switch and a safety interlock. All class 4 laser operation, particularly as part of installation testing when personnel from other subsystems will be present, will be documented at the consortium management level and activities will carefully coordinated with the safety experts.

\item[Liquid and gaseous cryogens used in module testing:] Full hazard analysis plans will be in place at the consortium management level for all module or module component testing involving cryogenic hazards, and these safety plans will be reviewed in the appropriate pre-production and production reviews.

\item[High voltage safety:] Some of the candidate \dshorts{sipm} may require bias voltages above \SI{50}{VDC} during warm testing (although not during cryogenic operation), which may be a regulated voltage as determined by specific laboratories and institutions.  Fabrication and testing plans will demonstrate compliance with local \dword{hv} safety requirements at the particular institution or laboratory where the testing or operation is performed, and this compliance will be reviewed as part of the standard review process.

\item[UV and \dword{vuv} light exposure:] Some \dshort{qa} and \dshort{qc} procedures used for module testing and qualification may require use of UV and/or \dword{vuv} light sources, which can be hazardous  to unprotected operators.  Full safety plans must be in place and reviewed by consortium management prior to beginning such testing.

\item[Working at heights, underground:]  Some aspects of \dshort{spvd} \dshort{pds} module fabrication, testing and installation will require working at heights or deep underground. Personnel safety will be an important factor in the design and planning for these operations, all procedures will be reviewed prior to implementation, and all applicable safety requirements at the relevant institutions will be observed at all times.

\end{description}

%%%%%%%%%%%%%%%%%%%%

\section{Organization and Management}
\label{sec_OrgMan}

The \dshort{pd} consortium benefits from the contributions of many institutions and facilities in Europe and North and South America.  Table~\ref{tab:sp-pds-institutes-i}
%and \ref{tab:sp-pds-institutes-ii} 
lists the member institutions. 

%%%%%%%%%%%%%%%%%%%%%%%%%%%%%%%%%%%
\subsection{Consortium Organization}
\label{sec:fdsp-pd-org-consortium}

\begin{longtable}
{ll}
\caption{PDS consortium institutions.}\\ \colhline
\rowcolor{dunetablecolor} Member Institute  &  Country       \\  \toprowrule
Federal University of ABC$^\dagger$ & Brazil \\ \colhline
Federal University of Alfenas Po\c{c}os de Caldas & Brazil \\ \colhline
%State University of Feira de Santana & Brazil \\ \colhline
Centro Brasileiro de Pesquisas F\'isicas & Brazil \\ \colhline
Federal University of Goi\'as & Brazil \\ \colhline
Brazilian Synchrotron Light Laboratory LNLS/CNPEM & Brazil \\ \colhline
University of Campinas$^\dagger$ & Brazil \\ \colhline
CTI Renato Archer & Brazil \\ \colhline
Federal Technological University of Paran\'a & Brazil \\ \colhline
Instituto Tecnol\'{o}gico de Aeron\'{a}utica$^\dagger$ & Brazil \\ \colhline
University Antonio Nari\~{n}o$^\dagger$ & Colombia \\ \colhline
Universidad del Atlantico$^\dagger$ & Colombia \\ \colhline
University EIA$^\dagger$ & Colombia \\ \colhline
Universidad Sergia Ablada & Colombia \\ \colhline
Institute of Physics CAS & Czech Republic \\ \colhline
Czech Technical University in Prague & Czech Republic \\ \colhline
Laboratoire APC$^\dagger$ & France \\ \colhline
University of Bologna and INFN$^\dagger$ & Italy \\ \colhline
University of Milano and INFN$^\dagger$ & Italy \\ \colhline
University of Milano Bicocca and INFN$^\dagger$ & Italy \\ \colhline
%University of Genova and INFN & Italy \\ \colhline
University of Insubria and INFN & Italy \\ \colhline
%University of Catania and INFN & Italy \\ \colhline
Laboratori Nazionali del Sud & Italy \\ \colhline
%University of Lecce and INFN & Italy \\ \colhline
University of Naples "Federico II" and INFN$^\dagger$ & Italy \\ \colhline
University of Ferrara and INFN$^\dagger$ & Italy \\ \colhline
INFN Padova & Italy \\  \colhline
INFN Pavia$^\dagger$ & Italy \\ \colhline
Nikhef$^\dagger$ & Netherlands \\ \colhline
Universidad Nacional de Assuncion$^\dagger$ & Paraguay \\ \colhline
Comisi\'{o}n Nacional de Investigaci\'{o}n y Desarrollo Aeroespacial$^\dagger$  & Per\'{u} \\ \colhline
Pontificia Universidad Catolica Per\'{u} & Per\'{u} \\ \colhline
Universidad Nacional de Ingineria$^\dagger$ & Per\'{u} \\ \colhline
Chung Ang University$^\dagger$ & South Korea \\ \colhline
CIEMAT$^\dagger$ & Spain \\ \colhline
IFIC (CSIC and University of Valencia)$^\dagger$ & Spain \\ \colhline
University of Granada$^\dagger$ & Spain \\ \colhline
Edinburgh University & UK \\ \colhline
Argonne National Laboratory$^\dagger$ & USA \\\colhline
Boston University$^\dagger$ & USA \\ \colhline
Brookhaven National Laboratory$^\dagger$ & USA \\ \colhline
%California Institute of Technology & USA \\ \colhline
University of California, Santa Barbara$^\dagger$ & USA \\ \colhline
Colorado State University$^\dagger$   &  USA  \\ \colhline
\dshort{fnal}$^\dagger$    &   USA    \\ \colhline
University of Illinois - Urbana-Champaign$^\dagger$ & USA \\ \colhline
Indiana University$^\dagger$ & USA \\ \colhline
University of Iowa$^\dagger$ & USA \\ \colhline
Lawrence Berkeley National Laboratory$^\dagger$ & USA \\ \colhline
University of Michigan$^\dagger$ & USA \\ \colhline
Northern Illinois University$^\dagger$ & USA \\ \colhline
South Dakota School of Mines and Technology$^\dagger$ & USA \\ \colhline
Stony Brook University$^\dagger$ & USA \\ \colhline
Syracuse University & USA \\ \colhline
\label{tab:sp-pds-institutes-i}
\end{longtable}

The %\single 
\dshort{pds} consortium has an organizational structure as follows:
\begin{itemize}
\item A consortium lead provides overall leadership for the effort and attends meetings of the \dshort{dune} Executive and Technical Boards.
\item A deputy consortium lead, who provides support to the consortium lead and has specific responsibility for oversight of the \dshort{spvd} \dshort{pds}.
\item Two technical leads, one with primary responsibility for mechanical systems, and one with primary responsibility for electronic and electrical systems. The technical leads provide technical support to the consortium lead and deputy, attend the Technical Board and other project meetings, oversee the project schedule and \dword{wbs}, and oversee the operation of the project working groups.  
\item A Project Management Board composed by the project leads from the participating countries, the consortium leadership team and a few {\it ad hoc} members. The Board maintains tight communication between the countries participating in the consortium construction activity.
\end{itemize}

Below the leadership, the consortium is divided up into six working groups, each led by two or three working group conveners. %(see Table~\ref{tbl:pds-wgs}).  
Each working group is charged with one primary area of responsibility within the consortium, and the conveners report directly to the Technical Lead regarding those responsibilities. Responsible institutions have been identified for all aspects of \dshort{pds} detector fabrication as indicated in Table~\ref{tab:sp-pds-institution-deliverables}.
A Memorandum of Understanding has been signed by representatives of all participating institutions and international supporting funding agencies with commitments for delivery on schedule. 

The working group conveners are appointed by the \dshort{pds} consortium lead and technical lead; the structure may evolve as the consortium matures and additional needs are identified.

\begin{longtable}
{p{0.20\textwidth}p{0.45\textwidth}p{0.20\textwidth}}
\caption{PDS deliverables and responsible institutions}\\ \colhline
\rowcolor{dunetablecolor} Subsystem & Description & Contributing Institutions  \\  \toprowrule

Photosensors (\dshort{sipm}) & 
Validation, procurement and component testing for all cathode and membrane modules. & 
INFN-MiB, INFN-Bo, CIEMAT, IFIC, U. Granada, U. Prague \\

\colhline
Dichroic Filters & 
Validation, procurement, primary wavelength shifting thin-film deposit, and component testing for all cathode and membrane modules. &
INFN-MiB, INFN-Na, CIEMAT, IFIC, U. Granada, UNICAMP, UFABC, ITA \\
\colhline

\dshort{wls} Plates & 
Validation, procurement, groove milling and component testing for all cathode and membrane modules. &
INFN-MiB, INFN-Na \\
\colhline

\dshort{xarapu} Mechanical Frames & 
Procurement, fabrication, construction. &
CSU, Iowa, NIU \\
\colhline

\dshort{sipm} Mounting and Ganging Cold Electronics & 
Validation, procurement, cold FlexBoards fabrication and testing, cold FlexBoard Integration (\dshort{sipm} mounting \& ganging electronics) for all cathode and membrane modules. &
UCSB, INFN-MiB \\
\colhline

Signal-Shaping Cold Electronics & 
Component down-select, validation, procurement, cold motherboard integration (signal conditioning stage) of all cathode and membrane modules. &
FNAL, UCSB, APC-Paris, LBL, U. Mich, Iowa, NIU
 \\
\colhline

\dshort{sof}: Electrical-to-Optical Conversion Cold Electronics & 
Component down-select, procurement, cold motherboard integration (SoF stage – analog signal transmitters and driver circuitry) for cathode modules. &
FNAL, UCSB,
APC-Paris
 \\
\colhline

\dshort{sof}: Optical-to-Electrical Conversion Warm Electronics & 
Component down-select, procurement, warm board fabrication, construction, integration (analog signal optical-to-electrical conversion stage) for cathode modules. &
APC-Paris \\
\colhline

Signal Digitization Warm Electronics & 
Warm board construction (signal digitizer, DAQ interface, online software) for cathode and membrane modules. &
INFN-MiB, INFN-Bo, CIEMAT, IFIC, U. Granada, SBU \\
\colhline

\dshort{pof}: Warm Transmitter and Cold Receiver & 
PoF Laser Module down-select, procurement, Warm Laser Box fabrication and integration, Photovoltaic power converter design, selection, procurement, cold motherboard integration (PoF stage – power optical-to-electrical conversion) for cathode modules. &
FNAL, UIUC, SDSMT \\
\colhline

\dshort{pof}: Step-Up Cold Electronics & 
\dshort{pof} Voltage StepUp Cold Electronics. &
FNAL, Iowa, LBL, BNL,
INFN-Mi \\
\colhline

Cathode PDS Fibers, Cables, Flanges & 
Procurement, fabrication, construction, integration of all fibers, flanges, and feedthroughs for cathode modules. &
FNAL, NIU, UMich, SDSMT, SBU, CAU \\
\colhline

Membrane PDS Fibers, Cables, Flanges & 
Procurement, fabrication, construction, integration of all cables, flanges and feedthroughs for membrane modules. &
CIEMAT, IFIC \\
\colhline

Cathode PDS Integration in Cathode systems & 
Procurement of electrical (cables \& balun) and mechanical cathode mounting solutions, interface mechanics, integration of cathode modules. &
BNL, FNAL, CSU, NIU, Iowa \\
\colhline

Membrane PDS Support Structure Mechanics & 
Procurement and fabrication of cryostat mounting solutions, interface mechanics, integration of membrane modules. &
CIEMAT \\
\colhline

Cathode PDS Response and Monitoring System & 
LED flashers and diffuser procurement, fabrication, integration of fibers, flanges, and feedthroughs for cathode response and monitoring system kit. &
ANL, CIEMAT, SDSMT \\
\colhline

Membrane PDS Response and Monitoring System  & 
Procurement, fabrication, construction, integration of all fibers, flanges and feedthroughs for the response and monitoring system kits. &
ANL, CIEMAT \\
\colhline

Detector Components Qualification in Cold & 
Test Stands (7) construction and operation: 

1.	Cold electronics components %30-yr lifetime CE qualification

2.	\dshort{sipm} cold FlexBoards

3.	PoF and fibers components

4.	Cold motherboards signal cables
&
BNL, SBU, INFN-MiB, UCSB, SDSMT, FNAL, NIU, UMich \\
\colhline

Production QA & 
\dshort{sipm}, \dshort{wls}, Dichroic Filter. &
INFN-Bo, INFN-FE, INFN-MiB, INFN-Na, CIEMAT, IFIC, U. Granada, UNICAMP \\
\colhline

\dshort{xarapu} Integrated Module and Electronics Assembly & 
Cathode and membrane modules. &
NIU, CIEMAT, U. Granada \\
\colhline

Production QC & 
Test Stand construction and operation:

1.	\dshort{xarapu} integrated cathode Modules

2.	\dshort{xarapu} integrated membrane Modules
 &
1.	CSU, FNAL, NIU

2.	CIEMAT, INFN-Na, INFN-Pv, INFN-MiB
 \\
\colhline

\label{tab:sp-pds-institution-deliverables}
\end{longtable}

\subsection{High-Level Schedule}
\label{sec:fdsp-pd-org-cs}

Table \ref{tab:PDS_milestones} lists key milestones in the design, validation, construction, and installation of the \dshort{spvd} \dshort{pds}.  This list includes external milestones indicating linkages to the main \dshort{dune} schedule (highlighted in color in the table), as well as internal milestones such as design validation and technical reviews.

In general, the flow of the schedule commences with a 60\% design review based on module performance testing at \dshort{pds} consortium test stands and integration testing at the \dshort{cern} \coldbox.  Additional similar design validation follows, leading to a final design review (FDR).  Following the FDR, 16 \dshort{xarapu}s and required electronics, cabling, fibers, and \dshort{pd} monitoring system components for \dshort{vdmod0} will be built, installed, and validated at \dshort{cern}.  Once the data from \dshort{vdmod0} have undergone initial analysis, production readiness reviews will be conducted and module fabrication will begin.

Some parts of the \dshort{spvd} \dshort{pds} system have a long procurement cycle and will require an accelerated design review process, as shown in the milestone table. This is the case for the \dshorts{sipm}, required in mid-2024 for flex \dshort{pcb} assembly.

\begin{dunetable}
[PDS milestones]
{p{0.8\textwidth}p{0.13\textwidth}}
{tab:PDS_milestones}
{Photon Detection System Milestones}
Milestone & Date   \\ \toprowrule
Ready to begin \dshort{pds} install at \dshort{cern} \dshort{vdmod0}	& Dec 2022\\ \colhline
\dshort{pds} installation complete at \dshort{cern} \dshort{vdmod0} & Mar 2023\\ \colhline
{\bf Ready for Final Design Review	}&{\bf  Apr 2023}\\ \colhline
Full Detection-to-Digitization Test complete & Oct 2023\\ \colhline
Ready for \dshort{sipm}s Production Readiness Review	& Jan 2024\\ \colhline
Ready for \dshort{xarapu} Production Readiness Review	& Mar 2024\\ \colhline
Ready for Warm Electronics Production Readiness Review	& May 2024\\ \colhline
Ready for Fibers, Cables, and Feedthroughs Production Readiness Review	& July 2024\\ \colhline
{\bf All Production Readiness Reviews complete	}&{\bf  Aug 2024}\\ \colhline
Warm electronics units construction complete	& Feb 2026\\ \colhline
Cold electronics (Sof \& PoF) construction complete	& Sep 2025\\ \colhline
Cathode and Membrane \dshort{xarapu} modules construction complete	& May 2027\\ \colhline
{\bf Photon Detection System Ready for Installation}	& May 2027\\ \colhline
Membrane \dshort{xarapu} and Response and Monitoring system installed	& Oct 2027\\ \colhline
Cathode \dshort{xarapu}  installed	& Feb 2028\\ \colhline
{\bf Photon Detection System commissioned}&{\bf  May 2028}\\\colhline
{\bf Photon Detection System project complete	}&{\bf  Jul 2028}\\ 
\end{dunetable}

A more detailed schedule for production and installation of the \dshort{spvd} is found in Figure~\ref{fig:pds_schedule}.
\begin{dunefigure}[Key \dshort{pds} milestones and activities toward 
\dshort{spvd}]{fig:pds_schedule}{
Key \dshort{pds} milestones and activities toward  the \dshort{spvd} in graphical format (Data from~\cite{docdb22261v28}).}
\includegraphics[width=1.0\textwidth]{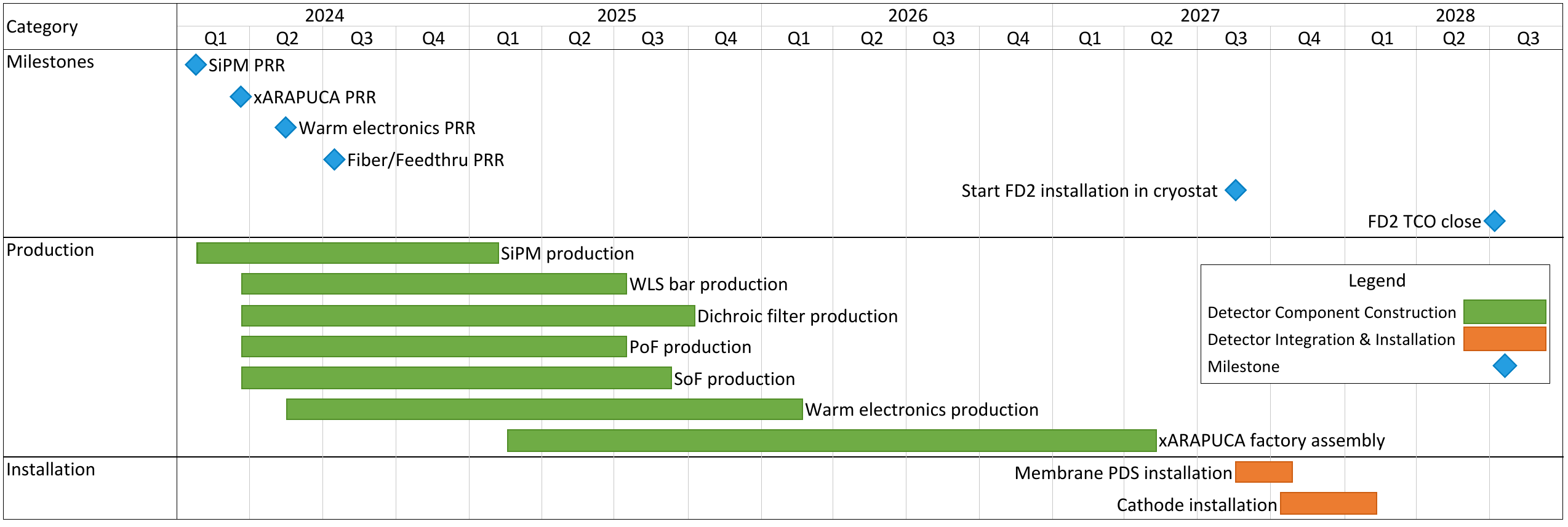}
\end{dunefigure}

%%%%

\chapter{Trigger and DAQ}
\label{sec:DAQ}

%%%%
\section{Introduction}
\label{subsec:INss}

The trigger and \dword{daq} (\dword{tdaq}) system is responsible for receiving,
processing, and recording data from the DUNE experiment.
For the \dword{spvd} this system:
\begin{itemize}
\item provides timing and synchronization to the detector electronics and calibration devices; 
\item receives and buffers data streaming from the \dword{tpc} top and bottom electronics and the \dword{pds}; 
\item extracts information from the data at a local level to subsequently form \dwords{trigdecision}; 
\item builds \dwords{trigrec}, defined as a collection of data and metadata from selected detector space-time volumes corresponding to a \dshort{trigdecision};
\item carries out additional data reduction and compression as needed; and
\item  relays \dshort{trigrec}s to permanent storage.
\end{itemize}

The main challenge for the DUNE \dshort{tdaq} lies in the development of effective, resilient software and firmware that optimize the performance of the underlying hardware. The design is driven not only by data rate and throughput considerations, but also---and predominantly---by the stringent uptime requirements of the experiment. 

The \dshort{tdaq} is subdivided into a set of subsystems.  Figure~\ref{fig:dq_subsys_view} shows the different subsystems and their relationships. All subsystems rely on the functionality provided by the \textit{\dword{daqccm}} and \textit{\dword{dqm}}, that is the glue of the overall \dshort{tdaq}, transforming the set of components into a coherent system.  
The \textit{\dword{daqtrs}} and \textit{Data filter}  are in charge of the selection and compression of data. The \textit{Dataflow} subsystem provides the communication layer to exchange data (i.e., it is used by the other subsystems intersecting it in the diagram). In addition, it implements the data collection functionality, i.e., the logic for building \dshort{trigrec}s as well as the organization of data into files. 
The \textit{\dword{daqros}} receives the data streams from the \dword{tpc} and \dword{pds}, processes them to extract information for the trigger, and buffers data while the trigger is forming a decision. The \textit{\dword{daqdts}} is in charge of distributing the clock, synchronizing the \dwords{detmodule}, as well as applying timestamps to hardware signals that may be used for triggering, such as calibration pulses. 
The \textit{Detector electronics software} is not part of the \dshort{tdaq} responsibility: it is shown in the diagram to indicate that detector experts will develop the software to configure, control, and monitor the electronics using the tools provided by the \textit{\dshort{daqccm}} subsystem.

%$$$$$$$$$$$$$$$  
\begin{dunefigure}
	[\dshort{tdaq} subsystems diagram]
	{fig:dq_subsys_view}
	{Diagram showing the relationships between \dshort{tdaq} subsystems. Overlaps indicate dependence on another subsystem, e.g., readout depends on the software communication layer provided by dataflow. All subsystems rely on control, configuration and monitoring libraries and framework provided by \dshort{daqccm}. It also shows, in grey, the detector electronics software, which is not part of \dshort{tdaq} but will be developed fully embedded into the \dshort{daqccm} environment.}
	\includegraphics[width=.6\textwidth]{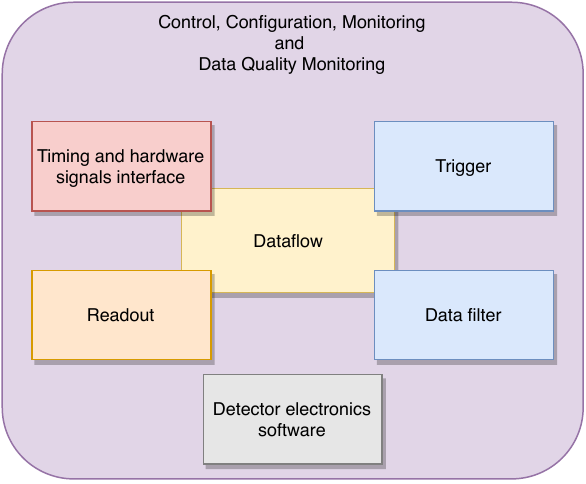}
\end{dunefigure}
%$$$$$$$$$$$$$$$ 

The physical components of the \dshort{tdaq} are primarily \dword{cots} components---servers, switches, fibers, etc.---as shown in Figure~\ref{fig:dq_phys_view}. A high performance Ethernet network interconnects all the elements and allows them to operate as a single, distributed system. At the output of the \dshort{tdaq} the high-bandwidth \dword{wan} allows the transfer of data from the \dword{surf} to \dword{fnal}. 

%$$$$$$$$$$$$$$$  
\begin{dunefigure}
	[Physical view diagram of the \dshort{tdaq} system. ]
	{fig:dq_phys_view}
	{Physical view diagram of the \dshort{tdaq} system. At \dword{surf}, the hardware components are distributed across the detector caverns and the surface \dshort{daq} room. The \dshort{tdaq} data are transferred to \dword{fnal} over a \dword{wan} connection.}
	\includegraphics[width=0.9\textwidth]{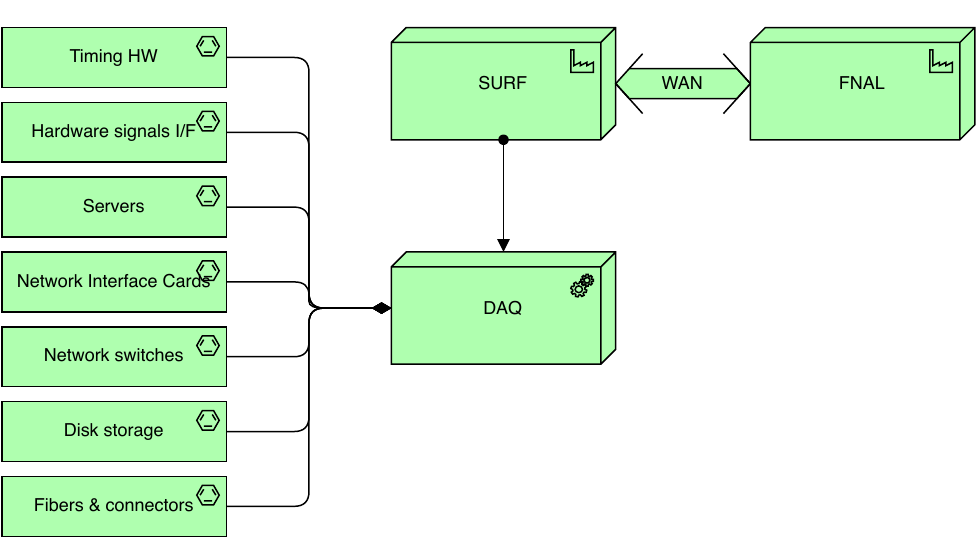}
\end{dunefigure}
%$$$$$$$$$$$$$$$ 

The \dshort{tdaq} for the whole of DUNE is designed and developed coherently by a joint team for both \dword{fd} and \dword{nd}. The \dshort{tdaq} systems for the \dword{nd} and the different \dword{fd} modules differ only in minor details so as to support the electronics and the data selection criteria for each. 

In particular, the \dword{spvd} module \dshort{tdaq} system will be very similar to that for the \dword{sphd} detector module, but with customizations for the top and bottom anode planes and the \dword{pd} electronics readout, as well as for the data selection algorithms, which will be tuned to the module's geometry. The main \dshort{tdaq} features for the \dword{fd} are described in Volume IV of the DUNE \dword{tdr}~\cite{Abi:2017aow}. 

The chapter begins with an overview of the \dshort{daq} design requirements (Section~\ref{sec:daq:spec}) that the design must meet and specifications for interfaces between the \dshort{daq} and other DUNE \dword{fd} systems. Subsequently, Section~\ref{subsec:OVss}, which makes up the majority this chapter, describes the design of the \dword{fd} \dshort{daq} in greater detail. Section~\ref{sec:tdaq:des-val-dev} describes design validation efforts to date, as well as future design development and validation plans. At the center of these efforts is the \dword{protodune} \dshort{daq} system (described in Section~\ref{subsec:tdaq:pduneii}), which has demonstrated several key aspects of the DUNE \dword{fd} \dshort{daq} design and continues to serve as a platform for further developing and validating the final design. The chapter finishes with Section ~\ref{sec:tdaq:org-mgmt},
which detail the management of the \dshort{daq} project, including the schedule for completing the design, production, and installation of the system, as well as %safety 
risk considerations.

%%%%%%%%%%%%%%%%%%%%%%%%%%
\section{Requirements and Specifications}
\label{sec:daq:spec}

The specifications for the \dword{tdaq} derive from higher-level requirements~\cite{edms-198204}. Table~\ref{tab:specs:SP-FD2} %tab:specs:FD2-VD-EBheld} 
lists the specifications that are critical to the \dword{fd} performance, including two relating to the \dshort{tdaq} (FD-22 and 23). 
These two and the additional \dshort{tdaq} specifications are listed in Table~\ref{tab:specs:SP-DAQ}. 
Note that the differences in \dshort{tdaq} requirements due to detector configuration and technology between the \dword{sphd} and \dword{spvd} are very limited, amounting only to a slightly larger total data rate of 1.8~TB/s for \dword{spvd} as opposed to 1.4~TB/s for \dword{sphd}, and the necessity of incorporating the \dword{tde}.  The requirement for data rate to tape of 30 PB/year is across all \dword{fd} modules.

% This file is generated, any edits may be lost.
\begin{footnotesize}
%\begin{longtable}{p{0.14\textwidth}p{0.13\textwidth}p{0.18\textwidth}p{0.22\textwidth}p{0.20\textwidth}}
\begin{longtable}{p{0.10\textwidth}p{0.22\textwidth}p{0.14\textwidth}p{0.23\textwidth}p{0.21\textwidth}}
\caption{TDAQ specifications %\fixmehl{ref \texttt{tab:spec:SP-DAQ}}
} \\
  \rowcolor{dunesky}
       Label & Description  & Specification \newline (Goal) & Rationale & Validation \\  \colhline

  \newtag{FD-22}{spec:data-rate-to-tape}& Data rate to tape  &  $<\,\SI{30}{PB/year}$ &  Cost.  Bandwidth. &  ProtoDUNE \\ \colhline

  \newtag{FD-23}{ spec:sn-trigger }  & Supernova trigger  & Efficiency for a SNB producing at least 60 interactions with a $\nu$ energy $>$\,10 MeV in 12\,kt of active detector mass during the first 10\,s of the burst. &  $>\,$95\% efficiency for SNB within 20 kpc &  Simulation and bench tests  %\fixme{C. Cuesta, is 12 kton correct? the fiducial/total mass (10/17 kton) of the FD2-VD could be in this table (AH 20 Dec '22: comes from EB held reqs; let's not change it)} 
  \\ \colhline

  \newtag{FD-DAQ-1}{ spec:DAQ-readout }  & DAQ readout throughput: The DAQ shall be able to accept the continuous data stream from the TPC and PDs.  &  1.8 TB/s per FD module &  Specification from TPC and PDS electronics &  Modular test on Proto\-DUNE; overall throughput scales linearly with number of APAs \\ \colhline
  % modified for spvd

  \newtag{FD-DAQ-2}{ spec:DAQ-throughput }  & DAQ storage throughput: The DAQ shall be able to store selected data at an average throughput of 10 Gb/s, with temporary peak throughput of 100 Gb/s.  &  10 Gb/s average storage throughput; 100 Gb/s peak temporary storage throughput per single phase detector module &  Average throughput estimated from physics and calibration requirements; peak throughput allowing for fast storage of SNB data ($\sim 10^4$ seconds to store 180 TB of data).  &  ProtoDUNE demonstrated steady storage at $\sim$ 40 Gb/s for a storage volume of 700 TB. Laboratory tests will allow to demonstrate the performance reach. \\ \colhline

  \newtag{FD-DAQ-3}{ spec:DAQ-readout-window }  & DAQ readout window: The DAQ shall support storing triggered data with a variable size readout window, from few $\mu$s (calibration) to 100 s (SNB), with a typical readout window for triggered interactions of 4.25 ms.  &  10 $\mu$s < readout window < 100 s &  Storage of the complete dataset for up to 100 s is required by the SNB physics studies; the typical readout window of 4.25 ms is defined by the drift time in the detector; calibration triggers can be configured to read out data over much shorter time intervals. &  Implementation techniques to be validated on the ProtoDUNE setup and in test labs. \\ \colhline
  % modified for spvd

  \newtag{FD-DAQ-4}{ spec:trigger-calibration }  & Calibration trigger: The DAQ shall provide the means to distribute time-synchronous commands to the calibration systems, in order to fire them, at a configurable rate and sequence and at configurable intervals in time. Those commands may be distributed during physics data taking or during special calibration data taking sessions. The DAQ shall trigger and acquire data at a fixed, configurable interval after the distribution of the commands, in order to capture the response of the detector to calibration stimuli.  &   &  Calibration is essential to attain required detector performance comprehension. &  %Techniques for doing this have been run successfully in MicroBooNE and ProtoDUNE.  
  Calibration techniques for these purposes have been operated successfully in MicroBooNE and ProtoDUNE.  
  \\ \colhline

  \newtag{FD-DAQ-5}{ spec:data-record }  & Data record: Corresponding to every trigger, the DAQ shall form a data record to be transferred to offline together with the metadata necessary for validation and processing.  &   &  Needed for offline analysis. &  Common experimental practice. \\ \colhline

  \newtag{FD-DAQ-6}{ spec:data-verification }  & Data verification: The DAQ shall check integrity of data at every data transfer step. It shall only delete data from the local storage after confirmation that data have been correctly recorded to permanent storage.  &   &  Data integrity checking is fundamental to ensure data quality. &   \\ \colhline

  \newtag{FD-DAQ-7}{ spec:trigger-high-energy }  & High-energy Trigger: The DAQ shall trigger and acquire data on visible energy deposition >100 MeV. Data acquisition may be limited to the area in which activity was detected.  &  $>$\SI{100}{\MeV} &  Driven by DUNE physics mission. &  Physics TDR. 100 MeV is an achievable parameter; lower thresholds are possible. \\ \colhline

  \newtag{FD-DAQ-8}{ spec:trigger-low-energy }  & Low-energy Trigger: The DAQ shall trigger and acquire data on visible energy deposition > 10 MeV of single neutrino interactions. Those triggers will normally be fired using a pre-scaling factor, in order to limit the data volume.  &  $>$\SI{10}{\MeV} &  Driven by DUNE physics mission. &  Physics TDR. 10 MeV is an achievable parameter; lower thresholds are possible. \\ \colhline

  \newtag{FD-DAQ-9}{ spec:daq-deadtime }  & DAQ deadtime: While taking data within the agreed conditions, the DAQ shall be able to trigger and acquire data without introducing any deadtime.  &  (Zero deadtime) &  Driven by DUNE physics mission. &  %Zero deadtime is an achievable inter-event deadtime but a small deadtime would not significantly compromise physics sensitivity.  
  Zero inter-event deadtime is achievable. A small deadtime would not significantly compromise physics sensitivity. 
  \\ \colhline

\label{tab:specs:SP-DAQ}
\end{longtable}
\end{footnotesize}

Putting these into context, these specifications 
can be viewed as falling into four categories:
\begin{itemize} 
	\item Synchronization: 
	\begin{itemize}
		\item The timing system shall provide a common timestamp to all \dword{spvd} detector systems.
		\item It shall be able to align the timestamp across the \dword{spvd} \dword{tpc} and \dword{pds} to better than \SI{10}{\nano\second} at all times (set by \dword{pds}).
		\item It shall be able to distribute synchronization, calibration, and control commands to \dword{spvd} systems. 
	   \end{itemize}
	\item Data selection: 
	
	  \begin{itemize} 
		
		\item The \dword{fd} shall be >90\% efficient for any interaction that leaves >100\,MeV of visible ionization energy inside the fiducial volume.
		\item The \dword{daqtrs} shall be capable of recognizing a \dword{snb} from a nearby supernova 		based on a threshold of more than 60 interactions within 10 seconds, with neutrino energy deposition above 10\,MeV each, with an efficiency $>$95\%.
		\item The \dword{fd} shall have high efficiency for any interaction leaving <100\,MeV of visible ionization energy inside the fiducial volume, and for single interactions with visible energy deposit >10\,MeV.  The lowest-energy of these may be pre-scaled. 
	 \end{itemize} 
	\item Data throughput: 
	  \begin{itemize}
		\item The \dshort{tdaq} shall be able to receive digitized data from the detector electronics ($\approx$1.8\,TB/s) and buffer them for up to 10 seconds awaiting a \dword{trigdecision}. 
		\item It shall be able to store data for up to one week without interrupting the \dword{daq}, in case of delays in the data transfer from \dword{surf} to \dword{fnal}. 
		\item It shall provide enough disk I/O capacity for driving the \dword{wan} link at 100\,G/s.
		\item It should be able to  transfer \dshort{snb} \dwords{trigrec} ($\approx$180\,TB) from the \dword{spvd} detector cavern to the \dshort{daq} components on the surface within one hour ($\approx$400\,Gb/s).
		\end{itemize} 
	\item Uptime: Each \dword{fd} module shall have an uptime of at least 95\%, and the \dword{fd} as a whole shall have an uptime of at least 98\%, during which at least one module is operational. 
	Since a few days per year of downtime for infrastructure maintenance cannot be avoided, and individual hardware failures (electronics, servers, disks, ...) may occur, 	this translates into a very strong requirement for the \dshort{tdaq}. The \dshort{tdaq} shall operate continuously, 	dynamically adjust to changing conditions, tolerate faults, and recover from errors autonomously. 
	\end{itemize} 

The electronics parameters essential for dimensioning the \dword{spvd} \dshort{tdaq} system are shown in Table~\ref{tab:table_detector_props}. 
The individual \dshort{tdaq} component counts for the \dword{spvd} are shown in Table~\ref{tab:table_daq_components}.

\begin{dunetable}
	[Detector parameters of \dshort{spvd} driving the \dshort{tdaq} design. ]
	{ll} 
	{tab:table_detector_props}
	{Summary of detector parameters driving the DAQ design.}
	\textbf{Parameter} & \textbf{Value} \\ \toprowrule
	\dword{tpc} channels  & 491520 \\ 
	\colhline
	TPC channel count per \dword{crp} & 3072 \\
	\colhline       
	TPC top electronics 40\,G links & 320 \\
	\colhline
	TPC bottom electronics 10\,G links & 960 \\
	\colhline
	TPC \dword{adc} sampling rate & 1.953125\,MHz \\
	\colhline
	TPC \dshort{adc} dynamic range  & 14\,bits \\ 
	\colhline
	\dshort{pds} channels  & 1280 \\
	\colhline       
	Max localized event record window (1 drift length)  & 4.25\,ms \\
	\colhline       
	Extended event record window & 100\,s \\
	\colhline
	Maximum size of TPC localized event record  (uncompressed) &  8\,GB\\
	\colhline
	Full size of TPC extended event record  (uncompressed)  &  180\,TB\\
	
\end{dunetable}

\begin{dunetable}
	[\dshort{tdaq} component counts for \dshort{spvd}]
	{lc} 
	{tab:table_daq_components}
	{\dshort{tdaq} component counts for \dshort{spvd}. All servers are interconnected via 10G/100G Ethernet network.}
	\textbf{Parameter} & \textbf{Count} \\ \toprowrule
	Clock speed & $62.5$~MHz \\ 
	\colhline
    Timing endpoints & $\approx{120}$ \\ 
     \colhline
	Readout switches (10G/100G)  & 20 \\ 
	  \colhline
	Readout switches (40G/100G)  & 8 \\ 
	\colhline
	TPC readout cards  & 80+80 \\ 
	\colhline      
	PDS readout cards & 4 \\
	\colhline
	Readout servers & 84 \\
	\colhline
	Trigger servers & 20 \\
	\colhline
	Data collection, storage servers & 8 \\ 
	\colhline
	Data filter servers  & 20 \\
	\colhline       
	\dshort{tdaq} control and monitoring servers & 25 \\
\end{dunetable}

\section{Interfaces}
\label{sec:daq:interf}

The overview diagram in Figure~\ref{fig:dq_overview} shows the main external interfaces of the \dword{tdaq}, with the external systems depicted in grey. Brief descriptions of these follow; they are detailed in referenced interface documents. The %technical details of the 
components' external interfaces are summarized in Table~\ref{tab:external}. 

%$$$$$$$$$$$$$$$  
\begin{dunefigure}
[Conceptual overview of \dshort{tdaq} system functionality]
{fig:dq_overview}
{Conceptual overview of \dshort{tdaq} system functionality for a single \dshort{fd} module. External systems are depicted in grey while the \dshort{tdaq} subsystems are represented using the same color scheme 
as in Figure~\ref{fig:dq_subsys_view}. The external interfaces are described in Section~\ref{sec:daq:interf} while the flow of data and messages is described in Section~\ref{subsec:OVss}. Acronyms used in the figure: HW=hardware, HS=hardware signal, TR=\dword{trigrec}, ICT=information and communication technologies}
\includegraphics[width=.9\textwidth]{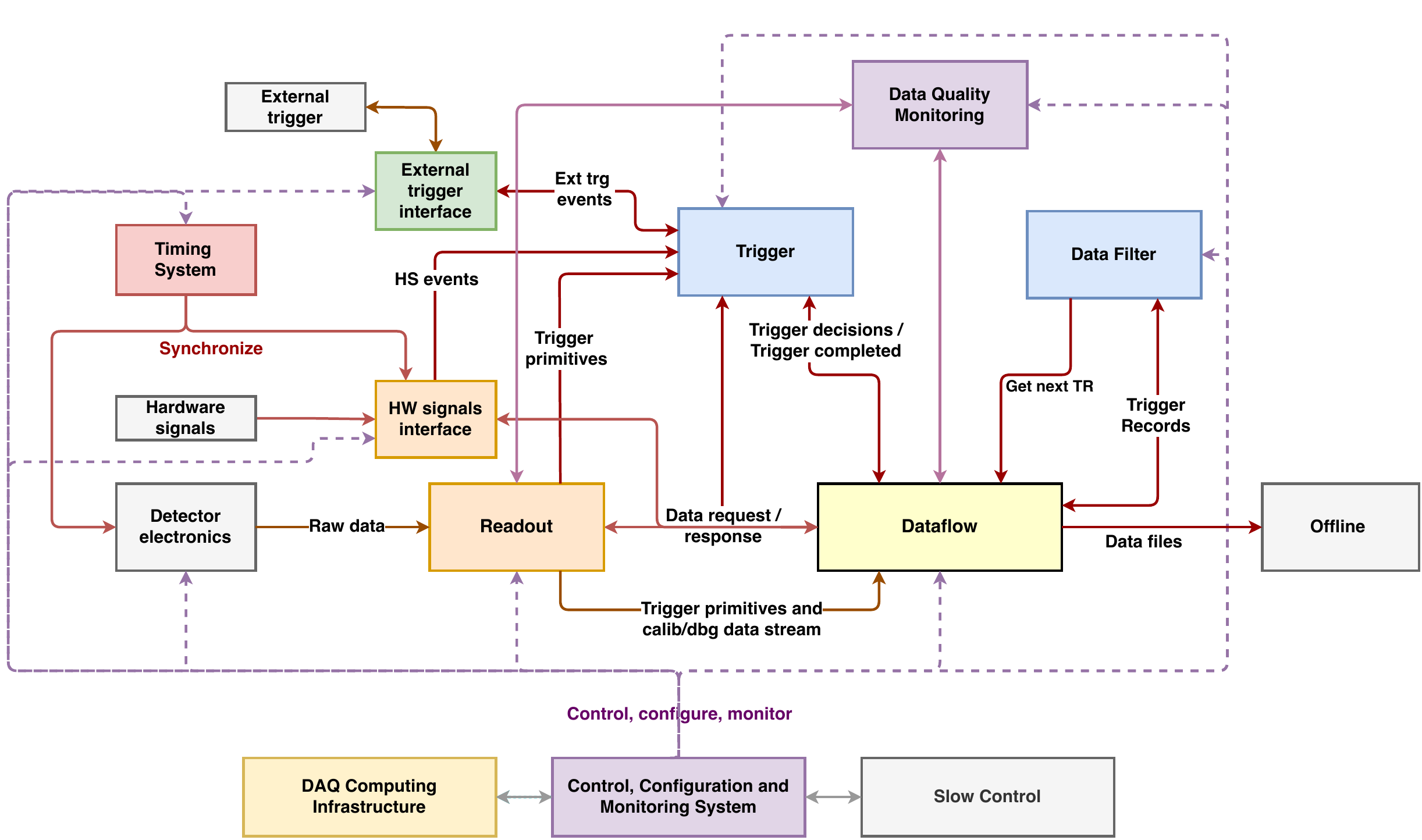}
\end{dunefigure}
%$$$$$$$$$$$$$$$ 
\begin{description}

\item[TPC Electronics and Photon Detection] The \dfirst{daqros} (Section~\ref{subsubsec:ROsss}) receives raw data from the various detector electronics devices. The interface to the top \dword{tpc} electronics is described in~\cite{daqinterfaceTDE} and the interface to the bottom \dword{tpc} electronics is in~\cite{daqinterfaceBDE}. The interface to the \dword{pds} electronics is in~\cite{daqinterfacePDS}. These documents describe the physical interface of the electronics to the \dshort{daq} system through optical links, as well as the expected data formats and transfer protocols. 

\item[Calibration] The hardware signals interface injects timestamped hardware signals, e.g., from calibration systems, into the \dshort{tdaq} data selection chain. The interface to calibration systems is through the \dword{daqdts} and the interfaces are described in~\cite{daqinterfacecalib}.

\item [External Triggers] The \dfirst{daqtrs} (Section~\ref{subsubsec:TRsss}) receives external timestamped trigger messages (e.g., from other \dword{fd} modules) % or the \dword{snews}) 
to initiate data collection irrespective of any activity inside the module.

\item [Computing] The data collection (Section~\ref{subsubsec:EBsss}) interfaces to the offline computing, for the purpose of transferring raw data and metadata for permanent storage. The interfaces are described in~\cite{daqinterfacecomput}.

\item [Slow Control] The \dshort{daq} relies on the Slow Control system to control and monitor the power distribution units in the server racks and to provide a user interface to remotely turn on/off computers (in addition to the network-based \dword{ipmi} interface). Furthermore, \dshort{daq} and Slow Control will exchange information about the status of different components. Details of the interaction are still to be worked out, but will occur purely on a software-data-exchange basis.

\item [Information and Communication Technologies] The \dshort{daq} transfers a large amount of data across networks within clusters of computers in certain locations, between computer cluster locations, and between the detector site and \dword{fnal} central computing. The responsibilities of the \dshort{daq} and of \dword{fnal} Networking are described in~\cite{daqinterfacenetworking}.

\end{description}

\begin{dunetable}
	[\dshort{tdaq} components' external interfaces]
	{p{0.45\textwidth}p{0.54\textwidth}} 
	{tab:external}
	{Brief description of \dshort{tdaq} components' external interfaces. References to interface documents are given in the text.}
	
%{    
\dshort{tdaq} Component & Interface description \\ \toprowrule
   Timing & 1000-BaseBX single mode fiber with custom protocol \\ \colhline
  \href{https://edms.cern.ch/document/2088713/7}{Readout - TPC bottom electronics} & 10\,G Ethernet, UDP/IP protocol \\ \colhline
  
   \href{https://edms.cern.ch/document/2618999/2}{Readout - TPC top electronics} & 40\,G Ethernet, UDP/IP protocol \\ \colhline
   
  \href{https://edms.cern.ch/document/2088726/4}{Readout - PDS} & 10\,G Ethernet, UDP/IP protocol 
  \\ \colhline
  
  \href{}{Trigger - external trigger sources/output} & Ethernet I/O of timestamped signals from calibration devices, or from/to other \dword{fd} modules or SNEWS \\ \colhline
  
  \href{https://edms.cern.ch/document/2620747/2}{Transient data store - offline computing} & WAN ethernet connection for transfer of HDF5 raw data files. HDF5 raw data format specifications.\\ \colhline
  
  \dword{daqccm} - offline computing & databases with the archive of run configurations and conditions \\ \colhline
  \dshort{daqccm} - slow control & Ethernet based inter process communication to exchange status information \\ \colhline
  \dshort{daqccm} - detectors software & Software libraries and tools for the implementation of the control, configuration and monitoring of the detector electronics \\ 
 
\end{dunetable}

%%%%%%%%%%%%%%%%%%%%%%%%%%
\section{Trigger and DAQ (TDAQ) System Design} %Overview}
\label{subsec:OVss}

The \dword{fd} \dword{tdaq} system is physically located at the \dword{fd} site, \dword{surf}. 
It uses space and power in each underground detector cavern (on top of the cryogenics mezzanine), and above-ground, in the \dword{mcr} within the Ross Dry building.
The upstream part of the system, responsible for
raw detector data reception, buffering, pre-processing, and triggering, resides underground. The back-end, which is responsible for
trigger-records-building, data storage, and data filtering, resides at the
surface.
Some elements of the \dshort{tdaq} span both locations, i.e., the timing, control, configuration, and monitoring systems (\dword{daqdts} and \dword{daqccm}). 
See Section~\ref{subsec:elect-inf} for information on the electrical infrastructure. 
The connectivity between all \dshort{tdaq} elements is provided through a distributed and redundant 10G/100G/400G Ethernet network, which is outside the scope of the DAQ consortium. The link between upstream and back-end elements is implemented over the redundant dual-fiber run in the Ross and Yates shafts described in Section~\ref{subsec:elect-inf}, ensuring the operation of the \dshort{tdaq} system even in the event of accidental damage to one of the two runs.
Data flow through the \dshort{daq} from upstream to the back-end and then offline. Most raw data are processed and buffered underground, thus controlling consumption of available data bandwidth to the surface.

All the detector modules and their subsystems are synchronized and timed against a common global clock, provided by the \dword{daqdts}. Cross-module communication and communication
to the outside world for data selection (trigger) purposes is facilitated through an \dword{daqeti}.

The \dshort{tdaq} system must acquire data from the both  the \dword{tpc} and the \dword{pds} in order to 
satisfy the requirements placed on it that derive from the experiment's physics objectives. 
Ionization charge measurement by the \dword{tpc} for any given activity in the \dword{fd} requires a
nominal recording of data over a time window determined by the drift speed of the ionization electrons in \dword{lar} and the detector dimension
along the drift direction (\maxdriftdist). Given a 
target drift \efield of \tgtdriftfield{}, the time window is set to 4.25\,ms.
The activity associated with beam, cosmic rays, and atmospheric neutrinos 
is localized in space and particularly in time; \dword{snb} neutrinos are associated with
activity that extends over the entirety of the detector and lasts between 10 and 100\,s. 

The detector data flow from the electronics output links, on top of the cryostat, through the \dshort{tdaq} system to permanent storage at \dword{fnal}. Two stages of data selection allow  reduction of the overall data volume from $\sim$1.8\,TB/s produced by the \dword{spvd} module to $\sim$30\,PB/year for all \dword{fd} modules: triggering the collection of data only for interesting detector regions and time windows, and 
applying data compression algorithms. Data selection strategies are defined in close collaboration with the DUNE physics groups, the offline processing team, and 
the detector subsystem and calibration experts.

 Figure~\ref{fig:dq_overview} depicts the flow of data within the \dshort{tdaq} system, left to right. Raw data produced by the detector electronics is streamed into the readout (\dword{daqros}). Here data are processed to produce \dword{trigprimitive} information, and stored for up to 10\,s. The trigger subsystem (\dword{daqtrs}) receives \dshort{trigprimitive}s, timestamped hardware signal events, and external trigger events, and combines this information to form a \dword{trigdecision}. 
 Trigger decision messages containing a list of time windows and detector locations are forwarded to the data collection component and an acknowledgment is sent back; data collection is in charge of requesting the relevant data from the \dword{daqros} and the \dshort{daqtrs}, forming \dwords{trigrec}, and storing them to disk. Trigger records are dispatched by data collection to the data filter for further data reduction. 
 The data filter returns the modified \dshort{trigrec}s to the data collection component for both storage and transfer to the offline computing system at \dword{fnal}. The whole \dshort{daq} process is orchestrated by the \dshort{daqccm}; the quality of the raw data is continuously monitored by the \dword{dqm} (Section~\ref{subsubsec:DQMsss}).
 
The \dshort{tdaq} can be partitioned---that is, multiple instances of the system can run on physically distinct regions of the detector, with each instance referred to as a \dword{daqpart}. This allows, for example, part of the module to be calibrated while other parts continue to collect physics data, maintaining the high uptime requirements of the system. 

In the following sections the different \dshort{tdaq} components shown in Figure~\ref{fig:dq_overview} and the external \dshort{tdaq} interfaces are described in more detail.

%%%%
\subsection{Timing and Clock Distribution}
\label{subsec:TCss}

The \dfirst{daqdts} provides services to allow the accurate synchronization of sampling and data processing for all elements of the \dword{fd}. The primary function is the distribution of clock and synchronization to \dword{fe} electronics, such that every data sample may be tagged with a 64-bit timestamp. These \dword{daqdts} timestamps, which are unique across the lifetime of the experiment, are used for data processing and event building within the \dshort{tdaq}, and are included within the recorded data sample to support event reconstruction within and across detector modules, and the correlation of DUNE data with external events (e.g., beam spills or astronomical observations).

All the electronics for the \dword{spvd} module are synchronized to a common clock, derived from
a single \dword{gps}-disciplined source. Individual detector channels are synchronized to a fraction of the  timestamp granularity. Initial bench tests show this to be <1~ns, and the expected final synchronization will be determined during the \dword{mod0} tests. % at \dword{protodune2}. 
Due to the extended physical size of the DUNE %detectors
\dshort{detmodule}s, consistency of timing alignment is ensured by both hardware- and software-based feedback loops, calibrating propagation delays on the timing links. The requirement on the absolute accuracy of timestamps with respect to \dword{tai} is one microsecond, though in practice is expected to be much better.

The consistent timing synchronization of all detector channels is continuously monitored and checked at multiple levels: within the timing distribution systems themselves; within the \dshort{daq} system against calibration signals; and offline against physics signals. As critical infrastructure for data-taking, the timing distribution systems are designed for robustness and fault-tolerance. All electronic components of the system for which failure would affect a large detector volume have redundant components contained in separate crates with redundant power and network links. Two independent \dword{gps} systems are used, with antennae at the Ross and Yates shaft entrances respectively, and are able to operate in a hot-spare configuration.

The \dword{daqdts} provides a common interface to timing ``endpoints'' in the readout systems. The endpoints provide a 62.5\,MHz master clock and 64\,bit timestamp. Synchronization information is transmitted on
single-mode optical fibers. Passive optical splitters are used to provide the ability to drive multiple timing endpoints from a single fiber. Passive optical combiners allow the use of two independent timing master systems. One timing master system is in operation at any one time, with the other available as a redundant hot-spare. The \dword{daqdts} measures, and compensates for, the propagation delays from timing master to endpoints in the readout systems.

The bottom drift electronics (\dword{bde}), and the \dword{pds}, have the same interface to the timing system as the corresponding systems in \dword{sphd} module. The top drift electronics were designed with an embedded timing distribution system based on the IEEE-1588/\dword{wr} standard, which, starting from a 10\,MHz clock and 1 pulse-per-second hardware signals distributes a timestamp to the digitization units in the \dword{utca} crates (Section~\ref{sec:TAROss}). Tests in the \dword{wr} lab at \dword{cern} have shown that it is possible to synchronize a \dword{wr} ``island'' from the \dword{daqdts}. Tests at larger scales will be done with the \dword{spvd} \coldbox{}es and %\dword{protodune2} 
\dword{mod0} detectors.

\subsection{Readout}
\label{subsubsec:ROsss}

The \dfirst{daqros} is responsible for receiving data from the electronics. It interfaces with the top and bottom drift \dword{tpc} electronics and the \dword{pds} electronics via optical fibers that connect the detector electronics on the cryostat roof with a set of Ethernet switches on the cryogenics mezzanine, inside the \dshort{daq} barrack. The switches aggregate traffic on 100G Ethernet links which are fed into the \dshort{daq} front-end (\dword{fe}) readout units.

Each top-drift electronics \dword{cro} \dshort{utca} crate aggregates raw data from twelve \dwords{amc} into a \SI{40}{\giga\bit/\second} optical link to the \dshort{daq}, for a total of 768 \dwords{crp} channels per crate, and of 320 links to the \dshort{daq} \dword{fe} readout. The typical effective throughput of a data link is expected to be around \SI{22}{\giga\bit/\second}.

Each bottom-drift electronics \dwords{wib} sends data via two \SI{10}{\giga\bit/\second} optical links to the \dshort{daq}. A total of 960 links is expected; the typical effective throughput of a data link is around \SI{7.2}{\giga\bit/\second}.

As the digitization rates of the charge electronics are directly streamed to readout, the rate of events does not change the bandwidth necessary between the electronics and the readout boards; the number of links is set to allow some redundancy in case of transmission failures. 

Each \dword{pds} electronics \dword{daphne} board sends data on one \SI{10}{\giga\bit/\second} optical link to the \dshort{daq}. A total of 40 links is expected; the typical effective throughput of a data link is estimated to be around \SI{3.5}{\giga\bit/\second}.

As the \dword{pds} electronics include zero-suppression, more overhead is allocated for the bandwidth of the system to account for variations in the event rate. 

The basic component of the readout subsystem is a \dword{daqrou}, %readout unit (RU), 
of which there are 84 responsible for reading out \dword{spvd}. A \dword{daqrou} consists of a high-end server hosting two \dword{pci}e cards which provide the high-bandwidth data connection to the \dword{fe}. Each \dword{daqrou} processes and buffers the data coming in from a portion of the detector (two \dshort{crp}s or 25\% of the \dword{pds}). 

Figure~\ref{fig:dq_subsys_readout} shows an expanded view of the readout subsystem in the context of the overall \dshort{daq}. The \dshort{daq} \dword{fe} readout units store the raw data stream from detectors in a circular memory buffer after data checking for at least \SI{10}{\second}, and perform a first stage of preprocessing. 
\Dwords{daqrou} respond to data requests from the event builder by providing the data fragments corresponding to the selected \dword{readout window} from the temporary memory buffers.

\begin{dunefigure}
	[\dshort{tdaq} subsystems diagram]
	{fig:dq_subsys_readout}
	{Diagram of the readout portion of the DAQ, showing the path of data movement and buffering, including SNB data. }
	\includegraphics[width=1.\textwidth]{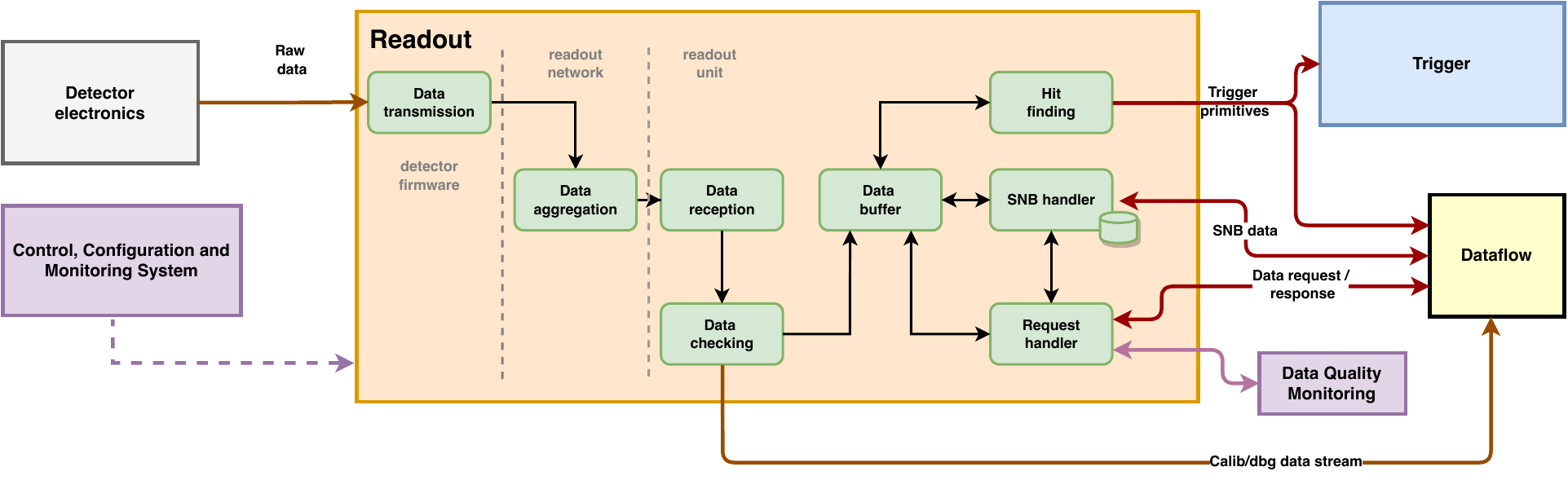}
\end{dunefigure}

In the event of an \dshort{snb} trigger firing, 
the \dwords{daqrou} will stream the content of the temporary memory buffer to local storage for a total duration of \SI{100}{second}.
Afterwards, each \dword{daqrou} makes the stored \dshort{snb} record available to the data collection system for transfer to the surface storage.

\Dwords{daqrou} perform the first stage of trigger processing on incoming data to generate local elementary trigger information (see \dshort{trigprimitive}s in Section~\ref{subsubsec:TRsss}), to minimize the raw data transfer over network. The \dshort{trigprimitive}s are streamed to both the \dshort{daqtrs} for further processing, and to data collection for semi-permanent storage. 

In addition to handling the main data path, \dshort{daq} \dword{fe} readout provides extra functions to support debugging, calibration and quality assessment. Specifically it:
\begin{itemize}
\item supports the \dword{bde} %bottom-drift \dword{ce} 
calibration with a dedicated raw data stream during calibration cycles; 
\item generates a reduced raw data stream for a configurable subset of channels, on demand, for debugging purposes; and
\item samples and extracts basic \dshort{dqm} information. 
\end{itemize}

At \dword{pdsp} a prototype implementation of the readout subsystem using a simplified communication protocol and custom \dword{felix} \dword{pci}e cards was demonstrated, and is described in~\cite{DUNE:2020txw}. 

The need of finding a common solution for reading out all the different detector electronics (including the \dword{tde}), justified a change in the implementation choice for the readout. 
Ethernet and the UPD/IP protocol are used as the underlying technologies for this change. Network switches and 100\,G smart network interface cards are the hardware components that allow to implement it. The switches multiplex several \dword{fe} data streams---320x40\,G links for the \dword{tde} and 960x10\,G links for the \dword{bde}--into 80 100\,G links for each type of \dword{fee} that are fed into the readout units. Thanks to recent technology advances it is possible to use exclusively commercial components for the implementation of this subsystem, thus reducing the complexity of (and the risks in) this area of the \dshort{tdaq}.

From a design point of view the system remains largely unchanged and much of the code developed for the \dshort{felix} based solution can be reused. In order to speed-up the evaluation of the readout implementation two prototypes have been developed, with a different balance of resources usage on the server or network interface cards. Those are being finalized and tested in a laboratory environment. The first integration with the \dword{bde} and \dword{tde} are planned before the \dshort{tdaq} final design review, while the integration with the \dword{pds} electronics will occur later. During the transition phase the existing \dshort{felix} based solution will continue being supported.

%%%%
\subsection{Trigger and Data Filter}
\label{subsubsec:TRsss}

\subsubsection{Strategies for Data Retention}
\label{subsubsec:TRsss:strat}

This section %briefly 
summarizes the strategies that have been devised to maximize the retention of interesting data while respecting the data volume that can be permanently stored long-term for the DUNE \dword{fd} ($\sim$30\,PB/y).  The goal is to be as inclusive as possible; at high energies (>\,100 MeV) 
it is important to accept all possible events with as high an efficiency and as wide a region-of-interest (\dword{roi}) in channel and time space as possible, regardless of event type. As 
energies approach 10\,MeV, when radiological backgrounds (including neutron captures, $^{42}$Ar, and $^{39}$Ar pileup) become dominant, the system should be semi-inclusive, leveraging the topological capabilities of the \dword{tpc} data to provide some discrimination of low-energy physics signals.

To facilitate partitioning, the \dshort{daq} trigger and data filter can be instantiated several times, and multiple instances can operate in parallel. Within any given \dword{daqpart}, the \dshort{daq} trigger and data filter will also be informed and aware of current detector configuration and conditions and apply certain masks and mapping on subdetectors or their fragments in its decision making. This information is delivered by the \dword{daqccm} system.

All digitized charge collection data are sent to the readout subsystem and processed. For each strip on a \dshort{crp} layer, a hit-finding algorithm allows identification of activity above the electronic noise; the threshold and time above threshold  will be configurable. With the expected noise of the \dword{tpc} electronics, the hit-finding threshold is expected to be such that hits will be generated for a large fraction of the \Ar39 decays; the spectrum endpoint is at 0.5\,MeV. Every hit generates a so-called \dshort{trigprimitive}.

Similarly, \dword{pds} electronics boards send waveform data for any channel that passed an internal threshold; \dshort{trigprimitive}s are also formed from these data.

The \dshort{trigprimitive}s serve two purposes: 
\begin{itemize}
	\item They are the basic elements 
	used to form a \dshort{trigdecision} in the \dshort{tdaq} system.
	\item They are stored as unbiased (at the ``event'' level) summary information that can be used for trigger, calibration, low-energy physics studies, etc.
\end{itemize}
To provide good sensitivity to different track topologies, each \dshort{trigprimitive} contains information such as the time-over-threshold of the waveform, its peak, and its total charge, as well as the timestamp of the start of the waveform.  To date, our \dword{tpc} trigger studies have been done using collection view only; exploitation of induction views is under development but is not seen as needed to satisfy any of DUNE's triggering requirements.

Trigger primitives generated from charge collection channels and the \dword{pds} will be stored on disk by the \dshort{tdaq} system. This data set is very important for carrying out trigger studies but can also be used for calibration purposes, as well as fast data analysis.  After compression and minimal clean-up, it is estimated that a few PB/year will be sufficient to store them. It is thus an option for DUNE to not only store the \dshort{trigprimitive}s temporarily (a few months) for specific studies, but to make them part of the data that will be stored permanently.
This data stream is particularly interesting in view of its role in potentially extending the low-energy physics reach of DUNE beyond its %main 
core program; it will also contain summary information for individual interactions with very low visible deposited energy.  Depending on the achieved \dword{s/n} on the collection strips, the \dshort{trigprimitive} threshold is expected to be around 250\,keV.

For triggering purposes, the \dshort{trigprimitive}s are the basic elements used by the \dshort{tdaq} to form a \dword{trigrec} and initiate the collection and storage of raw waveform data. The \dword{daqdsn} system takes \dshort{trigprimitive}s generated locally and looks for clusters in time and space. These clusters represent what is called ``trigger activity.''  Clusters of trigger activity are then passed to algorithms downstream, which determine whether any particular set of trigger activity clusters should be promoted to a \dword{trigcandidate}. Trigger candidates then are sent to \dshort{trigdecision} logic, which apply criteria that include both configuration parameters (e.g., which triggers are accepted in this data run) and dynamic decisions (e.g., does a \dword{tpc} \dshort{trigcandidate} come after an existing \dword{pds} \dshort{trigcandidate}?). The data selection work flow is shown in Figure~\ref{fig:dq_trigger}.

\begin{dunefigure}
	[Data selection work flow]
	{fig:dq_trigger}
	{The \dword{daqdsn} work flow. The trigger system is responsible for the decisions and algorithms in the blue boxes, and the dataflow system is responsible for drawing raw data and trigger system output into the tirgger records.}
	\includegraphics[width=.95\textwidth]{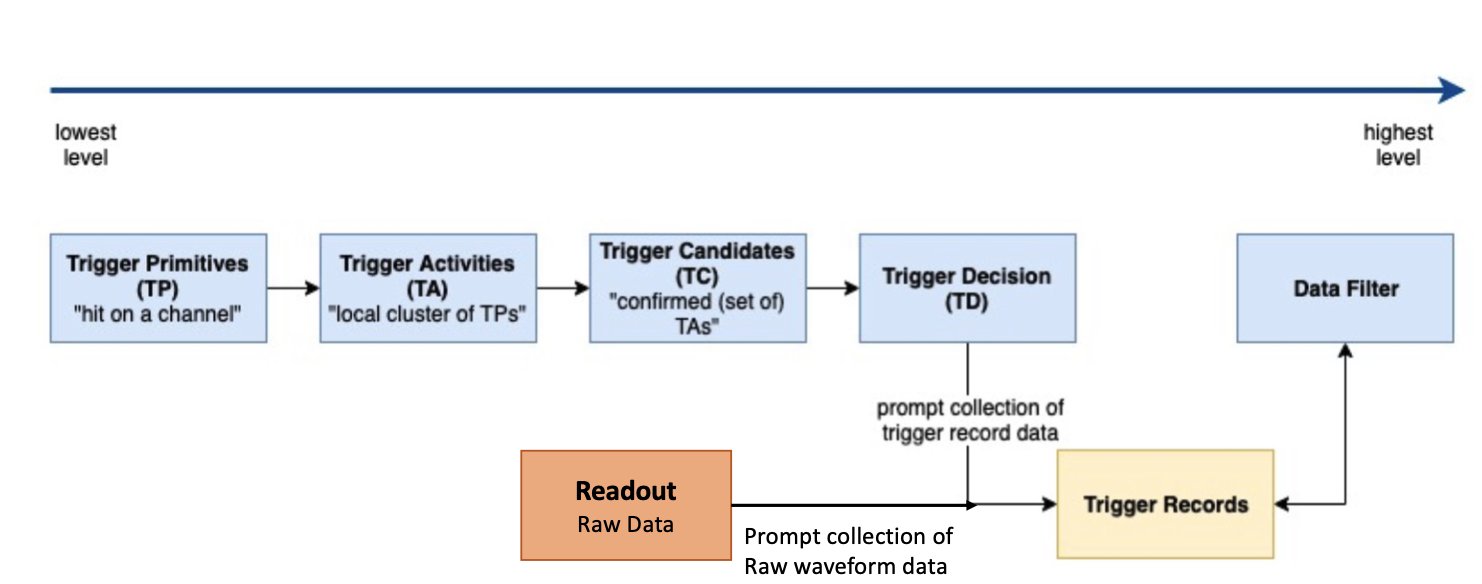}
\end{dunefigure}

There are two different raw data collection modes foreseen for the DUNE \dword{fd}:
\begin{itemize}
	\item  A \dshort{trigdecision} based on trigger activity consistent with a single interaction or internal decay that 
	includes a list of \dshort{crp} and/or \dword{pd} channels to be collected and their associated time window(s). The \dshort{tdaq} uses this information to collect the relevant raw data from its temporary buffers, form \dshort{trigrec}s, and store them persistently. The data files may be further trimmed or compressed through a data filter stage before being transmitted to \dword{fnal} for permanent storage. 
		\item When the trigger identifies several trigger activity clusters within a few seconds that are inconsistent with the expected fluctuations from background in rate and energy, 
		it fires a special \dshort{trigdecision} indicating 		a \dshort{snb} candidate.   
		For the \dword{spvd} the total collected raw data from a \dshort{snb} will be about $\sim$180\,TB of raw data. Thus, while 
		the effective burst threshold must be set low enough to satisfy DUNE's requirements on \dshort{snb} detection efficiency, 
		it is important to not fire too frequently on background fluctuations.   
		The present assumption is that the trigger conditions will be adjusted such that statistical fluctuations  cause on the order of one \dshort{snb} candidate trigger per month. It will take about one hour to transfer the data from this trigger event from the detector caverns to the storage on surface, and several additional hours to transfer those data to \dword{fnal}. Upon inspection, fake \dshort{snb} \dshort{trigrec}s will be discarded. 
\end{itemize}
 
Each trigger prompts the collection of data from the  \dword{daqros} to form a \dshort{trigrec}. 
In the extreme case of an \dshort{snb} trigger, data from the whole module is collected over a time window of \SI{100}{\second}. In most other cases, data from only a few \dshort{crp}s and \dshort{pd}s  will be collected over much shorter times ($\ll$\SI{10}{\milli\second}). 

The data filter acts on already-stored \dshort{trigrec}s. It processes them with the aim of further reducing the data volume to be transferred to \dword{fnal}.  Initially, the Data Filter was conceived as back-end ``insurance'' in the event that instrumental events (such as high-voltage ``streamers'') increased the data volume to unmanageable levels, and needed to be removed with more sophisticated algorithms than are possible upstream in the system.  It has become clear, however, that the Data Filter can also filter the data volume by imposing, if desired, various region-of-interest criteria on tracks and hits, thus reducing the data volume and allowing for more inclusive triggers.  Such regions-of-interest may be as simple as narrow (e.g., $\sim 100\mu$s) windows in time around hits, to rudimentary track reconstruction, or even machine learning approaches.  The interface between the data filter and the rest of the \dshort{daq} is specified so that events that get passed to it can equally well come off of the network, or from disk, so that algorithms can be tested offline and with little or not change, be included within the online Data Filter. Development of Data Filter algorithms will be ongoing as experience with prototypes and the as-installed detector progresses.

A final component to the trigger system, not shown in Figure~\ref{fig:dq_trigger}, is the %External Trigger Interface.  The 
\dword{daqeti}. It is intended to allow triggering of various modules off of one another within DUNE, triggering of modules due to external signals from \dword{lbnf} or calibration systems, publishing \dshort{snb} triggers to \dword{snews}, or even triggering by different detectors locally should various collaborations agree to this possibility. In the hierarchical design of the DUNE Data Selection system, the External Trigger thus represents the most ``global'' level.

Beam information from \dword{lbnf} or calibration systems can be used to judge if trigger candidates may come from beam or calibration activity, which may be helpful, for example, in allowing the Data Filter 
to decide on the \dword{roi} for the event. The beam time information can also be distributed to components of the calibration system to avoid producing activity in the detector that may interfere with activity from beam neutrinos.

The \dshort{daqeti} may also be useful as a way of increasing the sensitivity of \dshort{snb} triggering by requiring fewer events per module and summing events across all far detector modules. In the case of future \dword{fd} modules with complementary physics capabilities to \dword{spvd} and \dword{sphd}, the \dshort{daqeti} will enable this complementarity.

\subsubsection{Trigger Performance Studies}
\label{sec:tdaq:trigger}

Studies of the performance of the \dword{spvd} trigger have been done on simulated data as well as on %VD 
\coldbox data. Extensive performance studies have also been done for the \dword{sphd} %HD-FD1 
module, with which the \dword{trigprimitive}--trigger activity--\dword{trigcandidate}--\dword{trigdecision} chain was initially developed and validated.  Performance of trigger primitive generation on \dword{spvd} simulation has also been done, albeit with a relatively early version of the model of its channel signal and noise.  With this simulation, the \dword{spvd} sensitivity to triggering on a \dword{snb} neutrino flux has been studied, detailed in Section~\ref{sec:ph:le:triggerTPC}.  Further evaluation of the performance will be done with low-energy energy deposits in \dword{vdmod0} when data taking begins.  

\begin{figure}[h!]
	\includegraphics[width=0.3\textwidth]{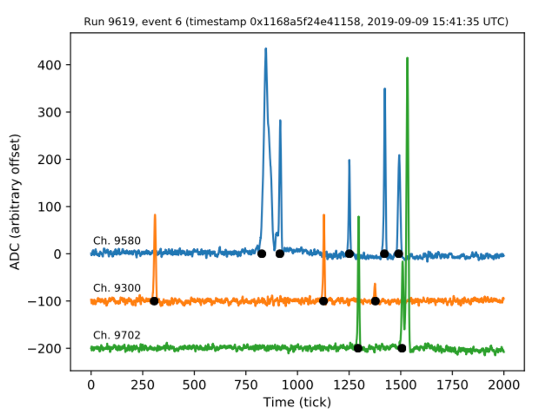}
	\includegraphics[width=0.7\textwidth]{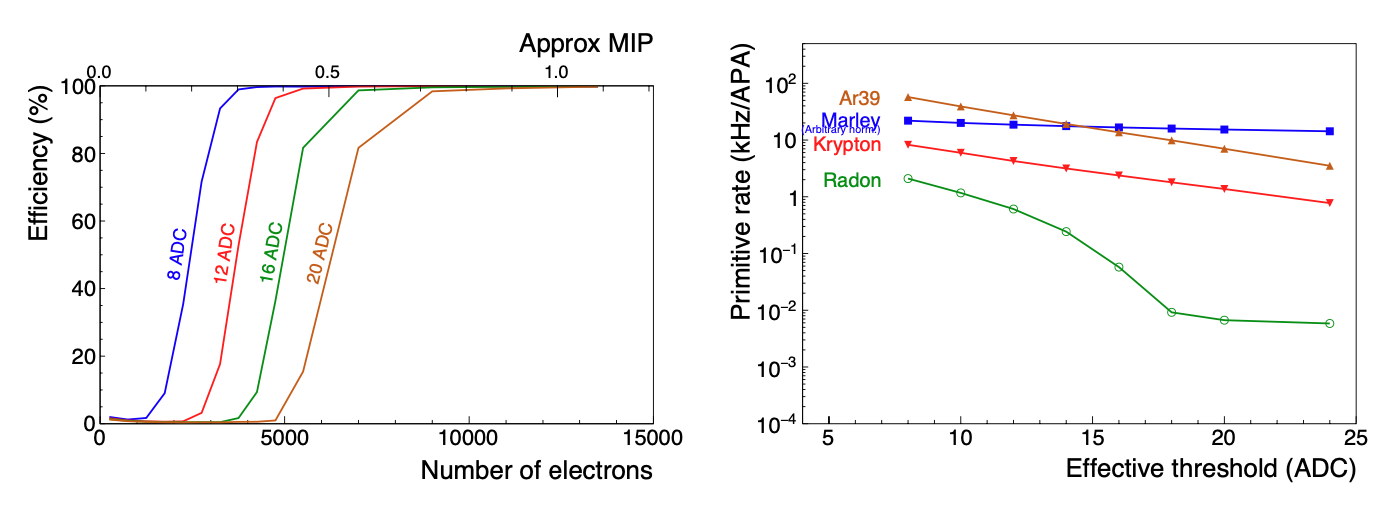}
	\caption[\dshort{trigprimitive}  hits in collection channels in \dshort{pdsp}]{(Left) Identification of \dshort{trigprimitive} hits in collection channels in \dword{pdsp}, using a software version of the \dshort{trigprimitive}  generation algorithm. (Middle) Efficiency curves for \dshort{trigprimitive}  identification as a function of threshold and number of electrons in a hit (or \dword{mip}-equivalent energy).  (Right) Rates of \dshort{trigprimitive}s  expected in \dword{sphd} for various radiological backgrounds and \dshort{snb} neutrino events (\dword{marley}) as a function of \dshort{trigprimitive}s threshold in \dshort{adc} counts. \label{fig:tpcurves}}
\end{figure}

For example, Figure~\ref{fig:tpcurves} shows the work done to validate the (software) \dword{trigprimitive} algorithm on \dword{pdsp} collection-channel data 
and on \dword{mc} simulations for %DUNE 
\dword{sphd} which performs similarly to \dword{spvd}.  As the figure shows, identification of hits to generate \dword{trigprimitive} is relatively easy with the algorithm, including adjustments for noise and baseline variations.  The middle plot of Figure~\ref{fig:tpcurves} shows the efficiency for detecting hits and generating \dshort{trigprimitive}s, as a function of threshold, for a gain that is roughly twice what the \dword{pdsp} channel gains were. The right side of Figure~\ref{fig:tpcurves} shows the \dshort{trigprimitive} rates for various physics sources: \dword{marley} events are hits from \dshort{snb} neutrinos, and the others are radiological backgrounds. At a threshold of about 7 \dword{adc} counts, the \dshort{trigprimitive} rate is entirely dominated by $^{39}$Ar decays, which in the \dshort{detmodule}s will be about $10$~MBq per 10\,kt module.   

\begin{figure}
	\includegraphics[width=0.9\textwidth]{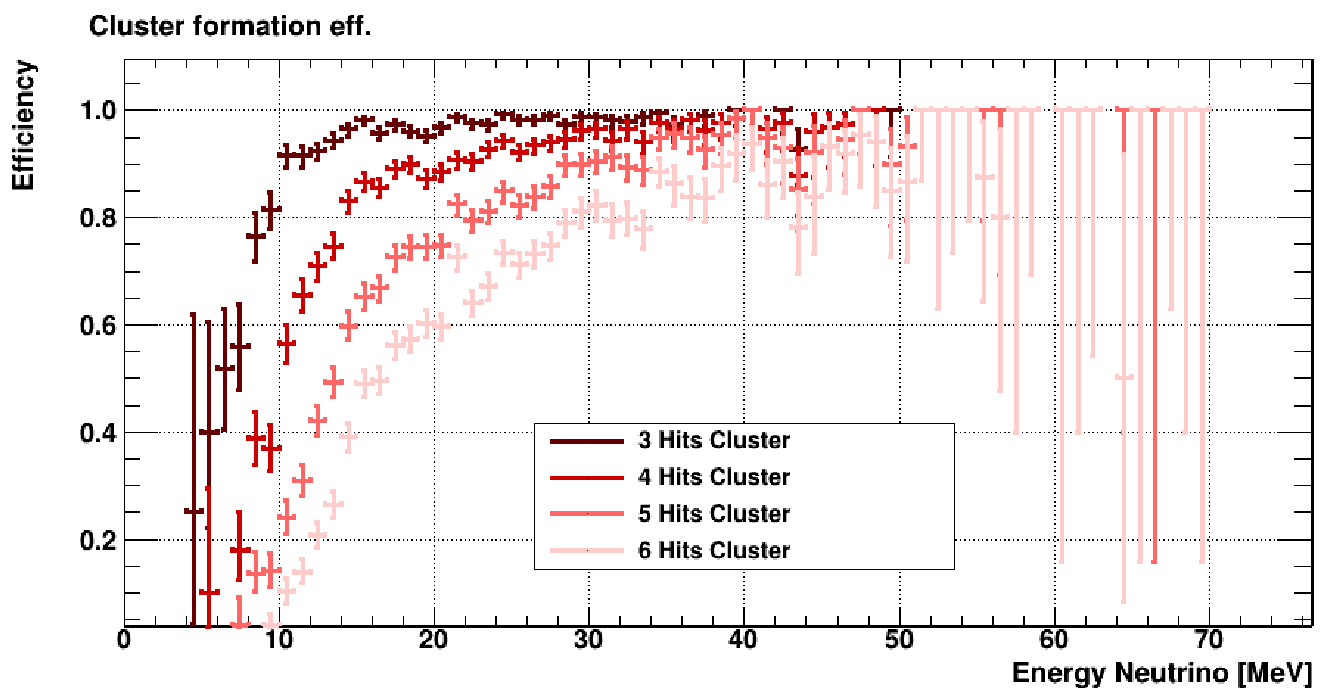}
	\includegraphics[width=0.9\textwidth]{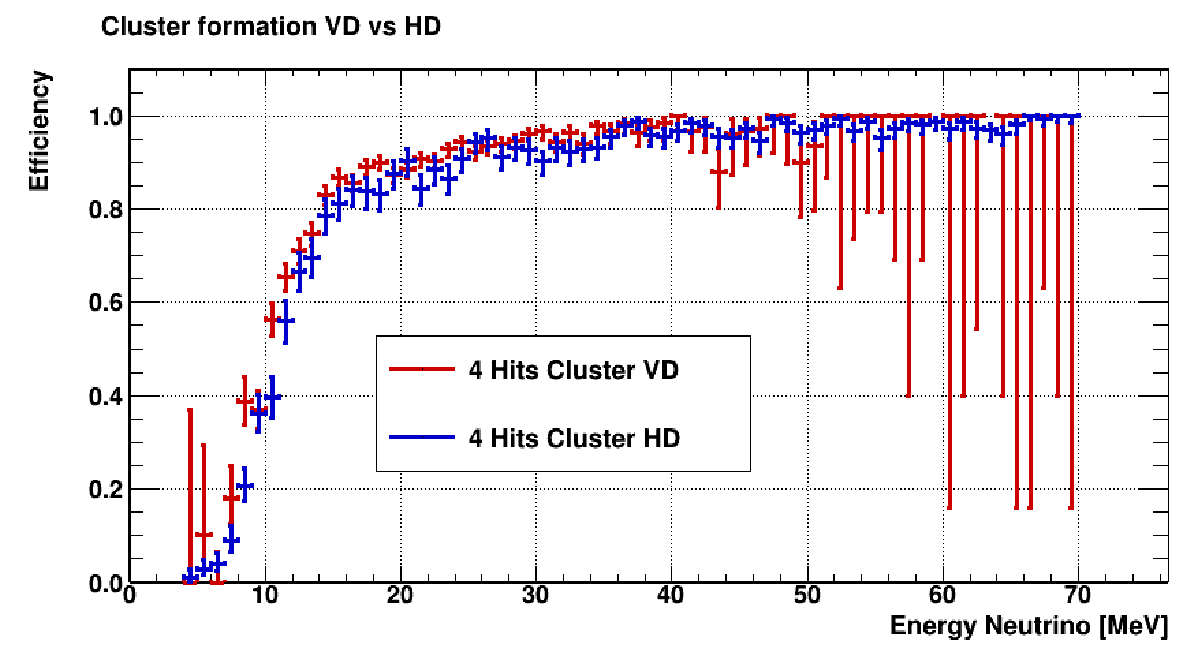}
	\caption[Trigger efficiency curves]{(Top) Trigger efficiency curves as a function of cluster size of \dshort{trigprimitive} hits for \dword{vd} and (bottom) a comparison of the trigger efficiency curves between \dshort{vd} and \dword{hd} simulations. 
	\label{fig:trigeff_comp}}

\end{figure}

Moving from identifying hits to identifying events, the effective energy threshold for the trigger is driven by the limits on total data volume. With a very conservative assumption that for every triggered event we write out an entire module's worth of data for twice the drift time, the event energy threshold is limited by the expected rates of background events, most notably neutrons, which yield roughly 6\,MeV of energy in a $\gamma$ cascade.  
As shown in Figure~\ref{fig:trigeff_comp}, with a purely inclusive trigger---cutting on windows of trigger activity by requiring clusters to have a minimum total and peak charge, a maximum time-over-threshold that exceeds a certain minimum, and a minimum hit multiplicity---the resultant efficiency curves pass through 50\% at 6~and 12\,MeV, depending on the requirement of hit multiplicity.  The agreement between  \dword{sphd} and \dword{spvd} is very good. 
This satisfies requirement SP-DAQ-8 (high efficiency above 10 MeV) and, with high efficiency above 40~MeV, requirement SP-DAQ-7 (>90\% efficiency above 100 MeV) is met for events more energetic than those observed in a \dshort{snb}. As shown in section~\ref{sec:ph:le:triggerTPC}, we achieve $>95\%$ triggering efficiency for 50 events in 10~kt of argon, performing better than requirement SP-FD-23. The effective threshold for \dshort{snb} events can be even lower due to burst structure of the neutrinos. 

Simple performance studies have also been done with data using the \dshort{vd} \coldbox.  For surface detectors like the \coldbox{}es and the \dwords{mod0}, an inclusive trigger does not make a lot of sense, because the cosmic ray rates are so high.  Fortunately, the software approach using \dword{tpc} topological information is naturally adaptable to creating exclusive triggers for low-rate event classes.  For the \dshort{vd} \coldbox, we implemented both a ``horizontal muon trigger'' (which essentially just counted collection-wire hits) and a trigger for Michel-like events. While efficiency studies in a configuration like the \dshort{vd} \coldbox are difficult to do (there is no precise normalization for the flux of horizontal cosmics, for example), the trigger performed well at selecting events with the expected topologies.  These tests were not just tests of selection purity, however, they also showed that the latencies are small and are accommodated easily within the system, and that the entire trigger generation chain %of 
functioned smoothly.

\begin{figure}
	\includegraphics[width=0.5\textwidth]{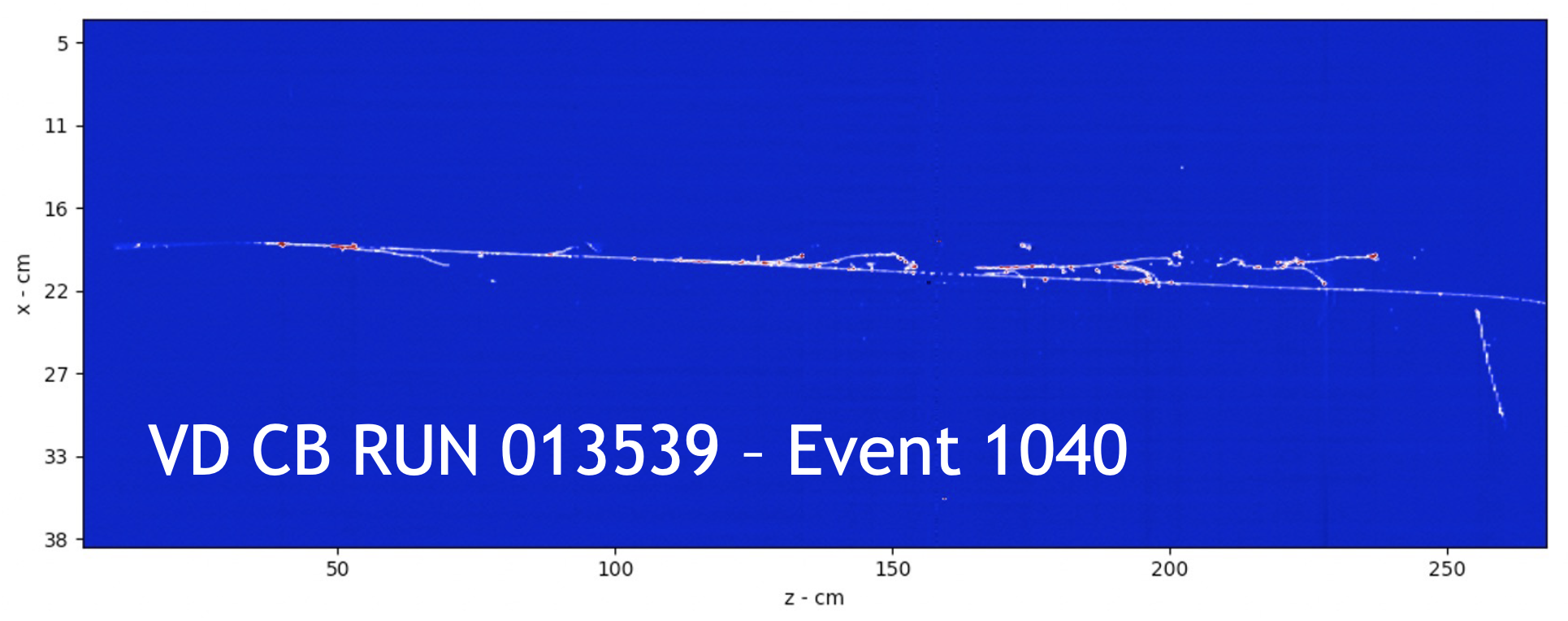}
	\includegraphics[width=0.5\textwidth]{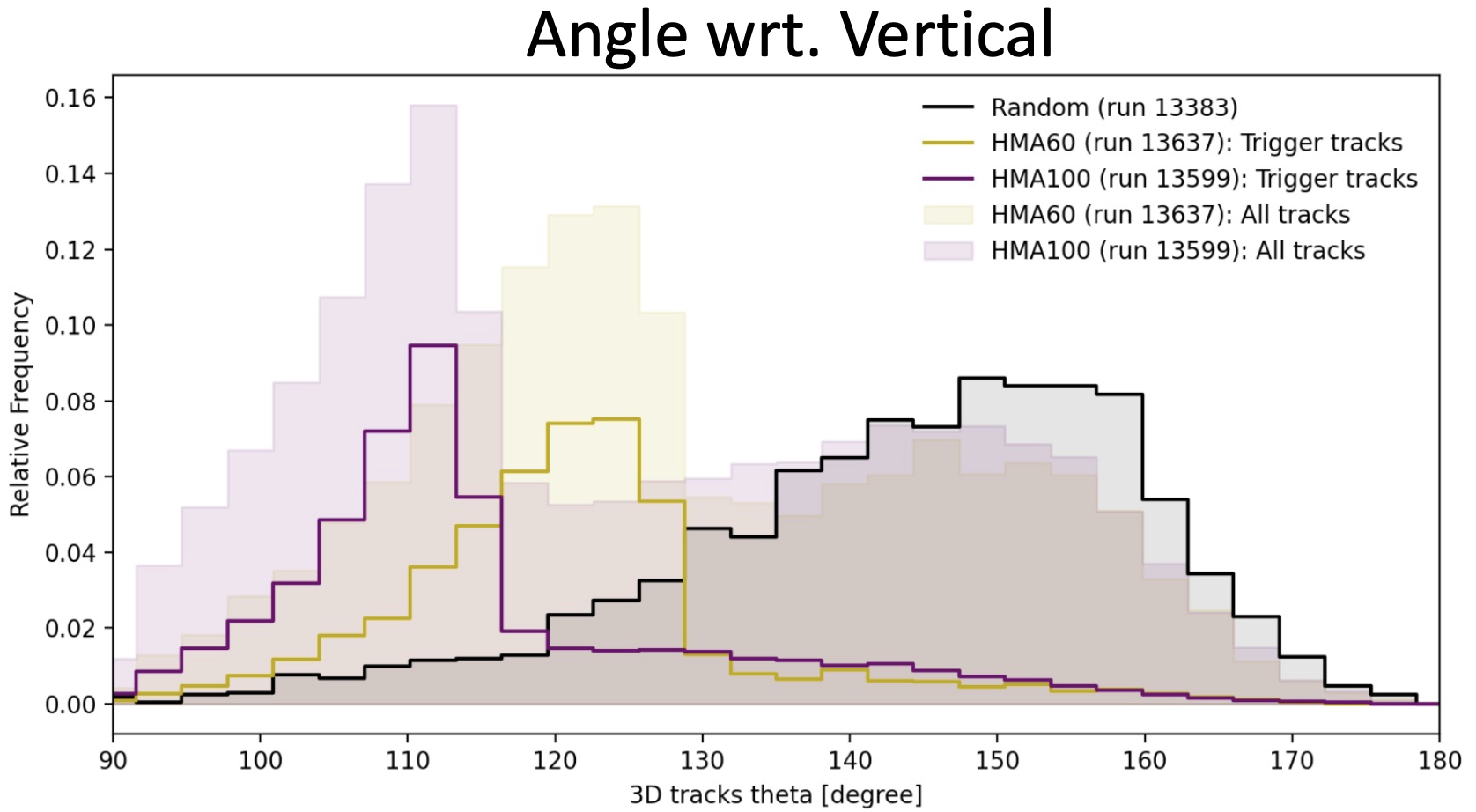}
	\caption[Muon trigger in \coldbox and reconstructed angles of muons]{(Left) A horizontal muon in the \dshort{vd} \coldbox captured by the very simple  hit-counting trigger. (Right) Reconstructed angles of muons for all tracks and tracks triggered by counting either 60 (HMA60) or 100 (HMA100) hits. Here, downward-going events have angles of $180^\circ$. As can be seen, the imposition of the higher hit multiplicity preferentially selects events closer to horizontal.
	\label{fig:hmatrig}}
\end{figure}

Figure~\ref{fig:hmatrig} shows both an event display of an event triggered by the Horizontal Muon Trigger (HMA) in the \dshort{vd} \coldbox, and distributions of the reconstructed angles of triggered tracks versus all tracks, for two different thresholds on the collection-wire hit multiplicity (60 and 100). As the right plot shows, increasing the trigger threshold enriches the data set with horizontal-going muons, despite the enormous flux of downward ($180^\circ$) cosmics.

\begin{figure}
	\includegraphics[width=0.5\textwidth]{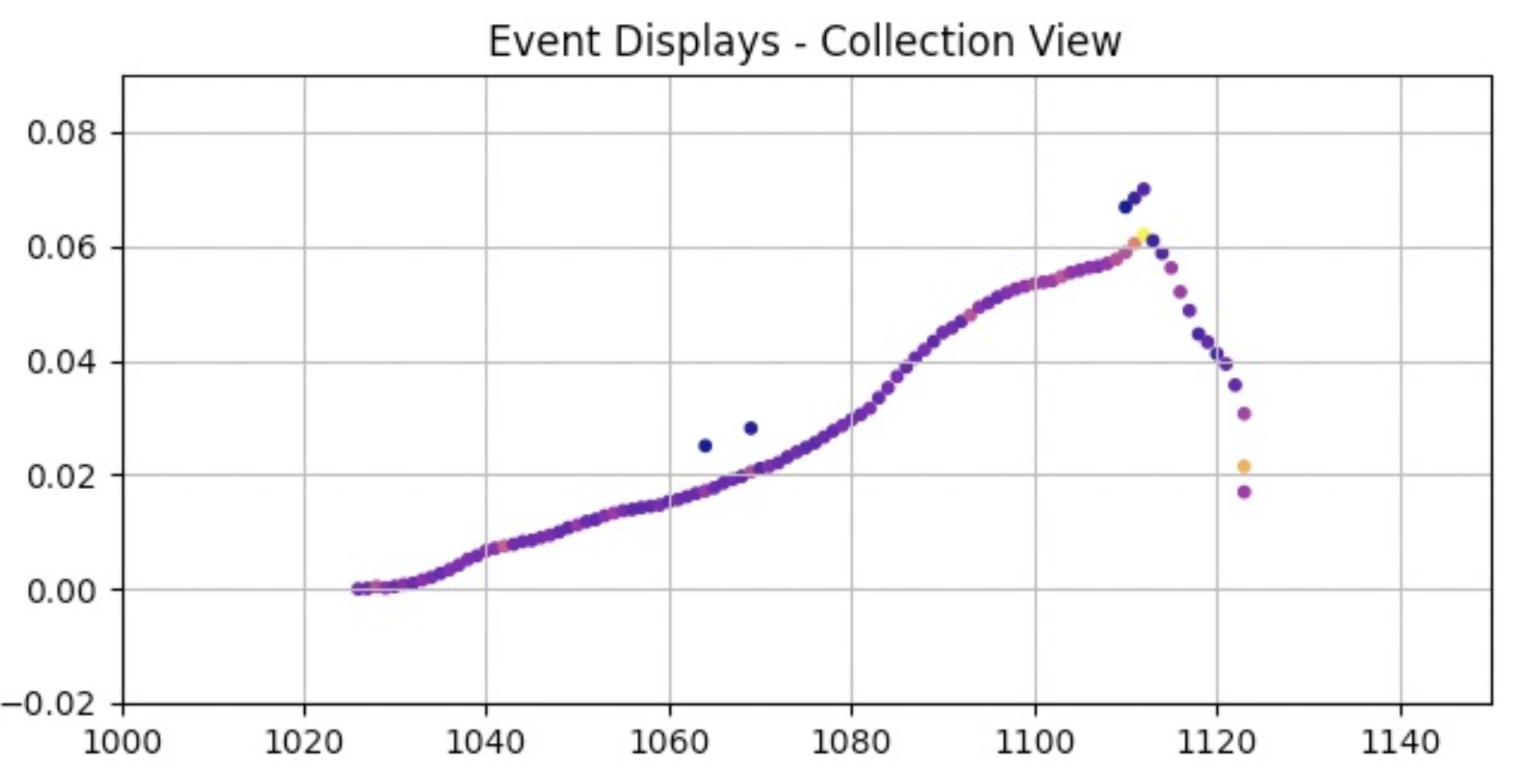}
	\includegraphics[width=0.5\textwidth]{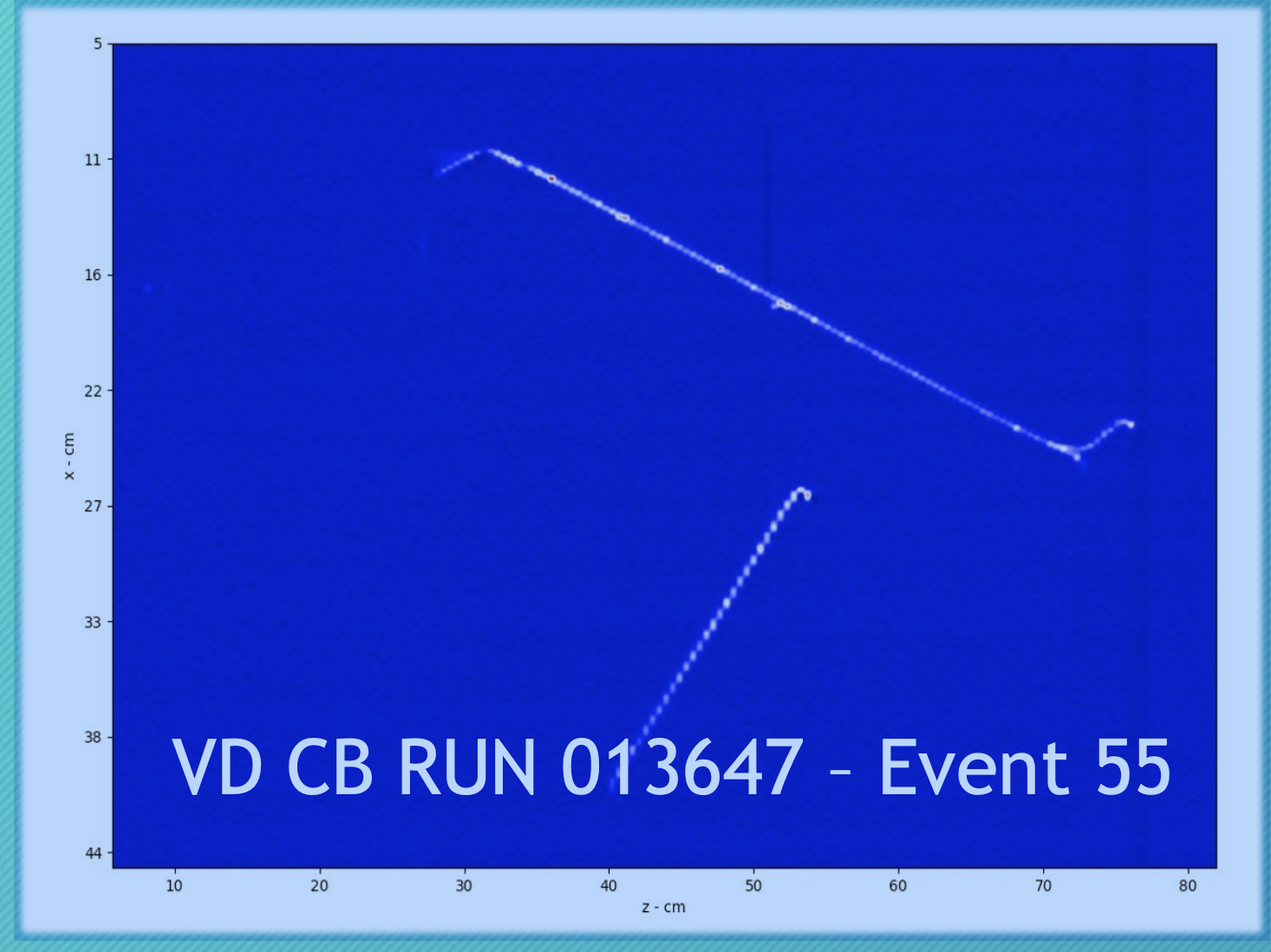}
		\caption[Michel events, simulation and data]{(Left) A Michel event triggered in simulation of the \dshort{vd} \coldbox (Right) A Michel candidate from the exclusive Michel trigger in \dshort{vd} \coldbox data.
	\label{fig:micheltrig}}
	\end{figure}

Figure~\ref{fig:micheltrig} shows the trigger catching a Michel in simulation on the left, and a Michel candidate from the \dshort{vd} \coldbox on the right.

%%%%
\subsection{Data Collection}
\label{subsubsec:EBsss}
The role of the data collection system is to handle the flow of data coherently from their sources to local storage, to provide inputs to data filtering, to collect the filtering result, and to interface with computing for data transfers to permanent storage. It is the only interface to the local \dshort{daq} storage at \gls{surf},  and is responsible for ensuring consistency, traceability and reliable handling of the data at all times until they are moved to offline computing at \dword{fnal}. Figure~\ref{fig:dq_subsys_data} shows a detailed diagram of the data collection flow. 

\begin{dunefigure}
	[\dshort{tdaq} subsystems diagram with data collection]
	{fig:dq_subsys_data}
	{A schematic diagram of the \dshort{daq} with detailed data collection flow, showing the different paths to local storage (cylinder) for triggered data, \dword{snb} triggers, and trigger primitive storage.}
	\includegraphics[width=0.95\textwidth]{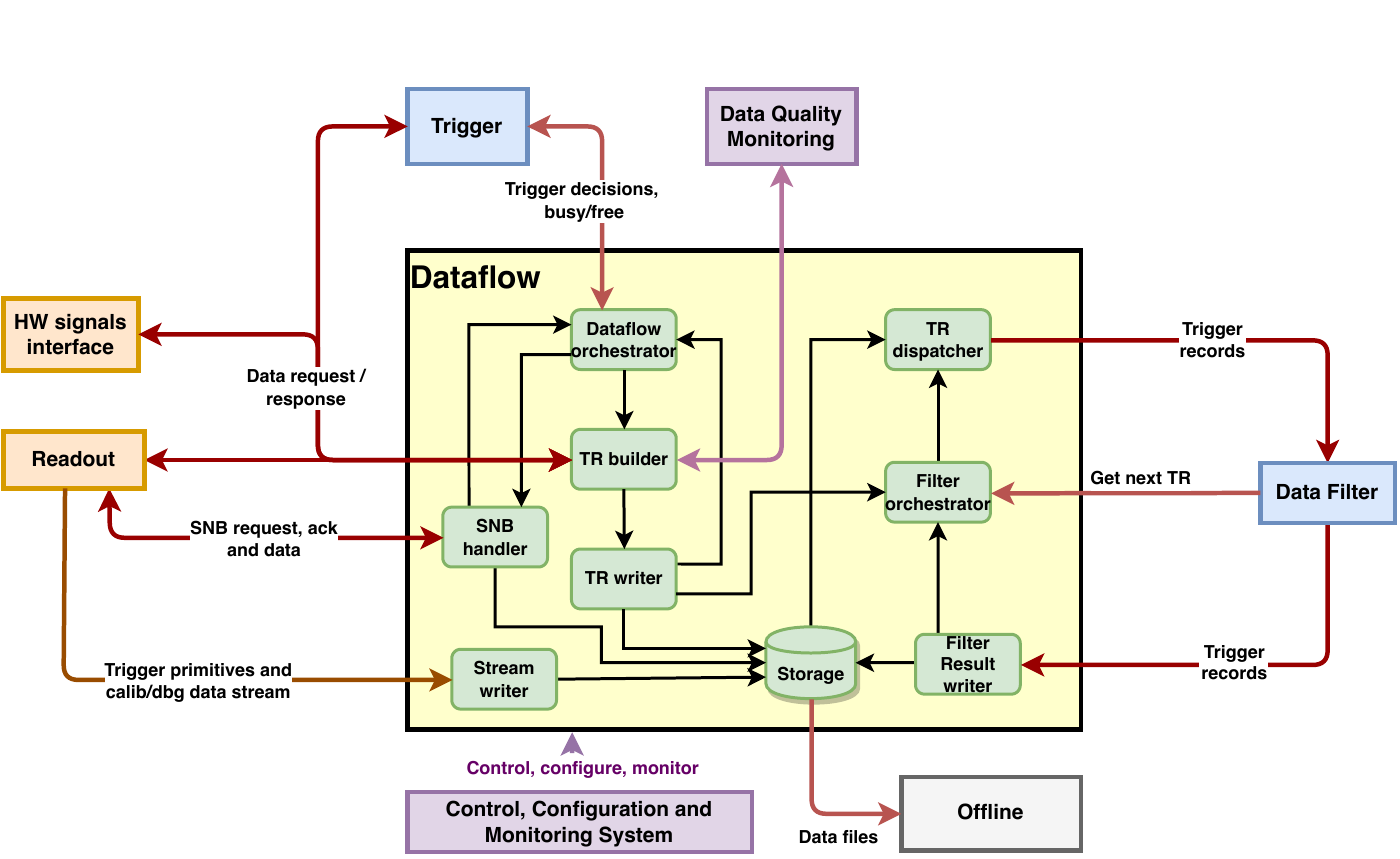}
\end{dunefigure}

The data collection strategy varies based on  the specific readout mode: interaction triggers, \dshort{snb} triggers, or streaming. 

Once an interaction \dshort{trigdecision} is formed, the data collection system gathers data fragments from the readout units into coherent \dwords{trigrec}. Intermediate trigger objects that contributed to the \dshort{trigdecision} are added to complement the information.
Trigger records are stored on disks prior to being served to the high-level filter system for further data reduction. The high-level filter forms modified \dshort{trigrec}s that the storage system  
saves on disk, awaiting their successful transfer to 
\dword{fnal}.

The \dshort{snb} records require a different approach due to their size: the data collection system instructs \dwords{daqrou} to initiate the raw streaming to local storage on the \dword{daqrou}. Afterwards, it transfers \dshort{snb} record fragments to the surface storage servers with high priority, compatibly with the available bandwidth and without interfering with the collection of interaction triggers, which continues in parallel.
From the surface storage, the \dshort{snb} record is transmitted to \dword{fnal} computing over the 100\,Gbps \dword{wan} link. To minimize the transfer time, which is expected to be a few hours, the data collection system will start uploading as soon as \dshort{snb} raw data reaches the surface buffer. 

The \dshort{trigprimitive}s are continuously streamed during data taking. The data collection system is responsible for keeping a copy of the stream in the surface temporary storage for a time period of at least a few months. The \dshort{trigprimitive} stream records will be made available on demand.

\subsection{Run Control, Configuration and Monitoring}
\label{subsubsec:RCsss}

The \dfirst{daqccm} subsystem is in charge of controlling, configuring, and monitoring the \dword{tdaq} system, as well as the detector components participating in data taking. It provides a central access point for the highly distributed \dshort{tdaq} components, allowing them to be treated and managed as a single coherent system through their corresponding subsystem interfaces. It is responsible for error handling and recovery, which is achieved through a robust and autonomous fault-tolerant control system. The main goal is to maximize system uptime, data-taking efficiency and data quality, taking into account that the system will encounter changes in data-taking conditions, both programmatic (e.g., calibrations) and unplanned (e.g., hardware failures or software faults). 

The \dshort{daqccm} provides an access point that delegates the user's actions, defined as any kind of human interaction, to internal function calls and procedures. It protects the direct access to detector and infrastructural resources. It also controls authentication and authorization, which limits different functionalities to certain groups and subsystems. As an example, only 
individuals authorized as detector experts can modify the \dword{fe} conﬁguration through the conﬁguration interfaces, or exclude a \dword{crp} from the readout.

The control component validates, distributes and executes commands on the \dshort{tdaq}, and is in charge of keeping the system in a coherent state. It consists of several components, such as access manager, process manager, resource manager and run control to carry out its tasks and also implements the intelligence required to automatically maintain the system in a properly functioning state or to alert operators if any parts  malfunction and cannot be automatically recovered. The smallest unit of a detector that can be controlled independently is referred to as a \dword{daqpart}.

The conﬁguration component provides several key elements for the conﬁguration management of the  \dshort{tdaq} components and detector \dword{fe} electronics. It provides  descriptions of system conﬁgurations, the ability to deﬁne and modify conﬁgurations, and graphical user interfaces for the human user to access the data. Data access libraries will hide the technology used for the database's implementation. The conﬁguration component  is also responsible for the archiving and bookkeeping of any used conﬁgurations.

Highly scalable and efficient operational monitoring is essential during data-taking periods. Any malfunctioning component of the experiment must be identified and reported as soon as possible. The monitoring component is intended to probe the \dshort{tdaq} components, services, and resources, collect and archive the obtained status information, and provide aggregation and visualization tools. 

The types of monitoring information vary greatly, ranging from log/error messages to metrics of different types. The monitoring infrastructure must therefore be flexible enough to seamlessly accommodate additions and modifications, and provide an aggregated view of the system behavior. The monitoring subsystem is a data source for the control subsystem that makes use of the information to automatically optimize the data taking conditions or recover from errors. Monitoring data is stored in databases and can be viewed through \dword{grafana} dashboards by the end user. The database structure continues to evolve as experience is gained in what is necessary to monitor.

The \dshort{daqccm} must also be flexible enough to be able to deploy the \dshort{daq} on a number of different test stands, prototypes, and production systems with varying sizes of server clusters, and manage the computing loads on these systems. 

The requirements on the \dshort{daqccm} led to an investigation of systematic solutions, and Kubernetes~\cite{k8s} was chosen as the path forward. Kubernetes is an open-source system for deploying and controlling applications running within containers, used very widely in the technology industry. There were several motivations to choose this approach for the DUNE \dshort{daq} system, driven, for example, by the location of the DUNE FD and the uptime requirements for the experiment. These include:

\begin{itemize}
    \item  Reliability engineering: Considering the stringent uptime requirement for the DUNE detector and looking to industry to see how comparable performance is achieved, cloud technology providers have similar or more stringent requirements for large data centers. The container orchestration paradigm is used throughout this industry, with Kubernetes the system of choice in almost all cases. 

    \item Provision of services: In addition to the DUNE \dshort{daq} applications for dataflow and triggering, the system relies on a large number of commercial and custom services, e.g., databases, visualization dashboards etc., which must be kept running reliably.  Kubernetes provides a robust and convenient way to define and deploy such services and to expose them in a simple way such that they may be used by other applications. 
    
    \item Test-stand support: In addition to the %experiment 
    \dword{fd} at \dword{surf} and larger test facilities, such as those based at \dword{fnal} and \dword{cern}, there is a need to provide \dshort{daq} functionality to small test stands located in labs at institutes around the world. A failure to provide coherent infrastructure for this use case %is expected to 
    would likely lead to homegrown testing scripts %being developed which 
    that duplicate some features of the \dshort{daq} software. 

    \item Resource management and networking: Kubernetes provides many tools that are required by the \dshort{daq} system, e.g., fail-safe process control, resource management to ensure validity of configuration and avoid clashes of hardware requirements, and a networking system that allows for improved flexibility compared to typical networking.

    \item System administration: given the location of the DUNE \dword{fd}, %complex at SURF 
    remote system administration is required. Running applications in containers gives some level of independence from the underlying operating systems on the \dshort{daq} nodes themselves, since the operating system and dependencies for applications can be distributed alongside the applications in their containers. This reduces % thus 
    the need for disruptive large-scale system upgrades. % is expected to be reduced.
\end{itemize}

Given the strong motivation to consider container-based systems, with Kubernetes as the most obvious choice, an R\&D program %of research and development 
was initiated in mid-2021. The starting point was the development of a single-node cluster configuration, providing some key services required by the \dshort{daq}, such as the  databases for storing operational monitoring and logging data, and the \dword{grafana} dashboard server for viewing this data. The development addressed the test-stand and provision-of-services requirements. 
After successful testing of the prototype %was tested and run 
by many \dshort{daq} developers, % was considered a success, and 
the decision was taken to continue this development %for 
to meet the full needs of the \dshort{daq}.

The next step was to set up a test cluster at the \dword{np04} facility at \dword{cern}, with about five nodes. This %system 
cluster has been used as the main development facility for over a year and has allowed us to scale up the early developments to an intermediate size and make use of realistic hardware systems. A key test was to benchmark the dataflow performance using the \dword{felix} hardware as an example of a hardware readout card 
to pass data to applications running within containers. 
Tests have shown that the performance is not affected by the use of containers, which was a potential concern, and have led to the cluster design shown in Figure~\ref{fig:daq_k8s}.

\begin{dunefigure}
	[\dshort{tdaq} subsystems diagram with data collection]
	{fig:daq_k8s}
	{A schematic diagram of cluster design based on the Kubernetes system. ``Pods'' are sets of containers running on servers (nodes); the pods run applications that carry out various functions of the \dshort{daq}, though not all nodes will run all applications simultaneously. Specialized nodes interface with readout hardware and with storage. Pods can connect to external services that can, e.g.,  provide logging services. Other services can be deployed on nodes natively. The run control interface allows the human user to configure and control apps across the cluster.}
	\includegraphics[width=1.15\textwidth, angle=90]{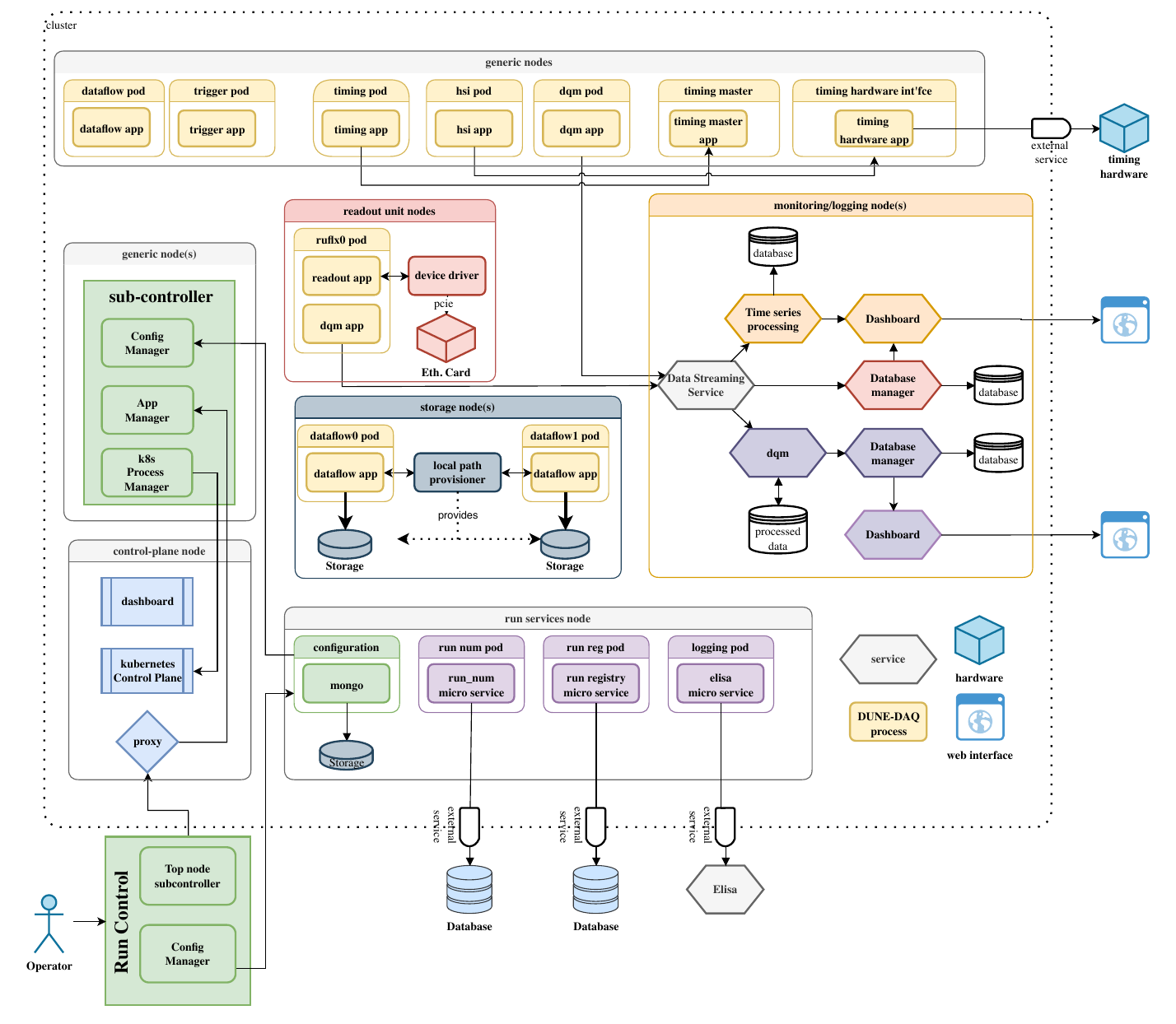}
\end{dunefigure}

Continuing work concentrates on scaling up the system from a small number of nodes to a larger, more production-like system, with input from multiple readout cards, output to multiple disk servers,  distribution of applications over a larger number of nodes, and corresponding monitoring of this larger system. Future plans include continuing to develop expertise in Kubernetes and leverage the wealth of features it provides, and  development of a supervisor control layer, likely based on an expert system, to provide automated responses to common problems.

\subsection{Data Quality Monitoring}
\label{subsubsec:DQMsss}

The \dword{dqm} subsystem is complementary to the monitoring component of the \dword{daqccm}. Instead of collecting counters, rates and logs, the \dshort{dqm} analyses samples of raw data and compares the results with the expectations, thus assessing the quality of the data themselves, rather than the quality of the \dshort{daq} process.

The \dshort{dqm} consists of three parts: the local processing module, the remote analysis module, and the web platform. Their interactions are shown in Figure~\ref{fig:daq_dqm}.

\begin{dunefigure}
	[DQM diagram]
	{fig:daq_dqm}
	{A schematic diagram of the \dshort{dqm} system showing the major components. %Light purple c
	Components at left plus the broadcast service are underground and %dark purple components 
	those at right are aboveground.}
	\includegraphics[width=0.7\textwidth]{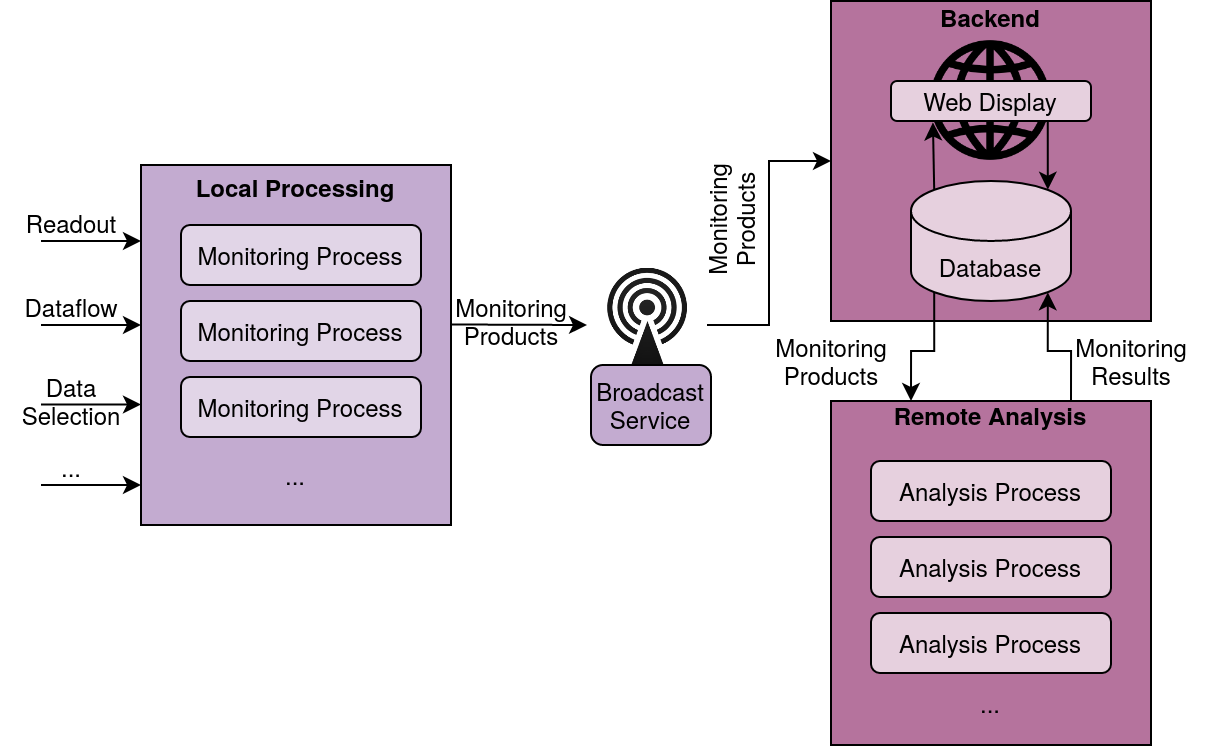}
\end{dunefigure}

The local processing module is directly connected to the rest of the \dshort{daq}, and will run underground at the \dword{fd} site. It samples data in parallel from across the \dshort{daq}, at configurable rates designed to keep processing requirements below 1 core per CRP. 
%\fixme{per \dword{crp}?} 
From these data, it produces monitoring products summarizing the data quality (such as histograms and Fourier transform), which are transmitted to the surface via an Apache Kafka broadcast service.

The web platform will be hosted on servers operating at the surface. It receives the monitoring products transmitted by the local processing, and indexes them in its own dedicated database. This database is connected to a web %UI
user interface, accessible to collaborators anywhere in the world, that allows users to navigate each detector subsystem and examine both live %data quality 
and historical data quality conditions. A data retention scheme will be developed to manage the total data volume in this database by down-sampling the time resolution of old run periods, keeping maximum granularity for current data quality conditions while ensuring that data quality records remain for the full run history of the experiment.

The remote analysis module will be hosted on the same surface-level servers as the web platform. It will monitor the \dshort{dqm} database as it is filled, and perform automated assessment of the monitoring products produced by the local processing. These assessment algorithms will generate alerts that can be passed directly to operators and experts through the web platform, or to the control subsystem in order to execute automated recovery actions. This module is designed to continuously evolve as understanding of the detector broadens during commissioning and early operation, and as detector conditions evolve and stabilize, allowing for the easy addition and removal of parallel analysis algorithms.

Prototyped versions of the local processing module and web platform have been tested at the \dword{cern} \coldbox{}es. Development of the algorithms for the remote analysis model will be developed with experience running detectors to define ``normal'' operations.

\section{Design Validation and Development}
\label{sec:tdaq:des-val-dev}

The validation and development of the \dshort{daq} design uses multiple test stands to prototype, develop, and test the performance of the system. This development is closely linked to the development of the electronics, and testing of the two systems goes hand-in-hand. 

\subsection{CERN \Coldbox}

As described in Chapter~\ref{ch:mod0} %Section~\textcolor{red}{chapter 8?} 
the readout electronics for \dword{spvd} are being tested at \dword{cern}'s \dword{np02}. This test provides the opportunity to test the integration of the %top electronics 
\dword{tde} with the \dshort{daq} to ensure that the interface between the electronics and the upstream readout units is working correctly, as well as the \dword{daqccm}.

\subsection{Module 0}%ProtoDUNE-II}
\label{subsec:tdaq:pduneii}

The primary large-scale test of the \dshort{daq} will be via the \dword{vdmod0}  activities, where the \dshort{daq} will be tested for both the \dword{sphd} and \dword{spvd} detectors. 

\section{Hardware Procurement and QA/QC}
\label{ch:qaqc}

The \dword{tdaq} system relies on \dword{cots} items, except for the custom built \dword{daqdts}. This section describes the approach to procurement as well as the \dword{qa}/\dword{qc} plans. The \dword{daqdts} hardware and the %other 
\dshort{cots} items are treated separately.

\subsection{Timing System}
\label{sec:hw:timing}
The timing system consists of three items of custom hardware, supported by  \dshort{cots} components. The custom components need to be produced, tested first individually and then as part of a system. Detailed procurement schedules and testing procedure documents are being developed, and will be available in the second quarter of 2023

\subsubsection{Custom Components}
\label{sec:hw:timing:custom}
Table \ref{tab:timing-custom-components-quantity} lists the custom components for the timing system that will be in operation in the first two \dshort{detmodule}s.

\begin{dunetable}
	[Custom items in \dshort{daqdts} ]
	{p{0.6\textwidth}p{0.15\textwidth}p{0.15\textwidth}} 
	{tab:timing-custom-components-quantity}
    {Summary of custom hardware items in \dword{daqdts}}
	\textbf{Description} & \textbf{Quantity in service} & \textbf{Quantity of spares}\\ \toprowrule
    GPS Interface Module. One at the top of each access shaft & 2 & 2 \\\colhline
    \dshort{utca} Interface Module. One per \dshort{utca} crate. Two crates per far detector module & 4 & 2\\\colhline
    Fiber Interface Module. Ten per \dshort{utca} crate & 40 & 5\\
	
\end{dunetable}

\subsubsection{COTS Items} % in Timing System}

Table \ref{tab:timing-cots-components-quantity} lists the \dword{cots} components for the timing system that will be in operation in the first two \dshort{detmodule}s. 

\begin{dunetable}
	[\dshort{cots} items in \dshort{daqdts} ]
	{p{0.6\textwidth}p{0.15\textwidth}p{0.15\textwidth}}
	{tab:timing-cots-components-quantity}
	{Summary of \dshort{cots} items in \dword{daqdts}}
	\textbf{Description} & \textbf{Quantity in service} & \textbf{Quantity of spares} \\ \toprowrule
\dshort{utca} crate: Either Vadatech or Schroff/nVent & 4 & 1 \\\colhline
 \dshort{utca} fan trays                               & 8 & 2 \\\colhline
 \dshort{utca} MCH: Crate controllers. One per crate   & 4 & 1 \\\colhline
 \dshort{utca} PSU: Power supply modules. Redundancy. Hot swappable & 8 & 2 \\\colhline
 \dshort{utca} JSM:  JTAG service modules (for reflashing FPGA firmware) & 4 & 1 \\\colhline
 \dshort{cots} FPGA based AMC that houses fiber interface board & 40 & 4\\\colhline
 GPS interface board PSU: 12V/3A output , 120V input external power supply  & 2 & 1 \\ 
	
\end{dunetable}

\subsubsection{Timing System QA/QC}

Production modules will be tested and ``burnt in'' in the UK before shipping to SURF, where they will undergo brief testing after installation. A plan for the testing of production modules will be documented in the second quarter of 2023, and will include a detailed testing procedure document, a daily testing schedule and manpower allocations specifying those responsible for various testing aspects.

The testing schedule provides one-year contingency to allow for another production batch in case some
components fail any of the specified tests. 
The fiber patch panels and passive optical spliters on the top of the cryostats will be installed by the infrastructure installation team, and tested together with the detector readout data and control fibers.

\subsection{TDAQ COTS Items}
\label{sec:cots}
The remaining \dshort{tdaq} hardware consists of computers and network switches, interconnected through either CAT6 copper cables or optical (\dword{om3}, \dword{om4} and Single Mode) fibers.

The computers themselves can be split into four categories as shown in Table~\ref{tab:daq-cots-components}.

\begin{dunetable}
	[\dshort{cots} items in \dshort{tdaq}]
	{p{0.1\textwidth}p{0.5\textwidth}p{0.13\textwidth}p{0.13\textwidth}}
	{tab:daq-cots-components}
	{Summary of \dshort{cots} items in \dshort{tdaq}}
	\textbf{Item} & \textbf{Description} & \textbf{Quantity in service} & \textbf{Quantity of spares} \\ \toprowrule
	Readout servers & high-end servers with 2x100G Ethernet \dwords{nic}, 2x10G Ethernet \dwords{nic}, at least 768\,GB of DDR5 RAM, at least 4 TB of fast \dword{nvme} storage and at least 32 physical CPU cores @2.5\,GHz & 167 & 8 \\ \colhline
Storage servers & used for data flow through the DAQ, with large storage and high performance I/O capacity & 12 & 1 \\  \colhline
Data processing servers & used for \dshort{daqtrs}, data flow, \dword{dqm}, with optimized CPU for processing & 102 & 5\\  \colhline
General purpose servers & used for \dword{daqccm}, \dshort{tdaq} infrastructure, \dword{daqeti}, etc., with a balanced configuration for I/O, processing and storage & 40 & 2 \\  \colhline
Readout switches & Ethernet aggregator switches to multiplex the detector links into 100G links for DAQ reception & 71 & 4 \\ 
\end{dunetable}

\subsection{Procurement Process}
The \dword{tdaq} infrastructure at the \dword{ehn1} is currently instrumental for the testing of hardware samples: it is used to finalize the hardware specifications prior to launching the procurement.

The main characteristics of the required hardware have been clarified through the prototype implementations of the \dshort{tdaq} components described in Sections~\ref{sec:hw:timing} and~\ref{sec:cots}, such that realistic cost estimates could be made.
Nevertheless, the pace of technology and products evolution is at present so high that detailed specifications for the computers and their peripherals will only be made closer to the purchasing date. For example, new motherboards supporting \dword{pci}e Gen 5 and \dword{dram} DDR5 have only recently been launched by both Intel and AMD, with important consequences on memory, storage and network performance.

The aim is to have a complete set of specifications in 2024, for procurement in 2024-2025 and installation of the pre-series hardware at the \dword{surf} in 2025-2026 for \dword{sphd} and 2027 for \dword{spvd}. The pre-series hardware makes up approximately 25\% of the final quantities, and will be used for detector integration and testing. The remaining hardware (approximetly 75\%) will be installed closer to the date scheduled for the filling of the detector with  \dword{lar}.

The specifications documents will be checked during a \dword{prr} prior to launching the procurement. The aim of this review is to give the final green light for the material spending.

Part of the procurement process is to test combinations of potential equipment together to eliminate compatibility problems in advance. Candidate hardware will thus be installed in the \dword{protodune} \dshort{tdaq} barracks at \dword{cern}, as part of the qualification step during the procurement process.

\subsection{Quality Assurance and Control}
Vendors of computing equipment have a set of standard \dword{qa}/\dword{qc} procedures to ensure the quality of their products. Nevertheless, during the tendering process it is possible to require additional tests to certify that the functionalities meet specifications. This is particularly important since the plan is for the  equipment to be shipped directly to the \dword{sdwf}, ready to be moved from the storage area to \dword{surf} when needed.

For the \dword{fd} \dshort{tdaq} computers the following tests (with the corresponding test result documents) will be required: 
\begin{itemize}
    \item installation of an agreed version of the BIOS firmware,
    \item adjustment of the BIOS settings,
    \item installation of a Linux operating system image provided by \dword{fnal},
    \item stress testing of CPUs using an agreed utility,
    \item stress testing of \dword{dram} using an agreed utility,
    \item stress testing of disks using an agreed utility adapted to the type of storage,
    \item stress testing of network I/O using an agreed utility,
\end{itemize}
The test reports will contain not only the information about pass/failed tests but also the monitoring values of temperatures and power consumption during the tests.

In order to catch problems that might have circumvented the manufacturer's QC process, acceptance testing will be done also {\it in situ}, at the \dword{surf}. 
This eliminates the additional time, logistics, and handling that would be needed if this were done elsewhere. To reduce the impact of items which fail the acceptance testing, spares (which are needed anyway) will be included in the initial order, allowing head room to return any defective hardware without impacting the installation schedule.

Acceptance testing for \dshort{cots} computing and networking hardware is part of the installation. A 24h burn-in procedure will be run to catch cases of infant mortality.  For computers, a standard Linux burn-in/test suite will be run (largely overlapping with the tests carried out at the manufacturer's premises).

The network switches will be tested by routing data through them at the full load, using industry standard tools.

For other \dshort{cots} components procured by the \dshort{tdaq}, the plan is to rely on the manufacturer's QA/QC standards only: this is in particular valid for optical transceivers, pre-terminated fibers and CAT6 cables, patch panels, passive optical splitters.

\section{Assembly, Installation, and Integration}
\label{sec:tdaq:assy-inst-integ}

The installation of the \dshort{daq} must be carefully timed to support the activities of the rest of the detector components, since the \dshort{daq} is required to check 
whether the installation and integration of the detector is proceeding well. The physical activities of the \dshort{daq} installation are not extensive in comparison to other subsystems, but the timing of the activities is critical to the overall success of the detector. 

The interface document describing the interface between the \dshort{daq} system and the facilities is in~\cite{daqinterfaceii}.

\subsection{DAQ Facilities and Infrastructure}

The \dshort{daq} for \dword{spvd} will inhabit similar space to that for  \dword{sphd}.  Note that the location of \dshort{daq} infrastructure has changed since the publication of the \dword{sphd} TDR\cite{Abi:2017aow}.  The \dshort{daq} computing infrastructure for each %individual module 
\dshort{detmodule} will be located on the cryogenics mezzanine on top of each detector module rather than all together in the \dword{cuc}.  Servers central to all modules will be in the Ross Dry basement at the surface, and all modules will connect to the fiber trunk running up the shaft in the \dword{cuc}'s \dword{mcr} room, where central networking equipment will be located.  Additionally, \dword{gps} antennae for the timing system will be located %on top of each headframe.  
at the top of both the Ross and Yates shafts.

\subsection{Ross Dry Basement}

The southwest corner of the Ross Dry basement area (Figure~\ref{fig:daq_rossdry}) at the surface will contain not only networking and central IT related gear, but eight 42U \dshort{daq} racks.  These will house the data collection servers and transient storage, the data filter farm, and servers for \dshort{daq} services common to all %four DUNE detector modules 
\dshort{detmodule}s (web services, databases, and so on).  The shaft fibers connect to this room, as does the connection to the \dword{wan}.  For this site, 50\,kVA of power and cooling, the racks, and cable trays will be installed by the fall of 2024, and the minimal installation of \dshort{daq} administrative machines needed to facilitate further work underground will be installed by the summer of 2025.

\begin{dunefigure}
	[DAQ room in Ross Dry]
	{fig:daq_rossdry}
	{Layout of the \dshort{daq} area in the Ross Dry Basement.}
	\includegraphics[width=1.0\textwidth]{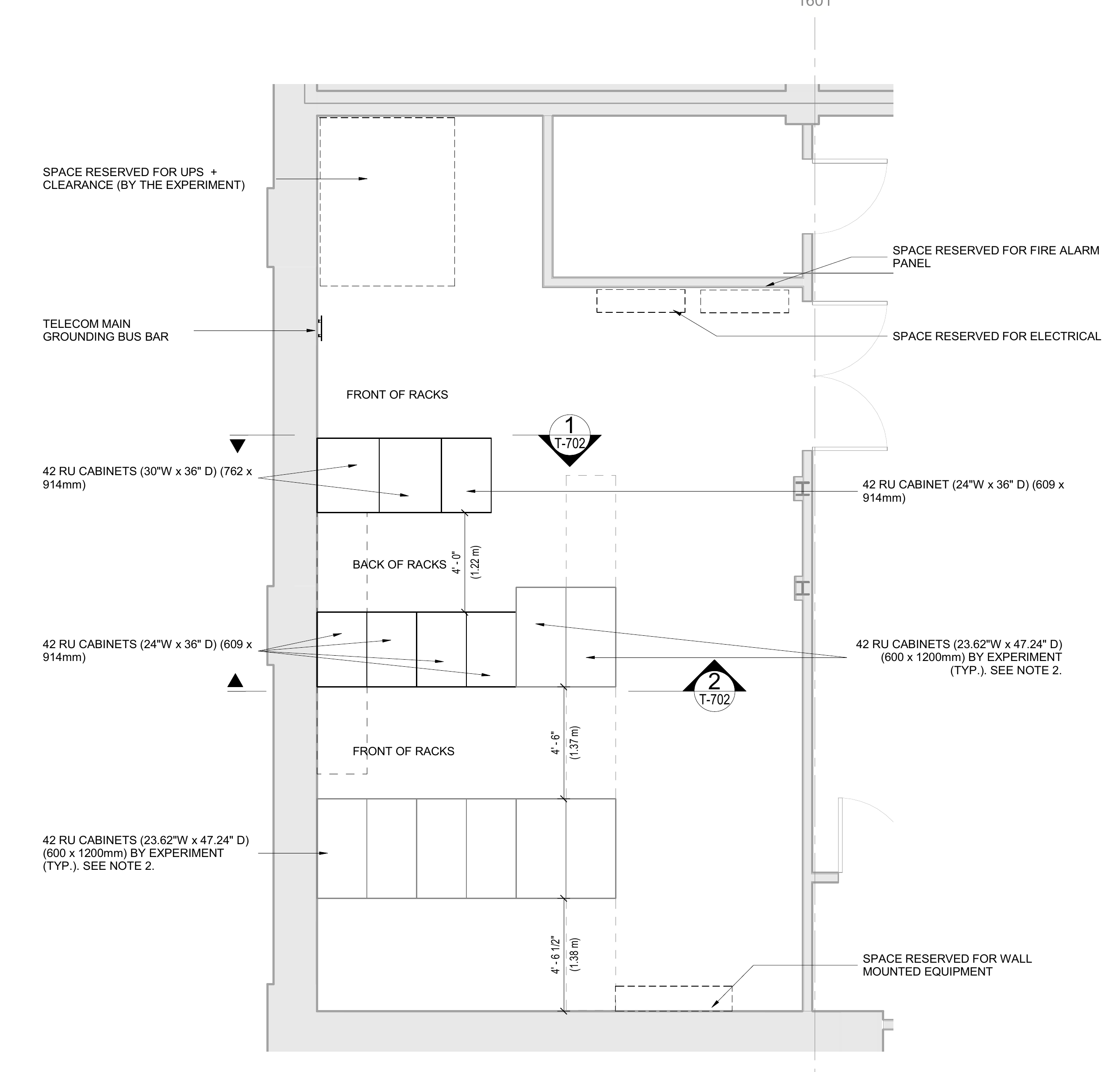}
\end{dunefigure}

\subsection{The \dshort{mcr} in the \dshort{cuc}}

The \dword{mcr} in the \dword{cuc} is the connectivity hub where all %four detector modules 
the \dshort{detmodule}s connect to the network out of the underground via \dword{fnal} central networking routers.  While this is facility space, it will be available to the \dshort{daq} consortium in the spring of 2025, well before either the surface or detector \dshort{daq} spaces will have need of it.

\subsection{The DAQ Barracks}

The main site for each detector module's \dshort{daq} infrastructure will be a barrack built on the cryogenics mezzanine above its cryostat. %each detector module.  
This space will contain 16 42U racks air cooled via a traditional data center hot/cold aisle arrangement.  Fibers will run from the warm electronics atop each detector module to patch panels in the barracks, then into the readout servers. This arrangement ensures that the electronically noisy \dshort{daq} computing infrastructure is galvanically isolated from the sensitive detector electronics. Data will then be transferred via fiber to the \dword{mcr} to be sent out to the surface components of the \dshort{daq}. 

Each \dshort{daq} barrack will have 150\,kVA of power and cooling.  Pallet staging and storage space is available on the cryogenics mezzanine floor nearby, and there is a separate small room suitable for %humans 
people to work outside the hostile environment of a server room.

Figure~\ref{fig:daq_barracks} shows the layout and \threed model of the barracks. 

\begin{dunefigure}
	[DAQ Barracks]
	{fig:daq_barracks}
	{Left: Overhead view of the DAQ barracks on the cryo mezzanine. Right: 3D model of the DAQ barracks. The server rooms and working areas are labeled.}
	\includegraphics[width=0.9\textwidth]{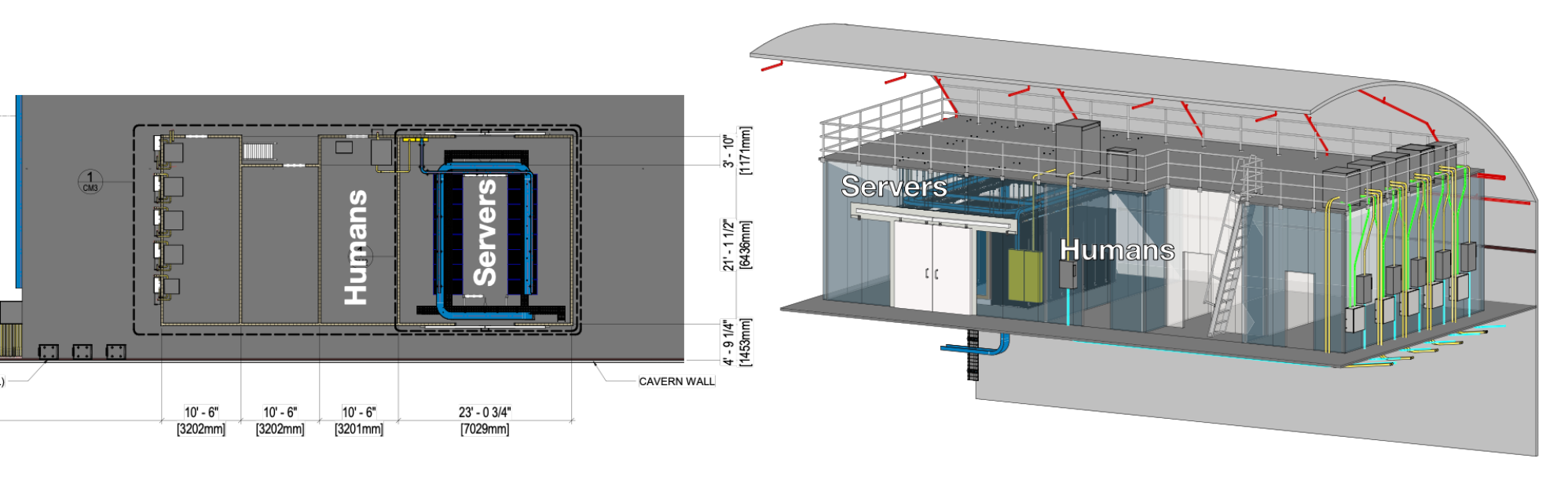}
\end{dunefigure}

\subsection{Installation Schedule}
\label{subsec:inst:sched}
The \dshort{daq} installation schedule is driven by when the warm electronics need to be connected to computers to test their readouts.  The \dword{sphd} barracks will be available complete with fiber connection to the \dword{mcr}, racks, cooling, and power early in 2026.  At that point servers can be installed in the racks and fibers laid in cable trays on the detector top to the warm electronics chimneys.  Three full months are foreseen in the schedule to finish the \dshort{daq} computing installation, configuration, and testing before it is needed to service detector electronics.

Both \dword{sphd} and \dword{spvd} fit into the same computing envelope, so the installation plan is the same for both detectors, offset by their installation schedules.  The \dword{spvd} barracks will be available for \dshort{daq} installation well after the \dword{sphd} \dshort{daq} installation is in the ``service \dword{apa} installation'' mode.

Items that go underground need to arrive at \dword{surf} at least three months before they are needed.  Thus, procurement is planned at present with ample headroom.

\subsection{Installation Scheme}

All three locations described here will need small crews of approximately six people for a limited time to rack in servers and switches, cable them up, and perform tests to validate the installation.  After that, a fractional  
\dword{fte} will be needed occasionally onsite to be hands-on, but the %installation is being designed 
system is designed to be operated remotely, with remote console logins and power distribution units. %This mode 
The remote operation is being developed and tested at \dword{cern}'s \dword{np04} site for  
\dword{hdmod0} and at \dword{fnal}'s \dword{iceberg} prototype, and will be replicated at test stands at multiple sites before the installation at \dword{surf}, by which point %it will be well-practiced.
any problems will have been solved.

\subsection{Networking}
The \dword{tdaq} relies largely on the network infrastructure provided by LBNF/DUNE at \dword{surf}. 

The Wide Area Network (WAN) for DUNE at SURF will be contracted and managed by Energy Sciences Network (ESNet), the network infrastructure and service provider funded by the Department of Energy Office of Science. The existing 10Gb/s VLAN circuit, provided by the Research, Education and Economic Development (REED) Network,will be upgraded to redundant links with a 100Gb/s primary path and at least 10Gb/s guaranteed bandwidth backup path provided by ESNet.  The redundant, full-bandwidth WAN will be available by the start of FD1 DAQ commissioning, June 2028.

DUNE networks at SURF share a high performance, redundant aggregation backbone provided by modular chassis switches/routers on the surface, in Ross Dry and Yates buildings, and underground in the MCR. The network equipment will be installed underground and in surface buildings. It will be configured, tested, managed and monitored by the Network Services group in Fermilab’s Core Computing Division and will have the capacity to support both FD1 and FD2 detectors.

DUNE DAQ requirements specify distinct, interconnected (via central hubs) networks with different performance and bandwidth needs: the Experiment and Technical networks. There is also a General Purpose network provisioned for other computers and user network access, and a separate management network infrastructure to manage all DUNE network devices. Fermilab Core Computing will configure all VLANs and provision ports for access switch uplinks for both FD1 and FD2 detectors before FD1 is ready for commissioning.

Core IT services such as DNS, DHCP, NTP, Kerberos and Active Directory/LDAP will run on a small virtual machine cluster co-located in the Ross Dry and Yates building computer rooms for redundancy. This infrastructure will be implemented in a fault tolerant scheme and is shared between FD1 and FD2. Access controls to the networks and to systems on the network will be applied on the basis of user roles and controlled centrally via Core IT applications managed by Fermilab Core Computing and will be built during the start of the initial DAQ surface installation, or before, in a standalone, locked rack temporarily placed outside the computer room in the Ross Dry lower level.

The DAQ consortium is responsible for the network used in the \dword{daqros}. Figure~\ref{fig:daq_network} illustrates the part of the overall network that will be used by \dshort{tdaq} for \dword{sphd} and \dword{spvd}. At the bottom of the diagram the readout switches are included.

\begin{dunefigure}
	[DAQ network.]
	{fig:daq_network}
	{Preliminary layout of the \dshort{tdaq} network.}
	\includegraphics[width=0.9\textwidth]{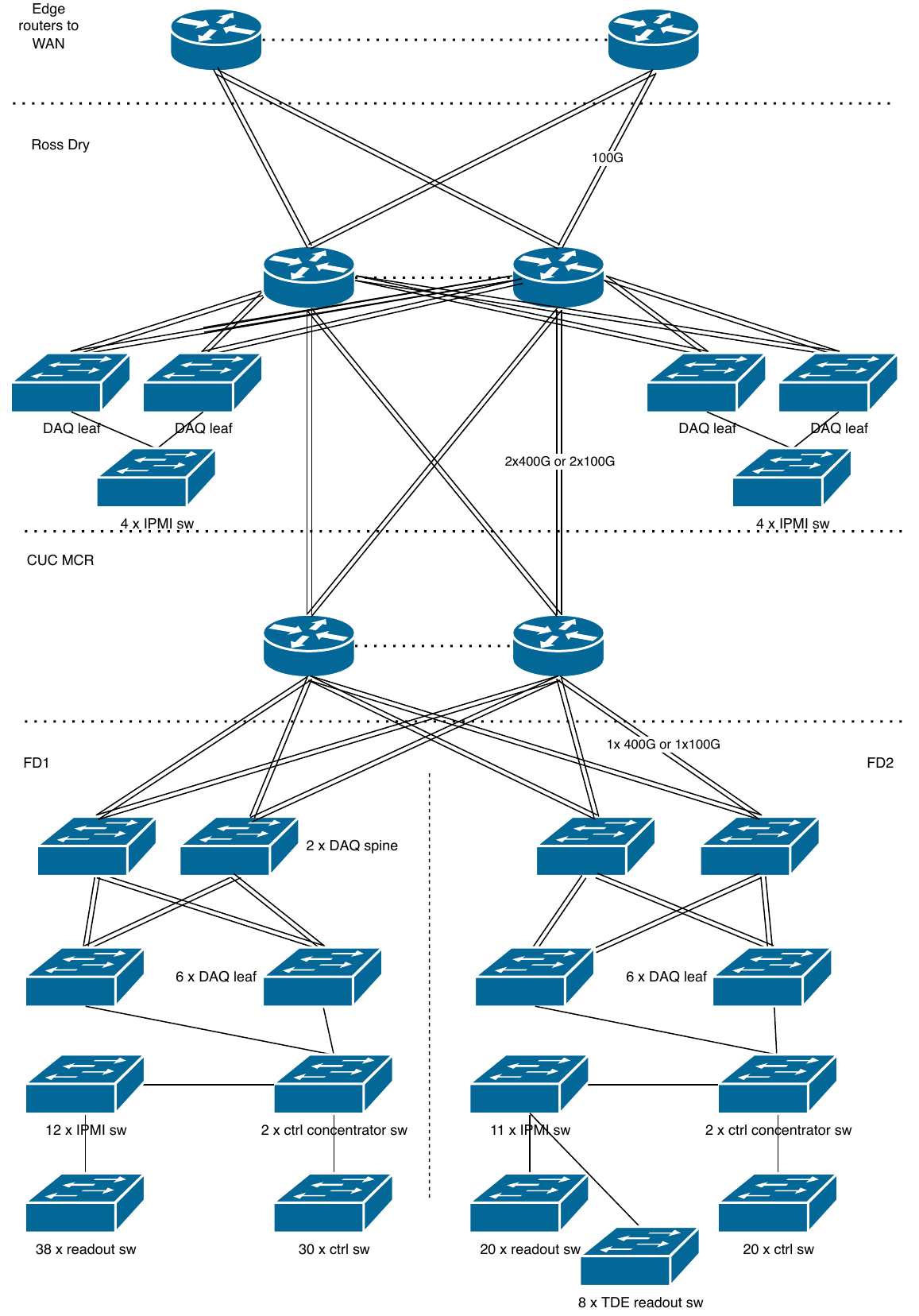}
\end{dunefigure}

\subsection{Rack Layout}

The rack layout and design is at a preliminary stage and shown in Figure~\ref{fig:daq_racks}, for the \dword{spvd} underground DAQ. Since the servers will be commodity hardware, specification and procurement of this hardware will happen closer to the time it is needed in order to manage costs and extend the lifetime of the equipment.
Once the final number, dimensions, and cooling needs for the servers are known, detailed rack design will be undertaken.

\begin{dunefigure}
	[DAQ racks for for \dshort{spvd}]
	{fig:daq_racks}
	{Preliminary layout of the underground racks for \dword{spvd}. Network devices use the same nomenclature as in Figure~\ref{fig:daq_network}. }
	\includegraphics[width=0.9\textwidth]{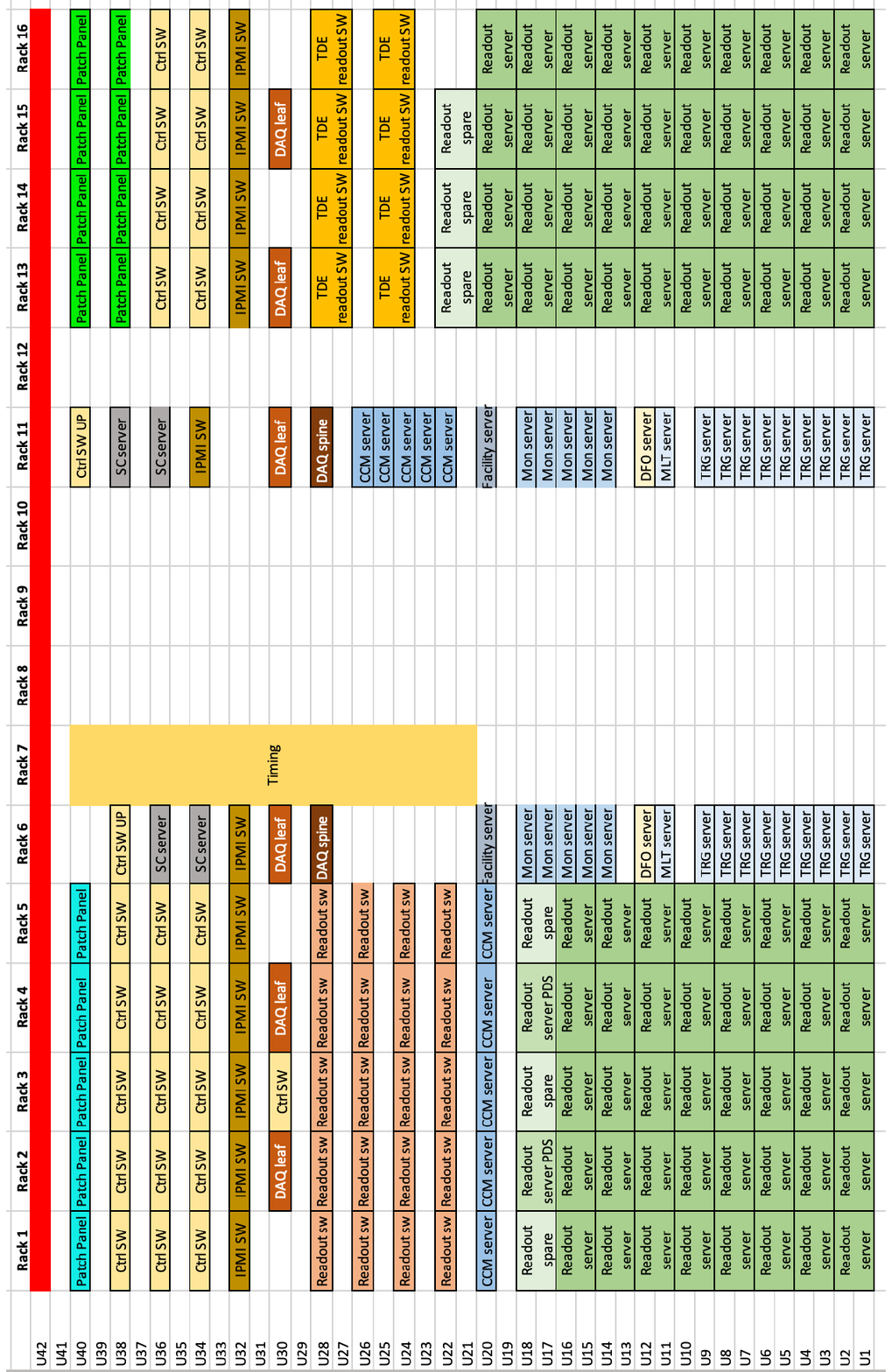}
\end{dunefigure}

\section{Organization and Schedule} %Project Management}
\label{sec:tdaq:org-mgmt}
\subsection{TDAQ Consortium}

The \dword{tdaq} consortium was inaugurated in 2017, with responsibility for delivery of the \dshort{daq} and trigger systems for all DUNE detector systems at the far and near sites. The management and working group structure of the consortium have evolved in the intervening time, and the current organization is shown in Figure~\ref{fig:daq_structure}. The working groups have responsibility for the \dshort{tdaq} subsystems described above. The data selection working groups have responsibility for both the design of the subsystem and the development of software and simulation tools to evaluate and optimize the physics performance of the overall system. Another working group is dedicated to the specification and design of the supporting infrastructure at \dword{surf} –– including computing services -- and planning the integration and installation steps required to deliver a working system.

\begin{dunefigure}
	[TDAQ Consortium Management Structure]
	{fig:daq_structure}
	{TDAQ Consortium Management Structure}
	\includegraphics[width=0.9\textwidth]{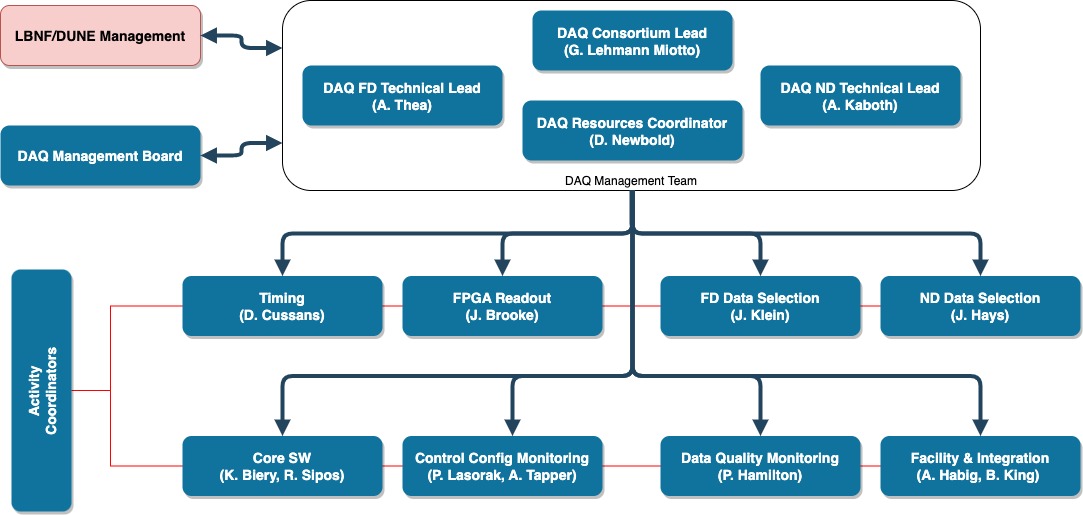}
\end{dunefigure}

The management team comprises the consortium leader, who is the %link person 
liaison to the DUNE \dword{exb}, and technical leads for far and near detector systems who are members of the DUNE \dword{tb}. They are supported by a resources coordinator with responsibility for resource and schedule management, liaising with the resources board. The funding agencies supporting the consortium are represented in the Management Board, which has ultimate responsibility for approving the project plan, monitoring progress, tracking changes, and ensuring that the necessary personnel and resources are committed to the project. 
A breakdown of consortium institutes and deliverables is shown in Table~\ref{tab:daq_institutes}. Many of these deliverables are common to all \dword{fd} modules, and so do not represent tasks specific to \dword{spvd}; however, the quantities of hardware components quoted below are those required for the \dword{spvd} system alone.

\begin{footnotesize}
\begin{longtable}{p{0.2\textwidth}p{0.38\textwidth}p{0.08\textwidth}p{0.14\textwidth}p{0.1\textwidth}}
\caption{TDAQ deliverables by institution} \\
\rowcolor{dunesky} Component & Description & Quantity (FD2) & Institutions & Funding Agencies \\  \colhline
TDAQ software \& firmware management & Administration of code repositories; \newline Development and maintenance of build tools; \newline Coordination, preparation, and distribution of SW/FW releases & & \dshort{fnal}, RAL & DOE, STFC\\ \colhline
Infrastructure servers & DAQ computing infrastructure (file servers, databases, etc) & 8 & \dshort{fnal} & DOE\\ \colhline
Computing administration and support & Development and testing of system configuration, management, and administration tools; \newline DAQ hardware inventory \newline system administration & & \dshort{fnal}, UMN Duluth & DOE\\ \colhline
Timing system & \dshort{utca}-based system to synchronize and distribute clock and timestamp to all electronics endpoints & 2 & Bristol, RAL & STFC\\ \colhline
External trigger interface & Server dedicated to the exchange of trigger messages with external systems (e.g. SNEWs) or other FD modules & 1 & Bristol, RAL & STFC\\ \colhline
Readout system cards & PCIe cards interfacing the DAQ to the detector electronics & 200 & Oxford, RAL & STFC\\ \colhline
Readout system SNB storage & Fast storage system for SNB data & 200 & Oxford & STFC\\ \colhline
Readout system servers & Servers hosting the readout cards and SNB storage, used to buffer data and transfer triggered data to the dataflow system & 100 & CERN & CERN\\ \colhline
Readout system firmware and software & Development and testing of FW and SW for the readout subsystem & & Toronto, CERN, Birmingham, Bristol, Oxford, STFC RAL, Sussex, UCL & Canada,
CERN, STFC\\ \colhline
Dataflow servers & Servers and storage to collect data and store them before final transfer to \dshort{fnal} & 8 & \dshort{fnal} & DOE\\ \colhline
Trigger servers & Servers to carry out the first stage of data selection & 20 & CERN & CERN\\ \colhline
Trigger software & Software development and testing to support the trigger algorithms execution & & Bristol, Oxford, RAL, Sussex & STFC\\ \colhline
Data filter servers & Servers to carry out the second stage of data selection; \newline software to support the filtering algorithms execution & 20 & \dshort{fnal} DOE\\ \colhline
Data filter software & Software development and testing to support the filtering algorithms execution & & CERN, Bristol, Oxford, RAL, Sussex & CERN, STFC\\ \colhline
CCM \& DQM servers & Servers to control, configure and monitor the system & 20 & RAL & STFC\\ \colhline
CCM SW & Development and testing of software to control, configure and monitor the system & & Birmingham, Imperial, Liverpool, STFC RAL, UCL, CERN, BNL, \dshort{fnal} & STFC, CERN, DOE\\ \colhline
DQM SW & Software development and testing to support the execution of data quality monitoring algorithms and visualize results & & Edinburgh, Imperial, UCL & STFC, CERN\\ \colhline
Algorithms, performance, simulation & Simulation and performance studies to devise optimal data selection strategies; development and testing of data selection and monitoring algorithms & & Toronto, Imperial, Oxford, Sussex, UCL, Columbia, Pennsylvania, BNL & Canada, STFC, DOE\\ \colhline
Fibers and cabling equipment & Patch panels, patch cords, timing connections, cable ties, labels &  & RAL & STFC\\ \colhline
HW for 
\dshort{mod0} & DAQ hardware for Module-0 in NP02 & & Toronto, RAL, CERN & Canada, STFC, CERN\\ \colhline
HW for FD2 CERN coldbox & Extensions to the %ProtoDUNE-I
\dword{pdsp} DAQ system for VD \coldbox & & Toronto, RAL, CERN & Canada, STFC, CERN\\ \colhline
Installation, integration, and support for %ProtoDUNE-II 
\dshort{mod0} & Effort to support the integration, commissioning, and operations of %ProtoDUNE-II 
\dshort{mod0} & & All TDAQ institutions\\ \colhline
Support for detector test setups & Effort to support the integration and commissioning of detector electronics and DAQ at detector test labs & & All TDAQ institutions\\ \colhline
DAQ hardware installation and commissioning at SURF & Effort to install DAQ hardware, test it and carry out integration and commissioning & & All TDAQ institutions\\ \colhline
DAQ integration and operations support at SURF & Effort to support the integration and commissioning of detector electronics and DAQ during the construction and commissioning phase & & All TDAQ institutions\\ \colhline

\label{tab:daq_institutes}
\end{longtable}
\end{footnotesize}

\subsection{%Project 
Schedule}

\subsubsection{Scheduling Approach}
The \dshort{tdaq} development schedule is driven by the requirements of other DUNE \dword{fd} subsystems through the various stages of their development, construction, installation and commissioning for physics.
\begin{itemize}
    \item During the development phase, primarily at test cryostats and at %\dword{protodune2}
    %ProtoDUNE-II 
the \dwords{mod0}, \dshort{tdaq} must support timing, readout and data storage functions adequate for reliable running and capture of data.
    \item Detector construction will be supported through provision of small-scale readout and timing systems at production sites as required.
    \item Early installation of \dshort{tdaq} components at the \dword{spvd} site will allow control, testing and evaluation of detector components and electronics during installation.
    \item Detector commissioning under warm and cold conditions will be supported by a fully functional \dshort{tdaq} system, with continuing development of data selection algorithms as the detector performance is understood and optimized.
\end{itemize}

The majority of resources in the \dshort{tdaq} project are devoted to software and firmware development, integration and testing. Moreover, the ``hard problems'' associated with development of a complex distributed real-time system lie to a large extent in the interfaces between components and the evolution of their specifications as the behavior and performance of the system is understood in detail.

For these deliverables, an agile-inspired approach to scheduling is therefore appropriate, and has been used with success during the development period leading up to %\dword{protodune2}
the \dwords{mod0}. This approach mandates the incremental development of software and firmware components in parallel, with emphasis on frequent integration leading to functional releases. The target feature set for each release is treated flexibly to account for the evolution of compontent specification. 
It is the responsibility of working group leaders to ensure that the overall schedule for the \dshort{tdaq} system is met, and moreover that the testing and development of hardware is adequately supported by mature software, as required. It is anticipated that continuous improvements will be made to \dshort{tdaq} software and firmware until DUNE physics running.

For hardware deliverables (either custom electronics or \dshort{cots} computing and networking), the development and evolution of the system stops at installation. Therefore, a more traditional scheduling approach is taken, involving steps of prototyping, integration and testing, pre-production, and production. Here the schedule is driven by the requirement to have all \dshort{tdaq} hardware available by the time of beneficial occupancy of the \dword{fd} site (surface and then underground areas respectively). Owing to risks associated with delays in hardware production, supply chain issues, or logistical problems, a conservative approach is taken, with explicit schedule contingency of up to six months for packaging and transatlantic shipping of goods, and three months allowed for logistics and handling at \dword{surf}.

\subsubsection{Milestones}

The \dshort{tdaq} project is a single common activity for all DUNE detector elements, and is planned so as to obtain maximum efficiency and cost saving from use of common technical approaches, choice of components, and integration and validation steps. The vast majority of development and integration activities are therefore common to \dword{sphd} and \dword{spvd}, and the project is planned around a single common set of milestones, also ensuring that no conflicts exist between the installation and commissioning steps for these two detector modules. The main \dword{spvd}-specific deliverables are:

\begin{itemize}
    \item Elements of online and offline software and timing system hardware which are required to support the top drift electronics;
    \item Data selection algorithms optimized for the \dword{spvd} charge readout and \dword{pds} geometry and performance;
    \item The installation campaign for \dword{spvd}.
\end{itemize}

Table~\ref{tab:daq_milestones} shows the top-level \dshort{tdaq}-specific milestones. These are largely driven by ready-by dates for infrastructure and facilities at \dword{surf}, and by need-by dates for installation.

\begin{footnotesize}
\begin{longtable}{p{0.4\textwidth}p{0.2\textwidth}p{0.25\textwidth}}
\caption{TDAQ Top-level milestones focused on \dshort{spvd}} \\
\rowcolor{dunesky} Milestone & Date & Driver\\ \colhline

Far detector final design review passed & March 2023 & Procurement schedule\\ \colhline
Ready for NP02 running & May 2023 & %ProtoDUNE-II  
\dshort{vdmod0} Schedule\\ \colhline
Procurement Readiness Review passed & September 2024 & Procurement schedule\\ \colhline
Initial DAQ surface installation complete & September 2025 & Surface beneficial occupancy\\ \colhline
Timing, readout, trigger, monitoring HW procurement complete & December 2026 & FD1/FD2 installation schedule \\ \colhline
Dataflow and Data Filter HW procurement complete & June 2027 & FD1/FD2 installation schedule \\ \colhline
FD2 DAQ underground installation complete & August 2027 & FD2 installation schedule\\ \colhline
FD2 DAQ surface installation complete & Oct 2027 & FD2 installation schedule\\ \colhline
Ready for FD2 commissioning & October 2028 & FD2 cool-down schedule\\ \colhline

\label{tab:daq_milestones}
\end{longtable}
\end{footnotesize}

A summary  schedule is shown in Figure~\ref{fig:daq_schedule}.
\begin{dunefigure}[Key DAQ milestones and activities toward 
\dshort{spvd}]{fig:daq_schedule}{
Key DAQ milestones and activities toward  the \dshort{spvd} in graphical format.}
\includegraphics[width=0.95\textwidth]{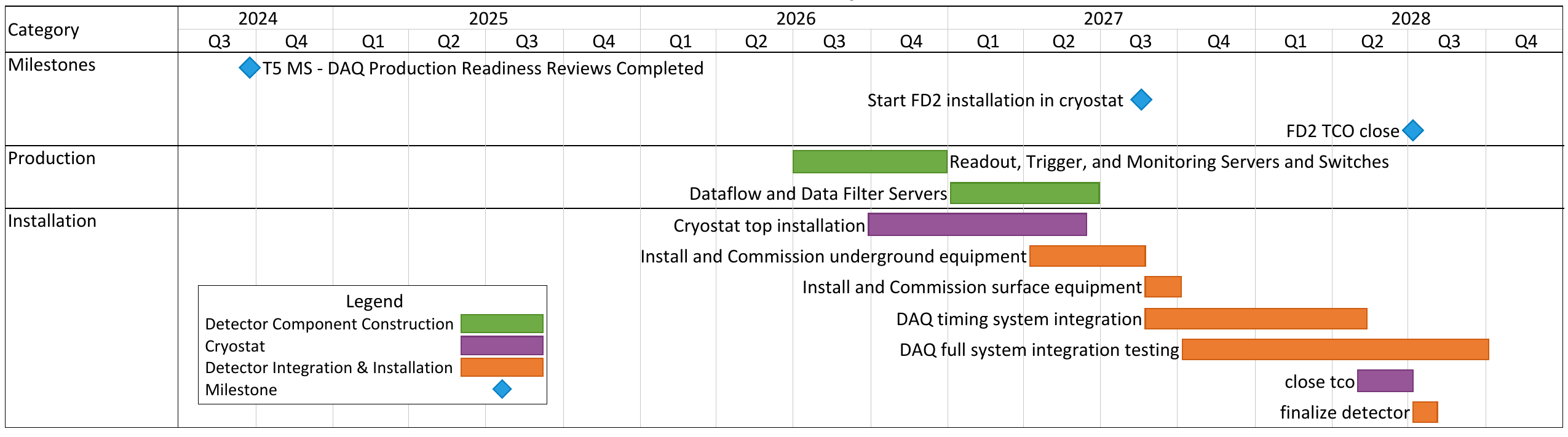}
\end{dunefigure}

%%%%

\chapter{\dshort{vdmod0}}
%\label{sec:PRTP}
\label{ch:mod0}
%\tableofcontents
\section{Introduction}
The \dword{vdmod0} detector is a full scale demonstrator of the %\Dword{vd} 
\dshort{dune} \dword{spvd}. Assembly of the detector in the \dword{np02} cryostat at the Neutrino Platform Facility at the \dword{cern} began in fall 2022 after decommissioning of the \dword{pddp} detector; Figure~\ref{fig:NP02-decommissioning} shows the inner volume of the \dshort{np02} cryostat during installation of the \dword{vdmod0} detector. There is currently some uncertainty as to when sufficient \dword{lar} will be available in Europe to fill the detector. The \dword{cern} cryogenics group is in monthly contact with all \dword{lar} vendors in Europe for the $\sim$kton of \dword{lar} that is needed. Our current understanding is that liquid oxygen production is very limited until steel production ramps back up. The price of delivery of \dword{lar} to \dword{cern} from the U.S. is quite high.
\begin{dunefigure}
[Inner volume of \dshort{np02} during installation of the \dshort{vdmod0} detector.]
{fig:NP02-decommissioning}
{The inner volume of \dshort{np02} during installation of the \dword{vdmod0} detector. The two top \dwords{crp} are visible, along with one cathode equipped with four \dwords{xarapu}, the downseam cathode is temporarily elevated and two \dwords{xarapu} on the upstream wall are visible behind the upstream field cage (the downstream membrane wall \dwords{xarapu} are not visible). All of the \dwords{xarapu} are covered by temporary bags.}
\includegraphics[angle=-90,width=0.8\linewidth]{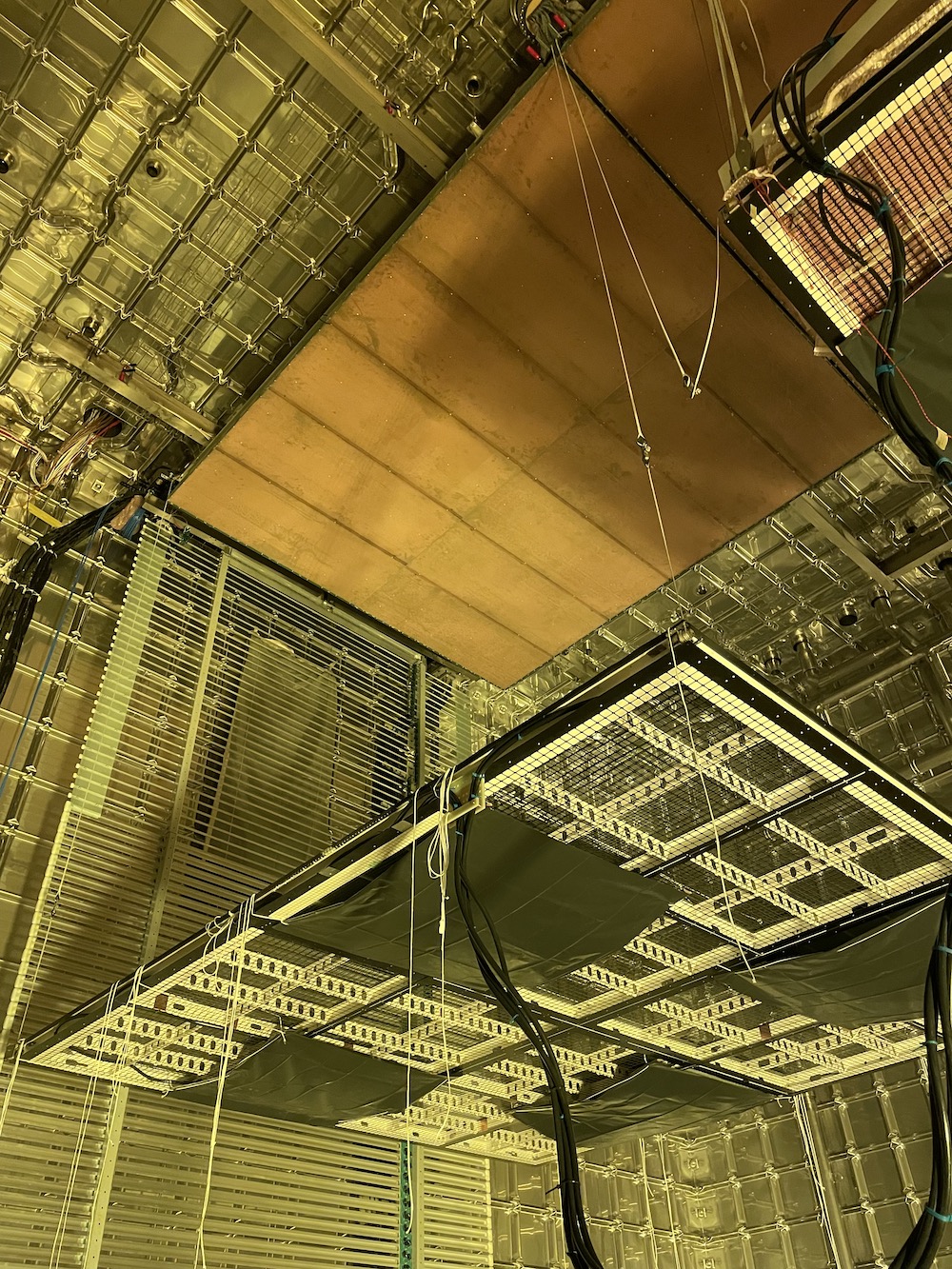}
\end{dunefigure}
The active components (\dwords{crp}, \dwords{pds}, cathode, etc.) are all  tested in smaller cryostats (the \coldbox or the dedicated \dword{pd}-\coldbox) to qualify the production process and detector performance. In the \coldbox, all \dword{tpc}  features (anode, cathode, \dshort{pds}, \efield) are reproduced at a reduced drift length. The \coldbox and its tests are presented in Section~\ref{sec:Cold box test facility}. More detailed discussion of \coldbox test results have been presented in the corresponding detector system chapters.

Section~\ref{sec:M0} introduces the \dshort{vdmod0} concept. Its goal is to demonstrate the high-level complete functional system integration of all \dshort{spvd} components. 
Prior to filling, tests at warm will verify all connectivity and grounding issues. The \coldbox tests have already demonstrated the performance of detector systems in an operational environment in \dshort{lar}. Given the uncertainty of \dshort{lar} availability in Europe in 2023, it is possible that full tests  at cold in \dshort{vdmod0} may not occur until 2024 and may be after the \dword{prr} process has been initiated.

Once filled, \dshort{vdmod0} is planned to be exposed to charged particle beams from the \dword{cern} \dword{h2} beamline. 
The beam test will provide valuable data to fully characterize the response and performance of the detector and complement the hadron-argon cross sections measured with \dword{pdsp}.

\section{\Coldbox Test Facility}
\label{sec:Cold box test facility}

Before \dwords{crp} or \dwords{pd} are installed in \dword{vdmod0}, they are tested in either the \coldbox located immediately next to the \dword{np02} cryostat or the dedicated \dword{pd}-\coldbox. The \coldbox makes use of the \dword{np02} cryogenics facility. The typical \coldbox installation requires 4--5 days. The top-cap of the \coldbox acts as its %the roof of the \coldbox, 
roof, and the \dword{crp} is attached to it. Before installation on the \coldbox, the top-cap, equipped with all necessary feedthroughs, is positioned in the dedicated Faraday cage, where all connections can be debugged and the grounding tested. This Faraday cage facility is also served by the \dword{daq} system, and therefore the entire \dword{crp} can be read out and qualified prior to installation. % into the \coldbox.
Figure~\ref{fig:Module0-1} shows a photograph of the \coldbox during an installation, where %. Visible is 
the roof of the \coldbox is shown with an attached \dword{crp} to be tested. The cathode (with an \dword{xarapu}) is visible at the bottom of the \coldbox.
%$$$$$$$$$$$$$$$ 
\begin{dunefigure}
[\Coldbox during installation of a \dshort{crp}]
{fig:Module0-1}
{The \coldbox during installation of a \dword{crp}. 
The cathode is visible on the bottom, with an \dword{xarapu} installed.}
\includegraphics[width=0.99\linewidth]{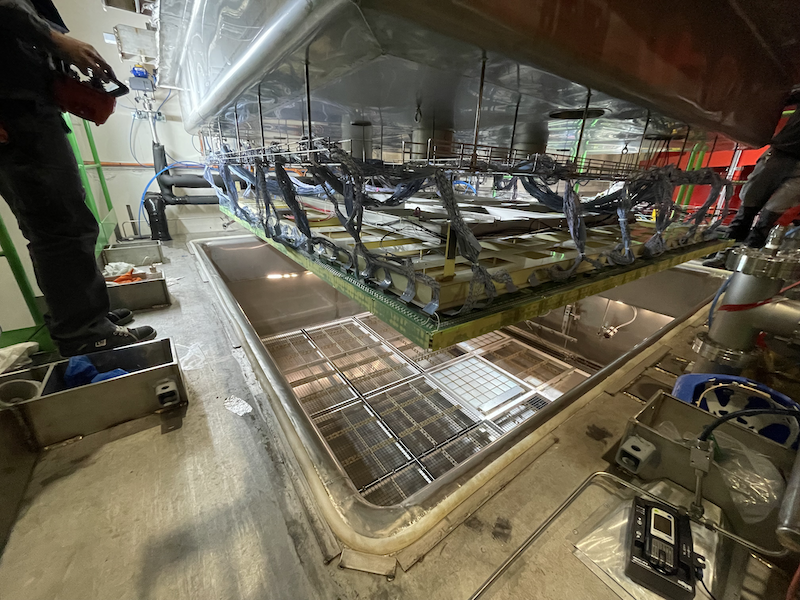}
\end{dunefigure}
%$$$$$$$$$$$$$$$ 

%$$$$$$$$$$$$$$$ 
\begin{dunefigure}
[Schematic layout of detector components inside the \coldbox]
{fig:Module0-3}
{Schematic layout of the detector components inside the \coldbox. The top-cap (cyan), the \dword{crp}, a drift of $\sim$25\,cm and the cathode (grey) with the \dword{xarapu} \dword{pd} on the bottom at HV.}
\includegraphics[width=0.8\linewidth]{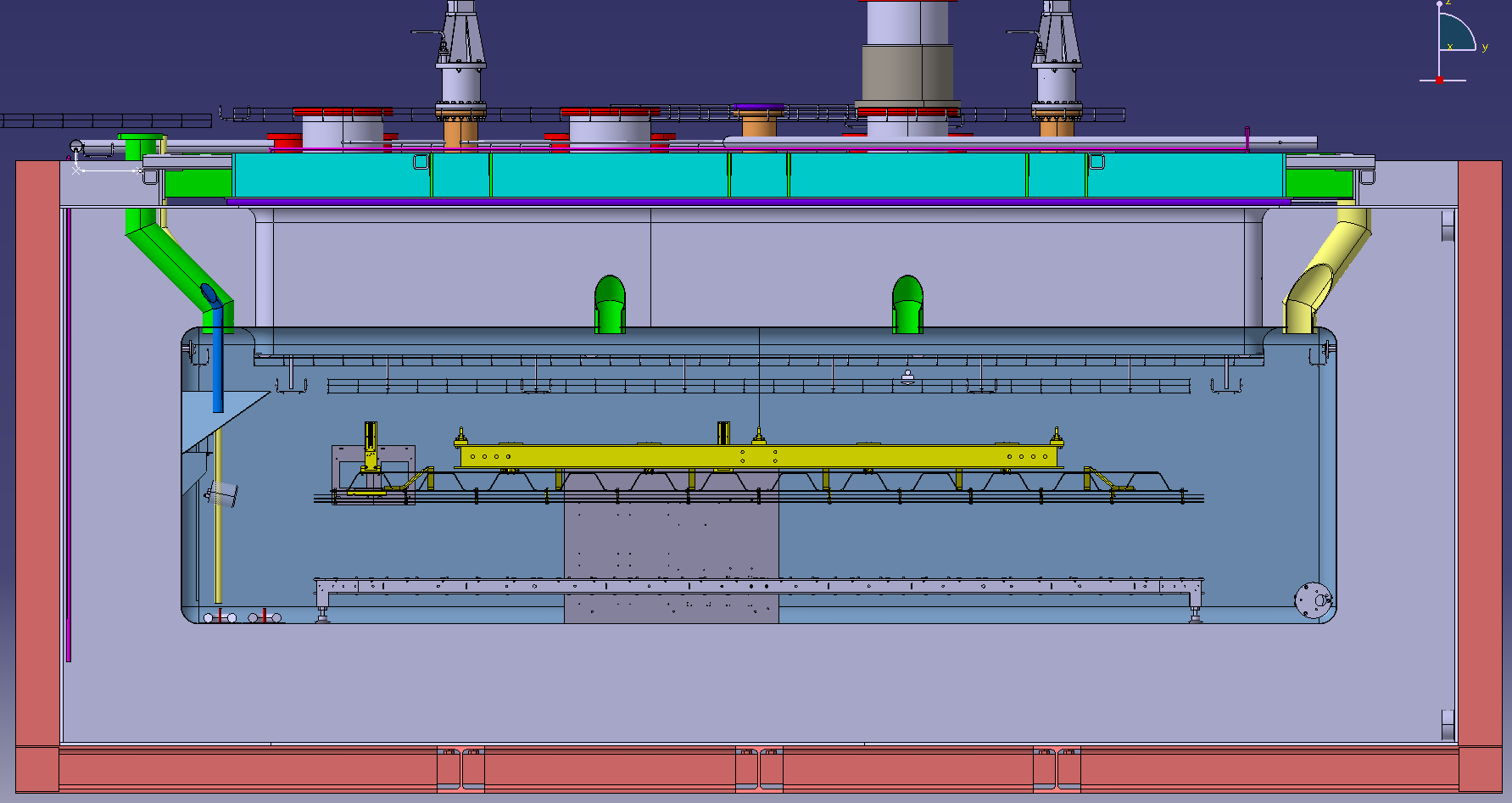}
\end{dunefigure}
%$$$$$$$$$$$$$$$

\begin{dunefigure}
[Typical raw data event display with reconstructed image]
{fig:Module0-4}
{Typical raw data event display showing the quality of the reconstructed image. The image shows a muon track crossing the full drift volume from anode (the first hit at low drift times) to cathode (the top most hit at higher drift times). The faint hit signals visible to the right of the main track are due to the ionization in the \dword{lar} above the \dword{crp}, where a weak residual \efield is present. The signals are weaker due to the higher electron-ion recombination, while the apparent track angle is steeper due to the slower drift velocity in the low \efield above the \dword{crp}.}
\includegraphics[width=0.8\linewidth]{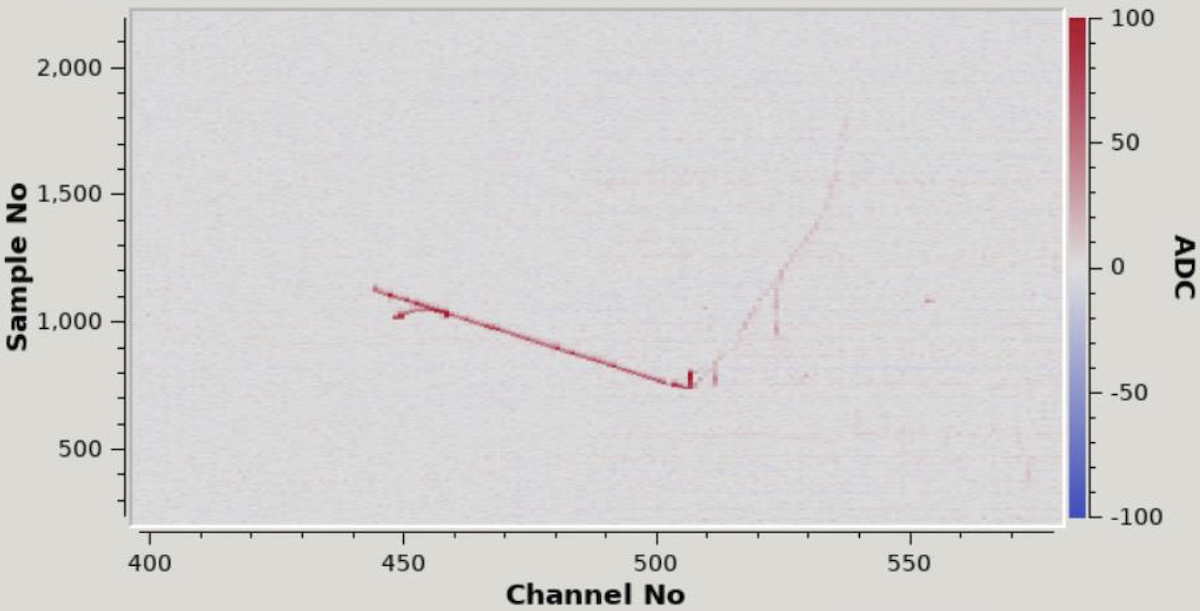}
\end{dunefigure}
%$$$$$$$$$$$$$$$

Figure~\ref{fig:Module0-3} shows the schematic layout of the \coldbox. Visible in yellow is the support structure of the \dword{crp} and services. Just below the \dword{crp} is an empty volume of about 25\,cm for drift, and on the floor of the \coldbox is the cathode with the \dword{pds}. 
Typically the cathode sits at $-$10\,kV.
Once the \coldbox is tested for leaks and closed, it is purged of air and filled with \dword{lar} above the yellow support structure. 
Then the \dword{lar} re-circulation is started, achieving a typical purity above 1.5\,ms. 
This purity corresponds to an attenuation length of about 2\,m, well beyond the 25\,cm drift length.
Once the \dword{daq} and slow control are activated, it is normal that cosmic muons appear immediately %visible 
in the event displays such as the one shown in Figure~\ref{fig:Module0-4}.
 
Five pre-production \dwords{crp} will have been tested through this process. The first hybrid CRP tested in 2021  (CRP-1) was equipped with both types of electronics (half top and half bottom  electronics --- \dword{tde} and \dword{bde}). A second test of CRP-1 in 2022 had improved grounding and some modification of the electronics layout. 
The second \dword{crp} (CRP-2) (top CRP with full \dword{tde} readout) was tested in summer 2022 and retested in early fall 2022 with a strip continuity issue resolved. The third \dword{crp} (CRP-3), (also full \dword{tde}) was tested in fall  2022. The first bottom \dword{crp} (CRP-5) (equipped with full cold \dword{bde}) was tested in February 2023 and the last one (CRP-4, with full \dword{bde}) was tested in March 2023.
 All \dwords{crp} tested reflect the layout and readout characteristics defined in the relevant chapters of this report.

 The preliminary tests in the \coldbox have required the following operational steps (time estimate given for each step): 
\begin{enumerate}
\item installing \dword{crp} under the top-cap of the \coldbox, cabling and preliminary testing of the connections: %estimated to take 
four to five days;
\item purging, cooling down and filling the \coldbox: three to four days;
\item turning on detector and running \dword{daq}; and
\item emptying, warming up, and opening the \coldbox: five days.
\end{enumerate}

In addition, a progression of \dword{xarapu} prototypes (v.1-v.4) toward the final design have been installed and tested in \coldbox runs throughout 2022--2023. Some runs were parasitic to \dword{crp} tests and some were dedicated to \dword{pds}. Several versions of cold \dword{pds} read-out motherboard (DCEm v.1.0 to v.2.0) including new generations of \dword{pof} receivers and \dword{sof} transmitters were tested and progressively optimized. A pulsed \dword{led} calibration system permanently installed in the \coldbox allowed a diagnostic, debugging and complete characterization of the response function of the new cold \dword{pds} readout system, toward the reference design presented in this report. More detailed description of these tests can be found in Chapter~\ref{chap:PDS}.

 The new dedicated \dword{pd}-\coldbox in a smaller cryostat has enabled much faster turnaround on  \dword{pds} testing and greatly expanded the testing capabilities for the \dword{pds}.  
 
 For the bottom \dword{crp} cold-tests in the U.S., a  cold-box was built at BNL.  As described in Section~\ref{subsubsec:CRP_5_productionTesting}, it has been used for the cold-tests of CRP-5a at BNL and the cold-tests of CRP-4 at Yale.

\section{\dshort{vdmod0}}
\label{sec:M0}

The \dword{vdmod0}  will validate the following aspects of the \dword{spvd} far detector module, which are not covered in either the \coldbox or the \dword{np02} HV tests:
\begin{itemize}
\item The bottom \dword{crp} in the design position with the supports on the cryostat floor;
\item Cathode suspension system;
\item Cathode modules in their final design as in \dword{spvd};
\item The $\sim$70\% transparent \dword{fc};
\item The symmetric top and bottom drifts;
\item The mechanical structures of \dwords{pds} on the cryostat wall and the cable layout testing and installation; and 
\item \dword{pds} operation at the nominal \dword{hv} on the cathode.
\end{itemize}

Figure~\ref{fig:Module0-2} shows a model of \dword{vdmod0} %detector 
inside the \dword{np02} cryostat.
%$$$$$$$$$$$$$$$ 
\begin{dunefigure}
[\threed model of \dshort{vdmod0}]
{fig:Module0-2}
{A \threed model of the \dword{vdmod0} layout to be installed inside the \dword{np02} cryostat.}
\includegraphics[width=0.8\linewidth]{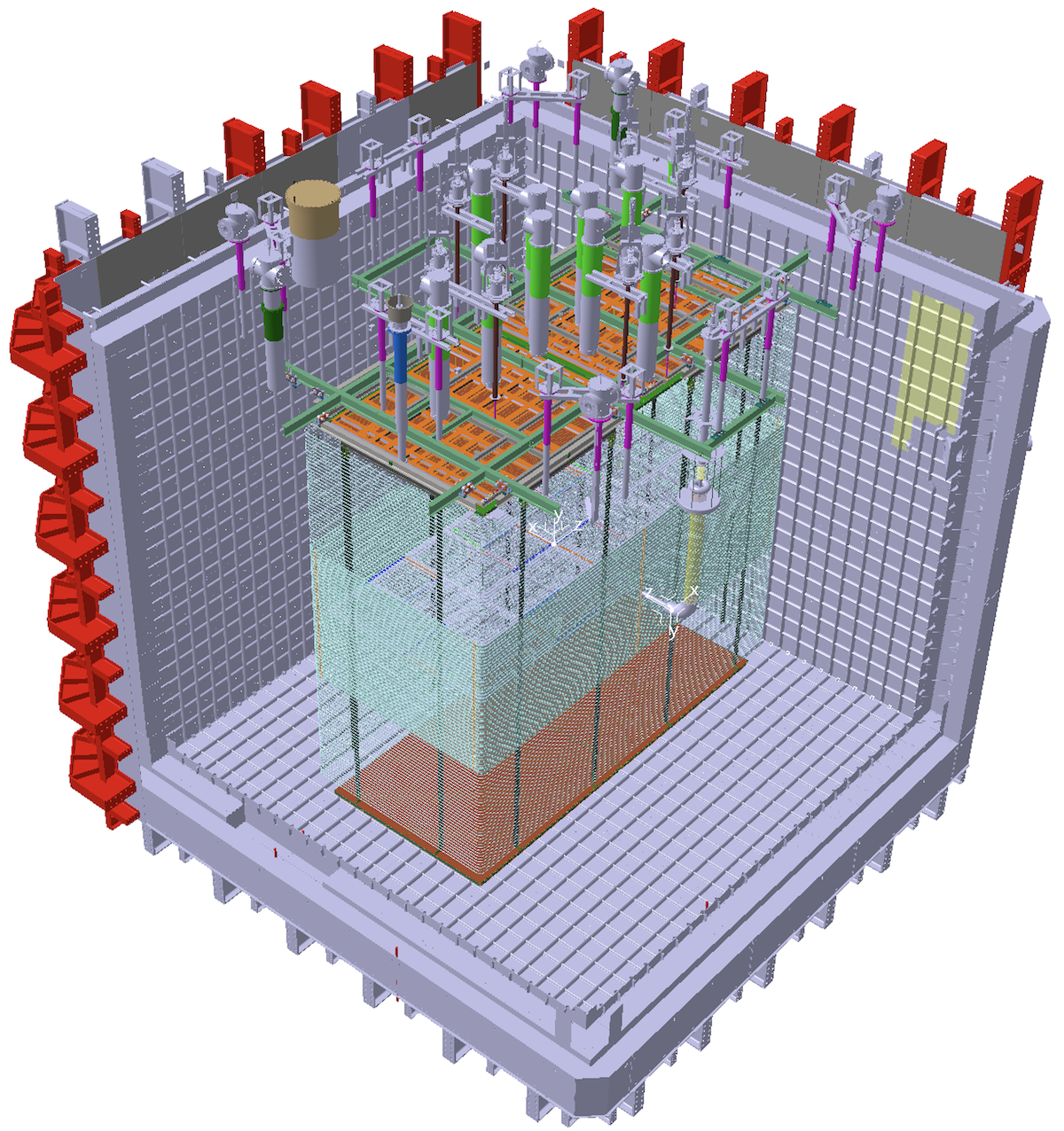}
\end{dunefigure}
%$$$$$$$$$$$$$$$

The \dword{vdmod0} detector design must accommodate the constraints set by the existing setup, most notably the \dword{np02} cryostat and its roof penetrations.
The interface structures are therefore required to match the positions of the various detector components with the positions of the penetrations.

\subsection{Charge-readout Planes (\dshort{crp}s)}
\label{sec:M0-crp}

\dword{vdmod0} contains an upper and a lower drift volume, each of which has two \dwords{crp} to read out the drift charge.

\subsubsection{Top \dshort{crp}s}
\label{sec:M0-tcrp}

The \dword{vdmod0} detector top anode plane consists of two top \dwords{crp} that have already been tested in the \coldbox.
They hang from dedicated stainless steel beams supported from roof penetrations (see Figure~\ref{fig:Module0-5}) as used by \dword{pddp}.
They are installed next to each other in the center of the cryostat, forming the \dword{tpc} active area of ${\rm 3m \times \sim 6.7m}$.
In this configuration, the distance between the \dword{crp} on the short side and the cryostat wall is about 73\,cm, close to the value in \dword{spvd}.

%$$$$$$$$$$$$$$$ 
\begin{dunefigure}
[\dshort{crp} mechanical support structure connected to the support feedthroughs]
{fig:Module0-5}
{Top \dword{crp} mechanical support structure connected to the support feedthroughs.}
\includegraphics[width=0.5\linewidth]{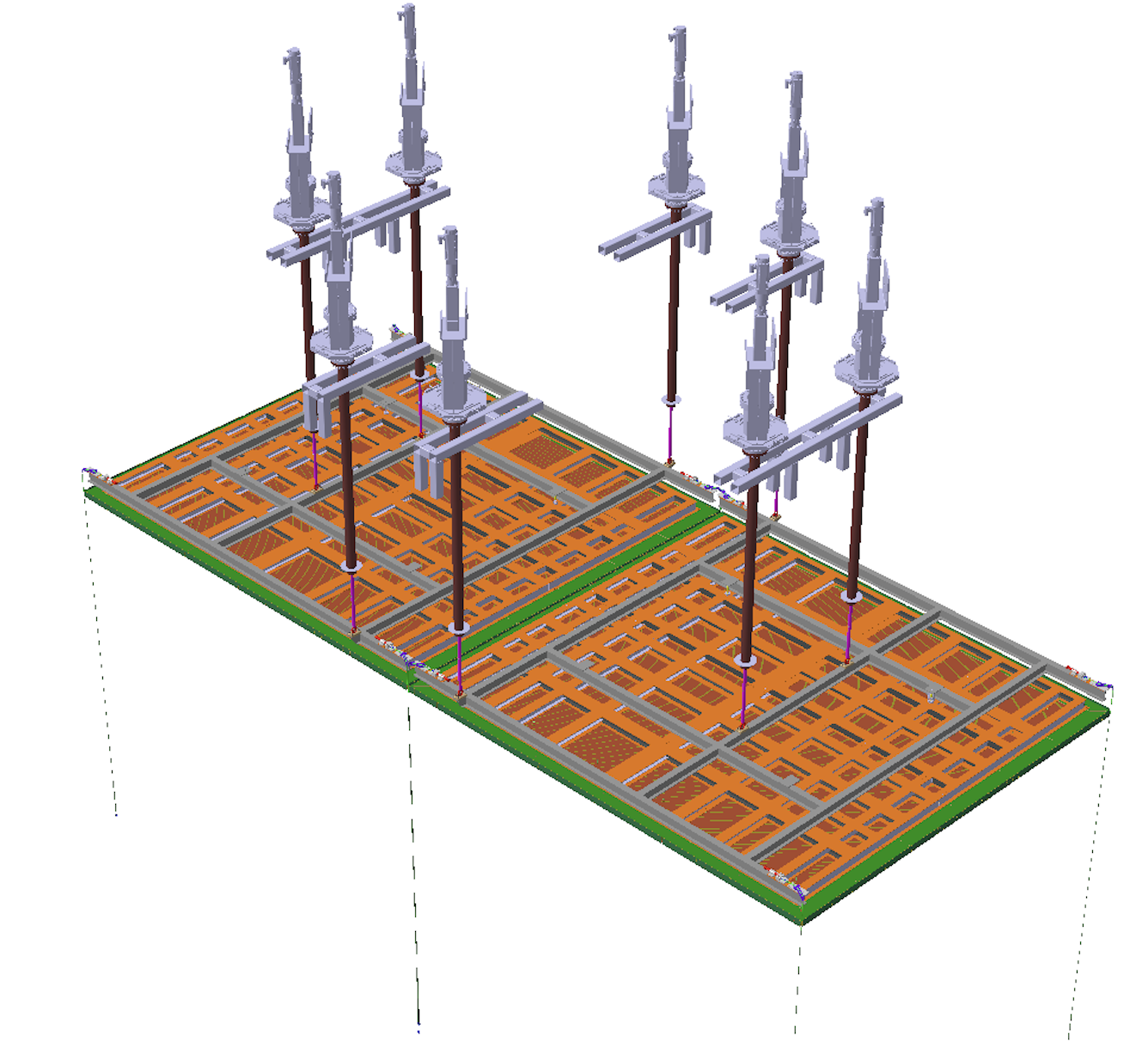}
\end{dunefigure}
%$$$$$$$$$$$$$$$

The \dword{tde} will use the signal feedthroughs (\dwords{sft}) built for \dword{pddp}, as shown in Figure~\ref{fig:Module0-6}.
They are conceptually identical %with respect 
to what is envisaged for \dword{spvd}, but they differ in the size of the flange separating the clean argon volume and the volume of the SGFT. This difference is dictated by the availability of the roof penetrations in \dword{np02}.
%$$$$$$$$$$$$$$$ 
\begin{dunefigure}
[Top \dshort{crp} signal feedthroughs (\dshort{sft}s)]
{fig:Module0-6}
{Top \dword{crp} signal feedthroughs (\dwords{sft})}
\includegraphics[width=0.5\linewidth]{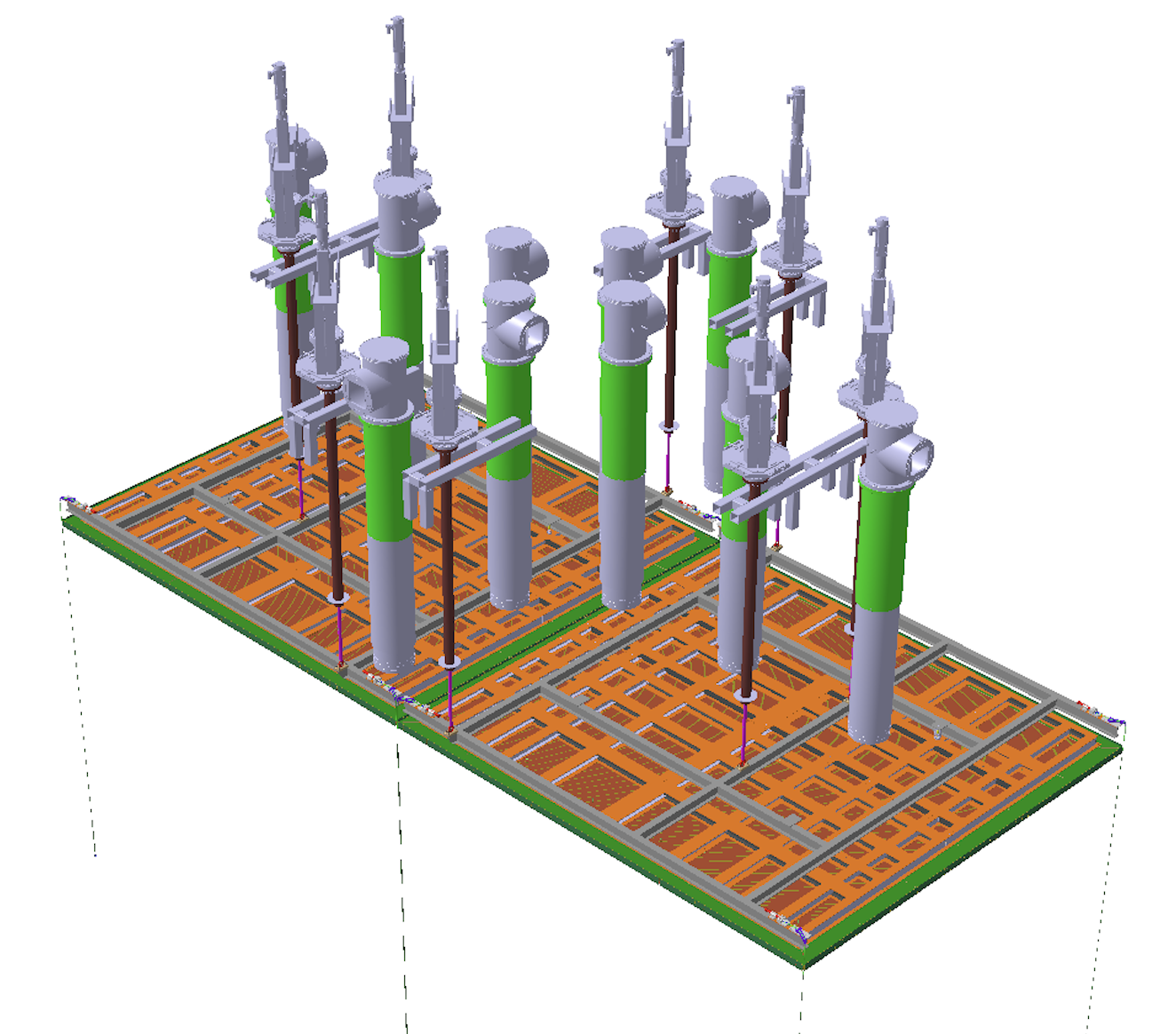}
\end{dunefigure}
%$$$$$$$$$$$$$$$

\subsubsection{Bottom \dshort{crp}s}
\label{sec:M0-bcrp}

Two bottom \dwords{crp} will be installed directly on the corrugated membrane of the cryostat, as envisaged for \dword{spvd}. 
The support structure is shown in Figure~\ref{fig:Module0-8}.
%$$$$$$$$$$$$$$$ 
\begin{dunefigure}
[Bottom \dshort{crp} supports]
{fig:Module0-8}
{Bottom \dword{crp} supports}
\includegraphics[width=0.7\linewidth]{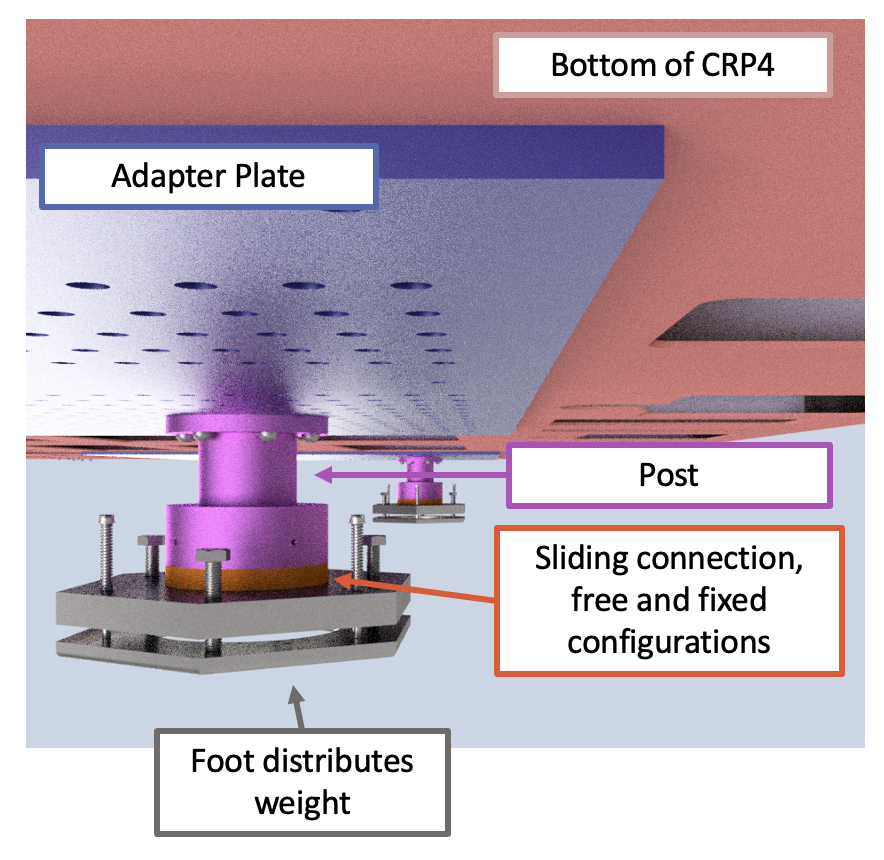}
\end{dunefigure}
%$$$$$$$$$$$$$$$
The bottom \dwords{crp} are identical in their layout and technology to the top \dwords{crp}. The difference lies in the readout electronics and interface boards connecting the  \dword{bde} \dwords{femb} to the the anodes strips.
The cables from the bottom \dwords{crp} will be routed on  cable trays along the walls of the cryostat as shown in Figure~\ref{fig:Module0-8-5}.
%$$$$$$$$$$$$$$$ 
\begin{dunefigure}
[Bottom \dshort{crp} cable routing in \dshort{vdmod0}]
{fig:Module0-8-5}
{Bottom \dword{crp} cable routing in \dshort{vdmod0}. The insert provides an expanded view of the cable routing near the roof. The \dshort{crp}s are not shown.}
\includegraphics[width=0.8\linewidth]{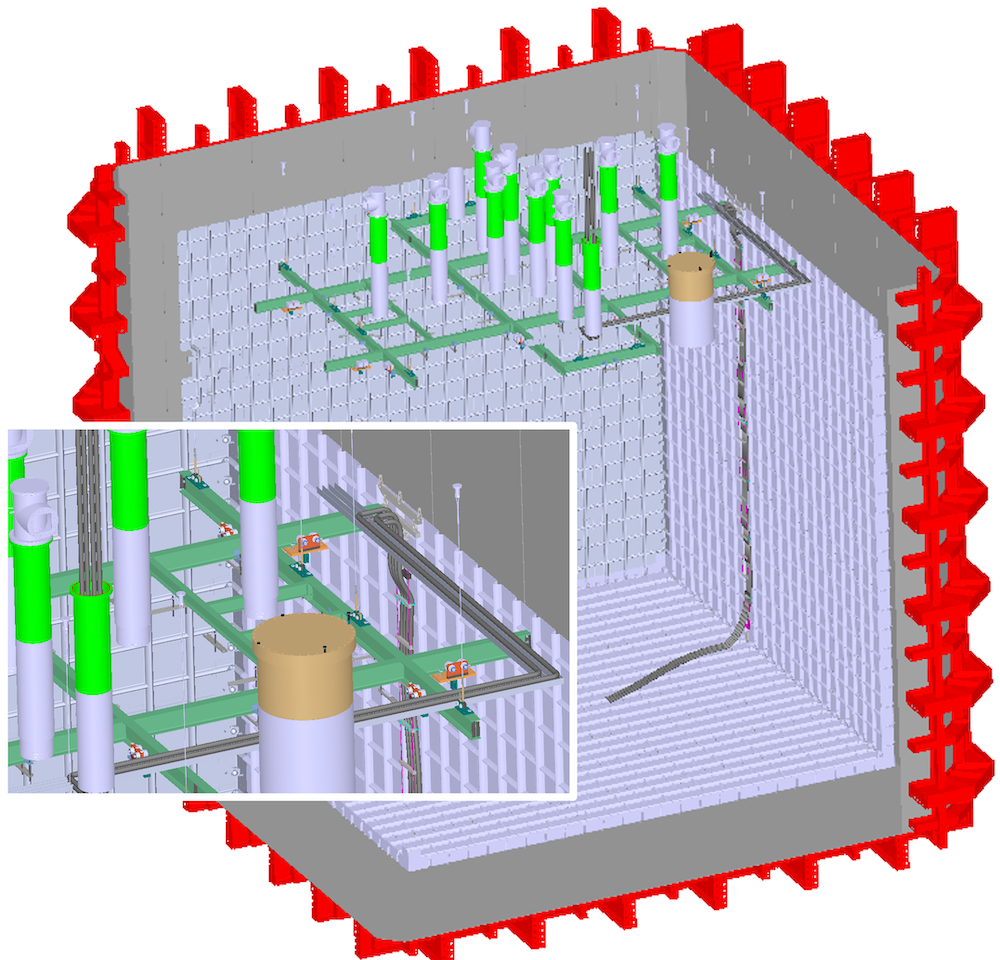}
\end{dunefigure}
%$$$$$$$$$$$$$$$

\subsection{HV System}
\label{sec:M0-hV system and field cage}

The \dword{hv} delivery system builds on the successful demonstration of $-$300\,kV in the \dword{np02} \dword{hv}  test.
The cathode consists of two modules (dimensions are the same as the \dword{crp}) suspended from the top \dword{crp} superstructure with supports at the four corners of each module at half-height of the active volume.
Each cathode module will be instrumented with four \dwords{xarapu}, operating at the cathode \dword{hv} and electrically isolated via \dword{pof} and \dword{sof}.
The two cathode modules will be electrically connected to the \dword{fc}. 
The \dword{fc} design is the same as that for \dword{spvd}, except for the support structure which is tailored for the \dword{vdmod0} cryostat roof penetrations. %\fixme{clarify}
The top and bottom drift will be symmetric, with the maximum drift length of 3.2\,m in both directions.
To achieve the nominal \efield in the drift region, half of the %\dword{dune} far detector 
\dword{spvd} voltage will be applied in \dword{vdmod0}.

\subsubsection{HV Delivery}

The nominal drift field of 450\,V/cm %500~V/cm 
will be achieved with approximately $-$160\,kV applied to the cathode. Operation at the full \dshort{dune} $-$300~kV was already achieved with the \dword{hv} stability test in the \dword{np02} \dword{hv} test and the \dword{vdmod0} is capable of operating the cathode at the full \dshort{dune} voltage to validate the operation at the nominal $-$300\,kV value of the fully integrated \dword{spvd}.
% Anne adding 7 mar 23:
Simulations indicate that the \dshort{vdmod0} is robust against \dshort{hv} breakdown at $-$300~kV. %there will be no fields above this value. 
\dshort{vdmod0} will provide an even more stringent test of the \dshort{fc} transition region than required for \dshort{spvd}.

For this reason --- and similar to the operation in the \dword{np02} \dword{hv} test --- the voltage will be provided by a commercial $-$300\,kV  power supply, fed into the cryostat through a vacuum-tight HV feedthrough (\dword{hvft}) rated for more than 300\,kV, and brought to the cathode depth by the \dword{hv} extender. 

The main improvements of the \dword{hv} delivery system with respect to the \dword{np02} \dword{hv} stability test, where the functionality of the HV extender was successfully demonstrated, are described in Chapter~\ref{ch:DFS}. In summary, they consists of a new, longer and wider \dword{hvft} to avoid  icing in the receptacle that receives the HV cable on the warm side, and a modified \dword{hvft}-to-extender coupling (replacing the spherical head with a cylinder and including a suspension disk adapted to support the new coupling) redesigned to mitigate the residual discharge events most likely due to the charging up of the supporting disk insulating surfaces. For the %NP02 
\dword{vdmod0} detector, the straight section of the extender will be shortened to about 3\,m to match the cathode depth in \dword{lar}.
A dedicated test of the new \dword{hvft} and the new extender coupling is planned to validate the design improvements. 

The installation of the \dword{hv} extender occurs after the completion of the \dword{tpc} with the procedure already tested in \dword{np02} in the HV stability test. It will be followed by the insertion of the \dword{hvft}, the electrical connection of the cathode to the HV bus and the electrical continuity checks.

\subsubsection{Cathode}

The two cathode modules are constructed and installed exactly as planned for \dword{spvd} as described in Chapter~\ref{ch:DFS}. Each module has the same foot print as the \dwords{crp} and consists of two 6\,cm thick \dword{frp} half frames that will be connected together on a dedicated cart in the \dword{np02} clean room.  This is followed by the mounting of the %resistive meshes 
perforated resistive panels on both surfaces of the frames and the metallic meshes in the locations where the \dword{xarapu}  are inserted. \Dword{xarapu} fiber routing through the frames follows the procedure depicted for \dword{spvd}.

As described in Chapter~\ref{ch:DFS}, the Length Adjusting Devices (LADs) are hosted in the four corners of the cathode frame. Being part of the suspension system, these devices receive the insulating suspension cables and allow for fine-tuning of the cathode level when hanging from the \dword{crp} support structure.

The other end of the suspension cables is connected to the Top Adjusting Device (TAD) on the top of the \dword{crp} superstructure. Both the adjusting devices and the insulation cables are being validated for cryogenic operation in extensive dedicated tests ongoing at IJCLab.

The suspension on the \dword{crp} structure for \dword{mod0} will follow closely the one depicted for the \dword{spvd}, except for the fact that each \dword{crp} supporting structure holds a single \dword{crp} and a single cathode module.

Figure~\ref{fig:NP02-Cathodes} shows the two cathode modules hanging from the \dword{crp} suspension structures. Figure~\ref{fig:NP02-Cathodes-detail} shows details of the cathode hanging device, the \dword{xarapu} location and the resistive %meshes 
panels with their supports to ensure planarity. 

%$$$$$$$$$$$$$$$ 
\begin{dunefigure}
[Suspended two-module cathode supermodule, with \dshort{hv} extender]
{fig:NP02-Cathodes}
{The %two-cathode modules 
two-module cathode supermodule hanging from the \dword{crp} suspension structures. The \dword{hv} extender is also shown.}
\includegraphics[width=0.7\linewidth]{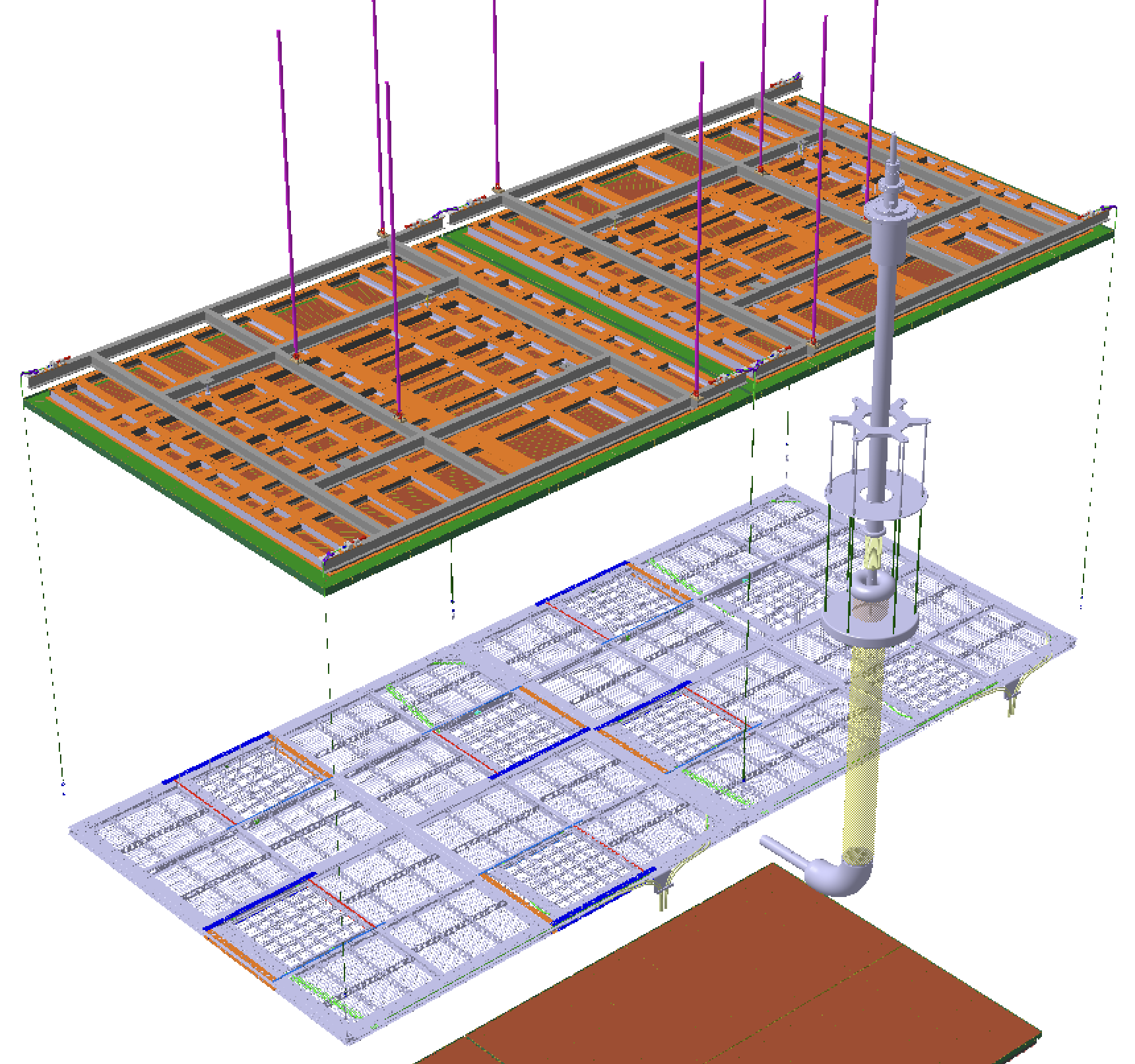}
\end{dunefigure}
%$$$$$$$$$$$$$$$

%$$$$$$$$$$$$$$$ 
\begin{dunefigure}
[Detail of the cathode hanging device and the \dshort{xarapu} locations]
{fig:NP02-Cathodes-detail}
{Detail of the cathode hanging device and the \dword{xarapu} locations.}
\includegraphics[width=0.66\linewidth]{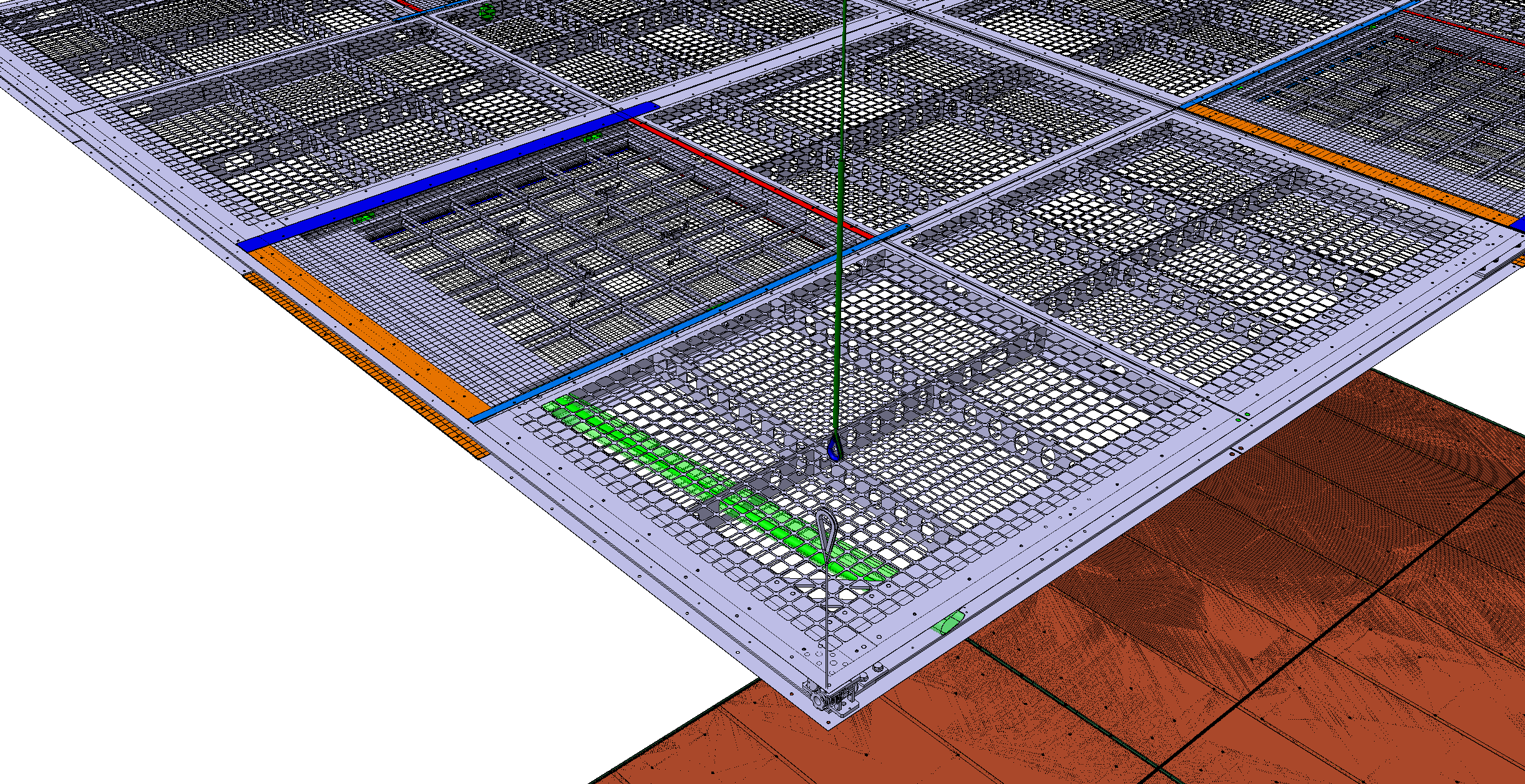}
\end{dunefigure}
%$$$$$$$$$$$$$$$

\subsubsection{Field Cage}

The \dword{vdmod0} \dword{fc} is designed to match as closely as possible that for \dword{spvd}, described in Chapter~\ref{ch:DFS}. It will consist of six independent columns, each of which is made of two \dword{fc} panels hanging one below the other and connected together at the cathode level. Figure~\ref{fig:Module0-7} gives a global view of the \dword{fc} layout for the \dword{vdmod0}.

%$$$$$$$$$$$$$$$ 
\begin{dunefigure}
[Field cage layout in \dshort{vdmod0} design]
{fig:Module0-7}
{The field cage layout in the \dword{vdmod0} design to provide two 3.2\,m drift regions, above and below the cathode, with the \efield{}s applied in opposite directions.}
\includegraphics[width=1.0\linewidth]{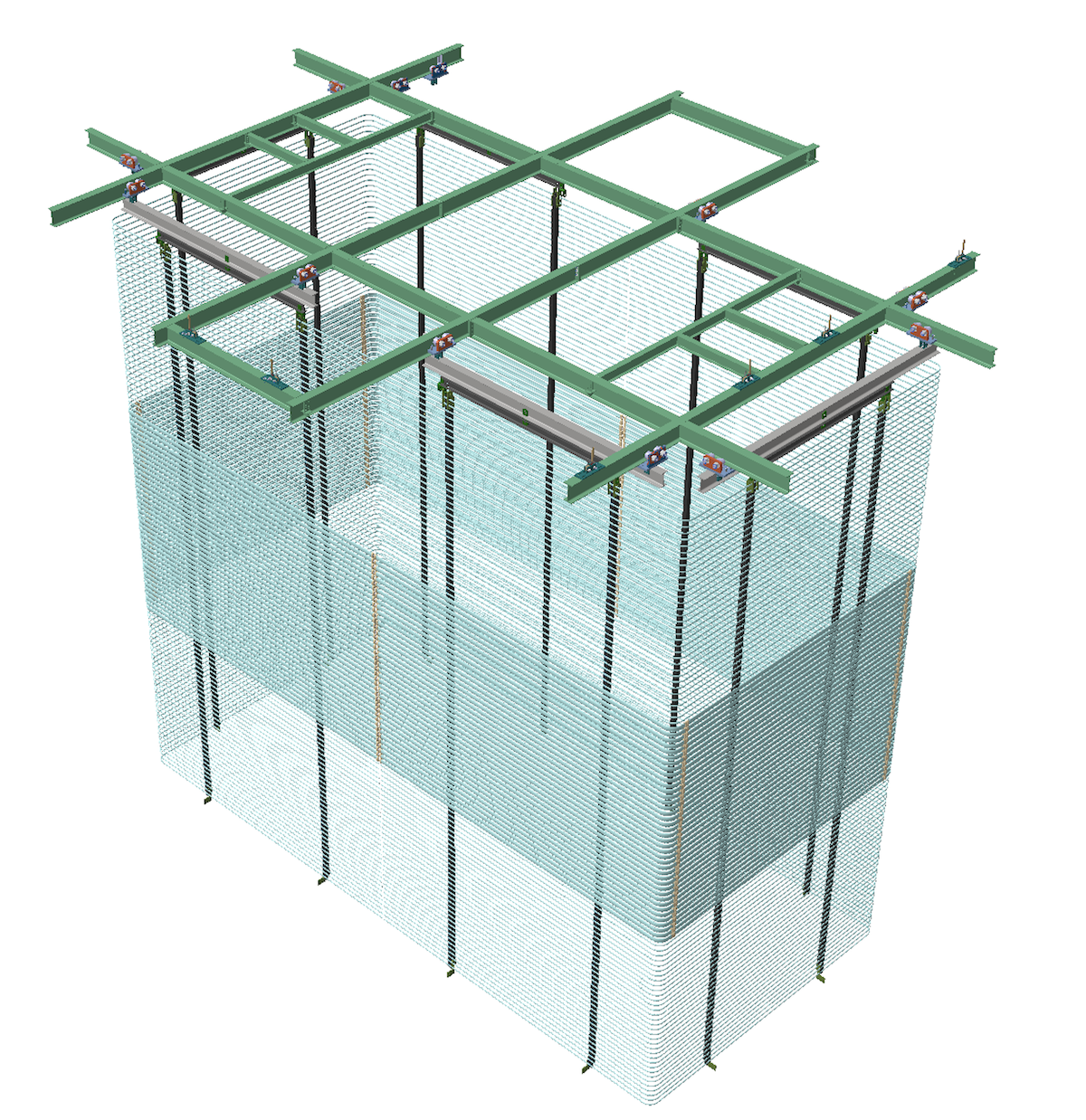}
\end{dunefigure}
%$$$$$$$$$$$$$$$

Each panel is 3.2\,m tall and hosts twenty-one 4.6\,cm thick profiles, starting from the cathode level, followed by 33 thinner profiles (1.5\,cm thick) to complete the \dword{fc} column to the \dword{crp}. The spacing between two neighboring \dword{fc} profiles is 6\,cm (center-to-center) for both the thin and %thick 
standard-thickness profiles. The profiles are kept in place with two \dword{frp} box beams (50\,mm$\times$50\,mm in cross section). This scheme allows for a 70\% optically transparent \dword{fc} in front of \dwords{pds} placed on the cryostat membrane, %and is %exactly 
%a scaled-down version of the \dword{spvd} layout. 
where the vertical height of the thick profile and thin profile regions are scaled by the ratio of heights of \dword{vdmod0} and \dword{spvd}.
The two columns on the short sides of \dword{vdmod0} are made of 3\,m long straight profiles, and their distance from the cryostat membrane is set to be $\sim$750\,mm as planned in \dword{spvd}.
The four columns composing the long sides of %the detector 
\dword{vdmod0} host 3.4\,m long profiles bent at $90^{\circ}$ on one end, to minimize the local \efield and to smoothly meet the profiles on the short sides. A \dword{fea} for the \dword{spvd} shows that this layout results in a uniform \efield including the corners of the active drift volume.  

Each profile will be terminated with \dword{uhmwpe} end caps. A \threed printed version, whose compatibility with \dword{lar} and \dword{hv} was previously tested, has been made for the thin profiles. 

Each \dword{fc} column is equipped with  high voltage divider boards (\dwords{hvdb}) where each step between two neighboring profiles is made of two 5\,G$\ohm$ resistors connected in parallel with a series of four varistors clamping at 1.74\,kV each. This scheme %has already been used for the NP02 Dual Phase detector. 
was used in \dword{pddp}. Most of the boards from \dword{pddp} test will be reused after being refurbished with new resistors. The implementation of four varistors instead of three, as planned for \dword{spvd} (see Section~\ref{subsubsec:HVDB}), is to allow testing the vertical drift layout with $-$300\,kV applied on the cathode (the ${\rm \Delta V}$ between the profiles will be as high as 5400\,V), as will be the case %planned 
in \dword{spvd}. 

Each \dword{fc} column is supported at the top by an aluminum yoke, from which the \dword{frp} box beams hang. At the bottom, a `\dword{fc} stabilizer'' that sits on the cryostat membrane is connected to the \dword{frp} beams to avoid uncontrolled swinging of the \dword{fc} columns (details are found in Section~\ref{subsubsec:FCsss}).
The yoke is connected at a single point to a stainless steel beam which in turn is supported by the \dword{fcss} by means of trolleys. 
This supporting scheme %allows to ensure 
ensures the verticality of the \dword{fc} columns even in case of roof/\dword{fcss} deformation. In addition, the use of trolleys allows for assembling the \dword{fc} columns away from the \dwords{crp} and cathode, and positioning them in their final location %when all QC's have been performed 
once \dword{qc} has been completed (including, possibly, the fiber routing of for the \dwords{xarapu} on the cathode).

At the %level of the 
cathode the \dword{fc} columns are connected together with a low-resistivity ``\dword{hv} bus'' (see Section~\ref{subsec:VDss}) to distribute the \dword{hv}, provided by the extender, to the outer boundary of the cathode plane. 

The total dry weight of each \dword{fc} column (including the yoke, the \dword{fcss} beam and the trolleys) will not exceed 200\,kg. The \dword{frp} connections between the yoke and the \dword{frp} beams have been tested to hold more than 1300\,kg without failing.

The assembly of the \dword{fc} panels will be performed in the \dword{np02} and \dword{np04} clean rooms. % of the space. 
%For this purpose, p
Prototypes of the assembly station and the related tooling, described in Section~\ref{subsubsec:fc-tools}, %is being presently constructed. 
were constructed in late 2022, %. This also includes 
including the profile-bending tool, adapted from that used for \dword{np04}, and the storage cart designed to transport the \dword{fc} panels inside the \dword{np02} cryostat.

\subsection{Photon Detectors}
\label{sec:M0 Photon detectors}

The \dword{pds} consortium has delivered eight \dwords{xarapu} with dedicated %power and readout over fiber f
\dword{pof} and \dword{sof} for the two cathode modules (4$+$4). The  detailed description is in Chapter~\ref{chap:PDS}.
Similarly, the \dword{pds} consortium has delivered four \dwords{xarapu} for the top membrane locations and plans to deliver four more \dwords{xarapu} for the bottom membrane locations (four covering the top drift and four covering the bottom drift).

\subsection{Data Acquisition (\dshort{daq})}
\label{sec:M0_DAQ}

The \dword{daq} system as developed for \dword{pdsp} has been used for the \coldbox operations and will be used for \dword{vdmod0}.

\subsection{Beam Plug}
\label{sec:The beam plug}

The dedicated tertiary beamline from the SPS-\dword{h2} line crosses the \dword{vdmod0} cryostat  diagonally (in the horizontal plane). It is inclined slightly  downward and is fully contained in the top drift volume.  It enters the \dword{fc} 96\,cm above the cathode and exits $\sim$36\,cm above it. Before entering the active volume, the total path length crossed by beam particle in inactive \dword{lar} (from the cryostat membrane to the \dword{fc}) is $\sim$4.3\,m.

To enable data-taking  with particles of well-known energy in the low-energy region of interest (\dword{roi}) to \dword{dune}, the present plan is to reduce the inactive \dword{lar} along the beamline inside the cryostat by inserting two consecutive ``beam plugs'' similar to the one built and installed in \dword{np04} for \dword{hdmod0}.  Figure~\ref{fig:BPModule0} shows a layout of the beam plug concept integrated into the \dword{np02} \dword{vdmod0} detector.  Figure~\ref{fig:BPNP04} shows the beam plug installed in the \dword{np04} cryostat.

The first beam plug, close to the cryostat membrane, could be made of a fully metallic vacuum pipe with metallic end caps. The length would be $\sim$2.5\,m and the diameter 25\,cm. The second one, in line with the first and reaching the \dword{fc} wall, will be a vacuum pipe made of insulating material with metallic end caps; its material, length, diameter and suspension system will be similar to those for the \dword{hdmod0} version.

The beam plug could be operated under vacuum or filled with nitrogen gas. In either case, the material budget encountered by the  beam particles consists essentially of the thickness of the stainless steel end caps ($\sim$4--5\,mm each). This thickness will ensure that a sizable fraction of the low-energy beam electrons will reach the active volume before showering.

%$$$$$$$$$$$$$$$ 
\begin{dunefigure}
[Beam plug concept for \dshort{vdmod0}]
%[Beam plug concept for the \dword{vdmod0} detector]
{fig:BPModule0}
{Beam plug concept for the \dword{vdmod0} detector}
\includegraphics[width=0.5\linewidth]{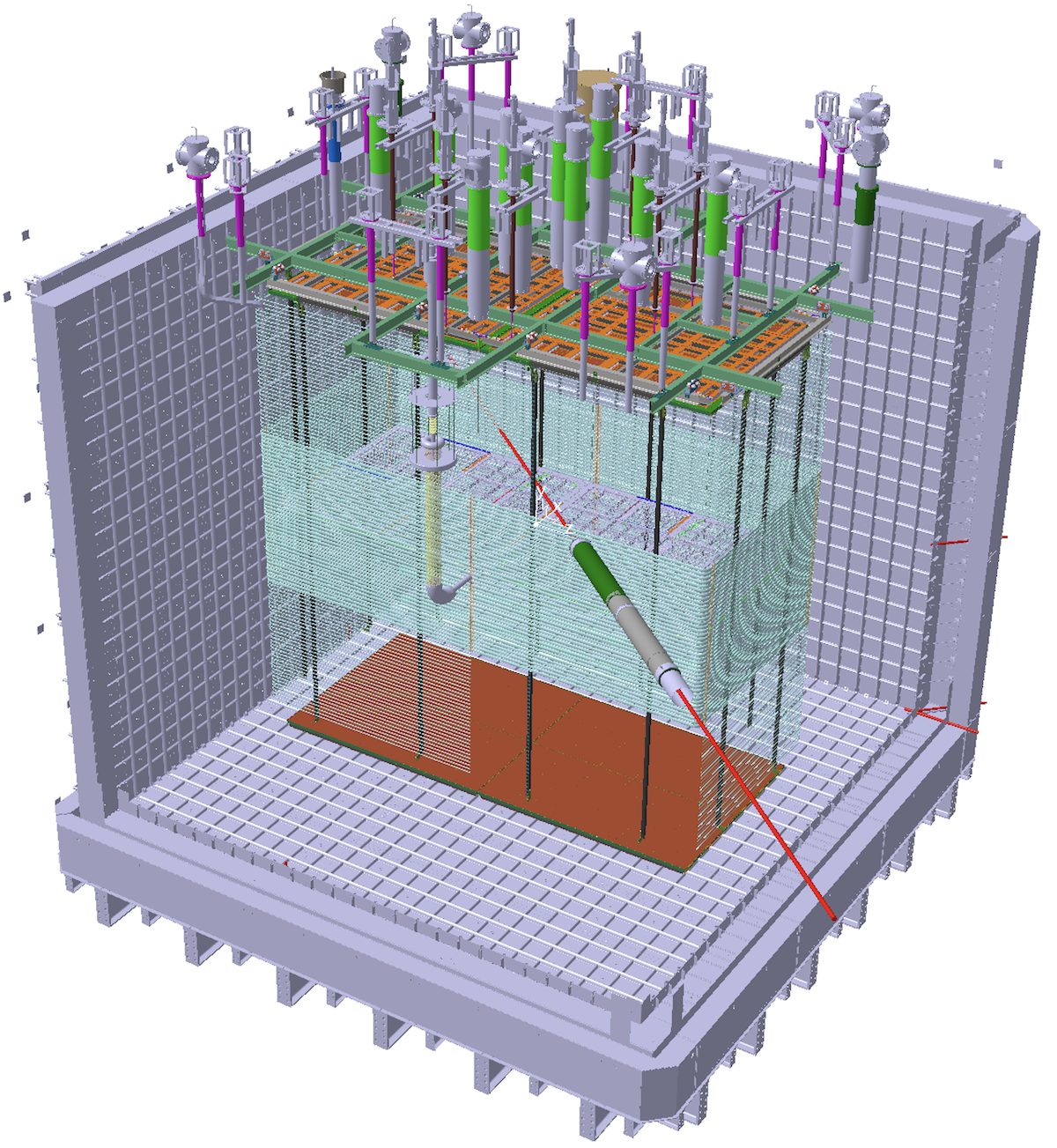}
\end{dunefigure}
%$$$$$$$$$$$$$$$

%$$$$$$$$$$$$$$$ 
\begin{dunefigure}
[Photo of the \dshort{hdmod0} beam plug]
%[Beam plug for the \dword{hdmod0} detector]
{fig:BPNP04}
{Photo of the beam plug realized and installed in the \dword{hdmod0}}
\includegraphics[width=0.5\linewidth]{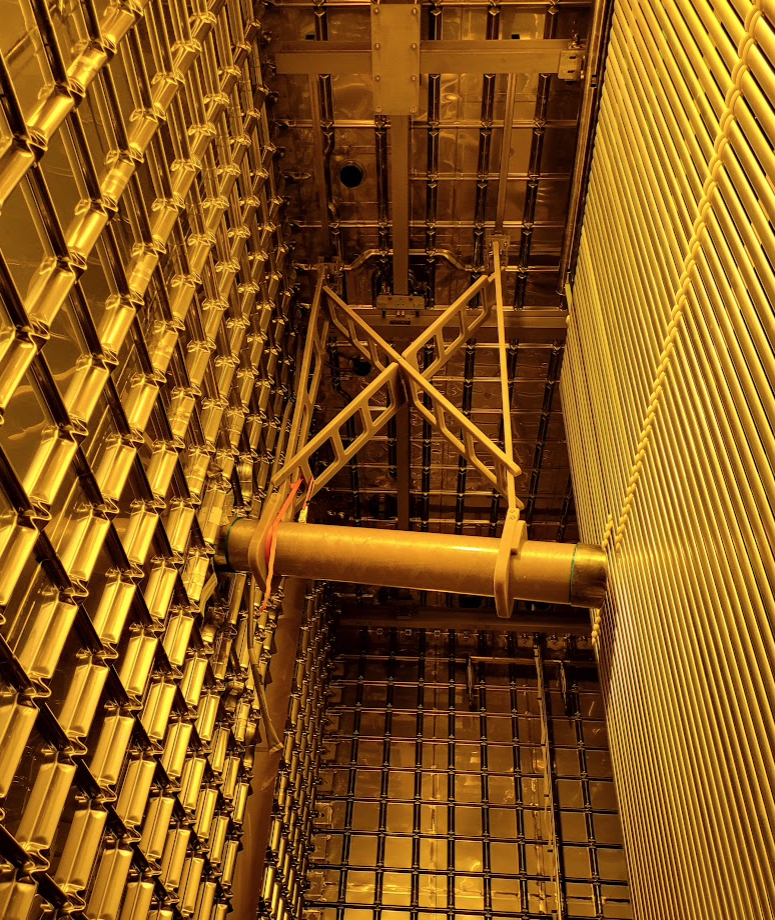}
\end{dunefigure}
%$$$$$$$$$$$$$$$

\subsection{Installation}

The decommissioning of the current \dword{hv} test setup in \dword{np02} and the opening of the cryostat were completed in fall 2022. 
Installation of the detector started in December 2022 and will continue through spring 2023.

The installation sequence is as follows:
\begin{enumerate}
\item{Install the \dword{fcss}.}
\item{Install top \dwords{crp}.}
\item{Connect the cables and bias to the existing feedthroughs.}
\item{Install the top four membrane \dwords{pd} and run the services to the feedthroughs.}
\item{Install the vertical cable trays.}
\item{Install the \dword{pds} in the two cathode modules.}
\item{Connect the two cathode modules to the suspension wires from the \dword{crp} suports.}
\item{Bring the \dword{pds} services to the cable trays and onward to the related feedthroughs.} 
\item{Install the bottom \dwords{crp} and run the cables via the vertical cable trays to the dedicated feedthroughs.}
\item{Install the \dword{fc} and the \dword{hv} system.}
\item{Install the four bottom membrane \dword{pds} and run the services to the feedthroughs.}
\item{Install the beam plug.}
\item{Clean the cryostat.}
\item{Close the \dword{tco}.}
\end{enumerate}

The slow control system remains is in place from \dword{pddp} and will be available for dedicated debugging activities after each component is installed.
Detector component tests are described in their respective chapters in this report.
All of the installation steps will go through the standard \dword{cern} planning and review, with dedicated \dword{esh} documentation to be prepared and accepted in advance for each step.
All components installed in \dword{vdmod0} are pre-production units. % as described in the various chapters of this TDR.

\subsection{Infrastructure}

The cryogenics is unchanged. The grounding scheme is based on full electrical isolation of all components on the cryostat and of the cryostat itself, with respect to the building ground.
\begin{itemize}
\item{Cryogenics instrumentation;}
\item{Grounding scheme;}
\item{Slow control;}
\item{\dword{tco} opening and closure approach;}
\item{Cryogenics scheme and operation.}
\end{itemize}

The \dword{tco} closing is similar to that of \dword{pdsp} and \dword{pddp}. The slow control is  essentially %basically 
unchanged. The cryogenics instrumentation (temperature and purity) is similar,  with some upgrades.

\subsection{Timeline and Strategy for Fill and Operation}

The filling schedule of \dword{vdmod0} depends on the market availability of \dword{lar} ($\sim$1\,kt) at reasonable cost. 
At the time of writing this report, no firm offer is available from any vendor in Europe. 
When \dword{lar}  becomes available, the present plan is to fill the \dword{np04} cryostat for \dword{hdmod0} first and then to transfer that \dword{lar}  to the \dword{np02} cryostat for \dshort{vdmod0} once \dshort{hdmod0} has been tested and qualified, probably five months after filling.
Discussions are ongoing as to whether to reverse the sequence and start with  \dword{np02}.  
The DUNE collaboration will make this decision once the availability of the necessary quantity of \dword{lar} is known. 
If the cost allows, it could even be possible to fill both cryostats at the same time.

The \dshort{vdmod0} is simpler to fill since no requirement exists on the maximum temperature gradient during filling, as is the case for \dshort{hdmod0}. The level of \dword{lar} should reach the height of the \dword{fcss} or above.

Slow control monitoring systems will follow the entire operation from the beginning. Dedicated dimensional and stress probes will be operational on the external wall of the cryostat to monitor abnormal behaviors. 
Once the \dword{np02} cryostat is full, the next step is to turn on the required \dword{hv} and the various biases to the electronics, %and the DAQ 
followed by the \dword{daq} systems.

Exposure to the dedicated \dword{h2} tertiary beam 
is planned. With the filling plan described above, this will most likely happen in 2024. The beam time request to \dword{cern} will be %perfected 
refined and submitted toward the end of 2023.  

%%%%

%\tableofcontents

\chapter{Integrated Engineering and Installation}
\label{ch:IEI}

%%%%%%%%%%%% 
\section{Introduction}
\label{ch:IEI:intro}

The \dword{usproj} \dword{fscfbsi} will provide facilities %and services, 
on the surface and underground at the \dword{surf} to house and support %integration and installation of 
the \dshort{dune} \dshort{detmodule}s, and once \dshort{dune} obtains \dword{aup}, the \dword{fnal} \dword{sdsd} will assist the %\dword{dune} 
\dshort{usproj} far site integration and installation (\dword{fsii}) team  with the integration and installation of the detector and its safe and productive operation, providing 
logistical, %cryogenics, 
electrical, mechanical, cyber, and environmental support.

The \dword{fdc} \dshort{fsii} team has responsibility for integration engineering at the far site. 
Section~\ref{ch:IEI:integ} describes the integration engineering  and 
it also describes the critical services the \dshort{fsii} team provides: warehousing, transportation, and logistics support. 
The \dshort{fdc} physical deliverables include the detector cryostat, cryogenics system, and installation infrastructure (Figure~\ref{fig:cryo-mezz}); \dword{fscfbsi} provides the initial electrical infrastructure. 
Section~\ref{ch:IEI:det-inst} outlines the installation process.

\begin{dunefigure}
[Cryostat and mezzanines]
{fig:cryo-mezz}
{Model of cryostat outer structure and the mezzanines above its roof. The cryogenics mezzanine (60.5\,m $\times$ 11.3\,m) appears in the foreground, the detector electronics mezzanine (61.5\,m $\times$ 3.6\,m) appears behind it.}
\includegraphics[width=0.95\textwidth]{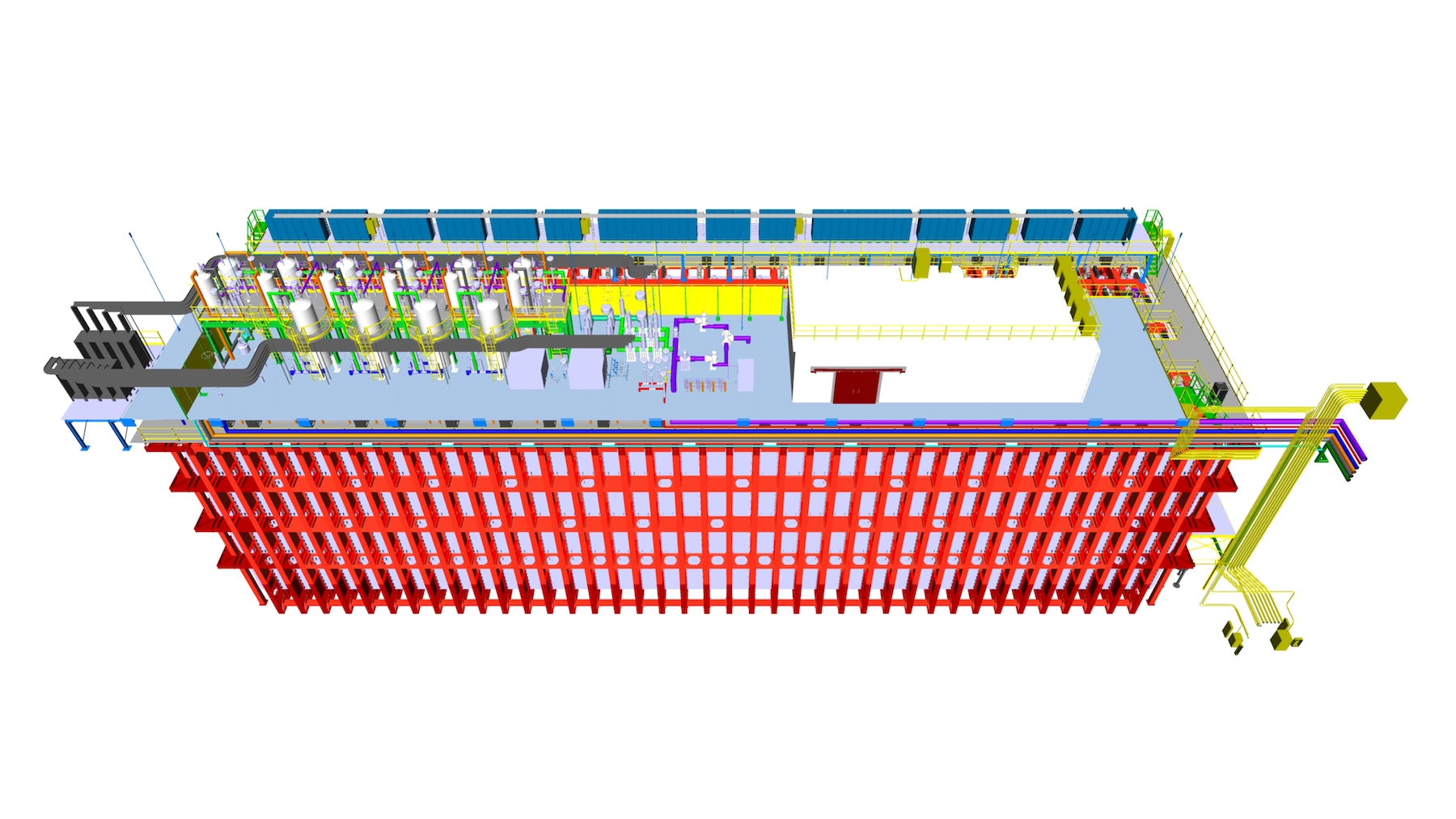}
\end{dunefigure}

%%%%%%%%%%%% 
\section{Requirements and Specifications}
\label{sec:fd2-ii-reqspec}

The principal requirements and specifications for these activites are listed in Table~\ref{tab:specs:SP-INST}.

% This file is generated, any edits may be lost. Not anymore!

\begin{footnotesize}
\begin{longtable}{p{0.12\textwidth}p{0.21\textwidth}p{0.17\textwidth}p{0.25\textwidth}p{0.13\textwidth}}
\caption{Installation specifications} \\ %\fixmehl{ref \texttt{tab:spec:SP-INST}} \\
  \rowcolor{dunesky}
       Label & Description  & Specification \newline (Goal) & Rationale & Validation \\  \colhline
   
  \newtag{INST-1}{ spec:logistics-material-handling }  & Compliance with the SURF Material Handling Specification for all material transported underground  &  SURF Material Handling Specification &  Loads must fit in the shaft be lifted safely. &  Visual and documentation check \\ \colhline

  \newtag{INST-2}{ spec:logistics-shipping-coord }  & Coordination of shipments with \dword{cmgc}; DUNE to schedule use of Ross Shaft  &  2 wk notice to CMGC &  Both DUNE and CMGC need to use Ross Shaft &  %Deliveries will be rejected 
  \\ \colhline

  \newtag{INST-3}{ spec:logistics-materials-buffer }  & Maintain materials buffer at logistics facility in SD   &  $>1$ month &  Prevent schedule delays in case of shipping or customs delays &  Documentatation and progress reporting \\ \colhline

  \newtag{INST-4}{ spec:cleanroom-specification }  & Installation cleanroom Specification  &  ISO 8 &  Reduce dust (contains U/Th) to prevent induced radiological background in detector &  Monitor air purity \\ \colhline
  % was INST-5 in HD

  \newtag{INST-5}{ spec:cleanroom-uv-filters }  & UV filter in installation cleanrooms for PDS sensor protection  &  filter $<\SI{400}{nm}$ for $>$ 2 wk exp; $<\SI{520}{nm}$ all else &  Prevent damage to PD coatings  &  Visual or spectrographic inspection \\ \colhline
    % was INST-6 in HD

\label{tab:specs:SP-INST}
\end{longtable}
\end{footnotesize}

The \dshort{fsii} team requires that all materials to be immersed in \dshort{lar} be validated in a \dword{lar} test stand in order to satisfy FD-5 in Table~\ref{tab:specs:SP-FD2}. This is the responsibility of the consortia.

%%%%%%%%%%%% 
 \section{Detector Integration}
 \label{ch:IEI:integ}
 
 \subsection{Responsibilities and Activities}
 \label{sec:IEI:resp-act}
 
The detector integration is the responsibility of the  \dword{fsii} team in cooperation with the various \dword{dune} consortium technical leads. It comprises the activities  listed below and involves several specialized engineering skills and groups within the %project, 
\dshort{usproj}, e.g., the \dword{ro}, the \dword{co}, and safety professionals. 
 \begin{itemize}
    \item \threed model integration,
    \item mechanical conflict detection,
    \item interface drawings,
    \item integration with \dshort{fscfbsi},
    \item interface to integration deliverables,
    \item warehousing and logistics, 
    \item scheduling and planning,
    \item detector slow control,
    \item material transport underground, 
    \item integration with cryostat, and
    \item management and oversight of underground work.
\end{itemize}
 
A configuration control office provides all \dword{cad} modeling and its verification, including all %relations
interfaces to the civil engineering activities of \dshort{fscfbsi}.

%A centralized 
The \dword{ro}, led by four senior physicists and engineers, will perform centralized technical reviews of %all (I'm always nervous about "all")
the various aspects of the detector throughout the design process, ending with \dwords{orr}.

%A dedicated 
The \dword{co} enforces %all 
the rules and codes %applicable for the 
that apply to the design of each component and verifies the validity of %all 
the structural analyses. 
The \dword{sdsd} is responsible for material and component receipt and logistics, and for the final delivery of %all the necessary 
materials to the appropriate underground cavern. This responsibility includes the hiring and management of %all technical crews, i.e.,  
underground work managers, a crew for basic services (electricity, ventilation, networking), and safety inspectors.

\subsection{Cryostat and Cryogenics Systems}
\label{ch:IEI:cryostat}

\subsubsection{Cryostat Design Overview}

The \dword{spvd} cryostat  
will be constructed using membrane cryostat technology, as for the \dword{sphd} and the two ProtoDUNE cryostats. 
In the cryostat design, described in~\cite{pdr-fs-cryo}, a \SI{1.2}{mm} thick, corrugated stainless steel membrane 
forms a sealed container for the \dshort{lar}, with surrounding layers of thermal insulation and vapor barriers. Outside these layers, a free-standing steel frame forms the outer (warm) vessel, the bottom and sides of which support the  hydrostatic load. The roof of the cryostat supports most of the components and equipment within the cryostat. The design differs between \dshort{sphd} and \dshort{spvd} largely in two areas: the number and location of the roof penetrations, and the size of the \dword{tco}.

\begin{dunefigure}
[Cryostat \dshort{tco} layout and dimensions]
{fig:TCO}
{View of part of cryostat endwall from exterior, showing \dshort{tco} layout and dimensions.}
\includegraphics[width=.45\textwidth]{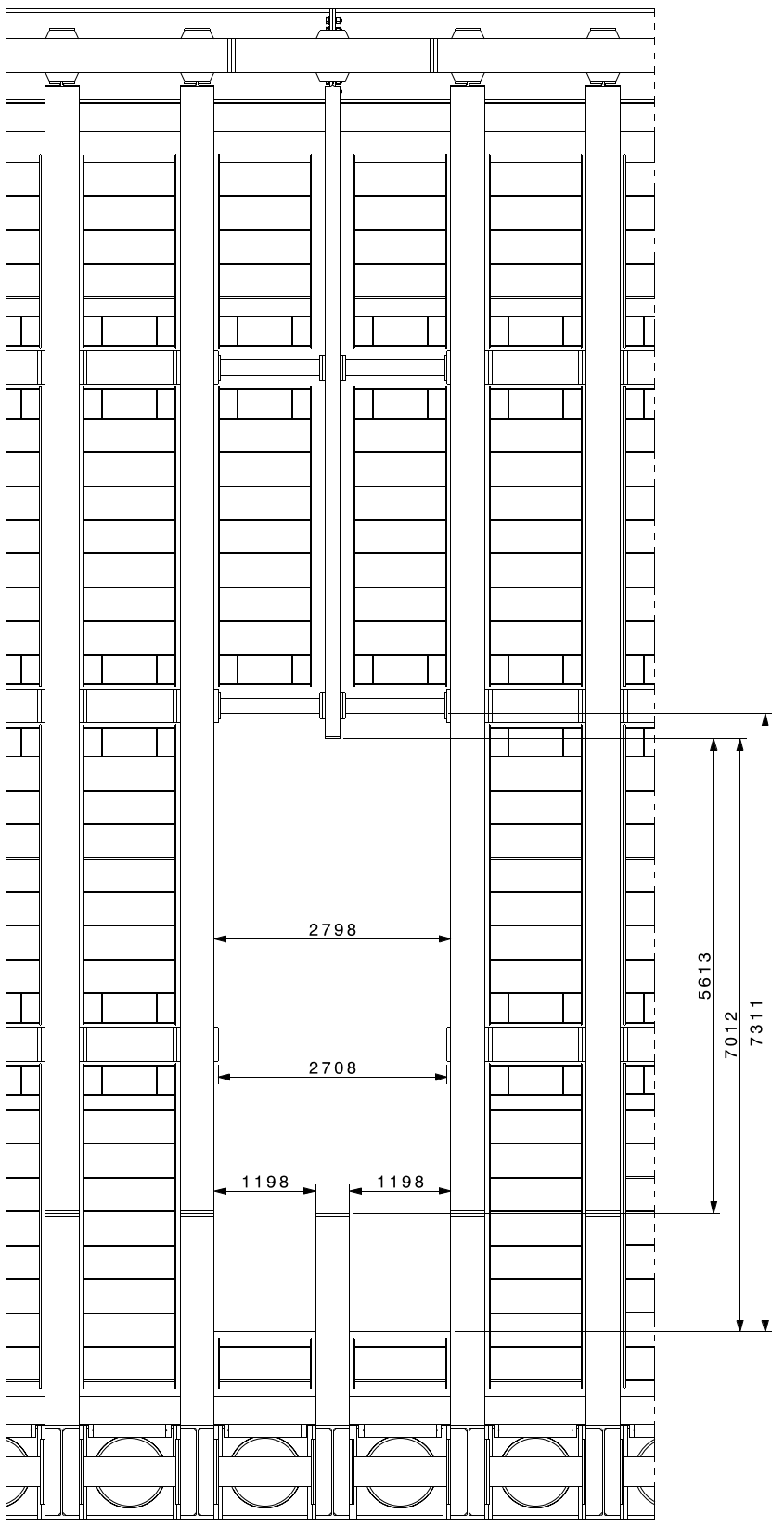}
\end{dunefigure}

The detector penetration details were finalized in November 2021. The cryostat engineering firm, GTT, provided an updated design for the cryostat in accordance with the new penetration  specifications, 
an updated list of materials to be procured,
and appropriate testing and installation procedures. The 
\dword{cern} holds the engineering contract with GTT for the cryostats, and this  contract accommodates these updates.

\subsubsection{Detector Assembly within Cryostat}

Given the sizes, relative robustness and modularity of the \dshort{spvd} 
components, much of the detector assembly work can be done inside the cryostat. The \dshort{tco} height can therefore be reduced by about half relative to the \dshort{sphd} design, to 7.3\,m  (see Figure~\ref{fig:TCO}), which will reduce the time needed to close it. Closing the \dshort{tco} from the exterior will be tested in 2023 on the \dword{np04} cryostat; experience gained from this will further simplify and shorten the procedure  for \dshort{spvd} as well as eliminate the risks and difficulties associated with performing work inside the cryostat after the detector is fully installed.

\subsubsection{Cryostat Roof Penetrations: Parameters and Uses}

The cryostat roof penetration locations and parameters are given in Figure~\ref{fig:penetration} and Table~\ref{tab:penetrationtable}. 
Together, the 293 penetrations on the cryostat roof accommodate all the cabling, wiring, and instrumentation needed by the \dword{crp} superstructures, the \dword{tpc} and \dword{pds} readout, and the interior cryogenics instrumentation. The 64 \dshort{crp} support penetrations (orange-colored in Figure~\ref{fig:penetration} and Table~\ref{tab:penetrationtable}) 
can be motorized in order to control the position and the planarity of the detector elements after \dshort{lar} filling.
Four of the penetrations (violet-colored in the diagram), one on each corner, are access holes that will allow continued access to the interior once the \dshort{tco} is closed and until everything is ready for the purge and \cooldown process to start.
During the detector installation, clean air will be injected into the cryostat through these access holes.

\begin{dunefigure}
[The layout of penetrations on the cryostat roof]
{fig:penetration}
{The layout of penetrations on the cryostat roof. The major dimensions (in meters) are 62.00 and 15.10 (primary membrane length and width), 60.02 and 13.50 (\dword{fc} length and width). See Figure~\ref{fig:penetrationdet} for a detail of the top right corner.}
\includegraphics[width=1.2\linewidth,keepaspectratio, angle=90]{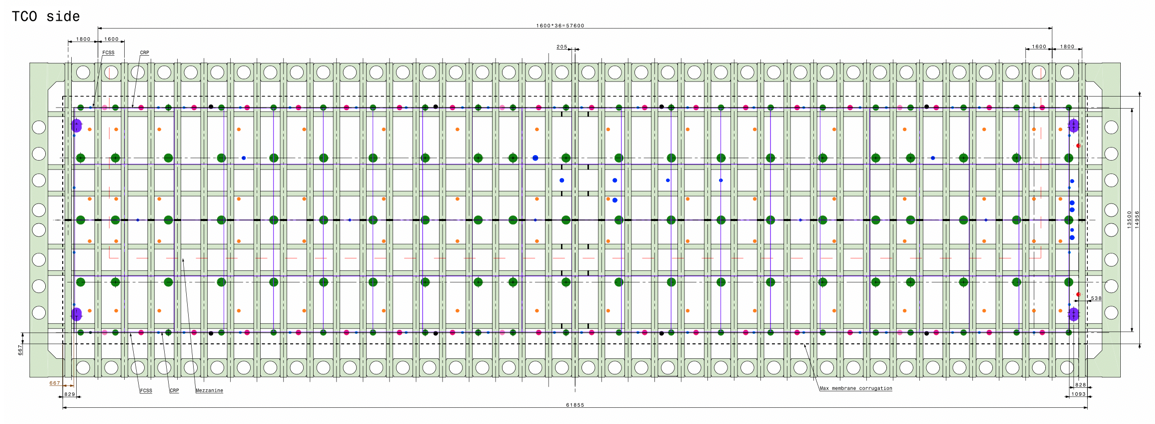}        
\end{dunefigure}

\begin{dunefigure}
[Layout of penetrations, detail]
{fig:penetrationdet}{Detail of the cryostat roof penetrations showing the end opposite from \dshort{tco} (top right corner of Figure~\ref{fig:penetration}).}
\includegraphics[width=.65\linewidth]{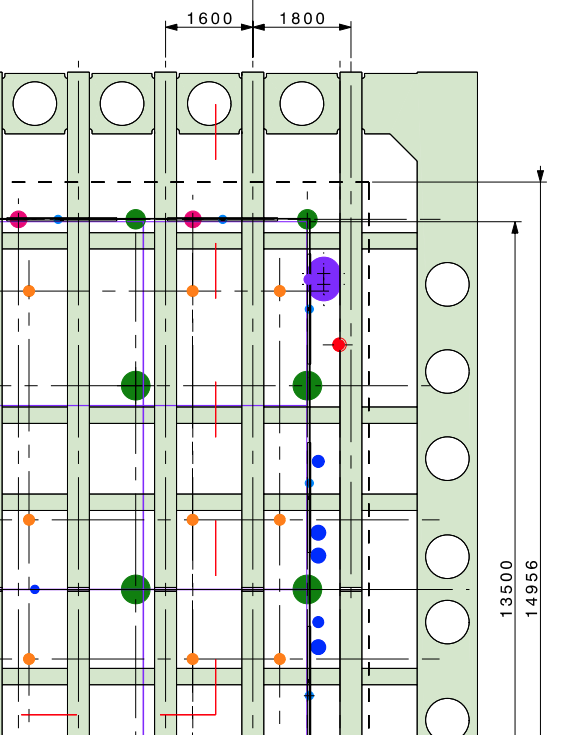}        
\end{dunefigure}

\begin{dunetable}
[Cryostat roof penetration parameters]
{lccl}
{tab:penetrationtable}
{Cryostat roof penetration parameters; colors correspond to Figures~\ref{fig:penetration} and~\ref{fig:penetrationdet}.}
Color & Diam. (mm) & Quantity & Description\\ \toprowrule
orange &200 & 64 & CRP supports \\ \colhline
green & 526 &63 & Top center CRP cables \\ \colhline
green & 355.6 & 42& Top side CRP cables \\ \colhline
dark pink & 304.8  & 34 &  Bottom CRP cables + PDS \\ \colhline
light pink & 304.8 & 6&  Bottom CRP cables \\ \colhline
red & 250& 2 & High voltage \\ \colhline
black &250 &8  & Laser \\ \colhline
violet & 800& 4& Access hatches \\ \colhline
light blue  &150 & 48&  FC supports\\ \colhline
light blue & 250 & 4 & \dword{calci} \\ \colhline
blue  &250 & 2&  Water trap\\ \colhline
blue  &200 & 3&  Spares\\ \colhline
blue  &152 & 5&  GAr controlled vent\\ \colhline
blue  &273 & 1&  GAr boiloff\\ \colhline
blue  &324 & 1&  \dword{psv} \\ \colhline
blue  &273 & 3&  LAr return\\ \colhline
blue  &219 & 3&  GAr purge; GAr make up\\ 
\end{dunetable}

\subsubsection{Cryogenics Infrastructure in the Cryostat}
\label{ch:IEI:cryoinfr}

The cryogenics system is described in detail in the Cryogenics Design Report~\cite{pdr-fs-cryo}. The cryogenics system has been scaled and designed to perform similarly to the \dword{protodune} systems at \dshort{cern}. Excellent purity was achieved in \dword{pdsp}, and simulations~\cite{voirin-sim-rpt} indicate that \dword{spvd} should perform in a similar manner, %aiming for impurity of 
limiting impurities to $<$30\,ppt, thereby reaching % as described in %requirement 
the specification goal listed in FD-5 (see Table~\ref{tab:specs:SP-FD2}). These simulations concluded that any possible effects from superheating of the \dshort{lar} would be confined to the top layer of the liquid.

As the cryostat dimensions and 
corresponding external heat loads are identical for the \dshort{sphd} and \dshort{spvd} detector modules,
and the heat load from the electrical power and the roof penetrations 
are similar, 
the primary cryogenics plant in the \dword{cuc} was designed for and will service both. 
This system is described in detail in~\cite{pdr-fs-cryo}. 

The \dshort{spvd} design does not require fine control of the temperature gradient during \cooldown{}, therefore no sprayers are needed on the short end walls, reducing the amount of internal cryogenics needed relative to the \dshort{sphd}. 
In addition, pipes for flushing \dword{gar} and distributing \dshort{lar} are needed only along the outer edges of the cryostat floor and up the long walls in this design, as shown in Figures~\ref{fig:Cryopipest} and~\ref{fig:Cryopipesb}. This allows flexibility in optimizing the vertical position of the detector within the cryostat.

\begin{dunefigure}
[Internal cryogenic piping at the top]
{fig:Cryopipest}
{Internal cryogenic piping at the top. The red pipe is the gas purge line and the two blue pipes are the liquid return lines.}
\includegraphics[width=0.95\linewidth]{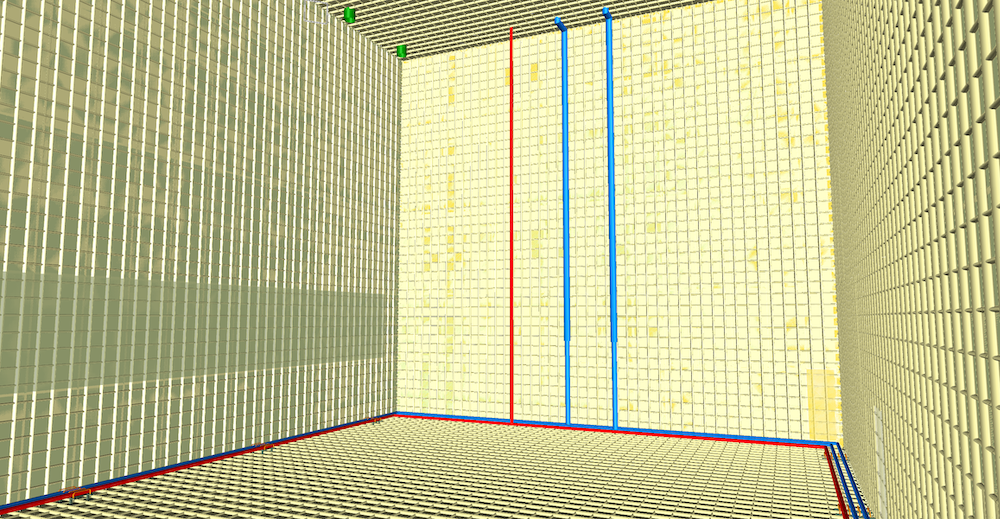}
\end{dunefigure}

\begin{dunefigure}
[Internal cryogenic piping and floor supports]
{fig:Cryopipesb}
{Internal cryogenic piping and floor supports.}
\includegraphics[width=0.7\linewidth]{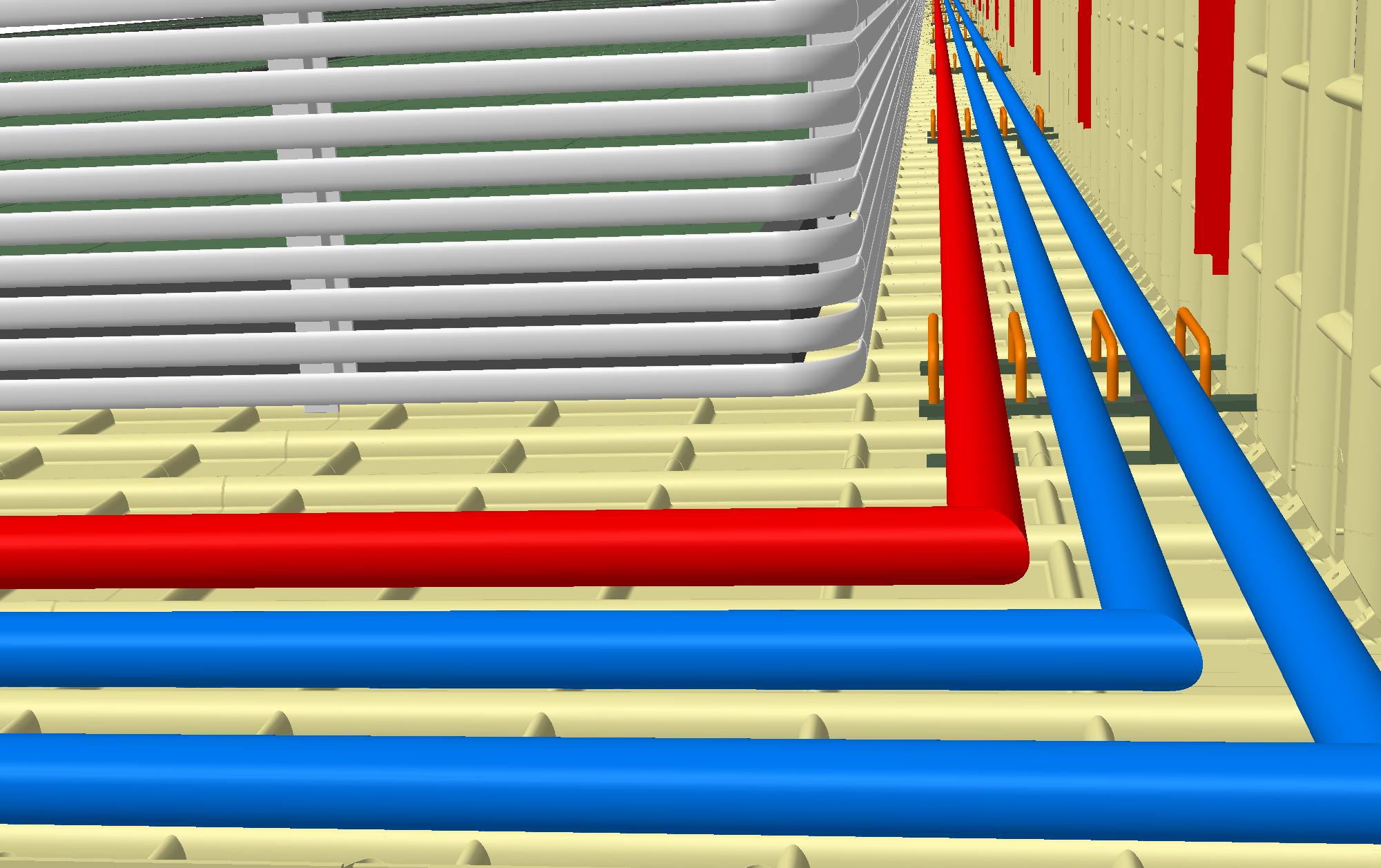}
\end{dunefigure}

%%%%
%\sub
\subsubsection{Cryogenic Instrumentation} % 11/15 11:15
\label{ch:IEI:cryoinst}

The \dword{spvd} cryogenics instrumentation is designed to provide real-time measurements of the state of the \dshort{lar} (i.e., purity, %electron lifetime,
temperature, level) and in particular, to provide prompt indications of any changes in its state.  As in the \dword{sphd} module, external purity monitors will be installed immediately downstream of the filtration system to signal any deterioration in the performance of the filters, and space is reserved for a purity monitor immediately upstream of the filtration system to measure the quality of the argon as it leaves the cryostat. 

\subsubsubsection{Purity Monitors}

Four purity monitors will be installed inside the cryostat, 
one pair on each of the upstream and downstream ends, one outside the upper drift volume and the other outside the lower.  
Each purity monitor requires quartz optical fibers that deliver xenon UV light to the gold photon cathode and are protected in a flexible fiber-bellow-hose, and traditional coaxial cables that provide \dword{hv} and signal readout.  
There are no dedicated roof penetrations for the purity monitors. Instead, the access hole covers will be used as feedthroughs for the monitor fibers and cables.  On each end of the detector, one (upper) purity monitor will be mounted to the bottom of the access hole radiation shielding (see Figure~\ref{fig:prm}), and the other will sit on a simple stand on the cryostat floor. The cables and fibers for both will be routed up through the %manhole 
radiation shielding, and for the lower monitor, they will continue 
down the corner of the cryostat and then across the floor to the lower monitor. % which will sit on a simple stand.

\begin{dunefigure}
[Purity monitors]
{fig:prm}
{Left: Photograph of a purity monitor. Right: 
\threed model of a purity monitor mounted under an access port.}
\includegraphics[width=0.35\linewidth]{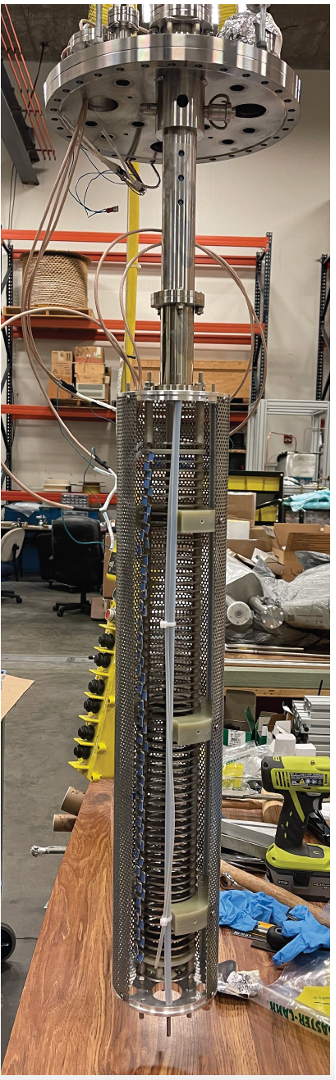}
\includegraphics[width=0.45\linewidth]{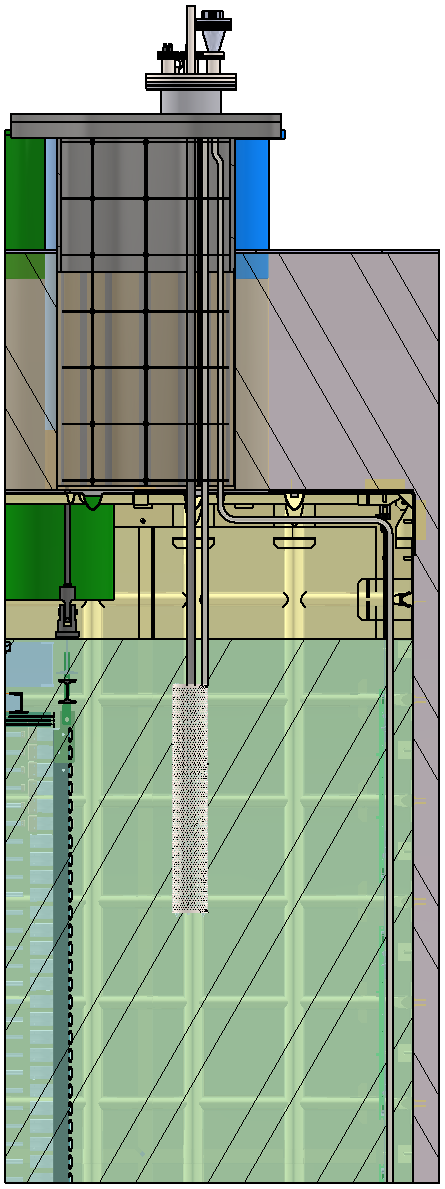}
\end{dunefigure}

The installation procedure of a purity monitor (PrM) assembly is as follows:
\begin{enumerate}
\item The cables and fibers are routed along the corner of the wall of the cryostat and placed near the roof so they can be reached at the completion of the detector installation.
\item After the top \dword{crp} and cathode are installed, but before installation of the east \dword{ewfc}, the east lower purity monitor is installed and connected to its cables. It will become inaccessible after this point. 
\item After the west %end wall 
\dword{ewfc} is in place, the west lower purity monitor is installed along with its stand, and cabled. % It can also them be cables.
\item Immediately before closing the \dword{tco}, the top purity monitors are mounted to the  radiation shields attached to the bottom of the sealing flange to the access ports, the cables inside the detector near the access holes are pulled up and attached to the purity monitor feedthrough that is welded to the access hole cover flange, and the access holes are closed.
\end{enumerate}

\subsubsubsection{Temperature Monitors}

Plans to measure the temperature of the \dshort{lar} entering the cryostat will be formulated based on the details of the cryogenic piping. The baseline design assumes a total of 16 precision \dshort{rtd}s, deployed in pairs, with four sensors per pipe at opposite cryostat corners. 

The temperature of the \dshort{lar} leaving the cryostat will be also monitored, with two sensors near each of the four pumps at the bottom of one of the short walls. Finally, one \dshort{rtd}s above and below each purity monitor will facilitate the study of correlations between temperature and purity. 

The temperature of the gas argon at the top of the cryostat (in the ullage) will be monitored using arrays of 18 sensors at ten different locations above the \dshort{crp}s. The top 14 sensors will be standard \dshort{rtd}s since the temperature gradient will be very large. The four sensors at the bottom, which are expected to be below the \dshort{lar} surface, will be precision \dshort{rtd}s. 

All systems described above have been prototyped in \dshort{pdsp}.

A prototype of %this 
a system to measure the \dshort{lar} vertical temperature gradient in the proximity of the active volume will be deployed in \dword{vdmod0}.

Vertical arrays of standard \dshort{rtd}s will be epoxied to the cryostat membrane in two opposite corners of the cryostat with the aim of monitoring the membrane temperature during \cooldown and filling.  

The thermometry is installed in the first phase of the detector installation when the cryostat false floor is in place but the walls are accessible using the scissor lifts.

\subsubsubsection{Alignment Monitors}
A laser-based alignment system is not part of the \dshort{spvd} baseline design, but active prototyping is ongoing. This system would provide artificial straight tracks used for mechanical alignment. 
Six penetrations will be available to accommodate a future laser-based alignment system and physical space is reserved.

\subsection{Electrical Infrastructure}
\label{subsec:elect-inf}

In addition to that provided by \dshort{fsii}, after gaining \dword{aup} of the caverns the \dshort{fdc} subproject will provide additional electrical infrastructure.  This includes the installation of detector power and ground, power and cooling for \dword{daq} racks, readout racks located both on the detector and on the detector rack mezzanine, optical fiber distribution for both data and network, and a \dword{ddss}. 

\subsubsection{Grounding and Detector Power} % section header added by Anne 9/9

The grounding strategy provides each detector module with an independent ground to minimize any environmental electrical noise that could couple into detector readout electronics either conductively or through emitted electromagnetic interference. 
The plan includes a separate detector ground, separated from the rest of the facility, for each of the four planned 
detector modules. The detector ground will primarily consist of the steel containment vessel, cryostat membrane, and connected readout electronics. 

For safety reasons, a saturable inductor must connect the detector ground to the facility ground.  This saturable inductor provides an impedance to any AC environmental noise on the facility ground as well as a safety path for any DC fault current that could damage equipment or harm personnel.

Power for the \dshort{spvd} will be provided through transformers located in the \dword{cuc} electrical room.   (\dword{fscfbsi} will supply a 1000\,kVA transformer for each cavern.)   
Power from these initial transformers will be run through special double-shielded transformers, which allows for separation of facility power and ground from detector power and ground.  The saturable inductor %described above 
will be located close to these double-shielded transformers and provide the safety ground. % described above.   

\begin{dunefigure}
[Detector power budget]
{fig:detpow}
{Power budget table.}
\includegraphics[width=0.95\linewidth]{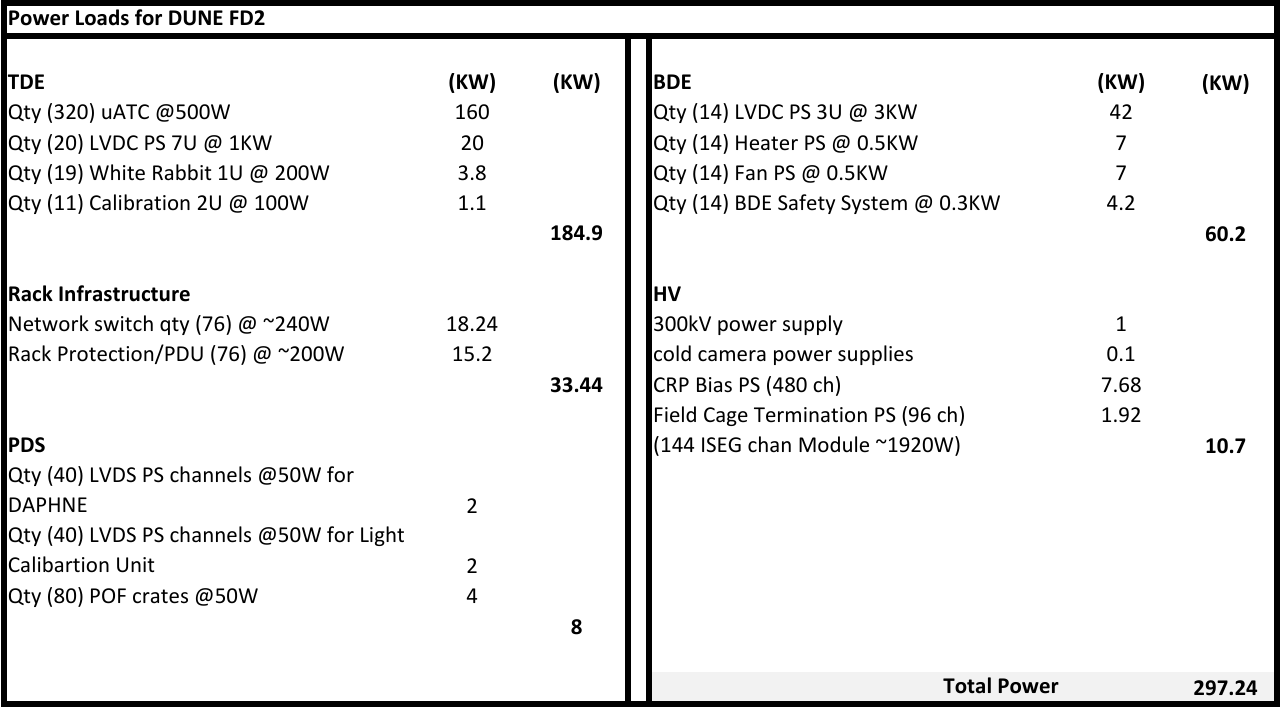}
\end{dunefigure}

%Individual detector module power 
Power to each detector module  will be de-rated to no more than 75\% of total power available (500\,kVA) at the electrical distribution panels. Consumption per detector module is planned to remain at or below 375\,kW.  AC power will be distributed from the \dshort{daq} barracks on the cryogenics mezzanine to approximately 80 detector racks %to be 
located on the detector rack mezzanine and to 320 \dword{tde} \dword{utca} crates distributed on top of the \dshort{spvd} cryostat. Figure~\ref{fig:detpow} shows the anticipated load of the %detector 
\dshort{spvd} to be approximately 300\,kW.  

\subsubsection{Detector Readout Electronics} 

A \dword{daq} barrack, to be located on top of the cryogenics mezzanine, will be referenced to facility ground.  Only fiber optic cables will be allowed to run between the detector electronics and the \dshort{daq} barrack in order to maintain the ground separation of facility and detector electronics.  \dshort{daq} will install 16 racks in this room with an expected load of 125\,kW.  A transformer to provide this power will be located on the roof of the barrack, along with a cooling unit to maintain the required temperature and humidity levels. All \dshort{daq} power will be UPS backed, with a requirement of 10 minutes of standby power  %This will 
to protect the \dshort{daq} servers. 

Detector electronics are installed on detector feedthroughs, in 27U racks close to feedthroughs, and in taller racks located on a detector mezzanine.  The detector rack mezzanine will hold approximately 80 racks and be referenced to detector ground, allowing easy access for maintenance and reducing complexity on top of the detector. %The detector racks provided 
These racks will include cooling fans, a rack protection system, smoke detection and a power distribution unit.  If smoke is detected within a rack, a hardware interlock will shut power off to the rack.

The DUNE experiment requires a number of fiber optic pairs to run between the surface and the \dword{4850l}.   These fibers serve as the pathways for data, networking, and slow controls.  A total of 96 fiber pairs, which will accommodate both DUNE and \dword{fscfbsi} needs, will be supplied through redundant paths with bundles of 96 pairs coming down both the Ross and Yates Shafts. The individual fibers are specified to allow for transmission of 100\,Gbps. The fibers run between the surface \dword{mcr} and a central location in the \dshort{cuc} and then to each \dshort{daq} barrack.  From the \dshort{mcr}, they connect to the WAN and ESnet to get to \dshort{fnal}. Figure~\ref{fig:opticalfiber}  shows the fiber plan for the detector.

\begin{dunefigure}
[Optical fiber plan]
{fig:opticalfiber}
{Summary of the number and routing of the optical fibers.}
\includegraphics[width=1.03\linewidth]{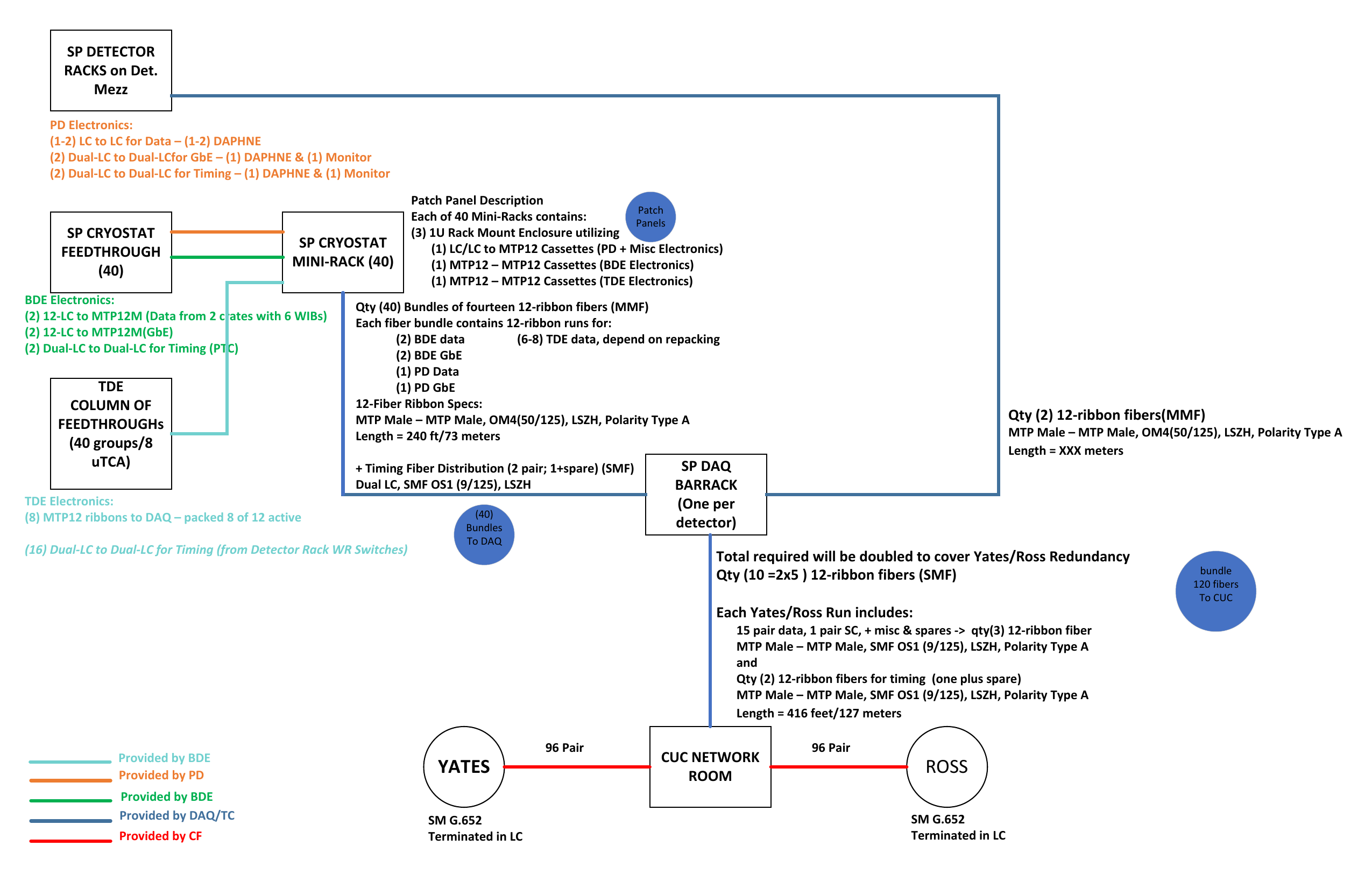}
%\par\bigskip
\end{dunefigure}

\section{Safety}
\label{sec:fdsp-safety}

The \dshort{lbnf-dune} Integrated \dword{esh} Plan~\cite{edms-2808692} outlines the requirements and regulations that \dword{dune} work must comply with, whether (1) at \dshort{fnal}, (2) in areas  leased by \dshort{fnal} or the \dword{doe}, (3) in leased space at \dshort{surf}, or (4) at collaborating institutions.
 
%%%%%%%%%%%%%%%%%%%%%%%%%%%%%%%%%
\subsection{Documentation Approval Process}

DUNE implements an engineering review and approval process for all required documentation, including structural calculations, assembly drawings, load tests, \dword{ha}, and procedural documents for a comprehensive set of identified individual tasks. 
As for \dword{pdsp}, all these documents are stored in the \dword{edms} at CERN. 
For the larger operations and systems like \dword{tpc} component factories, the \dword{dss}, cleanroom, and assembly infrastructure, DUNE safety also reviews the documentation then visits the site to conduct
 an \dword{orr}, which includes a demonstration of the final operations. The \dshorts{orr} are listed in project schedule. 
 
 Structural calculations, assembly drawings and proper documentation of  load tests, hazard analyses, and procedures for various items and activities will require review and approval before operational readiness is granted.

%%%%%%%%%%%%%%%%%%%%%%%%%%%%%%%%%
\subsection{Support and Responsibilities}

The \dword{esh} coordinator for each shift, who will report to the \dshort{dune} project \dshort{esh} manager, has overall \dshort{esh} oversight responsibility for the \dword{dune} activities at the  \dword{sdwf} and on the \dshort{surf} site. 
This person coordinates any \dshort{esh} activities and facilitates the resolution of any issues that are subject to the requirements of the \dword{doe} Workers Safety and Health Program, Title 10, Code Federal Regulations (CRF) Part 851 (10 CFR 851).  The on-site \dshort{esh} coordinator facilitates training and runs weekly safety meetings. This  person is also responsible for managing \dshort{esh}-related  documentation, including training records, \dword{ha} documents, weekly safety reports, records on materials-handling equipment, near-miss and accident reports, and equipment inspections. 

If the \dshort{esh} coordinator is absent, the shift supervisor acts in this capacity.
All workers have work stop authority in support of a safe working environment. 

%%%%%%%%%%%%%%%%%%%%%%%%%%%%%%%%%
\subsection{Safety Program}

The on-site \dshort{esh} coordinators will guide the \dword{fd} installation safety program, using the following:

\begin{enumerate}
\item	the \dword{feshm};
\item the \dshort{dune} Installation \dshort{esh} Plan~\cite{EDMS2858967}, 
which includes the fire evacuation plan, fire safety plan, lockdown plans, and the site plan;
\item	work planning and controls documentation which includes both hazard analysis and procedures; 
\item	Safety Data Sheets (SDS); 
\item	the respiratory plan, as required for chemical or \dword{odh} hazards; and 
\item	the training program, which covers required certifications and  training records.
\end{enumerate}

During the installation setup phase, as new equipment is being installed and tested, new employees and collaborators will be trained to access the facility and use the equipment. At the end of this phase, two shifts per day will be required.

The \dshort{dune} installation team
 will develop an  \dshort{esh} plan for detector  installation that defines  
the \dshort{esh} requirements and responsibilities for personnel during  assembly, installation, and construction of equipment at \dshort{surf}. It will cover at least the following areas:

{Work Planning and \dshort{ha}:} The goal of the work planning and \dshort{ha} process is to initiate thought about the hazards associated with work activities and plan how to perform the work. Work planning ensures the scope of the job is understood, appropriate materials and tools are available, all hazards are identified, mitigation efforts are established, and all affected employees understand what is expected of them. 
The work planning and \dshort{ha} program is documented in Chapter 2060 in the \dshort{feshm}.

The shift supervisor and the \dshort{esh} coordinator  will lead a work planning meeting at the start of each shift  to (1) coordinate the work activities, (2) notify the workers of potential safety issues, constraints, and hazard mitigations, (3) ensure that employees have the necessary \dshort{esh} training and \dword{ppe}, and (4) answer any questions.

{Access and training:}  All \dshort{dune} workers requiring access to the \dshort{surf} site must (1) register through the \dshort{fnal} Users Office to receive the necessary user training and a \dshort{fnal} identification number, and (2) they must apply for a \dshort{surf} identification badge. 
The workers will be required to complete \dshort{surf} surface and underground orientation classes. Workers accessing the underground must also complete 4850L and 4910L specific unescorted access training, and obtain a \dword{tap} for each trip to the underground area; this is required as part of \dshort{surf}'s Site Access Control Program. 
A properly trained guide will be stationed on all working levels. 

{\dword{ppe}:} 
The host laboratory is responsible for supplying appropriate \dshort{ppe} to all workers. 

{Emergency response team (\dword{ert}):} The \dword{sdsta} will maintain an emergency response incident command system and an \dshort{ert}.  The guides on each underground level will be trained as first responders to help in a medical emergency.
  
  Guides: The shift supervisor and lead workers will be trained as guides.
  
  {House cleaning:} All workers are responsible for keeping a clean organized work area. This is particularly important underground. Flammable items must be in proper storage cabinets, and items like empty shipping crates and boxes must be removed and 
transported back to the surface to make space.

{Equipment operation:} All overhead cranes, gantry cranes, fork lifts, motorized equipment, e.g., trains and carts, will be operated only by trained  operators. 
Other equipment, e.g., scissor lifts, pallet jacks, hand tools, and shop equipment, will be operated only by people trained
and certified for the particular piece of equipment. All installation equipment will be electrically powered.

\section{Detector Safety}

The DUNE detector safety system (\dword{ddss}) 
is intended to detect abnormal and potentially harmful operating conditions and provide interlocks, thereby protecting the experimental equipment. The system will recognize when conditions are not within the bounds of normal operating parameters and take pre-defined protective actions. % to protect equipment. 
Protective actions are hardware- or \dword{plc}-driven.  DUNE technical coordination works with the consortia to identify equipment hazards and ensure that harmful operating conditions can be prevented, or detected and mitigated.  %These 
Potential hazards include, e.g., smoke %detected 
in racks, \dword{odh}, % detection, 
a drop in the cryostat \dshort{lar} liquid level, %laser or radiation hazards in calibration systems, 
or over- or under-voltage conditions. %, and others to be determined.  
The \dshort{ddss} will receive alarms via cables from the various detector racks and 
provide input to the \dword{4850l} fire alarm system. %The 4850L fire 
This alarm system %\fixme{only alarm system or entire DDSS?}
is part of the life safety system and will play an integral role in detecting and responding to an event, as well as notifying occupants and emergency responders. This system is the responsibility of the host laboratory (\dshort{fnal}) and \dword{sdsta}.

\section{CRP and Cathode %Detector 
Support Structures}
\label{ch:IEI:dss}

The \dword{crp} superstructure, which supports the top \dshort{anodepln} and the cathode, is described in Section~\ref{sec:CRPMss}. It is designed by the \dshort{crp} consortium and the design will include an  engineering note 
with the \dword{qc} requirements. 
The \dword{fsii} team will be responsible for its fabrication and assembly, execution of the \dshort{qc}, and installation. 

The assembled \dshort{crp} superstructures are either 3\,m $\times$ 7\,m or
7\,m $\times$ 9\,m, i.e., too large to fit down the shaft in a single piece. The structures are designed to be brought underground in 3\,m $\times$ 3.4\,m  sub-assemblies, with final cleaning, assembly, and \dshort{qc} testing underground by \dshort{fsii}. 

\section{Installation Infrastructure}
\label{ch:IEI:inst-infr}

The installation infrastructure consists of the temporary equipment needed to install the detector, which is removed after the installation is complete. Installation infrastructure provided by the \dshort{fsii} team includes the \dword{greyrm} outside the cryostat with associated changing room, temporary cryostat and \dshort{greyrm} ventilation, the false floor inside the cryostat, temporary power and lighting in the cryostat and \dshort{greyrm}, scissor lifts capable of reaching 12\,m work height, material transport equipment from the \dshort{greyrm} to the cryostat, rigging equipment on the cryostat roof for the \dword{tde} and \dword{bde} equipment, and miscellaneous tools and rigging equipment. 
A crane will be in the cavern; there will be no fixed cranes in the \dshort{greyrm} or the cryostat except for the hoist beam through the \dshort{tco}. %Anne added last sentence 2/21/23 per comment 485

\begin{dunefigure}
[Space allocation in the cryostat and \dshort{greyrm} during \dshort{crp} installation]
{fig:cleanspace}
{Top: Image of the cryostat during week 8 of CRP installation, %when the CRPs are being installed. The figure shows 
showing the main space allocations inside the cryostat. Bottom: %Image of the 
\dshort{greyrm} %showing the 
work area and space allocations.}
\includegraphics[width=0.8\linewidth]{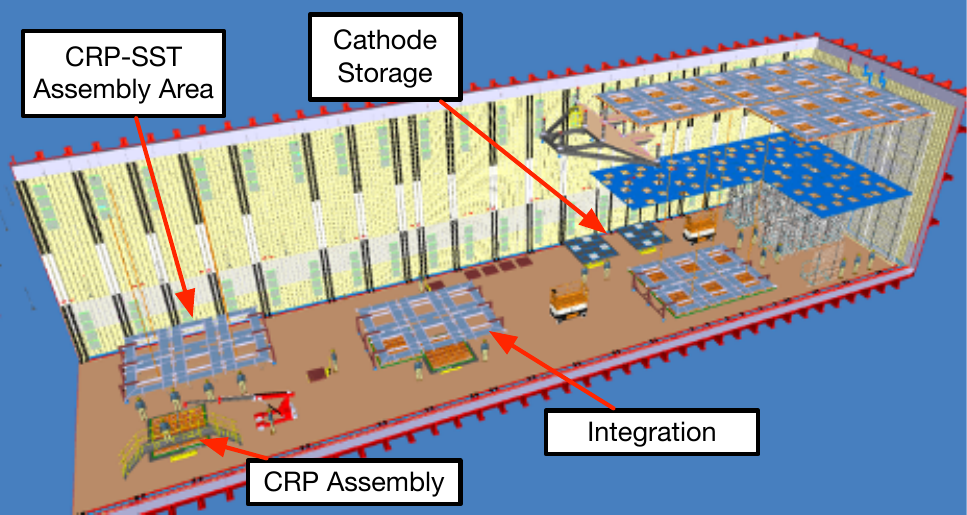}
\par\bigskip
\includegraphics[width=0.8\linewidth]{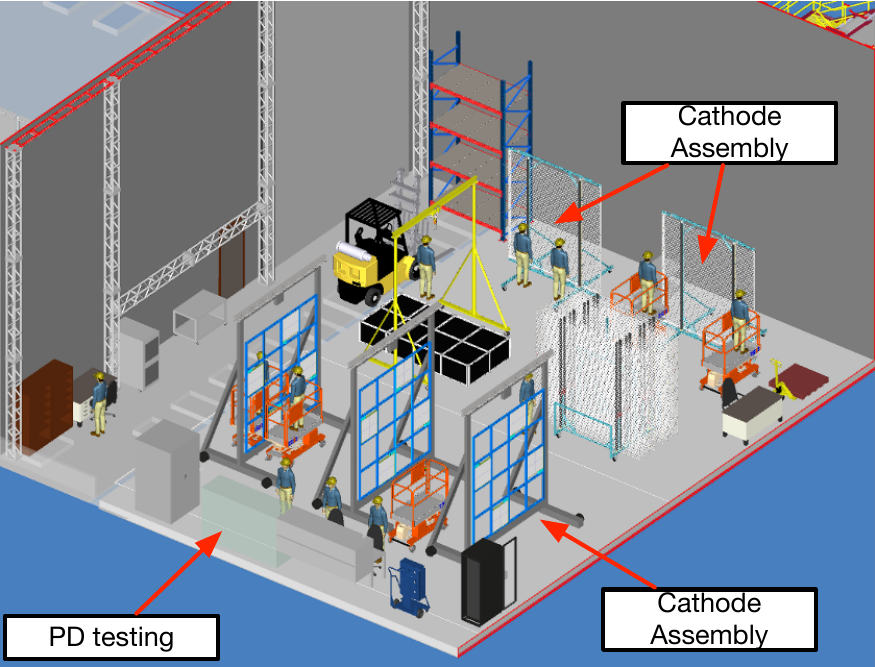}
\end{dunefigure}

The \dshort{fsii} team is also responsible for planning the material flow into the cryostat and budgeting the clean workspace provided by the cryostat and the \dshort{greyrm} for assembly, testing, and installation. 
Figure~\ref{fig:cleanspace} shows these spaces during 
the early phase of the \dshort{crp} installation, the period that requires the most clean workspace because of the activities going on in parallel.  
The floor of the cryostat measures 62\,m by 15.1\,m, offering a large available work area to be used for \dshort{crp} assembly and testing. 
The \dshort{fc} assembly area, the \dword{pd} testing area, and the cathode assembly area will be in the \dshort{greyrm}, 
as shown in the bottom of Figure~\ref{fig:cleanspace}. This set of activities determines the space requirements for the \dshort{greyrm}, which will be 15.5\,m deep by 17.45\,m wide. 
To enclose the cryostat's \dword{tco} (the bottom of which is 2\,m off the ground), the \dshort{greyrm} height is set at least 9.82\,m. 

\begin{dunefigure}
[\dshort{spvd} installation \dshort{greyrm}]
{fig:cleanroom}
{\dshort{greyrm} for the installation of the \dshort{spvd} detector module. The small structures to the left of the \dshort{greyrm} are the changing room and fenced areas for storage, and a small machine shop.}
\includegraphics[width=0.7\linewidth]{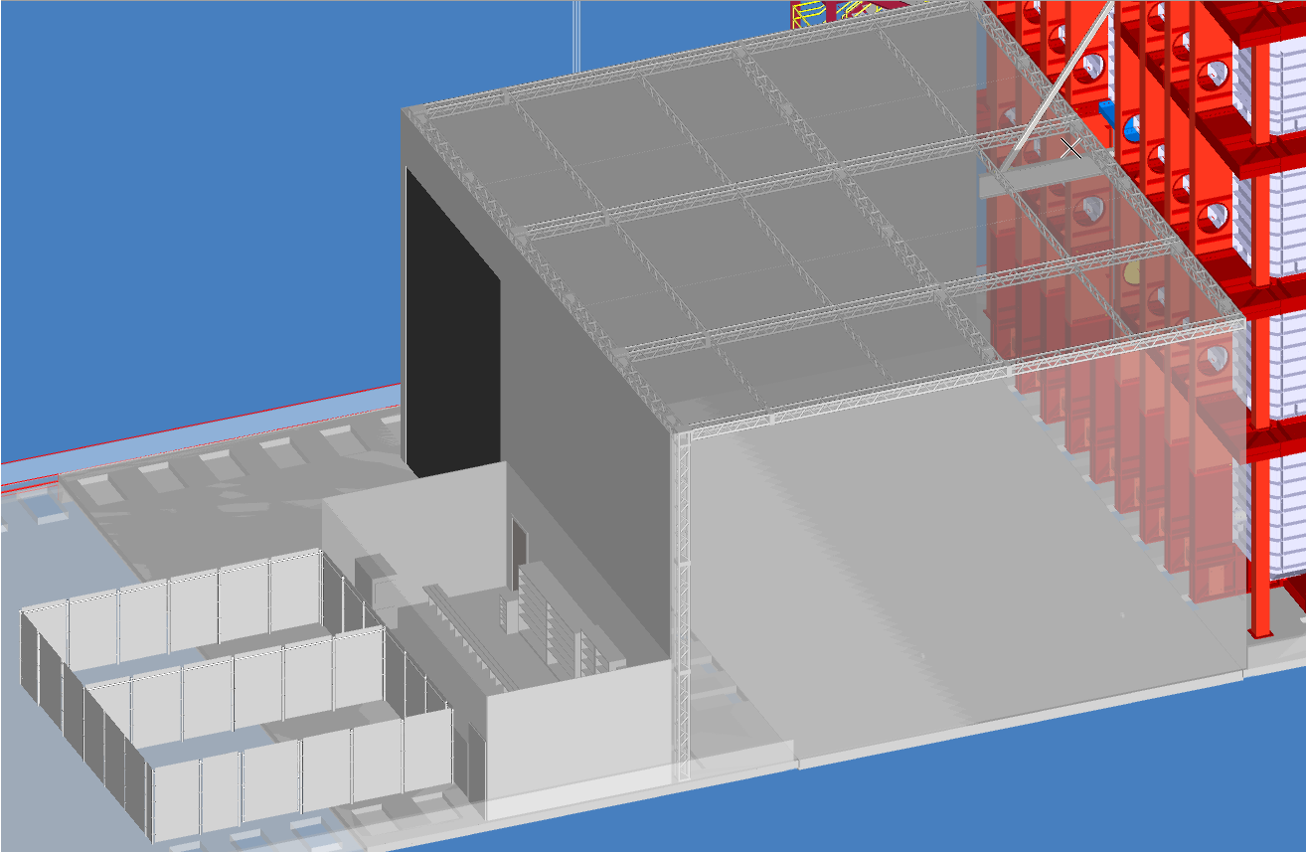}
\end{dunefigure}

A light-weight construction technique is used for the \dshort{greyrm} as there are no physical requirements beyond preventing dust migration and blocking UV light from entering. As shown in Figure~\ref{fig:cleanroom}, the \dshort{greyrm} support frame will be constructed of the same lightweight aluminum trusses that are used for the west wall of the \dshort{sphd} cleanroom. The frame will be covered with fiber-reinforced plastic sheets. Materials will enter the \dshort{greyrm} through a 6.1\,m by 8.3\,m opening in the west wall, which is normally closed with a pair of tarps. Since little sensitive electrical work will be performed inside the \dshort{greyrm} 
during this period, the floor will not be \dword{esd} protected. The \dshort{bde} consortium is planning to use this area for testing at the end of the installation and they will install temporary \dshort{esd} mats at that time. A small battery-operated forklift is used to move materials inside the \dshort{greyrm}. Temporary lighting and power are distributed in the area. In order to transport equipment from the \dshort{greyrm} through the \dshort{tco} into the cryostat a custom hoisting beam (\dshort{tco}--Beam) has been designed (Figure~\ref{fig:tcobeam}). As there are no mechanical connection points inside the cryostat this beam must be supported from the outer steel and cantilevered inside. To minimize the cantilever inside the cryostat, the area directly in front of the \dshort{tco} is kept free, and cryostat access during hoisting operations is via a fixed ladder. The hook coverage outside(inside) the cryostat is 2.2\,m(1.9\,m), which is sufficient to lift the longest piece of equipment (lifted from the center). A 2-ton (US) ultra-low-profile hoist %is planned to 
will give the maximum vertical hook clearance through the \dshort{tco} of 4.5\,m.

\begin{dunefigure}
[\dshort{tco} hoist system ]
{fig:tcobeam}
{The hoist system for transporting equipment from the \dshort{greyrm} into the cryostat through the \dshort{tco} allows objects up to 4\,m long to be brought in for installation.}
\includegraphics[width=0.9\linewidth]{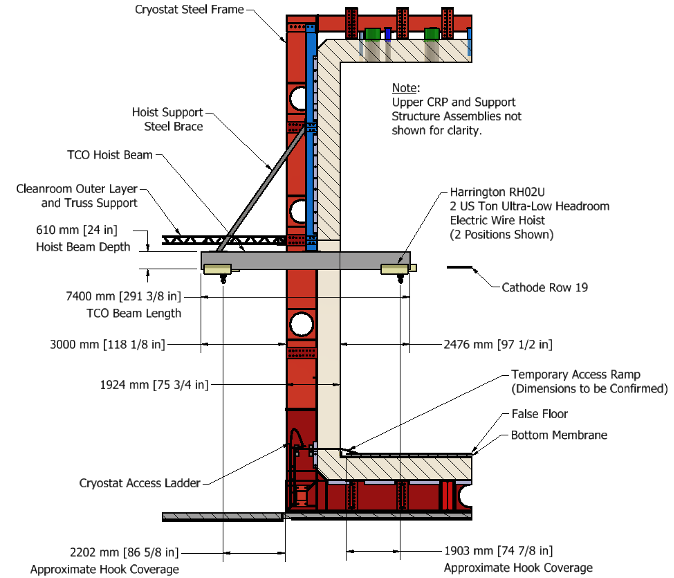}
\end{dunefigure}

Inside the cryostat, a temporary false floor %must be installed to 
will protect the cryostat's thin steel membrane and provide a flat surface for the scissor lifts during installation. Pedestals will be placed between the cryostat membrane corrugations to provide level support for the fire-retardant plywood flooring. Commercial floor pedestals have been identified with sufficient load rating for the scissor lifts.

The design of the cryostat false floor (Figure~\ref{fig:falsefloor}) %was designed, it had to take into 
 required consideration of the interfaces with the cryostat convolutions, the cryogenic piping along the wall, the bottom \dshort{crp} installation footprint, and the cables and fibers for the bottom \dshort{crp}s and the cathode \dshort{pd} modules, so it  %These constraints required that the 
  has to stop short of the vertical plane of the \dshort{fc} modules.   Figure~\ref{fig:falsefloor} shows a \threed model of the temporary floor and the floor tested at Ash River.

\begin{dunefigure}
[Cryostat temporary floor ]
{fig:falsefloor}
{Top: 3D model of the cryostat temporary floor with a section of the plywood removed to show the support pedestals and the underlying membrane. Bottom: Photograph of the flooring test at Ash River.}
\includegraphics[width=0.8\linewidth]{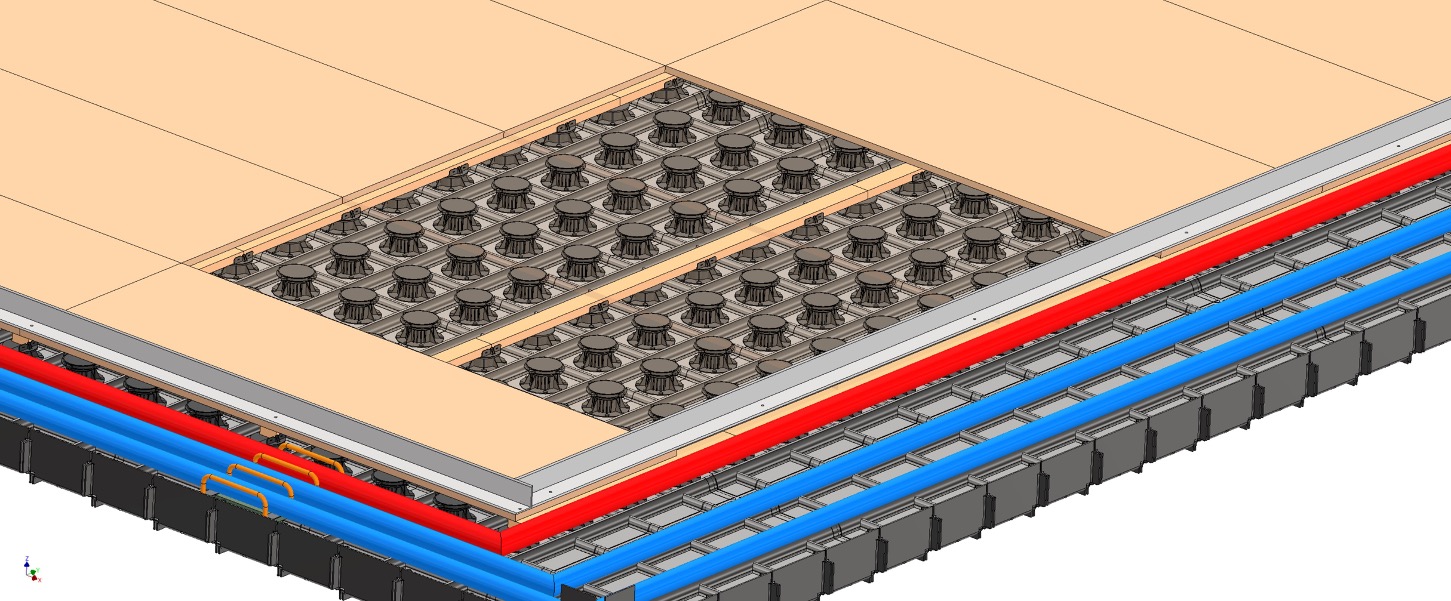}
\par\bigskip
\includegraphics[width=0.6\linewidth]{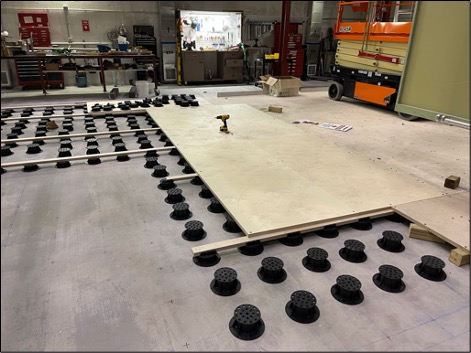}
\end{dunefigure}

Like the \dshort{sphd} detector module, the \dshort{spvd} module must be installed under relatively clean conditions. The radioactive decay of the naturally occurring Ar$^{39}$ isotope in the detector means that under normal operation the \dwords{tpc} will always have a steady background rate. Since the detector will be installed under ISO-8 (\dshort{greyrm}) conditions, the Ar$^{39}$ background will dominate. 
In order to keep the inside of the cryostat clean,  pure \dword{hepa}-filtered air is forced into the cryostat at the east end, which then flows along the cryostat and out the \dshort{tco} at the west end. This keeps the cryostat at an overpressure. The access hatches shown in Figure~\ref{fig:penetration} are circular, 
800\,mm in diameter,  
and are 1.2\,m in length. Using these ports as air ducts, an air flow of roughly 8,000\,CFM (13,600\,m$^3$/hr) through each is possible. 
The pressure in the cryostat will be sufficient ensure outward airflow while work is performed around the flanges on the roof, and covers will be added over the work area to prevent objects from falling in. 
The outer surfaces and cover will be cleaned just before closing.

A \dword{cfd} model of the air flow in the cryostat was performed and the average lifetime of the air in the cryostat computed.
Based on the results, the model was amended to assume eight additional portable air filter units placed in the cryostat at various locations and orientations, and the impact on the air stream age was re-computed.  The best results are shown in Figure~\ref{fig:airAge}. The 16,000 CFM input airflow combined with an additional 16,000 CFM of local filtration resulted in an average air age in the cryostat of 15 minutes, which is anticipated to be sufficient given the low expected occupancy. 
Note that cleanroom design guidelines do not directly apply to the %DUNE 
\dshort{spvd} cryostat since the dimensions and occupancy are very different from standard cleanrooms. (Under typical cleanroom conditions a minimum of five air exchanges per hour are recommended.) 
If in situ measurements indicate additional measure will be required, then workers in the cryostat may be required to wear traditional cleanroom garb to reduce dust production and/or additional portable air filters can be installed. Experience from \dshort{sphd} will also be taken into account. %clarify the number and placement for the \dshort{spvd}.

\begin{dunefigure}
[Air flow in the cryostat]
{fig:airAge}
{\Dword{fea} model of the air flow inside the cryostat with 16,000 CFM of pure filtered air entering through the east access hatches in the cryostat roof. The average duration of a given volume of air in the cryostat is 15 minutes. Plan view with \dshort{tco} at right.}
\includegraphics[width=0.9\linewidth]{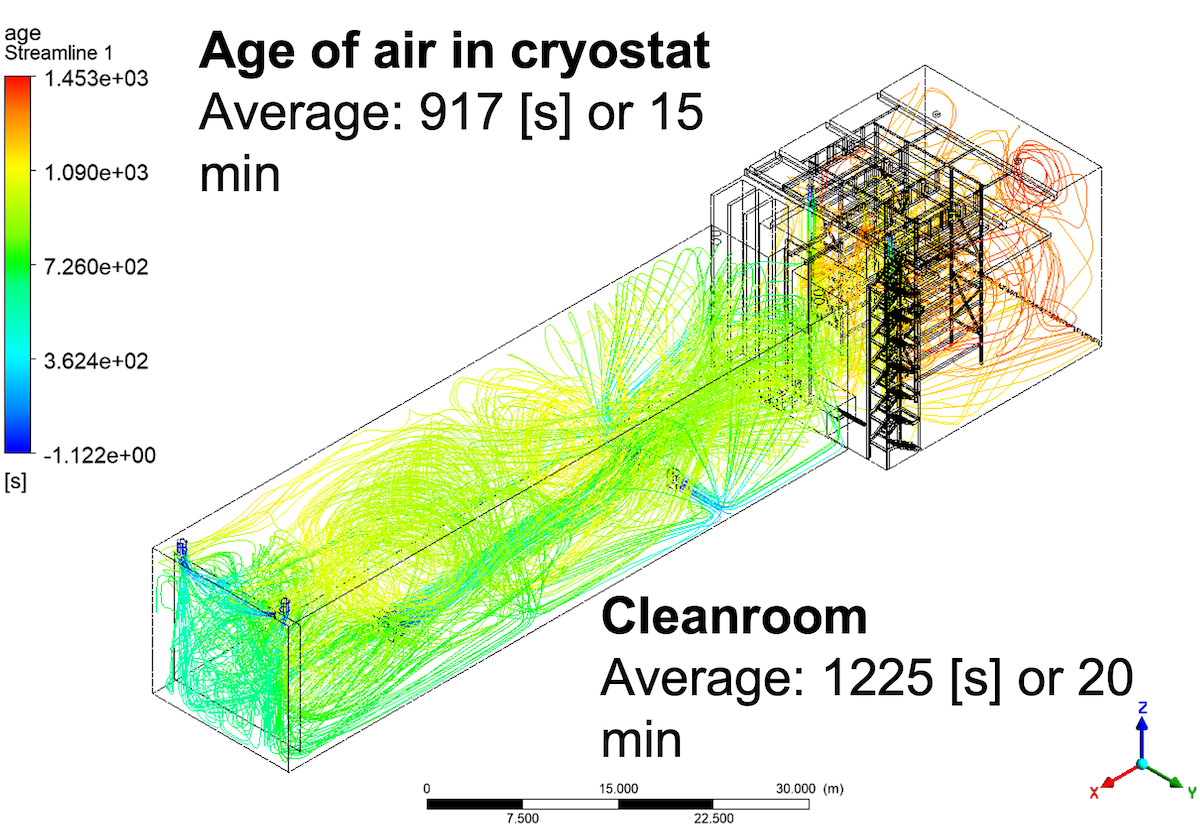}
\end{dunefigure}

The air flow through the \dshort{tco} was also investigated in the CFD model. The air velocity over a large majority of the surface exceeds 
0.2\,m/s, which, according to the ISO-14644-4 international standard, is sufficient to prevent dust from migrating from the \dshort{greyrm} into the cryostat. Figure~\ref{fig:TCO_air_velocity} shows the velocity of the air exiting the cryostat through the \dshort{tco}.

\begin{dunefigure}
[Air flow through the TCO]
{fig:TCO_air_velocity}
{CFD model of the air flowing out through the \dshort{tco} with 16,000 CFM of pure filtered air entering through the West access hatches in the cryostat roof. The velocity is consistently above v$_{air}$\,>\,0.2\,m/s.}
\includegraphics[width=0.7\linewidth]{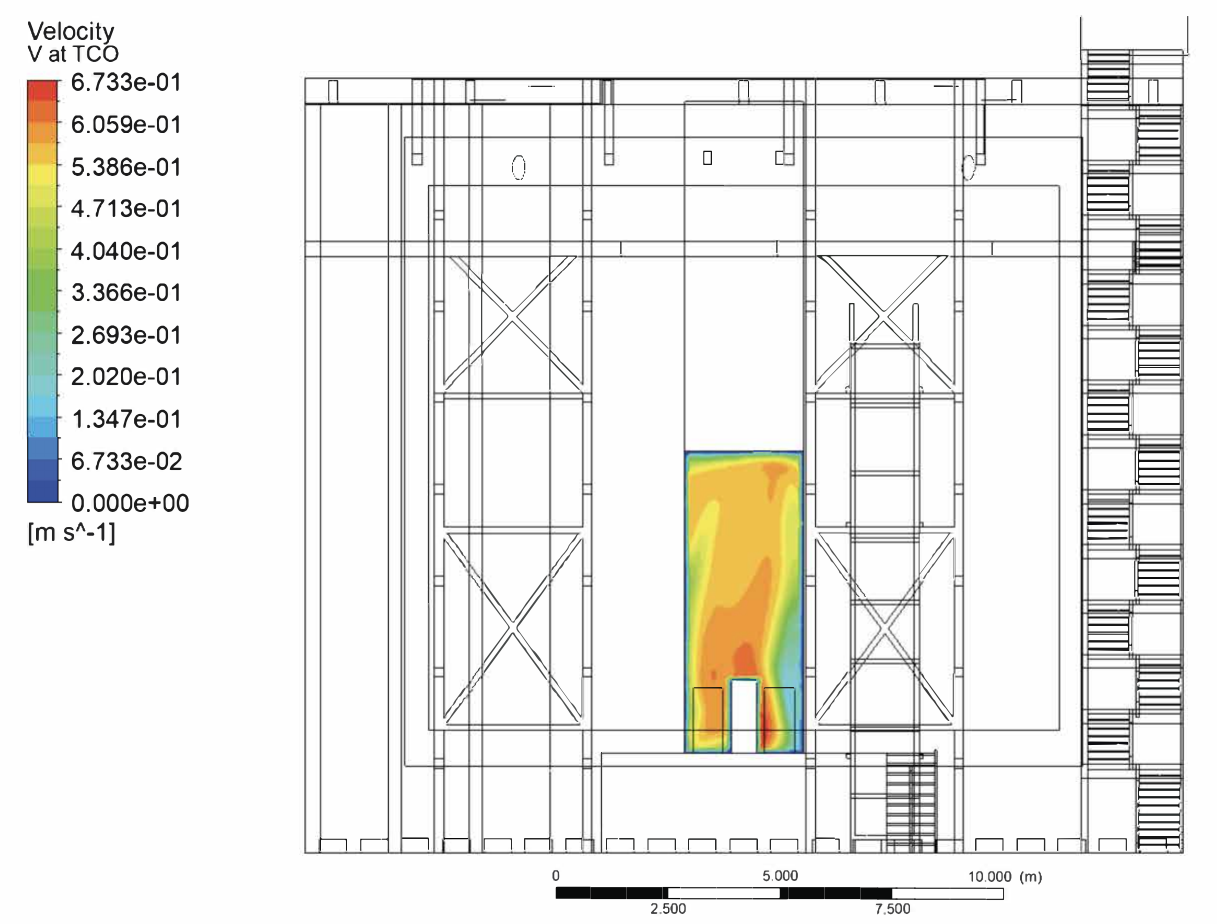}
\end{dunefigure}

Outside the \dshort{greyrm} a changing room is required to give personnel space to put on lab coats and clean shoes before entering. This will be a modular pre-fabricated room that can be erected and dismantled quickly.

The majority of the detector installation work takes place inside either the \dshort{greyrm} or the cryostat. However the installation of the electronics and readout related equipment takes place primarily on top of the cryostat. The 0.5\,m diameter \dshort{tde} chimneys weigh 340\,kg empty and are 2730\,mm long, requiring rigging equipment for transport and installation. Given the large number of  chimneys (105) and crosses (40), all of which are outside crane coverage, a custom gantry cart was designed to facilitate the installation. 

Each chimney will be loaded into the gantry cart in the horizontal orientation using the cavern crane, and moved across the cryostat roof to the correct port. The cart is designed to fit under the cryogenics and detector mezzanines.  It is equipped with a hydraulic cylinder that enables it to rotate the chimney to vertical, then integrated trolleys are used to adjust the position of the chimney over the flange.   The chimney is lowered  through the cryostat penetration using built-in hoists. These steps are shown in Figure~\ref{fig:chimneycart}. 

\begin{dunefigure}
[\dshort{tde} chimney installation gantry-cart]
{fig:chimneycart}
{Images showing a custom gantry cart used to install the \dshort{tde} chimney and the \dshort{bde} crosses.}
\includegraphics[width=0.4\linewidth]{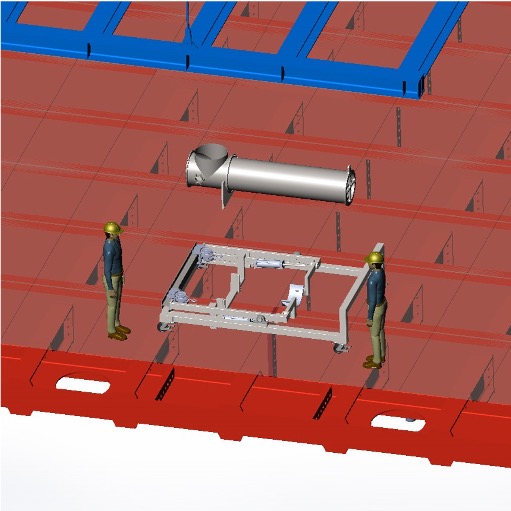}
\includegraphics[width=0.4\linewidth]{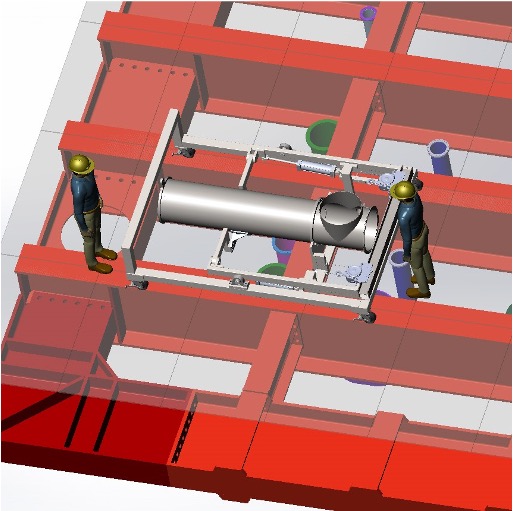}
\includegraphics[width=0.4\linewidth]{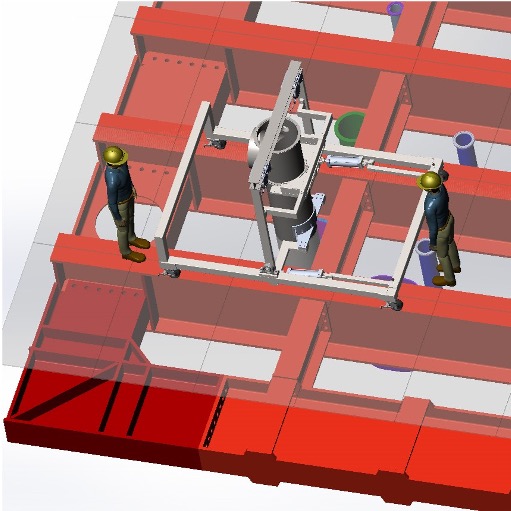}
\includegraphics[width=0.4\linewidth]{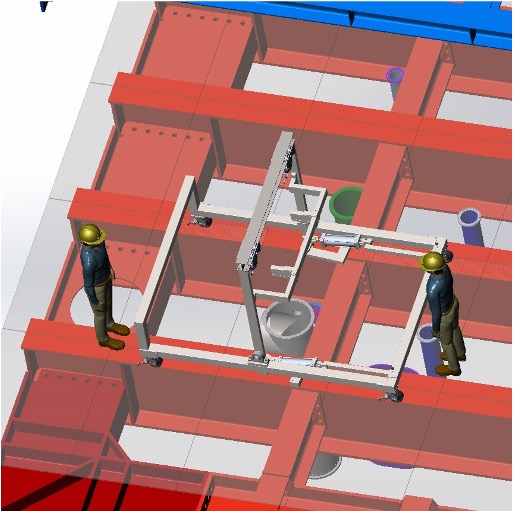}
\end{dunefigure}

Additional installation infrastructure provided by the \dshort{fsii} team includes temporary power and lighting in the cryostat, six scissor lifts capable of reaching a 12\,m work height, and miscellaneous tool and rigging equipment. This equipment is identical or very similar to what is required for the \dshort{sphd} installation. 
\FloatBarrier

\section{Detector Installation}
\label{ch:IEI:det-inst}

The installation of the \dword{spvd} detector is the last major exercise at the far site for this \dword{detmodule}. This makes it imperative that the work proceed as quickly as possible in order to minimize the installation cost and to bring the \dshort{spvd} detector online as fast as possible. The \dword{fsii} installation team  will complete the mechanical portions of the installation, and DUNE consortium members will complete the activities related to connectivity, testing, geometrical positioning, and surveying. It is planned to install the detector module using two 10 hour shifts per day, six days a week.

In planning the installation work it is assumed that roughly half the time is available for productive work and the rest is spent bringing people underground, breaks, entering the \dword{greyrm}, safety meetings, returning to the surface, or general inefficiency of underground work. The 50\% efficiency estimate is typical for underground work. The access to the underground area is limited on Sunday due to shaft maintenance and safety inspections.

\subsection{Installations on Cryostat Roof}

Work on installing the detector infrastructure and the equipment from the consortia begins in parallel with the installation of the cryostat insulation and stainless steel membrane. During this period the installation contractor will be working on the 4910 foot level both inside the cryostat and in the area outside. This leaves the roof of the cryostat free to begin installing the cryogenics and detector-related infrastructure. Installation of the cryostat cold structure (the membrane) will require one year of work, so the infrastructure on the roof can be completed before the cryostat work is complete. Both the cryogenics and detector mezzanines are installed at the beginning of the cryostat work, after which work on the proximity cryogenics system, barracks for the \dword{daq}, the power infrastructure, and all the racks is performed. The \dshort{daq} barracks shown in Figure~\ref{fig:daqbarracks} was designed to host the 16 \dshort{daq} electronics racks. The section view of the room shows one side of the double-row of racks with associated cable trays. The barracks will be outfitted with HVAC compatible with the computer heat load, and a dry agent fire extinguisher due to the high electrical load in the room and the large monetary value of the servers. It is planned to have the infrastructure for the \dshort{daq} available three months before the start of detector installation %so the there will be 
to ensure ample commissioning time (taking into account that there will already be a functioning \dshort{daq} in the north cavern).

\begin{dunefigure}
[DAQ barracks]
{fig:daqbarracks}
{Left: Isometric model of the DAQ barracks with racks installed. Right: Section view of the barracks showing the rack placement}
\includegraphics[width=0.45\linewidth]{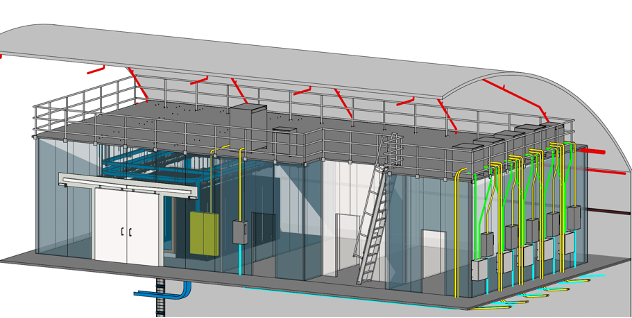}
\hspace{.1in}
\includegraphics[width=0.45\linewidth]{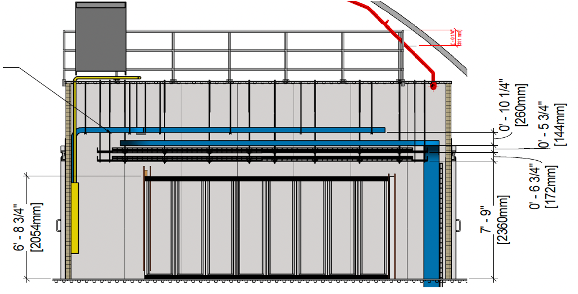}
\end{dunefigure}

On the cryostat roof itself, decking is installed for easy access, the \dword{gar} purge system and all the cable trays are installed, and the ladders for accessing the outside of the cryostat steel are installed.  This completes all the heavy work and associated welding so the installation of the delicate detector elements can begin. The electronics modules that are located in the racks on the detector mezzanine can be installed and connected to the \dword{ddss} and slow-control readout. Installation of cables and fibers on the cryostat roof and mezzanines will progress as the electronics installation proceeds. In the last few months before the cryostat is finished, the top and bottom \dword{tpc} electronics consortia can install the feedthroughs, chimneys, \dshort{bde} crosses, and the cable trays for the \dshort{crp}s, and connect the cables to the \dshort{daq}. 

Once the chimneys and crosses are installed the installation of the \dshort{tde} readout electronics and the \dshort{bde} \dword{wiec} can proceed. The installation of the electronics on the roof can progress in parallel with the detector installation work inside the cryostat.

\subsection{Preparing the Cryostat Interior}

Once the cryostat installation is complete and the cryostat has been leak-tested, the cleaning operation can begin. The hoisting beam in the \dword{tco} will be installed, 8,000 cfm (13,600 cubic m/hr) of purified air will be injected into the cryostat through the access hatches on the east end, the \dshort{greyrm} will be closed and cleaned, and the detector installation can begin inside the cryostat. The first steps are:
\begin{enumerate}
\item Rough clean the cryostat floor and install the false floor. 
\item Set up the temporary power and lighting in the cryostat.
\item Lift the five 12\,m tall scissor lifts  onto a temporary platform at the cryostat entrance using the hall crane and then drive them into the cryostat. 
\item Wipe down the cryostat interior; with five scissor lifts and a crew of 12 people per shift, the cryostat can be wiped down and most of the dust removed in two days.
\end{enumerate}

At this point detector installation can begin.

\subsection{Detector Component Installation}
\label{sec:detcompinst}
Installation begins on the east end wall, and moves west, toward the \dshort{tco}, in the order given here.
\begin{enumerate}[resume]
\item Remove the plastic sheet protecting the east wall and clean the membrane. 
\item  Install the two strings of membrane \dword{pd} modules and route cables to the first of the roof penetrations on the side wall of the detector. See Figure~\ref{fig:eastendwall}.
\item Install the HV extender and test it for continuity to the power supply. (The HV extender must be installed before the top \dshort{crp}s, % installation begins, %using the scissor lifts, 
as they will block access at the roof.) 
\end{enumerate}

\begin{dunefigure}
[East end wall PD and HV extender installation]
{fig:eastendwall}
{Model illustrating the installation of the PD modules on the east end wall and the HV extender.}
\includegraphics[width=0.5\linewidth]{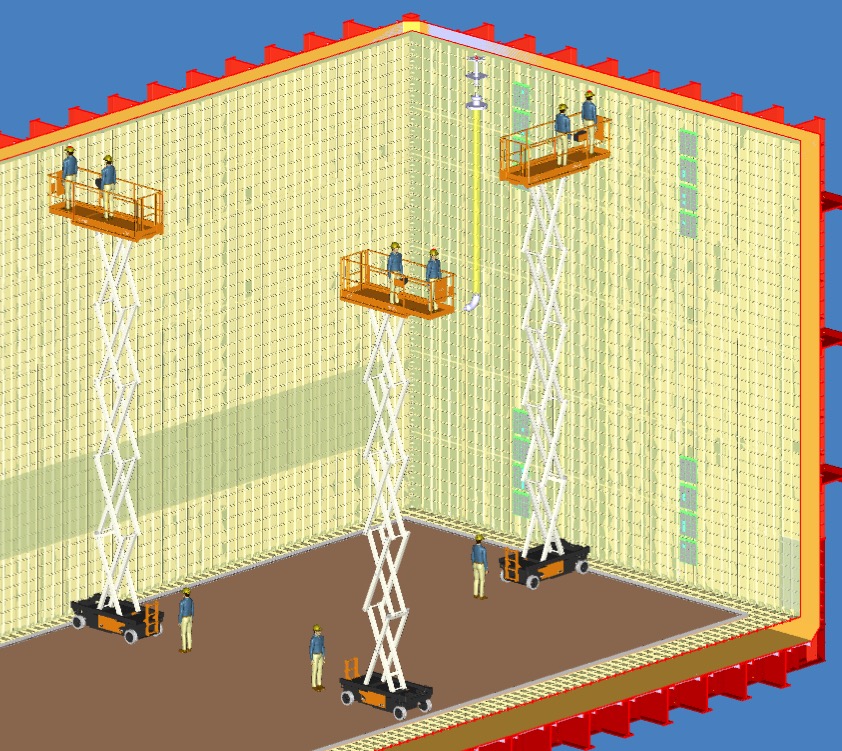}
\end{dunefigure}

 Next, installation of the cables for the \dword{bde}, fibers for the cathode \dshort{pd} modules, and the membrane \dshort{pd} modules begins along the north and south walls of the cryostat. The \dshort{bde} cables arrive underground on a large spool with the cable trays pre-attached. These cable trays are of a unique design based on a rope-ladder concept. 
 Access to the roof penetrations is still required. A manual chain hoist is brought into the cryostat for this installation work.  

Two crews, one on the north wall and one on the south wall, will work in parallel starting in the east corner and working west, to accomplish the following.
\begin{enumerate}[resume]
\item Remove the sheet protecting the cryostat membrane and clean the stainless steel surface including the roof.
\item Mount a hoisting fixture to the M10 corner bolts of the cryostat and hoist the cables up the wall.
\item Once in position, install a tensioning spring at the bottom and remove the hoist.
\item Route the cables through the roof penetrations and install the cathode-mount \dshort{pd} fibers down to floor level. 
\item In parallel, install the strings of membrane-mount \dwords{pd}. 
\item In the $\pm$2\,m region above and below the cathode, install a protective cover over the cable trays to prevent the cathode field from interacting with the thin conductors in the \dshort{bde} cables.
\end{enumerate}

Figure~\ref{fig:wallpdinstall} illustrates the \dshort{bde} cable and membrane \dshort{pd} installation.  The image on the left %in Figure \ref{fig:wallpdinstall} 
shows the relative orientation of the \dshort{pd} modules and the \dshort{bde} cables. Here the \dshort{pd} cables are not shown, but they will run from the \dshort{pd} modules to the same feedthrough. 

\begin{enumerate}[resume]
\item After the cables are in place, install temporary \dshort{bde} electronics  and test the entire readout chain through to the \dshort{daq}. Channel mapping, all electrical connections, and the warm readout are tested together.
\item After verification of the full functionality of the system, disconnect the \dshort{bde} cables and store them below the temporary floor.
\end{enumerate}

  The process of installing the false floor, cleaning  the walls, installing the cable trays, and installing the cables could be accomplished in eight weeks, but the work is spread over 11 weeks for resource-leveling reasons (once the \dshort{crp} installation begins, the crew gets split in order to accomplish the work in parallel.) %part of the installation is a single shift). 

\begin{dunefigure}
[\dshort{bde} cable and wall-mounted PD installation]
{fig:wallpdinstall}
{Left: Models of the installation of the wall PD modules and the \dshort{bde} cables along the cryostat long walls. Right: Closeup showing the relation of the \dshort{bde} cables and cable trays, the wall PD modules and the cryostat roof feedthrough.}
\includegraphics[width=0.45\linewidth]{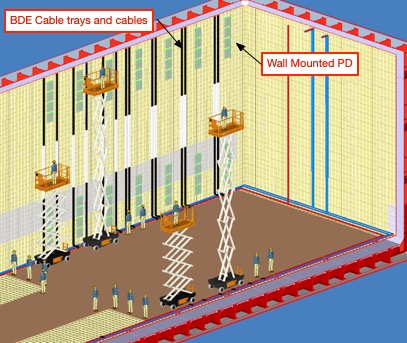}
\hspace{.1in}
\includegraphics[width=0.45\linewidth]{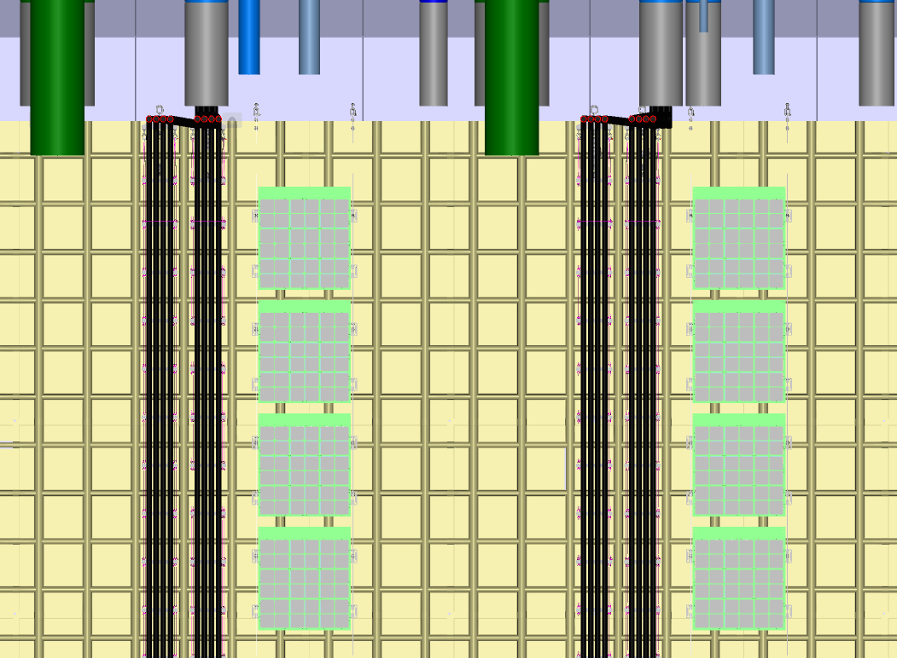}
\end{dunefigure}

Following the work on a portion of the north and south walls, but (for access considerations) before the \dshort{crp}s are in place, the circular cable trays for the top \dshort{crp}s for that area need to be installed. 
 The large \dword{tde} chimneys service 50 flat ribbon readout cables, making the cable density in this area high. As the pitch of the major I-beams on the cryostat roof do not match the pitch of the internal detectors, the locations of the chimneys relative to the \dshort{crp}s are not consistent. %different for each chimney. 
A circular cable tray, Figure~\ref{fig:circulartray}, was designed that mounts on the chimney to hold the cables and a jig was designed to route each cable %can be routed 
to the appropriate location. % for a given chimney. 
\begin{enumerate}[resume]
\item The cables will be attached first on the ground using a custom cabling jig.
\item The assembly is lifted into location and attached to the chimneys.
\item Once attached, the cables are connected to the chimneys.
\item The cables connecting the \dshort{crp}s to the readout electronics are installed. % first.
\end{enumerate}

Since the roof cabling work follows the wall \dshort{pd} installation, the cables will be in place before the top \dshort{crp}s are installed.

\begin{dunefigure}
[Top \dshort{crp} circular cable trays]
{fig:circulartray}
{Top: Model of %Images showing 
the installation of the circular cable trays for the top \dshort{crp}. Bottom: Installation jig used to pre-route the cables and to install the trays.}
\includegraphics[width=0.6\linewidth]{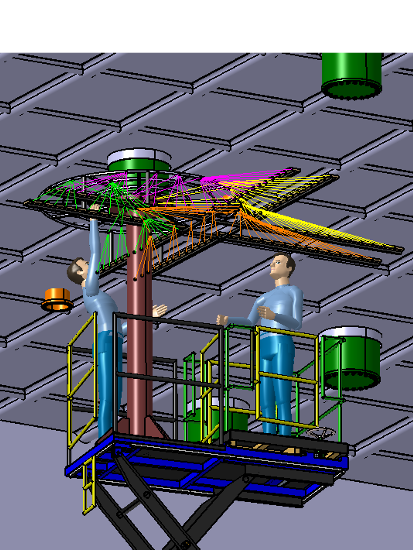}
\hspace{.1in}
\includegraphics[width=0.7\linewidth]{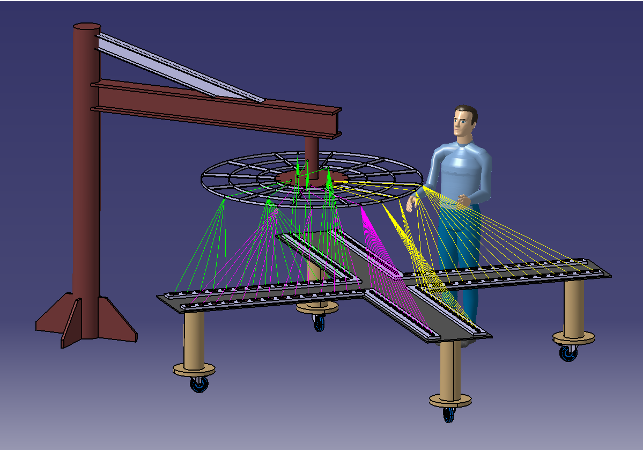}
\end{dunefigure}
 
After four weeks, enough of the wall-mount \dwords{pd} will be installed %so that there is 
to leave space %in the cryostat 
to begin installation of the top \dshort{crp}s. 
This procedure will differ in many respects from that used in \dword{pddp}, %and will be the same 
and will be prototyped in the \dword{vdmod0} test at CERN.  The \dword{spvd} top \dshort{crp}s are connected to an intermediary steel support frame that supports either two or six \dshort{crp} modules (see Section~\ref{sec:CRPMss}).
 This reduces the number of penetrations through the roof from three per \dshort{crp} for \dshort{pddp} to four per support structure, greatly simplifying the cryostat design. However, to connect the cables, the area above the \dshort{crp}s must remain accessible. A custom %access platform 
elevated workstation has therefore been designed that is supported through the cryostat roof and accessed from below via scissor lifts.  %and plays a key role in the top \dshort{crp} installation. 

Figure~\ref{fig:cleanspace} illustrates the layout of the \dshort{greyrm} in this phase of the installation, where the cathode and \dshort{pd} module components are brought for assembly and integration together. These activities occur in this order.
\begin{enumerate}[resume]
\item Bring the pieces of the \dshort{crp} support structures and the \dwords{cru}, which will be assembled into \dshort{crp} modules, directly into the cryostat and clean the components (Figure~\ref{fig:week8}. 
\item  Assemble the \dshort{crp} support structures and the \dshort{crp}s at their respective work stations.
\item Affix two or six \dshort{crp}s to a support structure using a series of linkages that must be adjusted to level the \dshort{crp}s and position them relative the to support structure. The \dshort{crp} support structures are equipped with temporary wheeled supports so they can be easily moved around inside the cryostat. (The 16 \dshort{crp} superstructures will be assembled and tested for dimensional tolerances before the start of installation.)
\item Hoist the %\dshort{crp} access platform 
elevated workstation and attach it to the cryostat roof using an unused upstream \dshort{crp} support feedthrough.
\item Wheel the set of two or six \dshort{crp}s into location and %installed. 
hoist it. 
\item Suspend the cathode plane from %this same 
the \dshort{crp} superstructure using several 3\,mm Dyneema cables.
\end{enumerate}

\begin{dunefigure}
[Cryostat layout during top \dshort{crp} installation]
{fig:week8}
{Layout of the equipment in the cryostat during top \dshort{crp} installation. The main work areas are labeled.
}
\includegraphics[width=0.9\linewidth]{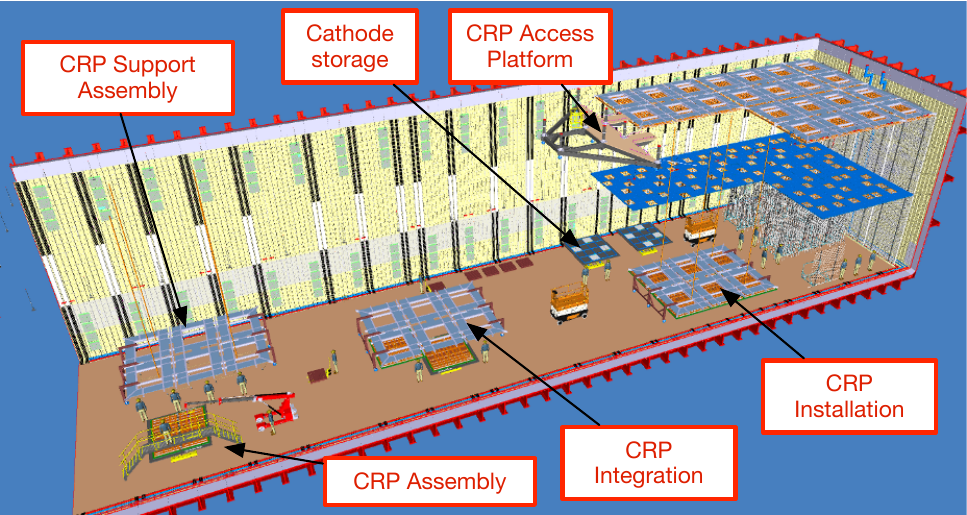}
\end{dunefigure}

In order to raise the superstructure, synchronized winches will be installed in place of the automated position system on top of the cryostat, as shown in Figure~\ref{fig:circulartray}. %fig:spft}. 
The external support structure is designed to allow switching between the manual winches and the automated suspension system while under load. 

Figure \ref{fig:Tcrp_install} illustrates the top \dshort{crp} and cathode installation process, which proceeds as follows. 
\begin{enumerate}[resume]
\item Install the hoisting system on the roof and lower the lifting cables through the \dshort{crp} support penetrations. 
\item When the cables are at roughly the correct height, move the assembly with the support structure and \dshort{crp} modules into position and connect it to the lifting cables. 
\item Remove the load from the support wheels, then remove the wheels.
\item Attach the cathode support cables and set them to the correct length.
\item Raise the \dshort{crp}s %are then raised until 
to where the cathode support cables are at the height of a cathode module on its installation table. 
\item Connect the support cables and set the gap to the \dshort{crp} with a laser range finder.
\item Connect the cathode PD fibers to the PD modules.
\item Make the interconnections between cathodes and perform continuity tests. % performed. 
\item Raise the \dshort{crp}s and cathode together to cabling height and attach the %access platform
elevated workstation. This workstation provides lateral support so the \dshort{crp} structure does not sway when people are working on top of it. 
\end{enumerate}

A 14 week schedule for the combined installation of the \dshort{crp}s and cathode modules has been developed.

 %$$$$$$$$$$$$$$$  
\begin{dunefigure}
[Top \dshort{crp} and cathode installation sequence]
{fig:Tcrp_install}
{The installation sequence for the top \dshort{crp} and the cathode modules is shown for a six-\dshort{crp} superstructure unit.  Top left: The \dshort{crp} support superstructure is moved into position and attached to the lifting cables.  Top right: The cathode support cables are attached and the \dshort{crp}s are lifted 6\,m.  Bottom Left: The cathode modules (blue) are installed under the \dshort{crp}s, and the cathode is suspended. Bottom Right: The entire assembly is lifted to cabling height where it is attached to the %access platform
elevated workstation.}
\includegraphics[width=0.48\linewidth]{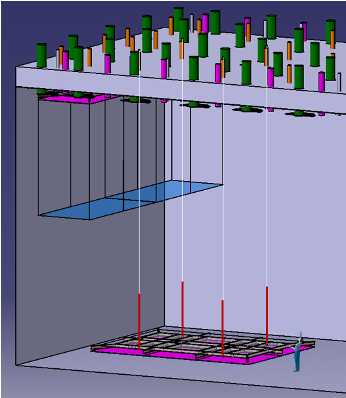}
\includegraphics[width=0.48\linewidth]{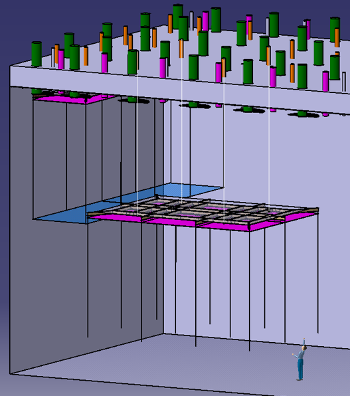}
\includegraphics[width=0.48\linewidth]{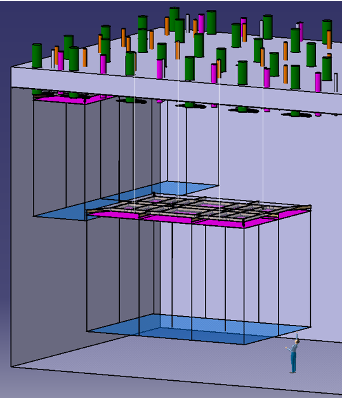}
\includegraphics[width=0.48\linewidth]{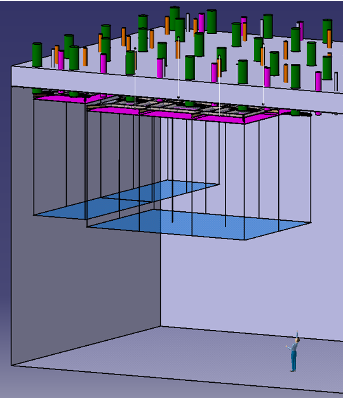}
\end{dunefigure}
%$$$$$$$$$$$$$$$ 

%$$$$$$$$$$$$$$$  
\begin{dunefigure}
[Cabling access top \dshort{crp}]
{fig:Tcrp_cabling}
{The cabling process for the top \dshort{crp}. Top: The \dshort{crp} support superstructure connected to the %access platform
elevated workstation. Middle: Crawl space for connecting the cables to the \dshort{crp}. Bottom: %Image showing 
The space near the roof of the cryostat after the \dshort{crp}s are raised to their final position.}

\includegraphics[width=0.7\linewidth]{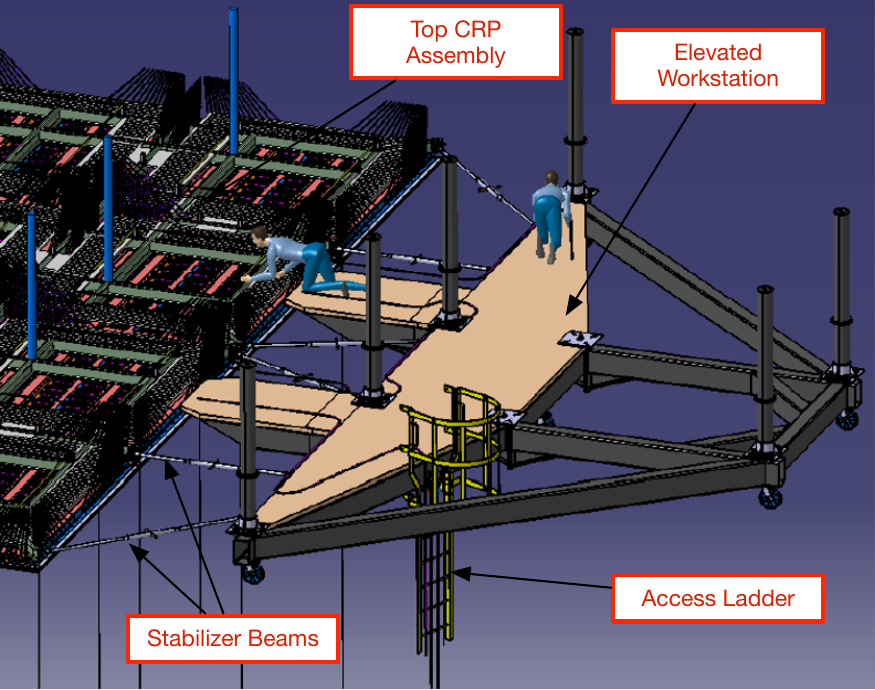}
\includegraphics[width=0.7\linewidth]{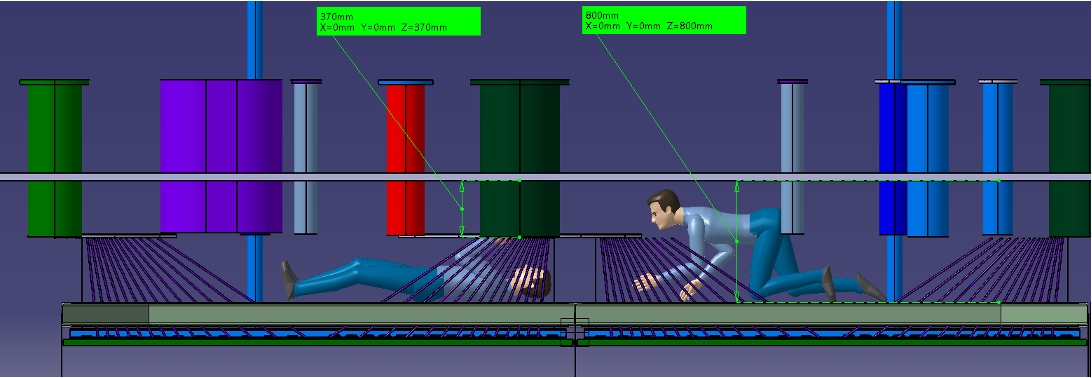}
\includegraphics[width=0.7\linewidth]{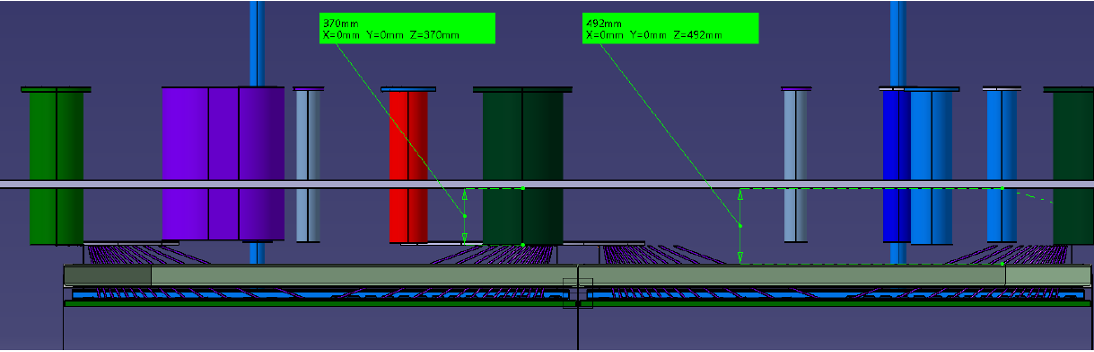}
\end{dunefigure}
%$$$$$$$$$$$$$$$ 

The top surface of the \dshort{crp} support structure is accessed by raising a scissor lift up to the access ladder attached to the elevated workstation (Figure~\ref{fig:Tcrp_cabling}). A system of fall restraint supports are integrated into the workstation and the \dshort{crp} support structures;  qualified personnel %can 
attach their fall restraint supports at the access ladder and climb %up on 
onto the workstation. 

The rails and attachment points are designed so a person can crawl out on the workstation and then onto the \dshort{crp} support structure. Figure~\ref{fig:Tcrp_cabling} shows the space between the cryostat roof and the work surface; it clearly illustrates the need to have the cables pre-positioned near the final connection point. Engineering related to the addition of a perimeter barrier along the outside edges of the \dshort{crp} supports and access platforms is actively being investigated. Upon conclusion, additional safety barriers may be added to the design.

\begin{enumerate}[resume]
\item Manually connect the cables to the \dshort{crp}s. 
\item Once all the connections are made, evacuate the area and test the readout.
\item Installation of the adjacent \dshort{crp} support structure can begin in parallel with the testing of the first set of \dshort{crp}s.
\item  When both sets of \dshort{crp}s have been installed and tested, disconnect the elevated workstation, lower it to the floor, and re-install it for the next pair of support structures.
\item Raise the \dshort{crp}s to the final height in the cryostat and monitor them regularly to ensure all the readout channels are functional.
\end{enumerate}

The installation of the \dshort{fc} can begin once the second pair of \dshort{crp} super structures is installed. %This schedule offset is sufficient to give enough 
This provides sufficient time and space under the installed \dshort{crp}s for the \dshort{fc} installation team to work, and corresponds to four weeks after the start of \dshort{crp} installation. Recall from Section~\ref{subsubsec:FCsss} that two columns of four \dshort{fc} modules each (eight total) form a supermodule, 6.0\,m(W) $\times$ 13\,m(H) for the long walls and 6.76\,m(W) $\times$ 13\,m(H) for the end walls.  
Each \dshort{fc} supermodule hangs from a 6.4\,m long \dshort{fc} support beam assembly %which is 
constructed of stainless steel I-beam and consists of a main spanning beam with two lift rods and two \dshort{fc} spanning yokes, as shown in Figure~\ref{fig:fc_beam}. The support beams are brought into the \dshort{greyrm} and cleaned and then transported into the cryostat.
The support beams are hoisted from the floor in the cryostat first to their  assembly heights and then to their final heights using winches and cables that raise %lift 
the \dshort{fc} lift rods. The flange and hoisting setup is designed so that while the FC support beam is lifted it is offset by 50\,mm from its final position. Since it will be lifted with hand winches that could be at slightly different levels,  the offset is critical, as it allows the modules supported from the beam to swing in-plane somewhat. 
When assembly in the cryostat is complete, the beams are shifted to the final position and placed on a spherical indent on the roof flange. Details of the roof alignment and support flange are shown in Figure~\ref{fig:fc_beam}. 

 %$$$$$$$$$$$$$$$  
\begin{dunefigure}
[Field cage support beams and roof positioning flange]
{fig:fc_beam}
{Left: \dshort{fc} support beam showing the major sub-assemblies. Right: \dshort{fc} roof penetration showing the $x$-$y$ position adjustment system, the penetration for lifting the support beam assembly, and the spherical socket for final positioning.}
\includegraphics[width=0.55\linewidth]{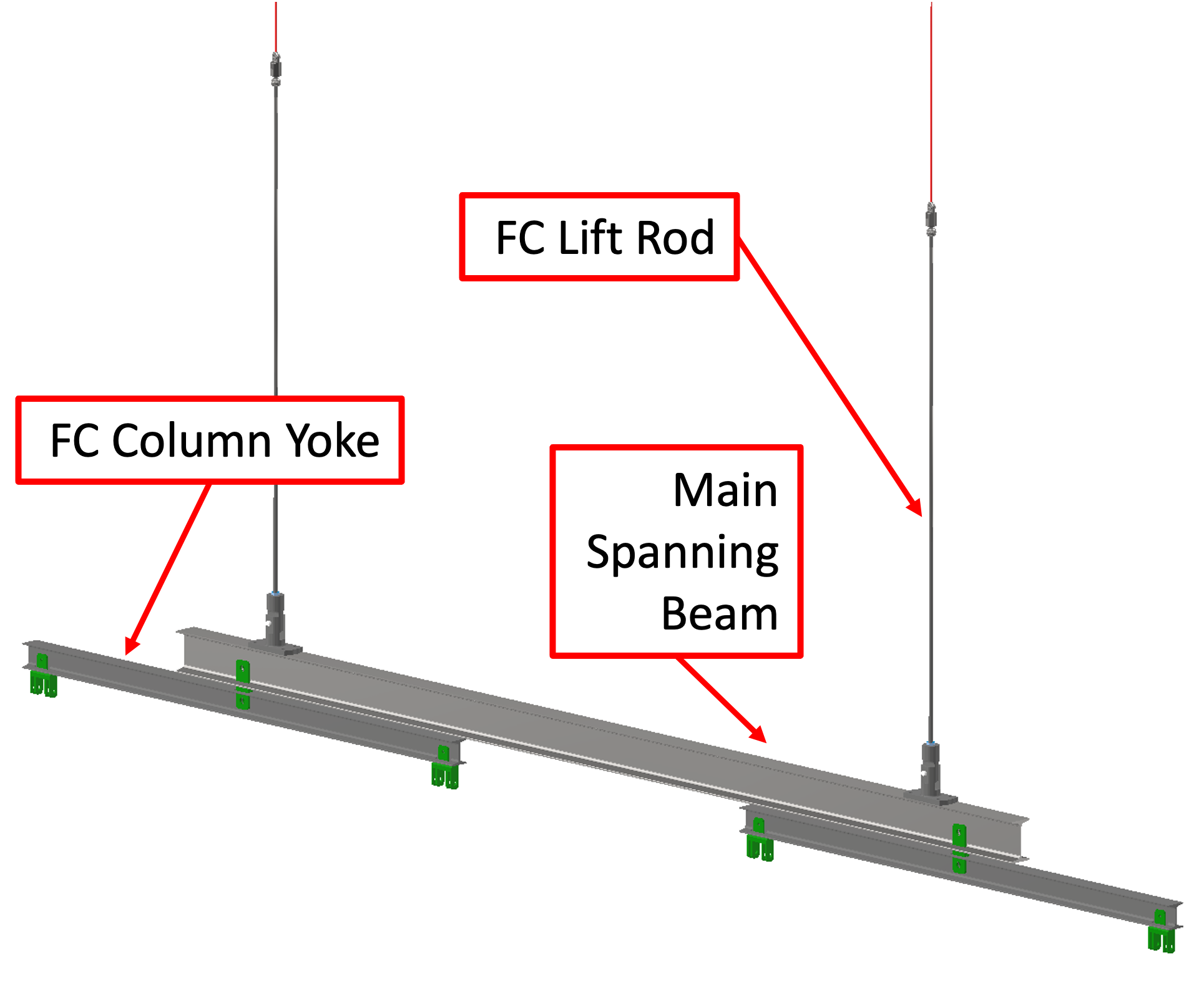}
\includegraphics[width=0.43\linewidth]{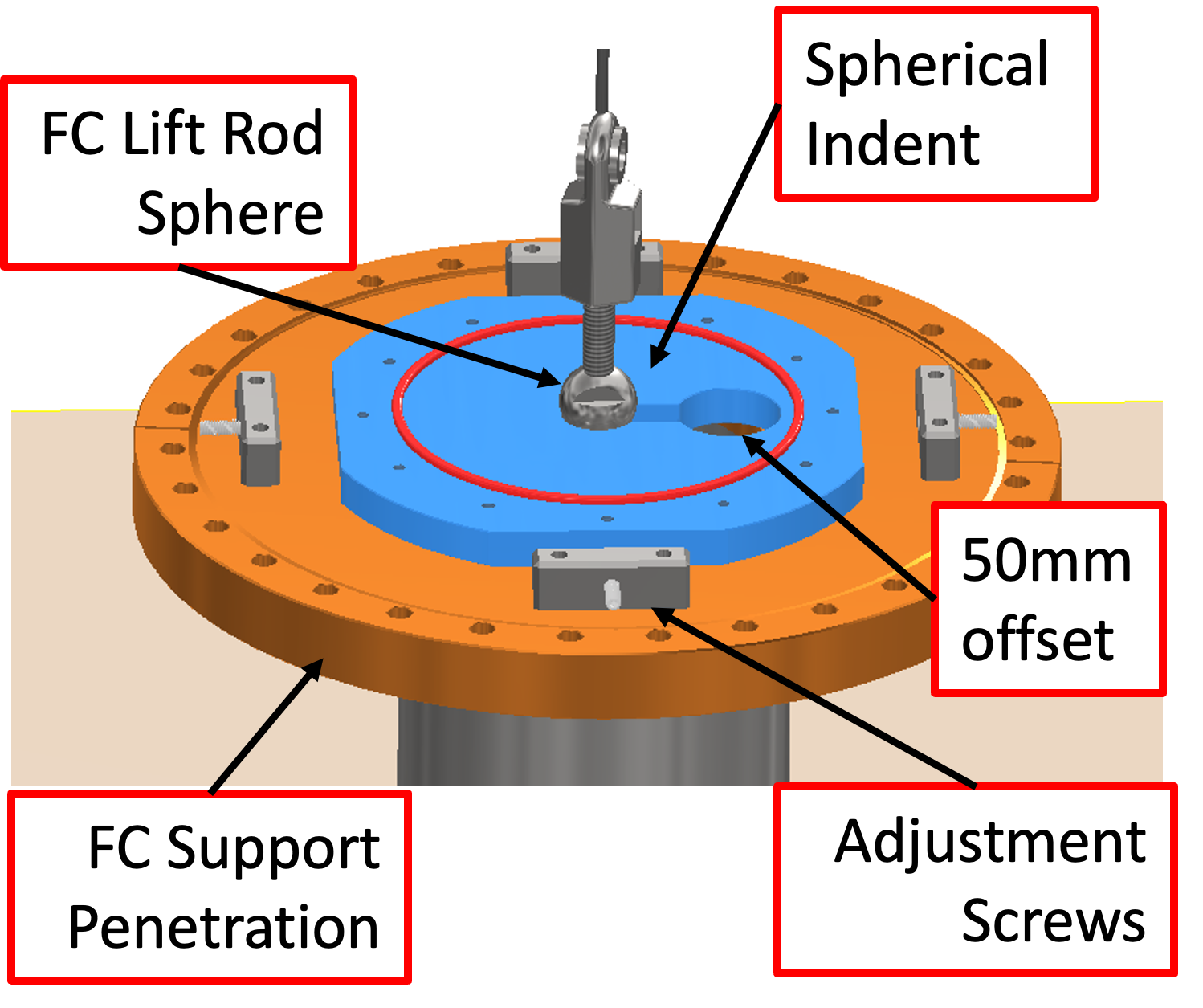}
\end{dunefigure}
%$$$$$$$$$$$$$$$

 The four walls surrounding the two active drift volumes are formed by ten \dshort{fc} supermodules each along the long walls, and two each along the end walls for a total of 24. % super modules. 
 The single \dshort{fc} modules will be assembled in the \dshort{greyrm} starting at the same time as \dshort{crp} installation. All the parts for the field cages and assembly tooling are cleaned as they are brought into the \dshort{greyrm}. As the modules are finished they will be transferred to custom storage carts (Figure~\ref{fig:fc_carts}) and hoisted into the cryostat. Once inside, the \dshort{fc} is constructed one supermodule at a time.  The \dshort{fc} installation process is substantially faster than that for the \dshort{crp}, so the \dshort{fc} team will be primarily fabricating modules and then installing them as space behind the \dshort{crp}s becomes available. 
%  \fixme{the false floor design text is now under infrastructure}
  In order to install the modules, a cantilevered cart was designed to hold a single \dshort{fc} module (Figure~\ref{fig:fc_carts}). During installation the single modules will either be lifted from the storage cart by hand and placed on the installation cart or a gantry crane will be used.   The \dshort{fc} installation sequence for the \dshort{fc} modules is shown in Figure~\ref{fig:fc_install} and enumerated below.

 %$$$$$$$$$$$$$$$  
\begin{dunefigure}
[Field cage carts]
{fig:fc_carts}
{Left: \dshort{fc} storage cart. Right: \dshort{fc} installation cart.}
\includegraphics[width=0.4\linewidth]{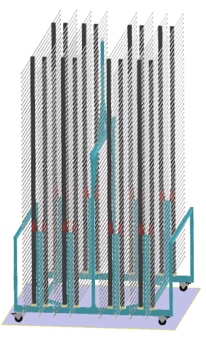}
\includegraphics[width=0.4\linewidth]{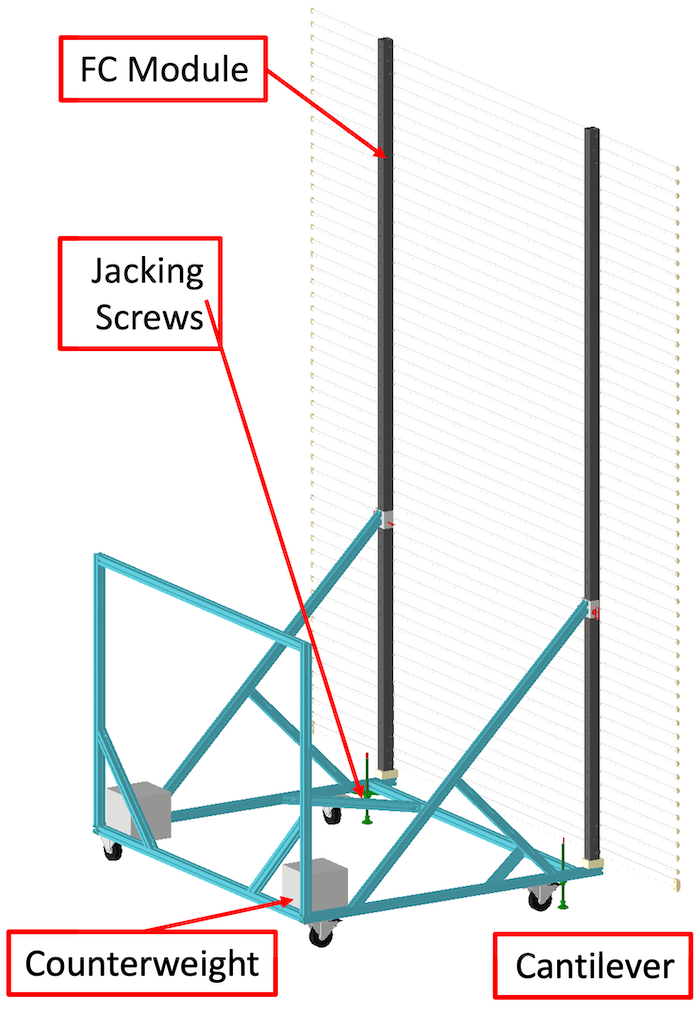}
\end{dunefigure}
%$$$$$$$$$$$$$$$

  %$$$$$$$$$$$$$$$  
\begin{dunefigure}
[Field cage installation sequence]
{fig:fc_install}
{\dshort{fc} installation sequence shown for an endwall panel. The top \dshort{crp} superstructure and cathode are in position. Top Left: The support I-Beam is installed with support rods and %FC panel 
\dshort{fc} module supports. The vertical lifting cables are not shown. Top Right: the first two %FC panels
\dshort{fc} modules (blue, at left) are installed. Bottom Left: %the condition after 
the second set of modules is installed. Bottom Right: %shows 
the full \dshort{fc}  supermodule is installed. }
\includegraphics[width=0.48\linewidth]{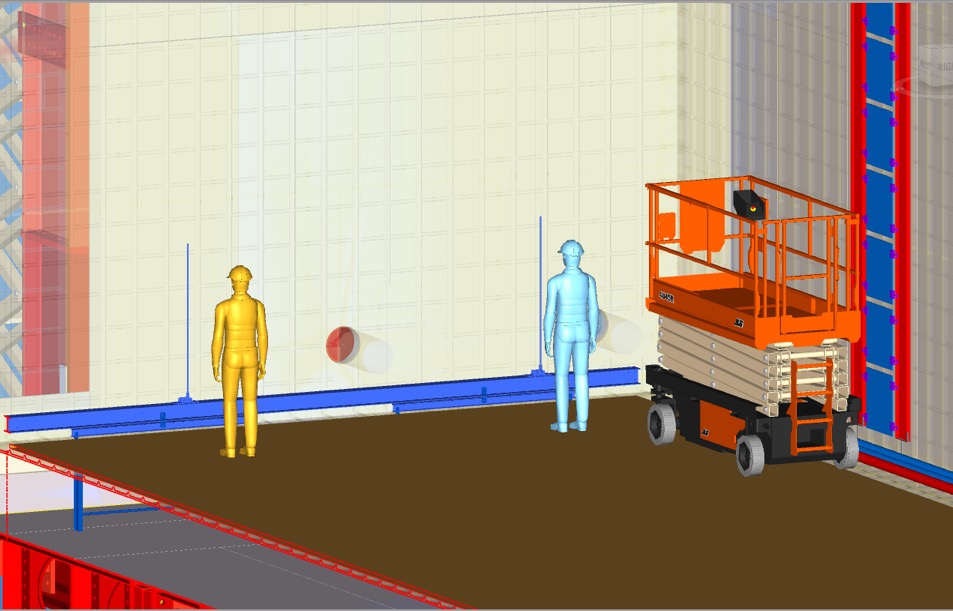} \hfill
\includegraphics[width=0.48\linewidth, trim={0 0 0 3.5cm},clip]{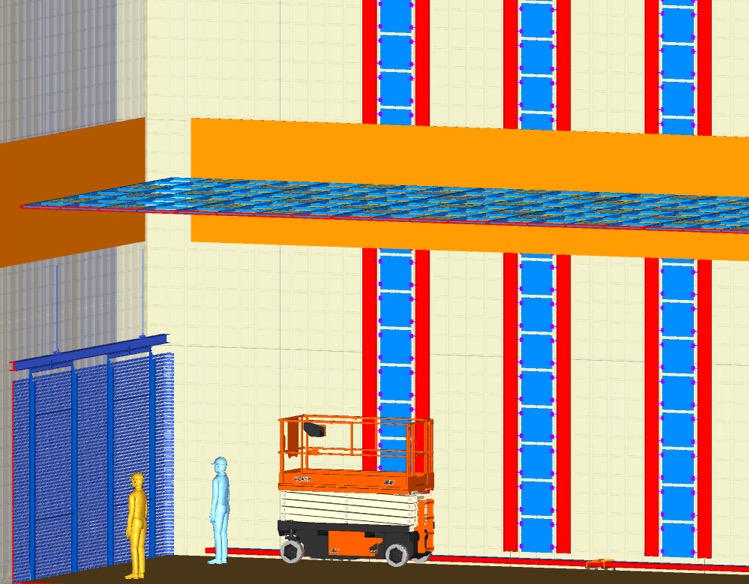}
\includegraphics[width=0.48\linewidth]{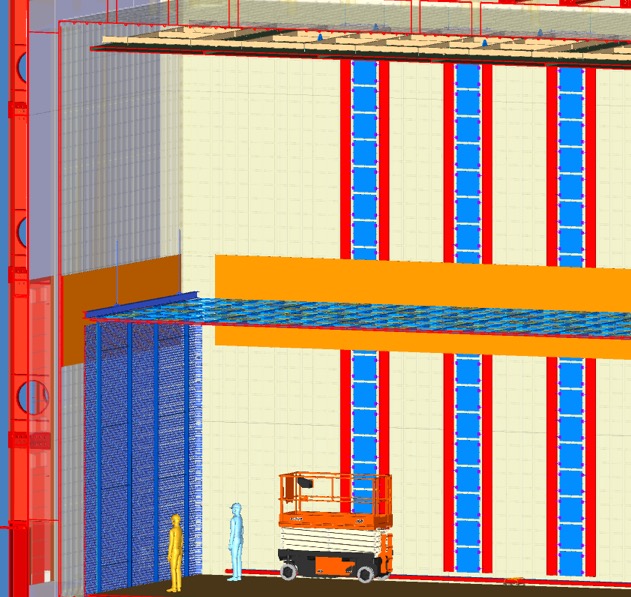} \hfill
\includegraphics[width=0.48\linewidth, trim={0 2.5cm 0 2cm},clip]{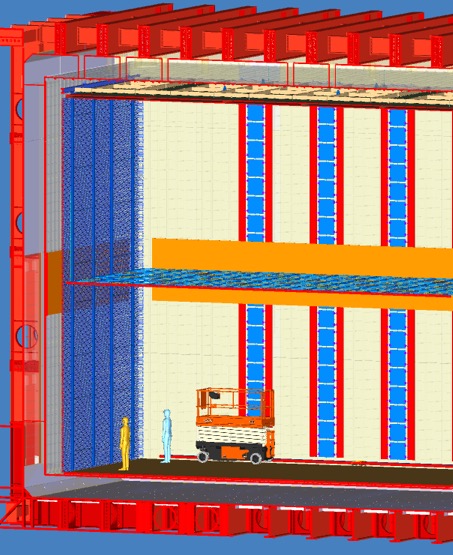}
\end{dunefigure}
%$$$$$$$$$$$$$$$ 
 
\begin{enumerate}[resume]
\item The installation of each \dshort{fc} supermodule  starts with moving one 6\,m stainless steel \dshort{fc} support I-beam into position.
\item Attach the stainless steel support rods and lower a pair of cables %and 
 from the cryostat roof  through  the \dshort{fc} roof penetrations. 
\item Three options exist for hoisting the modules; the use of (a) hoists on tripods just above the feedthroughs, %can be used, the gantry cart for 
  (b) the chimney installation gantry cart, % can be used, 
 or (c) a temporary I-beam, which can be mounted under the mezzanines with hoist attached. The simplest option %will be selected but this 
 may depend on %any 
 mechanical interferences in the local area.  The FC support I-beam is raised by about 3.5\,m, high enough to position two \dshort{fc} modules underneath it, side by side.
\item Once the two top modules are secured to the I-Beam, raise the partial supermodule again  by the same amount to mount the next set (row) of two \dshort{fc} modules under the top row, and to connect them, both mechanically and electrically. (These first two rows of the supermodule will be above the cathode).
\item Repeat this process for the third row of \dshort{fc} modules; the profiles at the boundary of the second and third rows are at the cathode level and at a later stage will be connected to the cathode on its bottom-facing side.
\item Once the third row is constructed, raise the nearly complete supermodule again, this time to its final position, and connect the fourth row of \dshort{fc} modules similarly to the others.
\item Transfer the load directly to the feedthrough flanges and remove the winch cables. 
\item On the cryostat roof place a vacuum-tight cover over the roof penetration flange, make the electrical connections through a side flange on the roof penetration, and test the electrical connections. At the bottom of the cryostat, install a support brace to stabilize the superstructure. The final roof penetration assembly and floor brace is shown in Figure~\ref{fig:fc_roof-assem}.
\item After the floor brace is secured, the \dword{pd} fibers that are hanging from the cathode are attached to the vertical \dshort{fc} \dword{frp} support beams. Any excess slack is stored under the cryogenic piping at the edge of the cryostat.
\item Test the readout and cathode \dshort{pd} modules in location.
\end{enumerate}

 %$$$$$$$$$$$$$$$  
\begin{dunefigure}
[Field cage roof penetration and floor support]
{fig:fc_roof-assem}
{Left: Fully assembled \dshort{fc} roof penetration. Right: Support brace to the cryostat floor.}
\includegraphics[width=0.4\linewidth]{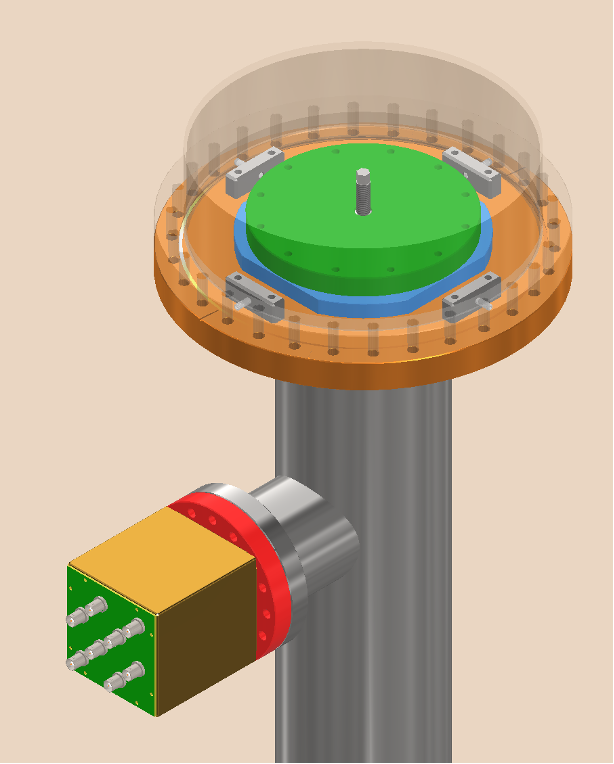}
\includegraphics[width=0.5\linewidth]{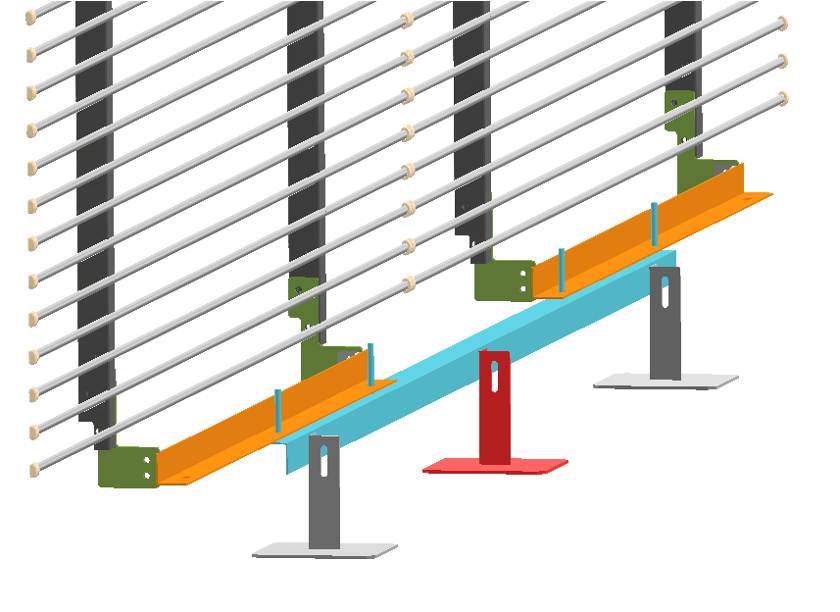}
\end{dunefigure}
%$$$$$$$$$$$$$$$ 

Once the next-to-final row of top \dshort{crp}s and north-south wall \dwords{fc} are installed, it will be necessary to remove the hoisting beam through the \dshort{tco} to make room to install the last \dshort{crp} row and the top $2/3$ of the west \dshort{fc} endwall. % can be installed.
Afterwards, \dshort{tco} hoisting beam is re-installed in a %lower 
position below the cathode plane in preparation for the bottom \dshort{crp} installation, the last major detector elements to be installed. 

\FloatBarrier

%%%% Bottom CRP installation
Although the bottom \dshort{crp}s
are mechanically very similar to the top ones, they are installed individually (no superstructure), making the installation process completely different. 
Tooling to rotate and lift the \dshort{crp}s will %therefore 
need to be designed and/or purchased. Each \dshort{crp} weighs roughly 200\,kg, making it too heavy to lift by hand, but this is well within the range of a light counterbalanced crane, many of which are on the market.

Like the top \dshort{crp}s, the bottom \dshort{crp}s are delivered to South Dakota as half-size \dwords{cru} in transport crates that can be rotated on-edge to fit down the Ross shaft. Unlike 
the top ones, however, the bottom \dshort{crp} are delivered with the electronics pre-installed and the internal cabling complete. Both top and bottom \dshort{crp} are bagged inside the shipping crates to prevent any dust accumulation. A dedicated area in the \dshort{greyrm} will be set up with \dword{esd} mats for acceptance testing. %In the \dshort{greyrm} 
The same lifting tooling in the \dshort{greyrm} can be used %as was used 
for both top and bottom \dshort{crp}s. 
The process is as follows.

\begin{enumerate}[resume]
\item Carry the shipping boxes into the \dshort{greyrm} with the anodes facing down and the composite frame up.
\item Execute the tests.
\item Install the final support feet.
\item Move the \dword{cru} to the \dshort{tco} using a simple cart, and hoist it into the cryostat.
\item Rotate the \dshort{cru}s to have the anodes face up.
\item Lower them onto an assembly table %(Figure~\ref{fig:crp-assembly}) 
and move them to an assembly area. Since one \dshort{bde} will have one leg and the adjacent one two, the former will require shimming on the table.
\item Assemble the \dshort{crp} in the cryostat by connecting the two sections of the composite frame and routing the downstream cables to the final patch panel. 
\end{enumerate}

 The installation of the bottom \dshort{crp}s will begin on the east end of the cryostat and single rows of bottom \dshort{crp}s will be installed at a time. %At the start of the installation of a
 
\begin{enumerate}[resume]
\item Before installing each new row of bottom \dshort{crp}s, remove the false floor and clean the region carefully. A protective bag over the cables to prevent dust contacting the cables is also removed.
\item Prepare the \dword{bde} cables that were stored under the floor for connection to the \dshort{crp}. 
\item Install a truss structure (Figure~\ref{fig:b-crp-tooling}) that pre-loads the membrane floor.
\item Lift each \dshort{crp} off the assembly table using a commercial counter-balanced crane with a custom 2.5\,m long fork.
\item Place the \dshort{crp} on the installation truss,  level it and set it to the correct height. 
\item Connect the \dshort{bde} cables to the pre-installed patch panels and test the unit.
\item Using the same crane, remove the load from the truss and %is then 
 disassemble it. 
\item Lower the \dshort{crp} onto the cryostat floor (Figure~\ref{fig:b-crp-install}).
\item Take measurements to determine the plane flatness and to verify the gaps between adjacent units.
\item Repeat this process until the installation reaches the last row in the detector, row 20.
\end{enumerate}

 %$$$$$$$$$$$$$$$  
\begin{dunefigure}
[Bottom \dshort{crp} tooling]
{fig:b-crp-tooling}
{Left: \dshort{crp} survey and leveling truss. Right: Crane and lifting fork.}
\includegraphics[width=0.45\linewidth]{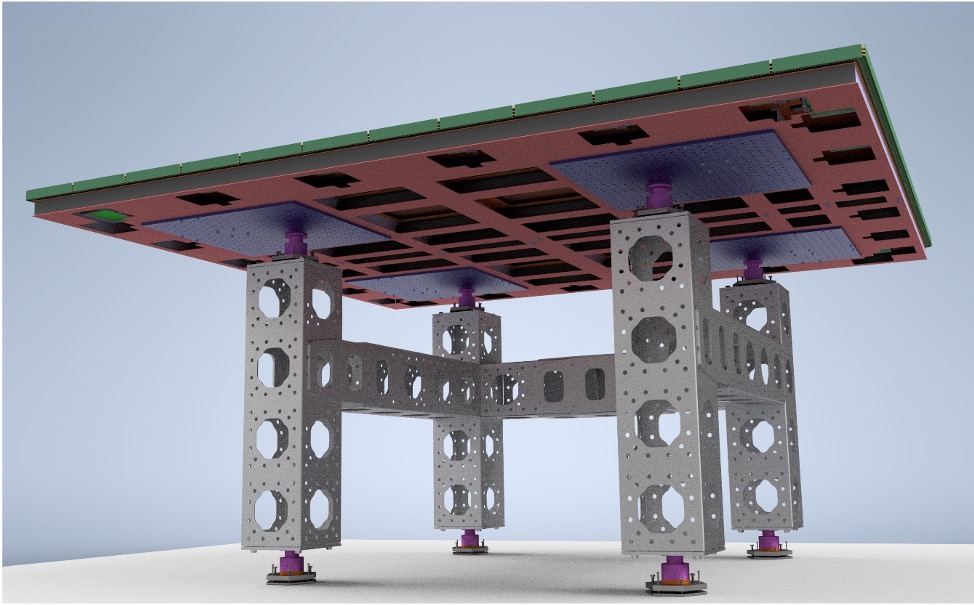}
\includegraphics[width=0.47\linewidth]{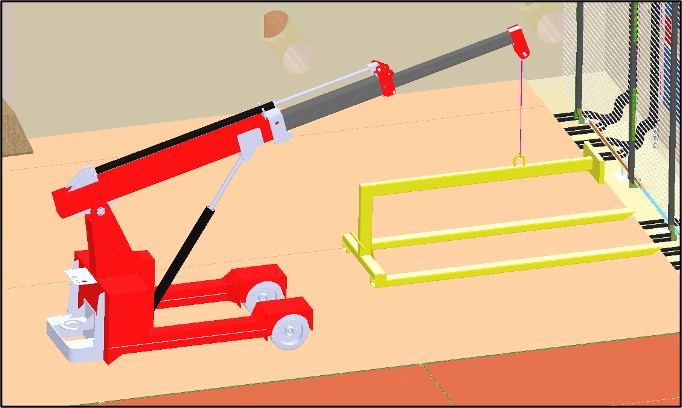}
\end{dunefigure}
%$$$$$$$$$$$$$$$ 

%$$$$$$$$$$$$$$$  
\begin{dunefigure}
[Bottom \dshort{crp} installation]
{fig:b-crp-install}
{Bottom \dshort{crp} being lowered into final position.}
\includegraphics[width=0.5\linewidth]{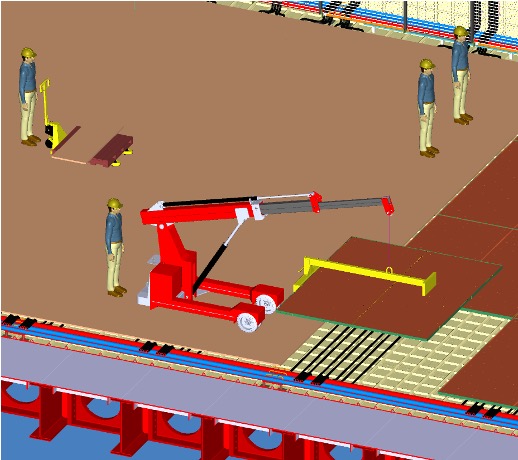}
\end{dunefigure}
%$$$$$$$$$$$$$$$

The last row of the detector will be challenging to install due to the limited space available. On the downstream end of the cryostat  only 700\,mm of space exists between the \dshort{fc} endwall and the cryostat membrane flat surface. This installation proceeds as follows.

\begin{enumerate}[resume]
\item Bring \dshort{fc} storage carts with enough modules for the top $2/3$ of the endwall into the cryostat.
\item Remove the \dshort{tco} hoist and install the endwall panels.
\item Make final HV connections between that cathode and the endwalls from the bottom
\item Remove the scissor lift from the cryostat.
\item Attach a temporary jib crane to the cryostat entrance (Figure~\ref{fig:jibcrane}) for installation of the final row of bottom \dshort{crp}s.
\item Position the \dshort{crp} assembly table directly in front of the \dshort{tco} opening, remove the floor under the north and south \dshort{crp}, and clean the membrane areas.
\item Move two \dshort{cru}s into the cryostat using the jib crane and place them on the assembly table. 
\item Assemble the two \dshort{bde}.
\item The tines of the bottom \dshort{crp}s lifting tool can then be placed under the \dshort{crp} to lift it. 
\item  Move the table out of the way, place the \dshort{crp} temporarily on the floor, and move the jib crane away. 
\item  Move the counterbalanced crane next to the \dshort{crp}, lift it, and move it to its final position. 
\item Repeat for the other outer \dshort{crp}.
\item Finally, bring the last \dshort{cru}s individually into the cryostat using the jib crane as shown in Figure~\ref{fig:last-cru}. At this time only 700\,mm of floor is left in the cryostat.
\item Carry in the last four \dshort{fc}  modules by hand. These modules have only a fraction of the aluminum profiles installed so as to keep the weight low.
\item Two technicians hold them in position while a third %technician 
connects the supports at the top.
\item Once the modules are freely hanging, install the floor support and the final profiles.
\item The remaining profiles are then installed on the lower field cage modules.
\item Install the west purity monitors.
\item  Remove the last of the floor, the protective covers over the end wall \dshort{pd} modules, and \dshort{lar} valves.
\item Clean all reachable surfaces.
\end{enumerate}

%$$$$$$$$$$$$$$$  
\begin{dunefigure}
[Jib crane for Row 20 installation]
{fig:jibcrane}
{Jib crane for Row 20 installation.}
\includegraphics[width=0.6\linewidth]{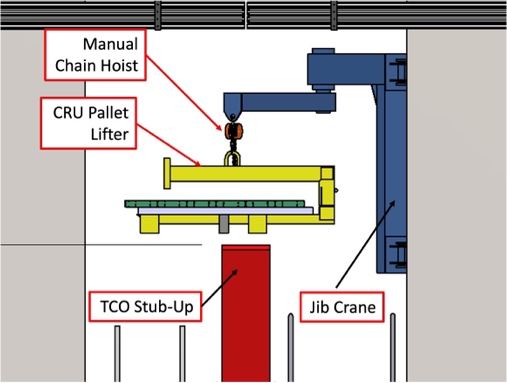}
\end{dunefigure}
%$$$$$$$$$$$$$$$

%$$$$$$$$$$$$$$$  
\begin{dunefigure}
[Final \dshort{crp} installation]
{fig:last-cru}
{Installation of the last \dshort{bde}.}
\includegraphics[width=0.7\linewidth]{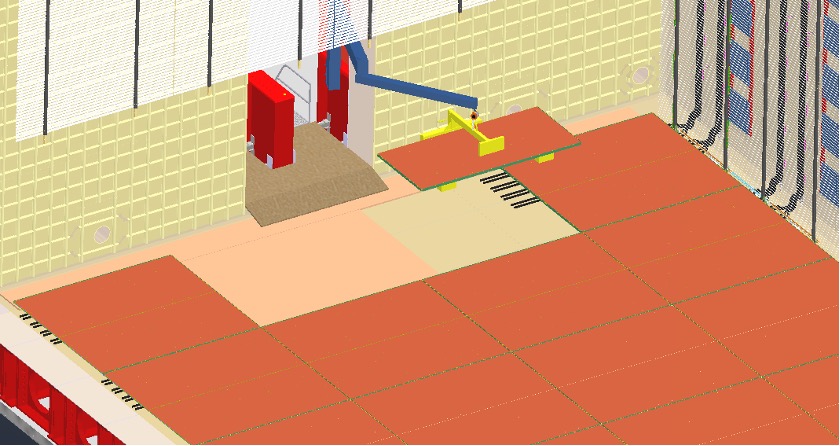}
\end{dunefigure}
%$$$$$$$$$$$$$$$

\FloatBarrier

Once detector installation completes, %and all the required equipment is in place inside the cryostat, 
the process of closing the \dshort{tco} %can 
begins.  The cryostat %installation company 
vendor will perform the mechanical closure of the \dshort{tco} as described.  
\begin{enumerate}[resume]
\item First weld the remaining membrane panels in position and clean the local area.
\item Working from the outside,  install the foam insulation and bolt the steel support structure in position with the tertiary 1\,cm membrane.
\item Weld the tertiary membrane, % is welded, 
sealing the \dshort{tco}.
\item Remove the cryostat ventilation from the four access hatches in the cryostat roof, and seal them.
\item This completes the detector module installation.
\end{enumerate}

%%%%%%%%%%%%%%%%%%%%%%%%%%%%
\section{Interfaces} %added by anne 30 sep

The interfaces between \dword{fsii} and the consortia are documented in a series of Interface Control Documents (ICD)s and interface drawings which are linked to left column in Table \ref{tbl:inst-interfaces}.

\begin{dunetable}
[Installation interface links]
{p{0.15\textwidth}p{0.8\textwidth}}
{tbl:inst-interfaces}
{Installation interface descriptions and links to full interface documents.}
Interfacing System & Description  \\ \toprowrule

\href{https://edms.cern.ch/document/2459132}{General} & \dshort{fsii} interfaces with all consortia in the same manner. This document defines the fundamental roles and responsibilities between \dshort{fsii} and every consortium. Consortia specific interfaces are defined in the interface appendices below.   \\ \colhline

\href{https://edms.cern.ch/document/2648559}{\dshort{crp}} & The CRP consortia designs the CRP supports. \dshort{fsii} will fabricate. CRP provides lifting fixtures and \dshort{fsii} provides commercial rigging equipment. CRP provides equipment to assemble the CRP from CRU in the cryostat.\\ \colhline

\href{https://edms.cern.ch/document/2736688}{\dshort{bde}} &  \dshort{fsii} provides the cryostat roof penetrations, the racks for \dshort{bde} modules and patch panels, cable trays and warm cables and optical fibers. \dshort{bde} provides all cold cables, all electronics, all electrical testing equipment.  \\ \colhline

\href{https://edms.cern.ch/document/2648556}{\dshort{tde}} &  \dshort{fsii} provides equipment to install the chimneys, \dshort{tde} provides tooling to connect to the chimneys. \dshort{fsii} provides the roof penetrations, cable trays, warm cables and optical fibers. \dshort{tde} provides all electronics and electrical testing equipment.\\ \colhline

\href{https://edms.cern.ch/document/2648558}{\dshort{hv}} &  \dshort{fsii} will provide a \dshort{greyrm} for field cage and cathode assembly. \dshort{hv} provides all equipment to do this work. \dshort{fsii} provides lifting equipment for raising the field cage support beams. Other equipment is provided by \dshort{hv}. \\ \colhline

\href{https://edms.cern.ch/document/2648555}{\dshort{pds}} & \dshort{fsii} provides the roof penetrations, racks, warm cables and cable trays. \dshort{pds} provides the detector modules and supports, the readout and testing equipment. \dshort{fsii} will provide a \dshort{greyrm} for assembling and testing the PD modules. \\  \colhline

\href{https://edms.cern.ch/document/2145183}{\dshort{daq}} &  \dshort{fsii} provides the barracks, racks, fire suppression, and cooling for the DAQ servers. \dshort{fsii} also provides the fibers running from the surface and fibers on the cryostat roof.  \\ \colhline

\href{https://edms.cern.ch/document/2391830}{\dshort{fscfbsi}} &  Interfaces to \dshort{fscfbsi} are described in the document linked to the left. These interfaces include: power, ventilation, cooling, fiber infrastructure, mechanical interfaces, and egress. \\ \colhline

\href{https://edms.cern.ch/project/CERN-0000231397}{Drawings} &  A collection of drawings defining the mechanical interfaces relevant to the installation process are collected in the folder linked to the right.\\ 

\end{dunetable}
\FloatBarrier

\section{Organization and Management}

\subsection{Contributing Institutions}

The \dword{dune} institutions contributing to the \dshort{fsii} scope are listed in Table~\ref{tab:FSii-institutes}.

\begin{longtable}
{ll}
\caption{\dshort{fsii} Institutions}\\ \colhline
\rowcolor{dunetablecolor} 
Member Institute  &  Country  \\  \toprowrule
BNL & USA\\ \colhline
CERN & Switzerland\\ \colhline
FNAL & USA \\ \colhline
U. Minn. & USA\\ \colhline
\label{tab:FSii-institutes}
\end{longtable}

\subsection{High-level Schedule}
\label{ch:inst:milestones}  

\begin{dunetable}
[Detector integration and installation schedule]
{p{0.73\textwidth}p{0.17\textwidth}}
{tab:IIsched}
{High level milestones and schedule for  \dshort{spvd} integration and installation.} 
Milestone & Date   \\ \toprowrule

\dshort{spvd} cryostat start of construction &   May 2025  \\ \colhline %June 2025

\dword{sdwf} ready to accept \dshort{spvd} shipments & February 2027 \\ \colhline %March 2025
\Dword{irr} all items for \dshort{spvd}& February 2027      \\ \colhline %October 2026
 
Top \dshort{tde}/\dshort{crp} assembly ready for installation &  April 2027 \\ \colhline

\dshort{spvd} cryostat complete; ready for detector installation & August 2027       \\ \colhline %July 2027
 
\dshort{spvd} fabrication and delivery of components to \dshort{surf} -- threshold \dword{kpp} (KPP)
 & August 2027      \\ \colhline

 Bottom \dshort{bde}/\dshort{crp} assembly ready for installation &  October 2027 \\ \colhline

\dshort{spvd} installation complete -- threshold KPP & August 2028    \\ \colhline

Final approval to operate cryogenics system \#2 & November 2028      \\ \colhline

All threshold KPPs met
& September 2029       \\ \colhline %August 2029

\dshort{spvd} commissioning test complete -- objective KPP met   (30\% filling)
& October 2029      \\ \colhline %August 2029 

\dshort{spvd} complete - objective KPP met
& November 2029      \\  %September 2029
\end{dunetable}

%%%%

\chapter{Project Management}
\label{ch:project}
%\tableofcontents
\section{Project Organization}
\label{sec:org}

%%%%%%%%%%%%%%%%%%%%%%%%
\subsection{DUNE Collaboration and Consortia}

The \dword{dune}, shown in Figure~\ref{fig:DUNE_consortia}, is led by two co-spokespersons overseeing the construction activities in three principal areas: \dword{sphd}, \dword{spvd} and \dword{nd}. 
An overall institutional board (\dword{ib}) is responsible for the governance of the collaboration. 
\dword{dune} is organized into several consortia, each of which is responsible for design, fabrication and commissioning of specific detector components and continue to maintain responsibility for these subsystems through the operations phase. 
\begin{dunefigure}[DUNE consortia]{fig:DUNE_consortia}
  {Organization of the \dword{dune} consortia.}
  \includegraphics[width=0.95\textwidth]{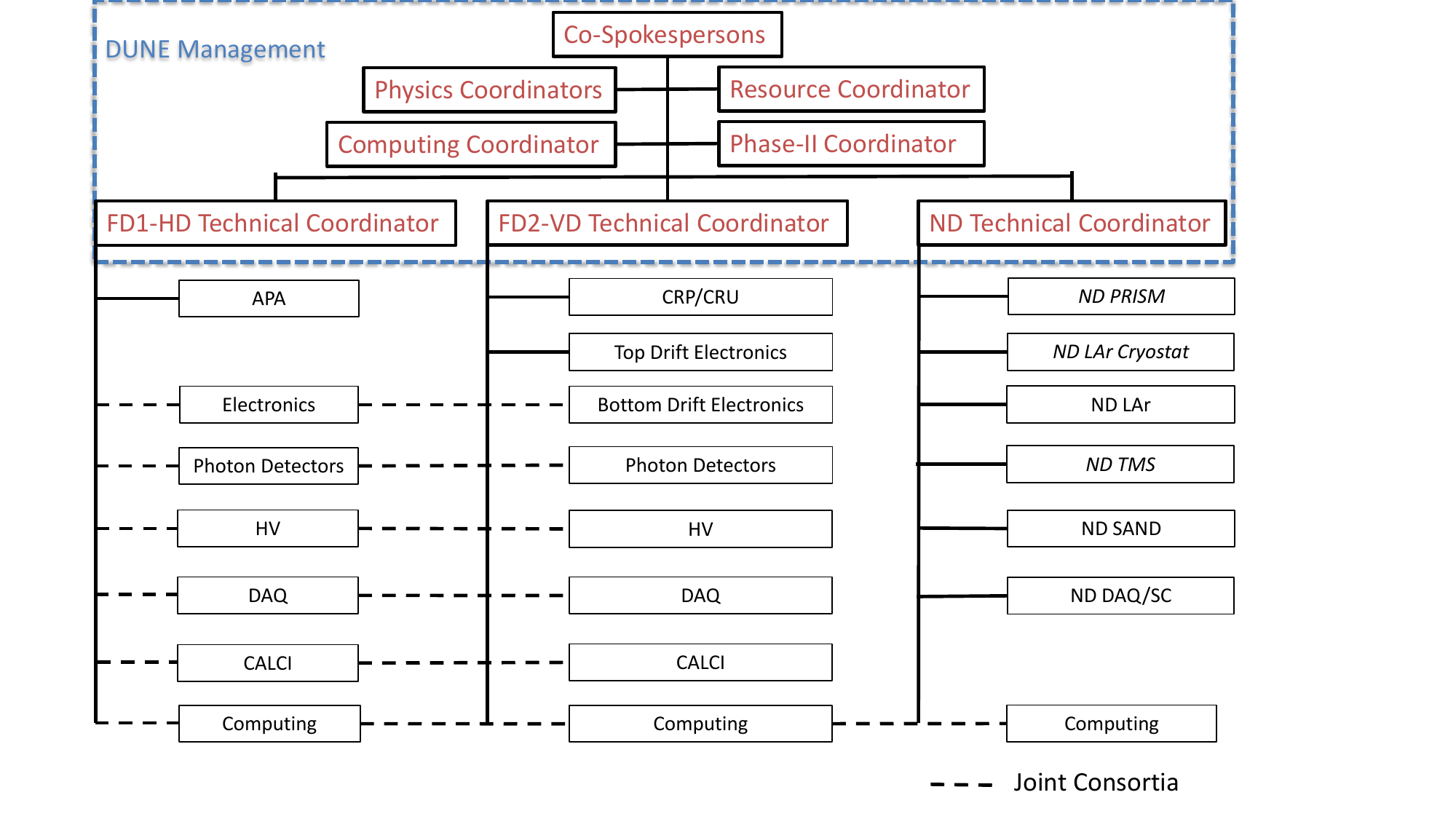}
\end{dunefigure}

The \dshort{dune} spokespersons serve on the \dword{efig} that coordinates between the 
conventional and beamline facilities and the \dshort{dune} detector subprojects. \dshort{dune} has three technical coordinators, one for each %detector (
of \dshort{sphd}, \dshort{spvd}, and \dshort{nd}, who coordinate the activities across their respective detector subsystem and provide the critical interface across the consortia. The technical coordinators are part of the DUNE management team as described in~\cite{edms-2808700,edms-2808699} along with the spokespersons, resource coordinator, physics coordinators, computing coordinator and upgrade coordinator.

There is a high degree of overlap between consortium deliverables in design, prototyping, and fabrication for \dshort{spvd} and \dshort{sphd}. In many cases entire subsystems are identical, and in others many of the components are identical. In these cases, a single consortia working on multiple detector systems has been instituted to provide maximum efficiency, and this is represented by the horizontal dashed lines in Figure~\ref{fig:DUNE_consortia}.

Each consortium is managed by a consortium leader and technical leads appointed by \dshort{dune} management. The consortium leader 
chairs a (consortium-specific) institutional board (\dshort{ib}) composed of one representative from each of the contributing institutions. Major consortium decisions such as technology selections and assignment of responsibilities to specific  
institutions pass through its \dshort{ib}. These decisions are then passed as recommendations to the overall \dshort{dune} \dword{exb}. (The \dshort{dune} \dword{exb} consists of the Spokespersons, \dshort{dune} \dshort{ib} chair, technical coordinators, resource coordinator, computing coordinator, physics coordinators, phase-II coordinator, and consortia leaders.)

Consortia manage the design and construction of subsystem deliverables that may be supported by multiple funding agencies, where each funding agency is responsible for its own deliverables. To ensure coordination between the separate internal projects contributing to the consortia, technical leads are responsible for chairing consortium project management boards that incorporate managers from each of the internal projects.

A total of nine \dword{fdc} consortia have been formed to cover the subsystems required for \dword{sphd} and \dword{spvd}, as shown in Figure~\ref{fig:DUNE_consortia}. 
In particular, two consortia focus exclusively on \dshort{spvd} (\dword{crp} and Top Drift Electronics) and another six consortia have responsibility for subsystems common to both detector technologies (Bottom Drift Electronics (combined with \dshort{sphd} Electronics), \dword{pds}, \dword{hv}, \dword{daq}, \dword{calci}, and Computing). One consortium, \dword{apa}, focuses exclusively on \dshort{sphd}. 

%%%%%%%%%%%%%%%%%%%%%%%%
\subsection{LBNF/DUNE Project}

The \dword{lbnf-dune} has been organized as shown in Figure~\ref{fig:DUNE_global}, with the \dword{lbnf-dune} project (and \dword{usproj} project) subdivided into five subprojects.

\begin{dunefigure}[Global project organization]{fig:DUNE_global}
  {\dshort{lbnf-dune} organization showing the \dshort{lbnf-dune} and \dshort{usproj} (top, center) as comprising five subprojects (SP), delineated with blue boxes. The DUNE Collaboration Management (top, left), \dshort{lbnf-dune} and \dshort{jpo} (top, right) work closely together. The exlusive US project organization, \dshort{fscf} includes two subprojects for excavation (EXE) and building and site infrastructure (BSI that include the utilities. The far detector and cryogenics (FDC) subproject (both US and international) includes both detectors FD1--2 and the cryostats and cryogenics. PM refers to project manager. CD refers to critical decision (part of the DOE system). "Sup \& Ser" refers to support and services. "Compnt PMs" refers to the different DUNE subsystems.}
  \includegraphics[width=0.95\textwidth]{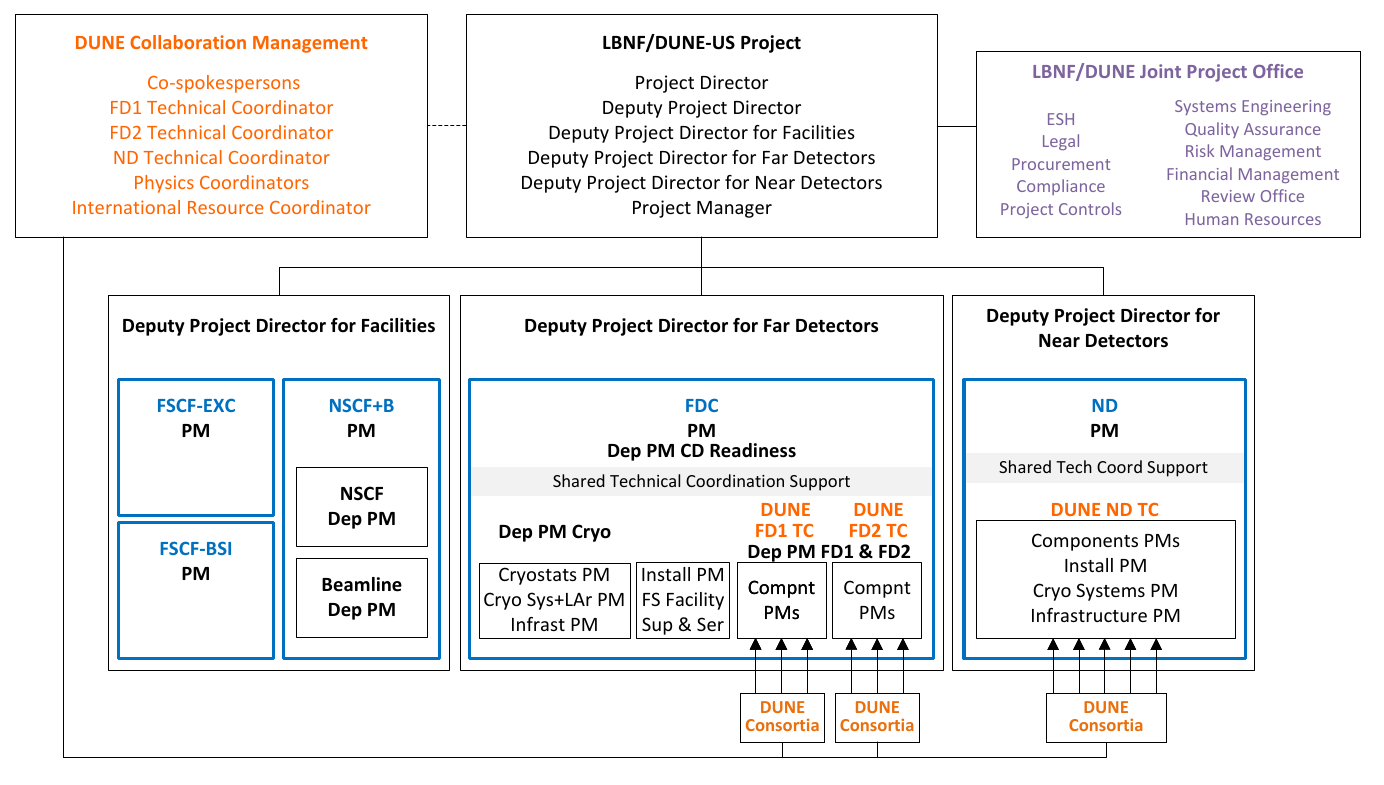}
\end{dunefigure}

\dshort{lbnf-dune} is led by a  Project Director. 
The five subprojects are managed by Deputy Project Directors who report to the Project Director, and engage the \dfirst{jpo} to  support various project functions. These management structures and functions are outlined in the \dshort{lbnf-dune} 
Project Management Plan~\cite{edms-2808695}. %protected in docdb 117
\dword{fscf} includes two subprojects for excavation (EXE) and building and site infrastructure (BSI that include the utilities. The far detector and cryogenics (FDC) subproject includes both detectors FD1--2 and the cryostats and cryogenics.

%%%%%%%%%%%%%%%%%%%
\subsection{Far Detector and Cryogenics (FDC) Subproject}

The international Far Detector and Cryogenics subproject (\dword{fdc}) is responsible for construction, assembly, installation and commissioning of the \dword{sphd} and \dword{spvd} \dshort{detmodule}s (including the supporting infrastructure: cryostats and cryogenics systems). Major parts of the detector are the responsibility of the \dshort{dune} consortia which are coordinated through the technical coordinators. 

The \dshort{fdc} subproject is overseen by the Deputy Project Director for Far Detectors (\dword{dpdfd}) and is organized as shown in Figure~\ref{fig:DUNE_fdc}.

\begin{dunefigure}[\dshort{fdc} organization]{fig:DUNE_fdc}  {Far detector and cryogenic (FDC) subproject organization. \dshort{cam} refers to control account manager, who is responsible for the budget and reporting.}
  \includegraphics[width=0.95\textwidth]{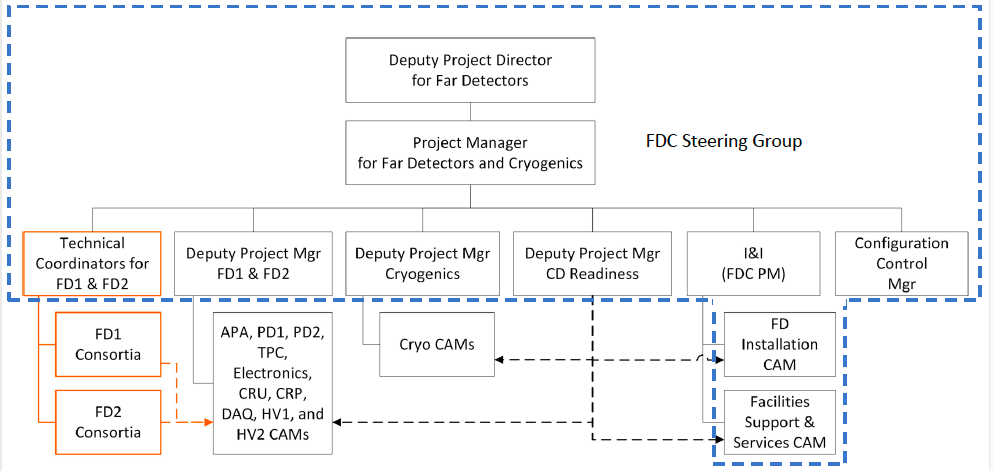}
\end{dunefigure}

The far site steering group (delineated by blue dashed line in Figure~\ref{fig:DUNE_fdc}) is chaired by and advises the \dshort{dpdfd}; its members are the 
\begin{itemize}
\item \dshort{dpdfd},
\item \dshort{sphd}/\dshort{spvd} %technical coordinators
\dwords{tcoord},
\item  \dshort{fdc} subproject manager and deputies,
\item  \dshort{sphd}/\dshort{spvd} installation managers,
\item  cryogenics manager,
\item  configuration control manager, 
\item  \dword{sdsd} manager, 
\item \dshort{usproj} Project Director,
\item \dshort{usproj} Project Manager,
\item Co-spokespersons.
\end{itemize}

The steering group oversees the entire \dshort{fdc} subproject, including the design and fabrication of the detector modules, cryostats, cryogenics and support systems and their installation and commissioning at \dword{surf} following \dword{aup}.

The \dshort{fdc} subproject maintains several management tools, including the international \dshort{fdc} \dword{p6} scheduling software and the \dshort{spvd} schedule. The \dshort{spvd} schedule was developed by the consortia contributing to \dshort{spvd} and is maintained by the  \dshort{fdc} project controls staff. The US \dshort{spvd} L2 \dwords{cam} are responsible for \dword{evms} on the US part of the schedule and track percent-complete and milestones with the consortia leaders for the rest of the schedule.  The \dshort{fdc} subproject coordinates with and oversees the \dshort{dune} consortia in the design, fabrication, installation and commissioning of their detector components and manages the interfaces between detector components and the detector support infrastructure (Interface documents can be found in {\tt https://edms.cern.ch/project/CERN-0000224744}.). It coordinates with the \dword{ro} and \dword{co} to ensure that all detector components meet specifications and are safe, and with the configuration control manager to ensure that the integrated detector model fits in the overall far site model, that installation envelopes are fully specified, and that detector component locations are understood when installed (at room temperature and dry) and during operations after the cryostat is filled with \dword{lar}. 
The \dshort{fdc} subproject coordinates with the consortia in the development of the \dshort{spvd} detector cost estimate, basis of estimate, risk register, contingency estimate, requirements, interfaces, \dshort{p6} schedule, milestones and for the U.S. scope of the project, \dshort{evms}. Requirements are managed at several levels and can be found in~\cite{EDMS214368}. 
The \dshort{fdc} subproject includes the \dword{fsii} organization that is responsible for all installation activities at the far site, including \dshort{sphd}, \dshort{spvd} and cryogenics, as discussed in Chapter~\ref{ch:IEI}.

 \dshort{spvd} maintains a set of tools to document, manage and track risks developed by the consortia together with the U.S. \dshort{cam}s. Risks are continually discussed and updated within each consortia and eventually approved by the \dshort{lbnf-dune} Risk Management Board and documented in the Fermilab Risk tool. A snapshot of the risk register is provided at~\cite{edms-2808697} and the technical risks are summaried in Section~\ref{sec:risks}. Regular risk workshops assess risks across the full scope of \dshort{fdc}.

%%%%%%%%%%%%%%%%%%%%%%%%
\subsection{Technical Boards}

Technical board meetings, chaired by the technical coordinators are the fora for resolving technical issues (e.g., how a \dword{fc} and \dword{crp} may interfere during the installation process). Minutes are taken and made available to the collaboration so that issues that impact science are brought to the attention of the \dword{exb}. Issues that impact project scope and schedule are brought to the consortia project management boards as needed. Two separate, but highly overlapping, \dword{fd} technical boards focus separately on \dword{sphd} and \dword{spvd}. The technical boards have expanded to incorporate additional project team members. These boards incorporate all parties needed to address issues arising across the full spectrum of detector-related activities and thereby ensure tight coordination between the consortia.
Each technical board consists of the following members:
\begin{itemize}
 \item technical coordinator (chair),
 \item consortium leadership teams,
 \item technical coordination engineering teams,
 \item installation coordinators (lead engineers),
  \item cryogenics coordinators (project managers and lead engineers),
 \item \dword{sdsd} (support service managers and logistics manager),
 \item configuration management team;
 \item project management (Deputy Directors for \dwords{fd} and detector subproject managers), and
 \item collaboration spokespersons and \dword{usproj} Project Director.    
\end{itemize}

%%%%%%%%%%%%%%%%%%%%%%%%
\subsection{Joint Project Office}

The \dword{jpo} reports to the \dword{lbnf-dune} Project Director as shown in Figure~\ref{fig:DUNE_global} and provides coordination across the \dshort{lbnf-dune} enterprize. Major functions provided by the \dshort{jpo} include \dword{esh}, \dword{qa}, systems engineering, and the review and compliance offices (\dword{ro} and \dword{co}).

\dshort{lbnf-dune} is committed to protecting the health and safety of staff, the community, and the environment, 
as well as to ensuring a safe work environment for \dshort{dune} workers at all institutions and protecting the public from hazards associated with constructing and operating \dshort{dune}. 
Accidents and injuries are preventable, and the \dshort{esh} team  works with the global \dshort{lbnf-dune} project and collaboration to establish an injury-free workplace. All work will be performed so as to preserve the quality of the environment and prevent property damage. The \dshort{lbnf-dune} \dshort{esh} program is described in detail in the LBNF/DUNE Integrated Safety and Health Management Plan~\cite{edms-2808692},
%protected in docdb 291 
and more briefly in~\cite{Abi:2020oxb}, Chapter 10.

The systems engineering group is responsible for the high-level project-wide management of requirements, interfaces and \dword{cad} integration, and oversees configuration management and change management. Systems engineering works with all \dshort{lbnf-dune} design teams to incorporate individual \dshort{cad} models into the global controlled model that is released in~\cite{edms-204070}. Configuration management establishes and maintains consistency across the entire \dshort{lbnf-dune} enterprize for project performance, function, 
and physical attributes via requirements, design models and drawing, and operational information for the duration of the project. The Systems Engineering Management Plan (\dword{semp})~\cite{edms-2808693} describes the  responsibilities and processes that support the design and implementation of configuration management. The main goal of the \dshort{semp} is to prevent unauthorized or uncontrolled changes to the engineering design or analysis, hardware, controlled documents, and controlled software. The latest official version of the requirements is stored in~\cite{edms-198204}. Change control processes are defined in the \dshort{semp}. 
Baseline change requests are catalogued and managed in the \dshort{lbnf-dune} change control tool according to thresholds defined in the Change Control Threshold and Authority Table~\cite{edms-2808695, edms-2808693}. The integrated \dshort{lbnf-dune} change control board provides oversight of changes. The \dshort{jpo} engineering team reviews subsystem component documentation in order to manage the detector configuration. 
The consortia provide engineering data for their detector subsystems to the \dshort{jpo} team for incorporation into global configuration files. The integration of the \dshort{spvd} detector is carried out by the \dword{tc} engineering team working with the consortia and in the broader framework of the \dshort{jpo} central engineering team. 

The \dshort{lbnf-dune} design review process, described in~\cite{edms-2173197}, is consistent with the \dshort{fnal} review process described in its \dshort{esh} manual~\cite{feshm}. Past and scheduled reviews are documented in~\cite{indico-586}. Review reports are saved in~\cite{edms-211273}). The review process is an important part of the \dshort{dune} \dshort{qa} process, both for design and production. 
The \dfirst{ro}, in coordination with the \dword{tc}, reviews all stages of development and works with each consortium or subsystem to arrange reviews at key stages of the design (\dword{cdr}, \dword{pdr} and \dword{fdr}) and production (\dword{prr} and \dword{ppr}). The office also conducts \dwords{irr} and \dwords{orr} of their subsystems. \dshort{cdr}, \dshort{pdr} and \dshort{fdr} are equated with 30\%, 60\% and 90\% design. 
The review charge for each is set by the \dword{ro} in coordination with \dword{tc}.

Full production of detector elements begins only after successful \dshort{prr}s. Regular production progress reviews will be held once production starts. The \dshort{prr}s will typically include a review of the first pre-production modules produced at each facility, which may include the production of \dword{mod0} components. Installation reviews will review the \dshort{dune} installation equipment, procedures, lifting fixtures and hazard analyses for the installation work.  The \dshorts{orr} will feed the \dshort{fnal} \dword{orc} process~\cite{feshm} and are the final safety checkout before equipment can be operated. Tracking review recommendations is part of the review process, with reports and recommendations maintained in~\cite{edms-204073}. 

The \dshort{jpo} \dfirst{co} has mechanical and electrical engineering experts from collaborating institutions and serves as the \dshort{lbnf-dune} engineering safety assurance team. The \dword{co} develops processes and procedures for evaluating engineering designs using accepted international safety standards. The codes and standards to which each system is designed will be reviewed as part of the \dword{pdr} and \dword{fdr} reviews. The \dshort{co} verifies that the structural analysis presented by the various engineering partners in the project is correct and of sufficient quality.

\dshort{dune} \dfirst{tc} monitors technical contributions from collaborating institutions and provides centralized project coordination functions with support from the \dshort{jpo}. 
Part of this project coordination is standardizing \dfirst{qa} and \dfirst{qc} practices, one facet of which is to assist the consortia in defining and implementing \dshort{qa}/\dshort plans that maintain uniform, high standards across the entire detector construction effort. The \dshort{lbnf-dune} \dshort{qa} program is described in detail in the project's Quality Assurance Plan~\cite{edms-2808696}  (\dword{qap}) 
and more briefly in~\cite{Abi:2020oxb}, Chapter 9.

%%%%%%%%%%%%%%%%%%%%%%%%
\section{Risks}
\label{sec:risks}

Table~\ref{tab:fd2-risks} lists the significant technical risks for the \dword{spvd}, along with the risk mitigations and their post-mitigation probabilities (P) and potential impacts on cost (C) and schedule (S). The impacts are designated as low (L), medium (M) or high (H) in each category. The impact values are assigned according to these criteria:

\begin{itemize}
    \item Probability (\%): L=0-19, M=20-39, H=$\geq$40;
    \item Cost (k\$ U.S.): L=0-100, M=100-500, H=$\geq$500;
    \item Schedule (months): L=0-2, M=2-6, H=$\geq$6.
\end{itemize}

Note that risks associated with delays, cost increases, insufficient personnel, or damages are not included. 
Two operational risks, BDE1 and HVS2, are included to emphasize that risk mitigation starts during the design phase, making design choices that go toward minimizing all operational risks. 

Most of the risks have been determined to be low probability (P); the few that are currently set to medium (M) will be reduced prior to the final design review (\dword{fdr}). This demonstrates the technical maturity of the design. The few risks designated high (H) (in cost or schedule only) are expected to decrease to low or medium prior to the \dword{prr}, once the production model is further understood and documented.

\begin{footnotesize}
\begin{longtable}
{P{0.06\textwidth}P{0.31\textwidth}P{0.43\textwidth}P{0.022\textwidth}P{0.022\textwidth}P{0.022\textwidth}} 
\caption[\dshort{spvd} Risks]{\dshort{spvd} Risks}
 \\
\rowcolor{dunesky}

ID& Description & Mitigation& P& C& S \\ \toprowrule

CRP1&
Bottom \dword{crp} support requires additional design consideration following \dword{mod0} test. & 
Engineering by an experienced group is planned for \dshort{mod0} supports. Prototype testing will be done before installation in \dshort{mod0}.&
L&M&L \\ \colhline

CRP2&Bottom CRP support requires additional design consideration following \dshort{mod0} test. &
Engineering by an experienced group is planned for \dshort{mod0} supports. Prototype testing will be done before installation in module 0.
&L&M&L\\ \colhline

CRP3&Bottom CRP require additional cold testing at far site due to damage on receipt &Extensive tests of shipping process during prototyping stage to minimize damage
&M&M&L \\ \colhline

CRP4&Additional design iteration of CRP adapter cards needed to address noise issues&CRP adapter cards are extensively tested during cold box tests prior to \dword{prr}
&L&L&M \\ \colhline

CRP5&Additional design iteration of edge boards needed to address mechanical connection issues prior to PRR. &Edge boards are extensively tested during prototyping stage; additional design considerations are explored to improve mechanical integrity. 
&L&L&M \\ \colhline

CRP6&Anode \dword{pcb} strip interconnect has some failure rate from handling and thermal cycle on the silver printing regions.&Identify alternative PCB panel building methods to eliminate the use of silver printing.&L&L&L \\ \colhline

%TDE1&Unexpected noise %issues due to grounding and shielding issues.&Several tests are performed in \coldbox and \dshort{protodune}  to assess, diagnose and attend to such issues. Ensure proper grounding and shielding of all electronics components.&L&L&L \\ \colhline

TDE1&Commercial components become obsolete  &Procure components with sufficient spares in advance. 
&L&L&L \\ \colhline

TDE2&Damage of \dword{fe} electronics due to HV discharges occurring at the level of CRP and/or field cage. &Design of suitable protection components on the FE card. Allow for replacement of cryogenic electronics via chimneys without contaminating LAr volume %and the need to switch 
or switching off the other cards in the same %chimney 
or other chimneys.&L&L&L \\ \colhline

TDE3& \dshort{sftchimney}s do not fit in roof penetrations & Jointly with I\&I, draft and execute validation procedure for chimney and penetration interface and tolerances.  
&L&L&L \\ \colhline

TDE4& Nitrogen leaks into cryostat via the \dshort{sftchimney}s & Before installation, leak-test each chimney by filling it with argon. Perform other tests in situ.
&L&L&L \\ \colhline

BDE1&Lifetime of TPC electronics components inside the cryostat.&
Engineering of \dwords{asic} follows design rules aimed at ensuring that the chips would work in \dword{lar}. The possibility of using redundant connections between the ASICs/\dwords{femb} and the \dwords{wib} is also being taken into account. Proper fabrication rules and QC tests for PCBs are being put in place based on the experience with other detectors that use cryogenic liquids.&L&L&H \\ \colhline

%BDE2&Cold ASIC(s) do not meet specifications&Extensive tests, including lifetime tests of ASICs during cold box tests and ahead of FDR to validate performance.&L&M&H \\ \colhline

BDE3&Incompatibilities with detector components provided by other consortia &Dcument and agree on proper design of interfaces; extensive integration tests during prototyping stage &L&M&M\\ \colhline

%BDE4&Excessive noise observed during detector commissioning &Perform multiple integration tests prior to FDR. Maintain enough flexibility in the connection of shielding of cables on top of the cryostat.&M&M&M\\ \colhline

BD4&Single source vendor for BDE&Use  fabrication technologies that are relatively recent and in widespread usage.  Investigate alternative vendors. Purchase all the components as soon as the design is finalized.&L&H&L\\ \colhline

HVS1&\dshort{sphd}/\dshort{spvd} \dword{hv} experience in %PD II 
\dshort{mod0} leads to a significant redesign&
If the HV performance during \dshort{mod0} does not improve upon \dword{protodune} performance, then there may be a significant redesign of the HV/cathode %CPA 
system
&L&M&M \\ \colhline

% Commented out 3/16 by Anne   HVS2& %Catastrophic 
%HV discharges causes damage to detector electronics&
%Operational risk; moved to a higher project level and not HVS ownership. &L&L&L \\ \colhline

HVS2&\Dword{fc} profiles need coating&A  %alternative 
method of applying a coating using chemical wipes has been found and can be implemented at SURF during \dshort{fc} assembly&L&L&L \\ \colhline
%HVS3&\Dword{fc} profiles need coating&Can apply a coating using chemical wipes at SURF during \dshort{fc} assembly&L&L&L \\ \colhline

PDS1&Insufficient \dword{pof} efficiency compared to baseline expectations.&
Explore and compare  PoF technologies from different vendors. Consider adding  PoF transmitters, receivers, and fiber penetrations in the cryostat.
&L&H&M \\ \colhline

PDS2&Simulations show additional detection efficiency required&
Explore and optimize geometries, optimize the number of \dwords{sipm} to maintain efficiency. Conclude studies before FDR.
&L&H&L \\ \colhline

PDS3&PDS Components fail 30-year cold validation testing&Extensive lifetime tests of indvidual components and integrated module prior to \dshort{fdr}.   Identify alternate vendor for several components in advance if issues arise during tests. &M&H&L \\ \colhline

PDS4&\Dword{pd} electronics generates noise on the TPC strips readout&
Reasonable shielding and simulations will be conducted to minimize risk of inducing noise on TPC at critical frequencies. Prototype and \dshort{mod0} completion will improve confidence if there is no indication of noise in neighboring detectors.
&L&M&M \\ \colhline

DAQ1&Inability to efficiently perform \dword{daq} work remotely&Tools developed to allow for remote access and for collaborative work between SURF personnel and remote DAQ experts; train SURF personnel to handle trouble-shooting issues independently &L&L&L \\ \colhline

DAQ2&SURF infrastructure stability impacts DAQ uptime and server lifetimes&
Communicate DAQ services requirements clearly, including those that affect stability. Develop a plan for operations support to monitor stability issues. Plan integration testing activities such that some instability will not cause a complete halt in work.
&L&L&L \\ \colhline

DAQ3&\dword{daqccm} and \dword{dqm} do not meet required specifications&Perform scaling tests at  \dshort{mod0}s to ensure CCM and DQM infrastructure can support full detectors with some headroom. Focus final development, integration, and review activities on achieving performance at scale.
&L&L&M \\ \colhline

DAQ4&Data filter is not optimized or deployed efficiently&Ensure data filter development work ends with full demonstration and integration with the DAQ well ahead of when it is needed for commissioning.&L&L&L \\ \colhline

I\&I1&Detector failure during \cooldown{} &Cold testing of all components before installation and continuous monitoring during \cooldown
&L&H&H\\ \colhline

I\&I2 &Missing scope due to poor interface definition at Far Site&Use systems engineering methodologies to identify and alleviate gaps in scope.  Ongoing inter-project interface meetings/discussions to ensure scope alignment and clarity of expectations.&L&H&L\\ \colhline

I\&I3 &Lack of agreement on international codes/standards affects partner design work&Use code equivalencies and international agreements to detail process for design and acceptance of non-U.S. components. Use design reviews to evaluate compliance to codes/standards applicable to the project;
establishment of the integration project office / compliance office to resolve issues ahead of execution.
&L&L&M \\ \colhline

I\&I4 &Grounding and shielding errors induce electronics noise requiring additional mitigation. &Document and implement the DUNE grounding and shielding plan. Test all instrumentation during \dshort{mod0}. &M&L&L \\ \colhline

I\&I5 &Additional Installation testing required after \dshort{mod0}&Provide for additional test setup during \dshort{mod0}
&M&L&M \\ \colhline

I\&I6 & I\&I Equipment requires major repair&Plan includes periodic maintenance and minor repair, including annual inspections, as appropriate.
&M&M&L \\ \colhline

\label{tab:fd2-risks}
\end{longtable}
\end{footnotesize}

%%%%%%%%%%%%%%%%%%%%%%%%
\section{Schedule}
\label{sec:sched}

The \dword{spvd} detector was added to the overall \dword{lbnf-dune} project in June 2021. The development steps of the \dshort{spvd} project are summarized in Figure~\ref{fig:FD2-development} as \dshort{spvd} advances from conceptual design through preliminary and final design, and into  production.
\begin{dunefigure}
[Project development plan for \dshort{spvd}]
{fig:FD2-development}
{Project development plan for \dshort{spvd}. The operation dates of \dshort{vdmod0} depend on the availability of \dshort{lar}. The dedicated \dshort{pd} \coldbox runs are not shown.}
\includegraphics[width=0.99\linewidth]{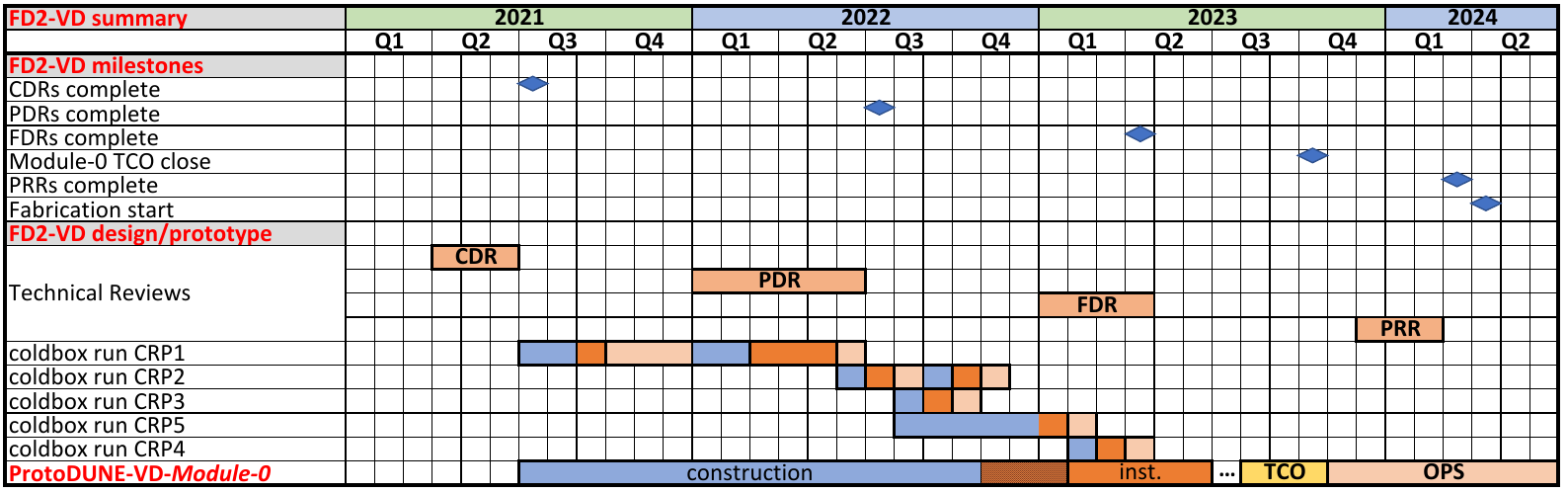}
\end{dunefigure}
The overall schedule includes the %vertical drift 
\dshort{spvd} phased R\&D plan starting with small-scale R\&D tests from  2020 through 2022, advancing to large-scale tests in 2021--2022 (as discussed in Chapter~\ref{ch:mod0}) and the complete \dshort{vdmod0} detector in 2023--2024. \dshort{vdmod0} is composed of pre-production components intended for production for %\dword{dune}
\dshort{spvd}. The conceptual design reviews (\dwords{cdr}) for each consortium were completed in spring 2021, followed by the conceptual design report (also abbreviated as \dshort{cdr}) that was endorsed by the \dword{lbnc} and approved by the \dshort{fnal} director in January 2022. Preliminary design reviews (\Dwords{pdr}) for each consortium were completed by June 2022. Final design reviews (\Dwords{fdr}) are expected in the first half of 2023 after \dshort{spvd} \coldbox runs have validated detector designs. The final full-scale pre-production components will be installed into \dshort{vdmod0} in early 2023. \Dwords{prr} for each consortium are expected in early 2024 in anticipation of full-scale production of components that is expected to start in 2024--2025 with installation at \dshort{surf} in 2026--2027. Near-term development milestones tracked by the \dshort{lbnc} are shown in Figure~\ref{fig:lbnc_milestones}. These milestones will be expanded and implemented into \dword{p6}.
\begin{dunefigure}
[Near-term FD2-VD milestones tracked by \dshort{lbnc}]
{fig:lbnc_milestones}
{FD2-VD milestones tracked by the \dshort{lbnc}. Green status indicates that it was accomplished before the milestone date and yellow means it was after the initial date.}
\includegraphics[width=0.9\linewidth]{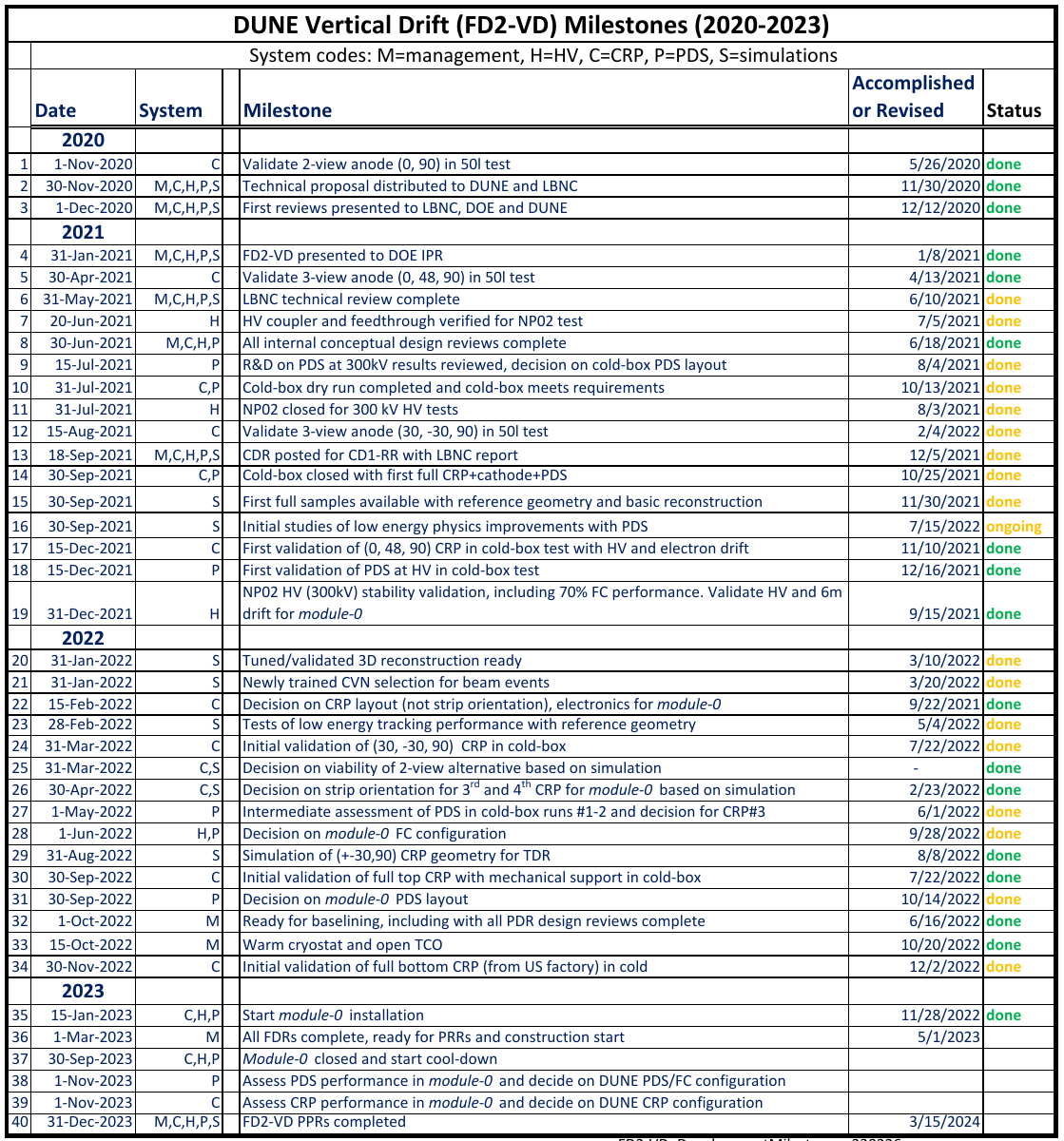}
\end{dunefigure}

\subsection{Prototyping Schedule}
\label{subsec:AAss}
The \coldbox tests of detector components are the key to demonstrating performance for the \dwords{fdr} (and U.S. \dword{doecd}-2/3). These tests demonstrate system prototype performance in an operational environment and give confidence that the components can deliver the necessary performance in \dword{spvd}. The required steps that follow before launching \dwords{prr} and final production include qualification of production sites and \dword{qc}, production processes and procedures. The \dshort{vdmod0} tests demonstrate a high level complete functional system integration of all \dshort{spvd} detector components.

As discussed in the \dshort{spvd} \dword{cdr}~\cite{FD2-VD-CDR}, several key areas of R\&D were advanced in small-scale test stands, including the 50\,L test stand at the \dword{cern}, \dshort{fnal}'s \dword{iceberg}, the \dword{bnl}/Yale \coldbox\, and the 2\,t cryostat at \dshort{cern}, and this development will continue. Larger-scale demonstrations focus on the \dshort{spvd} \coldbox and \dword{hv} test in the large \dword{np02} cryostat. The \dword{hvft}, coupler and \dword{hv} extender were tested in  test stands and in the large \dshort{np02} cryostat. This \dshort{np02} test has demonstrated that 300\,kV can be delivered to a cathode 6\,m under the \dshort{lar} surface and that ionization electronics can be drifted 6\,m; retiring significant risk for \dshort{spvd}. 
The \dword{crp} anode designs were tested in the 50\,L system and \coldbox at \dshort{cern}. This prototyping establishes performance parameters, refines designs, develops vendors, and progresses \dshort{spvd} to final design and \dword{mod0}. \dshort{vdmod0} installation started in early 2023 after  component testing in the \dshort{spvd} \coldbox. Operations are expected to start in 2024. The results may enable final adjustments for \dshort{spvd} production if there are outstanding issues from the \dwords{prr}. 

%%%%
\subsection{Production and Installation Schedule}
\label{subsec:CNss}

The \dwords{fdr} for \dword{spvd} components started in early 2023  and are expected to complete by May 2023. \Dwords{prr} are anticipated in early 2024 after \dword{usproj} obtains \dword{doecd}-3 approval. The production schedule is based on production estimates from each consortium as discussed in previous chapters; production is expected to occur in 2024--2026. The components for \dshort{spvd} should be delivered to \dshort{surf} in 2026--2027. The installation of \dshort{spvd}, the second \dshort{detmodule}, at \dshort{surf} is planned for 2026--2028. Additional details for the proposed installation procedure have been presented in Section~\ref{ch:IEI:det-inst}. The overall production and installation schedule is shown in Figure~\ref{fig:schedule_production}.
\begin{dunefigure}
[Vertical drift production and installation schedule]
{fig:schedule_production}
{Vertical drift production and installation schedule, based on \dshort{spvd} production estimates. Gap between TCO closing and purge/filling results from cooling power needs for FD1 filling. (Data from~\cite{docdb22261v28})}
\includegraphics[width=\linewidth]{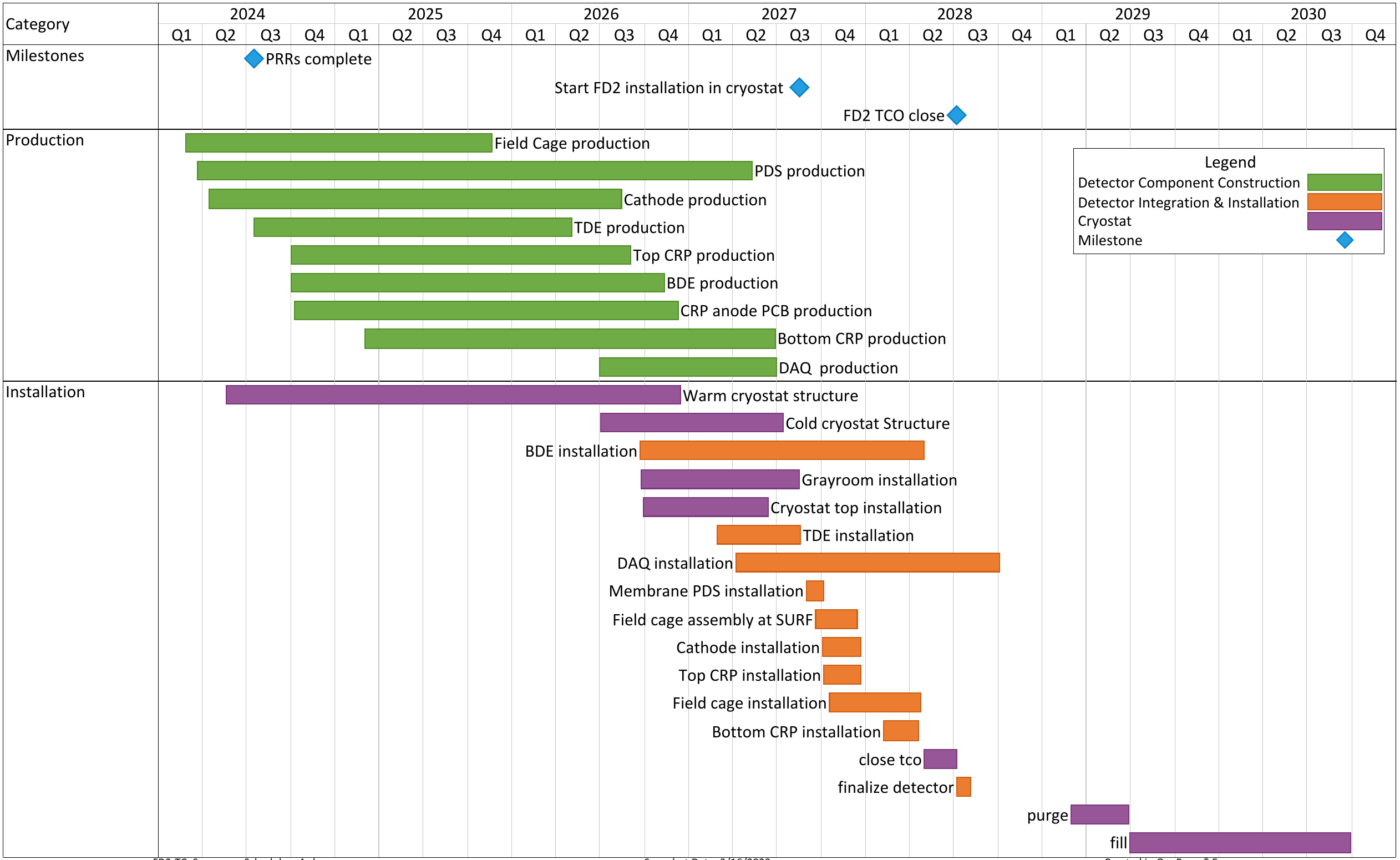}
\end{dunefigure}

%%%%

% this is added just after last chapter

\cleardoublepage
%\textsf{\listoffigures} %moving to end
%\textsf{\listoftables} %moving to end
\printglossaries

\cleardoublepage
%begin stuff from init per Thomas 6/2/2106
% end stuff from init
\cleardoublepage
\renewcommand{\bibname}{References}
\bibliographystyle{utphys} 
% To understand the style chosen, see:
% https://arxiv.org/hypertex/bibstyles/ (very bottom -- additions) and
% https://www.sharelatex.com/learn/Bibtex_bibliography_styles
% July 2017, AH and BV (and AM)
\bibliography{common/main} % main.bib

\end{document}